%% file: main.tex
\documentclass[a4paper]{llncs}

\usepackage{sosy-paper}

\newcommand{\smallsec}[1]{\smallskip\noindent\textbf{#1.}}
\newcommand{\smallersec}[1]{\smallskip\noindent\textit{#1.}}

\newcommand{\progconst}{\href{https://github.com/sosy-lab/sv-benchmarks/blob/svcomp19/c/loop-invariants/const_true-unreach-call_true-valid-memsafety_true-no-overflow_false-termination.c}{\texttt{const.c}}\xspace}
\newcommand{\progeqone}{\href{https://github.com/sosy-lab/sv-benchmarks/blob/svcomp19/c/loop-invariants/eq1_true-unreach-call_true-valid-memsafety_true-no-overflow_false-termination.c}{\texttt{eq1.c}}\xspace}
\newcommand{\progeqtwo}{\href{https://github.com/sosy-lab/sv-benchmarks/blob/svcomp19/c/loop-invariants/eq2_true-unreach-call_true-valid-memsafety_true-no-overflow_false-termination.c}{\texttt{eq2.c}}\xspace}
\newcommand{\progeven}{\href{https://github.com/sosy-lab/sv-benchmarks/blob/svcomp19/c/loop-invariants/even_true-unreach-call_true-valid-memsafety_true-no-overflow_false-termination.c}{\texttt{even.c}}\xspace}
\newcommand{\progodd}{\href{https://github.com/sosy-lab/sv-benchmarks/blob/svcomp19/c/loop-invariants/odd_true-unreach-call_true-valid-memsafety_true-no-overflow_false-termination.c}{\texttt{odd.c}}\xspace}
\newcommand{\progmodfour}{\href{https://github.com/sosy-lab/sv-benchmarks/blob/svcomp19/c/loop-invariants/mod4_true-unreach-call_true-valid-memsafety_true-no-overflow_false-termination.c}{\texttt{mod4.c}}\xspace}
\newcommand{\progbin}{\href{https://github.com/sosy-lab/sv-benchmarks/blob/svcomp19/c/loop-invariants/bin-suffix-5_true-unreach-call_true-valid-memsafety_true-no-overflow_false-termination.c}{\texttt{bin-suffix-5.c}}\xspace}

\title{Software Verification with \pdr:\\
       Implementation and Empirical Evaluation of the State of the Art}
\newcommand{\mypaperkeywords}{
Software verification,
Program analysis,
Invariant generation,
Property-directed reachability (\pdr),
IC3,
\kInduction,
VVT,
CPAchecker
}

\begin{document}
\author{Dirk Beyer and Matthias Dangl}
\institute{LMU Munich, Germany}

\maketitle

\begin{abstract}
\input{abstract}
\end{abstract}


\input{intro}

\input{background}
\input{approach}

\input{evaluation}

\input{conclusion}

\vspace{-3mm}
\bibliography{sw,dbeyer}

\end{document}

%% file: abstract.tex
Property-directed reachability (\pdr)
is a SAT/SMT-based reachability algorithm
that incrementally constructs inductive invariants.
After it was successfully applied to hardware model checking,
several adaptations to software model checking have been proposed.
We contribute a replicable and thorough comparative evaluation
of the state of the art: We
(1)~implemented a standalone \pdr algorithm and, as improvement,
a \pdr-based auxiliary-invariant generator for \kinduction, and
(2) performed an experimental study
on the largest publicly available benchmark set of C~verification tasks,
in which we explore the effectiveness and efficiency of software verification with \pdr.
The main contribution of our work is to establish a reproducible
baseline for ongoing research in the area by providing
a well-engineered reference implementation and
an experimental evaluation of the existing techniques.

%% file: intro.tex
\section{Introduction}

Automatic software verification~\cite{SoftwareModelChecking}
is a broad research area with many success stories
and large impact on technology that is applied in industry~\cite{SLAM,LDV,INFER}.
It nicely complements other general approaches to ensure functional correctness,
like software testing~\cite{ArtOfSoftwareTesting}
and interactive software verification~\cite{BeckertHaehnle14}.
One large sub-area of automatic software verification includes
algorithms and approaches that are based on SMT technology.
There are classic approaches like bounded model checking~\cite{BMC},
predicate abstraction~\cite{GrafSaidi97,BPR01},
and k-induction~\cite{kInduction,kInductionPrinciple,PKind}, which are well understood and evaluated;
a recent survey~\cite{AlgorithmComparison-JAR} provides a uniform overview and
sheds light on the differences of the algorithms.
Property-directed reachability (\pdr)~\cite{IC3} is a relatively recent (2011)
approach that is not yet included in surveys and comparative evaluations.
The approach was originally
applied to transition systems from hardware designs,
but was also adapted to software verification
in the last years~\cite{IC3,PropertyDirectedInvariants,SoftwareIC3,IC3-CFA,IC3-CFA-Improved,PredAbsPDR-FMSD,CTIGAR,PDR-kInduction}.

While in theory, the advantages and disadvantages of using \pdr seem clear,
we are interested in understanding the effect of applying \pdr to a large
set of verification tasks that were collected from academia and also from industrial software,
such as the Linux kernel.
To achieve this goal, we implemented one \pdr adaptation for software verification,
and another approach that integrates a \pdr-like invariant-generation module
into a k-induction approach.

\smallersec{\pdr Adaptation for Software Verification}
\pdr is a model-checking algorithm
that tries to construct an inductive safety invariant
by incrementally learning clauses
that are inductive relative to previously learned clauses.
The clause-learning strategy is guided by counterexamples to induction,
i.e., each time a proof of inductiveness fails,
the algorithm attempts to learn a new clause to avoid the same counterexample to induction in the future.
Originally, this algorithm was designed as a SAT-based technique for Boolean finite-state systems.
Every adaptation of \pdr to software verification
therefore needs to consider how to effectively and efficiently handle the infinite state space
and how to transfer the algorithm from SAT to SMT.
Furthermore, the adaptation to software has to deal with the program counter.

\smallersec{\pdr-like Invariant Generation}
Whenever an induction-proof attempt fails with a counterexample,
the counterexample describes a state $s$ that can transition into a bad state
(that violates the safety property),
which means that in order to make the proof succeed,
$s$ must be removed from consideration by an auxiliary invariant.
From this bad-state predecessor $s$,
the clause-learning strategy of \pdr
proceeds to generate such an auxiliary invariant
by applying the following two steps:
(1)~$s$ is first generalized to a set of states~$C$ that all transition into a bad state;
(2)~an invariant is constructed
that is (a)~inductive relative to previously found invariants%
\footnote{An assertion $F$ is said to be inductive relative to an invariant $\inv$
if $\inv$ can be used as an auxiliary invariant
for the proof of inductiveness
\mbox{$\forall s_j, s_{j+1} : F(s_j) \land T(s_j,s_{j+1}) \implies F(s_{j+1})$}
by conjoining~$\inv$ to the induction hypothesis~$F(s_j)$,
such that the modified induction query
\mbox{$\forall s_j, s_{j+1}: F(s_j) \boldsymbol{\land} \boldsymbol{\inv}\mathbf{(s_j)} \land T(s_j, s_{j+1}) \implies F(s_{j+1})$}
allows a proof by induction to succeed.~\cite{IC3}}
and (b)~at least strong enough to eliminate all states in~$C$.
If it fails to construct such an invariant and prove its inductiveness, then
the steps are recursively re-applied
to the counterexample obtained from the failed induction attempt.

We experimentally investigate two implementations of adaptations of \pdr
to software verification (\cpactigar and \vvtctigar),
as well as several combinations that use the \pdr-like invariant-generation module
that we designed and implemented for this study.

\lstset{
    breaklines=true,
    postbreak=\raisebox{0ex}[0ex][0ex]{\hspace{-3em}\ensuremath{\color{red}\hookrightarrow\space}},
    breakatwhitespace=true,
}

\begin{figure}[t]
\begin{lstlisting}[style=C]
extern void __VERIFIER_error() __attribute__ ((__noreturn__));
extern unsigned int __VERIFIER_nondet_uint(void);
void __VERIFIER_assert(int cond) {
  if (!(cond)) {
    ERROR: __VERIFIER_error();
  }
  return;
}
int main(void) {
  unsigned int w = __VERIFIER_nondet_uint();
  unsigned int x = w;
  unsigned int y = w + 1;
  unsigned int z = x + 1;
  while (__VERIFIER_nondet_uint()) {
    y++;
    z++;
  }
  __VERIFIER_assert(y == z);
  return 0;
}
\end{lstlisting}
  \vspace{-4mm}
  \caption{Example C program \progeqtwo}
  \label{fig:eq2.c}
  \vspace{-4mm}
\end{figure}

\lstset{
    breaklines=false
}

\smallsec{Example}
\Cref{fig:eq2.c} shows an example C program (\progeqtwo)
that contains four unsigned integer variables~\lstinline{w},~\lstinline{x},~\lstinline{y},~and~\lstinline{z}.
In line~\lstinline{10}, the variable~\lstinline{w} is initialized to an unknown value
via the input function~\lstinline{__VERIFIER_nondet_uint()};
then, its value is copied to~\lstinline{x} in line~\lstinline{11}.
In line~\lstinline{12}, variable~\lstinline{y} is initialized with the value of~\lstinline{w + 1},
and in line~\lstinline{13}, variable~\lstinline{z} is initialized with the value of~\lstinline{x + 1},
such that at this point,~\lstinline{w}~and~\lstinline{x} are equal to each other,
and~\lstinline{y}~and~\lstinline{z} are also equal to each other.
Then, from line~\lstinline{14} to line~\lstinline{17},
a loop with a nondeterministic exit condition
(and therefore an unknown number of iterations)
increments in each iteration both variables~\lstinline{y}~and~\lstinline{z}.
Lastly, line~\lstinline{18} asserts that after the loop,%
~\lstinline{y}~and~\lstinline{z} are (still) equal to each other.
Since~\lstinline{y}~and~\lstinline{z} are equal before the loop,
and are always incremented together within the loop,
the invariant~$y = z$ is inductive.
However, since there is no direct connection between~\lstinline{y}~and~\lstinline{z}
but only an indirect one via their shared dependency on~\lstinline{w},
na\"{i}ve data-flow-based techniques may fail to find this invariant.
In fact, we tried several configurations of the verification framework \cpachecker,
and found that many of them fail to prove this program:
\begin{itemize}
  \item Plain \kinduction without auxiliary-invariant generation fails,
        because it never checks if~$y = z$ is a loop invariant
	and instead only checks the reachability of the assertion failure (located after loop).
  The reachability of the assertion failure, in turn,
	depends on the nondeterministic loop-exit condition.
	Therefore we cannot conclude from
	``the assertion failure was not reached in $k$ previous iterations''
	that ``the assertion failure cannot be reached in the next iteration'':
  In the absence of auxiliary invariants,
	a valid counterexample to this induction hypothesis
	would always be that in the previous iterations the assertion \emph{condition}
	was in fact violated and an assertion \emph{failure}
	was not reached only \emph{because the loop was not exited}.
  \item A data-flow analysis based on the abstract domain of Boxes~\cite{Boxes}
        fails, because it is not able to track variable equalities.
  \item A data-flow analysis based on a template~$\mathsf{Eq}$
        for tracking the equality of pairs of variables
	fails, because while it detects the invariant $w = x$,
	it is unable to make the step to $y = z$
	due to the inequalities between~\lstinline{w}~and~\lstinline{y},
	and~\lstinline{x}~and~\lstinline{z}, respectively.
  \item For consistency with our evaluation,
        we also applied a data-flow analysis based on a template
	for tracking whether a variable is even or odd;
	obviously this is not useful for this program,
	and thus, this configuration also fails.
  \item Even combining the previous three techniques
        into a compound invariant generator that computes auxiliary invariants for \kinduction
	does not yield a successful configuration for this verification task.
  \item The invariant generator~\kipdr
        (the above-mentioned adaptation of \pdr to \kinduction, which we present
         in more detail in Sect.~\ref{sect:approach}), however,
	detects the invariant $y = z$
	and is therefore able to construct a proof by induction for this verification task.
\end{itemize}

We will now briefly sketch how \kipdr detects the invariant $y = z$ for the example verification task.
At first, \kipdr attempts to prove by induction that when line~\lstinline{18} is reached,
the assertion condition holds, which fails as discussed previously.
However, this failed induction attempt yields a counterexample to induction
where the values of~\lstinline{y}~and~\lstinline{z} differ from each other,
e.g., $y = 0 \land z = 1$, which is then generalized to~$y \neq z$,
i.e., a set of states that includes the concrete predecessor of a bad state from the counterexample,
as well as many other states that would violate the assertion,
if they were reachable themselves.
Then, \kipdr attempts to find an inductive invariant that eliminates all of these states,
and the attempt succeeds with the invariant~$y = z$.
Afterwards, \kipdr re-attempts its original induction proof
to show that the assertion is never violated,
which now succeeds due to the auxiliary invariant~$y = z$.

\smallsec{Contributions}
We present the following contributions:
\begin{itemize}
  \item We implement one adaptation of \pdr to software verification (based on~\cite{CTIGAR,VVT-COMP16})
        in the open-source verification framework \cpachecker, in order to
        establish a baseline for comparison with new ideas for improvement.
  \item We design and implement the algorithm \kipdr,
        as a new module for invariant generation
        that is based on ideas from \pdr
        and use this module
        as an extension to a state-of-the-art approach to \kinduction~\cite{kInduction}.
  \item We conduct a large experimental study
        to compare several tools and approaches to software verification using \pdr as a component,
        in order to highlight strengths and weaknesses of \pdr in the domain of software verification.
  \item We contribute a set of small examples that need invariants that are more difficult to obtain
        for standard data-flow-based approaches than the invariants necessary for programs in the large benchmark set.
\end{itemize}

\input{related}

%% file: related.tex
\smallsec{Related Work}
While \pdr (also known as IC3 for its first implementation~\cite{IC3})
was introduced as a SAT-based algorithm
for model checking finite-state Boolean transition systems~\cite{PropertyDirectedInvariants},
several approaches have since then been presented to extend it to SMT
and to apply it to the verification of software models:
\pdr has been suggested as an interpolation engine for \impact,
but experiments have shown that it is too expensive in the general case,
and is most effective if only applied as a fall-back engine
for cases where a cheaper interpolation engine
fails to produce useful interpolants~\cite{SoftwareIC3}.
It also has been proposed to improve this approach
by tracking control-flow locations explicitly
instead of symbolically~\cite{IC3-CFA},
thereby avoiding the problem that many iterations of the algorithm are spent
only to learn the control flow,
and this idea has later been extended
by several improvements to the generalization step of \pdr~\cite{IC3-CFA-Improved}.
Another approach is to model the program using a Boolean abstraction,
which has the advantage that it requires only few changes to the original algorithm,
but the disadvantage that a refinement procedure is necessary
to handle the spurious paths introduced by the abstraction:
One such approach uses infeasible error paths
(i.e., counterexample-guided abstraction refinement~(CEGAR)~\cite{ClarkeCEGAR})
to refine the abstraction~\cite{PredAbsPDR-FMSD},
while another (CTIGAR) uses counterexamples to induction~\cite{CTIGAR};
both of these refinement techniques use interpolation to obtain abstraction predicates;
the latter of the two techniques is used in
two of the configurations we compare in our evaluation (\cpactigar and \vvtctigar~\cite{VVT-COMP16}).
A different extension of \pdr to verify infinite-state systems
that does not require abstraction refinement is property-directed \kinduction~\cite{PDR-kInduction},
which increases the power of the induction checks used in \pdr
by applying $k$-induction instead of $1$-induction,
and which uses model-based generalization in addition to interpolation
to reason about potentially-infinite sets of states.
Unfortunately, support for effective model-based generalization is rare in SMT solvers\,%
\footnote{The implementation of this approach of property-directed \kinduction combines two SMT solvers, because neither of them supports all features required by the technique.},
making this approach impractical.
In contrast, our \kipdr algorithm presented in Sect.~\ref{sect:approach}
only requires support for interpolation,
which is available in several SMT solvers.

Despite this multitude of adaptations of \pdr to infinite-state systems,
most implementations in practice require their input
to be encoded as transition systems already.
The only available software verifiers that can be applied to actual C programs
and implement \pdr-based techniques
are \cpachecker~\cite{CPACHECKER}, \seahorn~\cite{SEAHORN}, and \vvt~\cite{VVT-COMP16}.

%% file: background.tex
\section{Background}
\label{sect:background}

In this section, we briefly introduce the algorithms \pdr and \kinduction,
which provide the core concepts on which we base our ideas.
In the following description of \pdr and \kinduction,
we use the following notation:
given the state variables $s$ and $s'$
within a state-transition system~$T$ that represents the program,
predicate $I(s)$ denotes that $s$ is an initial state,
$T(s,s')$ that a transition from $s$ to $s'$ exists,
and $P(s)$ that the safety property~$P$ holds for state~$s$.

In the context of the verification tasks we analyze,
we use the notion of a system's \emph{invariant}
to describe a logical assertion
that holds at every state of a system.
Similarly, we consider a \emph{loop invariant}
to be an assertion that holds at the entry point of a loop
and every time the control flow loops back to that point.
A loop invariant need not hold at every location within the loop,
but must be restored at the end of each loop iteration.
For simplicity,
when we reason about reachability properties,
we also use the term \emph{invariant}
for sets of states that can be described by an invariant,
i.e., as a shorthand for a set of program states
that contains all reachable states of the program.
In the context of the discussion of an algorithm,
we use the term \emph{invariant} to describe properties
the algorithm guarantees at every step,
i.e., these properties can be seen as system invariants of the algorithm.

\subsection{\pdr}

\pdr maintains a list of $k$~frames,
where a frame~$F_i$ is a predicate
that represents an overapproximation of all states
reachable within at most~$0 \leq i \leq k$ steps,
and a queue of proof obligations, which guide invariant discovery
towards invariants relevant to prove the correctness of a safety property~$P$.
For a given state $s$, the notation $F_i(s)$ means
that the predicate~$F_i$ holds for state~$s$.
The index~$i$ of a frame~$F_i$ is called its \emph{level},
and the frame $F_k$ is called the \emph{frontier},
because it represents the largest overapproximation of reachable states
computed by the algorithm~\cite{IC3}.
The algorithm maintains the following invariants:
\begin{enumerate}
  \item $F_0(s)=I(s)$, i.e., the first frame represents precisely the initial states.
  \item $\forall i \in \{0, \ldots, k\}: F_i(s) \implies P(s)$, i.e., every frame contains only states that satisfy the safety property.
  \item $\forall i \in \{0, \ldots, k-1\}: F_{i}(s) \implies F_{i+1}(s) $, i.e., a frame~$F_{i+1}$ represents in addition such states that are reachable with $i+1$~steps.
  \item $\forall i \in \{0, \ldots, k-1\}: F_i(s) \land T(s,s') \implies F_{i+1}(s')$, i.e., each frame is inductive relative to its predecessor.
\end{enumerate}
Using these data structures and algorithm invariants,
the algorithm attempts to find either a counterexample to~$P$
or a $1$\nobreakdash-inductive invariant~$F_i$
such that~$F_i(s) \Leftrightarrow F_{i+1}(s)$ for some level~$i \in \{0, \ldots, k-1\}$.
Until either of these potential outcomes is reached,
\pdr shifts back and forth between the following two phases:
\begin{enumerate}
  \item If the set of states represented by the frontier~$F_k$
        does not contain any predecessor states
        of $\lnot P$\nobreakdash-states
	(i.e.,~$\forall s_j, s_{j+1}: F_k(s_j) \land T(s_j,s_{j+1}) \implies P(s_{j+1})$,
	    called frontier-incrementation check),
        a new frontier~$F_{k+1}$ is created and initialized to~$P$.
        Subsequently, the algorithm attempts to push forward\,%
        \footnote{By ``push forward'', we mean to add a predicate~$c$ from frame~$F_{i}$ to frame~$F_{i+1}$~\cite{IC3}.}
        each predicate~$c$ of each frame~$F_i$ with $0 \leq i \leq k$
	      for which the consecution check
        ~$F_{i}(s_j) \land T(s_j,s_{j+1}) \implies c(s_{j+1})$ holds (see \cref{fig:consecution-check}).
	If, on the other hand, the frontier-incrementation check fails,
	\pdr extracts a~$\lnot P$\nobreakdash-predecessor~$t$ in~$F_k$,
	which represents a counterexample to induction (CTI),
	from the failed query as proof obligation~$\langle t, k - 1 \rangle$ (see \cref{fig:proof-obligation}, top).
  \item While the queue of proof obligations is not empty,
        \pdr processes the queue
        by trying to prove for each proof obligation~$\langle t, i \rangle$
        that the CTI\nobreakdash-state~$t$ is itself not reachable from~$F_i$
	and therefore does not need to be considered as a relevant~$\lnot P$\nobreakdash-predecessor.
  For this proof, \pdr chooses some predicate~$c \implies \lnot t$ with $\forall s: F_i(s) \implies c(s)$.
  \pdr then checks if~$c$ is inductive relative to~$F_i$
	by performing the consecution check~%
$F_i(s_j) \land c(s_j) \land T(s_j,s_{j+1}) \implies c(s_{j+1})$.
  If the consecution check succeeds,
	the frames~$F_1, \ldots, F_{i+1}$ can be strengthened by adding~$c$,
	thus ruling out the CTI~$t$ in these frames for the future (see \cref{fig:proof-obligation}, left).
  Also, unless $i=k$, we add a new proof obligation $\langle t, i+1\rangle$ to the queue
  as an optimization to initiate forward propagation,
  because we expect that the CTI\nobreakdash-state $s$
  would otherwise be rediscovered later at a higher level\cite{CTIGAR}.
	Otherwise, i.e., the consecution check does \emph{not} succeed for predicate~$c$,
  the algorithm extracts a predecessor~$u$~of~$t$ from the failed consecution check,
	which is added as a new proof obligation~$\langle u, i - 1 \rangle$ if~$i > 0$
	and~$t \land I$ is unsatisfiable (see \cref{fig:proof-obligation}, right).
	Otherwise,~$u$ represents the initial state of a real counterexample to~$P$.
\end{enumerate}
A more detailed presentation of \pdr can be found in the literature~\cite{IC3}.

\newcommand\consecutionCheckCaption{%
  Consecution check makes sure to only conjoin to frame~$F_{i+1}$ such $c_i$ from~$F_i$
  that are inductive relative to~$F_{i}$ w.r.t. transition relation~$T$%
}%
\begin{figure}
  \centering
  \input{figures/step-1-consecution-check}
  \caption{\consecutionCheckCaption}
  \label{fig:consecution-check}
\end{figure}

\newcommand\proofObligationCaption{%
  If phase~1 results in a proof obligation $\langle t, k-1 \rangle$ (top),
  then phase~2 resolves either by strengthening $F_{k}$ with $c$ (left),
  or by creating a new (backwards) proof obligation $\langle u, k-2 \rangle$ (right);
  if the chain of proof obligations propagates back to the initial states,
  then a feasible error path is found%
}%
\begin{figure}
  \centering
  \input{figures/step-2-proof-obligation}
  \caption{\proofObligationCaption}
  \label{fig:proof-obligation}
\end{figure}

\subsubsection{Example}
As an example, we will now show how \pdr is applied to a verification task with
\begin{itemize}
  \item the initial-state predicate~$I(s) = (x_s=2)$,
  \item the transition relation~$T(s,s') = (x_{s'}=2\cdot x_s - 1)$, and
  \item the safety property~$P(s) = (x_s>0)$.
\end{itemize}

Initially, we therefore have two frames~$F_0(s)=(x_s=2)$~and~$F_1(s)=(x_s>0)$,
and~$k=1$, i.e., $F_1$~is the frontier.
The algorithm begins in phase~$1$,
where we check whether the frontier contains any predecessors of~$\lnot P$\nobreakdash-states.
We find such a predecessor~$t$ with~$t = (x_t=0.5)$,
because $2 \cdot 0.5 - 1 = 0$, which violates~$P$.
From this CTI, we create the proof obligation~$\langle (x_s=0.5),0\rangle$,
add it to our queue of proof obligations, and switch to phase~$2$.
In phase~$2$, we try to find a predicate~$c \subseteq \lnot t$ with $\forall s:F_0(s) \implies c(s)$.
We can find a valid predicate~$c$ with $c(s)=(x_s \leq 0 \lor x_s \geq 1)$
that represents a subset of~$\lnot t$\nobreakdash-states
and a superset of the states represented by~$F_0$,
because $(x = 2) \implies (x \leq 0 \lor x \geq 1) \implies (x \neq 0.5)$,
and add this predicate to frames~$F_0$~and~$F_1$,
such that~$F_0(s)=(x_s=2 \land (x_s \leq 0 \lor x_s \geq 1))$
and~$F_1(s)=(x_s>0 \land (x_s \leq 0 \lor x_s \geq 1))$,
after which the algorithm switches back to phase~$1$.
Back in phase~$1$, we again check if the frontier, which is still~$F_1$,
contains any predecessors of~$\lnot P$\nobreakdash-states, which it now no longer does.
Hence, we create a new frontier~$F_2$ with~$F_2(s)=(x>0)$.
Then, we attempt to push forward each predicate of~$F_0$~and~$F_1$,
which succeeds for~$(x_s \leq 0 \lor x_s \geq 1)$~in~frame~$F_0$, which is already contained in~$F_1$,
and for~$(x_s \leq 0 \lor x_s \geq 1)$~in~frame~$F_1$, which we then add to~$F_2$.
Thus, our frames are now~$F_0(s)=(x_s=2 \land (x_s \leq 0 \lor x_s \geq 1))$,
${F_1(s)=(x_s>0 \land (x_s \leq 0 \lor x_s \geq 1))}$, and~$F_2(s)=(x_s>0 \land (x_s \leq 0 \lor x_s \geq 1))$,
which means that~$F_1(s) \Leftrightarrow F_2(s)$,
i.e., frame~$F_1$ represents an inductive invariant that implies~$P$
and therefore, the proof is complete.

\subsection{\kInduction}

Like \pdr, \kinduction attempts to prove a safety property~$P$ by applying induction.
However, while \pdr strengthens its induction hypothesis by using predicates
extracted from specific counterexamples to induction after failed induction attempts,
\kinduction strengthens its induction hypothesis
by increasing the length of the unrolling of the transition relation.

Starting with an initial value for the bound~$k$ (usually~$1$),
the \kinduction algorithm increases the value of~$k$ iteratively
after each unsuccessful attempt at
finding a specification violation (base case),
proving correctness via complete loop unrolling (forward condition), or
inductively proving correctness of the program (inductive-step case).

\subsubsection{Base Case}
The base case of \kinduction
consists of running BMC with the current bound~$k$.\,%
\footnote{We define the loop bound as the number of visits of the loop head,
that is, with loop bound~$k=1$, the loop head is visited once, but there
was not yet any unwinding of the loop body.
This nicely matches the intuition for $k$-induction:
$1$-inductiveness means that if the invariant holds for one state (without loop unrolling),
then it holds again after one loop unrolling in the successor state;
$k$-inductiveness means that if the invariant holds for $k$~states ($k-1$ loop unrollings),
then it holds again after one more loop unrolling in the successor state.}
This means that starting from all initial program states,
all states of the program reachable
within at most~$k-1$ unwindings of the transition relation
are explored.
If a~$\lnot P$\nobreakdash-state is found, the algorithm terminates.

\subsubsection{Forward Condition}
If no~$\lnot P$\nobreakdash-state is found by the BMC in the base case,
the algorithm continues by performing the forward-condition check,
which attempts to prove that BMC fully explored the state space of the program
by checking that no state with distance~$k' > k - 1$ to the initial state is reachable.
If this check is successful, the algorithm terminates.

\subsubsection{Inductive-Step Case}
The forward-condition check, however,
can only prove safety for programs with finite (and, in practice, short) loops.
To prove safety beyond the bound~$k$, the algorithm applies induction:
The inductive-step case attempts to prove that
after every sequence of $k$~unrollings of the transition relation
that did not reach a~$\lnot P$\nobreakdash-state,
there can also be no subsequent transition into a~$\lnot P$\nobreakdash-state
by unwinding the transition relation once more.
In the realm of model checking of software, however,
the safety property~$P$ is often not directly~$k$-inductive for any value of~$k$,
thus causing the inductive-step-case check to fail.
It is therefore state-of-the-art practice
to add auxiliary invariants to this check
to further strengthen the induction hypothesis
and make it more likely to succeed.
Thus, the inductive-step case proves a program safe if the following condition is unsatisfiable:
\vspace{-2mm}
\[
  \inv(s_n) \land \bigwedge\limits_{i=n}^{n+k-1} \left(P(s_i) \land T(s_i, s_{i+1})\right)  \land \lnot P(s_{n+k})
\]
\vspace{-3mm}\\
where $\inv$ is an auxiliary invariant,
and $s_n, \ldots, s_{n+k}$ is any sequence of states.
If this check fails, the induction attempt is inconclusive,
and the program is neither proved safe nor unsafe yet
with the current value of~$k$ and the given auxiliary invariant.
In this case, the algorithm increases the value of~$k$ and starts over.

A detailed presentation of \kinduction can be found in the literature~\cite{kInduction,AlgorithmComparison-JAR}.

%% file: figures/step-1-consecution-check.tex
\begin{tikzpicture}[>={Stealth[scale=1.5]}]

\newcommand{\hvn}[3]{
  \draw #1 -- +(1,0) -- +(1,1) -- +(0.5,1.5) -- +(0,1) --cycle;
  \node (#2) at #1 {};
  \node (#2-label) at ($#1 + (0.5,1.75)$) {#3};
}

\hvn{(0,0)}{Fi}{$F_i$}
\node (c1i) at (0.1,1.4) {$c_1$};
\node (c2i) at (0.9,1.4) {$c_2$};
\node (c3i) at (1.2,0.5) {$c_3$};
\node (c4i) at (0.5,-0.25) {$c_4$};
\node (c5i) at (-0.2,0.5) {$c_5$};

\node at (2,0.675) {\Large$\overset{T}{\Rightarrow}$};
\node (c4) at (3,0.5) {$c_4$};
\draw[->] (c4i) to[out=0,in=255] (c4);

\node at (5.5,1.75) {$F_{i+1}$};
\draw (5.5,0.65) ellipse (0.6 and 0.75);
\node (c4i1) at (5.5,-0.5) {$\land\ c_4$};

\draw[decoration={brace,mirror,raise=5pt},decorate]
  (3.5,-0.5) -- node[right=10pt] {\raisebox{10pt}{{\Large$\overset{T}{\Rightarrow}$}~}} (3.5,1.5);

\end{tikzpicture}

%% file: figures/step-2-proof-obligation.tex
\begin{tikzpicture}[>={Stealth[scale=1.5]}]

\newcommand{\hvn}[3]{
  \draw #1 -- +(1,0) -- +(1,1) -- +(0.5,1.5) -- +(0,1) --cycle;
  \node (#2) at #1 {};
  \node (#2-label) at ($#1 + (0.5,1.75)$) {#3};
}

\hvn{(0,0)}{Fk}{$F_k$}
\hvn{(1.5,0)}{P}{$P$}
\node[fill,circle,inner sep=1.5pt] (t) at ($(Fk) + (0.5,0.3)$) {};
\node[anchor=south east] at (t) {$t$};
\node[fill,circle,inner sep=1.5pt] (succT) at ($(P) + (0.5,-0.3)$) {};
\draw[->] (t) .. controls +(0.75,0.25) and +(-0.75,-0.25) .. (succT);

\node      at ($(t) - (0,1.0)$) {\Large$\Downarrow$};
\node (po) at ($(t) - (0,1.5)$) {$\langle t, k-1 \rangle$};

\hvn{(-4.5,-4)}{Fk1l}{$F_{k-1}$};
\hvn{(-3.0,-4)}{Fkl}{$F_{k}$};
\hvn{(-1.5,-4)}{Pl}{$P$};
\node[fill,circle,inner sep=1.5pt] (ul) at ($(Fk1l) + (0.5,0.3)$) {};
\node[fill,circle,inner sep=1.5pt] (tl) at ($(Fkl) + (0.5,0.3)$) {};
\node[anchor=south west] at (tl) {$t$};
\node[fill,circle,inner sep=1.5pt] (succTl) at ($(Pl) + (0.5,-0.3)$) {};
\draw[->] (ul) .. controls +(0.25,-1) and +(-0.25,-1) .. node[pos=0.5] (predTlTransition) {} (tl);
\draw[->] (tl) .. controls +(0.75,0.25) and +(-0.75,-0.25) .. (succTl);
\draw ($(predTlTransition) - (0.1,0.1)$) --($(predTlTransition) + (0.1,0.1)$);
\draw ($(predTlTransition) + (-0.1,0.1)$) --($(predTlTransition) + (0.1,-0.1)$);
\draw ($(Fkl) - (0.2,0.2)$) -- +(1.5,1.5);
\node (c) at ($(Fkl) + (1.2,0.95)$) {$\lnot c$};
\node (c) at ($(Fkl) + (1.,1.25)$) {$ c$};

\hvn{( 1.5,-4)}{Fk1r}{$F_{k-1}$};
\hvn{( 3.0,-4)}{Fkr}{$F_{k}$};
\hvn{( 4.5,-4)}{Pr}{$P$};
\node[fill,circle,inner sep=1.5pt] (ur) at ($(Fk1r) + (0.5,0.3)$) {};
\node[anchor=south west] at (ur) {$u$};
\node[fill,circle,inner sep=1.5pt] (tr) at ($(Fkr) + (0.5,0.3)$) {};
\node[anchor=south west] at (tr) {$t$};
\node[fill,circle,inner sep=1.5pt] (succTr) at ($(Pr) + (0.5,-0.3)$) {};
\draw[->] (ur) .. controls +(0.25,-1) and +(-0.25,-1) .. (tr);
\draw[->] (tr) .. controls +(0.75,0.25) and +(-0.75,-0.25) .. (succTr);

\node at ($(ur) - (0,1.0)$) {\Large$\Downarrow$};
\node at ($(ur) - (0,1.5)$) {$\langle u, k-2 \rangle$};

\draw[->] ($(po) - (0.2, 0.3)$) --($(Fkl-label) + ( 0.2,0.2)$);
\draw[->] ($(po) + (0.2,-0.3)$) --($(Fkr-label) + (-0.2,0.2)$);
\node at ($(po) - (0,0.5)$) {or};
\end{tikzpicture}

%% file: approach.tex
\section{Combining \kInduction with \pdr}
\label{sect:approach}

\newcommand{\intinv}{\mathit{InternalInv}}
\newcommand{\extinv}{\mathit{ExternalInv}}
\begin{algorithm}[t]
\renewcommand\algorithmicforall{\textbf{for each}}
\caption{Iterative-Deepening \kInduction with Property Direction}
\label{k-induction-pd-algo}
\resizebox{\linewidth}{!}{
\begin{minipage}{1.21\linewidth}
\begin{algorithmic}[1]
    \REQUIRE the initial value $k_{init} \geq 1$ for the bound $k$,\\
             an upper limit $k_{max}$ for the bound $k$,\\
             a function $\mathsf{inc}:\mathbb{N}\rightarrow\mathbb{N}$ with $\forall n\in\mathbb{N}: \mathsf{inc}(n) > n$,\\ %
             the initial states defined by the predicate $I$,\\
             the transfer relation defined by the predicate $T$,\\
             a safety property $P$,\\
	     a function $\mathsf{get\_currently\_known\_invariant}$ to obtain auxiliary invariants,\\
	     a Boolean $\mathit{pd}$ that enables or disables property direction,\\
             a function $\mathsf{lift} : \mathbb{N} \times (S \rightarrow \mathbb{B}) \times (S \rightarrow \mathbb{B}) \times S \rightarrow (S \rightarrow \mathbb{B})$, and\\
	     a function $\mathsf{strengthen} : \mathbb{N} \times (S \rightarrow \mathbb{B}) \times (S \rightarrow \mathbb{B}) \rightarrow (S \rightarrow \mathbb{B})$,\\
             where $S$ is the set of program states.

    \ENSURE \TRUE{} if $P$ holds, \FALSE{} otherwise

    \VARDECL the current bound $k := k_{init}$,\\
             the invariant $\intinv := \mathit{true}$ computed by this algorithm internally, and\\
             the set $O := \{\}$ of current proof obligations.
    \label{k-induction-algo-init-k}
    \WHILE{$k \leq k_{max}$}
    \STATE $O_\mathit{prev} := O$
    \STATE $O := \{\}$
    \label{k-induction-algo-max}
    \STATE $\mathit{base\_case} := I(s_0) \land \bigvee\limits_{n=0}^{k-1} \left(\bigwedge\limits_{i=0}^{n-1} T(s_i, s_{i+1}) \land \lnot P(s_n)\right)$
      \label{k-induction-algo-base-start}
      \IF{$\mathsf{sat}(\mathit{base\_case})$}
        \RETURN \FALSE
      \ENDIF
      \label{k-induction-algo-base-end}
\vspace*{0.2\baselineskip}
        \STATE $\mathit{forward\_condition} := I(s_0) \land \bigwedge\limits_{i=0}^{k-1} T(s_i, s_{i+1})$
        \label{k-induction-algo-fc-start}
        \IF{$\lnot\,\mathsf{sat}(\mathit{forward\_condition})$}
        \label{k-induction-algo-fc-check-end}
          \RETURN \TRUE
        \ENDIF
        \label{k-induction-algo-fc-end}
\vspace*{0.2\baselineskip}
	\IF{$\mathit{pd}$}
        \label{k-induction-algo-pd-start}
        \FORALL{$o \in O_\mathit{prev}$}
          \label{k-induction-algo-pd-choose-o}
          \STATE $\mathit{base\_case}_o := I(s_0) \land \bigvee\limits_{n=0}^{k-1} \left(\bigwedge\limits_{i=0}^{n-1} T(s_i, s_{i+1}) \land \lnot o(s_n)\right)$
          \label{k-induction-algo-pd-base-start}
	  \IF{$\mathsf{sat}(\mathit{base\_case_o})$}
          \label{k-induction-algo-pd-base-end}
	    \RETURN \FALSE
            \label{k-induction-algo-pd-base-sat}
	  \ELSE
	    \STATE $\begin{aligned}\mathit{step\_case_o}_n :=& \bigwedge\limits_{i=n}^{n+k-1} \left(o(s_i) \land T(s_i, s_{i+1})\right) \land \lnot o(s_{n+k})\end{aligned}$
            \label{k-induction-algo-pd-base-unsat-start}
              \STATE $\extinv := \mathsf{get\_currently\_known\_invariant}()$
              \label{k-induction-algo-pd-ind-aux-ext}
              \STATE $\inv := \intinv \land \extinv$
              \label{k-induction-algo-pd-ind-aux}
              \IF{$\mathsf{sat}(\inv(s_n) \land \mathit{step\_case_o}_n)$}
                \label{k-induction-algo-pd-ind-sat-start}
                \STATE $s_o := $ satisfying predecessor state
	        \STATE $O := O \cup \{ \lnot \mathsf{lift}(k, \inv, o, s_o) \}$
                \label{k-induction-algo-pd-ind-sat-end}
              \ELSE
                \label{k-induction-algo-pd-ind-unsat-start}
                \STATE $\intinv := \intinv \land \mathsf{strengthen}(k,\inv,o)$
                \label{k-induction-algo-pd-ind-unsat-end}
              \ENDIF
            \label{k-induction-algo-pd-base-unsat-end}
          \ENDIF
        \ENDFOR
	\ENDIF
        \label{k-induction-algo-pd-end}
\vspace*{0.4\baselineskip}
        \STATE $\begin{aligned}\mathit{step\_case}_n :=& \bigwedge\limits_{i=n}^{n+k-1} \left(P(s_i) \land T(s_i, s_{i+1})\right) \land \lnot P(s_{n+k})\end{aligned}$
        \label{k-induction-algo-ind-start}
          \STATE $\extinv := \mathsf{get\_currently\_known\_invariant}()~~~~$ \tikzmark{inv-consumer}
          \label{k-induction-algo-ind-aux-ext}
          \STATE $\inv := \intinv \land \extinv$
          \label{k-induction-algo-ind-aux}
          \IF{$\mathsf{sat}(\inv(s_n) \land \mathit{step\_case}_n)$}
            \label{k-induction-algo-ind-check}
	     \IF{$\mathit{pd}$}
               \label{k-induction-algo-ind-sat-start}
               \STATE $s := $ satisfying predecessor state
               \label{k-induction-algo-ind-sat-s}
               \STATE $O := O \cup \{ \lnot \mathsf{lift}(k, \inv, P, s) \}$
               \label{k-induction-algo-ind-sat-end}
	     \ENDIF
          \ELSE
            \RETURN \TRUE
            \label{k-induction-algo-ind-unsat}
          \ENDIF
        \label{k-induction-algo-repeat-ind}
        \label{k-induction-algo-ind-end}
\vspace*{0.2\baselineskip}
       \STATE $k$ := $\mathsf{inc}(k)$
       \label{k-induction-algo-inc}
    \ENDWHILE
    \RETURN \textbf{unknown}
\end{algorithmic}
\vspace{1mm}
\end{minipage}
}
\end{algorithm}

Algorithm~\ref{k-induction-pd-algo} shows an extension
of \kinduction with continuously-refined invariants~\cite{kInduction}
that applies \pdr's aspect of learning from counterexamples to induction and
that can be applied both as a main proof engine as well as an invariant generator.
This allows us to apply this extension of \kinduction
as an invariant generator to a main \kinduction procedure,
similar to the \kiki approach~\cite{kInduction}.

\subsubsection{Inputs}
The algorithm takes the following inputs:
The value~$k_{init}$ is used to initialize the unrolling bound~$k$,
whereas the function~$\mathsf{inc}$ is used to increase~$k$
in line~\ref{k-induction-algo-inc} after each major iteration of the algorithm,
up to an upper limit of~$k$ defined by the value~$k_{max}$ enforced in line~\ref{k-induction-algo-max}.
The set of initial program states is described by the predicate~$I$,
the possible state transitions are described by the transition relation~$T$,
and the set of safe states is described by the safety property~$P$.
The accessor~$\mathsf{get\_currently\_known\_invariant}$ is used to obtain
the strongest invariant currently available
via a concurrently running (external) auxiliary-invariant generator.
A Boolean flag~$\mathit{pd}$ (reminding of ``property-directed'') is used to control
whether or not failed induction checks are used to guide the algorithm
towards a sufficient strengthening of the safety property~$P$ to prove correctness;
if~$\mathit{pd}$ is set to~$\false$, the algorithm behaves exactly like standard \kinduction.
Given a failed attempt to prove some candidate invariant~$Q$\,%
\footnote{Depending on the step the algorithm is in,
$Q$~may be either the safety property~$P$ or a proof obligation~$o$.}
by induction,
the function~$\mathsf{lift}$ is used to obtain
from a concrete counterexample-to-induction (CTI) state
a set of CTI states described by a state predicate~$C$.
An implementation of the function $\mathsf{lift}$
needs to satisfy the condition that for
a CTI~$s \in S$ where~$S$ is the set of program states,
$k \in \mathbb{N}$,
$\inv \in  (S \rightarrow \mathbb{B})$,
$Q \in (S \rightarrow \mathbb{B})$,
and $C = \mathsf{lift}(k, \inv, Q, s)$,
the following holds:
\linebreak
{\thinmuskip=0mu\medmuskip=2mu\thickmuskip=4mu
$C(s) \land \left(\forall s_n \in S: C(s_n) ~~\implies~~ \inv(s_n) \land \bigwedge\limits_{i=n}^{n+k-1} \left(Q(s_i) \land T(s_i, s_{i+1})\right) \implies \lnot Q(s_{n+k})\right)$},
which means that the CTI~$s$ must be an element of the set of states described by the resulting predicate~$C$
and that all states in this set must be CTIs, i.e., they need to be $k$\nobreakdash-predecessors of~$\lnot Q$\nobreakdash-states,
or in other words,
each state in the set of states described by the predicate~$C$
must reach some~$\lnot Q$\nobreakdash-state via~$k$ unrollings of the transition relation~$T$.
We can implement~$\mathsf{lift}$ using Craig interpolation~\cite{Craig57,McMillanCraig}
between $A: s = s_n$ and $B: \inv(s_n) \land \bigwedge\limits_{i=n}^{n+k-1} \left(Q(s_i) \land T(s_i,s_{i+1})\right) \implies \lnot Q(s_{n+k})$,
because $s$ is a CTI, and therefore we know that $A \implies B$ holds.
\footnote{The formula~$C$ is called Craig interpolant for two formulas $A$ and $B$ with $A \implies B$,
if $A \implies C$, $C \implies B$, and all variables in~$C$ occur in both~$A$ and~$B$.}
Hence, the resulting interpolant satisfies the criteria for~$C$
to be a valid lifting of~$s$ according to the requirements towards the function~$\mathsf{lift}$
as outlined above.
The function~$\mathsf{strengthen}$
is used to obtain for a~$k$\nobreakdash-inductive invariant
a stronger~$k$\nobreakdash-inductive invariant,
i.e., its result needs to imply the input invariant,
and, just like the input invariant,
it must not be violated within~$k$ loop iterations
and must be $k$\nobreakdash-inductive.

\subsubsection{Algorithm}
Lines~\ref{k-induction-algo-base-start}~to~\ref{k-induction-algo-base-end}
show the base-case check (BMC)
and lines~\ref{k-induction-algo-fc-start}~to~\ref{k-induction-algo-fc-end}
show the forward-condition check,
both as described in Sect.~\ref{sect:background}.
If~$\mathit{pd}$ is set to~$\true$,
lines~\ref{k-induction-algo-pd-start}~to~\ref{k-induction-algo-pd-end}
attempt to prove each proof obligation using \kinduction:
Lines~\ref{k-induction-algo-pd-base-start}~to~\ref{k-induction-algo-pd-base-sat}
check the base case for a proof obligation~$o$.
If any violations of the proof obligation~$o$ are found,
this means that a predecessor state of a~$\lnot P$\nobreakdash-state,
and thus, transitively, a~$\lnot P$\nobreakdash-state, is reachable,
so we return~$\false$.
If, otherwise, no violation was found,
lines~\ref{k-induction-algo-pd-base-unsat-start}~to~\ref{k-induction-algo-pd-base-unsat-end}
check the inductive-step case to prove~$o$.%
\,\footnote{Note that we do not need to check the forward condition for proof obligations,
because the forward condition is unrelated to the safety property and the proof obligations,
and therefore only needs to be checked once in each major iteration (i.e., once after each increment of~$k$).}
We strengthen the induction hypothesis of the step-case check
by conjoining auxiliary invariants from an external invariant generator
(via a call to $\mathsf{get\_currently\_known\_invariant}$)
and the auxiliary invariant computed internally from proof obligations
that we successfully proved previously.
If the step-case check for~$o$ is unsuccessful, we extract the resulting CTI state,
$\mathsf{lift}$ it to a set of CTI states,
and construct a new proof obligation
so that we can later attempt to prove that these CTI states are unreachable.
If, on the other hand, the step-case check for~$o$ is successful,
we no longer track~$o$ in the set~$O$ of unproven proof obligations
(this case corresponds to line~\ref{k-induction-algo-pd-ind-unsat-start}).
We could now directly use the proof obligation as an invariant,
but instead, in line~\ref{k-induction-algo-pd-ind-unsat-end}
we first try to $\mathsf{strengthen}$ it into a stronger invariant
that removes even more unreachable states from future consideration
before conjoining it to our internally computed auxiliary invariant.
In our implementation, we implement $\mathsf{strengthen}$
by attempting to drop components from a (disjunctive) invariant
and checking if the remaining clause is still inductive.
In lines~\ref{k-induction-algo-ind-start}~to~\ref{k-induction-algo-ind-end},
we check the inductive-step case for the safety property~$P$.
This check is mostly analogous
to the inductive-step case check for the proof obligations described above,
except that if the check is successful,
we immediately return~$\true$.

Note that \cref{k-induction-pd-algo} eagerly increases~$k$, even if
the set~$O$ of proof obligations is not empty.
This heuristic prevents the \pdr part from iterating through
long chains of proof obligations, it rather delegates the unrolling to the $k$-induction part.

\subsubsection{Example}
We now give an example of applying Alg.~\ref{k-induction-pd-algo}
to the example introduced in Fig.~\ref{fig:eq2.c}.%
\footnote{For ease of presentation, we assume that
  $I$~encodes the beginning of the program (lines~10--13),
  $T$~encodes the loop body (lines~15--16), and
  $P$~encodes the rest of the program (which checks the property to verify).}
For this example, we will configure the algorithm as a property-directed invariant generator.
We choose~${k_{init} = 1}$ as the initial bound and~${k_{max} = \infty}$
to force the algorithm to increase the bound~$k$ until a proof is found or an alarm is raised,
and thus, the condition in line~\ref{k-induction-algo-init-k} of Alg.~\ref{k-induction-pd-algo}
will always evaluate to~$\mathit{true}$.
To increment~$k$ by a value of~$1$ in each major iteration, we define~${\mathsf{inc}(k) = k + 1}$.
The set of initial states and the transfer relation are given by the input program~\lstinline{eq2.c}.
As safety property~$P$ we require
that no call to the function~\lstinline{__VERIFIER_error()} is reachable.
We will not use an external auxiliary-invariant generator,
hence we define $\mathsf{get\_currently\_known\_invariant}() = \mathit{true}$.
Instead, we want to use the algorithm's capability of property-directed invariant generation
and set~${\mathit{pd} = \mathit{true}}$.
The most important steps and variable values at some algorithm locations
are summarized in Table~\ref{tab:example}.

\begin{table}[t]
  \centering
  \caption{Relevant steps and values within Alg.~\ref{k-induction-pd-algo}
           if applied to the example introduced in Fig.~\ref{fig:eq2.c}.}
  \label{tab:example}
  \begin{tabular}{l@{~}|@{~}p{.37\linewidth}@{~~~~}p{.37\linewidth}}
    \toprule
      Major iteration & $1$ & $2$ \\
    \midrule
      Line~\ref{k-induction-algo-init-k}       & $k = 1$
                                               & $k = 2$ \\
      Line~\ref{k-induction-algo-base-start}   & BMC finds no error in $0$ iterations
                                               & BMC finds no error in up to $1$~iteration \\
      Line~\ref{k-induction-algo-fc-check-end} & Forward-condition check fails
                                               & Forward-condition check fails \\
      Line~\ref{k-induction-algo-pd-start}     & $\mathit{pd} = \mathit{true}$
                                               & $\mathit{pd} = \mathit{true}$ \\
      Line~\ref{k-induction-algo-pd-choose-o}  & $O = \emptyset$
                                               & $O = \{\lstinline!y! = \lstinline!z!\}$,
                                                 $o = (\lstinline!y! = \lstinline!z!$) \\
      Line~\ref{k-induction-algo-pd-base-end}  &
                                               & BMC finds no counterexample
                                                 to $\lstinline!y! = \lstinline!z!$
                                                 in up to $1$ iteration \\
      Line~\ref{k-induction-algo-pd-ind-aux-ext} &
                                               & $\extinv = \mathit{true}$ \\
      Line~\ref{k-induction-algo-pd-ind-aux}   &
                                               & $\intinv = \mathit{true}$,
                                                 $\inv = \mathit{true}$\\
      Line~\ref{k-induction-algo-pd-ind-sat-start} &
                                               & Step-case check for $o$ succeeds,
                                                 because $\lstinline!y! = \lstinline!z!$
                                                 is inductive \\
      Line~\ref{k-induction-algo-pd-ind-unsat-start} &
                                               & $O = \emptyset$ \\
      Line~\ref{k-induction-algo-pd-ind-unsat-end} &
                                               & $\intinv = (\lstinline!y! = \lstinline!z!)$ \\
      Line~\ref{k-induction-algo-ind-aux-ext}  & $\extinv = \mathit{true}$
                                               & $\extinv = \mathit{true}$ \\
      Line~\ref{k-induction-algo-ind-aux}      & $\intinv = \mathit{true}$,
                                                 $\inv = \mathit{true}$
                                               & \mbox{$\intinv = (\lstinline!y! = \lstinline!z!)$},
                                                 \mbox{$\inv = ((\lstinline!y! = \lstinline!z!))$} \\
      Line~\ref{k-induction-algo-ind-check}    & Step-case check fails
                                               & Step-case check succeeds \\
      Line~\ref{k-induction-algo-ind-sat-start}& $\mathit{pd} = \mathit{true}$
                                               & \\
      Line~\ref{k-induction-algo-ind-sat-s}    & $s = (\lstinline!y! = 0 \land \lstinline!z! = 1)$
                                               & \\
      Line~\ref{k-induction-algo-ind-sat-end}  & $O = \{\lstinline!y! = \lstinline!z!\}$
                                               & \\
      Line~\ref{k-induction-algo-inc}          & $k = 2$
                                               & \\
    \bottomrule
  \end{tabular}
  \vspace{-4mm}
\end{table}

In the first iteration, with $k = 1$,
we first need to check the base case,
i.e., we need to check in lines~\ref{k-induction-algo-base-start} to~\ref{k-induction-algo-base-end} of the algorithm
whether the call to function~\lstinline{__VERIFIER_error()} in line~5 of the input program
is reachable
with zero loop iterations.
Since this is not the case,
we continue to lines~\ref{k-induction-algo-fc-start}~to~\ref{k-induction-algo-fc-end}
of the algorithm,
where we check the forward condition,
i.e., we check whether we completely unrolled the loop,
that is, not more than zero unrollings of the loop are possible.
Since the loop condition depends on a nondeterministic input value,
this check will always fail,
and so we continue to line~\ref{k-induction-algo-pd-start} of the algorithm,
where property-directed invariant generation begins,
if it is switched on.
Because~$\mathit{pd} = \mathit{true}$,
we attempt to generate invariants from our set~$O$ of proof obligations.
This set, however, is currently empty,
and so we jump to the step-case check
in lines~\ref{k-induction-algo-ind-start} to~\ref{k-induction-algo-ind-check} of the algorithm,
where we assume for any iteration~$n$
($k$~iterations from $n$~to~$n+k-1 = n$) that the safety property holds,
and from this assumption attempt to conclude that the safety property will also hold in the next iteration~$n+1$~($n+k$).
In the context of this example, this means that we assume for any iteration~$n$
that we did not reach a call to the function~\lstinline{__VERIFIER_error()},
and from this assumption try to conclude that~\lstinline{__VERIFIER_error()}
is also not reachable in the next iteration~$n+1$.
However, we have not yet computed any auxiliary invariants, so~$\intinv = \mathit{true}$
and we cannot obtain any invariants from an external invariant generator either,
because we defined $\mathsf{get\_currently\_known\_invariant}() = \mathit{true}$,
therefore also~$\inv = \mathit{true}$ in line~\ref{k-induction-algo-ind-aux}.
As described in the introduction,
checking the inductive-step case for the safety property~$P$
will always fail without auxiliary invariants
for this example,
because the SMT solver can provide a model
where~$\lstinline{y} \neq \lstinline{z}$ holds before this iteration,
but after this iteration,
the nondeterministic loop condition evaluates to~$\mathit{false}$
and the safety property is violated.
For this example, we assume the SMT solver produces a model
where before the iteration, $\lstinline{y} = 0$~and~$\lstinline{z} = 1$,
and this state becomes our satisfying predecessor state~$s$
in line~\ref{k-induction-algo-ind-sat-s} of the algorithm.
Next, we try to lift this state to a more abstract state
that still satisfies the property
that all of its successors violate the safety property.
For this example, we assume that lifting produces the abstract state~$\lstinline{y} \neq \lstinline{z}$.
We then negate this abstract state to obtain the proof obligation~$\lstinline{y} = \lstinline{z}$.
This means that we have learned that we should prove the invariant~$\lstinline{y} = \lstinline{z}$,
such that in future induction checks, we can remove all states where~$\lstinline{y} \neq \lstinline{z}$
from the set of predecessor states that need to be considered.
Therefore, we add the proof obligation to our set~$O$ of current proof obligations,
such that~$O := \{\lstinline{y} = \lstinline{z}\}$.
We then increment~$k$ using the function~$\mathsf{inc}(k) = k + 1$,
and continue into the next major iteration of our algorithm
with~$k = 2$.

Again, we check the base case in line~\ref{k-induction-algo-base-start},
and again, the check succeeds,
because we cannot reach a call to the function~\lstinline{__VERIFIER_error()}
in line~5 of the input program
within one loop iteration either.
As explained before,
the forward-condition check in line~\ref{k-induction-algo-fc-check-end} of the algorithm
must always fail for this input program,
and so we again continue with property-directed invariant generation.
This time, however, the set~$O$ of current proof obligations is not empty,
and so we choose from it
in line~\ref{k-induction-algo-pd-choose-o} of the algorithm
the only proof obligation~$o$ it currently contains,~$\lstinline{y} = \lstinline{z}$,
and try to prove it by induction.
Hence, we first check the proof-obligation base case in
lines~\ref{k-induction-algo-pd-base-start}~to~\ref{k-induction-algo-pd-base-end} of the algorithm.
Obviously, the check succeeds, because~$\lstinline{y} = \lstinline{z}$ holds
before and after the first loop iteration of the program:
the two variables are both initialized to~$0$,
and are both incremented once.
We still have not computed any auxiliary invariants yet, so~$\intinv = \mathit{true}$,
and we defined $\mathsf{get\_currently\_known\_invariant}() = \mathit{true}$,
therefore also~$\inv = \mathit{true}$ in line~\ref{k-induction-algo-pd-ind-aux}.
This means that when we check the inductive-step case in line~\ref{k-induction-algo-pd-ind-sat-start} of the algorithm, we have no auxiliary invariants.
Nevertheless, the check is successful,
because contrary to checking the safety property~$P$,
which concerns the reachability of a function call
outside the loop and therefore depends on the nondeterministic loop condition,
the condition that~$\lstinline{y} = \lstinline{z}$ before and after each loop iteration
does not depend on any nondeterministic input value.
Thus, we remove~$\lstinline{y} = \lstinline{z}$ from our set~$O$ of current proof obligations
in line~\ref{k-induction-algo-pd-ind-unsat-start} of the algorithm.
We then try to strengthen~$\lstinline{y} = \lstinline{z}$
to obtain a stronger~$k$\nobreakdash-inductive
(i.e., $2$\nobreakdash-inductive, since currently~$k = 2$)
invariant for this program.
However, since~$\lstinline{y} = \lstinline{z}$ is already a quite strong invariant
for this input program,
we are unlikely to find a stronger one easily.
For this example, we therefore assume that this step strengthens the invariant
to~$\lstinline{y} = \lstinline{z}$ again.
We then conjoin~$\lstinline{y} = \lstinline{z}$ to our current internally computed invariant~$\intinv$,
such that~$\intinv = (\lstinline{y} = \lstinline{z})$
in line~\ref{k-induction-algo-pd-ind-unsat-end} of the algorithm.
Next, we check the inductive-step case for the safety property~$P$
in lines~\ref{k-induction-algo-ind-start}~to~\ref{k-induction-algo-ind-check}.
This time, with~$\inv = (\lstinline{y} = \lstinline{z})$,
the check will succeed,
and the algorithm returns~$\mathit{true}$ in line~\ref{k-induction-algo-ind-unsat}.

%% file: evaluation.tex
\input{evaluation/data-commands}

\input{evaluation/new/data-commands}
\newcommand{\ntasksTrueNotSolvedByKinductionPlain}{\num{\PdrInvKinductionPlainTrueNotSolvedByKinductionPlainTotalCount}}

\input{evaluation/pathprograms/data-commands}
\newcommand{\npathprogramstasksTrueNotSolvedByKinductionPlain}{\num{\PdrInvPathprogramsKinductionPlainTrueNotSolvedByKinductionPlainTotalCount}}
\newcommand\npathprogramstasksTrueSolvedByKinductionPlain{\num{\the\numexpr (\npathprogramstruetasksplain - \PdrInvPathprogramsKinductionPlainTrueNotSolvedByKinductionPlainTotalCount)\relax}}

\newcommand{\plotpath}{evaluation}
\input{\plotpath/plot-defs.tex}

\section{Evaluation}

In this section, we present an extensive experimental study
on the effectiveness and efficiency of adaptations of \pdr to software verification.

\subsection{Compared Approaches}
We use the following abbreviations
to distinguish between the different techniques
that we evaluated:
\begin{description}[itemsep=0pt,parsep=1pt,leftmargin=11pt,topsep=1pt]
  \item[CTIGAR:]
    \ctigar~\cite{CTIGAR} is an adaptation of \pdr to software verification.
    Our evaluation compares two implementations of \ctigar,
    namely \vvtctigar from the tool \vvt and
    our own implementation \cpactigar.
    \vvt~\cite{VVT-COMP16} also provides a configuration
    that runs a parallel portfolio combination
    of \vvtctigar and bounded model checking,
    which we call \vvt-Portfolio.
  \item[\kind:]
    \kind~\cite{kInduction} denotes the plain \kinduction algorithm
    without property direction and without auxiliary invariants,
    i.e.,~we configure Alg.~\ref{k-induction-pd-algo}
    such that~$\mathit{pd} = \false$
    and $\mathsf{get\_currently\_known\_invariant()}$
    always returns~$\true$.
  \item[\kipdr:]
    \kipdr denotes a configuration of Alg.~\ref{k-induction-pd-algo}
    such that~$\mathit{pd} = \true$
    and $\mathsf{get\_currently\_known\_invariant()}$
    always returns~$\true$,
    i.e.,~\kinduction with property direction
    but without additional auxiliary-invariant generation.
    \kipdr is, like \ctigar, an adaptation of \pdr to software verification.
  \item[\kidf:]
    \kidf~\cite{kInduction} denotes a parallel combination of \kinduction
    (without property direction)
    with a data-flow-based auxiliary-invariant generator
    that continuously supplies the \kinduction procedure with invariants.
    Here, we configure Alg.~\ref{k-induction-pd-algo}
    such that~$\mathit{pd} = \false$
    and $\mathsf{get\_currently\_known\_invariant()}$
    always returns the most recent (strongest)
    invariant computed by the data-flow-based auxiliary-invariant generator.
  \item[\kikipdr:]
    Similarly to \kidf,
    \kikipdr denotes a parallel combination of \kinduction
    with an auxiliary-invariant generator
    --- in this case, \kipdr{} ---
    that continuously supplies invariants to the \kinduction procedure.
    Here, we configure one instance of Alg.~\ref{k-induction-pd-algo}
    such that~$\mathit{pd} = \false$
    and $\mathsf{get\_currently\_known\_invariant()}$
    always returns the most recent (strongest)
    invariant computed by \kipdr (a second instance of Alg.~\ref{k-induction-pd-algo}
    that is configured such that~$\mathit{pd} = \true$
    and $\mathsf{get\_currently\_known\_invariant()}$
    always returns~$\true$).
  \item[\kidfkipdr]
    \kidfkipdr denotes a parallel combination of \kinduction
    with an auxiliary-invariant generator
    that uses a sequential combination
    of a data-flow-based invariant generator
    and \kipdr to continuously supply \kinduction with auxiliary invariants.
    Here, we configure one instance of Alg.~\ref{k-induction-pd-algo}
    such that~$\mathit{pd} = \false$
    and $\mathsf{get\_currently\_known\_invariant()}$
    always returns the most recent (strongest)
    invariant computed by a sequential combination
    of the data-flow-based invariant generator
    and \kipdr (a second instance of Alg.~\ref{k-induction-pd-algo}
    that runs after the data-flow-based invariant generator finishes
    and is configured such that~$\mathit{pd} = \true$
    and $\mathsf{get\_currently\_known\_invariant()}$
    always returns~$\true$).
\end{description}
We do not evaluate the used invariant generators as standalone verification approaches,
because they are designed specifically to be used as auxiliary components
and do not perform well enough in isolation.
For example, data-flow based invariant-generation approaches
are often too imprecise to verify tasks by themselves,
whereas more precise techniques like \kipdr might spend too much time unnecessarily,
resulting in too many timeouts to be competitive.
Instead, we use the framework of \kinduction with continuously refined invariant generation,
which has been shown to be able to leverage the advantages
of both quick but imprecise and slow but precise techniques~\cite{kInduction}.

\subsection{Evaluation Goals}

The overall goals of our experimental evaluation are
to establish a baseline for experimental comparisons
involving \pdr in the context of software verification
and to determine whether we can combine the strengths of \pdr
with those of \kinduction.
We are interested in answers to the following research hypotheses.
We use the symbols \cmark{} and \xmark{} to indicate that a hypothesis will be confirmed
or refuted, respectively, by our experiments on the given benchmark sets.

\begin{description}[itemsep=0pt,parsep=1pt,leftmargin=11pt,topsep=1pt]
\item[Hypothesis 1:]
  \cpachecker is a suitable platform
  for implementing and evaluating \pdr-based techniques.
  That is, we are able to implement state-of-the-art \pdr techniques in this framework
  and obtain competitive results.
  \cmark
\item[Hypothesis 2:]
  By providing \kinduction with a \kipdr invariant generator,
  which utilizes the \pdr-aspect of guiding invariant discovery by leveraging failed induction attempts,
  we can improve the overall effectiveness of \kinduction.
  \xmark
\item[Hypothesis 3:]
  On small programs, such as path programs,
  \kipdr is a more effective invariant generator
  than data-flow-based techniques
  and is therefore well-suited
  for the approach of generating path invariants~\cite{PathInvariants}.
  \xmark
\item[Hypothesis 4:]
  While \kipdr is often outperformed by simpler, data-flow-based invariant-generation techniques,
  there exist programs that can not be solved using the considered data-flow-based invariant-generation techniques,
  but can be solved using \kipdr as auxiliary-invariant generator.
  Furthermore, \kikipdr can be more efficient than state-of-the-art verifiers from the \svcomp evaluation.
  \cmark
\item[Hypothesis 5:]
  Our conclusions are relevant,
  because our implementation of \kidfkipdr is competitive
  when compared to the best available implementations of \pdr technology for software verification.
  \cmark
\end{description}

\subsection{Benchmark Set}

The benchmark set we use in our experiments consists of verification tasks
from the International Competition on Software Verification (\svcomp)~\cite{SVCOMP17},
in particular, we use the benchmark categories as used for
\svcomp~2018\,\footnote{{\scriptsize\url{https://sv-comp.sosy-lab.org/2018/benchmarks.php}}}.
In addition, we use a set of new verification tasks, which are
(a)~generated path programs from existing verification tasks,
in order to obtain verification tasks that are small enough for the \pdr approach to handle, and
(b)~manually created programs that explore the expressive power of the \pdr approach.
All programs that were used in the evaluations are available on
the supplementary web page\,\footnote{{\scriptsize\url{https://www.sosy-lab.org/research/pdr-compare/}}}
and in the replication package~\cite{PDR-artifact}.

We consider only verification tasks
where the property to verify is the unreachability of a program location
(excluding the properties for
no-overflows,
memory safety,
and termination,
which are not in the scope of our evaluation).
From the resulting set of verification tasks, we excluded the categories
\textit{ReachSafety-Recursive} and
\textit{ConcurrencySafety},
each of which is not supported by at least one of the evaluated implementations.
The remaining set of categories consists of a total of~\ntasks~verification tasks from
the subcategory \mbox{\textit{DeviceDriversLinux64\_ReachSafety}}
of the category \textit{SoftwareSystems}
and from the following subcategories of the category \textit{ReachSafety}:
\textit{Arrays},
\textit{Bitvectors},
\textit{ControlFlow},
\textit{ECA},
\textit{Floats},
\textit{Heap},
\textit{Loops},
\textit{ProductLines}, and
\textit{Sequentialized}.
A total of \nfalsetasks~of these tasks
are known to contain a specification violation,
while the rest of the tasks are meant to satisfy their specification.

\subsection{Verification Tools and Algorithms}
We evaluate all verification tools that are
(1) publicly available,
(2) support C as an input-program language, and
(3) implement at least one \pdr-based approach.
There are three such verifiers available:
We use \cpachecker~\cite{CPACHECKER} in revision~\texttt{27\,742} from the trunk,
\seahorn~\cite{SEAHORN} in version \texttt{F16-0.1.0-rc3},
and the \vvt version used in the 2016 Competition on Software Verification (\svcomp~2016)~\cite{VVT-COMP16}.%
\,\footnote{We could not use the newer version of \vvt from GitHub, because it has a bug where its CTIGAR component does not report solved tasks and no fix was available.}

For implementing our own modules and extensions, we choose the
framework \cpachecker, because it
(a)~is a large open-source project that
    ``has a well established, mature codebase maintained by a large development team''\,%
    \footnote{Citation source: {\scriptsize\url{https://www.openhub.net/p/cpachecker}}}, and
(b)~seems to have efficient implementations of the core components\,%
    \footnote{\svcomp~2018: {\scriptsize\url{https://sv-comp.sosy-lab.org/2018/results/results-verified/}}}.
We configured \cpachecker to use \mathsat as an SMT solver
and to apply the SMT theories of equality with uninterpreted functions, bit vectors, and floats.

Unfortunately, we could include
neither the implementations of Cimatti and Grigio~\cite{SoftwareIC3},
nor that of Lange, Prinz, Neuh{\"{a}}u{\ss}er, Noll, and Katoen~\cite{IC3-CFA, IC3-CFA-Improved},
in our evaluation.
The former are only applicable to transition systems in SMT format
and control-flow graphs in SMTCFA format, respectively, not to C~programs,
and the latter is not publicly available.

\subsection{Experimental Setup}
For our experiments, we executed the chosen software verifiers on machines with
one 3.4\,GHz CPU (Intel Xeon E3-1230~v5) with 8~processing units and 33\,GB of RAM each.
The operating system was Ubuntu~16.04 (64~bit),
using Linux~4.4 and OpenJDK~1.8.
We limited each verification run to
two CPU cores,
a CPU run time of \SI{15}{min},
and a memory usage of \SI{15}{GB}.
To ensure reliable and accurate measurements,
we used the benchmarking framework \benchexec%
\footnote{{\scriptsize\url{\benchexecurl}}}~\cite{Benchmarking-STTT}
to conduct our experiments.

\subsection{Presentation}
All benchmarks, tools, and the full results
of our evaluation are available on
a supplementary web page\,%
\footnote{{\scriptsize\url{https://www.sosy-lab.org/research/pdr-compare/}}}
and in the replication package~\cite{PDR-artifact}.
All reported times are rounded to two significant digits.
Because it is sometimes difficult to compare results
in the presence of wrong alarms or wrong proofs,
we use the community-agreed schema of \svcomp
to assign quality values to each verification result, i.e.,
to calculate a score that quantifies the quality of the results for a verifier.
For every real bug found, 1~point is assigned,
for every correct safety proof, 2 points are assigned.
A score of~\num{16}~points is subtracted for every wrong alarm (false positive) reported by the tool,
and~\num{32}~points are subtracted for every wrong proof of safety (false negative).
While this scoring scheme may not match every use case,
it is not arbitrary: it follows a community consensus~\cite{SVCOMP17}
on the difficulty of verification versus falsification
and the importance of correct results,
where proving correctness (computing invariants) is considered more complicated than finding bugs (computing error paths),
and wrong answers are punished severely (by a factor of~16).
We consider this a good fit for evaluating approaches
such as \kinduction and \pdr,
which focus on producing safety proofs.

\subsection{Hypothesis 1: Suitability of \cpachecker for \pdr}

In our first set of experiments,
we establish that the \cpachecker framework
is a suitable platform for implementing \pdr-based techniques
by comparing the effectiveness and efficiency of our own implementation
of \pdr for software-model checking (\cpactigar)
to the only available verifier that implements a pure \pdr approach for software-model checking,
which is \vvtctigar~\cite{CTIGAR}.%
\footnote{We chose the direct comparison with \vvt over a comparison with \seahorn here,
because \cpactigar uses the same approach as \vvtctigar,
and because of the large number of incorrect results for \seahorn.}
The implementation in \vvt has been published previously~\cite{VVT-COMP16},
whereas the implementation in \cpachecker was newly developed for this article
to serve as a baseline to compare further implementations of adaptations of \pdr in \cpachecker against.
Columns two and three of Table~\ref{tab:main-configs-results} compare the results obtained
by running the two implementations of CTIGAR on the whole benchmark set,
and the last column of the table shows the results achieved with the standard configuration of~\vvt,
which runs not only CTIGAR, but a portfolio analysis of CTIGAR and bounded model checking.
The table lists the score achieved by each configuration
and breaks it down into the amount of correct and incorrect proofs and alarms.
It also lists for how many tasks a configuration
exceeded the time limit of \SI{15}{min}
or the memory limit of \SI{15}{GB},
as well as the amount of tasks a tool configuration
could not solve for other reasons,
such as failures in the parser, the SMT solver, or the analysis.
Furthermore, the table lists the total, mean, and median
CPU and wall times spent by a configuration.
We observe that while \cpactigar is more effective (i.e., solves more tasks correctly)
and has fewer wrong results than \vvtctigar,
\vvt appears to be more efficient (i.e., is faster).
We attribute the overall differences between the tools regarding correctly and incorrectly solved tasks
to the fact that \cpachecker as a framework is older and more mature than \vvt.
In fact, we see many cases where \vvt fails to parse the~C~code of a task,
likely due to limited support for some features of the~C~programming language.
We also see that only few real bugs are detected by \cpactigar~(\num{\PdrInvPdrCorrectFalseCount}) and \vvtctigar~(\num{\VvtCtigarCorrectFalseCount}),
while the \vvt portfolio analysis produces significantly more correct alarms~(\num{\VvtPortfolioCorrectFalseCount}).
This confirms our expectations,
because \pdr is mainly aimed at computing invariants (and thus finding proofs),
whereas bounded model checking is well-known as a good technique for finding bugs.
When considering the union of tasks solved by \cpactigar and \vvtctigar,
i.e., the tasks solved by either \cpactigar, \vvtctigar, or both (``virtual best''),
we find \num{\PdrInvOracleCorrectCount}~tasks solved correctly,
\num{\PdrInvOracleCorrectTrueCount}~of which constitute proofs, and
\num{\PdrInvOracleCorrectFalseCount}~of which constitute alarms.
This means that on the one hand,
for a large amount of tasks,
\cpactigar and \vvtctigar both produce the same correct results,
but on the other hand,
there are not only many tasks \cpactigar can solve correctly
that \vvtctigar cannot solve,
but also several tasks that \vvtctigar can solve but \cpactigar cannot.
Like the differences in the amounts of incorrect results,
we can attribute this observation to the way
the different frameworks model the tasks internally,
where for some types of tasks, \cpachecker takes a safer approach
to the trade-off between correctness and effectiveness than \vvt,
and where some language constructs supported by the older \cpachecker framework
are not supported by the younger \vvt.

\input{main-configs-table}

\begin{figure}[t]
  \centering
    \input{\plotpath/ctigar}
  \caption{Quantile plot
  for accumulated number of solved tasks (proofs and alarms)
  showing the CPU~time (linear scale below~\SI{1}{s}, logarithmic above)
  for the successful results
  of \cpactigar and \vvtctigar}
  \label{fig:ctigar-results}
\end{figure}
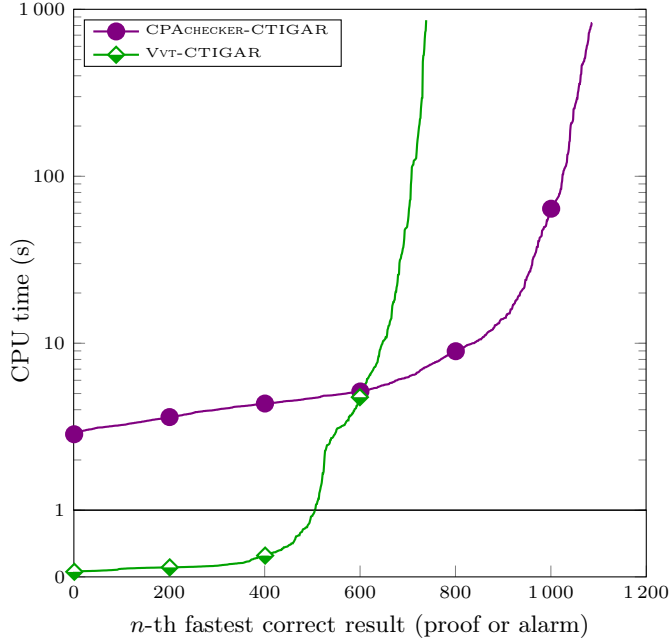

To further analyze the efficiency of \cpactigar and \vvtctigar,
the quantile plot in Fig.~\ref{fig:ctigar-results}
shows the CPU times that the two tool configurations spent on their correct results.
The plot shows that for both tools, there exist approximately~\num{610} tasks
for which each of the tasks can be solved in less than approximately~\SI{5.2}{\second},
and that in this range, \vvt is significantly faster than \cpachecker,
which can partially be explained by the approximately~\SI{3}{\second} startup time of the Java Virtual Machine that \cpachecker runs in.
Beyond that point, however, \cpactigar scales better than \vvtctigar.

In conclusion, our implementation was shown to be at least as good as (and even better than) the only
available implementation of \pdr for software model checking.
Therefore, the hypothesis that \cpachecker is a suitable platform
for implementing and evaluating adaptations of \pdr holds.
\cmark

\subsection{Hypothesis 2: Augmenting \kInduction with \kipdr}

We now attempt to establish that augmenting \kinduction with auxiliary invariants from \kipdr improves its overall effectiveness:
In the previous experiment, we presented \cpactigar,
which uses an adaptation of \pdr as its verification engine.
When we compare the results of \cpactigar from Table~\ref{tab:main-configs-results}
to the results from evaluations~\cite{kInduction,AlgorithmComparison-JAR}
of other techniques for previous versions of the same benchmark set,
we see that neither of the two CTIGAR implementations is competitive.
For example, running the \kinduction configuration of \cpachecker without auxiliary-invariant generation
on our benchmark set, we obtain~\num{\PdrInvKinductionPlainCorrectTrueCount}~correct proofs
and~\num{\PdrInvKinductionPlainCorrectFalseCount}~correct alarms,
as shown in the fourth column of Table~\ref{tab:main-configs-results}~(\kind).

In general, however, the strength of \pdr
is considered to be its capability for generating safety invariants,
so that it is more interesting to analyze its usefulness as an invariant generator.

\begin{figure}[t]
  \centering
    \input{\plotpath/kInduction-true-not-solved-by-kInduction-plain}
  \vspace{-3mm}
  \caption{Quantile plot
  of accumulated number of bug-free tasks proved correctly by
  different approaches for generating auxiliary invariants
  but not solved by \kinduction without auxiliary invariants
  }
  \label{fig:k-induction-true-not-solved-by-kInduction-plain-results}
  \vspace{-4mm}
\end{figure}
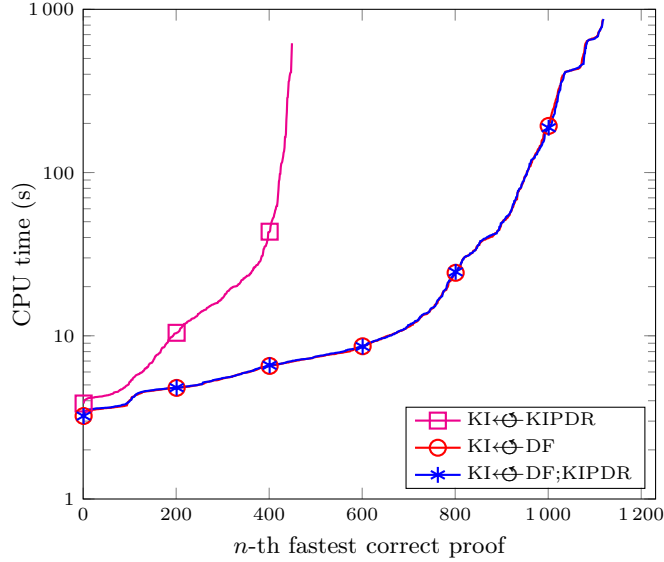

For our second experiment,
in which we solely compare configurations of \cpachecker to minimize confounding effects,
we therefore first took all~\ntasksTrueNotSolvedByKinductionPlain~tasks in our benchmark set
that do not contain bugs \textit{and} cannot be solved by \kinduction without an auxiliary-invariant generator,
i.e., all tasks where \kinduction might potentially benefit from auxiliary invariants.
On this set of tasks, we ran three configurations of \kinduction with auxiliary-invariant generation:
\kikipdr uses the property-directed extension of \kinduction described in Sect.~\ref{sect:approach}
as an auxiliary-invariant generator to a main \kinduction procedure,
\kidf uses a sequential combination of several data-flow analyses to generate auxiliary invariants,
and \kidfkipdr combines the invariant generators of the previous two.
Table~\ref{tab:kind-results-true-not-solved-by-k-induction-plain}
shows that for \kikipdr the property-directed invariant-generation component \kipdr
helps \kinduction to find proofs
for~\num{\PdrInvKinductionKipdrTrueNotSolvedByKinductionPlainButKipdrCorrectTrueCount}~tasks
that it could not solve without auxiliary invariants.
However, \kidf is much more efficient and effective,
and the combination of the two invariant generators
provided by \kidfkipdr does not contribute a significant improvement.
The corresponding quantile plot in Fig.~\ref{fig:k-induction-true-not-solved-by-kInduction-plain-results}
suggests that \kipdr is slower than \kidf (also compare the median CPU time),
and the fact that the graph of \kidfkipdr mostly overlaps with the graph of \kidf
suggests that there are only very few cases for which the behavior of the two configurations differs.
Therefore, on the currently available large benchmark set,
we \textit{cannot} confirm Hypothesis~2, which states that the invariant generator \kipdr
helps us improve the effectiveness of \kinduction beyond the current state of the art;
rather the null hypothesis holds, which states that there is no significant difference
(only two more solved instances, which is not a significant improvement).
\xmark

\input{k-induction-true-not-solved-by-kInduction-plain-table}

\subsection{Hypothesis 3: \kipdr as a Path-Invariant Generator}

While refuting Hypothesis~2,
we learned that \kipdr is a less efficient and effective invariant generator
than data-flow-based techniques on our set of benchmarks.
One explanation may be that \kipdr is too expensive to generate invariants for whole programs within the time limit.
There exists research, however,
that suggests that one solution for applying expensive invariant-generation techniques
is to extract a (smaller) so-called \textit{path program} from the (larger) original program,
apply the expensive techniques to the path program instead,
and then transfer the \textit{path invariants} computed for the path program
back to the analysis that attempts to solve the original verification task~\cite{PathInvariants}.

\input{pathprograms.k-induction-true-not-solved-by-kInduction-plain-table}

To determine if this approach provides a more suitable setting for \kipdr,
we extract from our original benchmark set~\num{\npathprogramstruetasksplain}~path programs
for which we are confident that they do not contain bugs.
Then, we take the same approach as in the previous experiment,
i.e., we exclude all~\npathprogramstasksTrueSolvedByKinductionPlain{} path programs
that can be solved by \kinduction without auxiliary invariants,
and on the remaining~\npathprogramstasksTrueNotSolvedByKinductionPlain{} tasks,
we run \kikipdr, \kidf, and \kidfkipdr.

Table~\ref{tab:pathprograms.kind-results-true-not-solved-by-k-induction-plain}
shows the results of this experiment.
We observe again that \kikipdr is the least effective configuration,
making \kipdr an ineffective invariant generator
even for this simplified set of tasks comprised of path programs.
Therefore, we consider Hypothesis~3 as refuted on the given benchmark set of path programs:
There is no evidence to suggest that \kipdr is an effective path-invariant generator
compared to state-of-the-art techniques
(only one more solved instance, which is not a significant improvement).
\xmark

\subsection{Hypothesis 4: \kipdr versus Data-Flow Techniques}

Now we show that the higher efficiency of data-flow-based techniques
is most likely due to the simple form of the invariants needed to prove the programs correct.
This explains why the conceptually more advanced \kipdr
is outperformed by the data-flow-based techniques.

For this experiment,
we selected from the previous set of benchmarks
those~\num{\PdrInvKinductionKipdrTrueNotSolvedByKinductionPlainButKipdrCorrectTrueCount}~tasks
that were solved by \kikipdr
and analyzed them using specific components
of the sequential combination of data-flow analyses used in \kidf.
Specifically, we used one configuration that uses the abstract domain of Boxes~\cite{Boxes},
one configuration using Boxes and the template~\mbox{$\mathsf{Eq} := (x = y)$}, and
one configuration using Boxes, the template~$\mathsf{Eq}$, and the template~\mbox{$\mathsf{Mod2}:=(|x|\ \%\ 2 = {c})$} with $x,y \in X$, where $X$ is the set of program variables, and $c \in {0,1}$.

\input{invariants-table}

The results are reported in Table~\ref{tab:kind-results-not-solved-by-k-induction-plain-but-kipdr}.
The first configuration, Boxes, is already sufficient
to verify~\num{\PdrInvKinductionDfStaticZeroZeroTTrueNotSolvedByKinductionPlainButKipdrCorrectTrueCount} of the~\num{\PdrInvKinductionKipdrTrueNotSolvedByKinductionPlainButKipdrCorrectTrueCount}~tasks
verified by \kikipdr.
Of the remaining~\num{12}~tasks,~\num{2}~can be verified using a combination of Boxes and $\mathsf{Eq}$
as an invariant generator.%
\,\footnote{However, due to the higher precision after adding the~$\mathsf{Eq}$~template,
there is also a task that is no longer solved,
thus, the total number of proofs increases only by one.}
Of the~\num{10}~tasks that are then left,
another~\num{3} can be verified when adding the $\mathsf{Mod2}$~template to the combination.
The flexibility of \kikipdr allows for another~\num{7}~tasks to be solved:
For two of these tasks,
namely \texttt{functions\_true-unreach-call1\_true-termination.i}
and \texttt{phases\_true-unreach-call1.i},
the key is still tracking whether variables are even or odd,
but the control flow is too complex for our simple data-flow analysis to succeed;
the task \texttt{ddlm2013\_true-unreach-call.i} requires
the disjunctive invariant $a = b \lor flag = 0$;
for \texttt{module\_get\_put-drivers-net-pppox\_false-termination.ko\_true-unreach-\allowbreak{}call.cil.out.i.pp.i},
\kikipdr succeeds by first proving the auxiliary invariant $ldv\_module\_refcounter > 0$
and then subsequently proving the stronger invariant $ldv\_module\_refcounter = 1$,
which in turn is a sufficiently strong auxiliary invariant to solve the task;
for each of the remaining three,
namely \texttt{s3\_srvr\_1\_true-unreach-call\_false-termination.cil.c},
\texttt{s3\_srvr\_2\_true-unreach-call\_false-termination.cil.c}, and
\texttt{s3\_srvr\_8\_true-unreach-call\_false-termination.cil.c},
\kikipdr constructs multiple large disjunctive invariants
that are too complex to manually dissect,
but each of which constrains the state space,
such that together, they suffice as an auxiliary invariant to solve the corresponding task.

The simple data-flow-based techniques, however, are much more efficient (see CPU times).
This explains why over the whole set of benchmarks,
which apparently contains many tasks for which simple invariants are sufficient,
the data-flow based techniques are more successful.

\begin{table}[t]
  \centering
  \caption{Results of four \kinduction-based configurations in \cpachecker
           with different approaches for generating auxiliary invariants
           for seven manually crafted verification tasks
           that do not contain bugs
           and are not solved by \kinduction without auxiliary invariants;
	   an entry ``T'' means that the CPU-time limit was exceeded,
	   an entry ``M'' means that the memory limit was exceeded,
	   and all other entries represent the CPU time a configuration spent
	   to correctly solve the task
           }
  \label{tab:hand-crafted-results}
  \newcommand\stoh[1]{\fpeval{#1/3600}}
  \newcommand\precnum[1]{\tablenum[round-mode=off,table-format=3.2]{#1}}
  \newcommand\rndnum[1]{\tablenum[round-mode=figures,round-precision=2,table-format=3.2]{#1}}
  \begin{tabular}{l|c|ccc|c}
    \toprule
    \multicolumn{1}{l|}{\multirow{2}{*}{Task}}
        & \multicolumn{1}{c|}{\multirow{2}{*}{Fig.}}
        & \multicolumn{3}{c|}{\static}
        & \multicolumn{1}{c}{\multirow{2}{*}{\kikipdr}}
        \\
             &
             & \multicolumn{1}{c}{Boxes}
             & \multicolumn{1}{@{~~~}c@{~~~}}{\begin{minipage}{5ex}Boxes,\\$\mathsf{Eq}$\end{minipage}}
             & \multicolumn{1}{@{~~}c@{~~~}|}{\begin{minipage}{5ex}Boxes,\\$\mathsf{Eq}$,\\$\mathsf{Mod2}$\end{minipage}} \\
    \midrule
    \progconst
      & \ref{fig:const.c}
      & \SI{3.3}{s}
      & \SI{3.3}{s}
      & \bfseries\SI{3.2}{s}
      & \SI{3.8}{s}\\
    \progeqone
      & \ref{fig:eq1.c}
      & T
      & \bfseries\SI{3.2}{s}
      & \SI{3.3}{s}
      & \SI{4.9}{s}\\
    \progeqtwo
      & \ref{fig:eq2.c}
      & M
      & M
      & M
      & \bfseries\SI{3.9}{s}\\
    \progeven
      & \ref{fig:even.c}
      & T
      & T
      & \bfseries\SI{3.5}{s}
      & \SI{3.9}{s}\\
    \progodd
      & \ref{fig:odd.c}
      & T
      & T
      & \bfseries\SI{3.4}{s}
      & \SI{4.1}{s}\\
    \progmodfour
      & \ref{fig:mod4.c}
      & T
      & T
      & T
      & \bfseries\SI{3.6}{s}\\
    \progbin
      & \ref{fig:bin-suffix-5.c}
      & M
      & M
      & M
      & \bfseries\SI{3.6}{s}\\
  \bottomrule
  \end{tabular}
\end{table}

\input{hand-crafted-examples}

To further explore the differences between the approaches regarding their expressive power,
we manually created seven additional verification tasks
that outline the strengths and weaknesses of the configurations discussed above.
We list all discussed example programs as figures in this section,
except for \progeqtwo, which is already listed in \cref{fig:eq2.c}.
We do not repeat lines~\lstinline{1}~to~\lstinline{9},
because they are the same as in \cref{fig:eq2.c} for all example programs.
Table~\ref{tab:hand-crafted-results} shows the results we obtained for these tasks.
All configurations were able to prove safety for the task~\progconst (\cref{fig:const.c}),
which was crafted such that it is sufficient to detect the invariant $s = 0$.
The task~\progeqone shown in \cref{fig:eq1.c}
requires the invariant generators to compute the invariant $w = x \land y = z$,
for which the abstract domain of Boxes is not strong enough by itself,
but which is trivial for the configurations using the~$\mathsf{Eq}$~template and for \kipdr.
The task~\progeqtwo shown in \cref{fig:eq2.c},
which has already been discussed as an example in the introduction,
requires a similarly simple equality invariant, $y = z$,
but the only way to prove it is to first prove that at least initially,%
~$x = w \land y = (w + 1) \land z = (x + 1)$ holds,
for which the simple abstract domains are too weak,
but which is once again easy for \kipdr.
The tasks~\progeven~(\cref{fig:even.c}) and~\progodd~(\cref{fig:odd.c})
were specifically designed
to be solved only using invariant generators
that could provide the invariants~$|x|\ \%\ 2 = 0$ and~$|x|\ \%\ 2 = 1$, respectively,
so it is not surprising that of the configurations using data-flow-based invariant generators
only the one using the $\mathsf{Mod2}$-template was able to solve the task.
The task~\progmodfour shown in \cref{fig:mod4.c},
however, while being a simple adaptation of \progeven
that requires the invariant~$|x|\ \%\ 4 = 0$,
is only solved by \kipdr,
as is the task~\progbin shown in \cref{fig:bin-suffix-5.c},
which requires an invariant~$(x\ \&\ 5) = 5$ where '$\&$' is the binary AND operator:
While these last two tasks are conceptually not more difficult to solve
than the tasks that are solved using templates,
each of them would require a new template to be specified (and implemented),
whereas the flexibility of \kipdr allows the invariant generator
to discover the invariant without any help.
This shows that while a fitting data-flow-based invariant generator may be efficient if it is available,
\kipdr is generally more powerful, because it does not require specific templates.

The conclusion of our study on hand-crafted verification tasks is that
there are programs for which \kipdr is the superior technique,
and that one of the following assumptions might be true:
(a)~the large benchmark set from the sv-benchmarks repository
is not diverse enough to represent situations that occur in practice, or,
(b)~the vast majority of programs needs only `simple' invariants
(that can be constructed using data-flow techniques).

On the chosen benchmark set, our experiments support the hypothesis
that \kipdr can be very strong and efficient on tasks that other approaches can not solve
(tasks with `interesting' invariants).
It is important to note that this is an `exists' statement and can not be generalized,
as shown by the results that \kipdr is often outperformed
by simpler, data-flow-based invariant-generation techniques.
\cmark

\subsubsection{Further Discussion}
The seven example programs\,%
\footnote{{\scriptsize\url{https://github.com/sosy-lab/sv-benchmarks/tree/svcomp19/c/loop-invariants/}}}
were added to the benchmark collection that was also used for \svcomp~2019,
and thus, results are available for all verifiers that participated in the competition\,%
\footnote{See the last seven rows in this table:\\
{\scriptsize\url{https://sv-comp.sosy-lab.org/2019/results/results-verified/ReachSafety-Loops.table.html}}}.
Table~\ref{tab:hand-crafted-results-svcomp} summarizes the results of the best six verifiers
in comparison with the \kikipdr approach that we created for the study in this paper.
Those verifiers are, in alphabetical order, \skink, \uautomizer, \ukojak, \utaipan, \veriabs, and \viap.
Please note that the results of those six verifiers were obtained in a slightly different environment
than the one for the evaluations we performed in this paper,
because while the same hardware configuration was used for both sets of results,
the operating-system version used in \svcomp~2019 was Ubuntu~18.04 based on Linux~4.15.

\begin{table}[t]
  \centering
  \caption{Results of \svcomp~2019 for the six verifiers that performed best
           on our seven manually crafted verification tasks,
           compared to the results of \kikipdr approach previously shown in \cref{tab:hand-crafted-results};
	   an entry ``T'' means that the CPU-time limit was exceeded,
	   an entry ``M'' means that the memory limit was exceeded,
	   an entry ``O'' means that the verifier gave up deliberately for other reasons,
	   and all other entries represent the CPU time a verifier configuration spent
	   to correctly solve the task;
     note that \svcomp~2019 used Ubuntu~18.04 based on Linux~4.15, whereas our evaluation of \kikipdr used Ubuntu~16.04 based on Linux~4.4; otherwise, the evaluation environment was the same
           }
  \label{tab:hand-crafted-results-svcomp}
  \newcommand\rndsec[1]{\tablenum[round-mode=figures,round-precision=2,table-format=3.1,table-space-text-post=]{#1}\,\si{\second}}
  \newcommand\rndsecv[1]{\tablenum[round-mode=figures,round-precision=2,table-format=3,table-space-text-post=]{#1}\,\si{\second}}
  \renewcommand{\uautomizer}{\tool{UAutomizer}\xspace}
  \renewcommand{\ukojak}{\tool{UKojak}\xspace}
  \renewcommand{\utaipan}{\tool{UTaipan}\xspace}
  \begin{tabular}{l|cccccc|c}
    \toprule
    \multicolumn{1}{l|}{\multirow{2}{*}{Task}}
        & \multicolumn{6}{c|}{\svcomp~2019}
        & \multicolumn{1}{c}{\multirow{2}{*}{\kikipdr}}
        \\
             & \multicolumn{1}{c}{\skink}
             & \multicolumn{1}{c}{\uautomizer}
             & \multicolumn{1}{c}{\ukojak}
             & \multicolumn{1}{c}{\utaipan}
             & \multicolumn{1}{c}{\veriabs}
             & \multicolumn{1}{c|}{\viap} \\
    \midrule
    \texttt{const.c}
      & \rndsec{4.2}
      & \rndsec{8.7}
      & \rndsec{9.1}
      & \rndsec{8.2}
      & \rndsecv{13}
      & \rndsec{110}
      & \bfseries\rndsec{3.8}\\
    \texttt{eq1.c}
      & \rndsec{290}
      & \rndsec{7.8}
      & \rndsec{7.6}
      & \rndsec{8.3}
      & \rndsecv{14}
      & \rndsec{57}
      & \bfseries\rndsec{4.9}\\
    \texttt{eq2.c}
      & \rndsec{4.1}
      & \rndsec{8.1}
      & \rndsec{8.6}
      & \rndsec{7.6}
      & \rndsecv{14}
      & \rndsec{4.7}
      & \bfseries\rndsec{3.9}\\
    \texttt{even.c}
      & \bfseries\rndsec{3.7}
      & \rndsec{7.4}
      & \rndsec{8.2}
      & \rndsec{8.6}
      & \rndsecv{140}
      & \rndsec{4.5}
      & \rndsec{3.9}\\
    \texttt{odd.c}
      & O
      & \rndsec{9.6}
      & T
      & \rndsec{11}
      & \rndsecv{140}
      & \rndsec{4.6}
      & \bfseries\rndsec{4.1}\\
    \texttt{mod4.c}
      & \rndsec{4.0}
      & \rndsec{8.4}
      & \rndsec{8.4}
      & \rndsec{7.7}
      & \rndsecv{140}
      & \rndsec{4.5}
      & \bfseries\rndsec{3.6}\\
    \texttt{bin-suffix-5.c}
      & O
      & \rndsec{14}
      & T
      & \rndsec{13}
      & \rndsecv{13}
      & \rndsec{4.7}
      & \bfseries\rndsec{3.6}\\
  \bottomrule
  \end{tabular}
\end{table}

As should be expected, we observe that there are several approaches other than \kikipdr
that can also solve the tasks we crafted.
For example, because each loop in our seven manually crafted examples can be replaced
by a single linear-arithmetic computation,
\veriabs is able to apply loop-acceleration to solve the tasks.
We also looked into the reasons why \uautomizer,
which uses automata-based trace abstraction~\cite{cav/HeizmannHP13}.
Interestingly, \uautomizer, which uses the SMT~solver~\zthree,
can solve all seven tasks,
a reimplementation of the same algorithm in \cpachecker
can also solve them if configured to use~\zthree,
but the same implementation cannot solve them using~\mathsat,
because no suitable interpolants are generated,
i.e.,~in the case of \uautomizer,
the results appear to be related more to the choice of SMT solver
rather than the algorithm itself.
\footnote{A thorough evaluation of the impact of the choice
of the SMT solver, and the used theory, on verification results
was done by Wendler~\cite{PhilippPredicateAnalysis}.}
While we already know from our previous experiments
that in general, data-flow analyses can be very efficient invariant generators,
this comparison between \kikipdr and the best verifiers from \svcomp~2019
reconfirms our observation from Table~\ref{tab:hand-crafted-results}
that there are tasks where other approaches are necessary and available,
and that one such approach is \kikipdr,
which performs at least as good as the best available verifiers for all seven examples,
in some cases even significantly better.

The results for Hypotheses~2 and~3 suggested to reconfirm the insight by Cimatti and Griggio
that \pdr is most effective if only applied as a fall-back engine
for cases where a cheaper interpolation engine
fails to produce useful interpolants~\cite{SoftwareIC3}.
The results of \cref{tab:hand-crafted-results-svcomp} draw a more optimistic picture.

\subsection{Hypothesis 5: Relevancy}

To conclude our evaluation, we establish the relevancy of our previous conclusions
by showing that the best of our configurations that use \pdr
is competitive when compared to the best available tool implementations
of adaptations of \pdr to software verification,
and by comparing \kikipdr against the best verifiers
in the subcategory \textit{ReachSafety-Loops} from \svcomp~2019,
a category that is well known to contain many tasks
that require effort to be spent on generating loop invariants.

\vspace{-4mm}
\subsubsection{Comparison against \pdr-Based Verification Tools}

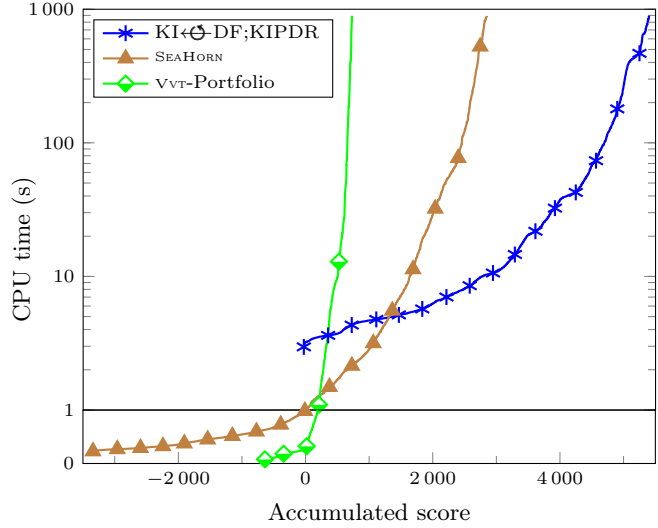
\begin{figure}[t]
  \centering
    \input{\plotpath/tools-score}
  \vspace{-2mm}
  \caption{Quantile plot
  for accumulated score of solved tasks
  (offset to the left by total penalty from wrong results)
  showing the CPU~time (linear scale below~\SI{1}{s}, logarithmic above)
  for the successful results of \kidfkipdr, \seahorn, and \vvt-Portfolio}
  \label{fig:tools-results}
  \vspace{-4mm}
\end{figure}

The last three columns of Table~\ref{tab:main-configs-results} give an overview
over the best configurations of three different software verifiers that use adaptations of \pdr:
For \cpachecker, we selected \kidfkipdr.
For \seahorn, we used the same configuration as submitted by the developers
to the 2016 Competition on Software Verification (\svcomp~2016)~\cite{SeaHorn-COMP15}.
For \vvt, we used the portfolio configuration.
We observe that \seahorn achieves the highest number of correct proofs,
but also has a significant amount of incorrect proofs.
\cpachecker is the slowest of the three tools
and finds fewer proofs than \seahorn,
but \cpachecker has no wrong proofs,
and also closely leads in the amount of found bugs.
The score-based quantile plot of these results displayed in Fig.~\ref{fig:tools-results}
visualizes the effects of incorrect results on the computed score.
While the graph for \seahorn is longer, i.e., shows that it solved the most tasks,
it is offset to the left by a total penalty of~\num{-3344}~points,
such that in the end, \kidfkipdr accumulates the highest score
because it has a smaller penalty of only~\num{-32}~points.
The plot also shows again, as in Fig.~\ref{fig:ctigar-results},
that the Java-based \cpachecker has a much higher startup time~(about~\SI{3}{s})
than the other two tools, which return results almost immediately for some tasks.

These results confirm our hypothesis that our previous conclusions are relevant,
because they are supported by an implementation
that is competitive when compared to the best available \pdr-based tool implementations.

\vspace{-4mm}
\subsubsection{Comparison against the Best Participants of \svcomp~2019 in the Category \textit{ReachSafety-Loops}}

Some of the previous experiments appeared to suggest that \kipdr is a slow invariant generator
that causes \kikipdr to be an inferier choice in many cases.
While \cref{tab:hand-crafted-results-svcomp} only shows a few anecdotal counterexamples
to this interpretation,
where \kikipdr is marginally faster than the best verifiers from \svcomp~2019
on six out of seven selected example tasks,
we will now consider how \kikipdr compares to the best verifiers from \svcomp~2019
in the subcategory \textit{ReachSafety-Loops},
which is known to contain many tasks
that require effort to be spent on generating loop invariants.
For a fair comparison, we re-executed our new implementations in the same environment,
i.e., not only on the same hardware configuration but also,
unlike our prior experiments, on Ubuntu~18.04 based on Linux~4.15.
In this newer execution environment, the Java~version used was OpenJDK~11.0.

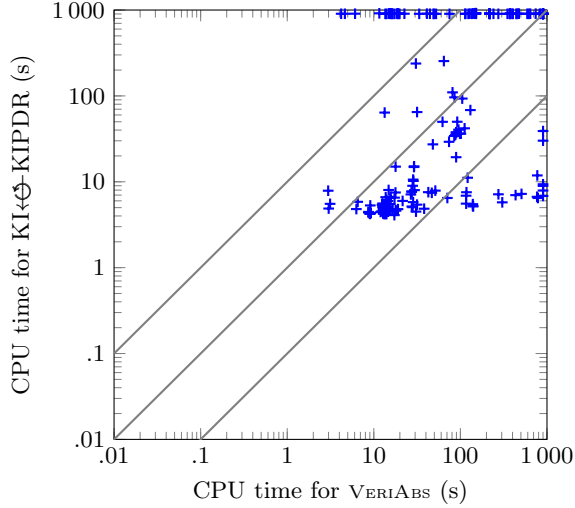
\begin{figure}[t]
  \vspace{4mm}
  \centering
    \input{\plotpath/svcomp19-loops.scatter}
  \vspace{-2mm}
  \caption{Scatter plot
  comparing the CPU times spent on tasks by \veriabs and \kikipdr}
  \label{fig:svcomp19-loops-scatter}
  \vspace{-3mm}
\end{figure}

\Cref{tab:svcomp19-loops-results} shows the results
for all~\nSvcompNineteenLoopsTasks~verification tasks
of the \svcomp~2019 subcategory \textit{ReachSafety-Loops},
\nSvcompNineteenLoopsFalsetasks~of which contain bugs,
while the other~\nSvcompNineteenLoopsTruetasks~are considered to be safe,
for the best three verifiers in that category,
namely \uautomizer, \utaipan, and \veriabs,
as well as for {\smaller\kikipdr} and {\smaller\kidfkipdr}.
While the \svcomp competitors clearly solve more tasks correctly
and obtain a significantly higher score than \kikipdr and \kidfkipdr,
we notice that for the tasks they can solve correctly,
the two \pdr\nobreakdash-based \kinduction configurations
have significantly lower median CPU times than the \svcomp competitors,
and that \kikipdr uses only about half as much CPU time in the arithmetic mean
than the other configurations.
For the same set of benchmarks,
i.e., the subcategory~\textit{ReachSafety-Loops} of \svcomp~2019,
\cref{fig:svcomp19-loops-scatter} directly compares
the CPU times spent on tasks by both \veriabs~(x-axis),
which was the best verifier in that subcategory,
and~\kikipdr~(y-axis),
to visualize the differences in efficiency between the two configurations.
We observe that there are more tasks solved by \veriabs
for which \kikipdr exceeds its time limit
than there are tasks solved by \kikipdr
for which \veriabs exceeds its time limit,
but we also see that for the majority of tasks that were solved by both verifiers,
\kikipdr is faster than \veriabs,
in a significant amount of cases even by more than an order of magnitude.
This shows that the invariant generator \kipdr
is not necessarily slower than other approaches,
and that on the contrary,
it can even be significantly faster than other approaches, depending on the benchmark set.

\input{svcomp19-loops-table}

In conclusion, compared to the \pdr\nobreakdash-based verifiers and compared to
the state-of-the-art verifiers that participated in \svcomp,
our \pdr\nobreakdash-based invariant generator obtains promising results.
\cmark

\subsection{Threats to Validity}

\noindent
\emph{External validity:}
We identify the following threats to the external validity of our experiments:
(1) We used the largest and most diverse publicly available benchmark for software verification,
but the risk remains that our conclusions are limited
to the kinds of programs represented by the benchmark.
As there were only two verifiers with support for \pdr participating in the competition,
this benchmark might contain few or no verification tasks
that are easy for \pdr and difficult for the other techniques.
We tried to limit this bias towards large C programs
by using path programs~\cite{PathInvariants}
(a way to reduce the complexity for the invariant generator),
and we also added a few ``hand-made'' programs that explore the boundaries of expressive power of the approaches.
(2) Our benchmark set contains only C programs,
because this is the only language supported by all of the evaluated tools.
Therefore, there is no guarantee that our results can be transferred
to programs written in other programming languages, for example software written in functional programming languages.
Moreover,
it might be possible that \pdr-based model checkers for hardware perform better on programs
that are converted from C to transition systems that those model checkers use.

\noindent
\emph{Internal validity:}
In our benchmarking environment, we execute up to four processes in parallel on each machine
to make our large experimental study feasible.
A drawback of the resulting shared usage of hardware resources
(such as caches or buses)~\cite{Benchmarking-STTT}
is that this setup may cause some minor noise in the measurements.
However, we use the benchmarking framework \benchexec
to properly account for the resources (CPU time and memory)
of each process and its subprocesses
and to ensure that for each physical core of the execution machines,
both of its processing units (virtual cores)
are allocated to the same process
to avoid the measurement errors
occurring from shared CPU resources
or from dynamic relocation of processes by the OS scheduler,
and to avoid performance influences from non-uniform memory access (NUMA)
by ensuring that all executions only use memory
belonging to their exclusively assigned CPU cores.

%% file: evaluation/data-commands.tex
\providecommand\StoreBenchExecResult[7]{\expandafter\newcommand\csname#1#2#3#4#5#6\endcsname{#7}}%
\StoreBenchExecResult{PdrInv}{KinductionDfStaticZeroZeroTTrueNotSolvedByKinductionPlainButKipdr}{Total}{}{Count}{}{449}%
\StoreBenchExecResult{PdrInv}{KinductionDfStaticZeroZeroTTrueNotSolvedByKinductionPlainButKipdr}{Total}{}{Cputime}{}{13659.135240813}%
\StoreBenchExecResult{PdrInv}{KinductionDfStaticZeroZeroTTrueNotSolvedByKinductionPlainButKipdr}{Total}{}{Cputime}{Avg}{30.42123661651002227171492205}%
\StoreBenchExecResult{PdrInv}{KinductionDfStaticZeroZeroTTrueNotSolvedByKinductionPlainButKipdr}{Total}{}{Cputime}{Median}{5.936393266}%
\StoreBenchExecResult{PdrInv}{KinductionDfStaticZeroZeroTTrueNotSolvedByKinductionPlainButKipdr}{Total}{}{Cputime}{Min}{3.399381636}%
\StoreBenchExecResult{PdrInv}{KinductionDfStaticZeroZeroTTrueNotSolvedByKinductionPlainButKipdr}{Total}{}{Cputime}{Max}{913.834354632}%
\StoreBenchExecResult{PdrInv}{KinductionDfStaticZeroZeroTTrueNotSolvedByKinductionPlainButKipdr}{Total}{}{Cputime}{Stdev}{143.5402462176682786660704708}%
\StoreBenchExecResult{PdrInv}{KinductionDfStaticZeroZeroTTrueNotSolvedByKinductionPlainButKipdr}{Total}{}{Walltime}{}{12114.00858592859}%
\StoreBenchExecResult{PdrInv}{KinductionDfStaticZeroZeroTTrueNotSolvedByKinductionPlainButKipdr}{Total}{}{Walltime}{Avg}{26.97997457890554565701559020}%
\StoreBenchExecResult{PdrInv}{KinductionDfStaticZeroZeroTTrueNotSolvedByKinductionPlainButKipdr}{Total}{}{Walltime}{Median}{3.15733909607}%
\StoreBenchExecResult{PdrInv}{KinductionDfStaticZeroZeroTTrueNotSolvedByKinductionPlainButKipdr}{Total}{}{Walltime}{Min}{1.87502193451}%
\StoreBenchExecResult{PdrInv}{KinductionDfStaticZeroZeroTTrueNotSolvedByKinductionPlainButKipdr}{Total}{}{Walltime}{Max}{897.71032691}%
\StoreBenchExecResult{PdrInv}{KinductionDfStaticZeroZeroTTrueNotSolvedByKinductionPlainButKipdr}{Total}{}{Walltime}{Stdev}{141.6288227494545033381207373}%
\StoreBenchExecResult{PdrInv}{KinductionDfStaticZeroZeroTTrueNotSolvedByKinductionPlainButKipdr}{Correct}{}{Count}{}{437}%
\StoreBenchExecResult{PdrInv}{KinductionDfStaticZeroZeroTTrueNotSolvedByKinductionPlainButKipdr}{Correct}{}{Cputime}{}{2911.214195619}%
\StoreBenchExecResult{PdrInv}{KinductionDfStaticZeroZeroTTrueNotSolvedByKinductionPlainButKipdr}{Correct}{}{Cputime}{Avg}{6.661817381279176201372997712}%
\StoreBenchExecResult{PdrInv}{KinductionDfStaticZeroZeroTTrueNotSolvedByKinductionPlainButKipdr}{Correct}{}{Cputime}{Median}{5.869181677}%
\StoreBenchExecResult{PdrInv}{KinductionDfStaticZeroZeroTTrueNotSolvedByKinductionPlainButKipdr}{Correct}{}{Cputime}{Min}{3.399381636}%
\StoreBenchExecResult{PdrInv}{KinductionDfStaticZeroZeroTTrueNotSolvedByKinductionPlainButKipdr}{Correct}{}{Cputime}{Max}{66.780305991}%
\StoreBenchExecResult{PdrInv}{KinductionDfStaticZeroZeroTTrueNotSolvedByKinductionPlainButKipdr}{Correct}{}{Cputime}{Stdev}{5.009111441138366994286176460}%
\StoreBenchExecResult{PdrInv}{KinductionDfStaticZeroZeroTTrueNotSolvedByKinductionPlainButKipdr}{Correct}{}{Walltime}{}{1541.01299667359}%
\StoreBenchExecResult{PdrInv}{KinductionDfStaticZeroZeroTTrueNotSolvedByKinductionPlainButKipdr}{Correct}{}{Walltime}{Avg}{3.526345530145514874141876430}%
\StoreBenchExecResult{PdrInv}{KinductionDfStaticZeroZeroTTrueNotSolvedByKinductionPlainButKipdr}{Correct}{}{Walltime}{Median}{3.121945858}%
\StoreBenchExecResult{PdrInv}{KinductionDfStaticZeroZeroTTrueNotSolvedByKinductionPlainButKipdr}{Correct}{}{Walltime}{Min}{1.87502193451}%
\StoreBenchExecResult{PdrInv}{KinductionDfStaticZeroZeroTTrueNotSolvedByKinductionPlainButKipdr}{Correct}{}{Walltime}{Max}{33.8622019291}%
\StoreBenchExecResult{PdrInv}{KinductionDfStaticZeroZeroTTrueNotSolvedByKinductionPlainButKipdr}{Correct}{}{Walltime}{Stdev}{2.523001096438479863048186274}%
\StoreBenchExecResult{PdrInv}{KinductionDfStaticZeroZeroTTrueNotSolvedByKinductionPlainButKipdr}{Correct}{True}{Count}{}{437}%
\StoreBenchExecResult{PdrInv}{KinductionDfStaticZeroZeroTTrueNotSolvedByKinductionPlainButKipdr}{Correct}{True}{Cputime}{}{2911.214195619}%
\StoreBenchExecResult{PdrInv}{KinductionDfStaticZeroZeroTTrueNotSolvedByKinductionPlainButKipdr}{Correct}{True}{Cputime}{Avg}{6.661817381279176201372997712}%
\StoreBenchExecResult{PdrInv}{KinductionDfStaticZeroZeroTTrueNotSolvedByKinductionPlainButKipdr}{Correct}{True}{Cputime}{Median}{5.869181677}%
\StoreBenchExecResult{PdrInv}{KinductionDfStaticZeroZeroTTrueNotSolvedByKinductionPlainButKipdr}{Correct}{True}{Cputime}{Min}{3.399381636}%
\StoreBenchExecResult{PdrInv}{KinductionDfStaticZeroZeroTTrueNotSolvedByKinductionPlainButKipdr}{Correct}{True}{Cputime}{Max}{66.780305991}%
\StoreBenchExecResult{PdrInv}{KinductionDfStaticZeroZeroTTrueNotSolvedByKinductionPlainButKipdr}{Correct}{True}{Cputime}{Stdev}{5.009111441138366994286176460}%
\StoreBenchExecResult{PdrInv}{KinductionDfStaticZeroZeroTTrueNotSolvedByKinductionPlainButKipdr}{Correct}{True}{Walltime}{}{1541.01299667359}%
\StoreBenchExecResult{PdrInv}{KinductionDfStaticZeroZeroTTrueNotSolvedByKinductionPlainButKipdr}{Correct}{True}{Walltime}{Avg}{3.526345530145514874141876430}%
\StoreBenchExecResult{PdrInv}{KinductionDfStaticZeroZeroTTrueNotSolvedByKinductionPlainButKipdr}{Correct}{True}{Walltime}{Median}{3.121945858}%
\StoreBenchExecResult{PdrInv}{KinductionDfStaticZeroZeroTTrueNotSolvedByKinductionPlainButKipdr}{Correct}{True}{Walltime}{Min}{1.87502193451}%
\StoreBenchExecResult{PdrInv}{KinductionDfStaticZeroZeroTTrueNotSolvedByKinductionPlainButKipdr}{Correct}{True}{Walltime}{Max}{33.8622019291}%
\StoreBenchExecResult{PdrInv}{KinductionDfStaticZeroZeroTTrueNotSolvedByKinductionPlainButKipdr}{Correct}{True}{Walltime}{Stdev}{2.523001096438479863048186274}%
\StoreBenchExecResult{PdrInv}{KinductionDfStaticZeroZeroTTrueNotSolvedByKinductionPlainButKipdr}{Wrong}{True}{Count}{}{0}%
\StoreBenchExecResult{PdrInv}{KinductionDfStaticZeroZeroTTrueNotSolvedByKinductionPlainButKipdr}{Wrong}{True}{Cputime}{}{0}%
\StoreBenchExecResult{PdrInv}{KinductionDfStaticZeroZeroTTrueNotSolvedByKinductionPlainButKipdr}{Wrong}{True}{Cputime}{Avg}{None}%
\StoreBenchExecResult{PdrInv}{KinductionDfStaticZeroZeroTTrueNotSolvedByKinductionPlainButKipdr}{Wrong}{True}{Cputime}{Median}{None}%
\StoreBenchExecResult{PdrInv}{KinductionDfStaticZeroZeroTTrueNotSolvedByKinductionPlainButKipdr}{Wrong}{True}{Cputime}{Min}{None}%
\StoreBenchExecResult{PdrInv}{KinductionDfStaticZeroZeroTTrueNotSolvedByKinductionPlainButKipdr}{Wrong}{True}{Cputime}{Max}{None}%
\StoreBenchExecResult{PdrInv}{KinductionDfStaticZeroZeroTTrueNotSolvedByKinductionPlainButKipdr}{Wrong}{True}{Cputime}{Stdev}{None}%
\StoreBenchExecResult{PdrInv}{KinductionDfStaticZeroZeroTTrueNotSolvedByKinductionPlainButKipdr}{Wrong}{True}{Walltime}{}{0}%
\StoreBenchExecResult{PdrInv}{KinductionDfStaticZeroZeroTTrueNotSolvedByKinductionPlainButKipdr}{Wrong}{True}{Walltime}{Avg}{None}%
\StoreBenchExecResult{PdrInv}{KinductionDfStaticZeroZeroTTrueNotSolvedByKinductionPlainButKipdr}{Wrong}{True}{Walltime}{Median}{None}%
\StoreBenchExecResult{PdrInv}{KinductionDfStaticZeroZeroTTrueNotSolvedByKinductionPlainButKipdr}{Wrong}{True}{Walltime}{Min}{None}%
\StoreBenchExecResult{PdrInv}{KinductionDfStaticZeroZeroTTrueNotSolvedByKinductionPlainButKipdr}{Wrong}{True}{Walltime}{Max}{None}%
\StoreBenchExecResult{PdrInv}{KinductionDfStaticZeroZeroTTrueNotSolvedByKinductionPlainButKipdr}{Wrong}{True}{Walltime}{Stdev}{None}%
\StoreBenchExecResult{PdrInv}{KinductionDfStaticZeroZeroTTrueNotSolvedByKinductionPlainButKipdr}{Error}{}{Count}{}{12}%
\StoreBenchExecResult{PdrInv}{KinductionDfStaticZeroZeroTTrueNotSolvedByKinductionPlainButKipdr}{Error}{}{Cputime}{}{10747.921045194}%
\StoreBenchExecResult{PdrInv}{KinductionDfStaticZeroZeroTTrueNotSolvedByKinductionPlainButKipdr}{Error}{}{Cputime}{Avg}{895.6600870995}%
\StoreBenchExecResult{PdrInv}{KinductionDfStaticZeroZeroTTrueNotSolvedByKinductionPlainButKipdr}{Error}{}{Cputime}{Median}{905.029497852}%
\StoreBenchExecResult{PdrInv}{KinductionDfStaticZeroZeroTTrueNotSolvedByKinductionPlainButKipdr}{Error}{}{Cputime}{Min}{811.329744803}%
\StoreBenchExecResult{PdrInv}{KinductionDfStaticZeroZeroTTrueNotSolvedByKinductionPlainButKipdr}{Error}{}{Cputime}{Max}{913.834354632}%
\StoreBenchExecResult{PdrInv}{KinductionDfStaticZeroZeroTTrueNotSolvedByKinductionPlainButKipdr}{Error}{}{Cputime}{Stdev}{28.56612310882484510418575915}%
\StoreBenchExecResult{PdrInv}{KinductionDfStaticZeroZeroTTrueNotSolvedByKinductionPlainButKipdr}{Error}{}{Walltime}{}{10572.995589255}%
\StoreBenchExecResult{PdrInv}{KinductionDfStaticZeroZeroTTrueNotSolvedByKinductionPlainButKipdr}{Error}{}{Walltime}{Avg}{881.08296577125}%
\StoreBenchExecResult{PdrInv}{KinductionDfStaticZeroZeroTTrueNotSolvedByKinductionPlainButKipdr}{Error}{}{Walltime}{Median}{891.264736056}%
\StoreBenchExecResult{PdrInv}{KinductionDfStaticZeroZeroTTrueNotSolvedByKinductionPlainButKipdr}{Error}{}{Walltime}{Min}{798.984657049}%
\StoreBenchExecResult{PdrInv}{KinductionDfStaticZeroZeroTTrueNotSolvedByKinductionPlainButKipdr}{Error}{}{Walltime}{Max}{897.71032691}%
\StoreBenchExecResult{PdrInv}{KinductionDfStaticZeroZeroTTrueNotSolvedByKinductionPlainButKipdr}{Error}{}{Walltime}{Stdev}{27.83922823009148553827706326}%
\StoreBenchExecResult{PdrInv}{KinductionDfStaticZeroZeroTTrueNotSolvedByKinductionPlainButKipdr}{Error}{OutOfMemory}{Count}{}{1}%
\StoreBenchExecResult{PdrInv}{KinductionDfStaticZeroZeroTTrueNotSolvedByKinductionPlainButKipdr}{Error}{OutOfMemory}{Cputime}{}{861.836319163}%
\StoreBenchExecResult{PdrInv}{KinductionDfStaticZeroZeroTTrueNotSolvedByKinductionPlainButKipdr}{Error}{OutOfMemory}{Cputime}{Avg}{861.836319163}%
\StoreBenchExecResult{PdrInv}{KinductionDfStaticZeroZeroTTrueNotSolvedByKinductionPlainButKipdr}{Error}{OutOfMemory}{Cputime}{Median}{861.836319163}%
\StoreBenchExecResult{PdrInv}{KinductionDfStaticZeroZeroTTrueNotSolvedByKinductionPlainButKipdr}{Error}{OutOfMemory}{Cputime}{Min}{861.836319163}%
\StoreBenchExecResult{PdrInv}{KinductionDfStaticZeroZeroTTrueNotSolvedByKinductionPlainButKipdr}{Error}{OutOfMemory}{Cputime}{Max}{861.836319163}%
\StoreBenchExecResult{PdrInv}{KinductionDfStaticZeroZeroTTrueNotSolvedByKinductionPlainButKipdr}{Error}{OutOfMemory}{Cputime}{Stdev}{0E-9}%
\StoreBenchExecResult{PdrInv}{KinductionDfStaticZeroZeroTTrueNotSolvedByKinductionPlainButKipdr}{Error}{OutOfMemory}{Walltime}{}{847.870331049}%
\StoreBenchExecResult{PdrInv}{KinductionDfStaticZeroZeroTTrueNotSolvedByKinductionPlainButKipdr}{Error}{OutOfMemory}{Walltime}{Avg}{847.870331049}%
\StoreBenchExecResult{PdrInv}{KinductionDfStaticZeroZeroTTrueNotSolvedByKinductionPlainButKipdr}{Error}{OutOfMemory}{Walltime}{Median}{847.870331049}%
\StoreBenchExecResult{PdrInv}{KinductionDfStaticZeroZeroTTrueNotSolvedByKinductionPlainButKipdr}{Error}{OutOfMemory}{Walltime}{Min}{847.870331049}%
\StoreBenchExecResult{PdrInv}{KinductionDfStaticZeroZeroTTrueNotSolvedByKinductionPlainButKipdr}{Error}{OutOfMemory}{Walltime}{Max}{847.870331049}%
\StoreBenchExecResult{PdrInv}{KinductionDfStaticZeroZeroTTrueNotSolvedByKinductionPlainButKipdr}{Error}{OutOfMemory}{Walltime}{Stdev}{0E-9}%
\StoreBenchExecResult{PdrInv}{KinductionDfStaticZeroZeroTTrueNotSolvedByKinductionPlainButKipdr}{Error}{OutOfNativeMemory}{Count}{}{1}%
\StoreBenchExecResult{PdrInv}{KinductionDfStaticZeroZeroTTrueNotSolvedByKinductionPlainButKipdr}{Error}{OutOfNativeMemory}{Cputime}{}{811.329744803}%
\StoreBenchExecResult{PdrInv}{KinductionDfStaticZeroZeroTTrueNotSolvedByKinductionPlainButKipdr}{Error}{OutOfNativeMemory}{Cputime}{Avg}{811.329744803}%
\StoreBenchExecResult{PdrInv}{KinductionDfStaticZeroZeroTTrueNotSolvedByKinductionPlainButKipdr}{Error}{OutOfNativeMemory}{Cputime}{Median}{811.329744803}%
\StoreBenchExecResult{PdrInv}{KinductionDfStaticZeroZeroTTrueNotSolvedByKinductionPlainButKipdr}{Error}{OutOfNativeMemory}{Cputime}{Min}{811.329744803}%
\StoreBenchExecResult{PdrInv}{KinductionDfStaticZeroZeroTTrueNotSolvedByKinductionPlainButKipdr}{Error}{OutOfNativeMemory}{Cputime}{Max}{811.329744803}%
\StoreBenchExecResult{PdrInv}{KinductionDfStaticZeroZeroTTrueNotSolvedByKinductionPlainButKipdr}{Error}{OutOfNativeMemory}{Cputime}{Stdev}{0E-9}%
\StoreBenchExecResult{PdrInv}{KinductionDfStaticZeroZeroTTrueNotSolvedByKinductionPlainButKipdr}{Error}{OutOfNativeMemory}{Walltime}{}{798.984657049}%
\StoreBenchExecResult{PdrInv}{KinductionDfStaticZeroZeroTTrueNotSolvedByKinductionPlainButKipdr}{Error}{OutOfNativeMemory}{Walltime}{Avg}{798.984657049}%
\StoreBenchExecResult{PdrInv}{KinductionDfStaticZeroZeroTTrueNotSolvedByKinductionPlainButKipdr}{Error}{OutOfNativeMemory}{Walltime}{Median}{798.984657049}%
\StoreBenchExecResult{PdrInv}{KinductionDfStaticZeroZeroTTrueNotSolvedByKinductionPlainButKipdr}{Error}{OutOfNativeMemory}{Walltime}{Min}{798.984657049}%
\StoreBenchExecResult{PdrInv}{KinductionDfStaticZeroZeroTTrueNotSolvedByKinductionPlainButKipdr}{Error}{OutOfNativeMemory}{Walltime}{Max}{798.984657049}%
\StoreBenchExecResult{PdrInv}{KinductionDfStaticZeroZeroTTrueNotSolvedByKinductionPlainButKipdr}{Error}{OutOfNativeMemory}{Walltime}{Stdev}{0E-9}%
\StoreBenchExecResult{PdrInv}{KinductionDfStaticZeroZeroTTrueNotSolvedByKinductionPlainButKipdr}{Error}{Timeout}{Count}{}{10}%
\StoreBenchExecResult{PdrInv}{KinductionDfStaticZeroZeroTTrueNotSolvedByKinductionPlainButKipdr}{Error}{Timeout}{Cputime}{}{9074.754981228}%
\StoreBenchExecResult{PdrInv}{KinductionDfStaticZeroZeroTTrueNotSolvedByKinductionPlainButKipdr}{Error}{Timeout}{Cputime}{Avg}{907.4754981228}%
\StoreBenchExecResult{PdrInv}{KinductionDfStaticZeroZeroTTrueNotSolvedByKinductionPlainButKipdr}{Error}{Timeout}{Cputime}{Median}{906.673867331}%
\StoreBenchExecResult{PdrInv}{KinductionDfStaticZeroZeroTTrueNotSolvedByKinductionPlainButKipdr}{Error}{Timeout}{Cputime}{Min}{902.898967409}%
\StoreBenchExecResult{PdrInv}{KinductionDfStaticZeroZeroTTrueNotSolvedByKinductionPlainButKipdr}{Error}{Timeout}{Cputime}{Max}{913.834354632}%
\StoreBenchExecResult{PdrInv}{KinductionDfStaticZeroZeroTTrueNotSolvedByKinductionPlainButKipdr}{Error}{Timeout}{Cputime}{Stdev}{3.749498471265031808048676777}%
\StoreBenchExecResult{PdrInv}{KinductionDfStaticZeroZeroTTrueNotSolvedByKinductionPlainButKipdr}{Error}{Timeout}{Walltime}{}{8926.140601157}%
\StoreBenchExecResult{PdrInv}{KinductionDfStaticZeroZeroTTrueNotSolvedByKinductionPlainButKipdr}{Error}{Timeout}{Walltime}{Avg}{892.6140601157}%
\StoreBenchExecResult{PdrInv}{KinductionDfStaticZeroZeroTTrueNotSolvedByKinductionPlainButKipdr}{Error}{Timeout}{Walltime}{Median}{892.715479970}%
\StoreBenchExecResult{PdrInv}{KinductionDfStaticZeroZeroTTrueNotSolvedByKinductionPlainButKipdr}{Error}{Timeout}{Walltime}{Min}{887.866760969}%
\StoreBenchExecResult{PdrInv}{KinductionDfStaticZeroZeroTTrueNotSolvedByKinductionPlainButKipdr}{Error}{Timeout}{Walltime}{Max}{897.71032691}%
\StoreBenchExecResult{PdrInv}{KinductionDfStaticZeroZeroTTrueNotSolvedByKinductionPlainButKipdr}{Error}{Timeout}{Walltime}{Stdev}{3.569296475664748737546116720}%
\providecommand\StoreBenchExecResult[7]{\expandafter\newcommand\csname#1#2#3#4#5#6\endcsname{#7}}%
\StoreBenchExecResult{PdrInv}{KinductionDfStaticZeroZeroTTrueNotSolvedByKinductionPlain}{Total}{}{Count}{}{2893}%
\StoreBenchExecResult{PdrInv}{KinductionDfStaticZeroZeroTTrueNotSolvedByKinductionPlain}{Total}{}{Cputime}{}{1640322.345285189}%
\StoreBenchExecResult{PdrInv}{KinductionDfStaticZeroZeroTTrueNotSolvedByKinductionPlain}{Total}{}{Cputime}{Avg}{566.9970083944656066367092983}%
\StoreBenchExecResult{PdrInv}{KinductionDfStaticZeroZeroTTrueNotSolvedByKinductionPlain}{Total}{}{Cputime}{Median}{901.1711571}%
\StoreBenchExecResult{PdrInv}{KinductionDfStaticZeroZeroTTrueNotSolvedByKinductionPlain}{Total}{}{Cputime}{Min}{2.446906053}%
\StoreBenchExecResult{PdrInv}{KinductionDfStaticZeroZeroTTrueNotSolvedByKinductionPlain}{Total}{}{Cputime}{Max}{1002.39260405}%
\StoreBenchExecResult{PdrInv}{KinductionDfStaticZeroZeroTTrueNotSolvedByKinductionPlain}{Total}{}{Cputime}{Stdev}{416.6652374680205811664607075}%
\StoreBenchExecResult{PdrInv}{KinductionDfStaticZeroZeroTTrueNotSolvedByKinductionPlain}{Total}{}{Walltime}{}{1151680.57774257515}%
\StoreBenchExecResult{PdrInv}{KinductionDfStaticZeroZeroTTrueNotSolvedByKinductionPlain}{Total}{}{Walltime}{Avg}{398.0921457803578119599032147}%
\StoreBenchExecResult{PdrInv}{KinductionDfStaticZeroZeroTTrueNotSolvedByKinductionPlain}{Total}{}{Walltime}{Median}{451.922544956}%
\StoreBenchExecResult{PdrInv}{KinductionDfStaticZeroZeroTTrueNotSolvedByKinductionPlain}{Total}{}{Walltime}{Min}{1.34633517265}%
\StoreBenchExecResult{PdrInv}{KinductionDfStaticZeroZeroTTrueNotSolvedByKinductionPlain}{Total}{}{Walltime}{Max}{908.435231209}%
\StoreBenchExecResult{PdrInv}{KinductionDfStaticZeroZeroTTrueNotSolvedByKinductionPlain}{Total}{}{Walltime}{Stdev}{331.6802488877884469935718733}%
\StoreBenchExecResult{PdrInv}{KinductionDfStaticZeroZeroTTrueNotSolvedByKinductionPlain}{Correct}{}{Count}{}{939}%
\StoreBenchExecResult{PdrInv}{KinductionDfStaticZeroZeroTTrueNotSolvedByKinductionPlain}{Correct}{}{Cputime}{}{40646.072668398}%
\StoreBenchExecResult{PdrInv}{KinductionDfStaticZeroZeroTTrueNotSolvedByKinductionPlain}{Correct}{}{Cputime}{Avg}{43.28655236251118210862619808}%
\StoreBenchExecResult{PdrInv}{KinductionDfStaticZeroZeroTTrueNotSolvedByKinductionPlain}{Correct}{}{Cputime}{Median}{6.973506103}%
\StoreBenchExecResult{PdrInv}{KinductionDfStaticZeroZeroTTrueNotSolvedByKinductionPlain}{Correct}{}{Cputime}{Min}{3.185602622}%
\StoreBenchExecResult{PdrInv}{KinductionDfStaticZeroZeroTTrueNotSolvedByKinductionPlain}{Correct}{}{Cputime}{Max}{839.210855366}%
\StoreBenchExecResult{PdrInv}{KinductionDfStaticZeroZeroTTrueNotSolvedByKinductionPlain}{Correct}{}{Cputime}{Stdev}{118.1397526106431932812213122}%
\StoreBenchExecResult{PdrInv}{KinductionDfStaticZeroZeroTTrueNotSolvedByKinductionPlain}{Correct}{}{Walltime}{}{22985.53274822025}%
\StoreBenchExecResult{PdrInv}{KinductionDfStaticZeroZeroTTrueNotSolvedByKinductionPlain}{Correct}{}{Walltime}{Avg}{24.47873562110782747603833866}%
\StoreBenchExecResult{PdrInv}{KinductionDfStaticZeroZeroTTrueNotSolvedByKinductionPlain}{Correct}{}{Walltime}{Median}{3.67091083527}%
\StoreBenchExecResult{PdrInv}{KinductionDfStaticZeroZeroTTrueNotSolvedByKinductionPlain}{Correct}{}{Walltime}{Min}{1.76809287071}%
\StoreBenchExecResult{PdrInv}{KinductionDfStaticZeroZeroTTrueNotSolvedByKinductionPlain}{Correct}{}{Walltime}{Max}{780.007678986}%
\StoreBenchExecResult{PdrInv}{KinductionDfStaticZeroZeroTTrueNotSolvedByKinductionPlain}{Correct}{}{Walltime}{Stdev}{74.21114884889057926415801341}%
\StoreBenchExecResult{PdrInv}{KinductionDfStaticZeroZeroTTrueNotSolvedByKinductionPlain}{Correct}{True}{Count}{}{939}%
\StoreBenchExecResult{PdrInv}{KinductionDfStaticZeroZeroTTrueNotSolvedByKinductionPlain}{Correct}{True}{Cputime}{}{40646.072668398}%
\StoreBenchExecResult{PdrInv}{KinductionDfStaticZeroZeroTTrueNotSolvedByKinductionPlain}{Correct}{True}{Cputime}{Avg}{43.28655236251118210862619808}%
\StoreBenchExecResult{PdrInv}{KinductionDfStaticZeroZeroTTrueNotSolvedByKinductionPlain}{Correct}{True}{Cputime}{Median}{6.973506103}%
\StoreBenchExecResult{PdrInv}{KinductionDfStaticZeroZeroTTrueNotSolvedByKinductionPlain}{Correct}{True}{Cputime}{Min}{3.185602622}%
\StoreBenchExecResult{PdrInv}{KinductionDfStaticZeroZeroTTrueNotSolvedByKinductionPlain}{Correct}{True}{Cputime}{Max}{839.210855366}%
\StoreBenchExecResult{PdrInv}{KinductionDfStaticZeroZeroTTrueNotSolvedByKinductionPlain}{Correct}{True}{Cputime}{Stdev}{118.1397526106431932812213122}%
\StoreBenchExecResult{PdrInv}{KinductionDfStaticZeroZeroTTrueNotSolvedByKinductionPlain}{Correct}{True}{Walltime}{}{22985.53274822025}%
\StoreBenchExecResult{PdrInv}{KinductionDfStaticZeroZeroTTrueNotSolvedByKinductionPlain}{Correct}{True}{Walltime}{Avg}{24.47873562110782747603833866}%
\StoreBenchExecResult{PdrInv}{KinductionDfStaticZeroZeroTTrueNotSolvedByKinductionPlain}{Correct}{True}{Walltime}{Median}{3.67091083527}%
\StoreBenchExecResult{PdrInv}{KinductionDfStaticZeroZeroTTrueNotSolvedByKinductionPlain}{Correct}{True}{Walltime}{Min}{1.76809287071}%
\StoreBenchExecResult{PdrInv}{KinductionDfStaticZeroZeroTTrueNotSolvedByKinductionPlain}{Correct}{True}{Walltime}{Max}{780.007678986}%
\StoreBenchExecResult{PdrInv}{KinductionDfStaticZeroZeroTTrueNotSolvedByKinductionPlain}{Correct}{True}{Walltime}{Stdev}{74.21114884889057926415801341}%
\StoreBenchExecResult{PdrInv}{KinductionDfStaticZeroZeroTTrueNotSolvedByKinductionPlain}{Wrong}{True}{Count}{}{0}%
\StoreBenchExecResult{PdrInv}{KinductionDfStaticZeroZeroTTrueNotSolvedByKinductionPlain}{Wrong}{True}{Cputime}{}{0}%
\StoreBenchExecResult{PdrInv}{KinductionDfStaticZeroZeroTTrueNotSolvedByKinductionPlain}{Wrong}{True}{Cputime}{Avg}{None}%
\StoreBenchExecResult{PdrInv}{KinductionDfStaticZeroZeroTTrueNotSolvedByKinductionPlain}{Wrong}{True}{Cputime}{Median}{None}%
\StoreBenchExecResult{PdrInv}{KinductionDfStaticZeroZeroTTrueNotSolvedByKinductionPlain}{Wrong}{True}{Cputime}{Min}{None}%
\StoreBenchExecResult{PdrInv}{KinductionDfStaticZeroZeroTTrueNotSolvedByKinductionPlain}{Wrong}{True}{Cputime}{Max}{None}%
\StoreBenchExecResult{PdrInv}{KinductionDfStaticZeroZeroTTrueNotSolvedByKinductionPlain}{Wrong}{True}{Cputime}{Stdev}{None}%
\StoreBenchExecResult{PdrInv}{KinductionDfStaticZeroZeroTTrueNotSolvedByKinductionPlain}{Wrong}{True}{Walltime}{}{0}%
\StoreBenchExecResult{PdrInv}{KinductionDfStaticZeroZeroTTrueNotSolvedByKinductionPlain}{Wrong}{True}{Walltime}{Avg}{None}%
\StoreBenchExecResult{PdrInv}{KinductionDfStaticZeroZeroTTrueNotSolvedByKinductionPlain}{Wrong}{True}{Walltime}{Median}{None}%
\StoreBenchExecResult{PdrInv}{KinductionDfStaticZeroZeroTTrueNotSolvedByKinductionPlain}{Wrong}{True}{Walltime}{Min}{None}%
\StoreBenchExecResult{PdrInv}{KinductionDfStaticZeroZeroTTrueNotSolvedByKinductionPlain}{Wrong}{True}{Walltime}{Max}{None}%
\StoreBenchExecResult{PdrInv}{KinductionDfStaticZeroZeroTTrueNotSolvedByKinductionPlain}{Wrong}{True}{Walltime}{Stdev}{None}%
\StoreBenchExecResult{PdrInv}{KinductionDfStaticZeroZeroTTrueNotSolvedByKinductionPlain}{Error}{}{Count}{}{1954}%
\StoreBenchExecResult{PdrInv}{KinductionDfStaticZeroZeroTTrueNotSolvedByKinductionPlain}{Error}{}{Cputime}{}{1599676.272616791}%
\StoreBenchExecResult{PdrInv}{KinductionDfStaticZeroZeroTTrueNotSolvedByKinductionPlain}{Error}{}{Cputime}{Avg}{818.6674885449288638689866940}%
\StoreBenchExecResult{PdrInv}{KinductionDfStaticZeroZeroTTrueNotSolvedByKinductionPlain}{Error}{}{Cputime}{Median}{901.8400761595}%
\StoreBenchExecResult{PdrInv}{KinductionDfStaticZeroZeroTTrueNotSolvedByKinductionPlain}{Error}{}{Cputime}{Min}{2.446906053}%
\StoreBenchExecResult{PdrInv}{KinductionDfStaticZeroZeroTTrueNotSolvedByKinductionPlain}{Error}{}{Cputime}{Max}{1002.39260405}%
\StoreBenchExecResult{PdrInv}{KinductionDfStaticZeroZeroTTrueNotSolvedByKinductionPlain}{Error}{}{Cputime}{Stdev}{234.9277992398295017325950902}%
\StoreBenchExecResult{PdrInv}{KinductionDfStaticZeroZeroTTrueNotSolvedByKinductionPlain}{Error}{}{Walltime}{}{1128695.04499435490}%
\StoreBenchExecResult{PdrInv}{KinductionDfStaticZeroZeroTTrueNotSolvedByKinductionPlain}{Error}{}{Walltime}{Avg}{577.6330834157394575230296827}%
\StoreBenchExecResult{PdrInv}{KinductionDfStaticZeroZeroTTrueNotSolvedByKinductionPlain}{Error}{}{Walltime}{Median}{456.1915575265}%
\StoreBenchExecResult{PdrInv}{KinductionDfStaticZeroZeroTTrueNotSolvedByKinductionPlain}{Error}{}{Walltime}{Min}{1.34633517265}%
\StoreBenchExecResult{PdrInv}{KinductionDfStaticZeroZeroTTrueNotSolvedByKinductionPlain}{Error}{}{Walltime}{Max}{908.435231209}%
\StoreBenchExecResult{PdrInv}{KinductionDfStaticZeroZeroTTrueNotSolvedByKinductionPlain}{Error}{}{Walltime}{Stdev}{246.8154264956752649864268645}%
\StoreBenchExecResult{PdrInv}{KinductionDfStaticZeroZeroTTrueNotSolvedByKinductionPlain}{Error}{Assertion}{Count}{}{2}%
\StoreBenchExecResult{PdrInv}{KinductionDfStaticZeroZeroTTrueNotSolvedByKinductionPlain}{Error}{Assertion}{Cputime}{}{6.448787321}%
\StoreBenchExecResult{PdrInv}{KinductionDfStaticZeroZeroTTrueNotSolvedByKinductionPlain}{Error}{Assertion}{Cputime}{Avg}{3.2243936605}%
\StoreBenchExecResult{PdrInv}{KinductionDfStaticZeroZeroTTrueNotSolvedByKinductionPlain}{Error}{Assertion}{Cputime}{Median}{3.2243936605}%
\StoreBenchExecResult{PdrInv}{KinductionDfStaticZeroZeroTTrueNotSolvedByKinductionPlain}{Error}{Assertion}{Cputime}{Min}{3.185055688}%
\StoreBenchExecResult{PdrInv}{KinductionDfStaticZeroZeroTTrueNotSolvedByKinductionPlain}{Error}{Assertion}{Cputime}{Max}{3.263731633}%
\StoreBenchExecResult{PdrInv}{KinductionDfStaticZeroZeroTTrueNotSolvedByKinductionPlain}{Error}{Assertion}{Cputime}{Stdev}{0.0393379725}%
\StoreBenchExecResult{PdrInv}{KinductionDfStaticZeroZeroTTrueNotSolvedByKinductionPlain}{Error}{Assertion}{Walltime}{}{3.56652212143}%
\StoreBenchExecResult{PdrInv}{KinductionDfStaticZeroZeroTTrueNotSolvedByKinductionPlain}{Error}{Assertion}{Walltime}{Avg}{1.783261060715}%
\StoreBenchExecResult{PdrInv}{KinductionDfStaticZeroZeroTTrueNotSolvedByKinductionPlain}{Error}{Assertion}{Walltime}{Median}{1.783261060715}%
\StoreBenchExecResult{PdrInv}{KinductionDfStaticZeroZeroTTrueNotSolvedByKinductionPlain}{Error}{Assertion}{Walltime}{Min}{1.76435208321}%
\StoreBenchExecResult{PdrInv}{KinductionDfStaticZeroZeroTTrueNotSolvedByKinductionPlain}{Error}{Assertion}{Walltime}{Max}{1.80217003822}%
\StoreBenchExecResult{PdrInv}{KinductionDfStaticZeroZeroTTrueNotSolvedByKinductionPlain}{Error}{Assertion}{Walltime}{Stdev}{0.018908977505}%
\StoreBenchExecResult{PdrInv}{KinductionDfStaticZeroZeroTTrueNotSolvedByKinductionPlain}{Error}{Error}{Count}{}{134}%
\StoreBenchExecResult{PdrInv}{KinductionDfStaticZeroZeroTTrueNotSolvedByKinductionPlain}{Error}{Error}{Cputime}{}{22716.174177127}%
\StoreBenchExecResult{PdrInv}{KinductionDfStaticZeroZeroTTrueNotSolvedByKinductionPlain}{Error}{Error}{Cputime}{Avg}{169.5236878890074626865671642}%
\StoreBenchExecResult{PdrInv}{KinductionDfStaticZeroZeroTTrueNotSolvedByKinductionPlain}{Error}{Error}{Cputime}{Median}{106.2288360695}%
\StoreBenchExecResult{PdrInv}{KinductionDfStaticZeroZeroTTrueNotSolvedByKinductionPlain}{Error}{Error}{Cputime}{Min}{2.446906053}%
\StoreBenchExecResult{PdrInv}{KinductionDfStaticZeroZeroTTrueNotSolvedByKinductionPlain}{Error}{Error}{Cputime}{Max}{898.008416126}%
\StoreBenchExecResult{PdrInv}{KinductionDfStaticZeroZeroTTrueNotSolvedByKinductionPlain}{Error}{Error}{Cputime}{Stdev}{197.1400823594661319500672426}%
\StoreBenchExecResult{PdrInv}{KinductionDfStaticZeroZeroTTrueNotSolvedByKinductionPlain}{Error}{Error}{Walltime}{}{19020.82942509676}%
\StoreBenchExecResult{PdrInv}{KinductionDfStaticZeroZeroTTrueNotSolvedByKinductionPlain}{Error}{Error}{Walltime}{Avg}{141.9464882469907462686567164}%
\StoreBenchExecResult{PdrInv}{KinductionDfStaticZeroZeroTTrueNotSolvedByKinductionPlain}{Error}{Error}{Walltime}{Median}{84.0375635624}%
\StoreBenchExecResult{PdrInv}{KinductionDfStaticZeroZeroTTrueNotSolvedByKinductionPlain}{Error}{Error}{Walltime}{Min}{1.34633517265}%
\StoreBenchExecResult{PdrInv}{KinductionDfStaticZeroZeroTTrueNotSolvedByKinductionPlain}{Error}{Error}{Walltime}{Max}{880.894071817}%
\StoreBenchExecResult{PdrInv}{KinductionDfStaticZeroZeroTTrueNotSolvedByKinductionPlain}{Error}{Error}{Walltime}{Stdev}{174.4029341359029823238108820}%
\StoreBenchExecResult{PdrInv}{KinductionDfStaticZeroZeroTTrueNotSolvedByKinductionPlain}{Error}{Exception}{Count}{}{7}%
\StoreBenchExecResult{PdrInv}{KinductionDfStaticZeroZeroTTrueNotSolvedByKinductionPlain}{Error}{Exception}{Cputime}{}{759.772362829}%
\StoreBenchExecResult{PdrInv}{KinductionDfStaticZeroZeroTTrueNotSolvedByKinductionPlain}{Error}{Exception}{Cputime}{Avg}{108.5389089755714285714285714}%
\StoreBenchExecResult{PdrInv}{KinductionDfStaticZeroZeroTTrueNotSolvedByKinductionPlain}{Error}{Exception}{Cputime}{Median}{78.164800942}%
\StoreBenchExecResult{PdrInv}{KinductionDfStaticZeroZeroTTrueNotSolvedByKinductionPlain}{Error}{Exception}{Cputime}{Min}{13.814084064}%
\StoreBenchExecResult{PdrInv}{KinductionDfStaticZeroZeroTTrueNotSolvedByKinductionPlain}{Error}{Exception}{Cputime}{Max}{231.490727806}%
\StoreBenchExecResult{PdrInv}{KinductionDfStaticZeroZeroTTrueNotSolvedByKinductionPlain}{Error}{Exception}{Cputime}{Stdev}{80.86673374883862890858846675}%
\StoreBenchExecResult{PdrInv}{KinductionDfStaticZeroZeroTTrueNotSolvedByKinductionPlain}{Error}{Exception}{Walltime}{}{396.00020146341}%
\StoreBenchExecResult{PdrInv}{KinductionDfStaticZeroZeroTTrueNotSolvedByKinductionPlain}{Error}{Exception}{Walltime}{Avg}{56.57145735191571428571428571}%
\StoreBenchExecResult{PdrInv}{KinductionDfStaticZeroZeroTTrueNotSolvedByKinductionPlain}{Error}{Exception}{Walltime}{Median}{39.3472690582}%
\StoreBenchExecResult{PdrInv}{KinductionDfStaticZeroZeroTTrueNotSolvedByKinductionPlain}{Error}{Exception}{Walltime}{Min}{7.07906699181}%
\StoreBenchExecResult{PdrInv}{KinductionDfStaticZeroZeroTTrueNotSolvedByKinductionPlain}{Error}{Exception}{Walltime}{Max}{116.446436167}%
\StoreBenchExecResult{PdrInv}{KinductionDfStaticZeroZeroTTrueNotSolvedByKinductionPlain}{Error}{Exception}{Walltime}{Stdev}{40.70577078591509998635292748}%
\StoreBenchExecResult{PdrInv}{KinductionDfStaticZeroZeroTTrueNotSolvedByKinductionPlain}{Error}{OutOfJavaMemory}{Count}{}{5}%
\StoreBenchExecResult{PdrInv}{KinductionDfStaticZeroZeroTTrueNotSolvedByKinductionPlain}{Error}{OutOfJavaMemory}{Cputime}{}{2055.963406224}%
\StoreBenchExecResult{PdrInv}{KinductionDfStaticZeroZeroTTrueNotSolvedByKinductionPlain}{Error}{OutOfJavaMemory}{Cputime}{Avg}{411.1926812448}%
\StoreBenchExecResult{PdrInv}{KinductionDfStaticZeroZeroTTrueNotSolvedByKinductionPlain}{Error}{OutOfJavaMemory}{Cputime}{Median}{328.093769619}%
\StoreBenchExecResult{PdrInv}{KinductionDfStaticZeroZeroTTrueNotSolvedByKinductionPlain}{Error}{OutOfJavaMemory}{Cputime}{Min}{168.105338919}%
\StoreBenchExecResult{PdrInv}{KinductionDfStaticZeroZeroTTrueNotSolvedByKinductionPlain}{Error}{OutOfJavaMemory}{Cputime}{Max}{730.703728756}%
\StoreBenchExecResult{PdrInv}{KinductionDfStaticZeroZeroTTrueNotSolvedByKinductionPlain}{Error}{OutOfJavaMemory}{Cputime}{Stdev}{205.0991584186664257297374765}%
\StoreBenchExecResult{PdrInv}{KinductionDfStaticZeroZeroTTrueNotSolvedByKinductionPlain}{Error}{OutOfJavaMemory}{Walltime}{}{1160.376106262}%
\StoreBenchExecResult{PdrInv}{KinductionDfStaticZeroZeroTTrueNotSolvedByKinductionPlain}{Error}{OutOfJavaMemory}{Walltime}{Avg}{232.0752212524}%
\StoreBenchExecResult{PdrInv}{KinductionDfStaticZeroZeroTTrueNotSolvedByKinductionPlain}{Error}{OutOfJavaMemory}{Walltime}{Median}{209.268599987}%
\StoreBenchExecResult{PdrInv}{KinductionDfStaticZeroZeroTTrueNotSolvedByKinductionPlain}{Error}{OutOfJavaMemory}{Walltime}{Min}{103.409188986}%
\StoreBenchExecResult{PdrInv}{KinductionDfStaticZeroZeroTTrueNotSolvedByKinductionPlain}{Error}{OutOfJavaMemory}{Walltime}{Max}{377.878635168}%
\StoreBenchExecResult{PdrInv}{KinductionDfStaticZeroZeroTTrueNotSolvedByKinductionPlain}{Error}{OutOfJavaMemory}{Walltime}{Stdev}{94.31462138851109448737205581}%
\StoreBenchExecResult{PdrInv}{KinductionDfStaticZeroZeroTTrueNotSolvedByKinductionPlain}{Error}{OutOfMemory}{Count}{}{127}%
\StoreBenchExecResult{PdrInv}{KinductionDfStaticZeroZeroTTrueNotSolvedByKinductionPlain}{Error}{OutOfMemory}{Cputime}{}{51985.806541265}%
\StoreBenchExecResult{PdrInv}{KinductionDfStaticZeroZeroTTrueNotSolvedByKinductionPlain}{Error}{OutOfMemory}{Cputime}{Avg}{409.3370593800393700787401575}%
\StoreBenchExecResult{PdrInv}{KinductionDfStaticZeroZeroTTrueNotSolvedByKinductionPlain}{Error}{OutOfMemory}{Cputime}{Median}{361.24282182}%
\StoreBenchExecResult{PdrInv}{KinductionDfStaticZeroZeroTTrueNotSolvedByKinductionPlain}{Error}{OutOfMemory}{Cputime}{Min}{140.343119534}%
\StoreBenchExecResult{PdrInv}{KinductionDfStaticZeroZeroTTrueNotSolvedByKinductionPlain}{Error}{OutOfMemory}{Cputime}{Max}{895.903635969}%
\StoreBenchExecResult{PdrInv}{KinductionDfStaticZeroZeroTTrueNotSolvedByKinductionPlain}{Error}{OutOfMemory}{Cputime}{Stdev}{208.5696336073739129970929652}%
\StoreBenchExecResult{PdrInv}{KinductionDfStaticZeroZeroTTrueNotSolvedByKinductionPlain}{Error}{OutOfMemory}{Walltime}{}{45784.3200325943}%
\StoreBenchExecResult{PdrInv}{KinductionDfStaticZeroZeroTTrueNotSolvedByKinductionPlain}{Error}{OutOfMemory}{Walltime}{Avg}{360.5064569495614173228346457}%
\StoreBenchExecResult{PdrInv}{KinductionDfStaticZeroZeroTTrueNotSolvedByKinductionPlain}{Error}{OutOfMemory}{Walltime}{Median}{283.133469105}%
\StoreBenchExecResult{PdrInv}{KinductionDfStaticZeroZeroTTrueNotSolvedByKinductionPlain}{Error}{OutOfMemory}{Walltime}{Min}{78.3415880203}%
\StoreBenchExecResult{PdrInv}{KinductionDfStaticZeroZeroTTrueNotSolvedByKinductionPlain}{Error}{OutOfMemory}{Walltime}{Max}{879.735621929}%
\StoreBenchExecResult{PdrInv}{KinductionDfStaticZeroZeroTTrueNotSolvedByKinductionPlain}{Error}{OutOfMemory}{Walltime}{Stdev}{226.5784370537721874541212506}%
\StoreBenchExecResult{PdrInv}{KinductionDfStaticZeroZeroTTrueNotSolvedByKinductionPlain}{Error}{OutOfNativeMemory}{Count}{}{1}%
\StoreBenchExecResult{PdrInv}{KinductionDfStaticZeroZeroTTrueNotSolvedByKinductionPlain}{Error}{OutOfNativeMemory}{Cputime}{}{811.329744803}%
\StoreBenchExecResult{PdrInv}{KinductionDfStaticZeroZeroTTrueNotSolvedByKinductionPlain}{Error}{OutOfNativeMemory}{Cputime}{Avg}{811.329744803}%
\StoreBenchExecResult{PdrInv}{KinductionDfStaticZeroZeroTTrueNotSolvedByKinductionPlain}{Error}{OutOfNativeMemory}{Cputime}{Median}{811.329744803}%
\StoreBenchExecResult{PdrInv}{KinductionDfStaticZeroZeroTTrueNotSolvedByKinductionPlain}{Error}{OutOfNativeMemory}{Cputime}{Min}{811.329744803}%
\StoreBenchExecResult{PdrInv}{KinductionDfStaticZeroZeroTTrueNotSolvedByKinductionPlain}{Error}{OutOfNativeMemory}{Cputime}{Max}{811.329744803}%
\StoreBenchExecResult{PdrInv}{KinductionDfStaticZeroZeroTTrueNotSolvedByKinductionPlain}{Error}{OutOfNativeMemory}{Cputime}{Stdev}{0E-9}%
\StoreBenchExecResult{PdrInv}{KinductionDfStaticZeroZeroTTrueNotSolvedByKinductionPlain}{Error}{OutOfNativeMemory}{Walltime}{}{798.984657049}%
\StoreBenchExecResult{PdrInv}{KinductionDfStaticZeroZeroTTrueNotSolvedByKinductionPlain}{Error}{OutOfNativeMemory}{Walltime}{Avg}{798.984657049}%
\StoreBenchExecResult{PdrInv}{KinductionDfStaticZeroZeroTTrueNotSolvedByKinductionPlain}{Error}{OutOfNativeMemory}{Walltime}{Median}{798.984657049}%
\StoreBenchExecResult{PdrInv}{KinductionDfStaticZeroZeroTTrueNotSolvedByKinductionPlain}{Error}{OutOfNativeMemory}{Walltime}{Min}{798.984657049}%
\StoreBenchExecResult{PdrInv}{KinductionDfStaticZeroZeroTTrueNotSolvedByKinductionPlain}{Error}{OutOfNativeMemory}{Walltime}{Max}{798.984657049}%
\StoreBenchExecResult{PdrInv}{KinductionDfStaticZeroZeroTTrueNotSolvedByKinductionPlain}{Error}{OutOfNativeMemory}{Walltime}{Stdev}{0E-9}%
\StoreBenchExecResult{PdrInv}{KinductionDfStaticZeroZeroTTrueNotSolvedByKinductionPlain}{Error}{Timeout}{Count}{}{1678}%
\StoreBenchExecResult{PdrInv}{KinductionDfStaticZeroZeroTTrueNotSolvedByKinductionPlain}{Error}{Timeout}{Cputime}{}{1521340.777597222}%
\StoreBenchExecResult{PdrInv}{KinductionDfStaticZeroZeroTTrueNotSolvedByKinductionPlain}{Error}{Timeout}{Cputime}{Avg}{906.6393191878557806912991657}%
\StoreBenchExecResult{PdrInv}{KinductionDfStaticZeroZeroTTrueNotSolvedByKinductionPlain}{Error}{Timeout}{Cputime}{Median}{902.229736642}%
\StoreBenchExecResult{PdrInv}{KinductionDfStaticZeroZeroTTrueNotSolvedByKinductionPlain}{Error}{Timeout}{Cputime}{Min}{900.91988852}%
\StoreBenchExecResult{PdrInv}{KinductionDfStaticZeroZeroTTrueNotSolvedByKinductionPlain}{Error}{Timeout}{Cputime}{Max}{1002.39260405}%
\StoreBenchExecResult{PdrInv}{KinductionDfStaticZeroZeroTTrueNotSolvedByKinductionPlain}{Error}{Timeout}{Cputime}{Stdev}{16.07461431170777469769003163}%
\StoreBenchExecResult{PdrInv}{KinductionDfStaticZeroZeroTTrueNotSolvedByKinductionPlain}{Error}{Timeout}{Walltime}{}{1061530.968049768}%
\StoreBenchExecResult{PdrInv}{KinductionDfStaticZeroZeroTTrueNotSolvedByKinductionPlain}{Error}{Timeout}{Walltime}{Avg}{632.6167866804338498212157330}%
\StoreBenchExecResult{PdrInv}{KinductionDfStaticZeroZeroTTrueNotSolvedByKinductionPlain}{Error}{Timeout}{Walltime}{Median}{463.5906124115}%
\StoreBenchExecResult{PdrInv}{KinductionDfStaticZeroZeroTTrueNotSolvedByKinductionPlain}{Error}{Timeout}{Walltime}{Min}{451.108106136}%
\StoreBenchExecResult{PdrInv}{KinductionDfStaticZeroZeroTTrueNotSolvedByKinductionPlain}{Error}{Timeout}{Walltime}{Max}{908.435231209}%
\StoreBenchExecResult{PdrInv}{KinductionDfStaticZeroZeroTTrueNotSolvedByKinductionPlain}{Error}{Timeout}{Walltime}{Stdev}{202.3041368542948525769403659}%
\providecommand\StoreBenchExecResult[7]{\expandafter\newcommand\csname#1#2#3#4#5#6\endcsname{#7}}%
\StoreBenchExecResult{PdrInv}{KinductionDfStaticZeroZeroT}{Total}{}{Count}{}{5591}%
\StoreBenchExecResult{PdrInv}{KinductionDfStaticZeroZeroT}{Total}{}{Cputime}{}{2228771.998109020}%
\StoreBenchExecResult{PdrInv}{KinductionDfStaticZeroZeroT}{Total}{}{Cputime}{Avg}{398.6356641225219102128420676}%
\StoreBenchExecResult{PdrInv}{KinductionDfStaticZeroZeroT}{Total}{}{Cputime}{Median}{119.248532199}%
\StoreBenchExecResult{PdrInv}{KinductionDfStaticZeroZeroT}{Total}{}{Cputime}{Min}{2.446906053}%
\StoreBenchExecResult{PdrInv}{KinductionDfStaticZeroZeroT}{Total}{}{Cputime}{Max}{1002.39260405}%
\StoreBenchExecResult{PdrInv}{KinductionDfStaticZeroZeroT}{Total}{}{Cputime}{Stdev}{420.2902082308360838728460608}%
\StoreBenchExecResult{PdrInv}{KinductionDfStaticZeroZeroT}{Total}{}{Walltime}{}{1591760.24884772267}%
\StoreBenchExecResult{PdrInv}{KinductionDfStaticZeroZeroT}{Total}{}{Walltime}{Avg}{284.7004558840498426041852978}%
\StoreBenchExecResult{PdrInv}{KinductionDfStaticZeroZeroT}{Total}{}{Walltime}{Median}{76.2768700123}%
\StoreBenchExecResult{PdrInv}{KinductionDfStaticZeroZeroT}{Total}{}{Walltime}{Min}{1.34633517265}%
\StoreBenchExecResult{PdrInv}{KinductionDfStaticZeroZeroT}{Total}{}{Walltime}{Max}{922.01857996}%
\StoreBenchExecResult{PdrInv}{KinductionDfStaticZeroZeroT}{Total}{}{Walltime}{Stdev}{325.8485988627558278832563158}%
\StoreBenchExecResult{PdrInv}{KinductionDfStaticZeroZeroT}{Correct}{}{Count}{}{2986}%
\StoreBenchExecResult{PdrInv}{KinductionDfStaticZeroZeroT}{Correct}{}{Cputime}{}{164092.272545182}%
\StoreBenchExecResult{PdrInv}{KinductionDfStaticZeroZeroT}{Correct}{}{Cputime}{Avg}{54.95387560119959812458137977}%
\StoreBenchExecResult{PdrInv}{KinductionDfStaticZeroZeroT}{Correct}{}{Cputime}{Median}{9.333570445}%
\StoreBenchExecResult{PdrInv}{KinductionDfStaticZeroZeroT}{Correct}{}{Cputime}{Min}{2.929669398}%
\StoreBenchExecResult{PdrInv}{KinductionDfStaticZeroZeroT}{Correct}{}{Cputime}{Max}{899.570086353}%
\StoreBenchExecResult{PdrInv}{KinductionDfStaticZeroZeroT}{Correct}{}{Cputime}{Stdev}{133.0905506911270231143433140}%
\StoreBenchExecResult{PdrInv}{KinductionDfStaticZeroZeroT}{Correct}{}{Walltime}{}{116123.82615589203}%
\StoreBenchExecResult{PdrInv}{KinductionDfStaticZeroZeroT}{Correct}{}{Walltime}{Avg}{38.88942604015138312123241795}%
\StoreBenchExecResult{PdrInv}{KinductionDfStaticZeroZeroT}{Correct}{}{Walltime}{Median}{4.91299700737}%
\StoreBenchExecResult{PdrInv}{KinductionDfStaticZeroZeroT}{Correct}{}{Walltime}{Min}{1.63654112816}%
\StoreBenchExecResult{PdrInv}{KinductionDfStaticZeroZeroT}{Correct}{}{Walltime}{Max}{884.368417025}%
\StoreBenchExecResult{PdrInv}{KinductionDfStaticZeroZeroT}{Correct}{}{Walltime}{Stdev}{109.9442154580803061282486468}%
\StoreBenchExecResult{PdrInv}{KinductionDfStaticZeroZeroT}{Correct}{False}{Count}{}{816}%
\StoreBenchExecResult{PdrInv}{KinductionDfStaticZeroZeroT}{Correct}{False}{Cputime}{}{71907.844576699}%
\StoreBenchExecResult{PdrInv}{KinductionDfStaticZeroZeroT}{Correct}{False}{Cputime}{Avg}{88.12235854987622549019607843}%
\StoreBenchExecResult{PdrInv}{KinductionDfStaticZeroZeroT}{Correct}{False}{Cputime}{Median}{21.3756631515}%
\StoreBenchExecResult{PdrInv}{KinductionDfStaticZeroZeroT}{Correct}{False}{Cputime}{Min}{3.18525768}%
\StoreBenchExecResult{PdrInv}{KinductionDfStaticZeroZeroT}{Correct}{False}{Cputime}{Max}{899.570086353}%
\StoreBenchExecResult{PdrInv}{KinductionDfStaticZeroZeroT}{Correct}{False}{Cputime}{Stdev}{175.1519608172406247816902107}%
\StoreBenchExecResult{PdrInv}{KinductionDfStaticZeroZeroT}{Correct}{False}{Walltime}{}{57836.85044979664}%
\StoreBenchExecResult{PdrInv}{KinductionDfStaticZeroZeroT}{Correct}{False}{Walltime}{Avg}{70.87849319828019607843137255}%
\StoreBenchExecResult{PdrInv}{KinductionDfStaticZeroZeroT}{Correct}{False}{Walltime}{Median}{11.8407919407}%
\StoreBenchExecResult{PdrInv}{KinductionDfStaticZeroZeroT}{Correct}{False}{Walltime}{Min}{1.77245593071}%
\StoreBenchExecResult{PdrInv}{KinductionDfStaticZeroZeroT}{Correct}{False}{Walltime}{Max}{884.368417025}%
\StoreBenchExecResult{PdrInv}{KinductionDfStaticZeroZeroT}{Correct}{False}{Walltime}{Stdev}{161.6807805732460739025935227}%
\StoreBenchExecResult{PdrInv}{KinductionDfStaticZeroZeroT}{Correct}{True}{Count}{}{2170}%
\StoreBenchExecResult{PdrInv}{KinductionDfStaticZeroZeroT}{Correct}{True}{Cputime}{}{92184.427968483}%
\StoreBenchExecResult{PdrInv}{KinductionDfStaticZeroZeroT}{Correct}{True}{Cputime}{Avg}{42.48130321128248847926267281}%
\StoreBenchExecResult{PdrInv}{KinductionDfStaticZeroZeroT}{Correct}{True}{Cputime}{Median}{7.032626742}%
\StoreBenchExecResult{PdrInv}{KinductionDfStaticZeroZeroT}{Correct}{True}{Cputime}{Min}{2.929669398}%
\StoreBenchExecResult{PdrInv}{KinductionDfStaticZeroZeroT}{Correct}{True}{Cputime}{Max}{880.007326739}%
\StoreBenchExecResult{PdrInv}{KinductionDfStaticZeroZeroT}{Correct}{True}{Cputime}{Stdev}{110.7631356307678788967444715}%
\StoreBenchExecResult{PdrInv}{KinductionDfStaticZeroZeroT}{Correct}{True}{Walltime}{}{58286.97570609539}%
\StoreBenchExecResult{PdrInv}{KinductionDfStaticZeroZeroT}{Correct}{True}{Walltime}{Avg}{26.86035746824672350230414747}%
\StoreBenchExecResult{PdrInv}{KinductionDfStaticZeroZeroT}{Correct}{True}{Walltime}{Median}{3.737602472305}%
\StoreBenchExecResult{PdrInv}{KinductionDfStaticZeroZeroT}{Correct}{True}{Walltime}{Min}{1.63654112816}%
\StoreBenchExecResult{PdrInv}{KinductionDfStaticZeroZeroT}{Correct}{True}{Walltime}{Max}{876.970759869}%
\StoreBenchExecResult{PdrInv}{KinductionDfStaticZeroZeroT}{Correct}{True}{Walltime}{Stdev}{79.20737468265552503046002007}%
\StoreBenchExecResult{PdrInv}{KinductionDfStaticZeroZeroT}{Wrong}{True}{Count}{}{0}%
\StoreBenchExecResult{PdrInv}{KinductionDfStaticZeroZeroT}{Wrong}{True}{Cputime}{}{0}%
\StoreBenchExecResult{PdrInv}{KinductionDfStaticZeroZeroT}{Wrong}{True}{Cputime}{Avg}{None}%
\StoreBenchExecResult{PdrInv}{KinductionDfStaticZeroZeroT}{Wrong}{True}{Cputime}{Median}{None}%
\StoreBenchExecResult{PdrInv}{KinductionDfStaticZeroZeroT}{Wrong}{True}{Cputime}{Min}{None}%
\StoreBenchExecResult{PdrInv}{KinductionDfStaticZeroZeroT}{Wrong}{True}{Cputime}{Max}{None}%
\StoreBenchExecResult{PdrInv}{KinductionDfStaticZeroZeroT}{Wrong}{True}{Cputime}{Stdev}{None}%
\StoreBenchExecResult{PdrInv}{KinductionDfStaticZeroZeroT}{Wrong}{True}{Walltime}{}{0}%
\StoreBenchExecResult{PdrInv}{KinductionDfStaticZeroZeroT}{Wrong}{True}{Walltime}{Avg}{None}%
\StoreBenchExecResult{PdrInv}{KinductionDfStaticZeroZeroT}{Wrong}{True}{Walltime}{Median}{None}%
\StoreBenchExecResult{PdrInv}{KinductionDfStaticZeroZeroT}{Wrong}{True}{Walltime}{Min}{None}%
\StoreBenchExecResult{PdrInv}{KinductionDfStaticZeroZeroT}{Wrong}{True}{Walltime}{Max}{None}%
\StoreBenchExecResult{PdrInv}{KinductionDfStaticZeroZeroT}{Wrong}{True}{Walltime}{Stdev}{None}%
\StoreBenchExecResult{PdrInv}{KinductionDfStaticZeroZeroT}{Error}{}{Count}{}{2603}%
\StoreBenchExecResult{PdrInv}{KinductionDfStaticZeroZeroT}{Error}{}{Cputime}{}{2064656.491657810}%
\StoreBenchExecResult{PdrInv}{KinductionDfStaticZeroZeroT}{Error}{}{Cputime}{Avg}{793.1834389772608528620822128}%
\StoreBenchExecResult{PdrInv}{KinductionDfStaticZeroZeroT}{Error}{}{Cputime}{Median}{901.848432732}%
\StoreBenchExecResult{PdrInv}{KinductionDfStaticZeroZeroT}{Error}{}{Cputime}{Min}{2.446906053}%
\StoreBenchExecResult{PdrInv}{KinductionDfStaticZeroZeroT}{Error}{}{Cputime}{Max}{1002.39260405}%
\StoreBenchExecResult{PdrInv}{KinductionDfStaticZeroZeroT}{Error}{}{Cputime}{Stdev}{260.4137531708260998579645010}%
\StoreBenchExecResult{PdrInv}{KinductionDfStaticZeroZeroT}{Error}{}{Walltime}{}{1475623.89269400502}%
\StoreBenchExecResult{PdrInv}{KinductionDfStaticZeroZeroT}{Error}{}{Walltime}{Avg}{566.8935431018075374567806377}%
\StoreBenchExecResult{PdrInv}{KinductionDfStaticZeroZeroT}{Error}{}{Walltime}{Median}{457.887514114}%
\StoreBenchExecResult{PdrInv}{KinductionDfStaticZeroZeroT}{Error}{}{Walltime}{Min}{1.34633517265}%
\StoreBenchExecResult{PdrInv}{KinductionDfStaticZeroZeroT}{Error}{}{Walltime}{Max}{922.01857996}%
\StoreBenchExecResult{PdrInv}{KinductionDfStaticZeroZeroT}{Error}{}{Walltime}{Stdev}{255.3164878197741728665071426}%
\StoreBenchExecResult{PdrInv}{KinductionDfStaticZeroZeroT}{Error}{Assertion}{Count}{}{4}%
\StoreBenchExecResult{PdrInv}{KinductionDfStaticZeroZeroT}{Error}{Assertion}{Cputime}{}{14.496794910}%
\StoreBenchExecResult{PdrInv}{KinductionDfStaticZeroZeroT}{Error}{Assertion}{Cputime}{Avg}{3.6241987275}%
\StoreBenchExecResult{PdrInv}{KinductionDfStaticZeroZeroT}{Error}{Assertion}{Cputime}{Median}{3.627673476}%
\StoreBenchExecResult{PdrInv}{KinductionDfStaticZeroZeroT}{Error}{Assertion}{Cputime}{Min}{3.185055688}%
\StoreBenchExecResult{PdrInv}{KinductionDfStaticZeroZeroT}{Error}{Assertion}{Cputime}{Max}{4.05639227}%
\StoreBenchExecResult{PdrInv}{KinductionDfStaticZeroZeroT}{Error}{Assertion}{Cputime}{Stdev}{0.4014253807517493247249002452}%
\StoreBenchExecResult{PdrInv}{KinductionDfStaticZeroZeroT}{Error}{Assertion}{Walltime}{}{8.01108312607}%
\StoreBenchExecResult{PdrInv}{KinductionDfStaticZeroZeroT}{Error}{Assertion}{Walltime}{Avg}{2.0027707815175}%
\StoreBenchExecResult{PdrInv}{KinductionDfStaticZeroZeroT}{Error}{Assertion}{Walltime}{Median}{2.00131046772}%
\StoreBenchExecResult{PdrInv}{KinductionDfStaticZeroZeroT}{Error}{Assertion}{Walltime}{Min}{1.76435208321}%
\StoreBenchExecResult{PdrInv}{KinductionDfStaticZeroZeroT}{Error}{Assertion}{Walltime}{Max}{2.24411010742}%
\StoreBenchExecResult{PdrInv}{KinductionDfStaticZeroZeroT}{Error}{Assertion}{Walltime}{Stdev}{0.2204576105997358673666008613}%
\StoreBenchExecResult{PdrInv}{KinductionDfStaticZeroZeroT}{Error}{Error}{Count}{}{203}%
\StoreBenchExecResult{PdrInv}{KinductionDfStaticZeroZeroT}{Error}{Error}{Cputime}{}{36647.134910167}%
\StoreBenchExecResult{PdrInv}{KinductionDfStaticZeroZeroT}{Error}{Error}{Cputime}{Avg}{180.5277581781625615763546798}%
\StoreBenchExecResult{PdrInv}{KinductionDfStaticZeroZeroT}{Error}{Error}{Cputime}{Median}{116.164938463}%
\StoreBenchExecResult{PdrInv}{KinductionDfStaticZeroZeroT}{Error}{Error}{Cputime}{Min}{2.446906053}%
\StoreBenchExecResult{PdrInv}{KinductionDfStaticZeroZeroT}{Error}{Error}{Cputime}{Max}{898.008416126}%
\StoreBenchExecResult{PdrInv}{KinductionDfStaticZeroZeroT}{Error}{Error}{Cputime}{Stdev}{194.2344508128652500347058607}%
\StoreBenchExecResult{PdrInv}{KinductionDfStaticZeroZeroT}{Error}{Error}{Walltime}{}{30273.41303301224}%
\StoreBenchExecResult{PdrInv}{KinductionDfStaticZeroZeroT}{Error}{Error}{Walltime}{Avg}{149.1301134631144827586206897}%
\StoreBenchExecResult{PdrInv}{KinductionDfStaticZeroZeroT}{Error}{Error}{Walltime}{Median}{87.2534880638}%
\StoreBenchExecResult{PdrInv}{KinductionDfStaticZeroZeroT}{Error}{Error}{Walltime}{Min}{1.34633517265}%
\StoreBenchExecResult{PdrInv}{KinductionDfStaticZeroZeroT}{Error}{Error}{Walltime}{Max}{880.894071817}%
\StoreBenchExecResult{PdrInv}{KinductionDfStaticZeroZeroT}{Error}{Error}{Walltime}{Stdev}{168.9415272912382817216386492}%
\StoreBenchExecResult{PdrInv}{KinductionDfStaticZeroZeroT}{Error}{Exception}{Count}{}{13}%
\StoreBenchExecResult{PdrInv}{KinductionDfStaticZeroZeroT}{Error}{Exception}{Cputime}{}{1401.424136434}%
\StoreBenchExecResult{PdrInv}{KinductionDfStaticZeroZeroT}{Error}{Exception}{Cputime}{Avg}{107.8018566487692307692307692}%
\StoreBenchExecResult{PdrInv}{KinductionDfStaticZeroZeroT}{Error}{Exception}{Cputime}{Median}{78.164800942}%
\StoreBenchExecResult{PdrInv}{KinductionDfStaticZeroZeroT}{Error}{Exception}{Cputime}{Min}{13.814084064}%
\StoreBenchExecResult{PdrInv}{KinductionDfStaticZeroZeroT}{Error}{Exception}{Cputime}{Max}{407.357064118}%
\StoreBenchExecResult{PdrInv}{KinductionDfStaticZeroZeroT}{Error}{Exception}{Cputime}{Stdev}{110.2595442011727280241044583}%
\StoreBenchExecResult{PdrInv}{KinductionDfStaticZeroZeroT}{Error}{Exception}{Walltime}{}{764.26113247861}%
\StoreBenchExecResult{PdrInv}{KinductionDfStaticZeroZeroT}{Error}{Exception}{Walltime}{Avg}{58.78931788297}%
\StoreBenchExecResult{PdrInv}{KinductionDfStaticZeroZeroT}{Error}{Exception}{Walltime}{Median}{39.3472690582}%
\StoreBenchExecResult{PdrInv}{KinductionDfStaticZeroZeroT}{Error}{Exception}{Walltime}{Min}{7.07906699181}%
\StoreBenchExecResult{PdrInv}{KinductionDfStaticZeroZeroT}{Error}{Exception}{Walltime}{Max}{242.534669161}%
\StoreBenchExecResult{PdrInv}{KinductionDfStaticZeroZeroT}{Error}{Exception}{Walltime}{Stdev}{63.45562960863017754052852563}%
\StoreBenchExecResult{PdrInv}{KinductionDfStaticZeroZeroT}{Error}{OutOfJavaMemory}{Count}{}{10}%
\StoreBenchExecResult{PdrInv}{KinductionDfStaticZeroZeroT}{Error}{OutOfJavaMemory}{Cputime}{}{5386.919930769}%
\StoreBenchExecResult{PdrInv}{KinductionDfStaticZeroZeroT}{Error}{OutOfJavaMemory}{Cputime}{Avg}{538.6919930769}%
\StoreBenchExecResult{PdrInv}{KinductionDfStaticZeroZeroT}{Error}{OutOfJavaMemory}{Cputime}{Median}{597.414178156}%
\StoreBenchExecResult{PdrInv}{KinductionDfStaticZeroZeroT}{Error}{OutOfJavaMemory}{Cputime}{Min}{168.105338919}%
\StoreBenchExecResult{PdrInv}{KinductionDfStaticZeroZeroT}{Error}{OutOfJavaMemory}{Cputime}{Max}{788.125840536}%
\StoreBenchExecResult{PdrInv}{KinductionDfStaticZeroZeroT}{Error}{OutOfJavaMemory}{Cputime}{Stdev}{205.8733675279559495905694428}%
\StoreBenchExecResult{PdrInv}{KinductionDfStaticZeroZeroT}{Error}{OutOfJavaMemory}{Walltime}{}{3290.023833275}%
\StoreBenchExecResult{PdrInv}{KinductionDfStaticZeroZeroT}{Error}{OutOfJavaMemory}{Walltime}{Avg}{329.0023833275}%
\StoreBenchExecResult{PdrInv}{KinductionDfStaticZeroZeroT}{Error}{OutOfJavaMemory}{Walltime}{Median}{345.4251089095}%
\StoreBenchExecResult{PdrInv}{KinductionDfStaticZeroZeroT}{Error}{OutOfJavaMemory}{Walltime}{Min}{103.409188986}%
\StoreBenchExecResult{PdrInv}{KinductionDfStaticZeroZeroT}{Error}{OutOfJavaMemory}{Walltime}{Max}{638.058216095}%
\StoreBenchExecResult{PdrInv}{KinductionDfStaticZeroZeroT}{Error}{OutOfJavaMemory}{Walltime}{Stdev}{141.5307965538915434206291333}%
\StoreBenchExecResult{PdrInv}{KinductionDfStaticZeroZeroT}{Error}{OutOfMemory}{Count}{}{257}%
\StoreBenchExecResult{PdrInv}{KinductionDfStaticZeroZeroT}{Error}{OutOfMemory}{Cputime}{}{97969.073735798}%
\StoreBenchExecResult{PdrInv}{KinductionDfStaticZeroZeroT}{Error}{OutOfMemory}{Cputime}{Avg}{381.2026215400700389105058366}%
\StoreBenchExecResult{PdrInv}{KinductionDfStaticZeroZeroT}{Error}{OutOfMemory}{Cputime}{Median}{311.44161181}%
\StoreBenchExecResult{PdrInv}{KinductionDfStaticZeroZeroT}{Error}{OutOfMemory}{Cputime}{Min}{139.758242099}%
\StoreBenchExecResult{PdrInv}{KinductionDfStaticZeroZeroT}{Error}{OutOfMemory}{Cputime}{Max}{895.903635969}%
\StoreBenchExecResult{PdrInv}{KinductionDfStaticZeroZeroT}{Error}{OutOfMemory}{Cputime}{Stdev}{196.8460335576480916080264428}%
\StoreBenchExecResult{PdrInv}{KinductionDfStaticZeroZeroT}{Error}{OutOfMemory}{Walltime}{}{85791.0328340471}%
\StoreBenchExecResult{PdrInv}{KinductionDfStaticZeroZeroT}{Error}{OutOfMemory}{Walltime}{Avg}{333.8172483815062256809338521}%
\StoreBenchExecResult{PdrInv}{KinductionDfStaticZeroZeroT}{Error}{OutOfMemory}{Walltime}{Median}{242.748559952}%
\StoreBenchExecResult{PdrInv}{KinductionDfStaticZeroZeroT}{Error}{OutOfMemory}{Walltime}{Min}{78.3415880203}%
\StoreBenchExecResult{PdrInv}{KinductionDfStaticZeroZeroT}{Error}{OutOfMemory}{Walltime}{Max}{879.735621929}%
\StoreBenchExecResult{PdrInv}{KinductionDfStaticZeroZeroT}{Error}{OutOfMemory}{Walltime}{Stdev}{212.0619403700497388597316501}%
\StoreBenchExecResult{PdrInv}{KinductionDfStaticZeroZeroT}{Error}{OutOfNativeMemory}{Count}{}{1}%
\StoreBenchExecResult{PdrInv}{KinductionDfStaticZeroZeroT}{Error}{OutOfNativeMemory}{Cputime}{}{811.329744803}%
\StoreBenchExecResult{PdrInv}{KinductionDfStaticZeroZeroT}{Error}{OutOfNativeMemory}{Cputime}{Avg}{811.329744803}%
\StoreBenchExecResult{PdrInv}{KinductionDfStaticZeroZeroT}{Error}{OutOfNativeMemory}{Cputime}{Median}{811.329744803}%
\StoreBenchExecResult{PdrInv}{KinductionDfStaticZeroZeroT}{Error}{OutOfNativeMemory}{Cputime}{Min}{811.329744803}%
\StoreBenchExecResult{PdrInv}{KinductionDfStaticZeroZeroT}{Error}{OutOfNativeMemory}{Cputime}{Max}{811.329744803}%
\StoreBenchExecResult{PdrInv}{KinductionDfStaticZeroZeroT}{Error}{OutOfNativeMemory}{Cputime}{Stdev}{0E-9}%
\StoreBenchExecResult{PdrInv}{KinductionDfStaticZeroZeroT}{Error}{OutOfNativeMemory}{Walltime}{}{798.984657049}%
\StoreBenchExecResult{PdrInv}{KinductionDfStaticZeroZeroT}{Error}{OutOfNativeMemory}{Walltime}{Avg}{798.984657049}%
\StoreBenchExecResult{PdrInv}{KinductionDfStaticZeroZeroT}{Error}{OutOfNativeMemory}{Walltime}{Median}{798.984657049}%
\StoreBenchExecResult{PdrInv}{KinductionDfStaticZeroZeroT}{Error}{OutOfNativeMemory}{Walltime}{Min}{798.984657049}%
\StoreBenchExecResult{PdrInv}{KinductionDfStaticZeroZeroT}{Error}{OutOfNativeMemory}{Walltime}{Max}{798.984657049}%
\StoreBenchExecResult{PdrInv}{KinductionDfStaticZeroZeroT}{Error}{OutOfNativeMemory}{Walltime}{Stdev}{0E-9}%
\StoreBenchExecResult{PdrInv}{KinductionDfStaticZeroZeroT}{Error}{Timeout}{Count}{}{2115}%
\StoreBenchExecResult{PdrInv}{KinductionDfStaticZeroZeroT}{Error}{Timeout}{Cputime}{}{1922426.112404929}%
\StoreBenchExecResult{PdrInv}{KinductionDfStaticZeroZeroT}{Error}{Timeout}{Cputime}{Avg}{908.9485165035125295508274232}%
\StoreBenchExecResult{PdrInv}{KinductionDfStaticZeroZeroT}{Error}{Timeout}{Cputime}{Median}{902.44943302}%
\StoreBenchExecResult{PdrInv}{KinductionDfStaticZeroZeroT}{Error}{Timeout}{Cputime}{Min}{900.91988852}%
\StoreBenchExecResult{PdrInv}{KinductionDfStaticZeroZeroT}{Error}{Timeout}{Cputime}{Max}{1002.39260405}%
\StoreBenchExecResult{PdrInv}{KinductionDfStaticZeroZeroT}{Error}{Timeout}{Cputime}{Stdev}{20.81432273029335947645812102}%
\StoreBenchExecResult{PdrInv}{KinductionDfStaticZeroZeroT}{Error}{Timeout}{Walltime}{}{1354698.166121017}%
\StoreBenchExecResult{PdrInv}{KinductionDfStaticZeroZeroT}{Error}{Timeout}{Walltime}{Avg}{640.5192274803862884160756501}%
\StoreBenchExecResult{PdrInv}{KinductionDfStaticZeroZeroT}{Error}{Timeout}{Walltime}{Median}{508.988873959}%
\StoreBenchExecResult{PdrInv}{KinductionDfStaticZeroZeroT}{Error}{Timeout}{Walltime}{Min}{451.108106136}%
\StoreBenchExecResult{PdrInv}{KinductionDfStaticZeroZeroT}{Error}{Timeout}{Walltime}{Max}{922.01857996}%
\StoreBenchExecResult{PdrInv}{KinductionDfStaticZeroZeroT}{Error}{Timeout}{Walltime}{Stdev}{201.6118008937117059659137554}%
\StoreBenchExecResult{PdrInv}{KinductionDfStaticZeroZeroT}{Wrong}{}{Count}{}{2}%
\StoreBenchExecResult{PdrInv}{KinductionDfStaticZeroZeroT}{Wrong}{}{Cputime}{}{23.233906028}%
\StoreBenchExecResult{PdrInv}{KinductionDfStaticZeroZeroT}{Wrong}{}{Cputime}{Avg}{11.616953014}%
\StoreBenchExecResult{PdrInv}{KinductionDfStaticZeroZeroT}{Wrong}{}{Cputime}{Median}{11.616953014}%
\StoreBenchExecResult{PdrInv}{KinductionDfStaticZeroZeroT}{Wrong}{}{Cputime}{Min}{3.762388799}%
\StoreBenchExecResult{PdrInv}{KinductionDfStaticZeroZeroT}{Wrong}{}{Cputime}{Max}{19.471517229}%
\StoreBenchExecResult{PdrInv}{KinductionDfStaticZeroZeroT}{Wrong}{}{Cputime}{Stdev}{7.854564215}%
\StoreBenchExecResult{PdrInv}{KinductionDfStaticZeroZeroT}{Wrong}{}{Walltime}{}{12.52999782562}%
\StoreBenchExecResult{PdrInv}{KinductionDfStaticZeroZeroT}{Wrong}{}{Walltime}{Avg}{6.26499891281}%
\StoreBenchExecResult{PdrInv}{KinductionDfStaticZeroZeroT}{Wrong}{}{Walltime}{Median}{6.26499891281}%
\StoreBenchExecResult{PdrInv}{KinductionDfStaticZeroZeroT}{Wrong}{}{Walltime}{Min}{2.08920884132}%
\StoreBenchExecResult{PdrInv}{KinductionDfStaticZeroZeroT}{Wrong}{}{Walltime}{Max}{10.4407889843}%
\StoreBenchExecResult{PdrInv}{KinductionDfStaticZeroZeroT}{Wrong}{}{Walltime}{Stdev}{4.17579007149}%
\StoreBenchExecResult{PdrInv}{KinductionDfStaticZeroZeroT}{Wrong}{False}{Count}{}{2}%
\StoreBenchExecResult{PdrInv}{KinductionDfStaticZeroZeroT}{Wrong}{False}{Cputime}{}{23.233906028}%
\StoreBenchExecResult{PdrInv}{KinductionDfStaticZeroZeroT}{Wrong}{False}{Cputime}{Avg}{11.616953014}%
\StoreBenchExecResult{PdrInv}{KinductionDfStaticZeroZeroT}{Wrong}{False}{Cputime}{Median}{11.616953014}%
\StoreBenchExecResult{PdrInv}{KinductionDfStaticZeroZeroT}{Wrong}{False}{Cputime}{Min}{3.762388799}%
\StoreBenchExecResult{PdrInv}{KinductionDfStaticZeroZeroT}{Wrong}{False}{Cputime}{Max}{19.471517229}%
\StoreBenchExecResult{PdrInv}{KinductionDfStaticZeroZeroT}{Wrong}{False}{Cputime}{Stdev}{7.854564215}%
\StoreBenchExecResult{PdrInv}{KinductionDfStaticZeroZeroT}{Wrong}{False}{Walltime}{}{12.52999782562}%
\StoreBenchExecResult{PdrInv}{KinductionDfStaticZeroZeroT}{Wrong}{False}{Walltime}{Avg}{6.26499891281}%
\StoreBenchExecResult{PdrInv}{KinductionDfStaticZeroZeroT}{Wrong}{False}{Walltime}{Median}{6.26499891281}%
\StoreBenchExecResult{PdrInv}{KinductionDfStaticZeroZeroT}{Wrong}{False}{Walltime}{Min}{2.08920884132}%
\StoreBenchExecResult{PdrInv}{KinductionDfStaticZeroZeroT}{Wrong}{False}{Walltime}{Max}{10.4407889843}%
\StoreBenchExecResult{PdrInv}{KinductionDfStaticZeroZeroT}{Wrong}{False}{Walltime}{Stdev}{4.17579007149}%
\providecommand\StoreBenchExecResult[7]{\expandafter\newcommand\csname#1#2#3#4#5#6\endcsname{#7}}%
\StoreBenchExecResult{PdrInv}{KinductionDfStaticZeroOneTFTrueNotSolvedByKinductionPlainButKipdr}{Total}{}{Count}{}{449}%
\StoreBenchExecResult{PdrInv}{KinductionDfStaticZeroOneTFTrueNotSolvedByKinductionPlainButKipdr}{Total}{}{Cputime}{}{13070.429615551}%
\StoreBenchExecResult{PdrInv}{KinductionDfStaticZeroOneTFTrueNotSolvedByKinductionPlainButKipdr}{Total}{}{Cputime}{Avg}{29.11008823062583518930957684}%
\StoreBenchExecResult{PdrInv}{KinductionDfStaticZeroOneTFTrueNotSolvedByKinductionPlainButKipdr}{Total}{}{Cputime}{Median}{6.070939062}%
\StoreBenchExecResult{PdrInv}{KinductionDfStaticZeroOneTFTrueNotSolvedByKinductionPlainButKipdr}{Total}{}{Cputime}{Min}{3.17534903}%
\StoreBenchExecResult{PdrInv}{KinductionDfStaticZeroOneTFTrueNotSolvedByKinductionPlainButKipdr}{Total}{}{Cputime}{Max}{913.517748961}%
\StoreBenchExecResult{PdrInv}{KinductionDfStaticZeroOneTFTrueNotSolvedByKinductionPlainButKipdr}{Total}{}{Cputime}{Stdev}{138.4201118189713258950161199}%
\StoreBenchExecResult{PdrInv}{KinductionDfStaticZeroOneTFTrueNotSolvedByKinductionPlainButKipdr}{Total}{}{Walltime}{}{10987.40954446767}%
\StoreBenchExecResult{PdrInv}{KinductionDfStaticZeroOneTFTrueNotSolvedByKinductionPlainButKipdr}{Total}{}{Walltime}{Avg}{24.47084531061841870824053452}%
\StoreBenchExecResult{PdrInv}{KinductionDfStaticZeroOneTFTrueNotSolvedByKinductionPlainButKipdr}{Total}{}{Walltime}{Median}{3.21363711357}%
\StoreBenchExecResult{PdrInv}{KinductionDfStaticZeroOneTFTrueNotSolvedByKinductionPlainButKipdr}{Total}{}{Walltime}{Min}{1.78259801865}%
\StoreBenchExecResult{PdrInv}{KinductionDfStaticZeroOneTFTrueNotSolvedByKinductionPlainButKipdr}{Total}{}{Walltime}{Max}{898.996891975}%
\StoreBenchExecResult{PdrInv}{KinductionDfStaticZeroOneTFTrueNotSolvedByKinductionPlainButKipdr}{Total}{}{Walltime}{Stdev}{131.9741719296530350055527773}%
\StoreBenchExecResult{PdrInv}{KinductionDfStaticZeroOneTFTrueNotSolvedByKinductionPlainButKipdr}{Correct}{}{Count}{}{438}%
\StoreBenchExecResult{PdrInv}{KinductionDfStaticZeroOneTFTrueNotSolvedByKinductionPlainButKipdr}{Correct}{}{Cputime}{}{3155.515533554}%
\StoreBenchExecResult{PdrInv}{KinductionDfStaticZeroOneTFTrueNotSolvedByKinductionPlainButKipdr}{Correct}{}{Cputime}{Avg}{7.204373364278538812785388128}%
\StoreBenchExecResult{PdrInv}{KinductionDfStaticZeroOneTFTrueNotSolvedByKinductionPlainButKipdr}{Correct}{}{Cputime}{Median}{5.986700053}%
\StoreBenchExecResult{PdrInv}{KinductionDfStaticZeroOneTFTrueNotSolvedByKinductionPlainButKipdr}{Correct}{}{Cputime}{Min}{3.17534903}%
\StoreBenchExecResult{PdrInv}{KinductionDfStaticZeroOneTFTrueNotSolvedByKinductionPlainButKipdr}{Correct}{}{Cputime}{Max}{79.246301155}%
\StoreBenchExecResult{PdrInv}{KinductionDfStaticZeroOneTFTrueNotSolvedByKinductionPlainButKipdr}{Correct}{}{Cputime}{Stdev}{6.738581134001495127348074013}%
\StoreBenchExecResult{PdrInv}{KinductionDfStaticZeroOneTFTrueNotSolvedByKinductionPlainButKipdr}{Correct}{}{Walltime}{}{1663.72902464867}%
\StoreBenchExecResult{PdrInv}{KinductionDfStaticZeroOneTFTrueNotSolvedByKinductionPlainButKipdr}{Correct}{}{Walltime}{Avg}{3.798468092805182648401826484}%
\StoreBenchExecResult{PdrInv}{KinductionDfStaticZeroOneTFTrueNotSolvedByKinductionPlainButKipdr}{Correct}{}{Walltime}{Median}{3.190543532375}%
\StoreBenchExecResult{PdrInv}{KinductionDfStaticZeroOneTFTrueNotSolvedByKinductionPlainButKipdr}{Correct}{}{Walltime}{Min}{1.78259801865}%
\StoreBenchExecResult{PdrInv}{KinductionDfStaticZeroOneTFTrueNotSolvedByKinductionPlainButKipdr}{Correct}{}{Walltime}{Max}{40.1153149605}%
\StoreBenchExecResult{PdrInv}{KinductionDfStaticZeroOneTFTrueNotSolvedByKinductionPlainButKipdr}{Correct}{}{Walltime}{Stdev}{3.396808986909482431606470797}%
\StoreBenchExecResult{PdrInv}{KinductionDfStaticZeroOneTFTrueNotSolvedByKinductionPlainButKipdr}{Correct}{True}{Count}{}{438}%
\StoreBenchExecResult{PdrInv}{KinductionDfStaticZeroOneTFTrueNotSolvedByKinductionPlainButKipdr}{Correct}{True}{Cputime}{}{3155.515533554}%
\StoreBenchExecResult{PdrInv}{KinductionDfStaticZeroOneTFTrueNotSolvedByKinductionPlainButKipdr}{Correct}{True}{Cputime}{Avg}{7.204373364278538812785388128}%
\StoreBenchExecResult{PdrInv}{KinductionDfStaticZeroOneTFTrueNotSolvedByKinductionPlainButKipdr}{Correct}{True}{Cputime}{Median}{5.986700053}%
\StoreBenchExecResult{PdrInv}{KinductionDfStaticZeroOneTFTrueNotSolvedByKinductionPlainButKipdr}{Correct}{True}{Cputime}{Min}{3.17534903}%
\StoreBenchExecResult{PdrInv}{KinductionDfStaticZeroOneTFTrueNotSolvedByKinductionPlainButKipdr}{Correct}{True}{Cputime}{Max}{79.246301155}%
\StoreBenchExecResult{PdrInv}{KinductionDfStaticZeroOneTFTrueNotSolvedByKinductionPlainButKipdr}{Correct}{True}{Cputime}{Stdev}{6.738581134001495127348074013}%
\StoreBenchExecResult{PdrInv}{KinductionDfStaticZeroOneTFTrueNotSolvedByKinductionPlainButKipdr}{Correct}{True}{Walltime}{}{1663.72902464867}%
\StoreBenchExecResult{PdrInv}{KinductionDfStaticZeroOneTFTrueNotSolvedByKinductionPlainButKipdr}{Correct}{True}{Walltime}{Avg}{3.798468092805182648401826484}%
\StoreBenchExecResult{PdrInv}{KinductionDfStaticZeroOneTFTrueNotSolvedByKinductionPlainButKipdr}{Correct}{True}{Walltime}{Median}{3.190543532375}%
\StoreBenchExecResult{PdrInv}{KinductionDfStaticZeroOneTFTrueNotSolvedByKinductionPlainButKipdr}{Correct}{True}{Walltime}{Min}{1.78259801865}%
\StoreBenchExecResult{PdrInv}{KinductionDfStaticZeroOneTFTrueNotSolvedByKinductionPlainButKipdr}{Correct}{True}{Walltime}{Max}{40.1153149605}%
\StoreBenchExecResult{PdrInv}{KinductionDfStaticZeroOneTFTrueNotSolvedByKinductionPlainButKipdr}{Correct}{True}{Walltime}{Stdev}{3.396808986909482431606470797}%
\StoreBenchExecResult{PdrInv}{KinductionDfStaticZeroOneTFTrueNotSolvedByKinductionPlainButKipdr}{Wrong}{True}{Count}{}{0}%
\StoreBenchExecResult{PdrInv}{KinductionDfStaticZeroOneTFTrueNotSolvedByKinductionPlainButKipdr}{Wrong}{True}{Cputime}{}{0}%
\StoreBenchExecResult{PdrInv}{KinductionDfStaticZeroOneTFTrueNotSolvedByKinductionPlainButKipdr}{Wrong}{True}{Cputime}{Avg}{None}%
\StoreBenchExecResult{PdrInv}{KinductionDfStaticZeroOneTFTrueNotSolvedByKinductionPlainButKipdr}{Wrong}{True}{Cputime}{Median}{None}%
\StoreBenchExecResult{PdrInv}{KinductionDfStaticZeroOneTFTrueNotSolvedByKinductionPlainButKipdr}{Wrong}{True}{Cputime}{Min}{None}%
\StoreBenchExecResult{PdrInv}{KinductionDfStaticZeroOneTFTrueNotSolvedByKinductionPlainButKipdr}{Wrong}{True}{Cputime}{Max}{None}%
\StoreBenchExecResult{PdrInv}{KinductionDfStaticZeroOneTFTrueNotSolvedByKinductionPlainButKipdr}{Wrong}{True}{Cputime}{Stdev}{None}%
\StoreBenchExecResult{PdrInv}{KinductionDfStaticZeroOneTFTrueNotSolvedByKinductionPlainButKipdr}{Wrong}{True}{Walltime}{}{0}%
\StoreBenchExecResult{PdrInv}{KinductionDfStaticZeroOneTFTrueNotSolvedByKinductionPlainButKipdr}{Wrong}{True}{Walltime}{Avg}{None}%
\StoreBenchExecResult{PdrInv}{KinductionDfStaticZeroOneTFTrueNotSolvedByKinductionPlainButKipdr}{Wrong}{True}{Walltime}{Median}{None}%
\StoreBenchExecResult{PdrInv}{KinductionDfStaticZeroOneTFTrueNotSolvedByKinductionPlainButKipdr}{Wrong}{True}{Walltime}{Min}{None}%
\StoreBenchExecResult{PdrInv}{KinductionDfStaticZeroOneTFTrueNotSolvedByKinductionPlainButKipdr}{Wrong}{True}{Walltime}{Max}{None}%
\StoreBenchExecResult{PdrInv}{KinductionDfStaticZeroOneTFTrueNotSolvedByKinductionPlainButKipdr}{Wrong}{True}{Walltime}{Stdev}{None}%
\StoreBenchExecResult{PdrInv}{KinductionDfStaticZeroOneTFTrueNotSolvedByKinductionPlainButKipdr}{Error}{}{Count}{}{11}%
\StoreBenchExecResult{PdrInv}{KinductionDfStaticZeroOneTFTrueNotSolvedByKinductionPlainButKipdr}{Error}{}{Cputime}{}{9914.914081997}%
\StoreBenchExecResult{PdrInv}{KinductionDfStaticZeroOneTFTrueNotSolvedByKinductionPlainButKipdr}{Error}{}{Cputime}{Avg}{901.3558256360909090909090909}%
\StoreBenchExecResult{PdrInv}{KinductionDfStaticZeroOneTFTrueNotSolvedByKinductionPlainButKipdr}{Error}{}{Cputime}{Median}{904.216327561}%
\StoreBenchExecResult{PdrInv}{KinductionDfStaticZeroOneTFTrueNotSolvedByKinductionPlainButKipdr}{Error}{}{Cputime}{Min}{842.997454308}%
\StoreBenchExecResult{PdrInv}{KinductionDfStaticZeroOneTFTrueNotSolvedByKinductionPlainButKipdr}{Error}{}{Cputime}{Max}{913.517748961}%
\StoreBenchExecResult{PdrInv}{KinductionDfStaticZeroOneTFTrueNotSolvedByKinductionPlainButKipdr}{Error}{}{Cputime}{Stdev}{18.80654926573657577719264499}%
\StoreBenchExecResult{PdrInv}{KinductionDfStaticZeroOneTFTrueNotSolvedByKinductionPlainButKipdr}{Error}{}{Walltime}{}{9323.680519819}%
\StoreBenchExecResult{PdrInv}{KinductionDfStaticZeroOneTFTrueNotSolvedByKinductionPlainButKipdr}{Error}{}{Walltime}{Avg}{847.6073199835454545454545455}%
\StoreBenchExecResult{PdrInv}{KinductionDfStaticZeroOneTFTrueNotSolvedByKinductionPlainButKipdr}{Error}{}{Walltime}{Median}{889.688685894}%
\StoreBenchExecResult{PdrInv}{KinductionDfStaticZeroOneTFTrueNotSolvedByKinductionPlainButKipdr}{Error}{}{Walltime}{Min}{453.09119606}%
\StoreBenchExecResult{PdrInv}{KinductionDfStaticZeroOneTFTrueNotSolvedByKinductionPlainButKipdr}{Error}{}{Walltime}{Max}{898.996891975}%
\StoreBenchExecResult{PdrInv}{KinductionDfStaticZeroOneTFTrueNotSolvedByKinductionPlainButKipdr}{Error}{}{Walltime}{Stdev}{126.1286078698833590091922770}%
\StoreBenchExecResult{PdrInv}{KinductionDfStaticZeroOneTFTrueNotSolvedByKinductionPlainButKipdr}{Error}{OutOfMemory}{Count}{}{1}%
\StoreBenchExecResult{PdrInv}{KinductionDfStaticZeroOneTFTrueNotSolvedByKinductionPlainButKipdr}{Error}{OutOfMemory}{Cputime}{}{842.997454308}%
\StoreBenchExecResult{PdrInv}{KinductionDfStaticZeroOneTFTrueNotSolvedByKinductionPlainButKipdr}{Error}{OutOfMemory}{Cputime}{Avg}{842.997454308}%
\StoreBenchExecResult{PdrInv}{KinductionDfStaticZeroOneTFTrueNotSolvedByKinductionPlainButKipdr}{Error}{OutOfMemory}{Cputime}{Median}{842.997454308}%
\StoreBenchExecResult{PdrInv}{KinductionDfStaticZeroOneTFTrueNotSolvedByKinductionPlainButKipdr}{Error}{OutOfMemory}{Cputime}{Min}{842.997454308}%
\StoreBenchExecResult{PdrInv}{KinductionDfStaticZeroOneTFTrueNotSolvedByKinductionPlainButKipdr}{Error}{OutOfMemory}{Cputime}{Max}{842.997454308}%
\StoreBenchExecResult{PdrInv}{KinductionDfStaticZeroOneTFTrueNotSolvedByKinductionPlainButKipdr}{Error}{OutOfMemory}{Cputime}{Stdev}{0E-9}%
\StoreBenchExecResult{PdrInv}{KinductionDfStaticZeroOneTFTrueNotSolvedByKinductionPlainButKipdr}{Error}{OutOfMemory}{Walltime}{}{829.883877993}%
\StoreBenchExecResult{PdrInv}{KinductionDfStaticZeroOneTFTrueNotSolvedByKinductionPlainButKipdr}{Error}{OutOfMemory}{Walltime}{Avg}{829.883877993}%
\StoreBenchExecResult{PdrInv}{KinductionDfStaticZeroOneTFTrueNotSolvedByKinductionPlainButKipdr}{Error}{OutOfMemory}{Walltime}{Median}{829.883877993}%
\StoreBenchExecResult{PdrInv}{KinductionDfStaticZeroOneTFTrueNotSolvedByKinductionPlainButKipdr}{Error}{OutOfMemory}{Walltime}{Min}{829.883877993}%
\StoreBenchExecResult{PdrInv}{KinductionDfStaticZeroOneTFTrueNotSolvedByKinductionPlainButKipdr}{Error}{OutOfMemory}{Walltime}{Max}{829.883877993}%
\StoreBenchExecResult{PdrInv}{KinductionDfStaticZeroOneTFTrueNotSolvedByKinductionPlainButKipdr}{Error}{OutOfMemory}{Walltime}{Stdev}{0E-9}%
\StoreBenchExecResult{PdrInv}{KinductionDfStaticZeroOneTFTrueNotSolvedByKinductionPlainButKipdr}{Error}{Timeout}{Count}{}{10}%
\StoreBenchExecResult{PdrInv}{KinductionDfStaticZeroOneTFTrueNotSolvedByKinductionPlainButKipdr}{Error}{Timeout}{Cputime}{}{9071.916627689}%
\StoreBenchExecResult{PdrInv}{KinductionDfStaticZeroOneTFTrueNotSolvedByKinductionPlainButKipdr}{Error}{Timeout}{Cputime}{Avg}{907.1916627689}%
\StoreBenchExecResult{PdrInv}{KinductionDfStaticZeroOneTFTrueNotSolvedByKinductionPlainButKipdr}{Error}{Timeout}{Cputime}{Median}{906.221807512}%
\StoreBenchExecResult{PdrInv}{KinductionDfStaticZeroOneTFTrueNotSolvedByKinductionPlainButKipdr}{Error}{Timeout}{Cputime}{Min}{903.193587886}%
\StoreBenchExecResult{PdrInv}{KinductionDfStaticZeroOneTFTrueNotSolvedByKinductionPlainButKipdr}{Error}{Timeout}{Cputime}{Max}{913.517748961}%
\StoreBenchExecResult{PdrInv}{KinductionDfStaticZeroOneTFTrueNotSolvedByKinductionPlainButKipdr}{Error}{Timeout}{Cputime}{Stdev}{3.798418008280143281930691571}%
\StoreBenchExecResult{PdrInv}{KinductionDfStaticZeroOneTFTrueNotSolvedByKinductionPlainButKipdr}{Error}{Timeout}{Walltime}{}{8493.796641826}%
\StoreBenchExecResult{PdrInv}{KinductionDfStaticZeroOneTFTrueNotSolvedByKinductionPlainButKipdr}{Error}{Timeout}{Walltime}{Avg}{849.3796641826}%
\StoreBenchExecResult{PdrInv}{KinductionDfStaticZeroOneTFTrueNotSolvedByKinductionPlainButKipdr}{Error}{Timeout}{Walltime}{Median}{891.869151950}%
\StoreBenchExecResult{PdrInv}{KinductionDfStaticZeroOneTFTrueNotSolvedByKinductionPlainButKipdr}{Error}{Timeout}{Walltime}{Min}{453.09119606}%
\StoreBenchExecResult{PdrInv}{KinductionDfStaticZeroOneTFTrueNotSolvedByKinductionPlainButKipdr}{Error}{Timeout}{Walltime}{Max}{898.996891975}%
\StoreBenchExecResult{PdrInv}{KinductionDfStaticZeroOneTFTrueNotSolvedByKinductionPlainButKipdr}{Error}{Timeout}{Walltime}{Stdev}{132.1541336922728927666525645}%
\providecommand\StoreBenchExecResult[7]{\expandafter\newcommand\csname#1#2#3#4#5#6\endcsname{#7}}%
\StoreBenchExecResult{PdrInv}{KinductionDfStaticZeroOneTFTrueNotSolvedByKinductionPlain}{Total}{}{Count}{}{2893}%
\StoreBenchExecResult{PdrInv}{KinductionDfStaticZeroOneTFTrueNotSolvedByKinductionPlain}{Total}{}{Cputime}{}{1628197.578319046}%
\StoreBenchExecResult{PdrInv}{KinductionDfStaticZeroOneTFTrueNotSolvedByKinductionPlain}{Total}{}{Cputime}{Avg}{562.8059378911323885240235050}%
\StoreBenchExecResult{PdrInv}{KinductionDfStaticZeroOneTFTrueNotSolvedByKinductionPlain}{Total}{}{Cputime}{Median}{901.139796496}%
\StoreBenchExecResult{PdrInv}{KinductionDfStaticZeroOneTFTrueNotSolvedByKinductionPlain}{Total}{}{Cputime}{Min}{2.941055393}%
\StoreBenchExecResult{PdrInv}{KinductionDfStaticZeroOneTFTrueNotSolvedByKinductionPlain}{Total}{}{Cputime}{Max}{1002.39740868}%
\StoreBenchExecResult{PdrInv}{KinductionDfStaticZeroOneTFTrueNotSolvedByKinductionPlain}{Total}{}{Cputime}{Stdev}{417.7577257857241917132527089}%
\StoreBenchExecResult{PdrInv}{KinductionDfStaticZeroOneTFTrueNotSolvedByKinductionPlain}{Total}{}{Walltime}{}{1124192.76795032096}%
\StoreBenchExecResult{PdrInv}{KinductionDfStaticZeroOneTFTrueNotSolvedByKinductionPlain}{Total}{}{Walltime}{Avg}{388.5906560491949395091600415}%
\StoreBenchExecResult{PdrInv}{KinductionDfStaticZeroOneTFTrueNotSolvedByKinductionPlain}{Total}{}{Walltime}{Median}{451.787154913}%
\StoreBenchExecResult{PdrInv}{KinductionDfStaticZeroOneTFTrueNotSolvedByKinductionPlain}{Total}{}{Walltime}{Min}{1.593998909}%
\StoreBenchExecResult{PdrInv}{KinductionDfStaticZeroOneTFTrueNotSolvedByKinductionPlain}{Total}{}{Walltime}{Max}{907.579365015}%
\StoreBenchExecResult{PdrInv}{KinductionDfStaticZeroOneTFTrueNotSolvedByKinductionPlain}{Total}{}{Walltime}{Stdev}{327.9400868145379962323306518}%
\StoreBenchExecResult{PdrInv}{KinductionDfStaticZeroOneTFTrueNotSolvedByKinductionPlain}{Correct}{}{Count}{}{946}%
\StoreBenchExecResult{PdrInv}{KinductionDfStaticZeroOneTFTrueNotSolvedByKinductionPlain}{Correct}{}{Cputime}{}{38877.789070207}%
\StoreBenchExecResult{PdrInv}{KinductionDfStaticZeroOneTFTrueNotSolvedByKinductionPlain}{Correct}{}{Cputime}{Avg}{41.09702861544080338266384778}%
\StoreBenchExecResult{PdrInv}{KinductionDfStaticZeroOneTFTrueNotSolvedByKinductionPlain}{Correct}{}{Cputime}{Median}{7.098176964}%
\StoreBenchExecResult{PdrInv}{KinductionDfStaticZeroOneTFTrueNotSolvedByKinductionPlain}{Correct}{}{Cputime}{Min}{3.165618484}%
\StoreBenchExecResult{PdrInv}{KinductionDfStaticZeroOneTFTrueNotSolvedByKinductionPlain}{Correct}{}{Cputime}{Max}{865.106067237}%
\StoreBenchExecResult{PdrInv}{KinductionDfStaticZeroOneTFTrueNotSolvedByKinductionPlain}{Correct}{}{Cputime}{Stdev}{112.5834349479032244276134650}%
\StoreBenchExecResult{PdrInv}{KinductionDfStaticZeroOneTFTrueNotSolvedByKinductionPlain}{Correct}{}{Walltime}{}{21252.22661876753}%
\StoreBenchExecResult{PdrInv}{KinductionDfStaticZeroOneTFTrueNotSolvedByKinductionPlain}{Correct}{}{Walltime}{Avg}{22.46535583379231501057082452}%
\StoreBenchExecResult{PdrInv}{KinductionDfStaticZeroOneTFTrueNotSolvedByKinductionPlain}{Correct}{}{Walltime}{Median}{3.76447558403}%
\StoreBenchExecResult{PdrInv}{KinductionDfStaticZeroOneTFTrueNotSolvedByKinductionPlain}{Correct}{}{Walltime}{Min}{1.76165986061}%
\StoreBenchExecResult{PdrInv}{KinductionDfStaticZeroOneTFTrueNotSolvedByKinductionPlain}{Correct}{}{Walltime}{Max}{739.456933022}%
\StoreBenchExecResult{PdrInv}{KinductionDfStaticZeroOneTFTrueNotSolvedByKinductionPlain}{Correct}{}{Walltime}{Stdev}{67.22690992902681272097840632}%
\StoreBenchExecResult{PdrInv}{KinductionDfStaticZeroOneTFTrueNotSolvedByKinductionPlain}{Correct}{True}{Count}{}{946}%
\StoreBenchExecResult{PdrInv}{KinductionDfStaticZeroOneTFTrueNotSolvedByKinductionPlain}{Correct}{True}{Cputime}{}{38877.789070207}%
\StoreBenchExecResult{PdrInv}{KinductionDfStaticZeroOneTFTrueNotSolvedByKinductionPlain}{Correct}{True}{Cputime}{Avg}{41.09702861544080338266384778}%
\StoreBenchExecResult{PdrInv}{KinductionDfStaticZeroOneTFTrueNotSolvedByKinductionPlain}{Correct}{True}{Cputime}{Median}{7.098176964}%
\StoreBenchExecResult{PdrInv}{KinductionDfStaticZeroOneTFTrueNotSolvedByKinductionPlain}{Correct}{True}{Cputime}{Min}{3.165618484}%
\StoreBenchExecResult{PdrInv}{KinductionDfStaticZeroOneTFTrueNotSolvedByKinductionPlain}{Correct}{True}{Cputime}{Max}{865.106067237}%
\StoreBenchExecResult{PdrInv}{KinductionDfStaticZeroOneTFTrueNotSolvedByKinductionPlain}{Correct}{True}{Cputime}{Stdev}{112.5834349479032244276134650}%
\StoreBenchExecResult{PdrInv}{KinductionDfStaticZeroOneTFTrueNotSolvedByKinductionPlain}{Correct}{True}{Walltime}{}{21252.22661876753}%
\StoreBenchExecResult{PdrInv}{KinductionDfStaticZeroOneTFTrueNotSolvedByKinductionPlain}{Correct}{True}{Walltime}{Avg}{22.46535583379231501057082452}%
\StoreBenchExecResult{PdrInv}{KinductionDfStaticZeroOneTFTrueNotSolvedByKinductionPlain}{Correct}{True}{Walltime}{Median}{3.76447558403}%
\StoreBenchExecResult{PdrInv}{KinductionDfStaticZeroOneTFTrueNotSolvedByKinductionPlain}{Correct}{True}{Walltime}{Min}{1.76165986061}%
\StoreBenchExecResult{PdrInv}{KinductionDfStaticZeroOneTFTrueNotSolvedByKinductionPlain}{Correct}{True}{Walltime}{Max}{739.456933022}%
\StoreBenchExecResult{PdrInv}{KinductionDfStaticZeroOneTFTrueNotSolvedByKinductionPlain}{Correct}{True}{Walltime}{Stdev}{67.22690992902681272097840632}%
\StoreBenchExecResult{PdrInv}{KinductionDfStaticZeroOneTFTrueNotSolvedByKinductionPlain}{Wrong}{True}{Count}{}{0}%
\StoreBenchExecResult{PdrInv}{KinductionDfStaticZeroOneTFTrueNotSolvedByKinductionPlain}{Wrong}{True}{Cputime}{}{0}%
\StoreBenchExecResult{PdrInv}{KinductionDfStaticZeroOneTFTrueNotSolvedByKinductionPlain}{Wrong}{True}{Cputime}{Avg}{None}%
\StoreBenchExecResult{PdrInv}{KinductionDfStaticZeroOneTFTrueNotSolvedByKinductionPlain}{Wrong}{True}{Cputime}{Median}{None}%
\StoreBenchExecResult{PdrInv}{KinductionDfStaticZeroOneTFTrueNotSolvedByKinductionPlain}{Wrong}{True}{Cputime}{Min}{None}%
\StoreBenchExecResult{PdrInv}{KinductionDfStaticZeroOneTFTrueNotSolvedByKinductionPlain}{Wrong}{True}{Cputime}{Max}{None}%
\StoreBenchExecResult{PdrInv}{KinductionDfStaticZeroOneTFTrueNotSolvedByKinductionPlain}{Wrong}{True}{Cputime}{Stdev}{None}%
\StoreBenchExecResult{PdrInv}{KinductionDfStaticZeroOneTFTrueNotSolvedByKinductionPlain}{Wrong}{True}{Walltime}{}{0}%
\StoreBenchExecResult{PdrInv}{KinductionDfStaticZeroOneTFTrueNotSolvedByKinductionPlain}{Wrong}{True}{Walltime}{Avg}{None}%
\StoreBenchExecResult{PdrInv}{KinductionDfStaticZeroOneTFTrueNotSolvedByKinductionPlain}{Wrong}{True}{Walltime}{Median}{None}%
\StoreBenchExecResult{PdrInv}{KinductionDfStaticZeroOneTFTrueNotSolvedByKinductionPlain}{Wrong}{True}{Walltime}{Min}{None}%
\StoreBenchExecResult{PdrInv}{KinductionDfStaticZeroOneTFTrueNotSolvedByKinductionPlain}{Wrong}{True}{Walltime}{Max}{None}%
\StoreBenchExecResult{PdrInv}{KinductionDfStaticZeroOneTFTrueNotSolvedByKinductionPlain}{Wrong}{True}{Walltime}{Stdev}{None}%
\StoreBenchExecResult{PdrInv}{KinductionDfStaticZeroOneTFTrueNotSolvedByKinductionPlain}{Error}{}{Count}{}{1947}%
\StoreBenchExecResult{PdrInv}{KinductionDfStaticZeroOneTFTrueNotSolvedByKinductionPlain}{Error}{}{Cputime}{}{1589319.789248839}%
\StoreBenchExecResult{PdrInv}{KinductionDfStaticZeroOneTFTrueNotSolvedByKinductionPlain}{Error}{}{Cputime}{Avg}{816.2916226239542886492039034}%
\StoreBenchExecResult{PdrInv}{KinductionDfStaticZeroOneTFTrueNotSolvedByKinductionPlain}{Error}{}{Cputime}{Median}{901.751534722}%
\StoreBenchExecResult{PdrInv}{KinductionDfStaticZeroOneTFTrueNotSolvedByKinductionPlain}{Error}{}{Cputime}{Min}{2.941055393}%
\StoreBenchExecResult{PdrInv}{KinductionDfStaticZeroOneTFTrueNotSolvedByKinductionPlain}{Error}{}{Cputime}{Max}{1002.39740868}%
\StoreBenchExecResult{PdrInv}{KinductionDfStaticZeroOneTFTrueNotSolvedByKinductionPlain}{Error}{}{Cputime}{Stdev}{238.0295390604570700044635638}%
\StoreBenchExecResult{PdrInv}{KinductionDfStaticZeroOneTFTrueNotSolvedByKinductionPlain}{Error}{}{Walltime}{}{1102940.54133155343}%
\StoreBenchExecResult{PdrInv}{KinductionDfStaticZeroOneTFTrueNotSolvedByKinductionPlain}{Error}{}{Walltime}{Avg}{566.4820448544188135593220339}%
\StoreBenchExecResult{PdrInv}{KinductionDfStaticZeroOneTFTrueNotSolvedByKinductionPlain}{Error}{}{Walltime}{Median}{454.666778088}%
\StoreBenchExecResult{PdrInv}{KinductionDfStaticZeroOneTFTrueNotSolvedByKinductionPlain}{Error}{}{Walltime}{Min}{1.593998909}%
\StoreBenchExecResult{PdrInv}{KinductionDfStaticZeroOneTFTrueNotSolvedByKinductionPlain}{Error}{}{Walltime}{Max}{907.579365015}%
\StoreBenchExecResult{PdrInv}{KinductionDfStaticZeroOneTFTrueNotSolvedByKinductionPlain}{Error}{}{Walltime}{Stdev}{246.6298419825546888488779948}%
\StoreBenchExecResult{PdrInv}{KinductionDfStaticZeroOneTFTrueNotSolvedByKinductionPlain}{Error}{Assertion}{Count}{}{2}%
\StoreBenchExecResult{PdrInv}{KinductionDfStaticZeroOneTFTrueNotSolvedByKinductionPlain}{Error}{Assertion}{Cputime}{}{7.153166457}%
\StoreBenchExecResult{PdrInv}{KinductionDfStaticZeroOneTFTrueNotSolvedByKinductionPlain}{Error}{Assertion}{Cputime}{Avg}{3.5765832285}%
\StoreBenchExecResult{PdrInv}{KinductionDfStaticZeroOneTFTrueNotSolvedByKinductionPlain}{Error}{Assertion}{Cputime}{Median}{3.5765832285}%
\StoreBenchExecResult{PdrInv}{KinductionDfStaticZeroOneTFTrueNotSolvedByKinductionPlain}{Error}{Assertion}{Cputime}{Min}{3.14358981}%
\StoreBenchExecResult{PdrInv}{KinductionDfStaticZeroOneTFTrueNotSolvedByKinductionPlain}{Error}{Assertion}{Cputime}{Max}{4.009576647}%
\StoreBenchExecResult{PdrInv}{KinductionDfStaticZeroOneTFTrueNotSolvedByKinductionPlain}{Error}{Assertion}{Cputime}{Stdev}{0.4329934185}%
\StoreBenchExecResult{PdrInv}{KinductionDfStaticZeroOneTFTrueNotSolvedByKinductionPlain}{Error}{Assertion}{Walltime}{}{3.99050474167}%
\StoreBenchExecResult{PdrInv}{KinductionDfStaticZeroOneTFTrueNotSolvedByKinductionPlain}{Error}{Assertion}{Walltime}{Avg}{1.995252370835}%
\StoreBenchExecResult{PdrInv}{KinductionDfStaticZeroOneTFTrueNotSolvedByKinductionPlain}{Error}{Assertion}{Walltime}{Median}{1.995252370835}%
\StoreBenchExecResult{PdrInv}{KinductionDfStaticZeroOneTFTrueNotSolvedByKinductionPlain}{Error}{Assertion}{Walltime}{Min}{1.7533288002}%
\StoreBenchExecResult{PdrInv}{KinductionDfStaticZeroOneTFTrueNotSolvedByKinductionPlain}{Error}{Assertion}{Walltime}{Max}{2.23717594147}%
\StoreBenchExecResult{PdrInv}{KinductionDfStaticZeroOneTFTrueNotSolvedByKinductionPlain}{Error}{Assertion}{Walltime}{Stdev}{0.241923570635}%
\StoreBenchExecResult{PdrInv}{KinductionDfStaticZeroOneTFTrueNotSolvedByKinductionPlain}{Error}{Error}{Count}{}{130}%
\StoreBenchExecResult{PdrInv}{KinductionDfStaticZeroOneTFTrueNotSolvedByKinductionPlain}{Error}{Error}{Cputime}{}{22276.789412242}%
\StoreBenchExecResult{PdrInv}{KinductionDfStaticZeroOneTFTrueNotSolvedByKinductionPlain}{Error}{Error}{Cputime}{Avg}{171.3599185557076923076923077}%
\StoreBenchExecResult{PdrInv}{KinductionDfStaticZeroOneTFTrueNotSolvedByKinductionPlain}{Error}{Error}{Cputime}{Median}{107.552999978}%
\StoreBenchExecResult{PdrInv}{KinductionDfStaticZeroOneTFTrueNotSolvedByKinductionPlain}{Error}{Error}{Cputime}{Min}{2.941055393}%
\StoreBenchExecResult{PdrInv}{KinductionDfStaticZeroOneTFTrueNotSolvedByKinductionPlain}{Error}{Error}{Cputime}{Max}{796.417582796}%
\StoreBenchExecResult{PdrInv}{KinductionDfStaticZeroOneTFTrueNotSolvedByKinductionPlain}{Error}{Error}{Cputime}{Stdev}{193.2905794854175251220384764}%
\StoreBenchExecResult{PdrInv}{KinductionDfStaticZeroOneTFTrueNotSolvedByKinductionPlain}{Error}{Error}{Walltime}{}{18778.88551354895}%
\StoreBenchExecResult{PdrInv}{KinductionDfStaticZeroOneTFTrueNotSolvedByKinductionPlain}{Error}{Error}{Walltime}{Avg}{144.4529654888380769230769231}%
\StoreBenchExecResult{PdrInv}{KinductionDfStaticZeroOneTFTrueNotSolvedByKinductionPlain}{Error}{Error}{Walltime}{Median}{77.89796197415}%
\StoreBenchExecResult{PdrInv}{KinductionDfStaticZeroOneTFTrueNotSolvedByKinductionPlain}{Error}{Error}{Walltime}{Min}{1.593998909}%
\StoreBenchExecResult{PdrInv}{KinductionDfStaticZeroOneTFTrueNotSolvedByKinductionPlain}{Error}{Error}{Walltime}{Max}{782.34778595}%
\StoreBenchExecResult{PdrInv}{KinductionDfStaticZeroOneTFTrueNotSolvedByKinductionPlain}{Error}{Error}{Walltime}{Stdev}{173.0318135813569634775581675}%
\StoreBenchExecResult{PdrInv}{KinductionDfStaticZeroOneTFTrueNotSolvedByKinductionPlain}{Error}{Exception}{Count}{}{7}%
\StoreBenchExecResult{PdrInv}{KinductionDfStaticZeroOneTFTrueNotSolvedByKinductionPlain}{Error}{Exception}{Cputime}{}{799.296850342}%
\StoreBenchExecResult{PdrInv}{KinductionDfStaticZeroOneTFTrueNotSolvedByKinductionPlain}{Error}{Exception}{Cputime}{Avg}{114.1852643345714285714285714}%
\StoreBenchExecResult{PdrInv}{KinductionDfStaticZeroOneTFTrueNotSolvedByKinductionPlain}{Error}{Exception}{Cputime}{Median}{90.927229316}%
\StoreBenchExecResult{PdrInv}{KinductionDfStaticZeroOneTFTrueNotSolvedByKinductionPlain}{Error}{Exception}{Cputime}{Min}{14.848690343}%
\StoreBenchExecResult{PdrInv}{KinductionDfStaticZeroOneTFTrueNotSolvedByKinductionPlain}{Error}{Exception}{Cputime}{Max}{240.247048821}%
\StoreBenchExecResult{PdrInv}{KinductionDfStaticZeroOneTFTrueNotSolvedByKinductionPlain}{Error}{Exception}{Cputime}{Stdev}{83.61990205654049584091144832}%
\StoreBenchExecResult{PdrInv}{KinductionDfStaticZeroOneTFTrueNotSolvedByKinductionPlain}{Error}{Exception}{Walltime}{}{417.48527002291}%
\StoreBenchExecResult{PdrInv}{KinductionDfStaticZeroOneTFTrueNotSolvedByKinductionPlain}{Error}{Exception}{Walltime}{Avg}{59.64075286041571428571428571}%
\StoreBenchExecResult{PdrInv}{KinductionDfStaticZeroOneTFTrueNotSolvedByKinductionPlain}{Error}{Exception}{Walltime}{Median}{45.7856659889}%
\StoreBenchExecResult{PdrInv}{KinductionDfStaticZeroOneTFTrueNotSolvedByKinductionPlain}{Error}{Exception}{Walltime}{Min}{7.60693311691}%
\StoreBenchExecResult{PdrInv}{KinductionDfStaticZeroOneTFTrueNotSolvedByKinductionPlain}{Error}{Exception}{Walltime}{Max}{120.636415958}%
\StoreBenchExecResult{PdrInv}{KinductionDfStaticZeroOneTFTrueNotSolvedByKinductionPlain}{Error}{Exception}{Walltime}{Stdev}{42.21174398039302091782027527}%
\StoreBenchExecResult{PdrInv}{KinductionDfStaticZeroOneTFTrueNotSolvedByKinductionPlain}{Error}{OutOfJavaMemory}{Count}{}{5}%
\StoreBenchExecResult{PdrInv}{KinductionDfStaticZeroOneTFTrueNotSolvedByKinductionPlain}{Error}{OutOfJavaMemory}{Cputime}{}{2114.729289184}%
\StoreBenchExecResult{PdrInv}{KinductionDfStaticZeroOneTFTrueNotSolvedByKinductionPlain}{Error}{OutOfJavaMemory}{Cputime}{Avg}{422.9458578368}%
\StoreBenchExecResult{PdrInv}{KinductionDfStaticZeroOneTFTrueNotSolvedByKinductionPlain}{Error}{OutOfJavaMemory}{Cputime}{Median}{334.648636291}%
\StoreBenchExecResult{PdrInv}{KinductionDfStaticZeroOneTFTrueNotSolvedByKinductionPlain}{Error}{OutOfJavaMemory}{Cputime}{Min}{187.515973605}%
\StoreBenchExecResult{PdrInv}{KinductionDfStaticZeroOneTFTrueNotSolvedByKinductionPlain}{Error}{OutOfJavaMemory}{Cputime}{Max}{858.725358526}%
\StoreBenchExecResult{PdrInv}{KinductionDfStaticZeroOneTFTrueNotSolvedByKinductionPlain}{Error}{OutOfJavaMemory}{Cputime}{Stdev}{236.9527661226046092026886696}%
\StoreBenchExecResult{PdrInv}{KinductionDfStaticZeroOneTFTrueNotSolvedByKinductionPlain}{Error}{OutOfJavaMemory}{Walltime}{}{1198.533137322}%
\StoreBenchExecResult{PdrInv}{KinductionDfStaticZeroOneTFTrueNotSolvedByKinductionPlain}{Error}{OutOfJavaMemory}{Walltime}{Avg}{239.7066274644}%
\StoreBenchExecResult{PdrInv}{KinductionDfStaticZeroOneTFTrueNotSolvedByKinductionPlain}{Error}{OutOfJavaMemory}{Walltime}{Median}{217.030892134}%
\StoreBenchExecResult{PdrInv}{KinductionDfStaticZeroOneTFTrueNotSolvedByKinductionPlain}{Error}{OutOfJavaMemory}{Walltime}{Min}{115.741152048}%
\StoreBenchExecResult{PdrInv}{KinductionDfStaticZeroOneTFTrueNotSolvedByKinductionPlain}{Error}{OutOfJavaMemory}{Walltime}{Max}{441.640638113}%
\StoreBenchExecResult{PdrInv}{KinductionDfStaticZeroOneTFTrueNotSolvedByKinductionPlain}{Error}{OutOfJavaMemory}{Walltime}{Stdev}{109.9539791830988033118249783}%
\StoreBenchExecResult{PdrInv}{KinductionDfStaticZeroOneTFTrueNotSolvedByKinductionPlain}{Error}{OutOfMemory}{Count}{}{132}%
\StoreBenchExecResult{PdrInv}{KinductionDfStaticZeroOneTFTrueNotSolvedByKinductionPlain}{Error}{OutOfMemory}{Cputime}{}{49770.766525138}%
\StoreBenchExecResult{PdrInv}{KinductionDfStaticZeroOneTFTrueNotSolvedByKinductionPlain}{Error}{OutOfMemory}{Cputime}{Avg}{377.0512615540757575757575758}%
\StoreBenchExecResult{PdrInv}{KinductionDfStaticZeroOneTFTrueNotSolvedByKinductionPlain}{Error}{OutOfMemory}{Cputime}{Median}{286.1648015945}%
\StoreBenchExecResult{PdrInv}{KinductionDfStaticZeroOneTFTrueNotSolvedByKinductionPlain}{Error}{OutOfMemory}{Cputime}{Min}{172.41826193}%
\StoreBenchExecResult{PdrInv}{KinductionDfStaticZeroOneTFTrueNotSolvedByKinductionPlain}{Error}{OutOfMemory}{Cputime}{Max}{882.926155716}%
\StoreBenchExecResult{PdrInv}{KinductionDfStaticZeroOneTFTrueNotSolvedByKinductionPlain}{Error}{OutOfMemory}{Cputime}{Stdev}{215.1154238749574972828409710}%
\StoreBenchExecResult{PdrInv}{KinductionDfStaticZeroOneTFTrueNotSolvedByKinductionPlain}{Error}{OutOfMemory}{Walltime}{}{42044.5961866379}%
\StoreBenchExecResult{PdrInv}{KinductionDfStaticZeroOneTFTrueNotSolvedByKinductionPlain}{Error}{OutOfMemory}{Walltime}{Avg}{318.5196680805901515151515152}%
\StoreBenchExecResult{PdrInv}{KinductionDfStaticZeroOneTFTrueNotSolvedByKinductionPlain}{Error}{OutOfMemory}{Walltime}{Median}{206.4880429505}%
\StoreBenchExecResult{PdrInv}{KinductionDfStaticZeroOneTFTrueNotSolvedByKinductionPlain}{Error}{OutOfMemory}{Walltime}{Min}{86.7566621304}%
\StoreBenchExecResult{PdrInv}{KinductionDfStaticZeroOneTFTrueNotSolvedByKinductionPlain}{Error}{OutOfMemory}{Walltime}{Max}{869.022175074}%
\StoreBenchExecResult{PdrInv}{KinductionDfStaticZeroOneTFTrueNotSolvedByKinductionPlain}{Error}{OutOfMemory}{Walltime}{Stdev}{240.6599652981393104476546955}%
\StoreBenchExecResult{PdrInv}{KinductionDfStaticZeroOneTFTrueNotSolvedByKinductionPlain}{Error}{Timeout}{Count}{}{1671}%
\StoreBenchExecResult{PdrInv}{KinductionDfStaticZeroOneTFTrueNotSolvedByKinductionPlain}{Error}{Timeout}{Cputime}{}{1514351.054005476}%
\StoreBenchExecResult{PdrInv}{KinductionDfStaticZeroOneTFTrueNotSolvedByKinductionPlain}{Error}{Timeout}{Cputime}{Avg}{906.2543710385852782764811490}%
\StoreBenchExecResult{PdrInv}{KinductionDfStaticZeroOneTFTrueNotSolvedByKinductionPlain}{Error}{Timeout}{Cputime}{Median}{902.089113485}%
\StoreBenchExecResult{PdrInv}{KinductionDfStaticZeroOneTFTrueNotSolvedByKinductionPlain}{Error}{Timeout}{Cputime}{Min}{900.818633269}%
\StoreBenchExecResult{PdrInv}{KinductionDfStaticZeroOneTFTrueNotSolvedByKinductionPlain}{Error}{Timeout}{Cputime}{Max}{1002.39740868}%
\StoreBenchExecResult{PdrInv}{KinductionDfStaticZeroOneTFTrueNotSolvedByKinductionPlain}{Error}{Timeout}{Cputime}{Stdev}{15.67033529895932739333547400}%
\StoreBenchExecResult{PdrInv}{KinductionDfStaticZeroOneTFTrueNotSolvedByKinductionPlain}{Error}{Timeout}{Walltime}{}{1040497.050719280}%
\StoreBenchExecResult{PdrInv}{KinductionDfStaticZeroOneTFTrueNotSolvedByKinductionPlain}{Error}{Timeout}{Walltime}{Avg}{622.6792643442728904847396768}%
\StoreBenchExecResult{PdrInv}{KinductionDfStaticZeroOneTFTrueNotSolvedByKinductionPlain}{Error}{Timeout}{Walltime}{Median}{458.536730051}%
\StoreBenchExecResult{PdrInv}{KinductionDfStaticZeroOneTFTrueNotSolvedByKinductionPlain}{Error}{Timeout}{Walltime}{Min}{451.09356904}%
\StoreBenchExecResult{PdrInv}{KinductionDfStaticZeroOneTFTrueNotSolvedByKinductionPlain}{Error}{Timeout}{Walltime}{Max}{907.579365015}%
\StoreBenchExecResult{PdrInv}{KinductionDfStaticZeroOneTFTrueNotSolvedByKinductionPlain}{Error}{Timeout}{Walltime}{Stdev}{200.6896308288325360304017845}%
\providecommand\StoreBenchExecResult[7]{\expandafter\newcommand\csname#1#2#3#4#5#6\endcsname{#7}}%
\StoreBenchExecResult{PdrInv}{KinductionDfStaticZeroOneTF}{Total}{}{Count}{}{5591}%
\StoreBenchExecResult{PdrInv}{KinductionDfStaticZeroOneTF}{Total}{}{Cputime}{}{2213866.663598018}%
\StoreBenchExecResult{PdrInv}{KinductionDfStaticZeroOneTF}{Total}{}{Cputime}{Avg}{395.9697126807401180468610266}%
\StoreBenchExecResult{PdrInv}{KinductionDfStaticZeroOneTF}{Total}{}{Cputime}{Median}{116.891387304}%
\StoreBenchExecResult{PdrInv}{KinductionDfStaticZeroOneTF}{Total}{}{Cputime}{Min}{2.941055393}%
\StoreBenchExecResult{PdrInv}{KinductionDfStaticZeroOneTF}{Total}{}{Cputime}{Max}{1002.39740868}%
\StoreBenchExecResult{PdrInv}{KinductionDfStaticZeroOneTF}{Total}{}{Cputime}{Stdev}{419.6692471857884121575994785}%
\StoreBenchExecResult{PdrInv}{KinductionDfStaticZeroOneTF}{Total}{}{Walltime}{}{1558322.87050130259}%
\StoreBenchExecResult{PdrInv}{KinductionDfStaticZeroOneTF}{Total}{}{Walltime}{Avg}{278.7198838313901967447683777}%
\StoreBenchExecResult{PdrInv}{KinductionDfStaticZeroOneTF}{Total}{}{Walltime}{Median}{74.8368251324}%
\StoreBenchExecResult{PdrInv}{KinductionDfStaticZeroOneTF}{Total}{}{Walltime}{Min}{1.593998909}%
\StoreBenchExecResult{PdrInv}{KinductionDfStaticZeroOneTF}{Total}{}{Walltime}{Max}{939.42276907}%
\StoreBenchExecResult{PdrInv}{KinductionDfStaticZeroOneTF}{Total}{}{Walltime}{Stdev}{321.5800669785694594870455930}%
\StoreBenchExecResult{PdrInv}{KinductionDfStaticZeroOneTF}{Correct}{}{Count}{}{2994}%
\StoreBenchExecResult{PdrInv}{KinductionDfStaticZeroOneTF}{Correct}{}{Cputime}{}{164570.426509634}%
\StoreBenchExecResult{PdrInv}{KinductionDfStaticZeroOneTF}{Correct}{}{Cputime}{Avg}{54.96674232118704074816299265}%
\StoreBenchExecResult{PdrInv}{KinductionDfStaticZeroOneTF}{Correct}{}{Cputime}{Median}{9.2760015205}%
\StoreBenchExecResult{PdrInv}{KinductionDfStaticZeroOneTF}{Correct}{}{Cputime}{Min}{3.002322542}%
\StoreBenchExecResult{PdrInv}{KinductionDfStaticZeroOneTF}{Correct}{}{Cputime}{Max}{890.658340113}%
\StoreBenchExecResult{PdrInv}{KinductionDfStaticZeroOneTF}{Correct}{}{Cputime}{Stdev}{133.0861807685385779472764545}%
\StoreBenchExecResult{PdrInv}{KinductionDfStaticZeroOneTF}{Correct}{}{Walltime}{}{115701.21230124868}%
\StoreBenchExecResult{PdrInv}{KinductionDfStaticZeroOneTF}{Correct}{}{Walltime}{Avg}{38.64435948605500334001336005}%
\StoreBenchExecResult{PdrInv}{KinductionDfStaticZeroOneTF}{Correct}{}{Walltime}{Median}{4.93507993221}%
\StoreBenchExecResult{PdrInv}{KinductionDfStaticZeroOneTF}{Correct}{}{Walltime}{Min}{1.66395783424}%
\StoreBenchExecResult{PdrInv}{KinductionDfStaticZeroOneTF}{Correct}{}{Walltime}{Max}{881.880240917}%
\StoreBenchExecResult{PdrInv}{KinductionDfStaticZeroOneTF}{Correct}{}{Walltime}{Stdev}{109.6015237047997842733887331}%
\StoreBenchExecResult{PdrInv}{KinductionDfStaticZeroOneTF}{Correct}{False}{Count}{}{817}%
\StoreBenchExecResult{PdrInv}{KinductionDfStaticZeroOneTF}{Correct}{False}{Cputime}{}{73242.803833919}%
\StoreBenchExecResult{PdrInv}{KinductionDfStaticZeroOneTF}{Correct}{False}{Cputime}{Avg}{89.64847470491921664626682987}%
\StoreBenchExecResult{PdrInv}{KinductionDfStaticZeroOneTF}{Correct}{False}{Cputime}{Median}{21.228473147}%
\StoreBenchExecResult{PdrInv}{KinductionDfStaticZeroOneTF}{Correct}{False}{Cputime}{Min}{3.142797848}%
\StoreBenchExecResult{PdrInv}{KinductionDfStaticZeroOneTF}{Correct}{False}{Cputime}{Max}{888.785830086}%
\StoreBenchExecResult{PdrInv}{KinductionDfStaticZeroOneTF}{Correct}{False}{Cputime}{Stdev}{177.7987690288642875357918023}%
\StoreBenchExecResult{PdrInv}{KinductionDfStaticZeroOneTF}{Correct}{False}{Walltime}{}{58386.17158960894}%
\StoreBenchExecResult{PdrInv}{KinductionDfStaticZeroOneTF}{Correct}{False}{Walltime}{Avg}{71.46410231286283965728274174}%
\StoreBenchExecResult{PdrInv}{KinductionDfStaticZeroOneTF}{Correct}{False}{Walltime}{Median}{11.812169075}%
\StoreBenchExecResult{PdrInv}{KinductionDfStaticZeroOneTF}{Correct}{False}{Walltime}{Min}{1.75175595284}%
\StoreBenchExecResult{PdrInv}{KinductionDfStaticZeroOneTF}{Correct}{False}{Walltime}{Max}{867.797919989}%
\StoreBenchExecResult{PdrInv}{KinductionDfStaticZeroOneTF}{Correct}{False}{Walltime}{Stdev}{162.1687803316727235483764804}%
\StoreBenchExecResult{PdrInv}{KinductionDfStaticZeroOneTF}{Correct}{True}{Count}{}{2177}%
\StoreBenchExecResult{PdrInv}{KinductionDfStaticZeroOneTF}{Correct}{True}{Cputime}{}{91327.622675715}%
\StoreBenchExecResult{PdrInv}{KinductionDfStaticZeroOneTF}{Correct}{True}{Cputime}{Avg}{41.95113581796738631143775838}%
\StoreBenchExecResult{PdrInv}{KinductionDfStaticZeroOneTF}{Correct}{True}{Cputime}{Median}{7.120131128}%
\StoreBenchExecResult{PdrInv}{KinductionDfStaticZeroOneTF}{Correct}{True}{Cputime}{Min}{3.002322542}%
\StoreBenchExecResult{PdrInv}{KinductionDfStaticZeroOneTF}{Correct}{True}{Cputime}{Max}{890.658340113}%
\StoreBenchExecResult{PdrInv}{KinductionDfStaticZeroOneTF}{Correct}{True}{Cputime}{Stdev}{108.9699704739653466433363409}%
\StoreBenchExecResult{PdrInv}{KinductionDfStaticZeroOneTF}{Correct}{True}{Walltime}{}{57315.04071163974}%
\StoreBenchExecResult{PdrInv}{KinductionDfStaticZeroOneTF}{Correct}{True}{Walltime}{Avg}{26.32753362960024804777216353}%
\StoreBenchExecResult{PdrInv}{KinductionDfStaticZeroOneTF}{Correct}{True}{Walltime}{Median}{3.79017782211}%
\StoreBenchExecResult{PdrInv}{KinductionDfStaticZeroOneTF}{Correct}{True}{Walltime}{Min}{1.66395783424}%
\StoreBenchExecResult{PdrInv}{KinductionDfStaticZeroOneTF}{Correct}{True}{Walltime}{Max}{881.880240917}%
\StoreBenchExecResult{PdrInv}{KinductionDfStaticZeroOneTF}{Correct}{True}{Walltime}{Stdev}{78.07125613672724684042932870}%
\StoreBenchExecResult{PdrInv}{KinductionDfStaticZeroOneTF}{Wrong}{True}{Count}{}{0}%
\StoreBenchExecResult{PdrInv}{KinductionDfStaticZeroOneTF}{Wrong}{True}{Cputime}{}{0}%
\StoreBenchExecResult{PdrInv}{KinductionDfStaticZeroOneTF}{Wrong}{True}{Cputime}{Avg}{None}%
\StoreBenchExecResult{PdrInv}{KinductionDfStaticZeroOneTF}{Wrong}{True}{Cputime}{Median}{None}%
\StoreBenchExecResult{PdrInv}{KinductionDfStaticZeroOneTF}{Wrong}{True}{Cputime}{Min}{None}%
\StoreBenchExecResult{PdrInv}{KinductionDfStaticZeroOneTF}{Wrong}{True}{Cputime}{Max}{None}%
\StoreBenchExecResult{PdrInv}{KinductionDfStaticZeroOneTF}{Wrong}{True}{Cputime}{Stdev}{None}%
\StoreBenchExecResult{PdrInv}{KinductionDfStaticZeroOneTF}{Wrong}{True}{Walltime}{}{0}%
\StoreBenchExecResult{PdrInv}{KinductionDfStaticZeroOneTF}{Wrong}{True}{Walltime}{Avg}{None}%
\StoreBenchExecResult{PdrInv}{KinductionDfStaticZeroOneTF}{Wrong}{True}{Walltime}{Median}{None}%
\StoreBenchExecResult{PdrInv}{KinductionDfStaticZeroOneTF}{Wrong}{True}{Walltime}{Min}{None}%
\StoreBenchExecResult{PdrInv}{KinductionDfStaticZeroOneTF}{Wrong}{True}{Walltime}{Max}{None}%
\StoreBenchExecResult{PdrInv}{KinductionDfStaticZeroOneTF}{Wrong}{True}{Walltime}{Stdev}{None}%
\StoreBenchExecResult{PdrInv}{KinductionDfStaticZeroOneTF}{Error}{}{Count}{}{2595}%
\StoreBenchExecResult{PdrInv}{KinductionDfStaticZeroOneTF}{Error}{}{Cputime}{}{2049273.346969025}%
\StoreBenchExecResult{PdrInv}{KinductionDfStaticZeroOneTF}{Error}{}{Cputime}{Avg}{789.7007117414354527938342967}%
\StoreBenchExecResult{PdrInv}{KinductionDfStaticZeroOneTF}{Error}{}{Cputime}{Median}{901.768376157}%
\StoreBenchExecResult{PdrInv}{KinductionDfStaticZeroOneTF}{Error}{}{Cputime}{Min}{2.941055393}%
\StoreBenchExecResult{PdrInv}{KinductionDfStaticZeroOneTF}{Error}{}{Cputime}{Max}{1002.39740868}%
\StoreBenchExecResult{PdrInv}{KinductionDfStaticZeroOneTF}{Error}{}{Cputime}{Stdev}{264.0541893413757914232840369}%
\StoreBenchExecResult{PdrInv}{KinductionDfStaticZeroOneTF}{Error}{}{Walltime}{}{1442609.25414088221}%
\StoreBenchExecResult{PdrInv}{KinductionDfStaticZeroOneTF}{Error}{}{Walltime}{Avg}{555.9187877228833179190751445}%
\StoreBenchExecResult{PdrInv}{KinductionDfStaticZeroOneTF}{Error}{}{Walltime}{Median}{455.724066973}%
\StoreBenchExecResult{PdrInv}{KinductionDfStaticZeroOneTF}{Error}{}{Walltime}{Min}{1.593998909}%
\StoreBenchExecResult{PdrInv}{KinductionDfStaticZeroOneTF}{Error}{}{Walltime}{Max}{939.42276907}%
\StoreBenchExecResult{PdrInv}{KinductionDfStaticZeroOneTF}{Error}{}{Walltime}{Stdev}{256.0342901578786113082287588}%
\StoreBenchExecResult{PdrInv}{KinductionDfStaticZeroOneTF}{Error}{Assertion}{Count}{}{4}%
\StoreBenchExecResult{PdrInv}{KinductionDfStaticZeroOneTF}{Error}{Assertion}{Cputime}{}{13.908514699}%
\StoreBenchExecResult{PdrInv}{KinductionDfStaticZeroOneTF}{Error}{Assertion}{Cputime}{Avg}{3.47712867475}%
\StoreBenchExecResult{PdrInv}{KinductionDfStaticZeroOneTF}{Error}{Assertion}{Cputime}{Median}{3.377674121}%
\StoreBenchExecResult{PdrInv}{KinductionDfStaticZeroOneTF}{Error}{Assertion}{Cputime}{Min}{3.14358981}%
\StoreBenchExecResult{PdrInv}{KinductionDfStaticZeroOneTF}{Error}{Assertion}{Cputime}{Max}{4.009576647}%
\StoreBenchExecResult{PdrInv}{KinductionDfStaticZeroOneTF}{Error}{Assertion}{Cputime}{Stdev}{0.3233887863538206733957542533}%
\StoreBenchExecResult{PdrInv}{KinductionDfStaticZeroOneTF}{Error}{Assertion}{Walltime}{}{7.71363568307}%
\StoreBenchExecResult{PdrInv}{KinductionDfStaticZeroOneTF}{Error}{Assertion}{Walltime}{Avg}{1.9284089207675}%
\StoreBenchExecResult{PdrInv}{KinductionDfStaticZeroOneTF}{Error}{Assertion}{Walltime}{Median}{1.86156547070}%
\StoreBenchExecResult{PdrInv}{KinductionDfStaticZeroOneTF}{Error}{Assertion}{Walltime}{Min}{1.7533288002}%
\StoreBenchExecResult{PdrInv}{KinductionDfStaticZeroOneTF}{Error}{Assertion}{Walltime}{Max}{2.23717594147}%
\StoreBenchExecResult{PdrInv}{KinductionDfStaticZeroOneTF}{Error}{Assertion}{Walltime}{Stdev}{0.1844176822140709304684308599}%
\StoreBenchExecResult{PdrInv}{KinductionDfStaticZeroOneTF}{Error}{Error}{Count}{}{198}%
\StoreBenchExecResult{PdrInv}{KinductionDfStaticZeroOneTF}{Error}{Error}{Cputime}{}{36833.575097642}%
\StoreBenchExecResult{PdrInv}{KinductionDfStaticZeroOneTF}{Error}{Error}{Cputime}{Avg}{186.0281570587979797979797980}%
\StoreBenchExecResult{PdrInv}{KinductionDfStaticZeroOneTF}{Error}{Error}{Cputime}{Median}{117.6340111745}%
\StoreBenchExecResult{PdrInv}{KinductionDfStaticZeroOneTF}{Error}{Error}{Cputime}{Min}{2.941055393}%
\StoreBenchExecResult{PdrInv}{KinductionDfStaticZeroOneTF}{Error}{Error}{Cputime}{Max}{796.417582796}%
\StoreBenchExecResult{PdrInv}{KinductionDfStaticZeroOneTF}{Error}{Error}{Cputime}{Stdev}{192.2147523792283768530625542}%
\StoreBenchExecResult{PdrInv}{KinductionDfStaticZeroOneTF}{Error}{Error}{Walltime}{}{30574.81831336463}%
\StoreBenchExecResult{PdrInv}{KinductionDfStaticZeroOneTF}{Error}{Error}{Walltime}{Avg}{154.4182743099223737373737374}%
\StoreBenchExecResult{PdrInv}{KinductionDfStaticZeroOneTF}{Error}{Error}{Walltime}{Median}{95.1683989763}%
\StoreBenchExecResult{PdrInv}{KinductionDfStaticZeroOneTF}{Error}{Error}{Walltime}{Min}{1.593998909}%
\StoreBenchExecResult{PdrInv}{KinductionDfStaticZeroOneTF}{Error}{Error}{Walltime}{Max}{782.34778595}%
\StoreBenchExecResult{PdrInv}{KinductionDfStaticZeroOneTF}{Error}{Error}{Walltime}{Stdev}{167.9311588297211845971004812}%
\StoreBenchExecResult{PdrInv}{KinductionDfStaticZeroOneTF}{Error}{Exception}{Count}{}{13}%
\StoreBenchExecResult{PdrInv}{KinductionDfStaticZeroOneTF}{Error}{Exception}{Cputime}{}{1486.648121027}%
\StoreBenchExecResult{PdrInv}{KinductionDfStaticZeroOneTF}{Error}{Exception}{Cputime}{Avg}{114.3575477713076923076923077}%
\StoreBenchExecResult{PdrInv}{KinductionDfStaticZeroOneTF}{Error}{Exception}{Cputime}{Median}{71.753267556}%
\StoreBenchExecResult{PdrInv}{KinductionDfStaticZeroOneTF}{Error}{Exception}{Cputime}{Min}{14.848690343}%
\StoreBenchExecResult{PdrInv}{KinductionDfStaticZeroOneTF}{Error}{Exception}{Cputime}{Max}{457.164451935}%
\StoreBenchExecResult{PdrInv}{KinductionDfStaticZeroOneTF}{Error}{Exception}{Cputime}{Stdev}{121.9201243415273467562653806}%
\StoreBenchExecResult{PdrInv}{KinductionDfStaticZeroOneTF}{Error}{Exception}{Walltime}{}{804.31326889961}%
\StoreBenchExecResult{PdrInv}{KinductionDfStaticZeroOneTF}{Error}{Exception}{Walltime}{Avg}{61.87025145381615384615384615}%
\StoreBenchExecResult{PdrInv}{KinductionDfStaticZeroOneTF}{Error}{Exception}{Walltime}{Median}{42.2126741409}%
\StoreBenchExecResult{PdrInv}{KinductionDfStaticZeroOneTF}{Error}{Exception}{Walltime}{Min}{7.60693311691}%
\StoreBenchExecResult{PdrInv}{KinductionDfStaticZeroOneTF}{Error}{Exception}{Walltime}{Max}{263.376106024}%
\StoreBenchExecResult{PdrInv}{KinductionDfStaticZeroOneTF}{Error}{Exception}{Walltime}{Stdev}{68.56374249661857185689079818}%
\StoreBenchExecResult{PdrInv}{KinductionDfStaticZeroOneTF}{Error}{OutOfJavaMemory}{Count}{}{10}%
\StoreBenchExecResult{PdrInv}{KinductionDfStaticZeroOneTF}{Error}{OutOfJavaMemory}{Cputime}{}{5400.931323796}%
\StoreBenchExecResult{PdrInv}{KinductionDfStaticZeroOneTF}{Error}{OutOfJavaMemory}{Cputime}{Avg}{540.0931323796}%
\StoreBenchExecResult{PdrInv}{KinductionDfStaticZeroOneTF}{Error}{OutOfJavaMemory}{Cputime}{Median}{533.5566705105}%
\StoreBenchExecResult{PdrInv}{KinductionDfStaticZeroOneTF}{Error}{OutOfJavaMemory}{Cputime}{Min}{187.515973605}%
\StoreBenchExecResult{PdrInv}{KinductionDfStaticZeroOneTF}{Error}{OutOfJavaMemory}{Cputime}{Max}{858.725358526}%
\StoreBenchExecResult{PdrInv}{KinductionDfStaticZeroOneTF}{Error}{OutOfJavaMemory}{Cputime}{Stdev}{219.5492666535895219222267335}%
\StoreBenchExecResult{PdrInv}{KinductionDfStaticZeroOneTF}{Error}{OutOfJavaMemory}{Walltime}{}{3267.862046004}%
\StoreBenchExecResult{PdrInv}{KinductionDfStaticZeroOneTF}{Error}{OutOfJavaMemory}{Walltime}{Avg}{326.7862046004}%
\StoreBenchExecResult{PdrInv}{KinductionDfStaticZeroOneTF}{Error}{OutOfJavaMemory}{Walltime}{Median}{328.248784900}%
\StoreBenchExecResult{PdrInv}{KinductionDfStaticZeroOneTF}{Error}{OutOfJavaMemory}{Walltime}{Min}{115.741152048}%
\StoreBenchExecResult{PdrInv}{KinductionDfStaticZeroOneTF}{Error}{OutOfJavaMemory}{Walltime}{Max}{569.054636955}%
\StoreBenchExecResult{PdrInv}{KinductionDfStaticZeroOneTF}{Error}{OutOfJavaMemory}{Walltime}{Stdev}{135.0884084704200503970881653}%
\StoreBenchExecResult{PdrInv}{KinductionDfStaticZeroOneTF}{Error}{OutOfMemory}{Count}{}{267}%
\StoreBenchExecResult{PdrInv}{KinductionDfStaticZeroOneTF}{Error}{OutOfMemory}{Cputime}{}{94900.146460749}%
\StoreBenchExecResult{PdrInv}{KinductionDfStaticZeroOneTF}{Error}{OutOfMemory}{Cputime}{Avg}{355.4312601526179775280898876}%
\StoreBenchExecResult{PdrInv}{KinductionDfStaticZeroOneTF}{Error}{OutOfMemory}{Cputime}{Median}{268.839962628}%
\StoreBenchExecResult{PdrInv}{KinductionDfStaticZeroOneTF}{Error}{OutOfMemory}{Cputime}{Min}{163.180241856}%
\StoreBenchExecResult{PdrInv}{KinductionDfStaticZeroOneTF}{Error}{OutOfMemory}{Cputime}{Max}{882.926155716}%
\StoreBenchExecResult{PdrInv}{KinductionDfStaticZeroOneTF}{Error}{OutOfMemory}{Cputime}{Stdev}{201.7252576971908151305145036}%
\StoreBenchExecResult{PdrInv}{KinductionDfStaticZeroOneTF}{Error}{OutOfMemory}{Walltime}{}{79894.0205557369}%
\StoreBenchExecResult{PdrInv}{KinductionDfStaticZeroOneTF}{Error}{OutOfMemory}{Walltime}{Avg}{299.2285414072543071161048689}%
\StoreBenchExecResult{PdrInv}{KinductionDfStaticZeroOneTF}{Error}{OutOfMemory}{Walltime}{Median}{200.211754084}%
\StoreBenchExecResult{PdrInv}{KinductionDfStaticZeroOneTF}{Error}{OutOfMemory}{Walltime}{Min}{86.5999338627}%
\StoreBenchExecResult{PdrInv}{KinductionDfStaticZeroOneTF}{Error}{OutOfMemory}{Walltime}{Max}{869.022175074}%
\StoreBenchExecResult{PdrInv}{KinductionDfStaticZeroOneTF}{Error}{OutOfMemory}{Walltime}{Stdev}{222.7490328764558969992704092}%
\StoreBenchExecResult{PdrInv}{KinductionDfStaticZeroOneTF}{Error}{Timeout}{Count}{}{2103}%
\StoreBenchExecResult{PdrInv}{KinductionDfStaticZeroOneTF}{Error}{Timeout}{Cputime}{}{1910638.137451112}%
\StoreBenchExecResult{PdrInv}{KinductionDfStaticZeroOneTF}{Error}{Timeout}{Cputime}{Avg}{908.5297848079467427484545887}%
\StoreBenchExecResult{PdrInv}{KinductionDfStaticZeroOneTF}{Error}{Timeout}{Cputime}{Median}{902.318294521}%
\StoreBenchExecResult{PdrInv}{KinductionDfStaticZeroOneTF}{Error}{Timeout}{Cputime}{Min}{900.818633269}%
\StoreBenchExecResult{PdrInv}{KinductionDfStaticZeroOneTF}{Error}{Timeout}{Cputime}{Max}{1002.39740868}%
\StoreBenchExecResult{PdrInv}{KinductionDfStaticZeroOneTF}{Error}{Timeout}{Cputime}{Stdev}{20.11612331718880463638965197}%
\StoreBenchExecResult{PdrInv}{KinductionDfStaticZeroOneTF}{Error}{Timeout}{Walltime}{}{1328060.526321194}%
\StoreBenchExecResult{PdrInv}{KinductionDfStaticZeroOneTF}{Error}{Timeout}{Walltime}{Avg}{631.5076206948140751307655730}%
\StoreBenchExecResult{PdrInv}{KinductionDfStaticZeroOneTF}{Error}{Timeout}{Walltime}{Median}{471.109051943}%
\StoreBenchExecResult{PdrInv}{KinductionDfStaticZeroOneTF}{Error}{Timeout}{Walltime}{Min}{451.09356904}%
\StoreBenchExecResult{PdrInv}{KinductionDfStaticZeroOneTF}{Error}{Timeout}{Walltime}{Max}{939.42276907}%
\StoreBenchExecResult{PdrInv}{KinductionDfStaticZeroOneTF}{Error}{Timeout}{Walltime}{Stdev}{200.5508650837991603526720436}%
\StoreBenchExecResult{PdrInv}{KinductionDfStaticZeroOneTF}{Wrong}{}{Count}{}{2}%
\StoreBenchExecResult{PdrInv}{KinductionDfStaticZeroOneTF}{Wrong}{}{Cputime}{}{22.890119359}%
\StoreBenchExecResult{PdrInv}{KinductionDfStaticZeroOneTF}{Wrong}{}{Cputime}{Avg}{11.4450596795}%
\StoreBenchExecResult{PdrInv}{KinductionDfStaticZeroOneTF}{Wrong}{}{Cputime}{Median}{11.4450596795}%
\StoreBenchExecResult{PdrInv}{KinductionDfStaticZeroOneTF}{Wrong}{}{Cputime}{Min}{3.826062888}%
\StoreBenchExecResult{PdrInv}{KinductionDfStaticZeroOneTF}{Wrong}{}{Cputime}{Max}{19.064056471}%
\StoreBenchExecResult{PdrInv}{KinductionDfStaticZeroOneTF}{Wrong}{}{Cputime}{Stdev}{7.6189967915}%
\StoreBenchExecResult{PdrInv}{KinductionDfStaticZeroOneTF}{Wrong}{}{Walltime}{}{12.4040591717}%
\StoreBenchExecResult{PdrInv}{KinductionDfStaticZeroOneTF}{Wrong}{}{Walltime}{Avg}{6.20202958585}%
\StoreBenchExecResult{PdrInv}{KinductionDfStaticZeroOneTF}{Wrong}{}{Walltime}{Median}{6.20202958585}%
\StoreBenchExecResult{PdrInv}{KinductionDfStaticZeroOneTF}{Wrong}{}{Walltime}{Min}{2.102643013}%
\StoreBenchExecResult{PdrInv}{KinductionDfStaticZeroOneTF}{Wrong}{}{Walltime}{Max}{10.3014161587}%
\StoreBenchExecResult{PdrInv}{KinductionDfStaticZeroOneTF}{Wrong}{}{Walltime}{Stdev}{4.09938657285}%
\StoreBenchExecResult{PdrInv}{KinductionDfStaticZeroOneTF}{Wrong}{False}{Count}{}{2}%
\StoreBenchExecResult{PdrInv}{KinductionDfStaticZeroOneTF}{Wrong}{False}{Cputime}{}{22.890119359}%
\StoreBenchExecResult{PdrInv}{KinductionDfStaticZeroOneTF}{Wrong}{False}{Cputime}{Avg}{11.4450596795}%
\StoreBenchExecResult{PdrInv}{KinductionDfStaticZeroOneTF}{Wrong}{False}{Cputime}{Median}{11.4450596795}%
\StoreBenchExecResult{PdrInv}{KinductionDfStaticZeroOneTF}{Wrong}{False}{Cputime}{Min}{3.826062888}%
\StoreBenchExecResult{PdrInv}{KinductionDfStaticZeroOneTF}{Wrong}{False}{Cputime}{Max}{19.064056471}%
\StoreBenchExecResult{PdrInv}{KinductionDfStaticZeroOneTF}{Wrong}{False}{Cputime}{Stdev}{7.6189967915}%
\StoreBenchExecResult{PdrInv}{KinductionDfStaticZeroOneTF}{Wrong}{False}{Walltime}{}{12.4040591717}%
\StoreBenchExecResult{PdrInv}{KinductionDfStaticZeroOneTF}{Wrong}{False}{Walltime}{Avg}{6.20202958585}%
\StoreBenchExecResult{PdrInv}{KinductionDfStaticZeroOneTF}{Wrong}{False}{Walltime}{Median}{6.20202958585}%
\StoreBenchExecResult{PdrInv}{KinductionDfStaticZeroOneTF}{Wrong}{False}{Walltime}{Min}{2.102643013}%
\StoreBenchExecResult{PdrInv}{KinductionDfStaticZeroOneTF}{Wrong}{False}{Walltime}{Max}{10.3014161587}%
\StoreBenchExecResult{PdrInv}{KinductionDfStaticZeroOneTF}{Wrong}{False}{Walltime}{Stdev}{4.09938657285}%
\providecommand\StoreBenchExecResult[7]{\expandafter\newcommand\csname#1#2#3#4#5#6\endcsname{#7}}%
\StoreBenchExecResult{PdrInv}{KinductionDfStaticZeroOneTTTrueNotSolvedByKinductionPlainButKipdr}{Total}{}{Count}{}{449}%
\StoreBenchExecResult{PdrInv}{KinductionDfStaticZeroOneTTTrueNotSolvedByKinductionPlainButKipdr}{Total}{}{Cputime}{}{10363.398440191}%
\StoreBenchExecResult{PdrInv}{KinductionDfStaticZeroOneTTTrueNotSolvedByKinductionPlainButKipdr}{Total}{}{Cputime}{Avg}{23.08106556835412026726057906}%
\StoreBenchExecResult{PdrInv}{KinductionDfStaticZeroOneTTTrueNotSolvedByKinductionPlainButKipdr}{Total}{}{Cputime}{Median}{6.010850061}%
\StoreBenchExecResult{PdrInv}{KinductionDfStaticZeroOneTTTrueNotSolvedByKinductionPlainButKipdr}{Total}{}{Cputime}{Min}{3.391066446}%
\StoreBenchExecResult{PdrInv}{KinductionDfStaticZeroOneTTTrueNotSolvedByKinductionPlainButKipdr}{Total}{}{Cputime}{Max}{914.638838463}%
\StoreBenchExecResult{PdrInv}{KinductionDfStaticZeroOneTTTrueNotSolvedByKinductionPlainButKipdr}{Total}{}{Cputime}{Stdev}{118.2808275414295504513583054}%
\StoreBenchExecResult{PdrInv}{KinductionDfStaticZeroOneTTTrueNotSolvedByKinductionPlainButKipdr}{Total}{}{Walltime}{}{8317.51362514363}%
\StoreBenchExecResult{PdrInv}{KinductionDfStaticZeroOneTTTrueNotSolvedByKinductionPlainButKipdr}{Total}{}{Walltime}{Avg}{18.52452923194572383073496659}%
\StoreBenchExecResult{PdrInv}{KinductionDfStaticZeroOneTTTrueNotSolvedByKinductionPlainButKipdr}{Total}{}{Walltime}{Median}{3.18773698807}%
\StoreBenchExecResult{PdrInv}{KinductionDfStaticZeroOneTTTrueNotSolvedByKinductionPlainButKipdr}{Total}{}{Walltime}{Min}{1.86370992661}%
\StoreBenchExecResult{PdrInv}{KinductionDfStaticZeroOneTTTrueNotSolvedByKinductionPlainButKipdr}{Total}{}{Walltime}{Max}{900.856547117}%
\StoreBenchExecResult{PdrInv}{KinductionDfStaticZeroOneTTTrueNotSolvedByKinductionPlainButKipdr}{Total}{}{Walltime}{Stdev}{111.1380546002709544615620701}%
\StoreBenchExecResult{PdrInv}{KinductionDfStaticZeroOneTTTrueNotSolvedByKinductionPlainButKipdr}{Correct}{}{Count}{}{441}%
\StoreBenchExecResult{PdrInv}{KinductionDfStaticZeroOneTTTrueNotSolvedByKinductionPlainButKipdr}{Correct}{}{Cputime}{}{3165.143430645}%
\StoreBenchExecResult{PdrInv}{KinductionDfStaticZeroOneTTTrueNotSolvedByKinductionPlainButKipdr}{Correct}{}{Cputime}{Avg}{7.177195987857142857142857143}%
\StoreBenchExecResult{PdrInv}{KinductionDfStaticZeroOneTTTrueNotSolvedByKinductionPlainButKipdr}{Correct}{}{Cputime}{Median}{5.99240848}%
\StoreBenchExecResult{PdrInv}{KinductionDfStaticZeroOneTTTrueNotSolvedByKinductionPlainButKipdr}{Correct}{}{Cputime}{Min}{3.391066446}%
\StoreBenchExecResult{PdrInv}{KinductionDfStaticZeroOneTTTrueNotSolvedByKinductionPlainButKipdr}{Correct}{}{Cputime}{Max}{89.900925884}%
\StoreBenchExecResult{PdrInv}{KinductionDfStaticZeroOneTTTrueNotSolvedByKinductionPlainButKipdr}{Correct}{}{Cputime}{Stdev}{6.623908016763101619144855118}%
\StoreBenchExecResult{PdrInv}{KinductionDfStaticZeroOneTTTrueNotSolvedByKinductionPlainButKipdr}{Correct}{}{Walltime}{}{1670.42976021763}%
\StoreBenchExecResult{PdrInv}{KinductionDfStaticZeroOneTTTrueNotSolvedByKinductionPlainButKipdr}{Correct}{}{Walltime}{Avg}{3.787822585527505668934240363}%
\StoreBenchExecResult{PdrInv}{KinductionDfStaticZeroOneTTTrueNotSolvedByKinductionPlainButKipdr}{Correct}{}{Walltime}{Median}{3.17253303528}%
\StoreBenchExecResult{PdrInv}{KinductionDfStaticZeroOneTTTrueNotSolvedByKinductionPlainButKipdr}{Correct}{}{Walltime}{Min}{1.86370992661}%
\StoreBenchExecResult{PdrInv}{KinductionDfStaticZeroOneTTTrueNotSolvedByKinductionPlainButKipdr}{Correct}{}{Walltime}{Max}{45.5330300331}%
\StoreBenchExecResult{PdrInv}{KinductionDfStaticZeroOneTTTrueNotSolvedByKinductionPlainButKipdr}{Correct}{}{Walltime}{Stdev}{3.341032479657021810752355016}%
\StoreBenchExecResult{PdrInv}{KinductionDfStaticZeroOneTTTrueNotSolvedByKinductionPlainButKipdr}{Correct}{True}{Count}{}{441}%
\StoreBenchExecResult{PdrInv}{KinductionDfStaticZeroOneTTTrueNotSolvedByKinductionPlainButKipdr}{Correct}{True}{Cputime}{}{3165.143430645}%
\StoreBenchExecResult{PdrInv}{KinductionDfStaticZeroOneTTTrueNotSolvedByKinductionPlainButKipdr}{Correct}{True}{Cputime}{Avg}{7.177195987857142857142857143}%
\StoreBenchExecResult{PdrInv}{KinductionDfStaticZeroOneTTTrueNotSolvedByKinductionPlainButKipdr}{Correct}{True}{Cputime}{Median}{5.99240848}%
\StoreBenchExecResult{PdrInv}{KinductionDfStaticZeroOneTTTrueNotSolvedByKinductionPlainButKipdr}{Correct}{True}{Cputime}{Min}{3.391066446}%
\StoreBenchExecResult{PdrInv}{KinductionDfStaticZeroOneTTTrueNotSolvedByKinductionPlainButKipdr}{Correct}{True}{Cputime}{Max}{89.900925884}%
\StoreBenchExecResult{PdrInv}{KinductionDfStaticZeroOneTTTrueNotSolvedByKinductionPlainButKipdr}{Correct}{True}{Cputime}{Stdev}{6.623908016763101619144855118}%
\StoreBenchExecResult{PdrInv}{KinductionDfStaticZeroOneTTTrueNotSolvedByKinductionPlainButKipdr}{Correct}{True}{Walltime}{}{1670.42976021763}%
\StoreBenchExecResult{PdrInv}{KinductionDfStaticZeroOneTTTrueNotSolvedByKinductionPlainButKipdr}{Correct}{True}{Walltime}{Avg}{3.787822585527505668934240363}%
\StoreBenchExecResult{PdrInv}{KinductionDfStaticZeroOneTTTrueNotSolvedByKinductionPlainButKipdr}{Correct}{True}{Walltime}{Median}{3.17253303528}%
\StoreBenchExecResult{PdrInv}{KinductionDfStaticZeroOneTTTrueNotSolvedByKinductionPlainButKipdr}{Correct}{True}{Walltime}{Min}{1.86370992661}%
\StoreBenchExecResult{PdrInv}{KinductionDfStaticZeroOneTTTrueNotSolvedByKinductionPlainButKipdr}{Correct}{True}{Walltime}{Max}{45.5330300331}%
\StoreBenchExecResult{PdrInv}{KinductionDfStaticZeroOneTTTrueNotSolvedByKinductionPlainButKipdr}{Correct}{True}{Walltime}{Stdev}{3.341032479657021810752355016}%
\StoreBenchExecResult{PdrInv}{KinductionDfStaticZeroOneTTTrueNotSolvedByKinductionPlainButKipdr}{Wrong}{True}{Count}{}{0}%
\StoreBenchExecResult{PdrInv}{KinductionDfStaticZeroOneTTTrueNotSolvedByKinductionPlainButKipdr}{Wrong}{True}{Cputime}{}{0}%
\StoreBenchExecResult{PdrInv}{KinductionDfStaticZeroOneTTTrueNotSolvedByKinductionPlainButKipdr}{Wrong}{True}{Cputime}{Avg}{None}%
\StoreBenchExecResult{PdrInv}{KinductionDfStaticZeroOneTTTrueNotSolvedByKinductionPlainButKipdr}{Wrong}{True}{Cputime}{Median}{None}%
\StoreBenchExecResult{PdrInv}{KinductionDfStaticZeroOneTTTrueNotSolvedByKinductionPlainButKipdr}{Wrong}{True}{Cputime}{Min}{None}%
\StoreBenchExecResult{PdrInv}{KinductionDfStaticZeroOneTTTrueNotSolvedByKinductionPlainButKipdr}{Wrong}{True}{Cputime}{Max}{None}%
\StoreBenchExecResult{PdrInv}{KinductionDfStaticZeroOneTTTrueNotSolvedByKinductionPlainButKipdr}{Wrong}{True}{Cputime}{Stdev}{None}%
\StoreBenchExecResult{PdrInv}{KinductionDfStaticZeroOneTTTrueNotSolvedByKinductionPlainButKipdr}{Wrong}{True}{Walltime}{}{0}%
\StoreBenchExecResult{PdrInv}{KinductionDfStaticZeroOneTTTrueNotSolvedByKinductionPlainButKipdr}{Wrong}{True}{Walltime}{Avg}{None}%
\StoreBenchExecResult{PdrInv}{KinductionDfStaticZeroOneTTTrueNotSolvedByKinductionPlainButKipdr}{Wrong}{True}{Walltime}{Median}{None}%
\StoreBenchExecResult{PdrInv}{KinductionDfStaticZeroOneTTTrueNotSolvedByKinductionPlainButKipdr}{Wrong}{True}{Walltime}{Min}{None}%
\StoreBenchExecResult{PdrInv}{KinductionDfStaticZeroOneTTTrueNotSolvedByKinductionPlainButKipdr}{Wrong}{True}{Walltime}{Max}{None}%
\StoreBenchExecResult{PdrInv}{KinductionDfStaticZeroOneTTTrueNotSolvedByKinductionPlainButKipdr}{Wrong}{True}{Walltime}{Stdev}{None}%
\StoreBenchExecResult{PdrInv}{KinductionDfStaticZeroOneTTTrueNotSolvedByKinductionPlainButKipdr}{Error}{}{Count}{}{8}%
\StoreBenchExecResult{PdrInv}{KinductionDfStaticZeroOneTTTrueNotSolvedByKinductionPlainButKipdr}{Error}{}{Cputime}{}{7198.255009546}%
\StoreBenchExecResult{PdrInv}{KinductionDfStaticZeroOneTTTrueNotSolvedByKinductionPlainButKipdr}{Error}{}{Cputime}{Avg}{899.78187619325}%
\StoreBenchExecResult{PdrInv}{KinductionDfStaticZeroOneTTTrueNotSolvedByKinductionPlainButKipdr}{Error}{}{Cputime}{Median}{904.0518159675}%
\StoreBenchExecResult{PdrInv}{KinductionDfStaticZeroOneTTTrueNotSolvedByKinductionPlainButKipdr}{Error}{}{Cputime}{Min}{859.673677721}%
\StoreBenchExecResult{PdrInv}{KinductionDfStaticZeroOneTTTrueNotSolvedByKinductionPlainButKipdr}{Error}{}{Cputime}{Max}{914.638838463}%
\StoreBenchExecResult{PdrInv}{KinductionDfStaticZeroOneTTTrueNotSolvedByKinductionPlainButKipdr}{Error}{}{Cputime}{Stdev}{15.57907464991836413204204126}%
\StoreBenchExecResult{PdrInv}{KinductionDfStaticZeroOneTTTrueNotSolvedByKinductionPlainButKipdr}{Error}{}{Walltime}{}{6647.083864926}%
\StoreBenchExecResult{PdrInv}{KinductionDfStaticZeroOneTTTrueNotSolvedByKinductionPlainButKipdr}{Error}{}{Walltime}{Avg}{830.88548311575}%
\StoreBenchExecResult{PdrInv}{KinductionDfStaticZeroOneTTTrueNotSolvedByKinductionPlainButKipdr}{Error}{}{Walltime}{Median}{888.669688463}%
\StoreBenchExecResult{PdrInv}{KinductionDfStaticZeroOneTTTrueNotSolvedByKinductionPlainButKipdr}{Error}{}{Walltime}{Min}{452.42462182}%
\StoreBenchExecResult{PdrInv}{KinductionDfStaticZeroOneTTTrueNotSolvedByKinductionPlainButKipdr}{Error}{}{Walltime}{Max}{900.856547117}%
\StoreBenchExecResult{PdrInv}{KinductionDfStaticZeroOneTTTrueNotSolvedByKinductionPlainButKipdr}{Error}{}{Walltime}{Stdev}{143.9449315845258333123041258}%
\StoreBenchExecResult{PdrInv}{KinductionDfStaticZeroOneTTTrueNotSolvedByKinductionPlainButKipdr}{Error}{OutOfMemory}{Count}{}{1}%
\StoreBenchExecResult{PdrInv}{KinductionDfStaticZeroOneTTTrueNotSolvedByKinductionPlainButKipdr}{Error}{OutOfMemory}{Cputime}{}{859.673677721}%
\StoreBenchExecResult{PdrInv}{KinductionDfStaticZeroOneTTTrueNotSolvedByKinductionPlainButKipdr}{Error}{OutOfMemory}{Cputime}{Avg}{859.673677721}%
\StoreBenchExecResult{PdrInv}{KinductionDfStaticZeroOneTTTrueNotSolvedByKinductionPlainButKipdr}{Error}{OutOfMemory}{Cputime}{Median}{859.673677721}%
\StoreBenchExecResult{PdrInv}{KinductionDfStaticZeroOneTTTrueNotSolvedByKinductionPlainButKipdr}{Error}{OutOfMemory}{Cputime}{Min}{859.673677721}%
\StoreBenchExecResult{PdrInv}{KinductionDfStaticZeroOneTTTrueNotSolvedByKinductionPlainButKipdr}{Error}{OutOfMemory}{Cputime}{Max}{859.673677721}%
\StoreBenchExecResult{PdrInv}{KinductionDfStaticZeroOneTTTrueNotSolvedByKinductionPlainButKipdr}{Error}{OutOfMemory}{Cputime}{Stdev}{0E-9}%
\StoreBenchExecResult{PdrInv}{KinductionDfStaticZeroOneTTTrueNotSolvedByKinductionPlainButKipdr}{Error}{OutOfMemory}{Walltime}{}{844.118904114}%
\StoreBenchExecResult{PdrInv}{KinductionDfStaticZeroOneTTTrueNotSolvedByKinductionPlainButKipdr}{Error}{OutOfMemory}{Walltime}{Avg}{844.118904114}%
\StoreBenchExecResult{PdrInv}{KinductionDfStaticZeroOneTTTrueNotSolvedByKinductionPlainButKipdr}{Error}{OutOfMemory}{Walltime}{Median}{844.118904114}%
\StoreBenchExecResult{PdrInv}{KinductionDfStaticZeroOneTTTrueNotSolvedByKinductionPlainButKipdr}{Error}{OutOfMemory}{Walltime}{Min}{844.118904114}%
\StoreBenchExecResult{PdrInv}{KinductionDfStaticZeroOneTTTrueNotSolvedByKinductionPlainButKipdr}{Error}{OutOfMemory}{Walltime}{Max}{844.118904114}%
\StoreBenchExecResult{PdrInv}{KinductionDfStaticZeroOneTTTrueNotSolvedByKinductionPlainButKipdr}{Error}{OutOfMemory}{Walltime}{Stdev}{0E-9}%
\StoreBenchExecResult{PdrInv}{KinductionDfStaticZeroOneTTTrueNotSolvedByKinductionPlainButKipdr}{Error}{Timeout}{Count}{}{7}%
\StoreBenchExecResult{PdrInv}{KinductionDfStaticZeroOneTTTrueNotSolvedByKinductionPlainButKipdr}{Error}{Timeout}{Cputime}{}{6338.581331825}%
\StoreBenchExecResult{PdrInv}{KinductionDfStaticZeroOneTTTrueNotSolvedByKinductionPlainButKipdr}{Error}{Timeout}{Cputime}{Avg}{905.5116188321428571428571429}%
\StoreBenchExecResult{PdrInv}{KinductionDfStaticZeroOneTTTrueNotSolvedByKinductionPlainButKipdr}{Error}{Timeout}{Cputime}{Median}{904.190325988}%
\StoreBenchExecResult{PdrInv}{KinductionDfStaticZeroOneTTTrueNotSolvedByKinductionPlainButKipdr}{Error}{Timeout}{Cputime}{Min}{902.299507982}%
\StoreBenchExecResult{PdrInv}{KinductionDfStaticZeroOneTTTrueNotSolvedByKinductionPlainButKipdr}{Error}{Timeout}{Cputime}{Max}{914.638838463}%
\StoreBenchExecResult{PdrInv}{KinductionDfStaticZeroOneTTTrueNotSolvedByKinductionPlainButKipdr}{Error}{Timeout}{Cputime}{Stdev}{3.839332083594127837845037329}%
\StoreBenchExecResult{PdrInv}{KinductionDfStaticZeroOneTTTrueNotSolvedByKinductionPlainButKipdr}{Error}{Timeout}{Walltime}{}{5802.964960812}%
\StoreBenchExecResult{PdrInv}{KinductionDfStaticZeroOneTTTrueNotSolvedByKinductionPlainButKipdr}{Error}{Timeout}{Walltime}{Avg}{828.9949944017142857142857143}%
\StoreBenchExecResult{PdrInv}{KinductionDfStaticZeroOneTTTrueNotSolvedByKinductionPlainButKipdr}{Error}{Timeout}{Walltime}{Median}{888.916290998}%
\StoreBenchExecResult{PdrInv}{KinductionDfStaticZeroOneTTTrueNotSolvedByKinductionPlainButKipdr}{Error}{Timeout}{Walltime}{Min}{452.42462182}%
\StoreBenchExecResult{PdrInv}{KinductionDfStaticZeroOneTTTrueNotSolvedByKinductionPlainButKipdr}{Error}{Timeout}{Walltime}{Max}{900.856547117}%
\StoreBenchExecResult{PdrInv}{KinductionDfStaticZeroOneTTTrueNotSolvedByKinductionPlainButKipdr}{Error}{Timeout}{Walltime}{Stdev}{153.7906766478516239872060819}%
\providecommand\StoreBenchExecResult[7]{\expandafter\newcommand\csname#1#2#3#4#5#6\endcsname{#7}}%
\StoreBenchExecResult{PdrInv}{KinductionDfStaticZeroOneTTTrueNotSolvedByKinductionPlain}{Total}{}{Count}{}{2893}%
\StoreBenchExecResult{PdrInv}{KinductionDfStaticZeroOneTTTrueNotSolvedByKinductionPlain}{Total}{}{Cputime}{}{1619607.959532677}%
\StoreBenchExecResult{PdrInv}{KinductionDfStaticZeroOneTTTrueNotSolvedByKinductionPlain}{Total}{}{Cputime}{Avg}{559.8368335750698237124092637}%
\StoreBenchExecResult{PdrInv}{KinductionDfStaticZeroOneTTTrueNotSolvedByKinductionPlain}{Total}{}{Cputime}{Median}{901.141315444}%
\StoreBenchExecResult{PdrInv}{KinductionDfStaticZeroOneTTTrueNotSolvedByKinductionPlain}{Total}{}{Cputime}{Min}{2.633111638}%
\StoreBenchExecResult{PdrInv}{KinductionDfStaticZeroOneTTTrueNotSolvedByKinductionPlain}{Total}{}{Cputime}{Max}{1002.33951193}%
\StoreBenchExecResult{PdrInv}{KinductionDfStaticZeroOneTTTrueNotSolvedByKinductionPlain}{Total}{}{Cputime}{Stdev}{418.0290008365795666439129419}%
\StoreBenchExecResult{PdrInv}{KinductionDfStaticZeroOneTTTrueNotSolvedByKinductionPlain}{Total}{}{Walltime}{}{1118102.06768175876}%
\StoreBenchExecResult{PdrInv}{KinductionDfStaticZeroOneTTTrueNotSolvedByKinductionPlain}{Total}{}{Walltime}{Avg}{386.4853327624468579329415831}%
\StoreBenchExecResult{PdrInv}{KinductionDfStaticZeroOneTTTrueNotSolvedByKinductionPlain}{Total}{}{Walltime}{Median}{451.810097933}%
\StoreBenchExecResult{PdrInv}{KinductionDfStaticZeroOneTTTrueNotSolvedByKinductionPlain}{Total}{}{Walltime}{Min}{1.4413330555}%
\StoreBenchExecResult{PdrInv}{KinductionDfStaticZeroOneTTTrueNotSolvedByKinductionPlain}{Total}{}{Walltime}{Max}{920.880501032}%
\StoreBenchExecResult{PdrInv}{KinductionDfStaticZeroOneTTTrueNotSolvedByKinductionPlain}{Total}{}{Walltime}{Stdev}{328.0026912066346960275298600}%
\StoreBenchExecResult{PdrInv}{KinductionDfStaticZeroOneTTTrueNotSolvedByKinductionPlain}{Correct}{}{Count}{}{950}%
\StoreBenchExecResult{PdrInv}{KinductionDfStaticZeroOneTTTrueNotSolvedByKinductionPlain}{Correct}{}{Cputime}{}{39080.406451603}%
\StoreBenchExecResult{PdrInv}{KinductionDfStaticZeroOneTTTrueNotSolvedByKinductionPlain}{Correct}{}{Cputime}{Avg}{41.13726994905578947368421053}%
\StoreBenchExecResult{PdrInv}{KinductionDfStaticZeroOneTTTrueNotSolvedByKinductionPlain}{Correct}{}{Cputime}{Median}{7.107670114}%
\StoreBenchExecResult{PdrInv}{KinductionDfStaticZeroOneTTTrueNotSolvedByKinductionPlain}{Correct}{}{Cputime}{Min}{3.351895368}%
\StoreBenchExecResult{PdrInv}{KinductionDfStaticZeroOneTTTrueNotSolvedByKinductionPlain}{Correct}{}{Cputime}{Max}{896.584723939}%
\StoreBenchExecResult{PdrInv}{KinductionDfStaticZeroOneTTTrueNotSolvedByKinductionPlain}{Correct}{}{Cputime}{Stdev}{113.1887613798698222461115770}%
\StoreBenchExecResult{PdrInv}{KinductionDfStaticZeroOneTTTrueNotSolvedByKinductionPlain}{Correct}{}{Walltime}{}{21387.28514075158}%
\StoreBenchExecResult{PdrInv}{KinductionDfStaticZeroOneTTTrueNotSolvedByKinductionPlain}{Correct}{}{Walltime}{Avg}{22.51293172710692631578947368}%
\StoreBenchExecResult{PdrInv}{KinductionDfStaticZeroOneTTTrueNotSolvedByKinductionPlain}{Correct}{}{Walltime}{Median}{3.74982059002}%
\StoreBenchExecResult{PdrInv}{KinductionDfStaticZeroOneTTTrueNotSolvedByKinductionPlain}{Correct}{}{Walltime}{Min}{1.86370992661}%
\StoreBenchExecResult{PdrInv}{KinductionDfStaticZeroOneTTTrueNotSolvedByKinductionPlain}{Correct}{}{Walltime}{Max}{780.438195944}%
\StoreBenchExecResult{PdrInv}{KinductionDfStaticZeroOneTTTrueNotSolvedByKinductionPlain}{Correct}{}{Walltime}{Stdev}{67.94019020714819522190392966}%
\StoreBenchExecResult{PdrInv}{KinductionDfStaticZeroOneTTTrueNotSolvedByKinductionPlain}{Correct}{True}{Count}{}{950}%
\StoreBenchExecResult{PdrInv}{KinductionDfStaticZeroOneTTTrueNotSolvedByKinductionPlain}{Correct}{True}{Cputime}{}{39080.406451603}%
\StoreBenchExecResult{PdrInv}{KinductionDfStaticZeroOneTTTrueNotSolvedByKinductionPlain}{Correct}{True}{Cputime}{Avg}{41.13726994905578947368421053}%
\StoreBenchExecResult{PdrInv}{KinductionDfStaticZeroOneTTTrueNotSolvedByKinductionPlain}{Correct}{True}{Cputime}{Median}{7.107670114}%
\StoreBenchExecResult{PdrInv}{KinductionDfStaticZeroOneTTTrueNotSolvedByKinductionPlain}{Correct}{True}{Cputime}{Min}{3.351895368}%
\StoreBenchExecResult{PdrInv}{KinductionDfStaticZeroOneTTTrueNotSolvedByKinductionPlain}{Correct}{True}{Cputime}{Max}{896.584723939}%
\StoreBenchExecResult{PdrInv}{KinductionDfStaticZeroOneTTTrueNotSolvedByKinductionPlain}{Correct}{True}{Cputime}{Stdev}{113.1887613798698222461115770}%
\StoreBenchExecResult{PdrInv}{KinductionDfStaticZeroOneTTTrueNotSolvedByKinductionPlain}{Correct}{True}{Walltime}{}{21387.28514075158}%
\StoreBenchExecResult{PdrInv}{KinductionDfStaticZeroOneTTTrueNotSolvedByKinductionPlain}{Correct}{True}{Walltime}{Avg}{22.51293172710692631578947368}%
\StoreBenchExecResult{PdrInv}{KinductionDfStaticZeroOneTTTrueNotSolvedByKinductionPlain}{Correct}{True}{Walltime}{Median}{3.74982059002}%
\StoreBenchExecResult{PdrInv}{KinductionDfStaticZeroOneTTTrueNotSolvedByKinductionPlain}{Correct}{True}{Walltime}{Min}{1.86370992661}%
\StoreBenchExecResult{PdrInv}{KinductionDfStaticZeroOneTTTrueNotSolvedByKinductionPlain}{Correct}{True}{Walltime}{Max}{780.438195944}%
\StoreBenchExecResult{PdrInv}{KinductionDfStaticZeroOneTTTrueNotSolvedByKinductionPlain}{Correct}{True}{Walltime}{Stdev}{67.94019020714819522190392966}%
\StoreBenchExecResult{PdrInv}{KinductionDfStaticZeroOneTTTrueNotSolvedByKinductionPlain}{Wrong}{True}{Count}{}{0}%
\StoreBenchExecResult{PdrInv}{KinductionDfStaticZeroOneTTTrueNotSolvedByKinductionPlain}{Wrong}{True}{Cputime}{}{0}%
\StoreBenchExecResult{PdrInv}{KinductionDfStaticZeroOneTTTrueNotSolvedByKinductionPlain}{Wrong}{True}{Cputime}{Avg}{None}%
\StoreBenchExecResult{PdrInv}{KinductionDfStaticZeroOneTTTrueNotSolvedByKinductionPlain}{Wrong}{True}{Cputime}{Median}{None}%
\StoreBenchExecResult{PdrInv}{KinductionDfStaticZeroOneTTTrueNotSolvedByKinductionPlain}{Wrong}{True}{Cputime}{Min}{None}%
\StoreBenchExecResult{PdrInv}{KinductionDfStaticZeroOneTTTrueNotSolvedByKinductionPlain}{Wrong}{True}{Cputime}{Max}{None}%
\StoreBenchExecResult{PdrInv}{KinductionDfStaticZeroOneTTTrueNotSolvedByKinductionPlain}{Wrong}{True}{Cputime}{Stdev}{None}%
\StoreBenchExecResult{PdrInv}{KinductionDfStaticZeroOneTTTrueNotSolvedByKinductionPlain}{Wrong}{True}{Walltime}{}{0}%
\StoreBenchExecResult{PdrInv}{KinductionDfStaticZeroOneTTTrueNotSolvedByKinductionPlain}{Wrong}{True}{Walltime}{Avg}{None}%
\StoreBenchExecResult{PdrInv}{KinductionDfStaticZeroOneTTTrueNotSolvedByKinductionPlain}{Wrong}{True}{Walltime}{Median}{None}%
\StoreBenchExecResult{PdrInv}{KinductionDfStaticZeroOneTTTrueNotSolvedByKinductionPlain}{Wrong}{True}{Walltime}{Min}{None}%
\StoreBenchExecResult{PdrInv}{KinductionDfStaticZeroOneTTTrueNotSolvedByKinductionPlain}{Wrong}{True}{Walltime}{Max}{None}%
\StoreBenchExecResult{PdrInv}{KinductionDfStaticZeroOneTTTrueNotSolvedByKinductionPlain}{Wrong}{True}{Walltime}{Stdev}{None}%
\StoreBenchExecResult{PdrInv}{KinductionDfStaticZeroOneTTTrueNotSolvedByKinductionPlain}{Error}{}{Count}{}{1943}%
\StoreBenchExecResult{PdrInv}{KinductionDfStaticZeroOneTTTrueNotSolvedByKinductionPlain}{Error}{}{Cputime}{}{1580527.553081074}%
\StoreBenchExecResult{PdrInv}{KinductionDfStaticZeroOneTTTrueNotSolvedByKinductionPlain}{Error}{}{Cputime}{Avg}{813.4470165111034482758620690}%
\StoreBenchExecResult{PdrInv}{KinductionDfStaticZeroOneTTTrueNotSolvedByKinductionPlain}{Error}{}{Cputime}{Median}{901.720366508}%
\StoreBenchExecResult{PdrInv}{KinductionDfStaticZeroOneTTTrueNotSolvedByKinductionPlain}{Error}{}{Cputime}{Min}{2.633111638}%
\StoreBenchExecResult{PdrInv}{KinductionDfStaticZeroOneTTTrueNotSolvedByKinductionPlain}{Error}{}{Cputime}{Max}{1002.33951193}%
\StoreBenchExecResult{PdrInv}{KinductionDfStaticZeroOneTTTrueNotSolvedByKinductionPlain}{Error}{}{Cputime}{Stdev}{240.9543891885713969412903515}%
\StoreBenchExecResult{PdrInv}{KinductionDfStaticZeroOneTTTrueNotSolvedByKinductionPlain}{Error}{}{Walltime}{}{1096714.78254100718}%
\StoreBenchExecResult{PdrInv}{KinductionDfStaticZeroOneTTTrueNotSolvedByKinductionPlain}{Error}{}{Walltime}{Avg}{564.4440465985626248069994853}%
\StoreBenchExecResult{PdrInv}{KinductionDfStaticZeroOneTTTrueNotSolvedByKinductionPlain}{Error}{}{Walltime}{Median}{454.932538033}%
\StoreBenchExecResult{PdrInv}{KinductionDfStaticZeroOneTTTrueNotSolvedByKinductionPlain}{Error}{}{Walltime}{Min}{1.4413330555}%
\StoreBenchExecResult{PdrInv}{KinductionDfStaticZeroOneTTTrueNotSolvedByKinductionPlain}{Error}{}{Walltime}{Max}{920.880501032}%
\StoreBenchExecResult{PdrInv}{KinductionDfStaticZeroOneTTTrueNotSolvedByKinductionPlain}{Error}{}{Walltime}{Stdev}{247.9716717376005737752776513}%
\StoreBenchExecResult{PdrInv}{KinductionDfStaticZeroOneTTTrueNotSolvedByKinductionPlain}{Error}{Assertion}{Count}{}{2}%
\StoreBenchExecResult{PdrInv}{KinductionDfStaticZeroOneTTTrueNotSolvedByKinductionPlain}{Error}{Assertion}{Cputime}{}{6.989160864}%
\StoreBenchExecResult{PdrInv}{KinductionDfStaticZeroOneTTTrueNotSolvedByKinductionPlain}{Error}{Assertion}{Cputime}{Avg}{3.494580432}%
\StoreBenchExecResult{PdrInv}{KinductionDfStaticZeroOneTTTrueNotSolvedByKinductionPlain}{Error}{Assertion}{Cputime}{Median}{3.494580432}%
\StoreBenchExecResult{PdrInv}{KinductionDfStaticZeroOneTTTrueNotSolvedByKinductionPlain}{Error}{Assertion}{Cputime}{Min}{3.254889053}%
\StoreBenchExecResult{PdrInv}{KinductionDfStaticZeroOneTTTrueNotSolvedByKinductionPlain}{Error}{Assertion}{Cputime}{Max}{3.734271811}%
\StoreBenchExecResult{PdrInv}{KinductionDfStaticZeroOneTTTrueNotSolvedByKinductionPlain}{Error}{Assertion}{Cputime}{Stdev}{0.239691379}%
\StoreBenchExecResult{PdrInv}{KinductionDfStaticZeroOneTTTrueNotSolvedByKinductionPlain}{Error}{Assertion}{Walltime}{}{3.87976098061}%
\StoreBenchExecResult{PdrInv}{KinductionDfStaticZeroOneTTTrueNotSolvedByKinductionPlain}{Error}{Assertion}{Walltime}{Avg}{1.939880490305}%
\StoreBenchExecResult{PdrInv}{KinductionDfStaticZeroOneTTTrueNotSolvedByKinductionPlain}{Error}{Assertion}{Walltime}{Median}{1.939880490305}%
\StoreBenchExecResult{PdrInv}{KinductionDfStaticZeroOneTTTrueNotSolvedByKinductionPlain}{Error}{Assertion}{Walltime}{Min}{1.7997341156}%
\StoreBenchExecResult{PdrInv}{KinductionDfStaticZeroOneTTTrueNotSolvedByKinductionPlain}{Error}{Assertion}{Walltime}{Max}{2.08002686501}%
\StoreBenchExecResult{PdrInv}{KinductionDfStaticZeroOneTTTrueNotSolvedByKinductionPlain}{Error}{Assertion}{Walltime}{Stdev}{0.140146374705}%
\StoreBenchExecResult{PdrInv}{KinductionDfStaticZeroOneTTTrueNotSolvedByKinductionPlain}{Error}{Error}{Count}{}{130}%
\StoreBenchExecResult{PdrInv}{KinductionDfStaticZeroOneTTTrueNotSolvedByKinductionPlain}{Error}{Error}{Cputime}{}{22374.184968762}%
\StoreBenchExecResult{PdrInv}{KinductionDfStaticZeroOneTTTrueNotSolvedByKinductionPlain}{Error}{Error}{Cputime}{Avg}{172.1091151443230769230769231}%
\StoreBenchExecResult{PdrInv}{KinductionDfStaticZeroOneTTTrueNotSolvedByKinductionPlain}{Error}{Error}{Cputime}{Median}{111.272374487}%
\StoreBenchExecResult{PdrInv}{KinductionDfStaticZeroOneTTTrueNotSolvedByKinductionPlain}{Error}{Error}{Cputime}{Min}{2.633111638}%
\StoreBenchExecResult{PdrInv}{KinductionDfStaticZeroOneTTTrueNotSolvedByKinductionPlain}{Error}{Error}{Cputime}{Max}{859.122657289}%
\StoreBenchExecResult{PdrInv}{KinductionDfStaticZeroOneTTTrueNotSolvedByKinductionPlain}{Error}{Error}{Cputime}{Stdev}{193.4155195178754855666892763}%
\StoreBenchExecResult{PdrInv}{KinductionDfStaticZeroOneTTTrueNotSolvedByKinductionPlain}{Error}{Error}{Walltime}{}{18822.95931529830}%
\StoreBenchExecResult{PdrInv}{KinductionDfStaticZeroOneTTTrueNotSolvedByKinductionPlain}{Error}{Error}{Walltime}{Avg}{144.7919947330638461538461538}%
\StoreBenchExecResult{PdrInv}{KinductionDfStaticZeroOneTTTrueNotSolvedByKinductionPlain}{Error}{Error}{Walltime}{Median}{85.8972519636}%
\StoreBenchExecResult{PdrInv}{KinductionDfStaticZeroOneTTTrueNotSolvedByKinductionPlain}{Error}{Error}{Walltime}{Min}{1.4413330555}%
\StoreBenchExecResult{PdrInv}{KinductionDfStaticZeroOneTTTrueNotSolvedByKinductionPlain}{Error}{Error}{Walltime}{Max}{841.78747201}%
\StoreBenchExecResult{PdrInv}{KinductionDfStaticZeroOneTTTrueNotSolvedByKinductionPlain}{Error}{Error}{Walltime}{Stdev}{172.6448903893023680695575260}%
\StoreBenchExecResult{PdrInv}{KinductionDfStaticZeroOneTTTrueNotSolvedByKinductionPlain}{Error}{Exception}{Count}{}{16}%
\StoreBenchExecResult{PdrInv}{KinductionDfStaticZeroOneTTTrueNotSolvedByKinductionPlain}{Error}{Exception}{Cputime}{}{3468.257278562}%
\StoreBenchExecResult{PdrInv}{KinductionDfStaticZeroOneTTTrueNotSolvedByKinductionPlain}{Error}{Exception}{Cputime}{Avg}{216.766079910125}%
\StoreBenchExecResult{PdrInv}{KinductionDfStaticZeroOneTTTrueNotSolvedByKinductionPlain}{Error}{Exception}{Cputime}{Median}{195.4111243865}%
\StoreBenchExecResult{PdrInv}{KinductionDfStaticZeroOneTTTrueNotSolvedByKinductionPlain}{Error}{Exception}{Cputime}{Min}{14.921760009}%
\StoreBenchExecResult{PdrInv}{KinductionDfStaticZeroOneTTTrueNotSolvedByKinductionPlain}{Error}{Exception}{Cputime}{Max}{605.497592878}%
\StoreBenchExecResult{PdrInv}{KinductionDfStaticZeroOneTTTrueNotSolvedByKinductionPlain}{Error}{Exception}{Cputime}{Stdev}{166.2337177423187469322115240}%
\StoreBenchExecResult{PdrInv}{KinductionDfStaticZeroOneTTTrueNotSolvedByKinductionPlain}{Error}{Exception}{Walltime}{}{1756.28703856487}%
\StoreBenchExecResult{PdrInv}{KinductionDfStaticZeroOneTTTrueNotSolvedByKinductionPlain}{Error}{Exception}{Walltime}{Avg}{109.767939910304375}%
\StoreBenchExecResult{PdrInv}{KinductionDfStaticZeroOneTTTrueNotSolvedByKinductionPlain}{Error}{Exception}{Walltime}{Median}{98.1585280896}%
\StoreBenchExecResult{PdrInv}{KinductionDfStaticZeroOneTTTrueNotSolvedByKinductionPlain}{Error}{Exception}{Walltime}{Min}{7.62086105347}%
\StoreBenchExecResult{PdrInv}{KinductionDfStaticZeroOneTTTrueNotSolvedByKinductionPlain}{Error}{Exception}{Walltime}{Max}{303.745280981}%
\StoreBenchExecResult{PdrInv}{KinductionDfStaticZeroOneTTTrueNotSolvedByKinductionPlain}{Error}{Exception}{Walltime}{Stdev}{82.86144293151853717429017562}%
\StoreBenchExecResult{PdrInv}{KinductionDfStaticZeroOneTTTrueNotSolvedByKinductionPlain}{Error}{OutOfJavaMemory}{Count}{}{5}%
\StoreBenchExecResult{PdrInv}{KinductionDfStaticZeroOneTTTrueNotSolvedByKinductionPlain}{Error}{OutOfJavaMemory}{Cputime}{}{1816.973756593}%
\StoreBenchExecResult{PdrInv}{KinductionDfStaticZeroOneTTTrueNotSolvedByKinductionPlain}{Error}{OutOfJavaMemory}{Cputime}{Avg}{363.3947513186}%
\StoreBenchExecResult{PdrInv}{KinductionDfStaticZeroOneTTTrueNotSolvedByKinductionPlain}{Error}{OutOfJavaMemory}{Cputime}{Median}{293.257636372}%
\StoreBenchExecResult{PdrInv}{KinductionDfStaticZeroOneTTTrueNotSolvedByKinductionPlain}{Error}{OutOfJavaMemory}{Cputime}{Min}{176.983807877}%
\StoreBenchExecResult{PdrInv}{KinductionDfStaticZeroOneTTTrueNotSolvedByKinductionPlain}{Error}{OutOfJavaMemory}{Cputime}{Max}{562.974894718}%
\StoreBenchExecResult{PdrInv}{KinductionDfStaticZeroOneTTTrueNotSolvedByKinductionPlain}{Error}{OutOfJavaMemory}{Cputime}{Stdev}{150.3912701454620946633744732}%
\StoreBenchExecResult{PdrInv}{KinductionDfStaticZeroOneTTTrueNotSolvedByKinductionPlain}{Error}{OutOfJavaMemory}{Walltime}{}{1112.236532211}%
\StoreBenchExecResult{PdrInv}{KinductionDfStaticZeroOneTTTrueNotSolvedByKinductionPlain}{Error}{OutOfJavaMemory}{Walltime}{Avg}{222.4473064422}%
\StoreBenchExecResult{PdrInv}{KinductionDfStaticZeroOneTTTrueNotSolvedByKinductionPlain}{Error}{OutOfJavaMemory}{Walltime}{Median}{195.936332941}%
\StoreBenchExecResult{PdrInv}{KinductionDfStaticZeroOneTTTrueNotSolvedByKinductionPlain}{Error}{OutOfJavaMemory}{Walltime}{Min}{106.547775984}%
\StoreBenchExecResult{PdrInv}{KinductionDfStaticZeroOneTTTrueNotSolvedByKinductionPlain}{Error}{OutOfJavaMemory}{Walltime}{Max}{359.09446311}%
\StoreBenchExecResult{PdrInv}{KinductionDfStaticZeroOneTTTrueNotSolvedByKinductionPlain}{Error}{OutOfJavaMemory}{Walltime}{Stdev}{85.59581734171590415197509505}%
\StoreBenchExecResult{PdrInv}{KinductionDfStaticZeroOneTTTrueNotSolvedByKinductionPlain}{Error}{OutOfMemory}{Count}{}{130}%
\StoreBenchExecResult{PdrInv}{KinductionDfStaticZeroOneTTTrueNotSolvedByKinductionPlain}{Error}{OutOfMemory}{Cputime}{}{48222.072390087}%
\StoreBenchExecResult{PdrInv}{KinductionDfStaticZeroOneTTTrueNotSolvedByKinductionPlain}{Error}{OutOfMemory}{Cputime}{Avg}{370.9390183852846153846153846}%
\StoreBenchExecResult{PdrInv}{KinductionDfStaticZeroOneTTTrueNotSolvedByKinductionPlain}{Error}{OutOfMemory}{Cputime}{Median}{311.090992006}%
\StoreBenchExecResult{PdrInv}{KinductionDfStaticZeroOneTTTrueNotSolvedByKinductionPlain}{Error}{OutOfMemory}{Cputime}{Min}{159.375222151}%
\StoreBenchExecResult{PdrInv}{KinductionDfStaticZeroOneTTTrueNotSolvedByKinductionPlain}{Error}{OutOfMemory}{Cputime}{Max}{891.59037469}%
\StoreBenchExecResult{PdrInv}{KinductionDfStaticZeroOneTTTrueNotSolvedByKinductionPlain}{Error}{OutOfMemory}{Cputime}{Stdev}{208.9565559223031491860889751}%
\StoreBenchExecResult{PdrInv}{KinductionDfStaticZeroOneTTTrueNotSolvedByKinductionPlain}{Error}{OutOfMemory}{Walltime}{}{40522.9398469924}%
\StoreBenchExecResult{PdrInv}{KinductionDfStaticZeroOneTTTrueNotSolvedByKinductionPlain}{Error}{OutOfMemory}{Walltime}{Avg}{311.7149218999415384615384615}%
\StoreBenchExecResult{PdrInv}{KinductionDfStaticZeroOneTTTrueNotSolvedByKinductionPlain}{Error}{OutOfMemory}{Walltime}{Median}{190.390333891}%
\StoreBenchExecResult{PdrInv}{KinductionDfStaticZeroOneTTTrueNotSolvedByKinductionPlain}{Error}{OutOfMemory}{Walltime}{Min}{85.8784229755}%
\StoreBenchExecResult{PdrInv}{KinductionDfStaticZeroOneTTTrueNotSolvedByKinductionPlain}{Error}{OutOfMemory}{Walltime}{Max}{866.566967964}%
\StoreBenchExecResult{PdrInv}{KinductionDfStaticZeroOneTTTrueNotSolvedByKinductionPlain}{Error}{OutOfMemory}{Walltime}{Stdev}{233.8710140417263807672318444}%
\StoreBenchExecResult{PdrInv}{KinductionDfStaticZeroOneTTTrueNotSolvedByKinductionPlain}{Error}{Timeout}{Count}{}{1660}%
\StoreBenchExecResult{PdrInv}{KinductionDfStaticZeroOneTTTrueNotSolvedByKinductionPlain}{Error}{Timeout}{Cputime}{}{1504639.075526206}%
\StoreBenchExecResult{PdrInv}{KinductionDfStaticZeroOneTTTrueNotSolvedByKinductionPlain}{Error}{Timeout}{Cputime}{Avg}{906.4090816422927710843373494}%
\StoreBenchExecResult{PdrInv}{KinductionDfStaticZeroOneTTTrueNotSolvedByKinductionPlain}{Error}{Timeout}{Cputime}{Median}{902.0866164735}%
\StoreBenchExecResult{PdrInv}{KinductionDfStaticZeroOneTTTrueNotSolvedByKinductionPlain}{Error}{Timeout}{Cputime}{Min}{900.911325191}%
\StoreBenchExecResult{PdrInv}{KinductionDfStaticZeroOneTTTrueNotSolvedByKinductionPlain}{Error}{Timeout}{Cputime}{Max}{1002.33951193}%
\StoreBenchExecResult{PdrInv}{KinductionDfStaticZeroOneTTTrueNotSolvedByKinductionPlain}{Error}{Timeout}{Cputime}{Stdev}{15.91731859390448994717804566}%
\StoreBenchExecResult{PdrInv}{KinductionDfStaticZeroOneTTTrueNotSolvedByKinductionPlain}{Error}{Timeout}{Walltime}{}{1034496.480046960}%
\StoreBenchExecResult{PdrInv}{KinductionDfStaticZeroOneTTTrueNotSolvedByKinductionPlain}{Error}{Timeout}{Walltime}{Avg}{623.1906506306987951807228916}%
\StoreBenchExecResult{PdrInv}{KinductionDfStaticZeroOneTTTrueNotSolvedByKinductionPlain}{Error}{Timeout}{Walltime}{Median}{459.2154369355}%
\StoreBenchExecResult{PdrInv}{KinductionDfStaticZeroOneTTTrueNotSolvedByKinductionPlain}{Error}{Timeout}{Walltime}{Min}{451.102283955}%
\StoreBenchExecResult{PdrInv}{KinductionDfStaticZeroOneTTTrueNotSolvedByKinductionPlain}{Error}{Timeout}{Walltime}{Max}{920.880501032}%
\StoreBenchExecResult{PdrInv}{KinductionDfStaticZeroOneTTTrueNotSolvedByKinductionPlain}{Error}{Timeout}{Walltime}{Stdev}{200.7390051217924126714160345}%
\providecommand\StoreBenchExecResult[7]{\expandafter\newcommand\csname#1#2#3#4#5#6\endcsname{#7}}%
\StoreBenchExecResult{PdrInv}{KinductionDfStaticZeroOneTT}{Total}{}{Count}{}{5591}%
\StoreBenchExecResult{PdrInv}{KinductionDfStaticZeroOneTT}{Total}{}{Cputime}{}{2203202.961078871}%
\StoreBenchExecResult{PdrInv}{KinductionDfStaticZeroOneTT}{Total}{}{Cputime}{Avg}{394.0624147878502951171525666}%
\StoreBenchExecResult{PdrInv}{KinductionDfStaticZeroOneTT}{Total}{}{Cputime}{Median}{113.542403566}%
\StoreBenchExecResult{PdrInv}{KinductionDfStaticZeroOneTT}{Total}{}{Cputime}{Min}{2.633111638}%
\StoreBenchExecResult{PdrInv}{KinductionDfStaticZeroOneTT}{Total}{}{Cputime}{Max}{1002.33951193}%
\StoreBenchExecResult{PdrInv}{KinductionDfStaticZeroOneTT}{Total}{}{Cputime}{Stdev}{419.0586414534723612926103041}%
\StoreBenchExecResult{PdrInv}{KinductionDfStaticZeroOneTT}{Total}{}{Walltime}{}{1550822.81704494393}%
\StoreBenchExecResult{PdrInv}{KinductionDfStaticZeroOneTT}{Total}{}{Walltime}{Avg}{277.3784326676701717045251297}%
\StoreBenchExecResult{PdrInv}{KinductionDfStaticZeroOneTT}{Total}{}{Walltime}{Median}{75.5665559769}%
\StoreBenchExecResult{PdrInv}{KinductionDfStaticZeroOneTT}{Total}{}{Walltime}{Min}{1.4413330555}%
\StoreBenchExecResult{PdrInv}{KinductionDfStaticZeroOneTT}{Total}{}{Walltime}{Max}{921.056605816}%
\StoreBenchExecResult{PdrInv}{KinductionDfStaticZeroOneTT}{Total}{}{Walltime}{Stdev}{321.2618115719674917293354119}%
\StoreBenchExecResult{PdrInv}{KinductionDfStaticZeroOneTT}{Correct}{}{Count}{}{2997}%
\StoreBenchExecResult{PdrInv}{KinductionDfStaticZeroOneTT}{Correct}{}{Cputime}{}{164736.483130135}%
\StoreBenchExecResult{PdrInv}{KinductionDfStaticZeroOneTT}{Correct}{}{Cputime}{Avg}{54.96712817154988321654988322}%
\StoreBenchExecResult{PdrInv}{KinductionDfStaticZeroOneTT}{Correct}{}{Cputime}{Median}{9.427844447}%
\StoreBenchExecResult{PdrInv}{KinductionDfStaticZeroOneTT}{Correct}{}{Cputime}{Min}{3.052078842}%
\StoreBenchExecResult{PdrInv}{KinductionDfStaticZeroOneTT}{Correct}{}{Cputime}{Max}{899.693369081}%
\StoreBenchExecResult{PdrInv}{KinductionDfStaticZeroOneTT}{Correct}{}{Cputime}{Stdev}{133.2573088385227525595107154}%
\StoreBenchExecResult{PdrInv}{KinductionDfStaticZeroOneTT}{Correct}{}{Walltime}{}{115449.44951200126}%
\StoreBenchExecResult{PdrInv}{KinductionDfStaticZeroOneTT}{Correct}{}{Walltime}{Avg}{38.52167150884259592926259593}%
\StoreBenchExecResult{PdrInv}{KinductionDfStaticZeroOneTT}{Correct}{}{Walltime}{Median}{4.98909091949}%
\StoreBenchExecResult{PdrInv}{KinductionDfStaticZeroOneTT}{Correct}{}{Walltime}{Min}{1.68862104416}%
\StoreBenchExecResult{PdrInv}{KinductionDfStaticZeroOneTT}{Correct}{}{Walltime}{Max}{877.609319925}%
\StoreBenchExecResult{PdrInv}{KinductionDfStaticZeroOneTT}{Correct}{}{Walltime}{Stdev}{109.4341137706197110951861814}%
\StoreBenchExecResult{PdrInv}{KinductionDfStaticZeroOneTT}{Correct}{False}{Count}{}{818}%
\StoreBenchExecResult{PdrInv}{KinductionDfStaticZeroOneTT}{Correct}{False}{Cputime}{}{74609.970337497}%
\StoreBenchExecResult{PdrInv}{KinductionDfStaticZeroOneTT}{Correct}{False}{Cputime}{Avg}{91.21023268642665036674816626}%
\StoreBenchExecResult{PdrInv}{KinductionDfStaticZeroOneTT}{Correct}{False}{Cputime}{Median}{21.753853096}%
\StoreBenchExecResult{PdrInv}{KinductionDfStaticZeroOneTT}{Correct}{False}{Cputime}{Min}{3.106383958}%
\StoreBenchExecResult{PdrInv}{KinductionDfStaticZeroOneTT}{Correct}{False}{Cputime}{Max}{899.693369081}%
\StoreBenchExecResult{PdrInv}{KinductionDfStaticZeroOneTT}{Correct}{False}{Cputime}{Stdev}{180.7808928845225085352432477}%
\StoreBenchExecResult{PdrInv}{KinductionDfStaticZeroOneTT}{Correct}{False}{Walltime}{}{59446.26409410979}%
\StoreBenchExecResult{PdrInv}{KinductionDfStaticZeroOneTT}{Correct}{False}{Walltime}{Avg}{72.67269449157676039119804401}%
\StoreBenchExecResult{PdrInv}{KinductionDfStaticZeroOneTT}{Correct}{False}{Walltime}{Median}{11.9742344618}%
\StoreBenchExecResult{PdrInv}{KinductionDfStaticZeroOneTT}{Correct}{False}{Walltime}{Min}{1.75132894516}%
\StoreBenchExecResult{PdrInv}{KinductionDfStaticZeroOneTT}{Correct}{False}{Walltime}{Max}{877.609319925}%
\StoreBenchExecResult{PdrInv}{KinductionDfStaticZeroOneTT}{Correct}{False}{Walltime}{Stdev}{164.7712095486097088115676964}%
\StoreBenchExecResult{PdrInv}{KinductionDfStaticZeroOneTT}{Correct}{True}{Count}{}{2179}%
\StoreBenchExecResult{PdrInv}{KinductionDfStaticZeroOneTT}{Correct}{True}{Cputime}{}{90126.512792638}%
\StoreBenchExecResult{PdrInv}{KinductionDfStaticZeroOneTT}{Correct}{True}{Cputime}{Avg}{41.36141018478109224414869206}%
\StoreBenchExecResult{PdrInv}{KinductionDfStaticZeroOneTT}{Correct}{True}{Cputime}{Median}{7.139609194}%
\StoreBenchExecResult{PdrInv}{KinductionDfStaticZeroOneTT}{Correct}{True}{Cputime}{Min}{3.052078842}%
\StoreBenchExecResult{PdrInv}{KinductionDfStaticZeroOneTT}{Correct}{True}{Cputime}{Max}{896.584723939}%
\StoreBenchExecResult{PdrInv}{KinductionDfStaticZeroOneTT}{Correct}{True}{Cputime}{Stdev}{107.1293868887125313546935307}%
\StoreBenchExecResult{PdrInv}{KinductionDfStaticZeroOneTT}{Correct}{True}{Walltime}{}{56003.18541789147}%
\StoreBenchExecResult{PdrInv}{KinductionDfStaticZeroOneTT}{Correct}{True}{Walltime}{Avg}{25.70132419361701239100504819}%
\StoreBenchExecResult{PdrInv}{KinductionDfStaticZeroOneTT}{Correct}{True}{Walltime}{Median}{3.79638004303}%
\StoreBenchExecResult{PdrInv}{KinductionDfStaticZeroOneTT}{Correct}{True}{Walltime}{Min}{1.68862104416}%
\StoreBenchExecResult{PdrInv}{KinductionDfStaticZeroOneTT}{Correct}{True}{Walltime}{Max}{845.474861145}%
\StoreBenchExecResult{PdrInv}{KinductionDfStaticZeroOneTT}{Correct}{True}{Walltime}{Stdev}{75.34849347605701889801348927}%
\StoreBenchExecResult{PdrInv}{KinductionDfStaticZeroOneTT}{Wrong}{True}{Count}{}{0}%
\StoreBenchExecResult{PdrInv}{KinductionDfStaticZeroOneTT}{Wrong}{True}{Cputime}{}{0}%
\StoreBenchExecResult{PdrInv}{KinductionDfStaticZeroOneTT}{Wrong}{True}{Cputime}{Avg}{None}%
\StoreBenchExecResult{PdrInv}{KinductionDfStaticZeroOneTT}{Wrong}{True}{Cputime}{Median}{None}%
\StoreBenchExecResult{PdrInv}{KinductionDfStaticZeroOneTT}{Wrong}{True}{Cputime}{Min}{None}%
\StoreBenchExecResult{PdrInv}{KinductionDfStaticZeroOneTT}{Wrong}{True}{Cputime}{Max}{None}%
\StoreBenchExecResult{PdrInv}{KinductionDfStaticZeroOneTT}{Wrong}{True}{Cputime}{Stdev}{None}%
\StoreBenchExecResult{PdrInv}{KinductionDfStaticZeroOneTT}{Wrong}{True}{Walltime}{}{0}%
\StoreBenchExecResult{PdrInv}{KinductionDfStaticZeroOneTT}{Wrong}{True}{Walltime}{Avg}{None}%
\StoreBenchExecResult{PdrInv}{KinductionDfStaticZeroOneTT}{Wrong}{True}{Walltime}{Median}{None}%
\StoreBenchExecResult{PdrInv}{KinductionDfStaticZeroOneTT}{Wrong}{True}{Walltime}{Min}{None}%
\StoreBenchExecResult{PdrInv}{KinductionDfStaticZeroOneTT}{Wrong}{True}{Walltime}{Max}{None}%
\StoreBenchExecResult{PdrInv}{KinductionDfStaticZeroOneTT}{Wrong}{True}{Walltime}{Stdev}{None}%
\StoreBenchExecResult{PdrInv}{KinductionDfStaticZeroOneTT}{Error}{}{Count}{}{2592}%
\StoreBenchExecResult{PdrInv}{KinductionDfStaticZeroOneTT}{Error}{}{Cputime}{}{2038443.598081342}%
\StoreBenchExecResult{PdrInv}{KinductionDfStaticZeroOneTT}{Error}{}{Cputime}{Avg}{786.4365733338510802469135802}%
\StoreBenchExecResult{PdrInv}{KinductionDfStaticZeroOneTT}{Error}{}{Cputime}{Median}{901.7299303795}%
\StoreBenchExecResult{PdrInv}{KinductionDfStaticZeroOneTT}{Error}{}{Cputime}{Min}{2.633111638}%
\StoreBenchExecResult{PdrInv}{KinductionDfStaticZeroOneTT}{Error}{}{Cputime}{Max}{1002.33951193}%
\StoreBenchExecResult{PdrInv}{KinductionDfStaticZeroOneTT}{Error}{}{Cputime}{Stdev}{266.9085061515243914092431933}%
\StoreBenchExecResult{PdrInv}{KinductionDfStaticZeroOneTT}{Error}{}{Walltime}{}{1435360.98295495314}%
\StoreBenchExecResult{PdrInv}{KinductionDfStaticZeroOneTT}{Error}{}{Walltime}{Avg}{553.7658113252134027777777778}%
\StoreBenchExecResult{PdrInv}{KinductionDfStaticZeroOneTT}{Error}{}{Walltime}{Median}{455.7373874185}%
\StoreBenchExecResult{PdrInv}{KinductionDfStaticZeroOneTT}{Error}{}{Walltime}{Min}{1.4413330555}%
\StoreBenchExecResult{PdrInv}{KinductionDfStaticZeroOneTT}{Error}{}{Walltime}{Max}{921.056605816}%
\StoreBenchExecResult{PdrInv}{KinductionDfStaticZeroOneTT}{Error}{}{Walltime}{Stdev}{257.6112701822149673146841316}%
\StoreBenchExecResult{PdrInv}{KinductionDfStaticZeroOneTT}{Error}{Assertion}{Count}{}{4}%
\StoreBenchExecResult{PdrInv}{KinductionDfStaticZeroOneTT}{Error}{Assertion}{Cputime}{}{13.313761120}%
\StoreBenchExecResult{PdrInv}{KinductionDfStaticZeroOneTT}{Error}{Assertion}{Cputime}{Avg}{3.328440280}%
\StoreBenchExecResult{PdrInv}{KinductionDfStaticZeroOneTT}{Error}{Assertion}{Cputime}{Median}{3.2577418075}%
\StoreBenchExecResult{PdrInv}{KinductionDfStaticZeroOneTT}{Error}{Assertion}{Cputime}{Min}{3.064005694}%
\StoreBenchExecResult{PdrInv}{KinductionDfStaticZeroOneTT}{Error}{Assertion}{Cputime}{Max}{3.734271811}%
\StoreBenchExecResult{PdrInv}{KinductionDfStaticZeroOneTT}{Error}{Assertion}{Cputime}{Stdev}{0.2473043197514635054459008457}%
\StoreBenchExecResult{PdrInv}{KinductionDfStaticZeroOneTT}{Error}{Assertion}{Walltime}{}{7.40536689758}%
\StoreBenchExecResult{PdrInv}{KinductionDfStaticZeroOneTT}{Error}{Assertion}{Walltime}{Avg}{1.851341724395}%
\StoreBenchExecResult{PdrInv}{KinductionDfStaticZeroOneTT}{Error}{Assertion}{Walltime}{Median}{1.811854481695}%
\StoreBenchExecResult{PdrInv}{KinductionDfStaticZeroOneTT}{Error}{Assertion}{Walltime}{Min}{1.70163106918}%
\StoreBenchExecResult{PdrInv}{KinductionDfStaticZeroOneTT}{Error}{Assertion}{Walltime}{Max}{2.08002686501}%
\StoreBenchExecResult{PdrInv}{KinductionDfStaticZeroOneTT}{Error}{Assertion}{Walltime}{Stdev}{0.1397519812414639677981385904}%
\StoreBenchExecResult{PdrInv}{KinductionDfStaticZeroOneTT}{Error}{Error}{Count}{}{198}%
\StoreBenchExecResult{PdrInv}{KinductionDfStaticZeroOneTT}{Error}{Error}{Cputime}{}{36890.456755785}%
\StoreBenchExecResult{PdrInv}{KinductionDfStaticZeroOneTT}{Error}{Error}{Cputime}{Avg}{186.3154381605303030303030303}%
\StoreBenchExecResult{PdrInv}{KinductionDfStaticZeroOneTT}{Error}{Error}{Cputime}{Median}{123.5362712405}%
\StoreBenchExecResult{PdrInv}{KinductionDfStaticZeroOneTT}{Error}{Error}{Cputime}{Min}{2.633111638}%
\StoreBenchExecResult{PdrInv}{KinductionDfStaticZeroOneTT}{Error}{Error}{Cputime}{Max}{859.122657289}%
\StoreBenchExecResult{PdrInv}{KinductionDfStaticZeroOneTT}{Error}{Error}{Cputime}{Stdev}{193.4284181716213241374687047}%
\StoreBenchExecResult{PdrInv}{KinductionDfStaticZeroOneTT}{Error}{Error}{Walltime}{}{30556.05902600013}%
\StoreBenchExecResult{PdrInv}{KinductionDfStaticZeroOneTT}{Error}{Error}{Walltime}{Avg}{154.3235304343440909090909091}%
\StoreBenchExecResult{PdrInv}{KinductionDfStaticZeroOneTT}{Error}{Error}{Walltime}{Median}{98.43613958355}%
\StoreBenchExecResult{PdrInv}{KinductionDfStaticZeroOneTT}{Error}{Error}{Walltime}{Min}{1.4413330555}%
\StoreBenchExecResult{PdrInv}{KinductionDfStaticZeroOneTT}{Error}{Error}{Walltime}{Max}{841.78747201}%
\StoreBenchExecResult{PdrInv}{KinductionDfStaticZeroOneTT}{Error}{Error}{Walltime}{Stdev}{168.3039118090603390819832232}%
\StoreBenchExecResult{PdrInv}{KinductionDfStaticZeroOneTT}{Error}{Exception}{Count}{}{25}%
\StoreBenchExecResult{PdrInv}{KinductionDfStaticZeroOneTT}{Error}{Exception}{Cputime}{}{4395.399357106}%
\StoreBenchExecResult{PdrInv}{KinductionDfStaticZeroOneTT}{Error}{Exception}{Cputime}{Avg}{175.81597428424}%
\StoreBenchExecResult{PdrInv}{KinductionDfStaticZeroOneTT}{Error}{Exception}{Cputime}{Median}{96.490323667}%
\StoreBenchExecResult{PdrInv}{KinductionDfStaticZeroOneTT}{Error}{Exception}{Cputime}{Min}{14.921760009}%
\StoreBenchExecResult{PdrInv}{KinductionDfStaticZeroOneTT}{Error}{Exception}{Cputime}{Max}{605.497592878}%
\StoreBenchExecResult{PdrInv}{KinductionDfStaticZeroOneTT}{Error}{Exception}{Cputime}{Stdev}{166.5069163680094635629201646}%
\StoreBenchExecResult{PdrInv}{KinductionDfStaticZeroOneTT}{Error}{Exception}{Walltime}{}{2263.51656246183}%
\StoreBenchExecResult{PdrInv}{KinductionDfStaticZeroOneTT}{Error}{Exception}{Walltime}{Avg}{90.5406624984732}%
\StoreBenchExecResult{PdrInv}{KinductionDfStaticZeroOneTT}{Error}{Exception}{Walltime}{Median}{48.6018331051}%
\StoreBenchExecResult{PdrInv}{KinductionDfStaticZeroOneTT}{Error}{Exception}{Walltime}{Min}{7.62086105347}%
\StoreBenchExecResult{PdrInv}{KinductionDfStaticZeroOneTT}{Error}{Exception}{Walltime}{Max}{303.745280981}%
\StoreBenchExecResult{PdrInv}{KinductionDfStaticZeroOneTT}{Error}{Exception}{Walltime}{Stdev}{85.88954560136705425557316435}%
\StoreBenchExecResult{PdrInv}{KinductionDfStaticZeroOneTT}{Error}{OutOfJavaMemory}{Count}{}{10}%
\StoreBenchExecResult{PdrInv}{KinductionDfStaticZeroOneTT}{Error}{OutOfJavaMemory}{Cputime}{}{4828.684167717}%
\StoreBenchExecResult{PdrInv}{KinductionDfStaticZeroOneTT}{Error}{OutOfJavaMemory}{Cputime}{Avg}{482.8684167717}%
\StoreBenchExecResult{PdrInv}{KinductionDfStaticZeroOneTT}{Error}{OutOfJavaMemory}{Cputime}{Median}{536.442143046}%
\StoreBenchExecResult{PdrInv}{KinductionDfStaticZeroOneTT}{Error}{OutOfJavaMemory}{Cputime}{Min}{176.983807877}%
\StoreBenchExecResult{PdrInv}{KinductionDfStaticZeroOneTT}{Error}{OutOfJavaMemory}{Cputime}{Max}{701.113484446}%
\StoreBenchExecResult{PdrInv}{KinductionDfStaticZeroOneTT}{Error}{OutOfJavaMemory}{Cputime}{Stdev}{170.9761863665121734060228620}%
\StoreBenchExecResult{PdrInv}{KinductionDfStaticZeroOneTT}{Error}{OutOfJavaMemory}{Walltime}{}{3040.559777975}%
\StoreBenchExecResult{PdrInv}{KinductionDfStaticZeroOneTT}{Error}{OutOfJavaMemory}{Walltime}{Avg}{304.0559777975}%
\StoreBenchExecResult{PdrInv}{KinductionDfStaticZeroOneTT}{Error}{OutOfJavaMemory}{Walltime}{Median}{296.1006685495}%
\StoreBenchExecResult{PdrInv}{KinductionDfStaticZeroOneTT}{Error}{OutOfJavaMemory}{Walltime}{Min}{106.547775984}%
\StoreBenchExecResult{PdrInv}{KinductionDfStaticZeroOneTT}{Error}{OutOfJavaMemory}{Walltime}{Max}{551.725485802}%
\StoreBenchExecResult{PdrInv}{KinductionDfStaticZeroOneTT}{Error}{OutOfJavaMemory}{Walltime}{Stdev}{122.4566052621858732691724463}%
\StoreBenchExecResult{PdrInv}{KinductionDfStaticZeroOneTT}{Error}{OutOfMemory}{Count}{}{264}%
\StoreBenchExecResult{PdrInv}{KinductionDfStaticZeroOneTT}{Error}{OutOfMemory}{Cputime}{}{92441.421009829}%
\StoreBenchExecResult{PdrInv}{KinductionDfStaticZeroOneTT}{Error}{OutOfMemory}{Cputime}{Avg}{350.1568977645037878787878788}%
\StoreBenchExecResult{PdrInv}{KinductionDfStaticZeroOneTT}{Error}{OutOfMemory}{Cputime}{Median}{269.3984416535}%
\StoreBenchExecResult{PdrInv}{KinductionDfStaticZeroOneTT}{Error}{OutOfMemory}{Cputime}{Min}{157.286603843}%
\StoreBenchExecResult{PdrInv}{KinductionDfStaticZeroOneTT}{Error}{OutOfMemory}{Cputime}{Max}{891.59037469}%
\StoreBenchExecResult{PdrInv}{KinductionDfStaticZeroOneTT}{Error}{OutOfMemory}{Cputime}{Stdev}{194.0139476192664922320246266}%
\StoreBenchExecResult{PdrInv}{KinductionDfStaticZeroOneTT}{Error}{OutOfMemory}{Walltime}{}{77203.6609396956}%
\StoreBenchExecResult{PdrInv}{KinductionDfStaticZeroOneTT}{Error}{OutOfMemory}{Walltime}{Avg}{292.4381096200590909090909091}%
\StoreBenchExecResult{PdrInv}{KinductionDfStaticZeroOneTT}{Error}{OutOfMemory}{Walltime}{Median}{192.7539035085}%
\StoreBenchExecResult{PdrInv}{KinductionDfStaticZeroOneTT}{Error}{OutOfMemory}{Walltime}{Min}{85.8784229755}%
\StoreBenchExecResult{PdrInv}{KinductionDfStaticZeroOneTT}{Error}{OutOfMemory}{Walltime}{Max}{866.566967964}%
\StoreBenchExecResult{PdrInv}{KinductionDfStaticZeroOneTT}{Error}{OutOfMemory}{Walltime}{Stdev}{214.8921485128454484545520802}%
\StoreBenchExecResult{PdrInv}{KinductionDfStaticZeroOneTT}{Error}{Timeout}{Count}{}{2091}%
\StoreBenchExecResult{PdrInv}{KinductionDfStaticZeroOneTT}{Error}{Timeout}{Cputime}{}{1899874.323029785}%
\StoreBenchExecResult{PdrInv}{KinductionDfStaticZeroOneTT}{Error}{Timeout}{Cputime}{Avg}{908.5960416211310377809660450}%
\StoreBenchExecResult{PdrInv}{KinductionDfStaticZeroOneTT}{Error}{Timeout}{Cputime}{Median}{902.252394723}%
\StoreBenchExecResult{PdrInv}{KinductionDfStaticZeroOneTT}{Error}{Timeout}{Cputime}{Min}{900.911325191}%
\StoreBenchExecResult{PdrInv}{KinductionDfStaticZeroOneTT}{Error}{Timeout}{Cputime}{Max}{1002.33951193}%
\StoreBenchExecResult{PdrInv}{KinductionDfStaticZeroOneTT}{Error}{Timeout}{Cputime}{Stdev}{20.34159192212590422140634277}%
\StoreBenchExecResult{PdrInv}{KinductionDfStaticZeroOneTT}{Error}{Timeout}{Walltime}{}{1322289.781281923}%
\StoreBenchExecResult{PdrInv}{KinductionDfStaticZeroOneTT}{Error}{Timeout}{Walltime}{Avg}{632.3719661797814442850310856}%
\StoreBenchExecResult{PdrInv}{KinductionDfStaticZeroOneTT}{Error}{Timeout}{Walltime}{Median}{471.011808872}%
\StoreBenchExecResult{PdrInv}{KinductionDfStaticZeroOneTT}{Error}{Timeout}{Walltime}{Min}{451.102283955}%
\StoreBenchExecResult{PdrInv}{KinductionDfStaticZeroOneTT}{Error}{Timeout}{Walltime}{Max}{921.056605816}%
\StoreBenchExecResult{PdrInv}{KinductionDfStaticZeroOneTT}{Error}{Timeout}{Walltime}{Stdev}{200.5883486027997472639797132}%
\StoreBenchExecResult{PdrInv}{KinductionDfStaticZeroOneTT}{Wrong}{}{Count}{}{2}%
\StoreBenchExecResult{PdrInv}{KinductionDfStaticZeroOneTT}{Wrong}{}{Cputime}{}{22.879867394}%
\StoreBenchExecResult{PdrInv}{KinductionDfStaticZeroOneTT}{Wrong}{}{Cputime}{Avg}{11.439933697}%
\StoreBenchExecResult{PdrInv}{KinductionDfStaticZeroOneTT}{Wrong}{}{Cputime}{Median}{11.439933697}%
\StoreBenchExecResult{PdrInv}{KinductionDfStaticZeroOneTT}{Wrong}{}{Cputime}{Min}{3.716800266}%
\StoreBenchExecResult{PdrInv}{KinductionDfStaticZeroOneTT}{Wrong}{}{Cputime}{Max}{19.163067128}%
\StoreBenchExecResult{PdrInv}{KinductionDfStaticZeroOneTT}{Wrong}{}{Cputime}{Stdev}{7.723133431}%
\StoreBenchExecResult{PdrInv}{KinductionDfStaticZeroOneTT}{Wrong}{}{Walltime}{}{12.38457798953}%
\StoreBenchExecResult{PdrInv}{KinductionDfStaticZeroOneTT}{Wrong}{}{Walltime}{Avg}{6.192288994765}%
\StoreBenchExecResult{PdrInv}{KinductionDfStaticZeroOneTT}{Wrong}{}{Walltime}{Median}{6.192288994765}%
\StoreBenchExecResult{PdrInv}{KinductionDfStaticZeroOneTT}{Wrong}{}{Walltime}{Min}{2.04899787903}%
\StoreBenchExecResult{PdrInv}{KinductionDfStaticZeroOneTT}{Wrong}{}{Walltime}{Max}{10.3355801105}%
\StoreBenchExecResult{PdrInv}{KinductionDfStaticZeroOneTT}{Wrong}{}{Walltime}{Stdev}{4.143291115735}%
\StoreBenchExecResult{PdrInv}{KinductionDfStaticZeroOneTT}{Wrong}{False}{Count}{}{2}%
\StoreBenchExecResult{PdrInv}{KinductionDfStaticZeroOneTT}{Wrong}{False}{Cputime}{}{22.879867394}%
\StoreBenchExecResult{PdrInv}{KinductionDfStaticZeroOneTT}{Wrong}{False}{Cputime}{Avg}{11.439933697}%
\StoreBenchExecResult{PdrInv}{KinductionDfStaticZeroOneTT}{Wrong}{False}{Cputime}{Median}{11.439933697}%
\StoreBenchExecResult{PdrInv}{KinductionDfStaticZeroOneTT}{Wrong}{False}{Cputime}{Min}{3.716800266}%
\StoreBenchExecResult{PdrInv}{KinductionDfStaticZeroOneTT}{Wrong}{False}{Cputime}{Max}{19.163067128}%
\StoreBenchExecResult{PdrInv}{KinductionDfStaticZeroOneTT}{Wrong}{False}{Cputime}{Stdev}{7.723133431}%
\StoreBenchExecResult{PdrInv}{KinductionDfStaticZeroOneTT}{Wrong}{False}{Walltime}{}{12.38457798953}%
\StoreBenchExecResult{PdrInv}{KinductionDfStaticZeroOneTT}{Wrong}{False}{Walltime}{Avg}{6.192288994765}%
\StoreBenchExecResult{PdrInv}{KinductionDfStaticZeroOneTT}{Wrong}{False}{Walltime}{Median}{6.192288994765}%
\StoreBenchExecResult{PdrInv}{KinductionDfStaticZeroOneTT}{Wrong}{False}{Walltime}{Min}{2.04899787903}%
\StoreBenchExecResult{PdrInv}{KinductionDfStaticZeroOneTT}{Wrong}{False}{Walltime}{Max}{10.3355801105}%
\StoreBenchExecResult{PdrInv}{KinductionDfStaticZeroOneTT}{Wrong}{False}{Walltime}{Stdev}{4.143291115735}%
\providecommand\StoreBenchExecResult[7]{\expandafter\newcommand\csname#1#2#3#4#5#6\endcsname{#7}}%
\StoreBenchExecResult{PdrInv}{KinductionDfStaticZeroTwoTFTrueNotSolvedByKinductionPlainButKipdr}{Total}{}{Count}{}{449}%
\StoreBenchExecResult{PdrInv}{KinductionDfStaticZeroTwoTFTrueNotSolvedByKinductionPlainButKipdr}{Total}{}{Cputime}{}{13070.424902700}%
\StoreBenchExecResult{PdrInv}{KinductionDfStaticZeroTwoTFTrueNotSolvedByKinductionPlainButKipdr}{Total}{}{Cputime}{Avg}{29.11007773429844097995545657}%
\StoreBenchExecResult{PdrInv}{KinductionDfStaticZeroTwoTFTrueNotSolvedByKinductionPlainButKipdr}{Total}{}{Cputime}{Median}{5.96597499}%
\StoreBenchExecResult{PdrInv}{KinductionDfStaticZeroTwoTFTrueNotSolvedByKinductionPlainButKipdr}{Total}{}{Cputime}{Min}{3.317091806}%
\StoreBenchExecResult{PdrInv}{KinductionDfStaticZeroTwoTFTrueNotSolvedByKinductionPlainButKipdr}{Total}{}{Cputime}{Max}{912.391570893}%
\StoreBenchExecResult{PdrInv}{KinductionDfStaticZeroTwoTFTrueNotSolvedByKinductionPlainButKipdr}{Total}{}{Cputime}{Stdev}{138.3413459837981623287931352}%
\StoreBenchExecResult{PdrInv}{KinductionDfStaticZeroTwoTFTrueNotSolvedByKinductionPlainButKipdr}{Total}{}{Walltime}{}{10982.28426432515}%
\StoreBenchExecResult{PdrInv}{KinductionDfStaticZeroTwoTFTrueNotSolvedByKinductionPlainButKipdr}{Total}{}{Walltime}{Avg}{24.45943043279543429844097996}%
\StoreBenchExecResult{PdrInv}{KinductionDfStaticZeroTwoTFTrueNotSolvedByKinductionPlainButKipdr}{Total}{}{Walltime}{Median}{3.17618989944}%
\StoreBenchExecResult{PdrInv}{KinductionDfStaticZeroTwoTFTrueNotSolvedByKinductionPlainButKipdr}{Total}{}{Walltime}{Min}{1.84105300903}%
\StoreBenchExecResult{PdrInv}{KinductionDfStaticZeroTwoTFTrueNotSolvedByKinductionPlainButKipdr}{Total}{}{Walltime}{Max}{898.350416899}%
\StoreBenchExecResult{PdrInv}{KinductionDfStaticZeroTwoTFTrueNotSolvedByKinductionPlainButKipdr}{Total}{}{Walltime}{Stdev}{131.8451328114365941328654344}%
\StoreBenchExecResult{PdrInv}{KinductionDfStaticZeroTwoTFTrueNotSolvedByKinductionPlainButKipdr}{Correct}{}{Count}{}{438}%
\StoreBenchExecResult{PdrInv}{KinductionDfStaticZeroTwoTFTrueNotSolvedByKinductionPlainButKipdr}{Correct}{}{Cputime}{}{3160.345373597}%
\StoreBenchExecResult{PdrInv}{KinductionDfStaticZeroTwoTFTrueNotSolvedByKinductionPlainButKipdr}{Correct}{}{Cputime}{Avg}{7.215400396340182648401826484}%
\StoreBenchExecResult{PdrInv}{KinductionDfStaticZeroTwoTFTrueNotSolvedByKinductionPlainButKipdr}{Correct}{}{Cputime}{Median}{5.9248691975}%
\StoreBenchExecResult{PdrInv}{KinductionDfStaticZeroTwoTFTrueNotSolvedByKinductionPlainButKipdr}{Correct}{}{Cputime}{Min}{3.317091806}%
\StoreBenchExecResult{PdrInv}{KinductionDfStaticZeroTwoTFTrueNotSolvedByKinductionPlainButKipdr}{Correct}{}{Cputime}{Max}{77.490850528}%
\StoreBenchExecResult{PdrInv}{KinductionDfStaticZeroTwoTFTrueNotSolvedByKinductionPlainButKipdr}{Correct}{}{Cputime}{Stdev}{6.804670292784557054522865292}%
\StoreBenchExecResult{PdrInv}{KinductionDfStaticZeroTwoTFTrueNotSolvedByKinductionPlainButKipdr}{Correct}{}{Walltime}{}{1666.93747162815}%
\StoreBenchExecResult{PdrInv}{KinductionDfStaticZeroTwoTFTrueNotSolvedByKinductionPlainButKipdr}{Correct}{}{Walltime}{Avg}{3.805793314219520547945205479}%
\StoreBenchExecResult{PdrInv}{KinductionDfStaticZeroTwoTFTrueNotSolvedByKinductionPlainButKipdr}{Correct}{}{Walltime}{Median}{3.132267951965}%
\StoreBenchExecResult{PdrInv}{KinductionDfStaticZeroTwoTFTrueNotSolvedByKinductionPlainButKipdr}{Correct}{}{Walltime}{Min}{1.84105300903}%
\StoreBenchExecResult{PdrInv}{KinductionDfStaticZeroTwoTFTrueNotSolvedByKinductionPlainButKipdr}{Correct}{}{Walltime}{Max}{39.2559719086}%
\StoreBenchExecResult{PdrInv}{KinductionDfStaticZeroTwoTFTrueNotSolvedByKinductionPlainButKipdr}{Correct}{}{Walltime}{Stdev}{3.434222586945868696540545045}%
\StoreBenchExecResult{PdrInv}{KinductionDfStaticZeroTwoTFTrueNotSolvedByKinductionPlainButKipdr}{Correct}{True}{Count}{}{438}%
\StoreBenchExecResult{PdrInv}{KinductionDfStaticZeroTwoTFTrueNotSolvedByKinductionPlainButKipdr}{Correct}{True}{Cputime}{}{3160.345373597}%
\StoreBenchExecResult{PdrInv}{KinductionDfStaticZeroTwoTFTrueNotSolvedByKinductionPlainButKipdr}{Correct}{True}{Cputime}{Avg}{7.215400396340182648401826484}%
\StoreBenchExecResult{PdrInv}{KinductionDfStaticZeroTwoTFTrueNotSolvedByKinductionPlainButKipdr}{Correct}{True}{Cputime}{Median}{5.9248691975}%
\StoreBenchExecResult{PdrInv}{KinductionDfStaticZeroTwoTFTrueNotSolvedByKinductionPlainButKipdr}{Correct}{True}{Cputime}{Min}{3.317091806}%
\StoreBenchExecResult{PdrInv}{KinductionDfStaticZeroTwoTFTrueNotSolvedByKinductionPlainButKipdr}{Correct}{True}{Cputime}{Max}{77.490850528}%
\StoreBenchExecResult{PdrInv}{KinductionDfStaticZeroTwoTFTrueNotSolvedByKinductionPlainButKipdr}{Correct}{True}{Cputime}{Stdev}{6.804670292784557054522865292}%
\StoreBenchExecResult{PdrInv}{KinductionDfStaticZeroTwoTFTrueNotSolvedByKinductionPlainButKipdr}{Correct}{True}{Walltime}{}{1666.93747162815}%
\StoreBenchExecResult{PdrInv}{KinductionDfStaticZeroTwoTFTrueNotSolvedByKinductionPlainButKipdr}{Correct}{True}{Walltime}{Avg}{3.805793314219520547945205479}%
\StoreBenchExecResult{PdrInv}{KinductionDfStaticZeroTwoTFTrueNotSolvedByKinductionPlainButKipdr}{Correct}{True}{Walltime}{Median}{3.132267951965}%
\StoreBenchExecResult{PdrInv}{KinductionDfStaticZeroTwoTFTrueNotSolvedByKinductionPlainButKipdr}{Correct}{True}{Walltime}{Min}{1.84105300903}%
\StoreBenchExecResult{PdrInv}{KinductionDfStaticZeroTwoTFTrueNotSolvedByKinductionPlainButKipdr}{Correct}{True}{Walltime}{Max}{39.2559719086}%
\StoreBenchExecResult{PdrInv}{KinductionDfStaticZeroTwoTFTrueNotSolvedByKinductionPlainButKipdr}{Correct}{True}{Walltime}{Stdev}{3.434222586945868696540545045}%
\StoreBenchExecResult{PdrInv}{KinductionDfStaticZeroTwoTFTrueNotSolvedByKinductionPlainButKipdr}{Wrong}{True}{Count}{}{0}%
\StoreBenchExecResult{PdrInv}{KinductionDfStaticZeroTwoTFTrueNotSolvedByKinductionPlainButKipdr}{Wrong}{True}{Cputime}{}{0}%
\StoreBenchExecResult{PdrInv}{KinductionDfStaticZeroTwoTFTrueNotSolvedByKinductionPlainButKipdr}{Wrong}{True}{Cputime}{Avg}{None}%
\StoreBenchExecResult{PdrInv}{KinductionDfStaticZeroTwoTFTrueNotSolvedByKinductionPlainButKipdr}{Wrong}{True}{Cputime}{Median}{None}%
\StoreBenchExecResult{PdrInv}{KinductionDfStaticZeroTwoTFTrueNotSolvedByKinductionPlainButKipdr}{Wrong}{True}{Cputime}{Min}{None}%
\StoreBenchExecResult{PdrInv}{KinductionDfStaticZeroTwoTFTrueNotSolvedByKinductionPlainButKipdr}{Wrong}{True}{Cputime}{Max}{None}%
\StoreBenchExecResult{PdrInv}{KinductionDfStaticZeroTwoTFTrueNotSolvedByKinductionPlainButKipdr}{Wrong}{True}{Cputime}{Stdev}{None}%
\StoreBenchExecResult{PdrInv}{KinductionDfStaticZeroTwoTFTrueNotSolvedByKinductionPlainButKipdr}{Wrong}{True}{Walltime}{}{0}%
\StoreBenchExecResult{PdrInv}{KinductionDfStaticZeroTwoTFTrueNotSolvedByKinductionPlainButKipdr}{Wrong}{True}{Walltime}{Avg}{None}%
\StoreBenchExecResult{PdrInv}{KinductionDfStaticZeroTwoTFTrueNotSolvedByKinductionPlainButKipdr}{Wrong}{True}{Walltime}{Median}{None}%
\StoreBenchExecResult{PdrInv}{KinductionDfStaticZeroTwoTFTrueNotSolvedByKinductionPlainButKipdr}{Wrong}{True}{Walltime}{Min}{None}%
\StoreBenchExecResult{PdrInv}{KinductionDfStaticZeroTwoTFTrueNotSolvedByKinductionPlainButKipdr}{Wrong}{True}{Walltime}{Max}{None}%
\StoreBenchExecResult{PdrInv}{KinductionDfStaticZeroTwoTFTrueNotSolvedByKinductionPlainButKipdr}{Wrong}{True}{Walltime}{Stdev}{None}%
\StoreBenchExecResult{PdrInv}{KinductionDfStaticZeroTwoTFTrueNotSolvedByKinductionPlainButKipdr}{Error}{}{Count}{}{11}%
\StoreBenchExecResult{PdrInv}{KinductionDfStaticZeroTwoTFTrueNotSolvedByKinductionPlainButKipdr}{Error}{}{Cputime}{}{9910.079529103}%
\StoreBenchExecResult{PdrInv}{KinductionDfStaticZeroTwoTFTrueNotSolvedByKinductionPlainButKipdr}{Error}{}{Cputime}{Avg}{900.9163208275454545454545455}%
\StoreBenchExecResult{PdrInv}{KinductionDfStaticZeroTwoTFTrueNotSolvedByKinductionPlainButKipdr}{Error}{}{Cputime}{Median}{903.864967405}%
\StoreBenchExecResult{PdrInv}{KinductionDfStaticZeroTwoTFTrueNotSolvedByKinductionPlainButKipdr}{Error}{}{Cputime}{Min}{856.412016135}%
\StoreBenchExecResult{PdrInv}{KinductionDfStaticZeroTwoTFTrueNotSolvedByKinductionPlainButKipdr}{Error}{}{Cputime}{Max}{912.391570893}%
\StoreBenchExecResult{PdrInv}{KinductionDfStaticZeroTwoTFTrueNotSolvedByKinductionPlainButKipdr}{Error}{}{Cputime}{Stdev}{14.62832097709200532477746686}%
\StoreBenchExecResult{PdrInv}{KinductionDfStaticZeroTwoTFTrueNotSolvedByKinductionPlainButKipdr}{Error}{}{Walltime}{}{9315.346792697}%
\StoreBenchExecResult{PdrInv}{KinductionDfStaticZeroTwoTFTrueNotSolvedByKinductionPlainButKipdr}{Error}{}{Walltime}{Avg}{846.849708427}%
\StoreBenchExecResult{PdrInv}{KinductionDfStaticZeroTwoTFTrueNotSolvedByKinductionPlainButKipdr}{Error}{}{Walltime}{Median}{888.318466902}%
\StoreBenchExecResult{PdrInv}{KinductionDfStaticZeroTwoTFTrueNotSolvedByKinductionPlainButKipdr}{Error}{}{Walltime}{Min}{452.431560993}%
\StoreBenchExecResult{PdrInv}{KinductionDfStaticZeroTwoTFTrueNotSolvedByKinductionPlainButKipdr}{Error}{}{Walltime}{Max}{898.350416899}%
\StoreBenchExecResult{PdrInv}{KinductionDfStaticZeroTwoTFTrueNotSolvedByKinductionPlainButKipdr}{Error}{}{Walltime}{Stdev}{125.5683056894862026843935800}%
\StoreBenchExecResult{PdrInv}{KinductionDfStaticZeroTwoTFTrueNotSolvedByKinductionPlainButKipdr}{Error}{OutOfMemory}{Count}{}{2}%
\StoreBenchExecResult{PdrInv}{KinductionDfStaticZeroTwoTFTrueNotSolvedByKinductionPlainButKipdr}{Error}{OutOfMemory}{Cputime}{}{1754.669582383}%
\StoreBenchExecResult{PdrInv}{KinductionDfStaticZeroTwoTFTrueNotSolvedByKinductionPlainButKipdr}{Error}{OutOfMemory}{Cputime}{Avg}{877.3347911915}%
\StoreBenchExecResult{PdrInv}{KinductionDfStaticZeroTwoTFTrueNotSolvedByKinductionPlainButKipdr}{Error}{OutOfMemory}{Cputime}{Median}{877.3347911915}%
\StoreBenchExecResult{PdrInv}{KinductionDfStaticZeroTwoTFTrueNotSolvedByKinductionPlainButKipdr}{Error}{OutOfMemory}{Cputime}{Min}{856.412016135}%
\StoreBenchExecResult{PdrInv}{KinductionDfStaticZeroTwoTFTrueNotSolvedByKinductionPlainButKipdr}{Error}{OutOfMemory}{Cputime}{Max}{898.257566248}%
\StoreBenchExecResult{PdrInv}{KinductionDfStaticZeroTwoTFTrueNotSolvedByKinductionPlainButKipdr}{Error}{OutOfMemory}{Cputime}{Stdev}{20.9227750565}%
\StoreBenchExecResult{PdrInv}{KinductionDfStaticZeroTwoTFTrueNotSolvedByKinductionPlainButKipdr}{Error}{OutOfMemory}{Walltime}{}{1727.410573005}%
\StoreBenchExecResult{PdrInv}{KinductionDfStaticZeroTwoTFTrueNotSolvedByKinductionPlainButKipdr}{Error}{OutOfMemory}{Walltime}{Avg}{863.7052865025}%
\StoreBenchExecResult{PdrInv}{KinductionDfStaticZeroTwoTFTrueNotSolvedByKinductionPlainButKipdr}{Error}{OutOfMemory}{Walltime}{Median}{863.7052865025}%
\StoreBenchExecResult{PdrInv}{KinductionDfStaticZeroTwoTFTrueNotSolvedByKinductionPlainButKipdr}{Error}{OutOfMemory}{Walltime}{Min}{842.484472036}%
\StoreBenchExecResult{PdrInv}{KinductionDfStaticZeroTwoTFTrueNotSolvedByKinductionPlainButKipdr}{Error}{OutOfMemory}{Walltime}{Max}{884.926100969}%
\StoreBenchExecResult{PdrInv}{KinductionDfStaticZeroTwoTFTrueNotSolvedByKinductionPlainButKipdr}{Error}{OutOfMemory}{Walltime}{Stdev}{21.2208144665}%
\StoreBenchExecResult{PdrInv}{KinductionDfStaticZeroTwoTFTrueNotSolvedByKinductionPlainButKipdr}{Error}{Timeout}{Count}{}{9}%
\StoreBenchExecResult{PdrInv}{KinductionDfStaticZeroTwoTFTrueNotSolvedByKinductionPlainButKipdr}{Error}{Timeout}{Cputime}{}{8155.409946720}%
\StoreBenchExecResult{PdrInv}{KinductionDfStaticZeroTwoTFTrueNotSolvedByKinductionPlainButKipdr}{Error}{Timeout}{Cputime}{Avg}{906.1566607466666666666666667}%
\StoreBenchExecResult{PdrInv}{KinductionDfStaticZeroTwoTFTrueNotSolvedByKinductionPlainButKipdr}{Error}{Timeout}{Cputime}{Median}{904.465593015}%
\StoreBenchExecResult{PdrInv}{KinductionDfStaticZeroTwoTFTrueNotSolvedByKinductionPlainButKipdr}{Error}{Timeout}{Cputime}{Min}{902.098148545}%
\StoreBenchExecResult{PdrInv}{KinductionDfStaticZeroTwoTFTrueNotSolvedByKinductionPlainButKipdr}{Error}{Timeout}{Cputime}{Max}{912.391570893}%
\StoreBenchExecResult{PdrInv}{KinductionDfStaticZeroTwoTFTrueNotSolvedByKinductionPlainButKipdr}{Error}{Timeout}{Cputime}{Stdev}{3.636435165325375431610631866}%
\StoreBenchExecResult{PdrInv}{KinductionDfStaticZeroTwoTFTrueNotSolvedByKinductionPlainButKipdr}{Error}{Timeout}{Walltime}{}{7587.936219692}%
\StoreBenchExecResult{PdrInv}{KinductionDfStaticZeroTwoTFTrueNotSolvedByKinductionPlainButKipdr}{Error}{Timeout}{Walltime}{Avg}{843.1040244102222222222222222}%
\StoreBenchExecResult{PdrInv}{KinductionDfStaticZeroTwoTFTrueNotSolvedByKinductionPlainButKipdr}{Error}{Timeout}{Walltime}{Median}{888.462490082}%
\StoreBenchExecResult{PdrInv}{KinductionDfStaticZeroTwoTFTrueNotSolvedByKinductionPlainButKipdr}{Error}{Timeout}{Walltime}{Min}{452.431560993}%
\StoreBenchExecResult{PdrInv}{KinductionDfStaticZeroTwoTFTrueNotSolvedByKinductionPlainButKipdr}{Error}{Timeout}{Walltime}{Max}{898.350416899}%
\StoreBenchExecResult{PdrInv}{KinductionDfStaticZeroTwoTFTrueNotSolvedByKinductionPlainButKipdr}{Error}{Timeout}{Walltime}{Stdev}{138.1811432039951354854891122}%
\providecommand\StoreBenchExecResult[7]{\expandafter\newcommand\csname#1#2#3#4#5#6\endcsname{#7}}%
\StoreBenchExecResult{PdrInv}{KinductionDfStaticZeroTwoTFTrueNotSolvedByKinductionPlain}{Total}{}{Count}{}{2893}%
\StoreBenchExecResult{PdrInv}{KinductionDfStaticZeroTwoTFTrueNotSolvedByKinductionPlain}{Total}{}{Cputime}{}{1629132.239869387}%
\StoreBenchExecResult{PdrInv}{KinductionDfStaticZeroTwoTFTrueNotSolvedByKinductionPlain}{Total}{}{Cputime}{Avg}{563.1290148183155893536121673}%
\StoreBenchExecResult{PdrInv}{KinductionDfStaticZeroTwoTFTrueNotSolvedByKinductionPlain}{Total}{}{Cputime}{Median}{901.145238562}%
\StoreBenchExecResult{PdrInv}{KinductionDfStaticZeroTwoTFTrueNotSolvedByKinductionPlain}{Total}{}{Cputime}{Min}{2.426216131}%
\StoreBenchExecResult{PdrInv}{KinductionDfStaticZeroTwoTFTrueNotSolvedByKinductionPlain}{Total}{}{Cputime}{Max}{1002.35417576}%
\StoreBenchExecResult{PdrInv}{KinductionDfStaticZeroTwoTFTrueNotSolvedByKinductionPlain}{Total}{}{Cputime}{Stdev}{417.8809691099819803409519528}%
\StoreBenchExecResult{PdrInv}{KinductionDfStaticZeroTwoTFTrueNotSolvedByKinductionPlain}{Total}{}{Walltime}{}{1125603.04263021290}%
\StoreBenchExecResult{PdrInv}{KinductionDfStaticZeroTwoTFTrueNotSolvedByKinductionPlain}{Total}{}{Walltime}{Avg}{389.0781343346743518838575873}%
\StoreBenchExecResult{PdrInv}{KinductionDfStaticZeroTwoTFTrueNotSolvedByKinductionPlain}{Total}{}{Walltime}{Median}{451.814517021}%
\StoreBenchExecResult{PdrInv}{KinductionDfStaticZeroTwoTFTrueNotSolvedByKinductionPlain}{Total}{}{Walltime}{Min}{1.34655117989}%
\StoreBenchExecResult{PdrInv}{KinductionDfStaticZeroTwoTFTrueNotSolvedByKinductionPlain}{Total}{}{Walltime}{Max}{927.069442034}%
\StoreBenchExecResult{PdrInv}{KinductionDfStaticZeroTwoTFTrueNotSolvedByKinductionPlain}{Total}{}{Walltime}{Stdev}{328.2500912335275724772018724}%
\StoreBenchExecResult{PdrInv}{KinductionDfStaticZeroTwoTFTrueNotSolvedByKinductionPlain}{Correct}{}{Count}{}{944}%
\StoreBenchExecResult{PdrInv}{KinductionDfStaticZeroTwoTFTrueNotSolvedByKinductionPlain}{Correct}{}{Cputime}{}{37572.566165707}%
\StoreBenchExecResult{PdrInv}{KinductionDfStaticZeroTwoTFTrueNotSolvedByKinductionPlain}{Correct}{}{Cputime}{Avg}{39.80144720943538135593220339}%
\StoreBenchExecResult{PdrInv}{KinductionDfStaticZeroTwoTFTrueNotSolvedByKinductionPlain}{Correct}{}{Cputime}{Median}{7.1149832515}%
\StoreBenchExecResult{PdrInv}{KinductionDfStaticZeroTwoTFTrueNotSolvedByKinductionPlain}{Correct}{}{Cputime}{Min}{3.290390684}%
\StoreBenchExecResult{PdrInv}{KinductionDfStaticZeroTwoTFTrueNotSolvedByKinductionPlain}{Correct}{}{Cputime}{Max}{888.220701356}%
\StoreBenchExecResult{PdrInv}{KinductionDfStaticZeroTwoTFTrueNotSolvedByKinductionPlain}{Correct}{}{Cputime}{Stdev}{109.4503944313724326742841710}%
\StoreBenchExecResult{PdrInv}{KinductionDfStaticZeroTwoTFTrueNotSolvedByKinductionPlain}{Correct}{}{Walltime}{}{20261.88592624841}%
\StoreBenchExecResult{PdrInv}{KinductionDfStaticZeroTwoTFTrueNotSolvedByKinductionPlain}{Correct}{}{Walltime}{Avg}{21.46386221000890889830508475}%
\StoreBenchExecResult{PdrInv}{KinductionDfStaticZeroTwoTFTrueNotSolvedByKinductionPlain}{Correct}{}{Walltime}{Median}{3.767378091815}%
\StoreBenchExecResult{PdrInv}{KinductionDfStaticZeroTwoTFTrueNotSolvedByKinductionPlain}{Correct}{}{Walltime}{Min}{1.82420611382}%
\StoreBenchExecResult{PdrInv}{KinductionDfStaticZeroTwoTFTrueNotSolvedByKinductionPlain}{Correct}{}{Walltime}{Max}{761.255792141}%
\StoreBenchExecResult{PdrInv}{KinductionDfStaticZeroTwoTFTrueNotSolvedByKinductionPlain}{Correct}{}{Walltime}{Stdev}{63.20030732434069114221641080}%
\StoreBenchExecResult{PdrInv}{KinductionDfStaticZeroTwoTFTrueNotSolvedByKinductionPlain}{Correct}{True}{Count}{}{944}%
\StoreBenchExecResult{PdrInv}{KinductionDfStaticZeroTwoTFTrueNotSolvedByKinductionPlain}{Correct}{True}{Cputime}{}{37572.566165707}%
\StoreBenchExecResult{PdrInv}{KinductionDfStaticZeroTwoTFTrueNotSolvedByKinductionPlain}{Correct}{True}{Cputime}{Avg}{39.80144720943538135593220339}%
\StoreBenchExecResult{PdrInv}{KinductionDfStaticZeroTwoTFTrueNotSolvedByKinductionPlain}{Correct}{True}{Cputime}{Median}{7.1149832515}%
\StoreBenchExecResult{PdrInv}{KinductionDfStaticZeroTwoTFTrueNotSolvedByKinductionPlain}{Correct}{True}{Cputime}{Min}{3.290390684}%
\StoreBenchExecResult{PdrInv}{KinductionDfStaticZeroTwoTFTrueNotSolvedByKinductionPlain}{Correct}{True}{Cputime}{Max}{888.220701356}%
\StoreBenchExecResult{PdrInv}{KinductionDfStaticZeroTwoTFTrueNotSolvedByKinductionPlain}{Correct}{True}{Cputime}{Stdev}{109.4503944313724326742841710}%
\StoreBenchExecResult{PdrInv}{KinductionDfStaticZeroTwoTFTrueNotSolvedByKinductionPlain}{Correct}{True}{Walltime}{}{20261.88592624841}%
\StoreBenchExecResult{PdrInv}{KinductionDfStaticZeroTwoTFTrueNotSolvedByKinductionPlain}{Correct}{True}{Walltime}{Avg}{21.46386221000890889830508475}%
\StoreBenchExecResult{PdrInv}{KinductionDfStaticZeroTwoTFTrueNotSolvedByKinductionPlain}{Correct}{True}{Walltime}{Median}{3.767378091815}%
\StoreBenchExecResult{PdrInv}{KinductionDfStaticZeroTwoTFTrueNotSolvedByKinductionPlain}{Correct}{True}{Walltime}{Min}{1.82420611382}%
\StoreBenchExecResult{PdrInv}{KinductionDfStaticZeroTwoTFTrueNotSolvedByKinductionPlain}{Correct}{True}{Walltime}{Max}{761.255792141}%
\StoreBenchExecResult{PdrInv}{KinductionDfStaticZeroTwoTFTrueNotSolvedByKinductionPlain}{Correct}{True}{Walltime}{Stdev}{63.20030732434069114221641080}%
\StoreBenchExecResult{PdrInv}{KinductionDfStaticZeroTwoTFTrueNotSolvedByKinductionPlain}{Wrong}{True}{Count}{}{0}%
\StoreBenchExecResult{PdrInv}{KinductionDfStaticZeroTwoTFTrueNotSolvedByKinductionPlain}{Wrong}{True}{Cputime}{}{0}%
\StoreBenchExecResult{PdrInv}{KinductionDfStaticZeroTwoTFTrueNotSolvedByKinductionPlain}{Wrong}{True}{Cputime}{Avg}{None}%
\StoreBenchExecResult{PdrInv}{KinductionDfStaticZeroTwoTFTrueNotSolvedByKinductionPlain}{Wrong}{True}{Cputime}{Median}{None}%
\StoreBenchExecResult{PdrInv}{KinductionDfStaticZeroTwoTFTrueNotSolvedByKinductionPlain}{Wrong}{True}{Cputime}{Min}{None}%
\StoreBenchExecResult{PdrInv}{KinductionDfStaticZeroTwoTFTrueNotSolvedByKinductionPlain}{Wrong}{True}{Cputime}{Max}{None}%
\StoreBenchExecResult{PdrInv}{KinductionDfStaticZeroTwoTFTrueNotSolvedByKinductionPlain}{Wrong}{True}{Cputime}{Stdev}{None}%
\StoreBenchExecResult{PdrInv}{KinductionDfStaticZeroTwoTFTrueNotSolvedByKinductionPlain}{Wrong}{True}{Walltime}{}{0}%
\StoreBenchExecResult{PdrInv}{KinductionDfStaticZeroTwoTFTrueNotSolvedByKinductionPlain}{Wrong}{True}{Walltime}{Avg}{None}%
\StoreBenchExecResult{PdrInv}{KinductionDfStaticZeroTwoTFTrueNotSolvedByKinductionPlain}{Wrong}{True}{Walltime}{Median}{None}%
\StoreBenchExecResult{PdrInv}{KinductionDfStaticZeroTwoTFTrueNotSolvedByKinductionPlain}{Wrong}{True}{Walltime}{Min}{None}%
\StoreBenchExecResult{PdrInv}{KinductionDfStaticZeroTwoTFTrueNotSolvedByKinductionPlain}{Wrong}{True}{Walltime}{Max}{None}%
\StoreBenchExecResult{PdrInv}{KinductionDfStaticZeroTwoTFTrueNotSolvedByKinductionPlain}{Wrong}{True}{Walltime}{Stdev}{None}%
\StoreBenchExecResult{PdrInv}{KinductionDfStaticZeroTwoTFTrueNotSolvedByKinductionPlain}{Error}{}{Count}{}{1949}%
\StoreBenchExecResult{PdrInv}{KinductionDfStaticZeroTwoTFTrueNotSolvedByKinductionPlain}{Error}{}{Cputime}{}{1591559.673703680}%
\StoreBenchExecResult{PdrInv}{KinductionDfStaticZeroTwoTFTrueNotSolvedByKinductionPlain}{Error}{}{Cputime}{Avg}{816.6032189346741918932786044}%
\StoreBenchExecResult{PdrInv}{KinductionDfStaticZeroTwoTFTrueNotSolvedByKinductionPlain}{Error}{}{Cputime}{Median}{901.752966754}%
\StoreBenchExecResult{PdrInv}{KinductionDfStaticZeroTwoTFTrueNotSolvedByKinductionPlain}{Error}{}{Cputime}{Min}{2.426216131}%
\StoreBenchExecResult{PdrInv}{KinductionDfStaticZeroTwoTFTrueNotSolvedByKinductionPlain}{Error}{}{Cputime}{Max}{1002.35417576}%
\StoreBenchExecResult{PdrInv}{KinductionDfStaticZeroTwoTFTrueNotSolvedByKinductionPlain}{Error}{}{Cputime}{Stdev}{237.7027757042616504621208745}%
\StoreBenchExecResult{PdrInv}{KinductionDfStaticZeroTwoTFTrueNotSolvedByKinductionPlain}{Error}{}{Walltime}{}{1105341.15670396449}%
\StoreBenchExecResult{PdrInv}{KinductionDfStaticZeroTwoTFTrueNotSolvedByKinductionPlain}{Error}{}{Walltime}{Avg}{567.1324559794584350949204720}%
\StoreBenchExecResult{PdrInv}{KinductionDfStaticZeroTwoTFTrueNotSolvedByKinductionPlain}{Error}{}{Walltime}{Median}{454.884833097}%
\StoreBenchExecResult{PdrInv}{KinductionDfStaticZeroTwoTFTrueNotSolvedByKinductionPlain}{Error}{}{Walltime}{Min}{1.34655117989}%
\StoreBenchExecResult{PdrInv}{KinductionDfStaticZeroTwoTFTrueNotSolvedByKinductionPlain}{Error}{}{Walltime}{Max}{927.069442034}%
\StoreBenchExecResult{PdrInv}{KinductionDfStaticZeroTwoTFTrueNotSolvedByKinductionPlain}{Error}{}{Walltime}{Stdev}{246.6632191772781221104622206}%
\StoreBenchExecResult{PdrInv}{KinductionDfStaticZeroTwoTFTrueNotSolvedByKinductionPlain}{Error}{Assertion}{Count}{}{2}%
\StoreBenchExecResult{PdrInv}{KinductionDfStaticZeroTwoTFTrueNotSolvedByKinductionPlain}{Error}{Assertion}{Cputime}{}{6.467041391}%
\StoreBenchExecResult{PdrInv}{KinductionDfStaticZeroTwoTFTrueNotSolvedByKinductionPlain}{Error}{Assertion}{Cputime}{Avg}{3.2335206955}%
\StoreBenchExecResult{PdrInv}{KinductionDfStaticZeroTwoTFTrueNotSolvedByKinductionPlain}{Error}{Assertion}{Cputime}{Median}{3.2335206955}%
\StoreBenchExecResult{PdrInv}{KinductionDfStaticZeroTwoTFTrueNotSolvedByKinductionPlain}{Error}{Assertion}{Cputime}{Min}{3.217962913}%
\StoreBenchExecResult{PdrInv}{KinductionDfStaticZeroTwoTFTrueNotSolvedByKinductionPlain}{Error}{Assertion}{Cputime}{Max}{3.249078478}%
\StoreBenchExecResult{PdrInv}{KinductionDfStaticZeroTwoTFTrueNotSolvedByKinductionPlain}{Error}{Assertion}{Cputime}{Stdev}{0.0155577825}%
\StoreBenchExecResult{PdrInv}{KinductionDfStaticZeroTwoTFTrueNotSolvedByKinductionPlain}{Error}{Assertion}{Walltime}{}{3.58185791970}%
\StoreBenchExecResult{PdrInv}{KinductionDfStaticZeroTwoTFTrueNotSolvedByKinductionPlain}{Error}{Assertion}{Walltime}{Avg}{1.79092895985}%
\StoreBenchExecResult{PdrInv}{KinductionDfStaticZeroTwoTFTrueNotSolvedByKinductionPlain}{Error}{Assertion}{Walltime}{Median}{1.79092895985}%
\StoreBenchExecResult{PdrInv}{KinductionDfStaticZeroTwoTFTrueNotSolvedByKinductionPlain}{Error}{Assertion}{Walltime}{Min}{1.78519392014}%
\StoreBenchExecResult{PdrInv}{KinductionDfStaticZeroTwoTFTrueNotSolvedByKinductionPlain}{Error}{Assertion}{Walltime}{Max}{1.79666399956}%
\StoreBenchExecResult{PdrInv}{KinductionDfStaticZeroTwoTFTrueNotSolvedByKinductionPlain}{Error}{Assertion}{Walltime}{Stdev}{0.00573503971}%
\StoreBenchExecResult{PdrInv}{KinductionDfStaticZeroTwoTFTrueNotSolvedByKinductionPlain}{Error}{Error}{Count}{}{130}%
\StoreBenchExecResult{PdrInv}{KinductionDfStaticZeroTwoTFTrueNotSolvedByKinductionPlain}{Error}{Error}{Cputime}{}{22380.268372900}%
\StoreBenchExecResult{PdrInv}{KinductionDfStaticZeroTwoTFTrueNotSolvedByKinductionPlain}{Error}{Error}{Cputime}{Avg}{172.1559105607692307692307692}%
\StoreBenchExecResult{PdrInv}{KinductionDfStaticZeroTwoTFTrueNotSolvedByKinductionPlain}{Error}{Error}{Cputime}{Median}{106.9104131135}%
\StoreBenchExecResult{PdrInv}{KinductionDfStaticZeroTwoTFTrueNotSolvedByKinductionPlain}{Error}{Error}{Cputime}{Min}{2.426216131}%
\StoreBenchExecResult{PdrInv}{KinductionDfStaticZeroTwoTFTrueNotSolvedByKinductionPlain}{Error}{Error}{Cputime}{Max}{826.46802186}%
\StoreBenchExecResult{PdrInv}{KinductionDfStaticZeroTwoTFTrueNotSolvedByKinductionPlain}{Error}{Error}{Cputime}{Stdev}{195.7635549930027334604113079}%
\StoreBenchExecResult{PdrInv}{KinductionDfStaticZeroTwoTFTrueNotSolvedByKinductionPlain}{Error}{Error}{Walltime}{}{18890.24246692569}%
\StoreBenchExecResult{PdrInv}{KinductionDfStaticZeroTwoTFTrueNotSolvedByKinductionPlain}{Error}{Error}{Walltime}{Avg}{145.3095574378899230769230769}%
\StoreBenchExecResult{PdrInv}{KinductionDfStaticZeroTwoTFTrueNotSolvedByKinductionPlain}{Error}{Error}{Walltime}{Median}{83.3389300108}%
\StoreBenchExecResult{PdrInv}{KinductionDfStaticZeroTwoTFTrueNotSolvedByKinductionPlain}{Error}{Error}{Walltime}{Min}{1.34655117989}%
\StoreBenchExecResult{PdrInv}{KinductionDfStaticZeroTwoTFTrueNotSolvedByKinductionPlain}{Error}{Error}{Walltime}{Max}{785.018320084}%
\StoreBenchExecResult{PdrInv}{KinductionDfStaticZeroTwoTFTrueNotSolvedByKinductionPlain}{Error}{Error}{Walltime}{Stdev}{175.6839896863705987374150024}%
\StoreBenchExecResult{PdrInv}{KinductionDfStaticZeroTwoTFTrueNotSolvedByKinductionPlain}{Error}{Exception}{Count}{}{7}%
\StoreBenchExecResult{PdrInv}{KinductionDfStaticZeroTwoTFTrueNotSolvedByKinductionPlain}{Error}{Exception}{Cputime}{}{799.633121341}%
\StoreBenchExecResult{PdrInv}{KinductionDfStaticZeroTwoTFTrueNotSolvedByKinductionPlain}{Error}{Exception}{Cputime}{Avg}{114.2333030487142857142857143}%
\StoreBenchExecResult{PdrInv}{KinductionDfStaticZeroTwoTFTrueNotSolvedByKinductionPlain}{Error}{Exception}{Cputime}{Median}{81.067961902}%
\StoreBenchExecResult{PdrInv}{KinductionDfStaticZeroTwoTFTrueNotSolvedByKinductionPlain}{Error}{Exception}{Cputime}{Min}{18.334883213}%
\StoreBenchExecResult{PdrInv}{KinductionDfStaticZeroTwoTFTrueNotSolvedByKinductionPlain}{Error}{Exception}{Cputime}{Max}{237.274632241}%
\StoreBenchExecResult{PdrInv}{KinductionDfStaticZeroTwoTFTrueNotSolvedByKinductionPlain}{Error}{Exception}{Cputime}{Stdev}{82.23471597806715185108169069}%
\StoreBenchExecResult{PdrInv}{KinductionDfStaticZeroTwoTFTrueNotSolvedByKinductionPlain}{Error}{Exception}{Walltime}{}{418.0184311872}%
\StoreBenchExecResult{PdrInv}{KinductionDfStaticZeroTwoTFTrueNotSolvedByKinductionPlain}{Error}{Exception}{Walltime}{Avg}{59.71691874102857142857142857}%
\StoreBenchExecResult{PdrInv}{KinductionDfStaticZeroTwoTFTrueNotSolvedByKinductionPlain}{Error}{Exception}{Walltime}{Median}{40.8400390148}%
\StoreBenchExecResult{PdrInv}{KinductionDfStaticZeroTwoTFTrueNotSolvedByKinductionPlain}{Error}{Exception}{Walltime}{Min}{9.4049270153}%
\StoreBenchExecResult{PdrInv}{KinductionDfStaticZeroTwoTFTrueNotSolvedByKinductionPlain}{Error}{Exception}{Walltime}{Max}{119.430325985}%
\StoreBenchExecResult{PdrInv}{KinductionDfStaticZeroTwoTFTrueNotSolvedByKinductionPlain}{Error}{Exception}{Walltime}{Stdev}{41.50622871485567083478085577}%
\StoreBenchExecResult{PdrInv}{KinductionDfStaticZeroTwoTFTrueNotSolvedByKinductionPlain}{Error}{OutOfJavaMemory}{Count}{}{6}%
\StoreBenchExecResult{PdrInv}{KinductionDfStaticZeroTwoTFTrueNotSolvedByKinductionPlain}{Error}{OutOfJavaMemory}{Cputime}{}{2539.129255168}%
\StoreBenchExecResult{PdrInv}{KinductionDfStaticZeroTwoTFTrueNotSolvedByKinductionPlain}{Error}{OutOfJavaMemory}{Cputime}{Avg}{423.1882091946666666666666667}%
\StoreBenchExecResult{PdrInv}{KinductionDfStaticZeroTwoTFTrueNotSolvedByKinductionPlain}{Error}{OutOfJavaMemory}{Cputime}{Median}{370.5141695255}%
\StoreBenchExecResult{PdrInv}{KinductionDfStaticZeroTwoTFTrueNotSolvedByKinductionPlain}{Error}{OutOfJavaMemory}{Cputime}{Min}{160.664889085}%
\StoreBenchExecResult{PdrInv}{KinductionDfStaticZeroTwoTFTrueNotSolvedByKinductionPlain}{Error}{OutOfJavaMemory}{Cputime}{Max}{850.760286023}%
\StoreBenchExecResult{PdrInv}{KinductionDfStaticZeroTwoTFTrueNotSolvedByKinductionPlain}{Error}{OutOfJavaMemory}{Cputime}{Stdev}{224.7841312482598039316855185}%
\StoreBenchExecResult{PdrInv}{KinductionDfStaticZeroTwoTFTrueNotSolvedByKinductionPlain}{Error}{OutOfJavaMemory}{Walltime}{}{1551.5270998474}%
\StoreBenchExecResult{PdrInv}{KinductionDfStaticZeroTwoTFTrueNotSolvedByKinductionPlain}{Error}{OutOfJavaMemory}{Walltime}{Avg}{258.5878499745666666666666667}%
\StoreBenchExecResult{PdrInv}{KinductionDfStaticZeroTwoTFTrueNotSolvedByKinductionPlain}{Error}{OutOfJavaMemory}{Walltime}{Median}{216.618752956}%
\StoreBenchExecResult{PdrInv}{KinductionDfStaticZeroTwoTFTrueNotSolvedByKinductionPlain}{Error}{OutOfJavaMemory}{Walltime}{Min}{99.0947179794}%
\StoreBenchExecResult{PdrInv}{KinductionDfStaticZeroTwoTFTrueNotSolvedByKinductionPlain}{Error}{OutOfJavaMemory}{Walltime}{Max}{568.973742962}%
\StoreBenchExecResult{PdrInv}{KinductionDfStaticZeroTwoTFTrueNotSolvedByKinductionPlain}{Error}{OutOfJavaMemory}{Walltime}{Stdev}{148.4131683202094598777126053}%
\StoreBenchExecResult{PdrInv}{KinductionDfStaticZeroTwoTFTrueNotSolvedByKinductionPlain}{Error}{OutOfMemory}{Count}{}{131}%
\StoreBenchExecResult{PdrInv}{KinductionDfStaticZeroTwoTFTrueNotSolvedByKinductionPlain}{Error}{OutOfMemory}{Cputime}{}{49575.331300857}%
\StoreBenchExecResult{PdrInv}{KinductionDfStaticZeroTwoTFTrueNotSolvedByKinductionPlain}{Error}{OutOfMemory}{Cputime}{Avg}{378.4376435179923664122137405}%
\StoreBenchExecResult{PdrInv}{KinductionDfStaticZeroTwoTFTrueNotSolvedByKinductionPlain}{Error}{OutOfMemory}{Cputime}{Median}{310.130740732}%
\StoreBenchExecResult{PdrInv}{KinductionDfStaticZeroTwoTFTrueNotSolvedByKinductionPlain}{Error}{OutOfMemory}{Cputime}{Min}{160.857776081}%
\StoreBenchExecResult{PdrInv}{KinductionDfStaticZeroTwoTFTrueNotSolvedByKinductionPlain}{Error}{OutOfMemory}{Cputime}{Max}{898.257566248}%
\StoreBenchExecResult{PdrInv}{KinductionDfStaticZeroTwoTFTrueNotSolvedByKinductionPlain}{Error}{OutOfMemory}{Cputime}{Stdev}{215.1959356441701496561683521}%
\StoreBenchExecResult{PdrInv}{KinductionDfStaticZeroTwoTFTrueNotSolvedByKinductionPlain}{Error}{OutOfMemory}{Walltime}{}{41981.0932400225}%
\StoreBenchExecResult{PdrInv}{KinductionDfStaticZeroTwoTFTrueNotSolvedByKinductionPlain}{Error}{OutOfMemory}{Walltime}{Avg}{320.4663606108587786259541985}%
\StoreBenchExecResult{PdrInv}{KinductionDfStaticZeroTwoTFTrueNotSolvedByKinductionPlain}{Error}{OutOfMemory}{Walltime}{Median}{216.001611948}%
\StoreBenchExecResult{PdrInv}{KinductionDfStaticZeroTwoTFTrueNotSolvedByKinductionPlain}{Error}{OutOfMemory}{Walltime}{Min}{86.2205598354}%
\StoreBenchExecResult{PdrInv}{KinductionDfStaticZeroTwoTFTrueNotSolvedByKinductionPlain}{Error}{OutOfMemory}{Walltime}{Max}{884.926100969}%
\StoreBenchExecResult{PdrInv}{KinductionDfStaticZeroTwoTFTrueNotSolvedByKinductionPlain}{Error}{OutOfMemory}{Walltime}{Stdev}{239.3317739680805739750275932}%
\StoreBenchExecResult{PdrInv}{KinductionDfStaticZeroTwoTFTrueNotSolvedByKinductionPlain}{Error}{Timeout}{Count}{}{1673}%
\StoreBenchExecResult{PdrInv}{KinductionDfStaticZeroTwoTFTrueNotSolvedByKinductionPlain}{Error}{Timeout}{Cputime}{}{1516258.844612023}%
\StoreBenchExecResult{PdrInv}{KinductionDfStaticZeroTwoTFTrueNotSolvedByKinductionPlain}{Error}{Timeout}{Cputime}{Avg}{906.3113237370131500298864316}%
\StoreBenchExecResult{PdrInv}{KinductionDfStaticZeroTwoTFTrueNotSolvedByKinductionPlain}{Error}{Timeout}{Cputime}{Median}{902.087651715}%
\StoreBenchExecResult{PdrInv}{KinductionDfStaticZeroTwoTFTrueNotSolvedByKinductionPlain}{Error}{Timeout}{Cputime}{Min}{900.470704413}%
\StoreBenchExecResult{PdrInv}{KinductionDfStaticZeroTwoTFTrueNotSolvedByKinductionPlain}{Error}{Timeout}{Cputime}{Max}{1002.35417576}%
\StoreBenchExecResult{PdrInv}{KinductionDfStaticZeroTwoTFTrueNotSolvedByKinductionPlain}{Error}{Timeout}{Cputime}{Stdev}{15.74519769802029142285820141}%
\StoreBenchExecResult{PdrInv}{KinductionDfStaticZeroTwoTFTrueNotSolvedByKinductionPlain}{Error}{Timeout}{Walltime}{}{1042496.693608062}%
\StoreBenchExecResult{PdrInv}{KinductionDfStaticZeroTwoTFTrueNotSolvedByKinductionPlain}{Error}{Timeout}{Walltime}{Avg}{623.1301217023682008368200837}%
\StoreBenchExecResult{PdrInv}{KinductionDfStaticZeroTwoTFTrueNotSolvedByKinductionPlain}{Error}{Timeout}{Walltime}{Median}{458.854245186}%
\StoreBenchExecResult{PdrInv}{KinductionDfStaticZeroTwoTFTrueNotSolvedByKinductionPlain}{Error}{Timeout}{Walltime}{Min}{450.96033597}%
\StoreBenchExecResult{PdrInv}{KinductionDfStaticZeroTwoTFTrueNotSolvedByKinductionPlain}{Error}{Timeout}{Walltime}{Max}{927.069442034}%
\StoreBenchExecResult{PdrInv}{KinductionDfStaticZeroTwoTFTrueNotSolvedByKinductionPlain}{Error}{Timeout}{Walltime}{Stdev}{200.9560735067899573270756291}%
\providecommand\StoreBenchExecResult[7]{\expandafter\newcommand\csname#1#2#3#4#5#6\endcsname{#7}}%
\StoreBenchExecResult{PdrInv}{KinductionDfStaticZeroTwoTF}{Total}{}{Count}{}{5591}%
\StoreBenchExecResult{PdrInv}{KinductionDfStaticZeroTwoTF}{Total}{}{Cputime}{}{2215390.037861008}%
\StoreBenchExecResult{PdrInv}{KinductionDfStaticZeroTwoTF}{Total}{}{Cputime}{Avg}{396.2421816957624754069039528}%
\StoreBenchExecResult{PdrInv}{KinductionDfStaticZeroTwoTF}{Total}{}{Cputime}{Median}{117.361944043}%
\StoreBenchExecResult{PdrInv}{KinductionDfStaticZeroTwoTF}{Total}{}{Cputime}{Min}{2.426216131}%
\StoreBenchExecResult{PdrInv}{KinductionDfStaticZeroTwoTF}{Total}{}{Cputime}{Max}{1002.35417576}%
\StoreBenchExecResult{PdrInv}{KinductionDfStaticZeroTwoTF}{Total}{}{Cputime}{Stdev}{419.6698512355003044518624647}%
\StoreBenchExecResult{PdrInv}{KinductionDfStaticZeroTwoTF}{Total}{}{Walltime}{}{1560250.61183501111}%
\StoreBenchExecResult{PdrInv}{KinductionDfStaticZeroTwoTF}{Total}{}{Walltime}{Avg}{279.0646774879290126989805044}%
\StoreBenchExecResult{PdrInv}{KinductionDfStaticZeroTwoTF}{Total}{}{Walltime}{Median}{78.8606309891}%
\StoreBenchExecResult{PdrInv}{KinductionDfStaticZeroTwoTF}{Total}{}{Walltime}{Min}{1.34655117989}%
\StoreBenchExecResult{PdrInv}{KinductionDfStaticZeroTwoTF}{Total}{}{Walltime}{Max}{929.81765008}%
\StoreBenchExecResult{PdrInv}{KinductionDfStaticZeroTwoTF}{Total}{}{Walltime}{Stdev}{321.7531216395019465642636488}%
\StoreBenchExecResult{PdrInv}{KinductionDfStaticZeroTwoTF}{Correct}{}{Count}{}{2991}%
\StoreBenchExecResult{PdrInv}{KinductionDfStaticZeroTwoTF}{Correct}{}{Cputime}{}{161809.170270126}%
\StoreBenchExecResult{PdrInv}{KinductionDfStaticZeroTwoTF}{Correct}{}{Cputime}{Avg}{54.09868614848746238716148445}%
\StoreBenchExecResult{PdrInv}{KinductionDfStaticZeroTwoTF}{Correct}{}{Cputime}{Median}{9.349687447}%
\StoreBenchExecResult{PdrInv}{KinductionDfStaticZeroTwoTF}{Correct}{}{Cputime}{Min}{3.012923053}%
\StoreBenchExecResult{PdrInv}{KinductionDfStaticZeroTwoTF}{Correct}{}{Cputime}{Max}{888.220701356}%
\StoreBenchExecResult{PdrInv}{KinductionDfStaticZeroTwoTF}{Correct}{}{Cputime}{Stdev}{130.3357802540340533782999657}%
\StoreBenchExecResult{PdrInv}{KinductionDfStaticZeroTwoTF}{Correct}{}{Walltime}{}{113844.20388746322}%
\StoreBenchExecResult{PdrInv}{KinductionDfStaticZeroTwoTF}{Correct}{}{Walltime}{Avg}{38.06225472666774322968906720}%
\StoreBenchExecResult{PdrInv}{KinductionDfStaticZeroTwoTF}{Correct}{}{Walltime}{Median}{4.95151805878}%
\StoreBenchExecResult{PdrInv}{KinductionDfStaticZeroTwoTF}{Correct}{}{Walltime}{Min}{1.6919260025}%
\StoreBenchExecResult{PdrInv}{KinductionDfStaticZeroTwoTF}{Correct}{}{Walltime}{Max}{866.346286058}%
\StoreBenchExecResult{PdrInv}{KinductionDfStaticZeroTwoTF}{Correct}{}{Walltime}{Stdev}{107.6499638423522556538736324}%
\StoreBenchExecResult{PdrInv}{KinductionDfStaticZeroTwoTF}{Correct}{False}{Count}{}{815}%
\StoreBenchExecResult{PdrInv}{KinductionDfStaticZeroTwoTF}{Correct}{False}{Cputime}{}{71337.492640770}%
\StoreBenchExecResult{PdrInv}{KinductionDfStaticZeroTwoTF}{Correct}{False}{Cputime}{Avg}{87.53066581689570552147239264}%
\StoreBenchExecResult{PdrInv}{KinductionDfStaticZeroTwoTF}{Correct}{False}{Cputime}{Median}{21.461250445}%
\StoreBenchExecResult{PdrInv}{KinductionDfStaticZeroTwoTF}{Correct}{False}{Cputime}{Min}{3.198651402}%
\StoreBenchExecResult{PdrInv}{KinductionDfStaticZeroTwoTF}{Correct}{False}{Cputime}{Max}{884.394865787}%
\StoreBenchExecResult{PdrInv}{KinductionDfStaticZeroTwoTF}{Correct}{False}{Cputime}{Stdev}{172.8164303087713420081436990}%
\StoreBenchExecResult{PdrInv}{KinductionDfStaticZeroTwoTF}{Correct}{False}{Walltime}{}{57190.06736445327}%
\StoreBenchExecResult{PdrInv}{KinductionDfStaticZeroTwoTF}{Correct}{False}{Walltime}{Avg}{70.17186179687517791411042945}%
\StoreBenchExecResult{PdrInv}{KinductionDfStaticZeroTwoTF}{Correct}{False}{Walltime}{Median}{11.8514611721}%
\StoreBenchExecResult{PdrInv}{KinductionDfStaticZeroTwoTF}{Correct}{False}{Walltime}{Min}{1.77485013008}%
\StoreBenchExecResult{PdrInv}{KinductionDfStaticZeroTwoTF}{Correct}{False}{Walltime}{Max}{866.346286058}%
\StoreBenchExecResult{PdrInv}{KinductionDfStaticZeroTwoTF}{Correct}{False}{Walltime}{Stdev}{159.8608173231593985584143787}%
\StoreBenchExecResult{PdrInv}{KinductionDfStaticZeroTwoTF}{Correct}{True}{Count}{}{2176}%
\StoreBenchExecResult{PdrInv}{KinductionDfStaticZeroTwoTF}{Correct}{True}{Cputime}{}{90471.677629356}%
\StoreBenchExecResult{PdrInv}{KinductionDfStaticZeroTwoTF}{Correct}{True}{Cputime}{Avg}{41.57705773407904411764705882}%
\StoreBenchExecResult{PdrInv}{KinductionDfStaticZeroTwoTF}{Correct}{True}{Cputime}{Median}{7.1441533045}%
\StoreBenchExecResult{PdrInv}{KinductionDfStaticZeroTwoTF}{Correct}{True}{Cputime}{Min}{3.012923053}%
\StoreBenchExecResult{PdrInv}{KinductionDfStaticZeroTwoTF}{Correct}{True}{Cputime}{Max}{888.220701356}%
\StoreBenchExecResult{PdrInv}{KinductionDfStaticZeroTwoTF}{Correct}{True}{Cputime}{Stdev}{107.6505050695271826214968529}%
\StoreBenchExecResult{PdrInv}{KinductionDfStaticZeroTwoTF}{Correct}{True}{Walltime}{}{56654.13652300995}%
\StoreBenchExecResult{PdrInv}{KinductionDfStaticZeroTwoTF}{Correct}{True}{Walltime}{Avg}{26.03590832858913143382352941}%
\StoreBenchExecResult{PdrInv}{KinductionDfStaticZeroTwoTF}{Correct}{True}{Walltime}{Median}{3.791200995445}%
\StoreBenchExecResult{PdrInv}{KinductionDfStaticZeroTwoTF}{Correct}{True}{Walltime}{Min}{1.6919260025}%
\StoreBenchExecResult{PdrInv}{KinductionDfStaticZeroTwoTF}{Correct}{True}{Walltime}{Max}{834.461910009}%
\StoreBenchExecResult{PdrInv}{KinductionDfStaticZeroTwoTF}{Correct}{True}{Walltime}{Stdev}{76.33168914551476511318900836}%
\StoreBenchExecResult{PdrInv}{KinductionDfStaticZeroTwoTF}{Wrong}{True}{Count}{}{0}%
\StoreBenchExecResult{PdrInv}{KinductionDfStaticZeroTwoTF}{Wrong}{True}{Cputime}{}{0}%
\StoreBenchExecResult{PdrInv}{KinductionDfStaticZeroTwoTF}{Wrong}{True}{Cputime}{Avg}{None}%
\StoreBenchExecResult{PdrInv}{KinductionDfStaticZeroTwoTF}{Wrong}{True}{Cputime}{Median}{None}%
\StoreBenchExecResult{PdrInv}{KinductionDfStaticZeroTwoTF}{Wrong}{True}{Cputime}{Min}{None}%
\StoreBenchExecResult{PdrInv}{KinductionDfStaticZeroTwoTF}{Wrong}{True}{Cputime}{Max}{None}%
\StoreBenchExecResult{PdrInv}{KinductionDfStaticZeroTwoTF}{Wrong}{True}{Cputime}{Stdev}{None}%
\StoreBenchExecResult{PdrInv}{KinductionDfStaticZeroTwoTF}{Wrong}{True}{Walltime}{}{0}%
\StoreBenchExecResult{PdrInv}{KinductionDfStaticZeroTwoTF}{Wrong}{True}{Walltime}{Avg}{None}%
\StoreBenchExecResult{PdrInv}{KinductionDfStaticZeroTwoTF}{Wrong}{True}{Walltime}{Median}{None}%
\StoreBenchExecResult{PdrInv}{KinductionDfStaticZeroTwoTF}{Wrong}{True}{Walltime}{Min}{None}%
\StoreBenchExecResult{PdrInv}{KinductionDfStaticZeroTwoTF}{Wrong}{True}{Walltime}{Max}{None}%
\StoreBenchExecResult{PdrInv}{KinductionDfStaticZeroTwoTF}{Wrong}{True}{Walltime}{Stdev}{None}%
\StoreBenchExecResult{PdrInv}{KinductionDfStaticZeroTwoTF}{Error}{}{Count}{}{2598}%
\StoreBenchExecResult{PdrInv}{KinductionDfStaticZeroTwoTF}{Error}{}{Cputime}{}{2053558.162546193}%
\StoreBenchExecResult{PdrInv}{KinductionDfStaticZeroTwoTF}{Error}{}{Cputime}{Avg}{790.4380918191658968437259430}%
\StoreBenchExecResult{PdrInv}{KinductionDfStaticZeroTwoTF}{Error}{}{Cputime}{Median}{901.763718728}%
\StoreBenchExecResult{PdrInv}{KinductionDfStaticZeroTwoTF}{Error}{}{Cputime}{Min}{2.426216131}%
\StoreBenchExecResult{PdrInv}{KinductionDfStaticZeroTwoTF}{Error}{}{Cputime}{Max}{1002.35417576}%
\StoreBenchExecResult{PdrInv}{KinductionDfStaticZeroTwoTF}{Error}{}{Cputime}{Stdev}{263.0427471811758278818628030}%
\StoreBenchExecResult{PdrInv}{KinductionDfStaticZeroTwoTF}{Error}{}{Walltime}{}{1446394.16124559208}%
\StoreBenchExecResult{PdrInv}{KinductionDfStaticZeroTwoTF}{Error}{}{Walltime}{Avg}{556.7337033277875596612779061}%
\StoreBenchExecResult{PdrInv}{KinductionDfStaticZeroTwoTF}{Error}{}{Walltime}{Median}{455.892175555}%
\StoreBenchExecResult{PdrInv}{KinductionDfStaticZeroTwoTF}{Error}{}{Walltime}{Min}{1.34655117989}%
\StoreBenchExecResult{PdrInv}{KinductionDfStaticZeroTwoTF}{Error}{}{Walltime}{Max}{929.81765008}%
\StoreBenchExecResult{PdrInv}{KinductionDfStaticZeroTwoTF}{Error}{}{Walltime}{Stdev}{255.7789418866312737745361605}%
\StoreBenchExecResult{PdrInv}{KinductionDfStaticZeroTwoTF}{Error}{Assertion}{Count}{}{4}%
\StoreBenchExecResult{PdrInv}{KinductionDfStaticZeroTwoTF}{Error}{Assertion}{Cputime}{}{12.760919498}%
\StoreBenchExecResult{PdrInv}{KinductionDfStaticZeroTwoTF}{Error}{Assertion}{Cputime}{Avg}{3.1902298745}%
\StoreBenchExecResult{PdrInv}{KinductionDfStaticZeroTwoTF}{Error}{Assertion}{Cputime}{Median}{3.208643984}%
\StoreBenchExecResult{PdrInv}{KinductionDfStaticZeroTwoTF}{Error}{Assertion}{Cputime}{Min}{3.094553052}%
\StoreBenchExecResult{PdrInv}{KinductionDfStaticZeroTwoTF}{Error}{Assertion}{Cputime}{Max}{3.249078478}%
\StoreBenchExecResult{PdrInv}{KinductionDfStaticZeroTwoTF}{Error}{Assertion}{Cputime}{Stdev}{0.05802813160521106464483206629}%
\StoreBenchExecResult{PdrInv}{KinductionDfStaticZeroTwoTF}{Error}{Assertion}{Walltime}{}{7.06547689438}%
\StoreBenchExecResult{PdrInv}{KinductionDfStaticZeroTwoTF}{Error}{Assertion}{Walltime}{Avg}{1.766369223595}%
\StoreBenchExecResult{PdrInv}{KinductionDfStaticZeroTwoTF}{Error}{Assertion}{Walltime}{Median}{1.768117547035}%
\StoreBenchExecResult{PdrInv}{KinductionDfStaticZeroTwoTF}{Error}{Assertion}{Walltime}{Min}{1.73257780075}%
\StoreBenchExecResult{PdrInv}{KinductionDfStaticZeroTwoTF}{Error}{Assertion}{Walltime}{Max}{1.79666399956}%
\StoreBenchExecResult{PdrInv}{KinductionDfStaticZeroTwoTF}{Error}{Assertion}{Walltime}{Stdev}{0.02573398538510490374536266170}%
\StoreBenchExecResult{PdrInv}{KinductionDfStaticZeroTwoTF}{Error}{Error}{Count}{}{198}%
\StoreBenchExecResult{PdrInv}{KinductionDfStaticZeroTwoTF}{Error}{Error}{Cputime}{}{36821.226120114}%
\StoreBenchExecResult{PdrInv}{KinductionDfStaticZeroTwoTF}{Error}{Error}{Cputime}{Avg}{185.9657884854242424242424242}%
\StoreBenchExecResult{PdrInv}{KinductionDfStaticZeroTwoTF}{Error}{Error}{Cputime}{Median}{120.4270870795}%
\StoreBenchExecResult{PdrInv}{KinductionDfStaticZeroTwoTF}{Error}{Error}{Cputime}{Min}{2.426216131}%
\StoreBenchExecResult{PdrInv}{KinductionDfStaticZeroTwoTF}{Error}{Error}{Cputime}{Max}{826.46802186}%
\StoreBenchExecResult{PdrInv}{KinductionDfStaticZeroTwoTF}{Error}{Error}{Cputime}{Stdev}{194.7754780034602015525265780}%
\StoreBenchExecResult{PdrInv}{KinductionDfStaticZeroTwoTF}{Error}{Error}{Walltime}{}{30612.41054248960}%
\StoreBenchExecResult{PdrInv}{KinductionDfStaticZeroTwoTF}{Error}{Error}{Walltime}{Avg}{154.6081340529777777777777778}%
\StoreBenchExecResult{PdrInv}{KinductionDfStaticZeroTwoTF}{Error}{Error}{Walltime}{Median}{93.9041539431}%
\StoreBenchExecResult{PdrInv}{KinductionDfStaticZeroTwoTF}{Error}{Error}{Walltime}{Min}{1.34655117989}%
\StoreBenchExecResult{PdrInv}{KinductionDfStaticZeroTwoTF}{Error}{Error}{Walltime}{Max}{785.018320084}%
\StoreBenchExecResult{PdrInv}{KinductionDfStaticZeroTwoTF}{Error}{Error}{Walltime}{Stdev}{170.8848651820777870465761717}%
\StoreBenchExecResult{PdrInv}{KinductionDfStaticZeroTwoTF}{Error}{Exception}{Count}{}{13}%
\StoreBenchExecResult{PdrInv}{KinductionDfStaticZeroTwoTF}{Error}{Exception}{Cputime}{}{1464.046378316}%
\StoreBenchExecResult{PdrInv}{KinductionDfStaticZeroTwoTF}{Error}{Exception}{Cputime}{Avg}{112.6189521781538461538461538}%
\StoreBenchExecResult{PdrInv}{KinductionDfStaticZeroTwoTF}{Error}{Exception}{Cputime}{Median}{76.469722898}%
\StoreBenchExecResult{PdrInv}{KinductionDfStaticZeroTwoTF}{Error}{Exception}{Cputime}{Min}{18.334883213}%
\StoreBenchExecResult{PdrInv}{KinductionDfStaticZeroTwoTF}{Error}{Exception}{Cputime}{Max}{444.065638215}%
\StoreBenchExecResult{PdrInv}{KinductionDfStaticZeroTwoTF}{Error}{Exception}{Cputime}{Stdev}{118.8845423070812076130878464}%
\StoreBenchExecResult{PdrInv}{KinductionDfStaticZeroTwoTF}{Error}{Exception}{Walltime}{}{792.8184678561}%
\StoreBenchExecResult{PdrInv}{KinductionDfStaticZeroTwoTF}{Error}{Exception}{Walltime}{Avg}{60.98603598893076923076923077}%
\StoreBenchExecResult{PdrInv}{KinductionDfStaticZeroTwoTF}{Error}{Exception}{Walltime}{Median}{39.6774880886}%
\StoreBenchExecResult{PdrInv}{KinductionDfStaticZeroTwoTF}{Error}{Exception}{Walltime}{Min}{9.4049270153}%
\StoreBenchExecResult{PdrInv}{KinductionDfStaticZeroTwoTF}{Error}{Exception}{Walltime}{Max}{256.358600855}%
\StoreBenchExecResult{PdrInv}{KinductionDfStaticZeroTwoTF}{Error}{Exception}{Walltime}{Stdev}{66.89772503084794263968114434}%
\StoreBenchExecResult{PdrInv}{KinductionDfStaticZeroTwoTF}{Error}{OutOfJavaMemory}{Count}{}{11}%
\StoreBenchExecResult{PdrInv}{KinductionDfStaticZeroTwoTF}{Error}{OutOfJavaMemory}{Cputime}{}{5862.495755797}%
\StoreBenchExecResult{PdrInv}{KinductionDfStaticZeroTwoTF}{Error}{OutOfJavaMemory}{Cputime}{Avg}{532.9541596179090909090909091}%
\StoreBenchExecResult{PdrInv}{KinductionDfStaticZeroTwoTF}{Error}{OutOfJavaMemory}{Cputime}{Median}{522.612025999}%
\StoreBenchExecResult{PdrInv}{KinductionDfStaticZeroTwoTF}{Error}{OutOfJavaMemory}{Cputime}{Min}{160.664889085}%
\StoreBenchExecResult{PdrInv}{KinductionDfStaticZeroTwoTF}{Error}{OutOfJavaMemory}{Cputime}{Max}{850.760286023}%
\StoreBenchExecResult{PdrInv}{KinductionDfStaticZeroTwoTF}{Error}{OutOfJavaMemory}{Cputime}{Stdev}{220.2466617501171656197196613}%
\StoreBenchExecResult{PdrInv}{KinductionDfStaticZeroTwoTF}{Error}{OutOfJavaMemory}{Walltime}{}{3642.0547437664}%
\StoreBenchExecResult{PdrInv}{KinductionDfStaticZeroTwoTF}{Error}{OutOfJavaMemory}{Walltime}{Avg}{331.0958857969454545454545455}%
\StoreBenchExecResult{PdrInv}{KinductionDfStaticZeroTwoTF}{Error}{OutOfJavaMemory}{Walltime}{Median}{297.590420008}%
\StoreBenchExecResult{PdrInv}{KinductionDfStaticZeroTwoTF}{Error}{OutOfJavaMemory}{Walltime}{Min}{99.0947179794}%
\StoreBenchExecResult{PdrInv}{KinductionDfStaticZeroTwoTF}{Error}{OutOfJavaMemory}{Walltime}{Max}{577.828119993}%
\StoreBenchExecResult{PdrInv}{KinductionDfStaticZeroTwoTF}{Error}{OutOfJavaMemory}{Walltime}{Stdev}{149.6835766905386975438495872}%
\StoreBenchExecResult{PdrInv}{KinductionDfStaticZeroTwoTF}{Error}{OutOfMemory}{Count}{}{264}%
\StoreBenchExecResult{PdrInv}{KinductionDfStaticZeroTwoTF}{Error}{OutOfMemory}{Cputime}{}{94641.228660051}%
\StoreBenchExecResult{PdrInv}{KinductionDfStaticZeroTwoTF}{Error}{OutOfMemory}{Cputime}{Avg}{358.4895025001931818181818182}%
\StoreBenchExecResult{PdrInv}{KinductionDfStaticZeroTwoTF}{Error}{OutOfMemory}{Cputime}{Median}{284.4751263105}%
\StoreBenchExecResult{PdrInv}{KinductionDfStaticZeroTwoTF}{Error}{OutOfMemory}{Cputime}{Min}{160.857776081}%
\StoreBenchExecResult{PdrInv}{KinductionDfStaticZeroTwoTF}{Error}{OutOfMemory}{Cputime}{Max}{898.257566248}%
\StoreBenchExecResult{PdrInv}{KinductionDfStaticZeroTwoTF}{Error}{OutOfMemory}{Cputime}{Stdev}{200.5696849683138659529558327}%
\StoreBenchExecResult{PdrInv}{KinductionDfStaticZeroTwoTF}{Error}{OutOfMemory}{Walltime}{}{79770.7905673996}%
\StoreBenchExecResult{PdrInv}{KinductionDfStaticZeroTwoTF}{Error}{OutOfMemory}{Walltime}{Avg}{302.1620854825742424242424242}%
\StoreBenchExecResult{PdrInv}{KinductionDfStaticZeroTwoTF}{Error}{OutOfMemory}{Walltime}{Median}{207.7582731245}%
\StoreBenchExecResult{PdrInv}{KinductionDfStaticZeroTwoTF}{Error}{OutOfMemory}{Walltime}{Min}{85.564209938}%
\StoreBenchExecResult{PdrInv}{KinductionDfStaticZeroTwoTF}{Error}{OutOfMemory}{Walltime}{Max}{884.926100969}%
\StoreBenchExecResult{PdrInv}{KinductionDfStaticZeroTwoTF}{Error}{OutOfMemory}{Walltime}{Stdev}{221.0142442799828818628745809}%
\StoreBenchExecResult{PdrInv}{KinductionDfStaticZeroTwoTF}{Error}{Timeout}{Count}{}{2108}%
\StoreBenchExecResult{PdrInv}{KinductionDfStaticZeroTwoTF}{Error}{Timeout}{Cputime}{}{1914756.404712417}%
\StoreBenchExecResult{PdrInv}{KinductionDfStaticZeroTwoTF}{Error}{Timeout}{Cputime}{Avg}{908.3284652335944022770398482}%
\StoreBenchExecResult{PdrInv}{KinductionDfStaticZeroTwoTF}{Error}{Timeout}{Cputime}{Median}{902.2969213535}%
\StoreBenchExecResult{PdrInv}{KinductionDfStaticZeroTwoTF}{Error}{Timeout}{Cputime}{Min}{900.470704413}%
\StoreBenchExecResult{PdrInv}{KinductionDfStaticZeroTwoTF}{Error}{Timeout}{Cputime}{Max}{1002.35417576}%
\StoreBenchExecResult{PdrInv}{KinductionDfStaticZeroTwoTF}{Error}{Timeout}{Cputime}{Stdev}{19.82367356177052038243138584}%
\StoreBenchExecResult{PdrInv}{KinductionDfStaticZeroTwoTF}{Error}{Timeout}{Walltime}{}{1331569.021447186}%
\StoreBenchExecResult{PdrInv}{KinductionDfStaticZeroTwoTF}{Error}{Timeout}{Walltime}{Avg}{631.6741088459136622390891841}%
\StoreBenchExecResult{PdrInv}{KinductionDfStaticZeroTwoTF}{Error}{Timeout}{Walltime}{Median}{470.433751583}%
\StoreBenchExecResult{PdrInv}{KinductionDfStaticZeroTwoTF}{Error}{Timeout}{Walltime}{Min}{450.96033597}%
\StoreBenchExecResult{PdrInv}{KinductionDfStaticZeroTwoTF}{Error}{Timeout}{Walltime}{Max}{929.81765008}%
\StoreBenchExecResult{PdrInv}{KinductionDfStaticZeroTwoTF}{Error}{Timeout}{Walltime}{Stdev}{200.8452213750952929248796870}%
\StoreBenchExecResult{PdrInv}{KinductionDfStaticZeroTwoTF}{Wrong}{}{Count}{}{2}%
\StoreBenchExecResult{PdrInv}{KinductionDfStaticZeroTwoTF}{Wrong}{}{Cputime}{}{22.705044689}%
\StoreBenchExecResult{PdrInv}{KinductionDfStaticZeroTwoTF}{Wrong}{}{Cputime}{Avg}{11.3525223445}%
\StoreBenchExecResult{PdrInv}{KinductionDfStaticZeroTwoTF}{Wrong}{}{Cputime}{Median}{11.3525223445}%
\StoreBenchExecResult{PdrInv}{KinductionDfStaticZeroTwoTF}{Wrong}{}{Cputime}{Min}{4.010559026}%
\StoreBenchExecResult{PdrInv}{KinductionDfStaticZeroTwoTF}{Wrong}{}{Cputime}{Max}{18.694485663}%
\StoreBenchExecResult{PdrInv}{KinductionDfStaticZeroTwoTF}{Wrong}{}{Cputime}{Stdev}{7.3419633185}%
\StoreBenchExecResult{PdrInv}{KinductionDfStaticZeroTwoTF}{Wrong}{}{Walltime}{}{12.24670195581}%
\StoreBenchExecResult{PdrInv}{KinductionDfStaticZeroTwoTF}{Wrong}{}{Walltime}{Avg}{6.123350977905}%
\StoreBenchExecResult{PdrInv}{KinductionDfStaticZeroTwoTF}{Wrong}{}{Walltime}{Median}{6.123350977905}%
\StoreBenchExecResult{PdrInv}{KinductionDfStaticZeroTwoTF}{Wrong}{}{Walltime}{Min}{2.17791795731}%
\StoreBenchExecResult{PdrInv}{KinductionDfStaticZeroTwoTF}{Wrong}{}{Walltime}{Max}{10.0687839985}%
\StoreBenchExecResult{PdrInv}{KinductionDfStaticZeroTwoTF}{Wrong}{}{Walltime}{Stdev}{3.945433020595}%
\StoreBenchExecResult{PdrInv}{KinductionDfStaticZeroTwoTF}{Wrong}{False}{Count}{}{2}%
\StoreBenchExecResult{PdrInv}{KinductionDfStaticZeroTwoTF}{Wrong}{False}{Cputime}{}{22.705044689}%
\StoreBenchExecResult{PdrInv}{KinductionDfStaticZeroTwoTF}{Wrong}{False}{Cputime}{Avg}{11.3525223445}%
\StoreBenchExecResult{PdrInv}{KinductionDfStaticZeroTwoTF}{Wrong}{False}{Cputime}{Median}{11.3525223445}%
\StoreBenchExecResult{PdrInv}{KinductionDfStaticZeroTwoTF}{Wrong}{False}{Cputime}{Min}{4.010559026}%
\StoreBenchExecResult{PdrInv}{KinductionDfStaticZeroTwoTF}{Wrong}{False}{Cputime}{Max}{18.694485663}%
\StoreBenchExecResult{PdrInv}{KinductionDfStaticZeroTwoTF}{Wrong}{False}{Cputime}{Stdev}{7.3419633185}%
\StoreBenchExecResult{PdrInv}{KinductionDfStaticZeroTwoTF}{Wrong}{False}{Walltime}{}{12.24670195581}%
\StoreBenchExecResult{PdrInv}{KinductionDfStaticZeroTwoTF}{Wrong}{False}{Walltime}{Avg}{6.123350977905}%
\StoreBenchExecResult{PdrInv}{KinductionDfStaticZeroTwoTF}{Wrong}{False}{Walltime}{Median}{6.123350977905}%
\StoreBenchExecResult{PdrInv}{KinductionDfStaticZeroTwoTF}{Wrong}{False}{Walltime}{Min}{2.17791795731}%
\StoreBenchExecResult{PdrInv}{KinductionDfStaticZeroTwoTF}{Wrong}{False}{Walltime}{Max}{10.0687839985}%
\StoreBenchExecResult{PdrInv}{KinductionDfStaticZeroTwoTF}{Wrong}{False}{Walltime}{Stdev}{3.945433020595}%
\providecommand\StoreBenchExecResult[7]{\expandafter\newcommand\csname#1#2#3#4#5#6\endcsname{#7}}%
\StoreBenchExecResult{PdrInv}{KinductionDfStaticZeroTwoTTTrueNotSolvedByKinductionPlainButKipdr}{Total}{}{Count}{}{449}%
\StoreBenchExecResult{PdrInv}{KinductionDfStaticZeroTwoTTTrueNotSolvedByKinductionPlainButKipdr}{Total}{}{Cputime}{}{10362.698904288}%
\StoreBenchExecResult{PdrInv}{KinductionDfStaticZeroTwoTTTrueNotSolvedByKinductionPlainButKipdr}{Total}{}{Cputime}{Avg}{23.07950758193318485523385301}%
\StoreBenchExecResult{PdrInv}{KinductionDfStaticZeroTwoTTTrueNotSolvedByKinductionPlainButKipdr}{Total}{}{Cputime}{Median}{5.979093145}%
\StoreBenchExecResult{PdrInv}{KinductionDfStaticZeroTwoTTTrueNotSolvedByKinductionPlainButKipdr}{Total}{}{Cputime}{Min}{3.375486491}%
\StoreBenchExecResult{PdrInv}{KinductionDfStaticZeroTwoTTTrueNotSolvedByKinductionPlainButKipdr}{Total}{}{Cputime}{Max}{915.528216184}%
\StoreBenchExecResult{PdrInv}{KinductionDfStaticZeroTwoTTTrueNotSolvedByKinductionPlainButKipdr}{Total}{}{Cputime}{Stdev}{118.1337568770116192608670781}%
\StoreBenchExecResult{PdrInv}{KinductionDfStaticZeroTwoTTTrueNotSolvedByKinductionPlainButKipdr}{Total}{}{Walltime}{}{8313.63321828735}%
\StoreBenchExecResult{PdrInv}{KinductionDfStaticZeroTwoTTTrueNotSolvedByKinductionPlainButKipdr}{Total}{}{Walltime}{Avg}{18.51588690041726057906458797}%
\StoreBenchExecResult{PdrInv}{KinductionDfStaticZeroTwoTTTrueNotSolvedByKinductionPlainButKipdr}{Total}{}{Walltime}{Median}{3.1916410923}%
\StoreBenchExecResult{PdrInv}{KinductionDfStaticZeroTwoTTTrueNotSolvedByKinductionPlainButKipdr}{Total}{}{Walltime}{Min}{1.87939119339}%
\StoreBenchExecResult{PdrInv}{KinductionDfStaticZeroTwoTTTrueNotSolvedByKinductionPlainButKipdr}{Total}{}{Walltime}{Max}{901.808320999}%
\StoreBenchExecResult{PdrInv}{KinductionDfStaticZeroTwoTTTrueNotSolvedByKinductionPlainButKipdr}{Total}{}{Walltime}{Stdev}{111.0036543602434896064288927}%
\StoreBenchExecResult{PdrInv}{KinductionDfStaticZeroTwoTTTrueNotSolvedByKinductionPlainButKipdr}{Correct}{}{Count}{}{441}%
\StoreBenchExecResult{PdrInv}{KinductionDfStaticZeroTwoTTTrueNotSolvedByKinductionPlainButKipdr}{Correct}{}{Cputime}{}{3174.597851019}%
\StoreBenchExecResult{PdrInv}{KinductionDfStaticZeroTwoTTTrueNotSolvedByKinductionPlainButKipdr}{Correct}{}{Cputime}{Avg}{7.198634582809523809523809524}%
\StoreBenchExecResult{PdrInv}{KinductionDfStaticZeroTwoTTTrueNotSolvedByKinductionPlainButKipdr}{Correct}{}{Cputime}{Median}{5.906411852}%
\StoreBenchExecResult{PdrInv}{KinductionDfStaticZeroTwoTTTrueNotSolvedByKinductionPlainButKipdr}{Correct}{}{Cputime}{Min}{3.375486491}%
\StoreBenchExecResult{PdrInv}{KinductionDfStaticZeroTwoTTTrueNotSolvedByKinductionPlainButKipdr}{Correct}{}{Cputime}{Max}{79.788350592}%
\StoreBenchExecResult{PdrInv}{KinductionDfStaticZeroTwoTTTrueNotSolvedByKinductionPlainButKipdr}{Correct}{}{Cputime}{Stdev}{6.803784787936250035251400917}%
\StoreBenchExecResult{PdrInv}{KinductionDfStaticZeroTwoTTTrueNotSolvedByKinductionPlainButKipdr}{Correct}{}{Walltime}{}{1675.10056376435}%
\StoreBenchExecResult{PdrInv}{KinductionDfStaticZeroTwoTTTrueNotSolvedByKinductionPlainButKipdr}{Correct}{}{Walltime}{Avg}{3.798413976789909297052154195}%
\StoreBenchExecResult{PdrInv}{KinductionDfStaticZeroTwoTTTrueNotSolvedByKinductionPlainButKipdr}{Correct}{}{Walltime}{Median}{3.13349199295}%
\StoreBenchExecResult{PdrInv}{KinductionDfStaticZeroTwoTTTrueNotSolvedByKinductionPlainButKipdr}{Correct}{}{Walltime}{Min}{1.87939119339}%
\StoreBenchExecResult{PdrInv}{KinductionDfStaticZeroTwoTTTrueNotSolvedByKinductionPlainButKipdr}{Correct}{}{Walltime}{Max}{40.3914949894}%
\StoreBenchExecResult{PdrInv}{KinductionDfStaticZeroTwoTTTrueNotSolvedByKinductionPlainButKipdr}{Correct}{}{Walltime}{Stdev}{3.431568396623904430622728355}%
\StoreBenchExecResult{PdrInv}{KinductionDfStaticZeroTwoTTTrueNotSolvedByKinductionPlainButKipdr}{Correct}{True}{Count}{}{441}%
\StoreBenchExecResult{PdrInv}{KinductionDfStaticZeroTwoTTTrueNotSolvedByKinductionPlainButKipdr}{Correct}{True}{Cputime}{}{3174.597851019}%
\StoreBenchExecResult{PdrInv}{KinductionDfStaticZeroTwoTTTrueNotSolvedByKinductionPlainButKipdr}{Correct}{True}{Cputime}{Avg}{7.198634582809523809523809524}%
\StoreBenchExecResult{PdrInv}{KinductionDfStaticZeroTwoTTTrueNotSolvedByKinductionPlainButKipdr}{Correct}{True}{Cputime}{Median}{5.906411852}%
\StoreBenchExecResult{PdrInv}{KinductionDfStaticZeroTwoTTTrueNotSolvedByKinductionPlainButKipdr}{Correct}{True}{Cputime}{Min}{3.375486491}%
\StoreBenchExecResult{PdrInv}{KinductionDfStaticZeroTwoTTTrueNotSolvedByKinductionPlainButKipdr}{Correct}{True}{Cputime}{Max}{79.788350592}%
\StoreBenchExecResult{PdrInv}{KinductionDfStaticZeroTwoTTTrueNotSolvedByKinductionPlainButKipdr}{Correct}{True}{Cputime}{Stdev}{6.803784787936250035251400917}%
\StoreBenchExecResult{PdrInv}{KinductionDfStaticZeroTwoTTTrueNotSolvedByKinductionPlainButKipdr}{Correct}{True}{Walltime}{}{1675.10056376435}%
\StoreBenchExecResult{PdrInv}{KinductionDfStaticZeroTwoTTTrueNotSolvedByKinductionPlainButKipdr}{Correct}{True}{Walltime}{Avg}{3.798413976789909297052154195}%
\StoreBenchExecResult{PdrInv}{KinductionDfStaticZeroTwoTTTrueNotSolvedByKinductionPlainButKipdr}{Correct}{True}{Walltime}{Median}{3.13349199295}%
\StoreBenchExecResult{PdrInv}{KinductionDfStaticZeroTwoTTTrueNotSolvedByKinductionPlainButKipdr}{Correct}{True}{Walltime}{Min}{1.87939119339}%
\StoreBenchExecResult{PdrInv}{KinductionDfStaticZeroTwoTTTrueNotSolvedByKinductionPlainButKipdr}{Correct}{True}{Walltime}{Max}{40.3914949894}%
\StoreBenchExecResult{PdrInv}{KinductionDfStaticZeroTwoTTTrueNotSolvedByKinductionPlainButKipdr}{Correct}{True}{Walltime}{Stdev}{3.431568396623904430622728355}%
\StoreBenchExecResult{PdrInv}{KinductionDfStaticZeroTwoTTTrueNotSolvedByKinductionPlainButKipdr}{Wrong}{True}{Count}{}{0}%
\StoreBenchExecResult{PdrInv}{KinductionDfStaticZeroTwoTTTrueNotSolvedByKinductionPlainButKipdr}{Wrong}{True}{Cputime}{}{0}%
\StoreBenchExecResult{PdrInv}{KinductionDfStaticZeroTwoTTTrueNotSolvedByKinductionPlainButKipdr}{Wrong}{True}{Cputime}{Avg}{None}%
\StoreBenchExecResult{PdrInv}{KinductionDfStaticZeroTwoTTTrueNotSolvedByKinductionPlainButKipdr}{Wrong}{True}{Cputime}{Median}{None}%
\StoreBenchExecResult{PdrInv}{KinductionDfStaticZeroTwoTTTrueNotSolvedByKinductionPlainButKipdr}{Wrong}{True}{Cputime}{Min}{None}%
\StoreBenchExecResult{PdrInv}{KinductionDfStaticZeroTwoTTTrueNotSolvedByKinductionPlainButKipdr}{Wrong}{True}{Cputime}{Max}{None}%
\StoreBenchExecResult{PdrInv}{KinductionDfStaticZeroTwoTTTrueNotSolvedByKinductionPlainButKipdr}{Wrong}{True}{Cputime}{Stdev}{None}%
\StoreBenchExecResult{PdrInv}{KinductionDfStaticZeroTwoTTTrueNotSolvedByKinductionPlainButKipdr}{Wrong}{True}{Walltime}{}{0}%
\StoreBenchExecResult{PdrInv}{KinductionDfStaticZeroTwoTTTrueNotSolvedByKinductionPlainButKipdr}{Wrong}{True}{Walltime}{Avg}{None}%
\StoreBenchExecResult{PdrInv}{KinductionDfStaticZeroTwoTTTrueNotSolvedByKinductionPlainButKipdr}{Wrong}{True}{Walltime}{Median}{None}%
\StoreBenchExecResult{PdrInv}{KinductionDfStaticZeroTwoTTTrueNotSolvedByKinductionPlainButKipdr}{Wrong}{True}{Walltime}{Min}{None}%
\StoreBenchExecResult{PdrInv}{KinductionDfStaticZeroTwoTTTrueNotSolvedByKinductionPlainButKipdr}{Wrong}{True}{Walltime}{Max}{None}%
\StoreBenchExecResult{PdrInv}{KinductionDfStaticZeroTwoTTTrueNotSolvedByKinductionPlainButKipdr}{Wrong}{True}{Walltime}{Stdev}{None}%
\StoreBenchExecResult{PdrInv}{KinductionDfStaticZeroTwoTTTrueNotSolvedByKinductionPlainButKipdr}{Error}{}{Count}{}{8}%
\StoreBenchExecResult{PdrInv}{KinductionDfStaticZeroTwoTTTrueNotSolvedByKinductionPlainButKipdr}{Error}{}{Cputime}{}{7188.101053269}%
\StoreBenchExecResult{PdrInv}{KinductionDfStaticZeroTwoTTTrueNotSolvedByKinductionPlainButKipdr}{Error}{}{Cputime}{Avg}{898.512631658625}%
\StoreBenchExecResult{PdrInv}{KinductionDfStaticZeroTwoTTTrueNotSolvedByKinductionPlainButKipdr}{Error}{}{Cputime}{Median}{903.8352687545}%
\StoreBenchExecResult{PdrInv}{KinductionDfStaticZeroTwoTTTrueNotSolvedByKinductionPlainButKipdr}{Error}{}{Cputime}{Min}{845.419291568}%
\StoreBenchExecResult{PdrInv}{KinductionDfStaticZeroTwoTTTrueNotSolvedByKinductionPlainButKipdr}{Error}{}{Cputime}{Max}{915.528216184}%
\StoreBenchExecResult{PdrInv}{KinductionDfStaticZeroTwoTTTrueNotSolvedByKinductionPlainButKipdr}{Error}{}{Cputime}{Stdev}{20.48330058402428394398412704}%
\StoreBenchExecResult{PdrInv}{KinductionDfStaticZeroTwoTTTrueNotSolvedByKinductionPlainButKipdr}{Error}{}{Walltime}{}{6638.532654523}%
\StoreBenchExecResult{PdrInv}{KinductionDfStaticZeroTwoTTTrueNotSolvedByKinductionPlainButKipdr}{Error}{}{Walltime}{Avg}{829.816581815375}%
\StoreBenchExecResult{PdrInv}{KinductionDfStaticZeroTwoTTTrueNotSolvedByKinductionPlainButKipdr}{Error}{}{Walltime}{Median}{888.3491876125}%
\StoreBenchExecResult{PdrInv}{KinductionDfStaticZeroTwoTTTrueNotSolvedByKinductionPlainButKipdr}{Error}{}{Walltime}{Min}{452.432250023}%
\StoreBenchExecResult{PdrInv}{KinductionDfStaticZeroTwoTTTrueNotSolvedByKinductionPlainButKipdr}{Error}{}{Walltime}{Max}{901.808320999}%
\StoreBenchExecResult{PdrInv}{KinductionDfStaticZeroTwoTTTrueNotSolvedByKinductionPlainButKipdr}{Error}{}{Walltime}{Stdev}{144.0951464681785866248907003}%
\StoreBenchExecResult{PdrInv}{KinductionDfStaticZeroTwoTTTrueNotSolvedByKinductionPlainButKipdr}{Error}{OutOfMemory}{Count}{}{1}%
\StoreBenchExecResult{PdrInv}{KinductionDfStaticZeroTwoTTTrueNotSolvedByKinductionPlainButKipdr}{Error}{OutOfMemory}{Cputime}{}{845.419291568}%
\StoreBenchExecResult{PdrInv}{KinductionDfStaticZeroTwoTTTrueNotSolvedByKinductionPlainButKipdr}{Error}{OutOfMemory}{Cputime}{Avg}{845.419291568}%
\StoreBenchExecResult{PdrInv}{KinductionDfStaticZeroTwoTTTrueNotSolvedByKinductionPlainButKipdr}{Error}{OutOfMemory}{Cputime}{Median}{845.419291568}%
\StoreBenchExecResult{PdrInv}{KinductionDfStaticZeroTwoTTTrueNotSolvedByKinductionPlainButKipdr}{Error}{OutOfMemory}{Cputime}{Min}{845.419291568}%
\StoreBenchExecResult{PdrInv}{KinductionDfStaticZeroTwoTTTrueNotSolvedByKinductionPlainButKipdr}{Error}{OutOfMemory}{Cputime}{Max}{845.419291568}%
\StoreBenchExecResult{PdrInv}{KinductionDfStaticZeroTwoTTTrueNotSolvedByKinductionPlainButKipdr}{Error}{OutOfMemory}{Cputime}{Stdev}{0E-9}%
\StoreBenchExecResult{PdrInv}{KinductionDfStaticZeroTwoTTTrueNotSolvedByKinductionPlainButKipdr}{Error}{OutOfMemory}{Walltime}{}{831.671846151}%
\StoreBenchExecResult{PdrInv}{KinductionDfStaticZeroTwoTTTrueNotSolvedByKinductionPlainButKipdr}{Error}{OutOfMemory}{Walltime}{Avg}{831.671846151}%
\StoreBenchExecResult{PdrInv}{KinductionDfStaticZeroTwoTTTrueNotSolvedByKinductionPlainButKipdr}{Error}{OutOfMemory}{Walltime}{Median}{831.671846151}%
\StoreBenchExecResult{PdrInv}{KinductionDfStaticZeroTwoTTTrueNotSolvedByKinductionPlainButKipdr}{Error}{OutOfMemory}{Walltime}{Min}{831.671846151}%
\StoreBenchExecResult{PdrInv}{KinductionDfStaticZeroTwoTTTrueNotSolvedByKinductionPlainButKipdr}{Error}{OutOfMemory}{Walltime}{Max}{831.671846151}%
\StoreBenchExecResult{PdrInv}{KinductionDfStaticZeroTwoTTTrueNotSolvedByKinductionPlainButKipdr}{Error}{OutOfMemory}{Walltime}{Stdev}{0E-9}%
\StoreBenchExecResult{PdrInv}{KinductionDfStaticZeroTwoTTTrueNotSolvedByKinductionPlainButKipdr}{Error}{Timeout}{Count}{}{7}%
\StoreBenchExecResult{PdrInv}{KinductionDfStaticZeroTwoTTTrueNotSolvedByKinductionPlainButKipdr}{Error}{Timeout}{Cputime}{}{6342.681761701}%
\StoreBenchExecResult{PdrInv}{KinductionDfStaticZeroTwoTTTrueNotSolvedByKinductionPlainButKipdr}{Error}{Timeout}{Cputime}{Avg}{906.0973945287142857142857143}%
\StoreBenchExecResult{PdrInv}{KinductionDfStaticZeroTwoTTTrueNotSolvedByKinductionPlainButKipdr}{Error}{Timeout}{Cputime}{Median}{903.968253896}%
\StoreBenchExecResult{PdrInv}{KinductionDfStaticZeroTwoTTTrueNotSolvedByKinductionPlainButKipdr}{Error}{Timeout}{Cputime}{Min}{902.397540097}%
\StoreBenchExecResult{PdrInv}{KinductionDfStaticZeroTwoTTTrueNotSolvedByKinductionPlainButKipdr}{Error}{Timeout}{Cputime}{Max}{915.528216184}%
\StoreBenchExecResult{PdrInv}{KinductionDfStaticZeroTwoTTTrueNotSolvedByKinductionPlainButKipdr}{Error}{Timeout}{Cputime}{Stdev}{4.390276040406201391809678529}%
\StoreBenchExecResult{PdrInv}{KinductionDfStaticZeroTwoTTTrueNotSolvedByKinductionPlainButKipdr}{Error}{Timeout}{Walltime}{}{5806.860808372}%
\StoreBenchExecResult{PdrInv}{KinductionDfStaticZeroTwoTTTrueNotSolvedByKinductionPlainButKipdr}{Error}{Timeout}{Walltime}{Avg}{829.5515440531428571428571429}%
\StoreBenchExecResult{PdrInv}{KinductionDfStaticZeroTwoTTTrueNotSolvedByKinductionPlainButKipdr}{Error}{Timeout}{Walltime}{Median}{888.496967077}%
\StoreBenchExecResult{PdrInv}{KinductionDfStaticZeroTwoTTTrueNotSolvedByKinductionPlainButKipdr}{Error}{Timeout}{Walltime}{Min}{452.432250023}%
\StoreBenchExecResult{PdrInv}{KinductionDfStaticZeroTwoTTTrueNotSolvedByKinductionPlainButKipdr}{Error}{Timeout}{Walltime}{Max}{901.808320999}%
\StoreBenchExecResult{PdrInv}{KinductionDfStaticZeroTwoTTTrueNotSolvedByKinductionPlainButKipdr}{Error}{Timeout}{Walltime}{Stdev}{154.0423671613122957298843356}%
\providecommand\StoreBenchExecResult[7]{\expandafter\newcommand\csname#1#2#3#4#5#6\endcsname{#7}}%
\StoreBenchExecResult{PdrInv}{KinductionDfStaticZeroTwoTTTrueNotSolvedByKinductionPlain}{Total}{}{Count}{}{2893}%
\StoreBenchExecResult{PdrInv}{KinductionDfStaticZeroTwoTTTrueNotSolvedByKinductionPlain}{Total}{}{Cputime}{}{1620514.460272006}%
\StoreBenchExecResult{PdrInv}{KinductionDfStaticZeroTwoTTTrueNotSolvedByKinductionPlain}{Total}{}{Cputime}{Avg}{560.1501763816128586242654684}%
\StoreBenchExecResult{PdrInv}{KinductionDfStaticZeroTwoTTTrueNotSolvedByKinductionPlain}{Total}{}{Cputime}{Median}{901.146746428}%
\StoreBenchExecResult{PdrInv}{KinductionDfStaticZeroTwoTTTrueNotSolvedByKinductionPlain}{Total}{}{Cputime}{Min}{2.474144852}%
\StoreBenchExecResult{PdrInv}{KinductionDfStaticZeroTwoTTTrueNotSolvedByKinductionPlain}{Total}{}{Cputime}{Max}{1002.37842465}%
\StoreBenchExecResult{PdrInv}{KinductionDfStaticZeroTwoTTTrueNotSolvedByKinductionPlain}{Total}{}{Cputime}{Stdev}{417.7757677463471764880566936}%
\StoreBenchExecResult{PdrInv}{KinductionDfStaticZeroTwoTTTrueNotSolvedByKinductionPlain}{Total}{}{Walltime}{}{1119093.90487647341}%
\StoreBenchExecResult{PdrInv}{KinductionDfStaticZeroTwoTTTrueNotSolvedByKinductionPlain}{Total}{}{Walltime}{Avg}{386.8281731339348116142412720}%
\StoreBenchExecResult{PdrInv}{KinductionDfStaticZeroTwoTTTrueNotSolvedByKinductionPlain}{Total}{}{Walltime}{Median}{451.81624794}%
\StoreBenchExecResult{PdrInv}{KinductionDfStaticZeroTwoTTTrueNotSolvedByKinductionPlain}{Total}{}{Walltime}{Min}{1.36245203018}%
\StoreBenchExecResult{PdrInv}{KinductionDfStaticZeroTwoTTTrueNotSolvedByKinductionPlain}{Total}{}{Walltime}{Max}{904.666206837}%
\StoreBenchExecResult{PdrInv}{KinductionDfStaticZeroTwoTTTrueNotSolvedByKinductionPlain}{Total}{}{Walltime}{Stdev}{327.9692643730342551938941569}%
\StoreBenchExecResult{PdrInv}{KinductionDfStaticZeroTwoTTTrueNotSolvedByKinductionPlain}{Correct}{}{Count}{}{949}%
\StoreBenchExecResult{PdrInv}{KinductionDfStaticZeroTwoTTTrueNotSolvedByKinductionPlain}{Correct}{}{Cputime}{}{38662.672376005}%
\StoreBenchExecResult{PdrInv}{KinductionDfStaticZeroTwoTTTrueNotSolvedByKinductionPlain}{Correct}{}{Cputime}{Avg}{40.74043453741306638566912540}%
\StoreBenchExecResult{PdrInv}{KinductionDfStaticZeroTwoTTTrueNotSolvedByKinductionPlain}{Correct}{}{Cputime}{Median}{7.191191815}%
\StoreBenchExecResult{PdrInv}{KinductionDfStaticZeroTwoTTTrueNotSolvedByKinductionPlain}{Correct}{}{Cputime}{Min}{3.331639646}%
\StoreBenchExecResult{PdrInv}{KinductionDfStaticZeroTwoTTTrueNotSolvedByKinductionPlain}{Correct}{}{Cputime}{Max}{883.567764281}%
\StoreBenchExecResult{PdrInv}{KinductionDfStaticZeroTwoTTTrueNotSolvedByKinductionPlain}{Correct}{}{Cputime}{Stdev}{112.4591597826857177073849575}%
\StoreBenchExecResult{PdrInv}{KinductionDfStaticZeroTwoTTTrueNotSolvedByKinductionPlain}{Correct}{}{Walltime}{}{21259.27748179263}%
\StoreBenchExecResult{PdrInv}{KinductionDfStaticZeroTwoTTTrueNotSolvedByKinductionPlain}{Correct}{}{Walltime}{Avg}{22.40176763097221285563751317}%
\StoreBenchExecResult{PdrInv}{KinductionDfStaticZeroTwoTTTrueNotSolvedByKinductionPlain}{Correct}{}{Walltime}{Median}{3.80509400368}%
\StoreBenchExecResult{PdrInv}{KinductionDfStaticZeroTwoTTTrueNotSolvedByKinductionPlain}{Correct}{}{Walltime}{Min}{1.82239103317}%
\StoreBenchExecResult{PdrInv}{KinductionDfStaticZeroTwoTTTrueNotSolvedByKinductionPlain}{Correct}{}{Walltime}{Max}{846.123702049}%
\StoreBenchExecResult{PdrInv}{KinductionDfStaticZeroTwoTTTrueNotSolvedByKinductionPlain}{Correct}{}{Walltime}{Stdev}{69.07920203947794604051853465}%
\StoreBenchExecResult{PdrInv}{KinductionDfStaticZeroTwoTTTrueNotSolvedByKinductionPlain}{Correct}{True}{Count}{}{949}%
\StoreBenchExecResult{PdrInv}{KinductionDfStaticZeroTwoTTTrueNotSolvedByKinductionPlain}{Correct}{True}{Cputime}{}{38662.672376005}%
\StoreBenchExecResult{PdrInv}{KinductionDfStaticZeroTwoTTTrueNotSolvedByKinductionPlain}{Correct}{True}{Cputime}{Avg}{40.74043453741306638566912540}%
\StoreBenchExecResult{PdrInv}{KinductionDfStaticZeroTwoTTTrueNotSolvedByKinductionPlain}{Correct}{True}{Cputime}{Median}{7.191191815}%
\StoreBenchExecResult{PdrInv}{KinductionDfStaticZeroTwoTTTrueNotSolvedByKinductionPlain}{Correct}{True}{Cputime}{Min}{3.331639646}%
\StoreBenchExecResult{PdrInv}{KinductionDfStaticZeroTwoTTTrueNotSolvedByKinductionPlain}{Correct}{True}{Cputime}{Max}{883.567764281}%
\StoreBenchExecResult{PdrInv}{KinductionDfStaticZeroTwoTTTrueNotSolvedByKinductionPlain}{Correct}{True}{Cputime}{Stdev}{112.4591597826857177073849575}%
\StoreBenchExecResult{PdrInv}{KinductionDfStaticZeroTwoTTTrueNotSolvedByKinductionPlain}{Correct}{True}{Walltime}{}{21259.27748179263}%
\StoreBenchExecResult{PdrInv}{KinductionDfStaticZeroTwoTTTrueNotSolvedByKinductionPlain}{Correct}{True}{Walltime}{Avg}{22.40176763097221285563751317}%
\StoreBenchExecResult{PdrInv}{KinductionDfStaticZeroTwoTTTrueNotSolvedByKinductionPlain}{Correct}{True}{Walltime}{Median}{3.80509400368}%
\StoreBenchExecResult{PdrInv}{KinductionDfStaticZeroTwoTTTrueNotSolvedByKinductionPlain}{Correct}{True}{Walltime}{Min}{1.82239103317}%
\StoreBenchExecResult{PdrInv}{KinductionDfStaticZeroTwoTTTrueNotSolvedByKinductionPlain}{Correct}{True}{Walltime}{Max}{846.123702049}%
\StoreBenchExecResult{PdrInv}{KinductionDfStaticZeroTwoTTTrueNotSolvedByKinductionPlain}{Correct}{True}{Walltime}{Stdev}{69.07920203947794604051853465}%
\StoreBenchExecResult{PdrInv}{KinductionDfStaticZeroTwoTTTrueNotSolvedByKinductionPlain}{Wrong}{True}{Count}{}{0}%
\StoreBenchExecResult{PdrInv}{KinductionDfStaticZeroTwoTTTrueNotSolvedByKinductionPlain}{Wrong}{True}{Cputime}{}{0}%
\StoreBenchExecResult{PdrInv}{KinductionDfStaticZeroTwoTTTrueNotSolvedByKinductionPlain}{Wrong}{True}{Cputime}{Avg}{None}%
\StoreBenchExecResult{PdrInv}{KinductionDfStaticZeroTwoTTTrueNotSolvedByKinductionPlain}{Wrong}{True}{Cputime}{Median}{None}%
\StoreBenchExecResult{PdrInv}{KinductionDfStaticZeroTwoTTTrueNotSolvedByKinductionPlain}{Wrong}{True}{Cputime}{Min}{None}%
\StoreBenchExecResult{PdrInv}{KinductionDfStaticZeroTwoTTTrueNotSolvedByKinductionPlain}{Wrong}{True}{Cputime}{Max}{None}%
\StoreBenchExecResult{PdrInv}{KinductionDfStaticZeroTwoTTTrueNotSolvedByKinductionPlain}{Wrong}{True}{Cputime}{Stdev}{None}%
\StoreBenchExecResult{PdrInv}{KinductionDfStaticZeroTwoTTTrueNotSolvedByKinductionPlain}{Wrong}{True}{Walltime}{}{0}%
\StoreBenchExecResult{PdrInv}{KinductionDfStaticZeroTwoTTTrueNotSolvedByKinductionPlain}{Wrong}{True}{Walltime}{Avg}{None}%
\StoreBenchExecResult{PdrInv}{KinductionDfStaticZeroTwoTTTrueNotSolvedByKinductionPlain}{Wrong}{True}{Walltime}{Median}{None}%
\StoreBenchExecResult{PdrInv}{KinductionDfStaticZeroTwoTTTrueNotSolvedByKinductionPlain}{Wrong}{True}{Walltime}{Min}{None}%
\StoreBenchExecResult{PdrInv}{KinductionDfStaticZeroTwoTTTrueNotSolvedByKinductionPlain}{Wrong}{True}{Walltime}{Max}{None}%
\StoreBenchExecResult{PdrInv}{KinductionDfStaticZeroTwoTTTrueNotSolvedByKinductionPlain}{Wrong}{True}{Walltime}{Stdev}{None}%
\StoreBenchExecResult{PdrInv}{KinductionDfStaticZeroTwoTTTrueNotSolvedByKinductionPlain}{Error}{}{Count}{}{1944}%
\StoreBenchExecResult{PdrInv}{KinductionDfStaticZeroTwoTTTrueNotSolvedByKinductionPlain}{Error}{}{Cputime}{}{1581851.787896001}%
\StoreBenchExecResult{PdrInv}{KinductionDfStaticZeroTwoTTTrueNotSolvedByKinductionPlain}{Error}{}{Cputime}{Avg}{813.7097674362145061728395062}%
\StoreBenchExecResult{PdrInv}{KinductionDfStaticZeroTwoTTTrueNotSolvedByKinductionPlain}{Error}{}{Cputime}{Median}{901.735510104}%
\StoreBenchExecResult{PdrInv}{KinductionDfStaticZeroTwoTTTrueNotSolvedByKinductionPlain}{Error}{}{Cputime}{Min}{2.474144852}%
\StoreBenchExecResult{PdrInv}{KinductionDfStaticZeroTwoTTTrueNotSolvedByKinductionPlain}{Error}{}{Cputime}{Max}{1002.37842465}%
\StoreBenchExecResult{PdrInv}{KinductionDfStaticZeroTwoTTTrueNotSolvedByKinductionPlain}{Error}{}{Cputime}{Stdev}{239.9420956341458248521785424}%
\StoreBenchExecResult{PdrInv}{KinductionDfStaticZeroTwoTTTrueNotSolvedByKinductionPlain}{Error}{}{Walltime}{}{1097834.62739468078}%
\StoreBenchExecResult{PdrInv}{KinductionDfStaticZeroTwoTTTrueNotSolvedByKinductionPlain}{Error}{}{Walltime}{Avg}{564.7297466022020473251028807}%
\StoreBenchExecResult{PdrInv}{KinductionDfStaticZeroTwoTTTrueNotSolvedByKinductionPlain}{Error}{}{Walltime}{Median}{454.7170020345}%
\StoreBenchExecResult{PdrInv}{KinductionDfStaticZeroTwoTTTrueNotSolvedByKinductionPlain}{Error}{}{Walltime}{Min}{1.36245203018}%
\StoreBenchExecResult{PdrInv}{KinductionDfStaticZeroTwoTTTrueNotSolvedByKinductionPlain}{Error}{}{Walltime}{Max}{904.666206837}%
\StoreBenchExecResult{PdrInv}{KinductionDfStaticZeroTwoTTTrueNotSolvedByKinductionPlain}{Error}{}{Walltime}{Stdev}{247.5128872424187190037642303}%
\StoreBenchExecResult{PdrInv}{KinductionDfStaticZeroTwoTTTrueNotSolvedByKinductionPlain}{Error}{Assertion}{Count}{}{2}%
\StoreBenchExecResult{PdrInv}{KinductionDfStaticZeroTwoTTTrueNotSolvedByKinductionPlain}{Error}{Assertion}{Cputime}{}{6.557490180}%
\StoreBenchExecResult{PdrInv}{KinductionDfStaticZeroTwoTTTrueNotSolvedByKinductionPlain}{Error}{Assertion}{Cputime}{Avg}{3.278745090}%
\StoreBenchExecResult{PdrInv}{KinductionDfStaticZeroTwoTTTrueNotSolvedByKinductionPlain}{Error}{Assertion}{Cputime}{Median}{3.278745090}%
\StoreBenchExecResult{PdrInv}{KinductionDfStaticZeroTwoTTTrueNotSolvedByKinductionPlain}{Error}{Assertion}{Cputime}{Min}{3.235285018}%
\StoreBenchExecResult{PdrInv}{KinductionDfStaticZeroTwoTTTrueNotSolvedByKinductionPlain}{Error}{Assertion}{Cputime}{Max}{3.322205162}%
\StoreBenchExecResult{PdrInv}{KinductionDfStaticZeroTwoTTTrueNotSolvedByKinductionPlain}{Error}{Assertion}{Cputime}{Stdev}{0.043460072}%
\StoreBenchExecResult{PdrInv}{KinductionDfStaticZeroTwoTTTrueNotSolvedByKinductionPlain}{Error}{Assertion}{Walltime}{}{3.61097598075}%
\StoreBenchExecResult{PdrInv}{KinductionDfStaticZeroTwoTTTrueNotSolvedByKinductionPlain}{Error}{Assertion}{Walltime}{Avg}{1.805487990375}%
\StoreBenchExecResult{PdrInv}{KinductionDfStaticZeroTwoTTTrueNotSolvedByKinductionPlain}{Error}{Assertion}{Walltime}{Median}{1.805487990375}%
\StoreBenchExecResult{PdrInv}{KinductionDfStaticZeroTwoTTTrueNotSolvedByKinductionPlain}{Error}{Assertion}{Walltime}{Min}{1.78283500671}%
\StoreBenchExecResult{PdrInv}{KinductionDfStaticZeroTwoTTTrueNotSolvedByKinductionPlain}{Error}{Assertion}{Walltime}{Max}{1.82814097404}%
\StoreBenchExecResult{PdrInv}{KinductionDfStaticZeroTwoTTTrueNotSolvedByKinductionPlain}{Error}{Assertion}{Walltime}{Stdev}{0.022652983665}%
\StoreBenchExecResult{PdrInv}{KinductionDfStaticZeroTwoTTTrueNotSolvedByKinductionPlain}{Error}{Error}{Count}{}{130}%
\StoreBenchExecResult{PdrInv}{KinductionDfStaticZeroTwoTTTrueNotSolvedByKinductionPlain}{Error}{Error}{Cputime}{}{22209.545631935}%
\StoreBenchExecResult{PdrInv}{KinductionDfStaticZeroTwoTTTrueNotSolvedByKinductionPlain}{Error}{Error}{Cputime}{Avg}{170.8426587071923076923076923}%
\StoreBenchExecResult{PdrInv}{KinductionDfStaticZeroTwoTTTrueNotSolvedByKinductionPlain}{Error}{Error}{Cputime}{Median}{102.401722072}%
\StoreBenchExecResult{PdrInv}{KinductionDfStaticZeroTwoTTTrueNotSolvedByKinductionPlain}{Error}{Error}{Cputime}{Min}{2.474144852}%
\StoreBenchExecResult{PdrInv}{KinductionDfStaticZeroTwoTTTrueNotSolvedByKinductionPlain}{Error}{Error}{Cputime}{Max}{851.506871282}%
\StoreBenchExecResult{PdrInv}{KinductionDfStaticZeroTwoTTTrueNotSolvedByKinductionPlain}{Error}{Error}{Cputime}{Stdev}{194.0570005291027984944209476}%
\StoreBenchExecResult{PdrInv}{KinductionDfStaticZeroTwoTTTrueNotSolvedByKinductionPlain}{Error}{Error}{Walltime}{}{18705.18491768873}%
\StoreBenchExecResult{PdrInv}{KinductionDfStaticZeroTwoTTTrueNotSolvedByKinductionPlain}{Error}{Error}{Walltime}{Avg}{143.8860378283748461538461538}%
\StoreBenchExecResult{PdrInv}{KinductionDfStaticZeroTwoTTTrueNotSolvedByKinductionPlain}{Error}{Error}{Walltime}{Median}{76.9642339945}%
\StoreBenchExecResult{PdrInv}{KinductionDfStaticZeroTwoTTTrueNotSolvedByKinductionPlain}{Error}{Error}{Walltime}{Min}{1.36245203018}%
\StoreBenchExecResult{PdrInv}{KinductionDfStaticZeroTwoTTTrueNotSolvedByKinductionPlain}{Error}{Error}{Walltime}{Max}{835.936573029}%
\StoreBenchExecResult{PdrInv}{KinductionDfStaticZeroTwoTTTrueNotSolvedByKinductionPlain}{Error}{Error}{Walltime}{Stdev}{173.4870232887416100657157260}%
\StoreBenchExecResult{PdrInv}{KinductionDfStaticZeroTwoTTTrueNotSolvedByKinductionPlain}{Error}{Exception}{Count}{}{17}%
\StoreBenchExecResult{PdrInv}{KinductionDfStaticZeroTwoTTTrueNotSolvedByKinductionPlain}{Error}{Exception}{Cputime}{}{4411.708540334}%
\StoreBenchExecResult{PdrInv}{KinductionDfStaticZeroTwoTTTrueNotSolvedByKinductionPlain}{Error}{Exception}{Cputime}{Avg}{259.5122670784705882352941176}%
\StoreBenchExecResult{PdrInv}{KinductionDfStaticZeroTwoTTTrueNotSolvedByKinductionPlain}{Error}{Exception}{Cputime}{Median}{236.89622936}%
\StoreBenchExecResult{PdrInv}{KinductionDfStaticZeroTwoTTTrueNotSolvedByKinductionPlain}{Error}{Exception}{Cputime}{Min}{15.007275125}%
\StoreBenchExecResult{PdrInv}{KinductionDfStaticZeroTwoTTTrueNotSolvedByKinductionPlain}{Error}{Exception}{Cputime}{Max}{899.809594714}%
\StoreBenchExecResult{PdrInv}{KinductionDfStaticZeroTwoTTTrueNotSolvedByKinductionPlain}{Error}{Exception}{Cputime}{Stdev}{221.3711191691447662963973729}%
\StoreBenchExecResult{PdrInv}{KinductionDfStaticZeroTwoTTTrueNotSolvedByKinductionPlain}{Error}{Exception}{Walltime}{}{2228.2231562150}%
\StoreBenchExecResult{PdrInv}{KinductionDfStaticZeroTwoTTTrueNotSolvedByKinductionPlain}{Error}{Exception}{Walltime}{Avg}{131.0719503655882352941176471}%
\StoreBenchExecResult{PdrInv}{KinductionDfStaticZeroTwoTTTrueNotSolvedByKinductionPlain}{Error}{Exception}{Walltime}{Median}{118.99434495}%
\StoreBenchExecResult{PdrInv}{KinductionDfStaticZeroTwoTTTrueNotSolvedByKinductionPlain}{Error}{Exception}{Walltime}{Min}{7.7244079113}%
\StoreBenchExecResult{PdrInv}{KinductionDfStaticZeroTwoTTTrueNotSolvedByKinductionPlain}{Error}{Exception}{Walltime}{Max}{450.515960932}%
\StoreBenchExecResult{PdrInv}{KinductionDfStaticZeroTwoTTTrueNotSolvedByKinductionPlain}{Error}{Exception}{Walltime}{Stdev}{110.3393252356436048120752835}%
\StoreBenchExecResult{PdrInv}{KinductionDfStaticZeroTwoTTTrueNotSolvedByKinductionPlain}{Error}{OutOfJavaMemory}{Count}{}{5}%
\StoreBenchExecResult{PdrInv}{KinductionDfStaticZeroTwoTTTrueNotSolvedByKinductionPlain}{Error}{OutOfJavaMemory}{Cputime}{}{1631.055618868}%
\StoreBenchExecResult{PdrInv}{KinductionDfStaticZeroTwoTTTrueNotSolvedByKinductionPlain}{Error}{OutOfJavaMemory}{Cputime}{Avg}{326.2111237736}%
\StoreBenchExecResult{PdrInv}{KinductionDfStaticZeroTwoTTTrueNotSolvedByKinductionPlain}{Error}{OutOfJavaMemory}{Cputime}{Median}{273.916928925}%
\StoreBenchExecResult{PdrInv}{KinductionDfStaticZeroTwoTTTrueNotSolvedByKinductionPlain}{Error}{OutOfJavaMemory}{Cputime}{Min}{161.513061861}%
\StoreBenchExecResult{PdrInv}{KinductionDfStaticZeroTwoTTTrueNotSolvedByKinductionPlain}{Error}{OutOfJavaMemory}{Cputime}{Max}{493.136683326}%
\StoreBenchExecResult{PdrInv}{KinductionDfStaticZeroTwoTTTrueNotSolvedByKinductionPlain}{Error}{OutOfJavaMemory}{Cputime}{Stdev}{125.1234423249478816530251550}%
\StoreBenchExecResult{PdrInv}{KinductionDfStaticZeroTwoTTTrueNotSolvedByKinductionPlain}{Error}{OutOfJavaMemory}{Walltime}{}{945.3858060846}%
\StoreBenchExecResult{PdrInv}{KinductionDfStaticZeroTwoTTTrueNotSolvedByKinductionPlain}{Error}{OutOfJavaMemory}{Walltime}{Avg}{189.07716121692}%
\StoreBenchExecResult{PdrInv}{KinductionDfStaticZeroTwoTTTrueNotSolvedByKinductionPlain}{Error}{OutOfJavaMemory}{Walltime}{Median}{184.599290848}%
\StoreBenchExecResult{PdrInv}{KinductionDfStaticZeroTwoTTTrueNotSolvedByKinductionPlain}{Error}{OutOfJavaMemory}{Walltime}{Min}{99.6581470966}%
\StoreBenchExecResult{PdrInv}{KinductionDfStaticZeroTwoTTTrueNotSolvedByKinductionPlain}{Error}{OutOfJavaMemory}{Walltime}{Max}{256.033595085}%
\StoreBenchExecResult{PdrInv}{KinductionDfStaticZeroTwoTTTrueNotSolvedByKinductionPlain}{Error}{OutOfJavaMemory}{Walltime}{Stdev}{54.48587170680636737870249549}%
\StoreBenchExecResult{PdrInv}{KinductionDfStaticZeroTwoTTTrueNotSolvedByKinductionPlain}{Error}{OutOfMemory}{Count}{}{131}%
\StoreBenchExecResult{PdrInv}{KinductionDfStaticZeroTwoTTTrueNotSolvedByKinductionPlain}{Error}{OutOfMemory}{Cputime}{}{50148.226843244}%
\StoreBenchExecResult{PdrInv}{KinductionDfStaticZeroTwoTTTrueNotSolvedByKinductionPlain}{Error}{OutOfMemory}{Cputime}{Avg}{382.8108919331603053435114504}%
\StoreBenchExecResult{PdrInv}{KinductionDfStaticZeroTwoTTTrueNotSolvedByKinductionPlain}{Error}{OutOfMemory}{Cputime}{Median}{329.860271779}%
\StoreBenchExecResult{PdrInv}{KinductionDfStaticZeroTwoTTTrueNotSolvedByKinductionPlain}{Error}{OutOfMemory}{Cputime}{Min}{164.375458015}%
\StoreBenchExecResult{PdrInv}{KinductionDfStaticZeroTwoTTTrueNotSolvedByKinductionPlain}{Error}{OutOfMemory}{Cputime}{Max}{898.549972212}%
\StoreBenchExecResult{PdrInv}{KinductionDfStaticZeroTwoTTTrueNotSolvedByKinductionPlain}{Error}{OutOfMemory}{Cputime}{Stdev}{209.3470629180238259684568049}%
\StoreBenchExecResult{PdrInv}{KinductionDfStaticZeroTwoTTTrueNotSolvedByKinductionPlain}{Error}{OutOfMemory}{Walltime}{}{42401.2165379497}%
\StoreBenchExecResult{PdrInv}{KinductionDfStaticZeroTwoTTTrueNotSolvedByKinductionPlain}{Error}{OutOfMemory}{Walltime}{Avg}{323.6734086866389312977099237}%
\StoreBenchExecResult{PdrInv}{KinductionDfStaticZeroTwoTTTrueNotSolvedByKinductionPlain}{Error}{OutOfMemory}{Walltime}{Median}{230.248093128}%
\StoreBenchExecResult{PdrInv}{KinductionDfStaticZeroTwoTTTrueNotSolvedByKinductionPlain}{Error}{OutOfMemory}{Walltime}{Min}{86.7613019943}%
\StoreBenchExecResult{PdrInv}{KinductionDfStaticZeroTwoTTTrueNotSolvedByKinductionPlain}{Error}{OutOfMemory}{Walltime}{Max}{884.884253025}%
\StoreBenchExecResult{PdrInv}{KinductionDfStaticZeroTwoTTTrueNotSolvedByKinductionPlain}{Error}{OutOfMemory}{Walltime}{Stdev}{236.0228050216872995907183243}%
\StoreBenchExecResult{PdrInv}{KinductionDfStaticZeroTwoTTTrueNotSolvedByKinductionPlain}{Error}{Timeout}{Count}{}{1659}%
\StoreBenchExecResult{PdrInv}{KinductionDfStaticZeroTwoTTTrueNotSolvedByKinductionPlain}{Error}{Timeout}{Cputime}{}{1503444.693771440}%
\StoreBenchExecResult{PdrInv}{KinductionDfStaticZeroTwoTTTrueNotSolvedByKinductionPlain}{Error}{Timeout}{Cputime}{Avg}{906.2354995608438818565400844}%
\StoreBenchExecResult{PdrInv}{KinductionDfStaticZeroTwoTTTrueNotSolvedByKinductionPlain}{Error}{Timeout}{Cputime}{Median}{902.078785767}%
\StoreBenchExecResult{PdrInv}{KinductionDfStaticZeroTwoTTTrueNotSolvedByKinductionPlain}{Error}{Timeout}{Cputime}{Min}{900.94451516}%
\StoreBenchExecResult{PdrInv}{KinductionDfStaticZeroTwoTTTrueNotSolvedByKinductionPlain}{Error}{Timeout}{Cputime}{Max}{1002.37842465}%
\StoreBenchExecResult{PdrInv}{KinductionDfStaticZeroTwoTTTrueNotSolvedByKinductionPlain}{Error}{Timeout}{Cputime}{Stdev}{15.46760181984277017671641098}%
\StoreBenchExecResult{PdrInv}{KinductionDfStaticZeroTwoTTTrueNotSolvedByKinductionPlain}{Error}{Timeout}{Walltime}{}{1033551.006000762}%
\StoreBenchExecResult{PdrInv}{KinductionDfStaticZeroTwoTTTrueNotSolvedByKinductionPlain}{Error}{Timeout}{Walltime}{Avg}{622.9963869805678119349005425}%
\StoreBenchExecResult{PdrInv}{KinductionDfStaticZeroTwoTTTrueNotSolvedByKinductionPlain}{Error}{Timeout}{Walltime}{Median}{458.401332855}%
\StoreBenchExecResult{PdrInv}{KinductionDfStaticZeroTwoTTTrueNotSolvedByKinductionPlain}{Error}{Timeout}{Walltime}{Min}{451.084649086}%
\StoreBenchExecResult{PdrInv}{KinductionDfStaticZeroTwoTTTrueNotSolvedByKinductionPlain}{Error}{Timeout}{Walltime}{Max}{904.666206837}%
\StoreBenchExecResult{PdrInv}{KinductionDfStaticZeroTwoTTTrueNotSolvedByKinductionPlain}{Error}{Timeout}{Walltime}{Stdev}{200.7481423743045394644221789}%
\providecommand\StoreBenchExecResult[7]{\expandafter\newcommand\csname#1#2#3#4#5#6\endcsname{#7}}%
\StoreBenchExecResult{PdrInv}{KinductionDfStaticZeroTwoTT}{Total}{}{Count}{}{5591}%
\StoreBenchExecResult{PdrInv}{KinductionDfStaticZeroTwoTT}{Total}{}{Cputime}{}{2203216.042288799}%
\StoreBenchExecResult{PdrInv}{KinductionDfStaticZeroTwoTT}{Total}{}{Cputime}{Avg}{394.0647544784115542836701842}%
\StoreBenchExecResult{PdrInv}{KinductionDfStaticZeroTwoTT}{Total}{}{Cputime}{Median}{112.638219981}%
\StoreBenchExecResult{PdrInv}{KinductionDfStaticZeroTwoTT}{Total}{}{Cputime}{Min}{2.474144852}%
\StoreBenchExecResult{PdrInv}{KinductionDfStaticZeroTwoTT}{Total}{}{Cputime}{Max}{1002.37842465}%
\StoreBenchExecResult{PdrInv}{KinductionDfStaticZeroTwoTT}{Total}{}{Cputime}{Stdev}{419.0127349067617139964847541}%
\StoreBenchExecResult{PdrInv}{KinductionDfStaticZeroTwoTT}{Total}{}{Walltime}{}{1551256.14478374512}%
\StoreBenchExecResult{PdrInv}{KinductionDfStaticZeroTwoTT}{Total}{}{Walltime}{Avg}{277.4559371818538937578250760}%
\StoreBenchExecResult{PdrInv}{KinductionDfStaticZeroTwoTT}{Total}{}{Walltime}{Median}{73.9484930038}%
\StoreBenchExecResult{PdrInv}{KinductionDfStaticZeroTwoTT}{Total}{}{Walltime}{Min}{1.36245203018}%
\StoreBenchExecResult{PdrInv}{KinductionDfStaticZeroTwoTT}{Total}{}{Walltime}{Max}{933.67899394}%
\StoreBenchExecResult{PdrInv}{KinductionDfStaticZeroTwoTT}{Total}{}{Walltime}{Stdev}{321.3356901797579147285610809}%
\StoreBenchExecResult{PdrInv}{KinductionDfStaticZeroTwoTT}{Correct}{}{Count}{}{2998}%
\StoreBenchExecResult{PdrInv}{KinductionDfStaticZeroTwoTT}{Correct}{}{Cputime}{}{165648.585772842}%
\StoreBenchExecResult{PdrInv}{KinductionDfStaticZeroTwoTT}{Correct}{}{Cputime}{Avg}{55.25303061135490326884589726}%
\StoreBenchExecResult{PdrInv}{KinductionDfStaticZeroTwoTT}{Correct}{}{Cputime}{Median}{9.48755703}%
\StoreBenchExecResult{PdrInv}{KinductionDfStaticZeroTwoTT}{Correct}{}{Cputime}{Min}{2.941830114}%
\StoreBenchExecResult{PdrInv}{KinductionDfStaticZeroTwoTT}{Correct}{}{Cputime}{Max}{895.8744127}%
\StoreBenchExecResult{PdrInv}{KinductionDfStaticZeroTwoTT}{Correct}{}{Cputime}{Stdev}{134.7132131438655990821048538}%
\StoreBenchExecResult{PdrInv}{KinductionDfStaticZeroTwoTT}{Correct}{}{Walltime}{}{116681.14835500662}%
\StoreBenchExecResult{PdrInv}{KinductionDfStaticZeroTwoTT}{Correct}{}{Walltime}{Avg}{38.91966256004223482321547698}%
\StoreBenchExecResult{PdrInv}{KinductionDfStaticZeroTwoTT}{Correct}{}{Walltime}{Median}{4.993619799615}%
\StoreBenchExecResult{PdrInv}{KinductionDfStaticZeroTwoTT}{Correct}{}{Walltime}{Min}{1.6507499218}%
\StoreBenchExecResult{PdrInv}{KinductionDfStaticZeroTwoTT}{Correct}{}{Walltime}{Max}{871.032688141}%
\StoreBenchExecResult{PdrInv}{KinductionDfStaticZeroTwoTT}{Correct}{}{Walltime}{Stdev}{111.3431851263497999399181598}%
\StoreBenchExecResult{PdrInv}{KinductionDfStaticZeroTwoTT}{Correct}{False}{Count}{}{819}%
\StoreBenchExecResult{PdrInv}{KinductionDfStaticZeroTwoTT}{Correct}{False}{Cputime}{}{74817.550117128}%
\StoreBenchExecResult{PdrInv}{KinductionDfStaticZeroTwoTT}{Correct}{False}{Cputime}{Avg}{91.35232004533333333333333333}%
\StoreBenchExecResult{PdrInv}{KinductionDfStaticZeroTwoTT}{Correct}{False}{Cputime}{Median}{21.821285749}%
\StoreBenchExecResult{PdrInv}{KinductionDfStaticZeroTwoTT}{Correct}{False}{Cputime}{Min}{3.119659875}%
\StoreBenchExecResult{PdrInv}{KinductionDfStaticZeroTwoTT}{Correct}{False}{Cputime}{Max}{895.8744127}%
\StoreBenchExecResult{PdrInv}{KinductionDfStaticZeroTwoTT}{Correct}{False}{Cputime}{Stdev}{181.6234268388716792126503880}%
\StoreBenchExecResult{PdrInv}{KinductionDfStaticZeroTwoTT}{Correct}{False}{Walltime}{}{59764.89520382683}%
\StoreBenchExecResult{PdrInv}{KinductionDfStaticZeroTwoTT}{Correct}{False}{Walltime}{Avg}{72.97301001688257631257631258}%
\StoreBenchExecResult{PdrInv}{KinductionDfStaticZeroTwoTT}{Correct}{False}{Walltime}{Median}{11.8863689899}%
\StoreBenchExecResult{PdrInv}{KinductionDfStaticZeroTwoTT}{Correct}{False}{Walltime}{Min}{1.76187109947}%
\StoreBenchExecResult{PdrInv}{KinductionDfStaticZeroTwoTT}{Correct}{False}{Walltime}{Max}{871.032688141}%
\StoreBenchExecResult{PdrInv}{KinductionDfStaticZeroTwoTT}{Correct}{False}{Walltime}{Stdev}{165.6564902358295958460300545}%
\StoreBenchExecResult{PdrInv}{KinductionDfStaticZeroTwoTT}{Correct}{True}{Count}{}{2179}%
\StoreBenchExecResult{PdrInv}{KinductionDfStaticZeroTwoTT}{Correct}{True}{Cputime}{}{90831.035655714}%
\StoreBenchExecResult{PdrInv}{KinductionDfStaticZeroTwoTT}{Correct}{True}{Cputime}{Avg}{41.68473412377879761358421294}%
\StoreBenchExecResult{PdrInv}{KinductionDfStaticZeroTwoTT}{Correct}{True}{Cputime}{Median}{7.228068766}%
\StoreBenchExecResult{PdrInv}{KinductionDfStaticZeroTwoTT}{Correct}{True}{Cputime}{Min}{2.941830114}%
\StoreBenchExecResult{PdrInv}{KinductionDfStaticZeroTwoTT}{Correct}{True}{Cputime}{Max}{883.567764281}%
\StoreBenchExecResult{PdrInv}{KinductionDfStaticZeroTwoTT}{Correct}{True}{Cputime}{Stdev}{109.0696744128755511759502150}%
\StoreBenchExecResult{PdrInv}{KinductionDfStaticZeroTwoTT}{Correct}{True}{Walltime}{}{56916.25315117979}%
\StoreBenchExecResult{PdrInv}{KinductionDfStaticZeroTwoTT}{Correct}{True}{Walltime}{Avg}{26.12035481926562184488297384}%
\StoreBenchExecResult{PdrInv}{KinductionDfStaticZeroTwoTT}{Correct}{True}{Walltime}{Median}{3.84789514542}%
\StoreBenchExecResult{PdrInv}{KinductionDfStaticZeroTwoTT}{Correct}{True}{Walltime}{Min}{1.6507499218}%
\StoreBenchExecResult{PdrInv}{KinductionDfStaticZeroTwoTT}{Correct}{True}{Walltime}{Max}{869.842550993}%
\StoreBenchExecResult{PdrInv}{KinductionDfStaticZeroTwoTT}{Correct}{True}{Walltime}{Stdev}{78.37658799961494269127984248}%
\StoreBenchExecResult{PdrInv}{KinductionDfStaticZeroTwoTT}{Wrong}{True}{Count}{}{0}%
\StoreBenchExecResult{PdrInv}{KinductionDfStaticZeroTwoTT}{Wrong}{True}{Cputime}{}{0}%
\StoreBenchExecResult{PdrInv}{KinductionDfStaticZeroTwoTT}{Wrong}{True}{Cputime}{Avg}{None}%
\StoreBenchExecResult{PdrInv}{KinductionDfStaticZeroTwoTT}{Wrong}{True}{Cputime}{Median}{None}%
\StoreBenchExecResult{PdrInv}{KinductionDfStaticZeroTwoTT}{Wrong}{True}{Cputime}{Min}{None}%
\StoreBenchExecResult{PdrInv}{KinductionDfStaticZeroTwoTT}{Wrong}{True}{Cputime}{Max}{None}%
\StoreBenchExecResult{PdrInv}{KinductionDfStaticZeroTwoTT}{Wrong}{True}{Cputime}{Stdev}{None}%
\StoreBenchExecResult{PdrInv}{KinductionDfStaticZeroTwoTT}{Wrong}{True}{Walltime}{}{0}%
\StoreBenchExecResult{PdrInv}{KinductionDfStaticZeroTwoTT}{Wrong}{True}{Walltime}{Avg}{None}%
\StoreBenchExecResult{PdrInv}{KinductionDfStaticZeroTwoTT}{Wrong}{True}{Walltime}{Median}{None}%
\StoreBenchExecResult{PdrInv}{KinductionDfStaticZeroTwoTT}{Wrong}{True}{Walltime}{Min}{None}%
\StoreBenchExecResult{PdrInv}{KinductionDfStaticZeroTwoTT}{Wrong}{True}{Walltime}{Max}{None}%
\StoreBenchExecResult{PdrInv}{KinductionDfStaticZeroTwoTT}{Wrong}{True}{Walltime}{Stdev}{None}%
\StoreBenchExecResult{PdrInv}{KinductionDfStaticZeroTwoTT}{Error}{}{Count}{}{2591}%
\StoreBenchExecResult{PdrInv}{KinductionDfStaticZeroTwoTT}{Error}{}{Cputime}{}{2037544.173657123}%
\StoreBenchExecResult{PdrInv}{KinductionDfStaticZeroTwoTT}{Error}{}{Cputime}{Avg}{786.3929655179942107294480895}%
\StoreBenchExecResult{PdrInv}{KinductionDfStaticZeroTwoTT}{Error}{}{Cputime}{Median}{901.754978771}%
\StoreBenchExecResult{PdrInv}{KinductionDfStaticZeroTwoTT}{Error}{}{Cputime}{Min}{2.474144852}%
\StoreBenchExecResult{PdrInv}{KinductionDfStaticZeroTwoTT}{Error}{}{Cputime}{Max}{1002.37842465}%
\StoreBenchExecResult{PdrInv}{KinductionDfStaticZeroTwoTT}{Error}{}{Cputime}{Stdev}{266.4582346112498264617718360}%
\StoreBenchExecResult{PdrInv}{KinductionDfStaticZeroTwoTT}{Error}{}{Walltime}{}{1434562.42211366736}%
\StoreBenchExecResult{PdrInv}{KinductionDfStaticZeroTwoTT}{Error}{}{Walltime}{Avg}{553.6713323479997529911231185}%
\StoreBenchExecResult{PdrInv}{KinductionDfStaticZeroTwoTT}{Error}{}{Walltime}{Median}{455.547183037}%
\StoreBenchExecResult{PdrInv}{KinductionDfStaticZeroTwoTT}{Error}{}{Walltime}{Min}{1.36245203018}%
\StoreBenchExecResult{PdrInv}{KinductionDfStaticZeroTwoTT}{Error}{}{Walltime}{Max}{933.67899394}%
\StoreBenchExecResult{PdrInv}{KinductionDfStaticZeroTwoTT}{Error}{}{Walltime}{Stdev}{257.4467992740422121772918181}%
\StoreBenchExecResult{PdrInv}{KinductionDfStaticZeroTwoTT}{Error}{Assertion}{Count}{}{4}%
\StoreBenchExecResult{PdrInv}{KinductionDfStaticZeroTwoTT}{Error}{Assertion}{Cputime}{}{13.035128682}%
\StoreBenchExecResult{PdrInv}{KinductionDfStaticZeroTwoTT}{Error}{Assertion}{Cputime}{Avg}{3.2587821705}%
\StoreBenchExecResult{PdrInv}{KinductionDfStaticZeroTwoTT}{Error}{Assertion}{Cputime}{Median}{3.2467014295}%
\StoreBenchExecResult{PdrInv}{KinductionDfStaticZeroTwoTT}{Error}{Assertion}{Cputime}{Min}{3.219520661}%
\StoreBenchExecResult{PdrInv}{KinductionDfStaticZeroTwoTT}{Error}{Assertion}{Cputime}{Max}{3.322205162}%
\StoreBenchExecResult{PdrInv}{KinductionDfStaticZeroTwoTT}{Error}{Assertion}{Cputime}{Stdev}{0.03910402628988634272550233448}%
\StoreBenchExecResult{PdrInv}{KinductionDfStaticZeroTwoTT}{Error}{Assertion}{Walltime}{}{7.22377181052}%
\StoreBenchExecResult{PdrInv}{KinductionDfStaticZeroTwoTT}{Error}{Assertion}{Walltime}{Avg}{1.80594295263}%
\StoreBenchExecResult{PdrInv}{KinductionDfStaticZeroTwoTT}{Error}{Assertion}{Walltime}{Median}{1.806397914885}%
\StoreBenchExecResult{PdrInv}{KinductionDfStaticZeroTwoTT}{Error}{Assertion}{Walltime}{Min}{1.78283500671}%
\StoreBenchExecResult{PdrInv}{KinductionDfStaticZeroTwoTT}{Error}{Assertion}{Walltime}{Max}{1.82814097404}%
\StoreBenchExecResult{PdrInv}{KinductionDfStaticZeroTwoTT}{Error}{Assertion}{Walltime}{Stdev}{0.01698851870831198789512209965}%
\StoreBenchExecResult{PdrInv}{KinductionDfStaticZeroTwoTT}{Error}{Error}{Count}{}{198}%
\StoreBenchExecResult{PdrInv}{KinductionDfStaticZeroTwoTT}{Error}{Error}{Cputime}{}{36885.324096903}%
\StoreBenchExecResult{PdrInv}{KinductionDfStaticZeroTwoTT}{Error}{Error}{Cputime}{Avg}{186.2895156409242424242424242}%
\StoreBenchExecResult{PdrInv}{KinductionDfStaticZeroTwoTT}{Error}{Error}{Cputime}{Median}{121.9785426265}%
\StoreBenchExecResult{PdrInv}{KinductionDfStaticZeroTwoTT}{Error}{Error}{Cputime}{Min}{2.474144852}%
\StoreBenchExecResult{PdrInv}{KinductionDfStaticZeroTwoTT}{Error}{Error}{Cputime}{Max}{851.506871282}%
\StoreBenchExecResult{PdrInv}{KinductionDfStaticZeroTwoTT}{Error}{Error}{Cputime}{Stdev}{193.3026342380959780364895678}%
\StoreBenchExecResult{PdrInv}{KinductionDfStaticZeroTwoTT}{Error}{Error}{Walltime}{}{30565.07850218080}%
\StoreBenchExecResult{PdrInv}{KinductionDfStaticZeroTwoTT}{Error}{Error}{Walltime}{Avg}{154.3690833443474747474747475}%
\StoreBenchExecResult{PdrInv}{KinductionDfStaticZeroTwoTT}{Error}{Error}{Walltime}{Median}{95.65421640875}%
\StoreBenchExecResult{PdrInv}{KinductionDfStaticZeroTwoTT}{Error}{Error}{Walltime}{Min}{1.36245203018}%
\StoreBenchExecResult{PdrInv}{KinductionDfStaticZeroTwoTT}{Error}{Error}{Walltime}{Max}{835.936573029}%
\StoreBenchExecResult{PdrInv}{KinductionDfStaticZeroTwoTT}{Error}{Error}{Walltime}{Stdev}{169.0325676022038708478339303}%
\StoreBenchExecResult{PdrInv}{KinductionDfStaticZeroTwoTT}{Error}{Exception}{Count}{}{26}%
\StoreBenchExecResult{PdrInv}{KinductionDfStaticZeroTwoTT}{Error}{Exception}{Cputime}{}{5279.788595874}%
\StoreBenchExecResult{PdrInv}{KinductionDfStaticZeroTwoTT}{Error}{Exception}{Cputime}{Avg}{203.068792149}%
\StoreBenchExecResult{PdrInv}{KinductionDfStaticZeroTwoTT}{Error}{Exception}{Cputime}{Median}{105.2068347605}%
\StoreBenchExecResult{PdrInv}{KinductionDfStaticZeroTwoTT}{Error}{Exception}{Cputime}{Min}{14.917365422}%
\StoreBenchExecResult{PdrInv}{KinductionDfStaticZeroTwoTT}{Error}{Exception}{Cputime}{Max}{899.809594714}%
\StoreBenchExecResult{PdrInv}{KinductionDfStaticZeroTwoTT}{Error}{Exception}{Cputime}{Stdev}{208.7732417731372040657201753}%
\StoreBenchExecResult{PdrInv}{KinductionDfStaticZeroTwoTT}{Error}{Exception}{Walltime}{}{2702.60543322714}%
\StoreBenchExecResult{PdrInv}{KinductionDfStaticZeroTwoTT}{Error}{Exception}{Walltime}{Avg}{103.9463628164284615384615385}%
\StoreBenchExecResult{PdrInv}{KinductionDfStaticZeroTwoTT}{Error}{Exception}{Walltime}{Median}{59.9230309725}%
\StoreBenchExecResult{PdrInv}{KinductionDfStaticZeroTwoTT}{Error}{Exception}{Walltime}{Min}{7.64925289154}%
\StoreBenchExecResult{PdrInv}{KinductionDfStaticZeroTwoTT}{Error}{Exception}{Walltime}{Max}{450.515960932}%
\StoreBenchExecResult{PdrInv}{KinductionDfStaticZeroTwoTT}{Error}{Exception}{Walltime}{Stdev}{105.7380108430962066106857856}%
\StoreBenchExecResult{PdrInv}{KinductionDfStaticZeroTwoTT}{Error}{OutOfJavaMemory}{Count}{}{10}%
\StoreBenchExecResult{PdrInv}{KinductionDfStaticZeroTwoTT}{Error}{OutOfJavaMemory}{Cputime}{}{4706.187813872}%
\StoreBenchExecResult{PdrInv}{KinductionDfStaticZeroTwoTT}{Error}{OutOfJavaMemory}{Cputime}{Avg}{470.6187813872}%
\StoreBenchExecResult{PdrInv}{KinductionDfStaticZeroTwoTT}{Error}{OutOfJavaMemory}{Cputime}{Median}{471.3100306775}%
\StoreBenchExecResult{PdrInv}{KinductionDfStaticZeroTwoTT}{Error}{OutOfJavaMemory}{Cputime}{Min}{161.513061861}%
\StoreBenchExecResult{PdrInv}{KinductionDfStaticZeroTwoTT}{Error}{OutOfJavaMemory}{Cputime}{Max}{724.672332486}%
\StoreBenchExecResult{PdrInv}{KinductionDfStaticZeroTwoTT}{Error}{OutOfJavaMemory}{Cputime}{Stdev}{185.4611141446172975739544982}%
\StoreBenchExecResult{PdrInv}{KinductionDfStaticZeroTwoTT}{Error}{OutOfJavaMemory}{Walltime}{}{2908.6453669076}%
\StoreBenchExecResult{PdrInv}{KinductionDfStaticZeroTwoTT}{Error}{OutOfJavaMemory}{Walltime}{Avg}{290.86453669076}%
\StoreBenchExecResult{PdrInv}{KinductionDfStaticZeroTwoTT}{Error}{OutOfJavaMemory}{Walltime}{Median}{272.5665160415}%
\StoreBenchExecResult{PdrInv}{KinductionDfStaticZeroTwoTT}{Error}{OutOfJavaMemory}{Walltime}{Min}{99.6581470966}%
\StoreBenchExecResult{PdrInv}{KinductionDfStaticZeroTwoTT}{Error}{OutOfJavaMemory}{Walltime}{Max}{581.468343973}%
\StoreBenchExecResult{PdrInv}{KinductionDfStaticZeroTwoTT}{Error}{OutOfJavaMemory}{Walltime}{Stdev}{132.6862558736948519200786251}%
\StoreBenchExecResult{PdrInv}{KinductionDfStaticZeroTwoTT}{Error}{OutOfMemory}{Count}{}{264}%
\StoreBenchExecResult{PdrInv}{KinductionDfStaticZeroTwoTT}{Error}{OutOfMemory}{Cputime}{}{93172.779412821}%
\StoreBenchExecResult{PdrInv}{KinductionDfStaticZeroTwoTT}{Error}{OutOfMemory}{Cputime}{Avg}{352.9271947455340909090909091}%
\StoreBenchExecResult{PdrInv}{KinductionDfStaticZeroTwoTT}{Error}{OutOfMemory}{Cputime}{Median}{273.8312881385}%
\StoreBenchExecResult{PdrInv}{KinductionDfStaticZeroTwoTT}{Error}{OutOfMemory}{Cputime}{Min}{153.733812818}%
\StoreBenchExecResult{PdrInv}{KinductionDfStaticZeroTwoTT}{Error}{OutOfMemory}{Cputime}{Max}{898.549972212}%
\StoreBenchExecResult{PdrInv}{KinductionDfStaticZeroTwoTT}{Error}{OutOfMemory}{Cputime}{Stdev}{194.7975818682180376918120644}%
\StoreBenchExecResult{PdrInv}{KinductionDfStaticZeroTwoTT}{Error}{OutOfMemory}{Walltime}{}{78045.6059927913}%
\StoreBenchExecResult{PdrInv}{KinductionDfStaticZeroTwoTT}{Error}{OutOfMemory}{Walltime}{Avg}{295.6272954272397727272727273}%
\StoreBenchExecResult{PdrInv}{KinductionDfStaticZeroTwoTT}{Error}{OutOfMemory}{Walltime}{Median}{197.792630553}%
\StoreBenchExecResult{PdrInv}{KinductionDfStaticZeroTwoTT}{Error}{OutOfMemory}{Walltime}{Min}{85.2494750023}%
\StoreBenchExecResult{PdrInv}{KinductionDfStaticZeroTwoTT}{Error}{OutOfMemory}{Walltime}{Max}{884.884253025}%
\StoreBenchExecResult{PdrInv}{KinductionDfStaticZeroTwoTT}{Error}{OutOfMemory}{Walltime}{Stdev}{216.1217665312378498351079472}%
\StoreBenchExecResult{PdrInv}{KinductionDfStaticZeroTwoTT}{Error}{Timeout}{Count}{}{2089}%
\StoreBenchExecResult{PdrInv}{KinductionDfStaticZeroTwoTT}{Error}{Timeout}{Cputime}{}{1897487.058608971}%
\StoreBenchExecResult{PdrInv}{KinductionDfStaticZeroTwoTT}{Error}{Timeout}{Cputime}{Avg}{908.3231491665730014360938248}%
\StoreBenchExecResult{PdrInv}{KinductionDfStaticZeroTwoTT}{Error}{Timeout}{Cputime}{Median}{902.284925152}%
\StoreBenchExecResult{PdrInv}{KinductionDfStaticZeroTwoTT}{Error}{Timeout}{Cputime}{Min}{900.218568273}%
\StoreBenchExecResult{PdrInv}{KinductionDfStaticZeroTwoTT}{Error}{Timeout}{Cputime}{Max}{1002.37842465}%
\StoreBenchExecResult{PdrInv}{KinductionDfStaticZeroTwoTT}{Error}{Timeout}{Cputime}{Stdev}{19.62989041125770084571693361}%
\StoreBenchExecResult{PdrInv}{KinductionDfStaticZeroTwoTT}{Error}{Timeout}{Walltime}{}{1320333.263046750}%
\StoreBenchExecResult{PdrInv}{KinductionDfStaticZeroTwoTT}{Error}{Timeout}{Walltime}{Avg}{632.0408152449736716132120632}%
\StoreBenchExecResult{PdrInv}{KinductionDfStaticZeroTwoTT}{Error}{Timeout}{Walltime}{Median}{472.239786863}%
\StoreBenchExecResult{PdrInv}{KinductionDfStaticZeroTwoTT}{Error}{Timeout}{Walltime}{Min}{451.084649086}%
\StoreBenchExecResult{PdrInv}{KinductionDfStaticZeroTwoTT}{Error}{Timeout}{Walltime}{Max}{933.67899394}%
\StoreBenchExecResult{PdrInv}{KinductionDfStaticZeroTwoTT}{Error}{Timeout}{Walltime}{Stdev}{200.6777996156000627876622718}%
\StoreBenchExecResult{PdrInv}{KinductionDfStaticZeroTwoTT}{Wrong}{}{Count}{}{2}%
\StoreBenchExecResult{PdrInv}{KinductionDfStaticZeroTwoTT}{Wrong}{}{Cputime}{}{23.282858834}%
\StoreBenchExecResult{PdrInv}{KinductionDfStaticZeroTwoTT}{Wrong}{}{Cputime}{Avg}{11.641429417}%
\StoreBenchExecResult{PdrInv}{KinductionDfStaticZeroTwoTT}{Wrong}{}{Cputime}{Median}{11.641429417}%
\StoreBenchExecResult{PdrInv}{KinductionDfStaticZeroTwoTT}{Wrong}{}{Cputime}{Min}{4.083448912}%
\StoreBenchExecResult{PdrInv}{KinductionDfStaticZeroTwoTT}{Wrong}{}{Cputime}{Max}{19.199409922}%
\StoreBenchExecResult{PdrInv}{KinductionDfStaticZeroTwoTT}{Wrong}{}{Cputime}{Stdev}{7.557980505}%
\StoreBenchExecResult{PdrInv}{KinductionDfStaticZeroTwoTT}{Wrong}{}{Walltime}{}{12.57431507114}%
\StoreBenchExecResult{PdrInv}{KinductionDfStaticZeroTwoTT}{Wrong}{}{Walltime}{Avg}{6.28715753557}%
\StoreBenchExecResult{PdrInv}{KinductionDfStaticZeroTwoTT}{Wrong}{}{Walltime}{Median}{6.28715753557}%
\StoreBenchExecResult{PdrInv}{KinductionDfStaticZeroTwoTT}{Wrong}{}{Walltime}{Min}{2.24120116234}%
\StoreBenchExecResult{PdrInv}{KinductionDfStaticZeroTwoTT}{Wrong}{}{Walltime}{Max}{10.3331139088}%
\StoreBenchExecResult{PdrInv}{KinductionDfStaticZeroTwoTT}{Wrong}{}{Walltime}{Stdev}{4.04595637323}%
\StoreBenchExecResult{PdrInv}{KinductionDfStaticZeroTwoTT}{Wrong}{False}{Count}{}{2}%
\StoreBenchExecResult{PdrInv}{KinductionDfStaticZeroTwoTT}{Wrong}{False}{Cputime}{}{23.282858834}%
\StoreBenchExecResult{PdrInv}{KinductionDfStaticZeroTwoTT}{Wrong}{False}{Cputime}{Avg}{11.641429417}%
\StoreBenchExecResult{PdrInv}{KinductionDfStaticZeroTwoTT}{Wrong}{False}{Cputime}{Median}{11.641429417}%
\StoreBenchExecResult{PdrInv}{KinductionDfStaticZeroTwoTT}{Wrong}{False}{Cputime}{Min}{4.083448912}%
\StoreBenchExecResult{PdrInv}{KinductionDfStaticZeroTwoTT}{Wrong}{False}{Cputime}{Max}{19.199409922}%
\StoreBenchExecResult{PdrInv}{KinductionDfStaticZeroTwoTT}{Wrong}{False}{Cputime}{Stdev}{7.557980505}%
\StoreBenchExecResult{PdrInv}{KinductionDfStaticZeroTwoTT}{Wrong}{False}{Walltime}{}{12.57431507114}%
\StoreBenchExecResult{PdrInv}{KinductionDfStaticZeroTwoTT}{Wrong}{False}{Walltime}{Avg}{6.28715753557}%
\StoreBenchExecResult{PdrInv}{KinductionDfStaticZeroTwoTT}{Wrong}{False}{Walltime}{Median}{6.28715753557}%
\StoreBenchExecResult{PdrInv}{KinductionDfStaticZeroTwoTT}{Wrong}{False}{Walltime}{Min}{2.24120116234}%
\StoreBenchExecResult{PdrInv}{KinductionDfStaticZeroTwoTT}{Wrong}{False}{Walltime}{Max}{10.3331139088}%
\StoreBenchExecResult{PdrInv}{KinductionDfStaticZeroTwoTT}{Wrong}{False}{Walltime}{Stdev}{4.04595637323}%
\providecommand\StoreBenchExecResult[7]{\expandafter\newcommand\csname#1#2#3#4#5#6\endcsname{#7}}%
\StoreBenchExecResult{PdrInv}{KinductionDfStaticSixteenTwoFTrueNotSolvedByKinductionPlainButKipdr}{Total}{}{Count}{}{449}%
\StoreBenchExecResult{PdrInv}{KinductionDfStaticSixteenTwoFTrueNotSolvedByKinductionPlainButKipdr}{Total}{}{Cputime}{}{15890.426689820}%
\StoreBenchExecResult{PdrInv}{KinductionDfStaticSixteenTwoFTrueNotSolvedByKinductionPlainButKipdr}{Total}{}{Cputime}{Avg}{35.39070532253897550111358575}%
\StoreBenchExecResult{PdrInv}{KinductionDfStaticSixteenTwoFTrueNotSolvedByKinductionPlainButKipdr}{Total}{}{Cputime}{Median}{5.960920655}%
\StoreBenchExecResult{PdrInv}{KinductionDfStaticSixteenTwoFTrueNotSolvedByKinductionPlainButKipdr}{Total}{}{Cputime}{Min}{3.165165323}%
\StoreBenchExecResult{PdrInv}{KinductionDfStaticSixteenTwoFTrueNotSolvedByKinductionPlainButKipdr}{Total}{}{Cputime}{Max}{989.241307688}%
\StoreBenchExecResult{PdrInv}{KinductionDfStaticSixteenTwoFTrueNotSolvedByKinductionPlainButKipdr}{Total}{}{Cputime}{Stdev}{153.2241272079781166539676259}%
\StoreBenchExecResult{PdrInv}{KinductionDfStaticSixteenTwoFTrueNotSolvedByKinductionPlainButKipdr}{Total}{}{Walltime}{}{8094.50286698334}%
\StoreBenchExecResult{PdrInv}{KinductionDfStaticSixteenTwoFTrueNotSolvedByKinductionPlainButKipdr}{Total}{}{Walltime}{Avg}{18.02784602891612472160356347}%
\StoreBenchExecResult{PdrInv}{KinductionDfStaticSixteenTwoFTrueNotSolvedByKinductionPlainButKipdr}{Total}{}{Walltime}{Median}{3.17353391647}%
\StoreBenchExecResult{PdrInv}{KinductionDfStaticSixteenTwoFTrueNotSolvedByKinductionPlainButKipdr}{Total}{}{Walltime}{Min}{1.75485801697}%
\StoreBenchExecResult{PdrInv}{KinductionDfStaticSixteenTwoFTrueNotSolvedByKinductionPlainButKipdr}{Total}{}{Walltime}{Max}{533.235811949}%
\StoreBenchExecResult{PdrInv}{KinductionDfStaticSixteenTwoFTrueNotSolvedByKinductionPlainButKipdr}{Total}{}{Walltime}{Stdev}{77.43469675742652026264603363}%
\StoreBenchExecResult{PdrInv}{KinductionDfStaticSixteenTwoFTrueNotSolvedByKinductionPlainButKipdr}{Correct}{}{Count}{}{437}%
\StoreBenchExecResult{PdrInv}{KinductionDfStaticSixteenTwoFTrueNotSolvedByKinductionPlainButKipdr}{Correct}{}{Cputime}{}{4968.345454077}%
\StoreBenchExecResult{PdrInv}{KinductionDfStaticSixteenTwoFTrueNotSolvedByKinductionPlainButKipdr}{Correct}{}{Cputime}{Avg}{11.36921156539359267734553776}%
\StoreBenchExecResult{PdrInv}{KinductionDfStaticSixteenTwoFTrueNotSolvedByKinductionPlainButKipdr}{Correct}{}{Cputime}{Median}{5.899811221}%
\StoreBenchExecResult{PdrInv}{KinductionDfStaticSixteenTwoFTrueNotSolvedByKinductionPlainButKipdr}{Correct}{}{Cputime}{Min}{3.165165323}%
\StoreBenchExecResult{PdrInv}{KinductionDfStaticSixteenTwoFTrueNotSolvedByKinductionPlainButKipdr}{Correct}{}{Cputime}{Max}{870.198310293}%
\StoreBenchExecResult{PdrInv}{KinductionDfStaticSixteenTwoFTrueNotSolvedByKinductionPlainButKipdr}{Correct}{}{Cputime}{Stdev}{50.16042975861097951185743561}%
\StoreBenchExecResult{PdrInv}{KinductionDfStaticSixteenTwoFTrueNotSolvedByKinductionPlainButKipdr}{Correct}{}{Walltime}{}{2572.88388180734}%
\StoreBenchExecResult{PdrInv}{KinductionDfStaticSixteenTwoFTrueNotSolvedByKinductionPlainButKipdr}{Correct}{}{Walltime}{Avg}{5.887606136858901601830663616}%
\StoreBenchExecResult{PdrInv}{KinductionDfStaticSixteenTwoFTrueNotSolvedByKinductionPlainButKipdr}{Correct}{}{Walltime}{Median}{3.13147902489}%
\StoreBenchExecResult{PdrInv}{KinductionDfStaticSixteenTwoFTrueNotSolvedByKinductionPlainButKipdr}{Correct}{}{Walltime}{Min}{1.75485801697}%
\StoreBenchExecResult{PdrInv}{KinductionDfStaticSixteenTwoFTrueNotSolvedByKinductionPlainButKipdr}{Correct}{}{Walltime}{Max}{436.343190908}%
\StoreBenchExecResult{PdrInv}{KinductionDfStaticSixteenTwoFTrueNotSolvedByKinductionPlainButKipdr}{Correct}{}{Walltime}{Stdev}{25.15433145974295163908529838}%
\StoreBenchExecResult{PdrInv}{KinductionDfStaticSixteenTwoFTrueNotSolvedByKinductionPlainButKipdr}{Correct}{True}{Count}{}{437}%
\StoreBenchExecResult{PdrInv}{KinductionDfStaticSixteenTwoFTrueNotSolvedByKinductionPlainButKipdr}{Correct}{True}{Cputime}{}{4968.345454077}%
\StoreBenchExecResult{PdrInv}{KinductionDfStaticSixteenTwoFTrueNotSolvedByKinductionPlainButKipdr}{Correct}{True}{Cputime}{Avg}{11.36921156539359267734553776}%
\StoreBenchExecResult{PdrInv}{KinductionDfStaticSixteenTwoFTrueNotSolvedByKinductionPlainButKipdr}{Correct}{True}{Cputime}{Median}{5.899811221}%
\StoreBenchExecResult{PdrInv}{KinductionDfStaticSixteenTwoFTrueNotSolvedByKinductionPlainButKipdr}{Correct}{True}{Cputime}{Min}{3.165165323}%
\StoreBenchExecResult{PdrInv}{KinductionDfStaticSixteenTwoFTrueNotSolvedByKinductionPlainButKipdr}{Correct}{True}{Cputime}{Max}{870.198310293}%
\StoreBenchExecResult{PdrInv}{KinductionDfStaticSixteenTwoFTrueNotSolvedByKinductionPlainButKipdr}{Correct}{True}{Cputime}{Stdev}{50.16042975861097951185743561}%
\StoreBenchExecResult{PdrInv}{KinductionDfStaticSixteenTwoFTrueNotSolvedByKinductionPlainButKipdr}{Correct}{True}{Walltime}{}{2572.88388180734}%
\StoreBenchExecResult{PdrInv}{KinductionDfStaticSixteenTwoFTrueNotSolvedByKinductionPlainButKipdr}{Correct}{True}{Walltime}{Avg}{5.887606136858901601830663616}%
\StoreBenchExecResult{PdrInv}{KinductionDfStaticSixteenTwoFTrueNotSolvedByKinductionPlainButKipdr}{Correct}{True}{Walltime}{Median}{3.13147902489}%
\StoreBenchExecResult{PdrInv}{KinductionDfStaticSixteenTwoFTrueNotSolvedByKinductionPlainButKipdr}{Correct}{True}{Walltime}{Min}{1.75485801697}%
\StoreBenchExecResult{PdrInv}{KinductionDfStaticSixteenTwoFTrueNotSolvedByKinductionPlainButKipdr}{Correct}{True}{Walltime}{Max}{436.343190908}%
\StoreBenchExecResult{PdrInv}{KinductionDfStaticSixteenTwoFTrueNotSolvedByKinductionPlainButKipdr}{Correct}{True}{Walltime}{Stdev}{25.15433145974295163908529838}%
\StoreBenchExecResult{PdrInv}{KinductionDfStaticSixteenTwoFTrueNotSolvedByKinductionPlainButKipdr}{Wrong}{True}{Count}{}{0}%
\StoreBenchExecResult{PdrInv}{KinductionDfStaticSixteenTwoFTrueNotSolvedByKinductionPlainButKipdr}{Wrong}{True}{Cputime}{}{0}%
\StoreBenchExecResult{PdrInv}{KinductionDfStaticSixteenTwoFTrueNotSolvedByKinductionPlainButKipdr}{Wrong}{True}{Cputime}{Avg}{None}%
\StoreBenchExecResult{PdrInv}{KinductionDfStaticSixteenTwoFTrueNotSolvedByKinductionPlainButKipdr}{Wrong}{True}{Cputime}{Median}{None}%
\StoreBenchExecResult{PdrInv}{KinductionDfStaticSixteenTwoFTrueNotSolvedByKinductionPlainButKipdr}{Wrong}{True}{Cputime}{Min}{None}%
\StoreBenchExecResult{PdrInv}{KinductionDfStaticSixteenTwoFTrueNotSolvedByKinductionPlainButKipdr}{Wrong}{True}{Cputime}{Max}{None}%
\StoreBenchExecResult{PdrInv}{KinductionDfStaticSixteenTwoFTrueNotSolvedByKinductionPlainButKipdr}{Wrong}{True}{Cputime}{Stdev}{None}%
\StoreBenchExecResult{PdrInv}{KinductionDfStaticSixteenTwoFTrueNotSolvedByKinductionPlainButKipdr}{Wrong}{True}{Walltime}{}{0}%
\StoreBenchExecResult{PdrInv}{KinductionDfStaticSixteenTwoFTrueNotSolvedByKinductionPlainButKipdr}{Wrong}{True}{Walltime}{Avg}{None}%
\StoreBenchExecResult{PdrInv}{KinductionDfStaticSixteenTwoFTrueNotSolvedByKinductionPlainButKipdr}{Wrong}{True}{Walltime}{Median}{None}%
\StoreBenchExecResult{PdrInv}{KinductionDfStaticSixteenTwoFTrueNotSolvedByKinductionPlainButKipdr}{Wrong}{True}{Walltime}{Min}{None}%
\StoreBenchExecResult{PdrInv}{KinductionDfStaticSixteenTwoFTrueNotSolvedByKinductionPlainButKipdr}{Wrong}{True}{Walltime}{Max}{None}%
\StoreBenchExecResult{PdrInv}{KinductionDfStaticSixteenTwoFTrueNotSolvedByKinductionPlainButKipdr}{Wrong}{True}{Walltime}{Stdev}{None}%
\StoreBenchExecResult{PdrInv}{KinductionDfStaticSixteenTwoFTrueNotSolvedByKinductionPlainButKipdr}{Error}{}{Count}{}{12}%
\StoreBenchExecResult{PdrInv}{KinductionDfStaticSixteenTwoFTrueNotSolvedByKinductionPlainButKipdr}{Error}{}{Cputime}{}{10922.081235743}%
\StoreBenchExecResult{PdrInv}{KinductionDfStaticSixteenTwoFTrueNotSolvedByKinductionPlainButKipdr}{Error}{}{Cputime}{Avg}{910.1734363119166666666666667}%
\StoreBenchExecResult{PdrInv}{KinductionDfStaticSixteenTwoFTrueNotSolvedByKinductionPlainButKipdr}{Error}{}{Cputime}{Median}{903.198342323}%
\StoreBenchExecResult{PdrInv}{KinductionDfStaticSixteenTwoFTrueNotSolvedByKinductionPlainButKipdr}{Error}{}{Cputime}{Min}{901.060319688}%
\StoreBenchExecResult{PdrInv}{KinductionDfStaticSixteenTwoFTrueNotSolvedByKinductionPlainButKipdr}{Error}{}{Cputime}{Max}{989.241307688}%
\StoreBenchExecResult{PdrInv}{KinductionDfStaticSixteenTwoFTrueNotSolvedByKinductionPlainButKipdr}{Error}{}{Cputime}{Stdev}{23.86470235209545855370128767}%
\StoreBenchExecResult{PdrInv}{KinductionDfStaticSixteenTwoFTrueNotSolvedByKinductionPlainButKipdr}{Error}{}{Walltime}{}{5521.618985176}%
\StoreBenchExecResult{PdrInv}{KinductionDfStaticSixteenTwoFTrueNotSolvedByKinductionPlainButKipdr}{Error}{}{Walltime}{Avg}{460.1349154313333333333333333}%
\StoreBenchExecResult{PdrInv}{KinductionDfStaticSixteenTwoFTrueNotSolvedByKinductionPlainButKipdr}{Error}{}{Walltime}{Median}{453.7513245345}%
\StoreBenchExecResult{PdrInv}{KinductionDfStaticSixteenTwoFTrueNotSolvedByKinductionPlainButKipdr}{Error}{}{Walltime}{Min}{451.800915956}%
\StoreBenchExecResult{PdrInv}{KinductionDfStaticSixteenTwoFTrueNotSolvedByKinductionPlainButKipdr}{Error}{}{Walltime}{Max}{533.235811949}%
\StoreBenchExecResult{PdrInv}{KinductionDfStaticSixteenTwoFTrueNotSolvedByKinductionPlainButKipdr}{Error}{}{Walltime}{Stdev}{22.06897199264134734729010305}%
\StoreBenchExecResult{PdrInv}{KinductionDfStaticSixteenTwoFTrueNotSolvedByKinductionPlainButKipdr}{Error}{Timeout}{Count}{}{12}%
\StoreBenchExecResult{PdrInv}{KinductionDfStaticSixteenTwoFTrueNotSolvedByKinductionPlainButKipdr}{Error}{Timeout}{Cputime}{}{10922.081235743}%
\StoreBenchExecResult{PdrInv}{KinductionDfStaticSixteenTwoFTrueNotSolvedByKinductionPlainButKipdr}{Error}{Timeout}{Cputime}{Avg}{910.1734363119166666666666667}%
\StoreBenchExecResult{PdrInv}{KinductionDfStaticSixteenTwoFTrueNotSolvedByKinductionPlainButKipdr}{Error}{Timeout}{Cputime}{Median}{903.198342323}%
\StoreBenchExecResult{PdrInv}{KinductionDfStaticSixteenTwoFTrueNotSolvedByKinductionPlainButKipdr}{Error}{Timeout}{Cputime}{Min}{901.060319688}%
\StoreBenchExecResult{PdrInv}{KinductionDfStaticSixteenTwoFTrueNotSolvedByKinductionPlainButKipdr}{Error}{Timeout}{Cputime}{Max}{989.241307688}%
\StoreBenchExecResult{PdrInv}{KinductionDfStaticSixteenTwoFTrueNotSolvedByKinductionPlainButKipdr}{Error}{Timeout}{Cputime}{Stdev}{23.86470235209545855370128767}%
\StoreBenchExecResult{PdrInv}{KinductionDfStaticSixteenTwoFTrueNotSolvedByKinductionPlainButKipdr}{Error}{Timeout}{Walltime}{}{5521.618985176}%
\StoreBenchExecResult{PdrInv}{KinductionDfStaticSixteenTwoFTrueNotSolvedByKinductionPlainButKipdr}{Error}{Timeout}{Walltime}{Avg}{460.1349154313333333333333333}%
\StoreBenchExecResult{PdrInv}{KinductionDfStaticSixteenTwoFTrueNotSolvedByKinductionPlainButKipdr}{Error}{Timeout}{Walltime}{Median}{453.7513245345}%
\StoreBenchExecResult{PdrInv}{KinductionDfStaticSixteenTwoFTrueNotSolvedByKinductionPlainButKipdr}{Error}{Timeout}{Walltime}{Min}{451.800915956}%
\StoreBenchExecResult{PdrInv}{KinductionDfStaticSixteenTwoFTrueNotSolvedByKinductionPlainButKipdr}{Error}{Timeout}{Walltime}{Max}{533.235811949}%
\StoreBenchExecResult{PdrInv}{KinductionDfStaticSixteenTwoFTrueNotSolvedByKinductionPlainButKipdr}{Error}{Timeout}{Walltime}{Stdev}{22.06897199264134734729010305}%
\providecommand\StoreBenchExecResult[7]{\expandafter\newcommand\csname#1#2#3#4#5#6\endcsname{#7}}%
\StoreBenchExecResult{PdrInv}{KinductionDfStaticSixteenTwoFTrueNotSolvedByKinductionPlain}{Total}{}{Count}{}{2893}%
\StoreBenchExecResult{PdrInv}{KinductionDfStaticSixteenTwoFTrueNotSolvedByKinductionPlain}{Total}{}{Cputime}{}{1670109.470878209}%
\StoreBenchExecResult{PdrInv}{KinductionDfStaticSixteenTwoFTrueNotSolvedByKinductionPlain}{Total}{}{Cputime}{Avg}{577.2932840920183200829588662}%
\StoreBenchExecResult{PdrInv}{KinductionDfStaticSixteenTwoFTrueNotSolvedByKinductionPlain}{Total}{}{Cputime}{Median}{901.109937136}%
\StoreBenchExecResult{PdrInv}{KinductionDfStaticSixteenTwoFTrueNotSolvedByKinductionPlain}{Total}{}{Cputime}{Min}{2.37078614}%
\StoreBenchExecResult{PdrInv}{KinductionDfStaticSixteenTwoFTrueNotSolvedByKinductionPlain}{Total}{}{Cputime}{Max}{1001.28896153}%
\StoreBenchExecResult{PdrInv}{KinductionDfStaticSixteenTwoFTrueNotSolvedByKinductionPlain}{Total}{}{Cputime}{Stdev}{413.7750776159398353093864119}%
\StoreBenchExecResult{PdrInv}{KinductionDfStaticSixteenTwoFTrueNotSolvedByKinductionPlain}{Total}{}{Walltime}{}{977513.19531535284}%
\StoreBenchExecResult{PdrInv}{KinductionDfStaticSixteenTwoFTrueNotSolvedByKinductionPlain}{Total}{}{Walltime}{Avg}{337.8891100295032284825440719}%
\StoreBenchExecResult{PdrInv}{KinductionDfStaticSixteenTwoFTrueNotSolvedByKinductionPlain}{Total}{}{Walltime}{Median}{451.63441515}%
\StoreBenchExecResult{PdrInv}{KinductionDfStaticSixteenTwoFTrueNotSolvedByKinductionPlain}{Total}{}{Walltime}{Min}{1.32805800438}%
\StoreBenchExecResult{PdrInv}{KinductionDfStaticSixteenTwoFTrueNotSolvedByKinductionPlain}{Total}{}{Walltime}{Max}{901.200150967}%
\StoreBenchExecResult{PdrInv}{KinductionDfStaticSixteenTwoFTrueNotSolvedByKinductionPlain}{Total}{}{Walltime}{Stdev}{272.4002071293838708143291813}%
\StoreBenchExecResult{PdrInv}{KinductionDfStaticSixteenTwoFTrueNotSolvedByKinductionPlain}{Correct}{}{Count}{}{929}%
\StoreBenchExecResult{PdrInv}{KinductionDfStaticSixteenTwoFTrueNotSolvedByKinductionPlain}{Correct}{}{Cputime}{}{42135.178476774}%
\StoreBenchExecResult{PdrInv}{KinductionDfStaticSixteenTwoFTrueNotSolvedByKinductionPlain}{Correct}{}{Cputime}{Avg}{45.35541278447147470398277718}%
\StoreBenchExecResult{PdrInv}{KinductionDfStaticSixteenTwoFTrueNotSolvedByKinductionPlain}{Correct}{}{Cputime}{Median}{7.050458974}%
\StoreBenchExecResult{PdrInv}{KinductionDfStaticSixteenTwoFTrueNotSolvedByKinductionPlain}{Correct}{}{Cputime}{Min}{3.165165323}%
\StoreBenchExecResult{PdrInv}{KinductionDfStaticSixteenTwoFTrueNotSolvedByKinductionPlain}{Correct}{}{Cputime}{Max}{870.198310293}%
\StoreBenchExecResult{PdrInv}{KinductionDfStaticSixteenTwoFTrueNotSolvedByKinductionPlain}{Correct}{}{Cputime}{Stdev}{109.3627316606266524406420816}%
\StoreBenchExecResult{PdrInv}{KinductionDfStaticSixteenTwoFTrueNotSolvedByKinductionPlain}{Correct}{}{Walltime}{}{22912.91866874647}%
\StoreBenchExecResult{PdrInv}{KinductionDfStaticSixteenTwoFTrueNotSolvedByKinductionPlain}{Correct}{}{Walltime}{Avg}{24.66406745828468245425188375}%
\StoreBenchExecResult{PdrInv}{KinductionDfStaticSixteenTwoFTrueNotSolvedByKinductionPlain}{Correct}{}{Walltime}{Median}{3.7198741436}%
\StoreBenchExecResult{PdrInv}{KinductionDfStaticSixteenTwoFTrueNotSolvedByKinductionPlain}{Correct}{}{Walltime}{Min}{1.75485801697}%
\StoreBenchExecResult{PdrInv}{KinductionDfStaticSixteenTwoFTrueNotSolvedByKinductionPlain}{Correct}{}{Walltime}{Max}{772.718392134}%
\StoreBenchExecResult{PdrInv}{KinductionDfStaticSixteenTwoFTrueNotSolvedByKinductionPlain}{Correct}{}{Walltime}{Stdev}{66.44357958492411084258166782}%
\StoreBenchExecResult{PdrInv}{KinductionDfStaticSixteenTwoFTrueNotSolvedByKinductionPlain}{Correct}{True}{Count}{}{929}%
\StoreBenchExecResult{PdrInv}{KinductionDfStaticSixteenTwoFTrueNotSolvedByKinductionPlain}{Correct}{True}{Cputime}{}{42135.178476774}%
\StoreBenchExecResult{PdrInv}{KinductionDfStaticSixteenTwoFTrueNotSolvedByKinductionPlain}{Correct}{True}{Cputime}{Avg}{45.35541278447147470398277718}%
\StoreBenchExecResult{PdrInv}{KinductionDfStaticSixteenTwoFTrueNotSolvedByKinductionPlain}{Correct}{True}{Cputime}{Median}{7.050458974}%
\StoreBenchExecResult{PdrInv}{KinductionDfStaticSixteenTwoFTrueNotSolvedByKinductionPlain}{Correct}{True}{Cputime}{Min}{3.165165323}%
\StoreBenchExecResult{PdrInv}{KinductionDfStaticSixteenTwoFTrueNotSolvedByKinductionPlain}{Correct}{True}{Cputime}{Max}{870.198310293}%
\StoreBenchExecResult{PdrInv}{KinductionDfStaticSixteenTwoFTrueNotSolvedByKinductionPlain}{Correct}{True}{Cputime}{Stdev}{109.3627316606266524406420816}%
\StoreBenchExecResult{PdrInv}{KinductionDfStaticSixteenTwoFTrueNotSolvedByKinductionPlain}{Correct}{True}{Walltime}{}{22912.91866874647}%
\StoreBenchExecResult{PdrInv}{KinductionDfStaticSixteenTwoFTrueNotSolvedByKinductionPlain}{Correct}{True}{Walltime}{Avg}{24.66406745828468245425188375}%
\StoreBenchExecResult{PdrInv}{KinductionDfStaticSixteenTwoFTrueNotSolvedByKinductionPlain}{Correct}{True}{Walltime}{Median}{3.7198741436}%
\StoreBenchExecResult{PdrInv}{KinductionDfStaticSixteenTwoFTrueNotSolvedByKinductionPlain}{Correct}{True}{Walltime}{Min}{1.75485801697}%
\StoreBenchExecResult{PdrInv}{KinductionDfStaticSixteenTwoFTrueNotSolvedByKinductionPlain}{Correct}{True}{Walltime}{Max}{772.718392134}%
\StoreBenchExecResult{PdrInv}{KinductionDfStaticSixteenTwoFTrueNotSolvedByKinductionPlain}{Correct}{True}{Walltime}{Stdev}{66.44357958492411084258166782}%
\StoreBenchExecResult{PdrInv}{KinductionDfStaticSixteenTwoFTrueNotSolvedByKinductionPlain}{Wrong}{True}{Count}{}{0}%
\StoreBenchExecResult{PdrInv}{KinductionDfStaticSixteenTwoFTrueNotSolvedByKinductionPlain}{Wrong}{True}{Cputime}{}{0}%
\StoreBenchExecResult{PdrInv}{KinductionDfStaticSixteenTwoFTrueNotSolvedByKinductionPlain}{Wrong}{True}{Cputime}{Avg}{None}%
\StoreBenchExecResult{PdrInv}{KinductionDfStaticSixteenTwoFTrueNotSolvedByKinductionPlain}{Wrong}{True}{Cputime}{Median}{None}%
\StoreBenchExecResult{PdrInv}{KinductionDfStaticSixteenTwoFTrueNotSolvedByKinductionPlain}{Wrong}{True}{Cputime}{Min}{None}%
\StoreBenchExecResult{PdrInv}{KinductionDfStaticSixteenTwoFTrueNotSolvedByKinductionPlain}{Wrong}{True}{Cputime}{Max}{None}%
\StoreBenchExecResult{PdrInv}{KinductionDfStaticSixteenTwoFTrueNotSolvedByKinductionPlain}{Wrong}{True}{Cputime}{Stdev}{None}%
\StoreBenchExecResult{PdrInv}{KinductionDfStaticSixteenTwoFTrueNotSolvedByKinductionPlain}{Wrong}{True}{Walltime}{}{0}%
\StoreBenchExecResult{PdrInv}{KinductionDfStaticSixteenTwoFTrueNotSolvedByKinductionPlain}{Wrong}{True}{Walltime}{Avg}{None}%
\StoreBenchExecResult{PdrInv}{KinductionDfStaticSixteenTwoFTrueNotSolvedByKinductionPlain}{Wrong}{True}{Walltime}{Median}{None}%
\StoreBenchExecResult{PdrInv}{KinductionDfStaticSixteenTwoFTrueNotSolvedByKinductionPlain}{Wrong}{True}{Walltime}{Min}{None}%
\StoreBenchExecResult{PdrInv}{KinductionDfStaticSixteenTwoFTrueNotSolvedByKinductionPlain}{Wrong}{True}{Walltime}{Max}{None}%
\StoreBenchExecResult{PdrInv}{KinductionDfStaticSixteenTwoFTrueNotSolvedByKinductionPlain}{Wrong}{True}{Walltime}{Stdev}{None}%
\StoreBenchExecResult{PdrInv}{KinductionDfStaticSixteenTwoFTrueNotSolvedByKinductionPlain}{Error}{}{Count}{}{1964}%
\StoreBenchExecResult{PdrInv}{KinductionDfStaticSixteenTwoFTrueNotSolvedByKinductionPlain}{Error}{}{Cputime}{}{1627974.292401435}%
\StoreBenchExecResult{PdrInv}{KinductionDfStaticSixteenTwoFTrueNotSolvedByKinductionPlain}{Error}{}{Cputime}{Avg}{828.9074808561278004073319756}%
\StoreBenchExecResult{PdrInv}{KinductionDfStaticSixteenTwoFTrueNotSolvedByKinductionPlain}{Error}{}{Cputime}{Median}{901.547633211}%
\StoreBenchExecResult{PdrInv}{KinductionDfStaticSixteenTwoFTrueNotSolvedByKinductionPlain}{Error}{}{Cputime}{Min}{2.37078614}%
\StoreBenchExecResult{PdrInv}{KinductionDfStaticSixteenTwoFTrueNotSolvedByKinductionPlain}{Error}{}{Cputime}{Max}{1001.28896153}%
\StoreBenchExecResult{PdrInv}{KinductionDfStaticSixteenTwoFTrueNotSolvedByKinductionPlain}{Error}{}{Cputime}{Stdev}{222.2258433806117087842342534}%
\StoreBenchExecResult{PdrInv}{KinductionDfStaticSixteenTwoFTrueNotSolvedByKinductionPlain}{Error}{}{Walltime}{}{954600.27664660637}%
\StoreBenchExecResult{PdrInv}{KinductionDfStaticSixteenTwoFTrueNotSolvedByKinductionPlain}{Error}{}{Walltime}{Avg}{486.0490206958280906313645621}%
\StoreBenchExecResult{PdrInv}{KinductionDfStaticSixteenTwoFTrueNotSolvedByKinductionPlain}{Error}{}{Walltime}{Median}{452.435005546}%
\StoreBenchExecResult{PdrInv}{KinductionDfStaticSixteenTwoFTrueNotSolvedByKinductionPlain}{Error}{}{Walltime}{Min}{1.32805800438}%
\StoreBenchExecResult{PdrInv}{KinductionDfStaticSixteenTwoFTrueNotSolvedByKinductionPlain}{Error}{}{Walltime}{Max}{901.200150967}%
\StoreBenchExecResult{PdrInv}{KinductionDfStaticSixteenTwoFTrueNotSolvedByKinductionPlain}{Error}{}{Walltime}{Stdev}{197.1127184369261436366402970}%
\StoreBenchExecResult{PdrInv}{KinductionDfStaticSixteenTwoFTrueNotSolvedByKinductionPlain}{Error}{Assertion}{Count}{}{2}%
\StoreBenchExecResult{PdrInv}{KinductionDfStaticSixteenTwoFTrueNotSolvedByKinductionPlain}{Error}{Assertion}{Cputime}{}{7.031062284}%
\StoreBenchExecResult{PdrInv}{KinductionDfStaticSixteenTwoFTrueNotSolvedByKinductionPlain}{Error}{Assertion}{Cputime}{Avg}{3.515531142}%
\StoreBenchExecResult{PdrInv}{KinductionDfStaticSixteenTwoFTrueNotSolvedByKinductionPlain}{Error}{Assertion}{Cputime}{Median}{3.515531142}%
\StoreBenchExecResult{PdrInv}{KinductionDfStaticSixteenTwoFTrueNotSolvedByKinductionPlain}{Error}{Assertion}{Cputime}{Min}{3.231633808}%
\StoreBenchExecResult{PdrInv}{KinductionDfStaticSixteenTwoFTrueNotSolvedByKinductionPlain}{Error}{Assertion}{Cputime}{Max}{3.799428476}%
\StoreBenchExecResult{PdrInv}{KinductionDfStaticSixteenTwoFTrueNotSolvedByKinductionPlain}{Error}{Assertion}{Cputime}{Stdev}{0.283897334}%
\StoreBenchExecResult{PdrInv}{KinductionDfStaticSixteenTwoFTrueNotSolvedByKinductionPlain}{Error}{Assertion}{Walltime}{}{3.88648891449}%
\StoreBenchExecResult{PdrInv}{KinductionDfStaticSixteenTwoFTrueNotSolvedByKinductionPlain}{Error}{Assertion}{Walltime}{Avg}{1.943244457245}%
\StoreBenchExecResult{PdrInv}{KinductionDfStaticSixteenTwoFTrueNotSolvedByKinductionPlain}{Error}{Assertion}{Walltime}{Median}{1.943244457245}%
\StoreBenchExecResult{PdrInv}{KinductionDfStaticSixteenTwoFTrueNotSolvedByKinductionPlain}{Error}{Assertion}{Walltime}{Min}{1.78250694275}%
\StoreBenchExecResult{PdrInv}{KinductionDfStaticSixteenTwoFTrueNotSolvedByKinductionPlain}{Error}{Assertion}{Walltime}{Max}{2.10398197174}%
\StoreBenchExecResult{PdrInv}{KinductionDfStaticSixteenTwoFTrueNotSolvedByKinductionPlain}{Error}{Assertion}{Walltime}{Stdev}{0.160737514495}%
\StoreBenchExecResult{PdrInv}{KinductionDfStaticSixteenTwoFTrueNotSolvedByKinductionPlain}{Error}{Error}{Count}{}{103}%
\StoreBenchExecResult{PdrInv}{KinductionDfStaticSixteenTwoFTrueNotSolvedByKinductionPlain}{Error}{Error}{Cputime}{}{14327.930064932}%
\StoreBenchExecResult{PdrInv}{KinductionDfStaticSixteenTwoFTrueNotSolvedByKinductionPlain}{Error}{Error}{Cputime}{Avg}{139.1061171352621359223300971}%
\StoreBenchExecResult{PdrInv}{KinductionDfStaticSixteenTwoFTrueNotSolvedByKinductionPlain}{Error}{Error}{Cputime}{Median}{64.944057882}%
\StoreBenchExecResult{PdrInv}{KinductionDfStaticSixteenTwoFTrueNotSolvedByKinductionPlain}{Error}{Error}{Cputime}{Min}{2.37078614}%
\StoreBenchExecResult{PdrInv}{KinductionDfStaticSixteenTwoFTrueNotSolvedByKinductionPlain}{Error}{Error}{Cputime}{Max}{843.661008287}%
\StoreBenchExecResult{PdrInv}{KinductionDfStaticSixteenTwoFTrueNotSolvedByKinductionPlain}{Error}{Error}{Cputime}{Stdev}{174.7504312467654652577503085}%
\StoreBenchExecResult{PdrInv}{KinductionDfStaticSixteenTwoFTrueNotSolvedByKinductionPlain}{Error}{Error}{Walltime}{}{12054.35421419150}%
\StoreBenchExecResult{PdrInv}{KinductionDfStaticSixteenTwoFTrueNotSolvedByKinductionPlain}{Error}{Error}{Walltime}{Avg}{117.0325651863252427184466019}%
\StoreBenchExecResult{PdrInv}{KinductionDfStaticSixteenTwoFTrueNotSolvedByKinductionPlain}{Error}{Error}{Walltime}{Median}{52.2554209232}%
\StoreBenchExecResult{PdrInv}{KinductionDfStaticSixteenTwoFTrueNotSolvedByKinductionPlain}{Error}{Error}{Walltime}{Min}{1.32805800438}%
\StoreBenchExecResult{PdrInv}{KinductionDfStaticSixteenTwoFTrueNotSolvedByKinductionPlain}{Error}{Error}{Walltime}{Max}{829.643739939}%
\StoreBenchExecResult{PdrInv}{KinductionDfStaticSixteenTwoFTrueNotSolvedByKinductionPlain}{Error}{Error}{Walltime}{Stdev}{158.9406027664754963846032287}%
\StoreBenchExecResult{PdrInv}{KinductionDfStaticSixteenTwoFTrueNotSolvedByKinductionPlain}{Error}{Exception}{Count}{}{5}%
\StoreBenchExecResult{PdrInv}{KinductionDfStaticSixteenTwoFTrueNotSolvedByKinductionPlain}{Error}{Exception}{Cputime}{}{637.479221069}%
\StoreBenchExecResult{PdrInv}{KinductionDfStaticSixteenTwoFTrueNotSolvedByKinductionPlain}{Error}{Exception}{Cputime}{Avg}{127.4958442138}%
\StoreBenchExecResult{PdrInv}{KinductionDfStaticSixteenTwoFTrueNotSolvedByKinductionPlain}{Error}{Exception}{Cputime}{Median}{70.647068768}%
\StoreBenchExecResult{PdrInv}{KinductionDfStaticSixteenTwoFTrueNotSolvedByKinductionPlain}{Error}{Exception}{Cputime}{Min}{17.502246669}%
\StoreBenchExecResult{PdrInv}{KinductionDfStaticSixteenTwoFTrueNotSolvedByKinductionPlain}{Error}{Exception}{Cputime}{Max}{309.746377123}%
\StoreBenchExecResult{PdrInv}{KinductionDfStaticSixteenTwoFTrueNotSolvedByKinductionPlain}{Error}{Exception}{Cputime}{Stdev}{117.7263608694235889435636292}%
\StoreBenchExecResult{PdrInv}{KinductionDfStaticSixteenTwoFTrueNotSolvedByKinductionPlain}{Error}{Exception}{Walltime}{}{320.71823024798}%
\StoreBenchExecResult{PdrInv}{KinductionDfStaticSixteenTwoFTrueNotSolvedByKinductionPlain}{Error}{Exception}{Walltime}{Avg}{64.143646049596}%
\StoreBenchExecResult{PdrInv}{KinductionDfStaticSixteenTwoFTrueNotSolvedByKinductionPlain}{Error}{Exception}{Walltime}{Median}{35.5427501202}%
\StoreBenchExecResult{PdrInv}{KinductionDfStaticSixteenTwoFTrueNotSolvedByKinductionPlain}{Error}{Exception}{Walltime}{Min}{8.98041892052}%
\StoreBenchExecResult{PdrInv}{KinductionDfStaticSixteenTwoFTrueNotSolvedByKinductionPlain}{Error}{Exception}{Walltime}{Max}{155.59612608}%
\StoreBenchExecResult{PdrInv}{KinductionDfStaticSixteenTwoFTrueNotSolvedByKinductionPlain}{Error}{Exception}{Walltime}{Stdev}{59.06282196240071875324933199}%
\StoreBenchExecResult{PdrInv}{KinductionDfStaticSixteenTwoFTrueNotSolvedByKinductionPlain}{Error}{OutOfJavaMemory}{Count}{}{4}%
\StoreBenchExecResult{PdrInv}{KinductionDfStaticSixteenTwoFTrueNotSolvedByKinductionPlain}{Error}{OutOfJavaMemory}{Cputime}{}{1524.522277843}%
\StoreBenchExecResult{PdrInv}{KinductionDfStaticSixteenTwoFTrueNotSolvedByKinductionPlain}{Error}{OutOfJavaMemory}{Cputime}{Avg}{381.13056946075}%
\StoreBenchExecResult{PdrInv}{KinductionDfStaticSixteenTwoFTrueNotSolvedByKinductionPlain}{Error}{OutOfJavaMemory}{Cputime}{Median}{356.365390115}%
\StoreBenchExecResult{PdrInv}{KinductionDfStaticSixteenTwoFTrueNotSolvedByKinductionPlain}{Error}{OutOfJavaMemory}{Cputime}{Min}{275.892520346}%
\StoreBenchExecResult{PdrInv}{KinductionDfStaticSixteenTwoFTrueNotSolvedByKinductionPlain}{Error}{OutOfJavaMemory}{Cputime}{Max}{535.898977267}%
\StoreBenchExecResult{PdrInv}{KinductionDfStaticSixteenTwoFTrueNotSolvedByKinductionPlain}{Error}{OutOfJavaMemory}{Cputime}{Stdev}{108.8990590327429041984333846}%
\StoreBenchExecResult{PdrInv}{KinductionDfStaticSixteenTwoFTrueNotSolvedByKinductionPlain}{Error}{OutOfJavaMemory}{Walltime}{}{806.256072999}%
\StoreBenchExecResult{PdrInv}{KinductionDfStaticSixteenTwoFTrueNotSolvedByKinductionPlain}{Error}{OutOfJavaMemory}{Walltime}{Avg}{201.56401824975}%
\StoreBenchExecResult{PdrInv}{KinductionDfStaticSixteenTwoFTrueNotSolvedByKinductionPlain}{Error}{OutOfJavaMemory}{Walltime}{Median}{195.5206454995}%
\StoreBenchExecResult{PdrInv}{KinductionDfStaticSixteenTwoFTrueNotSolvedByKinductionPlain}{Error}{OutOfJavaMemory}{Walltime}{Min}{141.277446032}%
\StoreBenchExecResult{PdrInv}{KinductionDfStaticSixteenTwoFTrueNotSolvedByKinductionPlain}{Error}{OutOfJavaMemory}{Walltime}{Max}{273.937335968}%
\StoreBenchExecResult{PdrInv}{KinductionDfStaticSixteenTwoFTrueNotSolvedByKinductionPlain}{Error}{OutOfJavaMemory}{Walltime}{Stdev}{51.44026946140919563442350665}%
\StoreBenchExecResult{PdrInv}{KinductionDfStaticSixteenTwoFTrueNotSolvedByKinductionPlain}{Error}{OutOfMemory}{Count}{}{122}%
\StoreBenchExecResult{PdrInv}{KinductionDfStaticSixteenTwoFTrueNotSolvedByKinductionPlain}{Error}{OutOfMemory}{Cputime}{}{44939.963607378}%
\StoreBenchExecResult{PdrInv}{KinductionDfStaticSixteenTwoFTrueNotSolvedByKinductionPlain}{Error}{OutOfMemory}{Cputime}{Avg}{368.3603574375245901639344262}%
\StoreBenchExecResult{PdrInv}{KinductionDfStaticSixteenTwoFTrueNotSolvedByKinductionPlain}{Error}{OutOfMemory}{Cputime}{Median}{348.8682152345}%
\StoreBenchExecResult{PdrInv}{KinductionDfStaticSixteenTwoFTrueNotSolvedByKinductionPlain}{Error}{OutOfMemory}{Cputime}{Min}{168.071531181}%
\StoreBenchExecResult{PdrInv}{KinductionDfStaticSixteenTwoFTrueNotSolvedByKinductionPlain}{Error}{OutOfMemory}{Cputime}{Max}{811.66161456}%
\StoreBenchExecResult{PdrInv}{KinductionDfStaticSixteenTwoFTrueNotSolvedByKinductionPlain}{Error}{OutOfMemory}{Cputime}{Stdev}{166.3792363268339429534587532}%
\StoreBenchExecResult{PdrInv}{KinductionDfStaticSixteenTwoFTrueNotSolvedByKinductionPlain}{Error}{OutOfMemory}{Walltime}{}{22557.8359644434}%
\StoreBenchExecResult{PdrInv}{KinductionDfStaticSixteenTwoFTrueNotSolvedByKinductionPlain}{Error}{OutOfMemory}{Walltime}{Avg}{184.9002947905196721311475410}%
\StoreBenchExecResult{PdrInv}{KinductionDfStaticSixteenTwoFTrueNotSolvedByKinductionPlain}{Error}{OutOfMemory}{Walltime}{Median}{175.2706849575}%
\StoreBenchExecResult{PdrInv}{KinductionDfStaticSixteenTwoFTrueNotSolvedByKinductionPlain}{Error}{OutOfMemory}{Walltime}{Min}{84.6973059177}%
\StoreBenchExecResult{PdrInv}{KinductionDfStaticSixteenTwoFTrueNotSolvedByKinductionPlain}{Error}{OutOfMemory}{Walltime}{Max}{406.786455154}%
\StoreBenchExecResult{PdrInv}{KinductionDfStaticSixteenTwoFTrueNotSolvedByKinductionPlain}{Error}{OutOfMemory}{Walltime}{Stdev}{83.27193045728746357769100793}%
\StoreBenchExecResult{PdrInv}{KinductionDfStaticSixteenTwoFTrueNotSolvedByKinductionPlain}{Error}{Timeout}{Count}{}{1728}%
\StoreBenchExecResult{PdrInv}{KinductionDfStaticSixteenTwoFTrueNotSolvedByKinductionPlain}{Error}{Timeout}{Cputime}{}{1566537.366167929}%
\StoreBenchExecResult{PdrInv}{KinductionDfStaticSixteenTwoFTrueNotSolvedByKinductionPlain}{Error}{Timeout}{Cputime}{Avg}{906.5609757916255787037037037}%
\StoreBenchExecResult{PdrInv}{KinductionDfStaticSixteenTwoFTrueNotSolvedByKinductionPlain}{Error}{Timeout}{Cputime}{Median}{901.7812157255}%
\StoreBenchExecResult{PdrInv}{KinductionDfStaticSixteenTwoFTrueNotSolvedByKinductionPlain}{Error}{Timeout}{Cputime}{Min}{900.779935944}%
\StoreBenchExecResult{PdrInv}{KinductionDfStaticSixteenTwoFTrueNotSolvedByKinductionPlain}{Error}{Timeout}{Cputime}{Max}{1001.28896153}%
\StoreBenchExecResult{PdrInv}{KinductionDfStaticSixteenTwoFTrueNotSolvedByKinductionPlain}{Error}{Timeout}{Cputime}{Stdev}{15.59704526969044213346444981}%
\StoreBenchExecResult{PdrInv}{KinductionDfStaticSixteenTwoFTrueNotSolvedByKinductionPlain}{Error}{Timeout}{Walltime}{}{918857.225675810}%
\StoreBenchExecResult{PdrInv}{KinductionDfStaticSixteenTwoFTrueNotSolvedByKinductionPlain}{Error}{Timeout}{Walltime}{Avg}{531.7460796735011574074074074}%
\StoreBenchExecResult{PdrInv}{KinductionDfStaticSixteenTwoFTrueNotSolvedByKinductionPlain}{Error}{Timeout}{Walltime}{Median}{452.870404959}%
\StoreBenchExecResult{PdrInv}{KinductionDfStaticSixteenTwoFTrueNotSolvedByKinductionPlain}{Error}{Timeout}{Walltime}{Min}{451.096029997}%
\StoreBenchExecResult{PdrInv}{KinductionDfStaticSixteenTwoFTrueNotSolvedByKinductionPlain}{Error}{Timeout}{Walltime}{Max}{901.200150967}%
\StoreBenchExecResult{PdrInv}{KinductionDfStaticSixteenTwoFTrueNotSolvedByKinductionPlain}{Error}{Timeout}{Walltime}{Stdev}{156.7374498191103680886548301}%
\providecommand\StoreBenchExecResult[7]{\expandafter\newcommand\csname#1#2#3#4#5#6\endcsname{#7}}%
\StoreBenchExecResult{PdrInv}{KinductionDfStaticSixteenTwoF}{Total}{}{Count}{}{5591}%
\StoreBenchExecResult{PdrInv}{KinductionDfStaticSixteenTwoF}{Total}{}{Cputime}{}{2325442.001353531}%
\StoreBenchExecResult{PdrInv}{KinductionDfStaticSixteenTwoF}{Total}{}{Cputime}{Avg}{415.9259526656288678232874262}%
\StoreBenchExecResult{PdrInv}{KinductionDfStaticSixteenTwoF}{Total}{}{Cputime}{Median}{170.820770977}%
\StoreBenchExecResult{PdrInv}{KinductionDfStaticSixteenTwoF}{Total}{}{Cputime}{Min}{2.37078614}%
\StoreBenchExecResult{PdrInv}{KinductionDfStaticSixteenTwoF}{Total}{}{Cputime}{Max}{1001.28896153}%
\StoreBenchExecResult{PdrInv}{KinductionDfStaticSixteenTwoF}{Total}{}{Cputime}{Stdev}{422.2561105382409144727094862}%
\StoreBenchExecResult{PdrInv}{KinductionDfStaticSixteenTwoF}{Total}{}{Walltime}{}{1368310.02336119829}%
\StoreBenchExecResult{PdrInv}{KinductionDfStaticSixteenTwoF}{Total}{}{Walltime}{Avg}{244.7343987410478071901269898}%
\StoreBenchExecResult{PdrInv}{KinductionDfStaticSixteenTwoF}{Total}{}{Walltime}{Median}{87.3208680153}%
\StoreBenchExecResult{PdrInv}{KinductionDfStaticSixteenTwoF}{Total}{}{Walltime}{Min}{1.32805800438}%
\StoreBenchExecResult{PdrInv}{KinductionDfStaticSixteenTwoF}{Total}{}{Walltime}{Max}{954.062977791}%
\StoreBenchExecResult{PdrInv}{KinductionDfStaticSixteenTwoF}{Total}{}{Walltime}{Stdev}{269.8815667440143663377833756}%
\StoreBenchExecResult{PdrInv}{KinductionDfStaticSixteenTwoF}{Correct}{}{Count}{}{2931}%
\StoreBenchExecResult{PdrInv}{KinductionDfStaticSixteenTwoF}{Correct}{}{Cputime}{}{162244.240582299}%
\StoreBenchExecResult{PdrInv}{KinductionDfStaticSixteenTwoF}{Correct}{}{Cputime}{Avg}{55.35456860535619242579324463}%
\StoreBenchExecResult{PdrInv}{KinductionDfStaticSixteenTwoF}{Correct}{}{Cputime}{Median}{9.487355698}%
\StoreBenchExecResult{PdrInv}{KinductionDfStaticSixteenTwoF}{Correct}{}{Cputime}{Min}{3.034831037}%
\StoreBenchExecResult{PdrInv}{KinductionDfStaticSixteenTwoF}{Correct}{}{Cputime}{Max}{888.462982153}%
\StoreBenchExecResult{PdrInv}{KinductionDfStaticSixteenTwoF}{Correct}{}{Cputime}{Stdev}{126.0615373252849297216294895}%
\StoreBenchExecResult{PdrInv}{KinductionDfStaticSixteenTwoF}{Correct}{}{Walltime}{}{97766.60730624420}%
\StoreBenchExecResult{PdrInv}{KinductionDfStaticSixteenTwoF}{Correct}{}{Walltime}{Avg}{33.35605844634738996929375640}%
\StoreBenchExecResult{PdrInv}{KinductionDfStaticSixteenTwoF}{Correct}{}{Walltime}{Median}{5.02267193794}%
\StoreBenchExecResult{PdrInv}{KinductionDfStaticSixteenTwoF}{Correct}{}{Walltime}{Min}{1.67546486855}%
\StoreBenchExecResult{PdrInv}{KinductionDfStaticSixteenTwoF}{Correct}{}{Walltime}{Max}{825.061019897}%
\StoreBenchExecResult{PdrInv}{KinductionDfStaticSixteenTwoF}{Correct}{}{Walltime}{Stdev}{88.60389448868703525530503482}%
\StoreBenchExecResult{PdrInv}{KinductionDfStaticSixteenTwoF}{Correct}{False}{Count}{}{786}%
\StoreBenchExecResult{PdrInv}{KinductionDfStaticSixteenTwoF}{Correct}{False}{Cputime}{}{64879.594336429}%
\StoreBenchExecResult{PdrInv}{KinductionDfStaticSixteenTwoF}{Correct}{False}{Cputime}{Avg}{82.54401315067302798982188295}%
\StoreBenchExecResult{PdrInv}{KinductionDfStaticSixteenTwoF}{Correct}{False}{Cputime}{Median}{23.7393956765}%
\StoreBenchExecResult{PdrInv}{KinductionDfStaticSixteenTwoF}{Correct}{False}{Cputime}{Min}{3.168040461}%
\StoreBenchExecResult{PdrInv}{KinductionDfStaticSixteenTwoF}{Correct}{False}{Cputime}{Max}{888.462982153}%
\StoreBenchExecResult{PdrInv}{KinductionDfStaticSixteenTwoF}{Correct}{False}{Cputime}{Stdev}{158.1152624008852502215243390}%
\StoreBenchExecResult{PdrInv}{KinductionDfStaticSixteenTwoF}{Correct}{False}{Walltime}{}{44028.99386334445}%
\StoreBenchExecResult{PdrInv}{KinductionDfStaticSixteenTwoF}{Correct}{False}{Walltime}{Avg}{56.01653163275375318066157761}%
\StoreBenchExecResult{PdrInv}{KinductionDfStaticSixteenTwoF}{Correct}{False}{Walltime}{Median}{12.3462090492}%
\StoreBenchExecResult{PdrInv}{KinductionDfStaticSixteenTwoF}{Correct}{False}{Walltime}{Min}{1.76473808289}%
\StoreBenchExecResult{PdrInv}{KinductionDfStaticSixteenTwoF}{Correct}{False}{Walltime}{Max}{814.212782145}%
\StoreBenchExecResult{PdrInv}{KinductionDfStaticSixteenTwoF}{Correct}{False}{Walltime}{Stdev}{125.7729971042611235205743914}%
\StoreBenchExecResult{PdrInv}{KinductionDfStaticSixteenTwoF}{Correct}{True}{Count}{}{2145}%
\StoreBenchExecResult{PdrInv}{KinductionDfStaticSixteenTwoF}{Correct}{True}{Cputime}{}{97364.646245870}%
\StoreBenchExecResult{PdrInv}{KinductionDfStaticSixteenTwoF}{Correct}{True}{Cputime}{Avg}{45.39144347126806526806526807}%
\StoreBenchExecResult{PdrInv}{KinductionDfStaticSixteenTwoF}{Correct}{True}{Cputime}{Median}{7.069380448}%
\StoreBenchExecResult{PdrInv}{KinductionDfStaticSixteenTwoF}{Correct}{True}{Cputime}{Min}{3.034831037}%
\StoreBenchExecResult{PdrInv}{KinductionDfStaticSixteenTwoF}{Correct}{True}{Cputime}{Max}{870.198310293}%
\StoreBenchExecResult{PdrInv}{KinductionDfStaticSixteenTwoF}{Correct}{True}{Cputime}{Stdev}{110.3790723542452380354808214}%
\StoreBenchExecResult{PdrInv}{KinductionDfStaticSixteenTwoF}{Correct}{True}{Walltime}{}{53737.61344289975}%
\StoreBenchExecResult{PdrInv}{KinductionDfStaticSixteenTwoF}{Correct}{True}{Walltime}{Avg}{25.05250043958030303030303030}%
\StoreBenchExecResult{PdrInv}{KinductionDfStaticSixteenTwoF}{Correct}{True}{Walltime}{Median}{3.74775791168}%
\StoreBenchExecResult{PdrInv}{KinductionDfStaticSixteenTwoF}{Correct}{True}{Walltime}{Min}{1.67546486855}%
\StoreBenchExecResult{PdrInv}{KinductionDfStaticSixteenTwoF}{Correct}{True}{Walltime}{Max}{825.061019897}%
\StoreBenchExecResult{PdrInv}{KinductionDfStaticSixteenTwoF}{Correct}{True}{Walltime}{Stdev}{68.36463823635426699543794984}%
\StoreBenchExecResult{PdrInv}{KinductionDfStaticSixteenTwoF}{Wrong}{True}{Count}{}{0}%
\StoreBenchExecResult{PdrInv}{KinductionDfStaticSixteenTwoF}{Wrong}{True}{Cputime}{}{0}%
\StoreBenchExecResult{PdrInv}{KinductionDfStaticSixteenTwoF}{Wrong}{True}{Cputime}{Avg}{None}%
\StoreBenchExecResult{PdrInv}{KinductionDfStaticSixteenTwoF}{Wrong}{True}{Cputime}{Median}{None}%
\StoreBenchExecResult{PdrInv}{KinductionDfStaticSixteenTwoF}{Wrong}{True}{Cputime}{Min}{None}%
\StoreBenchExecResult{PdrInv}{KinductionDfStaticSixteenTwoF}{Wrong}{True}{Cputime}{Max}{None}%
\StoreBenchExecResult{PdrInv}{KinductionDfStaticSixteenTwoF}{Wrong}{True}{Cputime}{Stdev}{None}%
\StoreBenchExecResult{PdrInv}{KinductionDfStaticSixteenTwoF}{Wrong}{True}{Walltime}{}{0}%
\StoreBenchExecResult{PdrInv}{KinductionDfStaticSixteenTwoF}{Wrong}{True}{Walltime}{Avg}{None}%
\StoreBenchExecResult{PdrInv}{KinductionDfStaticSixteenTwoF}{Wrong}{True}{Walltime}{Median}{None}%
\StoreBenchExecResult{PdrInv}{KinductionDfStaticSixteenTwoF}{Wrong}{True}{Walltime}{Min}{None}%
\StoreBenchExecResult{PdrInv}{KinductionDfStaticSixteenTwoF}{Wrong}{True}{Walltime}{Max}{None}%
\StoreBenchExecResult{PdrInv}{KinductionDfStaticSixteenTwoF}{Wrong}{True}{Walltime}{Stdev}{None}%
\StoreBenchExecResult{PdrInv}{KinductionDfStaticSixteenTwoF}{Error}{}{Count}{}{2658}%
\StoreBenchExecResult{PdrInv}{KinductionDfStaticSixteenTwoF}{Error}{}{Cputime}{}{2163172.119429213}%
\StoreBenchExecResult{PdrInv}{KinductionDfStaticSixteenTwoF}{Error}{}{Cputime}{Avg}{813.8345069334887133182844244}%
\StoreBenchExecResult{PdrInv}{KinductionDfStaticSixteenTwoF}{Error}{}{Cputime}{Median}{901.6137660235}%
\StoreBenchExecResult{PdrInv}{KinductionDfStaticSixteenTwoF}{Error}{}{Cputime}{Min}{2.37078614}%
\StoreBenchExecResult{PdrInv}{KinductionDfStaticSixteenTwoF}{Error}{}{Cputime}{Max}{1001.28896153}%
\StoreBenchExecResult{PdrInv}{KinductionDfStaticSixteenTwoF}{Error}{}{Cputime}{Stdev}{236.0196668458132507692663077}%
\StoreBenchExecResult{PdrInv}{KinductionDfStaticSixteenTwoF}{Error}{}{Walltime}{}{1270529.48870872458}%
\StoreBenchExecResult{PdrInv}{KinductionDfStaticSixteenTwoF}{Error}{}{Walltime}{Avg}{478.0020649769467945823927765}%
\StoreBenchExecResult{PdrInv}{KinductionDfStaticSixteenTwoF}{Error}{}{Walltime}{Median}{452.584066987}%
\StoreBenchExecResult{PdrInv}{KinductionDfStaticSixteenTwoF}{Error}{}{Walltime}{Min}{1.32805800438}%
\StoreBenchExecResult{PdrInv}{KinductionDfStaticSixteenTwoF}{Error}{}{Walltime}{Max}{954.062977791}%
\StoreBenchExecResult{PdrInv}{KinductionDfStaticSixteenTwoF}{Error}{}{Walltime}{Stdev}{202.0509173712563488460629747}%
\StoreBenchExecResult{PdrInv}{KinductionDfStaticSixteenTwoF}{Error}{Assertion}{Count}{}{4}%
\StoreBenchExecResult{PdrInv}{KinductionDfStaticSixteenTwoF}{Error}{Assertion}{Cputime}{}{13.571104040}%
\StoreBenchExecResult{PdrInv}{KinductionDfStaticSixteenTwoF}{Error}{Assertion}{Cputime}{Avg}{3.392776010}%
\StoreBenchExecResult{PdrInv}{KinductionDfStaticSixteenTwoF}{Error}{Assertion}{Cputime}{Median}{3.270020878}%
\StoreBenchExecResult{PdrInv}{KinductionDfStaticSixteenTwoF}{Error}{Assertion}{Cputime}{Min}{3.231633808}%
\StoreBenchExecResult{PdrInv}{KinductionDfStaticSixteenTwoF}{Error}{Assertion}{Cputime}{Max}{3.799428476}%
\StoreBenchExecResult{PdrInv}{KinductionDfStaticSixteenTwoF}{Error}{Assertion}{Cputime}{Stdev}{0.2353033627688911769752069203}%
\StoreBenchExecResult{PdrInv}{KinductionDfStaticSixteenTwoF}{Error}{Assertion}{Walltime}{}{7.51349973679}%
\StoreBenchExecResult{PdrInv}{KinductionDfStaticSixteenTwoF}{Error}{Assertion}{Walltime}{Avg}{1.8783749341975}%
\StoreBenchExecResult{PdrInv}{KinductionDfStaticSixteenTwoF}{Error}{Assertion}{Walltime}{Median}{1.81350541115}%
\StoreBenchExecResult{PdrInv}{KinductionDfStaticSixteenTwoF}{Error}{Assertion}{Walltime}{Min}{1.78250694275}%
\StoreBenchExecResult{PdrInv}{KinductionDfStaticSixteenTwoF}{Error}{Assertion}{Walltime}{Max}{2.10398197174}%
\StoreBenchExecResult{PdrInv}{KinductionDfStaticSixteenTwoF}{Error}{Assertion}{Walltime}{Stdev}{0.1311591894302750984327986721}%
\StoreBenchExecResult{PdrInv}{KinductionDfStaticSixteenTwoF}{Error}{Error}{Count}{}{139}%
\StoreBenchExecResult{PdrInv}{KinductionDfStaticSixteenTwoF}{Error}{Error}{Cputime}{}{20330.097463133}%
\StoreBenchExecResult{PdrInv}{KinductionDfStaticSixteenTwoF}{Error}{Error}{Cputime}{Avg}{146.2596939793741007194244604}%
\StoreBenchExecResult{PdrInv}{KinductionDfStaticSixteenTwoF}{Error}{Error}{Cputime}{Median}{77.621327437}%
\StoreBenchExecResult{PdrInv}{KinductionDfStaticSixteenTwoF}{Error}{Error}{Cputime}{Min}{2.37078614}%
\StoreBenchExecResult{PdrInv}{KinductionDfStaticSixteenTwoF}{Error}{Error}{Cputime}{Max}{843.661008287}%
\StoreBenchExecResult{PdrInv}{KinductionDfStaticSixteenTwoF}{Error}{Error}{Cputime}{Stdev}{177.2506487670829411325092188}%
\StoreBenchExecResult{PdrInv}{KinductionDfStaticSixteenTwoF}{Error}{Error}{Walltime}{}{17138.31728410661}%
\StoreBenchExecResult{PdrInv}{KinductionDfStaticSixteenTwoF}{Error}{Error}{Walltime}{Avg}{123.2972466482489928057553957}%
\StoreBenchExecResult{PdrInv}{KinductionDfStaticSixteenTwoF}{Error}{Error}{Walltime}{Median}{57.4175028801}%
\StoreBenchExecResult{PdrInv}{KinductionDfStaticSixteenTwoF}{Error}{Error}{Walltime}{Min}{1.32805800438}%
\StoreBenchExecResult{PdrInv}{KinductionDfStaticSixteenTwoF}{Error}{Error}{Walltime}{Max}{829.643739939}%
\StoreBenchExecResult{PdrInv}{KinductionDfStaticSixteenTwoF}{Error}{Error}{Walltime}{Stdev}{161.0082499773154662148540827}%
\StoreBenchExecResult{PdrInv}{KinductionDfStaticSixteenTwoF}{Error}{Exception}{Count}{}{8}%
\StoreBenchExecResult{PdrInv}{KinductionDfStaticSixteenTwoF}{Error}{Exception}{Cputime}{}{762.043020844}%
\StoreBenchExecResult{PdrInv}{KinductionDfStaticSixteenTwoF}{Error}{Exception}{Cputime}{Avg}{95.2553776055}%
\StoreBenchExecResult{PdrInv}{KinductionDfStaticSixteenTwoF}{Error}{Exception}{Cputime}{Median}{48.3061533055}%
\StoreBenchExecResult{PdrInv}{KinductionDfStaticSixteenTwoF}{Error}{Exception}{Cputime}{Min}{17.502246669}%
\StoreBenchExecResult{PdrInv}{KinductionDfStaticSixteenTwoF}{Error}{Exception}{Cputime}{Max}{309.746377123}%
\StoreBenchExecResult{PdrInv}{KinductionDfStaticSixteenTwoF}{Error}{Exception}{Cputime}{Stdev}{103.0418812587667941983269707}%
\StoreBenchExecResult{PdrInv}{KinductionDfStaticSixteenTwoF}{Error}{Exception}{Walltime}{}{383.66582846688}%
\StoreBenchExecResult{PdrInv}{KinductionDfStaticSixteenTwoF}{Error}{Exception}{Walltime}{Avg}{47.95822855836}%
\StoreBenchExecResult{PdrInv}{KinductionDfStaticSixteenTwoF}{Error}{Exception}{Walltime}{Median}{24.3608931303}%
\StoreBenchExecResult{PdrInv}{KinductionDfStaticSixteenTwoF}{Error}{Exception}{Walltime}{Min}{8.98041892052}%
\StoreBenchExecResult{PdrInv}{KinductionDfStaticSixteenTwoF}{Error}{Exception}{Walltime}{Max}{155.59612608}%
\StoreBenchExecResult{PdrInv}{KinductionDfStaticSixteenTwoF}{Error}{Exception}{Walltime}{Stdev}{51.70104399418535949819692090}%
\StoreBenchExecResult{PdrInv}{KinductionDfStaticSixteenTwoF}{Error}{OutOfJavaMemory}{Count}{}{8}%
\StoreBenchExecResult{PdrInv}{KinductionDfStaticSixteenTwoF}{Error}{OutOfJavaMemory}{Cputime}{}{4214.622595442}%
\StoreBenchExecResult{PdrInv}{KinductionDfStaticSixteenTwoF}{Error}{OutOfJavaMemory}{Cputime}{Avg}{526.82782443025}%
\StoreBenchExecResult{PdrInv}{KinductionDfStaticSixteenTwoF}{Error}{OutOfJavaMemory}{Cputime}{Median}{588.188674569}%
\StoreBenchExecResult{PdrInv}{KinductionDfStaticSixteenTwoF}{Error}{OutOfJavaMemory}{Cputime}{Min}{275.892520346}%
\StoreBenchExecResult{PdrInv}{KinductionDfStaticSixteenTwoF}{Error}{OutOfJavaMemory}{Cputime}{Max}{750.061671287}%
\StoreBenchExecResult{PdrInv}{KinductionDfStaticSixteenTwoF}{Error}{OutOfJavaMemory}{Cputime}{Stdev}{167.8776138895862693820645306}%
\StoreBenchExecResult{PdrInv}{KinductionDfStaticSixteenTwoF}{Error}{OutOfJavaMemory}{Walltime}{}{2192.380754950}%
\StoreBenchExecResult{PdrInv}{KinductionDfStaticSixteenTwoF}{Error}{OutOfJavaMemory}{Walltime}{Avg}{274.04759436875}%
\StoreBenchExecResult{PdrInv}{KinductionDfStaticSixteenTwoF}{Error}{OutOfJavaMemory}{Walltime}{Median}{298.924592972}%
\StoreBenchExecResult{PdrInv}{KinductionDfStaticSixteenTwoF}{Error}{OutOfJavaMemory}{Walltime}{Min}{141.277446032}%
\StoreBenchExecResult{PdrInv}{KinductionDfStaticSixteenTwoF}{Error}{OutOfJavaMemory}{Walltime}{Max}{404.074120045}%
\StoreBenchExecResult{PdrInv}{KinductionDfStaticSixteenTwoF}{Error}{OutOfJavaMemory}{Walltime}{Stdev}{84.47595073375275985134866735}%
\StoreBenchExecResult{PdrInv}{KinductionDfStaticSixteenTwoF}{Error}{OutOfMemory}{Count}{}{266}%
\StoreBenchExecResult{PdrInv}{KinductionDfStaticSixteenTwoF}{Error}{OutOfMemory}{Cputime}{}{108714.449581564}%
\StoreBenchExecResult{PdrInv}{KinductionDfStaticSixteenTwoF}{Error}{OutOfMemory}{Cputime}{Avg}{408.7009382765563909774436090}%
\StoreBenchExecResult{PdrInv}{KinductionDfStaticSixteenTwoF}{Error}{OutOfMemory}{Cputime}{Median}{381.4870377615}%
\StoreBenchExecResult{PdrInv}{KinductionDfStaticSixteenTwoF}{Error}{OutOfMemory}{Cputime}{Min}{168.071531181}%
\StoreBenchExecResult{PdrInv}{KinductionDfStaticSixteenTwoF}{Error}{OutOfMemory}{Cputime}{Max}{892.89828359}%
\StoreBenchExecResult{PdrInv}{KinductionDfStaticSixteenTwoF}{Error}{OutOfMemory}{Cputime}{Stdev}{187.9886357594161107691623076}%
\StoreBenchExecResult{PdrInv}{KinductionDfStaticSixteenTwoF}{Error}{OutOfMemory}{Walltime}{}{55052.4809734813}%
\StoreBenchExecResult{PdrInv}{KinductionDfStaticSixteenTwoF}{Error}{OutOfMemory}{Walltime}{Avg}{206.9642141860199248120300752}%
\StoreBenchExecResult{PdrInv}{KinductionDfStaticSixteenTwoF}{Error}{OutOfMemory}{Walltime}{Median}{191.539700508}%
\StoreBenchExecResult{PdrInv}{KinductionDfStaticSixteenTwoF}{Error}{OutOfMemory}{Walltime}{Min}{84.6973059177}%
\StoreBenchExecResult{PdrInv}{KinductionDfStaticSixteenTwoF}{Error}{OutOfMemory}{Walltime}{Max}{856.128892183}%
\StoreBenchExecResult{PdrInv}{KinductionDfStaticSixteenTwoF}{Error}{OutOfMemory}{Walltime}{Stdev}{100.8875684045516486377241791}%
\StoreBenchExecResult{PdrInv}{KinductionDfStaticSixteenTwoF}{Error}{Timeout}{Count}{}{2233}%
\StoreBenchExecResult{PdrInv}{KinductionDfStaticSixteenTwoF}{Error}{Timeout}{Cputime}{}{2029137.335664190}%
\StoreBenchExecResult{PdrInv}{KinductionDfStaticSixteenTwoF}{Error}{Timeout}{Cputime}{Avg}{908.7045838173712494402149575}%
\StoreBenchExecResult{PdrInv}{KinductionDfStaticSixteenTwoF}{Error}{Timeout}{Cputime}{Median}{902.029895154}%
\StoreBenchExecResult{PdrInv}{KinductionDfStaticSixteenTwoF}{Error}{Timeout}{Cputime}{Min}{900.779935944}%
\StoreBenchExecResult{PdrInv}{KinductionDfStaticSixteenTwoF}{Error}{Timeout}{Cputime}{Max}{1001.28896153}%
\StoreBenchExecResult{PdrInv}{KinductionDfStaticSixteenTwoF}{Error}{Timeout}{Cputime}{Stdev}{19.69685319664126737512595783}%
\StoreBenchExecResult{PdrInv}{KinductionDfStaticSixteenTwoF}{Error}{Timeout}{Walltime}{}{1195755.130367983}%
\StoreBenchExecResult{PdrInv}{KinductionDfStaticSixteenTwoF}{Error}{Timeout}{Walltime}{Avg}{535.4926692198759516345723242}%
\StoreBenchExecResult{PdrInv}{KinductionDfStaticSixteenTwoF}{Error}{Timeout}{Walltime}{Median}{453.578819036}%
\StoreBenchExecResult{PdrInv}{KinductionDfStaticSixteenTwoF}{Error}{Timeout}{Walltime}{Min}{451.096029997}%
\StoreBenchExecResult{PdrInv}{KinductionDfStaticSixteenTwoF}{Error}{Timeout}{Walltime}{Max}{954.062977791}%
\StoreBenchExecResult{PdrInv}{KinductionDfStaticSixteenTwoF}{Error}{Timeout}{Walltime}{Stdev}{156.9323688319183119567595766}%
\StoreBenchExecResult{PdrInv}{KinductionDfStaticSixteenTwoF}{Wrong}{}{Count}{}{2}%
\StoreBenchExecResult{PdrInv}{KinductionDfStaticSixteenTwoF}{Wrong}{}{Cputime}{}{25.641342019}%
\StoreBenchExecResult{PdrInv}{KinductionDfStaticSixteenTwoF}{Wrong}{}{Cputime}{Avg}{12.8206710095}%
\StoreBenchExecResult{PdrInv}{KinductionDfStaticSixteenTwoF}{Wrong}{}{Cputime}{Median}{12.8206710095}%
\StoreBenchExecResult{PdrInv}{KinductionDfStaticSixteenTwoF}{Wrong}{}{Cputime}{Min}{3.872432588}%
\StoreBenchExecResult{PdrInv}{KinductionDfStaticSixteenTwoF}{Wrong}{}{Cputime}{Max}{21.768909431}%
\StoreBenchExecResult{PdrInv}{KinductionDfStaticSixteenTwoF}{Wrong}{}{Cputime}{Stdev}{8.9482384215}%
\StoreBenchExecResult{PdrInv}{KinductionDfStaticSixteenTwoF}{Wrong}{}{Walltime}{}{13.92734622951}%
\StoreBenchExecResult{PdrInv}{KinductionDfStaticSixteenTwoF}{Wrong}{}{Walltime}{Avg}{6.963673114755}%
\StoreBenchExecResult{PdrInv}{KinductionDfStaticSixteenTwoF}{Wrong}{}{Walltime}{Median}{6.963673114755}%
\StoreBenchExecResult{PdrInv}{KinductionDfStaticSixteenTwoF}{Wrong}{}{Walltime}{Min}{2.12947106361}%
\StoreBenchExecResult{PdrInv}{KinductionDfStaticSixteenTwoF}{Wrong}{}{Walltime}{Max}{11.7978751659}%
\StoreBenchExecResult{PdrInv}{KinductionDfStaticSixteenTwoF}{Wrong}{}{Walltime}{Stdev}{4.834202051145}%
\StoreBenchExecResult{PdrInv}{KinductionDfStaticSixteenTwoF}{Wrong}{False}{Count}{}{2}%
\StoreBenchExecResult{PdrInv}{KinductionDfStaticSixteenTwoF}{Wrong}{False}{Cputime}{}{25.641342019}%
\StoreBenchExecResult{PdrInv}{KinductionDfStaticSixteenTwoF}{Wrong}{False}{Cputime}{Avg}{12.8206710095}%
\StoreBenchExecResult{PdrInv}{KinductionDfStaticSixteenTwoF}{Wrong}{False}{Cputime}{Median}{12.8206710095}%
\StoreBenchExecResult{PdrInv}{KinductionDfStaticSixteenTwoF}{Wrong}{False}{Cputime}{Min}{3.872432588}%
\StoreBenchExecResult{PdrInv}{KinductionDfStaticSixteenTwoF}{Wrong}{False}{Cputime}{Max}{21.768909431}%
\StoreBenchExecResult{PdrInv}{KinductionDfStaticSixteenTwoF}{Wrong}{False}{Cputime}{Stdev}{8.9482384215}%
\StoreBenchExecResult{PdrInv}{KinductionDfStaticSixteenTwoF}{Wrong}{False}{Walltime}{}{13.92734622951}%
\StoreBenchExecResult{PdrInv}{KinductionDfStaticSixteenTwoF}{Wrong}{False}{Walltime}{Avg}{6.963673114755}%
\StoreBenchExecResult{PdrInv}{KinductionDfStaticSixteenTwoF}{Wrong}{False}{Walltime}{Median}{6.963673114755}%
\StoreBenchExecResult{PdrInv}{KinductionDfStaticSixteenTwoF}{Wrong}{False}{Walltime}{Min}{2.12947106361}%
\StoreBenchExecResult{PdrInv}{KinductionDfStaticSixteenTwoF}{Wrong}{False}{Walltime}{Max}{11.7978751659}%
\StoreBenchExecResult{PdrInv}{KinductionDfStaticSixteenTwoF}{Wrong}{False}{Walltime}{Stdev}{4.834202051145}%
\providecommand\StoreBenchExecResult[7]{\expandafter\newcommand\csname#1#2#3#4#5#6\endcsname{#7}}%
\StoreBenchExecResult{PdrInv}{KinductionDfStaticSixteenTwoTTrueNotSolvedByKinductionPlainButKipdr}{Total}{}{Count}{}{449}%
\StoreBenchExecResult{PdrInv}{KinductionDfStaticSixteenTwoTTrueNotSolvedByKinductionPlainButKipdr}{Total}{}{Cputime}{}{14505.577133569}%
\StoreBenchExecResult{PdrInv}{KinductionDfStaticSixteenTwoTTrueNotSolvedByKinductionPlainButKipdr}{Total}{}{Cputime}{Avg}{32.30640786986414253897550111}%
\StoreBenchExecResult{PdrInv}{KinductionDfStaticSixteenTwoTTrueNotSolvedByKinductionPlainButKipdr}{Total}{}{Cputime}{Median}{6.263811156}%
\StoreBenchExecResult{PdrInv}{KinductionDfStaticSixteenTwoTTrueNotSolvedByKinductionPlainButKipdr}{Total}{}{Cputime}{Min}{3.280487786}%
\StoreBenchExecResult{PdrInv}{KinductionDfStaticSixteenTwoTTrueNotSolvedByKinductionPlainButKipdr}{Total}{}{Cputime}{Max}{912.637190922}%
\StoreBenchExecResult{PdrInv}{KinductionDfStaticSixteenTwoTTrueNotSolvedByKinductionPlainButKipdr}{Total}{}{Cputime}{Stdev}{145.1010106259151519380961585}%
\StoreBenchExecResult{PdrInv}{KinductionDfStaticSixteenTwoTTrueNotSolvedByKinductionPlainButKipdr}{Total}{}{Walltime}{}{9558.42117643135}%
\StoreBenchExecResult{PdrInv}{KinductionDfStaticSixteenTwoTTrueNotSolvedByKinductionPlainButKipdr}{Total}{}{Walltime}{Avg}{21.28824315463552338530066815}%
\StoreBenchExecResult{PdrInv}{KinductionDfStaticSixteenTwoTTrueNotSolvedByKinductionPlainButKipdr}{Total}{}{Walltime}{Median}{3.3344271183}%
\StoreBenchExecResult{PdrInv}{KinductionDfStaticSixteenTwoTTrueNotSolvedByKinductionPlainButKipdr}{Total}{}{Walltime}{Min}{1.82558703423}%
\StoreBenchExecResult{PdrInv}{KinductionDfStaticSixteenTwoTTrueNotSolvedByKinductionPlainButKipdr}{Total}{}{Walltime}{Max}{898.390832901}%
\StoreBenchExecResult{PdrInv}{KinductionDfStaticSixteenTwoTTrueNotSolvedByKinductionPlainButKipdr}{Total}{}{Walltime}{Stdev}{108.2886436711852928149089397}%
\StoreBenchExecResult{PdrInv}{KinductionDfStaticSixteenTwoTTrueNotSolvedByKinductionPlainButKipdr}{Correct}{}{Count}{}{437}%
\StoreBenchExecResult{PdrInv}{KinductionDfStaticSixteenTwoTTrueNotSolvedByKinductionPlainButKipdr}{Correct}{}{Cputime}{}{3637.555199286}%
\StoreBenchExecResult{PdrInv}{KinductionDfStaticSixteenTwoTTrueNotSolvedByKinductionPlainButKipdr}{Correct}{}{Cputime}{Avg}{8.323924941157894736842105263}%
\StoreBenchExecResult{PdrInv}{KinductionDfStaticSixteenTwoTTrueNotSolvedByKinductionPlainButKipdr}{Correct}{}{Cputime}{Median}{6.187354126}%
\StoreBenchExecResult{PdrInv}{KinductionDfStaticSixteenTwoTTrueNotSolvedByKinductionPlainButKipdr}{Correct}{}{Cputime}{Min}{3.280487786}%
\StoreBenchExecResult{PdrInv}{KinductionDfStaticSixteenTwoTTrueNotSolvedByKinductionPlainButKipdr}{Correct}{}{Cputime}{Max}{151.45430341}%
\StoreBenchExecResult{PdrInv}{KinductionDfStaticSixteenTwoTTrueNotSolvedByKinductionPlainButKipdr}{Correct}{}{Cputime}{Stdev}{10.55469151802187151223955885}%
\StoreBenchExecResult{PdrInv}{KinductionDfStaticSixteenTwoTTrueNotSolvedByKinductionPlainButKipdr}{Correct}{}{Walltime}{}{1909.11905813235}%
\StoreBenchExecResult{PdrInv}{KinductionDfStaticSixteenTwoTTrueNotSolvedByKinductionPlainButKipdr}{Correct}{}{Walltime}{Avg}{4.368693496870366132723112128}%
\StoreBenchExecResult{PdrInv}{KinductionDfStaticSixteenTwoTTrueNotSolvedByKinductionPlainButKipdr}{Correct}{}{Walltime}{Median}{3.26929783821}%
\StoreBenchExecResult{PdrInv}{KinductionDfStaticSixteenTwoTTrueNotSolvedByKinductionPlainButKipdr}{Correct}{}{Walltime}{Min}{1.82558703423}%
\StoreBenchExecResult{PdrInv}{KinductionDfStaticSixteenTwoTTrueNotSolvedByKinductionPlainButKipdr}{Correct}{}{Walltime}{Max}{78.4419739246}%
\StoreBenchExecResult{PdrInv}{KinductionDfStaticSixteenTwoTTrueNotSolvedByKinductionPlainButKipdr}{Correct}{}{Walltime}{Stdev}{5.381148760016992247206527633}%
\StoreBenchExecResult{PdrInv}{KinductionDfStaticSixteenTwoTTrueNotSolvedByKinductionPlainButKipdr}{Correct}{True}{Count}{}{437}%
\StoreBenchExecResult{PdrInv}{KinductionDfStaticSixteenTwoTTrueNotSolvedByKinductionPlainButKipdr}{Correct}{True}{Cputime}{}{3637.555199286}%
\StoreBenchExecResult{PdrInv}{KinductionDfStaticSixteenTwoTTrueNotSolvedByKinductionPlainButKipdr}{Correct}{True}{Cputime}{Avg}{8.323924941157894736842105263}%
\StoreBenchExecResult{PdrInv}{KinductionDfStaticSixteenTwoTTrueNotSolvedByKinductionPlainButKipdr}{Correct}{True}{Cputime}{Median}{6.187354126}%
\StoreBenchExecResult{PdrInv}{KinductionDfStaticSixteenTwoTTrueNotSolvedByKinductionPlainButKipdr}{Correct}{True}{Cputime}{Min}{3.280487786}%
\StoreBenchExecResult{PdrInv}{KinductionDfStaticSixteenTwoTTrueNotSolvedByKinductionPlainButKipdr}{Correct}{True}{Cputime}{Max}{151.45430341}%
\StoreBenchExecResult{PdrInv}{KinductionDfStaticSixteenTwoTTrueNotSolvedByKinductionPlainButKipdr}{Correct}{True}{Cputime}{Stdev}{10.55469151802187151223955885}%
\StoreBenchExecResult{PdrInv}{KinductionDfStaticSixteenTwoTTrueNotSolvedByKinductionPlainButKipdr}{Correct}{True}{Walltime}{}{1909.11905813235}%
\StoreBenchExecResult{PdrInv}{KinductionDfStaticSixteenTwoTTrueNotSolvedByKinductionPlainButKipdr}{Correct}{True}{Walltime}{Avg}{4.368693496870366132723112128}%
\StoreBenchExecResult{PdrInv}{KinductionDfStaticSixteenTwoTTrueNotSolvedByKinductionPlainButKipdr}{Correct}{True}{Walltime}{Median}{3.26929783821}%
\StoreBenchExecResult{PdrInv}{KinductionDfStaticSixteenTwoTTrueNotSolvedByKinductionPlainButKipdr}{Correct}{True}{Walltime}{Min}{1.82558703423}%
\StoreBenchExecResult{PdrInv}{KinductionDfStaticSixteenTwoTTrueNotSolvedByKinductionPlainButKipdr}{Correct}{True}{Walltime}{Max}{78.4419739246}%
\StoreBenchExecResult{PdrInv}{KinductionDfStaticSixteenTwoTTrueNotSolvedByKinductionPlainButKipdr}{Correct}{True}{Walltime}{Stdev}{5.381148760016992247206527633}%
\StoreBenchExecResult{PdrInv}{KinductionDfStaticSixteenTwoTTrueNotSolvedByKinductionPlainButKipdr}{Wrong}{True}{Count}{}{0}%
\StoreBenchExecResult{PdrInv}{KinductionDfStaticSixteenTwoTTrueNotSolvedByKinductionPlainButKipdr}{Wrong}{True}{Cputime}{}{0}%
\StoreBenchExecResult{PdrInv}{KinductionDfStaticSixteenTwoTTrueNotSolvedByKinductionPlainButKipdr}{Wrong}{True}{Cputime}{Avg}{None}%
\StoreBenchExecResult{PdrInv}{KinductionDfStaticSixteenTwoTTrueNotSolvedByKinductionPlainButKipdr}{Wrong}{True}{Cputime}{Median}{None}%
\StoreBenchExecResult{PdrInv}{KinductionDfStaticSixteenTwoTTrueNotSolvedByKinductionPlainButKipdr}{Wrong}{True}{Cputime}{Min}{None}%
\StoreBenchExecResult{PdrInv}{KinductionDfStaticSixteenTwoTTrueNotSolvedByKinductionPlainButKipdr}{Wrong}{True}{Cputime}{Max}{None}%
\StoreBenchExecResult{PdrInv}{KinductionDfStaticSixteenTwoTTrueNotSolvedByKinductionPlainButKipdr}{Wrong}{True}{Cputime}{Stdev}{None}%
\StoreBenchExecResult{PdrInv}{KinductionDfStaticSixteenTwoTTrueNotSolvedByKinductionPlainButKipdr}{Wrong}{True}{Walltime}{}{0}%
\StoreBenchExecResult{PdrInv}{KinductionDfStaticSixteenTwoTTrueNotSolvedByKinductionPlainButKipdr}{Wrong}{True}{Walltime}{Avg}{None}%
\StoreBenchExecResult{PdrInv}{KinductionDfStaticSixteenTwoTTrueNotSolvedByKinductionPlainButKipdr}{Wrong}{True}{Walltime}{Median}{None}%
\StoreBenchExecResult{PdrInv}{KinductionDfStaticSixteenTwoTTrueNotSolvedByKinductionPlainButKipdr}{Wrong}{True}{Walltime}{Min}{None}%
\StoreBenchExecResult{PdrInv}{KinductionDfStaticSixteenTwoTTrueNotSolvedByKinductionPlainButKipdr}{Wrong}{True}{Walltime}{Max}{None}%
\StoreBenchExecResult{PdrInv}{KinductionDfStaticSixteenTwoTTrueNotSolvedByKinductionPlainButKipdr}{Wrong}{True}{Walltime}{Stdev}{None}%
\StoreBenchExecResult{PdrInv}{KinductionDfStaticSixteenTwoTTrueNotSolvedByKinductionPlainButKipdr}{Error}{}{Count}{}{12}%
\StoreBenchExecResult{PdrInv}{KinductionDfStaticSixteenTwoTTrueNotSolvedByKinductionPlainButKipdr}{Error}{}{Cputime}{}{10868.021934283}%
\StoreBenchExecResult{PdrInv}{KinductionDfStaticSixteenTwoTTrueNotSolvedByKinductionPlainButKipdr}{Error}{}{Cputime}{Avg}{905.6684945235833333333333333}%
\StoreBenchExecResult{PdrInv}{KinductionDfStaticSixteenTwoTTrueNotSolvedByKinductionPlainButKipdr}{Error}{}{Cputime}{Median}{903.886263741}%
\StoreBenchExecResult{PdrInv}{KinductionDfStaticSixteenTwoTTrueNotSolvedByKinductionPlainButKipdr}{Error}{}{Cputime}{Min}{901.156437411}%
\StoreBenchExecResult{PdrInv}{KinductionDfStaticSixteenTwoTTrueNotSolvedByKinductionPlainButKipdr}{Error}{}{Cputime}{Max}{912.637190922}%
\StoreBenchExecResult{PdrInv}{KinductionDfStaticSixteenTwoTTrueNotSolvedByKinductionPlainButKipdr}{Error}{}{Cputime}{Stdev}{4.272048296595088137734338986}%
\StoreBenchExecResult{PdrInv}{KinductionDfStaticSixteenTwoTTrueNotSolvedByKinductionPlainButKipdr}{Error}{}{Walltime}{}{7649.302118299}%
\StoreBenchExecResult{PdrInv}{KinductionDfStaticSixteenTwoTTrueNotSolvedByKinductionPlainButKipdr}{Error}{}{Walltime}{Avg}{637.4418431915833333333333333}%
\StoreBenchExecResult{PdrInv}{KinductionDfStaticSixteenTwoTTrueNotSolvedByKinductionPlainButKipdr}{Error}{}{Walltime}{Median}{454.2377245425}%
\StoreBenchExecResult{PdrInv}{KinductionDfStaticSixteenTwoTTrueNotSolvedByKinductionPlainButKipdr}{Error}{}{Walltime}{Min}{451.771400928}%
\StoreBenchExecResult{PdrInv}{KinductionDfStaticSixteenTwoTTrueNotSolvedByKinductionPlainButKipdr}{Error}{}{Walltime}{Max}{898.390832901}%
\StoreBenchExecResult{PdrInv}{KinductionDfStaticSixteenTwoTTrueNotSolvedByKinductionPlainButKipdr}{Error}{}{Walltime}{Stdev}{218.2638918983577374420538454}%
\StoreBenchExecResult{PdrInv}{KinductionDfStaticSixteenTwoTTrueNotSolvedByKinductionPlainButKipdr}{Error}{Timeout}{Count}{}{12}%
\StoreBenchExecResult{PdrInv}{KinductionDfStaticSixteenTwoTTrueNotSolvedByKinductionPlainButKipdr}{Error}{Timeout}{Cputime}{}{10868.021934283}%
\StoreBenchExecResult{PdrInv}{KinductionDfStaticSixteenTwoTTrueNotSolvedByKinductionPlainButKipdr}{Error}{Timeout}{Cputime}{Avg}{905.6684945235833333333333333}%
\StoreBenchExecResult{PdrInv}{KinductionDfStaticSixteenTwoTTrueNotSolvedByKinductionPlainButKipdr}{Error}{Timeout}{Cputime}{Median}{903.886263741}%
\StoreBenchExecResult{PdrInv}{KinductionDfStaticSixteenTwoTTrueNotSolvedByKinductionPlainButKipdr}{Error}{Timeout}{Cputime}{Min}{901.156437411}%
\StoreBenchExecResult{PdrInv}{KinductionDfStaticSixteenTwoTTrueNotSolvedByKinductionPlainButKipdr}{Error}{Timeout}{Cputime}{Max}{912.637190922}%
\StoreBenchExecResult{PdrInv}{KinductionDfStaticSixteenTwoTTrueNotSolvedByKinductionPlainButKipdr}{Error}{Timeout}{Cputime}{Stdev}{4.272048296595088137734338986}%
\StoreBenchExecResult{PdrInv}{KinductionDfStaticSixteenTwoTTrueNotSolvedByKinductionPlainButKipdr}{Error}{Timeout}{Walltime}{}{7649.302118299}%
\StoreBenchExecResult{PdrInv}{KinductionDfStaticSixteenTwoTTrueNotSolvedByKinductionPlainButKipdr}{Error}{Timeout}{Walltime}{Avg}{637.4418431915833333333333333}%
\StoreBenchExecResult{PdrInv}{KinductionDfStaticSixteenTwoTTrueNotSolvedByKinductionPlainButKipdr}{Error}{Timeout}{Walltime}{Median}{454.2377245425}%
\StoreBenchExecResult{PdrInv}{KinductionDfStaticSixteenTwoTTrueNotSolvedByKinductionPlainButKipdr}{Error}{Timeout}{Walltime}{Min}{451.771400928}%
\StoreBenchExecResult{PdrInv}{KinductionDfStaticSixteenTwoTTrueNotSolvedByKinductionPlainButKipdr}{Error}{Timeout}{Walltime}{Max}{898.390832901}%
\StoreBenchExecResult{PdrInv}{KinductionDfStaticSixteenTwoTTrueNotSolvedByKinductionPlainButKipdr}{Error}{Timeout}{Walltime}{Stdev}{218.2638918983577374420538454}%
\providecommand\StoreBenchExecResult[7]{\expandafter\newcommand\csname#1#2#3#4#5#6\endcsname{#7}}%
\StoreBenchExecResult{PdrInv}{KinductionDfStaticSixteenTwoTTrueNotSolvedByKinductionPlain}{Total}{}{Count}{}{2893}%
\StoreBenchExecResult{PdrInv}{KinductionDfStaticSixteenTwoTTrueNotSolvedByKinductionPlain}{Total}{}{Cputime}{}{1605350.268591482}%
\StoreBenchExecResult{PdrInv}{KinductionDfStaticSixteenTwoTTrueNotSolvedByKinductionPlain}{Total}{}{Cputime}{Avg}{554.9084924270591081921880401}%
\StoreBenchExecResult{PdrInv}{KinductionDfStaticSixteenTwoTTrueNotSolvedByKinductionPlain}{Total}{}{Cputime}{Median}{901.111721868}%
\StoreBenchExecResult{PdrInv}{KinductionDfStaticSixteenTwoTTrueNotSolvedByKinductionPlain}{Total}{}{Cputime}{Min}{2.50398733}%
\StoreBenchExecResult{PdrInv}{KinductionDfStaticSixteenTwoTTrueNotSolvedByKinductionPlain}{Total}{}{Cputime}{Max}{1002.29711634}%
\StoreBenchExecResult{PdrInv}{KinductionDfStaticSixteenTwoTTrueNotSolvedByKinductionPlain}{Total}{}{Cputime}{Stdev}{414.7499855405361130022039517}%
\StoreBenchExecResult{PdrInv}{KinductionDfStaticSixteenTwoTTrueNotSolvedByKinductionPlain}{Total}{}{Walltime}{}{1014794.74451755016}%
\StoreBenchExecResult{PdrInv}{KinductionDfStaticSixteenTwoTTrueNotSolvedByKinductionPlain}{Total}{}{Walltime}{Avg}{350.7759227506222468026270308}%
\StoreBenchExecResult{PdrInv}{KinductionDfStaticSixteenTwoTTrueNotSolvedByKinductionPlain}{Total}{}{Walltime}{Median}{451.673289061}%
\StoreBenchExecResult{PdrInv}{KinductionDfStaticSixteenTwoTTrueNotSolvedByKinductionPlain}{Total}{}{Walltime}{Min}{1.38024616241}%
\StoreBenchExecResult{PdrInv}{KinductionDfStaticSixteenTwoTTrueNotSolvedByKinductionPlain}{Total}{}{Walltime}{Max}{981.297166109}%
\StoreBenchExecResult{PdrInv}{KinductionDfStaticSixteenTwoTTrueNotSolvedByKinductionPlain}{Total}{}{Walltime}{Stdev}{301.1040825320668892747770856}%
\StoreBenchExecResult{PdrInv}{KinductionDfStaticSixteenTwoTTrueNotSolvedByKinductionPlain}{Correct}{}{Count}{}{1020}%
\StoreBenchExecResult{PdrInv}{KinductionDfStaticSixteenTwoTTrueNotSolvedByKinductionPlain}{Correct}{}{Cputime}{}{67529.367130130}%
\StoreBenchExecResult{PdrInv}{KinductionDfStaticSixteenTwoTTrueNotSolvedByKinductionPlain}{Correct}{}{Cputime}{Avg}{66.20526189228431372549019608}%
\StoreBenchExecResult{PdrInv}{KinductionDfStaticSixteenTwoTTrueNotSolvedByKinductionPlain}{Correct}{}{Cputime}{Median}{8.978295397}%
\StoreBenchExecResult{PdrInv}{KinductionDfStaticSixteenTwoTTrueNotSolvedByKinductionPlain}{Correct}{}{Cputime}{Min}{3.280487786}%
\StoreBenchExecResult{PdrInv}{KinductionDfStaticSixteenTwoTTrueNotSolvedByKinductionPlain}{Correct}{}{Cputime}{Max}{887.480163708}%
\StoreBenchExecResult{PdrInv}{KinductionDfStaticSixteenTwoTTrueNotSolvedByKinductionPlain}{Correct}{}{Cputime}{Stdev}{141.6921311696175207766355130}%
\StoreBenchExecResult{PdrInv}{KinductionDfStaticSixteenTwoTTrueNotSolvedByKinductionPlain}{Correct}{}{Walltime}{}{35755.96707725180}%
\StoreBenchExecResult{PdrInv}{KinductionDfStaticSixteenTwoTTrueNotSolvedByKinductionPlain}{Correct}{}{Walltime}{Avg}{35.05486968358019607843137255}%
\StoreBenchExecResult{PdrInv}{KinductionDfStaticSixteenTwoTTrueNotSolvedByKinductionPlain}{Correct}{}{Walltime}{Median}{4.693468928335}%
\StoreBenchExecResult{PdrInv}{KinductionDfStaticSixteenTwoTTrueNotSolvedByKinductionPlain}{Correct}{}{Walltime}{Min}{1.82558703423}%
\StoreBenchExecResult{PdrInv}{KinductionDfStaticSixteenTwoTTrueNotSolvedByKinductionPlain}{Correct}{}{Walltime}{Max}{848.542956829}%
\StoreBenchExecResult{PdrInv}{KinductionDfStaticSixteenTwoTTrueNotSolvedByKinductionPlain}{Correct}{}{Walltime}{Stdev}{80.11332421818754571902907276}%
\StoreBenchExecResult{PdrInv}{KinductionDfStaticSixteenTwoTTrueNotSolvedByKinductionPlain}{Correct}{True}{Count}{}{1020}%
\StoreBenchExecResult{PdrInv}{KinductionDfStaticSixteenTwoTTrueNotSolvedByKinductionPlain}{Correct}{True}{Cputime}{}{67529.367130130}%
\StoreBenchExecResult{PdrInv}{KinductionDfStaticSixteenTwoTTrueNotSolvedByKinductionPlain}{Correct}{True}{Cputime}{Avg}{66.20526189228431372549019608}%
\StoreBenchExecResult{PdrInv}{KinductionDfStaticSixteenTwoTTrueNotSolvedByKinductionPlain}{Correct}{True}{Cputime}{Median}{8.978295397}%
\StoreBenchExecResult{PdrInv}{KinductionDfStaticSixteenTwoTTrueNotSolvedByKinductionPlain}{Correct}{True}{Cputime}{Min}{3.280487786}%
\StoreBenchExecResult{PdrInv}{KinductionDfStaticSixteenTwoTTrueNotSolvedByKinductionPlain}{Correct}{True}{Cputime}{Max}{887.480163708}%
\StoreBenchExecResult{PdrInv}{KinductionDfStaticSixteenTwoTTrueNotSolvedByKinductionPlain}{Correct}{True}{Cputime}{Stdev}{141.6921311696175207766355130}%
\StoreBenchExecResult{PdrInv}{KinductionDfStaticSixteenTwoTTrueNotSolvedByKinductionPlain}{Correct}{True}{Walltime}{}{35755.96707725180}%
\StoreBenchExecResult{PdrInv}{KinductionDfStaticSixteenTwoTTrueNotSolvedByKinductionPlain}{Correct}{True}{Walltime}{Avg}{35.05486968358019607843137255}%
\StoreBenchExecResult{PdrInv}{KinductionDfStaticSixteenTwoTTrueNotSolvedByKinductionPlain}{Correct}{True}{Walltime}{Median}{4.693468928335}%
\StoreBenchExecResult{PdrInv}{KinductionDfStaticSixteenTwoTTrueNotSolvedByKinductionPlain}{Correct}{True}{Walltime}{Min}{1.82558703423}%
\StoreBenchExecResult{PdrInv}{KinductionDfStaticSixteenTwoTTrueNotSolvedByKinductionPlain}{Correct}{True}{Walltime}{Max}{848.542956829}%
\StoreBenchExecResult{PdrInv}{KinductionDfStaticSixteenTwoTTrueNotSolvedByKinductionPlain}{Correct}{True}{Walltime}{Stdev}{80.11332421818754571902907276}%
\StoreBenchExecResult{PdrInv}{KinductionDfStaticSixteenTwoTTrueNotSolvedByKinductionPlain}{Wrong}{True}{Count}{}{0}%
\StoreBenchExecResult{PdrInv}{KinductionDfStaticSixteenTwoTTrueNotSolvedByKinductionPlain}{Wrong}{True}{Cputime}{}{0}%
\StoreBenchExecResult{PdrInv}{KinductionDfStaticSixteenTwoTTrueNotSolvedByKinductionPlain}{Wrong}{True}{Cputime}{Avg}{None}%
\StoreBenchExecResult{PdrInv}{KinductionDfStaticSixteenTwoTTrueNotSolvedByKinductionPlain}{Wrong}{True}{Cputime}{Median}{None}%
\StoreBenchExecResult{PdrInv}{KinductionDfStaticSixteenTwoTTrueNotSolvedByKinductionPlain}{Wrong}{True}{Cputime}{Min}{None}%
\StoreBenchExecResult{PdrInv}{KinductionDfStaticSixteenTwoTTrueNotSolvedByKinductionPlain}{Wrong}{True}{Cputime}{Max}{None}%
\StoreBenchExecResult{PdrInv}{KinductionDfStaticSixteenTwoTTrueNotSolvedByKinductionPlain}{Wrong}{True}{Cputime}{Stdev}{None}%
\StoreBenchExecResult{PdrInv}{KinductionDfStaticSixteenTwoTTrueNotSolvedByKinductionPlain}{Wrong}{True}{Walltime}{}{0}%
\StoreBenchExecResult{PdrInv}{KinductionDfStaticSixteenTwoTTrueNotSolvedByKinductionPlain}{Wrong}{True}{Walltime}{Avg}{None}%
\StoreBenchExecResult{PdrInv}{KinductionDfStaticSixteenTwoTTrueNotSolvedByKinductionPlain}{Wrong}{True}{Walltime}{Median}{None}%
\StoreBenchExecResult{PdrInv}{KinductionDfStaticSixteenTwoTTrueNotSolvedByKinductionPlain}{Wrong}{True}{Walltime}{Min}{None}%
\StoreBenchExecResult{PdrInv}{KinductionDfStaticSixteenTwoTTrueNotSolvedByKinductionPlain}{Wrong}{True}{Walltime}{Max}{None}%
\StoreBenchExecResult{PdrInv}{KinductionDfStaticSixteenTwoTTrueNotSolvedByKinductionPlain}{Wrong}{True}{Walltime}{Stdev}{None}%
\StoreBenchExecResult{PdrInv}{KinductionDfStaticSixteenTwoTTrueNotSolvedByKinductionPlain}{Error}{}{Count}{}{1873}%
\StoreBenchExecResult{PdrInv}{KinductionDfStaticSixteenTwoTTrueNotSolvedByKinductionPlain}{Error}{}{Cputime}{}{1537820.901461352}%
\StoreBenchExecResult{PdrInv}{KinductionDfStaticSixteenTwoTTrueNotSolvedByKinductionPlain}{Error}{}{Cputime}{Avg}{821.0469308389492792311799253}%
\StoreBenchExecResult{PdrInv}{KinductionDfStaticSixteenTwoTTrueNotSolvedByKinductionPlain}{Error}{}{Cputime}{Median}{901.601689525}%
\StoreBenchExecResult{PdrInv}{KinductionDfStaticSixteenTwoTTrueNotSolvedByKinductionPlain}{Error}{}{Cputime}{Min}{2.50398733}%
\StoreBenchExecResult{PdrInv}{KinductionDfStaticSixteenTwoTTrueNotSolvedByKinductionPlain}{Error}{}{Cputime}{Max}{1002.29711634}%
\StoreBenchExecResult{PdrInv}{KinductionDfStaticSixteenTwoTTrueNotSolvedByKinductionPlain}{Error}{}{Cputime}{Stdev}{232.0975490576682328748957652}%
\StoreBenchExecResult{PdrInv}{KinductionDfStaticSixteenTwoTTrueNotSolvedByKinductionPlain}{Error}{}{Walltime}{}{979038.77744029836}%
\StoreBenchExecResult{PdrInv}{KinductionDfStaticSixteenTwoTTrueNotSolvedByKinductionPlain}{Error}{}{Walltime}{Avg}{522.7115736467156219967965830}%
\StoreBenchExecResult{PdrInv}{KinductionDfStaticSixteenTwoTTrueNotSolvedByKinductionPlain}{Error}{}{Walltime}{Median}{453.158585787}%
\StoreBenchExecResult{PdrInv}{KinductionDfStaticSixteenTwoTTrueNotSolvedByKinductionPlain}{Error}{}{Walltime}{Min}{1.38024616241}%
\StoreBenchExecResult{PdrInv}{KinductionDfStaticSixteenTwoTTrueNotSolvedByKinductionPlain}{Error}{}{Walltime}{Max}{981.297166109}%
\StoreBenchExecResult{PdrInv}{KinductionDfStaticSixteenTwoTTrueNotSolvedByKinductionPlain}{Error}{}{Walltime}{Stdev}{229.5573952715370984563856477}%
\StoreBenchExecResult{PdrInv}{KinductionDfStaticSixteenTwoTTrueNotSolvedByKinductionPlain}{Error}{Assertion}{Count}{}{2}%
\StoreBenchExecResult{PdrInv}{KinductionDfStaticSixteenTwoTTrueNotSolvedByKinductionPlain}{Error}{Assertion}{Cputime}{}{6.477308002}%
\StoreBenchExecResult{PdrInv}{KinductionDfStaticSixteenTwoTTrueNotSolvedByKinductionPlain}{Error}{Assertion}{Cputime}{Avg}{3.238654001}%
\StoreBenchExecResult{PdrInv}{KinductionDfStaticSixteenTwoTTrueNotSolvedByKinductionPlain}{Error}{Assertion}{Cputime}{Median}{3.238654001}%
\StoreBenchExecResult{PdrInv}{KinductionDfStaticSixteenTwoTTrueNotSolvedByKinductionPlain}{Error}{Assertion}{Cputime}{Min}{3.214009309}%
\StoreBenchExecResult{PdrInv}{KinductionDfStaticSixteenTwoTTrueNotSolvedByKinductionPlain}{Error}{Assertion}{Cputime}{Max}{3.263298693}%
\StoreBenchExecResult{PdrInv}{KinductionDfStaticSixteenTwoTTrueNotSolvedByKinductionPlain}{Error}{Assertion}{Cputime}{Stdev}{0.024644692}%
\StoreBenchExecResult{PdrInv}{KinductionDfStaticSixteenTwoTTrueNotSolvedByKinductionPlain}{Error}{Assertion}{Walltime}{}{3.59253406525}%
\StoreBenchExecResult{PdrInv}{KinductionDfStaticSixteenTwoTTrueNotSolvedByKinductionPlain}{Error}{Assertion}{Walltime}{Avg}{1.796267032625}%
\StoreBenchExecResult{PdrInv}{KinductionDfStaticSixteenTwoTTrueNotSolvedByKinductionPlain}{Error}{Assertion}{Walltime}{Median}{1.796267032625}%
\StoreBenchExecResult{PdrInv}{KinductionDfStaticSixteenTwoTTrueNotSolvedByKinductionPlain}{Error}{Assertion}{Walltime}{Min}{1.79007005692}%
\StoreBenchExecResult{PdrInv}{KinductionDfStaticSixteenTwoTTrueNotSolvedByKinductionPlain}{Error}{Assertion}{Walltime}{Max}{1.80246400833}%
\StoreBenchExecResult{PdrInv}{KinductionDfStaticSixteenTwoTTrueNotSolvedByKinductionPlain}{Error}{Assertion}{Walltime}{Stdev}{0.006196975705}%
\StoreBenchExecResult{PdrInv}{KinductionDfStaticSixteenTwoTTrueNotSolvedByKinductionPlain}{Error}{Error}{Count}{}{128}%
\StoreBenchExecResult{PdrInv}{KinductionDfStaticSixteenTwoTTrueNotSolvedByKinductionPlain}{Error}{Error}{Cputime}{}{21505.857519088}%
\StoreBenchExecResult{PdrInv}{KinductionDfStaticSixteenTwoTTrueNotSolvedByKinductionPlain}{Error}{Error}{Cputime}{Avg}{168.014511867875}%
\StoreBenchExecResult{PdrInv}{KinductionDfStaticSixteenTwoTTrueNotSolvedByKinductionPlain}{Error}{Error}{Cputime}{Median}{102.9595954275}%
\StoreBenchExecResult{PdrInv}{KinductionDfStaticSixteenTwoTTrueNotSolvedByKinductionPlain}{Error}{Error}{Cputime}{Min}{2.50398733}%
\StoreBenchExecResult{PdrInv}{KinductionDfStaticSixteenTwoTTrueNotSolvedByKinductionPlain}{Error}{Error}{Cputime}{Max}{795.038907079}%
\StoreBenchExecResult{PdrInv}{KinductionDfStaticSixteenTwoTTrueNotSolvedByKinductionPlain}{Error}{Error}{Cputime}{Stdev}{190.7735753924141541130701114}%
\StoreBenchExecResult{PdrInv}{KinductionDfStaticSixteenTwoTTrueNotSolvedByKinductionPlain}{Error}{Error}{Walltime}{}{18043.70963573254}%
\StoreBenchExecResult{PdrInv}{KinductionDfStaticSixteenTwoTTrueNotSolvedByKinductionPlain}{Error}{Error}{Walltime}{Avg}{140.96648152916046875}%
\StoreBenchExecResult{PdrInv}{KinductionDfStaticSixteenTwoTTrueNotSolvedByKinductionPlain}{Error}{Error}{Walltime}{Median}{76.3887295723}%
\StoreBenchExecResult{PdrInv}{KinductionDfStaticSixteenTwoTTrueNotSolvedByKinductionPlain}{Error}{Error}{Walltime}{Min}{1.38024616241}%
\StoreBenchExecResult{PdrInv}{KinductionDfStaticSixteenTwoTTrueNotSolvedByKinductionPlain}{Error}{Error}{Walltime}{Max}{778.845774889}%
\StoreBenchExecResult{PdrInv}{KinductionDfStaticSixteenTwoTTrueNotSolvedByKinductionPlain}{Error}{Error}{Walltime}{Stdev}{169.1962457333500342447920130}%
\StoreBenchExecResult{PdrInv}{KinductionDfStaticSixteenTwoTTrueNotSolvedByKinductionPlain}{Error}{Exception}{Count}{}{6}%
\StoreBenchExecResult{PdrInv}{KinductionDfStaticSixteenTwoTTrueNotSolvedByKinductionPlain}{Error}{Exception}{Cputime}{}{734.281362559}%
\StoreBenchExecResult{PdrInv}{KinductionDfStaticSixteenTwoTTrueNotSolvedByKinductionPlain}{Error}{Exception}{Cputime}{Avg}{122.3802270931666666666666667}%
\StoreBenchExecResult{PdrInv}{KinductionDfStaticSixteenTwoTTrueNotSolvedByKinductionPlain}{Error}{Exception}{Cputime}{Median}{77.178988480}%
\StoreBenchExecResult{PdrInv}{KinductionDfStaticSixteenTwoTTrueNotSolvedByKinductionPlain}{Error}{Exception}{Cputime}{Min}{16.263779209}%
\StoreBenchExecResult{PdrInv}{KinductionDfStaticSixteenTwoTTrueNotSolvedByKinductionPlain}{Error}{Exception}{Cputime}{Max}{281.886836573}%
\StoreBenchExecResult{PdrInv}{KinductionDfStaticSixteenTwoTTrueNotSolvedByKinductionPlain}{Error}{Exception}{Cputime}{Stdev}{102.5505902694453010791846019}%
\StoreBenchExecResult{PdrInv}{KinductionDfStaticSixteenTwoTTrueNotSolvedByKinductionPlain}{Error}{Exception}{Walltime}{}{369.53014183127}%
\StoreBenchExecResult{PdrInv}{KinductionDfStaticSixteenTwoTTrueNotSolvedByKinductionPlain}{Error}{Exception}{Walltime}{Avg}{61.58835697187833333333333333}%
\StoreBenchExecResult{PdrInv}{KinductionDfStaticSixteenTwoTTrueNotSolvedByKinductionPlain}{Error}{Exception}{Walltime}{Median}{38.8762884140}%
\StoreBenchExecResult{PdrInv}{KinductionDfStaticSixteenTwoTTrueNotSolvedByKinductionPlain}{Error}{Exception}{Walltime}{Min}{8.31629300117}%
\StoreBenchExecResult{PdrInv}{KinductionDfStaticSixteenTwoTTrueNotSolvedByKinductionPlain}{Error}{Exception}{Walltime}{Max}{141.4968431}%
\StoreBenchExecResult{PdrInv}{KinductionDfStaticSixteenTwoTTrueNotSolvedByKinductionPlain}{Error}{Exception}{Walltime}{Stdev}{51.37172458070190118089204396}%
\StoreBenchExecResult{PdrInv}{KinductionDfStaticSixteenTwoTTrueNotSolvedByKinductionPlain}{Error}{OutOfJavaMemory}{Count}{}{6}%
\StoreBenchExecResult{PdrInv}{KinductionDfStaticSixteenTwoTTrueNotSolvedByKinductionPlain}{Error}{OutOfJavaMemory}{Cputime}{}{2599.613776811}%
\StoreBenchExecResult{PdrInv}{KinductionDfStaticSixteenTwoTTrueNotSolvedByKinductionPlain}{Error}{OutOfJavaMemory}{Cputime}{Avg}{433.2689628018333333333333333}%
\StoreBenchExecResult{PdrInv}{KinductionDfStaticSixteenTwoTTrueNotSolvedByKinductionPlain}{Error}{OutOfJavaMemory}{Cputime}{Median}{405.735498207}%
\StoreBenchExecResult{PdrInv}{KinductionDfStaticSixteenTwoTTrueNotSolvedByKinductionPlain}{Error}{OutOfJavaMemory}{Cputime}{Min}{202.520177461}%
\StoreBenchExecResult{PdrInv}{KinductionDfStaticSixteenTwoTTrueNotSolvedByKinductionPlain}{Error}{OutOfJavaMemory}{Cputime}{Max}{801.156967938}%
\StoreBenchExecResult{PdrInv}{KinductionDfStaticSixteenTwoTTrueNotSolvedByKinductionPlain}{Error}{OutOfJavaMemory}{Cputime}{Stdev}{193.6123585168342817924957319}%
\StoreBenchExecResult{PdrInv}{KinductionDfStaticSixteenTwoTTrueNotSolvedByKinductionPlain}{Error}{OutOfJavaMemory}{Walltime}{}{1512.897072315}%
\StoreBenchExecResult{PdrInv}{KinductionDfStaticSixteenTwoTTrueNotSolvedByKinductionPlain}{Error}{OutOfJavaMemory}{Walltime}{Avg}{252.1495120525}%
\StoreBenchExecResult{PdrInv}{KinductionDfStaticSixteenTwoTTrueNotSolvedByKinductionPlain}{Error}{OutOfJavaMemory}{Walltime}{Median}{232.9759271145}%
\StoreBenchExecResult{PdrInv}{KinductionDfStaticSixteenTwoTTrueNotSolvedByKinductionPlain}{Error}{OutOfJavaMemory}{Walltime}{Min}{115.817079067}%
\StoreBenchExecResult{PdrInv}{KinductionDfStaticSixteenTwoTTrueNotSolvedByKinductionPlain}{Error}{OutOfJavaMemory}{Walltime}{Max}{482.796853065}%
\StoreBenchExecResult{PdrInv}{KinductionDfStaticSixteenTwoTTrueNotSolvedByKinductionPlain}{Error}{OutOfJavaMemory}{Walltime}{Stdev}{113.2771312633968457188224735}%
\StoreBenchExecResult{PdrInv}{KinductionDfStaticSixteenTwoTTrueNotSolvedByKinductionPlain}{Error}{OutOfMemory}{Count}{}{106}%
\StoreBenchExecResult{PdrInv}{KinductionDfStaticSixteenTwoTTrueNotSolvedByKinductionPlain}{Error}{OutOfMemory}{Cputime}{}{42094.920889265}%
\StoreBenchExecResult{PdrInv}{KinductionDfStaticSixteenTwoTTrueNotSolvedByKinductionPlain}{Error}{OutOfMemory}{Cputime}{Avg}{397.1218951817452830188679245}%
\StoreBenchExecResult{PdrInv}{KinductionDfStaticSixteenTwoTTrueNotSolvedByKinductionPlain}{Error}{OutOfMemory}{Cputime}{Median}{329.231781912}%
\StoreBenchExecResult{PdrInv}{KinductionDfStaticSixteenTwoTTrueNotSolvedByKinductionPlain}{Error}{OutOfMemory}{Cputime}{Min}{168.885290968}%
\StoreBenchExecResult{PdrInv}{KinductionDfStaticSixteenTwoTTrueNotSolvedByKinductionPlain}{Error}{OutOfMemory}{Cputime}{Max}{899.769421909}%
\StoreBenchExecResult{PdrInv}{KinductionDfStaticSixteenTwoTTrueNotSolvedByKinductionPlain}{Error}{OutOfMemory}{Cputime}{Stdev}{226.3034778112724398420279687}%
\StoreBenchExecResult{PdrInv}{KinductionDfStaticSixteenTwoTTrueNotSolvedByKinductionPlain}{Error}{OutOfMemory}{Walltime}{}{22410.9524910423}%
\StoreBenchExecResult{PdrInv}{KinductionDfStaticSixteenTwoTTrueNotSolvedByKinductionPlain}{Error}{OutOfMemory}{Walltime}{Avg}{211.4240801041726415094339623}%
\StoreBenchExecResult{PdrInv}{KinductionDfStaticSixteenTwoTTrueNotSolvedByKinductionPlain}{Error}{OutOfMemory}{Walltime}{Median}{166.960020900}%
\StoreBenchExecResult{PdrInv}{KinductionDfStaticSixteenTwoTTrueNotSolvedByKinductionPlain}{Error}{OutOfMemory}{Walltime}{Min}{85.0336318016}%
\StoreBenchExecResult{PdrInv}{KinductionDfStaticSixteenTwoTTrueNotSolvedByKinductionPlain}{Error}{OutOfMemory}{Walltime}{Max}{885.447710991}%
\StoreBenchExecResult{PdrInv}{KinductionDfStaticSixteenTwoTTrueNotSolvedByKinductionPlain}{Error}{OutOfMemory}{Walltime}{Stdev}{148.3132850700440637726075043}%
\StoreBenchExecResult{PdrInv}{KinductionDfStaticSixteenTwoTTrueNotSolvedByKinductionPlain}{Error}{Timeout}{Count}{}{1625}%
\StoreBenchExecResult{PdrInv}{KinductionDfStaticSixteenTwoTTrueNotSolvedByKinductionPlain}{Error}{Timeout}{Cputime}{}{1470879.750605627}%
\StoreBenchExecResult{PdrInv}{KinductionDfStaticSixteenTwoTTrueNotSolvedByKinductionPlain}{Error}{Timeout}{Cputime}{Avg}{905.1567696034627692307692308}%
\StoreBenchExecResult{PdrInv}{KinductionDfStaticSixteenTwoTTrueNotSolvedByKinductionPlain}{Error}{Timeout}{Cputime}{Median}{902.042024851}%
\StoreBenchExecResult{PdrInv}{KinductionDfStaticSixteenTwoTTrueNotSolvedByKinductionPlain}{Error}{Timeout}{Cputime}{Min}{900.827584074}%
\StoreBenchExecResult{PdrInv}{KinductionDfStaticSixteenTwoTTrueNotSolvedByKinductionPlain}{Error}{Timeout}{Cputime}{Max}{1002.29711634}%
\StoreBenchExecResult{PdrInv}{KinductionDfStaticSixteenTwoTTrueNotSolvedByKinductionPlain}{Error}{Timeout}{Cputime}{Stdev}{11.68999260402978320687439626}%
\StoreBenchExecResult{PdrInv}{KinductionDfStaticSixteenTwoTTrueNotSolvedByKinductionPlain}{Error}{Timeout}{Walltime}{}{936698.095565312}%
\StoreBenchExecResult{PdrInv}{KinductionDfStaticSixteenTwoTTrueNotSolvedByKinductionPlain}{Error}{Timeout}{Walltime}{Avg}{576.4295972709612307692307692}%
\StoreBenchExecResult{PdrInv}{KinductionDfStaticSixteenTwoTTrueNotSolvedByKinductionPlain}{Error}{Timeout}{Walltime}{Median}{454.314702034}%
\StoreBenchExecResult{PdrInv}{KinductionDfStaticSixteenTwoTTrueNotSolvedByKinductionPlain}{Error}{Timeout}{Walltime}{Min}{451.101598978}%
\StoreBenchExecResult{PdrInv}{KinductionDfStaticSixteenTwoTTrueNotSolvedByKinductionPlain}{Error}{Timeout}{Walltime}{Max}{981.297166109}%
\StoreBenchExecResult{PdrInv}{KinductionDfStaticSixteenTwoTTrueNotSolvedByKinductionPlain}{Error}{Timeout}{Walltime}{Stdev}{186.8612235437034773130178048}%
\providecommand\StoreBenchExecResult[7]{\expandafter\newcommand\csname#1#2#3#4#5#6\endcsname{#7}}%
\StoreBenchExecResult{PdrInv}{KinductionDfStaticSixteenTwoT}{Total}{}{Count}{}{5591}%
\StoreBenchExecResult{PdrInv}{KinductionDfStaticSixteenTwoT}{Total}{}{Cputime}{}{2253908.613486173}%
\StoreBenchExecResult{PdrInv}{KinductionDfStaticSixteenTwoT}{Total}{}{Cputime}{Avg}{403.1315710045024145948846360}%
\StoreBenchExecResult{PdrInv}{KinductionDfStaticSixteenTwoT}{Total}{}{Cputime}{Median}{142.747244576}%
\StoreBenchExecResult{PdrInv}{KinductionDfStaticSixteenTwoT}{Total}{}{Cputime}{Min}{2.50398733}%
\StoreBenchExecResult{PdrInv}{KinductionDfStaticSixteenTwoT}{Total}{}{Cputime}{Max}{1002.29711634}%
\StoreBenchExecResult{PdrInv}{KinductionDfStaticSixteenTwoT}{Total}{}{Cputime}{Stdev}{418.2579840475603185181166207}%
\StoreBenchExecResult{PdrInv}{KinductionDfStaticSixteenTwoT}{Total}{}{Walltime}{}{1437336.90527724023}%
\StoreBenchExecResult{PdrInv}{KinductionDfStaticSixteenTwoT}{Total}{}{Walltime}{Avg}{257.0804695541477785727061349}%
\StoreBenchExecResult{PdrInv}{KinductionDfStaticSixteenTwoT}{Total}{}{Walltime}{Median}{76.8240571022}%
\StoreBenchExecResult{PdrInv}{KinductionDfStaticSixteenTwoT}{Total}{}{Walltime}{Min}{1.38024616241}%
\StoreBenchExecResult{PdrInv}{KinductionDfStaticSixteenTwoT}{Total}{}{Walltime}{Max}{981.305259943}%
\StoreBenchExecResult{PdrInv}{KinductionDfStaticSixteenTwoT}{Total}{}{Walltime}{Stdev}{295.6605684933201799684864332}%
\StoreBenchExecResult{PdrInv}{KinductionDfStaticSixteenTwoT}{Correct}{}{Count}{}{3017}%
\StoreBenchExecResult{PdrInv}{KinductionDfStaticSixteenTwoT}{Correct}{}{Cputime}{}{190382.714220707}%
\StoreBenchExecResult{PdrInv}{KinductionDfStaticSixteenTwoT}{Correct}{}{Cputime}{Avg}{63.10331926440404375207159430}%
\StoreBenchExecResult{PdrInv}{KinductionDfStaticSixteenTwoT}{Correct}{}{Cputime}{Median}{10.284590085}%
\StoreBenchExecResult{PdrInv}{KinductionDfStaticSixteenTwoT}{Correct}{}{Cputime}{Min}{2.967298488}%
\StoreBenchExecResult{PdrInv}{KinductionDfStaticSixteenTwoT}{Correct}{}{Cputime}{Max}{891.322198762}%
\StoreBenchExecResult{PdrInv}{KinductionDfStaticSixteenTwoT}{Correct}{}{Cputime}{Stdev}{141.5232957623061883276383892}%
\StoreBenchExecResult{PdrInv}{KinductionDfStaticSixteenTwoT}{Correct}{}{Walltime}{}{109427.71614623371}%
\StoreBenchExecResult{PdrInv}{KinductionDfStaticSixteenTwoT}{Correct}{}{Walltime}{Avg}{36.27037326689881007623467020}%
\StoreBenchExecResult{PdrInv}{KinductionDfStaticSixteenTwoT}{Correct}{}{Walltime}{Median}{5.36666417122}%
\StoreBenchExecResult{PdrInv}{KinductionDfStaticSixteenTwoT}{Correct}{}{Walltime}{Min}{1.66779017448}%
\StoreBenchExecResult{PdrInv}{KinductionDfStaticSixteenTwoT}{Correct}{}{Walltime}{Max}{848.542956829}%
\StoreBenchExecResult{PdrInv}{KinductionDfStaticSixteenTwoT}{Correct}{}{Walltime}{Stdev}{90.35273162913076311035830149}%
\StoreBenchExecResult{PdrInv}{KinductionDfStaticSixteenTwoT}{Correct}{False}{Count}{}{775}%
\StoreBenchExecResult{PdrInv}{KinductionDfStaticSixteenTwoT}{Correct}{False}{Cputime}{}{67202.881022572}%
\StoreBenchExecResult{PdrInv}{KinductionDfStaticSixteenTwoT}{Correct}{False}{Cputime}{Avg}{86.71339486783483870967741935}%
\StoreBenchExecResult{PdrInv}{KinductionDfStaticSixteenTwoT}{Correct}{False}{Cputime}{Median}{21.720433415}%
\StoreBenchExecResult{PdrInv}{KinductionDfStaticSixteenTwoT}{Correct}{False}{Cputime}{Min}{3.069289991}%
\StoreBenchExecResult{PdrInv}{KinductionDfStaticSixteenTwoT}{Correct}{False}{Cputime}{Max}{891.322198762}%
\StoreBenchExecResult{PdrInv}{KinductionDfStaticSixteenTwoT}{Correct}{False}{Cputime}{Stdev}{182.4430787550094812841630929}%
\StoreBenchExecResult{PdrInv}{KinductionDfStaticSixteenTwoT}{Correct}{False}{Walltime}{}{40541.05791497478}%
\StoreBenchExecResult{PdrInv}{KinductionDfStaticSixteenTwoT}{Correct}{False}{Walltime}{Avg}{52.3110424709352}%
\StoreBenchExecResult{PdrInv}{KinductionDfStaticSixteenTwoT}{Correct}{False}{Walltime}{Median}{11.736921072}%
\StoreBenchExecResult{PdrInv}{KinductionDfStaticSixteenTwoT}{Correct}{False}{Walltime}{Min}{1.73284602165}%
\StoreBenchExecResult{PdrInv}{KinductionDfStaticSixteenTwoT}{Correct}{False}{Walltime}{Max}{796.04433918}%
\StoreBenchExecResult{PdrInv}{KinductionDfStaticSixteenTwoT}{Correct}{False}{Walltime}{Stdev}{118.6415739396567246959288609}%
\StoreBenchExecResult{PdrInv}{KinductionDfStaticSixteenTwoT}{Correct}{True}{Count}{}{2242}%
\StoreBenchExecResult{PdrInv}{KinductionDfStaticSixteenTwoT}{Correct}{True}{Cputime}{}{123179.833198135}%
\StoreBenchExecResult{PdrInv}{KinductionDfStaticSixteenTwoT}{Correct}{True}{Cputime}{Avg}{54.94194165840098126672613738}%
\StoreBenchExecResult{PdrInv}{KinductionDfStaticSixteenTwoT}{Correct}{True}{Cputime}{Median}{8.7859332735}%
\StoreBenchExecResult{PdrInv}{KinductionDfStaticSixteenTwoT}{Correct}{True}{Cputime}{Min}{2.967298488}%
\StoreBenchExecResult{PdrInv}{KinductionDfStaticSixteenTwoT}{Correct}{True}{Cputime}{Max}{887.480163708}%
\StoreBenchExecResult{PdrInv}{KinductionDfStaticSixteenTwoT}{Correct}{True}{Cputime}{Stdev}{123.2358606735329040986931160}%
\StoreBenchExecResult{PdrInv}{KinductionDfStaticSixteenTwoT}{Correct}{True}{Walltime}{}{68886.65823125893}%
\StoreBenchExecResult{PdrInv}{KinductionDfStaticSixteenTwoT}{Correct}{True}{Walltime}{Avg}{30.72553890778721231043710972}%
\StoreBenchExecResult{PdrInv}{KinductionDfStaticSixteenTwoT}{Correct}{True}{Walltime}{Median}{4.635053396225}%
\StoreBenchExecResult{PdrInv}{KinductionDfStaticSixteenTwoT}{Correct}{True}{Walltime}{Min}{1.66779017448}%
\StoreBenchExecResult{PdrInv}{KinductionDfStaticSixteenTwoT}{Correct}{True}{Walltime}{Max}{848.542956829}%
\StoreBenchExecResult{PdrInv}{KinductionDfStaticSixteenTwoT}{Correct}{True}{Walltime}{Stdev}{77.46118231878286014256313839}%
\StoreBenchExecResult{PdrInv}{KinductionDfStaticSixteenTwoT}{Wrong}{True}{Count}{}{0}%
\StoreBenchExecResult{PdrInv}{KinductionDfStaticSixteenTwoT}{Wrong}{True}{Cputime}{}{0}%
\StoreBenchExecResult{PdrInv}{KinductionDfStaticSixteenTwoT}{Wrong}{True}{Cputime}{Avg}{None}%
\StoreBenchExecResult{PdrInv}{KinductionDfStaticSixteenTwoT}{Wrong}{True}{Cputime}{Median}{None}%
\StoreBenchExecResult{PdrInv}{KinductionDfStaticSixteenTwoT}{Wrong}{True}{Cputime}{Min}{None}%
\StoreBenchExecResult{PdrInv}{KinductionDfStaticSixteenTwoT}{Wrong}{True}{Cputime}{Max}{None}%
\StoreBenchExecResult{PdrInv}{KinductionDfStaticSixteenTwoT}{Wrong}{True}{Cputime}{Stdev}{None}%
\StoreBenchExecResult{PdrInv}{KinductionDfStaticSixteenTwoT}{Wrong}{True}{Walltime}{}{0}%
\StoreBenchExecResult{PdrInv}{KinductionDfStaticSixteenTwoT}{Wrong}{True}{Walltime}{Avg}{None}%
\StoreBenchExecResult{PdrInv}{KinductionDfStaticSixteenTwoT}{Wrong}{True}{Walltime}{Median}{None}%
\StoreBenchExecResult{PdrInv}{KinductionDfStaticSixteenTwoT}{Wrong}{True}{Walltime}{Min}{None}%
\StoreBenchExecResult{PdrInv}{KinductionDfStaticSixteenTwoT}{Wrong}{True}{Walltime}{Max}{None}%
\StoreBenchExecResult{PdrInv}{KinductionDfStaticSixteenTwoT}{Wrong}{True}{Walltime}{Stdev}{None}%
\StoreBenchExecResult{PdrInv}{KinductionDfStaticSixteenTwoT}{Error}{}{Count}{}{2572}%
\StoreBenchExecResult{PdrInv}{KinductionDfStaticSixteenTwoT}{Error}{}{Cputime}{}{2063502.772253593}%
\StoreBenchExecResult{PdrInv}{KinductionDfStaticSixteenTwoT}{Error}{}{Cputime}{Avg}{802.2950125402772161741835148}%
\StoreBenchExecResult{PdrInv}{KinductionDfStaticSixteenTwoT}{Error}{}{Cputime}{Median}{901.7511845065}%
\StoreBenchExecResult{PdrInv}{KinductionDfStaticSixteenTwoT}{Error}{}{Cputime}{Min}{2.50398733}%
\StoreBenchExecResult{PdrInv}{KinductionDfStaticSixteenTwoT}{Error}{}{Cputime}{Max}{1002.29711634}%
\StoreBenchExecResult{PdrInv}{KinductionDfStaticSixteenTwoT}{Error}{}{Cputime}{Stdev}{248.4247794878058588933795660}%
\StoreBenchExecResult{PdrInv}{KinductionDfStaticSixteenTwoT}{Error}{}{Walltime}{}{1327896.59111689021}%
\StoreBenchExecResult{PdrInv}{KinductionDfStaticSixteenTwoT}{Error}{}{Walltime}{Avg}{516.2894988790397395023328149}%
\StoreBenchExecResult{PdrInv}{KinductionDfStaticSixteenTwoT}{Error}{}{Walltime}{Median}{453.7642980815}%
\StoreBenchExecResult{PdrInv}{KinductionDfStaticSixteenTwoT}{Error}{}{Walltime}{Min}{1.38024616241}%
\StoreBenchExecResult{PdrInv}{KinductionDfStaticSixteenTwoT}{Error}{}{Walltime}{Max}{981.305259943}%
\StoreBenchExecResult{PdrInv}{KinductionDfStaticSixteenTwoT}{Error}{}{Walltime}{Stdev}{236.6757856405771359370824186}%
\StoreBenchExecResult{PdrInv}{KinductionDfStaticSixteenTwoT}{Error}{Assertion}{Count}{}{4}%
\StoreBenchExecResult{PdrInv}{KinductionDfStaticSixteenTwoT}{Error}{Assertion}{Cputime}{}{13.559490652}%
\StoreBenchExecResult{PdrInv}{KinductionDfStaticSixteenTwoT}{Error}{Assertion}{Cputime}{Avg}{3.389872663}%
\StoreBenchExecResult{PdrInv}{KinductionDfStaticSixteenTwoT}{Error}{Assertion}{Cputime}{Median}{3.238654001}%
\StoreBenchExecResult{PdrInv}{KinductionDfStaticSixteenTwoT}{Error}{Assertion}{Cputime}{Min}{3.114881842}%
\StoreBenchExecResult{PdrInv}{KinductionDfStaticSixteenTwoT}{Error}{Assertion}{Cputime}{Max}{3.967300808}%
\StoreBenchExecResult{PdrInv}{KinductionDfStaticSixteenTwoT}{Error}{Assertion}{Cputime}{Stdev}{0.3376359368588025768715243243}%
\StoreBenchExecResult{PdrInv}{KinductionDfStaticSixteenTwoT}{Error}{Assertion}{Walltime}{}{7.59280920029}%
\StoreBenchExecResult{PdrInv}{KinductionDfStaticSixteenTwoT}{Error}{Assertion}{Walltime}{Avg}{1.8982023000725}%
\StoreBenchExecResult{PdrInv}{KinductionDfStaticSixteenTwoT}{Error}{Assertion}{Walltime}{Median}{1.796267032625}%
\StoreBenchExecResult{PdrInv}{KinductionDfStaticSixteenTwoT}{Error}{Assertion}{Walltime}{Min}{1.7677590847}%
\StoreBenchExecResult{PdrInv}{KinductionDfStaticSixteenTwoT}{Error}{Assertion}{Walltime}{Max}{2.23251605034}%
\StoreBenchExecResult{PdrInv}{KinductionDfStaticSixteenTwoT}{Error}{Assertion}{Walltime}{Stdev}{0.1934163375788096419434004388}%
\StoreBenchExecResult{PdrInv}{KinductionDfStaticSixteenTwoT}{Error}{Error}{Count}{}{190}%
\StoreBenchExecResult{PdrInv}{KinductionDfStaticSixteenTwoT}{Error}{Error}{Cputime}{}{34129.909660955}%
\StoreBenchExecResult{PdrInv}{KinductionDfStaticSixteenTwoT}{Error}{Error}{Cputime}{Avg}{179.6311034787105263157894737}%
\StoreBenchExecResult{PdrInv}{KinductionDfStaticSixteenTwoT}{Error}{Error}{Cputime}{Median}{110.7314389595}%
\StoreBenchExecResult{PdrInv}{KinductionDfStaticSixteenTwoT}{Error}{Error}{Cputime}{Min}{2.50398733}%
\StoreBenchExecResult{PdrInv}{KinductionDfStaticSixteenTwoT}{Error}{Error}{Cputime}{Max}{795.038907079}%
\StoreBenchExecResult{PdrInv}{KinductionDfStaticSixteenTwoT}{Error}{Error}{Cputime}{Stdev}{189.3535013338417307215399571}%
\StoreBenchExecResult{PdrInv}{KinductionDfStaticSixteenTwoT}{Error}{Error}{Walltime}{}{28532.32504868355}%
\StoreBenchExecResult{PdrInv}{KinductionDfStaticSixteenTwoT}{Error}{Error}{Walltime}{Avg}{150.1701318351765789473684211}%
\StoreBenchExecResult{PdrInv}{KinductionDfStaticSixteenTwoT}{Error}{Error}{Walltime}{Median}{90.7152965069}%
\StoreBenchExecResult{PdrInv}{KinductionDfStaticSixteenTwoT}{Error}{Error}{Walltime}{Min}{1.38024616241}%
\StoreBenchExecResult{PdrInv}{KinductionDfStaticSixteenTwoT}{Error}{Error}{Walltime}{Max}{778.845774889}%
\StoreBenchExecResult{PdrInv}{KinductionDfStaticSixteenTwoT}{Error}{Error}{Walltime}{Stdev}{167.3225104874063018033642185}%
\StoreBenchExecResult{PdrInv}{KinductionDfStaticSixteenTwoT}{Error}{Exception}{Count}{}{13}%
\StoreBenchExecResult{PdrInv}{KinductionDfStaticSixteenTwoT}{Error}{Exception}{Cputime}{}{1635.030253637}%
\StoreBenchExecResult{PdrInv}{KinductionDfStaticSixteenTwoT}{Error}{Exception}{Cputime}{Avg}{125.7715579720769230769230769}%
\StoreBenchExecResult{PdrInv}{KinductionDfStaticSixteenTwoT}{Error}{Exception}{Cputime}{Median}{77.712693382}%
\StoreBenchExecResult{PdrInv}{KinductionDfStaticSixteenTwoT}{Error}{Exception}{Cputime}{Min}{16.263779209}%
\StoreBenchExecResult{PdrInv}{KinductionDfStaticSixteenTwoT}{Error}{Exception}{Cputime}{Max}{469.334782108}%
\StoreBenchExecResult{PdrInv}{KinductionDfStaticSixteenTwoT}{Error}{Exception}{Cputime}{Stdev}{130.0270771180987584948561862}%
\StoreBenchExecResult{PdrInv}{KinductionDfStaticSixteenTwoT}{Error}{Exception}{Walltime}{}{885.01162386087}%
\StoreBenchExecResult{PdrInv}{KinductionDfStaticSixteenTwoT}{Error}{Exception}{Walltime}{Avg}{68.07781722006692307692307692}%
\StoreBenchExecResult{PdrInv}{KinductionDfStaticSixteenTwoT}{Error}{Exception}{Walltime}{Median}{39.1346468925}%
\StoreBenchExecResult{PdrInv}{KinductionDfStaticSixteenTwoT}{Error}{Exception}{Walltime}{Min}{8.31629300117}%
\StoreBenchExecResult{PdrInv}{KinductionDfStaticSixteenTwoT}{Error}{Exception}{Walltime}{Max}{268.797929049}%
\StoreBenchExecResult{PdrInv}{KinductionDfStaticSixteenTwoT}{Error}{Exception}{Walltime}{Stdev}{72.96457681569164644161944142}%
\StoreBenchExecResult{PdrInv}{KinductionDfStaticSixteenTwoT}{Error}{OutOfJavaMemory}{Count}{}{9}%
\StoreBenchExecResult{PdrInv}{KinductionDfStaticSixteenTwoT}{Error}{OutOfJavaMemory}{Cputime}{}{4780.616627561}%
\StoreBenchExecResult{PdrInv}{KinductionDfStaticSixteenTwoT}{Error}{OutOfJavaMemory}{Cputime}{Avg}{531.1796252845555555555555556}%
\StoreBenchExecResult{PdrInv}{KinductionDfStaticSixteenTwoT}{Error}{OutOfJavaMemory}{Cputime}{Median}{504.204614107}%
\StoreBenchExecResult{PdrInv}{KinductionDfStaticSixteenTwoT}{Error}{OutOfJavaMemory}{Cputime}{Min}{202.520177461}%
\StoreBenchExecResult{PdrInv}{KinductionDfStaticSixteenTwoT}{Error}{OutOfJavaMemory}{Cputime}{Max}{801.156967938}%
\StoreBenchExecResult{PdrInv}{KinductionDfStaticSixteenTwoT}{Error}{OutOfJavaMemory}{Cputime}{Stdev}{211.5278318672056670610705599}%
\StoreBenchExecResult{PdrInv}{KinductionDfStaticSixteenTwoT}{Error}{OutOfJavaMemory}{Walltime}{}{2875.545161486}%
\StoreBenchExecResult{PdrInv}{KinductionDfStaticSixteenTwoT}{Error}{OutOfJavaMemory}{Walltime}{Avg}{319.5050179428888888888888889}%
\StoreBenchExecResult{PdrInv}{KinductionDfStaticSixteenTwoT}{Error}{OutOfJavaMemory}{Walltime}{Median}{261.618649006}%
\StoreBenchExecResult{PdrInv}{KinductionDfStaticSixteenTwoT}{Error}{OutOfJavaMemory}{Walltime}{Min}{115.817079067}%
\StoreBenchExecResult{PdrInv}{KinductionDfStaticSixteenTwoT}{Error}{OutOfJavaMemory}{Walltime}{Max}{654.607665062}%
\StoreBenchExecResult{PdrInv}{KinductionDfStaticSixteenTwoT}{Error}{OutOfJavaMemory}{Walltime}{Stdev}{155.9538611227563001740149664}%
\StoreBenchExecResult{PdrInv}{KinductionDfStaticSixteenTwoT}{Error}{OutOfMemory}{Count}{}{253}%
\StoreBenchExecResult{PdrInv}{KinductionDfStaticSixteenTwoT}{Error}{OutOfMemory}{Cputime}{}{114648.958621886}%
\StoreBenchExecResult{PdrInv}{KinductionDfStaticSixteenTwoT}{Error}{OutOfMemory}{Cputime}{Avg}{453.1579392169407114624505929}%
\StoreBenchExecResult{PdrInv}{KinductionDfStaticSixteenTwoT}{Error}{OutOfMemory}{Cputime}{Median}{364.277790003}%
\StoreBenchExecResult{PdrInv}{KinductionDfStaticSixteenTwoT}{Error}{OutOfMemory}{Cputime}{Min}{158.564928869}%
\StoreBenchExecResult{PdrInv}{KinductionDfStaticSixteenTwoT}{Error}{OutOfMemory}{Cputime}{Max}{899.769421909}%
\StoreBenchExecResult{PdrInv}{KinductionDfStaticSixteenTwoT}{Error}{OutOfMemory}{Cputime}{Stdev}{233.0455800453661493735313283}%
\StoreBenchExecResult{PdrInv}{KinductionDfStaticSixteenTwoT}{Error}{OutOfMemory}{Walltime}{}{61887.3745403235}%
\StoreBenchExecResult{PdrInv}{KinductionDfStaticSixteenTwoT}{Error}{OutOfMemory}{Walltime}{Avg}{244.6141286178794466403162055}%
\StoreBenchExecResult{PdrInv}{KinductionDfStaticSixteenTwoT}{Error}{OutOfMemory}{Walltime}{Median}{183.494902134}%
\StoreBenchExecResult{PdrInv}{KinductionDfStaticSixteenTwoT}{Error}{OutOfMemory}{Walltime}{Min}{85.0336318016}%
\StoreBenchExecResult{PdrInv}{KinductionDfStaticSixteenTwoT}{Error}{OutOfMemory}{Walltime}{Max}{885.447710991}%
\StoreBenchExecResult{PdrInv}{KinductionDfStaticSixteenTwoT}{Error}{OutOfMemory}{Walltime}{Stdev}{153.0761303367602108852956390}%
\StoreBenchExecResult{PdrInv}{KinductionDfStaticSixteenTwoT}{Error}{Timeout}{Count}{}{2103}%
\StoreBenchExecResult{PdrInv}{KinductionDfStaticSixteenTwoT}{Error}{Timeout}{Cputime}{}{1908294.697598902}%
\StoreBenchExecResult{PdrInv}{KinductionDfStaticSixteenTwoT}{Error}{Timeout}{Cputime}{Avg}{907.4154529714227294341417023}%
\StoreBenchExecResult{PdrInv}{KinductionDfStaticSixteenTwoT}{Error}{Timeout}{Cputime}{Median}{902.302296693}%
\StoreBenchExecResult{PdrInv}{KinductionDfStaticSixteenTwoT}{Error}{Timeout}{Cputime}{Min}{900.263400339}%
\StoreBenchExecResult{PdrInv}{KinductionDfStaticSixteenTwoT}{Error}{Timeout}{Cputime}{Max}{1002.29711634}%
\StoreBenchExecResult{PdrInv}{KinductionDfStaticSixteenTwoT}{Error}{Timeout}{Cputime}{Stdev}{17.54310214879217209308349934}%
\StoreBenchExecResult{PdrInv}{KinductionDfStaticSixteenTwoT}{Error}{Timeout}{Walltime}{}{1233708.741933336}%
\StoreBenchExecResult{PdrInv}{KinductionDfStaticSixteenTwoT}{Error}{Timeout}{Walltime}{Avg}{586.6422928831840228245363766}%
\StoreBenchExecResult{PdrInv}{KinductionDfStaticSixteenTwoT}{Error}{Timeout}{Walltime}{Median}{456.055104971}%
\StoreBenchExecResult{PdrInv}{KinductionDfStaticSixteenTwoT}{Error}{Timeout}{Walltime}{Min}{450.864520073}%
\StoreBenchExecResult{PdrInv}{KinductionDfStaticSixteenTwoT}{Error}{Timeout}{Walltime}{Max}{981.305259943}%
\StoreBenchExecResult{PdrInv}{KinductionDfStaticSixteenTwoT}{Error}{Timeout}{Walltime}{Stdev}{187.5422319065541820347283320}%
\StoreBenchExecResult{PdrInv}{KinductionDfStaticSixteenTwoT}{Wrong}{}{Count}{}{2}%
\StoreBenchExecResult{PdrInv}{KinductionDfStaticSixteenTwoT}{Wrong}{}{Cputime}{}{23.127011873}%
\StoreBenchExecResult{PdrInv}{KinductionDfStaticSixteenTwoT}{Wrong}{}{Cputime}{Avg}{11.5635059365}%
\StoreBenchExecResult{PdrInv}{KinductionDfStaticSixteenTwoT}{Wrong}{}{Cputime}{Median}{11.5635059365}%
\StoreBenchExecResult{PdrInv}{KinductionDfStaticSixteenTwoT}{Wrong}{}{Cputime}{Min}{4.020183712}%
\StoreBenchExecResult{PdrInv}{KinductionDfStaticSixteenTwoT}{Wrong}{}{Cputime}{Max}{19.106828161}%
\StoreBenchExecResult{PdrInv}{KinductionDfStaticSixteenTwoT}{Wrong}{}{Cputime}{Stdev}{7.5433222245}%
\StoreBenchExecResult{PdrInv}{KinductionDfStaticSixteenTwoT}{Wrong}{}{Walltime}{}{12.59801411631}%
\StoreBenchExecResult{PdrInv}{KinductionDfStaticSixteenTwoT}{Wrong}{}{Walltime}{Avg}{6.299007058155}%
\StoreBenchExecResult{PdrInv}{KinductionDfStaticSixteenTwoT}{Wrong}{}{Walltime}{Median}{6.299007058155}%
\StoreBenchExecResult{PdrInv}{KinductionDfStaticSixteenTwoT}{Wrong}{}{Walltime}{Min}{2.21149206161}%
\StoreBenchExecResult{PdrInv}{KinductionDfStaticSixteenTwoT}{Wrong}{}{Walltime}{Max}{10.3865220547}%
\StoreBenchExecResult{PdrInv}{KinductionDfStaticSixteenTwoT}{Wrong}{}{Walltime}{Stdev}{4.087514996545}%
\StoreBenchExecResult{PdrInv}{KinductionDfStaticSixteenTwoT}{Wrong}{False}{Count}{}{2}%
\StoreBenchExecResult{PdrInv}{KinductionDfStaticSixteenTwoT}{Wrong}{False}{Cputime}{}{23.127011873}%
\StoreBenchExecResult{PdrInv}{KinductionDfStaticSixteenTwoT}{Wrong}{False}{Cputime}{Avg}{11.5635059365}%
\StoreBenchExecResult{PdrInv}{KinductionDfStaticSixteenTwoT}{Wrong}{False}{Cputime}{Median}{11.5635059365}%
\StoreBenchExecResult{PdrInv}{KinductionDfStaticSixteenTwoT}{Wrong}{False}{Cputime}{Min}{4.020183712}%
\StoreBenchExecResult{PdrInv}{KinductionDfStaticSixteenTwoT}{Wrong}{False}{Cputime}{Max}{19.106828161}%
\StoreBenchExecResult{PdrInv}{KinductionDfStaticSixteenTwoT}{Wrong}{False}{Cputime}{Stdev}{7.5433222245}%
\StoreBenchExecResult{PdrInv}{KinductionDfStaticSixteenTwoT}{Wrong}{False}{Walltime}{}{12.59801411631}%
\StoreBenchExecResult{PdrInv}{KinductionDfStaticSixteenTwoT}{Wrong}{False}{Walltime}{Avg}{6.299007058155}%
\StoreBenchExecResult{PdrInv}{KinductionDfStaticSixteenTwoT}{Wrong}{False}{Walltime}{Median}{6.299007058155}%
\StoreBenchExecResult{PdrInv}{KinductionDfStaticSixteenTwoT}{Wrong}{False}{Walltime}{Min}{2.21149206161}%
\StoreBenchExecResult{PdrInv}{KinductionDfStaticSixteenTwoT}{Wrong}{False}{Walltime}{Max}{10.3865220547}%
\StoreBenchExecResult{PdrInv}{KinductionDfStaticSixteenTwoT}{Wrong}{False}{Walltime}{Stdev}{4.087514996545}%
\providecommand\StoreBenchExecResult[7]{\expandafter\newcommand\csname#1#2#3#4#5#6\endcsname{#7}}%
\StoreBenchExecResult{PdrInv}{KinductionDfStaticEightTwoTTrueNotSolvedByKinductionPlainButKipdr}{Total}{}{Count}{}{449}%
\StoreBenchExecResult{PdrInv}{KinductionDfStaticEightTwoTTrueNotSolvedByKinductionPlainButKipdr}{Total}{}{Cputime}{}{11090.893956728}%
\StoreBenchExecResult{PdrInv}{KinductionDfStaticEightTwoTTrueNotSolvedByKinductionPlainButKipdr}{Total}{}{Cputime}{Avg}{24.70132284349220489977728285}%
\StoreBenchExecResult{PdrInv}{KinductionDfStaticEightTwoTTrueNotSolvedByKinductionPlainButKipdr}{Total}{}{Cputime}{Median}{6.203414866}%
\StoreBenchExecResult{PdrInv}{KinductionDfStaticEightTwoTTrueNotSolvedByKinductionPlainButKipdr}{Total}{}{Cputime}{Min}{3.34601729}%
\StoreBenchExecResult{PdrInv}{KinductionDfStaticEightTwoTTrueNotSolvedByKinductionPlainButKipdr}{Total}{}{Cputime}{Max}{913.696656759}%
\StoreBenchExecResult{PdrInv}{KinductionDfStaticEightTwoTTrueNotSolvedByKinductionPlainButKipdr}{Total}{}{Cputime}{Stdev}{119.7870977035813147076926623}%
\StoreBenchExecResult{PdrInv}{KinductionDfStaticEightTwoTTrueNotSolvedByKinductionPlainButKipdr}{Total}{}{Walltime}{}{7845.45223331651}%
\StoreBenchExecResult{PdrInv}{KinductionDfStaticEightTwoTTrueNotSolvedByKinductionPlainButKipdr}{Total}{}{Walltime}{Avg}{17.47316755749779510022271715}%
\StoreBenchExecResult{PdrInv}{KinductionDfStaticEightTwoTTrueNotSolvedByKinductionPlainButKipdr}{Total}{}{Walltime}{Median}{3.29556703568}%
\StoreBenchExecResult{PdrInv}{KinductionDfStaticEightTwoTTrueNotSolvedByKinductionPlainButKipdr}{Total}{}{Walltime}{Min}{1.84488391876}%
\StoreBenchExecResult{PdrInv}{KinductionDfStaticEightTwoTTrueNotSolvedByKinductionPlainButKipdr}{Total}{}{Walltime}{Max}{899.615576982}%
\StoreBenchExecResult{PdrInv}{KinductionDfStaticEightTwoTTrueNotSolvedByKinductionPlainButKipdr}{Total}{}{Walltime}{Stdev}{100.4385668647929900109244816}%
\StoreBenchExecResult{PdrInv}{KinductionDfStaticEightTwoTTrueNotSolvedByKinductionPlainButKipdr}{Correct}{}{Count}{}{441}%
\StoreBenchExecResult{PdrInv}{KinductionDfStaticEightTwoTTrueNotSolvedByKinductionPlainButKipdr}{Correct}{}{Cputime}{}{3829.313815367}%
\StoreBenchExecResult{PdrInv}{KinductionDfStaticEightTwoTTrueNotSolvedByKinductionPlainButKipdr}{Correct}{}{Cputime}{Avg}{8.683251282011337868480725624}%
\StoreBenchExecResult{PdrInv}{KinductionDfStaticEightTwoTTrueNotSolvedByKinductionPlainButKipdr}{Correct}{}{Cputime}{Median}{6.152555209}%
\StoreBenchExecResult{PdrInv}{KinductionDfStaticEightTwoTTrueNotSolvedByKinductionPlainButKipdr}{Correct}{}{Cputime}{Min}{3.34601729}%
\StoreBenchExecResult{PdrInv}{KinductionDfStaticEightTwoTTrueNotSolvedByKinductionPlainButKipdr}{Correct}{}{Cputime}{Max}{158.10915202}%
\StoreBenchExecResult{PdrInv}{KinductionDfStaticEightTwoTTrueNotSolvedByKinductionPlainButKipdr}{Correct}{}{Cputime}{Stdev}{14.43891750419195141008611290}%
\StoreBenchExecResult{PdrInv}{KinductionDfStaticEightTwoTTrueNotSolvedByKinductionPlainButKipdr}{Correct}{}{Walltime}{}{2003.23324346551}%
\StoreBenchExecResult{PdrInv}{KinductionDfStaticEightTwoTTrueNotSolvedByKinductionPlainButKipdr}{Correct}{}{Walltime}{Avg}{4.542479010125873015873015873}%
\StoreBenchExecResult{PdrInv}{KinductionDfStaticEightTwoTTrueNotSolvedByKinductionPlainButKipdr}{Correct}{}{Walltime}{Median}{3.24802994728}%
\StoreBenchExecResult{PdrInv}{KinductionDfStaticEightTwoTTrueNotSolvedByKinductionPlainButKipdr}{Correct}{}{Walltime}{Min}{1.84488391876}%
\StoreBenchExecResult{PdrInv}{KinductionDfStaticEightTwoTTrueNotSolvedByKinductionPlainButKipdr}{Correct}{}{Walltime}{Max}{79.5911028385}%
\StoreBenchExecResult{PdrInv}{KinductionDfStaticEightTwoTTrueNotSolvedByKinductionPlainButKipdr}{Correct}{}{Walltime}{Stdev}{7.258505068010244135577758071}%
\StoreBenchExecResult{PdrInv}{KinductionDfStaticEightTwoTTrueNotSolvedByKinductionPlainButKipdr}{Correct}{True}{Count}{}{441}%
\StoreBenchExecResult{PdrInv}{KinductionDfStaticEightTwoTTrueNotSolvedByKinductionPlainButKipdr}{Correct}{True}{Cputime}{}{3829.313815367}%
\StoreBenchExecResult{PdrInv}{KinductionDfStaticEightTwoTTrueNotSolvedByKinductionPlainButKipdr}{Correct}{True}{Cputime}{Avg}{8.683251282011337868480725624}%
\StoreBenchExecResult{PdrInv}{KinductionDfStaticEightTwoTTrueNotSolvedByKinductionPlainButKipdr}{Correct}{True}{Cputime}{Median}{6.152555209}%
\StoreBenchExecResult{PdrInv}{KinductionDfStaticEightTwoTTrueNotSolvedByKinductionPlainButKipdr}{Correct}{True}{Cputime}{Min}{3.34601729}%
\StoreBenchExecResult{PdrInv}{KinductionDfStaticEightTwoTTrueNotSolvedByKinductionPlainButKipdr}{Correct}{True}{Cputime}{Max}{158.10915202}%
\StoreBenchExecResult{PdrInv}{KinductionDfStaticEightTwoTTrueNotSolvedByKinductionPlainButKipdr}{Correct}{True}{Cputime}{Stdev}{14.43891750419195141008611290}%
\StoreBenchExecResult{PdrInv}{KinductionDfStaticEightTwoTTrueNotSolvedByKinductionPlainButKipdr}{Correct}{True}{Walltime}{}{2003.23324346551}%
\StoreBenchExecResult{PdrInv}{KinductionDfStaticEightTwoTTrueNotSolvedByKinductionPlainButKipdr}{Correct}{True}{Walltime}{Avg}{4.542479010125873015873015873}%
\StoreBenchExecResult{PdrInv}{KinductionDfStaticEightTwoTTrueNotSolvedByKinductionPlainButKipdr}{Correct}{True}{Walltime}{Median}{3.24802994728}%
\StoreBenchExecResult{PdrInv}{KinductionDfStaticEightTwoTTrueNotSolvedByKinductionPlainButKipdr}{Correct}{True}{Walltime}{Min}{1.84488391876}%
\StoreBenchExecResult{PdrInv}{KinductionDfStaticEightTwoTTrueNotSolvedByKinductionPlainButKipdr}{Correct}{True}{Walltime}{Max}{79.5911028385}%
\StoreBenchExecResult{PdrInv}{KinductionDfStaticEightTwoTTrueNotSolvedByKinductionPlainButKipdr}{Correct}{True}{Walltime}{Stdev}{7.258505068010244135577758071}%
\StoreBenchExecResult{PdrInv}{KinductionDfStaticEightTwoTTrueNotSolvedByKinductionPlainButKipdr}{Wrong}{True}{Count}{}{0}%
\StoreBenchExecResult{PdrInv}{KinductionDfStaticEightTwoTTrueNotSolvedByKinductionPlainButKipdr}{Wrong}{True}{Cputime}{}{0}%
\StoreBenchExecResult{PdrInv}{KinductionDfStaticEightTwoTTrueNotSolvedByKinductionPlainButKipdr}{Wrong}{True}{Cputime}{Avg}{None}%
\StoreBenchExecResult{PdrInv}{KinductionDfStaticEightTwoTTrueNotSolvedByKinductionPlainButKipdr}{Wrong}{True}{Cputime}{Median}{None}%
\StoreBenchExecResult{PdrInv}{KinductionDfStaticEightTwoTTrueNotSolvedByKinductionPlainButKipdr}{Wrong}{True}{Cputime}{Min}{None}%
\StoreBenchExecResult{PdrInv}{KinductionDfStaticEightTwoTTrueNotSolvedByKinductionPlainButKipdr}{Wrong}{True}{Cputime}{Max}{None}%
\StoreBenchExecResult{PdrInv}{KinductionDfStaticEightTwoTTrueNotSolvedByKinductionPlainButKipdr}{Wrong}{True}{Cputime}{Stdev}{None}%
\StoreBenchExecResult{PdrInv}{KinductionDfStaticEightTwoTTrueNotSolvedByKinductionPlainButKipdr}{Wrong}{True}{Walltime}{}{0}%
\StoreBenchExecResult{PdrInv}{KinductionDfStaticEightTwoTTrueNotSolvedByKinductionPlainButKipdr}{Wrong}{True}{Walltime}{Avg}{None}%
\StoreBenchExecResult{PdrInv}{KinductionDfStaticEightTwoTTrueNotSolvedByKinductionPlainButKipdr}{Wrong}{True}{Walltime}{Median}{None}%
\StoreBenchExecResult{PdrInv}{KinductionDfStaticEightTwoTTrueNotSolvedByKinductionPlainButKipdr}{Wrong}{True}{Walltime}{Min}{None}%
\StoreBenchExecResult{PdrInv}{KinductionDfStaticEightTwoTTrueNotSolvedByKinductionPlainButKipdr}{Wrong}{True}{Walltime}{Max}{None}%
\StoreBenchExecResult{PdrInv}{KinductionDfStaticEightTwoTTrueNotSolvedByKinductionPlainButKipdr}{Wrong}{True}{Walltime}{Stdev}{None}%
\StoreBenchExecResult{PdrInv}{KinductionDfStaticEightTwoTTrueNotSolvedByKinductionPlainButKipdr}{Error}{}{Count}{}{8}%
\StoreBenchExecResult{PdrInv}{KinductionDfStaticEightTwoTTrueNotSolvedByKinductionPlainButKipdr}{Error}{}{Cputime}{}{7261.580141361}%
\StoreBenchExecResult{PdrInv}{KinductionDfStaticEightTwoTTrueNotSolvedByKinductionPlainButKipdr}{Error}{}{Cputime}{Avg}{907.697517670125}%
\StoreBenchExecResult{PdrInv}{KinductionDfStaticEightTwoTTrueNotSolvedByKinductionPlainButKipdr}{Error}{}{Cputime}{Median}{907.9385129965}%
\StoreBenchExecResult{PdrInv}{KinductionDfStaticEightTwoTTrueNotSolvedByKinductionPlainButKipdr}{Error}{}{Cputime}{Min}{901.742360748}%
\StoreBenchExecResult{PdrInv}{KinductionDfStaticEightTwoTTrueNotSolvedByKinductionPlainButKipdr}{Error}{}{Cputime}{Max}{913.696656759}%
\StoreBenchExecResult{PdrInv}{KinductionDfStaticEightTwoTTrueNotSolvedByKinductionPlainButKipdr}{Error}{}{Cputime}{Stdev}{3.998103409944779695162552801}%
\StoreBenchExecResult{PdrInv}{KinductionDfStaticEightTwoTTrueNotSolvedByKinductionPlainButKipdr}{Error}{}{Walltime}{}{5842.218989851}%
\StoreBenchExecResult{PdrInv}{KinductionDfStaticEightTwoTTrueNotSolvedByKinductionPlainButKipdr}{Error}{}{Walltime}{Avg}{730.277373731375}%
\StoreBenchExecResult{PdrInv}{KinductionDfStaticEightTwoTTrueNotSolvedByKinductionPlainButKipdr}{Error}{}{Walltime}{Median}{893.8723269705}%
\StoreBenchExecResult{PdrInv}{KinductionDfStaticEightTwoTTrueNotSolvedByKinductionPlainButKipdr}{Error}{}{Walltime}{Min}{452.297701836}%
\StoreBenchExecResult{PdrInv}{KinductionDfStaticEightTwoTTrueNotSolvedByKinductionPlainButKipdr}{Error}{}{Walltime}{Max}{899.615576982}%
\StoreBenchExecResult{PdrInv}{KinductionDfStaticEightTwoTTrueNotSolvedByKinductionPlainButKipdr}{Error}{}{Walltime}{Stdev}{214.4120161240012334251346913}%
\StoreBenchExecResult{PdrInv}{KinductionDfStaticEightTwoTTrueNotSolvedByKinductionPlainButKipdr}{Error}{Timeout}{Count}{}{8}%
\StoreBenchExecResult{PdrInv}{KinductionDfStaticEightTwoTTrueNotSolvedByKinductionPlainButKipdr}{Error}{Timeout}{Cputime}{}{7261.580141361}%
\StoreBenchExecResult{PdrInv}{KinductionDfStaticEightTwoTTrueNotSolvedByKinductionPlainButKipdr}{Error}{Timeout}{Cputime}{Avg}{907.697517670125}%
\StoreBenchExecResult{PdrInv}{KinductionDfStaticEightTwoTTrueNotSolvedByKinductionPlainButKipdr}{Error}{Timeout}{Cputime}{Median}{907.9385129965}%
\StoreBenchExecResult{PdrInv}{KinductionDfStaticEightTwoTTrueNotSolvedByKinductionPlainButKipdr}{Error}{Timeout}{Cputime}{Min}{901.742360748}%
\StoreBenchExecResult{PdrInv}{KinductionDfStaticEightTwoTTrueNotSolvedByKinductionPlainButKipdr}{Error}{Timeout}{Cputime}{Max}{913.696656759}%
\StoreBenchExecResult{PdrInv}{KinductionDfStaticEightTwoTTrueNotSolvedByKinductionPlainButKipdr}{Error}{Timeout}{Cputime}{Stdev}{3.998103409944779695162552801}%
\StoreBenchExecResult{PdrInv}{KinductionDfStaticEightTwoTTrueNotSolvedByKinductionPlainButKipdr}{Error}{Timeout}{Walltime}{}{5842.218989851}%
\StoreBenchExecResult{PdrInv}{KinductionDfStaticEightTwoTTrueNotSolvedByKinductionPlainButKipdr}{Error}{Timeout}{Walltime}{Avg}{730.277373731375}%
\StoreBenchExecResult{PdrInv}{KinductionDfStaticEightTwoTTrueNotSolvedByKinductionPlainButKipdr}{Error}{Timeout}{Walltime}{Median}{893.8723269705}%
\StoreBenchExecResult{PdrInv}{KinductionDfStaticEightTwoTTrueNotSolvedByKinductionPlainButKipdr}{Error}{Timeout}{Walltime}{Min}{452.297701836}%
\StoreBenchExecResult{PdrInv}{KinductionDfStaticEightTwoTTrueNotSolvedByKinductionPlainButKipdr}{Error}{Timeout}{Walltime}{Max}{899.615576982}%
\StoreBenchExecResult{PdrInv}{KinductionDfStaticEightTwoTTrueNotSolvedByKinductionPlainButKipdr}{Error}{Timeout}{Walltime}{Stdev}{214.4120161240012334251346913}%
\providecommand\StoreBenchExecResult[7]{\expandafter\newcommand\csname#1#2#3#4#5#6\endcsname{#7}}%
\StoreBenchExecResult{PdrInv}{KinductionDfStaticEightTwoTTrueNotSolvedByKinductionPlain}{Total}{}{Count}{}{2893}%
\StoreBenchExecResult{PdrInv}{KinductionDfStaticEightTwoTTrueNotSolvedByKinductionPlain}{Total}{}{Cputime}{}{1555476.246879197}%
\StoreBenchExecResult{PdrInv}{KinductionDfStaticEightTwoTTrueNotSolvedByKinductionPlain}{Total}{}{Cputime}{Avg}{537.6689411957127549256826823}%
\StoreBenchExecResult{PdrInv}{KinductionDfStaticEightTwoTTrueNotSolvedByKinductionPlain}{Total}{}{Cputime}{Median}{901.083666893}%
\StoreBenchExecResult{PdrInv}{KinductionDfStaticEightTwoTTrueNotSolvedByKinductionPlain}{Total}{}{Cputime}{Min}{2.502797224}%
\StoreBenchExecResult{PdrInv}{KinductionDfStaticEightTwoTTrueNotSolvedByKinductionPlain}{Total}{}{Cputime}{Max}{1001.30928116}%
\StoreBenchExecResult{PdrInv}{KinductionDfStaticEightTwoTTrueNotSolvedByKinductionPlain}{Total}{}{Cputime}{Stdev}{418.7373033484069086000161959}%
\StoreBenchExecResult{PdrInv}{KinductionDfStaticEightTwoTTrueNotSolvedByKinductionPlain}{Total}{}{Walltime}{}{993669.22632550336}%
\StoreBenchExecResult{PdrInv}{KinductionDfStaticEightTwoTTrueNotSolvedByKinductionPlain}{Total}{}{Walltime}{Avg}{343.4736350935027169028689941}%
\StoreBenchExecResult{PdrInv}{KinductionDfStaticEightTwoTTrueNotSolvedByKinductionPlain}{Total}{}{Walltime}{Median}{451.602358818}%
\StoreBenchExecResult{PdrInv}{KinductionDfStaticEightTwoTTrueNotSolvedByKinductionPlain}{Total}{}{Walltime}{Min}{1.36538696289}%
\StoreBenchExecResult{PdrInv}{KinductionDfStaticEightTwoTTrueNotSolvedByKinductionPlain}{Total}{}{Walltime}{Max}{975.006866932}%
\StoreBenchExecResult{PdrInv}{KinductionDfStaticEightTwoTTrueNotSolvedByKinductionPlain}{Total}{}{Walltime}{Stdev}{306.2438864747021046205818227}%
\StoreBenchExecResult{PdrInv}{KinductionDfStaticEightTwoTTrueNotSolvedByKinductionPlain}{Correct}{}{Count}{}{1073}%
\StoreBenchExecResult{PdrInv}{KinductionDfStaticEightTwoTTrueNotSolvedByKinductionPlain}{Correct}{}{Cputime}{}{64889.891544167}%
\StoreBenchExecResult{PdrInv}{KinductionDfStaticEightTwoTTrueNotSolvedByKinductionPlain}{Correct}{}{Cputime}{Avg}{60.47520181189841565703634669}%
\StoreBenchExecResult{PdrInv}{KinductionDfStaticEightTwoTTrueNotSolvedByKinductionPlain}{Correct}{}{Cputime}{Median}{8.131056508}%
\StoreBenchExecResult{PdrInv}{KinductionDfStaticEightTwoTTrueNotSolvedByKinductionPlain}{Correct}{}{Cputime}{Min}{3.34601729}%
\StoreBenchExecResult{PdrInv}{KinductionDfStaticEightTwoTTrueNotSolvedByKinductionPlain}{Correct}{}{Cputime}{Max}{890.091335986}%
\StoreBenchExecResult{PdrInv}{KinductionDfStaticEightTwoTTrueNotSolvedByKinductionPlain}{Correct}{}{Cputime}{Stdev}{129.9072053864073828257877568}%
\StoreBenchExecResult{PdrInv}{KinductionDfStaticEightTwoTTrueNotSolvedByKinductionPlain}{Correct}{}{Walltime}{}{34384.20136427723}%
\StoreBenchExecResult{PdrInv}{KinductionDfStaticEightTwoTTrueNotSolvedByKinductionPlain}{Correct}{}{Walltime}{Avg}{32.04492205431242311276794035}%
\StoreBenchExecResult{PdrInv}{KinductionDfStaticEightTwoTTrueNotSolvedByKinductionPlain}{Correct}{}{Walltime}{Median}{4.27121710777}%
\StoreBenchExecResult{PdrInv}{KinductionDfStaticEightTwoTTrueNotSolvedByKinductionPlain}{Correct}{}{Walltime}{Min}{1.84488391876}%
\StoreBenchExecResult{PdrInv}{KinductionDfStaticEightTwoTTrueNotSolvedByKinductionPlain}{Correct}{}{Walltime}{Max}{776.734863997}%
\StoreBenchExecResult{PdrInv}{KinductionDfStaticEightTwoTTrueNotSolvedByKinductionPlain}{Correct}{}{Walltime}{Stdev}{73.78426759361506186699123151}%
\StoreBenchExecResult{PdrInv}{KinductionDfStaticEightTwoTTrueNotSolvedByKinductionPlain}{Correct}{True}{Count}{}{1073}%
\StoreBenchExecResult{PdrInv}{KinductionDfStaticEightTwoTTrueNotSolvedByKinductionPlain}{Correct}{True}{Cputime}{}{64889.891544167}%
\StoreBenchExecResult{PdrInv}{KinductionDfStaticEightTwoTTrueNotSolvedByKinductionPlain}{Correct}{True}{Cputime}{Avg}{60.47520181189841565703634669}%
\StoreBenchExecResult{PdrInv}{KinductionDfStaticEightTwoTTrueNotSolvedByKinductionPlain}{Correct}{True}{Cputime}{Median}{8.131056508}%
\StoreBenchExecResult{PdrInv}{KinductionDfStaticEightTwoTTrueNotSolvedByKinductionPlain}{Correct}{True}{Cputime}{Min}{3.34601729}%
\StoreBenchExecResult{PdrInv}{KinductionDfStaticEightTwoTTrueNotSolvedByKinductionPlain}{Correct}{True}{Cputime}{Max}{890.091335986}%
\StoreBenchExecResult{PdrInv}{KinductionDfStaticEightTwoTTrueNotSolvedByKinductionPlain}{Correct}{True}{Cputime}{Stdev}{129.9072053864073828257877568}%
\StoreBenchExecResult{PdrInv}{KinductionDfStaticEightTwoTTrueNotSolvedByKinductionPlain}{Correct}{True}{Walltime}{}{34384.20136427723}%
\StoreBenchExecResult{PdrInv}{KinductionDfStaticEightTwoTTrueNotSolvedByKinductionPlain}{Correct}{True}{Walltime}{Avg}{32.04492205431242311276794035}%
\StoreBenchExecResult{PdrInv}{KinductionDfStaticEightTwoTTrueNotSolvedByKinductionPlain}{Correct}{True}{Walltime}{Median}{4.27121710777}%
\StoreBenchExecResult{PdrInv}{KinductionDfStaticEightTwoTTrueNotSolvedByKinductionPlain}{Correct}{True}{Walltime}{Min}{1.84488391876}%
\StoreBenchExecResult{PdrInv}{KinductionDfStaticEightTwoTTrueNotSolvedByKinductionPlain}{Correct}{True}{Walltime}{Max}{776.734863997}%
\StoreBenchExecResult{PdrInv}{KinductionDfStaticEightTwoTTrueNotSolvedByKinductionPlain}{Correct}{True}{Walltime}{Stdev}{73.78426759361506186699123151}%
\StoreBenchExecResult{PdrInv}{KinductionDfStaticEightTwoTTrueNotSolvedByKinductionPlain}{Wrong}{True}{Count}{}{0}%
\StoreBenchExecResult{PdrInv}{KinductionDfStaticEightTwoTTrueNotSolvedByKinductionPlain}{Wrong}{True}{Cputime}{}{0}%
\StoreBenchExecResult{PdrInv}{KinductionDfStaticEightTwoTTrueNotSolvedByKinductionPlain}{Wrong}{True}{Cputime}{Avg}{None}%
\StoreBenchExecResult{PdrInv}{KinductionDfStaticEightTwoTTrueNotSolvedByKinductionPlain}{Wrong}{True}{Cputime}{Median}{None}%
\StoreBenchExecResult{PdrInv}{KinductionDfStaticEightTwoTTrueNotSolvedByKinductionPlain}{Wrong}{True}{Cputime}{Min}{None}%
\StoreBenchExecResult{PdrInv}{KinductionDfStaticEightTwoTTrueNotSolvedByKinductionPlain}{Wrong}{True}{Cputime}{Max}{None}%
\StoreBenchExecResult{PdrInv}{KinductionDfStaticEightTwoTTrueNotSolvedByKinductionPlain}{Wrong}{True}{Cputime}{Stdev}{None}%
\StoreBenchExecResult{PdrInv}{KinductionDfStaticEightTwoTTrueNotSolvedByKinductionPlain}{Wrong}{True}{Walltime}{}{0}%
\StoreBenchExecResult{PdrInv}{KinductionDfStaticEightTwoTTrueNotSolvedByKinductionPlain}{Wrong}{True}{Walltime}{Avg}{None}%
\StoreBenchExecResult{PdrInv}{KinductionDfStaticEightTwoTTrueNotSolvedByKinductionPlain}{Wrong}{True}{Walltime}{Median}{None}%
\StoreBenchExecResult{PdrInv}{KinductionDfStaticEightTwoTTrueNotSolvedByKinductionPlain}{Wrong}{True}{Walltime}{Min}{None}%
\StoreBenchExecResult{PdrInv}{KinductionDfStaticEightTwoTTrueNotSolvedByKinductionPlain}{Wrong}{True}{Walltime}{Max}{None}%
\StoreBenchExecResult{PdrInv}{KinductionDfStaticEightTwoTTrueNotSolvedByKinductionPlain}{Wrong}{True}{Walltime}{Stdev}{None}%
\StoreBenchExecResult{PdrInv}{KinductionDfStaticEightTwoTTrueNotSolvedByKinductionPlain}{Error}{}{Count}{}{1820}%
\StoreBenchExecResult{PdrInv}{KinductionDfStaticEightTwoTTrueNotSolvedByKinductionPlain}{Error}{}{Cputime}{}{1490586.355335030}%
\StoreBenchExecResult{PdrInv}{KinductionDfStaticEightTwoTTrueNotSolvedByKinductionPlain}{Error}{}{Cputime}{Avg}{819.0034919423241758241758242}%
\StoreBenchExecResult{PdrInv}{KinductionDfStaticEightTwoTTrueNotSolvedByKinductionPlain}{Error}{}{Cputime}{Median}{901.6528868715}%
\StoreBenchExecResult{PdrInv}{KinductionDfStaticEightTwoTTrueNotSolvedByKinductionPlain}{Error}{}{Cputime}{Min}{2.502797224}%
\StoreBenchExecResult{PdrInv}{KinductionDfStaticEightTwoTTrueNotSolvedByKinductionPlain}{Error}{}{Cputime}{Max}{1001.30928116}%
\StoreBenchExecResult{PdrInv}{KinductionDfStaticEightTwoTTrueNotSolvedByKinductionPlain}{Error}{}{Cputime}{Stdev}{235.2986110620393480730940016}%
\StoreBenchExecResult{PdrInv}{KinductionDfStaticEightTwoTTrueNotSolvedByKinductionPlain}{Error}{}{Walltime}{}{959285.02496122613}%
\StoreBenchExecResult{PdrInv}{KinductionDfStaticEightTwoTTrueNotSolvedByKinductionPlain}{Error}{}{Walltime}{Avg}{527.0796840446297417582417582}%
\StoreBenchExecResult{PdrInv}{KinductionDfStaticEightTwoTTrueNotSolvedByKinductionPlain}{Error}{}{Walltime}{Median}{453.5218254325}%
\StoreBenchExecResult{PdrInv}{KinductionDfStaticEightTwoTTrueNotSolvedByKinductionPlain}{Error}{}{Walltime}{Min}{1.36538696289}%
\StoreBenchExecResult{PdrInv}{KinductionDfStaticEightTwoTTrueNotSolvedByKinductionPlain}{Error}{}{Walltime}{Max}{975.006866932}%
\StoreBenchExecResult{PdrInv}{KinductionDfStaticEightTwoTTrueNotSolvedByKinductionPlain}{Error}{}{Walltime}{Stdev}{234.4705053534018856873320162}%
\StoreBenchExecResult{PdrInv}{KinductionDfStaticEightTwoTTrueNotSolvedByKinductionPlain}{Error}{Assertion}{Count}{}{2}%
\StoreBenchExecResult{PdrInv}{KinductionDfStaticEightTwoTTrueNotSolvedByKinductionPlain}{Error}{Assertion}{Cputime}{}{6.750158056}%
\StoreBenchExecResult{PdrInv}{KinductionDfStaticEightTwoTTrueNotSolvedByKinductionPlain}{Error}{Assertion}{Cputime}{Avg}{3.375079028}%
\StoreBenchExecResult{PdrInv}{KinductionDfStaticEightTwoTTrueNotSolvedByKinductionPlain}{Error}{Assertion}{Cputime}{Median}{3.375079028}%
\StoreBenchExecResult{PdrInv}{KinductionDfStaticEightTwoTTrueNotSolvedByKinductionPlain}{Error}{Assertion}{Cputime}{Min}{3.231676692}%
\StoreBenchExecResult{PdrInv}{KinductionDfStaticEightTwoTTrueNotSolvedByKinductionPlain}{Error}{Assertion}{Cputime}{Max}{3.518481364}%
\StoreBenchExecResult{PdrInv}{KinductionDfStaticEightTwoTTrueNotSolvedByKinductionPlain}{Error}{Assertion}{Cputime}{Stdev}{0.143402336}%
\StoreBenchExecResult{PdrInv}{KinductionDfStaticEightTwoTTrueNotSolvedByKinductionPlain}{Error}{Assertion}{Walltime}{}{3.79572510719}%
\StoreBenchExecResult{PdrInv}{KinductionDfStaticEightTwoTTrueNotSolvedByKinductionPlain}{Error}{Assertion}{Walltime}{Avg}{1.897862553595}%
\StoreBenchExecResult{PdrInv}{KinductionDfStaticEightTwoTTrueNotSolvedByKinductionPlain}{Error}{Assertion}{Walltime}{Median}{1.897862553595}%
\StoreBenchExecResult{PdrInv}{KinductionDfStaticEightTwoTTrueNotSolvedByKinductionPlain}{Error}{Assertion}{Walltime}{Min}{1.79999709129}%
\StoreBenchExecResult{PdrInv}{KinductionDfStaticEightTwoTTrueNotSolvedByKinductionPlain}{Error}{Assertion}{Walltime}{Max}{1.9957280159}%
\StoreBenchExecResult{PdrInv}{KinductionDfStaticEightTwoTTrueNotSolvedByKinductionPlain}{Error}{Assertion}{Walltime}{Stdev}{0.097865462305}%
\StoreBenchExecResult{PdrInv}{KinductionDfStaticEightTwoTTrueNotSolvedByKinductionPlain}{Error}{Error}{Count}{}{129}%
\StoreBenchExecResult{PdrInv}{KinductionDfStaticEightTwoTTrueNotSolvedByKinductionPlain}{Error}{Error}{Cputime}{}{21743.532629970}%
\StoreBenchExecResult{PdrInv}{KinductionDfStaticEightTwoTTrueNotSolvedByKinductionPlain}{Error}{Error}{Cputime}{Avg}{168.5545165113953488372093023}%
\StoreBenchExecResult{PdrInv}{KinductionDfStaticEightTwoTTrueNotSolvedByKinductionPlain}{Error}{Error}{Cputime}{Median}{101.637451654}%
\StoreBenchExecResult{PdrInv}{KinductionDfStaticEightTwoTTrueNotSolvedByKinductionPlain}{Error}{Error}{Cputime}{Min}{2.502797224}%
\StoreBenchExecResult{PdrInv}{KinductionDfStaticEightTwoTTrueNotSolvedByKinductionPlain}{Error}{Error}{Cputime}{Max}{794.256551189}%
\StoreBenchExecResult{PdrInv}{KinductionDfStaticEightTwoTTrueNotSolvedByKinductionPlain}{Error}{Error}{Cputime}{Stdev}{187.1734517020029787882850634}%
\StoreBenchExecResult{PdrInv}{KinductionDfStaticEightTwoTTrueNotSolvedByKinductionPlain}{Error}{Error}{Walltime}{}{18269.20834183458}%
\StoreBenchExecResult{PdrInv}{KinductionDfStaticEightTwoTTrueNotSolvedByKinductionPlain}{Error}{Error}{Walltime}{Avg}{141.6217700917409302325581395}%
\StoreBenchExecResult{PdrInv}{KinductionDfStaticEightTwoTTrueNotSolvedByKinductionPlain}{Error}{Error}{Walltime}{Median}{74.2252981663}%
\StoreBenchExecResult{PdrInv}{KinductionDfStaticEightTwoTTrueNotSolvedByKinductionPlain}{Error}{Error}{Walltime}{Min}{1.36538696289}%
\StoreBenchExecResult{PdrInv}{KinductionDfStaticEightTwoTTrueNotSolvedByKinductionPlain}{Error}{Error}{Walltime}{Max}{778.652194023}%
\StoreBenchExecResult{PdrInv}{KinductionDfStaticEightTwoTTrueNotSolvedByKinductionPlain}{Error}{Error}{Walltime}{Stdev}{167.5509577057884714673993639}%
\StoreBenchExecResult{PdrInv}{KinductionDfStaticEightTwoTTrueNotSolvedByKinductionPlain}{Error}{Exception}{Count}{}{6}%
\StoreBenchExecResult{PdrInv}{KinductionDfStaticEightTwoTTrueNotSolvedByKinductionPlain}{Error}{Exception}{Cputime}{}{718.135361413}%
\StoreBenchExecResult{PdrInv}{KinductionDfStaticEightTwoTTrueNotSolvedByKinductionPlain}{Error}{Exception}{Cputime}{Avg}{119.6892269021666666666666667}%
\StoreBenchExecResult{PdrInv}{KinductionDfStaticEightTwoTTrueNotSolvedByKinductionPlain}{Error}{Exception}{Cputime}{Median}{80.770924803}%
\StoreBenchExecResult{PdrInv}{KinductionDfStaticEightTwoTTrueNotSolvedByKinductionPlain}{Error}{Exception}{Cputime}{Min}{14.587210578}%
\StoreBenchExecResult{PdrInv}{KinductionDfStaticEightTwoTTrueNotSolvedByKinductionPlain}{Error}{Exception}{Cputime}{Max}{279.945523639}%
\StoreBenchExecResult{PdrInv}{KinductionDfStaticEightTwoTTrueNotSolvedByKinductionPlain}{Error}{Exception}{Cputime}{Stdev}{97.50759513606682780472104384}%
\StoreBenchExecResult{PdrInv}{KinductionDfStaticEightTwoTTrueNotSolvedByKinductionPlain}{Error}{Exception}{Walltime}{}{361.45104861246}%
\StoreBenchExecResult{PdrInv}{KinductionDfStaticEightTwoTTrueNotSolvedByKinductionPlain}{Error}{Exception}{Walltime}{Avg}{60.24184143541}%
\StoreBenchExecResult{PdrInv}{KinductionDfStaticEightTwoTTrueNotSolvedByKinductionPlain}{Error}{Exception}{Walltime}{Median}{40.7244694233}%
\StoreBenchExecResult{PdrInv}{KinductionDfStaticEightTwoTTrueNotSolvedByKinductionPlain}{Error}{Exception}{Walltime}{Min}{7.51351094246}%
\StoreBenchExecResult{PdrInv}{KinductionDfStaticEightTwoTTrueNotSolvedByKinductionPlain}{Error}{Exception}{Walltime}{Max}{140.643676996}%
\StoreBenchExecResult{PdrInv}{KinductionDfStaticEightTwoTTrueNotSolvedByKinductionPlain}{Error}{Exception}{Walltime}{Stdev}{48.86834795175321720243089095}%
\StoreBenchExecResult{PdrInv}{KinductionDfStaticEightTwoTTrueNotSolvedByKinductionPlain}{Error}{OutOfJavaMemory}{Count}{}{5}%
\StoreBenchExecResult{PdrInv}{KinductionDfStaticEightTwoTTrueNotSolvedByKinductionPlain}{Error}{OutOfJavaMemory}{Cputime}{}{1759.149016130}%
\StoreBenchExecResult{PdrInv}{KinductionDfStaticEightTwoTTrueNotSolvedByKinductionPlain}{Error}{OutOfJavaMemory}{Cputime}{Avg}{351.829803226}%
\StoreBenchExecResult{PdrInv}{KinductionDfStaticEightTwoTTrueNotSolvedByKinductionPlain}{Error}{OutOfJavaMemory}{Cputime}{Median}{372.243618935}%
\StoreBenchExecResult{PdrInv}{KinductionDfStaticEightTwoTTrueNotSolvedByKinductionPlain}{Error}{OutOfJavaMemory}{Cputime}{Min}{189.622584203}%
\StoreBenchExecResult{PdrInv}{KinductionDfStaticEightTwoTTrueNotSolvedByKinductionPlain}{Error}{OutOfJavaMemory}{Cputime}{Max}{472.133994606}%
\StoreBenchExecResult{PdrInv}{KinductionDfStaticEightTwoTTrueNotSolvedByKinductionPlain}{Error}{OutOfJavaMemory}{Cputime}{Stdev}{100.2225898295508035957914850}%
\StoreBenchExecResult{PdrInv}{KinductionDfStaticEightTwoTTrueNotSolvedByKinductionPlain}{Error}{OutOfJavaMemory}{Walltime}{}{1014.754484176}%
\StoreBenchExecResult{PdrInv}{KinductionDfStaticEightTwoTTrueNotSolvedByKinductionPlain}{Error}{OutOfJavaMemory}{Walltime}{Avg}{202.9508968352}%
\StoreBenchExecResult{PdrInv}{KinductionDfStaticEightTwoTTrueNotSolvedByKinductionPlain}{Error}{OutOfJavaMemory}{Walltime}{Median}{222.909708023}%
\StoreBenchExecResult{PdrInv}{KinductionDfStaticEightTwoTTrueNotSolvedByKinductionPlain}{Error}{OutOfJavaMemory}{Walltime}{Min}{113.677829027}%
\StoreBenchExecResult{PdrInv}{KinductionDfStaticEightTwoTTrueNotSolvedByKinductionPlain}{Error}{OutOfJavaMemory}{Walltime}{Max}{244.924804926}%
\StoreBenchExecResult{PdrInv}{KinductionDfStaticEightTwoTTrueNotSolvedByKinductionPlain}{Error}{OutOfJavaMemory}{Walltime}{Stdev}{47.89107179699436265841440350}%
\StoreBenchExecResult{PdrInv}{KinductionDfStaticEightTwoTTrueNotSolvedByKinductionPlain}{Error}{OutOfMemory}{Count}{}{106}%
\StoreBenchExecResult{PdrInv}{KinductionDfStaticEightTwoTTrueNotSolvedByKinductionPlain}{Error}{OutOfMemory}{Cputime}{}{42250.331114836}%
\StoreBenchExecResult{PdrInv}{KinductionDfStaticEightTwoTTrueNotSolvedByKinductionPlain}{Error}{OutOfMemory}{Cputime}{Avg}{398.5880293852452830188679245}%
\StoreBenchExecResult{PdrInv}{KinductionDfStaticEightTwoTTrueNotSolvedByKinductionPlain}{Error}{OutOfMemory}{Cputime}{Median}{336.8256022045}%
\StoreBenchExecResult{PdrInv}{KinductionDfStaticEightTwoTTrueNotSolvedByKinductionPlain}{Error}{OutOfMemory}{Cputime}{Min}{167.139458137}%
\StoreBenchExecResult{PdrInv}{KinductionDfStaticEightTwoTTrueNotSolvedByKinductionPlain}{Error}{OutOfMemory}{Cputime}{Max}{880.806898157}%
\StoreBenchExecResult{PdrInv}{KinductionDfStaticEightTwoTTrueNotSolvedByKinductionPlain}{Error}{OutOfMemory}{Cputime}{Stdev}{223.3336400563398101088318913}%
\StoreBenchExecResult{PdrInv}{KinductionDfStaticEightTwoTTrueNotSolvedByKinductionPlain}{Error}{OutOfMemory}{Walltime}{}{22474.2947764419}%
\StoreBenchExecResult{PdrInv}{KinductionDfStaticEightTwoTTrueNotSolvedByKinductionPlain}{Error}{OutOfMemory}{Walltime}{Avg}{212.0216488343575471698113208}%
\StoreBenchExecResult{PdrInv}{KinductionDfStaticEightTwoTTrueNotSolvedByKinductionPlain}{Error}{OutOfMemory}{Walltime}{Median}{169.542504430}%
\StoreBenchExecResult{PdrInv}{KinductionDfStaticEightTwoTTrueNotSolvedByKinductionPlain}{Error}{OutOfMemory}{Walltime}{Min}{84.0819849968}%
\StoreBenchExecResult{PdrInv}{KinductionDfStaticEightTwoTTrueNotSolvedByKinductionPlain}{Error}{OutOfMemory}{Walltime}{Max}{840.760004997}%
\StoreBenchExecResult{PdrInv}{KinductionDfStaticEightTwoTTrueNotSolvedByKinductionPlain}{Error}{OutOfMemory}{Walltime}{Stdev}{145.3371698570629781992760646}%
\StoreBenchExecResult{PdrInv}{KinductionDfStaticEightTwoTTrueNotSolvedByKinductionPlain}{Error}{Timeout}{Count}{}{1572}%
\StoreBenchExecResult{PdrInv}{KinductionDfStaticEightTwoTTrueNotSolvedByKinductionPlain}{Error}{Timeout}{Cputime}{}{1424108.457054625}%
\StoreBenchExecResult{PdrInv}{KinductionDfStaticEightTwoTTrueNotSolvedByKinductionPlain}{Error}{Timeout}{Cputime}{Avg}{905.9214103400922391857506361}%
\StoreBenchExecResult{PdrInv}{KinductionDfStaticEightTwoTTrueNotSolvedByKinductionPlain}{Error}{Timeout}{Cputime}{Median}{902.047815224}%
\StoreBenchExecResult{PdrInv}{KinductionDfStaticEightTwoTTrueNotSolvedByKinductionPlain}{Error}{Timeout}{Cputime}{Min}{900.824878562}%
\StoreBenchExecResult{PdrInv}{KinductionDfStaticEightTwoTTrueNotSolvedByKinductionPlain}{Error}{Timeout}{Cputime}{Max}{1001.30928116}%
\StoreBenchExecResult{PdrInv}{KinductionDfStaticEightTwoTTrueNotSolvedByKinductionPlain}{Error}{Timeout}{Cputime}{Stdev}{13.95506109321471765368566839}%
\StoreBenchExecResult{PdrInv}{KinductionDfStaticEightTwoTTrueNotSolvedByKinductionPlain}{Error}{Timeout}{Walltime}{}{917161.520585054}%
\StoreBenchExecResult{PdrInv}{KinductionDfStaticEightTwoTTrueNotSolvedByKinductionPlain}{Error}{Timeout}{Walltime}{Avg}{583.4360817971081424936386768}%
\StoreBenchExecResult{PdrInv}{KinductionDfStaticEightTwoTTrueNotSolvedByKinductionPlain}{Error}{Timeout}{Walltime}{Median}{455.000699520}%
\StoreBenchExecResult{PdrInv}{KinductionDfStaticEightTwoTTrueNotSolvedByKinductionPlain}{Error}{Timeout}{Walltime}{Min}{451.098812819}%
\StoreBenchExecResult{PdrInv}{KinductionDfStaticEightTwoTTrueNotSolvedByKinductionPlain}{Error}{Timeout}{Walltime}{Max}{975.006866932}%
\StoreBenchExecResult{PdrInv}{KinductionDfStaticEightTwoTTrueNotSolvedByKinductionPlain}{Error}{Timeout}{Walltime}{Stdev}{190.5950921254572417309567905}%
\providecommand\StoreBenchExecResult[7]{\expandafter\newcommand\csname#1#2#3#4#5#6\endcsname{#7}}%
\StoreBenchExecResult{PdrInv}{KinductionDfStaticEightTwoT}{Total}{}{Count}{}{5591}%
\StoreBenchExecResult{PdrInv}{KinductionDfStaticEightTwoT}{Total}{}{Cputime}{}{2196149.479597208}%
\StoreBenchExecResult{PdrInv}{KinductionDfStaticEightTwoT}{Total}{}{Cputime}{Avg}{392.8008369875170810230727956}%
\StoreBenchExecResult{PdrInv}{KinductionDfStaticEightTwoT}{Total}{}{Cputime}{Median}{113.369613682}%
\StoreBenchExecResult{PdrInv}{KinductionDfStaticEightTwoT}{Total}{}{Cputime}{Min}{2.502797224}%
\StoreBenchExecResult{PdrInv}{KinductionDfStaticEightTwoT}{Total}{}{Cputime}{Max}{1002.31487952}%
\StoreBenchExecResult{PdrInv}{KinductionDfStaticEightTwoT}{Total}{}{Cputime}{Stdev}{416.0794904450780209737098437}%
\StoreBenchExecResult{PdrInv}{KinductionDfStaticEightTwoT}{Total}{}{Walltime}{}{1420661.43854975187}%
\StoreBenchExecResult{PdrInv}{KinductionDfStaticEightTwoT}{Total}{}{Walltime}{Avg}{254.0979142460654390985512431}%
\StoreBenchExecResult{PdrInv}{KinductionDfStaticEightTwoT}{Total}{}{Walltime}{Median}{66.9046721458}%
\StoreBenchExecResult{PdrInv}{KinductionDfStaticEightTwoT}{Total}{}{Walltime}{Min}{1.36538696289}%
\StoreBenchExecResult{PdrInv}{KinductionDfStaticEightTwoT}{Total}{}{Walltime}{Max}{979.662005901}%
\StoreBenchExecResult{PdrInv}{KinductionDfStaticEightTwoT}{Total}{}{Walltime}{Stdev}{298.6684797289346038207592467}%
\StoreBenchExecResult{PdrInv}{KinductionDfStaticEightTwoT}{Correct}{}{Count}{}{3084}%
\StoreBenchExecResult{PdrInv}{KinductionDfStaticEightTwoT}{Correct}{}{Cputime}{}{194596.101874585}%
\StoreBenchExecResult{PdrInv}{KinductionDfStaticEightTwoT}{Correct}{}{Cputime}{Avg}{63.09860631471627756160830091}%
\StoreBenchExecResult{PdrInv}{KinductionDfStaticEightTwoT}{Correct}{}{Cputime}{Median}{10.027357559}%
\StoreBenchExecResult{PdrInv}{KinductionDfStaticEightTwoT}{Correct}{}{Cputime}{Min}{2.958535013}%
\StoreBenchExecResult{PdrInv}{KinductionDfStaticEightTwoT}{Correct}{}{Cputime}{Max}{896.232896503}%
\StoreBenchExecResult{PdrInv}{KinductionDfStaticEightTwoT}{Correct}{}{Cputime}{Stdev}{139.4080697238763925357393269}%
\StoreBenchExecResult{PdrInv}{KinductionDfStaticEightTwoT}{Correct}{}{Walltime}{}{118450.05748152226}%
\StoreBenchExecResult{PdrInv}{KinductionDfStaticEightTwoT}{Correct}{}{Walltime}{Avg}{38.40793044147933203631647211}%
\StoreBenchExecResult{PdrInv}{KinductionDfStaticEightTwoT}{Correct}{}{Walltime}{Median}{5.249336600305}%
\StoreBenchExecResult{PdrInv}{KinductionDfStaticEightTwoT}{Correct}{}{Walltime}{Min}{1.64178109169}%
\StoreBenchExecResult{PdrInv}{KinductionDfStaticEightTwoT}{Correct}{}{Walltime}{Max}{867.906393051}%
\StoreBenchExecResult{PdrInv}{KinductionDfStaticEightTwoT}{Correct}{}{Walltime}{Stdev}{99.73574208135708460476096801}%
\StoreBenchExecResult{PdrInv}{KinductionDfStaticEightTwoT}{Correct}{False}{Count}{}{788}%
\StoreBenchExecResult{PdrInv}{KinductionDfStaticEightTwoT}{Correct}{False}{Cputime}{}{74340.962064144}%
\StoreBenchExecResult{PdrInv}{KinductionDfStaticEightTwoT}{Correct}{False}{Cputime}{Avg}{94.34132241642639593908629442}%
\StoreBenchExecResult{PdrInv}{KinductionDfStaticEightTwoT}{Correct}{False}{Cputime}{Median}{22.3636106285}%
\StoreBenchExecResult{PdrInv}{KinductionDfStaticEightTwoT}{Correct}{False}{Cputime}{Min}{3.095316337}%
\StoreBenchExecResult{PdrInv}{KinductionDfStaticEightTwoT}{Correct}{False}{Cputime}{Max}{896.232896503}%
\StoreBenchExecResult{PdrInv}{KinductionDfStaticEightTwoT}{Correct}{False}{Cputime}{Stdev}{186.3334938091157496137996914}%
\StoreBenchExecResult{PdrInv}{KinductionDfStaticEightTwoT}{Correct}{False}{Walltime}{}{51091.01714348764}%
\StoreBenchExecResult{PdrInv}{KinductionDfStaticEightTwoT}{Correct}{False}{Walltime}{Avg}{64.83631617193862944162436548}%
\StoreBenchExecResult{PdrInv}{KinductionDfStaticEightTwoT}{Correct}{False}{Walltime}{Median}{12.1339185238}%
\StoreBenchExecResult{PdrInv}{KinductionDfStaticEightTwoT}{Correct}{False}{Walltime}{Min}{1.73207998276}%
\StoreBenchExecResult{PdrInv}{KinductionDfStaticEightTwoT}{Correct}{False}{Walltime}{Max}{867.906393051}%
\StoreBenchExecResult{PdrInv}{KinductionDfStaticEightTwoT}{Correct}{False}{Walltime}{Stdev}{147.1718776338455548264720238}%
\StoreBenchExecResult{PdrInv}{KinductionDfStaticEightTwoT}{Correct}{True}{Count}{}{2296}%
\StoreBenchExecResult{PdrInv}{KinductionDfStaticEightTwoT}{Correct}{True}{Cputime}{}{120255.139810441}%
\StoreBenchExecResult{PdrInv}{KinductionDfStaticEightTwoT}{Correct}{True}{Cputime}{Avg}{52.37593197318858885017421603}%
\StoreBenchExecResult{PdrInv}{KinductionDfStaticEightTwoT}{Correct}{True}{Cputime}{Median}{8.044652705}%
\StoreBenchExecResult{PdrInv}{KinductionDfStaticEightTwoT}{Correct}{True}{Cputime}{Min}{2.958535013}%
\StoreBenchExecResult{PdrInv}{KinductionDfStaticEightTwoT}{Correct}{True}{Cputime}{Max}{890.091335986}%
\StoreBenchExecResult{PdrInv}{KinductionDfStaticEightTwoT}{Correct}{True}{Cputime}{Stdev}{117.2115136586916609617919118}%
\StoreBenchExecResult{PdrInv}{KinductionDfStaticEightTwoT}{Correct}{True}{Walltime}{}{67359.04033803462}%
\StoreBenchExecResult{PdrInv}{KinductionDfStaticEightTwoT}{Correct}{True}{Walltime}{Avg}{29.33756112283737804878048780}%
\StoreBenchExecResult{PdrInv}{KinductionDfStaticEightTwoT}{Correct}{True}{Walltime}{Median}{4.255840063095}%
\StoreBenchExecResult{PdrInv}{KinductionDfStaticEightTwoT}{Correct}{True}{Walltime}{Min}{1.64178109169}%
\StoreBenchExecResult{PdrInv}{KinductionDfStaticEightTwoT}{Correct}{True}{Walltime}{Max}{827.768306017}%
\StoreBenchExecResult{PdrInv}{KinductionDfStaticEightTwoT}{Correct}{True}{Walltime}{Stdev}{74.86981974016746056754584900}%
\StoreBenchExecResult{PdrInv}{KinductionDfStaticEightTwoT}{Wrong}{True}{Count}{}{0}%
\StoreBenchExecResult{PdrInv}{KinductionDfStaticEightTwoT}{Wrong}{True}{Cputime}{}{0}%
\StoreBenchExecResult{PdrInv}{KinductionDfStaticEightTwoT}{Wrong}{True}{Cputime}{Avg}{None}%
\StoreBenchExecResult{PdrInv}{KinductionDfStaticEightTwoT}{Wrong}{True}{Cputime}{Median}{None}%
\StoreBenchExecResult{PdrInv}{KinductionDfStaticEightTwoT}{Wrong}{True}{Cputime}{Min}{None}%
\StoreBenchExecResult{PdrInv}{KinductionDfStaticEightTwoT}{Wrong}{True}{Cputime}{Max}{None}%
\StoreBenchExecResult{PdrInv}{KinductionDfStaticEightTwoT}{Wrong}{True}{Cputime}{Stdev}{None}%
\StoreBenchExecResult{PdrInv}{KinductionDfStaticEightTwoT}{Wrong}{True}{Walltime}{}{0}%
\StoreBenchExecResult{PdrInv}{KinductionDfStaticEightTwoT}{Wrong}{True}{Walltime}{Avg}{None}%
\StoreBenchExecResult{PdrInv}{KinductionDfStaticEightTwoT}{Wrong}{True}{Walltime}{Median}{None}%
\StoreBenchExecResult{PdrInv}{KinductionDfStaticEightTwoT}{Wrong}{True}{Walltime}{Min}{None}%
\StoreBenchExecResult{PdrInv}{KinductionDfStaticEightTwoT}{Wrong}{True}{Walltime}{Max}{None}%
\StoreBenchExecResult{PdrInv}{KinductionDfStaticEightTwoT}{Wrong}{True}{Walltime}{Stdev}{None}%
\StoreBenchExecResult{PdrInv}{KinductionDfStaticEightTwoT}{Error}{}{Count}{}{2505}%
\StoreBenchExecResult{PdrInv}{KinductionDfStaticEightTwoT}{Error}{}{Cputime}{}{2001528.628832067}%
\StoreBenchExecResult{PdrInv}{KinductionDfStaticEightTwoT}{Error}{}{Cputime}{Avg}{799.0134246834598802395209581}%
\StoreBenchExecResult{PdrInv}{KinductionDfStaticEightTwoT}{Error}{}{Cputime}{Median}{901.708126507}%
\StoreBenchExecResult{PdrInv}{KinductionDfStaticEightTwoT}{Error}{}{Cputime}{Min}{2.502797224}%
\StoreBenchExecResult{PdrInv}{KinductionDfStaticEightTwoT}{Error}{}{Cputime}{Max}{1002.31487952}%
\StoreBenchExecResult{PdrInv}{KinductionDfStaticEightTwoT}{Error}{}{Cputime}{Stdev}{252.0269198791784842200735097}%
\StoreBenchExecResult{PdrInv}{KinductionDfStaticEightTwoT}{Error}{}{Walltime}{}{1302198.03992795936}%
\StoreBenchExecResult{PdrInv}{KinductionDfStaticEightTwoT}{Error}{}{Walltime}{Avg}{519.8395368973889660678642715}%
\StoreBenchExecResult{PdrInv}{KinductionDfStaticEightTwoT}{Error}{}{Walltime}{Median}{453.995809078}%
\StoreBenchExecResult{PdrInv}{KinductionDfStaticEightTwoT}{Error}{}{Walltime}{Min}{1.36538696289}%
\StoreBenchExecResult{PdrInv}{KinductionDfStaticEightTwoT}{Error}{}{Walltime}{Max}{979.662005901}%
\StoreBenchExecResult{PdrInv}{KinductionDfStaticEightTwoT}{Error}{}{Walltime}{Stdev}{242.7055693201915622712236044}%
\StoreBenchExecResult{PdrInv}{KinductionDfStaticEightTwoT}{Error}{Assertion}{Count}{}{4}%
\StoreBenchExecResult{PdrInv}{KinductionDfStaticEightTwoT}{Error}{Assertion}{Cputime}{}{14.791480667}%
\StoreBenchExecResult{PdrInv}{KinductionDfStaticEightTwoT}{Error}{Assertion}{Cputime}{Avg}{3.69787016675}%
\StoreBenchExecResult{PdrInv}{KinductionDfStaticEightTwoT}{Error}{Assertion}{Cputime}{Median}{3.6794155785}%
\StoreBenchExecResult{PdrInv}{KinductionDfStaticEightTwoT}{Error}{Assertion}{Cputime}{Min}{3.231676692}%
\StoreBenchExecResult{PdrInv}{KinductionDfStaticEightTwoT}{Error}{Assertion}{Cputime}{Max}{4.200972818}%
\StoreBenchExecResult{PdrInv}{KinductionDfStaticEightTwoT}{Error}{Assertion}{Cputime}{Stdev}{0.3615692948947255776744224036}%
\StoreBenchExecResult{PdrInv}{KinductionDfStaticEightTwoT}{Error}{Assertion}{Walltime}{}{8.24579715729}%
\StoreBenchExecResult{PdrInv}{KinductionDfStaticEightTwoT}{Error}{Assertion}{Walltime}{Avg}{2.0614492893225}%
\StoreBenchExecResult{PdrInv}{KinductionDfStaticEightTwoT}{Error}{Assertion}{Walltime}{Median}{2.068484544755}%
\StoreBenchExecResult{PdrInv}{KinductionDfStaticEightTwoT}{Error}{Assertion}{Walltime}{Min}{1.79999709129}%
\StoreBenchExecResult{PdrInv}{KinductionDfStaticEightTwoT}{Error}{Assertion}{Walltime}{Max}{2.30883097649}%
\StoreBenchExecResult{PdrInv}{KinductionDfStaticEightTwoT}{Error}{Assertion}{Walltime}{Stdev}{0.1872438020482402828619171734}%
\StoreBenchExecResult{PdrInv}{KinductionDfStaticEightTwoT}{Error}{Error}{Count}{}{193}%
\StoreBenchExecResult{PdrInv}{KinductionDfStaticEightTwoT}{Error}{Error}{Cputime}{}{34527.717594068}%
\StoreBenchExecResult{PdrInv}{KinductionDfStaticEightTwoT}{Error}{Error}{Cputime}{Avg}{178.9000911609740932642487047}%
\StoreBenchExecResult{PdrInv}{KinductionDfStaticEightTwoT}{Error}{Error}{Cputime}{Median}{113.203940359}%
\StoreBenchExecResult{PdrInv}{KinductionDfStaticEightTwoT}{Error}{Error}{Cputime}{Min}{2.502797224}%
\StoreBenchExecResult{PdrInv}{KinductionDfStaticEightTwoT}{Error}{Error}{Cputime}{Max}{794.256551189}%
\StoreBenchExecResult{PdrInv}{KinductionDfStaticEightTwoT}{Error}{Error}{Cputime}{Stdev}{186.6019902651484391674633569}%
\StoreBenchExecResult{PdrInv}{KinductionDfStaticEightTwoT}{Error}{Error}{Walltime}{}{28880.69359755201}%
\StoreBenchExecResult{PdrInv}{KinductionDfStaticEightTwoT}{Error}{Error}{Walltime}{Avg}{149.6408994691813989637305699}%
\StoreBenchExecResult{PdrInv}{KinductionDfStaticEightTwoT}{Error}{Error}{Walltime}{Median}{91.1960031986}%
\StoreBenchExecResult{PdrInv}{KinductionDfStaticEightTwoT}{Error}{Error}{Walltime}{Min}{1.36538696289}%
\StoreBenchExecResult{PdrInv}{KinductionDfStaticEightTwoT}{Error}{Error}{Walltime}{Max}{778.652194023}%
\StoreBenchExecResult{PdrInv}{KinductionDfStaticEightTwoT}{Error}{Error}{Walltime}{Stdev}{166.1254330062943349579332038}%
\StoreBenchExecResult{PdrInv}{KinductionDfStaticEightTwoT}{Error}{Exception}{Count}{}{12}%
\StoreBenchExecResult{PdrInv}{KinductionDfStaticEightTwoT}{Error}{Exception}{Cputime}{}{1426.221680367}%
\StoreBenchExecResult{PdrInv}{KinductionDfStaticEightTwoT}{Error}{Exception}{Cputime}{Avg}{118.85180669725}%
\StoreBenchExecResult{PdrInv}{KinductionDfStaticEightTwoT}{Error}{Exception}{Cputime}{Median}{80.770924803}%
\StoreBenchExecResult{PdrInv}{KinductionDfStaticEightTwoT}{Error}{Exception}{Cputime}{Min}{14.587210578}%
\StoreBenchExecResult{PdrInv}{KinductionDfStaticEightTwoT}{Error}{Exception}{Cputime}{Max}{441.781428059}%
\StoreBenchExecResult{PdrInv}{KinductionDfStaticEightTwoT}{Error}{Exception}{Cputime}{Stdev}{125.4474392675971912924374807}%
\StoreBenchExecResult{PdrInv}{KinductionDfStaticEightTwoT}{Error}{Exception}{Walltime}{}{756.97566390066}%
\StoreBenchExecResult{PdrInv}{KinductionDfStaticEightTwoT}{Error}{Exception}{Walltime}{Avg}{63.081305325055}%
\StoreBenchExecResult{PdrInv}{KinductionDfStaticEightTwoT}{Error}{Exception}{Walltime}{Median}{40.7244694233}%
\StoreBenchExecResult{PdrInv}{KinductionDfStaticEightTwoT}{Error}{Exception}{Walltime}{Min}{7.51351094246}%
\StoreBenchExecResult{PdrInv}{KinductionDfStaticEightTwoT}{Error}{Exception}{Walltime}{Max}{253.052404881}%
\StoreBenchExecResult{PdrInv}{KinductionDfStaticEightTwoT}{Error}{Exception}{Walltime}{Stdev}{69.71279629496273431341327904}%
\StoreBenchExecResult{PdrInv}{KinductionDfStaticEightTwoT}{Error}{OutOfJavaMemory}{Count}{}{9}%
\StoreBenchExecResult{PdrInv}{KinductionDfStaticEightTwoT}{Error}{OutOfJavaMemory}{Cputime}{}{4295.322274724}%
\StoreBenchExecResult{PdrInv}{KinductionDfStaticEightTwoT}{Error}{OutOfJavaMemory}{Cputime}{Avg}{477.2580305248888888888888889}%
\StoreBenchExecResult{PdrInv}{KinductionDfStaticEightTwoT}{Error}{OutOfJavaMemory}{Cputime}{Median}{472.133994606}%
\StoreBenchExecResult{PdrInv}{KinductionDfStaticEightTwoT}{Error}{OutOfJavaMemory}{Cputime}{Min}{189.622584203}%
\StoreBenchExecResult{PdrInv}{KinductionDfStaticEightTwoT}{Error}{OutOfJavaMemory}{Cputime}{Max}{696.759175759}%
\StoreBenchExecResult{PdrInv}{KinductionDfStaticEightTwoT}{Error}{OutOfJavaMemory}{Cputime}{Stdev}{163.7194953981294087855761581}%
\StoreBenchExecResult{PdrInv}{KinductionDfStaticEightTwoT}{Error}{OutOfJavaMemory}{Walltime}{}{2537.254302025}%
\StoreBenchExecResult{PdrInv}{KinductionDfStaticEightTwoT}{Error}{OutOfJavaMemory}{Walltime}{Avg}{281.9171446694444444444444444}%
\StoreBenchExecResult{PdrInv}{KinductionDfStaticEightTwoT}{Error}{OutOfJavaMemory}{Walltime}{Median}{244.924804926}%
\StoreBenchExecResult{PdrInv}{KinductionDfStaticEightTwoT}{Error}{OutOfJavaMemory}{Walltime}{Min}{113.677829027}%
\StoreBenchExecResult{PdrInv}{KinductionDfStaticEightTwoT}{Error}{OutOfJavaMemory}{Walltime}{Max}{460.266738892}%
\StoreBenchExecResult{PdrInv}{KinductionDfStaticEightTwoT}{Error}{OutOfJavaMemory}{Walltime}{Stdev}{104.8716442265648176709636023}%
\StoreBenchExecResult{PdrInv}{KinductionDfStaticEightTwoT}{Error}{OutOfMemory}{Count}{}{260}%
\StoreBenchExecResult{PdrInv}{KinductionDfStaticEightTwoT}{Error}{OutOfMemory}{Cputime}{}{120710.447601292}%
\StoreBenchExecResult{PdrInv}{KinductionDfStaticEightTwoT}{Error}{OutOfMemory}{Cputime}{Avg}{464.2709523126615384615384615}%
\StoreBenchExecResult{PdrInv}{KinductionDfStaticEightTwoT}{Error}{OutOfMemory}{Cputime}{Median}{387.8615494835}%
\StoreBenchExecResult{PdrInv}{KinductionDfStaticEightTwoT}{Error}{OutOfMemory}{Cputime}{Min}{154.990550731}%
\StoreBenchExecResult{PdrInv}{KinductionDfStaticEightTwoT}{Error}{OutOfMemory}{Cputime}{Max}{899.007156656}%
\StoreBenchExecResult{PdrInv}{KinductionDfStaticEightTwoT}{Error}{OutOfMemory}{Cputime}{Stdev}{238.5244597698032553861137109}%
\StoreBenchExecResult{PdrInv}{KinductionDfStaticEightTwoT}{Error}{OutOfMemory}{Walltime}{}{65279.0576193434}%
\StoreBenchExecResult{PdrInv}{KinductionDfStaticEightTwoT}{Error}{OutOfMemory}{Walltime}{Avg}{251.0732985359361538461538462}%
\StoreBenchExecResult{PdrInv}{KinductionDfStaticEightTwoT}{Error}{OutOfMemory}{Walltime}{Median}{194.6069840195}%
\StoreBenchExecResult{PdrInv}{KinductionDfStaticEightTwoT}{Error}{OutOfMemory}{Walltime}{Min}{84.0819849968}%
\StoreBenchExecResult{PdrInv}{KinductionDfStaticEightTwoT}{Error}{OutOfMemory}{Walltime}{Max}{850.07557106}%
\StoreBenchExecResult{PdrInv}{KinductionDfStaticEightTwoT}{Error}{OutOfMemory}{Walltime}{Stdev}{156.8807536612681923731029980}%
\StoreBenchExecResult{PdrInv}{KinductionDfStaticEightTwoT}{Error}{Timeout}{Count}{}{2027}%
\StoreBenchExecResult{PdrInv}{KinductionDfStaticEightTwoT}{Error}{Timeout}{Cputime}{}{1840554.128200949}%
\StoreBenchExecResult{PdrInv}{KinductionDfStaticEightTwoT}{Error}{Timeout}{Cputime}{Avg}{908.0188101632703502713369512}%
\StoreBenchExecResult{PdrInv}{KinductionDfStaticEightTwoT}{Error}{Timeout}{Cputime}{Median}{902.255944045}%
\StoreBenchExecResult{PdrInv}{KinductionDfStaticEightTwoT}{Error}{Timeout}{Cputime}{Min}{900.428906238}%
\StoreBenchExecResult{PdrInv}{KinductionDfStaticEightTwoT}{Error}{Timeout}{Cputime}{Max}{1002.31487952}%
\StoreBenchExecResult{PdrInv}{KinductionDfStaticEightTwoT}{Error}{Timeout}{Cputime}{Stdev}{18.85024521732988234316715227}%
\StoreBenchExecResult{PdrInv}{KinductionDfStaticEightTwoT}{Error}{Timeout}{Walltime}{}{1204735.812947981}%
\StoreBenchExecResult{PdrInv}{KinductionDfStaticEightTwoT}{Error}{Timeout}{Walltime}{Avg}{594.3442589777903305377405032}%
\StoreBenchExecResult{PdrInv}{KinductionDfStaticEightTwoT}{Error}{Timeout}{Walltime}{Median}{456.286587}%
\StoreBenchExecResult{PdrInv}{KinductionDfStaticEightTwoT}{Error}{Timeout}{Walltime}{Min}{451.098812819}%
\StoreBenchExecResult{PdrInv}{KinductionDfStaticEightTwoT}{Error}{Timeout}{Walltime}{Max}{979.662005901}%
\StoreBenchExecResult{PdrInv}{KinductionDfStaticEightTwoT}{Error}{Timeout}{Walltime}{Stdev}{192.4941362118159911517353245}%
\StoreBenchExecResult{PdrInv}{KinductionDfStaticEightTwoT}{Wrong}{}{Count}{}{2}%
\StoreBenchExecResult{PdrInv}{KinductionDfStaticEightTwoT}{Wrong}{}{Cputime}{}{24.748890556}%
\StoreBenchExecResult{PdrInv}{KinductionDfStaticEightTwoT}{Wrong}{}{Cputime}{Avg}{12.374445278}%
\StoreBenchExecResult{PdrInv}{KinductionDfStaticEightTwoT}{Wrong}{}{Cputime}{Median}{12.374445278}%
\StoreBenchExecResult{PdrInv}{KinductionDfStaticEightTwoT}{Wrong}{}{Cputime}{Min}{4.868655329}%
\StoreBenchExecResult{PdrInv}{KinductionDfStaticEightTwoT}{Wrong}{}{Cputime}{Max}{19.880235227}%
\StoreBenchExecResult{PdrInv}{KinductionDfStaticEightTwoT}{Wrong}{}{Cputime}{Stdev}{7.505789949}%
\StoreBenchExecResult{PdrInv}{KinductionDfStaticEightTwoT}{Wrong}{}{Walltime}{}{13.34114027025}%
\StoreBenchExecResult{PdrInv}{KinductionDfStaticEightTwoT}{Wrong}{}{Walltime}{Avg}{6.670570135125}%
\StoreBenchExecResult{PdrInv}{KinductionDfStaticEightTwoT}{Wrong}{}{Walltime}{Median}{6.670570135125}%
\StoreBenchExecResult{PdrInv}{KinductionDfStaticEightTwoT}{Wrong}{}{Walltime}{Min}{2.67388010025}%
\StoreBenchExecResult{PdrInv}{KinductionDfStaticEightTwoT}{Wrong}{}{Walltime}{Max}{10.66726017}%
\StoreBenchExecResult{PdrInv}{KinductionDfStaticEightTwoT}{Wrong}{}{Walltime}{Stdev}{3.996690034875}%
\StoreBenchExecResult{PdrInv}{KinductionDfStaticEightTwoT}{Wrong}{False}{Count}{}{2}%
\StoreBenchExecResult{PdrInv}{KinductionDfStaticEightTwoT}{Wrong}{False}{Cputime}{}{24.748890556}%
\StoreBenchExecResult{PdrInv}{KinductionDfStaticEightTwoT}{Wrong}{False}{Cputime}{Avg}{12.374445278}%
\StoreBenchExecResult{PdrInv}{KinductionDfStaticEightTwoT}{Wrong}{False}{Cputime}{Median}{12.374445278}%
\StoreBenchExecResult{PdrInv}{KinductionDfStaticEightTwoT}{Wrong}{False}{Cputime}{Min}{4.868655329}%
\StoreBenchExecResult{PdrInv}{KinductionDfStaticEightTwoT}{Wrong}{False}{Cputime}{Max}{19.880235227}%
\StoreBenchExecResult{PdrInv}{KinductionDfStaticEightTwoT}{Wrong}{False}{Cputime}{Stdev}{7.505789949}%
\StoreBenchExecResult{PdrInv}{KinductionDfStaticEightTwoT}{Wrong}{False}{Walltime}{}{13.34114027025}%
\StoreBenchExecResult{PdrInv}{KinductionDfStaticEightTwoT}{Wrong}{False}{Walltime}{Avg}{6.670570135125}%
\StoreBenchExecResult{PdrInv}{KinductionDfStaticEightTwoT}{Wrong}{False}{Walltime}{Median}{6.670570135125}%
\StoreBenchExecResult{PdrInv}{KinductionDfStaticEightTwoT}{Wrong}{False}{Walltime}{Min}{2.67388010025}%
\StoreBenchExecResult{PdrInv}{KinductionDfStaticEightTwoT}{Wrong}{False}{Walltime}{Max}{10.66726017}%
\StoreBenchExecResult{PdrInv}{KinductionDfStaticEightTwoT}{Wrong}{False}{Walltime}{Stdev}{3.996690034875}%
\providecommand\StoreBenchExecResult[7]{\expandafter\newcommand\csname#1#2#3#4#5#6\endcsname{#7}}%
\StoreBenchExecResult{PdrInv}{KinductionDfTrueNotSolvedByKinductionPlainButKipdr}{Total}{}{Count}{}{449}%
\StoreBenchExecResult{PdrInv}{KinductionDfTrueNotSolvedByKinductionPlainButKipdr}{Total}{}{Cputime}{}{7618.969198167}%
\StoreBenchExecResult{PdrInv}{KinductionDfTrueNotSolvedByKinductionPlainButKipdr}{Total}{}{Cputime}{Avg}{16.96875099814476614699331849}%
\StoreBenchExecResult{PdrInv}{KinductionDfTrueNotSolvedByKinductionPlainButKipdr}{Total}{}{Cputime}{Median}{5.828812832}%
\StoreBenchExecResult{PdrInv}{KinductionDfTrueNotSolvedByKinductionPlainButKipdr}{Total}{}{Cputime}{Min}{3.318034525}%
\StoreBenchExecResult{PdrInv}{KinductionDfTrueNotSolvedByKinductionPlainButKipdr}{Total}{}{Cputime}{Max}{905.201109481}%
\StoreBenchExecResult{PdrInv}{KinductionDfTrueNotSolvedByKinductionPlainButKipdr}{Total}{}{Cputime}{Stdev}{94.30936579071097349378233008}%
\StoreBenchExecResult{PdrInv}{KinductionDfTrueNotSolvedByKinductionPlainButKipdr}{Total}{}{Walltime}{}{3905.66689348265}%
\StoreBenchExecResult{PdrInv}{KinductionDfTrueNotSolvedByKinductionPlainButKipdr}{Total}{}{Walltime}{Avg}{8.698589963213028953229398664}%
\StoreBenchExecResult{PdrInv}{KinductionDfTrueNotSolvedByKinductionPlainButKipdr}{Total}{}{Walltime}{Median}{3.08683681488}%
\StoreBenchExecResult{PdrInv}{KinductionDfTrueNotSolvedByKinductionPlainButKipdr}{Total}{}{Walltime}{Min}{1.81842899323}%
\StoreBenchExecResult{PdrInv}{KinductionDfTrueNotSolvedByKinductionPlainButKipdr}{Total}{}{Walltime}{Max}{455.971984148}%
\StoreBenchExecResult{PdrInv}{KinductionDfTrueNotSolvedByKinductionPlainButKipdr}{Total}{}{Walltime}{Stdev}{47.37483516002843883513989991}%
\StoreBenchExecResult{PdrInv}{KinductionDfTrueNotSolvedByKinductionPlainButKipdr}{Correct}{}{Count}{}{444}%
\StoreBenchExecResult{PdrInv}{KinductionDfTrueNotSolvedByKinductionPlainButKipdr}{Correct}{}{Cputime}{}{3100.003927112}%
\StoreBenchExecResult{PdrInv}{KinductionDfTrueNotSolvedByKinductionPlainButKipdr}{Correct}{}{Cputime}{Avg}{6.981990826828828828828828829}%
\StoreBenchExecResult{PdrInv}{KinductionDfTrueNotSolvedByKinductionPlainButKipdr}{Correct}{}{Cputime}{Median}{5.7965014135}%
\StoreBenchExecResult{PdrInv}{KinductionDfTrueNotSolvedByKinductionPlainButKipdr}{Correct}{}{Cputime}{Min}{3.318034525}%
\StoreBenchExecResult{PdrInv}{KinductionDfTrueNotSolvedByKinductionPlainButKipdr}{Correct}{}{Cputime}{Max}{75.890612022}%
\StoreBenchExecResult{PdrInv}{KinductionDfTrueNotSolvedByKinductionPlainButKipdr}{Correct}{}{Cputime}{Stdev}{6.177208362615215328964496607}%
\StoreBenchExecResult{PdrInv}{KinductionDfTrueNotSolvedByKinductionPlainButKipdr}{Correct}{}{Walltime}{}{1634.81321454065}%
\StoreBenchExecResult{PdrInv}{KinductionDfTrueNotSolvedByKinductionPlainButKipdr}{Correct}{}{Walltime}{Avg}{3.682011744460923423423423423}%
\StoreBenchExecResult{PdrInv}{KinductionDfTrueNotSolvedByKinductionPlainButKipdr}{Correct}{}{Walltime}{Median}{3.072164416315}%
\StoreBenchExecResult{PdrInv}{KinductionDfTrueNotSolvedByKinductionPlainButKipdr}{Correct}{}{Walltime}{Min}{1.81842899323}%
\StoreBenchExecResult{PdrInv}{KinductionDfTrueNotSolvedByKinductionPlainButKipdr}{Correct}{}{Walltime}{Max}{38.4671201706}%
\StoreBenchExecResult{PdrInv}{KinductionDfTrueNotSolvedByKinductionPlainButKipdr}{Correct}{}{Walltime}{Stdev}{3.116305077045772264511415604}%
\StoreBenchExecResult{PdrInv}{KinductionDfTrueNotSolvedByKinductionPlainButKipdr}{Correct}{True}{Count}{}{444}%
\StoreBenchExecResult{PdrInv}{KinductionDfTrueNotSolvedByKinductionPlainButKipdr}{Correct}{True}{Cputime}{}{3100.003927112}%
\StoreBenchExecResult{PdrInv}{KinductionDfTrueNotSolvedByKinductionPlainButKipdr}{Correct}{True}{Cputime}{Avg}{6.981990826828828828828828829}%
\StoreBenchExecResult{PdrInv}{KinductionDfTrueNotSolvedByKinductionPlainButKipdr}{Correct}{True}{Cputime}{Median}{5.7965014135}%
\StoreBenchExecResult{PdrInv}{KinductionDfTrueNotSolvedByKinductionPlainButKipdr}{Correct}{True}{Cputime}{Min}{3.318034525}%
\StoreBenchExecResult{PdrInv}{KinductionDfTrueNotSolvedByKinductionPlainButKipdr}{Correct}{True}{Cputime}{Max}{75.890612022}%
\StoreBenchExecResult{PdrInv}{KinductionDfTrueNotSolvedByKinductionPlainButKipdr}{Correct}{True}{Cputime}{Stdev}{6.177208362615215328964496607}%
\StoreBenchExecResult{PdrInv}{KinductionDfTrueNotSolvedByKinductionPlainButKipdr}{Correct}{True}{Walltime}{}{1634.81321454065}%
\StoreBenchExecResult{PdrInv}{KinductionDfTrueNotSolvedByKinductionPlainButKipdr}{Correct}{True}{Walltime}{Avg}{3.682011744460923423423423423}%
\StoreBenchExecResult{PdrInv}{KinductionDfTrueNotSolvedByKinductionPlainButKipdr}{Correct}{True}{Walltime}{Median}{3.072164416315}%
\StoreBenchExecResult{PdrInv}{KinductionDfTrueNotSolvedByKinductionPlainButKipdr}{Correct}{True}{Walltime}{Min}{1.81842899323}%
\StoreBenchExecResult{PdrInv}{KinductionDfTrueNotSolvedByKinductionPlainButKipdr}{Correct}{True}{Walltime}{Max}{38.4671201706}%
\StoreBenchExecResult{PdrInv}{KinductionDfTrueNotSolvedByKinductionPlainButKipdr}{Correct}{True}{Walltime}{Stdev}{3.116305077045772264511415604}%
\StoreBenchExecResult{PdrInv}{KinductionDfTrueNotSolvedByKinductionPlainButKipdr}{Wrong}{True}{Count}{}{0}%
\StoreBenchExecResult{PdrInv}{KinductionDfTrueNotSolvedByKinductionPlainButKipdr}{Wrong}{True}{Cputime}{}{0}%
\StoreBenchExecResult{PdrInv}{KinductionDfTrueNotSolvedByKinductionPlainButKipdr}{Wrong}{True}{Cputime}{Avg}{None}%
\StoreBenchExecResult{PdrInv}{KinductionDfTrueNotSolvedByKinductionPlainButKipdr}{Wrong}{True}{Cputime}{Median}{None}%
\StoreBenchExecResult{PdrInv}{KinductionDfTrueNotSolvedByKinductionPlainButKipdr}{Wrong}{True}{Cputime}{Min}{None}%
\StoreBenchExecResult{PdrInv}{KinductionDfTrueNotSolvedByKinductionPlainButKipdr}{Wrong}{True}{Cputime}{Max}{None}%
\StoreBenchExecResult{PdrInv}{KinductionDfTrueNotSolvedByKinductionPlainButKipdr}{Wrong}{True}{Cputime}{Stdev}{None}%
\StoreBenchExecResult{PdrInv}{KinductionDfTrueNotSolvedByKinductionPlainButKipdr}{Wrong}{True}{Walltime}{}{0}%
\StoreBenchExecResult{PdrInv}{KinductionDfTrueNotSolvedByKinductionPlainButKipdr}{Wrong}{True}{Walltime}{Avg}{None}%
\StoreBenchExecResult{PdrInv}{KinductionDfTrueNotSolvedByKinductionPlainButKipdr}{Wrong}{True}{Walltime}{Median}{None}%
\StoreBenchExecResult{PdrInv}{KinductionDfTrueNotSolvedByKinductionPlainButKipdr}{Wrong}{True}{Walltime}{Min}{None}%
\StoreBenchExecResult{PdrInv}{KinductionDfTrueNotSolvedByKinductionPlainButKipdr}{Wrong}{True}{Walltime}{Max}{None}%
\StoreBenchExecResult{PdrInv}{KinductionDfTrueNotSolvedByKinductionPlainButKipdr}{Wrong}{True}{Walltime}{Stdev}{None}%
\StoreBenchExecResult{PdrInv}{KinductionDfTrueNotSolvedByKinductionPlainButKipdr}{Error}{}{Count}{}{5}%
\StoreBenchExecResult{PdrInv}{KinductionDfTrueNotSolvedByKinductionPlainButKipdr}{Error}{}{Cputime}{}{4518.965271055}%
\StoreBenchExecResult{PdrInv}{KinductionDfTrueNotSolvedByKinductionPlainButKipdr}{Error}{}{Cputime}{Avg}{903.793054211}%
\StoreBenchExecResult{PdrInv}{KinductionDfTrueNotSolvedByKinductionPlainButKipdr}{Error}{}{Cputime}{Median}{904.58514465}%
\StoreBenchExecResult{PdrInv}{KinductionDfTrueNotSolvedByKinductionPlainButKipdr}{Error}{}{Cputime}{Min}{901.846845232}%
\StoreBenchExecResult{PdrInv}{KinductionDfTrueNotSolvedByKinductionPlainButKipdr}{Error}{}{Cputime}{Max}{905.201109481}%
\StoreBenchExecResult{PdrInv}{KinductionDfTrueNotSolvedByKinductionPlainButKipdr}{Error}{}{Cputime}{Stdev}{1.401189336311343766144343914}%
\StoreBenchExecResult{PdrInv}{KinductionDfTrueNotSolvedByKinductionPlainButKipdr}{Error}{}{Walltime}{}{2270.853678942}%
\StoreBenchExecResult{PdrInv}{KinductionDfTrueNotSolvedByKinductionPlainButKipdr}{Error}{}{Walltime}{Avg}{454.1707357884}%
\StoreBenchExecResult{PdrInv}{KinductionDfTrueNotSolvedByKinductionPlainButKipdr}{Error}{}{Walltime}{Median}{454.964095831}%
\StoreBenchExecResult{PdrInv}{KinductionDfTrueNotSolvedByKinductionPlainButKipdr}{Error}{}{Walltime}{Min}{452.335769892}%
\StoreBenchExecResult{PdrInv}{KinductionDfTrueNotSolvedByKinductionPlainButKipdr}{Error}{}{Walltime}{Max}{455.971984148}%
\StoreBenchExecResult{PdrInv}{KinductionDfTrueNotSolvedByKinductionPlainButKipdr}{Error}{}{Walltime}{Stdev}{1.525934862944779930765940808}%
\StoreBenchExecResult{PdrInv}{KinductionDfTrueNotSolvedByKinductionPlainButKipdr}{Error}{Timeout}{Count}{}{5}%
\StoreBenchExecResult{PdrInv}{KinductionDfTrueNotSolvedByKinductionPlainButKipdr}{Error}{Timeout}{Cputime}{}{4518.965271055}%
\StoreBenchExecResult{PdrInv}{KinductionDfTrueNotSolvedByKinductionPlainButKipdr}{Error}{Timeout}{Cputime}{Avg}{903.793054211}%
\StoreBenchExecResult{PdrInv}{KinductionDfTrueNotSolvedByKinductionPlainButKipdr}{Error}{Timeout}{Cputime}{Median}{904.58514465}%
\StoreBenchExecResult{PdrInv}{KinductionDfTrueNotSolvedByKinductionPlainButKipdr}{Error}{Timeout}{Cputime}{Min}{901.846845232}%
\StoreBenchExecResult{PdrInv}{KinductionDfTrueNotSolvedByKinductionPlainButKipdr}{Error}{Timeout}{Cputime}{Max}{905.201109481}%
\StoreBenchExecResult{PdrInv}{KinductionDfTrueNotSolvedByKinductionPlainButKipdr}{Error}{Timeout}{Cputime}{Stdev}{1.401189336311343766144343914}%
\StoreBenchExecResult{PdrInv}{KinductionDfTrueNotSolvedByKinductionPlainButKipdr}{Error}{Timeout}{Walltime}{}{2270.853678942}%
\StoreBenchExecResult{PdrInv}{KinductionDfTrueNotSolvedByKinductionPlainButKipdr}{Error}{Timeout}{Walltime}{Avg}{454.1707357884}%
\StoreBenchExecResult{PdrInv}{KinductionDfTrueNotSolvedByKinductionPlainButKipdr}{Error}{Timeout}{Walltime}{Median}{454.964095831}%
\StoreBenchExecResult{PdrInv}{KinductionDfTrueNotSolvedByKinductionPlainButKipdr}{Error}{Timeout}{Walltime}{Min}{452.335769892}%
\StoreBenchExecResult{PdrInv}{KinductionDfTrueNotSolvedByKinductionPlainButKipdr}{Error}{Timeout}{Walltime}{Max}{455.971984148}%
\StoreBenchExecResult{PdrInv}{KinductionDfTrueNotSolvedByKinductionPlainButKipdr}{Error}{Timeout}{Walltime}{Stdev}{1.525934862944779930765940808}%
\providecommand\StoreBenchExecResult[7]{\expandafter\newcommand\csname#1#2#3#4#5#6\endcsname{#7}}%
\StoreBenchExecResult{PdrInv}{KinductionDfTrueNotSolvedByKinductionPlain}{Total}{}{Count}{}{2893}%
\StoreBenchExecResult{PdrInv}{KinductionDfTrueNotSolvedByKinductionPlain}{Total}{}{Cputime}{}{1524526.305246742}%
\StoreBenchExecResult{PdrInv}{KinductionDfTrueNotSolvedByKinductionPlain}{Total}{}{Cputime}{Avg}{526.9707242470591081921880401}%
\StoreBenchExecResult{PdrInv}{KinductionDfTrueNotSolvedByKinductionPlain}{Total}{}{Cputime}{Median}{901.031151997}%
\StoreBenchExecResult{PdrInv}{KinductionDfTrueNotSolvedByKinductionPlain}{Total}{}{Cputime}{Min}{2.440895303}%
\StoreBenchExecResult{PdrInv}{KinductionDfTrueNotSolvedByKinductionPlain}{Total}{}{Cputime}{Max}{1002.34088193}%
\StoreBenchExecResult{PdrInv}{KinductionDfTrueNotSolvedByKinductionPlain}{Total}{}{Cputime}{Stdev}{420.6670138160805154971524799}%
\StoreBenchExecResult{PdrInv}{KinductionDfTrueNotSolvedByKinductionPlain}{Total}{}{Walltime}{}{876223.57532666498}%
\StoreBenchExecResult{PdrInv}{KinductionDfTrueNotSolvedByKinductionPlain}{Total}{}{Walltime}{Avg}{302.8771432169598963014172140}%
\StoreBenchExecResult{PdrInv}{KinductionDfTrueNotSolvedByKinductionPlain}{Total}{}{Walltime}{Median}{451.393793821}%
\StoreBenchExecResult{PdrInv}{KinductionDfTrueNotSolvedByKinductionPlain}{Total}{}{Walltime}{Min}{1.32832694054}%
\StoreBenchExecResult{PdrInv}{KinductionDfTrueNotSolvedByKinductionPlain}{Total}{}{Walltime}{Max}{919.826648951}%
\StoreBenchExecResult{PdrInv}{KinductionDfTrueNotSolvedByKinductionPlain}{Total}{}{Walltime}{Stdev}{265.8726845906066136029365897}%
\StoreBenchExecResult{PdrInv}{KinductionDfTrueNotSolvedByKinductionPlain}{Correct}{}{Count}{}{1117}%
\StoreBenchExecResult{PdrInv}{KinductionDfTrueNotSolvedByKinductionPlain}{Correct}{}{Cputime}{}{75697.941508254}%
\StoreBenchExecResult{PdrInv}{KinductionDfTrueNotSolvedByKinductionPlain}{Correct}{}{Cputime}{Avg}{67.76897180685228290062667860}%
\StoreBenchExecResult{PdrInv}{KinductionDfTrueNotSolvedByKinductionPlain}{Correct}{}{Cputime}{Median}{7.892559581}%
\StoreBenchExecResult{PdrInv}{KinductionDfTrueNotSolvedByKinductionPlain}{Correct}{}{Cputime}{Min}{3.224740351}%
\StoreBenchExecResult{PdrInv}{KinductionDfTrueNotSolvedByKinductionPlain}{Correct}{}{Cputime}{Max}{864.33111854}%
\StoreBenchExecResult{PdrInv}{KinductionDfTrueNotSolvedByKinductionPlain}{Correct}{}{Cputime}{Stdev}{153.4050038171952280970716228}%
\StoreBenchExecResult{PdrInv}{KinductionDfTrueNotSolvedByKinductionPlain}{Correct}{}{Walltime}{}{39430.53829098090}%
\StoreBenchExecResult{PdrInv}{KinductionDfTrueNotSolvedByKinductionPlain}{Correct}{}{Walltime}{Avg}{35.30039238225684870188003581}%
\StoreBenchExecResult{PdrInv}{KinductionDfTrueNotSolvedByKinductionPlain}{Correct}{}{Walltime}{Median}{4.16023015976}%
\StoreBenchExecResult{PdrInv}{KinductionDfTrueNotSolvedByKinductionPlain}{Correct}{}{Walltime}{Min}{1.8099091053}%
\StoreBenchExecResult{PdrInv}{KinductionDfTrueNotSolvedByKinductionPlain}{Correct}{}{Walltime}{Max}{744.50601697}%
\StoreBenchExecResult{PdrInv}{KinductionDfTrueNotSolvedByKinductionPlain}{Correct}{}{Walltime}{Stdev}{81.74323731384320106139925564}%
\StoreBenchExecResult{PdrInv}{KinductionDfTrueNotSolvedByKinductionPlain}{Correct}{True}{Count}{}{1117}%
\StoreBenchExecResult{PdrInv}{KinductionDfTrueNotSolvedByKinductionPlain}{Correct}{True}{Cputime}{}{75697.941508254}%
\StoreBenchExecResult{PdrInv}{KinductionDfTrueNotSolvedByKinductionPlain}{Correct}{True}{Cputime}{Avg}{67.76897180685228290062667860}%
\StoreBenchExecResult{PdrInv}{KinductionDfTrueNotSolvedByKinductionPlain}{Correct}{True}{Cputime}{Median}{7.892559581}%
\StoreBenchExecResult{PdrInv}{KinductionDfTrueNotSolvedByKinductionPlain}{Correct}{True}{Cputime}{Min}{3.224740351}%
\StoreBenchExecResult{PdrInv}{KinductionDfTrueNotSolvedByKinductionPlain}{Correct}{True}{Cputime}{Max}{864.33111854}%
\StoreBenchExecResult{PdrInv}{KinductionDfTrueNotSolvedByKinductionPlain}{Correct}{True}{Cputime}{Stdev}{153.4050038171952280970716228}%
\StoreBenchExecResult{PdrInv}{KinductionDfTrueNotSolvedByKinductionPlain}{Correct}{True}{Walltime}{}{39430.53829098090}%
\StoreBenchExecResult{PdrInv}{KinductionDfTrueNotSolvedByKinductionPlain}{Correct}{True}{Walltime}{Avg}{35.30039238225684870188003581}%
\StoreBenchExecResult{PdrInv}{KinductionDfTrueNotSolvedByKinductionPlain}{Correct}{True}{Walltime}{Median}{4.16023015976}%
\StoreBenchExecResult{PdrInv}{KinductionDfTrueNotSolvedByKinductionPlain}{Correct}{True}{Walltime}{Min}{1.8099091053}%
\StoreBenchExecResult{PdrInv}{KinductionDfTrueNotSolvedByKinductionPlain}{Correct}{True}{Walltime}{Max}{744.50601697}%
\StoreBenchExecResult{PdrInv}{KinductionDfTrueNotSolvedByKinductionPlain}{Correct}{True}{Walltime}{Stdev}{81.74323731384320106139925564}%
\StoreBenchExecResult{PdrInv}{KinductionDfTrueNotSolvedByKinductionPlain}{Wrong}{True}{Count}{}{0}%
\StoreBenchExecResult{PdrInv}{KinductionDfTrueNotSolvedByKinductionPlain}{Wrong}{True}{Cputime}{}{0}%
\StoreBenchExecResult{PdrInv}{KinductionDfTrueNotSolvedByKinductionPlain}{Wrong}{True}{Cputime}{Avg}{None}%
\StoreBenchExecResult{PdrInv}{KinductionDfTrueNotSolvedByKinductionPlain}{Wrong}{True}{Cputime}{Median}{None}%
\StoreBenchExecResult{PdrInv}{KinductionDfTrueNotSolvedByKinductionPlain}{Wrong}{True}{Cputime}{Min}{None}%
\StoreBenchExecResult{PdrInv}{KinductionDfTrueNotSolvedByKinductionPlain}{Wrong}{True}{Cputime}{Max}{None}%
\StoreBenchExecResult{PdrInv}{KinductionDfTrueNotSolvedByKinductionPlain}{Wrong}{True}{Cputime}{Stdev}{None}%
\StoreBenchExecResult{PdrInv}{KinductionDfTrueNotSolvedByKinductionPlain}{Wrong}{True}{Walltime}{}{0}%
\StoreBenchExecResult{PdrInv}{KinductionDfTrueNotSolvedByKinductionPlain}{Wrong}{True}{Walltime}{Avg}{None}%
\StoreBenchExecResult{PdrInv}{KinductionDfTrueNotSolvedByKinductionPlain}{Wrong}{True}{Walltime}{Median}{None}%
\StoreBenchExecResult{PdrInv}{KinductionDfTrueNotSolvedByKinductionPlain}{Wrong}{True}{Walltime}{Min}{None}%
\StoreBenchExecResult{PdrInv}{KinductionDfTrueNotSolvedByKinductionPlain}{Wrong}{True}{Walltime}{Max}{None}%
\StoreBenchExecResult{PdrInv}{KinductionDfTrueNotSolvedByKinductionPlain}{Wrong}{True}{Walltime}{Stdev}{None}%
\StoreBenchExecResult{PdrInv}{KinductionDfTrueNotSolvedByKinductionPlain}{Error}{}{Count}{}{1776}%
\StoreBenchExecResult{PdrInv}{KinductionDfTrueNotSolvedByKinductionPlain}{Error}{}{Cputime}{}{1448828.363738488}%
\StoreBenchExecResult{PdrInv}{KinductionDfTrueNotSolvedByKinductionPlain}{Error}{}{Cputime}{Avg}{815.7817363392387387387387387}%
\StoreBenchExecResult{PdrInv}{KinductionDfTrueNotSolvedByKinductionPlain}{Error}{}{Cputime}{Median}{901.502483225}%
\StoreBenchExecResult{PdrInv}{KinductionDfTrueNotSolvedByKinductionPlain}{Error}{}{Cputime}{Min}{2.440895303}%
\StoreBenchExecResult{PdrInv}{KinductionDfTrueNotSolvedByKinductionPlain}{Error}{}{Cputime}{Max}{1002.34088193}%
\StoreBenchExecResult{PdrInv}{KinductionDfTrueNotSolvedByKinductionPlain}{Error}{}{Cputime}{Stdev}{239.6318318112386757603580439}%
\StoreBenchExecResult{PdrInv}{KinductionDfTrueNotSolvedByKinductionPlain}{Error}{}{Walltime}{}{836793.03703568408}%
\StoreBenchExecResult{PdrInv}{KinductionDfTrueNotSolvedByKinductionPlain}{Error}{}{Walltime}{Avg}{471.1672505831554504504504505}%
\StoreBenchExecResult{PdrInv}{KinductionDfTrueNotSolvedByKinductionPlain}{Error}{}{Walltime}{Median}{452.317220926}%
\StoreBenchExecResult{PdrInv}{KinductionDfTrueNotSolvedByKinductionPlain}{Error}{}{Walltime}{Min}{1.32832694054}%
\StoreBenchExecResult{PdrInv}{KinductionDfTrueNotSolvedByKinductionPlain}{Error}{}{Walltime}{Max}{919.826648951}%
\StoreBenchExecResult{PdrInv}{KinductionDfTrueNotSolvedByKinductionPlain}{Error}{}{Walltime}{Stdev}{193.8876890559128332095222057}%
\StoreBenchExecResult{PdrInv}{KinductionDfTrueNotSolvedByKinductionPlain}{Error}{Assertion}{Count}{}{2}%
\StoreBenchExecResult{PdrInv}{KinductionDfTrueNotSolvedByKinductionPlain}{Error}{Assertion}{Cputime}{}{6.608858744}%
\StoreBenchExecResult{PdrInv}{KinductionDfTrueNotSolvedByKinductionPlain}{Error}{Assertion}{Cputime}{Avg}{3.304429372}%
\StoreBenchExecResult{PdrInv}{KinductionDfTrueNotSolvedByKinductionPlain}{Error}{Assertion}{Cputime}{Median}{3.304429372}%
\StoreBenchExecResult{PdrInv}{KinductionDfTrueNotSolvedByKinductionPlain}{Error}{Assertion}{Cputime}{Min}{3.252592312}%
\StoreBenchExecResult{PdrInv}{KinductionDfTrueNotSolvedByKinductionPlain}{Error}{Assertion}{Cputime}{Max}{3.356266432}%
\StoreBenchExecResult{PdrInv}{KinductionDfTrueNotSolvedByKinductionPlain}{Error}{Assertion}{Cputime}{Stdev}{0.051837060}%
\StoreBenchExecResult{PdrInv}{KinductionDfTrueNotSolvedByKinductionPlain}{Error}{Assertion}{Walltime}{}{3.65375494956}%
\StoreBenchExecResult{PdrInv}{KinductionDfTrueNotSolvedByKinductionPlain}{Error}{Assertion}{Walltime}{Avg}{1.82687747478}%
\StoreBenchExecResult{PdrInv}{KinductionDfTrueNotSolvedByKinductionPlain}{Error}{Assertion}{Walltime}{Median}{1.82687747478}%
\StoreBenchExecResult{PdrInv}{KinductionDfTrueNotSolvedByKinductionPlain}{Error}{Assertion}{Walltime}{Min}{1.78716897964}%
\StoreBenchExecResult{PdrInv}{KinductionDfTrueNotSolvedByKinductionPlain}{Error}{Assertion}{Walltime}{Max}{1.86658596992}%
\StoreBenchExecResult{PdrInv}{KinductionDfTrueNotSolvedByKinductionPlain}{Error}{Assertion}{Walltime}{Stdev}{0.03970849514}%
\StoreBenchExecResult{PdrInv}{KinductionDfTrueNotSolvedByKinductionPlain}{Error}{Error}{Count}{}{128}%
\StoreBenchExecResult{PdrInv}{KinductionDfTrueNotSolvedByKinductionPlain}{Error}{Error}{Cputime}{}{21366.041696426}%
\StoreBenchExecResult{PdrInv}{KinductionDfTrueNotSolvedByKinductionPlain}{Error}{Error}{Cputime}{Avg}{166.922200753328125}%
\StoreBenchExecResult{PdrInv}{KinductionDfTrueNotSolvedByKinductionPlain}{Error}{Error}{Cputime}{Median}{102.4599692875}%
\StoreBenchExecResult{PdrInv}{KinductionDfTrueNotSolvedByKinductionPlain}{Error}{Error}{Cputime}{Min}{2.440895303}%
\StoreBenchExecResult{PdrInv}{KinductionDfTrueNotSolvedByKinductionPlain}{Error}{Error}{Cputime}{Max}{796.361291406}%
\StoreBenchExecResult{PdrInv}{KinductionDfTrueNotSolvedByKinductionPlain}{Error}{Error}{Cputime}{Stdev}{189.0821436773064756548232362}%
\StoreBenchExecResult{PdrInv}{KinductionDfTrueNotSolvedByKinductionPlain}{Error}{Error}{Walltime}{}{17984.09465742279}%
\StoreBenchExecResult{PdrInv}{KinductionDfTrueNotSolvedByKinductionPlain}{Error}{Error}{Walltime}{Avg}{140.500739511115546875}%
\StoreBenchExecResult{PdrInv}{KinductionDfTrueNotSolvedByKinductionPlain}{Error}{Error}{Walltime}{Median}{79.29778289795}%
\StoreBenchExecResult{PdrInv}{KinductionDfTrueNotSolvedByKinductionPlain}{Error}{Error}{Walltime}{Min}{1.32832694054}%
\StoreBenchExecResult{PdrInv}{KinductionDfTrueNotSolvedByKinductionPlain}{Error}{Error}{Walltime}{Max}{780.546742201}%
\StoreBenchExecResult{PdrInv}{KinductionDfTrueNotSolvedByKinductionPlain}{Error}{Error}{Walltime}{Stdev}{169.1837837076091150713353421}%
\StoreBenchExecResult{PdrInv}{KinductionDfTrueNotSolvedByKinductionPlain}{Error}{Exception}{Count}{}{15}%
\StoreBenchExecResult{PdrInv}{KinductionDfTrueNotSolvedByKinductionPlain}{Error}{Exception}{Cputime}{}{3418.257013971}%
\StoreBenchExecResult{PdrInv}{KinductionDfTrueNotSolvedByKinductionPlain}{Error}{Exception}{Cputime}{Avg}{227.8838009314}%
\StoreBenchExecResult{PdrInv}{KinductionDfTrueNotSolvedByKinductionPlain}{Error}{Exception}{Cputime}{Median}{220.848681684}%
\StoreBenchExecResult{PdrInv}{KinductionDfTrueNotSolvedByKinductionPlain}{Error}{Exception}{Cputime}{Min}{15.185833627}%
\StoreBenchExecResult{PdrInv}{KinductionDfTrueNotSolvedByKinductionPlain}{Error}{Exception}{Cputime}{Max}{602.333519643}%
\StoreBenchExecResult{PdrInv}{KinductionDfTrueNotSolvedByKinductionPlain}{Error}{Exception}{Cputime}{Stdev}{163.9639494580446150382579038}%
\StoreBenchExecResult{PdrInv}{KinductionDfTrueNotSolvedByKinductionPlain}{Error}{Exception}{Walltime}{}{1716.14370465263}%
\StoreBenchExecResult{PdrInv}{KinductionDfTrueNotSolvedByKinductionPlain}{Error}{Exception}{Walltime}{Avg}{114.4095803101753333333333333}%
\StoreBenchExecResult{PdrInv}{KinductionDfTrueNotSolvedByKinductionPlain}{Error}{Exception}{Walltime}{Median}{110.90247798}%
\StoreBenchExecResult{PdrInv}{KinductionDfTrueNotSolvedByKinductionPlain}{Error}{Exception}{Walltime}{Min}{7.76194119453}%
\StoreBenchExecResult{PdrInv}{KinductionDfTrueNotSolvedByKinductionPlain}{Error}{Exception}{Walltime}{Max}{301.968492985}%
\StoreBenchExecResult{PdrInv}{KinductionDfTrueNotSolvedByKinductionPlain}{Error}{Exception}{Walltime}{Stdev}{82.17714061210816701154183429}%
\StoreBenchExecResult{PdrInv}{KinductionDfTrueNotSolvedByKinductionPlain}{Error}{OutOfJavaMemory}{Count}{}{4}%
\StoreBenchExecResult{PdrInv}{KinductionDfTrueNotSolvedByKinductionPlain}{Error}{OutOfJavaMemory}{Cputime}{}{1759.748167518}%
\StoreBenchExecResult{PdrInv}{KinductionDfTrueNotSolvedByKinductionPlain}{Error}{OutOfJavaMemory}{Cputime}{Avg}{439.9370418795}%
\StoreBenchExecResult{PdrInv}{KinductionDfTrueNotSolvedByKinductionPlain}{Error}{OutOfJavaMemory}{Cputime}{Median}{458.535837567}%
\StoreBenchExecResult{PdrInv}{KinductionDfTrueNotSolvedByKinductionPlain}{Error}{OutOfJavaMemory}{Cputime}{Min}{270.351151354}%
\StoreBenchExecResult{PdrInv}{KinductionDfTrueNotSolvedByKinductionPlain}{Error}{OutOfJavaMemory}{Cputime}{Max}{572.32534103}%
\StoreBenchExecResult{PdrInv}{KinductionDfTrueNotSolvedByKinductionPlain}{Error}{OutOfJavaMemory}{Cputime}{Stdev}{110.0058052944893796124611574}%
\StoreBenchExecResult{PdrInv}{KinductionDfTrueNotSolvedByKinductionPlain}{Error}{OutOfJavaMemory}{Walltime}{}{905.664759875}%
\StoreBenchExecResult{PdrInv}{KinductionDfTrueNotSolvedByKinductionPlain}{Error}{OutOfJavaMemory}{Walltime}{Avg}{226.41618996875}%
\StoreBenchExecResult{PdrInv}{KinductionDfTrueNotSolvedByKinductionPlain}{Error}{OutOfJavaMemory}{Walltime}{Median}{238.0941244365}%
\StoreBenchExecResult{PdrInv}{KinductionDfTrueNotSolvedByKinductionPlain}{Error}{OutOfJavaMemory}{Walltime}{Min}{138.367254019}%
\StoreBenchExecResult{PdrInv}{KinductionDfTrueNotSolvedByKinductionPlain}{Error}{OutOfJavaMemory}{Walltime}{Max}{291.109256983}%
\StoreBenchExecResult{PdrInv}{KinductionDfTrueNotSolvedByKinductionPlain}{Error}{OutOfJavaMemory}{Walltime}{Stdev}{56.09251916907131929507459356}%
\StoreBenchExecResult{PdrInv}{KinductionDfTrueNotSolvedByKinductionPlain}{Error}{OutOfMemory}{Count}{}{98}%
\StoreBenchExecResult{PdrInv}{KinductionDfTrueNotSolvedByKinductionPlain}{Error}{OutOfMemory}{Cputime}{}{36779.601877869}%
\StoreBenchExecResult{PdrInv}{KinductionDfTrueNotSolvedByKinductionPlain}{Error}{OutOfMemory}{Cputime}{Avg}{375.3020599782551020408163265}%
\StoreBenchExecResult{PdrInv}{KinductionDfTrueNotSolvedByKinductionPlain}{Error}{OutOfMemory}{Cputime}{Median}{332.987818841}%
\StoreBenchExecResult{PdrInv}{KinductionDfTrueNotSolvedByKinductionPlain}{Error}{OutOfMemory}{Cputime}{Min}{168.857488142}%
\StoreBenchExecResult{PdrInv}{KinductionDfTrueNotSolvedByKinductionPlain}{Error}{OutOfMemory}{Cputime}{Max}{876.101535739}%
\StoreBenchExecResult{PdrInv}{KinductionDfTrueNotSolvedByKinductionPlain}{Error}{OutOfMemory}{Cputime}{Stdev}{204.2788477893602296102941373}%
\StoreBenchExecResult{PdrInv}{KinductionDfTrueNotSolvedByKinductionPlain}{Error}{OutOfMemory}{Walltime}{}{18452.2281267611}%
\StoreBenchExecResult{PdrInv}{KinductionDfTrueNotSolvedByKinductionPlain}{Error}{OutOfMemory}{Walltime}{Avg}{188.2880421098071428571428571}%
\StoreBenchExecResult{PdrInv}{KinductionDfTrueNotSolvedByKinductionPlain}{Error}{OutOfMemory}{Walltime}{Median}{167.0533549785}%
\StoreBenchExecResult{PdrInv}{KinductionDfTrueNotSolvedByKinductionPlain}{Error}{OutOfMemory}{Walltime}{Min}{84.9250471592}%
\StoreBenchExecResult{PdrInv}{KinductionDfTrueNotSolvedByKinductionPlain}{Error}{OutOfMemory}{Walltime}{Max}{438.882339001}%
\StoreBenchExecResult{PdrInv}{KinductionDfTrueNotSolvedByKinductionPlain}{Error}{OutOfMemory}{Walltime}{Stdev}{102.2007021092830555531190697}%
\StoreBenchExecResult{PdrInv}{KinductionDfTrueNotSolvedByKinductionPlain}{Error}{Timeout}{Count}{}{1529}%
\StoreBenchExecResult{PdrInv}{KinductionDfTrueNotSolvedByKinductionPlain}{Error}{Timeout}{Cputime}{}{1385498.106123960}%
\StoreBenchExecResult{PdrInv}{KinductionDfTrueNotSolvedByKinductionPlain}{Error}{Timeout}{Cputime}{Avg}{906.1465703884630477436232832}%
\StoreBenchExecResult{PdrInv}{KinductionDfTrueNotSolvedByKinductionPlain}{Error}{Timeout}{Cputime}{Median}{901.759522333}%
\StoreBenchExecResult{PdrInv}{KinductionDfTrueNotSolvedByKinductionPlain}{Error}{Timeout}{Cputime}{Min}{900.537943417}%
\StoreBenchExecResult{PdrInv}{KinductionDfTrueNotSolvedByKinductionPlain}{Error}{Timeout}{Cputime}{Max}{1002.34088193}%
\StoreBenchExecResult{PdrInv}{KinductionDfTrueNotSolvedByKinductionPlain}{Error}{Timeout}{Cputime}{Stdev}{16.31116986371776220322481127}%
\StoreBenchExecResult{PdrInv}{KinductionDfTrueNotSolvedByKinductionPlain}{Error}{Timeout}{Walltime}{}{797731.252032023}%
\StoreBenchExecResult{PdrInv}{KinductionDfTrueNotSolvedByKinductionPlain}{Error}{Timeout}{Walltime}{Avg}{521.7339777841877043819489863}%
\StoreBenchExecResult{PdrInv}{KinductionDfTrueNotSolvedByKinductionPlain}{Error}{Timeout}{Walltime}{Median}{452.817712069}%
\StoreBenchExecResult{PdrInv}{KinductionDfTrueNotSolvedByKinductionPlain}{Error}{Timeout}{Walltime}{Min}{450.988446951}%
\StoreBenchExecResult{PdrInv}{KinductionDfTrueNotSolvedByKinductionPlain}{Error}{Timeout}{Walltime}{Max}{919.826648951}%
\StoreBenchExecResult{PdrInv}{KinductionDfTrueNotSolvedByKinductionPlain}{Error}{Timeout}{Walltime}{Stdev}{148.2983228973945968765835646}%
\providecommand\StoreBenchExecResult[7]{\expandafter\newcommand\csname#1#2#3#4#5#6\endcsname{#7}}%
\StoreBenchExecResult{PdrInv}{KinductionDf}{Total}{}{Count}{}{5591}%
\StoreBenchExecResult{PdrInv}{KinductionDf}{Total}{}{Cputime}{}{2181535.870814876}%
\StoreBenchExecResult{PdrInv}{KinductionDf}{Total}{}{Cputime}{Avg}{390.1870632829325702021105348}%
\StoreBenchExecResult{PdrInv}{KinductionDf}{Total}{}{Cputime}{Median}{100.519299937}%
\StoreBenchExecResult{PdrInv}{KinductionDf}{Total}{}{Cputime}{Min}{2.440895303}%
\StoreBenchExecResult{PdrInv}{KinductionDf}{Total}{}{Cputime}{Max}{1002.34088193}%
\StoreBenchExecResult{PdrInv}{KinductionDf}{Total}{}{Cputime}{Stdev}{416.3306121906758638808897688}%
\StoreBenchExecResult{PdrInv}{KinductionDf}{Total}{}{Walltime}{}{1238304.89027211869}%
\StoreBenchExecResult{PdrInv}{KinductionDf}{Total}{}{Walltime}{Avg}{221.4818261978391504203183688}%
\StoreBenchExecResult{PdrInv}{KinductionDf}{Total}{}{Walltime}{Median}{56.157900095}%
\StoreBenchExecResult{PdrInv}{KinductionDf}{Total}{}{Walltime}{Min}{1.32832694054}%
\StoreBenchExecResult{PdrInv}{KinductionDf}{Total}{}{Walltime}{Max}{919.826648951}%
\StoreBenchExecResult{PdrInv}{KinductionDf}{Total}{}{Walltime}{Stdev}{252.9561802590267723010230080}%
\StoreBenchExecResult{PdrInv}{KinductionDf}{Correct}{}{Count}{}{3099}%
\StoreBenchExecResult{PdrInv}{KinductionDf}{Correct}{}{Cputime}{}{193656.889718340}%
\StoreBenchExecResult{PdrInv}{KinductionDf}{Correct}{}{Cputime}{Avg}{62.49012252931268151016456922}%
\StoreBenchExecResult{PdrInv}{KinductionDf}{Correct}{}{Cputime}{Median}{10.115170818}%
\StoreBenchExecResult{PdrInv}{KinductionDf}{Correct}{}{Cputime}{Min}{2.956727101}%
\StoreBenchExecResult{PdrInv}{KinductionDf}{Correct}{}{Cputime}{Max}{896.316567446}%
\StoreBenchExecResult{PdrInv}{KinductionDf}{Correct}{}{Cputime}{Stdev}{140.3537972146904451045577815}%
\StoreBenchExecResult{PdrInv}{KinductionDf}{Correct}{}{Walltime}{}{102937.92361998765}%
\StoreBenchExecResult{PdrInv}{KinductionDf}{Correct}{}{Walltime}{Avg}{33.21649681187081316553727009}%
\StoreBenchExecResult{PdrInv}{KinductionDf}{Correct}{}{Walltime}{Median}{5.2597630024}%
\StoreBenchExecResult{PdrInv}{KinductionDf}{Correct}{}{Walltime}{Min}{1.65714097023}%
\StoreBenchExecResult{PdrInv}{KinductionDf}{Correct}{}{Walltime}{Max}{863.815379143}%
\StoreBenchExecResult{PdrInv}{KinductionDf}{Correct}{}{Walltime}{Stdev}{77.45612348717077348470017489}%
\StoreBenchExecResult{PdrInv}{KinductionDf}{Correct}{False}{Count}{}{763}%
\StoreBenchExecResult{PdrInv}{KinductionDf}{Correct}{False}{Cputime}{}{62666.109383617}%
\StoreBenchExecResult{PdrInv}{KinductionDf}{Correct}{False}{Cputime}{Avg}{82.13120495886893840104849279}%
\StoreBenchExecResult{PdrInv}{KinductionDf}{Correct}{False}{Cputime}{Median}{23.916158259}%
\StoreBenchExecResult{PdrInv}{KinductionDf}{Correct}{False}{Cputime}{Min}{3.204232574}%
\StoreBenchExecResult{PdrInv}{KinductionDf}{Correct}{False}{Cputime}{Max}{896.316567446}%
\StoreBenchExecResult{PdrInv}{KinductionDf}{Correct}{False}{Cputime}{Stdev}{166.5408014615533496128472698}%
\StoreBenchExecResult{PdrInv}{KinductionDf}{Correct}{False}{Walltime}{}{32975.84839296634}%
\StoreBenchExecResult{PdrInv}{KinductionDf}{Correct}{False}{Walltime}{Avg}{43.21867417164657929226736566}%
\StoreBenchExecResult{PdrInv}{KinductionDf}{Correct}{False}{Walltime}{Median}{12.3396089077}%
\StoreBenchExecResult{PdrInv}{KinductionDf}{Correct}{False}{Walltime}{Min}{1.78456401825}%
\StoreBenchExecResult{PdrInv}{KinductionDf}{Correct}{False}{Walltime}{Max}{594.768676043}%
\StoreBenchExecResult{PdrInv}{KinductionDf}{Correct}{False}{Walltime}{Stdev}{87.31669060520710745077178169}%
\StoreBenchExecResult{PdrInv}{KinductionDf}{Correct}{True}{Count}{}{2336}%
\StoreBenchExecResult{PdrInv}{KinductionDf}{Correct}{True}{Cputime}{}{130990.780334723}%
\StoreBenchExecResult{PdrInv}{KinductionDf}{Correct}{True}{Cputime}{Avg}{56.07482034876840753424657534}%
\StoreBenchExecResult{PdrInv}{KinductionDf}{Correct}{True}{Cputime}{Median}{7.8623468795}%
\StoreBenchExecResult{PdrInv}{KinductionDf}{Correct}{True}{Cputime}{Min}{2.956727101}%
\StoreBenchExecResult{PdrInv}{KinductionDf}{Correct}{True}{Cputime}{Max}{883.606012348}%
\StoreBenchExecResult{PdrInv}{KinductionDf}{Correct}{True}{Cputime}{Stdev}{130.0270902237170862015207082}%
\StoreBenchExecResult{PdrInv}{KinductionDf}{Correct}{True}{Walltime}{}{69962.07522702131}%
\StoreBenchExecResult{PdrInv}{KinductionDf}{Correct}{True}{Walltime}{Avg}{29.94951850471802654109589041}%
\StoreBenchExecResult{PdrInv}{KinductionDf}{Correct}{True}{Walltime}{Median}{4.167878985405}%
\StoreBenchExecResult{PdrInv}{KinductionDf}{Correct}{True}{Walltime}{Min}{1.65714097023}%
\StoreBenchExecResult{PdrInv}{KinductionDf}{Correct}{True}{Walltime}{Max}{863.815379143}%
\StoreBenchExecResult{PdrInv}{KinductionDf}{Correct}{True}{Walltime}{Stdev}{73.65740399292191132248495055}%
\StoreBenchExecResult{PdrInv}{KinductionDf}{Wrong}{True}{Count}{}{0}%
\StoreBenchExecResult{PdrInv}{KinductionDf}{Wrong}{True}{Cputime}{}{0}%
\StoreBenchExecResult{PdrInv}{KinductionDf}{Wrong}{True}{Cputime}{Avg}{None}%
\StoreBenchExecResult{PdrInv}{KinductionDf}{Wrong}{True}{Cputime}{Median}{None}%
\StoreBenchExecResult{PdrInv}{KinductionDf}{Wrong}{True}{Cputime}{Min}{None}%
\StoreBenchExecResult{PdrInv}{KinductionDf}{Wrong}{True}{Cputime}{Max}{None}%
\StoreBenchExecResult{PdrInv}{KinductionDf}{Wrong}{True}{Cputime}{Stdev}{None}%
\StoreBenchExecResult{PdrInv}{KinductionDf}{Wrong}{True}{Walltime}{}{0}%
\StoreBenchExecResult{PdrInv}{KinductionDf}{Wrong}{True}{Walltime}{Avg}{None}%
\StoreBenchExecResult{PdrInv}{KinductionDf}{Wrong}{True}{Walltime}{Median}{None}%
\StoreBenchExecResult{PdrInv}{KinductionDf}{Wrong}{True}{Walltime}{Min}{None}%
\StoreBenchExecResult{PdrInv}{KinductionDf}{Wrong}{True}{Walltime}{Max}{None}%
\StoreBenchExecResult{PdrInv}{KinductionDf}{Wrong}{True}{Walltime}{Stdev}{None}%
\StoreBenchExecResult{PdrInv}{KinductionDf}{Error}{}{Count}{}{2490}%
\StoreBenchExecResult{PdrInv}{KinductionDf}{Error}{}{Cputime}{}{1987854.523547427}%
\StoreBenchExecResult{PdrInv}{KinductionDf}{Error}{}{Cputime}{Avg}{798.3351500190469879518072289}%
\StoreBenchExecResult{PdrInv}{KinductionDf}{Error}{}{Cputime}{Median}{901.5633291945}%
\StoreBenchExecResult{PdrInv}{KinductionDf}{Error}{}{Cputime}{Min}{2.440895303}%
\StoreBenchExecResult{PdrInv}{KinductionDf}{Error}{}{Cputime}{Max}{1002.34088193}%
\StoreBenchExecResult{PdrInv}{KinductionDf}{Error}{}{Cputime}{Stdev}{253.6304417937316888548929673}%
\StoreBenchExecResult{PdrInv}{KinductionDf}{Error}{}{Walltime}{}{1135354.03644416324}%
\StoreBenchExecResult{PdrInv}{KinductionDf}{Error}{}{Walltime}{Avg}{455.9654764835996947791164659}%
\StoreBenchExecResult{PdrInv}{KinductionDf}{Error}{}{Walltime}{Median}{452.357358932}%
\StoreBenchExecResult{PdrInv}{KinductionDf}{Error}{}{Walltime}{Min}{1.32832694054}%
\StoreBenchExecResult{PdrInv}{KinductionDf}{Error}{}{Walltime}{Max}{919.826648951}%
\StoreBenchExecResult{PdrInv}{KinductionDf}{Error}{}{Walltime}{Stdev}{192.5505164997201524035109663}%
\StoreBenchExecResult{PdrInv}{KinductionDf}{Error}{Assertion}{Count}{}{4}%
\StoreBenchExecResult{PdrInv}{KinductionDf}{Error}{Assertion}{Cputime}{}{13.068867928}%
\StoreBenchExecResult{PdrInv}{KinductionDf}{Error}{Assertion}{Cputime}{Avg}{3.267216982}%
\StoreBenchExecResult{PdrInv}{KinductionDf}{Error}{Assertion}{Cputime}{Median}{3.2528048485}%
\StoreBenchExecResult{PdrInv}{KinductionDf}{Error}{Assertion}{Cputime}{Min}{3.206991799}%
\StoreBenchExecResult{PdrInv}{KinductionDf}{Error}{Assertion}{Cputime}{Max}{3.356266432}%
\StoreBenchExecResult{PdrInv}{KinductionDf}{Error}{Assertion}{Cputime}{Stdev}{0.05470920110058567195929732730}%
\StoreBenchExecResult{PdrInv}{KinductionDf}{Error}{Assertion}{Walltime}{}{7.25183296203}%
\StoreBenchExecResult{PdrInv}{KinductionDf}{Error}{Assertion}{Walltime}{Avg}{1.8129582405075}%
\StoreBenchExecResult{PdrInv}{KinductionDf}{Error}{Assertion}{Walltime}{Median}{1.806653022765}%
\StoreBenchExecResult{PdrInv}{KinductionDf}{Error}{Assertion}{Walltime}{Min}{1.77194094658}%
\StoreBenchExecResult{PdrInv}{KinductionDf}{Error}{Assertion}{Walltime}{Max}{1.86658596992}%
\StoreBenchExecResult{PdrInv}{KinductionDf}{Error}{Assertion}{Walltime}{Stdev}{0.03673254407752626346215536383}%
\StoreBenchExecResult{PdrInv}{KinductionDf}{Error}{Error}{Count}{}{186}%
\StoreBenchExecResult{PdrInv}{KinductionDf}{Error}{Error}{Cputime}{}{32508.647000973}%
\StoreBenchExecResult{PdrInv}{KinductionDf}{Error}{Error}{Cputime}{Avg}{174.7776720482419354838709677}%
\StoreBenchExecResult{PdrInv}{KinductionDf}{Error}{Error}{Cputime}{Median}{111.130864952}%
\StoreBenchExecResult{PdrInv}{KinductionDf}{Error}{Error}{Cputime}{Min}{2.440895303}%
\StoreBenchExecResult{PdrInv}{KinductionDf}{Error}{Error}{Cputime}{Max}{796.361291406}%
\StoreBenchExecResult{PdrInv}{KinductionDf}{Error}{Error}{Cputime}{Stdev}{183.9777135698659867191546662}%
\StoreBenchExecResult{PdrInv}{KinductionDf}{Error}{Error}{Walltime}{}{27211.99485564359}%
\StoreBenchExecResult{PdrInv}{KinductionDf}{Error}{Error}{Walltime}{Avg}{146.3010476109870430107526882}%
\StoreBenchExecResult{PdrInv}{KinductionDf}{Error}{Error}{Walltime}{Median}{92.17062962055}%
\StoreBenchExecResult{PdrInv}{KinductionDf}{Error}{Error}{Walltime}{Min}{1.32832694054}%
\StoreBenchExecResult{PdrInv}{KinductionDf}{Error}{Error}{Walltime}{Max}{780.546742201}%
\StoreBenchExecResult{PdrInv}{KinductionDf}{Error}{Error}{Walltime}{Stdev}{163.2196362193402897707404294}%
\StoreBenchExecResult{PdrInv}{KinductionDf}{Error}{Exception}{Count}{}{25}%
\StoreBenchExecResult{PdrInv}{KinductionDf}{Error}{Exception}{Cputime}{}{4727.364941668}%
\StoreBenchExecResult{PdrInv}{KinductionDf}{Error}{Exception}{Cputime}{Avg}{189.09459766672}%
\StoreBenchExecResult{PdrInv}{KinductionDf}{Error}{Exception}{Cputime}{Median}{121.695664132}%
\StoreBenchExecResult{PdrInv}{KinductionDf}{Error}{Exception}{Cputime}{Min}{15.081006021}%
\StoreBenchExecResult{PdrInv}{KinductionDf}{Error}{Exception}{Cputime}{Max}{632.075608302}%
\StoreBenchExecResult{PdrInv}{KinductionDf}{Error}{Exception}{Cputime}{Stdev}{175.5892988323298453472961380}%
\StoreBenchExecResult{PdrInv}{KinductionDf}{Error}{Exception}{Walltime}{}{2374.33172249782}%
\StoreBenchExecResult{PdrInv}{KinductionDf}{Error}{Exception}{Walltime}{Avg}{94.9732688999128}%
\StoreBenchExecResult{PdrInv}{KinductionDf}{Error}{Exception}{Walltime}{Median}{61.3966090679}%
\StoreBenchExecResult{PdrInv}{KinductionDf}{Error}{Exception}{Walltime}{Min}{7.74311518669}%
\StoreBenchExecResult{PdrInv}{KinductionDf}{Error}{Exception}{Walltime}{Max}{316.888850927}%
\StoreBenchExecResult{PdrInv}{KinductionDf}{Error}{Exception}{Walltime}{Stdev}{87.99231327396314944743067395}%
\StoreBenchExecResult{PdrInv}{KinductionDf}{Error}{OutOfJavaMemory}{Count}{}{6}%
\StoreBenchExecResult{PdrInv}{KinductionDf}{Error}{OutOfJavaMemory}{Cputime}{}{2898.322881882}%
\StoreBenchExecResult{PdrInv}{KinductionDf}{Error}{OutOfJavaMemory}{Cputime}{Avg}{483.053813647}%
\StoreBenchExecResult{PdrInv}{KinductionDf}{Error}{OutOfJavaMemory}{Cputime}{Median}{525.5022827005}%
\StoreBenchExecResult{PdrInv}{KinductionDf}{Error}{OutOfJavaMemory}{Cputime}{Min}{270.351151354}%
\StoreBenchExecResult{PdrInv}{KinductionDf}{Error}{OutOfJavaMemory}{Cputime}{Max}{572.819675667}%
\StoreBenchExecResult{PdrInv}{KinductionDf}{Error}{OutOfJavaMemory}{Cputime}{Stdev}{108.5807959029615856405501614}%
\StoreBenchExecResult{PdrInv}{KinductionDf}{Error}{OutOfJavaMemory}{Walltime}{}{1483.665053845}%
\StoreBenchExecResult{PdrInv}{KinductionDf}{Error}{OutOfJavaMemory}{Walltime}{Avg}{247.2775089741666666666666667}%
\StoreBenchExecResult{PdrInv}{KinductionDf}{Error}{OutOfJavaMemory}{Walltime}{Median}{269.4279044865}%
\StoreBenchExecResult{PdrInv}{KinductionDf}{Error}{OutOfJavaMemory}{Walltime}{Min}{138.367254019}%
\StoreBenchExecResult{PdrInv}{KinductionDf}{Error}{OutOfJavaMemory}{Walltime}{Max}{291.109256983}%
\StoreBenchExecResult{PdrInv}{KinductionDf}{Error}{OutOfJavaMemory}{Walltime}{Stdev}{54.49047258509416747526632494}%
\StoreBenchExecResult{PdrInv}{KinductionDf}{Error}{OutOfMemory}{Count}{}{232}%
\StoreBenchExecResult{PdrInv}{KinductionDf}{Error}{OutOfMemory}{Cputime}{}{97858.994345196}%
\StoreBenchExecResult{PdrInv}{KinductionDf}{Error}{OutOfMemory}{Cputime}{Avg}{421.8060101086034482758620690}%
\StoreBenchExecResult{PdrInv}{KinductionDf}{Error}{OutOfMemory}{Cputime}{Median}{363.398569078}%
\StoreBenchExecResult{PdrInv}{KinductionDf}{Error}{OutOfMemory}{Cputime}{Min}{162.991121275}%
\StoreBenchExecResult{PdrInv}{KinductionDf}{Error}{OutOfMemory}{Cputime}{Max}{898.166072766}%
\StoreBenchExecResult{PdrInv}{KinductionDf}{Error}{OutOfMemory}{Cputime}{Stdev}{207.9915709776489077805032467}%
\StoreBenchExecResult{PdrInv}{KinductionDf}{Error}{OutOfMemory}{Walltime}{}{49150.6944339218}%
\StoreBenchExecResult{PdrInv}{KinductionDf}{Error}{OutOfMemory}{Walltime}{Avg}{211.856441525525}%
\StoreBenchExecResult{PdrInv}{KinductionDf}{Error}{OutOfMemory}{Walltime}{Median}{182.2998884915}%
\StoreBenchExecResult{PdrInv}{KinductionDf}{Error}{OutOfMemory}{Walltime}{Min}{84.9250471592}%
\StoreBenchExecResult{PdrInv}{KinductionDf}{Error}{OutOfMemory}{Walltime}{Max}{449.641111135}%
\StoreBenchExecResult{PdrInv}{KinductionDf}{Error}{OutOfMemory}{Walltime}{Stdev}{103.7642219013148227976490676}%
\StoreBenchExecResult{PdrInv}{KinductionDf}{Error}{Timeout}{Count}{}{2037}%
\StoreBenchExecResult{PdrInv}{KinductionDf}{Error}{Timeout}{Cputime}{}{1849848.125509780}%
\StoreBenchExecResult{PdrInv}{KinductionDf}{Error}{Timeout}{Cputime}{Avg}{908.1237729552184585174275896}%
\StoreBenchExecResult{PdrInv}{KinductionDf}{Error}{Timeout}{Cputime}{Median}{901.962375633}%
\StoreBenchExecResult{PdrInv}{KinductionDf}{Error}{Timeout}{Cputime}{Min}{900.195720149}%
\StoreBenchExecResult{PdrInv}{KinductionDf}{Error}{Timeout}{Cputime}{Max}{1002.34088193}%
\StoreBenchExecResult{PdrInv}{KinductionDf}{Error}{Timeout}{Cputime}{Stdev}{20.23231917595519031922362970}%
\StoreBenchExecResult{PdrInv}{KinductionDf}{Error}{Timeout}{Walltime}{}{1055126.098545293}%
\StoreBenchExecResult{PdrInv}{KinductionDf}{Error}{Timeout}{Walltime}{Avg}{517.9804116569921453117329406}%
\StoreBenchExecResult{PdrInv}{KinductionDf}{Error}{Timeout}{Walltime}{Median}{453.229801893}%
\StoreBenchExecResult{PdrInv}{KinductionDf}{Error}{Timeout}{Walltime}{Min}{450.988446951}%
\StoreBenchExecResult{PdrInv}{KinductionDf}{Error}{Timeout}{Walltime}{Max}{919.826648951}%
\StoreBenchExecResult{PdrInv}{KinductionDf}{Error}{Timeout}{Walltime}{Stdev}{141.5517786148441011771055185}%
\StoreBenchExecResult{PdrInv}{KinductionDf}{Wrong}{}{Count}{}{2}%
\StoreBenchExecResult{PdrInv}{KinductionDf}{Wrong}{}{Cputime}{}{24.457549109}%
\StoreBenchExecResult{PdrInv}{KinductionDf}{Wrong}{}{Cputime}{Avg}{12.2287745545}%
\StoreBenchExecResult{PdrInv}{KinductionDf}{Wrong}{}{Cputime}{Median}{12.2287745545}%
\StoreBenchExecResult{PdrInv}{KinductionDf}{Wrong}{}{Cputime}{Min}{4.804479699}%
\StoreBenchExecResult{PdrInv}{KinductionDf}{Wrong}{}{Cputime}{Max}{19.65306941}%
\StoreBenchExecResult{PdrInv}{KinductionDf}{Wrong}{}{Cputime}{Stdev}{7.4242948555}%
\StoreBenchExecResult{PdrInv}{KinductionDf}{Wrong}{}{Walltime}{}{12.9302079678}%
\StoreBenchExecResult{PdrInv}{KinductionDf}{Wrong}{}{Walltime}{Avg}{6.4651039839}%
\StoreBenchExecResult{PdrInv}{KinductionDf}{Wrong}{}{Walltime}{Median}{6.4651039839}%
\StoreBenchExecResult{PdrInv}{KinductionDf}{Wrong}{}{Walltime}{Min}{2.6051170826}%
\StoreBenchExecResult{PdrInv}{KinductionDf}{Wrong}{}{Walltime}{Max}{10.3250908852}%
\StoreBenchExecResult{PdrInv}{KinductionDf}{Wrong}{}{Walltime}{Stdev}{3.8599869013}%
\StoreBenchExecResult{PdrInv}{KinductionDf}{Wrong}{False}{Count}{}{2}%
\StoreBenchExecResult{PdrInv}{KinductionDf}{Wrong}{False}{Cputime}{}{24.457549109}%
\StoreBenchExecResult{PdrInv}{KinductionDf}{Wrong}{False}{Cputime}{Avg}{12.2287745545}%
\StoreBenchExecResult{PdrInv}{KinductionDf}{Wrong}{False}{Cputime}{Median}{12.2287745545}%
\StoreBenchExecResult{PdrInv}{KinductionDf}{Wrong}{False}{Cputime}{Min}{4.804479699}%
\StoreBenchExecResult{PdrInv}{KinductionDf}{Wrong}{False}{Cputime}{Max}{19.65306941}%
\StoreBenchExecResult{PdrInv}{KinductionDf}{Wrong}{False}{Cputime}{Stdev}{7.4242948555}%
\StoreBenchExecResult{PdrInv}{KinductionDf}{Wrong}{False}{Walltime}{}{12.9302079678}%
\StoreBenchExecResult{PdrInv}{KinductionDf}{Wrong}{False}{Walltime}{Avg}{6.4651039839}%
\StoreBenchExecResult{PdrInv}{KinductionDf}{Wrong}{False}{Walltime}{Median}{6.4651039839}%
\StoreBenchExecResult{PdrInv}{KinductionDf}{Wrong}{False}{Walltime}{Min}{2.6051170826}%
\StoreBenchExecResult{PdrInv}{KinductionDf}{Wrong}{False}{Walltime}{Max}{10.3250908852}%
\StoreBenchExecResult{PdrInv}{KinductionDf}{Wrong}{False}{Walltime}{Stdev}{3.8599869013}%
\providecommand\StoreBenchExecResult[7]{\expandafter\newcommand\csname#1#2#3#4#5#6\endcsname{#7}}%
\StoreBenchExecResult{PdrInv}{KinductionKipdrdfTrueNotSolvedByKinductionPlainButKipdr}{Total}{}{Count}{}{449}%
\StoreBenchExecResult{PdrInv}{KinductionKipdrdfTrueNotSolvedByKinductionPlainButKipdr}{Total}{}{Cputime}{}{5880.343924916}%
\StoreBenchExecResult{PdrInv}{KinductionKipdrdfTrueNotSolvedByKinductionPlainButKipdr}{Total}{}{Cputime}{Avg}{13.09653435393318485523385301}%
\StoreBenchExecResult{PdrInv}{KinductionKipdrdfTrueNotSolvedByKinductionPlainButKipdr}{Total}{}{Cputime}{Median}{5.818751292}%
\StoreBenchExecResult{PdrInv}{KinductionKipdrdfTrueNotSolvedByKinductionPlainButKipdr}{Total}{}{Cputime}{Min}{3.231039296}%
\StoreBenchExecResult{PdrInv}{KinductionKipdrdfTrueNotSolvedByKinductionPlainButKipdr}{Total}{}{Cputime}{Max}{904.698984463}%
\StoreBenchExecResult{PdrInv}{KinductionKipdrdfTrueNotSolvedByKinductionPlainButKipdr}{Total}{}{Cputime}{Stdev}{73.29713430528883659361024023}%
\StoreBenchExecResult{PdrInv}{KinductionKipdrdfTrueNotSolvedByKinductionPlainButKipdr}{Total}{}{Walltime}{}{3031.63344478611}%
\StoreBenchExecResult{PdrInv}{KinductionKipdrdfTrueNotSolvedByKinductionPlainButKipdr}{Total}{}{Walltime}{Avg}{6.751967583042561247216035635}%
\StoreBenchExecResult{PdrInv}{KinductionKipdrdfTrueNotSolvedByKinductionPlainButKipdr}{Total}{}{Walltime}{Median}{3.09030294418}%
\StoreBenchExecResult{PdrInv}{KinductionKipdrdfTrueNotSolvedByKinductionPlainButKipdr}{Total}{}{Walltime}{Min}{1.80096197128}%
\StoreBenchExecResult{PdrInv}{KinductionKipdrdfTrueNotSolvedByKinductionPlainButKipdr}{Total}{}{Walltime}{Max}{455.224486113}%
\StoreBenchExecResult{PdrInv}{KinductionKipdrdfTrueNotSolvedByKinductionPlainButKipdr}{Total}{}{Walltime}{Stdev}{36.79375255528257383555734964}%
\StoreBenchExecResult{PdrInv}{KinductionKipdrdfTrueNotSolvedByKinductionPlainButKipdr}{Correct}{}{Count}{}{446}%
\StoreBenchExecResult{PdrInv}{KinductionKipdrdfTrueNotSolvedByKinductionPlainButKipdr}{Correct}{}{Cputime}{}{3170.644247536}%
\StoreBenchExecResult{PdrInv}{KinductionKipdrdfTrueNotSolvedByKinductionPlainButKipdr}{Correct}{}{Cputime}{Avg}{7.109067819587443946188340807}%
\StoreBenchExecResult{PdrInv}{KinductionKipdrdfTrueNotSolvedByKinductionPlainButKipdr}{Correct}{}{Cputime}{Median}{5.8088309995}%
\StoreBenchExecResult{PdrInv}{KinductionKipdrdfTrueNotSolvedByKinductionPlainButKipdr}{Correct}{}{Cputime}{Min}{3.231039296}%
\StoreBenchExecResult{PdrInv}{KinductionKipdrdfTrueNotSolvedByKinductionPlainButKipdr}{Correct}{}{Cputime}{Max}{85.555877857}%
\StoreBenchExecResult{PdrInv}{KinductionKipdrdfTrueNotSolvedByKinductionPlainButKipdr}{Correct}{}{Cputime}{Stdev}{6.564027017295492445783263231}%
\StoreBenchExecResult{PdrInv}{KinductionKipdrdfTrueNotSolvedByKinductionPlainButKipdr}{Correct}{}{Walltime}{}{1670.93439483611}%
\StoreBenchExecResult{PdrInv}{KinductionKipdrdfTrueNotSolvedByKinductionPlainButKipdr}{Correct}{}{Walltime}{Avg}{3.746489674520426008968609865}%
\StoreBenchExecResult{PdrInv}{KinductionKipdrdfTrueNotSolvedByKinductionPlainButKipdr}{Correct}{}{Walltime}{Median}{3.08919751644}%
\StoreBenchExecResult{PdrInv}{KinductionKipdrdfTrueNotSolvedByKinductionPlainButKipdr}{Correct}{}{Walltime}{Min}{1.80096197128}%
\StoreBenchExecResult{PdrInv}{KinductionKipdrdfTrueNotSolvedByKinductionPlainButKipdr}{Correct}{}{Walltime}{Max}{43.2875180244}%
\StoreBenchExecResult{PdrInv}{KinductionKipdrdfTrueNotSolvedByKinductionPlainButKipdr}{Correct}{}{Walltime}{Stdev}{3.309593798710138424579738271}%
\StoreBenchExecResult{PdrInv}{KinductionKipdrdfTrueNotSolvedByKinductionPlainButKipdr}{Correct}{True}{Count}{}{446}%
\StoreBenchExecResult{PdrInv}{KinductionKipdrdfTrueNotSolvedByKinductionPlainButKipdr}{Correct}{True}{Cputime}{}{3170.644247536}%
\StoreBenchExecResult{PdrInv}{KinductionKipdrdfTrueNotSolvedByKinductionPlainButKipdr}{Correct}{True}{Cputime}{Avg}{7.109067819587443946188340807}%
\StoreBenchExecResult{PdrInv}{KinductionKipdrdfTrueNotSolvedByKinductionPlainButKipdr}{Correct}{True}{Cputime}{Median}{5.8088309995}%
\StoreBenchExecResult{PdrInv}{KinductionKipdrdfTrueNotSolvedByKinductionPlainButKipdr}{Correct}{True}{Cputime}{Min}{3.231039296}%
\StoreBenchExecResult{PdrInv}{KinductionKipdrdfTrueNotSolvedByKinductionPlainButKipdr}{Correct}{True}{Cputime}{Max}{85.555877857}%
\StoreBenchExecResult{PdrInv}{KinductionKipdrdfTrueNotSolvedByKinductionPlainButKipdr}{Correct}{True}{Cputime}{Stdev}{6.564027017295492445783263231}%
\StoreBenchExecResult{PdrInv}{KinductionKipdrdfTrueNotSolvedByKinductionPlainButKipdr}{Correct}{True}{Walltime}{}{1670.93439483611}%
\StoreBenchExecResult{PdrInv}{KinductionKipdrdfTrueNotSolvedByKinductionPlainButKipdr}{Correct}{True}{Walltime}{Avg}{3.746489674520426008968609865}%
\StoreBenchExecResult{PdrInv}{KinductionKipdrdfTrueNotSolvedByKinductionPlainButKipdr}{Correct}{True}{Walltime}{Median}{3.08919751644}%
\StoreBenchExecResult{PdrInv}{KinductionKipdrdfTrueNotSolvedByKinductionPlainButKipdr}{Correct}{True}{Walltime}{Min}{1.80096197128}%
\StoreBenchExecResult{PdrInv}{KinductionKipdrdfTrueNotSolvedByKinductionPlainButKipdr}{Correct}{True}{Walltime}{Max}{43.2875180244}%
\StoreBenchExecResult{PdrInv}{KinductionKipdrdfTrueNotSolvedByKinductionPlainButKipdr}{Correct}{True}{Walltime}{Stdev}{3.309593798710138424579738271}%
\StoreBenchExecResult{PdrInv}{KinductionKipdrdfTrueNotSolvedByKinductionPlainButKipdr}{Wrong}{True}{Count}{}{0}%
\StoreBenchExecResult{PdrInv}{KinductionKipdrdfTrueNotSolvedByKinductionPlainButKipdr}{Wrong}{True}{Cputime}{}{0}%
\StoreBenchExecResult{PdrInv}{KinductionKipdrdfTrueNotSolvedByKinductionPlainButKipdr}{Wrong}{True}{Cputime}{Avg}{None}%
\StoreBenchExecResult{PdrInv}{KinductionKipdrdfTrueNotSolvedByKinductionPlainButKipdr}{Wrong}{True}{Cputime}{Median}{None}%
\StoreBenchExecResult{PdrInv}{KinductionKipdrdfTrueNotSolvedByKinductionPlainButKipdr}{Wrong}{True}{Cputime}{Min}{None}%
\StoreBenchExecResult{PdrInv}{KinductionKipdrdfTrueNotSolvedByKinductionPlainButKipdr}{Wrong}{True}{Cputime}{Max}{None}%
\StoreBenchExecResult{PdrInv}{KinductionKipdrdfTrueNotSolvedByKinductionPlainButKipdr}{Wrong}{True}{Cputime}{Stdev}{None}%
\StoreBenchExecResult{PdrInv}{KinductionKipdrdfTrueNotSolvedByKinductionPlainButKipdr}{Wrong}{True}{Walltime}{}{0}%
\StoreBenchExecResult{PdrInv}{KinductionKipdrdfTrueNotSolvedByKinductionPlainButKipdr}{Wrong}{True}{Walltime}{Avg}{None}%
\StoreBenchExecResult{PdrInv}{KinductionKipdrdfTrueNotSolvedByKinductionPlainButKipdr}{Wrong}{True}{Walltime}{Median}{None}%
\StoreBenchExecResult{PdrInv}{KinductionKipdrdfTrueNotSolvedByKinductionPlainButKipdr}{Wrong}{True}{Walltime}{Min}{None}%
\StoreBenchExecResult{PdrInv}{KinductionKipdrdfTrueNotSolvedByKinductionPlainButKipdr}{Wrong}{True}{Walltime}{Max}{None}%
\StoreBenchExecResult{PdrInv}{KinductionKipdrdfTrueNotSolvedByKinductionPlainButKipdr}{Wrong}{True}{Walltime}{Stdev}{None}%
\StoreBenchExecResult{PdrInv}{KinductionKipdrdfTrueNotSolvedByKinductionPlainButKipdr}{Error}{}{Count}{}{3}%
\StoreBenchExecResult{PdrInv}{KinductionKipdrdfTrueNotSolvedByKinductionPlainButKipdr}{Error}{}{Cputime}{}{2709.699677380}%
\StoreBenchExecResult{PdrInv}{KinductionKipdrdfTrueNotSolvedByKinductionPlainButKipdr}{Error}{}{Cputime}{Avg}{903.2332257933333333333333333}%
\StoreBenchExecResult{PdrInv}{KinductionKipdrdfTrueNotSolvedByKinductionPlainButKipdr}{Error}{}{Cputime}{Median}{902.835717402}%
\StoreBenchExecResult{PdrInv}{KinductionKipdrdfTrueNotSolvedByKinductionPlainButKipdr}{Error}{}{Cputime}{Min}{902.164975515}%
\StoreBenchExecResult{PdrInv}{KinductionKipdrdfTrueNotSolvedByKinductionPlainButKipdr}{Error}{}{Cputime}{Max}{904.698984463}%
\StoreBenchExecResult{PdrInv}{KinductionKipdrdfTrueNotSolvedByKinductionPlainButKipdr}{Error}{}{Cputime}{Stdev}{1.072010580800185800210071111}%
\StoreBenchExecResult{PdrInv}{KinductionKipdrdfTrueNotSolvedByKinductionPlainButKipdr}{Error}{}{Walltime}{}{1360.699049950}%
\StoreBenchExecResult{PdrInv}{KinductionKipdrdfTrueNotSolvedByKinductionPlainButKipdr}{Error}{}{Walltime}{Avg}{453.5663499833333333333333333}%
\StoreBenchExecResult{PdrInv}{KinductionKipdrdfTrueNotSolvedByKinductionPlainButKipdr}{Error}{}{Walltime}{Median}{452.854034901}%
\StoreBenchExecResult{PdrInv}{KinductionKipdrdfTrueNotSolvedByKinductionPlainButKipdr}{Error}{}{Walltime}{Min}{452.620528936}%
\StoreBenchExecResult{PdrInv}{KinductionKipdrdfTrueNotSolvedByKinductionPlainButKipdr}{Error}{}{Walltime}{Max}{455.224486113}%
\StoreBenchExecResult{PdrInv}{KinductionKipdrdfTrueNotSolvedByKinductionPlainButKipdr}{Error}{}{Walltime}{Stdev}{1.176348255492992436855294755}%
\StoreBenchExecResult{PdrInv}{KinductionKipdrdfTrueNotSolvedByKinductionPlainButKipdr}{Error}{Timeout}{Count}{}{3}%
\StoreBenchExecResult{PdrInv}{KinductionKipdrdfTrueNotSolvedByKinductionPlainButKipdr}{Error}{Timeout}{Cputime}{}{2709.699677380}%
\StoreBenchExecResult{PdrInv}{KinductionKipdrdfTrueNotSolvedByKinductionPlainButKipdr}{Error}{Timeout}{Cputime}{Avg}{903.2332257933333333333333333}%
\StoreBenchExecResult{PdrInv}{KinductionKipdrdfTrueNotSolvedByKinductionPlainButKipdr}{Error}{Timeout}{Cputime}{Median}{902.835717402}%
\StoreBenchExecResult{PdrInv}{KinductionKipdrdfTrueNotSolvedByKinductionPlainButKipdr}{Error}{Timeout}{Cputime}{Min}{902.164975515}%
\StoreBenchExecResult{PdrInv}{KinductionKipdrdfTrueNotSolvedByKinductionPlainButKipdr}{Error}{Timeout}{Cputime}{Max}{904.698984463}%
\StoreBenchExecResult{PdrInv}{KinductionKipdrdfTrueNotSolvedByKinductionPlainButKipdr}{Error}{Timeout}{Cputime}{Stdev}{1.072010580800185800210071111}%
\StoreBenchExecResult{PdrInv}{KinductionKipdrdfTrueNotSolvedByKinductionPlainButKipdr}{Error}{Timeout}{Walltime}{}{1360.699049950}%
\StoreBenchExecResult{PdrInv}{KinductionKipdrdfTrueNotSolvedByKinductionPlainButKipdr}{Error}{Timeout}{Walltime}{Avg}{453.5663499833333333333333333}%
\StoreBenchExecResult{PdrInv}{KinductionKipdrdfTrueNotSolvedByKinductionPlainButKipdr}{Error}{Timeout}{Walltime}{Median}{452.854034901}%
\StoreBenchExecResult{PdrInv}{KinductionKipdrdfTrueNotSolvedByKinductionPlainButKipdr}{Error}{Timeout}{Walltime}{Min}{452.620528936}%
\StoreBenchExecResult{PdrInv}{KinductionKipdrdfTrueNotSolvedByKinductionPlainButKipdr}{Error}{Timeout}{Walltime}{Max}{455.224486113}%
\StoreBenchExecResult{PdrInv}{KinductionKipdrdfTrueNotSolvedByKinductionPlainButKipdr}{Error}{Timeout}{Walltime}{Stdev}{1.176348255492992436855294755}%
\providecommand\StoreBenchExecResult[7]{\expandafter\newcommand\csname#1#2#3#4#5#6\endcsname{#7}}%
\StoreBenchExecResult{PdrInv}{KinductionKipdrdfTrueNotSolvedByKinductionPlain}{Total}{}{Count}{}{2893}%
\StoreBenchExecResult{PdrInv}{KinductionKipdrdfTrueNotSolvedByKinductionPlain}{Total}{}{Cputime}{}{1509948.996897475}%
\StoreBenchExecResult{PdrInv}{KinductionKipdrdfTrueNotSolvedByKinductionPlain}{Total}{}{Cputime}{Avg}{521.9319035248790183200829589}%
\StoreBenchExecResult{PdrInv}{KinductionKipdrdfTrueNotSolvedByKinductionPlain}{Total}{}{Cputime}{Median}{901.013394247}%
\StoreBenchExecResult{PdrInv}{KinductionKipdrdfTrueNotSolvedByKinductionPlain}{Total}{}{Cputime}{Min}{2.333845157}%
\StoreBenchExecResult{PdrInv}{KinductionKipdrdfTrueNotSolvedByKinductionPlain}{Total}{}{Cputime}{Max}{1002.29981558}%
\StoreBenchExecResult{PdrInv}{KinductionKipdrdfTrueNotSolvedByKinductionPlain}{Total}{}{Cputime}{Stdev}{421.3879665171923677953011939}%
\StoreBenchExecResult{PdrInv}{KinductionKipdrdfTrueNotSolvedByKinductionPlain}{Total}{}{Walltime}{}{833867.89249252894}%
\StoreBenchExecResult{PdrInv}{KinductionKipdrdfTrueNotSolvedByKinductionPlain}{Total}{}{Walltime}{Avg}{288.2363956075108676114759765}%
\StoreBenchExecResult{PdrInv}{KinductionKipdrdfTrueNotSolvedByKinductionPlain}{Total}{}{Walltime}{Median}{451.34181118}%
\StoreBenchExecResult{PdrInv}{KinductionKipdrdfTrueNotSolvedByKinductionPlain}{Total}{}{Walltime}{Min}{1.29316186905}%
\StoreBenchExecResult{PdrInv}{KinductionKipdrdfTrueNotSolvedByKinductionPlain}{Total}{}{Walltime}{Max}{898.137360096}%
\StoreBenchExecResult{PdrInv}{KinductionKipdrdfTrueNotSolvedByKinductionPlain}{Total}{}{Walltime}{Stdev}{249.0243051760393717454108101}%
\StoreBenchExecResult{PdrInv}{KinductionKipdrdfTrueNotSolvedByKinductionPlain}{Correct}{}{Count}{}{1119}%
\StoreBenchExecResult{PdrInv}{KinductionKipdrdfTrueNotSolvedByKinductionPlain}{Correct}{}{Cputime}{}{76213.475371731}%
\StoreBenchExecResult{PdrInv}{KinductionKipdrdfTrueNotSolvedByKinductionPlain}{Correct}{}{Cputime}{Avg}{68.10855707929490616621983914}%
\StoreBenchExecResult{PdrInv}{KinductionKipdrdfTrueNotSolvedByKinductionPlain}{Correct}{}{Cputime}{Median}{8.016751553}%
\StoreBenchExecResult{PdrInv}{KinductionKipdrdfTrueNotSolvedByKinductionPlain}{Correct}{}{Cputime}{Min}{3.222336252}%
\StoreBenchExecResult{PdrInv}{KinductionKipdrdfTrueNotSolvedByKinductionPlain}{Correct}{}{Cputime}{Max}{872.592645571}%
\StoreBenchExecResult{PdrInv}{KinductionKipdrdfTrueNotSolvedByKinductionPlain}{Correct}{}{Cputime}{Stdev}{154.3641666561558942125628959}%
\StoreBenchExecResult{PdrInv}{KinductionKipdrdfTrueNotSolvedByKinductionPlain}{Correct}{}{Walltime}{}{40054.14290452222}%
\StoreBenchExecResult{PdrInv}{KinductionKipdrdfTrueNotSolvedByKinductionPlain}{Correct}{}{Walltime}{Avg}{35.79458704604309204647006256}%
\StoreBenchExecResult{PdrInv}{KinductionKipdrdfTrueNotSolvedByKinductionPlain}{Correct}{}{Walltime}{Median}{4.21357297897}%
\StoreBenchExecResult{PdrInv}{KinductionKipdrdfTrueNotSolvedByKinductionPlain}{Correct}{}{Walltime}{Min}{1.79158210754}%
\StoreBenchExecResult{PdrInv}{KinductionKipdrdfTrueNotSolvedByKinductionPlain}{Correct}{}{Walltime}{Max}{801.756588936}%
\StoreBenchExecResult{PdrInv}{KinductionKipdrdfTrueNotSolvedByKinductionPlain}{Correct}{}{Walltime}{Stdev}{84.50188543129154865722711913}%
\StoreBenchExecResult{PdrInv}{KinductionKipdrdfTrueNotSolvedByKinductionPlain}{Correct}{True}{Count}{}{1119}%
\StoreBenchExecResult{PdrInv}{KinductionKipdrdfTrueNotSolvedByKinductionPlain}{Correct}{True}{Cputime}{}{76213.475371731}%
\StoreBenchExecResult{PdrInv}{KinductionKipdrdfTrueNotSolvedByKinductionPlain}{Correct}{True}{Cputime}{Avg}{68.10855707929490616621983914}%
\StoreBenchExecResult{PdrInv}{KinductionKipdrdfTrueNotSolvedByKinductionPlain}{Correct}{True}{Cputime}{Median}{8.016751553}%
\StoreBenchExecResult{PdrInv}{KinductionKipdrdfTrueNotSolvedByKinductionPlain}{Correct}{True}{Cputime}{Min}{3.222336252}%
\StoreBenchExecResult{PdrInv}{KinductionKipdrdfTrueNotSolvedByKinductionPlain}{Correct}{True}{Cputime}{Max}{872.592645571}%
\StoreBenchExecResult{PdrInv}{KinductionKipdrdfTrueNotSolvedByKinductionPlain}{Correct}{True}{Cputime}{Stdev}{154.3641666561558942125628959}%
\StoreBenchExecResult{PdrInv}{KinductionKipdrdfTrueNotSolvedByKinductionPlain}{Correct}{True}{Walltime}{}{40054.14290452222}%
\StoreBenchExecResult{PdrInv}{KinductionKipdrdfTrueNotSolvedByKinductionPlain}{Correct}{True}{Walltime}{Avg}{35.79458704604309204647006256}%
\StoreBenchExecResult{PdrInv}{KinductionKipdrdfTrueNotSolvedByKinductionPlain}{Correct}{True}{Walltime}{Median}{4.21357297897}%
\StoreBenchExecResult{PdrInv}{KinductionKipdrdfTrueNotSolvedByKinductionPlain}{Correct}{True}{Walltime}{Min}{1.79158210754}%
\StoreBenchExecResult{PdrInv}{KinductionKipdrdfTrueNotSolvedByKinductionPlain}{Correct}{True}{Walltime}{Max}{801.756588936}%
\StoreBenchExecResult{PdrInv}{KinductionKipdrdfTrueNotSolvedByKinductionPlain}{Correct}{True}{Walltime}{Stdev}{84.50188543129154865722711913}%
\StoreBenchExecResult{PdrInv}{KinductionKipdrdfTrueNotSolvedByKinductionPlain}{Wrong}{True}{Count}{}{0}%
\StoreBenchExecResult{PdrInv}{KinductionKipdrdfTrueNotSolvedByKinductionPlain}{Wrong}{True}{Cputime}{}{0}%
\StoreBenchExecResult{PdrInv}{KinductionKipdrdfTrueNotSolvedByKinductionPlain}{Wrong}{True}{Cputime}{Avg}{None}%
\StoreBenchExecResult{PdrInv}{KinductionKipdrdfTrueNotSolvedByKinductionPlain}{Wrong}{True}{Cputime}{Median}{None}%
\StoreBenchExecResult{PdrInv}{KinductionKipdrdfTrueNotSolvedByKinductionPlain}{Wrong}{True}{Cputime}{Min}{None}%
\StoreBenchExecResult{PdrInv}{KinductionKipdrdfTrueNotSolvedByKinductionPlain}{Wrong}{True}{Cputime}{Max}{None}%
\StoreBenchExecResult{PdrInv}{KinductionKipdrdfTrueNotSolvedByKinductionPlain}{Wrong}{True}{Cputime}{Stdev}{None}%
\StoreBenchExecResult{PdrInv}{KinductionKipdrdfTrueNotSolvedByKinductionPlain}{Wrong}{True}{Walltime}{}{0}%
\StoreBenchExecResult{PdrInv}{KinductionKipdrdfTrueNotSolvedByKinductionPlain}{Wrong}{True}{Walltime}{Avg}{None}%
\StoreBenchExecResult{PdrInv}{KinductionKipdrdfTrueNotSolvedByKinductionPlain}{Wrong}{True}{Walltime}{Median}{None}%
\StoreBenchExecResult{PdrInv}{KinductionKipdrdfTrueNotSolvedByKinductionPlain}{Wrong}{True}{Walltime}{Min}{None}%
\StoreBenchExecResult{PdrInv}{KinductionKipdrdfTrueNotSolvedByKinductionPlain}{Wrong}{True}{Walltime}{Max}{None}%
\StoreBenchExecResult{PdrInv}{KinductionKipdrdfTrueNotSolvedByKinductionPlain}{Wrong}{True}{Walltime}{Stdev}{None}%
\StoreBenchExecResult{PdrInv}{KinductionKipdrdfTrueNotSolvedByKinductionPlain}{Error}{}{Count}{}{1774}%
\StoreBenchExecResult{PdrInv}{KinductionKipdrdfTrueNotSolvedByKinductionPlain}{Error}{}{Cputime}{}{1433735.521525744}%
\StoreBenchExecResult{PdrInv}{KinductionKipdrdfTrueNotSolvedByKinductionPlain}{Error}{}{Cputime}{Avg}{808.1936423482209695603156708}%
\StoreBenchExecResult{PdrInv}{KinductionKipdrdfTrueNotSolvedByKinductionPlain}{Error}{}{Cputime}{Median}{901.541527149}%
\StoreBenchExecResult{PdrInv}{KinductionKipdrdfTrueNotSolvedByKinductionPlain}{Error}{}{Cputime}{Min}{2.333845157}%
\StoreBenchExecResult{PdrInv}{KinductionKipdrdfTrueNotSolvedByKinductionPlain}{Error}{}{Cputime}{Max}{1002.29981558}%
\StoreBenchExecResult{PdrInv}{KinductionKipdrdfTrueNotSolvedByKinductionPlain}{Error}{}{Cputime}{Stdev}{250.3702627705511428457292474}%
\StoreBenchExecResult{PdrInv}{KinductionKipdrdfTrueNotSolvedByKinductionPlain}{Error}{}{Walltime}{}{793813.74958800672}%
\StoreBenchExecResult{PdrInv}{KinductionKipdrdfTrueNotSolvedByKinductionPlain}{Error}{}{Walltime}{Avg}{447.4711102525404284103720406}%
\StoreBenchExecResult{PdrInv}{KinductionKipdrdfTrueNotSolvedByKinductionPlain}{Error}{}{Walltime}{Median}{452.2161884305}%
\StoreBenchExecResult{PdrInv}{KinductionKipdrdfTrueNotSolvedByKinductionPlain}{Error}{}{Walltime}{Min}{1.29316186905}%
\StoreBenchExecResult{PdrInv}{KinductionKipdrdfTrueNotSolvedByKinductionPlain}{Error}{}{Walltime}{Max}{898.137360096}%
\StoreBenchExecResult{PdrInv}{KinductionKipdrdfTrueNotSolvedByKinductionPlain}{Error}{}{Walltime}{Stdev}{176.2733502724707737623383521}%
\StoreBenchExecResult{PdrInv}{KinductionKipdrdfTrueNotSolvedByKinductionPlain}{Error}{Assertion}{Count}{}{2}%
\StoreBenchExecResult{PdrInv}{KinductionKipdrdfTrueNotSolvedByKinductionPlain}{Error}{Assertion}{Cputime}{}{6.983796226}%
\StoreBenchExecResult{PdrInv}{KinductionKipdrdfTrueNotSolvedByKinductionPlain}{Error}{Assertion}{Cputime}{Avg}{3.491898113}%
\StoreBenchExecResult{PdrInv}{KinductionKipdrdfTrueNotSolvedByKinductionPlain}{Error}{Assertion}{Cputime}{Median}{3.491898113}%
\StoreBenchExecResult{PdrInv}{KinductionKipdrdfTrueNotSolvedByKinductionPlain}{Error}{Assertion}{Cputime}{Min}{3.368334957}%
\StoreBenchExecResult{PdrInv}{KinductionKipdrdfTrueNotSolvedByKinductionPlain}{Error}{Assertion}{Cputime}{Max}{3.615461269}%
\StoreBenchExecResult{PdrInv}{KinductionKipdrdfTrueNotSolvedByKinductionPlain}{Error}{Assertion}{Cputime}{Stdev}{0.123563156}%
\StoreBenchExecResult{PdrInv}{KinductionKipdrdfTrueNotSolvedByKinductionPlain}{Error}{Assertion}{Walltime}{}{3.87283825874}%
\StoreBenchExecResult{PdrInv}{KinductionKipdrdfTrueNotSolvedByKinductionPlain}{Error}{Assertion}{Walltime}{Avg}{1.93641912937}%
\StoreBenchExecResult{PdrInv}{KinductionKipdrdfTrueNotSolvedByKinductionPlain}{Error}{Assertion}{Walltime}{Median}{1.93641912937}%
\StoreBenchExecResult{PdrInv}{KinductionKipdrdfTrueNotSolvedByKinductionPlain}{Error}{Assertion}{Walltime}{Min}{1.87764120102}%
\StoreBenchExecResult{PdrInv}{KinductionKipdrdfTrueNotSolvedByKinductionPlain}{Error}{Assertion}{Walltime}{Max}{1.99519705772}%
\StoreBenchExecResult{PdrInv}{KinductionKipdrdfTrueNotSolvedByKinductionPlain}{Error}{Assertion}{Walltime}{Stdev}{0.05877792835}%
\StoreBenchExecResult{PdrInv}{KinductionKipdrdfTrueNotSolvedByKinductionPlain}{Error}{Error}{Count}{}{128}%
\StoreBenchExecResult{PdrInv}{KinductionKipdrdfTrueNotSolvedByKinductionPlain}{Error}{Error}{Cputime}{}{21469.538604572}%
\StoreBenchExecResult{PdrInv}{KinductionKipdrdfTrueNotSolvedByKinductionPlain}{Error}{Error}{Cputime}{Avg}{167.73077034821875}%
\StoreBenchExecResult{PdrInv}{KinductionKipdrdfTrueNotSolvedByKinductionPlain}{Error}{Error}{Cputime}{Median}{106.682116089}%
\StoreBenchExecResult{PdrInv}{KinductionKipdrdfTrueNotSolvedByKinductionPlain}{Error}{Error}{Cputime}{Min}{2.333845157}%
\StoreBenchExecResult{PdrInv}{KinductionKipdrdfTrueNotSolvedByKinductionPlain}{Error}{Error}{Cputime}{Max}{788.647328849}%
\StoreBenchExecResult{PdrInv}{KinductionKipdrdfTrueNotSolvedByKinductionPlain}{Error}{Error}{Cputime}{Stdev}{190.3470966469567674877889651}%
\StoreBenchExecResult{PdrInv}{KinductionKipdrdfTrueNotSolvedByKinductionPlain}{Error}{Error}{Walltime}{}{18070.83810401099}%
\StoreBenchExecResult{PdrInv}{KinductionKipdrdfTrueNotSolvedByKinductionPlain}{Error}{Error}{Walltime}{Avg}{141.178422687585859375}%
\StoreBenchExecResult{PdrInv}{KinductionKipdrdfTrueNotSolvedByKinductionPlain}{Error}{Error}{Walltime}{Median}{79.6765950918}%
\StoreBenchExecResult{PdrInv}{KinductionKipdrdfTrueNotSolvedByKinductionPlain}{Error}{Error}{Walltime}{Min}{1.29316186905}%
\StoreBenchExecResult{PdrInv}{KinductionKipdrdfTrueNotSolvedByKinductionPlain}{Error}{Error}{Walltime}{Max}{774.052371025}%
\StoreBenchExecResult{PdrInv}{KinductionKipdrdfTrueNotSolvedByKinductionPlain}{Error}{Error}{Walltime}{Stdev}{170.7430203287248198718881995}%
\StoreBenchExecResult{PdrInv}{KinductionKipdrdfTrueNotSolvedByKinductionPlain}{Error}{Exception}{Count}{}{16}%
\StoreBenchExecResult{PdrInv}{KinductionKipdrdfTrueNotSolvedByKinductionPlain}{Error}{Exception}{Cputime}{}{3473.504349252}%
\StoreBenchExecResult{PdrInv}{KinductionKipdrdfTrueNotSolvedByKinductionPlain}{Error}{Exception}{Cputime}{Avg}{217.09402182825}%
\StoreBenchExecResult{PdrInv}{KinductionKipdrdfTrueNotSolvedByKinductionPlain}{Error}{Exception}{Cputime}{Median}{191.667659044}%
\StoreBenchExecResult{PdrInv}{KinductionKipdrdfTrueNotSolvedByKinductionPlain}{Error}{Exception}{Cputime}{Min}{15.540618942}%
\StoreBenchExecResult{PdrInv}{KinductionKipdrdfTrueNotSolvedByKinductionPlain}{Error}{Exception}{Cputime}{Max}{546.507850636}%
\StoreBenchExecResult{PdrInv}{KinductionKipdrdfTrueNotSolvedByKinductionPlain}{Error}{Exception}{Cputime}{Stdev}{146.8282843745132260482814436}%
\StoreBenchExecResult{PdrInv}{KinductionKipdrdfTrueNotSolvedByKinductionPlain}{Error}{Exception}{Walltime}{}{1744.02903151501}%
\StoreBenchExecResult{PdrInv}{KinductionKipdrdfTrueNotSolvedByKinductionPlain}{Error}{Exception}{Walltime}{Avg}{109.001814469688125}%
\StoreBenchExecResult{PdrInv}{KinductionKipdrdfTrueNotSolvedByKinductionPlain}{Error}{Exception}{Walltime}{Median}{96.2818005086}%
\StoreBenchExecResult{PdrInv}{KinductionKipdrdfTrueNotSolvedByKinductionPlain}{Error}{Exception}{Walltime}{Min}{7.95783209801}%
\StoreBenchExecResult{PdrInv}{KinductionKipdrdfTrueNotSolvedByKinductionPlain}{Error}{Exception}{Walltime}{Max}{273.986470938}%
\StoreBenchExecResult{PdrInv}{KinductionKipdrdfTrueNotSolvedByKinductionPlain}{Error}{Exception}{Walltime}{Stdev}{73.58170055271908671051577824}%
\StoreBenchExecResult{PdrInv}{KinductionKipdrdfTrueNotSolvedByKinductionPlain}{Error}{OutOfJavaMemory}{Count}{}{4}%
\StoreBenchExecResult{PdrInv}{KinductionKipdrdfTrueNotSolvedByKinductionPlain}{Error}{OutOfJavaMemory}{Cputime}{}{2110.630689999}%
\StoreBenchExecResult{PdrInv}{KinductionKipdrdfTrueNotSolvedByKinductionPlain}{Error}{OutOfJavaMemory}{Cputime}{Avg}{527.65767249975}%
\StoreBenchExecResult{PdrInv}{KinductionKipdrdfTrueNotSolvedByKinductionPlain}{Error}{OutOfJavaMemory}{Cputime}{Median}{481.086194167}%
\StoreBenchExecResult{PdrInv}{KinductionKipdrdfTrueNotSolvedByKinductionPlain}{Error}{OutOfJavaMemory}{Cputime}{Min}{251.848360092}%
\StoreBenchExecResult{PdrInv}{KinductionKipdrdfTrueNotSolvedByKinductionPlain}{Error}{OutOfJavaMemory}{Cputime}{Max}{896.609941573}%
\StoreBenchExecResult{PdrInv}{KinductionKipdrdfTrueNotSolvedByKinductionPlain}{Error}{OutOfJavaMemory}{Cputime}{Stdev}{233.0307049556052303596210464}%
\StoreBenchExecResult{PdrInv}{KinductionKipdrdfTrueNotSolvedByKinductionPlain}{Error}{OutOfJavaMemory}{Walltime}{}{1083.567172051}%
\StoreBenchExecResult{PdrInv}{KinductionKipdrdfTrueNotSolvedByKinductionPlain}{Error}{OutOfJavaMemory}{Walltime}{Avg}{270.89179301275}%
\StoreBenchExecResult{PdrInv}{KinductionKipdrdfTrueNotSolvedByKinductionPlain}{Error}{OutOfJavaMemory}{Walltime}{Median}{249.4470169545}%
\StoreBenchExecResult{PdrInv}{KinductionKipdrdfTrueNotSolvedByKinductionPlain}{Error}{OutOfJavaMemory}{Walltime}{Min}{128.864484072}%
\StoreBenchExecResult{PdrInv}{KinductionKipdrdfTrueNotSolvedByKinductionPlain}{Error}{OutOfJavaMemory}{Walltime}{Max}{455.80865407}%
\StoreBenchExecResult{PdrInv}{KinductionKipdrdfTrueNotSolvedByKinductionPlain}{Error}{OutOfJavaMemory}{Walltime}{Stdev}{117.7615194763728266935976662}%
\StoreBenchExecResult{PdrInv}{KinductionKipdrdfTrueNotSolvedByKinductionPlain}{Error}{OutOfMemory}{Count}{}{106}%
\StoreBenchExecResult{PdrInv}{KinductionKipdrdfTrueNotSolvedByKinductionPlain}{Error}{OutOfMemory}{Cputime}{}{43322.253558575}%
\StoreBenchExecResult{PdrInv}{KinductionKipdrdfTrueNotSolvedByKinductionPlain}{Error}{OutOfMemory}{Cputime}{Avg}{408.7005052695754716981132075}%
\StoreBenchExecResult{PdrInv}{KinductionKipdrdfTrueNotSolvedByKinductionPlain}{Error}{OutOfMemory}{Cputime}{Median}{340.032344923}%
\StoreBenchExecResult{PdrInv}{KinductionKipdrdfTrueNotSolvedByKinductionPlain}{Error}{OutOfMemory}{Cputime}{Min}{168.557623561}%
\StoreBenchExecResult{PdrInv}{KinductionKipdrdfTrueNotSolvedByKinductionPlain}{Error}{OutOfMemory}{Cputime}{Max}{893.315293101}%
\StoreBenchExecResult{PdrInv}{KinductionKipdrdfTrueNotSolvedByKinductionPlain}{Error}{OutOfMemory}{Cputime}{Stdev}{231.1537518135104995123483160}%
\StoreBenchExecResult{PdrInv}{KinductionKipdrdfTrueNotSolvedByKinductionPlain}{Error}{OutOfMemory}{Walltime}{}{21727.4545679062}%
\StoreBenchExecResult{PdrInv}{KinductionKipdrdfTrueNotSolvedByKinductionPlain}{Error}{OutOfMemory}{Walltime}{Avg}{204.9759864896811320754716981}%
\StoreBenchExecResult{PdrInv}{KinductionKipdrdfTrueNotSolvedByKinductionPlain}{Error}{OutOfMemory}{Walltime}{Median}{170.6053814885}%
\StoreBenchExecResult{PdrInv}{KinductionKipdrdfTrueNotSolvedByKinductionPlain}{Error}{OutOfMemory}{Walltime}{Min}{84.8142340183}%
\StoreBenchExecResult{PdrInv}{KinductionKipdrdfTrueNotSolvedByKinductionPlain}{Error}{OutOfMemory}{Walltime}{Max}{447.369602919}%
\StoreBenchExecResult{PdrInv}{KinductionKipdrdfTrueNotSolvedByKinductionPlain}{Error}{OutOfMemory}{Walltime}{Stdev}{115.6336981656141442826311949}%
\StoreBenchExecResult{PdrInv}{KinductionKipdrdfTrueNotSolvedByKinductionPlain}{Error}{SegmentationFault}{Count}{}{15}%
\StoreBenchExecResult{PdrInv}{KinductionKipdrdfTrueNotSolvedByKinductionPlain}{Error}{SegmentationFault}{Cputime}{}{855.873219127}%
\StoreBenchExecResult{PdrInv}{KinductionKipdrdfTrueNotSolvedByKinductionPlain}{Error}{SegmentationFault}{Cputime}{Avg}{57.05821460846666666666666667}%
\StoreBenchExecResult{PdrInv}{KinductionKipdrdfTrueNotSolvedByKinductionPlain}{Error}{SegmentationFault}{Cputime}{Median}{12.636805861}%
\StoreBenchExecResult{PdrInv}{KinductionKipdrdfTrueNotSolvedByKinductionPlain}{Error}{SegmentationFault}{Cputime}{Min}{7.624064345}%
\StoreBenchExecResult{PdrInv}{KinductionKipdrdfTrueNotSolvedByKinductionPlain}{Error}{SegmentationFault}{Cputime}{Max}{605.084929953}%
\StoreBenchExecResult{PdrInv}{KinductionKipdrdfTrueNotSolvedByKinductionPlain}{Error}{SegmentationFault}{Cputime}{Stdev}{147.2814503274155312539784161}%
\StoreBenchExecResult{PdrInv}{KinductionKipdrdfTrueNotSolvedByKinductionPlain}{Error}{SegmentationFault}{Walltime}{}{431.42628693578}%
\StoreBenchExecResult{PdrInv}{KinductionKipdrdfTrueNotSolvedByKinductionPlain}{Error}{SegmentationFault}{Walltime}{Avg}{28.76175246238533333333333333}%
\StoreBenchExecResult{PdrInv}{KinductionKipdrdfTrueNotSolvedByKinductionPlain}{Error}{SegmentationFault}{Walltime}{Median}{6.52975416183}%
\StoreBenchExecResult{PdrInv}{KinductionKipdrdfTrueNotSolvedByKinductionPlain}{Error}{SegmentationFault}{Walltime}{Min}{4.00342702866}%
\StoreBenchExecResult{PdrInv}{KinductionKipdrdfTrueNotSolvedByKinductionPlain}{Error}{SegmentationFault}{Walltime}{Max}{303.038462162}%
\StoreBenchExecResult{PdrInv}{KinductionKipdrdfTrueNotSolvedByKinductionPlain}{Error}{SegmentationFault}{Walltime}{Stdev}{73.71353401438307302455556241}%
\StoreBenchExecResult{PdrInv}{KinductionKipdrdfTrueNotSolvedByKinductionPlain}{Error}{Timeout}{Count}{}{1503}%
\StoreBenchExecResult{PdrInv}{KinductionKipdrdfTrueNotSolvedByKinductionPlain}{Error}{Timeout}{Cputime}{}{1362496.737307993}%
\StoreBenchExecResult{PdrInv}{KinductionKipdrdfTrueNotSolvedByKinductionPlain}{Error}{Timeout}{Cputime}{Avg}{906.5181219614058549567531603}%
\StoreBenchExecResult{PdrInv}{KinductionKipdrdfTrueNotSolvedByKinductionPlain}{Error}{Timeout}{Cputime}{Median}{901.903838051}%
\StoreBenchExecResult{PdrInv}{KinductionKipdrdfTrueNotSolvedByKinductionPlain}{Error}{Timeout}{Cputime}{Min}{900.831185784}%
\StoreBenchExecResult{PdrInv}{KinductionKipdrdfTrueNotSolvedByKinductionPlain}{Error}{Timeout}{Cputime}{Max}{1002.29981558}%
\StoreBenchExecResult{PdrInv}{KinductionKipdrdfTrueNotSolvedByKinductionPlain}{Error}{Timeout}{Cputime}{Stdev}{16.29397334494988245096955848}%
\StoreBenchExecResult{PdrInv}{KinductionKipdrdfTrueNotSolvedByKinductionPlain}{Error}{Timeout}{Walltime}{}{750752.561587329}%
\StoreBenchExecResult{PdrInv}{KinductionKipdrdfTrueNotSolvedByKinductionPlain}{Error}{Timeout}{Walltime}{Avg}{499.5027023202455089820359281}%
\StoreBenchExecResult{PdrInv}{KinductionKipdrdfTrueNotSolvedByKinductionPlain}{Error}{Timeout}{Walltime}{Median}{452.55604291}%
\StoreBenchExecResult{PdrInv}{KinductionKipdrdfTrueNotSolvedByKinductionPlain}{Error}{Timeout}{Walltime}{Min}{451.090511799}%
\StoreBenchExecResult{PdrInv}{KinductionKipdrdfTrueNotSolvedByKinductionPlain}{Error}{Timeout}{Walltime}{Max}{898.137360096}%
\StoreBenchExecResult{PdrInv}{KinductionKipdrdfTrueNotSolvedByKinductionPlain}{Error}{Timeout}{Walltime}{Stdev}{122.2287368404636663721720980}%
\providecommand\StoreBenchExecResult[7]{\expandafter\newcommand\csname#1#2#3#4#5#6\endcsname{#7}}%
\StoreBenchExecResult{PdrInv}{KinductionKipdrdf}{Total}{}{Count}{}{5591}%
\StoreBenchExecResult{PdrInv}{KinductionKipdrdf}{Total}{}{Cputime}{}{2165000.073686068}%
\StoreBenchExecResult{PdrInv}{KinductionKipdrdf}{Total}{}{Cputime}{Avg}{387.2294891228882131997853693}%
\StoreBenchExecResult{PdrInv}{KinductionKipdrdf}{Total}{}{Cputime}{Median}{95.384247913}%
\StoreBenchExecResult{PdrInv}{KinductionKipdrdf}{Total}{}{Cputime}{Min}{2.333845157}%
\StoreBenchExecResult{PdrInv}{KinductionKipdrdf}{Total}{}{Cputime}{Max}{1002.3192275}%
\StoreBenchExecResult{PdrInv}{KinductionKipdrdf}{Total}{}{Cputime}{Stdev}{415.6003665725482723412748271}%
\StoreBenchExecResult{PdrInv}{KinductionKipdrdf}{Total}{}{Walltime}{}{1184843.57164883817}%
\StoreBenchExecResult{PdrInv}{KinductionKipdrdf}{Total}{}{Walltime}{Avg}{211.9197946071969540332677517}%
\StoreBenchExecResult{PdrInv}{KinductionKipdrdf}{Total}{}{Walltime}{Median}{51.6288340092}%
\StoreBenchExecResult{PdrInv}{KinductionKipdrdf}{Total}{}{Walltime}{Min}{1.29316186905}%
\StoreBenchExecResult{PdrInv}{KinductionKipdrdf}{Total}{}{Walltime}{Max}{898.137360096}%
\StoreBenchExecResult{PdrInv}{KinductionKipdrdf}{Total}{}{Walltime}{Stdev}{238.4861278476651870851661458}%
\StoreBenchExecResult{PdrInv}{KinductionKipdrdf}{Correct}{}{Count}{}{3095}%
\StoreBenchExecResult{PdrInv}{KinductionKipdrdf}{Correct}{}{Cputime}{}{194703.546133703}%
\StoreBenchExecResult{PdrInv}{KinductionKipdrdf}{Correct}{}{Cputime}{Avg}{62.90906175563909531502423263}%
\StoreBenchExecResult{PdrInv}{KinductionKipdrdf}{Correct}{}{Cputime}{Median}{10.32600199}%
\StoreBenchExecResult{PdrInv}{KinductionKipdrdf}{Correct}{}{Cputime}{Min}{2.968020149}%
\StoreBenchExecResult{PdrInv}{KinductionKipdrdf}{Correct}{}{Cputime}{Max}{895.474978613}%
\StoreBenchExecResult{PdrInv}{KinductionKipdrdf}{Correct}{}{Cputime}{Stdev}{141.0404137005318992744670659}%
\StoreBenchExecResult{PdrInv}{KinductionKipdrdf}{Correct}{}{Walltime}{}{102712.03205610076}%
\StoreBenchExecResult{PdrInv}{KinductionKipdrdf}{Correct}{}{Walltime}{Avg}{33.18644008274661066235864297}%
\StoreBenchExecResult{PdrInv}{KinductionKipdrdf}{Correct}{}{Walltime}{Median}{5.37669706345}%
\StoreBenchExecResult{PdrInv}{KinductionKipdrdf}{Correct}{}{Walltime}{Min}{1.65646290779}%
\StoreBenchExecResult{PdrInv}{KinductionKipdrdf}{Correct}{}{Walltime}{Max}{808.819927931}%
\StoreBenchExecResult{PdrInv}{KinductionKipdrdf}{Correct}{}{Walltime}{Stdev}{77.25890459264403219784220176}%
\StoreBenchExecResult{PdrInv}{KinductionKipdrdf}{Correct}{False}{Count}{}{760}%
\StoreBenchExecResult{PdrInv}{KinductionKipdrdf}{Correct}{False}{Cputime}{}{62188.585590622}%
\StoreBenchExecResult{PdrInv}{KinductionKipdrdf}{Correct}{False}{Cputime}{Avg}{81.82708630345}%
\StoreBenchExecResult{PdrInv}{KinductionKipdrdf}{Correct}{False}{Cputime}{Median}{24.146890478}%
\StoreBenchExecResult{PdrInv}{KinductionKipdrdf}{Correct}{False}{Cputime}{Min}{3.108073613}%
\StoreBenchExecResult{PdrInv}{KinductionKipdrdf}{Correct}{False}{Cputime}{Max}{895.474978613}%
\StoreBenchExecResult{PdrInv}{KinductionKipdrdf}{Correct}{False}{Cputime}{Stdev}{165.0970560441322587820532438}%
\StoreBenchExecResult{PdrInv}{KinductionKipdrdf}{Correct}{False}{Walltime}{}{32433.87466836532}%
\StoreBenchExecResult{PdrInv}{KinductionKipdrdf}{Correct}{False}{Walltime}{Avg}{42.67615087942805263157894737}%
\StoreBenchExecResult{PdrInv}{KinductionKipdrdf}{Correct}{False}{Walltime}{Median}{12.58785104755}%
\StoreBenchExecResult{PdrInv}{KinductionKipdrdf}{Correct}{False}{Walltime}{Min}{1.76118993759}%
\StoreBenchExecResult{PdrInv}{KinductionKipdrdf}{Correct}{False}{Walltime}{Max}{635.433564186}%
\StoreBenchExecResult{PdrInv}{KinductionKipdrdf}{Correct}{False}{Walltime}{Stdev}{85.91785168620934404905822478}%
\StoreBenchExecResult{PdrInv}{KinductionKipdrdf}{Correct}{True}{Count}{}{2335}%
\StoreBenchExecResult{PdrInv}{KinductionKipdrdf}{Correct}{True}{Cputime}{}{132514.960543081}%
\StoreBenchExecResult{PdrInv}{KinductionKipdrdf}{Correct}{True}{Cputime}{Avg}{56.75158909767922912205567452}%
\StoreBenchExecResult{PdrInv}{KinductionKipdrdf}{Correct}{True}{Cputime}{Median}{7.992936812}%
\StoreBenchExecResult{PdrInv}{KinductionKipdrdf}{Correct}{True}{Cputime}{Min}{2.968020149}%
\StoreBenchExecResult{PdrInv}{KinductionKipdrdf}{Correct}{True}{Cputime}{Max}{872.592645571}%
\StoreBenchExecResult{PdrInv}{KinductionKipdrdf}{Correct}{True}{Cputime}{Stdev}{131.6850046844931944317457518}%
\StoreBenchExecResult{PdrInv}{KinductionKipdrdf}{Correct}{True}{Walltime}{}{70278.15738773544}%
\StoreBenchExecResult{PdrInv}{KinductionKipdrdf}{Correct}{True}{Walltime}{Avg}{30.09771194335564882226980728}%
\StoreBenchExecResult{PdrInv}{KinductionKipdrdf}{Correct}{True}{Walltime}{Median}{4.21352815628}%
\StoreBenchExecResult{PdrInv}{KinductionKipdrdf}{Correct}{True}{Walltime}{Min}{1.65646290779}%
\StoreBenchExecResult{PdrInv}{KinductionKipdrdf}{Correct}{True}{Walltime}{Max}{808.819927931}%
\StoreBenchExecResult{PdrInv}{KinductionKipdrdf}{Correct}{True}{Walltime}{Stdev}{73.96080763535478018850007886}%
\StoreBenchExecResult{PdrInv}{KinductionKipdrdf}{Wrong}{True}{Count}{}{0}%
\StoreBenchExecResult{PdrInv}{KinductionKipdrdf}{Wrong}{True}{Cputime}{}{0}%
\StoreBenchExecResult{PdrInv}{KinductionKipdrdf}{Wrong}{True}{Cputime}{Avg}{None}%
\StoreBenchExecResult{PdrInv}{KinductionKipdrdf}{Wrong}{True}{Cputime}{Median}{None}%
\StoreBenchExecResult{PdrInv}{KinductionKipdrdf}{Wrong}{True}{Cputime}{Min}{None}%
\StoreBenchExecResult{PdrInv}{KinductionKipdrdf}{Wrong}{True}{Cputime}{Max}{None}%
\StoreBenchExecResult{PdrInv}{KinductionKipdrdf}{Wrong}{True}{Cputime}{Stdev}{None}%
\StoreBenchExecResult{PdrInv}{KinductionKipdrdf}{Wrong}{True}{Walltime}{}{0}%
\StoreBenchExecResult{PdrInv}{KinductionKipdrdf}{Wrong}{True}{Walltime}{Avg}{None}%
\StoreBenchExecResult{PdrInv}{KinductionKipdrdf}{Wrong}{True}{Walltime}{Median}{None}%
\StoreBenchExecResult{PdrInv}{KinductionKipdrdf}{Wrong}{True}{Walltime}{Min}{None}%
\StoreBenchExecResult{PdrInv}{KinductionKipdrdf}{Wrong}{True}{Walltime}{Max}{None}%
\StoreBenchExecResult{PdrInv}{KinductionKipdrdf}{Wrong}{True}{Walltime}{Stdev}{None}%
\StoreBenchExecResult{PdrInv}{KinductionKipdrdf}{Error}{}{Count}{}{2494}%
\StoreBenchExecResult{PdrInv}{KinductionKipdrdf}{Error}{}{Cputime}{}{1970271.771058227}%
\StoreBenchExecResult{PdrInv}{KinductionKipdrdf}{Error}{}{Cputime}{Avg}{790.0047197506924619085805934}%
\StoreBenchExecResult{PdrInv}{KinductionKipdrdf}{Error}{}{Cputime}{Median}{901.6171193875}%
\StoreBenchExecResult{PdrInv}{KinductionKipdrdf}{Error}{}{Cputime}{Min}{2.333845157}%
\StoreBenchExecResult{PdrInv}{KinductionKipdrdf}{Error}{}{Cputime}{Max}{1002.3192275}%
\StoreBenchExecResult{PdrInv}{KinductionKipdrdf}{Error}{}{Cputime}{Stdev}{263.9149594407958154803038006}%
\StoreBenchExecResult{PdrInv}{KinductionKipdrdf}{Error}{}{Walltime}{}{1082118.43627499979}%
\StoreBenchExecResult{PdrInv}{KinductionKipdrdf}{Error}{}{Walltime}{Avg}{433.8887074077785846030473136}%
\StoreBenchExecResult{PdrInv}{KinductionKipdrdf}{Error}{}{Walltime}{Median}{452.222868562}%
\StoreBenchExecResult{PdrInv}{KinductionKipdrdf}{Error}{}{Walltime}{Min}{1.29316186905}%
\StoreBenchExecResult{PdrInv}{KinductionKipdrdf}{Error}{}{Walltime}{Max}{898.137360096}%
\StoreBenchExecResult{PdrInv}{KinductionKipdrdf}{Error}{}{Walltime}{Stdev}{176.4865487992586285953197282}%
\StoreBenchExecResult{PdrInv}{KinductionKipdrdf}{Error}{Assertion}{Count}{}{4}%
\StoreBenchExecResult{PdrInv}{KinductionKipdrdf}{Error}{Assertion}{Cputime}{}{13.736281712}%
\StoreBenchExecResult{PdrInv}{KinductionKipdrdf}{Error}{Assertion}{Cputime}{Avg}{3.434070428}%
\StoreBenchExecResult{PdrInv}{KinductionKipdrdf}{Error}{Assertion}{Cputime}{Median}{3.4181100355}%
\StoreBenchExecResult{PdrInv}{KinductionKipdrdf}{Error}{Assertion}{Cputime}{Min}{3.284600372}%
\StoreBenchExecResult{PdrInv}{KinductionKipdrdf}{Error}{Assertion}{Cputime}{Max}{3.615461269}%
\StoreBenchExecResult{PdrInv}{KinductionKipdrdf}{Error}{Assertion}{Cputime}{Stdev}{0.1231954950233579503027054932}%
\StoreBenchExecResult{PdrInv}{KinductionKipdrdf}{Error}{Assertion}{Walltime}{}{7.61115527152}%
\StoreBenchExecResult{PdrInv}{KinductionKipdrdf}{Error}{Assertion}{Walltime}{Avg}{1.90278881788}%
\StoreBenchExecResult{PdrInv}{KinductionKipdrdf}{Error}{Assertion}{Walltime}{Median}{1.892352104185}%
\StoreBenchExecResult{PdrInv}{KinductionKipdrdf}{Error}{Assertion}{Walltime}{Min}{1.83125400543}%
\StoreBenchExecResult{PdrInv}{KinductionKipdrdf}{Error}{Assertion}{Walltime}{Max}{1.99519705772}%
\StoreBenchExecResult{PdrInv}{KinductionKipdrdf}{Error}{Assertion}{Walltime}{Stdev}{0.05980631971611874181157970637}%
\StoreBenchExecResult{PdrInv}{KinductionKipdrdf}{Error}{Error}{Count}{}{186}%
\StoreBenchExecResult{PdrInv}{KinductionKipdrdf}{Error}{Error}{Cputime}{}{32736.744683723}%
\StoreBenchExecResult{PdrInv}{KinductionKipdrdf}{Error}{Error}{Cputime}{Avg}{176.0040036759301075268817204}%
\StoreBenchExecResult{PdrInv}{KinductionKipdrdf}{Error}{Error}{Cputime}{Median}{110.0270978525}%
\StoreBenchExecResult{PdrInv}{KinductionKipdrdf}{Error}{Error}{Cputime}{Min}{2.333845157}%
\StoreBenchExecResult{PdrInv}{KinductionKipdrdf}{Error}{Error}{Cputime}{Max}{788.647328849}%
\StoreBenchExecResult{PdrInv}{KinductionKipdrdf}{Error}{Error}{Cputime}{Stdev}{184.8708488879188217315027654}%
\StoreBenchExecResult{PdrInv}{KinductionKipdrdf}{Error}{Error}{Walltime}{}{27414.39753794473}%
\StoreBenchExecResult{PdrInv}{KinductionKipdrdf}{Error}{Error}{Walltime}{Avg}{147.3892340749716666666666667}%
\StoreBenchExecResult{PdrInv}{KinductionKipdrdf}{Error}{Error}{Walltime}{Median}{94.1581196785}%
\StoreBenchExecResult{PdrInv}{KinductionKipdrdf}{Error}{Error}{Walltime}{Min}{1.29316186905}%
\StoreBenchExecResult{PdrInv}{KinductionKipdrdf}{Error}{Error}{Walltime}{Max}{774.052371025}%
\StoreBenchExecResult{PdrInv}{KinductionKipdrdf}{Error}{Error}{Walltime}{Stdev}{164.6252431924701088956965927}%
\StoreBenchExecResult{PdrInv}{KinductionKipdrdf}{Error}{Exception}{Count}{}{26}%
\StoreBenchExecResult{PdrInv}{KinductionKipdrdf}{Error}{Exception}{Cputime}{}{4765.143612361}%
\StoreBenchExecResult{PdrInv}{KinductionKipdrdf}{Error}{Exception}{Cputime}{Avg}{183.2747543215769230769230769}%
\StoreBenchExecResult{PdrInv}{KinductionKipdrdf}{Error}{Exception}{Cputime}{Median}{140.984029561}%
\StoreBenchExecResult{PdrInv}{KinductionKipdrdf}{Error}{Exception}{Cputime}{Min}{14.628540045}%
\StoreBenchExecResult{PdrInv}{KinductionKipdrdf}{Error}{Exception}{Cputime}{Max}{614.494016919}%
\StoreBenchExecResult{PdrInv}{KinductionKipdrdf}{Error}{Exception}{Cputime}{Stdev}{162.2439040973790469704906919}%
\StoreBenchExecResult{PdrInv}{KinductionKipdrdf}{Error}{Exception}{Walltime}{}{2393.37700176283}%
\StoreBenchExecResult{PdrInv}{KinductionKipdrdf}{Error}{Exception}{Walltime}{Avg}{92.05296160626269230769230769}%
\StoreBenchExecResult{PdrInv}{KinductionKipdrdf}{Error}{Exception}{Walltime}{Median}{70.91715812685}%
\StoreBenchExecResult{PdrInv}{KinductionKipdrdf}{Error}{Exception}{Walltime}{Min}{7.50129508972}%
\StoreBenchExecResult{PdrInv}{KinductionKipdrdf}{Error}{Exception}{Walltime}{Max}{308.067233086}%
\StoreBenchExecResult{PdrInv}{KinductionKipdrdf}{Error}{Exception}{Walltime}{Stdev}{81.30218963721160996577977026}%
\StoreBenchExecResult{PdrInv}{KinductionKipdrdf}{Error}{OutOfJavaMemory}{Count}{}{6}%
\StoreBenchExecResult{PdrInv}{KinductionKipdrdf}{Error}{OutOfJavaMemory}{Cputime}{}{3326.766798223}%
\StoreBenchExecResult{PdrInv}{KinductionKipdrdf}{Error}{OutOfJavaMemory}{Cputime}{Avg}{554.4611330371666666666666667}%
\StoreBenchExecResult{PdrInv}{KinductionKipdrdf}{Error}{OutOfJavaMemory}{Cputime}{Median}{535.181593427}%
\StoreBenchExecResult{PdrInv}{KinductionKipdrdf}{Error}{OutOfJavaMemory}{Cputime}{Min}{251.848360092}%
\StoreBenchExecResult{PdrInv}{KinductionKipdrdf}{Error}{OutOfJavaMemory}{Cputime}{Max}{896.609941573}%
\StoreBenchExecResult{PdrInv}{KinductionKipdrdf}{Error}{OutOfJavaMemory}{Cputime}{Stdev}{195.1940265472677211531194956}%
\StoreBenchExecResult{PdrInv}{KinductionKipdrdf}{Error}{OutOfJavaMemory}{Walltime}{}{1700.655373097}%
\StoreBenchExecResult{PdrInv}{KinductionKipdrdf}{Error}{OutOfJavaMemory}{Walltime}{Avg}{283.4425621828333333333333333}%
\StoreBenchExecResult{PdrInv}{KinductionKipdrdf}{Error}{OutOfJavaMemory}{Walltime}{Median}{274.336109519}%
\StoreBenchExecResult{PdrInv}{KinductionKipdrdf}{Error}{OutOfJavaMemory}{Walltime}{Min}{128.864484072}%
\StoreBenchExecResult{PdrInv}{KinductionKipdrdf}{Error}{OutOfJavaMemory}{Walltime}{Max}{455.80865407}%
\StoreBenchExecResult{PdrInv}{KinductionKipdrdf}{Error}{OutOfJavaMemory}{Walltime}{Stdev}{98.38633692604885596530559854}%
\StoreBenchExecResult{PdrInv}{KinductionKipdrdf}{Error}{OutOfMemory}{Count}{}{243}%
\StoreBenchExecResult{PdrInv}{KinductionKipdrdf}{Error}{OutOfMemory}{Cputime}{}{106068.526754605}%
\StoreBenchExecResult{PdrInv}{KinductionKipdrdf}{Error}{OutOfMemory}{Cputime}{Avg}{436.4959948749176954732510288}%
\StoreBenchExecResult{PdrInv}{KinductionKipdrdf}{Error}{OutOfMemory}{Cputime}{Median}{368.047714642}%
\StoreBenchExecResult{PdrInv}{KinductionKipdrdf}{Error}{OutOfMemory}{Cputime}{Min}{168.557623561}%
\StoreBenchExecResult{PdrInv}{KinductionKipdrdf}{Error}{OutOfMemory}{Cputime}{Max}{893.315293101}%
\StoreBenchExecResult{PdrInv}{KinductionKipdrdf}{Error}{OutOfMemory}{Cputime}{Stdev}{215.9318729914774991871776521}%
\StoreBenchExecResult{PdrInv}{KinductionKipdrdf}{Error}{OutOfMemory}{Walltime}{}{53191.9258704164}%
\StoreBenchExecResult{PdrInv}{KinductionKipdrdf}{Error}{OutOfMemory}{Walltime}{Avg}{218.8968142815489711934156379}%
\StoreBenchExecResult{PdrInv}{KinductionKipdrdf}{Error}{OutOfMemory}{Walltime}{Median}{184.591503143}%
\StoreBenchExecResult{PdrInv}{KinductionKipdrdf}{Error}{OutOfMemory}{Walltime}{Min}{84.8142340183}%
\StoreBenchExecResult{PdrInv}{KinductionKipdrdf}{Error}{OutOfMemory}{Walltime}{Max}{447.369602919}%
\StoreBenchExecResult{PdrInv}{KinductionKipdrdf}{Error}{OutOfMemory}{Walltime}{Stdev}{108.0247935533929620084726200}%
\StoreBenchExecResult{PdrInv}{KinductionKipdrdf}{Error}{SegmentationFault}{Count}{}{23}%
\StoreBenchExecResult{PdrInv}{KinductionKipdrdf}{Error}{SegmentationFault}{Cputime}{}{936.742642977}%
\StoreBenchExecResult{PdrInv}{KinductionKipdrdf}{Error}{SegmentationFault}{Cputime}{Avg}{40.727940999}%
\StoreBenchExecResult{PdrInv}{KinductionKipdrdf}{Error}{SegmentationFault}{Cputime}{Median}{11.205163869}%
\StoreBenchExecResult{PdrInv}{KinductionKipdrdf}{Error}{SegmentationFault}{Cputime}{Min}{5.991557589}%
\StoreBenchExecResult{PdrInv}{KinductionKipdrdf}{Error}{SegmentationFault}{Cputime}{Max}{605.084929953}%
\StoreBenchExecResult{PdrInv}{KinductionKipdrdf}{Error}{SegmentationFault}{Cputime}{Stdev}{121.0335533991299803140903582}%
\StoreBenchExecResult{PdrInv}{KinductionKipdrdf}{Error}{SegmentationFault}{Walltime}{}{473.45646071431}%
\StoreBenchExecResult{PdrInv}{KinductionKipdrdf}{Error}{SegmentationFault}{Walltime}{Avg}{20.58506350931782608695652174}%
\StoreBenchExecResult{PdrInv}{KinductionKipdrdf}{Error}{SegmentationFault}{Walltime}{Median}{5.79922199249}%
\StoreBenchExecResult{PdrInv}{KinductionKipdrdf}{Error}{SegmentationFault}{Walltime}{Min}{3.16496491432}%
\StoreBenchExecResult{PdrInv}{KinductionKipdrdf}{Error}{SegmentationFault}{Walltime}{Max}{303.038462162}%
\StoreBenchExecResult{PdrInv}{KinductionKipdrdf}{Error}{SegmentationFault}{Walltime}{Stdev}{60.57760239580449724720561594}%
\StoreBenchExecResult{PdrInv}{KinductionKipdrdf}{Error}{Timeout}{Count}{}{2006}%
\StoreBenchExecResult{PdrInv}{KinductionKipdrdf}{Error}{Timeout}{Cputime}{}{1822424.110284626}%
\StoreBenchExecResult{PdrInv}{KinductionKipdrdf}{Error}{Timeout}{Cputime}{Avg}{908.4865953562442671984047856}%
\StoreBenchExecResult{PdrInv}{KinductionKipdrdf}{Error}{Timeout}{Cputime}{Median}{902.1249176135}%
\StoreBenchExecResult{PdrInv}{KinductionKipdrdf}{Error}{Timeout}{Cputime}{Min}{900.699639385}%
\StoreBenchExecResult{PdrInv}{KinductionKipdrdf}{Error}{Timeout}{Cputime}{Max}{1002.3192275}%
\StoreBenchExecResult{PdrInv}{KinductionKipdrdf}{Error}{Timeout}{Cputime}{Stdev}{20.37102172153899036996495813}%
\StoreBenchExecResult{PdrInv}{KinductionKipdrdf}{Error}{Timeout}{Walltime}{}{996937.012875793}%
\StoreBenchExecResult{PdrInv}{KinductionKipdrdf}{Error}{Timeout}{Walltime}{Avg}{496.9775737167462612163509472}%
\StoreBenchExecResult{PdrInv}{KinductionKipdrdf}{Error}{Timeout}{Walltime}{Median}{452.871804476}%
\StoreBenchExecResult{PdrInv}{KinductionKipdrdf}{Error}{Timeout}{Walltime}{Min}{451.071259022}%
\StoreBenchExecResult{PdrInv}{KinductionKipdrdf}{Error}{Timeout}{Walltime}{Max}{898.137360096}%
\StoreBenchExecResult{PdrInv}{KinductionKipdrdf}{Error}{Timeout}{Walltime}{Stdev}{116.3553917092617503577606470}%
\StoreBenchExecResult{PdrInv}{KinductionKipdrdf}{Wrong}{}{Count}{}{2}%
\StoreBenchExecResult{PdrInv}{KinductionKipdrdf}{Wrong}{}{Cputime}{}{24.756494138}%
\StoreBenchExecResult{PdrInv}{KinductionKipdrdf}{Wrong}{}{Cputime}{Avg}{12.378247069}%
\StoreBenchExecResult{PdrInv}{KinductionKipdrdf}{Wrong}{}{Cputime}{Median}{12.378247069}%
\StoreBenchExecResult{PdrInv}{KinductionKipdrdf}{Wrong}{}{Cputime}{Min}{4.347590996}%
\StoreBenchExecResult{PdrInv}{KinductionKipdrdf}{Wrong}{}{Cputime}{Max}{20.408903142}%
\StoreBenchExecResult{PdrInv}{KinductionKipdrdf}{Wrong}{}{Cputime}{Stdev}{8.030656073}%
\StoreBenchExecResult{PdrInv}{KinductionKipdrdf}{Wrong}{}{Walltime}{}{13.10331773762}%
\StoreBenchExecResult{PdrInv}{KinductionKipdrdf}{Wrong}{}{Walltime}{Avg}{6.55165886881}%
\StoreBenchExecResult{PdrInv}{KinductionKipdrdf}{Wrong}{}{Walltime}{Median}{6.55165886881}%
\StoreBenchExecResult{PdrInv}{KinductionKipdrdf}{Wrong}{}{Walltime}{Min}{2.34495782852}%
\StoreBenchExecResult{PdrInv}{KinductionKipdrdf}{Wrong}{}{Walltime}{Max}{10.7583599091}%
\StoreBenchExecResult{PdrInv}{KinductionKipdrdf}{Wrong}{}{Walltime}{Stdev}{4.20670104029}%
\StoreBenchExecResult{PdrInv}{KinductionKipdrdf}{Wrong}{False}{Count}{}{2}%
\StoreBenchExecResult{PdrInv}{KinductionKipdrdf}{Wrong}{False}{Cputime}{}{24.756494138}%
\StoreBenchExecResult{PdrInv}{KinductionKipdrdf}{Wrong}{False}{Cputime}{Avg}{12.378247069}%
\StoreBenchExecResult{PdrInv}{KinductionKipdrdf}{Wrong}{False}{Cputime}{Median}{12.378247069}%
\StoreBenchExecResult{PdrInv}{KinductionKipdrdf}{Wrong}{False}{Cputime}{Min}{4.347590996}%
\StoreBenchExecResult{PdrInv}{KinductionKipdrdf}{Wrong}{False}{Cputime}{Max}{20.408903142}%
\StoreBenchExecResult{PdrInv}{KinductionKipdrdf}{Wrong}{False}{Cputime}{Stdev}{8.030656073}%
\StoreBenchExecResult{PdrInv}{KinductionKipdrdf}{Wrong}{False}{Walltime}{}{13.10331773762}%
\StoreBenchExecResult{PdrInv}{KinductionKipdrdf}{Wrong}{False}{Walltime}{Avg}{6.55165886881}%
\StoreBenchExecResult{PdrInv}{KinductionKipdrdf}{Wrong}{False}{Walltime}{Median}{6.55165886881}%
\StoreBenchExecResult{PdrInv}{KinductionKipdrdf}{Wrong}{False}{Walltime}{Min}{2.34495782852}%
\StoreBenchExecResult{PdrInv}{KinductionKipdrdf}{Wrong}{False}{Walltime}{Max}{10.7583599091}%
\StoreBenchExecResult{PdrInv}{KinductionKipdrdf}{Wrong}{False}{Walltime}{Stdev}{4.20670104029}%
\providecommand\StoreBenchExecResult[7]{\expandafter\newcommand\csname#1#2#3#4#5#6\endcsname{#7}}%
\StoreBenchExecResult{PdrInv}{KinductionKipdrTrueNotSolvedByKinductionPlainButKipdr}{Total}{}{Count}{}{449}%
\StoreBenchExecResult{PdrInv}{KinductionKipdrTrueNotSolvedByKinductionPlainButKipdr}{Total}{}{Cputime}{}{12941.840760226}%
\StoreBenchExecResult{PdrInv}{KinductionKipdrTrueNotSolvedByKinductionPlainButKipdr}{Total}{}{Cputime}{Avg}{28.82369879783073496659242762}%
\StoreBenchExecResult{PdrInv}{KinductionKipdrTrueNotSolvedByKinductionPlainButKipdr}{Total}{}{Cputime}{Median}{12.227945546}%
\StoreBenchExecResult{PdrInv}{KinductionKipdrTrueNotSolvedByKinductionPlainButKipdr}{Total}{}{Cputime}{Min}{3.847777203}%
\StoreBenchExecResult{PdrInv}{KinductionKipdrTrueNotSolvedByKinductionPlainButKipdr}{Total}{}{Cputime}{Max}{619.638703043}%
\StoreBenchExecResult{PdrInv}{KinductionKipdrTrueNotSolvedByKinductionPlainButKipdr}{Total}{}{Cputime}{Stdev}{66.14018380173883425962175482}%
\StoreBenchExecResult{PdrInv}{KinductionKipdrTrueNotSolvedByKinductionPlainButKipdr}{Total}{}{Walltime}{}{7020.46713399863}%
\StoreBenchExecResult{PdrInv}{KinductionKipdrTrueNotSolvedByKinductionPlainButKipdr}{Total}{}{Walltime}{Avg}{15.63578426280318485523385301}%
\StoreBenchExecResult{PdrInv}{KinductionKipdrTrueNotSolvedByKinductionPlainButKipdr}{Total}{}{Walltime}{Median}{6.35933184624}%
\StoreBenchExecResult{PdrInv}{KinductionKipdrTrueNotSolvedByKinductionPlainButKipdr}{Total}{}{Walltime}{Min}{2.1211669445}%
\StoreBenchExecResult{PdrInv}{KinductionKipdrTrueNotSolvedByKinductionPlainButKipdr}{Total}{}{Walltime}{Max}{341.991599083}%
\StoreBenchExecResult{PdrInv}{KinductionKipdrTrueNotSolvedByKinductionPlainButKipdr}{Total}{}{Walltime}{Stdev}{36.71837211358080385420770211}%
\StoreBenchExecResult{PdrInv}{KinductionKipdrTrueNotSolvedByKinductionPlainButKipdr}{Correct}{}{Count}{}{449}%
\StoreBenchExecResult{PdrInv}{KinductionKipdrTrueNotSolvedByKinductionPlainButKipdr}{Correct}{}{Cputime}{}{12941.840760226}%
\StoreBenchExecResult{PdrInv}{KinductionKipdrTrueNotSolvedByKinductionPlainButKipdr}{Correct}{}{Cputime}{Avg}{28.82369879783073496659242762}%
\StoreBenchExecResult{PdrInv}{KinductionKipdrTrueNotSolvedByKinductionPlainButKipdr}{Correct}{}{Cputime}{Median}{12.227945546}%
\StoreBenchExecResult{PdrInv}{KinductionKipdrTrueNotSolvedByKinductionPlainButKipdr}{Correct}{}{Cputime}{Min}{3.847777203}%
\StoreBenchExecResult{PdrInv}{KinductionKipdrTrueNotSolvedByKinductionPlainButKipdr}{Correct}{}{Cputime}{Max}{619.638703043}%
\StoreBenchExecResult{PdrInv}{KinductionKipdrTrueNotSolvedByKinductionPlainButKipdr}{Correct}{}{Cputime}{Stdev}{66.14018380173883425962175482}%
\StoreBenchExecResult{PdrInv}{KinductionKipdrTrueNotSolvedByKinductionPlainButKipdr}{Correct}{}{Walltime}{}{7020.46713399863}%
\StoreBenchExecResult{PdrInv}{KinductionKipdrTrueNotSolvedByKinductionPlainButKipdr}{Correct}{}{Walltime}{Avg}{15.63578426280318485523385301}%
\StoreBenchExecResult{PdrInv}{KinductionKipdrTrueNotSolvedByKinductionPlainButKipdr}{Correct}{}{Walltime}{Median}{6.35933184624}%
\StoreBenchExecResult{PdrInv}{KinductionKipdrTrueNotSolvedByKinductionPlainButKipdr}{Correct}{}{Walltime}{Min}{2.1211669445}%
\StoreBenchExecResult{PdrInv}{KinductionKipdrTrueNotSolvedByKinductionPlainButKipdr}{Correct}{}{Walltime}{Max}{341.991599083}%
\StoreBenchExecResult{PdrInv}{KinductionKipdrTrueNotSolvedByKinductionPlainButKipdr}{Correct}{}{Walltime}{Stdev}{36.71837211358080385420770211}%
\StoreBenchExecResult{PdrInv}{KinductionKipdrTrueNotSolvedByKinductionPlainButKipdr}{Correct}{True}{Count}{}{449}%
\StoreBenchExecResult{PdrInv}{KinductionKipdrTrueNotSolvedByKinductionPlainButKipdr}{Correct}{True}{Cputime}{}{12941.840760226}%
\StoreBenchExecResult{PdrInv}{KinductionKipdrTrueNotSolvedByKinductionPlainButKipdr}{Correct}{True}{Cputime}{Avg}{28.82369879783073496659242762}%
\StoreBenchExecResult{PdrInv}{KinductionKipdrTrueNotSolvedByKinductionPlainButKipdr}{Correct}{True}{Cputime}{Median}{12.227945546}%
\StoreBenchExecResult{PdrInv}{KinductionKipdrTrueNotSolvedByKinductionPlainButKipdr}{Correct}{True}{Cputime}{Min}{3.847777203}%
\StoreBenchExecResult{PdrInv}{KinductionKipdrTrueNotSolvedByKinductionPlainButKipdr}{Correct}{True}{Cputime}{Max}{619.638703043}%
\StoreBenchExecResult{PdrInv}{KinductionKipdrTrueNotSolvedByKinductionPlainButKipdr}{Correct}{True}{Cputime}{Stdev}{66.14018380173883425962175482}%
\StoreBenchExecResult{PdrInv}{KinductionKipdrTrueNotSolvedByKinductionPlainButKipdr}{Correct}{True}{Walltime}{}{7020.46713399863}%
\StoreBenchExecResult{PdrInv}{KinductionKipdrTrueNotSolvedByKinductionPlainButKipdr}{Correct}{True}{Walltime}{Avg}{15.63578426280318485523385301}%
\StoreBenchExecResult{PdrInv}{KinductionKipdrTrueNotSolvedByKinductionPlainButKipdr}{Correct}{True}{Walltime}{Median}{6.35933184624}%
\StoreBenchExecResult{PdrInv}{KinductionKipdrTrueNotSolvedByKinductionPlainButKipdr}{Correct}{True}{Walltime}{Min}{2.1211669445}%
\StoreBenchExecResult{PdrInv}{KinductionKipdrTrueNotSolvedByKinductionPlainButKipdr}{Correct}{True}{Walltime}{Max}{341.991599083}%
\StoreBenchExecResult{PdrInv}{KinductionKipdrTrueNotSolvedByKinductionPlainButKipdr}{Correct}{True}{Walltime}{Stdev}{36.71837211358080385420770211}%
\StoreBenchExecResult{PdrInv}{KinductionKipdrTrueNotSolvedByKinductionPlainButKipdr}{Wrong}{True}{Count}{}{0}%
\StoreBenchExecResult{PdrInv}{KinductionKipdrTrueNotSolvedByKinductionPlainButKipdr}{Wrong}{True}{Cputime}{}{0}%
\StoreBenchExecResult{PdrInv}{KinductionKipdrTrueNotSolvedByKinductionPlainButKipdr}{Wrong}{True}{Cputime}{Avg}{None}%
\StoreBenchExecResult{PdrInv}{KinductionKipdrTrueNotSolvedByKinductionPlainButKipdr}{Wrong}{True}{Cputime}{Median}{None}%
\StoreBenchExecResult{PdrInv}{KinductionKipdrTrueNotSolvedByKinductionPlainButKipdr}{Wrong}{True}{Cputime}{Min}{None}%
\StoreBenchExecResult{PdrInv}{KinductionKipdrTrueNotSolvedByKinductionPlainButKipdr}{Wrong}{True}{Cputime}{Max}{None}%
\StoreBenchExecResult{PdrInv}{KinductionKipdrTrueNotSolvedByKinductionPlainButKipdr}{Wrong}{True}{Cputime}{Stdev}{None}%
\StoreBenchExecResult{PdrInv}{KinductionKipdrTrueNotSolvedByKinductionPlainButKipdr}{Wrong}{True}{Walltime}{}{0}%
\StoreBenchExecResult{PdrInv}{KinductionKipdrTrueNotSolvedByKinductionPlainButKipdr}{Wrong}{True}{Walltime}{Avg}{None}%
\StoreBenchExecResult{PdrInv}{KinductionKipdrTrueNotSolvedByKinductionPlainButKipdr}{Wrong}{True}{Walltime}{Median}{None}%
\StoreBenchExecResult{PdrInv}{KinductionKipdrTrueNotSolvedByKinductionPlainButKipdr}{Wrong}{True}{Walltime}{Min}{None}%
\StoreBenchExecResult{PdrInv}{KinductionKipdrTrueNotSolvedByKinductionPlainButKipdr}{Wrong}{True}{Walltime}{Max}{None}%
\StoreBenchExecResult{PdrInv}{KinductionKipdrTrueNotSolvedByKinductionPlainButKipdr}{Wrong}{True}{Walltime}{Stdev}{None}%
\providecommand\StoreBenchExecResult[7]{\expandafter\newcommand\csname#1#2#3#4#5#6\endcsname{#7}}%
\StoreBenchExecResult{PdrInv}{KinductionKipdrTrueNotSolvedByKinductionPlain}{Total}{}{Count}{}{2893}%
\StoreBenchExecResult{PdrInv}{KinductionKipdrTrueNotSolvedByKinductionPlain}{Total}{}{Cputime}{}{1884905.658050719}%
\StoreBenchExecResult{PdrInv}{KinductionKipdrTrueNotSolvedByKinductionPlain}{Total}{}{Cputime}{Avg}{651.5401514174624956792257172}%
\StoreBenchExecResult{PdrInv}{KinductionKipdrTrueNotSolvedByKinductionPlain}{Total}{}{Cputime}{Median}{901.724237697}%
\StoreBenchExecResult{PdrInv}{KinductionKipdrTrueNotSolvedByKinductionPlain}{Total}{}{Cputime}{Min}{3.056082965}%
\StoreBenchExecResult{PdrInv}{KinductionKipdrTrueNotSolvedByKinductionPlain}{Total}{}{Cputime}{Max}{1002.39868112}%
\StoreBenchExecResult{PdrInv}{KinductionKipdrTrueNotSolvedByKinductionPlain}{Total}{}{Cputime}{Stdev}{389.2671674233688364471813979}%
\StoreBenchExecResult{PdrInv}{KinductionKipdrTrueNotSolvedByKinductionPlain}{Total}{}{Walltime}{}{974852.53084445820}%
\StoreBenchExecResult{PdrInv}{KinductionKipdrTrueNotSolvedByKinductionPlain}{Total}{}{Walltime}{Avg}{336.9694195798334600760456274}%
\StoreBenchExecResult{PdrInv}{KinductionKipdrTrueNotSolvedByKinductionPlain}{Total}{}{Walltime}{Median}{452.282715797}%
\StoreBenchExecResult{PdrInv}{KinductionKipdrTrueNotSolvedByKinductionPlain}{Total}{}{Walltime}{Min}{1.68593192101}%
\StoreBenchExecResult{PdrInv}{KinductionKipdrTrueNotSolvedByKinductionPlain}{Total}{}{Walltime}{Max}{900.688721895}%
\StoreBenchExecResult{PdrInv}{KinductionKipdrTrueNotSolvedByKinductionPlain}{Total}{}{Walltime}{Stdev}{208.7265250384730428584698727}%
\StoreBenchExecResult{PdrInv}{KinductionKipdrTrueNotSolvedByKinductionPlain}{Correct}{}{Count}{}{449}%
\StoreBenchExecResult{PdrInv}{KinductionKipdrTrueNotSolvedByKinductionPlain}{Correct}{}{Cputime}{}{12941.840760226}%
\StoreBenchExecResult{PdrInv}{KinductionKipdrTrueNotSolvedByKinductionPlain}{Correct}{}{Cputime}{Avg}{28.82369879783073496659242762}%
\StoreBenchExecResult{PdrInv}{KinductionKipdrTrueNotSolvedByKinductionPlain}{Correct}{}{Cputime}{Median}{12.227945546}%
\StoreBenchExecResult{PdrInv}{KinductionKipdrTrueNotSolvedByKinductionPlain}{Correct}{}{Cputime}{Min}{3.847777203}%
\StoreBenchExecResult{PdrInv}{KinductionKipdrTrueNotSolvedByKinductionPlain}{Correct}{}{Cputime}{Max}{619.638703043}%
\StoreBenchExecResult{PdrInv}{KinductionKipdrTrueNotSolvedByKinductionPlain}{Correct}{}{Cputime}{Stdev}{66.14018380173883425962175482}%
\StoreBenchExecResult{PdrInv}{KinductionKipdrTrueNotSolvedByKinductionPlain}{Correct}{}{Walltime}{}{7020.46713399863}%
\StoreBenchExecResult{PdrInv}{KinductionKipdrTrueNotSolvedByKinductionPlain}{Correct}{}{Walltime}{Avg}{15.63578426280318485523385301}%
\StoreBenchExecResult{PdrInv}{KinductionKipdrTrueNotSolvedByKinductionPlain}{Correct}{}{Walltime}{Median}{6.35933184624}%
\StoreBenchExecResult{PdrInv}{KinductionKipdrTrueNotSolvedByKinductionPlain}{Correct}{}{Walltime}{Min}{2.1211669445}%
\StoreBenchExecResult{PdrInv}{KinductionKipdrTrueNotSolvedByKinductionPlain}{Correct}{}{Walltime}{Max}{341.991599083}%
\StoreBenchExecResult{PdrInv}{KinductionKipdrTrueNotSolvedByKinductionPlain}{Correct}{}{Walltime}{Stdev}{36.71837211358080385420770211}%
\StoreBenchExecResult{PdrInv}{KinductionKipdrTrueNotSolvedByKinductionPlain}{Correct}{True}{Count}{}{449}%
\StoreBenchExecResult{PdrInv}{KinductionKipdrTrueNotSolvedByKinductionPlain}{Correct}{True}{Cputime}{}{12941.840760226}%
\StoreBenchExecResult{PdrInv}{KinductionKipdrTrueNotSolvedByKinductionPlain}{Correct}{True}{Cputime}{Avg}{28.82369879783073496659242762}%
\StoreBenchExecResult{PdrInv}{KinductionKipdrTrueNotSolvedByKinductionPlain}{Correct}{True}{Cputime}{Median}{12.227945546}%
\StoreBenchExecResult{PdrInv}{KinductionKipdrTrueNotSolvedByKinductionPlain}{Correct}{True}{Cputime}{Min}{3.847777203}%
\StoreBenchExecResult{PdrInv}{KinductionKipdrTrueNotSolvedByKinductionPlain}{Correct}{True}{Cputime}{Max}{619.638703043}%
\StoreBenchExecResult{PdrInv}{KinductionKipdrTrueNotSolvedByKinductionPlain}{Correct}{True}{Cputime}{Stdev}{66.14018380173883425962175482}%
\StoreBenchExecResult{PdrInv}{KinductionKipdrTrueNotSolvedByKinductionPlain}{Correct}{True}{Walltime}{}{7020.46713399863}%
\StoreBenchExecResult{PdrInv}{KinductionKipdrTrueNotSolvedByKinductionPlain}{Correct}{True}{Walltime}{Avg}{15.63578426280318485523385301}%
\StoreBenchExecResult{PdrInv}{KinductionKipdrTrueNotSolvedByKinductionPlain}{Correct}{True}{Walltime}{Median}{6.35933184624}%
\StoreBenchExecResult{PdrInv}{KinductionKipdrTrueNotSolvedByKinductionPlain}{Correct}{True}{Walltime}{Min}{2.1211669445}%
\StoreBenchExecResult{PdrInv}{KinductionKipdrTrueNotSolvedByKinductionPlain}{Correct}{True}{Walltime}{Max}{341.991599083}%
\StoreBenchExecResult{PdrInv}{KinductionKipdrTrueNotSolvedByKinductionPlain}{Correct}{True}{Walltime}{Stdev}{36.71837211358080385420770211}%
\StoreBenchExecResult{PdrInv}{KinductionKipdrTrueNotSolvedByKinductionPlain}{Wrong}{True}{Count}{}{0}%
\StoreBenchExecResult{PdrInv}{KinductionKipdrTrueNotSolvedByKinductionPlain}{Wrong}{True}{Cputime}{}{0}%
\StoreBenchExecResult{PdrInv}{KinductionKipdrTrueNotSolvedByKinductionPlain}{Wrong}{True}{Cputime}{Avg}{None}%
\StoreBenchExecResult{PdrInv}{KinductionKipdrTrueNotSolvedByKinductionPlain}{Wrong}{True}{Cputime}{Median}{None}%
\StoreBenchExecResult{PdrInv}{KinductionKipdrTrueNotSolvedByKinductionPlain}{Wrong}{True}{Cputime}{Min}{None}%
\StoreBenchExecResult{PdrInv}{KinductionKipdrTrueNotSolvedByKinductionPlain}{Wrong}{True}{Cputime}{Max}{None}%
\StoreBenchExecResult{PdrInv}{KinductionKipdrTrueNotSolvedByKinductionPlain}{Wrong}{True}{Cputime}{Stdev}{None}%
\StoreBenchExecResult{PdrInv}{KinductionKipdrTrueNotSolvedByKinductionPlain}{Wrong}{True}{Walltime}{}{0}%
\StoreBenchExecResult{PdrInv}{KinductionKipdrTrueNotSolvedByKinductionPlain}{Wrong}{True}{Walltime}{Avg}{None}%
\StoreBenchExecResult{PdrInv}{KinductionKipdrTrueNotSolvedByKinductionPlain}{Wrong}{True}{Walltime}{Median}{None}%
\StoreBenchExecResult{PdrInv}{KinductionKipdrTrueNotSolvedByKinductionPlain}{Wrong}{True}{Walltime}{Min}{None}%
\StoreBenchExecResult{PdrInv}{KinductionKipdrTrueNotSolvedByKinductionPlain}{Wrong}{True}{Walltime}{Max}{None}%
\StoreBenchExecResult{PdrInv}{KinductionKipdrTrueNotSolvedByKinductionPlain}{Wrong}{True}{Walltime}{Stdev}{None}%
\StoreBenchExecResult{PdrInv}{KinductionKipdrTrueNotSolvedByKinductionPlain}{Error}{}{Count}{}{2444}%
\StoreBenchExecResult{PdrInv}{KinductionKipdrTrueNotSolvedByKinductionPlain}{Error}{}{Cputime}{}{1871963.817290493}%
\StoreBenchExecResult{PdrInv}{KinductionKipdrTrueNotSolvedByKinductionPlain}{Error}{}{Cputime}{Avg}{765.9426420992197217675941080}%
\StoreBenchExecResult{PdrInv}{KinductionKipdrTrueNotSolvedByKinductionPlain}{Error}{}{Cputime}{Median}{902.442271916}%
\StoreBenchExecResult{PdrInv}{KinductionKipdrTrueNotSolvedByKinductionPlain}{Error}{}{Cputime}{Min}{3.056082965}%
\StoreBenchExecResult{PdrInv}{KinductionKipdrTrueNotSolvedByKinductionPlain}{Error}{}{Cputime}{Max}{1002.39868112}%
\StoreBenchExecResult{PdrInv}{KinductionKipdrTrueNotSolvedByKinductionPlain}{Error}{}{Cputime}{Stdev}{306.9775046882022785460008869}%
\StoreBenchExecResult{PdrInv}{KinductionKipdrTrueNotSolvedByKinductionPlain}{Error}{}{Walltime}{}{967832.06371045957}%
\StoreBenchExecResult{PdrInv}{KinductionKipdrTrueNotSolvedByKinductionPlain}{Error}{}{Walltime}{Avg}{396.0032993905317389525368249}%
\StoreBenchExecResult{PdrInv}{KinductionKipdrTrueNotSolvedByKinductionPlain}{Error}{}{Walltime}{Median}{452.740108967}%
\StoreBenchExecResult{PdrInv}{KinductionKipdrTrueNotSolvedByKinductionPlain}{Error}{}{Walltime}{Min}{1.68593192101}%
\StoreBenchExecResult{PdrInv}{KinductionKipdrTrueNotSolvedByKinductionPlain}{Error}{}{Walltime}{Max}{900.688721895}%
\StoreBenchExecResult{PdrInv}{KinductionKipdrTrueNotSolvedByKinductionPlain}{Error}{}{Walltime}{Stdev}{169.9069670084281617706729869}%
\StoreBenchExecResult{PdrInv}{KinductionKipdrTrueNotSolvedByKinductionPlain}{Error}{Assertion}{Count}{}{2}%
\StoreBenchExecResult{PdrInv}{KinductionKipdrTrueNotSolvedByKinductionPlain}{Error}{Assertion}{Cputime}{}{6.265707171}%
\StoreBenchExecResult{PdrInv}{KinductionKipdrTrueNotSolvedByKinductionPlain}{Error}{Assertion}{Cputime}{Avg}{3.1328535855}%
\StoreBenchExecResult{PdrInv}{KinductionKipdrTrueNotSolvedByKinductionPlain}{Error}{Assertion}{Cputime}{Median}{3.1328535855}%
\StoreBenchExecResult{PdrInv}{KinductionKipdrTrueNotSolvedByKinductionPlain}{Error}{Assertion}{Cputime}{Min}{3.10621539}%
\StoreBenchExecResult{PdrInv}{KinductionKipdrTrueNotSolvedByKinductionPlain}{Error}{Assertion}{Cputime}{Max}{3.159491781}%
\StoreBenchExecResult{PdrInv}{KinductionKipdrTrueNotSolvedByKinductionPlain}{Error}{Assertion}{Cputime}{Stdev}{0.0266381955}%
\StoreBenchExecResult{PdrInv}{KinductionKipdrTrueNotSolvedByKinductionPlain}{Error}{Assertion}{Walltime}{}{3.54407525062}%
\StoreBenchExecResult{PdrInv}{KinductionKipdrTrueNotSolvedByKinductionPlain}{Error}{Assertion}{Walltime}{Avg}{1.77203762531}%
\StoreBenchExecResult{PdrInv}{KinductionKipdrTrueNotSolvedByKinductionPlain}{Error}{Assertion}{Walltime}{Median}{1.77203762531}%
\StoreBenchExecResult{PdrInv}{KinductionKipdrTrueNotSolvedByKinductionPlain}{Error}{Assertion}{Walltime}{Min}{1.75860214233}%
\StoreBenchExecResult{PdrInv}{KinductionKipdrTrueNotSolvedByKinductionPlain}{Error}{Assertion}{Walltime}{Max}{1.78547310829}%
\StoreBenchExecResult{PdrInv}{KinductionKipdrTrueNotSolvedByKinductionPlain}{Error}{Assertion}{Walltime}{Stdev}{0.01343548298}%
\StoreBenchExecResult{PdrInv}{KinductionKipdrTrueNotSolvedByKinductionPlain}{Error}{Error}{Count}{}{282}%
\StoreBenchExecResult{PdrInv}{KinductionKipdrTrueNotSolvedByKinductionPlain}{Error}{Error}{Cputime}{}{19384.465826336}%
\StoreBenchExecResult{PdrInv}{KinductionKipdrTrueNotSolvedByKinductionPlain}{Error}{Error}{Cputime}{Avg}{68.73924051892198581560283688}%
\StoreBenchExecResult{PdrInv}{KinductionKipdrTrueNotSolvedByKinductionPlain}{Error}{Error}{Cputime}{Median}{24.4460062545}%
\StoreBenchExecResult{PdrInv}{KinductionKipdrTrueNotSolvedByKinductionPlain}{Error}{Error}{Cputime}{Min}{3.056082965}%
\StoreBenchExecResult{PdrInv}{KinductionKipdrTrueNotSolvedByKinductionPlain}{Error}{Error}{Cputime}{Max}{896.99756731}%
\StoreBenchExecResult{PdrInv}{KinductionKipdrTrueNotSolvedByKinductionPlain}{Error}{Error}{Cputime}{Stdev}{120.5482260810088167103874032}%
\StoreBenchExecResult{PdrInv}{KinductionKipdrTrueNotSolvedByKinductionPlain}{Error}{Error}{Walltime}{}{9888.52026176375}%
\StoreBenchExecResult{PdrInv}{KinductionKipdrTrueNotSolvedByKinductionPlain}{Error}{Error}{Walltime}{Avg}{35.06567468710549645390070922}%
\StoreBenchExecResult{PdrInv}{KinductionKipdrTrueNotSolvedByKinductionPlain}{Error}{Error}{Walltime}{Median}{12.64394700525}%
\StoreBenchExecResult{PdrInv}{KinductionKipdrTrueNotSolvedByKinductionPlain}{Error}{Error}{Walltime}{Min}{1.68593192101}%
\StoreBenchExecResult{PdrInv}{KinductionKipdrTrueNotSolvedByKinductionPlain}{Error}{Error}{Walltime}{Max}{455.590052128}%
\StoreBenchExecResult{PdrInv}{KinductionKipdrTrueNotSolvedByKinductionPlain}{Error}{Error}{Walltime}{Stdev}{61.43427915057459575549947953}%
\StoreBenchExecResult{PdrInv}{KinductionKipdrTrueNotSolvedByKinductionPlain}{Error}{Exception}{Count}{}{7}%
\StoreBenchExecResult{PdrInv}{KinductionKipdrTrueNotSolvedByKinductionPlain}{Error}{Exception}{Cputime}{}{1346.830511540}%
\StoreBenchExecResult{PdrInv}{KinductionKipdrTrueNotSolvedByKinductionPlain}{Error}{Exception}{Cputime}{Avg}{192.4043587914285714285714286}%
\StoreBenchExecResult{PdrInv}{KinductionKipdrTrueNotSolvedByKinductionPlain}{Error}{Exception}{Cputime}{Median}{52.858417422}%
\StoreBenchExecResult{PdrInv}{KinductionKipdrTrueNotSolvedByKinductionPlain}{Error}{Exception}{Cputime}{Min}{8.437501733}%
\StoreBenchExecResult{PdrInv}{KinductionKipdrTrueNotSolvedByKinductionPlain}{Error}{Exception}{Cputime}{Max}{876.752511221}%
\StoreBenchExecResult{PdrInv}{KinductionKipdrTrueNotSolvedByKinductionPlain}{Error}{Exception}{Cputime}{Stdev}{289.3766999511600076747067306}%
\StoreBenchExecResult{PdrInv}{KinductionKipdrTrueNotSolvedByKinductionPlain}{Error}{Exception}{Walltime}{}{701.61910819989}%
\StoreBenchExecResult{PdrInv}{KinductionKipdrTrueNotSolvedByKinductionPlain}{Error}{Exception}{Walltime}{Avg}{100.2313011714128571428571429}%
\StoreBenchExecResult{PdrInv}{KinductionKipdrTrueNotSolvedByKinductionPlain}{Error}{Exception}{Walltime}{Median}{26.6583139896}%
\StoreBenchExecResult{PdrInv}{KinductionKipdrTrueNotSolvedByKinductionPlain}{Error}{Exception}{Walltime}{Min}{4.4456179142}%
\StoreBenchExecResult{PdrInv}{KinductionKipdrTrueNotSolvedByKinductionPlain}{Error}{Exception}{Walltime}{Max}{439.963095188}%
\StoreBenchExecResult{PdrInv}{KinductionKipdrTrueNotSolvedByKinductionPlain}{Error}{Exception}{Walltime}{Stdev}{145.2348753299295525084619328}%
\StoreBenchExecResult{PdrInv}{KinductionKipdrTrueNotSolvedByKinductionPlain}{Error}{OutOfJavaMemory}{Count}{}{9}%
\StoreBenchExecResult{PdrInv}{KinductionKipdrTrueNotSolvedByKinductionPlain}{Error}{OutOfJavaMemory}{Cputime}{}{5824.482985355}%
\StoreBenchExecResult{PdrInv}{KinductionKipdrTrueNotSolvedByKinductionPlain}{Error}{OutOfJavaMemory}{Cputime}{Avg}{647.1647761505555555555555556}%
\StoreBenchExecResult{PdrInv}{KinductionKipdrTrueNotSolvedByKinductionPlain}{Error}{OutOfJavaMemory}{Cputime}{Median}{610.14201869}%
\StoreBenchExecResult{PdrInv}{KinductionKipdrTrueNotSolvedByKinductionPlain}{Error}{OutOfJavaMemory}{Cputime}{Min}{420.95326554}%
\StoreBenchExecResult{PdrInv}{KinductionKipdrTrueNotSolvedByKinductionPlain}{Error}{OutOfJavaMemory}{Cputime}{Max}{828.502859018}%
\StoreBenchExecResult{PdrInv}{KinductionKipdrTrueNotSolvedByKinductionPlain}{Error}{OutOfJavaMemory}{Cputime}{Stdev}{124.7837284289202695083844247}%
\StoreBenchExecResult{PdrInv}{KinductionKipdrTrueNotSolvedByKinductionPlain}{Error}{OutOfJavaMemory}{Walltime}{}{3090.865936520}%
\StoreBenchExecResult{PdrInv}{KinductionKipdrTrueNotSolvedByKinductionPlain}{Error}{OutOfJavaMemory}{Walltime}{Avg}{343.4295485022222222222222222}%
\StoreBenchExecResult{PdrInv}{KinductionKipdrTrueNotSolvedByKinductionPlain}{Error}{OutOfJavaMemory}{Walltime}{Median}{312.166846991}%
\StoreBenchExecResult{PdrInv}{KinductionKipdrTrueNotSolvedByKinductionPlain}{Error}{OutOfJavaMemory}{Walltime}{Min}{219.030888796}%
\StoreBenchExecResult{PdrInv}{KinductionKipdrTrueNotSolvedByKinductionPlain}{Error}{OutOfJavaMemory}{Walltime}{Max}{478.753119946}%
\StoreBenchExecResult{PdrInv}{KinductionKipdrTrueNotSolvedByKinductionPlain}{Error}{OutOfJavaMemory}{Walltime}{Stdev}{78.42253227110764254192303570}%
\StoreBenchExecResult{PdrInv}{KinductionKipdrTrueNotSolvedByKinductionPlain}{Error}{OutOfMemory}{Count}{}{143}%
\StoreBenchExecResult{PdrInv}{KinductionKipdrTrueNotSolvedByKinductionPlain}{Error}{OutOfMemory}{Cputime}{}{46349.471527955}%
\StoreBenchExecResult{PdrInv}{KinductionKipdrTrueNotSolvedByKinductionPlain}{Error}{OutOfMemory}{Cputime}{Avg}{324.1221785171678321678321678}%
\StoreBenchExecResult{PdrInv}{KinductionKipdrTrueNotSolvedByKinductionPlain}{Error}{OutOfMemory}{Cputime}{Median}{289.066459238}%
\StoreBenchExecResult{PdrInv}{KinductionKipdrTrueNotSolvedByKinductionPlain}{Error}{OutOfMemory}{Cputime}{Min}{108.686632731}%
\StoreBenchExecResult{PdrInv}{KinductionKipdrTrueNotSolvedByKinductionPlain}{Error}{OutOfMemory}{Cputime}{Max}{895.701690324}%
\StoreBenchExecResult{PdrInv}{KinductionKipdrTrueNotSolvedByKinductionPlain}{Error}{OutOfMemory}{Cputime}{Stdev}{201.0566545413328491167219689}%
\StoreBenchExecResult{PdrInv}{KinductionKipdrTrueNotSolvedByKinductionPlain}{Error}{OutOfMemory}{Walltime}{}{23267.5901927990}%
\StoreBenchExecResult{PdrInv}{KinductionKipdrTrueNotSolvedByKinductionPlain}{Error}{OutOfMemory}{Walltime}{Avg}{162.7104209286643356643356643}%
\StoreBenchExecResult{PdrInv}{KinductionKipdrTrueNotSolvedByKinductionPlain}{Error}{OutOfMemory}{Walltime}{Median}{145.09813714}%
\StoreBenchExecResult{PdrInv}{KinductionKipdrTrueNotSolvedByKinductionPlain}{Error}{OutOfMemory}{Walltime}{Min}{54.8755509853}%
\StoreBenchExecResult{PdrInv}{KinductionKipdrTrueNotSolvedByKinductionPlain}{Error}{OutOfMemory}{Walltime}{Max}{448.484357119}%
\StoreBenchExecResult{PdrInv}{KinductionKipdrTrueNotSolvedByKinductionPlain}{Error}{OutOfMemory}{Walltime}{Stdev}{100.5491775261484976622313354}%
\StoreBenchExecResult{PdrInv}{KinductionKipdrTrueNotSolvedByKinductionPlain}{Error}{SegmentationFault}{Count}{}{19}%
\StoreBenchExecResult{PdrInv}{KinductionKipdrTrueNotSolvedByKinductionPlain}{Error}{SegmentationFault}{Cputime}{}{135.095559582}%
\StoreBenchExecResult{PdrInv}{KinductionKipdrTrueNotSolvedByKinductionPlain}{Error}{SegmentationFault}{Cputime}{Avg}{7.110292609578947368421052632}%
\StoreBenchExecResult{PdrInv}{KinductionKipdrTrueNotSolvedByKinductionPlain}{Error}{SegmentationFault}{Cputime}{Median}{4.670452608}%
\StoreBenchExecResult{PdrInv}{KinductionKipdrTrueNotSolvedByKinductionPlain}{Error}{SegmentationFault}{Cputime}{Min}{3.791459989}%
\StoreBenchExecResult{PdrInv}{KinductionKipdrTrueNotSolvedByKinductionPlain}{Error}{SegmentationFault}{Cputime}{Max}{35.085368594}%
\StoreBenchExecResult{PdrInv}{KinductionKipdrTrueNotSolvedByKinductionPlain}{Error}{SegmentationFault}{Cputime}{Stdev}{7.197306836017248806050580649}%
\StoreBenchExecResult{PdrInv}{KinductionKipdrTrueNotSolvedByKinductionPlain}{Error}{SegmentationFault}{Walltime}{}{71.73991274831}%
\StoreBenchExecResult{PdrInv}{KinductionKipdrTrueNotSolvedByKinductionPlain}{Error}{SegmentationFault}{Walltime}{Avg}{3.77578488149}%
\StoreBenchExecResult{PdrInv}{KinductionKipdrTrueNotSolvedByKinductionPlain}{Error}{SegmentationFault}{Walltime}{Median}{2.54403686523}%
\StoreBenchExecResult{PdrInv}{KinductionKipdrTrueNotSolvedByKinductionPlain}{Error}{SegmentationFault}{Walltime}{Min}{2.123939991}%
\StoreBenchExecResult{PdrInv}{KinductionKipdrTrueNotSolvedByKinductionPlain}{Error}{SegmentationFault}{Walltime}{Max}{17.7801301479}%
\StoreBenchExecResult{PdrInv}{KinductionKipdrTrueNotSolvedByKinductionPlain}{Error}{SegmentationFault}{Walltime}{Stdev}{3.603226417991945489441539135}%
\StoreBenchExecResult{PdrInv}{KinductionKipdrTrueNotSolvedByKinductionPlain}{Error}{Timeout}{Count}{}{1982}%
\StoreBenchExecResult{PdrInv}{KinductionKipdrTrueNotSolvedByKinductionPlain}{Error}{Timeout}{Cputime}{}{1798917.205172554}%
\StoreBenchExecResult{PdrInv}{KinductionKipdrTrueNotSolvedByKinductionPlain}{Error}{Timeout}{Cputime}{Avg}{907.6272478166266397578203835}%
\StoreBenchExecResult{PdrInv}{KinductionKipdrTrueNotSolvedByKinductionPlain}{Error}{Timeout}{Cputime}{Median}{903.5831404585}%
\StoreBenchExecResult{PdrInv}{KinductionKipdrTrueNotSolvedByKinductionPlain}{Error}{Timeout}{Cputime}{Min}{900.842433901}%
\StoreBenchExecResult{PdrInv}{KinductionKipdrTrueNotSolvedByKinductionPlain}{Error}{Timeout}{Cputime}{Max}{1002.39868112}%
\StoreBenchExecResult{PdrInv}{KinductionKipdrTrueNotSolvedByKinductionPlain}{Error}{Timeout}{Cputime}{Stdev}{14.03963640468253650481776986}%
\StoreBenchExecResult{PdrInv}{KinductionKipdrTrueNotSolvedByKinductionPlain}{Error}{Timeout}{Walltime}{}{930808.184223178}%
\StoreBenchExecResult{PdrInv}{KinductionKipdrTrueNotSolvedByKinductionPlain}{Error}{Timeout}{Walltime}{Avg}{469.6307690328849646821392533}%
\StoreBenchExecResult{PdrInv}{KinductionKipdrTrueNotSolvedByKinductionPlain}{Error}{Timeout}{Walltime}{Median}{453.494397402}%
\StoreBenchExecResult{PdrInv}{KinductionKipdrTrueNotSolvedByKinductionPlain}{Error}{Timeout}{Walltime}{Min}{451.057632923}%
\StoreBenchExecResult{PdrInv}{KinductionKipdrTrueNotSolvedByKinductionPlain}{Error}{Timeout}{Walltime}{Max}{900.688721895}%
\StoreBenchExecResult{PdrInv}{KinductionKipdrTrueNotSolvedByKinductionPlain}{Error}{Timeout}{Walltime}{Stdev}{66.27162958180224847681650798}%
\providecommand\StoreBenchExecResult[7]{\expandafter\newcommand\csname#1#2#3#4#5#6\endcsname{#7}}%
\StoreBenchExecResult{PdrInv}{KinductionKipdr}{Total}{}{Count}{}{5591}%
\StoreBenchExecResult{PdrInv}{KinductionKipdr}{Total}{}{Cputime}{}{2481436.545486000}%
\StoreBenchExecResult{PdrInv}{KinductionKipdr}{Total}{}{Cputime}{Avg}{443.8269621688427830441781434}%
\StoreBenchExecResult{PdrInv}{KinductionKipdr}{Total}{}{Cputime}{Median}{270.401949488}%
\StoreBenchExecResult{PdrInv}{KinductionKipdr}{Total}{}{Cputime}{Min}{2.882157566}%
\StoreBenchExecResult{PdrInv}{KinductionKipdr}{Total}{}{Cputime}{Max}{1002.39868112}%
\StoreBenchExecResult{PdrInv}{KinductionKipdr}{Total}{}{Cputime}{Stdev}{427.6516633592446177994777559}%
\StoreBenchExecResult{PdrInv}{KinductionKipdr}{Total}{}{Walltime}{}{1285097.98105764629}%
\StoreBenchExecResult{PdrInv}{KinductionKipdr}{Total}{}{Walltime}{Avg}{229.8511860235461080307637274}%
\StoreBenchExecResult{PdrInv}{KinductionKipdr}{Total}{}{Walltime}{Median}{136.919751167}%
\StoreBenchExecResult{PdrInv}{KinductionKipdr}{Total}{}{Walltime}{Min}{1.64817690849}%
\StoreBenchExecResult{PdrInv}{KinductionKipdr}{Total}{}{Walltime}{Max}{972.360032082}%
\StoreBenchExecResult{PdrInv}{KinductionKipdr}{Total}{}{Walltime}{Stdev}{225.5875112703174158518727422}%
\StoreBenchExecResult{PdrInv}{KinductionKipdr}{Correct}{}{Count}{}{2428}%
\StoreBenchExecResult{PdrInv}{KinductionKipdr}{Correct}{}{Cputime}{}{121145.244972360}%
\StoreBenchExecResult{PdrInv}{KinductionKipdr}{Correct}{}{Cputime}{Avg}{49.89507618301482701812191104}%
\StoreBenchExecResult{PdrInv}{KinductionKipdr}{Correct}{}{Cputime}{Median}{11.121483797}%
\StoreBenchExecResult{PdrInv}{KinductionKipdr}{Correct}{}{Cputime}{Min}{2.882157566}%
\StoreBenchExecResult{PdrInv}{KinductionKipdr}{Correct}{}{Cputime}{Max}{869.839255515}%
\StoreBenchExecResult{PdrInv}{KinductionKipdr}{Correct}{}{Cputime}{Stdev}{116.3472632307657868915259094}%
\StoreBenchExecResult{PdrInv}{KinductionKipdr}{Correct}{}{Walltime}{}{66181.64514875047}%
\StoreBenchExecResult{PdrInv}{KinductionKipdr}{Correct}{}{Walltime}{Avg}{27.25767922106691515650741351}%
\StoreBenchExecResult{PdrInv}{KinductionKipdr}{Correct}{}{Walltime}{Median}{5.854636430745}%
\StoreBenchExecResult{PdrInv}{KinductionKipdr}{Correct}{}{Walltime}{Min}{1.64817690849}%
\StoreBenchExecResult{PdrInv}{KinductionKipdr}{Correct}{}{Walltime}{Max}{784.89930582}%
\StoreBenchExecResult{PdrInv}{KinductionKipdr}{Correct}{}{Walltime}{Stdev}{67.46829329763848431461369248}%
\StoreBenchExecResult{PdrInv}{KinductionKipdr}{Correct}{False}{Count}{}{766}%
\StoreBenchExecResult{PdrInv}{KinductionKipdr}{Correct}{False}{Cputime}{}{57840.005138869}%
\StoreBenchExecResult{PdrInv}{KinductionKipdr}{Correct}{False}{Cputime}{Avg}{75.50914508990731070496083551}%
\StoreBenchExecResult{PdrInv}{KinductionKipdr}{Correct}{False}{Cputime}{Median}{20.522011159}%
\StoreBenchExecResult{PdrInv}{KinductionKipdr}{Correct}{False}{Cputime}{Min}{2.970052697}%
\StoreBenchExecResult{PdrInv}{KinductionKipdr}{Correct}{False}{Cputime}{Max}{869.839255515}%
\StoreBenchExecResult{PdrInv}{KinductionKipdr}{Correct}{False}{Cputime}{Stdev}{158.0203489911577637385809726}%
\StoreBenchExecResult{PdrInv}{KinductionKipdr}{Correct}{False}{Walltime}{}{31398.50290489123}%
\StoreBenchExecResult{PdrInv}{KinductionKipdr}{Correct}{False}{Walltime}{Avg}{40.99021266957079634464751958}%
\StoreBenchExecResult{PdrInv}{KinductionKipdr}{Correct}{False}{Walltime}{Median}{10.8510448933}%
\StoreBenchExecResult{PdrInv}{KinductionKipdr}{Correct}{False}{Walltime}{Min}{1.70375013351}%
\StoreBenchExecResult{PdrInv}{KinductionKipdr}{Correct}{False}{Walltime}{Max}{626.602530956}%
\StoreBenchExecResult{PdrInv}{KinductionKipdr}{Correct}{False}{Walltime}{Stdev}{87.99173480682004298575744936}%
\StoreBenchExecResult{PdrInv}{KinductionKipdr}{Correct}{True}{Count}{}{1662}%
\StoreBenchExecResult{PdrInv}{KinductionKipdr}{Correct}{True}{Cputime}{}{63305.239833491}%
\StoreBenchExecResult{PdrInv}{KinductionKipdr}{Correct}{True}{Cputime}{Avg}{38.08979532701022864019253911}%
\StoreBenchExecResult{PdrInv}{KinductionKipdr}{Correct}{True}{Cputime}{Median}{9.121692615}%
\StoreBenchExecResult{PdrInv}{KinductionKipdr}{Correct}{True}{Cputime}{Min}{2.882157566}%
\StoreBenchExecResult{PdrInv}{KinductionKipdr}{Correct}{True}{Cputime}{Max}{787.313178234}%
\StoreBenchExecResult{PdrInv}{KinductionKipdr}{Correct}{True}{Cputime}{Stdev}{88.46036026363276483999394313}%
\StoreBenchExecResult{PdrInv}{KinductionKipdr}{Correct}{True}{Walltime}{}{34783.14224385924}%
\StoreBenchExecResult{PdrInv}{KinductionKipdr}{Correct}{True}{Walltime}{Avg}{20.92848510460844765342960289}%
\StoreBenchExecResult{PdrInv}{KinductionKipdr}{Correct}{True}{Walltime}{Median}{4.78512954712}%
\StoreBenchExecResult{PdrInv}{KinductionKipdr}{Correct}{True}{Walltime}{Min}{1.64817690849}%
\StoreBenchExecResult{PdrInv}{KinductionKipdr}{Correct}{True}{Walltime}{Max}{784.89930582}%
\StoreBenchExecResult{PdrInv}{KinductionKipdr}{Correct}{True}{Walltime}{Stdev}{54.35522460199165847820364943}%
\StoreBenchExecResult{PdrInv}{KinductionKipdr}{Wrong}{True}{Count}{}{0}%
\StoreBenchExecResult{PdrInv}{KinductionKipdr}{Wrong}{True}{Cputime}{}{0}%
\StoreBenchExecResult{PdrInv}{KinductionKipdr}{Wrong}{True}{Cputime}{Avg}{None}%
\StoreBenchExecResult{PdrInv}{KinductionKipdr}{Wrong}{True}{Cputime}{Median}{None}%
\StoreBenchExecResult{PdrInv}{KinductionKipdr}{Wrong}{True}{Cputime}{Min}{None}%
\StoreBenchExecResult{PdrInv}{KinductionKipdr}{Wrong}{True}{Cputime}{Max}{None}%
\StoreBenchExecResult{PdrInv}{KinductionKipdr}{Wrong}{True}{Cputime}{Stdev}{None}%
\StoreBenchExecResult{PdrInv}{KinductionKipdr}{Wrong}{True}{Walltime}{}{0}%
\StoreBenchExecResult{PdrInv}{KinductionKipdr}{Wrong}{True}{Walltime}{Avg}{None}%
\StoreBenchExecResult{PdrInv}{KinductionKipdr}{Wrong}{True}{Walltime}{Median}{None}%
\StoreBenchExecResult{PdrInv}{KinductionKipdr}{Wrong}{True}{Walltime}{Min}{None}%
\StoreBenchExecResult{PdrInv}{KinductionKipdr}{Wrong}{True}{Walltime}{Max}{None}%
\StoreBenchExecResult{PdrInv}{KinductionKipdr}{Wrong}{True}{Walltime}{Stdev}{None}%
\StoreBenchExecResult{PdrInv}{KinductionKipdr}{Error}{}{Count}{}{3161}%
\StoreBenchExecResult{PdrInv}{KinductionKipdr}{Error}{}{Cputime}{}{2360256.000045351}%
\StoreBenchExecResult{PdrInv}{KinductionKipdr}{Error}{}{Cputime}{Avg}{746.6801645192505536222714331}%
\StoreBenchExecResult{PdrInv}{KinductionKipdr}{Error}{}{Cputime}{Median}{902.431812701}%
\StoreBenchExecResult{PdrInv}{KinductionKipdr}{Error}{}{Cputime}{Min}{3.056082965}%
\StoreBenchExecResult{PdrInv}{KinductionKipdr}{Error}{}{Cputime}{Max}{1002.39868112}%
\StoreBenchExecResult{PdrInv}{KinductionKipdr}{Error}{}{Cputime}{Stdev}{319.4500474766149756042266067}%
\StoreBenchExecResult{PdrInv}{KinductionKipdr}{Error}{}{Walltime}{}{1218898.07415104519}%
\StoreBenchExecResult{PdrInv}{KinductionKipdr}{Error}{}{Walltime}{Avg}{385.6052116896694685226194242}%
\StoreBenchExecResult{PdrInv}{KinductionKipdr}{Error}{}{Walltime}{Median}{452.695104837}%
\StoreBenchExecResult{PdrInv}{KinductionKipdr}{Error}{}{Walltime}{Min}{1.68593192101}%
\StoreBenchExecResult{PdrInv}{KinductionKipdr}{Error}{}{Walltime}{Max}{972.360032082}%
\StoreBenchExecResult{PdrInv}{KinductionKipdr}{Error}{}{Walltime}{Stdev}{175.2080048057573017839954019}%
\StoreBenchExecResult{PdrInv}{KinductionKipdr}{Error}{Assertion}{Count}{}{4}%
\StoreBenchExecResult{PdrInv}{KinductionKipdr}{Error}{Assertion}{Cputime}{}{12.474104259}%
\StoreBenchExecResult{PdrInv}{KinductionKipdr}{Error}{Assertion}{Cputime}{Avg}{3.11852606475}%
\StoreBenchExecResult{PdrInv}{KinductionKipdr}{Error}{Assertion}{Cputime}{Median}{3.1052178185}%
\StoreBenchExecResult{PdrInv}{KinductionKipdr}{Error}{Assertion}{Cputime}{Min}{3.104176841}%
\StoreBenchExecResult{PdrInv}{KinductionKipdr}{Error}{Assertion}{Cputime}{Max}{3.159491781}%
\StoreBenchExecResult{PdrInv}{KinductionKipdr}{Error}{Assertion}{Cputime}{Stdev}{0.02366589985823262938240279533}%
\StoreBenchExecResult{PdrInv}{KinductionKipdr}{Error}{Assertion}{Walltime}{}{6.99939918517}%
\StoreBenchExecResult{PdrInv}{KinductionKipdr}{Error}{Assertion}{Walltime}{Avg}{1.7498497962925}%
\StoreBenchExecResult{PdrInv}{KinductionKipdr}{Error}{Assertion}{Walltime}{Median}{1.74595403671}%
\StoreBenchExecResult{PdrInv}{KinductionKipdr}{Error}{Assertion}{Walltime}{Min}{1.72201800346}%
\StoreBenchExecResult{PdrInv}{KinductionKipdr}{Error}{Assertion}{Walltime}{Max}{1.78547310829}%
\StoreBenchExecResult{PdrInv}{KinductionKipdr}{Error}{Assertion}{Walltime}{Stdev}{0.02446391264158831426844539406}%
\StoreBenchExecResult{PdrInv}{KinductionKipdr}{Error}{Error}{Count}{}{373}%
\StoreBenchExecResult{PdrInv}{KinductionKipdr}{Error}{Error}{Cputime}{}{25981.778185043}%
\StoreBenchExecResult{PdrInv}{KinductionKipdr}{Error}{Error}{Cputime}{Avg}{69.65624178295710455764075067}%
\StoreBenchExecResult{PdrInv}{KinductionKipdr}{Error}{Error}{Cputime}{Median}{23.880296916}%
\StoreBenchExecResult{PdrInv}{KinductionKipdr}{Error}{Error}{Cputime}{Min}{3.056082965}%
\StoreBenchExecResult{PdrInv}{KinductionKipdr}{Error}{Error}{Cputime}{Max}{896.99756731}%
\StoreBenchExecResult{PdrInv}{KinductionKipdr}{Error}{Error}{Cputime}{Stdev}{125.2428339759032998570052483}%
\StoreBenchExecResult{PdrInv}{KinductionKipdr}{Error}{Error}{Walltime}{}{13418.75828886107}%
\StoreBenchExecResult{PdrInv}{KinductionKipdr}{Error}{Error}{Walltime}{Avg}{35.97522329453369973190348525}%
\StoreBenchExecResult{PdrInv}{KinductionKipdr}{Error}{Error}{Walltime}{Median}{12.2018101215}%
\StoreBenchExecResult{PdrInv}{KinductionKipdr}{Error}{Error}{Walltime}{Min}{1.68593192101}%
\StoreBenchExecResult{PdrInv}{KinductionKipdr}{Error}{Error}{Walltime}{Max}{455.590052128}%
\StoreBenchExecResult{PdrInv}{KinductionKipdr}{Error}{Error}{Walltime}{Stdev}{65.00752790960140673192333235}%
\StoreBenchExecResult{PdrInv}{KinductionKipdr}{Error}{Exception}{Count}{}{10}%
\StoreBenchExecResult{PdrInv}{KinductionKipdr}{Error}{Exception}{Cputime}{}{1451.089607971}%
\StoreBenchExecResult{PdrInv}{KinductionKipdr}{Error}{Exception}{Cputime}{Avg}{145.1089607971}%
\StoreBenchExecResult{PdrInv}{KinductionKipdr}{Error}{Exception}{Cputime}{Median}{38.1154347355}%
\StoreBenchExecResult{PdrInv}{KinductionKipdr}{Error}{Exception}{Cputime}{Min}{8.437501733}%
\StoreBenchExecResult{PdrInv}{KinductionKipdr}{Error}{Exception}{Cputime}{Max}{876.752511221}%
\StoreBenchExecResult{PdrInv}{KinductionKipdr}{Error}{Exception}{Cputime}{Stdev}{252.8203473689266746922691288}%
\StoreBenchExecResult{PdrInv}{KinductionKipdr}{Error}{Exception}{Walltime}{}{755.01407289489}%
\StoreBenchExecResult{PdrInv}{KinductionKipdr}{Error}{Exception}{Walltime}{Avg}{75.501407289489}%
\StoreBenchExecResult{PdrInv}{KinductionKipdr}{Error}{Exception}{Walltime}{Median}{19.26314795015}%
\StoreBenchExecResult{PdrInv}{KinductionKipdr}{Error}{Exception}{Walltime}{Min}{4.4456179142}%
\StoreBenchExecResult{PdrInv}{KinductionKipdr}{Error}{Exception}{Walltime}{Max}{439.963095188}%
\StoreBenchExecResult{PdrInv}{KinductionKipdr}{Error}{Exception}{Walltime}{Stdev}{127.3349995751057719521251394}%
\StoreBenchExecResult{PdrInv}{KinductionKipdr}{Error}{OutOfJavaMemory}{Count}{}{13}%
\StoreBenchExecResult{PdrInv}{KinductionKipdr}{Error}{OutOfJavaMemory}{Cputime}{}{9102.882355307}%
\StoreBenchExecResult{PdrInv}{KinductionKipdr}{Error}{OutOfJavaMemory}{Cputime}{Avg}{700.221719639}%
\StoreBenchExecResult{PdrInv}{KinductionKipdr}{Error}{OutOfJavaMemory}{Cputime}{Median}{719.305648838}%
\StoreBenchExecResult{PdrInv}{KinductionKipdr}{Error}{OutOfJavaMemory}{Cputime}{Min}{420.95326554}%
\StoreBenchExecResult{PdrInv}{KinductionKipdr}{Error}{OutOfJavaMemory}{Cputime}{Max}{891.17454337}%
\StoreBenchExecResult{PdrInv}{KinductionKipdr}{Error}{OutOfJavaMemory}{Cputime}{Stdev}{136.7143754176299770270434338}%
\StoreBenchExecResult{PdrInv}{KinductionKipdr}{Error}{OutOfJavaMemory}{Walltime}{}{4754.403002742}%
\StoreBenchExecResult{PdrInv}{KinductionKipdr}{Error}{OutOfJavaMemory}{Walltime}{Avg}{365.7233079032307692307692308}%
\StoreBenchExecResult{PdrInv}{KinductionKipdr}{Error}{OutOfJavaMemory}{Walltime}{Median}{365.932016134}%
\StoreBenchExecResult{PdrInv}{KinductionKipdr}{Error}{OutOfJavaMemory}{Walltime}{Min}{219.030888796}%
\StoreBenchExecResult{PdrInv}{KinductionKipdr}{Error}{OutOfJavaMemory}{Walltime}{Max}{478.753119946}%
\StoreBenchExecResult{PdrInv}{KinductionKipdr}{Error}{OutOfJavaMemory}{Walltime}{Stdev}{75.99454274108263933558903048}%
\StoreBenchExecResult{PdrInv}{KinductionKipdr}{Error}{OutOfMemory}{Count}{}{273}%
\StoreBenchExecResult{PdrInv}{KinductionKipdr}{Error}{OutOfMemory}{Cputime}{}{89639.145605115}%
\StoreBenchExecResult{PdrInv}{KinductionKipdr}{Error}{OutOfMemory}{Cputime}{Avg}{328.3485187000549450549450549}%
\StoreBenchExecResult{PdrInv}{KinductionKipdr}{Error}{OutOfMemory}{Cputime}{Median}{287.730304895}%
\StoreBenchExecResult{PdrInv}{KinductionKipdr}{Error}{OutOfMemory}{Cputime}{Min}{108.686632731}%
\StoreBenchExecResult{PdrInv}{KinductionKipdr}{Error}{OutOfMemory}{Cputime}{Max}{895.701690324}%
\StoreBenchExecResult{PdrInv}{KinductionKipdr}{Error}{OutOfMemory}{Cputime}{Stdev}{195.7970403507343381133352247}%
\StoreBenchExecResult{PdrInv}{KinductionKipdr}{Error}{OutOfMemory}{Walltime}{}{44989.7063510456}%
\StoreBenchExecResult{PdrInv}{KinductionKipdr}{Error}{OutOfMemory}{Walltime}{Avg}{164.7974591613391941391941392}%
\StoreBenchExecResult{PdrInv}{KinductionKipdr}{Error}{OutOfMemory}{Walltime}{Median}{144.436069012}%
\StoreBenchExecResult{PdrInv}{KinductionKipdr}{Error}{OutOfMemory}{Walltime}{Min}{54.8755509853}%
\StoreBenchExecResult{PdrInv}{KinductionKipdr}{Error}{OutOfMemory}{Walltime}{Max}{448.484357119}%
\StoreBenchExecResult{PdrInv}{KinductionKipdr}{Error}{OutOfMemory}{Walltime}{Stdev}{97.90578555443208301811995414}%
\StoreBenchExecResult{PdrInv}{KinductionKipdr}{Error}{SegmentationFault}{Count}{}{29}%
\StoreBenchExecResult{PdrInv}{KinductionKipdr}{Error}{SegmentationFault}{Cputime}{}{221.205721438}%
\StoreBenchExecResult{PdrInv}{KinductionKipdr}{Error}{SegmentationFault}{Cputime}{Avg}{7.627783497862068965517241379}%
\StoreBenchExecResult{PdrInv}{KinductionKipdr}{Error}{SegmentationFault}{Cputime}{Median}{5.402467645}%
\StoreBenchExecResult{PdrInv}{KinductionKipdr}{Error}{SegmentationFault}{Cputime}{Min}{3.791459989}%
\StoreBenchExecResult{PdrInv}{KinductionKipdr}{Error}{SegmentationFault}{Cputime}{Max}{35.085368594}%
\StoreBenchExecResult{PdrInv}{KinductionKipdr}{Error}{SegmentationFault}{Cputime}{Stdev}{7.384773976027750444733929762}%
\StoreBenchExecResult{PdrInv}{KinductionKipdr}{Error}{SegmentationFault}{Walltime}{}{116.80447530746}%
\StoreBenchExecResult{PdrInv}{KinductionKipdr}{Error}{SegmentationFault}{Walltime}{Avg}{4.027740527843448275862068966}%
\StoreBenchExecResult{PdrInv}{KinductionKipdr}{Error}{SegmentationFault}{Walltime}{Median}{2.91926598549}%
\StoreBenchExecResult{PdrInv}{KinductionKipdr}{Error}{SegmentationFault}{Walltime}{Min}{2.123939991}%
\StoreBenchExecResult{PdrInv}{KinductionKipdr}{Error}{SegmentationFault}{Walltime}{Max}{17.7801301479}%
\StoreBenchExecResult{PdrInv}{KinductionKipdr}{Error}{SegmentationFault}{Walltime}{Stdev}{3.695168421116672656679474950}%
\StoreBenchExecResult{PdrInv}{KinductionKipdr}{Error}{Timeout}{Count}{}{2459}%
\StoreBenchExecResult{PdrInv}{KinductionKipdr}{Error}{Timeout}{Cputime}{}{2233847.424466218}%
\StoreBenchExecResult{PdrInv}{KinductionKipdr}{Error}{Timeout}{Cputime}{Avg}{908.4373421985433102887352582}%
\StoreBenchExecResult{PdrInv}{KinductionKipdr}{Error}{Timeout}{Cputime}{Median}{903.883282024}%
\StoreBenchExecResult{PdrInv}{KinductionKipdr}{Error}{Timeout}{Cputime}{Min}{900.842433901}%
\StoreBenchExecResult{PdrInv}{KinductionKipdr}{Error}{Timeout}{Cputime}{Max}{1002.39868112}%
\StoreBenchExecResult{PdrInv}{KinductionKipdr}{Error}{Timeout}{Cputime}{Stdev}{15.50695911529719650350396128}%
\StoreBenchExecResult{PdrInv}{KinductionKipdr}{Error}{Timeout}{Walltime}{}{1154856.388561009}%
\StoreBenchExecResult{PdrInv}{KinductionKipdr}{Error}{Timeout}{Walltime}{Avg}{469.6447289796701911346075641}%
\StoreBenchExecResult{PdrInv}{KinductionKipdr}{Error}{Timeout}{Walltime}{Median}{453.687979937}%
\StoreBenchExecResult{PdrInv}{KinductionKipdr}{Error}{Timeout}{Walltime}{Min}{451.049824953}%
\StoreBenchExecResult{PdrInv}{KinductionKipdr}{Error}{Timeout}{Walltime}{Max}{972.360032082}%
\StoreBenchExecResult{PdrInv}{KinductionKipdr}{Error}{Timeout}{Walltime}{Stdev}{65.51062693018710008586409280}%
\StoreBenchExecResult{PdrInv}{KinductionKipdr}{Wrong}{}{Count}{}{2}%
\StoreBenchExecResult{PdrInv}{KinductionKipdr}{Wrong}{}{Cputime}{}{35.300468289}%
\StoreBenchExecResult{PdrInv}{KinductionKipdr}{Wrong}{}{Cputime}{Avg}{17.6502341445}%
\StoreBenchExecResult{PdrInv}{KinductionKipdr}{Wrong}{}{Cputime}{Median}{17.6502341445}%
\StoreBenchExecResult{PdrInv}{KinductionKipdr}{Wrong}{}{Cputime}{Min}{3.549080658}%
\StoreBenchExecResult{PdrInv}{KinductionKipdr}{Wrong}{}{Cputime}{Max}{31.751387631}%
\StoreBenchExecResult{PdrInv}{KinductionKipdr}{Wrong}{}{Cputime}{Stdev}{14.1011534865}%
\StoreBenchExecResult{PdrInv}{KinductionKipdr}{Wrong}{}{Walltime}{}{18.26175785063}%
\StoreBenchExecResult{PdrInv}{KinductionKipdr}{Wrong}{}{Walltime}{Avg}{9.130878925315}%
\StoreBenchExecResult{PdrInv}{KinductionKipdr}{Wrong}{}{Walltime}{Median}{9.130878925315}%
\StoreBenchExecResult{PdrInv}{KinductionKipdr}{Wrong}{}{Walltime}{Min}{1.99939084053}%
\StoreBenchExecResult{PdrInv}{KinductionKipdr}{Wrong}{}{Walltime}{Max}{16.2623670101}%
\StoreBenchExecResult{PdrInv}{KinductionKipdr}{Wrong}{}{Walltime}{Stdev}{7.131488084785}%
\StoreBenchExecResult{PdrInv}{KinductionKipdr}{Wrong}{False}{Count}{}{2}%
\StoreBenchExecResult{PdrInv}{KinductionKipdr}{Wrong}{False}{Cputime}{}{35.300468289}%
\StoreBenchExecResult{PdrInv}{KinductionKipdr}{Wrong}{False}{Cputime}{Avg}{17.6502341445}%
\StoreBenchExecResult{PdrInv}{KinductionKipdr}{Wrong}{False}{Cputime}{Median}{17.6502341445}%
\StoreBenchExecResult{PdrInv}{KinductionKipdr}{Wrong}{False}{Cputime}{Min}{3.549080658}%
\StoreBenchExecResult{PdrInv}{KinductionKipdr}{Wrong}{False}{Cputime}{Max}{31.751387631}%
\StoreBenchExecResult{PdrInv}{KinductionKipdr}{Wrong}{False}{Cputime}{Stdev}{14.1011534865}%
\StoreBenchExecResult{PdrInv}{KinductionKipdr}{Wrong}{False}{Walltime}{}{18.26175785063}%
\StoreBenchExecResult{PdrInv}{KinductionKipdr}{Wrong}{False}{Walltime}{Avg}{9.130878925315}%
\StoreBenchExecResult{PdrInv}{KinductionKipdr}{Wrong}{False}{Walltime}{Median}{9.130878925315}%
\StoreBenchExecResult{PdrInv}{KinductionKipdr}{Wrong}{False}{Walltime}{Min}{1.99939084053}%
\StoreBenchExecResult{PdrInv}{KinductionKipdr}{Wrong}{False}{Walltime}{Max}{16.2623670101}%
\StoreBenchExecResult{PdrInv}{KinductionKipdr}{Wrong}{False}{Walltime}{Stdev}{7.131488084785}%
\providecommand\StoreBenchExecResult[7]{\expandafter\newcommand\csname#1#2#3#4#5#6\endcsname{#7}}%
\StoreBenchExecResult{PdrInv}{KinductionPlainTrueNotSolvedByKinductionPlainButKipdr}{Total}{}{Count}{}{449}%
\StoreBenchExecResult{PdrInv}{KinductionPlainTrueNotSolvedByKinductionPlainButKipdr}{Total}{}{Cputime}{}{404427.354180886}%
\StoreBenchExecResult{PdrInv}{KinductionPlainTrueNotSolvedByKinductionPlainButKipdr}{Total}{}{Cputime}{Avg}{900.7290738995233853006681514}%
\StoreBenchExecResult{PdrInv}{KinductionPlainTrueNotSolvedByKinductionPlainButKipdr}{Total}{}{Cputime}{Median}{907.161545194}%
\StoreBenchExecResult{PdrInv}{KinductionPlainTrueNotSolvedByKinductionPlainButKipdr}{Total}{}{Cputime}{Min}{634.573466834}%
\StoreBenchExecResult{PdrInv}{KinductionPlainTrueNotSolvedByKinductionPlainButKipdr}{Total}{}{Cputime}{Max}{938.082131514}%
\StoreBenchExecResult{PdrInv}{KinductionPlainTrueNotSolvedByKinductionPlainButKipdr}{Total}{}{Cputime}{Stdev}{29.53941275196986586690884446}%
\StoreBenchExecResult{PdrInv}{KinductionPlainTrueNotSolvedByKinductionPlainButKipdr}{Total}{}{Walltime}{}{397002.321395388}%
\StoreBenchExecResult{PdrInv}{KinductionPlainTrueNotSolvedByKinductionPlainButKipdr}{Total}{}{Walltime}{Avg}{884.1922525509755011135857461}%
\StoreBenchExecResult{PdrInv}{KinductionPlainTrueNotSolvedByKinductionPlainButKipdr}{Total}{}{Walltime}{Median}{893.099745989}%
\StoreBenchExecResult{PdrInv}{KinductionPlainTrueNotSolvedByKinductionPlainButKipdr}{Total}{}{Walltime}{Min}{622.163383007}%
\StoreBenchExecResult{PdrInv}{KinductionPlainTrueNotSolvedByKinductionPlainButKipdr}{Total}{}{Walltime}{Max}{902.229701042}%
\StoreBenchExecResult{PdrInv}{KinductionPlainTrueNotSolvedByKinductionPlainButKipdr}{Total}{}{Walltime}{Stdev}{32.00546432119713003221452350}%
\StoreBenchExecResult{PdrInv}{KinductionPlainTrueNotSolvedByKinductionPlainButKipdr}{Error}{}{Count}{}{449}%
\StoreBenchExecResult{PdrInv}{KinductionPlainTrueNotSolvedByKinductionPlainButKipdr}{Error}{}{Cputime}{}{404427.354180886}%
\StoreBenchExecResult{PdrInv}{KinductionPlainTrueNotSolvedByKinductionPlainButKipdr}{Error}{}{Cputime}{Avg}{900.7290738995233853006681514}%
\StoreBenchExecResult{PdrInv}{KinductionPlainTrueNotSolvedByKinductionPlainButKipdr}{Error}{}{Cputime}{Median}{907.161545194}%
\StoreBenchExecResult{PdrInv}{KinductionPlainTrueNotSolvedByKinductionPlainButKipdr}{Error}{}{Cputime}{Min}{634.573466834}%
\StoreBenchExecResult{PdrInv}{KinductionPlainTrueNotSolvedByKinductionPlainButKipdr}{Error}{}{Cputime}{Max}{938.082131514}%
\StoreBenchExecResult{PdrInv}{KinductionPlainTrueNotSolvedByKinductionPlainButKipdr}{Error}{}{Cputime}{Stdev}{29.53941275196986586690884446}%
\StoreBenchExecResult{PdrInv}{KinductionPlainTrueNotSolvedByKinductionPlainButKipdr}{Error}{}{Walltime}{}{397002.321395388}%
\StoreBenchExecResult{PdrInv}{KinductionPlainTrueNotSolvedByKinductionPlainButKipdr}{Error}{}{Walltime}{Avg}{884.1922525509755011135857461}%
\StoreBenchExecResult{PdrInv}{KinductionPlainTrueNotSolvedByKinductionPlainButKipdr}{Error}{}{Walltime}{Median}{893.099745989}%
\StoreBenchExecResult{PdrInv}{KinductionPlainTrueNotSolvedByKinductionPlainButKipdr}{Error}{}{Walltime}{Min}{622.163383007}%
\StoreBenchExecResult{PdrInv}{KinductionPlainTrueNotSolvedByKinductionPlainButKipdr}{Error}{}{Walltime}{Max}{902.229701042}%
\StoreBenchExecResult{PdrInv}{KinductionPlainTrueNotSolvedByKinductionPlainButKipdr}{Error}{}{Walltime}{Stdev}{32.00546432119713003221452350}%
\StoreBenchExecResult{PdrInv}{KinductionPlainTrueNotSolvedByKinductionPlainButKipdr}{Error}{OutOfMemory}{Count}{}{41}%
\StoreBenchExecResult{PdrInv}{KinductionPlainTrueNotSolvedByKinductionPlainButKipdr}{Error}{OutOfMemory}{Cputime}{}{33956.071849734}%
\StoreBenchExecResult{PdrInv}{KinductionPlainTrueNotSolvedByKinductionPlainButKipdr}{Error}{OutOfMemory}{Cputime}{Avg}{828.1968743837560975609756098}%
\StoreBenchExecResult{PdrInv}{KinductionPlainTrueNotSolvedByKinductionPlainButKipdr}{Error}{OutOfMemory}{Cputime}{Median}{837.779709833}%
\StoreBenchExecResult{PdrInv}{KinductionPlainTrueNotSolvedByKinductionPlainButKipdr}{Error}{OutOfMemory}{Cputime}{Min}{634.573466834}%
\StoreBenchExecResult{PdrInv}{KinductionPlainTrueNotSolvedByKinductionPlainButKipdr}{Error}{OutOfMemory}{Cputime}{Max}{899.806149114}%
\StoreBenchExecResult{PdrInv}{KinductionPlainTrueNotSolvedByKinductionPlainButKipdr}{Error}{OutOfMemory}{Cputime}{Stdev}{60.14744835840922854180178690}%
\StoreBenchExecResult{PdrInv}{KinductionPlainTrueNotSolvedByKinductionPlainButKipdr}{Error}{OutOfMemory}{Walltime}{}{33327.866990090}%
\StoreBenchExecResult{PdrInv}{KinductionPlainTrueNotSolvedByKinductionPlainButKipdr}{Error}{OutOfMemory}{Walltime}{Avg}{812.8748046363414634146341463}%
\StoreBenchExecResult{PdrInv}{KinductionPlainTrueNotSolvedByKinductionPlainButKipdr}{Error}{OutOfMemory}{Walltime}{Median}{822.730746984}%
\StoreBenchExecResult{PdrInv}{KinductionPlainTrueNotSolvedByKinductionPlainButKipdr}{Error}{OutOfMemory}{Walltime}{Min}{622.163383007}%
\StoreBenchExecResult{PdrInv}{KinductionPlainTrueNotSolvedByKinductionPlainButKipdr}{Error}{OutOfMemory}{Walltime}{Max}{881.583776951}%
\StoreBenchExecResult{PdrInv}{KinductionPlainTrueNotSolvedByKinductionPlainButKipdr}{Error}{OutOfMemory}{Walltime}{Stdev}{58.50982443520732262635351112}%
\StoreBenchExecResult{PdrInv}{KinductionPlainTrueNotSolvedByKinductionPlainButKipdr}{Error}{Timeout}{Count}{}{408}%
\StoreBenchExecResult{PdrInv}{KinductionPlainTrueNotSolvedByKinductionPlainButKipdr}{Error}{Timeout}{Cputime}{}{370471.282331152}%
\StoreBenchExecResult{PdrInv}{KinductionPlainTrueNotSolvedByKinductionPlainButKipdr}{Error}{Timeout}{Cputime}{Avg}{908.0178488508627450980392157}%
\StoreBenchExecResult{PdrInv}{KinductionPlainTrueNotSolvedByKinductionPlainButKipdr}{Error}{Timeout}{Cputime}{Median}{907.9285634005}%
\StoreBenchExecResult{PdrInv}{KinductionPlainTrueNotSolvedByKinductionPlainButKipdr}{Error}{Timeout}{Cputime}{Min}{900.180142923}%
\StoreBenchExecResult{PdrInv}{KinductionPlainTrueNotSolvedByKinductionPlainButKipdr}{Error}{Timeout}{Cputime}{Max}{938.082131514}%
\StoreBenchExecResult{PdrInv}{KinductionPlainTrueNotSolvedByKinductionPlainButKipdr}{Error}{Timeout}{Cputime}{Stdev}{3.862676069647131785217614641}%
\StoreBenchExecResult{PdrInv}{KinductionPlainTrueNotSolvedByKinductionPlainButKipdr}{Error}{Timeout}{Walltime}{}{363674.454405298}%
\StoreBenchExecResult{PdrInv}{KinductionPlainTrueNotSolvedByKinductionPlainButKipdr}{Error}{Timeout}{Walltime}{Avg}{891.3589568757303921568627451}%
\StoreBenchExecResult{PdrInv}{KinductionPlainTrueNotSolvedByKinductionPlainButKipdr}{Error}{Timeout}{Walltime}{Median}{893.3847715855}%
\StoreBenchExecResult{PdrInv}{KinductionPlainTrueNotSolvedByKinductionPlainButKipdr}{Error}{Timeout}{Walltime}{Min}{702.269698858}%
\StoreBenchExecResult{PdrInv}{KinductionPlainTrueNotSolvedByKinductionPlainButKipdr}{Error}{Timeout}{Walltime}{Max}{902.229701042}%
\StoreBenchExecResult{PdrInv}{KinductionPlainTrueNotSolvedByKinductionPlainButKipdr}{Error}{Timeout}{Walltime}{Stdev}{14.85920749656561051032816000}%
\providecommand\StoreBenchExecResult[7]{\expandafter\newcommand\csname#1#2#3#4#5#6\endcsname{#7}}%
\StoreBenchExecResult{PdrInv}{KinductionPlainTrueNotSolvedByKinductionPlain}{Total}{}{Count}{}{2893}%
\StoreBenchExecResult{PdrInv}{KinductionPlainTrueNotSolvedByKinductionPlain}{Total}{}{Cputime}{}{2286188.539811423}%
\StoreBenchExecResult{PdrInv}{KinductionPlainTrueNotSolvedByKinductionPlain}{Total}{}{Cputime}{Avg}{790.2483718670663670929830626}%
\StoreBenchExecResult{PdrInv}{KinductionPlainTrueNotSolvedByKinductionPlain}{Total}{}{Cputime}{Median}{903.196108197}%
\StoreBenchExecResult{PdrInv}{KinductionPlainTrueNotSolvedByKinductionPlain}{Total}{}{Cputime}{Min}{2.524657944}%
\StoreBenchExecResult{PdrInv}{KinductionPlainTrueNotSolvedByKinductionPlain}{Total}{}{Cputime}{Max}{980.03014337}%
\StoreBenchExecResult{PdrInv}{KinductionPlainTrueNotSolvedByKinductionPlain}{Total}{}{Cputime}{Stdev}{282.6171076763802904204595638}%
\StoreBenchExecResult{PdrInv}{KinductionPlainTrueNotSolvedByKinductionPlain}{Total}{}{Walltime}{}{2093708.93828677189}%
\StoreBenchExecResult{PdrInv}{KinductionPlainTrueNotSolvedByKinductionPlain}{Total}{}{Walltime}{Avg}{723.7154988893093294158313170}%
\StoreBenchExecResult{PdrInv}{KinductionPlainTrueNotSolvedByKinductionPlain}{Total}{}{Walltime}{Median}{848.12104702}%
\StoreBenchExecResult{PdrInv}{KinductionPlainTrueNotSolvedByKinductionPlain}{Total}{}{Walltime}{Min}{1.36754894257}%
\StoreBenchExecResult{PdrInv}{KinductionPlainTrueNotSolvedByKinductionPlain}{Total}{}{Walltime}{Max}{903.080847025}%
\StoreBenchExecResult{PdrInv}{KinductionPlainTrueNotSolvedByKinductionPlain}{Total}{}{Walltime}{Stdev}{271.3172914654583917314264084}%
\StoreBenchExecResult{PdrInv}{KinductionPlainTrueNotSolvedByKinductionPlain}{Error}{}{Count}{}{2893}%
\StoreBenchExecResult{PdrInv}{KinductionPlainTrueNotSolvedByKinductionPlain}{Error}{}{Cputime}{}{2286188.539811423}%
\StoreBenchExecResult{PdrInv}{KinductionPlainTrueNotSolvedByKinductionPlain}{Error}{}{Cputime}{Avg}{790.2483718670663670929830626}%
\StoreBenchExecResult{PdrInv}{KinductionPlainTrueNotSolvedByKinductionPlain}{Error}{}{Cputime}{Median}{903.196108197}%
\StoreBenchExecResult{PdrInv}{KinductionPlainTrueNotSolvedByKinductionPlain}{Error}{}{Cputime}{Min}{2.524657944}%
\StoreBenchExecResult{PdrInv}{KinductionPlainTrueNotSolvedByKinductionPlain}{Error}{}{Cputime}{Max}{980.03014337}%
\StoreBenchExecResult{PdrInv}{KinductionPlainTrueNotSolvedByKinductionPlain}{Error}{}{Cputime}{Stdev}{282.6171076763802904204595638}%
\StoreBenchExecResult{PdrInv}{KinductionPlainTrueNotSolvedByKinductionPlain}{Error}{}{Walltime}{}{2093708.93828677189}%
\StoreBenchExecResult{PdrInv}{KinductionPlainTrueNotSolvedByKinductionPlain}{Error}{}{Walltime}{Avg}{723.7154988893093294158313170}%
\StoreBenchExecResult{PdrInv}{KinductionPlainTrueNotSolvedByKinductionPlain}{Error}{}{Walltime}{Median}{848.12104702}%
\StoreBenchExecResult{PdrInv}{KinductionPlainTrueNotSolvedByKinductionPlain}{Error}{}{Walltime}{Min}{1.36754894257}%
\StoreBenchExecResult{PdrInv}{KinductionPlainTrueNotSolvedByKinductionPlain}{Error}{}{Walltime}{Max}{903.080847025}%
\StoreBenchExecResult{PdrInv}{KinductionPlainTrueNotSolvedByKinductionPlain}{Error}{}{Walltime}{Stdev}{271.3172914654583917314264084}%
\StoreBenchExecResult{PdrInv}{KinductionPlainTrueNotSolvedByKinductionPlain}{Error}{Assertion}{Count}{}{2}%
\StoreBenchExecResult{PdrInv}{KinductionPlainTrueNotSolvedByKinductionPlain}{Error}{Assertion}{Cputime}{}{6.168755808}%
\StoreBenchExecResult{PdrInv}{KinductionPlainTrueNotSolvedByKinductionPlain}{Error}{Assertion}{Cputime}{Avg}{3.084377904}%
\StoreBenchExecResult{PdrInv}{KinductionPlainTrueNotSolvedByKinductionPlain}{Error}{Assertion}{Cputime}{Median}{3.084377904}%
\StoreBenchExecResult{PdrInv}{KinductionPlainTrueNotSolvedByKinductionPlain}{Error}{Assertion}{Cputime}{Min}{3.070959123}%
\StoreBenchExecResult{PdrInv}{KinductionPlainTrueNotSolvedByKinductionPlain}{Error}{Assertion}{Cputime}{Max}{3.097796685}%
\StoreBenchExecResult{PdrInv}{KinductionPlainTrueNotSolvedByKinductionPlain}{Error}{Assertion}{Cputime}{Stdev}{0.013418781}%
\StoreBenchExecResult{PdrInv}{KinductionPlainTrueNotSolvedByKinductionPlain}{Error}{Assertion}{Walltime}{}{3.46690201759}%
\StoreBenchExecResult{PdrInv}{KinductionPlainTrueNotSolvedByKinductionPlain}{Error}{Assertion}{Walltime}{Avg}{1.733451008795}%
\StoreBenchExecResult{PdrInv}{KinductionPlainTrueNotSolvedByKinductionPlain}{Error}{Assertion}{Walltime}{Median}{1.733451008795}%
\StoreBenchExecResult{PdrInv}{KinductionPlainTrueNotSolvedByKinductionPlain}{Error}{Assertion}{Walltime}{Min}{1.71145105362}%
\StoreBenchExecResult{PdrInv}{KinductionPlainTrueNotSolvedByKinductionPlain}{Error}{Assertion}{Walltime}{Max}{1.75545096397}%
\StoreBenchExecResult{PdrInv}{KinductionPlainTrueNotSolvedByKinductionPlain}{Error}{Assertion}{Walltime}{Stdev}{0.021999955175}%
\StoreBenchExecResult{PdrInv}{KinductionPlainTrueNotSolvedByKinductionPlain}{Error}{Error}{Count}{}{299}%
\StoreBenchExecResult{PdrInv}{KinductionPlainTrueNotSolvedByKinductionPlain}{Error}{Error}{Cputime}{}{20184.785202878}%
\StoreBenchExecResult{PdrInv}{KinductionPlainTrueNotSolvedByKinductionPlain}{Error}{Error}{Cputime}{Avg}{67.50764281898996655518394649}%
\StoreBenchExecResult{PdrInv}{KinductionPlainTrueNotSolvedByKinductionPlain}{Error}{Error}{Cputime}{Median}{21.748242734}%
\StoreBenchExecResult{PdrInv}{KinductionPlainTrueNotSolvedByKinductionPlain}{Error}{Error}{Cputime}{Min}{2.524657944}%
\StoreBenchExecResult{PdrInv}{KinductionPlainTrueNotSolvedByKinductionPlain}{Error}{Error}{Cputime}{Max}{890.352562609}%
\StoreBenchExecResult{PdrInv}{KinductionPlainTrueNotSolvedByKinductionPlain}{Error}{Error}{Cputime}{Stdev}{147.5425889384626312371945183}%
\StoreBenchExecResult{PdrInv}{KinductionPlainTrueNotSolvedByKinductionPlain}{Error}{Error}{Walltime}{}{15337.47282647909}%
\StoreBenchExecResult{PdrInv}{KinductionPlainTrueNotSolvedByKinductionPlain}{Error}{Error}{Walltime}{Avg}{51.29589574073274247491638796}%
\StoreBenchExecResult{PdrInv}{KinductionPlainTrueNotSolvedByKinductionPlain}{Error}{Error}{Walltime}{Median}{12.1307199001}%
\StoreBenchExecResult{PdrInv}{KinductionPlainTrueNotSolvedByKinductionPlain}{Error}{Error}{Walltime}{Min}{1.36754894257}%
\StoreBenchExecResult{PdrInv}{KinductionPlainTrueNotSolvedByKinductionPlain}{Error}{Error}{Walltime}{Max}{822.59779191}%
\StoreBenchExecResult{PdrInv}{KinductionPlainTrueNotSolvedByKinductionPlain}{Error}{Error}{Walltime}{Stdev}{124.4882629136126827691036805}%
\StoreBenchExecResult{PdrInv}{KinductionPlainTrueNotSolvedByKinductionPlain}{Error}{Exception}{Count}{}{13}%
\StoreBenchExecResult{PdrInv}{KinductionPlainTrueNotSolvedByKinductionPlain}{Error}{Exception}{Cputime}{}{4811.339779132}%
\StoreBenchExecResult{PdrInv}{KinductionPlainTrueNotSolvedByKinductionPlain}{Error}{Exception}{Cputime}{Avg}{370.1030599332307692307692308}%
\StoreBenchExecResult{PdrInv}{KinductionPlainTrueNotSolvedByKinductionPlain}{Error}{Exception}{Cputime}{Median}{301.547945442}%
\StoreBenchExecResult{PdrInv}{KinductionPlainTrueNotSolvedByKinductionPlain}{Error}{Exception}{Cputime}{Min}{6.568480705}%
\StoreBenchExecResult{PdrInv}{KinductionPlainTrueNotSolvedByKinductionPlain}{Error}{Exception}{Cputime}{Max}{798.996830673}%
\StoreBenchExecResult{PdrInv}{KinductionPlainTrueNotSolvedByKinductionPlain}{Error}{Exception}{Cputime}{Stdev}{326.3010253845477313831097501}%
\StoreBenchExecResult{PdrInv}{KinductionPlainTrueNotSolvedByKinductionPlain}{Error}{Exception}{Walltime}{}{4138.03609681141}%
\StoreBenchExecResult{PdrInv}{KinductionPlainTrueNotSolvedByKinductionPlain}{Error}{Exception}{Walltime}{Avg}{318.3104689854930769230769231}%
\StoreBenchExecResult{PdrInv}{KinductionPlainTrueNotSolvedByKinductionPlain}{Error}{Exception}{Walltime}{Median}{279.604038}%
\StoreBenchExecResult{PdrInv}{KinductionPlainTrueNotSolvedByKinductionPlain}{Error}{Exception}{Walltime}{Min}{3.80832004547}%
\StoreBenchExecResult{PdrInv}{KinductionPlainTrueNotSolvedByKinductionPlain}{Error}{Exception}{Walltime}{Max}{716.377207041}%
\StoreBenchExecResult{PdrInv}{KinductionPlainTrueNotSolvedByKinductionPlain}{Error}{Exception}{Walltime}{Stdev}{283.0937405762332617031500959}%
\StoreBenchExecResult{PdrInv}{KinductionPlainTrueNotSolvedByKinductionPlain}{Error}{OutOfJavaMemory}{Count}{}{14}%
\StoreBenchExecResult{PdrInv}{KinductionPlainTrueNotSolvedByKinductionPlain}{Error}{OutOfJavaMemory}{Cputime}{}{8128.308213167}%
\StoreBenchExecResult{PdrInv}{KinductionPlainTrueNotSolvedByKinductionPlain}{Error}{OutOfJavaMemory}{Cputime}{Avg}{580.5934437976428571428571429}%
\StoreBenchExecResult{PdrInv}{KinductionPlainTrueNotSolvedByKinductionPlain}{Error}{OutOfJavaMemory}{Cputime}{Median}{631.200137720}%
\StoreBenchExecResult{PdrInv}{KinductionPlainTrueNotSolvedByKinductionPlain}{Error}{OutOfJavaMemory}{Cputime}{Min}{161.19237286}%
\StoreBenchExecResult{PdrInv}{KinductionPlainTrueNotSolvedByKinductionPlain}{Error}{OutOfJavaMemory}{Cputime}{Max}{881.51517178}%
\StoreBenchExecResult{PdrInv}{KinductionPlainTrueNotSolvedByKinductionPlain}{Error}{OutOfJavaMemory}{Cputime}{Stdev}{251.4657326877878607316811596}%
\StoreBenchExecResult{PdrInv}{KinductionPlainTrueNotSolvedByKinductionPlain}{Error}{OutOfJavaMemory}{Walltime}{}{6029.4704337118}%
\StoreBenchExecResult{PdrInv}{KinductionPlainTrueNotSolvedByKinductionPlain}{Error}{OutOfJavaMemory}{Walltime}{Avg}{430.6764595508428571428571429}%
\StoreBenchExecResult{PdrInv}{KinductionPlainTrueNotSolvedByKinductionPlain}{Error}{OutOfJavaMemory}{Walltime}{Median}{442.9773864745}%
\StoreBenchExecResult{PdrInv}{KinductionPlainTrueNotSolvedByKinductionPlain}{Error}{OutOfJavaMemory}{Walltime}{Min}{99.9721179008}%
\StoreBenchExecResult{PdrInv}{KinductionPlainTrueNotSolvedByKinductionPlain}{Error}{OutOfJavaMemory}{Walltime}{Max}{754.459775925}%
\StoreBenchExecResult{PdrInv}{KinductionPlainTrueNotSolvedByKinductionPlain}{Error}{OutOfJavaMemory}{Walltime}{Stdev}{211.6783655000350846008954759}%
\StoreBenchExecResult{PdrInv}{KinductionPlainTrueNotSolvedByKinductionPlain}{Error}{OutOfMemory}{Count}{}{184}%
\StoreBenchExecResult{PdrInv}{KinductionPlainTrueNotSolvedByKinductionPlain}{Error}{OutOfMemory}{Cputime}{}{93509.735844753}%
\StoreBenchExecResult{PdrInv}{KinductionPlainTrueNotSolvedByKinductionPlain}{Error}{OutOfMemory}{Cputime}{Avg}{508.2050861127880434782608696}%
\StoreBenchExecResult{PdrInv}{KinductionPlainTrueNotSolvedByKinductionPlain}{Error}{OutOfMemory}{Cputime}{Median}{509.0511092935}%
\StoreBenchExecResult{PdrInv}{KinductionPlainTrueNotSolvedByKinductionPlain}{Error}{OutOfMemory}{Cputime}{Min}{125.179554468}%
\StoreBenchExecResult{PdrInv}{KinductionPlainTrueNotSolvedByKinductionPlain}{Error}{OutOfMemory}{Cputime}{Max}{899.806149114}%
\StoreBenchExecResult{PdrInv}{KinductionPlainTrueNotSolvedByKinductionPlain}{Error}{OutOfMemory}{Cputime}{Stdev}{282.5784910954514830261127438}%
\StoreBenchExecResult{PdrInv}{KinductionPlainTrueNotSolvedByKinductionPlain}{Error}{OutOfMemory}{Walltime}{}{90391.452912329}%
\StoreBenchExecResult{PdrInv}{KinductionPlainTrueNotSolvedByKinductionPlain}{Error}{OutOfMemory}{Walltime}{Avg}{491.2578962626576086956521739}%
\StoreBenchExecResult{PdrInv}{KinductionPlainTrueNotSolvedByKinductionPlain}{Error}{OutOfMemory}{Walltime}{Median}{495.241203904}%
\StoreBenchExecResult{PdrInv}{KinductionPlainTrueNotSolvedByKinductionPlain}{Error}{OutOfMemory}{Walltime}{Min}{112.996083975}%
\StoreBenchExecResult{PdrInv}{KinductionPlainTrueNotSolvedByKinductionPlain}{Error}{OutOfMemory}{Walltime}{Max}{885.898652077}%
\StoreBenchExecResult{PdrInv}{KinductionPlainTrueNotSolvedByKinductionPlain}{Error}{OutOfMemory}{Walltime}{Stdev}{280.9577733513922433993479493}%
\StoreBenchExecResult{PdrInv}{KinductionPlainTrueNotSolvedByKinductionPlain}{Error}{Timeout}{Count}{}{2381}%
\StoreBenchExecResult{PdrInv}{KinductionPlainTrueNotSolvedByKinductionPlain}{Error}{Timeout}{Cputime}{}{2159548.202015685}%
\StoreBenchExecResult{PdrInv}{KinductionPlainTrueNotSolvedByKinductionPlain}{Error}{Timeout}{Cputime}{Avg}{906.9921050044876102477950441}%
\StoreBenchExecResult{PdrInv}{KinductionPlainTrueNotSolvedByKinductionPlain}{Error}{Timeout}{Cputime}{Median}{904.007466608}%
\StoreBenchExecResult{PdrInv}{KinductionPlainTrueNotSolvedByKinductionPlain}{Error}{Timeout}{Cputime}{Min}{900.180142923}%
\StoreBenchExecResult{PdrInv}{KinductionPlainTrueNotSolvedByKinductionPlain}{Error}{Timeout}{Cputime}{Max}{980.03014337}%
\StoreBenchExecResult{PdrInv}{KinductionPlainTrueNotSolvedByKinductionPlain}{Error}{Timeout}{Cputime}{Stdev}{8.671012958742172048522517556}%
\StoreBenchExecResult{PdrInv}{KinductionPlainTrueNotSolvedByKinductionPlain}{Error}{Timeout}{Walltime}{}{1977809.039115423}%
\StoreBenchExecResult{PdrInv}{KinductionPlainTrueNotSolvedByKinductionPlain}{Error}{Timeout}{Walltime}{Avg}{830.6631831648143637127257455}%
\StoreBenchExecResult{PdrInv}{KinductionPlainTrueNotSolvedByKinductionPlain}{Error}{Timeout}{Walltime}{Median}{877.557717085}%
\StoreBenchExecResult{PdrInv}{KinductionPlainTrueNotSolvedByKinductionPlain}{Error}{Timeout}{Walltime}{Min}{492.846322775}%
\StoreBenchExecResult{PdrInv}{KinductionPlainTrueNotSolvedByKinductionPlain}{Error}{Timeout}{Walltime}{Max}{903.080847025}%
\StoreBenchExecResult{PdrInv}{KinductionPlainTrueNotSolvedByKinductionPlain}{Error}{Timeout}{Walltime}{Stdev}{80.38536031843398635399626305}%
\providecommand\StoreBenchExecResult[7]{\expandafter\newcommand\csname#1#2#3#4#5#6\endcsname{#7}}%
\StoreBenchExecResult{PdrInv}{KinductionPlain}{Total}{}{Count}{}{5591}%
\StoreBenchExecResult{PdrInv}{KinductionPlain}{Total}{}{Cputime}{}{2798679.639984491}%
\StoreBenchExecResult{PdrInv}{KinductionPlain}{Total}{}{Cputime}{Avg}{500.5687068475211947773206940}%
\StoreBenchExecResult{PdrInv}{KinductionPlain}{Total}{}{Cputime}{Median}{868.685246897}%
\StoreBenchExecResult{PdrInv}{KinductionPlain}{Total}{}{Cputime}{Min}{2.524657944}%
\StoreBenchExecResult{PdrInv}{KinductionPlain}{Total}{}{Cputime}{Max}{1002.21123876}%
\StoreBenchExecResult{PdrInv}{KinductionPlain}{Total}{}{Cputime}{Stdev}{427.5461309750189673458525244}%
\StoreBenchExecResult{PdrInv}{KinductionPlain}{Total}{}{Walltime}{}{2561599.35869453786}%
\StoreBenchExecResult{PdrInv}{KinductionPlain}{Total}{}{Walltime}{Avg}{458.1647931844997066714362368}%
\StoreBenchExecResult{PdrInv}{KinductionPlain}{Total}{}{Walltime}{Median}{680.856317043}%
\StoreBenchExecResult{PdrInv}{KinductionPlain}{Total}{}{Walltime}{Min}{1.36754894257}%
\StoreBenchExecResult{PdrInv}{KinductionPlain}{Total}{}{Walltime}{Max}{963.118584871}%
\StoreBenchExecResult{PdrInv}{KinductionPlain}{Total}{}{Walltime}{Stdev}{398.6618569580856927022647113}%
\StoreBenchExecResult{PdrInv}{KinductionPlain}{Correct}{}{Count}{}{2075}%
\StoreBenchExecResult{PdrInv}{KinductionPlain}{Correct}{}{Cputime}{}{108878.492537876}%
\StoreBenchExecResult{PdrInv}{KinductionPlain}{Correct}{}{Cputime}{Avg}{52.47156266885590361445783133}%
\StoreBenchExecResult{PdrInv}{KinductionPlain}{Correct}{}{Cputime}{Median}{9.84852058}%
\StoreBenchExecResult{PdrInv}{KinductionPlain}{Correct}{}{Cputime}{Min}{2.932446386}%
\StoreBenchExecResult{PdrInv}{KinductionPlain}{Correct}{}{Cputime}{Max}{896.753930027}%
\StoreBenchExecResult{PdrInv}{KinductionPlain}{Correct}{}{Cputime}{Stdev}{135.1527842788481576084871880}%
\StoreBenchExecResult{PdrInv}{KinductionPlain}{Correct}{}{Walltime}{}{95192.33876013816}%
\StoreBenchExecResult{PdrInv}{KinductionPlain}{Correct}{}{Walltime}{Avg}{45.87582590850031807228915663}%
\StoreBenchExecResult{PdrInv}{KinductionPlain}{Correct}{}{Walltime}{Median}{5.34067893028}%
\StoreBenchExecResult{PdrInv}{KinductionPlain}{Correct}{}{Walltime}{Min}{1.65026211739}%
\StoreBenchExecResult{PdrInv}{KinductionPlain}{Correct}{}{Walltime}{Max}{881.709371805}%
\StoreBenchExecResult{PdrInv}{KinductionPlain}{Correct}{}{Walltime}{Stdev}{130.8503758471020355575892495}%
\StoreBenchExecResult{PdrInv}{KinductionPlain}{Correct}{False}{Count}{}{836}%
\StoreBenchExecResult{PdrInv}{KinductionPlain}{Correct}{False}{Cputime}{}{68550.709753055}%
\StoreBenchExecResult{PdrInv}{KinductionPlain}{Correct}{False}{Cputime}{Avg}{81.99845664241028708133971292}%
\StoreBenchExecResult{PdrInv}{KinductionPlain}{Correct}{False}{Cputime}{Median}{16.1491332775}%
\StoreBenchExecResult{PdrInv}{KinductionPlain}{Correct}{False}{Cputime}{Min}{3.016743131}%
\StoreBenchExecResult{PdrInv}{KinductionPlain}{Correct}{False}{Cputime}{Max}{896.753930027}%
\StoreBenchExecResult{PdrInv}{KinductionPlain}{Correct}{False}{Cputime}{Stdev}{176.8080134234912776799649557}%
\StoreBenchExecResult{PdrInv}{KinductionPlain}{Correct}{False}{Walltime}{}{61483.96026110621}%
\StoreBenchExecResult{PdrInv}{KinductionPlain}{Correct}{False}{Walltime}{Avg}{73.54540701089259569377990431}%
\StoreBenchExecResult{PdrInv}{KinductionPlain}{Correct}{False}{Walltime}{Median}{9.20947456360}%
\StoreBenchExecResult{PdrInv}{KinductionPlain}{Correct}{False}{Walltime}{Min}{1.71778702736}%
\StoreBenchExecResult{PdrInv}{KinductionPlain}{Correct}{False}{Walltime}{Max}{881.709371805}%
\StoreBenchExecResult{PdrInv}{KinductionPlain}{Correct}{False}{Walltime}{Stdev}{171.6972677803019092082721423}%
\StoreBenchExecResult{PdrInv}{KinductionPlain}{Correct}{True}{Count}{}{1239}%
\StoreBenchExecResult{PdrInv}{KinductionPlain}{Correct}{True}{Cputime}{}{40327.782784821}%
\StoreBenchExecResult{PdrInv}{KinductionPlain}{Correct}{True}{Cputime}{Avg}{32.54865438645762711864406780}%
\StoreBenchExecResult{PdrInv}{KinductionPlain}{Correct}{True}{Cputime}{Median}{7.214652128}%
\StoreBenchExecResult{PdrInv}{KinductionPlain}{Correct}{True}{Cputime}{Min}{2.932446386}%
\StoreBenchExecResult{PdrInv}{KinductionPlain}{Correct}{True}{Cputime}{Max}{893.777407096}%
\StoreBenchExecResult{PdrInv}{KinductionPlain}{Correct}{True}{Cputime}{Stdev}{92.26597894771223151763608956}%
\StoreBenchExecResult{PdrInv}{KinductionPlain}{Correct}{True}{Walltime}{}{33708.37849903195}%
\StoreBenchExecResult{PdrInv}{KinductionPlain}{Correct}{True}{Walltime}{Avg}{27.20611662553022598870056497}%
\StoreBenchExecResult{PdrInv}{KinductionPlain}{Correct}{True}{Walltime}{Median}{3.98854088783}%
\StoreBenchExecResult{PdrInv}{KinductionPlain}{Correct}{True}{Walltime}{Min}{1.65026211739}%
\StoreBenchExecResult{PdrInv}{KinductionPlain}{Correct}{True}{Walltime}{Max}{875.484183788}%
\StoreBenchExecResult{PdrInv}{KinductionPlain}{Correct}{True}{Walltime}{Stdev}{88.98417308599047139788059498}%
\StoreBenchExecResult{PdrInv}{KinductionPlain}{Wrong}{True}{Count}{}{0}%
\StoreBenchExecResult{PdrInv}{KinductionPlain}{Wrong}{True}{Cputime}{}{0}%
\StoreBenchExecResult{PdrInv}{KinductionPlain}{Wrong}{True}{Cputime}{Avg}{None}%
\StoreBenchExecResult{PdrInv}{KinductionPlain}{Wrong}{True}{Cputime}{Median}{None}%
\StoreBenchExecResult{PdrInv}{KinductionPlain}{Wrong}{True}{Cputime}{Min}{None}%
\StoreBenchExecResult{PdrInv}{KinductionPlain}{Wrong}{True}{Cputime}{Max}{None}%
\StoreBenchExecResult{PdrInv}{KinductionPlain}{Wrong}{True}{Cputime}{Stdev}{None}%
\StoreBenchExecResult{PdrInv}{KinductionPlain}{Wrong}{True}{Walltime}{}{0}%
\StoreBenchExecResult{PdrInv}{KinductionPlain}{Wrong}{True}{Walltime}{Avg}{None}%
\StoreBenchExecResult{PdrInv}{KinductionPlain}{Wrong}{True}{Walltime}{Median}{None}%
\StoreBenchExecResult{PdrInv}{KinductionPlain}{Wrong}{True}{Walltime}{Min}{None}%
\StoreBenchExecResult{PdrInv}{KinductionPlain}{Wrong}{True}{Walltime}{Max}{None}%
\StoreBenchExecResult{PdrInv}{KinductionPlain}{Wrong}{True}{Walltime}{Stdev}{None}%
\StoreBenchExecResult{PdrInv}{KinductionPlain}{Error}{}{Count}{}{3514}%
\StoreBenchExecResult{PdrInv}{KinductionPlain}{Error}{}{Cputime}{}{2689778.308291708}%
\StoreBenchExecResult{PdrInv}{KinductionPlain}{Error}{}{Cputime}{Avg}{765.4463028718577120091064314}%
\StoreBenchExecResult{PdrInv}{KinductionPlain}{Error}{}{Cputime}{Median}{902.9980651325}%
\StoreBenchExecResult{PdrInv}{KinductionPlain}{Error}{}{Cputime}{Min}{2.524657944}%
\StoreBenchExecResult{PdrInv}{KinductionPlain}{Error}{}{Cputime}{Max}{1002.21123876}%
\StoreBenchExecResult{PdrInv}{KinductionPlain}{Error}{}{Cputime}{Stdev}{301.9785816305871535791045679}%
\StoreBenchExecResult{PdrInv}{KinductionPlain}{Error}{}{Walltime}{}{2466394.65063713370}%
\StoreBenchExecResult{PdrInv}{KinductionPlain}{Error}{}{Walltime}{Avg}{701.8766791796054923164484917}%
\StoreBenchExecResult{PdrInv}{KinductionPlain}{Error}{}{Walltime}{Median}{836.4787650105}%
\StoreBenchExecResult{PdrInv}{KinductionPlain}{Error}{}{Walltime}{Min}{1.36754894257}%
\StoreBenchExecResult{PdrInv}{KinductionPlain}{Error}{}{Walltime}{Max}{963.118584871}%
\StoreBenchExecResult{PdrInv}{KinductionPlain}{Error}{}{Walltime}{Stdev}{287.8786290390889883896993991}%
\StoreBenchExecResult{PdrInv}{KinductionPlain}{Error}{Assertion}{Count}{}{4}%
\StoreBenchExecResult{PdrInv}{KinductionPlain}{Error}{Assertion}{Cputime}{}{12.515762964}%
\StoreBenchExecResult{PdrInv}{KinductionPlain}{Error}{Assertion}{Cputime}{Avg}{3.128940741}%
\StoreBenchExecResult{PdrInv}{KinductionPlain}{Error}{Assertion}{Cputime}{Median}{3.113573300}%
\StoreBenchExecResult{PdrInv}{KinductionPlain}{Error}{Assertion}{Cputime}{Min}{3.070959123}%
\StoreBenchExecResult{PdrInv}{KinductionPlain}{Error}{Assertion}{Cputime}{Max}{3.217657241}%
\StoreBenchExecResult{PdrInv}{KinductionPlain}{Error}{Assertion}{Cputime}{Stdev}{0.05523270101498145998486979531}%
\StoreBenchExecResult{PdrInv}{KinductionPlain}{Error}{Assertion}{Walltime}{}{7.03489279747}%
\StoreBenchExecResult{PdrInv}{KinductionPlain}{Error}{Assertion}{Walltime}{Avg}{1.7587231993675}%
\StoreBenchExecResult{PdrInv}{KinductionPlain}{Error}{Assertion}{Walltime}{Median}{1.757482409475}%
\StoreBenchExecResult{PdrInv}{KinductionPlain}{Error}{Assertion}{Walltime}{Min}{1.71145105362}%
\StoreBenchExecResult{PdrInv}{KinductionPlain}{Error}{Assertion}{Walltime}{Max}{1.8084769249}%
\StoreBenchExecResult{PdrInv}{KinductionPlain}{Error}{Assertion}{Walltime}{Stdev}{0.03435630083671135888880426452}%
\StoreBenchExecResult{PdrInv}{KinductionPlain}{Error}{Error}{Count}{}{390}%
\StoreBenchExecResult{PdrInv}{KinductionPlain}{Error}{Error}{Cputime}{}{24050.459487729}%
\StoreBenchExecResult{PdrInv}{KinductionPlain}{Error}{Error}{Cputime}{Avg}{61.66784484033076923076923077}%
\StoreBenchExecResult{PdrInv}{KinductionPlain}{Error}{Error}{Cputime}{Median}{19.7010947875}%
\StoreBenchExecResult{PdrInv}{KinductionPlain}{Error}{Error}{Cputime}{Min}{2.524657944}%
\StoreBenchExecResult{PdrInv}{KinductionPlain}{Error}{Error}{Cputime}{Max}{890.352562609}%
\StoreBenchExecResult{PdrInv}{KinductionPlain}{Error}{Error}{Cputime}{Stdev}{134.6887479181063960963451007}%
\StoreBenchExecResult{PdrInv}{KinductionPlain}{Error}{Error}{Walltime}{}{17960.16732144223}%
\StoreBenchExecResult{PdrInv}{KinductionPlain}{Error}{Error}{Walltime}{Avg}{46.05171108062110256410256410}%
\StoreBenchExecResult{PdrInv}{KinductionPlain}{Error}{Error}{Walltime}{Median}{10.4599869251}%
\StoreBenchExecResult{PdrInv}{KinductionPlain}{Error}{Error}{Walltime}{Min}{1.36754894257}%
\StoreBenchExecResult{PdrInv}{KinductionPlain}{Error}{Error}{Walltime}{Max}{822.59779191}%
\StoreBenchExecResult{PdrInv}{KinductionPlain}{Error}{Error}{Walltime}{Stdev}{112.5933597444059666192889487}%
\StoreBenchExecResult{PdrInv}{KinductionPlain}{Error}{Exception}{Count}{}{16}%
\StoreBenchExecResult{PdrInv}{KinductionPlain}{Error}{Exception}{Cputime}{}{4873.809355125}%
\StoreBenchExecResult{PdrInv}{KinductionPlain}{Error}{Exception}{Cputime}{Avg}{304.6130846953125}%
\StoreBenchExecResult{PdrInv}{KinductionPlain}{Error}{Exception}{Cputime}{Median}{93.754688825}%
\StoreBenchExecResult{PdrInv}{KinductionPlain}{Error}{Exception}{Cputime}{Min}{6.568480705}%
\StoreBenchExecResult{PdrInv}{KinductionPlain}{Error}{Exception}{Cputime}{Max}{798.996830673}%
\StoreBenchExecResult{PdrInv}{KinductionPlain}{Error}{Exception}{Cputime}{Stdev}{324.1952026524832109973862279}%
\StoreBenchExecResult{PdrInv}{KinductionPlain}{Error}{Exception}{Walltime}{}{4176.42934179320}%
\StoreBenchExecResult{PdrInv}{KinductionPlain}{Error}{Exception}{Walltime}{Avg}{261.026833862075}%
\StoreBenchExecResult{PdrInv}{KinductionPlain}{Error}{Exception}{Walltime}{Median}{74.43587601185}%
\StoreBenchExecResult{PdrInv}{KinductionPlain}{Error}{Exception}{Walltime}{Min}{3.80832004547}%
\StoreBenchExecResult{PdrInv}{KinductionPlain}{Error}{Exception}{Walltime}{Max}{716.377207041}%
\StoreBenchExecResult{PdrInv}{KinductionPlain}{Error}{Exception}{Walltime}{Stdev}{281.6756912224033343795941558}%
\StoreBenchExecResult{PdrInv}{KinductionPlain}{Error}{OutOfJavaMemory}{Count}{}{25}%
\StoreBenchExecResult{PdrInv}{KinductionPlain}{Error}{OutOfJavaMemory}{Cputime}{}{15171.244784099}%
\StoreBenchExecResult{PdrInv}{KinductionPlain}{Error}{OutOfJavaMemory}{Cputime}{Avg}{606.84979136396}%
\StoreBenchExecResult{PdrInv}{KinductionPlain}{Error}{OutOfJavaMemory}{Cputime}{Median}{664.870016737}%
\StoreBenchExecResult{PdrInv}{KinductionPlain}{Error}{OutOfJavaMemory}{Cputime}{Min}{161.19237286}%
\StoreBenchExecResult{PdrInv}{KinductionPlain}{Error}{OutOfJavaMemory}{Cputime}{Max}{886.137726391}%
\StoreBenchExecResult{PdrInv}{KinductionPlain}{Error}{OutOfJavaMemory}{Cputime}{Stdev}{238.7844394772508533279806844}%
\StoreBenchExecResult{PdrInv}{KinductionPlain}{Error}{OutOfJavaMemory}{Walltime}{}{10865.8108403698}%
\StoreBenchExecResult{PdrInv}{KinductionPlain}{Error}{OutOfJavaMemory}{Walltime}{Avg}{434.632433614792}%
\StoreBenchExecResult{PdrInv}{KinductionPlain}{Error}{OutOfJavaMemory}{Walltime}{Median}{447.082789898}%
\StoreBenchExecResult{PdrInv}{KinductionPlain}{Error}{OutOfJavaMemory}{Walltime}{Min}{99.9721179008}%
\StoreBenchExecResult{PdrInv}{KinductionPlain}{Error}{OutOfJavaMemory}{Walltime}{Max}{754.459775925}%
\StoreBenchExecResult{PdrInv}{KinductionPlain}{Error}{OutOfJavaMemory}{Walltime}{Stdev}{197.8941326190758679840764860}%
\StoreBenchExecResult{PdrInv}{KinductionPlain}{Error}{OutOfMemory}{Count}{}{315}%
\StoreBenchExecResult{PdrInv}{KinductionPlain}{Error}{OutOfMemory}{Cputime}{}{138308.710233129}%
\StoreBenchExecResult{PdrInv}{KinductionPlain}{Error}{OutOfMemory}{Cputime}{Avg}{439.0752705813619047619047619}%
\StoreBenchExecResult{PdrInv}{KinductionPlain}{Error}{OutOfMemory}{Cputime}{Median}{375.163127341}%
\StoreBenchExecResult{PdrInv}{KinductionPlain}{Error}{OutOfMemory}{Cputime}{Min}{125.179554468}%
\StoreBenchExecResult{PdrInv}{KinductionPlain}{Error}{OutOfMemory}{Cputime}{Max}{899.806149114}%
\StoreBenchExecResult{PdrInv}{KinductionPlain}{Error}{OutOfMemory}{Cputime}{Stdev}{260.9800670332385826660187508}%
\StoreBenchExecResult{PdrInv}{KinductionPlain}{Error}{OutOfMemory}{Walltime}{}{133334.550362345}%
\StoreBenchExecResult{PdrInv}{KinductionPlain}{Error}{OutOfMemory}{Walltime}{Avg}{423.2842868645873015873015873}%
\StoreBenchExecResult{PdrInv}{KinductionPlain}{Error}{OutOfMemory}{Walltime}{Median}{364.179842949}%
\StoreBenchExecResult{PdrInv}{KinductionPlain}{Error}{OutOfMemory}{Walltime}{Min}{112.996083975}%
\StoreBenchExecResult{PdrInv}{KinductionPlain}{Error}{OutOfMemory}{Walltime}{Max}{885.898652077}%
\StoreBenchExecResult{PdrInv}{KinductionPlain}{Error}{OutOfMemory}{Walltime}{Stdev}{259.6502232702249039295710231}%
\StoreBenchExecResult{PdrInv}{KinductionPlain}{Error}{Timeout}{Count}{}{2764}%
\StoreBenchExecResult{PdrInv}{KinductionPlain}{Error}{Timeout}{Cputime}{}{2507361.568668662}%
\StoreBenchExecResult{PdrInv}{KinductionPlain}{Error}{Timeout}{Cputime}{Avg}{907.1496268699934876989869754}%
\StoreBenchExecResult{PdrInv}{KinductionPlain}{Error}{Timeout}{Cputime}{Median}{904.0668178215}%
\StoreBenchExecResult{PdrInv}{KinductionPlain}{Error}{Timeout}{Cputime}{Min}{900.180142923}%
\StoreBenchExecResult{PdrInv}{KinductionPlain}{Error}{Timeout}{Cputime}{Max}{1002.21123876}%
\StoreBenchExecResult{PdrInv}{KinductionPlain}{Error}{Timeout}{Cputime}{Stdev}{9.216810758806584093493923280}%
\StoreBenchExecResult{PdrInv}{KinductionPlain}{Error}{Timeout}{Walltime}{}{2300050.657878386}%
\StoreBenchExecResult{PdrInv}{KinductionPlain}{Error}{Timeout}{Walltime}{Avg}{832.1456794060730824891461650}%
\StoreBenchExecResult{PdrInv}{KinductionPlain}{Error}{Timeout}{Walltime}{Median}{877.0268926625}%
\StoreBenchExecResult{PdrInv}{KinductionPlain}{Error}{Timeout}{Walltime}{Min}{492.846322775}%
\StoreBenchExecResult{PdrInv}{KinductionPlain}{Error}{Timeout}{Walltime}{Max}{963.118584871}%
\StoreBenchExecResult{PdrInv}{KinductionPlain}{Error}{Timeout}{Walltime}{Stdev}{78.05421852020561357971021639}%
\StoreBenchExecResult{PdrInv}{KinductionPlain}{Wrong}{}{Count}{}{2}%
\StoreBenchExecResult{PdrInv}{KinductionPlain}{Wrong}{}{Cputime}{}{22.839154907}%
\StoreBenchExecResult{PdrInv}{KinductionPlain}{Wrong}{}{Cputime}{Avg}{11.4195774535}%
\StoreBenchExecResult{PdrInv}{KinductionPlain}{Wrong}{}{Cputime}{Median}{11.4195774535}%
\StoreBenchExecResult{PdrInv}{KinductionPlain}{Wrong}{}{Cputime}{Min}{4.068832737}%
\StoreBenchExecResult{PdrInv}{KinductionPlain}{Wrong}{}{Cputime}{Max}{18.77032217}%
\StoreBenchExecResult{PdrInv}{KinductionPlain}{Wrong}{}{Cputime}{Stdev}{7.3507447165}%
\StoreBenchExecResult{PdrInv}{KinductionPlain}{Wrong}{}{Walltime}{}{12.3692972660}%
\StoreBenchExecResult{PdrInv}{KinductionPlain}{Wrong}{}{Walltime}{Avg}{6.1846486330}%
\StoreBenchExecResult{PdrInv}{KinductionPlain}{Wrong}{}{Walltime}{Median}{6.1846486330}%
\StoreBenchExecResult{PdrInv}{KinductionPlain}{Wrong}{}{Walltime}{Min}{2.2132871151}%
\StoreBenchExecResult{PdrInv}{KinductionPlain}{Wrong}{}{Walltime}{Max}{10.1560101509}%
\StoreBenchExecResult{PdrInv}{KinductionPlain}{Wrong}{}{Walltime}{Stdev}{3.9713615179}%
\StoreBenchExecResult{PdrInv}{KinductionPlain}{Wrong}{False}{Count}{}{2}%
\StoreBenchExecResult{PdrInv}{KinductionPlain}{Wrong}{False}{Cputime}{}{22.839154907}%
\StoreBenchExecResult{PdrInv}{KinductionPlain}{Wrong}{False}{Cputime}{Avg}{11.4195774535}%
\StoreBenchExecResult{PdrInv}{KinductionPlain}{Wrong}{False}{Cputime}{Median}{11.4195774535}%
\StoreBenchExecResult{PdrInv}{KinductionPlain}{Wrong}{False}{Cputime}{Min}{4.068832737}%
\StoreBenchExecResult{PdrInv}{KinductionPlain}{Wrong}{False}{Cputime}{Max}{18.77032217}%
\StoreBenchExecResult{PdrInv}{KinductionPlain}{Wrong}{False}{Cputime}{Stdev}{7.3507447165}%
\StoreBenchExecResult{PdrInv}{KinductionPlain}{Wrong}{False}{Walltime}{}{12.3692972660}%
\StoreBenchExecResult{PdrInv}{KinductionPlain}{Wrong}{False}{Walltime}{Avg}{6.1846486330}%
\StoreBenchExecResult{PdrInv}{KinductionPlain}{Wrong}{False}{Walltime}{Median}{6.1846486330}%
\StoreBenchExecResult{PdrInv}{KinductionPlain}{Wrong}{False}{Walltime}{Min}{2.2132871151}%
\StoreBenchExecResult{PdrInv}{KinductionPlain}{Wrong}{False}{Walltime}{Max}{10.1560101509}%
\StoreBenchExecResult{PdrInv}{KinductionPlain}{Wrong}{False}{Walltime}{Stdev}{3.9713615179}%
\providecommand\StoreBenchExecResult[7]{\expandafter\newcommand\csname#1#2#3#4#5#6\endcsname{#7}}%
\StoreBenchExecResult{PdrInv}{OracleTrueNotSolvedByKinductionPlainButKipdr}{Total}{}{Count}{}{449}%
\StoreBenchExecResult{PdrInv}{OracleTrueNotSolvedByKinductionPlainButKipdr}{Total}{}{Cputime}{}{2256.977390167}%
\StoreBenchExecResult{PdrInv}{OracleTrueNotSolvedByKinductionPlainButKipdr}{Total}{}{Cputime}{Avg}{5.026675701930957683741648107}%
\StoreBenchExecResult{PdrInv}{OracleTrueNotSolvedByKinductionPlainButKipdr}{Total}{}{Cputime}{Median}{0.104885966}%
\StoreBenchExecResult{PdrInv}{OracleTrueNotSolvedByKinductionPlainButKipdr}{Total}{}{Cputime}{Min}{0.054004265}%
\StoreBenchExecResult{PdrInv}{OracleTrueNotSolvedByKinductionPlainButKipdr}{Total}{}{Cputime}{Max}{902.614005656}%
\StoreBenchExecResult{PdrInv}{OracleTrueNotSolvedByKinductionPlainButKipdr}{Total}{}{Cputime}{Stdev}{55.99624827983418201188624939}%
\StoreBenchExecResult{PdrInv}{OracleTrueNotSolvedByKinductionPlainButKipdr}{Total}{}{Walltime}{}{2120.3604922301559}%
\StoreBenchExecResult{PdrInv}{OracleTrueNotSolvedByKinductionPlainButKipdr}{Total}{}{Walltime}{Avg}{4.722406441492552115812917595}%
\StoreBenchExecResult{PdrInv}{OracleTrueNotSolvedByKinductionPlainButKipdr}{Total}{}{Walltime}{Median}{0.0655188560486}%
\StoreBenchExecResult{PdrInv}{OracleTrueNotSolvedByKinductionPlainButKipdr}{Total}{}{Walltime}{Min}{0.0359630584717}%
\StoreBenchExecResult{PdrInv}{OracleTrueNotSolvedByKinductionPlainButKipdr}{Total}{}{Walltime}{Max}{888.099666834}%
\StoreBenchExecResult{PdrInv}{OracleTrueNotSolvedByKinductionPlainButKipdr}{Total}{}{Walltime}{Stdev}{54.67223411509272327150952409}%
\StoreBenchExecResult{PdrInv}{OracleTrueNotSolvedByKinductionPlainButKipdr}{Correct}{}{Count}{}{22}%
\StoreBenchExecResult{PdrInv}{OracleTrueNotSolvedByKinductionPlainButKipdr}{Correct}{}{Cputime}{}{1309.653318220}%
\StoreBenchExecResult{PdrInv}{OracleTrueNotSolvedByKinductionPlainButKipdr}{Correct}{}{Cputime}{Avg}{59.52969628272727272727272727}%
\StoreBenchExecResult{PdrInv}{OracleTrueNotSolvedByKinductionPlainButKipdr}{Correct}{}{Cputime}{Median}{6.791211289}%
\StoreBenchExecResult{PdrInv}{OracleTrueNotSolvedByKinductionPlainButKipdr}{Correct}{}{Cputime}{Min}{0.196609262}%
\StoreBenchExecResult{PdrInv}{OracleTrueNotSolvedByKinductionPlainButKipdr}{Correct}{}{Cputime}{Max}{720.680305649}%
\StoreBenchExecResult{PdrInv}{OracleTrueNotSolvedByKinductionPlainButKipdr}{Correct}{}{Cputime}{Stdev}{154.7049942502910344631301538}%
\StoreBenchExecResult{PdrInv}{OracleTrueNotSolvedByKinductionPlainButKipdr}{Correct}{}{Walltime}{}{1203.475627661324}%
\StoreBenchExecResult{PdrInv}{OracleTrueNotSolvedByKinductionPlainButKipdr}{Correct}{}{Walltime}{Avg}{54.70343762096927272727272727}%
\StoreBenchExecResult{PdrInv}{OracleTrueNotSolvedByKinductionPlainButKipdr}{Correct}{}{Walltime}{Median}{4.09995150566}%
\StoreBenchExecResult{PdrInv}{OracleTrueNotSolvedByKinductionPlainButKipdr}{Correct}{}{Walltime}{Min}{0.126837968826}%
\StoreBenchExecResult{PdrInv}{OracleTrueNotSolvedByKinductionPlainButKipdr}{Correct}{}{Walltime}{Max}{702.018512011}%
\StoreBenchExecResult{PdrInv}{OracleTrueNotSolvedByKinductionPlainButKipdr}{Correct}{}{Walltime}{Stdev}{150.3844349065699746974893521}%
\StoreBenchExecResult{PdrInv}{OracleTrueNotSolvedByKinductionPlainButKipdr}{Correct}{True}{Count}{}{22}%
\StoreBenchExecResult{PdrInv}{OracleTrueNotSolvedByKinductionPlainButKipdr}{Correct}{True}{Cputime}{}{1309.653318220}%
\StoreBenchExecResult{PdrInv}{OracleTrueNotSolvedByKinductionPlainButKipdr}{Correct}{True}{Cputime}{Avg}{59.52969628272727272727272727}%
\StoreBenchExecResult{PdrInv}{OracleTrueNotSolvedByKinductionPlainButKipdr}{Correct}{True}{Cputime}{Median}{6.791211289}%
\StoreBenchExecResult{PdrInv}{OracleTrueNotSolvedByKinductionPlainButKipdr}{Correct}{True}{Cputime}{Min}{0.196609262}%
\StoreBenchExecResult{PdrInv}{OracleTrueNotSolvedByKinductionPlainButKipdr}{Correct}{True}{Cputime}{Max}{720.680305649}%
\StoreBenchExecResult{PdrInv}{OracleTrueNotSolvedByKinductionPlainButKipdr}{Correct}{True}{Cputime}{Stdev}{154.7049942502910344631301538}%
\StoreBenchExecResult{PdrInv}{OracleTrueNotSolvedByKinductionPlainButKipdr}{Correct}{True}{Walltime}{}{1203.475627661324}%
\StoreBenchExecResult{PdrInv}{OracleTrueNotSolvedByKinductionPlainButKipdr}{Correct}{True}{Walltime}{Avg}{54.70343762096927272727272727}%
\StoreBenchExecResult{PdrInv}{OracleTrueNotSolvedByKinductionPlainButKipdr}{Correct}{True}{Walltime}{Median}{4.09995150566}%
\StoreBenchExecResult{PdrInv}{OracleTrueNotSolvedByKinductionPlainButKipdr}{Correct}{True}{Walltime}{Min}{0.126837968826}%
\StoreBenchExecResult{PdrInv}{OracleTrueNotSolvedByKinductionPlainButKipdr}{Correct}{True}{Walltime}{Max}{702.018512011}%
\StoreBenchExecResult{PdrInv}{OracleTrueNotSolvedByKinductionPlainButKipdr}{Correct}{True}{Walltime}{Stdev}{150.3844349065699746974893521}%
\StoreBenchExecResult{PdrInv}{OracleTrueNotSolvedByKinductionPlainButKipdr}{Wrong}{True}{Count}{}{0}%
\StoreBenchExecResult{PdrInv}{OracleTrueNotSolvedByKinductionPlainButKipdr}{Wrong}{True}{Cputime}{}{0}%
\StoreBenchExecResult{PdrInv}{OracleTrueNotSolvedByKinductionPlainButKipdr}{Wrong}{True}{Cputime}{Avg}{None}%
\StoreBenchExecResult{PdrInv}{OracleTrueNotSolvedByKinductionPlainButKipdr}{Wrong}{True}{Cputime}{Median}{None}%
\StoreBenchExecResult{PdrInv}{OracleTrueNotSolvedByKinductionPlainButKipdr}{Wrong}{True}{Cputime}{Min}{None}%
\StoreBenchExecResult{PdrInv}{OracleTrueNotSolvedByKinductionPlainButKipdr}{Wrong}{True}{Cputime}{Max}{None}%
\StoreBenchExecResult{PdrInv}{OracleTrueNotSolvedByKinductionPlainButKipdr}{Wrong}{True}{Cputime}{Stdev}{None}%
\StoreBenchExecResult{PdrInv}{OracleTrueNotSolvedByKinductionPlainButKipdr}{Wrong}{True}{Walltime}{}{0}%
\StoreBenchExecResult{PdrInv}{OracleTrueNotSolvedByKinductionPlainButKipdr}{Wrong}{True}{Walltime}{Avg}{None}%
\StoreBenchExecResult{PdrInv}{OracleTrueNotSolvedByKinductionPlainButKipdr}{Wrong}{True}{Walltime}{Median}{None}%
\StoreBenchExecResult{PdrInv}{OracleTrueNotSolvedByKinductionPlainButKipdr}{Wrong}{True}{Walltime}{Min}{None}%
\StoreBenchExecResult{PdrInv}{OracleTrueNotSolvedByKinductionPlainButKipdr}{Wrong}{True}{Walltime}{Max}{None}%
\StoreBenchExecResult{PdrInv}{OracleTrueNotSolvedByKinductionPlainButKipdr}{Wrong}{True}{Walltime}{Stdev}{None}%
\StoreBenchExecResult{PdrInv}{OracleTrueNotSolvedByKinductionPlainButKipdr}{Error}{}{Count}{}{1}%
\StoreBenchExecResult{PdrInv}{OracleTrueNotSolvedByKinductionPlainButKipdr}{Error}{}{Cputime}{}{902.614005656}%
\StoreBenchExecResult{PdrInv}{OracleTrueNotSolvedByKinductionPlainButKipdr}{Error}{}{Cputime}{Avg}{902.614005656}%
\StoreBenchExecResult{PdrInv}{OracleTrueNotSolvedByKinductionPlainButKipdr}{Error}{}{Cputime}{Median}{902.614005656}%
\StoreBenchExecResult{PdrInv}{OracleTrueNotSolvedByKinductionPlainButKipdr}{Error}{}{Cputime}{Min}{902.614005656}%
\StoreBenchExecResult{PdrInv}{OracleTrueNotSolvedByKinductionPlainButKipdr}{Error}{}{Cputime}{Max}{902.614005656}%
\StoreBenchExecResult{PdrInv}{OracleTrueNotSolvedByKinductionPlainButKipdr}{Error}{}{Cputime}{Stdev}{0E-9}%
\StoreBenchExecResult{PdrInv}{OracleTrueNotSolvedByKinductionPlainButKipdr}{Error}{}{Walltime}{}{888.099666834}%
\StoreBenchExecResult{PdrInv}{OracleTrueNotSolvedByKinductionPlainButKipdr}{Error}{}{Walltime}{Avg}{888.099666834}%
\StoreBenchExecResult{PdrInv}{OracleTrueNotSolvedByKinductionPlainButKipdr}{Error}{}{Walltime}{Median}{888.099666834}%
\StoreBenchExecResult{PdrInv}{OracleTrueNotSolvedByKinductionPlainButKipdr}{Error}{}{Walltime}{Min}{888.099666834}%
\StoreBenchExecResult{PdrInv}{OracleTrueNotSolvedByKinductionPlainButKipdr}{Error}{}{Walltime}{Max}{888.099666834}%
\StoreBenchExecResult{PdrInv}{OracleTrueNotSolvedByKinductionPlainButKipdr}{Error}{}{Walltime}{Stdev}{0E-9}%
\StoreBenchExecResult{PdrInv}{OracleTrueNotSolvedByKinductionPlainButKipdr}{Error}{Timeout}{Count}{}{1}%
\StoreBenchExecResult{PdrInv}{OracleTrueNotSolvedByKinductionPlainButKipdr}{Error}{Timeout}{Cputime}{}{902.614005656}%
\StoreBenchExecResult{PdrInv}{OracleTrueNotSolvedByKinductionPlainButKipdr}{Error}{Timeout}{Cputime}{Avg}{902.614005656}%
\StoreBenchExecResult{PdrInv}{OracleTrueNotSolvedByKinductionPlainButKipdr}{Error}{Timeout}{Cputime}{Median}{902.614005656}%
\StoreBenchExecResult{PdrInv}{OracleTrueNotSolvedByKinductionPlainButKipdr}{Error}{Timeout}{Cputime}{Min}{902.614005656}%
\StoreBenchExecResult{PdrInv}{OracleTrueNotSolvedByKinductionPlainButKipdr}{Error}{Timeout}{Cputime}{Max}{902.614005656}%
\StoreBenchExecResult{PdrInv}{OracleTrueNotSolvedByKinductionPlainButKipdr}{Error}{Timeout}{Cputime}{Stdev}{0E-9}%
\StoreBenchExecResult{PdrInv}{OracleTrueNotSolvedByKinductionPlainButKipdr}{Error}{Timeout}{Walltime}{}{888.099666834}%
\StoreBenchExecResult{PdrInv}{OracleTrueNotSolvedByKinductionPlainButKipdr}{Error}{Timeout}{Walltime}{Avg}{888.099666834}%
\StoreBenchExecResult{PdrInv}{OracleTrueNotSolvedByKinductionPlainButKipdr}{Error}{Timeout}{Walltime}{Median}{888.099666834}%
\StoreBenchExecResult{PdrInv}{OracleTrueNotSolvedByKinductionPlainButKipdr}{Error}{Timeout}{Walltime}{Min}{888.099666834}%
\StoreBenchExecResult{PdrInv}{OracleTrueNotSolvedByKinductionPlainButKipdr}{Error}{Timeout}{Walltime}{Max}{888.099666834}%
\StoreBenchExecResult{PdrInv}{OracleTrueNotSolvedByKinductionPlainButKipdr}{Error}{Timeout}{Walltime}{Stdev}{0E-9}%
\StoreBenchExecResult{PdrInv}{OracleTrueNotSolvedByKinductionPlainButKipdr}{Unknown}{}{Count}{}{426}%
\StoreBenchExecResult{PdrInv}{OracleTrueNotSolvedByKinductionPlainButKipdr}{Unknown}{}{Cputime}{}{44.710066291}%
\StoreBenchExecResult{PdrInv}{OracleTrueNotSolvedByKinductionPlainButKipdr}{Unknown}{}{Cputime}{Avg}{0.1049532072558685446009389671}%
\StoreBenchExecResult{PdrInv}{OracleTrueNotSolvedByKinductionPlainButKipdr}{Unknown}{}{Cputime}{Median}{0.1029969065}%
\StoreBenchExecResult{PdrInv}{OracleTrueNotSolvedByKinductionPlainButKipdr}{Unknown}{}{Cputime}{Min}{0.054004265}%
\StoreBenchExecResult{PdrInv}{OracleTrueNotSolvedByKinductionPlainButKipdr}{Unknown}{}{Cputime}{Max}{0.261340774}%
\StoreBenchExecResult{PdrInv}{OracleTrueNotSolvedByKinductionPlainButKipdr}{Unknown}{}{Cputime}{Stdev}{0.03278436357931169744070023063}%
\StoreBenchExecResult{PdrInv}{OracleTrueNotSolvedByKinductionPlainButKipdr}{Unknown}{}{Walltime}{}{28.7851977348319}%
\StoreBenchExecResult{PdrInv}{OracleTrueNotSolvedByKinductionPlainButKipdr}{Unknown}{}{Walltime}{Avg}{0.06757088670148333333333333333}%
\StoreBenchExecResult{PdrInv}{OracleTrueNotSolvedByKinductionPlainButKipdr}{Unknown}{}{Walltime}{Median}{0.06429135799405}%
\StoreBenchExecResult{PdrInv}{OracleTrueNotSolvedByKinductionPlainButKipdr}{Unknown}{}{Walltime}{Min}{0.0359630584717}%
\StoreBenchExecResult{PdrInv}{OracleTrueNotSolvedByKinductionPlainButKipdr}{Unknown}{}{Walltime}{Max}{0.417335033417}%
\StoreBenchExecResult{PdrInv}{OracleTrueNotSolvedByKinductionPlainButKipdr}{Unknown}{}{Walltime}{Stdev}{0.02842950297024843683208457592}%
\StoreBenchExecResult{PdrInv}{OracleTrueNotSolvedByKinductionPlainButKipdr}{Unknown}{Unknown}{Count}{}{426}%
\StoreBenchExecResult{PdrInv}{OracleTrueNotSolvedByKinductionPlainButKipdr}{Unknown}{Unknown}{Cputime}{}{44.710066291}%
\StoreBenchExecResult{PdrInv}{OracleTrueNotSolvedByKinductionPlainButKipdr}{Unknown}{Unknown}{Cputime}{Avg}{0.1049532072558685446009389671}%
\StoreBenchExecResult{PdrInv}{OracleTrueNotSolvedByKinductionPlainButKipdr}{Unknown}{Unknown}{Cputime}{Median}{0.1029969065}%
\StoreBenchExecResult{PdrInv}{OracleTrueNotSolvedByKinductionPlainButKipdr}{Unknown}{Unknown}{Cputime}{Min}{0.054004265}%
\StoreBenchExecResult{PdrInv}{OracleTrueNotSolvedByKinductionPlainButKipdr}{Unknown}{Unknown}{Cputime}{Max}{0.261340774}%
\StoreBenchExecResult{PdrInv}{OracleTrueNotSolvedByKinductionPlainButKipdr}{Unknown}{Unknown}{Cputime}{Stdev}{0.03278436357931169744070023063}%
\StoreBenchExecResult{PdrInv}{OracleTrueNotSolvedByKinductionPlainButKipdr}{Unknown}{Unknown}{Walltime}{}{28.7851977348319}%
\StoreBenchExecResult{PdrInv}{OracleTrueNotSolvedByKinductionPlainButKipdr}{Unknown}{Unknown}{Walltime}{Avg}{0.06757088670148333333333333333}%
\StoreBenchExecResult{PdrInv}{OracleTrueNotSolvedByKinductionPlainButKipdr}{Unknown}{Unknown}{Walltime}{Median}{0.06429135799405}%
\StoreBenchExecResult{PdrInv}{OracleTrueNotSolvedByKinductionPlainButKipdr}{Unknown}{Unknown}{Walltime}{Min}{0.0359630584717}%
\StoreBenchExecResult{PdrInv}{OracleTrueNotSolvedByKinductionPlainButKipdr}{Unknown}{Unknown}{Walltime}{Max}{0.417335033417}%
\StoreBenchExecResult{PdrInv}{OracleTrueNotSolvedByKinductionPlainButKipdr}{Unknown}{Unknown}{Walltime}{Stdev}{0.02842950297024843683208457592}%
\providecommand\StoreBenchExecResult[7]{\expandafter\newcommand\csname#1#2#3#4#5#6\endcsname{#7}}%
\StoreBenchExecResult{PdrInv}{OracleTrueNotSolvedByKinductionPlain}{Total}{}{Count}{}{2893}%
\StoreBenchExecResult{PdrInv}{OracleTrueNotSolvedByKinductionPlain}{Total}{}{Cputime}{}{96023.839087920}%
\StoreBenchExecResult{PdrInv}{OracleTrueNotSolvedByKinductionPlain}{Total}{}{Cputime}{Avg}{33.19178675697200138264777048}%
\StoreBenchExecResult{PdrInv}{OracleTrueNotSolvedByKinductionPlain}{Total}{}{Cputime}{Median}{0.214115227}%
\StoreBenchExecResult{PdrInv}{OracleTrueNotSolvedByKinductionPlain}{Total}{}{Cputime}{Min}{0.054004265}%
\StoreBenchExecResult{PdrInv}{OracleTrueNotSolvedByKinductionPlain}{Total}{}{Cputime}{Max}{908.843729296}%
\StoreBenchExecResult{PdrInv}{OracleTrueNotSolvedByKinductionPlain}{Total}{}{Cputime}{Stdev}{158.4531674761677900774409918}%
\StoreBenchExecResult{PdrInv}{OracleTrueNotSolvedByKinductionPlain}{Total}{}{Walltime}{}{96032.5656492804242}%
\StoreBenchExecResult{PdrInv}{OracleTrueNotSolvedByKinductionPlain}{Total}{}{Walltime}{Avg}{33.19480319712423926719668165}%
\StoreBenchExecResult{PdrInv}{OracleTrueNotSolvedByKinductionPlain}{Total}{}{Walltime}{Median}{0.17103600502}%
\StoreBenchExecResult{PdrInv}{OracleTrueNotSolvedByKinductionPlain}{Total}{}{Walltime}{Min}{0.0354268550873}%
\StoreBenchExecResult{PdrInv}{OracleTrueNotSolvedByKinductionPlain}{Total}{}{Walltime}{Max}{1031.00247908}%
\StoreBenchExecResult{PdrInv}{OracleTrueNotSolvedByKinductionPlain}{Total}{}{Walltime}{Stdev}{158.3808491069342319502230264}%
\StoreBenchExecResult{PdrInv}{OracleTrueNotSolvedByKinductionPlain}{Correct}{}{Count}{}{291}%
\StoreBenchExecResult{PdrInv}{OracleTrueNotSolvedByKinductionPlain}{Correct}{}{Cputime}{}{9458.923519358}%
\StoreBenchExecResult{PdrInv}{OracleTrueNotSolvedByKinductionPlain}{Correct}{}{Cputime}{Avg}{32.50489181909965635738831615}%
\StoreBenchExecResult{PdrInv}{OracleTrueNotSolvedByKinductionPlain}{Correct}{}{Cputime}{Median}{0.278327706}%
\StoreBenchExecResult{PdrInv}{OracleTrueNotSolvedByKinductionPlain}{Correct}{}{Cputime}{Min}{0.086225714}%
\StoreBenchExecResult{PdrInv}{OracleTrueNotSolvedByKinductionPlain}{Correct}{}{Cputime}{Max}{829.069057036}%
\StoreBenchExecResult{PdrInv}{OracleTrueNotSolvedByKinductionPlain}{Correct}{}{Cputime}{Stdev}{120.1691540835832324160668386}%
\StoreBenchExecResult{PdrInv}{OracleTrueNotSolvedByKinductionPlain}{Correct}{}{Walltime}{}{8535.0242125992498}%
\StoreBenchExecResult{PdrInv}{OracleTrueNotSolvedByKinductionPlain}{Correct}{}{Walltime}{Avg}{29.32998011202491340206185567}%
\StoreBenchExecResult{PdrInv}{OracleTrueNotSolvedByKinductionPlain}{Correct}{}{Walltime}{Median}{0.201776981354}%
\StoreBenchExecResult{PdrInv}{OracleTrueNotSolvedByKinductionPlain}{Correct}{}{Walltime}{Min}{0.0573229789734}%
\StoreBenchExecResult{PdrInv}{OracleTrueNotSolvedByKinductionPlain}{Correct}{}{Walltime}{Max}{826.533210993}%
\StoreBenchExecResult{PdrInv}{OracleTrueNotSolvedByKinductionPlain}{Correct}{}{Walltime}{Stdev}{113.0481302086461304684397476}%
\StoreBenchExecResult{PdrInv}{OracleTrueNotSolvedByKinductionPlain}{Correct}{True}{Count}{}{291}%
\StoreBenchExecResult{PdrInv}{OracleTrueNotSolvedByKinductionPlain}{Correct}{True}{Cputime}{}{9458.923519358}%
\StoreBenchExecResult{PdrInv}{OracleTrueNotSolvedByKinductionPlain}{Correct}{True}{Cputime}{Avg}{32.50489181909965635738831615}%
\StoreBenchExecResult{PdrInv}{OracleTrueNotSolvedByKinductionPlain}{Correct}{True}{Cputime}{Median}{0.278327706}%
\StoreBenchExecResult{PdrInv}{OracleTrueNotSolvedByKinductionPlain}{Correct}{True}{Cputime}{Min}{0.086225714}%
\StoreBenchExecResult{PdrInv}{OracleTrueNotSolvedByKinductionPlain}{Correct}{True}{Cputime}{Max}{829.069057036}%
\StoreBenchExecResult{PdrInv}{OracleTrueNotSolvedByKinductionPlain}{Correct}{True}{Cputime}{Stdev}{120.1691540835832324160668386}%
\StoreBenchExecResult{PdrInv}{OracleTrueNotSolvedByKinductionPlain}{Correct}{True}{Walltime}{}{8535.0242125992498}%
\StoreBenchExecResult{PdrInv}{OracleTrueNotSolvedByKinductionPlain}{Correct}{True}{Walltime}{Avg}{29.32998011202491340206185567}%
\StoreBenchExecResult{PdrInv}{OracleTrueNotSolvedByKinductionPlain}{Correct}{True}{Walltime}{Median}{0.201776981354}%
\StoreBenchExecResult{PdrInv}{OracleTrueNotSolvedByKinductionPlain}{Correct}{True}{Walltime}{Min}{0.0573229789734}%
\StoreBenchExecResult{PdrInv}{OracleTrueNotSolvedByKinductionPlain}{Correct}{True}{Walltime}{Max}{826.533210993}%
\StoreBenchExecResult{PdrInv}{OracleTrueNotSolvedByKinductionPlain}{Correct}{True}{Walltime}{Stdev}{113.0481302086461304684397476}%
\StoreBenchExecResult{PdrInv}{OracleTrueNotSolvedByKinductionPlain}{Wrong}{True}{Count}{}{0}%
\StoreBenchExecResult{PdrInv}{OracleTrueNotSolvedByKinductionPlain}{Wrong}{True}{Cputime}{}{0}%
\StoreBenchExecResult{PdrInv}{OracleTrueNotSolvedByKinductionPlain}{Wrong}{True}{Cputime}{Avg}{None}%
\StoreBenchExecResult{PdrInv}{OracleTrueNotSolvedByKinductionPlain}{Wrong}{True}{Cputime}{Median}{None}%
\StoreBenchExecResult{PdrInv}{OracleTrueNotSolvedByKinductionPlain}{Wrong}{True}{Cputime}{Min}{None}%
\StoreBenchExecResult{PdrInv}{OracleTrueNotSolvedByKinductionPlain}{Wrong}{True}{Cputime}{Max}{None}%
\StoreBenchExecResult{PdrInv}{OracleTrueNotSolvedByKinductionPlain}{Wrong}{True}{Cputime}{Stdev}{None}%
\StoreBenchExecResult{PdrInv}{OracleTrueNotSolvedByKinductionPlain}{Wrong}{True}{Walltime}{}{0}%
\StoreBenchExecResult{PdrInv}{OracleTrueNotSolvedByKinductionPlain}{Wrong}{True}{Walltime}{Avg}{None}%
\StoreBenchExecResult{PdrInv}{OracleTrueNotSolvedByKinductionPlain}{Wrong}{True}{Walltime}{Median}{None}%
\StoreBenchExecResult{PdrInv}{OracleTrueNotSolvedByKinductionPlain}{Wrong}{True}{Walltime}{Min}{None}%
\StoreBenchExecResult{PdrInv}{OracleTrueNotSolvedByKinductionPlain}{Wrong}{True}{Walltime}{Max}{None}%
\StoreBenchExecResult{PdrInv}{OracleTrueNotSolvedByKinductionPlain}{Wrong}{True}{Walltime}{Stdev}{None}%
\StoreBenchExecResult{PdrInv}{OracleTrueNotSolvedByKinductionPlain}{Error}{}{Count}{}{102}%
\StoreBenchExecResult{PdrInv}{OracleTrueNotSolvedByKinductionPlain}{Error}{}{Cputime}{}{81317.895876982}%
\StoreBenchExecResult{PdrInv}{OracleTrueNotSolvedByKinductionPlain}{Error}{}{Cputime}{Avg}{797.2342733037450980392156863}%
\StoreBenchExecResult{PdrInv}{OracleTrueNotSolvedByKinductionPlain}{Error}{}{Cputime}{Median}{901.9342677285}%
\StoreBenchExecResult{PdrInv}{OracleTrueNotSolvedByKinductionPlain}{Error}{}{Cputime}{Min}{3.343430506}%
\StoreBenchExecResult{PdrInv}{OracleTrueNotSolvedByKinductionPlain}{Error}{}{Cputime}{Max}{908.843729296}%
\StoreBenchExecResult{PdrInv}{OracleTrueNotSolvedByKinductionPlain}{Error}{}{Cputime}{Stdev}{250.2466211648216211601276365}%
\StoreBenchExecResult{PdrInv}{OracleTrueNotSolvedByKinductionPlain}{Error}{}{Walltime}{}{82430.84674216167}%
\StoreBenchExecResult{PdrInv}{OracleTrueNotSolvedByKinductionPlain}{Error}{}{Walltime}{Avg}{808.1455562957026470588235294}%
\StoreBenchExecResult{PdrInv}{OracleTrueNotSolvedByKinductionPlain}{Error}{}{Walltime}{Median}{890.090693831}%
\StoreBenchExecResult{PdrInv}{OracleTrueNotSolvedByKinductionPlain}{Error}{}{Walltime}{Min}{2.17186689377}%
\StoreBenchExecResult{PdrInv}{OracleTrueNotSolvedByKinductionPlain}{Error}{}{Walltime}{Max}{1031.00247908}%
\StoreBenchExecResult{PdrInv}{OracleTrueNotSolvedByKinductionPlain}{Error}{}{Walltime}{Stdev}{223.1494151962686410848195349}%
\StoreBenchExecResult{PdrInv}{OracleTrueNotSolvedByKinductionPlain}{Error}{Error}{Count}{}{1}%
\StoreBenchExecResult{PdrInv}{OracleTrueNotSolvedByKinductionPlain}{Error}{Error}{Cputime}{}{4.028347458}%
\StoreBenchExecResult{PdrInv}{OracleTrueNotSolvedByKinductionPlain}{Error}{Error}{Cputime}{Avg}{4.028347458}%
\StoreBenchExecResult{PdrInv}{OracleTrueNotSolvedByKinductionPlain}{Error}{Error}{Cputime}{Median}{4.028347458}%
\StoreBenchExecResult{PdrInv}{OracleTrueNotSolvedByKinductionPlain}{Error}{Error}{Cputime}{Min}{4.028347458}%
\StoreBenchExecResult{PdrInv}{OracleTrueNotSolvedByKinductionPlain}{Error}{Error}{Cputime}{Max}{4.028347458}%
\StoreBenchExecResult{PdrInv}{OracleTrueNotSolvedByKinductionPlain}{Error}{Error}{Cputime}{Stdev}{0E-9}%
\StoreBenchExecResult{PdrInv}{OracleTrueNotSolvedByKinductionPlain}{Error}{Error}{Walltime}{}{2.17186689377}%
\StoreBenchExecResult{PdrInv}{OracleTrueNotSolvedByKinductionPlain}{Error}{Error}{Walltime}{Avg}{2.17186689377}%
\StoreBenchExecResult{PdrInv}{OracleTrueNotSolvedByKinductionPlain}{Error}{Error}{Walltime}{Median}{2.17186689377}%
\StoreBenchExecResult{PdrInv}{OracleTrueNotSolvedByKinductionPlain}{Error}{Error}{Walltime}{Min}{2.17186689377}%
\StoreBenchExecResult{PdrInv}{OracleTrueNotSolvedByKinductionPlain}{Error}{Error}{Walltime}{Max}{2.17186689377}%
\StoreBenchExecResult{PdrInv}{OracleTrueNotSolvedByKinductionPlain}{Error}{Error}{Walltime}{Stdev}{0E-11}%
\StoreBenchExecResult{PdrInv}{OracleTrueNotSolvedByKinductionPlain}{Error}{OutOfMemory}{Count}{}{14}%
\StoreBenchExecResult{PdrInv}{OracleTrueNotSolvedByKinductionPlain}{Error}{OutOfMemory}{Cputime}{}{4593.057838206}%
\StoreBenchExecResult{PdrInv}{OracleTrueNotSolvedByKinductionPlain}{Error}{OutOfMemory}{Cputime}{Avg}{328.0755598718571428571428571}%
\StoreBenchExecResult{PdrInv}{OracleTrueNotSolvedByKinductionPlain}{Error}{OutOfMemory}{Cputime}{Median}{365.072017050}%
\StoreBenchExecResult{PdrInv}{OracleTrueNotSolvedByKinductionPlain}{Error}{OutOfMemory}{Cputime}{Min}{64.436629677}%
\StoreBenchExecResult{PdrInv}{OracleTrueNotSolvedByKinductionPlain}{Error}{OutOfMemory}{Cputime}{Max}{604.451123066}%
\StoreBenchExecResult{PdrInv}{OracleTrueNotSolvedByKinductionPlain}{Error}{OutOfMemory}{Cputime}{Stdev}{184.2560748558316428022860695}%
\StoreBenchExecResult{PdrInv}{OracleTrueNotSolvedByKinductionPlain}{Error}{OutOfMemory}{Walltime}{}{4556.9123659149}%
\StoreBenchExecResult{PdrInv}{OracleTrueNotSolvedByKinductionPlain}{Error}{OutOfMemory}{Walltime}{Avg}{325.4937404224928571428571429}%
\StoreBenchExecResult{PdrInv}{OracleTrueNotSolvedByKinductionPlain}{Error}{OutOfMemory}{Walltime}{Median}{362.180272102}%
\StoreBenchExecResult{PdrInv}{OracleTrueNotSolvedByKinductionPlain}{Error}{OutOfMemory}{Walltime}{Min}{63.6780879498}%
\StoreBenchExecResult{PdrInv}{OracleTrueNotSolvedByKinductionPlain}{Error}{OutOfMemory}{Walltime}{Max}{601.009039164}%
\StoreBenchExecResult{PdrInv}{OracleTrueNotSolvedByKinductionPlain}{Error}{OutOfMemory}{Walltime}{Stdev}{183.5827960599662929188988893}%
\StoreBenchExecResult{PdrInv}{OracleTrueNotSolvedByKinductionPlain}{Error}{Timeout}{Count}{}{87}%
\StoreBenchExecResult{PdrInv}{OracleTrueNotSolvedByKinductionPlain}{Error}{Timeout}{Cputime}{}{76720.809691318}%
\StoreBenchExecResult{PdrInv}{OracleTrueNotSolvedByKinductionPlain}{Error}{Timeout}{Cputime}{Avg}{881.8483872565287356321839080}%
\StoreBenchExecResult{PdrInv}{OracleTrueNotSolvedByKinductionPlain}{Error}{Timeout}{Cputime}{Median}{902.071967623}%
\StoreBenchExecResult{PdrInv}{OracleTrueNotSolvedByKinductionPlain}{Error}{Timeout}{Cputime}{Min}{3.343430506}%
\StoreBenchExecResult{PdrInv}{OracleTrueNotSolvedByKinductionPlain}{Error}{Timeout}{Cputime}{Max}{908.843729296}%
\StoreBenchExecResult{PdrInv}{OracleTrueNotSolvedByKinductionPlain}{Error}{Timeout}{Cputime}{Stdev}{134.7063536438931200875814182}%
\StoreBenchExecResult{PdrInv}{OracleTrueNotSolvedByKinductionPlain}{Error}{Timeout}{Walltime}{}{77871.762509353}%
\StoreBenchExecResult{PdrInv}{OracleTrueNotSolvedByKinductionPlain}{Error}{Timeout}{Walltime}{Avg}{895.0777299925632183908045977}%
\StoreBenchExecResult{PdrInv}{OracleTrueNotSolvedByKinductionPlain}{Error}{Timeout}{Walltime}{Median}{890.552845001}%
\StoreBenchExecResult{PdrInv}{OracleTrueNotSolvedByKinductionPlain}{Error}{Timeout}{Walltime}{Min}{883.549846888}%
\StoreBenchExecResult{PdrInv}{OracleTrueNotSolvedByKinductionPlain}{Error}{Timeout}{Walltime}{Max}{1031.00247908}%
\StoreBenchExecResult{PdrInv}{OracleTrueNotSolvedByKinductionPlain}{Error}{Timeout}{Walltime}{Stdev}{21.14845782079812208710744675}%
\StoreBenchExecResult{PdrInv}{OracleTrueNotSolvedByKinductionPlain}{Unknown}{}{Count}{}{2500}%
\StoreBenchExecResult{PdrInv}{OracleTrueNotSolvedByKinductionPlain}{Unknown}{}{Cputime}{}{5247.019691580}%
\StoreBenchExecResult{PdrInv}{OracleTrueNotSolvedByKinductionPlain}{Unknown}{}{Cputime}{Avg}{2.098807876632}%
\StoreBenchExecResult{PdrInv}{OracleTrueNotSolvedByKinductionPlain}{Unknown}{}{Cputime}{Median}{0.2028879305}%
\StoreBenchExecResult{PdrInv}{OracleTrueNotSolvedByKinductionPlain}{Unknown}{}{Cputime}{Min}{0.054004265}%
\StoreBenchExecResult{PdrInv}{OracleTrueNotSolvedByKinductionPlain}{Unknown}{}{Cputime}{Max}{111.544145802}%
\StoreBenchExecResult{PdrInv}{OracleTrueNotSolvedByKinductionPlain}{Unknown}{}{Cputime}{Stdev}{5.839035504688835100781781597}%
\StoreBenchExecResult{PdrInv}{OracleTrueNotSolvedByKinductionPlain}{Unknown}{}{Walltime}{}{5066.6946945195044}%
\StoreBenchExecResult{PdrInv}{OracleTrueNotSolvedByKinductionPlain}{Unknown}{}{Walltime}{Avg}{2.02667787780780176}%
\StoreBenchExecResult{PdrInv}{OracleTrueNotSolvedByKinductionPlain}{Unknown}{}{Walltime}{Median}{0.162702560425}%
\StoreBenchExecResult{PdrInv}{OracleTrueNotSolvedByKinductionPlain}{Unknown}{}{Walltime}{Min}{0.0354268550873}%
\StoreBenchExecResult{PdrInv}{OracleTrueNotSolvedByKinductionPlain}{Unknown}{}{Walltime}{Max}{109.348944902}%
\StoreBenchExecResult{PdrInv}{OracleTrueNotSolvedByKinductionPlain}{Unknown}{}{Walltime}{Stdev}{5.740933559064151247053458205}%
\StoreBenchExecResult{PdrInv}{OracleTrueNotSolvedByKinductionPlain}{Unknown}{Unknown}{Count}{}{2500}%
\StoreBenchExecResult{PdrInv}{OracleTrueNotSolvedByKinductionPlain}{Unknown}{Unknown}{Cputime}{}{5247.019691580}%
\StoreBenchExecResult{PdrInv}{OracleTrueNotSolvedByKinductionPlain}{Unknown}{Unknown}{Cputime}{Avg}{2.098807876632}%
\StoreBenchExecResult{PdrInv}{OracleTrueNotSolvedByKinductionPlain}{Unknown}{Unknown}{Cputime}{Median}{0.2028879305}%
\StoreBenchExecResult{PdrInv}{OracleTrueNotSolvedByKinductionPlain}{Unknown}{Unknown}{Cputime}{Min}{0.054004265}%
\StoreBenchExecResult{PdrInv}{OracleTrueNotSolvedByKinductionPlain}{Unknown}{Unknown}{Cputime}{Max}{111.544145802}%
\StoreBenchExecResult{PdrInv}{OracleTrueNotSolvedByKinductionPlain}{Unknown}{Unknown}{Cputime}{Stdev}{5.839035504688835100781781597}%
\StoreBenchExecResult{PdrInv}{OracleTrueNotSolvedByKinductionPlain}{Unknown}{Unknown}{Walltime}{}{5066.6946945195044}%
\StoreBenchExecResult{PdrInv}{OracleTrueNotSolvedByKinductionPlain}{Unknown}{Unknown}{Walltime}{Avg}{2.02667787780780176}%
\StoreBenchExecResult{PdrInv}{OracleTrueNotSolvedByKinductionPlain}{Unknown}{Unknown}{Walltime}{Median}{0.162702560425}%
\StoreBenchExecResult{PdrInv}{OracleTrueNotSolvedByKinductionPlain}{Unknown}{Unknown}{Walltime}{Min}{0.0354268550873}%
\StoreBenchExecResult{PdrInv}{OracleTrueNotSolvedByKinductionPlain}{Unknown}{Unknown}{Walltime}{Max}{109.348944902}%
\StoreBenchExecResult{PdrInv}{OracleTrueNotSolvedByKinductionPlain}{Unknown}{Unknown}{Walltime}{Stdev}{5.740933559064151247053458205}%
\providecommand\StoreBenchExecResult[7]{\expandafter\newcommand\csname#1#2#3#4#5#6\endcsname{#7}}%
\StoreBenchExecResult{PdrInv}{Oracle}{Total}{}{Count}{}{5591}%
\StoreBenchExecResult{PdrInv}{Oracle}{Total}{}{Cputime}{}{161478.730534274}%
\StoreBenchExecResult{PdrInv}{Oracle}{Total}{}{Cputime}{Avg}{28.88190494263530674297978895}%
\StoreBenchExecResult{PdrInv}{Oracle}{Total}{}{Cputime}{Median}{0.3193526}%
\StoreBenchExecResult{PdrInv}{Oracle}{Total}{}{Cputime}{Min}{0.054004265}%
\StoreBenchExecResult{PdrInv}{Oracle}{Total}{}{Cputime}{Max}{908.843729296}%
\StoreBenchExecResult{PdrInv}{Oracle}{Total}{}{Cputime}{Stdev}{137.7103490122047685692816118}%
\StoreBenchExecResult{PdrInv}{Oracle}{Total}{}{Walltime}{}{157614.4810469216820}%
\StoreBenchExecResult{PdrInv}{Oracle}{Total}{}{Walltime}{Avg}{28.19074960595987873367912717}%
\StoreBenchExecResult{PdrInv}{Oracle}{Total}{}{Walltime}{Median}{0.276914834976}%
\StoreBenchExecResult{PdrInv}{Oracle}{Total}{}{Walltime}{Min}{0.0354268550873}%
\StoreBenchExecResult{PdrInv}{Oracle}{Total}{}{Walltime}{Max}{1031.00247908}%
\StoreBenchExecResult{PdrInv}{Oracle}{Total}{}{Walltime}{Stdev}{136.9662421024864791586512062}%
\StoreBenchExecResult{PdrInv}{Oracle}{Correct}{}{Count}{}{1474}%
\StoreBenchExecResult{PdrInv}{Oracle}{Correct}{}{Cputime}{}{36381.504309437}%
\StoreBenchExecResult{PdrInv}{Oracle}{Correct}{}{Cputime}{Avg}{24.68216031847829036635006784}%
\StoreBenchExecResult{PdrInv}{Oracle}{Correct}{}{Cputime}{Median}{3.7834749815}%
\StoreBenchExecResult{PdrInv}{Oracle}{Correct}{}{Cputime}{Min}{0.076494197}%
\StoreBenchExecResult{PdrInv}{Oracle}{Correct}{}{Cputime}{Max}{860.835249649}%
\StoreBenchExecResult{PdrInv}{Oracle}{Correct}{}{Cputime}{Stdev}{94.00181235346768330538104912}%
\StoreBenchExecResult{PdrInv}{Oracle}{Correct}{}{Walltime}{}{32253.1044096934359}%
\StoreBenchExecResult{PdrInv}{Oracle}{Correct}{}{Walltime}{Avg}{21.88134627523299586160108548}%
\StoreBenchExecResult{PdrInv}{Oracle}{Correct}{}{Walltime}{Median}{2.21906208992}%
\StoreBenchExecResult{PdrInv}{Oracle}{Correct}{}{Walltime}{Min}{0.0550191402435}%
\StoreBenchExecResult{PdrInv}{Oracle}{Correct}{}{Walltime}{Max}{826.533210993}%
\StoreBenchExecResult{PdrInv}{Oracle}{Correct}{}{Walltime}{Stdev}{90.38636635348692370437587516}%
\StoreBenchExecResult{PdrInv}{Oracle}{Correct}{False}{Count}{}{350}%
\StoreBenchExecResult{PdrInv}{Oracle}{Correct}{False}{Cputime}{}{10382.620283796}%
\StoreBenchExecResult{PdrInv}{Oracle}{Correct}{False}{Cputime}{Avg}{29.66462938227428571428571429}%
\StoreBenchExecResult{PdrInv}{Oracle}{Correct}{False}{Cputime}{Median}{3.598937949}%
\StoreBenchExecResult{PdrInv}{Oracle}{Correct}{False}{Cputime}{Min}{0.076494197}%
\StoreBenchExecResult{PdrInv}{Oracle}{Correct}{False}{Cputime}{Max}{860.835249649}%
\StoreBenchExecResult{PdrInv}{Oracle}{Correct}{False}{Cputime}{Stdev}{90.77879204676246017568151475}%
\StoreBenchExecResult{PdrInv}{Oracle}{Correct}{False}{Walltime}{}{8901.5746548180578}%
\StoreBenchExecResult{PdrInv}{Oracle}{Correct}{False}{Walltime}{Avg}{25.433070442337308}%
\StoreBenchExecResult{PdrInv}{Oracle}{Correct}{False}{Walltime}{Median}{2.28490996361}%
\StoreBenchExecResult{PdrInv}{Oracle}{Correct}{False}{Walltime}{Min}{0.0559029579163}%
\StoreBenchExecResult{PdrInv}{Oracle}{Correct}{False}{Walltime}{Max}{809.225548029}%
\StoreBenchExecResult{PdrInv}{Oracle}{Correct}{False}{Walltime}{Stdev}{85.53833285881724108847713979}%
\StoreBenchExecResult{PdrInv}{Oracle}{Wrong}{False}{Count}{}{0}%
\StoreBenchExecResult{PdrInv}{Oracle}{Wrong}{False}{Cputime}{}{0}%
\StoreBenchExecResult{PdrInv}{Oracle}{Wrong}{False}{Cputime}{Avg}{None}%
\StoreBenchExecResult{PdrInv}{Oracle}{Wrong}{False}{Cputime}{Median}{None}%
\StoreBenchExecResult{PdrInv}{Oracle}{Wrong}{False}{Cputime}{Min}{None}%
\StoreBenchExecResult{PdrInv}{Oracle}{Wrong}{False}{Cputime}{Max}{None}%
\StoreBenchExecResult{PdrInv}{Oracle}{Wrong}{False}{Cputime}{Stdev}{None}%
\StoreBenchExecResult{PdrInv}{Oracle}{Wrong}{False}{Walltime}{}{0}%
\StoreBenchExecResult{PdrInv}{Oracle}{Wrong}{False}{Walltime}{Avg}{None}%
\StoreBenchExecResult{PdrInv}{Oracle}{Wrong}{False}{Walltime}{Median}{None}%
\StoreBenchExecResult{PdrInv}{Oracle}{Wrong}{False}{Walltime}{Min}{None}%
\StoreBenchExecResult{PdrInv}{Oracle}{Wrong}{False}{Walltime}{Max}{None}%
\StoreBenchExecResult{PdrInv}{Oracle}{Wrong}{False}{Walltime}{Stdev}{None}%
\StoreBenchExecResult{PdrInv}{Oracle}{Correct}{True}{Count}{}{1124}%
\StoreBenchExecResult{PdrInv}{Oracle}{Correct}{True}{Cputime}{}{25998.884025641}%
\StoreBenchExecResult{PdrInv}{Oracle}{Correct}{True}{Cputime}{Avg}{23.13067973811476868327402135}%
\StoreBenchExecResult{PdrInv}{Oracle}{Correct}{True}{Cputime}{Median}{3.7976293375}%
\StoreBenchExecResult{PdrInv}{Oracle}{Correct}{True}{Cputime}{Min}{0.07783832}%
\StoreBenchExecResult{PdrInv}{Oracle}{Correct}{True}{Cputime}{Max}{829.069057036}%
\StoreBenchExecResult{PdrInv}{Oracle}{Correct}{True}{Cputime}{Stdev}{94.92971613023428714551390415}%
\StoreBenchExecResult{PdrInv}{Oracle}{Correct}{True}{Walltime}{}{23351.5297548753781}%
\StoreBenchExecResult{PdrInv}{Oracle}{Correct}{True}{Walltime}{Avg}{20.77538234419517624555160142}%
\StoreBenchExecResult{PdrInv}{Oracle}{Correct}{True}{Walltime}{Median}{2.213792562485}%
\StoreBenchExecResult{PdrInv}{Oracle}{Correct}{True}{Walltime}{Min}{0.0550191402435}%
\StoreBenchExecResult{PdrInv}{Oracle}{Correct}{True}{Walltime}{Max}{826.533210993}%
\StoreBenchExecResult{PdrInv}{Oracle}{Correct}{True}{Walltime}{Stdev}{91.81570263797118170593441140}%
\StoreBenchExecResult{PdrInv}{Oracle}{Wrong}{True}{Count}{}{0}%
\StoreBenchExecResult{PdrInv}{Oracle}{Wrong}{True}{Cputime}{}{0}%
\StoreBenchExecResult{PdrInv}{Oracle}{Wrong}{True}{Cputime}{Avg}{None}%
\StoreBenchExecResult{PdrInv}{Oracle}{Wrong}{True}{Cputime}{Median}{None}%
\StoreBenchExecResult{PdrInv}{Oracle}{Wrong}{True}{Cputime}{Min}{None}%
\StoreBenchExecResult{PdrInv}{Oracle}{Wrong}{True}{Cputime}{Max}{None}%
\StoreBenchExecResult{PdrInv}{Oracle}{Wrong}{True}{Cputime}{Stdev}{None}%
\StoreBenchExecResult{PdrInv}{Oracle}{Wrong}{True}{Walltime}{}{0}%
\StoreBenchExecResult{PdrInv}{Oracle}{Wrong}{True}{Walltime}{Avg}{None}%
\StoreBenchExecResult{PdrInv}{Oracle}{Wrong}{True}{Walltime}{Median}{None}%
\StoreBenchExecResult{PdrInv}{Oracle}{Wrong}{True}{Walltime}{Min}{None}%
\StoreBenchExecResult{PdrInv}{Oracle}{Wrong}{True}{Walltime}{Max}{None}%
\StoreBenchExecResult{PdrInv}{Oracle}{Wrong}{True}{Walltime}{Stdev}{None}%
\StoreBenchExecResult{PdrInv}{Oracle}{Error}{}{Count}{}{145}%
\StoreBenchExecResult{PdrInv}{Oracle}{Error}{}{Cputime}{}{111033.739254929}%
\StoreBenchExecResult{PdrInv}{Oracle}{Error}{}{Cputime}{Avg}{765.7499258960620689655172414}%
\StoreBenchExecResult{PdrInv}{Oracle}{Error}{}{Cputime}{Median}{901.865513129}%
\StoreBenchExecResult{PdrInv}{Oracle}{Error}{}{Cputime}{Min}{3.343430506}%
\StoreBenchExecResult{PdrInv}{Oracle}{Error}{}{Cputime}{Max}{908.843729296}%
\StoreBenchExecResult{PdrInv}{Oracle}{Error}{}{Cputime}{Stdev}{281.3221924295873361112329777}%
\StoreBenchExecResult{PdrInv}{Oracle}{Error}{}{Walltime}{}{111713.64219761657}%
\StoreBenchExecResult{PdrInv}{Oracle}{Error}{}{Walltime}{Avg}{770.4389117077004827586206897}%
\StoreBenchExecResult{PdrInv}{Oracle}{Error}{}{Walltime}{Median}{889.905979156}%
\StoreBenchExecResult{PdrInv}{Oracle}{Error}{}{Walltime}{Min}{2.17186689377}%
\StoreBenchExecResult{PdrInv}{Oracle}{Error}{}{Walltime}{Max}{1031.00247908}%
\StoreBenchExecResult{PdrInv}{Oracle}{Error}{}{Walltime}{Stdev}{265.9325379047241477013285938}%
\StoreBenchExecResult{PdrInv}{Oracle}{Error}{Error}{Count}{}{1}%
\StoreBenchExecResult{PdrInv}{Oracle}{Error}{Error}{Cputime}{}{4.028347458}%
\StoreBenchExecResult{PdrInv}{Oracle}{Error}{Error}{Cputime}{Avg}{4.028347458}%
\StoreBenchExecResult{PdrInv}{Oracle}{Error}{Error}{Cputime}{Median}{4.028347458}%
\StoreBenchExecResult{PdrInv}{Oracle}{Error}{Error}{Cputime}{Min}{4.028347458}%
\StoreBenchExecResult{PdrInv}{Oracle}{Error}{Error}{Cputime}{Max}{4.028347458}%
\StoreBenchExecResult{PdrInv}{Oracle}{Error}{Error}{Cputime}{Stdev}{0E-9}%
\StoreBenchExecResult{PdrInv}{Oracle}{Error}{Error}{Walltime}{}{2.17186689377}%
\StoreBenchExecResult{PdrInv}{Oracle}{Error}{Error}{Walltime}{Avg}{2.17186689377}%
\StoreBenchExecResult{PdrInv}{Oracle}{Error}{Error}{Walltime}{Median}{2.17186689377}%
\StoreBenchExecResult{PdrInv}{Oracle}{Error}{Error}{Walltime}{Min}{2.17186689377}%
\StoreBenchExecResult{PdrInv}{Oracle}{Error}{Error}{Walltime}{Max}{2.17186689377}%
\StoreBenchExecResult{PdrInv}{Oracle}{Error}{Error}{Walltime}{Stdev}{0E-11}%
\StoreBenchExecResult{PdrInv}{Oracle}{Error}{OutOfMemory}{Count}{}{27}%
\StoreBenchExecResult{PdrInv}{Oracle}{Error}{OutOfMemory}{Cputime}{}{7241.305324225}%
\StoreBenchExecResult{PdrInv}{Oracle}{Error}{OutOfMemory}{Cputime}{Avg}{268.1964934898148148148148148}%
\StoreBenchExecResult{PdrInv}{Oracle}{Error}{OutOfMemory}{Cputime}{Median}{243.462884178}%
\StoreBenchExecResult{PdrInv}{Oracle}{Error}{OutOfMemory}{Cputime}{Min}{55.067234255}%
\StoreBenchExecResult{PdrInv}{Oracle}{Error}{OutOfMemory}{Cputime}{Max}{604.945274521}%
\StoreBenchExecResult{PdrInv}{Oracle}{Error}{OutOfMemory}{Cputime}{Stdev}{182.6684203788573515705938021}%
\StoreBenchExecResult{PdrInv}{Oracle}{Error}{OutOfMemory}{Walltime}{}{7113.9039719128}%
\StoreBenchExecResult{PdrInv}{Oracle}{Error}{OutOfMemory}{Walltime}{Avg}{263.4779248856592592592592593}%
\StoreBenchExecResult{PdrInv}{Oracle}{Error}{OutOfMemory}{Walltime}{Median}{218.099846125}%
\StoreBenchExecResult{PdrInv}{Oracle}{Error}{OutOfMemory}{Walltime}{Min}{54.4378728867}%
\StoreBenchExecResult{PdrInv}{Oracle}{Error}{OutOfMemory}{Walltime}{Max}{604.252864122}%
\StoreBenchExecResult{PdrInv}{Oracle}{Error}{OutOfMemory}{Walltime}{Stdev}{182.4997506332440137471009970}%
\StoreBenchExecResult{PdrInv}{Oracle}{Error}{Timeout}{Count}{}{117}%
\StoreBenchExecResult{PdrInv}{Oracle}{Error}{Timeout}{Cputime}{}{103788.405583246}%
\StoreBenchExecResult{PdrInv}{Oracle}{Error}{Timeout}{Cputime}{Avg}{887.0803896003931623931623932}%
\StoreBenchExecResult{PdrInv}{Oracle}{Error}{Timeout}{Cputime}{Median}{902.024250211}%
\StoreBenchExecResult{PdrInv}{Oracle}{Error}{Timeout}{Cputime}{Min}{3.343430506}%
\StoreBenchExecResult{PdrInv}{Oracle}{Error}{Timeout}{Cputime}{Max}{908.843729296}%
\StoreBenchExecResult{PdrInv}{Oracle}{Error}{Timeout}{Cputime}{Stdev}{116.5015548242598120553524644}%
\StoreBenchExecResult{PdrInv}{Oracle}{Error}{Timeout}{Walltime}{}{104597.566358810}%
\StoreBenchExecResult{PdrInv}{Oracle}{Error}{Timeout}{Walltime}{Avg}{893.9962936650427350427350427}%
\StoreBenchExecResult{PdrInv}{Oracle}{Error}{Timeout}{Walltime}{Median}{890.230445862}%
\StoreBenchExecResult{PdrInv}{Oracle}{Error}{Timeout}{Walltime}{Min}{883.549846888}%
\StoreBenchExecResult{PdrInv}{Oracle}{Error}{Timeout}{Walltime}{Max}{1031.00247908}%
\StoreBenchExecResult{PdrInv}{Oracle}{Error}{Timeout}{Walltime}{Stdev}{18.37157654765284563761739017}%
\StoreBenchExecResult{PdrInv}{Oracle}{Unknown}{}{Count}{}{3972}%
\StoreBenchExecResult{PdrInv}{Oracle}{Unknown}{}{Cputime}{}{14063.486969908}%
\StoreBenchExecResult{PdrInv}{Oracle}{Unknown}{}{Cputime}{Avg}{3.540656336834843907351460222}%
\StoreBenchExecResult{PdrInv}{Oracle}{Unknown}{}{Cputime}{Median}{0.2337717815}%
\StoreBenchExecResult{PdrInv}{Oracle}{Unknown}{}{Cputime}{Min}{0.054004265}%
\StoreBenchExecResult{PdrInv}{Oracle}{Unknown}{}{Cputime}{Max}{111.544145802}%
\StoreBenchExecResult{PdrInv}{Oracle}{Unknown}{}{Cputime}{Stdev}{7.443880307628795931912435538}%
\StoreBenchExecResult{PdrInv}{Oracle}{Unknown}{}{Walltime}{}{13647.7344396116761}%
\StoreBenchExecResult{PdrInv}{Oracle}{Unknown}{}{Walltime}{Avg}{3.435985508462154103726082578}%
\StoreBenchExecResult{PdrInv}{Oracle}{Unknown}{}{Walltime}{Median}{0.1926209926605}%
\StoreBenchExecResult{PdrInv}{Oracle}{Unknown}{}{Walltime}{Min}{0.0354268550873}%
\StoreBenchExecResult{PdrInv}{Oracle}{Unknown}{}{Walltime}{Max}{109.348944902}%
\StoreBenchExecResult{PdrInv}{Oracle}{Unknown}{}{Walltime}{Stdev}{7.294029697373493578397228124}%
\StoreBenchExecResult{PdrInv}{Oracle}{Unknown}{Unknown}{Count}{}{3972}%
\StoreBenchExecResult{PdrInv}{Oracle}{Unknown}{Unknown}{Cputime}{}{14063.486969908}%
\StoreBenchExecResult{PdrInv}{Oracle}{Unknown}{Unknown}{Cputime}{Avg}{3.540656336834843907351460222}%
\StoreBenchExecResult{PdrInv}{Oracle}{Unknown}{Unknown}{Cputime}{Median}{0.2337717815}%
\StoreBenchExecResult{PdrInv}{Oracle}{Unknown}{Unknown}{Cputime}{Min}{0.054004265}%
\StoreBenchExecResult{PdrInv}{Oracle}{Unknown}{Unknown}{Cputime}{Max}{111.544145802}%
\StoreBenchExecResult{PdrInv}{Oracle}{Unknown}{Unknown}{Cputime}{Stdev}{7.443880307628795931912435538}%
\StoreBenchExecResult{PdrInv}{Oracle}{Unknown}{Unknown}{Walltime}{}{13647.7344396116761}%
\StoreBenchExecResult{PdrInv}{Oracle}{Unknown}{Unknown}{Walltime}{Avg}{3.435985508462154103726082578}%
\StoreBenchExecResult{PdrInv}{Oracle}{Unknown}{Unknown}{Walltime}{Median}{0.1926209926605}%
\StoreBenchExecResult{PdrInv}{Oracle}{Unknown}{Unknown}{Walltime}{Min}{0.0354268550873}%
\StoreBenchExecResult{PdrInv}{Oracle}{Unknown}{Unknown}{Walltime}{Max}{109.348944902}%
\StoreBenchExecResult{PdrInv}{Oracle}{Unknown}{Unknown}{Walltime}{Stdev}{7.294029697373493578397228124}%
\providecommand\StoreBenchExecResult[7]{\expandafter\newcommand\csname#1#2#3#4#5#6\endcsname{#7}}%
\StoreBenchExecResult{PdrInv}{PdrTrueNotSolvedByKinductionPlainButKipdr}{Total}{}{Count}{}{449}%
\StoreBenchExecResult{PdrInv}{PdrTrueNotSolvedByKinductionPlainButKipdr}{Total}{}{Cputime}{}{389293.710233819}%
\StoreBenchExecResult{PdrInv}{PdrTrueNotSolvedByKinductionPlainButKipdr}{Total}{}{Cputime}{Avg}{867.0238535274365256124721604}%
\StoreBenchExecResult{PdrInv}{PdrTrueNotSolvedByKinductionPlainButKipdr}{Total}{}{Cputime}{Median}{902.621964676}%
\StoreBenchExecResult{PdrInv}{PdrTrueNotSolvedByKinductionPlainButKipdr}{Total}{}{Cputime}{Min}{3.09830843}%
\StoreBenchExecResult{PdrInv}{PdrTrueNotSolvedByKinductionPlainButKipdr}{Total}{}{Cputime}{Max}{912.264088922}%
\StoreBenchExecResult{PdrInv}{PdrTrueNotSolvedByKinductionPlainButKipdr}{Total}{}{Cputime}{Stdev}{175.2680975607713234429561739}%
\StoreBenchExecResult{PdrInv}{PdrTrueNotSolvedByKinductionPlainButKipdr}{Total}{}{Walltime}{}{383882.04715657844}%
\StoreBenchExecResult{PdrInv}{PdrTrueNotSolvedByKinductionPlainButKipdr}{Total}{}{Walltime}{Avg}{854.9711517963885077951002227}%
\StoreBenchExecResult{PdrInv}{PdrTrueNotSolvedByKinductionPlainButKipdr}{Total}{}{Walltime}{Median}{892.022969007}%
\StoreBenchExecResult{PdrInv}{PdrTrueNotSolvedByKinductionPlainButKipdr}{Total}{}{Walltime}{Min}{1.73475503922}%
\StoreBenchExecResult{PdrInv}{PdrTrueNotSolvedByKinductionPlainButKipdr}{Total}{}{Walltime}{Max}{898.839938164}%
\StoreBenchExecResult{PdrInv}{PdrTrueNotSolvedByKinductionPlainButKipdr}{Total}{}{Walltime}{Stdev}{173.4774121322924106016412381}%
\StoreBenchExecResult{PdrInv}{PdrTrueNotSolvedByKinductionPlainButKipdr}{Correct}{}{Count}{}{19}%
\StoreBenchExecResult{PdrInv}{PdrTrueNotSolvedByKinductionPlainButKipdr}{Correct}{}{Cputime}{}{909.691916116}%
\StoreBenchExecResult{PdrInv}{PdrTrueNotSolvedByKinductionPlainButKipdr}{Correct}{}{Cputime}{Avg}{47.87852190084210526315789474}%
\StoreBenchExecResult{PdrInv}{PdrTrueNotSolvedByKinductionPlainButKipdr}{Correct}{}{Cputime}{Median}{6.510913152}%
\StoreBenchExecResult{PdrInv}{PdrTrueNotSolvedByKinductionPlainButKipdr}{Correct}{}{Cputime}{Min}{3.09830843}%
\StoreBenchExecResult{PdrInv}{PdrTrueNotSolvedByKinductionPlainButKipdr}{Correct}{}{Cputime}{Max}{720.680305649}%
\StoreBenchExecResult{PdrInv}{PdrTrueNotSolvedByKinductionPlainButKipdr}{Correct}{}{Cputime}{Stdev}{158.8944736309152998480951652}%
\StoreBenchExecResult{PdrInv}{PdrTrueNotSolvedByKinductionPlainButKipdr}{Correct}{}{Walltime}{}{836.16679596944}%
\StoreBenchExecResult{PdrInv}{PdrTrueNotSolvedByKinductionPlainButKipdr}{Correct}{}{Walltime}{Avg}{44.00877873523368421052631579}%
\StoreBenchExecResult{PdrInv}{PdrTrueNotSolvedByKinductionPlainButKipdr}{Correct}{}{Walltime}{Median}{3.91548705101}%
\StoreBenchExecResult{PdrInv}{PdrTrueNotSolvedByKinductionPlainButKipdr}{Correct}{}{Walltime}{Min}{1.73475503922}%
\StoreBenchExecResult{PdrInv}{PdrTrueNotSolvedByKinductionPlainButKipdr}{Correct}{}{Walltime}{Max}{702.018512011}%
\StoreBenchExecResult{PdrInv}{PdrTrueNotSolvedByKinductionPlainButKipdr}{Correct}{}{Walltime}{Stdev}{155.3390110753679457913992950}%
\StoreBenchExecResult{PdrInv}{PdrTrueNotSolvedByKinductionPlainButKipdr}{Correct}{True}{Count}{}{19}%
\StoreBenchExecResult{PdrInv}{PdrTrueNotSolvedByKinductionPlainButKipdr}{Correct}{True}{Cputime}{}{909.691916116}%
\StoreBenchExecResult{PdrInv}{PdrTrueNotSolvedByKinductionPlainButKipdr}{Correct}{True}{Cputime}{Avg}{47.87852190084210526315789474}%
\StoreBenchExecResult{PdrInv}{PdrTrueNotSolvedByKinductionPlainButKipdr}{Correct}{True}{Cputime}{Median}{6.510913152}%
\StoreBenchExecResult{PdrInv}{PdrTrueNotSolvedByKinductionPlainButKipdr}{Correct}{True}{Cputime}{Min}{3.09830843}%
\StoreBenchExecResult{PdrInv}{PdrTrueNotSolvedByKinductionPlainButKipdr}{Correct}{True}{Cputime}{Max}{720.680305649}%
\StoreBenchExecResult{PdrInv}{PdrTrueNotSolvedByKinductionPlainButKipdr}{Correct}{True}{Cputime}{Stdev}{158.8944736309152998480951652}%
\StoreBenchExecResult{PdrInv}{PdrTrueNotSolvedByKinductionPlainButKipdr}{Correct}{True}{Walltime}{}{836.16679596944}%
\StoreBenchExecResult{PdrInv}{PdrTrueNotSolvedByKinductionPlainButKipdr}{Correct}{True}{Walltime}{Avg}{44.00877873523368421052631579}%
\StoreBenchExecResult{PdrInv}{PdrTrueNotSolvedByKinductionPlainButKipdr}{Correct}{True}{Walltime}{Median}{3.91548705101}%
\StoreBenchExecResult{PdrInv}{PdrTrueNotSolvedByKinductionPlainButKipdr}{Correct}{True}{Walltime}{Min}{1.73475503922}%
\StoreBenchExecResult{PdrInv}{PdrTrueNotSolvedByKinductionPlainButKipdr}{Correct}{True}{Walltime}{Max}{702.018512011}%
\StoreBenchExecResult{PdrInv}{PdrTrueNotSolvedByKinductionPlainButKipdr}{Correct}{True}{Walltime}{Stdev}{155.3390110753679457913992950}%
\StoreBenchExecResult{PdrInv}{PdrTrueNotSolvedByKinductionPlainButKipdr}{Wrong}{True}{Count}{}{0}%
\StoreBenchExecResult{PdrInv}{PdrTrueNotSolvedByKinductionPlainButKipdr}{Wrong}{True}{Cputime}{}{0}%
\StoreBenchExecResult{PdrInv}{PdrTrueNotSolvedByKinductionPlainButKipdr}{Wrong}{True}{Cputime}{Avg}{None}%
\StoreBenchExecResult{PdrInv}{PdrTrueNotSolvedByKinductionPlainButKipdr}{Wrong}{True}{Cputime}{Median}{None}%
\StoreBenchExecResult{PdrInv}{PdrTrueNotSolvedByKinductionPlainButKipdr}{Wrong}{True}{Cputime}{Min}{None}%
\StoreBenchExecResult{PdrInv}{PdrTrueNotSolvedByKinductionPlainButKipdr}{Wrong}{True}{Cputime}{Max}{None}%
\StoreBenchExecResult{PdrInv}{PdrTrueNotSolvedByKinductionPlainButKipdr}{Wrong}{True}{Cputime}{Stdev}{None}%
\StoreBenchExecResult{PdrInv}{PdrTrueNotSolvedByKinductionPlainButKipdr}{Wrong}{True}{Walltime}{}{0}%
\StoreBenchExecResult{PdrInv}{PdrTrueNotSolvedByKinductionPlainButKipdr}{Wrong}{True}{Walltime}{Avg}{None}%
\StoreBenchExecResult{PdrInv}{PdrTrueNotSolvedByKinductionPlainButKipdr}{Wrong}{True}{Walltime}{Median}{None}%
\StoreBenchExecResult{PdrInv}{PdrTrueNotSolvedByKinductionPlainButKipdr}{Wrong}{True}{Walltime}{Min}{None}%
\StoreBenchExecResult{PdrInv}{PdrTrueNotSolvedByKinductionPlainButKipdr}{Wrong}{True}{Walltime}{Max}{None}%
\StoreBenchExecResult{PdrInv}{PdrTrueNotSolvedByKinductionPlainButKipdr}{Wrong}{True}{Walltime}{Stdev}{None}%
\StoreBenchExecResult{PdrInv}{PdrTrueNotSolvedByKinductionPlainButKipdr}{Error}{}{Count}{}{430}%
\StoreBenchExecResult{PdrInv}{PdrTrueNotSolvedByKinductionPlainButKipdr}{Error}{}{Cputime}{}{388384.018317703}%
\StoreBenchExecResult{PdrInv}{PdrTrueNotSolvedByKinductionPlainButKipdr}{Error}{}{Cputime}{Avg}{903.2186472504720930232558140}%
\StoreBenchExecResult{PdrInv}{PdrTrueNotSolvedByKinductionPlainButKipdr}{Error}{}{Cputime}{Median}{902.6888988145}%
\StoreBenchExecResult{PdrInv}{PdrTrueNotSolvedByKinductionPlainButKipdr}{Error}{}{Cputime}{Min}{901.158180785}%
\StoreBenchExecResult{PdrInv}{PdrTrueNotSolvedByKinductionPlainButKipdr}{Error}{}{Cputime}{Max}{912.264088922}%
\StoreBenchExecResult{PdrInv}{PdrTrueNotSolvedByKinductionPlainButKipdr}{Error}{}{Cputime}{Stdev}{1.344792582200654941666816149}%
\StoreBenchExecResult{PdrInv}{PdrTrueNotSolvedByKinductionPlainButKipdr}{Error}{}{Walltime}{}{383045.880360609}%
\StoreBenchExecResult{PdrInv}{PdrTrueNotSolvedByKinductionPlainButKipdr}{Error}{}{Walltime}{Avg}{890.8043729316488372093023256}%
\StoreBenchExecResult{PdrInv}{PdrTrueNotSolvedByKinductionPlainButKipdr}{Error}{}{Walltime}{Median}{892.126722932}%
\StoreBenchExecResult{PdrInv}{PdrTrueNotSolvedByKinductionPlainButKipdr}{Error}{}{Walltime}{Min}{879.288122892}%
\StoreBenchExecResult{PdrInv}{PdrTrueNotSolvedByKinductionPlainButKipdr}{Error}{}{Walltime}{Max}{898.839938164}%
\StoreBenchExecResult{PdrInv}{PdrTrueNotSolvedByKinductionPlainButKipdr}{Error}{}{Walltime}{Stdev}{3.812261697986267528481593727}%
\StoreBenchExecResult{PdrInv}{PdrTrueNotSolvedByKinductionPlainButKipdr}{Error}{Timeout}{Count}{}{430}%
\StoreBenchExecResult{PdrInv}{PdrTrueNotSolvedByKinductionPlainButKipdr}{Error}{Timeout}{Cputime}{}{388384.018317703}%
\StoreBenchExecResult{PdrInv}{PdrTrueNotSolvedByKinductionPlainButKipdr}{Error}{Timeout}{Cputime}{Avg}{903.2186472504720930232558140}%
\StoreBenchExecResult{PdrInv}{PdrTrueNotSolvedByKinductionPlainButKipdr}{Error}{Timeout}{Cputime}{Median}{902.6888988145}%
\StoreBenchExecResult{PdrInv}{PdrTrueNotSolvedByKinductionPlainButKipdr}{Error}{Timeout}{Cputime}{Min}{901.158180785}%
\StoreBenchExecResult{PdrInv}{PdrTrueNotSolvedByKinductionPlainButKipdr}{Error}{Timeout}{Cputime}{Max}{912.264088922}%
\StoreBenchExecResult{PdrInv}{PdrTrueNotSolvedByKinductionPlainButKipdr}{Error}{Timeout}{Cputime}{Stdev}{1.344792582200654941666816149}%
\StoreBenchExecResult{PdrInv}{PdrTrueNotSolvedByKinductionPlainButKipdr}{Error}{Timeout}{Walltime}{}{383045.880360609}%
\StoreBenchExecResult{PdrInv}{PdrTrueNotSolvedByKinductionPlainButKipdr}{Error}{Timeout}{Walltime}{Avg}{890.8043729316488372093023256}%
\StoreBenchExecResult{PdrInv}{PdrTrueNotSolvedByKinductionPlainButKipdr}{Error}{Timeout}{Walltime}{Median}{892.126722932}%
\StoreBenchExecResult{PdrInv}{PdrTrueNotSolvedByKinductionPlainButKipdr}{Error}{Timeout}{Walltime}{Min}{879.288122892}%
\StoreBenchExecResult{PdrInv}{PdrTrueNotSolvedByKinductionPlainButKipdr}{Error}{Timeout}{Walltime}{Max}{898.839938164}%
\StoreBenchExecResult{PdrInv}{PdrTrueNotSolvedByKinductionPlainButKipdr}{Error}{Timeout}{Walltime}{Stdev}{3.812261697986267528481593727}%
\providecommand\StoreBenchExecResult[7]{\expandafter\newcommand\csname#1#2#3#4#5#6\endcsname{#7}}%
\StoreBenchExecResult{PdrInv}{PdrTrueNotSolvedByKinductionPlain}{Total}{}{Count}{}{2893}%
\StoreBenchExecResult{PdrInv}{PdrTrueNotSolvedByKinductionPlain}{Total}{}{Cputime}{}{2256330.289093411}%
\StoreBenchExecResult{PdrInv}{PdrTrueNotSolvedByKinductionPlain}{Total}{}{Cputime}{Avg}{779.9275109206398202557898375}%
\StoreBenchExecResult{PdrInv}{PdrTrueNotSolvedByKinductionPlain}{Total}{}{Cputime}{Median}{902.054289505}%
\StoreBenchExecResult{PdrInv}{PdrTrueNotSolvedByKinductionPlain}{Total}{}{Cputime}{Min}{2.839431595}%
\StoreBenchExecResult{PdrInv}{PdrTrueNotSolvedByKinductionPlain}{Total}{}{Cputime}{Max}{1001.0468483}%
\StoreBenchExecResult{PdrInv}{PdrTrueNotSolvedByKinductionPlain}{Total}{}{Cputime}{Stdev}{293.1743742381131650948304576}%
\StoreBenchExecResult{PdrInv}{PdrTrueNotSolvedByKinductionPlain}{Total}{}{Walltime}{}{2127071.95189810609}%
\StoreBenchExecResult{PdrInv}{PdrTrueNotSolvedByKinductionPlain}{Total}{}{Walltime}{Avg}{735.2478229858645316280677497}%
\StoreBenchExecResult{PdrInv}{PdrTrueNotSolvedByKinductionPlain}{Total}{}{Walltime}{Median}{885.163808107}%
\StoreBenchExecResult{PdrInv}{PdrTrueNotSolvedByKinductionPlain}{Total}{}{Walltime}{Min}{1.56016206741}%
\StoreBenchExecResult{PdrInv}{PdrTrueNotSolvedByKinductionPlain}{Total}{}{Walltime}{Max}{906.216305971}%
\StoreBenchExecResult{PdrInv}{PdrTrueNotSolvedByKinductionPlain}{Total}{}{Walltime}{Stdev}{289.3531132669351987857768250}%
\StoreBenchExecResult{PdrInv}{PdrTrueNotSolvedByKinductionPlain}{Correct}{}{Count}{}{67}%
\StoreBenchExecResult{PdrInv}{PdrTrueNotSolvedByKinductionPlain}{Correct}{}{Cputime}{}{7994.506788147}%
\StoreBenchExecResult{PdrInv}{PdrTrueNotSolvedByKinductionPlain}{Correct}{}{Cputime}{Avg}{119.3209968380149253731343284}%
\StoreBenchExecResult{PdrInv}{PdrTrueNotSolvedByKinductionPlain}{Correct}{}{Cputime}{Median}{43.691618057}%
\StoreBenchExecResult{PdrInv}{PdrTrueNotSolvedByKinductionPlain}{Correct}{}{Cputime}{Min}{3.09830843}%
\StoreBenchExecResult{PdrInv}{PdrTrueNotSolvedByKinductionPlain}{Correct}{}{Cputime}{Max}{829.069057036}%
\StoreBenchExecResult{PdrInv}{PdrTrueNotSolvedByKinductionPlain}{Correct}{}{Cputime}{Stdev}{187.5258572687791836730331924}%
\StoreBenchExecResult{PdrInv}{PdrTrueNotSolvedByKinductionPlain}{Correct}{}{Walltime}{}{7229.54784846313}%
\StoreBenchExecResult{PdrInv}{PdrTrueNotSolvedByKinductionPlain}{Correct}{}{Walltime}{Avg}{107.9036992307929850746268657}%
\StoreBenchExecResult{PdrInv}{PdrTrueNotSolvedByKinductionPlain}{Correct}{}{Walltime}{Median}{37.4351379871}%
\StoreBenchExecResult{PdrInv}{PdrTrueNotSolvedByKinductionPlain}{Correct}{}{Walltime}{Min}{1.73475503922}%
\StoreBenchExecResult{PdrInv}{PdrTrueNotSolvedByKinductionPlain}{Correct}{}{Walltime}{Max}{826.533210993}%
\StoreBenchExecResult{PdrInv}{PdrTrueNotSolvedByKinductionPlain}{Correct}{}{Walltime}{Stdev}{181.0212021520069676352710995}%
\StoreBenchExecResult{PdrInv}{PdrTrueNotSolvedByKinductionPlain}{Correct}{True}{Count}{}{67}%
\StoreBenchExecResult{PdrInv}{PdrTrueNotSolvedByKinductionPlain}{Correct}{True}{Cputime}{}{7994.506788147}%
\StoreBenchExecResult{PdrInv}{PdrTrueNotSolvedByKinductionPlain}{Correct}{True}{Cputime}{Avg}{119.3209968380149253731343284}%
\StoreBenchExecResult{PdrInv}{PdrTrueNotSolvedByKinductionPlain}{Correct}{True}{Cputime}{Median}{43.691618057}%
\StoreBenchExecResult{PdrInv}{PdrTrueNotSolvedByKinductionPlain}{Correct}{True}{Cputime}{Min}{3.09830843}%
\StoreBenchExecResult{PdrInv}{PdrTrueNotSolvedByKinductionPlain}{Correct}{True}{Cputime}{Max}{829.069057036}%
\StoreBenchExecResult{PdrInv}{PdrTrueNotSolvedByKinductionPlain}{Correct}{True}{Cputime}{Stdev}{187.5258572687791836730331924}%
\StoreBenchExecResult{PdrInv}{PdrTrueNotSolvedByKinductionPlain}{Correct}{True}{Walltime}{}{7229.54784846313}%
\StoreBenchExecResult{PdrInv}{PdrTrueNotSolvedByKinductionPlain}{Correct}{True}{Walltime}{Avg}{107.9036992307929850746268657}%
\StoreBenchExecResult{PdrInv}{PdrTrueNotSolvedByKinductionPlain}{Correct}{True}{Walltime}{Median}{37.4351379871}%
\StoreBenchExecResult{PdrInv}{PdrTrueNotSolvedByKinductionPlain}{Correct}{True}{Walltime}{Min}{1.73475503922}%
\StoreBenchExecResult{PdrInv}{PdrTrueNotSolvedByKinductionPlain}{Correct}{True}{Walltime}{Max}{826.533210993}%
\StoreBenchExecResult{PdrInv}{PdrTrueNotSolvedByKinductionPlain}{Correct}{True}{Walltime}{Stdev}{181.0212021520069676352710995}%
\StoreBenchExecResult{PdrInv}{PdrTrueNotSolvedByKinductionPlain}{Wrong}{True}{Count}{}{0}%
\StoreBenchExecResult{PdrInv}{PdrTrueNotSolvedByKinductionPlain}{Wrong}{True}{Cputime}{}{0}%
\StoreBenchExecResult{PdrInv}{PdrTrueNotSolvedByKinductionPlain}{Wrong}{True}{Cputime}{Avg}{None}%
\StoreBenchExecResult{PdrInv}{PdrTrueNotSolvedByKinductionPlain}{Wrong}{True}{Cputime}{Median}{None}%
\StoreBenchExecResult{PdrInv}{PdrTrueNotSolvedByKinductionPlain}{Wrong}{True}{Cputime}{Min}{None}%
\StoreBenchExecResult{PdrInv}{PdrTrueNotSolvedByKinductionPlain}{Wrong}{True}{Cputime}{Max}{None}%
\StoreBenchExecResult{PdrInv}{PdrTrueNotSolvedByKinductionPlain}{Wrong}{True}{Cputime}{Stdev}{None}%
\StoreBenchExecResult{PdrInv}{PdrTrueNotSolvedByKinductionPlain}{Wrong}{True}{Walltime}{}{0}%
\StoreBenchExecResult{PdrInv}{PdrTrueNotSolvedByKinductionPlain}{Wrong}{True}{Walltime}{Avg}{None}%
\StoreBenchExecResult{PdrInv}{PdrTrueNotSolvedByKinductionPlain}{Wrong}{True}{Walltime}{Median}{None}%
\StoreBenchExecResult{PdrInv}{PdrTrueNotSolvedByKinductionPlain}{Wrong}{True}{Walltime}{Min}{None}%
\StoreBenchExecResult{PdrInv}{PdrTrueNotSolvedByKinductionPlain}{Wrong}{True}{Walltime}{Max}{None}%
\StoreBenchExecResult{PdrInv}{PdrTrueNotSolvedByKinductionPlain}{Wrong}{True}{Walltime}{Stdev}{None}%
\StoreBenchExecResult{PdrInv}{PdrTrueNotSolvedByKinductionPlain}{Error}{}{Count}{}{2826}%
\StoreBenchExecResult{PdrInv}{PdrTrueNotSolvedByKinductionPlain}{Error}{}{Cputime}{}{2248335.782305264}%
\StoreBenchExecResult{PdrInv}{PdrTrueNotSolvedByKinductionPlain}{Error}{}{Cputime}{Avg}{795.5894487987487615003538570}%
\StoreBenchExecResult{PdrInv}{PdrTrueNotSolvedByKinductionPlain}{Error}{}{Cputime}{Median}{902.097027905}%
\StoreBenchExecResult{PdrInv}{PdrTrueNotSolvedByKinductionPlain}{Error}{}{Cputime}{Min}{2.839431595}%
\StoreBenchExecResult{PdrInv}{PdrTrueNotSolvedByKinductionPlain}{Error}{}{Cputime}{Max}{1001.0468483}%
\StoreBenchExecResult{PdrInv}{PdrTrueNotSolvedByKinductionPlain}{Error}{}{Cputime}{Stdev}{276.7012427641297595829647404}%
\StoreBenchExecResult{PdrInv}{PdrTrueNotSolvedByKinductionPlain}{Error}{}{Walltime}{}{2119842.40404964296}%
\StoreBenchExecResult{PdrInv}{PdrTrueNotSolvedByKinductionPlain}{Error}{}{Walltime}{Avg}{750.1211620840916348195329087}%
\StoreBenchExecResult{PdrInv}{PdrTrueNotSolvedByKinductionPlain}{Error}{}{Walltime}{Median}{885.7157969475}%
\StoreBenchExecResult{PdrInv}{PdrTrueNotSolvedByKinductionPlain}{Error}{}{Walltime}{Min}{1.56016206741}%
\StoreBenchExecResult{PdrInv}{PdrTrueNotSolvedByKinductionPlain}{Error}{}{Walltime}{Max}{906.216305971}%
\StoreBenchExecResult{PdrInv}{PdrTrueNotSolvedByKinductionPlain}{Error}{}{Walltime}{Stdev}{274.5567435036259783688710434}%
\StoreBenchExecResult{PdrInv}{PdrTrueNotSolvedByKinductionPlain}{Error}{Assertion}{Count}{}{2}%
\StoreBenchExecResult{PdrInv}{PdrTrueNotSolvedByKinductionPlain}{Error}{Assertion}{Cputime}{}{6.055434170}%
\StoreBenchExecResult{PdrInv}{PdrTrueNotSolvedByKinductionPlain}{Error}{Assertion}{Cputime}{Avg}{3.027717085}%
\StoreBenchExecResult{PdrInv}{PdrTrueNotSolvedByKinductionPlain}{Error}{Assertion}{Cputime}{Median}{3.027717085}%
\StoreBenchExecResult{PdrInv}{PdrTrueNotSolvedByKinductionPlain}{Error}{Assertion}{Cputime}{Min}{2.839431595}%
\StoreBenchExecResult{PdrInv}{PdrTrueNotSolvedByKinductionPlain}{Error}{Assertion}{Cputime}{Max}{3.216002575}%
\StoreBenchExecResult{PdrInv}{PdrTrueNotSolvedByKinductionPlain}{Error}{Assertion}{Cputime}{Stdev}{0.188285490}%
\StoreBenchExecResult{PdrInv}{PdrTrueNotSolvedByKinductionPlain}{Error}{Assertion}{Walltime}{}{3.38070201874}%
\StoreBenchExecResult{PdrInv}{PdrTrueNotSolvedByKinductionPlain}{Error}{Assertion}{Walltime}{Avg}{1.69035100937}%
\StoreBenchExecResult{PdrInv}{PdrTrueNotSolvedByKinductionPlain}{Error}{Assertion}{Walltime}{Median}{1.69035100937}%
\StoreBenchExecResult{PdrInv}{PdrTrueNotSolvedByKinductionPlain}{Error}{Assertion}{Walltime}{Min}{1.59299302101}%
\StoreBenchExecResult{PdrInv}{PdrTrueNotSolvedByKinductionPlain}{Error}{Assertion}{Walltime}{Max}{1.78770899773}%
\StoreBenchExecResult{PdrInv}{PdrTrueNotSolvedByKinductionPlain}{Error}{Assertion}{Walltime}{Stdev}{0.09735798836}%
\StoreBenchExecResult{PdrInv}{PdrTrueNotSolvedByKinductionPlain}{Error}{Error}{Count}{}{361}%
\StoreBenchExecResult{PdrInv}{PdrTrueNotSolvedByKinductionPlain}{Error}{Error}{Cputime}{}{42380.110039127}%
\StoreBenchExecResult{PdrInv}{PdrTrueNotSolvedByKinductionPlain}{Error}{Error}{Cputime}{Avg}{117.3964267011828254847645429}%
\StoreBenchExecResult{PdrInv}{PdrTrueNotSolvedByKinductionPlain}{Error}{Error}{Cputime}{Median}{47.109695394}%
\StoreBenchExecResult{PdrInv}{PdrTrueNotSolvedByKinductionPlain}{Error}{Error}{Cputime}{Min}{2.879662591}%
\StoreBenchExecResult{PdrInv}{PdrTrueNotSolvedByKinductionPlain}{Error}{Error}{Cputime}{Max}{896.467967052}%
\StoreBenchExecResult{PdrInv}{PdrTrueNotSolvedByKinductionPlain}{Error}{Error}{Cputime}{Stdev}{172.9687138074897669252149095}%
\StoreBenchExecResult{PdrInv}{PdrTrueNotSolvedByKinductionPlain}{Error}{Error}{Walltime}{}{35661.39786315006}%
\StoreBenchExecResult{PdrInv}{PdrTrueNotSolvedByKinductionPlain}{Error}{Error}{Walltime}{Avg}{98.78503563199462603878116343}%
\StoreBenchExecResult{PdrInv}{PdrTrueNotSolvedByKinductionPlain}{Error}{Error}{Walltime}{Median}{34.351336956}%
\StoreBenchExecResult{PdrInv}{PdrTrueNotSolvedByKinductionPlain}{Error}{Error}{Walltime}{Min}{1.56016206741}%
\StoreBenchExecResult{PdrInv}{PdrTrueNotSolvedByKinductionPlain}{Error}{Error}{Walltime}{Max}{798.802060127}%
\StoreBenchExecResult{PdrInv}{PdrTrueNotSolvedByKinductionPlain}{Error}{Error}{Walltime}{Stdev}{155.5291024935041228142531387}%
\StoreBenchExecResult{PdrInv}{PdrTrueNotSolvedByKinductionPlain}{Error}{Exception}{Count}{}{15}%
\StoreBenchExecResult{PdrInv}{PdrTrueNotSolvedByKinductionPlain}{Error}{Exception}{Cputime}{}{5185.588075852}%
\StoreBenchExecResult{PdrInv}{PdrTrueNotSolvedByKinductionPlain}{Error}{Exception}{Cputime}{Avg}{345.7058717234666666666666667}%
\StoreBenchExecResult{PdrInv}{PdrTrueNotSolvedByKinductionPlain}{Error}{Exception}{Cputime}{Median}{497.379835534}%
\StoreBenchExecResult{PdrInv}{PdrTrueNotSolvedByKinductionPlain}{Error}{Exception}{Cputime}{Min}{5.292909095}%
\StoreBenchExecResult{PdrInv}{PdrTrueNotSolvedByKinductionPlain}{Error}{Exception}{Cputime}{Max}{530.049401185}%
\StoreBenchExecResult{PdrInv}{PdrTrueNotSolvedByKinductionPlain}{Error}{Exception}{Cputime}{Stdev}{230.6259309614624597845750779}%
\StoreBenchExecResult{PdrInv}{PdrTrueNotSolvedByKinductionPlain}{Error}{Exception}{Walltime}{}{4349.04301333416}%
\StoreBenchExecResult{PdrInv}{PdrTrueNotSolvedByKinductionPlain}{Error}{Exception}{Walltime}{Avg}{289.936200888944}%
\StoreBenchExecResult{PdrInv}{PdrTrueNotSolvedByKinductionPlain}{Error}{Exception}{Walltime}{Median}{416.363477945}%
\StoreBenchExecResult{PdrInv}{PdrTrueNotSolvedByKinductionPlain}{Error}{Exception}{Walltime}{Min}{2.82928800583}%
\StoreBenchExecResult{PdrInv}{PdrTrueNotSolvedByKinductionPlain}{Error}{Exception}{Walltime}{Max}{447.364850998}%
\StoreBenchExecResult{PdrInv}{PdrTrueNotSolvedByKinductionPlain}{Error}{Exception}{Walltime}{Stdev}{195.9372824168882446577704698}%
\StoreBenchExecResult{PdrInv}{PdrTrueNotSolvedByKinductionPlain}{Error}{OutOfJavaMemory}{Count}{}{8}%
\StoreBenchExecResult{PdrInv}{PdrTrueNotSolvedByKinductionPlain}{Error}{OutOfJavaMemory}{Cputime}{}{3737.601849047}%
\StoreBenchExecResult{PdrInv}{PdrTrueNotSolvedByKinductionPlain}{Error}{OutOfJavaMemory}{Cputime}{Avg}{467.200231130875}%
\StoreBenchExecResult{PdrInv}{PdrTrueNotSolvedByKinductionPlain}{Error}{OutOfJavaMemory}{Cputime}{Median}{509.7144821585}%
\StoreBenchExecResult{PdrInv}{PdrTrueNotSolvedByKinductionPlain}{Error}{OutOfJavaMemory}{Cputime}{Min}{237.563009364}%
\StoreBenchExecResult{PdrInv}{PdrTrueNotSolvedByKinductionPlain}{Error}{OutOfJavaMemory}{Cputime}{Max}{542.556565247}%
\StoreBenchExecResult{PdrInv}{PdrTrueNotSolvedByKinductionPlain}{Error}{OutOfJavaMemory}{Cputime}{Stdev}{93.60009201187221336388969532}%
\StoreBenchExecResult{PdrInv}{PdrTrueNotSolvedByKinductionPlain}{Error}{OutOfJavaMemory}{Walltime}{}{2558.139458179}%
\StoreBenchExecResult{PdrInv}{PdrTrueNotSolvedByKinductionPlain}{Error}{OutOfJavaMemory}{Walltime}{Avg}{319.767432272375}%
\StoreBenchExecResult{PdrInv}{PdrTrueNotSolvedByKinductionPlain}{Error}{OutOfJavaMemory}{Walltime}{Median}{348.3717814685}%
\StoreBenchExecResult{PdrInv}{PdrTrueNotSolvedByKinductionPlain}{Error}{OutOfJavaMemory}{Walltime}{Min}{153.061067104}%
\StoreBenchExecResult{PdrInv}{PdrTrueNotSolvedByKinductionPlain}{Error}{OutOfJavaMemory}{Walltime}{Max}{363.610453129}%
\StoreBenchExecResult{PdrInv}{PdrTrueNotSolvedByKinductionPlain}{Error}{OutOfJavaMemory}{Walltime}{Stdev}{65.81995292651458814382911071}%
\StoreBenchExecResult{PdrInv}{PdrTrueNotSolvedByKinductionPlain}{Error}{OutOfMemory}{Count}{}{12}%
\StoreBenchExecResult{PdrInv}{PdrTrueNotSolvedByKinductionPlain}{Error}{OutOfMemory}{Cputime}{}{2302.905175416}%
\StoreBenchExecResult{PdrInv}{PdrTrueNotSolvedByKinductionPlain}{Error}{OutOfMemory}{Cputime}{Avg}{191.908764618}%
\StoreBenchExecResult{PdrInv}{PdrTrueNotSolvedByKinductionPlain}{Error}{OutOfMemory}{Cputime}{Median}{136.2481361165}%
\StoreBenchExecResult{PdrInv}{PdrTrueNotSolvedByKinductionPlain}{Error}{OutOfMemory}{Cputime}{Min}{129.325401862}%
\StoreBenchExecResult{PdrInv}{PdrTrueNotSolvedByKinductionPlain}{Error}{OutOfMemory}{Cputime}{Max}{766.707137278}%
\StoreBenchExecResult{PdrInv}{PdrTrueNotSolvedByKinductionPlain}{Error}{OutOfMemory}{Cputime}{Stdev}{173.6834475183998464350417704}%
\StoreBenchExecResult{PdrInv}{PdrTrueNotSolvedByKinductionPlain}{Error}{OutOfMemory}{Walltime}{}{1963.570679426}%
\StoreBenchExecResult{PdrInv}{PdrTrueNotSolvedByKinductionPlain}{Error}{OutOfMemory}{Walltime}{Avg}{163.6308899521666666666666667}%
\StoreBenchExecResult{PdrInv}{PdrTrueNotSolvedByKinductionPlain}{Error}{OutOfMemory}{Walltime}{Median}{119.3377144335}%
\StoreBenchExecResult{PdrInv}{PdrTrueNotSolvedByKinductionPlain}{Error}{OutOfMemory}{Walltime}{Min}{114.130378008}%
\StoreBenchExecResult{PdrInv}{PdrTrueNotSolvedByKinductionPlain}{Error}{OutOfMemory}{Walltime}{Max}{616.824223042}%
\StoreBenchExecResult{PdrInv}{PdrTrueNotSolvedByKinductionPlain}{Error}{OutOfMemory}{Walltime}{Stdev}{137.0021617987140022162335875}%
\StoreBenchExecResult{PdrInv}{PdrTrueNotSolvedByKinductionPlain}{Error}{SegmentationFault}{Count}{}{1}%
\StoreBenchExecResult{PdrInv}{PdrTrueNotSolvedByKinductionPlain}{Error}{SegmentationFault}{Cputime}{}{477.87720091}%
\StoreBenchExecResult{PdrInv}{PdrTrueNotSolvedByKinductionPlain}{Error}{SegmentationFault}{Cputime}{Avg}{477.87720091}%
\StoreBenchExecResult{PdrInv}{PdrTrueNotSolvedByKinductionPlain}{Error}{SegmentationFault}{Cputime}{Median}{477.87720091}%
\StoreBenchExecResult{PdrInv}{PdrTrueNotSolvedByKinductionPlain}{Error}{SegmentationFault}{Cputime}{Min}{477.87720091}%
\StoreBenchExecResult{PdrInv}{PdrTrueNotSolvedByKinductionPlain}{Error}{SegmentationFault}{Cputime}{Max}{477.87720091}%
\StoreBenchExecResult{PdrInv}{PdrTrueNotSolvedByKinductionPlain}{Error}{SegmentationFault}{Cputime}{Stdev}{0E-8}%
\StoreBenchExecResult{PdrInv}{PdrTrueNotSolvedByKinductionPlain}{Error}{SegmentationFault}{Walltime}{}{453.881695986}%
\StoreBenchExecResult{PdrInv}{PdrTrueNotSolvedByKinductionPlain}{Error}{SegmentationFault}{Walltime}{Avg}{453.881695986}%
\StoreBenchExecResult{PdrInv}{PdrTrueNotSolvedByKinductionPlain}{Error}{SegmentationFault}{Walltime}{Median}{453.881695986}%
\StoreBenchExecResult{PdrInv}{PdrTrueNotSolvedByKinductionPlain}{Error}{SegmentationFault}{Walltime}{Min}{453.881695986}%
\StoreBenchExecResult{PdrInv}{PdrTrueNotSolvedByKinductionPlain}{Error}{SegmentationFault}{Walltime}{Max}{453.881695986}%
\StoreBenchExecResult{PdrInv}{PdrTrueNotSolvedByKinductionPlain}{Error}{SegmentationFault}{Walltime}{Stdev}{0E-9}%
\StoreBenchExecResult{PdrInv}{PdrTrueNotSolvedByKinductionPlain}{Error}{Timeout}{Count}{}{2427}%
\StoreBenchExecResult{PdrInv}{PdrTrueNotSolvedByKinductionPlain}{Error}{Timeout}{Cputime}{}{2194245.644530742}%
\StoreBenchExecResult{PdrInv}{PdrTrueNotSolvedByKinductionPlain}{Error}{Timeout}{Cputime}{Avg}{904.0979169883568191182529872}%
\StoreBenchExecResult{PdrInv}{PdrTrueNotSolvedByKinductionPlain}{Error}{Timeout}{Cputime}{Median}{902.406219297}%
\StoreBenchExecResult{PdrInv}{PdrTrueNotSolvedByKinductionPlain}{Error}{Timeout}{Cputime}{Min}{900.73232011}%
\StoreBenchExecResult{PdrInv}{PdrTrueNotSolvedByKinductionPlain}{Error}{Timeout}{Cputime}{Max}{1001.0468483}%
\StoreBenchExecResult{PdrInv}{PdrTrueNotSolvedByKinductionPlain}{Error}{Timeout}{Cputime}{Stdev}{6.193939727156855294445259590}%
\StoreBenchExecResult{PdrInv}{PdrTrueNotSolvedByKinductionPlain}{Error}{Timeout}{Walltime}{}{2074852.990637549}%
\StoreBenchExecResult{PdrInv}{PdrTrueNotSolvedByKinductionPlain}{Error}{Timeout}{Walltime}{Avg}{854.9044048774408735063864854}%
\StoreBenchExecResult{PdrInv}{PdrTrueNotSolvedByKinductionPlain}{Error}{Timeout}{Walltime}{Median}{888.504296064}%
\StoreBenchExecResult{PdrInv}{PdrTrueNotSolvedByKinductionPlain}{Error}{Timeout}{Walltime}{Min}{486.869479895}%
\StoreBenchExecResult{PdrInv}{PdrTrueNotSolvedByKinductionPlain}{Error}{Timeout}{Walltime}{Max}{906.216305971}%
\StoreBenchExecResult{PdrInv}{PdrTrueNotSolvedByKinductionPlain}{Error}{Timeout}{Walltime}{Stdev}{75.04525025218917664955233417}%
\providecommand\StoreBenchExecResult[7]{\expandafter\newcommand\csname#1#2#3#4#5#6\endcsname{#7}}%
\StoreBenchExecResult{PdrInv}{Pdr}{Total}{}{Count}{}{5591}%
\StoreBenchExecResult{PdrInv}{Pdr}{Total}{}{Cputime}{}{3699791.292638809}%
\StoreBenchExecResult{PdrInv}{Pdr}{Total}{}{Cputime}{Avg}{661.7405281056714362368091576}%
\StoreBenchExecResult{PdrInv}{Pdr}{Total}{}{Cputime}{Median}{901.72994722}%
\StoreBenchExecResult{PdrInv}{Pdr}{Total}{}{Cputime}{Min}{2.793063979}%
\StoreBenchExecResult{PdrInv}{Pdr}{Total}{}{Cputime}{Max}{1001.0468483}%
\StoreBenchExecResult{PdrInv}{Pdr}{Total}{}{Cputime}{Stdev}{388.0946858217790410615056008}%
\StoreBenchExecResult{PdrInv}{Pdr}{Total}{}{Walltime}{}{3532637.62646293088}%
\StoreBenchExecResult{PdrInv}{Pdr}{Total}{}{Walltime}{Avg}{631.8436105281579109282775890}%
\StoreBenchExecResult{PdrInv}{Pdr}{Total}{}{Walltime}{Median}{883.891388178}%
\StoreBenchExecResult{PdrInv}{Pdr}{Total}{}{Walltime}{Min}{1.56016206741}%
\StoreBenchExecResult{PdrInv}{Pdr}{Total}{}{Walltime}{Max}{908.306071997}%
\StoreBenchExecResult{PdrInv}{Pdr}{Total}{}{Walltime}{Stdev}{378.0844172384772949364405977}%
\StoreBenchExecResult{PdrInv}{Pdr}{Correct}{}{Count}{}{1087}%
\StoreBenchExecResult{PdrInv}{Pdr}{Correct}{}{Cputime}{}{32579.620928755}%
\StoreBenchExecResult{PdrInv}{Pdr}{Correct}{}{Cputime}{Avg}{29.97205237235970561177552898}%
\StoreBenchExecResult{PdrInv}{Pdr}{Correct}{}{Cputime}{Median}{4.884578527}%
\StoreBenchExecResult{PdrInv}{Pdr}{Correct}{}{Cputime}{Min}{2.852027404}%
\StoreBenchExecResult{PdrInv}{Pdr}{Correct}{}{Cputime}{Max}{829.069057036}%
\StoreBenchExecResult{PdrInv}{Pdr}{Correct}{}{Cputime}{Stdev}{99.58710752205702694378020158}%
\StoreBenchExecResult{PdrInv}{Pdr}{Correct}{}{Walltime}{}{28342.01476311627}%
\StoreBenchExecResult{PdrInv}{Pdr}{Correct}{}{Walltime}{Avg}{26.07361063764146274149034039}%
\StoreBenchExecResult{PdrInv}{Pdr}{Correct}{}{Walltime}{Median}{2.64180493355}%
\StoreBenchExecResult{PdrInv}{Pdr}{Correct}{}{Walltime}{Min}{1.57658696175}%
\StoreBenchExecResult{PdrInv}{Pdr}{Correct}{}{Walltime}{Max}{826.533210993}%
\StoreBenchExecResult{PdrInv}{Pdr}{Correct}{}{Walltime}{Stdev}{97.16613162077675724472527826}%
\StoreBenchExecResult{PdrInv}{Pdr}{Correct}{False}{Count}{}{255}%
\StoreBenchExecResult{PdrInv}{Pdr}{Correct}{False}{Cputime}{}{7689.553875740}%
\StoreBenchExecResult{PdrInv}{Pdr}{Correct}{False}{Cputime}{Avg}{30.15511323819607843137254902}%
\StoreBenchExecResult{PdrInv}{Pdr}{Correct}{False}{Cputime}{Median}{5.494926053}%
\StoreBenchExecResult{PdrInv}{Pdr}{Correct}{False}{Cputime}{Min}{2.914757653}%
\StoreBenchExecResult{PdrInv}{Pdr}{Correct}{False}{Cputime}{Max}{658.544573689}%
\StoreBenchExecResult{PdrInv}{Pdr}{Correct}{False}{Cputime}{Stdev}{84.84810300053408625523382229}%
\StoreBenchExecResult{PdrInv}{Pdr}{Correct}{False}{Walltime}{}{6241.85171031995}%
\StoreBenchExecResult{PdrInv}{Pdr}{Correct}{False}{Walltime}{Avg}{24.47784984439196078431372549}%
\StoreBenchExecResult{PdrInv}{Pdr}{Correct}{False}{Walltime}{Median}{2.93016290665}%
\StoreBenchExecResult{PdrInv}{Pdr}{Correct}{False}{Walltime}{Min}{1.63834285736}%
\StoreBenchExecResult{PdrInv}{Pdr}{Correct}{False}{Walltime}{Max}{628.296149969}%
\StoreBenchExecResult{PdrInv}{Pdr}{Correct}{False}{Walltime}{Stdev}{80.60404966570848603434823765}%
\StoreBenchExecResult{PdrInv}{Pdr}{Correct}{True}{Count}{}{832}%
\StoreBenchExecResult{PdrInv}{Pdr}{Correct}{True}{Cputime}{}{24890.067053015}%
\StoreBenchExecResult{PdrInv}{Pdr}{Correct}{True}{Cputime}{Avg}{29.91594597718149038461538462}%
\StoreBenchExecResult{PdrInv}{Pdr}{Correct}{True}{Cputime}{Median}{4.723256050}%
\StoreBenchExecResult{PdrInv}{Pdr}{Correct}{True}{Cputime}{Min}{2.852027404}%
\StoreBenchExecResult{PdrInv}{Pdr}{Correct}{True}{Cputime}{Max}{829.069057036}%
\StoreBenchExecResult{PdrInv}{Pdr}{Correct}{True}{Cputime}{Stdev}{103.6857723705427604539452922}%
\StoreBenchExecResult{PdrInv}{Pdr}{Correct}{True}{Walltime}{}{22100.16305279632}%
\StoreBenchExecResult{PdrInv}{Pdr}{Correct}{True}{Walltime}{Avg}{26.56269597691865384615384615}%
\StoreBenchExecResult{PdrInv}{Pdr}{Correct}{True}{Walltime}{Median}{2.56928896904}%
\StoreBenchExecResult{PdrInv}{Pdr}{Correct}{True}{Walltime}{Min}{1.57658696175}%
\StoreBenchExecResult{PdrInv}{Pdr}{Correct}{True}{Walltime}{Max}{826.533210993}%
\StoreBenchExecResult{PdrInv}{Pdr}{Correct}{True}{Walltime}{Stdev}{101.6986727442518923602918523}%
\StoreBenchExecResult{PdrInv}{Pdr}{Wrong}{True}{Count}{}{0}%
\StoreBenchExecResult{PdrInv}{Pdr}{Wrong}{True}{Cputime}{}{0}%
\StoreBenchExecResult{PdrInv}{Pdr}{Wrong}{True}{Cputime}{Avg}{None}%
\StoreBenchExecResult{PdrInv}{Pdr}{Wrong}{True}{Cputime}{Median}{None}%
\StoreBenchExecResult{PdrInv}{Pdr}{Wrong}{True}{Cputime}{Min}{None}%
\StoreBenchExecResult{PdrInv}{Pdr}{Wrong}{True}{Cputime}{Max}{None}%
\StoreBenchExecResult{PdrInv}{Pdr}{Wrong}{True}{Cputime}{Stdev}{None}%
\StoreBenchExecResult{PdrInv}{Pdr}{Wrong}{True}{Walltime}{}{0}%
\StoreBenchExecResult{PdrInv}{Pdr}{Wrong}{True}{Walltime}{Avg}{None}%
\StoreBenchExecResult{PdrInv}{Pdr}{Wrong}{True}{Walltime}{Median}{None}%
\StoreBenchExecResult{PdrInv}{Pdr}{Wrong}{True}{Walltime}{Min}{None}%
\StoreBenchExecResult{PdrInv}{Pdr}{Wrong}{True}{Walltime}{Max}{None}%
\StoreBenchExecResult{PdrInv}{Pdr}{Wrong}{True}{Walltime}{Stdev}{None}%
\StoreBenchExecResult{PdrInv}{Pdr}{Error}{}{Count}{}{4503}%
\StoreBenchExecResult{PdrInv}{Pdr}{Error}{}{Cputime}{}{3667208.261903227}%
\StoreBenchExecResult{PdrInv}{Pdr}{Error}{}{Cputime}{Avg}{814.3922411510608483233399956}%
\StoreBenchExecResult{PdrInv}{Pdr}{Error}{}{Cputime}{Median}{902.123542648}%
\StoreBenchExecResult{PdrInv}{Pdr}{Error}{}{Cputime}{Min}{2.793063979}%
\StoreBenchExecResult{PdrInv}{Pdr}{Error}{}{Cputime}{Max}{1001.0468483}%
\StoreBenchExecResult{PdrInv}{Pdr}{Error}{}{Cputime}{Stdev}{254.6922899556973601183727935}%
\StoreBenchExecResult{PdrInv}{Pdr}{Error}{}{Walltime}{}{3504293.67539476853}%
\StoreBenchExecResult{PdrInv}{Pdr}{Error}{}{Walltime}{Avg}{778.2131191194245014434821230}%
\StoreBenchExecResult{PdrInv}{Pdr}{Error}{}{Walltime}{Median}{889.153652906}%
\StoreBenchExecResult{PdrInv}{Pdr}{Error}{}{Walltime}{Min}{1.56016206741}%
\StoreBenchExecResult{PdrInv}{Pdr}{Error}{}{Walltime}{Max}{908.306071997}%
\StoreBenchExecResult{PdrInv}{Pdr}{Error}{}{Walltime}{Stdev}{255.1738098325407376532115799}%
\StoreBenchExecResult{PdrInv}{Pdr}{Error}{Assertion}{Count}{}{4}%
\StoreBenchExecResult{PdrInv}{Pdr}{Error}{Assertion}{Cputime}{}{12.282966073}%
\StoreBenchExecResult{PdrInv}{Pdr}{Error}{Assertion}{Cputime}{Avg}{3.07074151825}%
\StoreBenchExecResult{PdrInv}{Pdr}{Error}{Assertion}{Cputime}{Median}{3.027717085}%
\StoreBenchExecResult{PdrInv}{Pdr}{Error}{Assertion}{Cputime}{Min}{2.793063979}%
\StoreBenchExecResult{PdrInv}{Pdr}{Error}{Assertion}{Cputime}{Max}{3.434467924}%
\StoreBenchExecResult{PdrInv}{Pdr}{Error}{Assertion}{Cputime}{Stdev}{0.2664614274364098907588722161}%
\StoreBenchExecResult{PdrInv}{Pdr}{Error}{Assertion}{Walltime}{}{6.91748523712}%
\StoreBenchExecResult{PdrInv}{Pdr}{Error}{Assertion}{Walltime}{Avg}{1.72937130928}%
\StoreBenchExecResult{PdrInv}{Pdr}{Error}{Assertion}{Walltime}{Median}{1.69035100937}%
\StoreBenchExecResult{PdrInv}{Pdr}{Error}{Assertion}{Walltime}{Min}{1.58088111877}%
\StoreBenchExecResult{PdrInv}{Pdr}{Error}{Assertion}{Walltime}{Max}{1.95590209961}%
\StoreBenchExecResult{PdrInv}{Pdr}{Error}{Assertion}{Walltime}{Stdev}{0.1544084348816273367031014613}%
\StoreBenchExecResult{PdrInv}{Pdr}{Error}{Error}{Count}{}{458}%
\StoreBenchExecResult{PdrInv}{Pdr}{Error}{Error}{Cputime}{}{51175.947430211}%
\StoreBenchExecResult{PdrInv}{Pdr}{Error}{Error}{Cputime}{Avg}{111.7378764851768558951965066}%
\StoreBenchExecResult{PdrInv}{Pdr}{Error}{Error}{Cputime}{Median}{43.4640290455}%
\StoreBenchExecResult{PdrInv}{Pdr}{Error}{Error}{Cputime}{Min}{2.879662591}%
\StoreBenchExecResult{PdrInv}{Pdr}{Error}{Error}{Cputime}{Max}{896.467967052}%
\StoreBenchExecResult{PdrInv}{Pdr}{Error}{Error}{Cputime}{Stdev}{168.6513372148348746452554575}%
\StoreBenchExecResult{PdrInv}{Pdr}{Error}{Error}{Walltime}{}{42671.86094784973}%
\StoreBenchExecResult{PdrInv}{Pdr}{Error}{Error}{Walltime}{Avg}{93.17000206954089519650655022}%
\StoreBenchExecResult{PdrInv}{Pdr}{Error}{Error}{Walltime}{Median}{29.37152194975}%
\StoreBenchExecResult{PdrInv}{Pdr}{Error}{Error}{Walltime}{Min}{1.56016206741}%
\StoreBenchExecResult{PdrInv}{Pdr}{Error}{Error}{Walltime}{Max}{798.802060127}%
\StoreBenchExecResult{PdrInv}{Pdr}{Error}{Error}{Walltime}{Stdev}{151.1174153984642685468035899}%
\StoreBenchExecResult{PdrInv}{Pdr}{Error}{Exception}{Count}{}{18}%
\StoreBenchExecResult{PdrInv}{Pdr}{Error}{Exception}{Cputime}{}{5268.824620302}%
\StoreBenchExecResult{PdrInv}{Pdr}{Error}{Exception}{Cputime}{Avg}{292.7124789056666666666666667}%
\StoreBenchExecResult{PdrInv}{Pdr}{Error}{Exception}{Cputime}{Median}{494.042621287}%
\StoreBenchExecResult{PdrInv}{Pdr}{Error}{Exception}{Cputime}{Min}{5.292909095}%
\StoreBenchExecResult{PdrInv}{Pdr}{Error}{Exception}{Cputime}{Max}{530.049401185}%
\StoreBenchExecResult{PdrInv}{Pdr}{Error}{Exception}{Cputime}{Stdev}{241.6097832827361551812955077}%
\StoreBenchExecResult{PdrInv}{Pdr}{Error}{Exception}{Walltime}{}{4400.27697825418}%
\StoreBenchExecResult{PdrInv}{Pdr}{Error}{Exception}{Walltime}{Avg}{244.4598321252322222222222222}%
\StoreBenchExecResult{PdrInv}{Pdr}{Error}{Exception}{Walltime}{Median}{413.4257229565}%
\StoreBenchExecResult{PdrInv}{Pdr}{Error}{Exception}{Walltime}{Min}{2.82928800583}%
\StoreBenchExecResult{PdrInv}{Pdr}{Error}{Exception}{Walltime}{Max}{447.364850998}%
\StoreBenchExecResult{PdrInv}{Pdr}{Error}{Exception}{Walltime}{Stdev}{205.7687340844525534015656748}%
\StoreBenchExecResult{PdrInv}{Pdr}{Error}{OutOfJavaMemory}{Count}{}{12}%
\StoreBenchExecResult{PdrInv}{Pdr}{Error}{OutOfJavaMemory}{Cputime}{}{5950.850835172}%
\StoreBenchExecResult{PdrInv}{Pdr}{Error}{OutOfJavaMemory}{Cputime}{Avg}{495.9042362643333333333333333}%
\StoreBenchExecResult{PdrInv}{Pdr}{Error}{OutOfJavaMemory}{Cputime}{Median}{516.10543668}%
\StoreBenchExecResult{PdrInv}{Pdr}{Error}{OutOfJavaMemory}{Cputime}{Min}{237.563009364}%
\StoreBenchExecResult{PdrInv}{Pdr}{Error}{OutOfJavaMemory}{Cputime}{Max}{765.693484479}%
\StoreBenchExecResult{PdrInv}{Pdr}{Error}{OutOfJavaMemory}{Cputime}{Stdev}{129.1659620274220061287899096}%
\StoreBenchExecResult{PdrInv}{Pdr}{Error}{OutOfJavaMemory}{Walltime}{}{4028.433263302}%
\StoreBenchExecResult{PdrInv}{Pdr}{Error}{OutOfJavaMemory}{Walltime}{Avg}{335.7027719418333333333333333}%
\StoreBenchExecResult{PdrInv}{Pdr}{Error}{OutOfJavaMemory}{Walltime}{Median}{352.9782434705}%
\StoreBenchExecResult{PdrInv}{Pdr}{Error}{OutOfJavaMemory}{Walltime}{Min}{153.061067104}%
\StoreBenchExecResult{PdrInv}{Pdr}{Error}{OutOfJavaMemory}{Walltime}{Max}{520.892397165}%
\StoreBenchExecResult{PdrInv}{Pdr}{Error}{OutOfJavaMemory}{Walltime}{Stdev}{88.10613758371058978565393764}%
\StoreBenchExecResult{PdrInv}{Pdr}{Error}{OutOfMemory}{Count}{}{23}%
\StoreBenchExecResult{PdrInv}{Pdr}{Error}{OutOfMemory}{Cputime}{}{3831.932666986}%
\StoreBenchExecResult{PdrInv}{Pdr}{Error}{OutOfMemory}{Cputime}{Avg}{166.6057681298260869565217391}%
\StoreBenchExecResult{PdrInv}{Pdr}{Error}{OutOfMemory}{Cputime}{Median}{138.303607833}%
\StoreBenchExecResult{PdrInv}{Pdr}{Error}{OutOfMemory}{Cputime}{Min}{129.325401862}%
\StoreBenchExecResult{PdrInv}{Pdr}{Error}{OutOfMemory}{Cputime}{Max}{766.707137278}%
\StoreBenchExecResult{PdrInv}{Pdr}{Error}{OutOfMemory}{Cputime}{Stdev}{128.2491389047114038529655436}%
\StoreBenchExecResult{PdrInv}{Pdr}{Error}{OutOfMemory}{Walltime}{}{3299.771935939}%
\StoreBenchExecResult{PdrInv}{Pdr}{Error}{OutOfMemory}{Walltime}{Avg}{143.4683450408260869565217391}%
\StoreBenchExecResult{PdrInv}{Pdr}{Error}{OutOfMemory}{Walltime}{Median}{120.053154945}%
\StoreBenchExecResult{PdrInv}{Pdr}{Error}{OutOfMemory}{Walltime}{Min}{114.130378008}%
\StoreBenchExecResult{PdrInv}{Pdr}{Error}{OutOfMemory}{Walltime}{Max}{616.824223042}%
\StoreBenchExecResult{PdrInv}{Pdr}{Error}{OutOfMemory}{Walltime}{Stdev}{101.2124483211764178815909650}%
\StoreBenchExecResult{PdrInv}{Pdr}{Error}{SegmentationFault}{Count}{}{6}%
\StoreBenchExecResult{PdrInv}{Pdr}{Error}{SegmentationFault}{Cputime}{}{2707.065732113}%
\StoreBenchExecResult{PdrInv}{Pdr}{Error}{SegmentationFault}{Cputime}{Avg}{451.1776220188333333333333333}%
\StoreBenchExecResult{PdrInv}{Pdr}{Error}{SegmentationFault}{Cputime}{Median}{470.291700992}%
\StoreBenchExecResult{PdrInv}{Pdr}{Error}{SegmentationFault}{Cputime}{Min}{16.843765714}%
\StoreBenchExecResult{PdrInv}{Pdr}{Error}{SegmentationFault}{Cputime}{Max}{722.529932062}%
\StoreBenchExecResult{PdrInv}{Pdr}{Error}{SegmentationFault}{Cputime}{Stdev}{227.9577281776415075483372814}%
\StoreBenchExecResult{PdrInv}{Pdr}{Error}{SegmentationFault}{Walltime}{}{2658.2326490875}%
\StoreBenchExecResult{PdrInv}{Pdr}{Error}{SegmentationFault}{Walltime}{Avg}{443.0387748479166666666666667}%
\StoreBenchExecResult{PdrInv}{Pdr}{Error}{SegmentationFault}{Walltime}{Median}{456.040765047}%
\StoreBenchExecResult{PdrInv}{Pdr}{Error}{SegmentationFault}{Walltime}{Min}{13.3237121105}%
\StoreBenchExecResult{PdrInv}{Pdr}{Error}{SegmentationFault}{Walltime}{Max}{715.523455858}%
\StoreBenchExecResult{PdrInv}{Pdr}{Error}{SegmentationFault}{Walltime}{Stdev}{226.7907989938180545113376746}%
\StoreBenchExecResult{PdrInv}{Pdr}{Error}{Timeout}{Count}{}{3982}%
\StoreBenchExecResult{PdrInv}{Pdr}{Error}{Timeout}{Cputime}{}{3598261.357652370}%
\StoreBenchExecResult{PdrInv}{Pdr}{Error}{Timeout}{Cputime}{Avg}{903.6316819820115519839276745}%
\StoreBenchExecResult{PdrInv}{Pdr}{Error}{Timeout}{Cputime}{Median}{902.3555764315}%
\StoreBenchExecResult{PdrInv}{Pdr}{Error}{Timeout}{Cputime}{Min}{900.73232011}%
\StoreBenchExecResult{PdrInv}{Pdr}{Error}{Timeout}{Cputime}{Max}{1001.0468483}%
\StoreBenchExecResult{PdrInv}{Pdr}{Error}{Timeout}{Cputime}{Stdev}{5.131053514807181973900558664}%
\StoreBenchExecResult{PdrInv}{Pdr}{Error}{Timeout}{Walltime}{}{3447228.182135099}%
\StoreBenchExecResult{PdrInv}{Pdr}{Error}{Timeout}{Walltime}{Avg}{865.7027077185080361627322953}%
\StoreBenchExecResult{PdrInv}{Pdr}{Error}{Timeout}{Walltime}{Median}{890.882922530}%
\StoreBenchExecResult{PdrInv}{Pdr}{Error}{Timeout}{Walltime}{Min}{486.869479895}%
\StoreBenchExecResult{PdrInv}{Pdr}{Error}{Timeout}{Walltime}{Max}{908.306071997}%
\StoreBenchExecResult{PdrInv}{Pdr}{Error}{Timeout}{Walltime}{Stdev}{63.59775691470617237615370222}%
\StoreBenchExecResult{PdrInv}{Pdr}{Wrong}{}{Count}{}{1}%
\StoreBenchExecResult{PdrInv}{Pdr}{Wrong}{}{Cputime}{}{3.409806827}%
\StoreBenchExecResult{PdrInv}{Pdr}{Wrong}{}{Cputime}{Avg}{3.409806827}%
\StoreBenchExecResult{PdrInv}{Pdr}{Wrong}{}{Cputime}{Median}{3.409806827}%
\StoreBenchExecResult{PdrInv}{Pdr}{Wrong}{}{Cputime}{Min}{3.409806827}%
\StoreBenchExecResult{PdrInv}{Pdr}{Wrong}{}{Cputime}{Max}{3.409806827}%
\StoreBenchExecResult{PdrInv}{Pdr}{Wrong}{}{Cputime}{Stdev}{0E-9}%
\StoreBenchExecResult{PdrInv}{Pdr}{Wrong}{}{Walltime}{}{1.93630504608}%
\StoreBenchExecResult{PdrInv}{Pdr}{Wrong}{}{Walltime}{Avg}{1.93630504608}%
\StoreBenchExecResult{PdrInv}{Pdr}{Wrong}{}{Walltime}{Median}{1.93630504608}%
\StoreBenchExecResult{PdrInv}{Pdr}{Wrong}{}{Walltime}{Min}{1.93630504608}%
\StoreBenchExecResult{PdrInv}{Pdr}{Wrong}{}{Walltime}{Max}{1.93630504608}%
\StoreBenchExecResult{PdrInv}{Pdr}{Wrong}{}{Walltime}{Stdev}{0E-11}%
\StoreBenchExecResult{PdrInv}{Pdr}{Wrong}{False}{Count}{}{1}%
\StoreBenchExecResult{PdrInv}{Pdr}{Wrong}{False}{Cputime}{}{3.409806827}%
\StoreBenchExecResult{PdrInv}{Pdr}{Wrong}{False}{Cputime}{Avg}{3.409806827}%
\StoreBenchExecResult{PdrInv}{Pdr}{Wrong}{False}{Cputime}{Median}{3.409806827}%
\StoreBenchExecResult{PdrInv}{Pdr}{Wrong}{False}{Cputime}{Min}{3.409806827}%
\StoreBenchExecResult{PdrInv}{Pdr}{Wrong}{False}{Cputime}{Max}{3.409806827}%
\StoreBenchExecResult{PdrInv}{Pdr}{Wrong}{False}{Cputime}{Stdev}{0E-9}%
\StoreBenchExecResult{PdrInv}{Pdr}{Wrong}{False}{Walltime}{}{1.93630504608}%
\StoreBenchExecResult{PdrInv}{Pdr}{Wrong}{False}{Walltime}{Avg}{1.93630504608}%
\StoreBenchExecResult{PdrInv}{Pdr}{Wrong}{False}{Walltime}{Median}{1.93630504608}%
\StoreBenchExecResult{PdrInv}{Pdr}{Wrong}{False}{Walltime}{Min}{1.93630504608}%
\StoreBenchExecResult{PdrInv}{Pdr}{Wrong}{False}{Walltime}{Max}{1.93630504608}%
\StoreBenchExecResult{PdrInv}{Pdr}{Wrong}{False}{Walltime}{Stdev}{0E-11}%
\providecommand\StoreBenchExecResult[7]{\expandafter\newcommand\csname#1#2#3#4#5#6\endcsname{#7}}%
\StoreBenchExecResult{Seahorn}{SeahornTrueNotSolvedByKinductionPlainButKipdr}{Total}{}{Count}{}{449}%
\StoreBenchExecResult{Seahorn}{SeahornTrueNotSolvedByKinductionPlainButKipdr}{Total}{}{Cputime}{}{1578.114586736}%
\StoreBenchExecResult{Seahorn}{SeahornTrueNotSolvedByKinductionPlainButKipdr}{Total}{}{Cputime}{Avg}{3.514731819011135857461024499}%
\StoreBenchExecResult{Seahorn}{SeahornTrueNotSolvedByKinductionPlainButKipdr}{Total}{}{Cputime}{Median}{0.473701265}%
\StoreBenchExecResult{Seahorn}{SeahornTrueNotSolvedByKinductionPlainButKipdr}{Total}{}{Cputime}{Min}{0.241837657}%
\StoreBenchExecResult{Seahorn}{SeahornTrueNotSolvedByKinductionPlainButKipdr}{Total}{}{Cputime}{Max}{1001.86650094}%
\StoreBenchExecResult{Seahorn}{SeahornTrueNotSolvedByKinductionPlainButKipdr}{Total}{}{Cputime}{Stdev}{48.73478440179888426613573047}%
\StoreBenchExecResult{Seahorn}{SeahornTrueNotSolvedByKinductionPlainButKipdr}{Total}{}{Walltime}{}{856.079966306089}%
\StoreBenchExecResult{Seahorn}{SeahornTrueNotSolvedByKinductionPlainButKipdr}{Total}{}{Walltime}{Avg}{1.906636896004652561247216036}%
\StoreBenchExecResult{Seahorn}{SeahornTrueNotSolvedByKinductionPlainButKipdr}{Total}{}{Walltime}{Median}{0.257434844971}%
\StoreBenchExecResult{Seahorn}{SeahornTrueNotSolvedByKinductionPlainButKipdr}{Total}{}{Walltime}{Min}{0.141296863556}%
\StoreBenchExecResult{Seahorn}{SeahornTrueNotSolvedByKinductionPlainButKipdr}{Total}{}{Walltime}{Max}{501.004492998}%
\StoreBenchExecResult{Seahorn}{SeahornTrueNotSolvedByKinductionPlainButKipdr}{Total}{}{Walltime}{Stdev}{24.78931321719413176913688904}%
\StoreBenchExecResult{Seahorn}{SeahornTrueNotSolvedByKinductionPlainButKipdr}{Correct}{}{Count}{}{448}%
\StoreBenchExecResult{Seahorn}{SeahornTrueNotSolvedByKinductionPlainButKipdr}{Correct}{}{Cputime}{}{576.248085796}%
\StoreBenchExecResult{Seahorn}{SeahornTrueNotSolvedByKinductionPlainButKipdr}{Correct}{}{Cputime}{Avg}{1.286268048651785714285714286}%
\StoreBenchExecResult{Seahorn}{SeahornTrueNotSolvedByKinductionPlainButKipdr}{Correct}{}{Cputime}{Median}{0.473601278}%
\StoreBenchExecResult{Seahorn}{SeahornTrueNotSolvedByKinductionPlainButKipdr}{Correct}{}{Cputime}{Min}{0.241837657}%
\StoreBenchExecResult{Seahorn}{SeahornTrueNotSolvedByKinductionPlainButKipdr}{Correct}{}{Cputime}{Max}{235.332648727}%
\StoreBenchExecResult{Seahorn}{SeahornTrueNotSolvedByKinductionPlainButKipdr}{Correct}{}{Cputime}{Stdev}{12.27289405352377016197657340}%
\StoreBenchExecResult{Seahorn}{SeahornTrueNotSolvedByKinductionPlainButKipdr}{Correct}{}{Walltime}{}{355.075473308089}%
\StoreBenchExecResult{Seahorn}{SeahornTrueNotSolvedByKinductionPlainButKipdr}{Correct}{}{Walltime}{Avg}{0.7925791814912700892857142857}%
\StoreBenchExecResult{Seahorn}{SeahornTrueNotSolvedByKinductionPlainButKipdr}{Correct}{}{Walltime}{Median}{0.2574124336245}%
\StoreBenchExecResult{Seahorn}{SeahornTrueNotSolvedByKinductionPlainButKipdr}{Correct}{}{Walltime}{Min}{0.141296863556}%
\StoreBenchExecResult{Seahorn}{SeahornTrueNotSolvedByKinductionPlainButKipdr}{Correct}{}{Walltime}{Max}{117.695271015}%
\StoreBenchExecResult{Seahorn}{SeahornTrueNotSolvedByKinductionPlainButKipdr}{Correct}{}{Walltime}{Stdev}{7.656159776260562842673104847}%
\StoreBenchExecResult{Seahorn}{SeahornTrueNotSolvedByKinductionPlainButKipdr}{Correct}{True}{Count}{}{448}%
\StoreBenchExecResult{Seahorn}{SeahornTrueNotSolvedByKinductionPlainButKipdr}{Correct}{True}{Cputime}{}{576.248085796}%
\StoreBenchExecResult{Seahorn}{SeahornTrueNotSolvedByKinductionPlainButKipdr}{Correct}{True}{Cputime}{Avg}{1.286268048651785714285714286}%
\StoreBenchExecResult{Seahorn}{SeahornTrueNotSolvedByKinductionPlainButKipdr}{Correct}{True}{Cputime}{Median}{0.473601278}%
\StoreBenchExecResult{Seahorn}{SeahornTrueNotSolvedByKinductionPlainButKipdr}{Correct}{True}{Cputime}{Min}{0.241837657}%
\StoreBenchExecResult{Seahorn}{SeahornTrueNotSolvedByKinductionPlainButKipdr}{Correct}{True}{Cputime}{Max}{235.332648727}%
\StoreBenchExecResult{Seahorn}{SeahornTrueNotSolvedByKinductionPlainButKipdr}{Correct}{True}{Cputime}{Stdev}{12.27289405352377016197657340}%
\StoreBenchExecResult{Seahorn}{SeahornTrueNotSolvedByKinductionPlainButKipdr}{Correct}{True}{Walltime}{}{355.075473308089}%
\StoreBenchExecResult{Seahorn}{SeahornTrueNotSolvedByKinductionPlainButKipdr}{Correct}{True}{Walltime}{Avg}{0.7925791814912700892857142857}%
\StoreBenchExecResult{Seahorn}{SeahornTrueNotSolvedByKinductionPlainButKipdr}{Correct}{True}{Walltime}{Median}{0.2574124336245}%
\StoreBenchExecResult{Seahorn}{SeahornTrueNotSolvedByKinductionPlainButKipdr}{Correct}{True}{Walltime}{Min}{0.141296863556}%
\StoreBenchExecResult{Seahorn}{SeahornTrueNotSolvedByKinductionPlainButKipdr}{Correct}{True}{Walltime}{Max}{117.695271015}%
\StoreBenchExecResult{Seahorn}{SeahornTrueNotSolvedByKinductionPlainButKipdr}{Correct}{True}{Walltime}{Stdev}{7.656159776260562842673104847}%
\StoreBenchExecResult{Seahorn}{SeahornTrueNotSolvedByKinductionPlainButKipdr}{Wrong}{True}{Count}{}{0}%
\StoreBenchExecResult{Seahorn}{SeahornTrueNotSolvedByKinductionPlainButKipdr}{Wrong}{True}{Cputime}{}{0}%
\StoreBenchExecResult{Seahorn}{SeahornTrueNotSolvedByKinductionPlainButKipdr}{Wrong}{True}{Cputime}{Avg}{None}%
\StoreBenchExecResult{Seahorn}{SeahornTrueNotSolvedByKinductionPlainButKipdr}{Wrong}{True}{Cputime}{Median}{None}%
\StoreBenchExecResult{Seahorn}{SeahornTrueNotSolvedByKinductionPlainButKipdr}{Wrong}{True}{Cputime}{Min}{None}%
\StoreBenchExecResult{Seahorn}{SeahornTrueNotSolvedByKinductionPlainButKipdr}{Wrong}{True}{Cputime}{Max}{None}%
\StoreBenchExecResult{Seahorn}{SeahornTrueNotSolvedByKinductionPlainButKipdr}{Wrong}{True}{Cputime}{Stdev}{None}%
\StoreBenchExecResult{Seahorn}{SeahornTrueNotSolvedByKinductionPlainButKipdr}{Wrong}{True}{Walltime}{}{0}%
\StoreBenchExecResult{Seahorn}{SeahornTrueNotSolvedByKinductionPlainButKipdr}{Wrong}{True}{Walltime}{Avg}{None}%
\StoreBenchExecResult{Seahorn}{SeahornTrueNotSolvedByKinductionPlainButKipdr}{Wrong}{True}{Walltime}{Median}{None}%
\StoreBenchExecResult{Seahorn}{SeahornTrueNotSolvedByKinductionPlainButKipdr}{Wrong}{True}{Walltime}{Min}{None}%
\StoreBenchExecResult{Seahorn}{SeahornTrueNotSolvedByKinductionPlainButKipdr}{Wrong}{True}{Walltime}{Max}{None}%
\StoreBenchExecResult{Seahorn}{SeahornTrueNotSolvedByKinductionPlainButKipdr}{Wrong}{True}{Walltime}{Stdev}{None}%
\StoreBenchExecResult{Seahorn}{SeahornTrueNotSolvedByKinductionPlainButKipdr}{Error}{}{Count}{}{1}%
\StoreBenchExecResult{Seahorn}{SeahornTrueNotSolvedByKinductionPlainButKipdr}{Error}{}{Cputime}{}{1001.86650094}%
\StoreBenchExecResult{Seahorn}{SeahornTrueNotSolvedByKinductionPlainButKipdr}{Error}{}{Cputime}{Avg}{1001.86650094}%
\StoreBenchExecResult{Seahorn}{SeahornTrueNotSolvedByKinductionPlainButKipdr}{Error}{}{Cputime}{Median}{1001.86650094}%
\StoreBenchExecResult{Seahorn}{SeahornTrueNotSolvedByKinductionPlainButKipdr}{Error}{}{Cputime}{Min}{1001.86650094}%
\StoreBenchExecResult{Seahorn}{SeahornTrueNotSolvedByKinductionPlainButKipdr}{Error}{}{Cputime}{Max}{1001.86650094}%
\StoreBenchExecResult{Seahorn}{SeahornTrueNotSolvedByKinductionPlainButKipdr}{Error}{}{Cputime}{Stdev}{0E-8}%
\StoreBenchExecResult{Seahorn}{SeahornTrueNotSolvedByKinductionPlainButKipdr}{Error}{}{Walltime}{}{501.004492998}%
\StoreBenchExecResult{Seahorn}{SeahornTrueNotSolvedByKinductionPlainButKipdr}{Error}{}{Walltime}{Avg}{501.004492998}%
\StoreBenchExecResult{Seahorn}{SeahornTrueNotSolvedByKinductionPlainButKipdr}{Error}{}{Walltime}{Median}{501.004492998}%
\StoreBenchExecResult{Seahorn}{SeahornTrueNotSolvedByKinductionPlainButKipdr}{Error}{}{Walltime}{Min}{501.004492998}%
\StoreBenchExecResult{Seahorn}{SeahornTrueNotSolvedByKinductionPlainButKipdr}{Error}{}{Walltime}{Max}{501.004492998}%
\StoreBenchExecResult{Seahorn}{SeahornTrueNotSolvedByKinductionPlainButKipdr}{Error}{}{Walltime}{Stdev}{0E-9}%
\StoreBenchExecResult{Seahorn}{SeahornTrueNotSolvedByKinductionPlainButKipdr}{Error}{Timeout}{Count}{}{1}%
\StoreBenchExecResult{Seahorn}{SeahornTrueNotSolvedByKinductionPlainButKipdr}{Error}{Timeout}{Cputime}{}{1001.86650094}%
\StoreBenchExecResult{Seahorn}{SeahornTrueNotSolvedByKinductionPlainButKipdr}{Error}{Timeout}{Cputime}{Avg}{1001.86650094}%
\StoreBenchExecResult{Seahorn}{SeahornTrueNotSolvedByKinductionPlainButKipdr}{Error}{Timeout}{Cputime}{Median}{1001.86650094}%
\StoreBenchExecResult{Seahorn}{SeahornTrueNotSolvedByKinductionPlainButKipdr}{Error}{Timeout}{Cputime}{Min}{1001.86650094}%
\StoreBenchExecResult{Seahorn}{SeahornTrueNotSolvedByKinductionPlainButKipdr}{Error}{Timeout}{Cputime}{Max}{1001.86650094}%
\StoreBenchExecResult{Seahorn}{SeahornTrueNotSolvedByKinductionPlainButKipdr}{Error}{Timeout}{Cputime}{Stdev}{0E-8}%
\StoreBenchExecResult{Seahorn}{SeahornTrueNotSolvedByKinductionPlainButKipdr}{Error}{Timeout}{Walltime}{}{501.004492998}%
\StoreBenchExecResult{Seahorn}{SeahornTrueNotSolvedByKinductionPlainButKipdr}{Error}{Timeout}{Walltime}{Avg}{501.004492998}%
\StoreBenchExecResult{Seahorn}{SeahornTrueNotSolvedByKinductionPlainButKipdr}{Error}{Timeout}{Walltime}{Median}{501.004492998}%
\StoreBenchExecResult{Seahorn}{SeahornTrueNotSolvedByKinductionPlainButKipdr}{Error}{Timeout}{Walltime}{Min}{501.004492998}%
\StoreBenchExecResult{Seahorn}{SeahornTrueNotSolvedByKinductionPlainButKipdr}{Error}{Timeout}{Walltime}{Max}{501.004492998}%
\StoreBenchExecResult{Seahorn}{SeahornTrueNotSolvedByKinductionPlainButKipdr}{Error}{Timeout}{Walltime}{Stdev}{0E-9}%
\providecommand\StoreBenchExecResult[7]{\expandafter\newcommand\csname#1#2#3#4#5#6\endcsname{#7}}%
\StoreBenchExecResult{Seahorn}{SeahornTrueNotSolvedByKinductionPlain}{Total}{}{Count}{}{2893}%
\StoreBenchExecResult{Seahorn}{SeahornTrueNotSolvedByKinductionPlain}{Total}{}{Cputime}{}{949303.496433844}%
\StoreBenchExecResult{Seahorn}{SeahornTrueNotSolvedByKinductionPlain}{Total}{}{Cputime}{Avg}{328.1380907133923263048738334}%
\StoreBenchExecResult{Seahorn}{SeahornTrueNotSolvedByKinductionPlain}{Total}{}{Cputime}{Median}{8.184512613}%
\StoreBenchExecResult{Seahorn}{SeahornTrueNotSolvedByKinductionPlain}{Total}{}{Cputime}{Min}{0.019925741}%
\StoreBenchExecResult{Seahorn}{SeahornTrueNotSolvedByKinductionPlain}{Total}{}{Cputime}{Max}{1001.94246017}%
\StoreBenchExecResult{Seahorn}{SeahornTrueNotSolvedByKinductionPlain}{Total}{}{Cputime}{Stdev}{443.3752776185665212088371950}%
\StoreBenchExecResult{Seahorn}{SeahornTrueNotSolvedByKinductionPlain}{Total}{}{Walltime}{}{734832.6836233138459}%
\StoreBenchExecResult{Seahorn}{SeahornTrueNotSolvedByKinductionPlain}{Total}{}{Walltime}{Avg}{254.0036929219888855513307985}%
\StoreBenchExecResult{Seahorn}{SeahornTrueNotSolvedByKinductionPlain}{Total}{}{Walltime}{Median}{5.79015994072}%
\StoreBenchExecResult{Seahorn}{SeahornTrueNotSolvedByKinductionPlain}{Total}{}{Walltime}{Min}{0.0202729701996}%
\StoreBenchExecResult{Seahorn}{SeahornTrueNotSolvedByKinductionPlain}{Total}{}{Walltime}{Max}{1000.61071706}%
\StoreBenchExecResult{Seahorn}{SeahornTrueNotSolvedByKinductionPlain}{Total}{}{Walltime}{Stdev}{364.2190316686873064964586622}%
\StoreBenchExecResult{Seahorn}{SeahornTrueNotSolvedByKinductionPlain}{Correct}{}{Count}{}{1683}%
\StoreBenchExecResult{Seahorn}{SeahornTrueNotSolvedByKinductionPlain}{Correct}{}{Cputime}{}{60776.974655244}%
\StoreBenchExecResult{Seahorn}{SeahornTrueNotSolvedByKinductionPlain}{Correct}{}{Cputime}{Avg}{36.11228440596791443850267380}%
\StoreBenchExecResult{Seahorn}{SeahornTrueNotSolvedByKinductionPlain}{Correct}{}{Cputime}{Median}{0.788996236}%
\StoreBenchExecResult{Seahorn}{SeahornTrueNotSolvedByKinductionPlain}{Correct}{}{Cputime}{Min}{0.239288082}%
\StoreBenchExecResult{Seahorn}{SeahornTrueNotSolvedByKinductionPlain}{Correct}{}{Cputime}{Max}{891.349400462}%
\StoreBenchExecResult{Seahorn}{SeahornTrueNotSolvedByKinductionPlain}{Correct}{}{Cputime}{Stdev}{122.3603871516301405396347694}%
\StoreBenchExecResult{Seahorn}{SeahornTrueNotSolvedByKinductionPlain}{Correct}{}{Walltime}{}{46351.430865996280}%
\StoreBenchExecResult{Seahorn}{SeahornTrueNotSolvedByKinductionPlain}{Correct}{}{Walltime}{Avg}{27.54095713962939988116458705}%
\StoreBenchExecResult{Seahorn}{SeahornTrueNotSolvedByKinductionPlain}{Correct}{}{Walltime}{Median}{0.432631969452}%
\StoreBenchExecResult{Seahorn}{SeahornTrueNotSolvedByKinductionPlain}{Correct}{}{Walltime}{Min}{0.138706207275}%
\StoreBenchExecResult{Seahorn}{SeahornTrueNotSolvedByKinductionPlain}{Correct}{}{Walltime}{Max}{871.729703903}%
\StoreBenchExecResult{Seahorn}{SeahornTrueNotSolvedByKinductionPlain}{Correct}{}{Walltime}{Stdev}{93.95944650280780955168753332}%
\StoreBenchExecResult{Seahorn}{SeahornTrueNotSolvedByKinductionPlain}{Correct}{True}{Count}{}{1683}%
\StoreBenchExecResult{Seahorn}{SeahornTrueNotSolvedByKinductionPlain}{Correct}{True}{Cputime}{}{60776.974655244}%
\StoreBenchExecResult{Seahorn}{SeahornTrueNotSolvedByKinductionPlain}{Correct}{True}{Cputime}{Avg}{36.11228440596791443850267380}%
\StoreBenchExecResult{Seahorn}{SeahornTrueNotSolvedByKinductionPlain}{Correct}{True}{Cputime}{Median}{0.788996236}%
\StoreBenchExecResult{Seahorn}{SeahornTrueNotSolvedByKinductionPlain}{Correct}{True}{Cputime}{Min}{0.239288082}%
\StoreBenchExecResult{Seahorn}{SeahornTrueNotSolvedByKinductionPlain}{Correct}{True}{Cputime}{Max}{891.349400462}%
\StoreBenchExecResult{Seahorn}{SeahornTrueNotSolvedByKinductionPlain}{Correct}{True}{Cputime}{Stdev}{122.3603871516301405396347694}%
\StoreBenchExecResult{Seahorn}{SeahornTrueNotSolvedByKinductionPlain}{Correct}{True}{Walltime}{}{46351.430865996280}%
\StoreBenchExecResult{Seahorn}{SeahornTrueNotSolvedByKinductionPlain}{Correct}{True}{Walltime}{Avg}{27.54095713962939988116458705}%
\StoreBenchExecResult{Seahorn}{SeahornTrueNotSolvedByKinductionPlain}{Correct}{True}{Walltime}{Median}{0.432631969452}%
\StoreBenchExecResult{Seahorn}{SeahornTrueNotSolvedByKinductionPlain}{Correct}{True}{Walltime}{Min}{0.138706207275}%
\StoreBenchExecResult{Seahorn}{SeahornTrueNotSolvedByKinductionPlain}{Correct}{True}{Walltime}{Max}{871.729703903}%
\StoreBenchExecResult{Seahorn}{SeahornTrueNotSolvedByKinductionPlain}{Correct}{True}{Walltime}{Stdev}{93.95944650280780955168753332}%
\StoreBenchExecResult{Seahorn}{SeahornTrueNotSolvedByKinductionPlain}{Wrong}{True}{Count}{}{0}%
\StoreBenchExecResult{Seahorn}{SeahornTrueNotSolvedByKinductionPlain}{Wrong}{True}{Cputime}{}{0}%
\StoreBenchExecResult{Seahorn}{SeahornTrueNotSolvedByKinductionPlain}{Wrong}{True}{Cputime}{Avg}{None}%
\StoreBenchExecResult{Seahorn}{SeahornTrueNotSolvedByKinductionPlain}{Wrong}{True}{Cputime}{Median}{None}%
\StoreBenchExecResult{Seahorn}{SeahornTrueNotSolvedByKinductionPlain}{Wrong}{True}{Cputime}{Min}{None}%
\StoreBenchExecResult{Seahorn}{SeahornTrueNotSolvedByKinductionPlain}{Wrong}{True}{Cputime}{Max}{None}%
\StoreBenchExecResult{Seahorn}{SeahornTrueNotSolvedByKinductionPlain}{Wrong}{True}{Cputime}{Stdev}{None}%
\StoreBenchExecResult{Seahorn}{SeahornTrueNotSolvedByKinductionPlain}{Wrong}{True}{Walltime}{}{0}%
\StoreBenchExecResult{Seahorn}{SeahornTrueNotSolvedByKinductionPlain}{Wrong}{True}{Walltime}{Avg}{None}%
\StoreBenchExecResult{Seahorn}{SeahornTrueNotSolvedByKinductionPlain}{Wrong}{True}{Walltime}{Median}{None}%
\StoreBenchExecResult{Seahorn}{SeahornTrueNotSolvedByKinductionPlain}{Wrong}{True}{Walltime}{Min}{None}%
\StoreBenchExecResult{Seahorn}{SeahornTrueNotSolvedByKinductionPlain}{Wrong}{True}{Walltime}{Max}{None}%
\StoreBenchExecResult{Seahorn}{SeahornTrueNotSolvedByKinductionPlain}{Wrong}{True}{Walltime}{Stdev}{None}%
\StoreBenchExecResult{Seahorn}{SeahornTrueNotSolvedByKinductionPlain}{Error}{}{Count}{}{1123}%
\StoreBenchExecResult{Seahorn}{SeahornTrueNotSolvedByKinductionPlain}{Error}{}{Cputime}{}{885893.088814230}%
\StoreBenchExecResult{Seahorn}{SeahornTrueNotSolvedByKinductionPlain}{Error}{}{Cputime}{Avg}{788.8629464062600178094390027}%
\StoreBenchExecResult{Seahorn}{SeahornTrueNotSolvedByKinductionPlain}{Error}{}{Cputime}{Median}{1000.10097824}%
\StoreBenchExecResult{Seahorn}{SeahornTrueNotSolvedByKinductionPlain}{Error}{}{Cputime}{Min}{0.019925741}%
\StoreBenchExecResult{Seahorn}{SeahornTrueNotSolvedByKinductionPlain}{Error}{}{Cputime}{Max}{1001.94246017}%
\StoreBenchExecResult{Seahorn}{SeahornTrueNotSolvedByKinductionPlain}{Error}{}{Cputime}{Stdev}{368.7669710573466940152859289}%
\StoreBenchExecResult{Seahorn}{SeahornTrueNotSolvedByKinductionPlain}{Error}{}{Walltime}{}{685910.5581641265949}%
\StoreBenchExecResult{Seahorn}{SeahornTrueNotSolvedByKinductionPlain}{Error}{}{Walltime}{Avg}{610.7841123456158458593054319}%
\StoreBenchExecResult{Seahorn}{SeahornTrueNotSolvedByKinductionPlain}{Error}{}{Walltime}{Median}{501.004588842}%
\StoreBenchExecResult{Seahorn}{SeahornTrueNotSolvedByKinductionPlain}{Error}{}{Walltime}{Min}{0.0202729701996}%
\StoreBenchExecResult{Seahorn}{SeahornTrueNotSolvedByKinductionPlain}{Error}{}{Walltime}{Max}{1000.61071706}%
\StoreBenchExecResult{Seahorn}{SeahornTrueNotSolvedByKinductionPlain}{Error}{}{Walltime}{Stdev}{345.5493476784825071606643289}%
\StoreBenchExecResult{Seahorn}{SeahornTrueNotSolvedByKinductionPlain}{Error}{Error}{Count}{}{69}%
\StoreBenchExecResult{Seahorn}{SeahornTrueNotSolvedByKinductionPlain}{Error}{Error}{Cputime}{}{6070.704394890}%
\StoreBenchExecResult{Seahorn}{SeahornTrueNotSolvedByKinductionPlain}{Error}{Error}{Cputime}{Avg}{87.98122311434782608695652174}%
\StoreBenchExecResult{Seahorn}{SeahornTrueNotSolvedByKinductionPlain}{Error}{Error}{Cputime}{Median}{29.10440541}%
\StoreBenchExecResult{Seahorn}{SeahornTrueNotSolvedByKinductionPlain}{Error}{Error}{Cputime}{Min}{1.117989155}%
\StoreBenchExecResult{Seahorn}{SeahornTrueNotSolvedByKinductionPlain}{Error}{Error}{Cputime}{Max}{597.055218492}%
\StoreBenchExecResult{Seahorn}{SeahornTrueNotSolvedByKinductionPlain}{Error}{Error}{Cputime}{Stdev}{138.7756197126176534005612460}%
\StoreBenchExecResult{Seahorn}{SeahornTrueNotSolvedByKinductionPlain}{Error}{Error}{Walltime}{}{5998.439326525272}%
\StoreBenchExecResult{Seahorn}{SeahornTrueNotSolvedByKinductionPlain}{Error}{Error}{Walltime}{Avg}{86.93390328297495652173913043}%
\StoreBenchExecResult{Seahorn}{SeahornTrueNotSolvedByKinductionPlain}{Error}{Error}{Walltime}{Median}{27.8889348507}%
\StoreBenchExecResult{Seahorn}{SeahornTrueNotSolvedByKinductionPlain}{Error}{Error}{Walltime}{Min}{0.665085792542}%
\StoreBenchExecResult{Seahorn}{SeahornTrueNotSolvedByKinductionPlain}{Error}{Error}{Walltime}{Max}{596.154284}%
\StoreBenchExecResult{Seahorn}{SeahornTrueNotSolvedByKinductionPlain}{Error}{Error}{Walltime}{Stdev}{138.7680606286361356623176767}%
\StoreBenchExecResult{Seahorn}{SeahornTrueNotSolvedByKinductionPlain}{Error}{Failure}{Count}{}{58}%
\StoreBenchExecResult{Seahorn}{SeahornTrueNotSolvedByKinductionPlain}{Error}{Failure}{Cputime}{}{1.398382791}%
\StoreBenchExecResult{Seahorn}{SeahornTrueNotSolvedByKinductionPlain}{Error}{Failure}{Cputime}{Avg}{0.02411004812068965517241379310}%
\StoreBenchExecResult{Seahorn}{SeahornTrueNotSolvedByKinductionPlain}{Error}{Failure}{Cputime}{Median}{0.021358194}%
\StoreBenchExecResult{Seahorn}{SeahornTrueNotSolvedByKinductionPlain}{Error}{Failure}{Cputime}{Min}{0.019925741}%
\StoreBenchExecResult{Seahorn}{SeahornTrueNotSolvedByKinductionPlain}{Error}{Failure}{Cputime}{Max}{0.051342763}%
\StoreBenchExecResult{Seahorn}{SeahornTrueNotSolvedByKinductionPlain}{Error}{Failure}{Cputime}{Stdev}{0.006540096589307971474133625633}%
\StoreBenchExecResult{Seahorn}{SeahornTrueNotSolvedByKinductionPlain}{Error}{Failure}{Walltime}{}{1.4120249748229}%
\StoreBenchExecResult{Seahorn}{SeahornTrueNotSolvedByKinductionPlain}{Error}{Failure}{Walltime}{Avg}{0.02434525818660172413793103448}%
\StoreBenchExecResult{Seahorn}{SeahornTrueNotSolvedByKinductionPlain}{Error}{Failure}{Walltime}{Median}{0.0216649770737}%
\StoreBenchExecResult{Seahorn}{SeahornTrueNotSolvedByKinductionPlain}{Error}{Failure}{Walltime}{Min}{0.0202729701996}%
\StoreBenchExecResult{Seahorn}{SeahornTrueNotSolvedByKinductionPlain}{Error}{Failure}{Walltime}{Max}{0.051558971405}%
\StoreBenchExecResult{Seahorn}{SeahornTrueNotSolvedByKinductionPlain}{Error}{Failure}{Walltime}{Stdev}{0.006567936360008522801066992212}%
\StoreBenchExecResult{Seahorn}{SeahornTrueNotSolvedByKinductionPlain}{Error}{OutOfMemory}{Count}{}{184}%
\StoreBenchExecResult{Seahorn}{SeahornTrueNotSolvedByKinductionPlain}{Error}{OutOfMemory}{Cputime}{}{68516.555233517}%
\StoreBenchExecResult{Seahorn}{SeahornTrueNotSolvedByKinductionPlain}{Error}{OutOfMemory}{Cputime}{Avg}{372.3725827908532608695652174}%
\StoreBenchExecResult{Seahorn}{SeahornTrueNotSolvedByKinductionPlain}{Error}{OutOfMemory}{Cputime}{Median}{259.414880485}%
\StoreBenchExecResult{Seahorn}{SeahornTrueNotSolvedByKinductionPlain}{Error}{OutOfMemory}{Cputime}{Min}{39.865694467}%
\StoreBenchExecResult{Seahorn}{SeahornTrueNotSolvedByKinductionPlain}{Error}{OutOfMemory}{Cputime}{Max}{895.393229374}%
\StoreBenchExecResult{Seahorn}{SeahornTrueNotSolvedByKinductionPlain}{Error}{OutOfMemory}{Cputime}{Stdev}{270.7072878729757420136965390}%
\StoreBenchExecResult{Seahorn}{SeahornTrueNotSolvedByKinductionPlain}{Error}{OutOfMemory}{Walltime}{}{66449.9506607105}%
\StoreBenchExecResult{Seahorn}{SeahornTrueNotSolvedByKinductionPlain}{Error}{OutOfMemory}{Walltime}{Avg}{361.1410361995135869565217391}%
\StoreBenchExecResult{Seahorn}{SeahornTrueNotSolvedByKinductionPlain}{Error}{OutOfMemory}{Walltime}{Median}{240.5157344345}%
\StoreBenchExecResult{Seahorn}{SeahornTrueNotSolvedByKinductionPlain}{Error}{OutOfMemory}{Walltime}{Min}{38.3606710434}%
\StoreBenchExecResult{Seahorn}{SeahornTrueNotSolvedByKinductionPlain}{Error}{OutOfMemory}{Walltime}{Max}{894.667023897}%
\StoreBenchExecResult{Seahorn}{SeahornTrueNotSolvedByKinductionPlain}{Error}{OutOfMemory}{Walltime}{Stdev}{273.3728064885876239926892811}%
\StoreBenchExecResult{Seahorn}{SeahornTrueNotSolvedByKinductionPlain}{Error}{Timeout}{Count}{}{812}%
\StoreBenchExecResult{Seahorn}{SeahornTrueNotSolvedByKinductionPlain}{Error}{Timeout}{Cputime}{}{811304.430803032}%
\StoreBenchExecResult{Seahorn}{SeahornTrueNotSolvedByKinductionPlain}{Error}{Timeout}{Cputime}{Avg}{999.1433876884630541871921182}%
\StoreBenchExecResult{Seahorn}{SeahornTrueNotSolvedByKinductionPlain}{Error}{Timeout}{Cputime}{Median}{1000.211214035}%
\StoreBenchExecResult{Seahorn}{SeahornTrueNotSolvedByKinductionPlain}{Error}{Timeout}{Cputime}{Min}{901.616882766}%
\StoreBenchExecResult{Seahorn}{SeahornTrueNotSolvedByKinductionPlain}{Error}{Timeout}{Cputime}{Max}{1001.94246017}%
\StoreBenchExecResult{Seahorn}{SeahornTrueNotSolvedByKinductionPlain}{Error}{Timeout}{Cputime}{Stdev}{10.75357683789759131287730506}%
\StoreBenchExecResult{Seahorn}{SeahornTrueNotSolvedByKinductionPlain}{Error}{Timeout}{Walltime}{}{613460.756151916}%
\StoreBenchExecResult{Seahorn}{SeahornTrueNotSolvedByKinductionPlain}{Error}{Timeout}{Walltime}{Avg}{755.4935420590098522167487685}%
\StoreBenchExecResult{Seahorn}{SeahornTrueNotSolvedByKinductionPlain}{Error}{Timeout}{Walltime}{Median}{946.3787635565}%
\StoreBenchExecResult{Seahorn}{SeahornTrueNotSolvedByKinductionPlain}{Error}{Timeout}{Walltime}{Min}{457.19666481}%
\StoreBenchExecResult{Seahorn}{SeahornTrueNotSolvedByKinductionPlain}{Error}{Timeout}{Walltime}{Max}{1000.61071706}%
\StoreBenchExecResult{Seahorn}{SeahornTrueNotSolvedByKinductionPlain}{Error}{Timeout}{Walltime}{Stdev}{248.0712001068825631382151185}%
\StoreBenchExecResult{Seahorn}{SeahornTrueNotSolvedByKinductionPlain}{Wrong}{}{Count}{}{87}%
\StoreBenchExecResult{Seahorn}{SeahornTrueNotSolvedByKinductionPlain}{Wrong}{}{Cputime}{}{2633.432964370}%
\StoreBenchExecResult{Seahorn}{SeahornTrueNotSolvedByKinductionPlain}{Wrong}{}{Cputime}{Avg}{30.26934441804597701149425287}%
\StoreBenchExecResult{Seahorn}{SeahornTrueNotSolvedByKinductionPlain}{Wrong}{}{Cputime}{Median}{0.547554896}%
\StoreBenchExecResult{Seahorn}{SeahornTrueNotSolvedByKinductionPlain}{Wrong}{}{Cputime}{Min}{0.283909742}%
\StoreBenchExecResult{Seahorn}{SeahornTrueNotSolvedByKinductionPlain}{Wrong}{}{Cputime}{Max}{884.206391999}%
\StoreBenchExecResult{Seahorn}{SeahornTrueNotSolvedByKinductionPlain}{Wrong}{}{Cputime}{Stdev}{116.3057914293032242146288364}%
\StoreBenchExecResult{Seahorn}{SeahornTrueNotSolvedByKinductionPlain}{Wrong}{}{Walltime}{}{2570.694593190971}%
\StoreBenchExecResult{Seahorn}{SeahornTrueNotSolvedByKinductionPlain}{Wrong}{}{Walltime}{Avg}{29.54821371483874712643678161}%
\StoreBenchExecResult{Seahorn}{SeahornTrueNotSolvedByKinductionPlain}{Wrong}{}{Walltime}{Median}{0.311375141144}%
\StoreBenchExecResult{Seahorn}{SeahornTrueNotSolvedByKinductionPlain}{Wrong}{}{Walltime}{Min}{0.163145065308}%
\StoreBenchExecResult{Seahorn}{SeahornTrueNotSolvedByKinductionPlain}{Wrong}{}{Walltime}{Max}{883.923076153}%
\StoreBenchExecResult{Seahorn}{SeahornTrueNotSolvedByKinductionPlain}{Wrong}{}{Walltime}{Stdev}{116.3400459418719684426182232}%
\StoreBenchExecResult{Seahorn}{SeahornTrueNotSolvedByKinductionPlain}{Wrong}{False}{Count}{}{87}%
\StoreBenchExecResult{Seahorn}{SeahornTrueNotSolvedByKinductionPlain}{Wrong}{False}{Cputime}{}{2633.432964370}%
\StoreBenchExecResult{Seahorn}{SeahornTrueNotSolvedByKinductionPlain}{Wrong}{False}{Cputime}{Avg}{30.26934441804597701149425287}%
\StoreBenchExecResult{Seahorn}{SeahornTrueNotSolvedByKinductionPlain}{Wrong}{False}{Cputime}{Median}{0.547554896}%
\StoreBenchExecResult{Seahorn}{SeahornTrueNotSolvedByKinductionPlain}{Wrong}{False}{Cputime}{Min}{0.283909742}%
\StoreBenchExecResult{Seahorn}{SeahornTrueNotSolvedByKinductionPlain}{Wrong}{False}{Cputime}{Max}{884.206391999}%
\StoreBenchExecResult{Seahorn}{SeahornTrueNotSolvedByKinductionPlain}{Wrong}{False}{Cputime}{Stdev}{116.3057914293032242146288364}%
\StoreBenchExecResult{Seahorn}{SeahornTrueNotSolvedByKinductionPlain}{Wrong}{False}{Walltime}{}{2570.694593190971}%
\StoreBenchExecResult{Seahorn}{SeahornTrueNotSolvedByKinductionPlain}{Wrong}{False}{Walltime}{Avg}{29.54821371483874712643678161}%
\StoreBenchExecResult{Seahorn}{SeahornTrueNotSolvedByKinductionPlain}{Wrong}{False}{Walltime}{Median}{0.311375141144}%
\StoreBenchExecResult{Seahorn}{SeahornTrueNotSolvedByKinductionPlain}{Wrong}{False}{Walltime}{Min}{0.163145065308}%
\StoreBenchExecResult{Seahorn}{SeahornTrueNotSolvedByKinductionPlain}{Wrong}{False}{Walltime}{Max}{883.923076153}%
\StoreBenchExecResult{Seahorn}{SeahornTrueNotSolvedByKinductionPlain}{Wrong}{False}{Walltime}{Stdev}{116.3400459418719684426182232}%
\StoreBenchExecResult{Seahorn}{SeahornTrueNotSolvedByKinductionPlain}{Correct}{False}{Count}{}{0}%
\StoreBenchExecResult{Seahorn}{SeahornTrueNotSolvedByKinductionPlain}{Correct}{False}{Cputime}{}{0}%
\StoreBenchExecResult{Seahorn}{SeahornTrueNotSolvedByKinductionPlain}{Correct}{False}{Cputime}{Avg}{None}%
\StoreBenchExecResult{Seahorn}{SeahornTrueNotSolvedByKinductionPlain}{Correct}{False}{Cputime}{Median}{None}%
\StoreBenchExecResult{Seahorn}{SeahornTrueNotSolvedByKinductionPlain}{Correct}{False}{Cputime}{Min}{None}%
\StoreBenchExecResult{Seahorn}{SeahornTrueNotSolvedByKinductionPlain}{Correct}{False}{Cputime}{Max}{None}%
\StoreBenchExecResult{Seahorn}{SeahornTrueNotSolvedByKinductionPlain}{Correct}{False}{Cputime}{Stdev}{None}%
\StoreBenchExecResult{Seahorn}{SeahornTrueNotSolvedByKinductionPlain}{Correct}{False}{Walltime}{}{0}%
\StoreBenchExecResult{Seahorn}{SeahornTrueNotSolvedByKinductionPlain}{Correct}{False}{Walltime}{Avg}{None}%
\StoreBenchExecResult{Seahorn}{SeahornTrueNotSolvedByKinductionPlain}{Correct}{False}{Walltime}{Median}{None}%
\StoreBenchExecResult{Seahorn}{SeahornTrueNotSolvedByKinductionPlain}{Correct}{False}{Walltime}{Min}{None}%
\StoreBenchExecResult{Seahorn}{SeahornTrueNotSolvedByKinductionPlain}{Correct}{False}{Walltime}{Max}{None}%
\StoreBenchExecResult{Seahorn}{SeahornTrueNotSolvedByKinductionPlain}{Correct}{False}{Walltime}{Stdev}{None}%
\providecommand\StoreBenchExecResult[7]{\expandafter\newcommand\csname#1#2#3#4#5#6\endcsname{#7}}%
\StoreBenchExecResult{Seahorn}{Seahorn}{Total}{}{Count}{}{5591}%
\StoreBenchExecResult{Seahorn}{Seahorn}{Total}{}{Cputime}{}{1681346.818685811}%
\StoreBenchExecResult{Seahorn}{Seahorn}{Total}{}{Cputime}{Avg}{300.7238094590969415131461277}%
\StoreBenchExecResult{Seahorn}{Seahorn}{Total}{}{Cputime}{Median}{4.072451462}%
\StoreBenchExecResult{Seahorn}{Seahorn}{Total}{}{Cputime}{Min}{0.019825803}%
\StoreBenchExecResult{Seahorn}{Seahorn}{Total}{}{Cputime}{Max}{1002.11823391}%
\StoreBenchExecResult{Seahorn}{Seahorn}{Total}{}{Cputime}{Stdev}{437.3238673364031623147909285}%
\StoreBenchExecResult{Seahorn}{Seahorn}{Total}{}{Walltime}{}{1181682.1175198614848}%
\StoreBenchExecResult{Seahorn}{Seahorn}{Total}{}{Walltime}{Avg}{211.3543404614311366124128063}%
\StoreBenchExecResult{Seahorn}{Seahorn}{Total}{}{Walltime}{Median}{2.97129702568}%
\StoreBenchExecResult{Seahorn}{Seahorn}{Total}{}{Walltime}{Min}{0.0200588703156}%
\StoreBenchExecResult{Seahorn}{Seahorn}{Total}{}{Walltime}{Max}{1000.70675588}%
\StoreBenchExecResult{Seahorn}{Seahorn}{Total}{}{Walltime}{Stdev}{327.8378503737180463716747136}%
\StoreBenchExecResult{Seahorn}{Seahorn}{Correct}{}{Count}{}{3468}%
\StoreBenchExecResult{Seahorn}{Seahorn}{Correct}{}{Cputime}{}{106036.234361517}%
\StoreBenchExecResult{Seahorn}{Seahorn}{Correct}{}{Cputime}{Avg}{30.57561544449740484429065744}%
\StoreBenchExecResult{Seahorn}{Seahorn}{Correct}{}{Cputime}{Median}{0.886301730}%
\StoreBenchExecResult{Seahorn}{Seahorn}{Correct}{}{Cputime}{Min}{0.228106627}%
\StoreBenchExecResult{Seahorn}{Seahorn}{Correct}{}{Cputime}{Max}{891.349400462}%
\StoreBenchExecResult{Seahorn}{Seahorn}{Correct}{}{Cputime}{Stdev}{109.2120680764612604589895799}%
\StoreBenchExecResult{Seahorn}{Seahorn}{Correct}{}{Walltime}{}{77916.473235363066}%
\StoreBenchExecResult{Seahorn}{Seahorn}{Correct}{}{Walltime}{Avg}{22.46726448539880795847750865}%
\StoreBenchExecResult{Seahorn}{Seahorn}{Correct}{}{Walltime}{Median}{0.4959733486175}%
\StoreBenchExecResult{Seahorn}{Seahorn}{Correct}{}{Walltime}{Min}{0.134125947952}%
\StoreBenchExecResult{Seahorn}{Seahorn}{Correct}{}{Walltime}{Max}{871.729703903}%
\StoreBenchExecResult{Seahorn}{Seahorn}{Correct}{}{Walltime}{Stdev}{83.71282690359337505051319341}%
\StoreBenchExecResult{Seahorn}{Seahorn}{Correct}{False}{Count}{}{744}%
\StoreBenchExecResult{Seahorn}{Seahorn}{Correct}{False}{Cputime}{}{27922.365755521}%
\StoreBenchExecResult{Seahorn}{Seahorn}{Correct}{False}{Cputime}{Avg}{37.53006149935618279569892473}%
\StoreBenchExecResult{Seahorn}{Seahorn}{Correct}{False}{Cputime}{Median}{2.422115510}%
\StoreBenchExecResult{Seahorn}{Seahorn}{Correct}{False}{Cputime}{Min}{0.245356156}%
\StoreBenchExecResult{Seahorn}{Seahorn}{Correct}{False}{Cputime}{Max}{825.873894751}%
\StoreBenchExecResult{Seahorn}{Seahorn}{Correct}{False}{Cputime}{Stdev}{115.8162565058359750761109818}%
\StoreBenchExecResult{Seahorn}{Seahorn}{Correct}{False}{Walltime}{}{22831.164708854083}%
\StoreBenchExecResult{Seahorn}{Seahorn}{Correct}{False}{Walltime}{Avg}{30.68704933985763844086021505}%
\StoreBenchExecResult{Seahorn}{Seahorn}{Correct}{False}{Walltime}{Median}{1.93488752842}%
\StoreBenchExecResult{Seahorn}{Seahorn}{Correct}{False}{Walltime}{Min}{0.144863843918}%
\StoreBenchExecResult{Seahorn}{Seahorn}{Correct}{False}{Walltime}{Max}{825.266508102}%
\StoreBenchExecResult{Seahorn}{Seahorn}{Correct}{False}{Walltime}{Stdev}{101.5352850739914974952062143}%
\StoreBenchExecResult{Seahorn}{Seahorn}{Correct}{True}{Count}{}{2724}%
\StoreBenchExecResult{Seahorn}{Seahorn}{Correct}{True}{Cputime}{}{78113.868605996}%
\StoreBenchExecResult{Seahorn}{Seahorn}{Correct}{True}{Cputime}{Avg}{28.67616321806020558002936858}%
\StoreBenchExecResult{Seahorn}{Seahorn}{Correct}{True}{Cputime}{Median}{0.738812959}%
\StoreBenchExecResult{Seahorn}{Seahorn}{Correct}{True}{Cputime}{Min}{0.228106627}%
\StoreBenchExecResult{Seahorn}{Seahorn}{Correct}{True}{Cputime}{Max}{891.349400462}%
\StoreBenchExecResult{Seahorn}{Seahorn}{Correct}{True}{Cputime}{Stdev}{107.2592889072644339884131768}%
\StoreBenchExecResult{Seahorn}{Seahorn}{Correct}{True}{Walltime}{}{55085.308526508983}%
\StoreBenchExecResult{Seahorn}{Seahorn}{Correct}{True}{Walltime}{Avg}{20.22221311545851064610866373}%
\StoreBenchExecResult{Seahorn}{Seahorn}{Correct}{True}{Walltime}{Median}{0.409259557724}%
\StoreBenchExecResult{Seahorn}{Seahorn}{Correct}{True}{Walltime}{Min}{0.134125947952}%
\StoreBenchExecResult{Seahorn}{Seahorn}{Correct}{True}{Walltime}{Max}{871.729703903}%
\StoreBenchExecResult{Seahorn}{Seahorn}{Correct}{True}{Walltime}{Stdev}{77.99096780730255235680112539}%
\StoreBenchExecResult{Seahorn}{Seahorn}{Error}{}{Count}{}{1960}%
\StoreBenchExecResult{Seahorn}{Seahorn}{Error}{}{Cputime}{}{1568841.463908659}%
\StoreBenchExecResult{Seahorn}{Seahorn}{Error}{}{Cputime}{Avg}{800.4293183207443877551020408}%
\StoreBenchExecResult{Seahorn}{Seahorn}{Error}{}{Cputime}{Median}{1000.179943155}%
\StoreBenchExecResult{Seahorn}{Seahorn}{Error}{}{Cputime}{Min}{0.019825803}%
\StoreBenchExecResult{Seahorn}{Seahorn}{Error}{}{Cputime}{Max}{1002.11823391}%
\StoreBenchExecResult{Seahorn}{Seahorn}{Error}{}{Cputime}{Stdev}{372.4855821821003809148896740}%
\StoreBenchExecResult{Seahorn}{Seahorn}{Error}{}{Walltime}{}{1097734.9214768531408}%
\StoreBenchExecResult{Seahorn}{Seahorn}{Error}{}{Walltime}{Avg}{560.0688374881903779591836735}%
\StoreBenchExecResult{Seahorn}{Seahorn}{Error}{}{Walltime}{Median}{501.004539490}%
\StoreBenchExecResult{Seahorn}{Seahorn}{Error}{}{Walltime}{Min}{0.0200588703156}%
\StoreBenchExecResult{Seahorn}{Seahorn}{Error}{}{Walltime}{Max}{1000.70675588}%
\StoreBenchExecResult{Seahorn}{Seahorn}{Error}{}{Walltime}{Stdev}{325.2047515985598686133860735}%
\StoreBenchExecResult{Seahorn}{Seahorn}{Error}{Error}{Count}{}{70}%
\StoreBenchExecResult{Seahorn}{Seahorn}{Error}{Error}{Cputime}{}{6071.499603259}%
\StoreBenchExecResult{Seahorn}{Seahorn}{Error}{Error}{Cputime}{Avg}{86.73570861798571428571428571}%
\StoreBenchExecResult{Seahorn}{Seahorn}{Error}{Error}{Cputime}{Median}{28.8472221605}%
\StoreBenchExecResult{Seahorn}{Seahorn}{Error}{Error}{Cputime}{Min}{0.795208369}%
\StoreBenchExecResult{Seahorn}{Seahorn}{Error}{Error}{Cputime}{Max}{597.055218492}%
\StoreBenchExecResult{Seahorn}{Seahorn}{Error}{Error}{Cputime}{Stdev}{138.1686971453367272463853933}%
\StoreBenchExecResult{Seahorn}{Seahorn}{Error}{Error}{Walltime}{}{5998.894587517322}%
\StoreBenchExecResult{Seahorn}{Seahorn}{Error}{Error}{Walltime}{Avg}{85.69849410739031428571428571}%
\StoreBenchExecResult{Seahorn}{Seahorn}{Error}{Error}{Walltime}{Median}{27.7043699026}%
\StoreBenchExecResult{Seahorn}{Seahorn}{Error}{Error}{Walltime}{Min}{0.45526099205}%
\StoreBenchExecResult{Seahorn}{Seahorn}{Error}{Error}{Walltime}{Max}{596.154284}%
\StoreBenchExecResult{Seahorn}{Seahorn}{Error}{Error}{Walltime}{Stdev}{138.1549528587359670996335084}%
\StoreBenchExecResult{Seahorn}{Seahorn}{Error}{Failure}{Count}{}{183}%
\StoreBenchExecResult{Seahorn}{Seahorn}{Error}{Failure}{Cputime}{}{4.477797387}%
\StoreBenchExecResult{Seahorn}{Seahorn}{Error}{Failure}{Cputime}{Avg}{0.02446883818032786885245901639}%
\StoreBenchExecResult{Seahorn}{Seahorn}{Error}{Failure}{Cputime}{Median}{0.02152828}%
\StoreBenchExecResult{Seahorn}{Seahorn}{Error}{Failure}{Cputime}{Min}{0.019825803}%
\StoreBenchExecResult{Seahorn}{Seahorn}{Error}{Failure}{Cputime}{Max}{0.053875123}%
\StoreBenchExecResult{Seahorn}{Seahorn}{Error}{Failure}{Cputime}{Stdev}{0.007230620919241951050039353546}%
\StoreBenchExecResult{Seahorn}{Seahorn}{Error}{Failure}{Walltime}{}{4.5235249996188}%
\StoreBenchExecResult{Seahorn}{Seahorn}{Error}{Failure}{Walltime}{Avg}{0.02471871584491147540983606557}%
\StoreBenchExecResult{Seahorn}{Seahorn}{Error}{Failure}{Walltime}{Median}{0.0218451023102}%
\StoreBenchExecResult{Seahorn}{Seahorn}{Error}{Failure}{Walltime}{Min}{0.0200588703156}%
\StoreBenchExecResult{Seahorn}{Seahorn}{Error}{Failure}{Walltime}{Max}{0.0554180145264}%
\StoreBenchExecResult{Seahorn}{Seahorn}{Error}{Failure}{Walltime}{Stdev}{0.007294602034953762372633413648}%
\StoreBenchExecResult{Seahorn}{Seahorn}{Error}{OutOfMemory}{Count}{}{231}%
\StoreBenchExecResult{Seahorn}{Seahorn}{Error}{OutOfMemory}{Cputime}{}{86756.893379552}%
\StoreBenchExecResult{Seahorn}{Seahorn}{Error}{OutOfMemory}{Cputime}{Avg}{375.5709670110476190476190476}%
\StoreBenchExecResult{Seahorn}{Seahorn}{Error}{OutOfMemory}{Cputime}{Median}{269.500605528}%
\StoreBenchExecResult{Seahorn}{Seahorn}{Error}{OutOfMemory}{Cputime}{Min}{39.865694467}%
\StoreBenchExecResult{Seahorn}{Seahorn}{Error}{OutOfMemory}{Cputime}{Max}{895.393229374}%
\StoreBenchExecResult{Seahorn}{Seahorn}{Error}{OutOfMemory}{Cputime}{Stdev}{272.1854632533923345475139241}%
\StoreBenchExecResult{Seahorn}{Seahorn}{Error}{OutOfMemory}{Walltime}{}{83957.4358105712}%
\StoreBenchExecResult{Seahorn}{Seahorn}{Error}{OutOfMemory}{Walltime}{Avg}{363.4521030760658008658008658}%
\StoreBenchExecResult{Seahorn}{Seahorn}{Error}{OutOfMemory}{Walltime}{Median}{255.622237921}%
\StoreBenchExecResult{Seahorn}{Seahorn}{Error}{OutOfMemory}{Walltime}{Min}{38.3606710434}%
\StoreBenchExecResult{Seahorn}{Seahorn}{Error}{OutOfMemory}{Walltime}{Max}{894.667023897}%
\StoreBenchExecResult{Seahorn}{Seahorn}{Error}{OutOfMemory}{Walltime}{Stdev}{272.8782704584789952903299721}%
\StoreBenchExecResult{Seahorn}{Seahorn}{Error}{Timeout}{Count}{}{1476}%
\StoreBenchExecResult{Seahorn}{Seahorn}{Error}{Timeout}{Cputime}{}{1476008.593128461}%
\StoreBenchExecResult{Seahorn}{Seahorn}{Error}{Timeout}{Cputime}{Avg}{1000.005821902751355013550136}%
\StoreBenchExecResult{Seahorn}{Seahorn}{Error}{Timeout}{Cputime}{Median}{1001.431024195}%
\StoreBenchExecResult{Seahorn}{Seahorn}{Error}{Timeout}{Cputime}{Min}{901.616882766}%
\StoreBenchExecResult{Seahorn}{Seahorn}{Error}{Timeout}{Cputime}{Max}{1002.11823391}%
\StoreBenchExecResult{Seahorn}{Seahorn}{Error}{Timeout}{Cputime}{Stdev}{8.662960848969381451048543123}%
\StoreBenchExecResult{Seahorn}{Seahorn}{Error}{Timeout}{Walltime}{}{1007774.067553765}%
\StoreBenchExecResult{Seahorn}{Seahorn}{Error}{Timeout}{Walltime}{Avg}{682.7737585052608401084010840}%
\StoreBenchExecResult{Seahorn}{Seahorn}{Error}{Timeout}{Walltime}{Median}{501.004587054}%
\StoreBenchExecResult{Seahorn}{Seahorn}{Error}{Timeout}{Walltime}{Min}{457.19666481}%
\StoreBenchExecResult{Seahorn}{Seahorn}{Error}{Timeout}{Walltime}{Max}{1000.70675588}%
\StoreBenchExecResult{Seahorn}{Seahorn}{Error}{Timeout}{Walltime}{Stdev}{239.1916175180410992949637341}%
\StoreBenchExecResult{Seahorn}{Seahorn}{Wrong}{}{Count}{}{163}%
\StoreBenchExecResult{Seahorn}{Seahorn}{Wrong}{}{Cputime}{}{6469.120415635}%
\StoreBenchExecResult{Seahorn}{Seahorn}{Wrong}{}{Cputime}{Avg}{39.68785531064417177914110429}%
\StoreBenchExecResult{Seahorn}{Seahorn}{Wrong}{}{Cputime}{Median}{1.089477407}%
\StoreBenchExecResult{Seahorn}{Seahorn}{Wrong}{}{Cputime}{Min}{0.282116012}%
\StoreBenchExecResult{Seahorn}{Seahorn}{Wrong}{}{Cputime}{Max}{884.206391999}%
\StoreBenchExecResult{Seahorn}{Seahorn}{Wrong}{}{Cputime}{Stdev}{120.3602044620867755498566552}%
\StoreBenchExecResult{Seahorn}{Seahorn}{Wrong}{}{Walltime}{}{6030.722807645278}%
\StoreBenchExecResult{Seahorn}{Seahorn}{Wrong}{}{Walltime}{Avg}{36.99829943340661349693251534}%
\StoreBenchExecResult{Seahorn}{Seahorn}{Wrong}{}{Walltime}{Median}{0.657075166702}%
\StoreBenchExecResult{Seahorn}{Seahorn}{Wrong}{}{Walltime}{Min}{0.158488035202}%
\StoreBenchExecResult{Seahorn}{Seahorn}{Wrong}{}{Walltime}{Max}{883.923076153}%
\StoreBenchExecResult{Seahorn}{Seahorn}{Wrong}{}{Walltime}{Stdev}{118.5621019458117758740803954}%
\StoreBenchExecResult{Seahorn}{Seahorn}{Wrong}{False}{Count}{}{117}%
\StoreBenchExecResult{Seahorn}{Seahorn}{Wrong}{False}{Cputime}{}{2694.394104547}%
\StoreBenchExecResult{Seahorn}{Seahorn}{Wrong}{False}{Cputime}{Avg}{23.02900944057264957264957265}%
\StoreBenchExecResult{Seahorn}{Seahorn}{Wrong}{False}{Cputime}{Median}{0.593322365}%
\StoreBenchExecResult{Seahorn}{Seahorn}{Wrong}{False}{Cputime}{Min}{0.282116012}%
\StoreBenchExecResult{Seahorn}{Seahorn}{Wrong}{False}{Cputime}{Max}{884.206391999}%
\StoreBenchExecResult{Seahorn}{Seahorn}{Wrong}{False}{Cputime}{Stdev}{101.0526480397852404583151152}%
\StoreBenchExecResult{Seahorn}{Seahorn}{Wrong}{False}{Walltime}{}{2623.501950740642}%
\StoreBenchExecResult{Seahorn}{Seahorn}{Wrong}{False}{Walltime}{Avg}{22.42309359607386324786324786}%
\StoreBenchExecResult{Seahorn}{Seahorn}{Wrong}{False}{Walltime}{Median}{0.345902919769}%
\StoreBenchExecResult{Seahorn}{Seahorn}{Wrong}{False}{Walltime}{Min}{0.158488035202}%
\StoreBenchExecResult{Seahorn}{Seahorn}{Wrong}{False}{Walltime}{Max}{883.923076153}%
\StoreBenchExecResult{Seahorn}{Seahorn}{Wrong}{False}{Walltime}{Stdev}{101.0578557590927956323319425}%
\StoreBenchExecResult{Seahorn}{Seahorn}{Wrong}{True}{Count}{}{46}%
\StoreBenchExecResult{Seahorn}{Seahorn}{Wrong}{True}{Cputime}{}{3774.726311088}%
\StoreBenchExecResult{Seahorn}{Seahorn}{Wrong}{True}{Cputime}{Avg}{82.05926763234782608695652174}%
\StoreBenchExecResult{Seahorn}{Seahorn}{Wrong}{True}{Cputime}{Median}{19.8141720315}%
\StoreBenchExecResult{Seahorn}{Seahorn}{Wrong}{True}{Cputime}{Min}{0.31919214}%
\StoreBenchExecResult{Seahorn}{Seahorn}{Wrong}{True}{Cputime}{Max}{802.346984221}%
\StoreBenchExecResult{Seahorn}{Seahorn}{Wrong}{True}{Cputime}{Stdev}{151.1906189907251717555431453}%
\StoreBenchExecResult{Seahorn}{Seahorn}{Wrong}{True}{Walltime}{}{3407.220856904636}%
\StoreBenchExecResult{Seahorn}{Seahorn}{Wrong}{True}{Walltime}{Avg}{74.07001862836165217391304348}%
\StoreBenchExecResult{Seahorn}{Seahorn}{Wrong}{True}{Walltime}{Median}{15.16221010685}%
\StoreBenchExecResult{Seahorn}{Seahorn}{Wrong}{True}{Walltime}{Min}{0.171535015106}%
\StoreBenchExecResult{Seahorn}{Seahorn}{Wrong}{True}{Walltime}{Max}{801.994575977}%
\StoreBenchExecResult{Seahorn}{Seahorn}{Wrong}{True}{Walltime}{Stdev}{148.0546484273853941924020404}%
\providecommand\StoreBenchExecResult[7]{\expandafter\newcommand\csname#1#2#3#4#5#6\endcsname{#7}}%
\StoreBenchExecResult{Vvt}{CtigarTrueNotSolvedByKinductionPlainButKipdr}{Total}{}{Count}{}{449}%
\StoreBenchExecResult{Vvt}{CtigarTrueNotSolvedByKinductionPlainButKipdr}{Total}{}{Cputime}{}{1450.615757307}%
\StoreBenchExecResult{Vvt}{CtigarTrueNotSolvedByKinductionPlainButKipdr}{Total}{}{Cputime}{Avg}{3.230770060817371937639198218}%
\StoreBenchExecResult{Vvt}{CtigarTrueNotSolvedByKinductionPlainButKipdr}{Total}{}{Cputime}{Median}{0.103523717}%
\StoreBenchExecResult{Vvt}{CtigarTrueNotSolvedByKinductionPlainButKipdr}{Total}{}{Cputime}{Min}{0.054004265}%
\StoreBenchExecResult{Vvt}{CtigarTrueNotSolvedByKinductionPlainButKipdr}{Total}{}{Cputime}{Max}{1000.32944058}%
\StoreBenchExecResult{Vvt}{CtigarTrueNotSolvedByKinductionPlainButKipdr}{Total}{}{Cputime}{Stdev}{48.98155527094958464886327032}%
\StoreBenchExecResult{Vvt}{CtigarTrueNotSolvedByKinductionPlainButKipdr}{Total}{}{Walltime}{}{1362.5356786256670}%
\StoreBenchExecResult{Vvt}{CtigarTrueNotSolvedByKinductionPlainButKipdr}{Total}{}{Walltime}{Avg}{3.034600620547142538975501114}%
\StoreBenchExecResult{Vvt}{CtigarTrueNotSolvedByKinductionPlainButKipdr}{Total}{}{Walltime}{Median}{0.0650970935822}%
\StoreBenchExecResult{Vvt}{CtigarTrueNotSolvedByKinductionPlainButKipdr}{Total}{}{Walltime}{Min}{0.0359630584717}%
\StoreBenchExecResult{Vvt}{CtigarTrueNotSolvedByKinductionPlainButKipdr}{Total}{}{Walltime}{Max}{961.512840033}%
\StoreBenchExecResult{Vvt}{CtigarTrueNotSolvedByKinductionPlainButKipdr}{Total}{}{Walltime}{Stdev}{46.92033859844549642355046828}%
\StoreBenchExecResult{Vvt}{CtigarTrueNotSolvedByKinductionPlainButKipdr}{Correct}{}{Count}{}{4}%
\StoreBenchExecResult{Vvt}{CtigarTrueNotSolvedByKinductionPlainButKipdr}{Correct}{}{Cputime}{}{403.547358136}%
\StoreBenchExecResult{Vvt}{CtigarTrueNotSolvedByKinductionPlainButKipdr}{Correct}{}{Cputime}{Avg}{100.886839534}%
\StoreBenchExecResult{Vvt}{CtigarTrueNotSolvedByKinductionPlainButKipdr}{Correct}{}{Cputime}{Median}{99.1055789765}%
\StoreBenchExecResult{Vvt}{CtigarTrueNotSolvedByKinductionPlainButKipdr}{Correct}{}{Cputime}{Min}{0.196609262}%
\StoreBenchExecResult{Vvt}{CtigarTrueNotSolvedByKinductionPlainButKipdr}{Correct}{}{Cputime}{Max}{205.139590921}%
\StoreBenchExecResult{Vvt}{CtigarTrueNotSolvedByKinductionPlainButKipdr}{Correct}{}{Cputime}{Stdev}{100.6816214729636167569844611}%
\StoreBenchExecResult{Vvt}{CtigarTrueNotSolvedByKinductionPlainButKipdr}{Correct}{}{Walltime}{}{369.354816913744}%
\StoreBenchExecResult{Vvt}{CtigarTrueNotSolvedByKinductionPlainButKipdr}{Correct}{}{Walltime}{Avg}{92.338704228436}%
\StoreBenchExecResult{Vvt}{CtigarTrueNotSolvedByKinductionPlainButKipdr}{Correct}{}{Walltime}{Median}{90.721166968459}%
\StoreBenchExecResult{Vvt}{CtigarTrueNotSolvedByKinductionPlainButKipdr}{Correct}{}{Walltime}{Min}{0.126837968826}%
\StoreBenchExecResult{Vvt}{CtigarTrueNotSolvedByKinductionPlainButKipdr}{Correct}{}{Walltime}{Max}{187.785645008}%
\StoreBenchExecResult{Vvt}{CtigarTrueNotSolvedByKinductionPlainButKipdr}{Correct}{}{Walltime}{Stdev}{92.19622846116512659818640392}%
\StoreBenchExecResult{Vvt}{CtigarTrueNotSolvedByKinductionPlainButKipdr}{Correct}{True}{Count}{}{4}%
\StoreBenchExecResult{Vvt}{CtigarTrueNotSolvedByKinductionPlainButKipdr}{Correct}{True}{Cputime}{}{403.547358136}%
\StoreBenchExecResult{Vvt}{CtigarTrueNotSolvedByKinductionPlainButKipdr}{Correct}{True}{Cputime}{Avg}{100.886839534}%
\StoreBenchExecResult{Vvt}{CtigarTrueNotSolvedByKinductionPlainButKipdr}{Correct}{True}{Cputime}{Median}{99.1055789765}%
\StoreBenchExecResult{Vvt}{CtigarTrueNotSolvedByKinductionPlainButKipdr}{Correct}{True}{Cputime}{Min}{0.196609262}%
\StoreBenchExecResult{Vvt}{CtigarTrueNotSolvedByKinductionPlainButKipdr}{Correct}{True}{Cputime}{Max}{205.139590921}%
\StoreBenchExecResult{Vvt}{CtigarTrueNotSolvedByKinductionPlainButKipdr}{Correct}{True}{Cputime}{Stdev}{100.6816214729636167569844611}%
\StoreBenchExecResult{Vvt}{CtigarTrueNotSolvedByKinductionPlainButKipdr}{Correct}{True}{Walltime}{}{369.354816913744}%
\StoreBenchExecResult{Vvt}{CtigarTrueNotSolvedByKinductionPlainButKipdr}{Correct}{True}{Walltime}{Avg}{92.338704228436}%
\StoreBenchExecResult{Vvt}{CtigarTrueNotSolvedByKinductionPlainButKipdr}{Correct}{True}{Walltime}{Median}{90.721166968459}%
\StoreBenchExecResult{Vvt}{CtigarTrueNotSolvedByKinductionPlainButKipdr}{Correct}{True}{Walltime}{Min}{0.126837968826}%
\StoreBenchExecResult{Vvt}{CtigarTrueNotSolvedByKinductionPlainButKipdr}{Correct}{True}{Walltime}{Max}{187.785645008}%
\StoreBenchExecResult{Vvt}{CtigarTrueNotSolvedByKinductionPlainButKipdr}{Correct}{True}{Walltime}{Stdev}{92.19622846116512659818640392}%
\StoreBenchExecResult{Vvt}{CtigarTrueNotSolvedByKinductionPlainButKipdr}{Wrong}{True}{Count}{}{0}%
\StoreBenchExecResult{Vvt}{CtigarTrueNotSolvedByKinductionPlainButKipdr}{Wrong}{True}{Cputime}{}{0}%
\StoreBenchExecResult{Vvt}{CtigarTrueNotSolvedByKinductionPlainButKipdr}{Wrong}{True}{Cputime}{Avg}{None}%
\StoreBenchExecResult{Vvt}{CtigarTrueNotSolvedByKinductionPlainButKipdr}{Wrong}{True}{Cputime}{Median}{None}%
\StoreBenchExecResult{Vvt}{CtigarTrueNotSolvedByKinductionPlainButKipdr}{Wrong}{True}{Cputime}{Min}{None}%
\StoreBenchExecResult{Vvt}{CtigarTrueNotSolvedByKinductionPlainButKipdr}{Wrong}{True}{Cputime}{Max}{None}%
\StoreBenchExecResult{Vvt}{CtigarTrueNotSolvedByKinductionPlainButKipdr}{Wrong}{True}{Cputime}{Stdev}{None}%
\StoreBenchExecResult{Vvt}{CtigarTrueNotSolvedByKinductionPlainButKipdr}{Wrong}{True}{Walltime}{}{0}%
\StoreBenchExecResult{Vvt}{CtigarTrueNotSolvedByKinductionPlainButKipdr}{Wrong}{True}{Walltime}{Avg}{None}%
\StoreBenchExecResult{Vvt}{CtigarTrueNotSolvedByKinductionPlainButKipdr}{Wrong}{True}{Walltime}{Median}{None}%
\StoreBenchExecResult{Vvt}{CtigarTrueNotSolvedByKinductionPlainButKipdr}{Wrong}{True}{Walltime}{Min}{None}%
\StoreBenchExecResult{Vvt}{CtigarTrueNotSolvedByKinductionPlainButKipdr}{Wrong}{True}{Walltime}{Max}{None}%
\StoreBenchExecResult{Vvt}{CtigarTrueNotSolvedByKinductionPlainButKipdr}{Wrong}{True}{Walltime}{Stdev}{None}%
\StoreBenchExecResult{Vvt}{CtigarTrueNotSolvedByKinductionPlainButKipdr}{Error}{}{Count}{}{1}%
\StoreBenchExecResult{Vvt}{CtigarTrueNotSolvedByKinductionPlainButKipdr}{Error}{}{Cputime}{}{1000.32944058}%
\StoreBenchExecResult{Vvt}{CtigarTrueNotSolvedByKinductionPlainButKipdr}{Error}{}{Cputime}{Avg}{1000.32944058}%
\StoreBenchExecResult{Vvt}{CtigarTrueNotSolvedByKinductionPlainButKipdr}{Error}{}{Cputime}{Median}{1000.32944058}%
\StoreBenchExecResult{Vvt}{CtigarTrueNotSolvedByKinductionPlainButKipdr}{Error}{}{Cputime}{Min}{1000.32944058}%
\StoreBenchExecResult{Vvt}{CtigarTrueNotSolvedByKinductionPlainButKipdr}{Error}{}{Cputime}{Max}{1000.32944058}%
\StoreBenchExecResult{Vvt}{CtigarTrueNotSolvedByKinductionPlainButKipdr}{Error}{}{Cputime}{Stdev}{0E-8}%
\StoreBenchExecResult{Vvt}{CtigarTrueNotSolvedByKinductionPlainButKipdr}{Error}{}{Walltime}{}{961.512840033}%
\StoreBenchExecResult{Vvt}{CtigarTrueNotSolvedByKinductionPlainButKipdr}{Error}{}{Walltime}{Avg}{961.512840033}%
\StoreBenchExecResult{Vvt}{CtigarTrueNotSolvedByKinductionPlainButKipdr}{Error}{}{Walltime}{Median}{961.512840033}%
\StoreBenchExecResult{Vvt}{CtigarTrueNotSolvedByKinductionPlainButKipdr}{Error}{}{Walltime}{Min}{961.512840033}%
\StoreBenchExecResult{Vvt}{CtigarTrueNotSolvedByKinductionPlainButKipdr}{Error}{}{Walltime}{Max}{961.512840033}%
\StoreBenchExecResult{Vvt}{CtigarTrueNotSolvedByKinductionPlainButKipdr}{Error}{}{Walltime}{Stdev}{0E-9}%
\StoreBenchExecResult{Vvt}{CtigarTrueNotSolvedByKinductionPlainButKipdr}{Error}{Timeout}{Count}{}{1}%
\StoreBenchExecResult{Vvt}{CtigarTrueNotSolvedByKinductionPlainButKipdr}{Error}{Timeout}{Cputime}{}{1000.32944058}%
\StoreBenchExecResult{Vvt}{CtigarTrueNotSolvedByKinductionPlainButKipdr}{Error}{Timeout}{Cputime}{Avg}{1000.32944058}%
\StoreBenchExecResult{Vvt}{CtigarTrueNotSolvedByKinductionPlainButKipdr}{Error}{Timeout}{Cputime}{Median}{1000.32944058}%
\StoreBenchExecResult{Vvt}{CtigarTrueNotSolvedByKinductionPlainButKipdr}{Error}{Timeout}{Cputime}{Min}{1000.32944058}%
\StoreBenchExecResult{Vvt}{CtigarTrueNotSolvedByKinductionPlainButKipdr}{Error}{Timeout}{Cputime}{Max}{1000.32944058}%
\StoreBenchExecResult{Vvt}{CtigarTrueNotSolvedByKinductionPlainButKipdr}{Error}{Timeout}{Cputime}{Stdev}{0E-8}%
\StoreBenchExecResult{Vvt}{CtigarTrueNotSolvedByKinductionPlainButKipdr}{Error}{Timeout}{Walltime}{}{961.512840033}%
\StoreBenchExecResult{Vvt}{CtigarTrueNotSolvedByKinductionPlainButKipdr}{Error}{Timeout}{Walltime}{Avg}{961.512840033}%
\StoreBenchExecResult{Vvt}{CtigarTrueNotSolvedByKinductionPlainButKipdr}{Error}{Timeout}{Walltime}{Median}{961.512840033}%
\StoreBenchExecResult{Vvt}{CtigarTrueNotSolvedByKinductionPlainButKipdr}{Error}{Timeout}{Walltime}{Min}{961.512840033}%
\StoreBenchExecResult{Vvt}{CtigarTrueNotSolvedByKinductionPlainButKipdr}{Error}{Timeout}{Walltime}{Max}{961.512840033}%
\StoreBenchExecResult{Vvt}{CtigarTrueNotSolvedByKinductionPlainButKipdr}{Error}{Timeout}{Walltime}{Stdev}{0E-9}%
\StoreBenchExecResult{Vvt}{CtigarTrueNotSolvedByKinductionPlainButKipdr}{Unknown}{}{Count}{}{444}%
\StoreBenchExecResult{Vvt}{CtigarTrueNotSolvedByKinductionPlainButKipdr}{Unknown}{}{Cputime}{}{46.738958591}%
\StoreBenchExecResult{Vvt}{CtigarTrueNotSolvedByKinductionPlainButKipdr}{Unknown}{}{Cputime}{Avg}{0.1052679247545045045045045045}%
\StoreBenchExecResult{Vvt}{CtigarTrueNotSolvedByKinductionPlainButKipdr}{Unknown}{}{Cputime}{Median}{0.1033867055}%
\StoreBenchExecResult{Vvt}{CtigarTrueNotSolvedByKinductionPlainButKipdr}{Unknown}{}{Cputime}{Min}{0.054004265}%
\StoreBenchExecResult{Vvt}{CtigarTrueNotSolvedByKinductionPlainButKipdr}{Unknown}{}{Cputime}{Max}{0.261340774}%
\StoreBenchExecResult{Vvt}{CtigarTrueNotSolvedByKinductionPlainButKipdr}{Unknown}{}{Cputime}{Stdev}{0.03256198447278204559880997106}%
\StoreBenchExecResult{Vvt}{CtigarTrueNotSolvedByKinductionPlainButKipdr}{Unknown}{}{Walltime}{}{31.6680216789230}%
\StoreBenchExecResult{Vvt}{CtigarTrueNotSolvedByKinductionPlainButKipdr}{Unknown}{}{Walltime}{Avg}{0.07132437315072747747747747748}%
\StoreBenchExecResult{Vvt}{CtigarTrueNotSolvedByKinductionPlainButKipdr}{Unknown}{}{Walltime}{Median}{0.0645134449005}%
\StoreBenchExecResult{Vvt}{CtigarTrueNotSolvedByKinductionPlainButKipdr}{Unknown}{}{Walltime}{Min}{0.0359630584717}%
\StoreBenchExecResult{Vvt}{CtigarTrueNotSolvedByKinductionPlainButKipdr}{Unknown}{}{Walltime}{Max}{0.417335033417}%
\StoreBenchExecResult{Vvt}{CtigarTrueNotSolvedByKinductionPlainButKipdr}{Unknown}{}{Walltime}{Stdev}{0.04472675072618591013530912523}%
\StoreBenchExecResult{Vvt}{CtigarTrueNotSolvedByKinductionPlainButKipdr}{Unknown}{Unknown}{Count}{}{444}%
\StoreBenchExecResult{Vvt}{CtigarTrueNotSolvedByKinductionPlainButKipdr}{Unknown}{Unknown}{Cputime}{}{46.738958591}%
\StoreBenchExecResult{Vvt}{CtigarTrueNotSolvedByKinductionPlainButKipdr}{Unknown}{Unknown}{Cputime}{Avg}{0.1052679247545045045045045045}%
\StoreBenchExecResult{Vvt}{CtigarTrueNotSolvedByKinductionPlainButKipdr}{Unknown}{Unknown}{Cputime}{Median}{0.1033867055}%
\StoreBenchExecResult{Vvt}{CtigarTrueNotSolvedByKinductionPlainButKipdr}{Unknown}{Unknown}{Cputime}{Min}{0.054004265}%
\StoreBenchExecResult{Vvt}{CtigarTrueNotSolvedByKinductionPlainButKipdr}{Unknown}{Unknown}{Cputime}{Max}{0.261340774}%
\StoreBenchExecResult{Vvt}{CtigarTrueNotSolvedByKinductionPlainButKipdr}{Unknown}{Unknown}{Cputime}{Stdev}{0.03256198447278204559880997106}%
\StoreBenchExecResult{Vvt}{CtigarTrueNotSolvedByKinductionPlainButKipdr}{Unknown}{Unknown}{Walltime}{}{31.6680216789230}%
\StoreBenchExecResult{Vvt}{CtigarTrueNotSolvedByKinductionPlainButKipdr}{Unknown}{Unknown}{Walltime}{Avg}{0.07132437315072747747747747748}%
\StoreBenchExecResult{Vvt}{CtigarTrueNotSolvedByKinductionPlainButKipdr}{Unknown}{Unknown}{Walltime}{Median}{0.0645134449005}%
\StoreBenchExecResult{Vvt}{CtigarTrueNotSolvedByKinductionPlainButKipdr}{Unknown}{Unknown}{Walltime}{Min}{0.0359630584717}%
\StoreBenchExecResult{Vvt}{CtigarTrueNotSolvedByKinductionPlainButKipdr}{Unknown}{Unknown}{Walltime}{Max}{0.417335033417}%
\StoreBenchExecResult{Vvt}{CtigarTrueNotSolvedByKinductionPlainButKipdr}{Unknown}{Unknown}{Walltime}{Stdev}{0.04472675072618591013530912523}%
\providecommand\StoreBenchExecResult[7]{\expandafter\newcommand\csname#1#2#3#4#5#6\endcsname{#7}}%
\StoreBenchExecResult{Vvt}{CtigarTrueNotSolvedByKinductionPlain}{Total}{}{Count}{}{2893}%
\StoreBenchExecResult{Vvt}{CtigarTrueNotSolvedByKinductionPlain}{Total}{}{Cputime}{}{90686.622551634}%
\StoreBenchExecResult{Vvt}{CtigarTrueNotSolvedByKinductionPlain}{Total}{}{Cputime}{Avg}{31.34691412085516764604217076}%
\StoreBenchExecResult{Vvt}{CtigarTrueNotSolvedByKinductionPlain}{Total}{}{Cputime}{Median}{0.207384898}%
\StoreBenchExecResult{Vvt}{CtigarTrueNotSolvedByKinductionPlain}{Total}{}{Cputime}{Min}{0.054004265}%
\StoreBenchExecResult{Vvt}{CtigarTrueNotSolvedByKinductionPlain}{Total}{}{Cputime}{Max}{1001.27562188}%
\StoreBenchExecResult{Vvt}{CtigarTrueNotSolvedByKinductionPlain}{Total}{}{Cputime}{Stdev}{163.4375218236953353812661141}%
\StoreBenchExecResult{Vvt}{CtigarTrueNotSolvedByKinductionPlain}{Total}{}{Walltime}{}{87903.7755239095764}%
\StoreBenchExecResult{Vvt}{CtigarTrueNotSolvedByKinductionPlain}{Total}{}{Walltime}{Avg}{30.38498981123732333218112686}%
\StoreBenchExecResult{Vvt}{CtigarTrueNotSolvedByKinductionPlain}{Total}{}{Walltime}{Median}{0.166859865189}%
\StoreBenchExecResult{Vvt}{CtigarTrueNotSolvedByKinductionPlain}{Total}{}{Walltime}{Min}{0.0354268550873}%
\StoreBenchExecResult{Vvt}{CtigarTrueNotSolvedByKinductionPlain}{Total}{}{Walltime}{Max}{1031.00247908}%
\StoreBenchExecResult{Vvt}{CtigarTrueNotSolvedByKinductionPlain}{Total}{}{Walltime}{Stdev}{156.8384494251762501868106001}%
\StoreBenchExecResult{Vvt}{CtigarTrueNotSolvedByKinductionPlain}{Correct}{}{Count}{}{252}%
\StoreBenchExecResult{Vvt}{CtigarTrueNotSolvedByKinductionPlain}{Correct}{}{Cputime}{}{4814.566922242}%
\StoreBenchExecResult{Vvt}{CtigarTrueNotSolvedByKinductionPlain}{Correct}{}{Cputime}{Avg}{19.10542429461111111111111111}%
\StoreBenchExecResult{Vvt}{CtigarTrueNotSolvedByKinductionPlain}{Correct}{}{Cputime}{Median}{0.1993733435}%
\StoreBenchExecResult{Vvt}{CtigarTrueNotSolvedByKinductionPlain}{Correct}{}{Cputime}{Min}{0.086225714}%
\StoreBenchExecResult{Vvt}{CtigarTrueNotSolvedByKinductionPlain}{Correct}{}{Cputime}{Max}{746.436011968}%
\StoreBenchExecResult{Vvt}{CtigarTrueNotSolvedByKinductionPlain}{Correct}{}{Cputime}{Stdev}{88.46452358750945471487164277}%
\StoreBenchExecResult{Vvt}{CtigarTrueNotSolvedByKinductionPlain}{Correct}{}{Walltime}{}{4434.1854977599398}%
\StoreBenchExecResult{Vvt}{CtigarTrueNotSolvedByKinductionPlain}{Correct}{}{Walltime}{Avg}{17.59597419746007857142857143}%
\StoreBenchExecResult{Vvt}{CtigarTrueNotSolvedByKinductionPlain}{Correct}{}{Walltime}{Median}{0.1316640377045}%
\StoreBenchExecResult{Vvt}{CtigarTrueNotSolvedByKinductionPlain}{Correct}{}{Walltime}{Min}{0.0573229789734}%
\StoreBenchExecResult{Vvt}{CtigarTrueNotSolvedByKinductionPlain}{Correct}{}{Walltime}{Max}{702.74341917}%
\StoreBenchExecResult{Vvt}{CtigarTrueNotSolvedByKinductionPlain}{Correct}{}{Walltime}{Stdev}{82.69314738702365943423809226}%
\StoreBenchExecResult{Vvt}{CtigarTrueNotSolvedByKinductionPlain}{Correct}{True}{Count}{}{252}%
\StoreBenchExecResult{Vvt}{CtigarTrueNotSolvedByKinductionPlain}{Correct}{True}{Cputime}{}{4814.566922242}%
\StoreBenchExecResult{Vvt}{CtigarTrueNotSolvedByKinductionPlain}{Correct}{True}{Cputime}{Avg}{19.10542429461111111111111111}%
\StoreBenchExecResult{Vvt}{CtigarTrueNotSolvedByKinductionPlain}{Correct}{True}{Cputime}{Median}{0.1993733435}%
\StoreBenchExecResult{Vvt}{CtigarTrueNotSolvedByKinductionPlain}{Correct}{True}{Cputime}{Min}{0.086225714}%
\StoreBenchExecResult{Vvt}{CtigarTrueNotSolvedByKinductionPlain}{Correct}{True}{Cputime}{Max}{746.436011968}%
\StoreBenchExecResult{Vvt}{CtigarTrueNotSolvedByKinductionPlain}{Correct}{True}{Cputime}{Stdev}{88.46452358750945471487164277}%
\StoreBenchExecResult{Vvt}{CtigarTrueNotSolvedByKinductionPlain}{Correct}{True}{Walltime}{}{4434.1854977599398}%
\StoreBenchExecResult{Vvt}{CtigarTrueNotSolvedByKinductionPlain}{Correct}{True}{Walltime}{Avg}{17.59597419746007857142857143}%
\StoreBenchExecResult{Vvt}{CtigarTrueNotSolvedByKinductionPlain}{Correct}{True}{Walltime}{Median}{0.1316640377045}%
\StoreBenchExecResult{Vvt}{CtigarTrueNotSolvedByKinductionPlain}{Correct}{True}{Walltime}{Min}{0.0573229789734}%
\StoreBenchExecResult{Vvt}{CtigarTrueNotSolvedByKinductionPlain}{Correct}{True}{Walltime}{Max}{702.74341917}%
\StoreBenchExecResult{Vvt}{CtigarTrueNotSolvedByKinductionPlain}{Correct}{True}{Walltime}{Stdev}{82.69314738702365943423809226}%
\StoreBenchExecResult{Vvt}{CtigarTrueNotSolvedByKinductionPlain}{Wrong}{True}{Count}{}{0}%
\StoreBenchExecResult{Vvt}{CtigarTrueNotSolvedByKinductionPlain}{Wrong}{True}{Cputime}{}{0}%
\StoreBenchExecResult{Vvt}{CtigarTrueNotSolvedByKinductionPlain}{Wrong}{True}{Cputime}{Avg}{None}%
\StoreBenchExecResult{Vvt}{CtigarTrueNotSolvedByKinductionPlain}{Wrong}{True}{Cputime}{Median}{None}%
\StoreBenchExecResult{Vvt}{CtigarTrueNotSolvedByKinductionPlain}{Wrong}{True}{Cputime}{Min}{None}%
\StoreBenchExecResult{Vvt}{CtigarTrueNotSolvedByKinductionPlain}{Wrong}{True}{Cputime}{Max}{None}%
\StoreBenchExecResult{Vvt}{CtigarTrueNotSolvedByKinductionPlain}{Wrong}{True}{Cputime}{Stdev}{None}%
\StoreBenchExecResult{Vvt}{CtigarTrueNotSolvedByKinductionPlain}{Wrong}{True}{Walltime}{}{0}%
\StoreBenchExecResult{Vvt}{CtigarTrueNotSolvedByKinductionPlain}{Wrong}{True}{Walltime}{Avg}{None}%
\StoreBenchExecResult{Vvt}{CtigarTrueNotSolvedByKinductionPlain}{Wrong}{True}{Walltime}{Median}{None}%
\StoreBenchExecResult{Vvt}{CtigarTrueNotSolvedByKinductionPlain}{Wrong}{True}{Walltime}{Min}{None}%
\StoreBenchExecResult{Vvt}{CtigarTrueNotSolvedByKinductionPlain}{Wrong}{True}{Walltime}{Max}{None}%
\StoreBenchExecResult{Vvt}{CtigarTrueNotSolvedByKinductionPlain}{Wrong}{True}{Walltime}{Stdev}{None}%
\StoreBenchExecResult{Vvt}{CtigarTrueNotSolvedByKinductionPlain}{Error}{}{Count}{}{92}%
\StoreBenchExecResult{Vvt}{CtigarTrueNotSolvedByKinductionPlain}{Error}{}{Cputime}{}{80587.312698369}%
\StoreBenchExecResult{Vvt}{CtigarTrueNotSolvedByKinductionPlain}{Error}{}{Cputime}{Avg}{875.9490510692282608695652174}%
\StoreBenchExecResult{Vvt}{CtigarTrueNotSolvedByKinductionPlain}{Error}{}{Cputime}{Median}{1000.56362794}%
\StoreBenchExecResult{Vvt}{CtigarTrueNotSolvedByKinductionPlain}{Error}{}{Cputime}{Min}{3.343430506}%
\StoreBenchExecResult{Vvt}{CtigarTrueNotSolvedByKinductionPlain}{Error}{}{Cputime}{Max}{1001.27562188}%
\StoreBenchExecResult{Vvt}{CtigarTrueNotSolvedByKinductionPlain}{Error}{}{Cputime}{Stdev}{283.0165398474017398568370801}%
\StoreBenchExecResult{Vvt}{CtigarTrueNotSolvedByKinductionPlain}{Error}{}{Walltime}{}{78367.8930440039}%
\StoreBenchExecResult{Vvt}{CtigarTrueNotSolvedByKinductionPlain}{Error}{}{Walltime}{Avg}{851.8249243913467391304347826}%
\StoreBenchExecResult{Vvt}{CtigarTrueNotSolvedByKinductionPlain}{Error}{}{Walltime}{Median}{957.496505022}%
\StoreBenchExecResult{Vvt}{CtigarTrueNotSolvedByKinductionPlain}{Error}{}{Walltime}{Min}{63.6780879498}%
\StoreBenchExecResult{Vvt}{CtigarTrueNotSolvedByKinductionPlain}{Error}{}{Walltime}{Max}{1031.00247908}%
\StoreBenchExecResult{Vvt}{CtigarTrueNotSolvedByKinductionPlain}{Error}{}{Walltime}{Stdev}{237.3766993631103349314946042}%
\StoreBenchExecResult{Vvt}{CtigarTrueNotSolvedByKinductionPlain}{Error}{OutOfMemory}{Count}{}{14}%
\StoreBenchExecResult{Vvt}{CtigarTrueNotSolvedByKinductionPlain}{Error}{OutOfMemory}{Cputime}{}{4593.057838206}%
\StoreBenchExecResult{Vvt}{CtigarTrueNotSolvedByKinductionPlain}{Error}{OutOfMemory}{Cputime}{Avg}{328.0755598718571428571428571}%
\StoreBenchExecResult{Vvt}{CtigarTrueNotSolvedByKinductionPlain}{Error}{OutOfMemory}{Cputime}{Median}{365.072017050}%
\StoreBenchExecResult{Vvt}{CtigarTrueNotSolvedByKinductionPlain}{Error}{OutOfMemory}{Cputime}{Min}{64.436629677}%
\StoreBenchExecResult{Vvt}{CtigarTrueNotSolvedByKinductionPlain}{Error}{OutOfMemory}{Cputime}{Max}{604.451123066}%
\StoreBenchExecResult{Vvt}{CtigarTrueNotSolvedByKinductionPlain}{Error}{OutOfMemory}{Cputime}{Stdev}{184.2560748558316428022860695}%
\StoreBenchExecResult{Vvt}{CtigarTrueNotSolvedByKinductionPlain}{Error}{OutOfMemory}{Walltime}{}{4556.9123659149}%
\StoreBenchExecResult{Vvt}{CtigarTrueNotSolvedByKinductionPlain}{Error}{OutOfMemory}{Walltime}{Avg}{325.4937404224928571428571429}%
\StoreBenchExecResult{Vvt}{CtigarTrueNotSolvedByKinductionPlain}{Error}{OutOfMemory}{Walltime}{Median}{362.180272102}%
\StoreBenchExecResult{Vvt}{CtigarTrueNotSolvedByKinductionPlain}{Error}{OutOfMemory}{Walltime}{Min}{63.6780879498}%
\StoreBenchExecResult{Vvt}{CtigarTrueNotSolvedByKinductionPlain}{Error}{OutOfMemory}{Walltime}{Max}{601.009039164}%
\StoreBenchExecResult{Vvt}{CtigarTrueNotSolvedByKinductionPlain}{Error}{OutOfMemory}{Walltime}{Stdev}{183.5827960599662929188988893}%
\StoreBenchExecResult{Vvt}{CtigarTrueNotSolvedByKinductionPlain}{Error}{Timeout}{Count}{}{78}%
\StoreBenchExecResult{Vvt}{CtigarTrueNotSolvedByKinductionPlain}{Error}{Timeout}{Cputime}{}{75994.254860163}%
\StoreBenchExecResult{Vvt}{CtigarTrueNotSolvedByKinductionPlain}{Error}{Timeout}{Cputime}{Avg}{974.2853187200384615384615385}%
\StoreBenchExecResult{Vvt}{CtigarTrueNotSolvedByKinductionPlain}{Error}{Timeout}{Cputime}{Median}{1000.575889335}%
\StoreBenchExecResult{Vvt}{CtigarTrueNotSolvedByKinductionPlain}{Error}{Timeout}{Cputime}{Min}{3.343430506}%
\StoreBenchExecResult{Vvt}{CtigarTrueNotSolvedByKinductionPlain}{Error}{Timeout}{Cputime}{Max}{1001.27562188}%
\StoreBenchExecResult{Vvt}{CtigarTrueNotSolvedByKinductionPlain}{Error}{Timeout}{Cputime}{Stdev}{157.5928021117844102025489540}%
\StoreBenchExecResult{Vvt}{CtigarTrueNotSolvedByKinductionPlain}{Error}{Timeout}{Walltime}{}{73810.980678089}%
\StoreBenchExecResult{Vvt}{CtigarTrueNotSolvedByKinductionPlain}{Error}{Timeout}{Walltime}{Avg}{946.2946240780641025641025641}%
\StoreBenchExecResult{Vvt}{CtigarTrueNotSolvedByKinductionPlain}{Error}{Timeout}{Walltime}{Median}{962.3372205495}%
\StoreBenchExecResult{Vvt}{CtigarTrueNotSolvedByKinductionPlain}{Error}{Timeout}{Walltime}{Min}{806.528889894}%
\StoreBenchExecResult{Vvt}{CtigarTrueNotSolvedByKinductionPlain}{Error}{Timeout}{Walltime}{Max}{1031.00247908}%
\StoreBenchExecResult{Vvt}{CtigarTrueNotSolvedByKinductionPlain}{Error}{Timeout}{Walltime}{Stdev}{42.01572679803775232721117394}%
\StoreBenchExecResult{Vvt}{CtigarTrueNotSolvedByKinductionPlain}{Unknown}{}{Count}{}{2539}%
\StoreBenchExecResult{Vvt}{CtigarTrueNotSolvedByKinductionPlain}{Unknown}{}{Cputime}{}{5264.369615725}%
\StoreBenchExecResult{Vvt}{CtigarTrueNotSolvedByKinductionPlain}{Unknown}{}{Cputime}{Avg}{2.073402763184324537219377708}%
\StoreBenchExecResult{Vvt}{CtigarTrueNotSolvedByKinductionPlain}{Unknown}{}{Cputime}{Median}{0.200246707}%
\StoreBenchExecResult{Vvt}{CtigarTrueNotSolvedByKinductionPlain}{Unknown}{}{Cputime}{Min}{0.054004265}%
\StoreBenchExecResult{Vvt}{CtigarTrueNotSolvedByKinductionPlain}{Unknown}{}{Cputime}{Max}{111.544145802}%
\StoreBenchExecResult{Vvt}{CtigarTrueNotSolvedByKinductionPlain}{Unknown}{}{Cputime}{Stdev}{5.802139460582965941629511149}%
\StoreBenchExecResult{Vvt}{CtigarTrueNotSolvedByKinductionPlain}{Unknown}{}{Walltime}{}{5084.3607614044636}%
\StoreBenchExecResult{Vvt}{CtigarTrueNotSolvedByKinductionPlain}{Unknown}{}{Walltime}{Avg}{2.002505223081710752264671130}%
\StoreBenchExecResult{Vvt}{CtigarTrueNotSolvedByKinductionPlain}{Unknown}{}{Walltime}{Median}{0.160934925079}%
\StoreBenchExecResult{Vvt}{CtigarTrueNotSolvedByKinductionPlain}{Unknown}{}{Walltime}{Min}{0.0354268550873}%
\StoreBenchExecResult{Vvt}{CtigarTrueNotSolvedByKinductionPlain}{Unknown}{}{Walltime}{Max}{109.348944902}%
\StoreBenchExecResult{Vvt}{CtigarTrueNotSolvedByKinductionPlain}{Unknown}{}{Walltime}{Stdev}{5.704586303570797871811669313}%
\StoreBenchExecResult{Vvt}{CtigarTrueNotSolvedByKinductionPlain}{Unknown}{Unknown}{Count}{}{2539}%
\StoreBenchExecResult{Vvt}{CtigarTrueNotSolvedByKinductionPlain}{Unknown}{Unknown}{Cputime}{}{5264.369615725}%
\StoreBenchExecResult{Vvt}{CtigarTrueNotSolvedByKinductionPlain}{Unknown}{Unknown}{Cputime}{Avg}{2.073402763184324537219377708}%
\StoreBenchExecResult{Vvt}{CtigarTrueNotSolvedByKinductionPlain}{Unknown}{Unknown}{Cputime}{Median}{0.200246707}%
\StoreBenchExecResult{Vvt}{CtigarTrueNotSolvedByKinductionPlain}{Unknown}{Unknown}{Cputime}{Min}{0.054004265}%
\StoreBenchExecResult{Vvt}{CtigarTrueNotSolvedByKinductionPlain}{Unknown}{Unknown}{Cputime}{Max}{111.544145802}%
\StoreBenchExecResult{Vvt}{CtigarTrueNotSolvedByKinductionPlain}{Unknown}{Unknown}{Cputime}{Stdev}{5.802139460582965941629511149}%
\StoreBenchExecResult{Vvt}{CtigarTrueNotSolvedByKinductionPlain}{Unknown}{Unknown}{Walltime}{}{5084.3607614044636}%
\StoreBenchExecResult{Vvt}{CtigarTrueNotSolvedByKinductionPlain}{Unknown}{Unknown}{Walltime}{Avg}{2.002505223081710752264671130}%
\StoreBenchExecResult{Vvt}{CtigarTrueNotSolvedByKinductionPlain}{Unknown}{Unknown}{Walltime}{Median}{0.160934925079}%
\StoreBenchExecResult{Vvt}{CtigarTrueNotSolvedByKinductionPlain}{Unknown}{Unknown}{Walltime}{Min}{0.0354268550873}%
\StoreBenchExecResult{Vvt}{CtigarTrueNotSolvedByKinductionPlain}{Unknown}{Unknown}{Walltime}{Max}{109.348944902}%
\StoreBenchExecResult{Vvt}{CtigarTrueNotSolvedByKinductionPlain}{Unknown}{Unknown}{Walltime}{Stdev}{5.704586303570797871811669313}%
\StoreBenchExecResult{Vvt}{CtigarTrueNotSolvedByKinductionPlain}{Wrong}{}{Count}{}{10}%
\StoreBenchExecResult{Vvt}{CtigarTrueNotSolvedByKinductionPlain}{Wrong}{}{Cputime}{}{20.373315298}%
\StoreBenchExecResult{Vvt}{CtigarTrueNotSolvedByKinductionPlain}{Wrong}{}{Cputime}{Avg}{2.0373315298}%
\StoreBenchExecResult{Vvt}{CtigarTrueNotSolvedByKinductionPlain}{Wrong}{}{Cputime}{Median}{0.4427419715}%
\StoreBenchExecResult{Vvt}{CtigarTrueNotSolvedByKinductionPlain}{Wrong}{}{Cputime}{Min}{0.186305606}%
\StoreBenchExecResult{Vvt}{CtigarTrueNotSolvedByKinductionPlain}{Wrong}{}{Cputime}{Max}{6.282368072}%
\StoreBenchExecResult{Vvt}{CtigarTrueNotSolvedByKinductionPlain}{Wrong}{}{Cputime}{Stdev}{2.469306520348652481353032156}%
\StoreBenchExecResult{Vvt}{CtigarTrueNotSolvedByKinductionPlain}{Wrong}{}{Walltime}{}{17.336220741273}%
\StoreBenchExecResult{Vvt}{CtigarTrueNotSolvedByKinductionPlain}{Wrong}{}{Walltime}{Avg}{1.7336220741273}%
\StoreBenchExecResult{Vvt}{CtigarTrueNotSolvedByKinductionPlain}{Wrong}{}{Walltime}{Median}{0.339862465858}%
\StoreBenchExecResult{Vvt}{CtigarTrueNotSolvedByKinductionPlain}{Wrong}{}{Walltime}{Min}{0.137578010559}%
\StoreBenchExecResult{Vvt}{CtigarTrueNotSolvedByKinductionPlain}{Wrong}{}{Walltime}{Max}{5.45112490654}%
\StoreBenchExecResult{Vvt}{CtigarTrueNotSolvedByKinductionPlain}{Wrong}{}{Walltime}{Stdev}{2.156970798388403880925974834}%
\StoreBenchExecResult{Vvt}{CtigarTrueNotSolvedByKinductionPlain}{Wrong}{False}{Count}{}{10}%
\StoreBenchExecResult{Vvt}{CtigarTrueNotSolvedByKinductionPlain}{Wrong}{False}{Cputime}{}{20.373315298}%
\StoreBenchExecResult{Vvt}{CtigarTrueNotSolvedByKinductionPlain}{Wrong}{False}{Cputime}{Avg}{2.0373315298}%
\StoreBenchExecResult{Vvt}{CtigarTrueNotSolvedByKinductionPlain}{Wrong}{False}{Cputime}{Median}{0.4427419715}%
\StoreBenchExecResult{Vvt}{CtigarTrueNotSolvedByKinductionPlain}{Wrong}{False}{Cputime}{Min}{0.186305606}%
\StoreBenchExecResult{Vvt}{CtigarTrueNotSolvedByKinductionPlain}{Wrong}{False}{Cputime}{Max}{6.282368072}%
\StoreBenchExecResult{Vvt}{CtigarTrueNotSolvedByKinductionPlain}{Wrong}{False}{Cputime}{Stdev}{2.469306520348652481353032156}%
\StoreBenchExecResult{Vvt}{CtigarTrueNotSolvedByKinductionPlain}{Wrong}{False}{Walltime}{}{17.336220741273}%
\StoreBenchExecResult{Vvt}{CtigarTrueNotSolvedByKinductionPlain}{Wrong}{False}{Walltime}{Avg}{1.7336220741273}%
\StoreBenchExecResult{Vvt}{CtigarTrueNotSolvedByKinductionPlain}{Wrong}{False}{Walltime}{Median}{0.339862465858}%
\StoreBenchExecResult{Vvt}{CtigarTrueNotSolvedByKinductionPlain}{Wrong}{False}{Walltime}{Min}{0.137578010559}%
\StoreBenchExecResult{Vvt}{CtigarTrueNotSolvedByKinductionPlain}{Wrong}{False}{Walltime}{Max}{5.45112490654}%
\StoreBenchExecResult{Vvt}{CtigarTrueNotSolvedByKinductionPlain}{Wrong}{False}{Walltime}{Stdev}{2.156970798388403880925974834}%
\StoreBenchExecResult{Vvt}{CtigarTrueNotSolvedByKinductionPlain}{Correct}{False}{Count}{}{0}%
\StoreBenchExecResult{Vvt}{CtigarTrueNotSolvedByKinductionPlain}{Correct}{False}{Cputime}{}{0}%
\StoreBenchExecResult{Vvt}{CtigarTrueNotSolvedByKinductionPlain}{Correct}{False}{Cputime}{Avg}{None}%
\StoreBenchExecResult{Vvt}{CtigarTrueNotSolvedByKinductionPlain}{Correct}{False}{Cputime}{Median}{None}%
\StoreBenchExecResult{Vvt}{CtigarTrueNotSolvedByKinductionPlain}{Correct}{False}{Cputime}{Min}{None}%
\StoreBenchExecResult{Vvt}{CtigarTrueNotSolvedByKinductionPlain}{Correct}{False}{Cputime}{Max}{None}%
\StoreBenchExecResult{Vvt}{CtigarTrueNotSolvedByKinductionPlain}{Correct}{False}{Cputime}{Stdev}{None}%
\StoreBenchExecResult{Vvt}{CtigarTrueNotSolvedByKinductionPlain}{Correct}{False}{Walltime}{}{0}%
\StoreBenchExecResult{Vvt}{CtigarTrueNotSolvedByKinductionPlain}{Correct}{False}{Walltime}{Avg}{None}%
\StoreBenchExecResult{Vvt}{CtigarTrueNotSolvedByKinductionPlain}{Correct}{False}{Walltime}{Median}{None}%
\StoreBenchExecResult{Vvt}{CtigarTrueNotSolvedByKinductionPlain}{Correct}{False}{Walltime}{Min}{None}%
\StoreBenchExecResult{Vvt}{CtigarTrueNotSolvedByKinductionPlain}{Correct}{False}{Walltime}{Max}{None}%
\StoreBenchExecResult{Vvt}{CtigarTrueNotSolvedByKinductionPlain}{Correct}{False}{Walltime}{Stdev}{None}%
\providecommand\StoreBenchExecResult[7]{\expandafter\newcommand\csname#1#2#3#4#5#6\endcsname{#7}}%
\StoreBenchExecResult{Vvt}{Ctigar}{Total}{}{Count}{}{5591}%
\StoreBenchExecResult{Vvt}{Ctigar}{Total}{}{Cputime}{}{141883.264579039}%
\StoreBenchExecResult{Vvt}{Ctigar}{Total}{}{Cputime}{Avg}{25.37708184207458415310320157}%
\StoreBenchExecResult{Vvt}{Ctigar}{Total}{}{Cputime}{Median}{0.207082674}%
\StoreBenchExecResult{Vvt}{Ctigar}{Total}{}{Cputime}{Min}{0.053039498}%
\StoreBenchExecResult{Vvt}{Ctigar}{Total}{}{Cputime}{Max}{1001.27562188}%
\StoreBenchExecResult{Vvt}{Ctigar}{Total}{}{Cputime}{Stdev}{141.4928026469089237534259134}%
\StoreBenchExecResult{Vvt}{Ctigar}{Total}{}{Walltime}{}{136367.8761286824379}%
\StoreBenchExecResult{Vvt}{Ctigar}{Total}{}{Walltime}{Avg}{24.39060563918483954569844393}%
\StoreBenchExecResult{Vvt}{Ctigar}{Total}{}{Walltime}{Median}{0.165588855743}%
\StoreBenchExecResult{Vvt}{Ctigar}{Total}{}{Walltime}{Min}{0.035012960434}%
\StoreBenchExecResult{Vvt}{Ctigar}{Total}{}{Walltime}{Max}{1031.00247908}%
\StoreBenchExecResult{Vvt}{Ctigar}{Total}{}{Walltime}{Stdev}{135.2321364239393844078882252}%
\StoreBenchExecResult{Vvt}{Ctigar}{Correct}{}{Count}{}{739}%
\StoreBenchExecResult{Vvt}{Ctigar}{Correct}{}{Cputime}{}{11692.403406317}%
\StoreBenchExecResult{Vvt}{Ctigar}{Correct}{}{Cputime}{Avg}{15.82192612492151556156968877}%
\StoreBenchExecResult{Vvt}{Ctigar}{Correct}{}{Cputime}{Median}{0.243684199}%
\StoreBenchExecResult{Vvt}{Ctigar}{Correct}{}{Cputime}{Min}{0.076494197}%
\StoreBenchExecResult{Vvt}{Ctigar}{Correct}{}{Cputime}{Max}{860.835249649}%
\StoreBenchExecResult{Vvt}{Ctigar}{Correct}{}{Cputime}{Stdev}{74.31267281688482602362123518}%
\StoreBenchExecResult{Vvt}{Ctigar}{Correct}{}{Walltime}{}{10699.2074549183559}%
\StoreBenchExecResult{Vvt}{Ctigar}{Correct}{}{Walltime}{Avg}{14.47795325428735575101488498}%
\StoreBenchExecResult{Vvt}{Ctigar}{Correct}{}{Walltime}{Median}{0.183952093124}%
\StoreBenchExecResult{Vvt}{Ctigar}{Correct}{}{Walltime}{Min}{0.0550191402435}%
\StoreBenchExecResult{Vvt}{Ctigar}{Correct}{}{Walltime}{Max}{809.225548029}%
\StoreBenchExecResult{Vvt}{Ctigar}{Correct}{}{Walltime}{Stdev}{69.58058286003833026371667328}%
\StoreBenchExecResult{Vvt}{Ctigar}{Correct}{False}{Count}{}{215}%
\StoreBenchExecResult{Vvt}{Ctigar}{Correct}{False}{Cputime}{}{3352.996590112}%
\StoreBenchExecResult{Vvt}{Ctigar}{Correct}{False}{Cputime}{Avg}{15.59533297726511627906976744}%
\StoreBenchExecResult{Vvt}{Ctigar}{Correct}{False}{Cputime}{Median}{0.229246393}%
\StoreBenchExecResult{Vvt}{Ctigar}{Correct}{False}{Cputime}{Min}{0.076494197}%
\StoreBenchExecResult{Vvt}{Ctigar}{Correct}{False}{Cputime}{Max}{860.835249649}%
\StoreBenchExecResult{Vvt}{Ctigar}{Correct}{False}{Cputime}{Stdev}{70.79811938273224582267956472}%
\StoreBenchExecResult{Vvt}{Ctigar}{Correct}{False}{Walltime}{}{3034.7787117958578}%
\StoreBenchExecResult{Vvt}{Ctigar}{Correct}{False}{Walltime}{Avg}{14.11524982230631534883720930}%
\StoreBenchExecResult{Vvt}{Ctigar}{Correct}{False}{Walltime}{Median}{0.156408786774}%
\StoreBenchExecResult{Vvt}{Ctigar}{Correct}{False}{Walltime}{Min}{0.0559029579163}%
\StoreBenchExecResult{Vvt}{Ctigar}{Correct}{False}{Walltime}{Max}{809.225548029}%
\StoreBenchExecResult{Vvt}{Ctigar}{Correct}{False}{Walltime}{Stdev}{66.02312976873192096952102136}%
\StoreBenchExecResult{Vvt}{Ctigar}{Correct}{True}{Count}{}{524}%
\StoreBenchExecResult{Vvt}{Ctigar}{Correct}{True}{Cputime}{}{8339.406816205}%
\StoreBenchExecResult{Vvt}{Ctigar}{Correct}{True}{Cputime}{Avg}{15.91489850420801526717557252}%
\StoreBenchExecResult{Vvt}{Ctigar}{Correct}{True}{Cputime}{Median}{0.2475773545}%
\StoreBenchExecResult{Vvt}{Ctigar}{Correct}{True}{Cputime}{Min}{0.07783832}%
\StoreBenchExecResult{Vvt}{Ctigar}{Correct}{True}{Cputime}{Max}{746.436011968}%
\StoreBenchExecResult{Vvt}{Ctigar}{Correct}{True}{Cputime}{Stdev}{75.70732596687374276337717859}%
\StoreBenchExecResult{Vvt}{Ctigar}{Correct}{True}{Walltime}{}{7664.4287431224981}%
\StoreBenchExecResult{Vvt}{Ctigar}{Correct}{True}{Walltime}{Avg}{14.62677241053911851145038168}%
\StoreBenchExecResult{Vvt}{Ctigar}{Correct}{True}{Walltime}{Median}{0.1845409870145}%
\StoreBenchExecResult{Vvt}{Ctigar}{Correct}{True}{Walltime}{Min}{0.0550191402435}%
\StoreBenchExecResult{Vvt}{Ctigar}{Correct}{True}{Walltime}{Max}{702.74341917}%
\StoreBenchExecResult{Vvt}{Ctigar}{Correct}{True}{Walltime}{Stdev}{70.98812761869082630231915822}%
\StoreBenchExecResult{Vvt}{Ctigar}{Error}{}{Count}{}{138}%
\StoreBenchExecResult{Vvt}{Ctigar}{Error}{}{Cputime}{}{115625.401200960}%
\StoreBenchExecResult{Vvt}{Ctigar}{Error}{}{Cputime}{Avg}{837.8652260939130434782608696}%
\StoreBenchExecResult{Vvt}{Ctigar}{Error}{}{Cputime}{Median}{1000.555964295}%
\StoreBenchExecResult{Vvt}{Ctigar}{Error}{}{Cputime}{Min}{3.343430506}%
\StoreBenchExecResult{Vvt}{Ctigar}{Error}{}{Cputime}{Max}{1001.27562188}%
\StoreBenchExecResult{Vvt}{Ctigar}{Error}{}{Cputime}{Stdev}{319.8010488563299056654674774}%
\StoreBenchExecResult{Vvt}{Ctigar}{Error}{}{Walltime}{}{111554.9846413238}%
\StoreBenchExecResult{Vvt}{Ctigar}{Error}{}{Walltime}{Avg}{808.3694539226362318840579710}%
\StoreBenchExecResult{Vvt}{Ctigar}{Error}{}{Walltime}{Median}{960.5560184715}%
\StoreBenchExecResult{Vvt}{Ctigar}{Error}{}{Walltime}{Min}{54.4378728867}%
\StoreBenchExecResult{Vvt}{Ctigar}{Error}{}{Walltime}{Max}{1031.00247908}%
\StoreBenchExecResult{Vvt}{Ctigar}{Error}{}{Walltime}{Stdev}{287.4061730813215841005400133}%
\StoreBenchExecResult{Vvt}{Ctigar}{Error}{OutOfMemory}{Count}{}{28}%
\StoreBenchExecResult{Vvt}{Ctigar}{Error}{OutOfMemory}{Cputime}{}{7615.075606323}%
\StoreBenchExecResult{Vvt}{Ctigar}{Error}{OutOfMemory}{Cputime}{Avg}{271.9669859401071428571428571}%
\StoreBenchExecResult{Vvt}{Ctigar}{Error}{OutOfMemory}{Cputime}{Median}{246.0190798385}%
\StoreBenchExecResult{Vvt}{Ctigar}{Error}{OutOfMemory}{Cputime}{Min}{55.067234255}%
\StoreBenchExecResult{Vvt}{Ctigar}{Error}{OutOfMemory}{Cputime}{Max}{604.945274521}%
\StoreBenchExecResult{Vvt}{Ctigar}{Error}{OutOfMemory}{Cputime}{Stdev}{180.4436059873671288201997042}%
\StoreBenchExecResult{Vvt}{Ctigar}{Error}{OutOfMemory}{Walltime}{}{7484.3773150468}%
\StoreBenchExecResult{Vvt}{Ctigar}{Error}{OutOfMemory}{Walltime}{Avg}{267.2991898231}%
\StoreBenchExecResult{Vvt}{Ctigar}{Error}{OutOfMemory}{Walltime}{Median}{218.3346276285}%
\StoreBenchExecResult{Vvt}{Ctigar}{Error}{OutOfMemory}{Walltime}{Min}{54.4378728867}%
\StoreBenchExecResult{Vvt}{Ctigar}{Error}{OutOfMemory}{Walltime}{Max}{604.252864122}%
\StoreBenchExecResult{Vvt}{Ctigar}{Error}{OutOfMemory}{Walltime}{Stdev}{180.3078175725985000501387319}%
\StoreBenchExecResult{Vvt}{Ctigar}{Error}{Timeout}{Count}{}{110}%
\StoreBenchExecResult{Vvt}{Ctigar}{Error}{Timeout}{Cputime}{}{108010.325594637}%
\StoreBenchExecResult{Vvt}{Ctigar}{Error}{Timeout}{Cputime}{Avg}{981.9120508603363636363636364}%
\StoreBenchExecResult{Vvt}{Ctigar}{Error}{Timeout}{Cputime}{Median}{1000.573411185}%
\StoreBenchExecResult{Vvt}{Ctigar}{Error}{Timeout}{Cputime}{Min}{3.343430506}%
\StoreBenchExecResult{Vvt}{Ctigar}{Error}{Timeout}{Cputime}{Max}{1001.27562188}%
\StoreBenchExecResult{Vvt}{Ctigar}{Error}{Timeout}{Cputime}{Stdev}{133.2383537205915746687159911}%
\StoreBenchExecResult{Vvt}{Ctigar}{Error}{Timeout}{Walltime}{}{104070.607326277}%
\StoreBenchExecResult{Vvt}{Ctigar}{Error}{Timeout}{Walltime}{Avg}{946.0964302388818181818181818}%
\StoreBenchExecResult{Vvt}{Ctigar}{Error}{Timeout}{Walltime}{Median}{962.882804632}%
\StoreBenchExecResult{Vvt}{Ctigar}{Error}{Timeout}{Walltime}{Min}{806.528889894}%
\StoreBenchExecResult{Vvt}{Ctigar}{Error}{Timeout}{Walltime}{Max}{1031.00247908}%
\StoreBenchExecResult{Vvt}{Ctigar}{Error}{Timeout}{Walltime}{Stdev}{43.17585970464453774459701912}%
\StoreBenchExecResult{Vvt}{Ctigar}{Unknown}{}{Count}{}{4695}%
\StoreBenchExecResult{Vvt}{Ctigar}{Unknown}{}{Cputime}{}{14538.507916760}%
\StoreBenchExecResult{Vvt}{Ctigar}{Unknown}{}{Cputime}{Avg}{3.096593805486687965921192758}%
\StoreBenchExecResult{Vvt}{Ctigar}{Unknown}{}{Cputime}{Median}{0.198579888}%
\StoreBenchExecResult{Vvt}{Ctigar}{Unknown}{}{Cputime}{Min}{0.053039498}%
\StoreBenchExecResult{Vvt}{Ctigar}{Unknown}{}{Cputime}{Max}{111.544145802}%
\StoreBenchExecResult{Vvt}{Ctigar}{Unknown}{}{Cputime}{Stdev}{7.003407037732431846566677251}%
\StoreBenchExecResult{Vvt}{Ctigar}{Unknown}{}{Walltime}{}{14091.0023288727785}%
\StoreBenchExecResult{Vvt}{Ctigar}{Unknown}{}{Walltime}{Avg}{3.001278451304106176783812567}%
\StoreBenchExecResult{Vvt}{Ctigar}{Unknown}{}{Walltime}{Median}{0.158845186234}%
\StoreBenchExecResult{Vvt}{Ctigar}{Unknown}{}{Walltime}{Min}{0.035012960434}%
\StoreBenchExecResult{Vvt}{Ctigar}{Unknown}{}{Walltime}{Max}{109.348944902}%
\StoreBenchExecResult{Vvt}{Ctigar}{Unknown}{}{Walltime}{Stdev}{6.859474421850403972079871641}%
\StoreBenchExecResult{Vvt}{Ctigar}{Unknown}{Unknown}{Count}{}{4695}%
\StoreBenchExecResult{Vvt}{Ctigar}{Unknown}{Unknown}{Cputime}{}{14538.507916760}%
\StoreBenchExecResult{Vvt}{Ctigar}{Unknown}{Unknown}{Cputime}{Avg}{3.096593805486687965921192758}%
\StoreBenchExecResult{Vvt}{Ctigar}{Unknown}{Unknown}{Cputime}{Median}{0.198579888}%
\StoreBenchExecResult{Vvt}{Ctigar}{Unknown}{Unknown}{Cputime}{Min}{0.053039498}%
\StoreBenchExecResult{Vvt}{Ctigar}{Unknown}{Unknown}{Cputime}{Max}{111.544145802}%
\StoreBenchExecResult{Vvt}{Ctigar}{Unknown}{Unknown}{Cputime}{Stdev}{7.003407037732431846566677251}%
\StoreBenchExecResult{Vvt}{Ctigar}{Unknown}{Unknown}{Walltime}{}{14091.0023288727785}%
\StoreBenchExecResult{Vvt}{Ctigar}{Unknown}{Unknown}{Walltime}{Avg}{3.001278451304106176783812567}%
\StoreBenchExecResult{Vvt}{Ctigar}{Unknown}{Unknown}{Walltime}{Median}{0.158845186234}%
\StoreBenchExecResult{Vvt}{Ctigar}{Unknown}{Unknown}{Walltime}{Min}{0.035012960434}%
\StoreBenchExecResult{Vvt}{Ctigar}{Unknown}{Unknown}{Walltime}{Max}{109.348944902}%
\StoreBenchExecResult{Vvt}{Ctigar}{Unknown}{Unknown}{Walltime}{Stdev}{6.859474421850403972079871641}%
\StoreBenchExecResult{Vvt}{Ctigar}{Wrong}{}{Count}{}{19}%
\StoreBenchExecResult{Vvt}{Ctigar}{Wrong}{}{Cputime}{}{26.952055002}%
\StoreBenchExecResult{Vvt}{Ctigar}{Wrong}{}{Cputime}{Avg}{1.418529210631578947368421053}%
\StoreBenchExecResult{Vvt}{Ctigar}{Wrong}{}{Cputime}{Median}{0.336493258}%
\StoreBenchExecResult{Vvt}{Ctigar}{Wrong}{}{Cputime}{Min}{0.097389841}%
\StoreBenchExecResult{Vvt}{Ctigar}{Wrong}{}{Cputime}{Max}{6.282368072}%
\StoreBenchExecResult{Vvt}{Ctigar}{Wrong}{}{Cputime}{Stdev}{2.187039335997554454936387778}%
\StoreBenchExecResult{Vvt}{Ctigar}{Wrong}{}{Walltime}{}{22.6817035675035}%
\StoreBenchExecResult{Vvt}{Ctigar}{Wrong}{}{Walltime}{Avg}{1.193773871973868421052631579}%
\StoreBenchExecResult{Vvt}{Ctigar}{Wrong}{}{Walltime}{Median}{0.256167173386}%
\StoreBenchExecResult{Vvt}{Ctigar}{Wrong}{}{Walltime}{Min}{0.0632410049438}%
\StoreBenchExecResult{Vvt}{Ctigar}{Wrong}{}{Walltime}{Max}{5.45112490654}%
\StoreBenchExecResult{Vvt}{Ctigar}{Wrong}{}{Walltime}{Stdev}{1.894530770652144493134934885}%
\StoreBenchExecResult{Vvt}{Ctigar}{Wrong}{False}{Count}{}{14}%
\StoreBenchExecResult{Vvt}{Ctigar}{Wrong}{False}{Cputime}{}{20.845044473}%
\StoreBenchExecResult{Vvt}{Ctigar}{Wrong}{False}{Cputime}{Avg}{1.488931748071428571428571429}%
\StoreBenchExecResult{Vvt}{Ctigar}{Wrong}{False}{Cputime}{Median}{0.3494185425}%
\StoreBenchExecResult{Vvt}{Ctigar}{Wrong}{False}{Cputime}{Min}{0.098044077}%
\StoreBenchExecResult{Vvt}{Ctigar}{Wrong}{False}{Cputime}{Max}{6.282368072}%
\StoreBenchExecResult{Vvt}{Ctigar}{Wrong}{False}{Cputime}{Stdev}{2.259940775177406114902785931}%
\StoreBenchExecResult{Vvt}{Ctigar}{Wrong}{False}{Walltime}{}{17.6405134201059}%
\StoreBenchExecResult{Vvt}{Ctigar}{Wrong}{False}{Walltime}{Avg}{1.260036672864707142857142857}%
\StoreBenchExecResult{Vvt}{Ctigar}{Wrong}{False}{Walltime}{Median}{0.2593730688095}%
\StoreBenchExecResult{Vvt}{Ctigar}{Wrong}{False}{Walltime}{Min}{0.0632410049438}%
\StoreBenchExecResult{Vvt}{Ctigar}{Wrong}{False}{Walltime}{Max}{5.45112490654}%
\StoreBenchExecResult{Vvt}{Ctigar}{Wrong}{False}{Walltime}{Stdev}{1.970780887295361812827998199}%
\StoreBenchExecResult{Vvt}{Ctigar}{Wrong}{True}{Count}{}{5}%
\StoreBenchExecResult{Vvt}{Ctigar}{Wrong}{True}{Cputime}{}{6.107010529}%
\StoreBenchExecResult{Vvt}{Ctigar}{Wrong}{True}{Cputime}{Avg}{1.2214021058}%
\StoreBenchExecResult{Vvt}{Ctigar}{Wrong}{True}{Cputime}{Median}{0.178759051}%
\StoreBenchExecResult{Vvt}{Ctigar}{Wrong}{True}{Cputime}{Min}{0.097389841}%
\StoreBenchExecResult{Vvt}{Ctigar}{Wrong}{True}{Cputime}{Max}{5.114596914}%
\StoreBenchExecResult{Vvt}{Ctigar}{Wrong}{True}{Cputime}{Stdev}{1.955164497697905050562333462}%
\StoreBenchExecResult{Vvt}{Ctigar}{Wrong}{True}{Walltime}{}{5.0411901473976}%
\StoreBenchExecResult{Vvt}{Ctigar}{Wrong}{True}{Walltime}{Avg}{1.00823802947952}%
\StoreBenchExecResult{Vvt}{Ctigar}{Wrong}{True}{Walltime}{Median}{0.123603105545}%
\StoreBenchExecResult{Vvt}{Ctigar}{Wrong}{True}{Walltime}{Min}{0.0681519508362}%
\StoreBenchExecResult{Vvt}{Ctigar}{Wrong}{True}{Walltime}{Max}{4.29117298126}%
\StoreBenchExecResult{Vvt}{Ctigar}{Wrong}{True}{Walltime}{Stdev}{1.648418595539957380768740830}%
\providecommand\StoreBenchExecResult[7]{\expandafter\newcommand\csname#1#2#3#4#5#6\endcsname{#7}}%
\StoreBenchExecResult{Vvt}{PortfolioTrueNotSolvedByKinductionPlainButKipdr}{Total}{}{Count}{}{449}%
\StoreBenchExecResult{Vvt}{PortfolioTrueNotSolvedByKinductionPlainButKipdr}{Total}{}{Cputime}{}{6669.649391430}%
\StoreBenchExecResult{Vvt}{PortfolioTrueNotSolvedByKinductionPlainButKipdr}{Total}{}{Cputime}{Avg}{14.85445298759465478841870824}%
\StoreBenchExecResult{Vvt}{PortfolioTrueNotSolvedByKinductionPlainButKipdr}{Total}{}{Cputime}{Median}{0.105628006}%
\StoreBenchExecResult{Vvt}{PortfolioTrueNotSolvedByKinductionPlainButKipdr}{Total}{}{Cputime}{Min}{0.054837838}%
\StoreBenchExecResult{Vvt}{PortfolioTrueNotSolvedByKinductionPlainButKipdr}{Total}{}{Cputime}{Max}{1000.45007986}%
\StoreBenchExecResult{Vvt}{PortfolioTrueNotSolvedByKinductionPlainButKipdr}{Total}{}{Cputime}{Stdev}{113.0060635155493753989947816}%
\StoreBenchExecResult{Vvt}{PortfolioTrueNotSolvedByKinductionPlainButKipdr}{Total}{}{Walltime}{}{5720.5560154921315}%
\StoreBenchExecResult{Vvt}{PortfolioTrueNotSolvedByKinductionPlainButKipdr}{Total}{}{Walltime}{Avg}{12.74065927726532628062360802}%
\StoreBenchExecResult{Vvt}{PortfolioTrueNotSolvedByKinductionPlainButKipdr}{Total}{}{Walltime}{Median}{0.0640480518341}%
\StoreBenchExecResult{Vvt}{PortfolioTrueNotSolvedByKinductionPlainButKipdr}{Total}{}{Walltime}{Min}{0.0367290973663}%
\StoreBenchExecResult{Vvt}{PortfolioTrueNotSolvedByKinductionPlainButKipdr}{Total}{}{Walltime}{Max}{999.208019972}%
\StoreBenchExecResult{Vvt}{PortfolioTrueNotSolvedByKinductionPlainButKipdr}{Total}{}{Walltime}{Stdev}{101.1145458472747393284893832}%
\StoreBenchExecResult{Vvt}{PortfolioTrueNotSolvedByKinductionPlainButKipdr}{Correct}{}{Count}{}{6}%
\StoreBenchExecResult{Vvt}{PortfolioTrueNotSolvedByKinductionPlainButKipdr}{Correct}{}{Cputime}{}{1623.493824491}%
\StoreBenchExecResult{Vvt}{PortfolioTrueNotSolvedByKinductionPlainButKipdr}{Correct}{}{Cputime}{Avg}{270.5823040818333333333333333}%
\StoreBenchExecResult{Vvt}{PortfolioTrueNotSolvedByKinductionPlainButKipdr}{Correct}{}{Cputime}{Median}{257.974186555}%
\StoreBenchExecResult{Vvt}{PortfolioTrueNotSolvedByKinductionPlainButKipdr}{Correct}{}{Cputime}{Min}{0.246094047}%
\StoreBenchExecResult{Vvt}{PortfolioTrueNotSolvedByKinductionPlainButKipdr}{Correct}{}{Cputime}{Max}{701.714423218}%
\StoreBenchExecResult{Vvt}{PortfolioTrueNotSolvedByKinductionPlainButKipdr}{Correct}{}{Cputime}{Stdev}{255.5793976309136871163720179}%
\StoreBenchExecResult{Vvt}{PortfolioTrueNotSolvedByKinductionPlainButKipdr}{Correct}{}{Walltime}{}{1207.816850899928}%
\StoreBenchExecResult{Vvt}{PortfolioTrueNotSolvedByKinductionPlainButKipdr}{Correct}{}{Walltime}{Avg}{201.3028084833213333333333333}%
\StoreBenchExecResult{Vvt}{PortfolioTrueNotSolvedByKinductionPlainButKipdr}{Correct}{}{Walltime}{Median}{204.331848502}%
\StoreBenchExecResult{Vvt}{PortfolioTrueNotSolvedByKinductionPlainButKipdr}{Correct}{}{Walltime}{Min}{0.133152008057}%
\StoreBenchExecResult{Vvt}{PortfolioTrueNotSolvedByKinductionPlainButKipdr}{Correct}{}{Walltime}{Max}{400.924507856}%
\StoreBenchExecResult{Vvt}{PortfolioTrueNotSolvedByKinductionPlainButKipdr}{Correct}{}{Walltime}{Stdev}{183.8429961214454294418150403}%
\StoreBenchExecResult{Vvt}{PortfolioTrueNotSolvedByKinductionPlainButKipdr}{Correct}{True}{Count}{}{6}%
\StoreBenchExecResult{Vvt}{PortfolioTrueNotSolvedByKinductionPlainButKipdr}{Correct}{True}{Cputime}{}{1623.493824491}%
\StoreBenchExecResult{Vvt}{PortfolioTrueNotSolvedByKinductionPlainButKipdr}{Correct}{True}{Cputime}{Avg}{270.5823040818333333333333333}%
\StoreBenchExecResult{Vvt}{PortfolioTrueNotSolvedByKinductionPlainButKipdr}{Correct}{True}{Cputime}{Median}{257.974186555}%
\StoreBenchExecResult{Vvt}{PortfolioTrueNotSolvedByKinductionPlainButKipdr}{Correct}{True}{Cputime}{Min}{0.246094047}%
\StoreBenchExecResult{Vvt}{PortfolioTrueNotSolvedByKinductionPlainButKipdr}{Correct}{True}{Cputime}{Max}{701.714423218}%
\StoreBenchExecResult{Vvt}{PortfolioTrueNotSolvedByKinductionPlainButKipdr}{Correct}{True}{Cputime}{Stdev}{255.5793976309136871163720179}%
\StoreBenchExecResult{Vvt}{PortfolioTrueNotSolvedByKinductionPlainButKipdr}{Correct}{True}{Walltime}{}{1207.816850899928}%
\StoreBenchExecResult{Vvt}{PortfolioTrueNotSolvedByKinductionPlainButKipdr}{Correct}{True}{Walltime}{Avg}{201.3028084833213333333333333}%
\StoreBenchExecResult{Vvt}{PortfolioTrueNotSolvedByKinductionPlainButKipdr}{Correct}{True}{Walltime}{Median}{204.331848502}%
\StoreBenchExecResult{Vvt}{PortfolioTrueNotSolvedByKinductionPlainButKipdr}{Correct}{True}{Walltime}{Min}{0.133152008057}%
\StoreBenchExecResult{Vvt}{PortfolioTrueNotSolvedByKinductionPlainButKipdr}{Correct}{True}{Walltime}{Max}{400.924507856}%
\StoreBenchExecResult{Vvt}{PortfolioTrueNotSolvedByKinductionPlainButKipdr}{Correct}{True}{Walltime}{Stdev}{183.8429961214454294418150403}%
\StoreBenchExecResult{Vvt}{PortfolioTrueNotSolvedByKinductionPlainButKipdr}{Wrong}{True}{Count}{}{0}%
\StoreBenchExecResult{Vvt}{PortfolioTrueNotSolvedByKinductionPlainButKipdr}{Wrong}{True}{Cputime}{}{0}%
\StoreBenchExecResult{Vvt}{PortfolioTrueNotSolvedByKinductionPlainButKipdr}{Wrong}{True}{Cputime}{Avg}{None}%
\StoreBenchExecResult{Vvt}{PortfolioTrueNotSolvedByKinductionPlainButKipdr}{Wrong}{True}{Cputime}{Median}{None}%
\StoreBenchExecResult{Vvt}{PortfolioTrueNotSolvedByKinductionPlainButKipdr}{Wrong}{True}{Cputime}{Min}{None}%
\StoreBenchExecResult{Vvt}{PortfolioTrueNotSolvedByKinductionPlainButKipdr}{Wrong}{True}{Cputime}{Max}{None}%
\StoreBenchExecResult{Vvt}{PortfolioTrueNotSolvedByKinductionPlainButKipdr}{Wrong}{True}{Cputime}{Stdev}{None}%
\StoreBenchExecResult{Vvt}{PortfolioTrueNotSolvedByKinductionPlainButKipdr}{Wrong}{True}{Walltime}{}{0}%
\StoreBenchExecResult{Vvt}{PortfolioTrueNotSolvedByKinductionPlainButKipdr}{Wrong}{True}{Walltime}{Avg}{None}%
\StoreBenchExecResult{Vvt}{PortfolioTrueNotSolvedByKinductionPlainButKipdr}{Wrong}{True}{Walltime}{Median}{None}%
\StoreBenchExecResult{Vvt}{PortfolioTrueNotSolvedByKinductionPlainButKipdr}{Wrong}{True}{Walltime}{Min}{None}%
\StoreBenchExecResult{Vvt}{PortfolioTrueNotSolvedByKinductionPlainButKipdr}{Wrong}{True}{Walltime}{Max}{None}%
\StoreBenchExecResult{Vvt}{PortfolioTrueNotSolvedByKinductionPlainButKipdr}{Wrong}{True}{Walltime}{Stdev}{None}%
\StoreBenchExecResult{Vvt}{PortfolioTrueNotSolvedByKinductionPlainButKipdr}{Error}{}{Count}{}{5}%
\StoreBenchExecResult{Vvt}{PortfolioTrueNotSolvedByKinductionPlainButKipdr}{Error}{}{Cputime}{}{5000.87568017}%
\StoreBenchExecResult{Vvt}{PortfolioTrueNotSolvedByKinductionPlainButKipdr}{Error}{}{Cputime}{Avg}{1000.175136034}%
\StoreBenchExecResult{Vvt}{PortfolioTrueNotSolvedByKinductionPlainButKipdr}{Error}{}{Cputime}{Median}{1000.13080526}%
\StoreBenchExecResult{Vvt}{PortfolioTrueNotSolvedByKinductionPlainButKipdr}{Error}{}{Cputime}{Min}{1000.05689599}%
\StoreBenchExecResult{Vvt}{PortfolioTrueNotSolvedByKinductionPlainButKipdr}{Error}{}{Cputime}{Max}{1000.45007986}%
\StoreBenchExecResult{Vvt}{PortfolioTrueNotSolvedByKinductionPlainButKipdr}{Error}{}{Cputime}{Stdev}{0.1426768927596970903419326338}%
\StoreBenchExecResult{Vvt}{PortfolioTrueNotSolvedByKinductionPlainButKipdr}{Error}{}{Walltime}{}{4484.249081613}%
\StoreBenchExecResult{Vvt}{PortfolioTrueNotSolvedByKinductionPlainButKipdr}{Error}{}{Walltime}{Avg}{896.8498163226}%
\StoreBenchExecResult{Vvt}{PortfolioTrueNotSolvedByKinductionPlainButKipdr}{Error}{}{Walltime}{Median}{993.437015057}%
\StoreBenchExecResult{Vvt}{PortfolioTrueNotSolvedByKinductionPlainButKipdr}{Error}{}{Walltime}{Min}{501.004195929}%
\StoreBenchExecResult{Vvt}{PortfolioTrueNotSolvedByKinductionPlainButKipdr}{Error}{}{Walltime}{Max}{999.208019972}%
\StoreBenchExecResult{Vvt}{PortfolioTrueNotSolvedByKinductionPlainButKipdr}{Error}{}{Walltime}{Stdev}{197.9399918175575129483277410}%
\StoreBenchExecResult{Vvt}{PortfolioTrueNotSolvedByKinductionPlainButKipdr}{Error}{Timeout}{Count}{}{5}%
\StoreBenchExecResult{Vvt}{PortfolioTrueNotSolvedByKinductionPlainButKipdr}{Error}{Timeout}{Cputime}{}{5000.87568017}%
\StoreBenchExecResult{Vvt}{PortfolioTrueNotSolvedByKinductionPlainButKipdr}{Error}{Timeout}{Cputime}{Avg}{1000.175136034}%
\StoreBenchExecResult{Vvt}{PortfolioTrueNotSolvedByKinductionPlainButKipdr}{Error}{Timeout}{Cputime}{Median}{1000.13080526}%
\StoreBenchExecResult{Vvt}{PortfolioTrueNotSolvedByKinductionPlainButKipdr}{Error}{Timeout}{Cputime}{Min}{1000.05689599}%
\StoreBenchExecResult{Vvt}{PortfolioTrueNotSolvedByKinductionPlainButKipdr}{Error}{Timeout}{Cputime}{Max}{1000.45007986}%
\StoreBenchExecResult{Vvt}{PortfolioTrueNotSolvedByKinductionPlainButKipdr}{Error}{Timeout}{Cputime}{Stdev}{0.1426768927596970903419326338}%
\StoreBenchExecResult{Vvt}{PortfolioTrueNotSolvedByKinductionPlainButKipdr}{Error}{Timeout}{Walltime}{}{4484.249081613}%
\StoreBenchExecResult{Vvt}{PortfolioTrueNotSolvedByKinductionPlainButKipdr}{Error}{Timeout}{Walltime}{Avg}{896.8498163226}%
\StoreBenchExecResult{Vvt}{PortfolioTrueNotSolvedByKinductionPlainButKipdr}{Error}{Timeout}{Walltime}{Median}{993.437015057}%
\StoreBenchExecResult{Vvt}{PortfolioTrueNotSolvedByKinductionPlainButKipdr}{Error}{Timeout}{Walltime}{Min}{501.004195929}%
\StoreBenchExecResult{Vvt}{PortfolioTrueNotSolvedByKinductionPlainButKipdr}{Error}{Timeout}{Walltime}{Max}{999.208019972}%
\StoreBenchExecResult{Vvt}{PortfolioTrueNotSolvedByKinductionPlainButKipdr}{Error}{Timeout}{Walltime}{Stdev}{197.9399918175575129483277410}%
\StoreBenchExecResult{Vvt}{PortfolioTrueNotSolvedByKinductionPlainButKipdr}{Unknown}{}{Count}{}{438}%
\StoreBenchExecResult{Vvt}{PortfolioTrueNotSolvedByKinductionPlainButKipdr}{Unknown}{}{Cputime}{}{45.279886769}%
\StoreBenchExecResult{Vvt}{PortfolioTrueNotSolvedByKinductionPlainButKipdr}{Unknown}{}{Cputime}{Avg}{0.1033787369155251141552511416}%
\StoreBenchExecResult{Vvt}{PortfolioTrueNotSolvedByKinductionPlainButKipdr}{Unknown}{}{Cputime}{Median}{0.105319557}%
\StoreBenchExecResult{Vvt}{PortfolioTrueNotSolvedByKinductionPlainButKipdr}{Unknown}{}{Cputime}{Min}{0.054837838}%
\StoreBenchExecResult{Vvt}{PortfolioTrueNotSolvedByKinductionPlainButKipdr}{Unknown}{}{Cputime}{Max}{0.260040124}%
\StoreBenchExecResult{Vvt}{PortfolioTrueNotSolvedByKinductionPlainButKipdr}{Unknown}{}{Cputime}{Stdev}{0.02752731344829647712985223425}%
\StoreBenchExecResult{Vvt}{PortfolioTrueNotSolvedByKinductionPlainButKipdr}{Unknown}{}{Walltime}{}{28.4900829792035}%
\StoreBenchExecResult{Vvt}{PortfolioTrueNotSolvedByKinductionPlainButKipdr}{Unknown}{}{Walltime}{Avg}{0.06504585155069292237442922374}%
\StoreBenchExecResult{Vvt}{PortfolioTrueNotSolvedByKinductionPlainButKipdr}{Unknown}{}{Walltime}{Median}{0.06342804431915}%
\StoreBenchExecResult{Vvt}{PortfolioTrueNotSolvedByKinductionPlainButKipdr}{Unknown}{}{Walltime}{Min}{0.0367290973663}%
\StoreBenchExecResult{Vvt}{PortfolioTrueNotSolvedByKinductionPlainButKipdr}{Unknown}{}{Walltime}{Max}{0.21385717392}%
\StoreBenchExecResult{Vvt}{PortfolioTrueNotSolvedByKinductionPlainButKipdr}{Unknown}{}{Walltime}{Stdev}{0.02125252564359967296866877075}%
\StoreBenchExecResult{Vvt}{PortfolioTrueNotSolvedByKinductionPlainButKipdr}{Unknown}{Unknown}{Count}{}{438}%
\StoreBenchExecResult{Vvt}{PortfolioTrueNotSolvedByKinductionPlainButKipdr}{Unknown}{Unknown}{Cputime}{}{45.279886769}%
\StoreBenchExecResult{Vvt}{PortfolioTrueNotSolvedByKinductionPlainButKipdr}{Unknown}{Unknown}{Cputime}{Avg}{0.1033787369155251141552511416}%
\StoreBenchExecResult{Vvt}{PortfolioTrueNotSolvedByKinductionPlainButKipdr}{Unknown}{Unknown}{Cputime}{Median}{0.105319557}%
\StoreBenchExecResult{Vvt}{PortfolioTrueNotSolvedByKinductionPlainButKipdr}{Unknown}{Unknown}{Cputime}{Min}{0.054837838}%
\StoreBenchExecResult{Vvt}{PortfolioTrueNotSolvedByKinductionPlainButKipdr}{Unknown}{Unknown}{Cputime}{Max}{0.260040124}%
\StoreBenchExecResult{Vvt}{PortfolioTrueNotSolvedByKinductionPlainButKipdr}{Unknown}{Unknown}{Cputime}{Stdev}{0.02752731344829647712985223425}%
\StoreBenchExecResult{Vvt}{PortfolioTrueNotSolvedByKinductionPlainButKipdr}{Unknown}{Unknown}{Walltime}{}{28.4900829792035}%
\StoreBenchExecResult{Vvt}{PortfolioTrueNotSolvedByKinductionPlainButKipdr}{Unknown}{Unknown}{Walltime}{Avg}{0.06504585155069292237442922374}%
\StoreBenchExecResult{Vvt}{PortfolioTrueNotSolvedByKinductionPlainButKipdr}{Unknown}{Unknown}{Walltime}{Median}{0.06342804431915}%
\StoreBenchExecResult{Vvt}{PortfolioTrueNotSolvedByKinductionPlainButKipdr}{Unknown}{Unknown}{Walltime}{Min}{0.0367290973663}%
\StoreBenchExecResult{Vvt}{PortfolioTrueNotSolvedByKinductionPlainButKipdr}{Unknown}{Unknown}{Walltime}{Max}{0.21385717392}%
\StoreBenchExecResult{Vvt}{PortfolioTrueNotSolvedByKinductionPlainButKipdr}{Unknown}{Unknown}{Walltime}{Stdev}{0.02125252564359967296866877075}%
\providecommand\StoreBenchExecResult[7]{\expandafter\newcommand\csname#1#2#3#4#5#6\endcsname{#7}}%
\StoreBenchExecResult{Vvt}{PortfolioTrueNotSolvedByKinductionPlain}{Total}{}{Count}{}{2893}%
\StoreBenchExecResult{Vvt}{PortfolioTrueNotSolvedByKinductionPlain}{Total}{}{Cputime}{}{201589.599472373}%
\StoreBenchExecResult{Vvt}{PortfolioTrueNotSolvedByKinductionPlain}{Total}{}{Cputime}{Avg}{69.68185256563187003110957484}%
\StoreBenchExecResult{Vvt}{PortfolioTrueNotSolvedByKinductionPlain}{Total}{}{Cputime}{Median}{0.220568137}%
\StoreBenchExecResult{Vvt}{PortfolioTrueNotSolvedByKinductionPlain}{Total}{}{Cputime}{Min}{0.054837838}%
\StoreBenchExecResult{Vvt}{PortfolioTrueNotSolvedByKinductionPlain}{Total}{}{Cputime}{Max}{1002.26186594}%
\StoreBenchExecResult{Vvt}{PortfolioTrueNotSolvedByKinductionPlain}{Total}{}{Cputime}{Stdev}{246.9369736133791448477233172}%
\StoreBenchExecResult{Vvt}{PortfolioTrueNotSolvedByKinductionPlain}{Total}{}{Walltime}{}{161259.2574553498861}%
\StoreBenchExecResult{Vvt}{PortfolioTrueNotSolvedByKinductionPlain}{Total}{}{Walltime}{Avg}{55.74118819749391154510888351}%
\StoreBenchExecResult{Vvt}{PortfolioTrueNotSolvedByKinductionPlain}{Total}{}{Walltime}{Median}{0.165946960449}%
\StoreBenchExecResult{Vvt}{PortfolioTrueNotSolvedByKinductionPlain}{Total}{}{Walltime}{Min}{0.0361199378967}%
\StoreBenchExecResult{Vvt}{PortfolioTrueNotSolvedByKinductionPlain}{Total}{}{Walltime}{Max}{999.976505041}%
\StoreBenchExecResult{Vvt}{PortfolioTrueNotSolvedByKinductionPlain}{Total}{}{Walltime}{Stdev}{206.6566354954300755535119905}%
\StoreBenchExecResult{Vvt}{PortfolioTrueNotSolvedByKinductionPlain}{Correct}{}{Count}{}{252}%
\StoreBenchExecResult{Vvt}{PortfolioTrueNotSolvedByKinductionPlain}{Correct}{}{Cputime}{}{7553.185461951}%
\StoreBenchExecResult{Vvt}{PortfolioTrueNotSolvedByKinductionPlain}{Correct}{}{Cputime}{Avg}{29.97295818234523809523809524}%
\StoreBenchExecResult{Vvt}{PortfolioTrueNotSolvedByKinductionPlain}{Correct}{}{Cputime}{Median}{0.273213132}%
\StoreBenchExecResult{Vvt}{PortfolioTrueNotSolvedByKinductionPlain}{Correct}{}{Cputime}{Min}{0.091354582}%
\StoreBenchExecResult{Vvt}{PortfolioTrueNotSolvedByKinductionPlain}{Correct}{}{Cputime}{Max}{892.573646031}%
\StoreBenchExecResult{Vvt}{PortfolioTrueNotSolvedByKinductionPlain}{Correct}{}{Cputime}{Stdev}{128.4599015054108048705259095}%
\StoreBenchExecResult{Vvt}{PortfolioTrueNotSolvedByKinductionPlain}{Correct}{}{Walltime}{}{4188.8224837774022}%
\StoreBenchExecResult{Vvt}{PortfolioTrueNotSolvedByKinductionPlain}{Correct}{}{Walltime}{Avg}{16.62231144356111984126984127}%
\StoreBenchExecResult{Vvt}{PortfolioTrueNotSolvedByKinductionPlain}{Correct}{}{Walltime}{Median}{0.155259013176}%
\StoreBenchExecResult{Vvt}{PortfolioTrueNotSolvedByKinductionPlain}{Correct}{}{Walltime}{Min}{0.049252986908}%
\StoreBenchExecResult{Vvt}{PortfolioTrueNotSolvedByKinductionPlain}{Correct}{}{Walltime}{Max}{450.800184965}%
\StoreBenchExecResult{Vvt}{PortfolioTrueNotSolvedByKinductionPlain}{Correct}{}{Walltime}{Stdev}{71.03890100532506803982468461}%
\StoreBenchExecResult{Vvt}{PortfolioTrueNotSolvedByKinductionPlain}{Correct}{True}{Count}{}{252}%
\StoreBenchExecResult{Vvt}{PortfolioTrueNotSolvedByKinductionPlain}{Correct}{True}{Cputime}{}{7553.185461951}%
\StoreBenchExecResult{Vvt}{PortfolioTrueNotSolvedByKinductionPlain}{Correct}{True}{Cputime}{Avg}{29.97295818234523809523809524}%
\StoreBenchExecResult{Vvt}{PortfolioTrueNotSolvedByKinductionPlain}{Correct}{True}{Cputime}{Median}{0.273213132}%
\StoreBenchExecResult{Vvt}{PortfolioTrueNotSolvedByKinductionPlain}{Correct}{True}{Cputime}{Min}{0.091354582}%
\StoreBenchExecResult{Vvt}{PortfolioTrueNotSolvedByKinductionPlain}{Correct}{True}{Cputime}{Max}{892.573646031}%
\StoreBenchExecResult{Vvt}{PortfolioTrueNotSolvedByKinductionPlain}{Correct}{True}{Cputime}{Stdev}{128.4599015054108048705259095}%
\StoreBenchExecResult{Vvt}{PortfolioTrueNotSolvedByKinductionPlain}{Correct}{True}{Walltime}{}{4188.8224837774022}%
\StoreBenchExecResult{Vvt}{PortfolioTrueNotSolvedByKinductionPlain}{Correct}{True}{Walltime}{Avg}{16.62231144356111984126984127}%
\StoreBenchExecResult{Vvt}{PortfolioTrueNotSolvedByKinductionPlain}{Correct}{True}{Walltime}{Median}{0.155259013176}%
\StoreBenchExecResult{Vvt}{PortfolioTrueNotSolvedByKinductionPlain}{Correct}{True}{Walltime}{Min}{0.049252986908}%
\StoreBenchExecResult{Vvt}{PortfolioTrueNotSolvedByKinductionPlain}{Correct}{True}{Walltime}{Max}{450.800184965}%
\StoreBenchExecResult{Vvt}{PortfolioTrueNotSolvedByKinductionPlain}{Correct}{True}{Walltime}{Stdev}{71.03890100532506803982468461}%
\StoreBenchExecResult{Vvt}{PortfolioTrueNotSolvedByKinductionPlain}{Wrong}{True}{Count}{}{0}%
\StoreBenchExecResult{Vvt}{PortfolioTrueNotSolvedByKinductionPlain}{Wrong}{True}{Cputime}{}{0}%
\StoreBenchExecResult{Vvt}{PortfolioTrueNotSolvedByKinductionPlain}{Wrong}{True}{Cputime}{Avg}{None}%
\StoreBenchExecResult{Vvt}{PortfolioTrueNotSolvedByKinductionPlain}{Wrong}{True}{Cputime}{Median}{None}%
\StoreBenchExecResult{Vvt}{PortfolioTrueNotSolvedByKinductionPlain}{Wrong}{True}{Cputime}{Min}{None}%
\StoreBenchExecResult{Vvt}{PortfolioTrueNotSolvedByKinductionPlain}{Wrong}{True}{Cputime}{Max}{None}%
\StoreBenchExecResult{Vvt}{PortfolioTrueNotSolvedByKinductionPlain}{Wrong}{True}{Cputime}{Stdev}{None}%
\StoreBenchExecResult{Vvt}{PortfolioTrueNotSolvedByKinductionPlain}{Wrong}{True}{Walltime}{}{0}%
\StoreBenchExecResult{Vvt}{PortfolioTrueNotSolvedByKinductionPlain}{Wrong}{True}{Walltime}{Avg}{None}%
\StoreBenchExecResult{Vvt}{PortfolioTrueNotSolvedByKinductionPlain}{Wrong}{True}{Walltime}{Median}{None}%
\StoreBenchExecResult{Vvt}{PortfolioTrueNotSolvedByKinductionPlain}{Wrong}{True}{Walltime}{Min}{None}%
\StoreBenchExecResult{Vvt}{PortfolioTrueNotSolvedByKinductionPlain}{Wrong}{True}{Walltime}{Max}{None}%
\StoreBenchExecResult{Vvt}{PortfolioTrueNotSolvedByKinductionPlain}{Wrong}{True}{Walltime}{Stdev}{None}%
\StoreBenchExecResult{Vvt}{PortfolioTrueNotSolvedByKinductionPlain}{Error}{}{Count}{}{198}%
\StoreBenchExecResult{Vvt}{PortfolioTrueNotSolvedByKinductionPlain}{Error}{}{Cputime}{}{189228.342938299}%
\StoreBenchExecResult{Vvt}{PortfolioTrueNotSolvedByKinductionPlain}{Error}{}{Cputime}{Avg}{955.6987017085808080808080808}%
\StoreBenchExecResult{Vvt}{PortfolioTrueNotSolvedByKinductionPlain}{Error}{}{Cputime}{Median}{1000.06292950}%
\StoreBenchExecResult{Vvt}{PortfolioTrueNotSolvedByKinductionPlain}{Error}{}{Cputime}{Min}{113.217950864}%
\StoreBenchExecResult{Vvt}{PortfolioTrueNotSolvedByKinductionPlain}{Error}{}{Cputime}{Max}{1002.26186594}%
\StoreBenchExecResult{Vvt}{PortfolioTrueNotSolvedByKinductionPlain}{Error}{}{Cputime}{Stdev}{160.8517265570084561057681971}%
\StoreBenchExecResult{Vvt}{PortfolioTrueNotSolvedByKinductionPlain}{Error}{}{Walltime}{}{152423.3196239486}%
\StoreBenchExecResult{Vvt}{PortfolioTrueNotSolvedByKinductionPlain}{Error}{}{Walltime}{Avg}{769.8147455754979797979797980}%
\StoreBenchExecResult{Vvt}{PortfolioTrueNotSolvedByKinductionPlain}{Error}{}{Walltime}{Median}{994.904842019}%
\StoreBenchExecResult{Vvt}{PortfolioTrueNotSolvedByKinductionPlain}{Error}{}{Walltime}{Min}{79.8744199276}%
\StoreBenchExecResult{Vvt}{PortfolioTrueNotSolvedByKinductionPlain}{Error}{}{Walltime}{Max}{999.976505041}%
\StoreBenchExecResult{Vvt}{PortfolioTrueNotSolvedByKinductionPlain}{Error}{}{Walltime}{Stdev}{263.6314674764290531163710794}%
\StoreBenchExecResult{Vvt}{PortfolioTrueNotSolvedByKinductionPlain}{Error}{OutOfMemory}{Count}{}{15}%
\StoreBenchExecResult{Vvt}{PortfolioTrueNotSolvedByKinductionPlain}{Error}{OutOfMemory}{Cputime}{}{6193.171628395}%
\StoreBenchExecResult{Vvt}{PortfolioTrueNotSolvedByKinductionPlain}{Error}{OutOfMemory}{Cputime}{Avg}{412.8781085596666666666666667}%
\StoreBenchExecResult{Vvt}{PortfolioTrueNotSolvedByKinductionPlain}{Error}{OutOfMemory}{Cputime}{Median}{371.290789317}%
\StoreBenchExecResult{Vvt}{PortfolioTrueNotSolvedByKinductionPlain}{Error}{OutOfMemory}{Cputime}{Min}{113.217950864}%
\StoreBenchExecResult{Vvt}{PortfolioTrueNotSolvedByKinductionPlain}{Error}{OutOfMemory}{Cputime}{Max}{697.51567418}%
\StoreBenchExecResult{Vvt}{PortfolioTrueNotSolvedByKinductionPlain}{Error}{OutOfMemory}{Cputime}{Stdev}{149.0151530647092771151374060}%
\StoreBenchExecResult{Vvt}{PortfolioTrueNotSolvedByKinductionPlain}{Error}{OutOfMemory}{Walltime}{}{5231.1458907126}%
\StoreBenchExecResult{Vvt}{PortfolioTrueNotSolvedByKinductionPlain}{Error}{OutOfMemory}{Walltime}{Avg}{348.74305938084}%
\StoreBenchExecResult{Vvt}{PortfolioTrueNotSolvedByKinductionPlain}{Error}{OutOfMemory}{Walltime}{Median}{364.109511852}%
\StoreBenchExecResult{Vvt}{PortfolioTrueNotSolvedByKinductionPlain}{Error}{OutOfMemory}{Walltime}{Min}{79.8744199276}%
\StoreBenchExecResult{Vvt}{PortfolioTrueNotSolvedByKinductionPlain}{Error}{OutOfMemory}{Walltime}{Max}{694.541884899}%
\StoreBenchExecResult{Vvt}{PortfolioTrueNotSolvedByKinductionPlain}{Error}{OutOfMemory}{Walltime}{Stdev}{174.0424976113480493581636777}%
\StoreBenchExecResult{Vvt}{PortfolioTrueNotSolvedByKinductionPlain}{Error}{Timeout}{Count}{}{183}%
\StoreBenchExecResult{Vvt}{PortfolioTrueNotSolvedByKinductionPlain}{Error}{Timeout}{Cputime}{}{183035.171309904}%
\StoreBenchExecResult{Vvt}{PortfolioTrueNotSolvedByKinductionPlain}{Error}{Timeout}{Cputime}{Avg}{1000.192192950295081967213115}%
\StoreBenchExecResult{Vvt}{PortfolioTrueNotSolvedByKinductionPlain}{Error}{Timeout}{Cputime}{Median}{1000.07462347}%
\StoreBenchExecResult{Vvt}{PortfolioTrueNotSolvedByKinductionPlain}{Error}{Timeout}{Cputime}{Min}{913.226467424}%
\StoreBenchExecResult{Vvt}{PortfolioTrueNotSolvedByKinductionPlain}{Error}{Timeout}{Cputime}{Max}{1002.26186594}%
\StoreBenchExecResult{Vvt}{PortfolioTrueNotSolvedByKinductionPlain}{Error}{Timeout}{Cputime}{Stdev}{6.500777151278354720221642893}%
\StoreBenchExecResult{Vvt}{PortfolioTrueNotSolvedByKinductionPlain}{Error}{Timeout}{Walltime}{}{147192.173733236}%
\StoreBenchExecResult{Vvt}{PortfolioTrueNotSolvedByKinductionPlain}{Error}{Timeout}{Walltime}{Avg}{804.3288182144043715846994536}%
\StoreBenchExecResult{Vvt}{PortfolioTrueNotSolvedByKinductionPlain}{Error}{Timeout}{Walltime}{Median}{996.228054047}%
\StoreBenchExecResult{Vvt}{PortfolioTrueNotSolvedByKinductionPlain}{Error}{Timeout}{Walltime}{Min}{457.325178862}%
\StoreBenchExecResult{Vvt}{PortfolioTrueNotSolvedByKinductionPlain}{Error}{Timeout}{Walltime}{Max}{999.976505041}%
\StoreBenchExecResult{Vvt}{PortfolioTrueNotSolvedByKinductionPlain}{Error}{Timeout}{Walltime}{Stdev}{238.7287736413882244821497660}%
\StoreBenchExecResult{Vvt}{PortfolioTrueNotSolvedByKinductionPlain}{Unknown}{}{Count}{}{2427}%
\StoreBenchExecResult{Vvt}{PortfolioTrueNotSolvedByKinductionPlain}{Unknown}{}{Cputime}{}{4715.527787764}%
\StoreBenchExecResult{Vvt}{PortfolioTrueNotSolvedByKinductionPlain}{Unknown}{}{Cputime}{Avg}{1.942945112387309435517099300}%
\StoreBenchExecResult{Vvt}{PortfolioTrueNotSolvedByKinductionPlain}{Unknown}{}{Cputime}{Median}{0.195849496}%
\StoreBenchExecResult{Vvt}{PortfolioTrueNotSolvedByKinductionPlain}{Unknown}{}{Cputime}{Min}{0.054837838}%
\StoreBenchExecResult{Vvt}{PortfolioTrueNotSolvedByKinductionPlain}{Unknown}{}{Cputime}{Max}{123.080849066}%
\StoreBenchExecResult{Vvt}{PortfolioTrueNotSolvedByKinductionPlain}{Unknown}{}{Cputime}{Stdev}{6.125325367640133153914002722}%
\StoreBenchExecResult{Vvt}{PortfolioTrueNotSolvedByKinductionPlain}{Unknown}{}{Walltime}{}{4562.8399333954491}%
\StoreBenchExecResult{Vvt}{PortfolioTrueNotSolvedByKinductionPlain}{Unknown}{}{Walltime}{Avg}{1.880032935061989740420271941}%
\StoreBenchExecResult{Vvt}{PortfolioTrueNotSolvedByKinductionPlain}{Unknown}{}{Walltime}{Median}{0.152294874191}%
\StoreBenchExecResult{Vvt}{PortfolioTrueNotSolvedByKinductionPlain}{Unknown}{}{Walltime}{Min}{0.0361199378967}%
\StoreBenchExecResult{Vvt}{PortfolioTrueNotSolvedByKinductionPlain}{Unknown}{}{Walltime}{Max}{121.012732029}%
\StoreBenchExecResult{Vvt}{PortfolioTrueNotSolvedByKinductionPlain}{Unknown}{}{Walltime}{Stdev}{6.043181923007583816712660625}%
\StoreBenchExecResult{Vvt}{PortfolioTrueNotSolvedByKinductionPlain}{Unknown}{Unknown}{Count}{}{2427}%
\StoreBenchExecResult{Vvt}{PortfolioTrueNotSolvedByKinductionPlain}{Unknown}{Unknown}{Cputime}{}{4715.527787764}%
\StoreBenchExecResult{Vvt}{PortfolioTrueNotSolvedByKinductionPlain}{Unknown}{Unknown}{Cputime}{Avg}{1.942945112387309435517099300}%
\StoreBenchExecResult{Vvt}{PortfolioTrueNotSolvedByKinductionPlain}{Unknown}{Unknown}{Cputime}{Median}{0.195849496}%
\StoreBenchExecResult{Vvt}{PortfolioTrueNotSolvedByKinductionPlain}{Unknown}{Unknown}{Cputime}{Min}{0.054837838}%
\StoreBenchExecResult{Vvt}{PortfolioTrueNotSolvedByKinductionPlain}{Unknown}{Unknown}{Cputime}{Max}{123.080849066}%
\StoreBenchExecResult{Vvt}{PortfolioTrueNotSolvedByKinductionPlain}{Unknown}{Unknown}{Cputime}{Stdev}{6.125325367640133153914002722}%
\StoreBenchExecResult{Vvt}{PortfolioTrueNotSolvedByKinductionPlain}{Unknown}{Unknown}{Walltime}{}{4562.8399333954491}%
\StoreBenchExecResult{Vvt}{PortfolioTrueNotSolvedByKinductionPlain}{Unknown}{Unknown}{Walltime}{Avg}{1.880032935061989740420271941}%
\StoreBenchExecResult{Vvt}{PortfolioTrueNotSolvedByKinductionPlain}{Unknown}{Unknown}{Walltime}{Median}{0.152294874191}%
\StoreBenchExecResult{Vvt}{PortfolioTrueNotSolvedByKinductionPlain}{Unknown}{Unknown}{Walltime}{Min}{0.0361199378967}%
\StoreBenchExecResult{Vvt}{PortfolioTrueNotSolvedByKinductionPlain}{Unknown}{Unknown}{Walltime}{Max}{121.012732029}%
\StoreBenchExecResult{Vvt}{PortfolioTrueNotSolvedByKinductionPlain}{Unknown}{Unknown}{Walltime}{Stdev}{6.043181923007583816712660625}%
\StoreBenchExecResult{Vvt}{PortfolioTrueNotSolvedByKinductionPlain}{Wrong}{}{Count}{}{16}%
\StoreBenchExecResult{Vvt}{PortfolioTrueNotSolvedByKinductionPlain}{Wrong}{}{Cputime}{}{92.543284359}%
\StoreBenchExecResult{Vvt}{PortfolioTrueNotSolvedByKinductionPlain}{Wrong}{}{Cputime}{Avg}{5.7839552724375}%
\StoreBenchExecResult{Vvt}{PortfolioTrueNotSolvedByKinductionPlain}{Wrong}{}{Cputime}{Median}{0.230086557}%
\StoreBenchExecResult{Vvt}{PortfolioTrueNotSolvedByKinductionPlain}{Wrong}{}{Cputime}{Min}{0.137341469}%
\StoreBenchExecResult{Vvt}{PortfolioTrueNotSolvedByKinductionPlain}{Wrong}{}{Cputime}{Max}{42.42557717}%
\StoreBenchExecResult{Vvt}{PortfolioTrueNotSolvedByKinductionPlain}{Wrong}{}{Cputime}{Stdev}{11.94186226086306751131884217}%
\StoreBenchExecResult{Vvt}{PortfolioTrueNotSolvedByKinductionPlain}{Wrong}{}{Walltime}{}{84.2754142284348}%
\StoreBenchExecResult{Vvt}{PortfolioTrueNotSolvedByKinductionPlain}{Wrong}{}{Walltime}{Avg}{5.267213389277175}%
\StoreBenchExecResult{Vvt}{PortfolioTrueNotSolvedByKinductionPlain}{Wrong}{}{Walltime}{Median}{0.1383640766145}%
\StoreBenchExecResult{Vvt}{PortfolioTrueNotSolvedByKinductionPlain}{Wrong}{}{Walltime}{Min}{0.0825860500336}%
\StoreBenchExecResult{Vvt}{PortfolioTrueNotSolvedByKinductionPlain}{Wrong}{}{Walltime}{Max}{39.8229458332}%
\StoreBenchExecResult{Vvt}{PortfolioTrueNotSolvedByKinductionPlain}{Wrong}{}{Walltime}{Stdev}{11.17372662594733931699860549}%
\StoreBenchExecResult{Vvt}{PortfolioTrueNotSolvedByKinductionPlain}{Wrong}{False}{Count}{}{16}%
\StoreBenchExecResult{Vvt}{PortfolioTrueNotSolvedByKinductionPlain}{Wrong}{False}{Cputime}{}{92.543284359}%
\StoreBenchExecResult{Vvt}{PortfolioTrueNotSolvedByKinductionPlain}{Wrong}{False}{Cputime}{Avg}{5.7839552724375}%
\StoreBenchExecResult{Vvt}{PortfolioTrueNotSolvedByKinductionPlain}{Wrong}{False}{Cputime}{Median}{0.230086557}%
\StoreBenchExecResult{Vvt}{PortfolioTrueNotSolvedByKinductionPlain}{Wrong}{False}{Cputime}{Min}{0.137341469}%
\StoreBenchExecResult{Vvt}{PortfolioTrueNotSolvedByKinductionPlain}{Wrong}{False}{Cputime}{Max}{42.42557717}%
\StoreBenchExecResult{Vvt}{PortfolioTrueNotSolvedByKinductionPlain}{Wrong}{False}{Cputime}{Stdev}{11.94186226086306751131884217}%
\StoreBenchExecResult{Vvt}{PortfolioTrueNotSolvedByKinductionPlain}{Wrong}{False}{Walltime}{}{84.2754142284348}%
\StoreBenchExecResult{Vvt}{PortfolioTrueNotSolvedByKinductionPlain}{Wrong}{False}{Walltime}{Avg}{5.267213389277175}%
\StoreBenchExecResult{Vvt}{PortfolioTrueNotSolvedByKinductionPlain}{Wrong}{False}{Walltime}{Median}{0.1383640766145}%
\StoreBenchExecResult{Vvt}{PortfolioTrueNotSolvedByKinductionPlain}{Wrong}{False}{Walltime}{Min}{0.0825860500336}%
\StoreBenchExecResult{Vvt}{PortfolioTrueNotSolvedByKinductionPlain}{Wrong}{False}{Walltime}{Max}{39.8229458332}%
\StoreBenchExecResult{Vvt}{PortfolioTrueNotSolvedByKinductionPlain}{Wrong}{False}{Walltime}{Stdev}{11.17372662594733931699860549}%
\StoreBenchExecResult{Vvt}{PortfolioTrueNotSolvedByKinductionPlain}{Correct}{False}{Count}{}{0}%
\StoreBenchExecResult{Vvt}{PortfolioTrueNotSolvedByKinductionPlain}{Correct}{False}{Cputime}{}{0}%
\StoreBenchExecResult{Vvt}{PortfolioTrueNotSolvedByKinductionPlain}{Correct}{False}{Cputime}{Avg}{None}%
\StoreBenchExecResult{Vvt}{PortfolioTrueNotSolvedByKinductionPlain}{Correct}{False}{Cputime}{Median}{None}%
\StoreBenchExecResult{Vvt}{PortfolioTrueNotSolvedByKinductionPlain}{Correct}{False}{Cputime}{Min}{None}%
\StoreBenchExecResult{Vvt}{PortfolioTrueNotSolvedByKinductionPlain}{Correct}{False}{Cputime}{Max}{None}%
\StoreBenchExecResult{Vvt}{PortfolioTrueNotSolvedByKinductionPlain}{Correct}{False}{Cputime}{Stdev}{None}%
\StoreBenchExecResult{Vvt}{PortfolioTrueNotSolvedByKinductionPlain}{Correct}{False}{Walltime}{}{0}%
\StoreBenchExecResult{Vvt}{PortfolioTrueNotSolvedByKinductionPlain}{Correct}{False}{Walltime}{Avg}{None}%
\StoreBenchExecResult{Vvt}{PortfolioTrueNotSolvedByKinductionPlain}{Correct}{False}{Walltime}{Median}{None}%
\StoreBenchExecResult{Vvt}{PortfolioTrueNotSolvedByKinductionPlain}{Correct}{False}{Walltime}{Min}{None}%
\StoreBenchExecResult{Vvt}{PortfolioTrueNotSolvedByKinductionPlain}{Correct}{False}{Walltime}{Max}{None}%
\StoreBenchExecResult{Vvt}{PortfolioTrueNotSolvedByKinductionPlain}{Correct}{False}{Walltime}{Stdev}{None}%
\providecommand\StoreBenchExecResult[7]{\expandafter\newcommand\csname#1#2#3#4#5#6\endcsname{#7}}%
\StoreBenchExecResult{Vvt}{Portfolio}{Total}{}{Count}{}{5591}%
\StoreBenchExecResult{Vvt}{Portfolio}{Total}{}{Cputime}{}{566757.688293365}%
\StoreBenchExecResult{Vvt}{Portfolio}{Total}{}{Cputime}{Avg}{101.3696455541700947952065820}%
\StoreBenchExecResult{Vvt}{Portfolio}{Total}{}{Cputime}{Median}{0.217362877}%
\StoreBenchExecResult{Vvt}{Portfolio}{Total}{}{Cputime}{Min}{0.054837838}%
\StoreBenchExecResult{Vvt}{Portfolio}{Total}{}{Cputime}{Max}{1002.26186594}%
\StoreBenchExecResult{Vvt}{Portfolio}{Total}{}{Cputime}{Stdev}{292.8653271262807638039913710}%
\StoreBenchExecResult{Vvt}{Portfolio}{Total}{}{Walltime}{}{517207.5143973746557}%
\StoreBenchExecResult{Vvt}{Portfolio}{Total}{}{Walltime}{Avg}{92.50715693031204716508674656}%
\StoreBenchExecResult{Vvt}{Portfolio}{Total}{}{Walltime}{Median}{0.162916898727}%
\StoreBenchExecResult{Vvt}{Portfolio}{Total}{}{Walltime}{Min}{0.0356471538544}%
\StoreBenchExecResult{Vvt}{Portfolio}{Total}{}{Walltime}{Max}{1000.40836501}%
\StoreBenchExecResult{Vvt}{Portfolio}{Total}{}{Walltime}{Stdev}{274.0049745050687186119668937}%
\StoreBenchExecResult{Vvt}{Portfolio}{Correct}{}{Count}{}{839}%
\StoreBenchExecResult{Vvt}{Portfolio}{Correct}{}{Cputime}{}{20594.952595098}%
\StoreBenchExecResult{Vvt}{Portfolio}{Correct}{}{Cputime}{Avg}{24.54702335530154946364719905}%
\StoreBenchExecResult{Vvt}{Portfolio}{Correct}{}{Cputime}{Median}{0.45135025}%
\StoreBenchExecResult{Vvt}{Portfolio}{Correct}{}{Cputime}{Min}{0.077574035}%
\StoreBenchExecResult{Vvt}{Portfolio}{Correct}{}{Cputime}{Max}{892.573646031}%
\StoreBenchExecResult{Vvt}{Portfolio}{Correct}{}{Cputime}{Stdev}{92.99653776584251043108566230}%
\StoreBenchExecResult{Vvt}{Portfolio}{Correct}{}{Walltime}{}{15583.2403891090848}%
\StoreBenchExecResult{Vvt}{Portfolio}{Correct}{}{Walltime}{Avg}{18.57358806806803909415971395}%
\StoreBenchExecResult{Vvt}{Portfolio}{Correct}{}{Walltime}{Median}{0.254212141037}%
\StoreBenchExecResult{Vvt}{Portfolio}{Correct}{}{Walltime}{Min}{0.0468108654022}%
\StoreBenchExecResult{Vvt}{Portfolio}{Correct}{}{Walltime}{Max}{766.448214054}%
\StoreBenchExecResult{Vvt}{Portfolio}{Correct}{}{Walltime}{Stdev}{70.89384349018322237767292994}%
\StoreBenchExecResult{Vvt}{Portfolio}{Correct}{False}{Count}{}{311}%
\StoreBenchExecResult{Vvt}{Portfolio}{Correct}{False}{Cputime}{}{9422.681857155}%
\StoreBenchExecResult{Vvt}{Portfolio}{Correct}{False}{Cputime}{Avg}{30.29801240242765273311897106}%
\StoreBenchExecResult{Vvt}{Portfolio}{Correct}{False}{Cputime}{Median}{0.522202119}%
\StoreBenchExecResult{Vvt}{Portfolio}{Correct}{False}{Cputime}{Min}{0.088101543}%
\StoreBenchExecResult{Vvt}{Portfolio}{Correct}{False}{Cputime}{Max}{767.911330292}%
\StoreBenchExecResult{Vvt}{Portfolio}{Correct}{False}{Cputime}{Stdev}{89.43303239856282000433791789}%
\StoreBenchExecResult{Vvt}{Portfolio}{Correct}{False}{Walltime}{}{9084.4517166621157}%
\StoreBenchExecResult{Vvt}{Portfolio}{Correct}{False}{Walltime}{Avg}{29.21045568058558102893890675}%
\StoreBenchExecResult{Vvt}{Portfolio}{Correct}{False}{Walltime}{Median}{0.362382888794}%
\StoreBenchExecResult{Vvt}{Portfolio}{Correct}{False}{Walltime}{Min}{0.0493488311768}%
\StoreBenchExecResult{Vvt}{Portfolio}{Correct}{False}{Walltime}{Max}{766.448214054}%
\StoreBenchExecResult{Vvt}{Portfolio}{Correct}{False}{Walltime}{Stdev}{88.77428399698858051011203099}%
\StoreBenchExecResult{Vvt}{Portfolio}{Correct}{True}{Count}{}{528}%
\StoreBenchExecResult{Vvt}{Portfolio}{Correct}{True}{Cputime}{}{11172.270737943}%
\StoreBenchExecResult{Vvt}{Portfolio}{Correct}{True}{Cputime}{Avg}{21.15960367034659090909090909}%
\StoreBenchExecResult{Vvt}{Portfolio}{Correct}{True}{Cputime}{Median}{0.4408726625}%
\StoreBenchExecResult{Vvt}{Portfolio}{Correct}{True}{Cputime}{Min}{0.077574035}%
\StoreBenchExecResult{Vvt}{Portfolio}{Correct}{True}{Cputime}{Max}{892.573646031}%
\StoreBenchExecResult{Vvt}{Portfolio}{Correct}{True}{Cputime}{Stdev}{94.86997699913932648848749771}%
\StoreBenchExecResult{Vvt}{Portfolio}{Correct}{True}{Walltime}{}{6498.7886724469691}%
\StoreBenchExecResult{Vvt}{Portfolio}{Correct}{True}{Walltime}{Avg}{12.30831187963441117424242424}%
\StoreBenchExecResult{Vvt}{Portfolio}{Correct}{True}{Walltime}{Median}{0.241463422775}%
\StoreBenchExecResult{Vvt}{Portfolio}{Correct}{True}{Walltime}{Min}{0.0468108654022}%
\StoreBenchExecResult{Vvt}{Portfolio}{Correct}{True}{Walltime}{Max}{488.931025982}%
\StoreBenchExecResult{Vvt}{Portfolio}{Correct}{True}{Walltime}{Stdev}{56.90728966912478709469420789}%
\StoreBenchExecResult{Vvt}{Portfolio}{Error}{}{Count}{}{546}%
\StoreBenchExecResult{Vvt}{Portfolio}{Error}{}{Cputime}{}{532751.055141085}%
\StoreBenchExecResult{Vvt}{Portfolio}{Error}{}{Cputime}{Avg}{975.7345332254304029304029304}%
\StoreBenchExecResult{Vvt}{Portfolio}{Error}{}{Cputime}{Median}{1000.055754145}%
\StoreBenchExecResult{Vvt}{Portfolio}{Error}{}{Cputime}{Min}{113.217950864}%
\StoreBenchExecResult{Vvt}{Portfolio}{Error}{}{Cputime}{Max}{1002.26186594}%
\StoreBenchExecResult{Vvt}{Portfolio}{Error}{}{Cputime}{Stdev}{123.1435671139726910394818115}%
\StoreBenchExecResult{Vvt}{Portfolio}{Error}{}{Walltime}{}{488574.8420691404}%
\StoreBenchExecResult{Vvt}{Portfolio}{Error}{}{Walltime}{Avg}{894.8257180753487179487179487}%
\StoreBenchExecResult{Vvt}{Portfolio}{Error}{}{Walltime}{Median}{997.7120400665}%
\StoreBenchExecResult{Vvt}{Portfolio}{Error}{}{Walltime}{Min}{79.8744199276}%
\StoreBenchExecResult{Vvt}{Portfolio}{Error}{}{Walltime}{Max}{1000.40836501}%
\StoreBenchExecResult{Vvt}{Portfolio}{Error}{}{Walltime}{Stdev}{213.1878471971383750951504593}%
\StoreBenchExecResult{Vvt}{Portfolio}{Error}{OutOfMemory}{Count}{}{22}%
\StoreBenchExecResult{Vvt}{Portfolio}{Error}{OutOfMemory}{Cputime}{}{8673.638364591}%
\StoreBenchExecResult{Vvt}{Portfolio}{Error}{OutOfMemory}{Cputime}{Avg}{394.2562892995909090909090909}%
\StoreBenchExecResult{Vvt}{Portfolio}{Error}{OutOfMemory}{Cputime}{Median}{369.265465066}%
\StoreBenchExecResult{Vvt}{Portfolio}{Error}{OutOfMemory}{Cputime}{Min}{113.217950864}%
\StoreBenchExecResult{Vvt}{Portfolio}{Error}{OutOfMemory}{Cputime}{Max}{697.51567418}%
\StoreBenchExecResult{Vvt}{Portfolio}{Error}{OutOfMemory}{Cputime}{Stdev}{153.8997533847279064983842486}%
\StoreBenchExecResult{Vvt}{Portfolio}{Error}{OutOfMemory}{Walltime}{}{7260.7060036654}%
\StoreBenchExecResult{Vvt}{Portfolio}{Error}{OutOfMemory}{Walltime}{Avg}{330.0320910757}%
\StoreBenchExecResult{Vvt}{Portfolio}{Error}{OutOfMemory}{Walltime}{Median}{363.239386439}%
\StoreBenchExecResult{Vvt}{Portfolio}{Error}{OutOfMemory}{Walltime}{Min}{79.8744199276}%
\StoreBenchExecResult{Vvt}{Portfolio}{Error}{OutOfMemory}{Walltime}{Max}{694.541884899}%
\StoreBenchExecResult{Vvt}{Portfolio}{Error}{OutOfMemory}{Walltime}{Stdev}{158.4869059836717167107443646}%
\StoreBenchExecResult{Vvt}{Portfolio}{Error}{Timeout}{Count}{}{524}%
\StoreBenchExecResult{Vvt}{Portfolio}{Error}{Timeout}{Cputime}{}{524077.416776494}%
\StoreBenchExecResult{Vvt}{Portfolio}{Error}{Timeout}{Cputime}{Avg}{1000.147741939874045801526718}%
\StoreBenchExecResult{Vvt}{Portfolio}{Error}{Timeout}{Cputime}{Median}{1000.056074105}%
\StoreBenchExecResult{Vvt}{Portfolio}{Error}{Timeout}{Cputime}{Min}{913.226467424}%
\StoreBenchExecResult{Vvt}{Portfolio}{Error}{Timeout}{Cputime}{Max}{1002.26186594}%
\StoreBenchExecResult{Vvt}{Portfolio}{Error}{Timeout}{Cputime}{Stdev}{3.852309396089440885378275393}%
\StoreBenchExecResult{Vvt}{Portfolio}{Error}{Timeout}{Walltime}{}{481314.136065475}%
\StoreBenchExecResult{Vvt}{Portfolio}{Error}{Timeout}{Walltime}{Avg}{918.5384276058683206106870229}%
\StoreBenchExecResult{Vvt}{Portfolio}{Error}{Timeout}{Walltime}{Median}{997.9254604575}%
\StoreBenchExecResult{Vvt}{Portfolio}{Error}{Timeout}{Walltime}{Min}{457.325178862}%
\StoreBenchExecResult{Vvt}{Portfolio}{Error}{Timeout}{Walltime}{Max}{1000.40836501}%
\StoreBenchExecResult{Vvt}{Portfolio}{Error}{Timeout}{Walltime}{Stdev}{179.8542972848907747933615811}%
\StoreBenchExecResult{Vvt}{Portfolio}{Unknown}{}{Count}{}{4175}%
\StoreBenchExecResult{Vvt}{Portfolio}{Unknown}{}{Cputime}{}{11403.974684807}%
\StoreBenchExecResult{Vvt}{Portfolio}{Unknown}{}{Cputime}{Avg}{2.731490942468742514970059880}%
\StoreBenchExecResult{Vvt}{Portfolio}{Unknown}{}{Cputime}{Median}{0.179372175}%
\StoreBenchExecResult{Vvt}{Portfolio}{Unknown}{}{Cputime}{Min}{0.054837838}%
\StoreBenchExecResult{Vvt}{Portfolio}{Unknown}{}{Cputime}{Max}{123.080849066}%
\StoreBenchExecResult{Vvt}{Portfolio}{Unknown}{}{Cputime}{Stdev}{7.318248912564787717948756483}%
\StoreBenchExecResult{Vvt}{Portfolio}{Unknown}{}{Walltime}{}{11080.9528553488616}%
\StoreBenchExecResult{Vvt}{Portfolio}{Unknown}{}{Walltime}{Avg}{2.654120444394936910179640719}%
\StoreBenchExecResult{Vvt}{Portfolio}{Unknown}{}{Walltime}{Median}{0.134914875031}%
\StoreBenchExecResult{Vvt}{Portfolio}{Unknown}{}{Walltime}{Min}{0.0356471538544}%
\StoreBenchExecResult{Vvt}{Portfolio}{Unknown}{}{Walltime}{Max}{121.012732029}%
\StoreBenchExecResult{Vvt}{Portfolio}{Unknown}{}{Walltime}{Stdev}{7.200566965995347481761287894}%
\StoreBenchExecResult{Vvt}{Portfolio}{Unknown}{Unknown}{Count}{}{4175}%
\StoreBenchExecResult{Vvt}{Portfolio}{Unknown}{Unknown}{Cputime}{}{11403.974684807}%
\StoreBenchExecResult{Vvt}{Portfolio}{Unknown}{Unknown}{Cputime}{Avg}{2.731490942468742514970059880}%
\StoreBenchExecResult{Vvt}{Portfolio}{Unknown}{Unknown}{Cputime}{Median}{0.179372175}%
\StoreBenchExecResult{Vvt}{Portfolio}{Unknown}{Unknown}{Cputime}{Min}{0.054837838}%
\StoreBenchExecResult{Vvt}{Portfolio}{Unknown}{Unknown}{Cputime}{Max}{123.080849066}%
\StoreBenchExecResult{Vvt}{Portfolio}{Unknown}{Unknown}{Cputime}{Stdev}{7.318248912564787717948756483}%
\StoreBenchExecResult{Vvt}{Portfolio}{Unknown}{Unknown}{Walltime}{}{11080.9528553488616}%
\StoreBenchExecResult{Vvt}{Portfolio}{Unknown}{Unknown}{Walltime}{Avg}{2.654120444394936910179640719}%
\StoreBenchExecResult{Vvt}{Portfolio}{Unknown}{Unknown}{Walltime}{Median}{0.134914875031}%
\StoreBenchExecResult{Vvt}{Portfolio}{Unknown}{Unknown}{Walltime}{Min}{0.0356471538544}%
\StoreBenchExecResult{Vvt}{Portfolio}{Unknown}{Unknown}{Walltime}{Max}{121.012732029}%
\StoreBenchExecResult{Vvt}{Portfolio}{Unknown}{Unknown}{Walltime}{Stdev}{7.200566965995347481761287894}%
\StoreBenchExecResult{Vvt}{Portfolio}{Wrong}{}{Count}{}{31}%
\StoreBenchExecResult{Vvt}{Portfolio}{Wrong}{}{Cputime}{}{2007.705872375}%
\StoreBenchExecResult{Vvt}{Portfolio}{Wrong}{}{Cputime}{Avg}{64.76470556048387096774193548}%
\StoreBenchExecResult{Vvt}{Portfolio}{Wrong}{}{Cputime}{Median}{0.283826447}%
\StoreBenchExecResult{Vvt}{Portfolio}{Wrong}{}{Cputime}{Min}{0.098004762}%
\StoreBenchExecResult{Vvt}{Portfolio}{Wrong}{}{Cputime}{Max}{512.637972405}%
\StoreBenchExecResult{Vvt}{Portfolio}{Wrong}{}{Cputime}{Stdev}{154.8183203387929317784988208}%
\StoreBenchExecResult{Vvt}{Portfolio}{Wrong}{}{Walltime}{}{1968.4790837763093}%
\StoreBenchExecResult{Vvt}{Portfolio}{Wrong}{}{Walltime}{Avg}{63.49932528310675161290322581}%
\StoreBenchExecResult{Vvt}{Portfolio}{Wrong}{}{Walltime}{Median}{0.166478157043}%
\StoreBenchExecResult{Vvt}{Portfolio}{Wrong}{}{Walltime}{Min}{0.0577459335327}%
\StoreBenchExecResult{Vvt}{Portfolio}{Wrong}{}{Walltime}{Max}{507.390437841}%
\StoreBenchExecResult{Vvt}{Portfolio}{Wrong}{}{Walltime}{Stdev}{153.3458216018608743732258898}%
\StoreBenchExecResult{Vvt}{Portfolio}{Wrong}{False}{Count}{}{22}%
\StoreBenchExecResult{Vvt}{Portfolio}{Wrong}{False}{Cputime}{}{136.265259603}%
\StoreBenchExecResult{Vvt}{Portfolio}{Wrong}{False}{Cputime}{Avg}{6.1938754365}%
\StoreBenchExecResult{Vvt}{Portfolio}{Wrong}{False}{Cputime}{Median}{0.1789548885}%
\StoreBenchExecResult{Vvt}{Portfolio}{Wrong}{False}{Cputime}{Min}{0.109771843}%
\StoreBenchExecResult{Vvt}{Portfolio}{Wrong}{False}{Cputime}{Max}{42.42557717}%
\StoreBenchExecResult{Vvt}{Portfolio}{Wrong}{False}{Cputime}{Stdev}{12.93132467090578230398841105}%
\StoreBenchExecResult{Vvt}{Portfolio}{Wrong}{False}{Walltime}{}{125.7506589889339}%
\StoreBenchExecResult{Vvt}{Portfolio}{Wrong}{False}{Walltime}{Avg}{5.715939044951540909090909091}%
\StoreBenchExecResult{Vvt}{Portfolio}{Wrong}{False}{Walltime}{Median}{0.110868096352}%
\StoreBenchExecResult{Vvt}{Portfolio}{Wrong}{False}{Walltime}{Min}{0.058522939682}%
\StoreBenchExecResult{Vvt}{Portfolio}{Wrong}{False}{Walltime}{Max}{39.8229458332}%
\StoreBenchExecResult{Vvt}{Portfolio}{Wrong}{False}{Walltime}{Stdev}{12.24776009839099025758730502}%
\StoreBenchExecResult{Vvt}{Portfolio}{Wrong}{True}{Count}{}{9}%
\StoreBenchExecResult{Vvt}{Portfolio}{Wrong}{True}{Cputime}{}{1871.440612772}%
\StoreBenchExecResult{Vvt}{Portfolio}{Wrong}{True}{Cputime}{Avg}{207.9378458635555555555555556}%
\StoreBenchExecResult{Vvt}{Portfolio}{Wrong}{True}{Cputime}{Median}{17.360425255}%
\StoreBenchExecResult{Vvt}{Portfolio}{Wrong}{True}{Cputime}{Min}{0.098004762}%
\StoreBenchExecResult{Vvt}{Portfolio}{Wrong}{True}{Cputime}{Max}{512.637972405}%
\StoreBenchExecResult{Vvt}{Portfolio}{Wrong}{True}{Cputime}{Stdev}{230.7938955885960188365681475}%
\StoreBenchExecResult{Vvt}{Portfolio}{Wrong}{True}{Walltime}{}{1842.7284247873754}%
\StoreBenchExecResult{Vvt}{Portfolio}{Wrong}{True}{Walltime}{Avg}{204.7476027541528222222222222}%
\StoreBenchExecResult{Vvt}{Portfolio}{Wrong}{True}{Walltime}{Median}{8.77038788795}%
\StoreBenchExecResult{Vvt}{Portfolio}{Wrong}{True}{Walltime}{Min}{0.0577459335327}%
\StoreBenchExecResult{Vvt}{Portfolio}{Wrong}{True}{Walltime}{Max}{507.390437841}%
\StoreBenchExecResult{Vvt}{Portfolio}{Wrong}{True}{Walltime}{Stdev}{229.1644451605636817900785342}%
\ifdefined\PdrInvKinductionPlainTotalCount\else\edef\PdrInvKinductionPlainTotalCount{0}\fi
\ifdefined\PdrInvKinductionPlainCorrectCount\else\edef\PdrInvKinductionPlainCorrectCount{0}\fi
\ifdefined\PdrInvKinductionPlainCorrectTrueCount\else\edef\PdrInvKinductionPlainCorrectTrueCount{0}\fi
\ifdefined\PdrInvKinductionPlainCorrectFalseCount\else\edef\PdrInvKinductionPlainCorrectFalseCount{0}\fi
\ifdefined\PdrInvKinductionPlainWrongTrueCount\else\edef\PdrInvKinductionPlainWrongTrueCount{0}\fi
\ifdefined\PdrInvKinductionPlainWrongFalseCount\else\edef\PdrInvKinductionPlainWrongFalseCount{0}\fi
\ifdefined\PdrInvKinductionPlainErrorTimeoutCount\else\edef\PdrInvKinductionPlainErrorTimeoutCount{0}\fi
\ifdefined\PdrInvKinductionPlainErrorOutOfMemoryCount\else\edef\PdrInvKinductionPlainErrorOutOfMemoryCount{0}\fi
\ifdefined\PdrInvKinductionPlainCorrectCputime\else\edef\PdrInvKinductionPlainCorrectCputime{0}\fi
\ifdefined\PdrInvKinductionPlainCorrectCputimeAvg\else\edef\PdrInvKinductionPlainCorrectCputimeAvg{None}\fi
\ifdefined\PdrInvKinductionPlainCorrectWalltime\else\edef\PdrInvKinductionPlainCorrectWalltime{0}\fi
\ifdefined\PdrInvKinductionPlainCorrectWalltimeAvg\else\edef\PdrInvKinductionPlainCorrectWalltimeAvg{None}\fi
\ifdefined\PdrInvKinductionDfStaticZeroZeroTTotalCount\else\edef\PdrInvKinductionDfStaticZeroZeroTTotalCount{0}\fi
\ifdefined\PdrInvKinductionDfStaticZeroZeroTCorrectCount\else\edef\PdrInvKinductionDfStaticZeroZeroTCorrectCount{0}\fi
\ifdefined\PdrInvKinductionDfStaticZeroZeroTCorrectTrueCount\else\edef\PdrInvKinductionDfStaticZeroZeroTCorrectTrueCount{0}\fi
\ifdefined\PdrInvKinductionDfStaticZeroZeroTCorrectFalseCount\else\edef\PdrInvKinductionDfStaticZeroZeroTCorrectFalseCount{0}\fi
\ifdefined\PdrInvKinductionDfStaticZeroZeroTWrongTrueCount\else\edef\PdrInvKinductionDfStaticZeroZeroTWrongTrueCount{0}\fi
\ifdefined\PdrInvKinductionDfStaticZeroZeroTWrongFalseCount\else\edef\PdrInvKinductionDfStaticZeroZeroTWrongFalseCount{0}\fi
\ifdefined\PdrInvKinductionDfStaticZeroZeroTErrorTimeoutCount\else\edef\PdrInvKinductionDfStaticZeroZeroTErrorTimeoutCount{0}\fi
\ifdefined\PdrInvKinductionDfStaticZeroZeroTErrorOutOfMemoryCount\else\edef\PdrInvKinductionDfStaticZeroZeroTErrorOutOfMemoryCount{0}\fi
\ifdefined\PdrInvKinductionDfStaticZeroZeroTCorrectCputime\else\edef\PdrInvKinductionDfStaticZeroZeroTCorrectCputime{0}\fi
\ifdefined\PdrInvKinductionDfStaticZeroZeroTCorrectCputimeAvg\else\edef\PdrInvKinductionDfStaticZeroZeroTCorrectCputimeAvg{None}\fi
\ifdefined\PdrInvKinductionDfStaticZeroZeroTCorrectWalltime\else\edef\PdrInvKinductionDfStaticZeroZeroTCorrectWalltime{0}\fi
\ifdefined\PdrInvKinductionDfStaticZeroZeroTCorrectWalltimeAvg\else\edef\PdrInvKinductionDfStaticZeroZeroTCorrectWalltimeAvg{None}\fi
\ifdefined\PdrInvKinductionDfStaticZeroOneTTTotalCount\else\edef\PdrInvKinductionDfStaticZeroOneTTTotalCount{0}\fi
\ifdefined\PdrInvKinductionDfStaticZeroOneTTCorrectCount\else\edef\PdrInvKinductionDfStaticZeroOneTTCorrectCount{0}\fi
\ifdefined\PdrInvKinductionDfStaticZeroOneTTCorrectTrueCount\else\edef\PdrInvKinductionDfStaticZeroOneTTCorrectTrueCount{0}\fi
\ifdefined\PdrInvKinductionDfStaticZeroOneTTCorrectFalseCount\else\edef\PdrInvKinductionDfStaticZeroOneTTCorrectFalseCount{0}\fi
\ifdefined\PdrInvKinductionDfStaticZeroOneTTWrongTrueCount\else\edef\PdrInvKinductionDfStaticZeroOneTTWrongTrueCount{0}\fi
\ifdefined\PdrInvKinductionDfStaticZeroOneTTWrongFalseCount\else\edef\PdrInvKinductionDfStaticZeroOneTTWrongFalseCount{0}\fi
\ifdefined\PdrInvKinductionDfStaticZeroOneTTErrorTimeoutCount\else\edef\PdrInvKinductionDfStaticZeroOneTTErrorTimeoutCount{0}\fi
\ifdefined\PdrInvKinductionDfStaticZeroOneTTErrorOutOfMemoryCount\else\edef\PdrInvKinductionDfStaticZeroOneTTErrorOutOfMemoryCount{0}\fi
\ifdefined\PdrInvKinductionDfStaticZeroOneTTCorrectCputime\else\edef\PdrInvKinductionDfStaticZeroOneTTCorrectCputime{0}\fi
\ifdefined\PdrInvKinductionDfStaticZeroOneTTCorrectCputimeAvg\else\edef\PdrInvKinductionDfStaticZeroOneTTCorrectCputimeAvg{None}\fi
\ifdefined\PdrInvKinductionDfStaticZeroOneTTCorrectWalltime\else\edef\PdrInvKinductionDfStaticZeroOneTTCorrectWalltime{0}\fi
\ifdefined\PdrInvKinductionDfStaticZeroOneTTCorrectWalltimeAvg\else\edef\PdrInvKinductionDfStaticZeroOneTTCorrectWalltimeAvg{None}\fi
\ifdefined\PdrInvKinductionDfStaticZeroOneTFTotalCount\else\edef\PdrInvKinductionDfStaticZeroOneTFTotalCount{0}\fi
\ifdefined\PdrInvKinductionDfStaticZeroOneTFCorrectCount\else\edef\PdrInvKinductionDfStaticZeroOneTFCorrectCount{0}\fi
\ifdefined\PdrInvKinductionDfStaticZeroOneTFCorrectTrueCount\else\edef\PdrInvKinductionDfStaticZeroOneTFCorrectTrueCount{0}\fi
\ifdefined\PdrInvKinductionDfStaticZeroOneTFCorrectFalseCount\else\edef\PdrInvKinductionDfStaticZeroOneTFCorrectFalseCount{0}\fi
\ifdefined\PdrInvKinductionDfStaticZeroOneTFWrongTrueCount\else\edef\PdrInvKinductionDfStaticZeroOneTFWrongTrueCount{0}\fi
\ifdefined\PdrInvKinductionDfStaticZeroOneTFWrongFalseCount\else\edef\PdrInvKinductionDfStaticZeroOneTFWrongFalseCount{0}\fi
\ifdefined\PdrInvKinductionDfStaticZeroOneTFErrorTimeoutCount\else\edef\PdrInvKinductionDfStaticZeroOneTFErrorTimeoutCount{0}\fi
\ifdefined\PdrInvKinductionDfStaticZeroOneTFErrorOutOfMemoryCount\else\edef\PdrInvKinductionDfStaticZeroOneTFErrorOutOfMemoryCount{0}\fi
\ifdefined\PdrInvKinductionDfStaticZeroOneTFCorrectCputime\else\edef\PdrInvKinductionDfStaticZeroOneTFCorrectCputime{0}\fi
\ifdefined\PdrInvKinductionDfStaticZeroOneTFCorrectCputimeAvg\else\edef\PdrInvKinductionDfStaticZeroOneTFCorrectCputimeAvg{None}\fi
\ifdefined\PdrInvKinductionDfStaticZeroOneTFCorrectWalltime\else\edef\PdrInvKinductionDfStaticZeroOneTFCorrectWalltime{0}\fi
\ifdefined\PdrInvKinductionDfStaticZeroOneTFCorrectWalltimeAvg\else\edef\PdrInvKinductionDfStaticZeroOneTFCorrectWalltimeAvg{None}\fi
\ifdefined\PdrInvKinductionDfStaticZeroTwoTTTotalCount\else\edef\PdrInvKinductionDfStaticZeroTwoTTTotalCount{0}\fi
\ifdefined\PdrInvKinductionDfStaticZeroTwoTTCorrectCount\else\edef\PdrInvKinductionDfStaticZeroTwoTTCorrectCount{0}\fi
\ifdefined\PdrInvKinductionDfStaticZeroTwoTTCorrectTrueCount\else\edef\PdrInvKinductionDfStaticZeroTwoTTCorrectTrueCount{0}\fi
\ifdefined\PdrInvKinductionDfStaticZeroTwoTTCorrectFalseCount\else\edef\PdrInvKinductionDfStaticZeroTwoTTCorrectFalseCount{0}\fi
\ifdefined\PdrInvKinductionDfStaticZeroTwoTTWrongTrueCount\else\edef\PdrInvKinductionDfStaticZeroTwoTTWrongTrueCount{0}\fi
\ifdefined\PdrInvKinductionDfStaticZeroTwoTTWrongFalseCount\else\edef\PdrInvKinductionDfStaticZeroTwoTTWrongFalseCount{0}\fi
\ifdefined\PdrInvKinductionDfStaticZeroTwoTTErrorTimeoutCount\else\edef\PdrInvKinductionDfStaticZeroTwoTTErrorTimeoutCount{0}\fi
\ifdefined\PdrInvKinductionDfStaticZeroTwoTTErrorOutOfMemoryCount\else\edef\PdrInvKinductionDfStaticZeroTwoTTErrorOutOfMemoryCount{0}\fi
\ifdefined\PdrInvKinductionDfStaticZeroTwoTTCorrectCputime\else\edef\PdrInvKinductionDfStaticZeroTwoTTCorrectCputime{0}\fi
\ifdefined\PdrInvKinductionDfStaticZeroTwoTTCorrectCputimeAvg\else\edef\PdrInvKinductionDfStaticZeroTwoTTCorrectCputimeAvg{None}\fi
\ifdefined\PdrInvKinductionDfStaticZeroTwoTTCorrectWalltime\else\edef\PdrInvKinductionDfStaticZeroTwoTTCorrectWalltime{0}\fi
\ifdefined\PdrInvKinductionDfStaticZeroTwoTTCorrectWalltimeAvg\else\edef\PdrInvKinductionDfStaticZeroTwoTTCorrectWalltimeAvg{None}\fi
\ifdefined\PdrInvKinductionDfStaticZeroTwoTFTotalCount\else\edef\PdrInvKinductionDfStaticZeroTwoTFTotalCount{0}\fi
\ifdefined\PdrInvKinductionDfStaticZeroTwoTFCorrectCount\else\edef\PdrInvKinductionDfStaticZeroTwoTFCorrectCount{0}\fi
\ifdefined\PdrInvKinductionDfStaticZeroTwoTFCorrectTrueCount\else\edef\PdrInvKinductionDfStaticZeroTwoTFCorrectTrueCount{0}\fi
\ifdefined\PdrInvKinductionDfStaticZeroTwoTFCorrectFalseCount\else\edef\PdrInvKinductionDfStaticZeroTwoTFCorrectFalseCount{0}\fi
\ifdefined\PdrInvKinductionDfStaticZeroTwoTFWrongTrueCount\else\edef\PdrInvKinductionDfStaticZeroTwoTFWrongTrueCount{0}\fi
\ifdefined\PdrInvKinductionDfStaticZeroTwoTFWrongFalseCount\else\edef\PdrInvKinductionDfStaticZeroTwoTFWrongFalseCount{0}\fi
\ifdefined\PdrInvKinductionDfStaticZeroTwoTFErrorTimeoutCount\else\edef\PdrInvKinductionDfStaticZeroTwoTFErrorTimeoutCount{0}\fi
\ifdefined\PdrInvKinductionDfStaticZeroTwoTFErrorOutOfMemoryCount\else\edef\PdrInvKinductionDfStaticZeroTwoTFErrorOutOfMemoryCount{0}\fi
\ifdefined\PdrInvKinductionDfStaticZeroTwoTFCorrectCputime\else\edef\PdrInvKinductionDfStaticZeroTwoTFCorrectCputime{0}\fi
\ifdefined\PdrInvKinductionDfStaticZeroTwoTFCorrectCputimeAvg\else\edef\PdrInvKinductionDfStaticZeroTwoTFCorrectCputimeAvg{None}\fi
\ifdefined\PdrInvKinductionDfStaticZeroTwoTFCorrectWalltime\else\edef\PdrInvKinductionDfStaticZeroTwoTFCorrectWalltime{0}\fi
\ifdefined\PdrInvKinductionDfStaticZeroTwoTFCorrectWalltimeAvg\else\edef\PdrInvKinductionDfStaticZeroTwoTFCorrectWalltimeAvg{None}\fi
\ifdefined\PdrInvKinductionDfStaticEightTwoTTotalCount\else\edef\PdrInvKinductionDfStaticEightTwoTTotalCount{0}\fi
\ifdefined\PdrInvKinductionDfStaticEightTwoTCorrectCount\else\edef\PdrInvKinductionDfStaticEightTwoTCorrectCount{0}\fi
\ifdefined\PdrInvKinductionDfStaticEightTwoTCorrectTrueCount\else\edef\PdrInvKinductionDfStaticEightTwoTCorrectTrueCount{0}\fi
\ifdefined\PdrInvKinductionDfStaticEightTwoTCorrectFalseCount\else\edef\PdrInvKinductionDfStaticEightTwoTCorrectFalseCount{0}\fi
\ifdefined\PdrInvKinductionDfStaticEightTwoTWrongTrueCount\else\edef\PdrInvKinductionDfStaticEightTwoTWrongTrueCount{0}\fi
\ifdefined\PdrInvKinductionDfStaticEightTwoTWrongFalseCount\else\edef\PdrInvKinductionDfStaticEightTwoTWrongFalseCount{0}\fi
\ifdefined\PdrInvKinductionDfStaticEightTwoTErrorTimeoutCount\else\edef\PdrInvKinductionDfStaticEightTwoTErrorTimeoutCount{0}\fi
\ifdefined\PdrInvKinductionDfStaticEightTwoTErrorOutOfMemoryCount\else\edef\PdrInvKinductionDfStaticEightTwoTErrorOutOfMemoryCount{0}\fi
\ifdefined\PdrInvKinductionDfStaticEightTwoTCorrectCputime\else\edef\PdrInvKinductionDfStaticEightTwoTCorrectCputime{0}\fi
\ifdefined\PdrInvKinductionDfStaticEightTwoTCorrectCputimeAvg\else\edef\PdrInvKinductionDfStaticEightTwoTCorrectCputimeAvg{None}\fi
\ifdefined\PdrInvKinductionDfStaticEightTwoTCorrectWalltime\else\edef\PdrInvKinductionDfStaticEightTwoTCorrectWalltime{0}\fi
\ifdefined\PdrInvKinductionDfStaticEightTwoTCorrectWalltimeAvg\else\edef\PdrInvKinductionDfStaticEightTwoTCorrectWalltimeAvg{None}\fi
\ifdefined\PdrInvKinductionDfStaticSixteenTwoTTotalCount\else\edef\PdrInvKinductionDfStaticSixteenTwoTTotalCount{0}\fi
\ifdefined\PdrInvKinductionDfStaticSixteenTwoTCorrectCount\else\edef\PdrInvKinductionDfStaticSixteenTwoTCorrectCount{0}\fi
\ifdefined\PdrInvKinductionDfStaticSixteenTwoTCorrectTrueCount\else\edef\PdrInvKinductionDfStaticSixteenTwoTCorrectTrueCount{0}\fi
\ifdefined\PdrInvKinductionDfStaticSixteenTwoTCorrectFalseCount\else\edef\PdrInvKinductionDfStaticSixteenTwoTCorrectFalseCount{0}\fi
\ifdefined\PdrInvKinductionDfStaticSixteenTwoTWrongTrueCount\else\edef\PdrInvKinductionDfStaticSixteenTwoTWrongTrueCount{0}\fi
\ifdefined\PdrInvKinductionDfStaticSixteenTwoTWrongFalseCount\else\edef\PdrInvKinductionDfStaticSixteenTwoTWrongFalseCount{0}\fi
\ifdefined\PdrInvKinductionDfStaticSixteenTwoTErrorTimeoutCount\else\edef\PdrInvKinductionDfStaticSixteenTwoTErrorTimeoutCount{0}\fi
\ifdefined\PdrInvKinductionDfStaticSixteenTwoTErrorOutOfMemoryCount\else\edef\PdrInvKinductionDfStaticSixteenTwoTErrorOutOfMemoryCount{0}\fi
\ifdefined\PdrInvKinductionDfStaticSixteenTwoTCorrectCputime\else\edef\PdrInvKinductionDfStaticSixteenTwoTCorrectCputime{0}\fi
\ifdefined\PdrInvKinductionDfStaticSixteenTwoTCorrectCputimeAvg\else\edef\PdrInvKinductionDfStaticSixteenTwoTCorrectCputimeAvg{None}\fi
\ifdefined\PdrInvKinductionDfStaticSixteenTwoTCorrectWalltime\else\edef\PdrInvKinductionDfStaticSixteenTwoTCorrectWalltime{0}\fi
\ifdefined\PdrInvKinductionDfStaticSixteenTwoTCorrectWalltimeAvg\else\edef\PdrInvKinductionDfStaticSixteenTwoTCorrectWalltimeAvg{None}\fi
\ifdefined\PdrInvKinductionDfStaticSixteenTwoFTotalCount\else\edef\PdrInvKinductionDfStaticSixteenTwoFTotalCount{0}\fi
\ifdefined\PdrInvKinductionDfStaticSixteenTwoFCorrectCount\else\edef\PdrInvKinductionDfStaticSixteenTwoFCorrectCount{0}\fi
\ifdefined\PdrInvKinductionDfStaticSixteenTwoFCorrectTrueCount\else\edef\PdrInvKinductionDfStaticSixteenTwoFCorrectTrueCount{0}\fi
\ifdefined\PdrInvKinductionDfStaticSixteenTwoFCorrectFalseCount\else\edef\PdrInvKinductionDfStaticSixteenTwoFCorrectFalseCount{0}\fi
\ifdefined\PdrInvKinductionDfStaticSixteenTwoFWrongTrueCount\else\edef\PdrInvKinductionDfStaticSixteenTwoFWrongTrueCount{0}\fi
\ifdefined\PdrInvKinductionDfStaticSixteenTwoFWrongFalseCount\else\edef\PdrInvKinductionDfStaticSixteenTwoFWrongFalseCount{0}\fi
\ifdefined\PdrInvKinductionDfStaticSixteenTwoFErrorTimeoutCount\else\edef\PdrInvKinductionDfStaticSixteenTwoFErrorTimeoutCount{0}\fi
\ifdefined\PdrInvKinductionDfStaticSixteenTwoFErrorOutOfMemoryCount\else\edef\PdrInvKinductionDfStaticSixteenTwoFErrorOutOfMemoryCount{0}\fi
\ifdefined\PdrInvKinductionDfStaticSixteenTwoFCorrectCputime\else\edef\PdrInvKinductionDfStaticSixteenTwoFCorrectCputime{0}\fi
\ifdefined\PdrInvKinductionDfStaticSixteenTwoFCorrectCputimeAvg\else\edef\PdrInvKinductionDfStaticSixteenTwoFCorrectCputimeAvg{None}\fi
\ifdefined\PdrInvKinductionDfStaticSixteenTwoFCorrectWalltime\else\edef\PdrInvKinductionDfStaticSixteenTwoFCorrectWalltime{0}\fi
\ifdefined\PdrInvKinductionDfStaticSixteenTwoFCorrectWalltimeAvg\else\edef\PdrInvKinductionDfStaticSixteenTwoFCorrectWalltimeAvg{None}\fi
\ifdefined\PdrInvKinductionDfTotalCount\else\edef\PdrInvKinductionDfTotalCount{0}\fi
\ifdefined\PdrInvKinductionDfCorrectCount\else\edef\PdrInvKinductionDfCorrectCount{0}\fi
\ifdefined\PdrInvKinductionDfCorrectTrueCount\else\edef\PdrInvKinductionDfCorrectTrueCount{0}\fi
\ifdefined\PdrInvKinductionDfCorrectFalseCount\else\edef\PdrInvKinductionDfCorrectFalseCount{0}\fi
\ifdefined\PdrInvKinductionDfWrongTrueCount\else\edef\PdrInvKinductionDfWrongTrueCount{0}\fi
\ifdefined\PdrInvKinductionDfWrongFalseCount\else\edef\PdrInvKinductionDfWrongFalseCount{0}\fi
\ifdefined\PdrInvKinductionDfErrorTimeoutCount\else\edef\PdrInvKinductionDfErrorTimeoutCount{0}\fi
\ifdefined\PdrInvKinductionDfErrorOutOfMemoryCount\else\edef\PdrInvKinductionDfErrorOutOfMemoryCount{0}\fi
\ifdefined\PdrInvKinductionDfCorrectCputime\else\edef\PdrInvKinductionDfCorrectCputime{0}\fi
\ifdefined\PdrInvKinductionDfCorrectCputimeAvg\else\edef\PdrInvKinductionDfCorrectCputimeAvg{None}\fi
\ifdefined\PdrInvKinductionDfCorrectWalltime\else\edef\PdrInvKinductionDfCorrectWalltime{0}\fi
\ifdefined\PdrInvKinductionDfCorrectWalltimeAvg\else\edef\PdrInvKinductionDfCorrectWalltimeAvg{None}\fi
\ifdefined\PdrInvKinductionKipdrTotalCount\else\edef\PdrInvKinductionKipdrTotalCount{0}\fi
\ifdefined\PdrInvKinductionKipdrCorrectCount\else\edef\PdrInvKinductionKipdrCorrectCount{0}\fi
\ifdefined\PdrInvKinductionKipdrCorrectTrueCount\else\edef\PdrInvKinductionKipdrCorrectTrueCount{0}\fi
\ifdefined\PdrInvKinductionKipdrCorrectFalseCount\else\edef\PdrInvKinductionKipdrCorrectFalseCount{0}\fi
\ifdefined\PdrInvKinductionKipdrWrongTrueCount\else\edef\PdrInvKinductionKipdrWrongTrueCount{0}\fi
\ifdefined\PdrInvKinductionKipdrWrongFalseCount\else\edef\PdrInvKinductionKipdrWrongFalseCount{0}\fi
\ifdefined\PdrInvKinductionKipdrErrorTimeoutCount\else\edef\PdrInvKinductionKipdrErrorTimeoutCount{0}\fi
\ifdefined\PdrInvKinductionKipdrErrorOutOfMemoryCount\else\edef\PdrInvKinductionKipdrErrorOutOfMemoryCount{0}\fi
\ifdefined\PdrInvKinductionKipdrCorrectCputime\else\edef\PdrInvKinductionKipdrCorrectCputime{0}\fi
\ifdefined\PdrInvKinductionKipdrCorrectCputimeAvg\else\edef\PdrInvKinductionKipdrCorrectCputimeAvg{None}\fi
\ifdefined\PdrInvKinductionKipdrCorrectWalltime\else\edef\PdrInvKinductionKipdrCorrectWalltime{0}\fi
\ifdefined\PdrInvKinductionKipdrCorrectWalltimeAvg\else\edef\PdrInvKinductionKipdrCorrectWalltimeAvg{None}\fi
\ifdefined\PdrInvKinductionKipdrdfTotalCount\else\edef\PdrInvKinductionKipdrdfTotalCount{0}\fi
\ifdefined\PdrInvKinductionKipdrdfCorrectCount\else\edef\PdrInvKinductionKipdrdfCorrectCount{0}\fi
\ifdefined\PdrInvKinductionKipdrdfCorrectTrueCount\else\edef\PdrInvKinductionKipdrdfCorrectTrueCount{0}\fi
\ifdefined\PdrInvKinductionKipdrdfCorrectFalseCount\else\edef\PdrInvKinductionKipdrdfCorrectFalseCount{0}\fi
\ifdefined\PdrInvKinductionKipdrdfWrongTrueCount\else\edef\PdrInvKinductionKipdrdfWrongTrueCount{0}\fi
\ifdefined\PdrInvKinductionKipdrdfWrongFalseCount\else\edef\PdrInvKinductionKipdrdfWrongFalseCount{0}\fi
\ifdefined\PdrInvKinductionKipdrdfErrorTimeoutCount\else\edef\PdrInvKinductionKipdrdfErrorTimeoutCount{0}\fi
\ifdefined\PdrInvKinductionKipdrdfErrorOutOfMemoryCount\else\edef\PdrInvKinductionKipdrdfErrorOutOfMemoryCount{0}\fi
\ifdefined\PdrInvKinductionKipdrdfCorrectCputime\else\edef\PdrInvKinductionKipdrdfCorrectCputime{0}\fi
\ifdefined\PdrInvKinductionKipdrdfCorrectCputimeAvg\else\edef\PdrInvKinductionKipdrdfCorrectCputimeAvg{None}\fi
\ifdefined\PdrInvKinductionKipdrdfCorrectWalltime\else\edef\PdrInvKinductionKipdrdfCorrectWalltime{0}\fi
\ifdefined\PdrInvKinductionKipdrdfCorrectWalltimeAvg\else\edef\PdrInvKinductionKipdrdfCorrectWalltimeAvg{None}\fi
\ifdefined\PdrInvPdrTotalCount\else\edef\PdrInvPdrTotalCount{0}\fi
\ifdefined\PdrInvPdrCorrectCount\else\edef\PdrInvPdrCorrectCount{0}\fi
\ifdefined\PdrInvPdrCorrectTrueCount\else\edef\PdrInvPdrCorrectTrueCount{0}\fi
\ifdefined\PdrInvPdrCorrectFalseCount\else\edef\PdrInvPdrCorrectFalseCount{0}\fi
\ifdefined\PdrInvPdrWrongTrueCount\else\edef\PdrInvPdrWrongTrueCount{0}\fi
\ifdefined\PdrInvPdrWrongFalseCount\else\edef\PdrInvPdrWrongFalseCount{0}\fi
\ifdefined\PdrInvPdrErrorTimeoutCount\else\edef\PdrInvPdrErrorTimeoutCount{0}\fi
\ifdefined\PdrInvPdrErrorOutOfMemoryCount\else\edef\PdrInvPdrErrorOutOfMemoryCount{0}\fi
\ifdefined\PdrInvPdrCorrectCputime\else\edef\PdrInvPdrCorrectCputime{0}\fi
\ifdefined\PdrInvPdrCorrectCputimeAvg\else\edef\PdrInvPdrCorrectCputimeAvg{None}\fi
\ifdefined\PdrInvPdrCorrectWalltime\else\edef\PdrInvPdrCorrectWalltime{0}\fi
\ifdefined\PdrInvPdrCorrectWalltimeAvg\else\edef\PdrInvPdrCorrectWalltimeAvg{None}\fi
\ifdefined\PdrInvOracleTotalCount\else\edef\PdrInvOracleTotalCount{0}\fi
\ifdefined\PdrInvOracleCorrectCount\else\edef\PdrInvOracleCorrectCount{0}\fi
\ifdefined\PdrInvOracleCorrectTrueCount\else\edef\PdrInvOracleCorrectTrueCount{0}\fi
\ifdefined\PdrInvOracleCorrectFalseCount\else\edef\PdrInvOracleCorrectFalseCount{0}\fi
\ifdefined\PdrInvOracleWrongTrueCount\else\edef\PdrInvOracleWrongTrueCount{0}\fi
\ifdefined\PdrInvOracleWrongFalseCount\else\edef\PdrInvOracleWrongFalseCount{0}\fi
\ifdefined\PdrInvOracleErrorTimeoutCount\else\edef\PdrInvOracleErrorTimeoutCount{0}\fi
\ifdefined\PdrInvOracleErrorOutOfMemoryCount\else\edef\PdrInvOracleErrorOutOfMemoryCount{0}\fi
\ifdefined\PdrInvOracleCorrectCputime\else\edef\PdrInvOracleCorrectCputime{0}\fi
\ifdefined\PdrInvOracleCorrectCputimeAvg\else\edef\PdrInvOracleCorrectCputimeAvg{None}\fi
\ifdefined\PdrInvOracleCorrectWalltime\else\edef\PdrInvOracleCorrectWalltime{0}\fi
\ifdefined\PdrInvOracleCorrectWalltimeAvg\else\edef\PdrInvOracleCorrectWalltimeAvg{None}\fi
\ifdefined\SeahornSeahornTotalCount\else\edef\SeahornSeahornTotalCount{0}\fi
\ifdefined\SeahornSeahornCorrectCount\else\edef\SeahornSeahornCorrectCount{0}\fi
\ifdefined\SeahornSeahornCorrectTrueCount\else\edef\SeahornSeahornCorrectTrueCount{0}\fi
\ifdefined\SeahornSeahornCorrectFalseCount\else\edef\SeahornSeahornCorrectFalseCount{0}\fi
\ifdefined\SeahornSeahornWrongTrueCount\else\edef\SeahornSeahornWrongTrueCount{0}\fi
\ifdefined\SeahornSeahornWrongFalseCount\else\edef\SeahornSeahornWrongFalseCount{0}\fi
\ifdefined\SeahornSeahornErrorTimeoutCount\else\edef\SeahornSeahornErrorTimeoutCount{0}\fi
\ifdefined\SeahornSeahornErrorOutOfMemoryCount\else\edef\SeahornSeahornErrorOutOfMemoryCount{0}\fi
\ifdefined\SeahornSeahornCorrectCputime\else\edef\SeahornSeahornCorrectCputime{0}\fi
\ifdefined\SeahornSeahornCorrectCputimeAvg\else\edef\SeahornSeahornCorrectCputimeAvg{None}\fi
\ifdefined\SeahornSeahornCorrectWalltime\else\edef\SeahornSeahornCorrectWalltime{0}\fi
\ifdefined\SeahornSeahornCorrectWalltimeAvg\else\edef\SeahornSeahornCorrectWalltimeAvg{None}\fi
\ifdefined\VvtCtigarTotalCount\else\edef\VvtCtigarTotalCount{0}\fi
\ifdefined\VvtCtigarCorrectCount\else\edef\VvtCtigarCorrectCount{0}\fi
\ifdefined\VvtCtigarCorrectTrueCount\else\edef\VvtCtigarCorrectTrueCount{0}\fi
\ifdefined\VvtCtigarCorrectFalseCount\else\edef\VvtCtigarCorrectFalseCount{0}\fi
\ifdefined\VvtCtigarWrongTrueCount\else\edef\VvtCtigarWrongTrueCount{0}\fi
\ifdefined\VvtCtigarWrongFalseCount\else\edef\VvtCtigarWrongFalseCount{0}\fi
\ifdefined\VvtCtigarErrorTimeoutCount\else\edef\VvtCtigarErrorTimeoutCount{0}\fi
\ifdefined\VvtCtigarErrorOutOfMemoryCount\else\edef\VvtCtigarErrorOutOfMemoryCount{0}\fi
\ifdefined\VvtCtigarCorrectCputime\else\edef\VvtCtigarCorrectCputime{0}\fi
\ifdefined\VvtCtigarCorrectCputimeAvg\else\edef\VvtCtigarCorrectCputimeAvg{None}\fi
\ifdefined\VvtCtigarCorrectWalltime\else\edef\VvtCtigarCorrectWalltime{0}\fi
\ifdefined\VvtCtigarCorrectWalltimeAvg\else\edef\VvtCtigarCorrectWalltimeAvg{None}\fi
\ifdefined\VvtPortfolioTotalCount\else\edef\VvtPortfolioTotalCount{0}\fi
\ifdefined\VvtPortfolioCorrectCount\else\edef\VvtPortfolioCorrectCount{0}\fi
\ifdefined\VvtPortfolioCorrectTrueCount\else\edef\VvtPortfolioCorrectTrueCount{0}\fi
\ifdefined\VvtPortfolioCorrectFalseCount\else\edef\VvtPortfolioCorrectFalseCount{0}\fi
\ifdefined\VvtPortfolioWrongTrueCount\else\edef\VvtPortfolioWrongTrueCount{0}\fi
\ifdefined\VvtPortfolioWrongFalseCount\else\edef\VvtPortfolioWrongFalseCount{0}\fi
\ifdefined\VvtPortfolioErrorTimeoutCount\else\edef\VvtPortfolioErrorTimeoutCount{0}\fi
\ifdefined\VvtPortfolioErrorOutOfMemoryCount\else\edef\VvtPortfolioErrorOutOfMemoryCount{0}\fi
\ifdefined\VvtPortfolioCorrectCputime\else\edef\VvtPortfolioCorrectCputime{0}\fi
\ifdefined\VvtPortfolioCorrectCputimeAvg\else\edef\VvtPortfolioCorrectCputimeAvg{None}\fi
\ifdefined\VvtPortfolioCorrectWalltime\else\edef\VvtPortfolioCorrectWalltime{0}\fi
\ifdefined\VvtPortfolioCorrectWalltimeAvg\else\edef\VvtPortfolioCorrectWalltimeAvg{None}\fi
\ifdefined\PdrInvKinductionPlainTrueNotSolvedByKinductionPlainTotalCount\else\edef\PdrInvKinductionPlainTrueNotSolvedByKinductionPlainTotalCount{0}\fi
\ifdefined\PdrInvKinductionPlainTrueNotSolvedByKinductionPlainCorrectCount\else\edef\PdrInvKinductionPlainTrueNotSolvedByKinductionPlainCorrectCount{0}\fi
\ifdefined\PdrInvKinductionPlainTrueNotSolvedByKinductionPlainCorrectTrueCount\else\edef\PdrInvKinductionPlainTrueNotSolvedByKinductionPlainCorrectTrueCount{0}\fi
\ifdefined\PdrInvKinductionPlainTrueNotSolvedByKinductionPlainCorrectFalseCount\else\edef\PdrInvKinductionPlainTrueNotSolvedByKinductionPlainCorrectFalseCount{0}\fi
\ifdefined\PdrInvKinductionPlainTrueNotSolvedByKinductionPlainWrongTrueCount\else\edef\PdrInvKinductionPlainTrueNotSolvedByKinductionPlainWrongTrueCount{0}\fi
\ifdefined\PdrInvKinductionPlainTrueNotSolvedByKinductionPlainWrongFalseCount\else\edef\PdrInvKinductionPlainTrueNotSolvedByKinductionPlainWrongFalseCount{0}\fi
\ifdefined\PdrInvKinductionPlainTrueNotSolvedByKinductionPlainErrorTimeoutCount\else\edef\PdrInvKinductionPlainTrueNotSolvedByKinductionPlainErrorTimeoutCount{0}\fi
\ifdefined\PdrInvKinductionPlainTrueNotSolvedByKinductionPlainErrorOutOfMemoryCount\else\edef\PdrInvKinductionPlainTrueNotSolvedByKinductionPlainErrorOutOfMemoryCount{0}\fi
\ifdefined\PdrInvKinductionPlainTrueNotSolvedByKinductionPlainCorrectCputime\else\edef\PdrInvKinductionPlainTrueNotSolvedByKinductionPlainCorrectCputime{0}\fi
\ifdefined\PdrInvKinductionPlainTrueNotSolvedByKinductionPlainCorrectCputimeAvg\else\edef\PdrInvKinductionPlainTrueNotSolvedByKinductionPlainCorrectCputimeAvg{None}\fi
\ifdefined\PdrInvKinductionPlainTrueNotSolvedByKinductionPlainCorrectWalltime\else\edef\PdrInvKinductionPlainTrueNotSolvedByKinductionPlainCorrectWalltime{0}\fi
\ifdefined\PdrInvKinductionPlainTrueNotSolvedByKinductionPlainCorrectWalltimeAvg\else\edef\PdrInvKinductionPlainTrueNotSolvedByKinductionPlainCorrectWalltimeAvg{None}\fi
\ifdefined\PdrInvKinductionDfStaticZeroZeroTTrueNotSolvedByKinductionPlainTotalCount\else\edef\PdrInvKinductionDfStaticZeroZeroTTrueNotSolvedByKinductionPlainTotalCount{0}\fi
\ifdefined\PdrInvKinductionDfStaticZeroZeroTTrueNotSolvedByKinductionPlainCorrectCount\else\edef\PdrInvKinductionDfStaticZeroZeroTTrueNotSolvedByKinductionPlainCorrectCount{0}\fi
\ifdefined\PdrInvKinductionDfStaticZeroZeroTTrueNotSolvedByKinductionPlainCorrectTrueCount\else\edef\PdrInvKinductionDfStaticZeroZeroTTrueNotSolvedByKinductionPlainCorrectTrueCount{0}\fi
\ifdefined\PdrInvKinductionDfStaticZeroZeroTTrueNotSolvedByKinductionPlainCorrectFalseCount\else\edef\PdrInvKinductionDfStaticZeroZeroTTrueNotSolvedByKinductionPlainCorrectFalseCount{0}\fi
\ifdefined\PdrInvKinductionDfStaticZeroZeroTTrueNotSolvedByKinductionPlainWrongTrueCount\else\edef\PdrInvKinductionDfStaticZeroZeroTTrueNotSolvedByKinductionPlainWrongTrueCount{0}\fi
\ifdefined\PdrInvKinductionDfStaticZeroZeroTTrueNotSolvedByKinductionPlainWrongFalseCount\else\edef\PdrInvKinductionDfStaticZeroZeroTTrueNotSolvedByKinductionPlainWrongFalseCount{0}\fi
\ifdefined\PdrInvKinductionDfStaticZeroZeroTTrueNotSolvedByKinductionPlainErrorTimeoutCount\else\edef\PdrInvKinductionDfStaticZeroZeroTTrueNotSolvedByKinductionPlainErrorTimeoutCount{0}\fi
\ifdefined\PdrInvKinductionDfStaticZeroZeroTTrueNotSolvedByKinductionPlainErrorOutOfMemoryCount\else\edef\PdrInvKinductionDfStaticZeroZeroTTrueNotSolvedByKinductionPlainErrorOutOfMemoryCount{0}\fi
\ifdefined\PdrInvKinductionDfStaticZeroZeroTTrueNotSolvedByKinductionPlainCorrectCputime\else\edef\PdrInvKinductionDfStaticZeroZeroTTrueNotSolvedByKinductionPlainCorrectCputime{0}\fi
\ifdefined\PdrInvKinductionDfStaticZeroZeroTTrueNotSolvedByKinductionPlainCorrectCputimeAvg\else\edef\PdrInvKinductionDfStaticZeroZeroTTrueNotSolvedByKinductionPlainCorrectCputimeAvg{None}\fi
\ifdefined\PdrInvKinductionDfStaticZeroZeroTTrueNotSolvedByKinductionPlainCorrectWalltime\else\edef\PdrInvKinductionDfStaticZeroZeroTTrueNotSolvedByKinductionPlainCorrectWalltime{0}\fi
\ifdefined\PdrInvKinductionDfStaticZeroZeroTTrueNotSolvedByKinductionPlainCorrectWalltimeAvg\else\edef\PdrInvKinductionDfStaticZeroZeroTTrueNotSolvedByKinductionPlainCorrectWalltimeAvg{None}\fi
\ifdefined\PdrInvKinductionDfStaticZeroOneTTTrueNotSolvedByKinductionPlainTotalCount\else\edef\PdrInvKinductionDfStaticZeroOneTTTrueNotSolvedByKinductionPlainTotalCount{0}\fi
\ifdefined\PdrInvKinductionDfStaticZeroOneTTTrueNotSolvedByKinductionPlainCorrectCount\else\edef\PdrInvKinductionDfStaticZeroOneTTTrueNotSolvedByKinductionPlainCorrectCount{0}\fi
\ifdefined\PdrInvKinductionDfStaticZeroOneTTTrueNotSolvedByKinductionPlainCorrectTrueCount\else\edef\PdrInvKinductionDfStaticZeroOneTTTrueNotSolvedByKinductionPlainCorrectTrueCount{0}\fi
\ifdefined\PdrInvKinductionDfStaticZeroOneTTTrueNotSolvedByKinductionPlainCorrectFalseCount\else\edef\PdrInvKinductionDfStaticZeroOneTTTrueNotSolvedByKinductionPlainCorrectFalseCount{0}\fi
\ifdefined\PdrInvKinductionDfStaticZeroOneTTTrueNotSolvedByKinductionPlainWrongTrueCount\else\edef\PdrInvKinductionDfStaticZeroOneTTTrueNotSolvedByKinductionPlainWrongTrueCount{0}\fi
\ifdefined\PdrInvKinductionDfStaticZeroOneTTTrueNotSolvedByKinductionPlainWrongFalseCount\else\edef\PdrInvKinductionDfStaticZeroOneTTTrueNotSolvedByKinductionPlainWrongFalseCount{0}\fi
\ifdefined\PdrInvKinductionDfStaticZeroOneTTTrueNotSolvedByKinductionPlainErrorTimeoutCount\else\edef\PdrInvKinductionDfStaticZeroOneTTTrueNotSolvedByKinductionPlainErrorTimeoutCount{0}\fi
\ifdefined\PdrInvKinductionDfStaticZeroOneTTTrueNotSolvedByKinductionPlainErrorOutOfMemoryCount\else\edef\PdrInvKinductionDfStaticZeroOneTTTrueNotSolvedByKinductionPlainErrorOutOfMemoryCount{0}\fi
\ifdefined\PdrInvKinductionDfStaticZeroOneTTTrueNotSolvedByKinductionPlainCorrectCputime\else\edef\PdrInvKinductionDfStaticZeroOneTTTrueNotSolvedByKinductionPlainCorrectCputime{0}\fi
\ifdefined\PdrInvKinductionDfStaticZeroOneTTTrueNotSolvedByKinductionPlainCorrectCputimeAvg\else\edef\PdrInvKinductionDfStaticZeroOneTTTrueNotSolvedByKinductionPlainCorrectCputimeAvg{None}\fi
\ifdefined\PdrInvKinductionDfStaticZeroOneTTTrueNotSolvedByKinductionPlainCorrectWalltime\else\edef\PdrInvKinductionDfStaticZeroOneTTTrueNotSolvedByKinductionPlainCorrectWalltime{0}\fi
\ifdefined\PdrInvKinductionDfStaticZeroOneTTTrueNotSolvedByKinductionPlainCorrectWalltimeAvg\else\edef\PdrInvKinductionDfStaticZeroOneTTTrueNotSolvedByKinductionPlainCorrectWalltimeAvg{None}\fi
\ifdefined\PdrInvKinductionDfStaticZeroOneTFTrueNotSolvedByKinductionPlainTotalCount\else\edef\PdrInvKinductionDfStaticZeroOneTFTrueNotSolvedByKinductionPlainTotalCount{0}\fi
\ifdefined\PdrInvKinductionDfStaticZeroOneTFTrueNotSolvedByKinductionPlainCorrectCount\else\edef\PdrInvKinductionDfStaticZeroOneTFTrueNotSolvedByKinductionPlainCorrectCount{0}\fi
\ifdefined\PdrInvKinductionDfStaticZeroOneTFTrueNotSolvedByKinductionPlainCorrectTrueCount\else\edef\PdrInvKinductionDfStaticZeroOneTFTrueNotSolvedByKinductionPlainCorrectTrueCount{0}\fi
\ifdefined\PdrInvKinductionDfStaticZeroOneTFTrueNotSolvedByKinductionPlainCorrectFalseCount\else\edef\PdrInvKinductionDfStaticZeroOneTFTrueNotSolvedByKinductionPlainCorrectFalseCount{0}\fi
\ifdefined\PdrInvKinductionDfStaticZeroOneTFTrueNotSolvedByKinductionPlainWrongTrueCount\else\edef\PdrInvKinductionDfStaticZeroOneTFTrueNotSolvedByKinductionPlainWrongTrueCount{0}\fi
\ifdefined\PdrInvKinductionDfStaticZeroOneTFTrueNotSolvedByKinductionPlainWrongFalseCount\else\edef\PdrInvKinductionDfStaticZeroOneTFTrueNotSolvedByKinductionPlainWrongFalseCount{0}\fi
\ifdefined\PdrInvKinductionDfStaticZeroOneTFTrueNotSolvedByKinductionPlainErrorTimeoutCount\else\edef\PdrInvKinductionDfStaticZeroOneTFTrueNotSolvedByKinductionPlainErrorTimeoutCount{0}\fi
\ifdefined\PdrInvKinductionDfStaticZeroOneTFTrueNotSolvedByKinductionPlainErrorOutOfMemoryCount\else\edef\PdrInvKinductionDfStaticZeroOneTFTrueNotSolvedByKinductionPlainErrorOutOfMemoryCount{0}\fi
\ifdefined\PdrInvKinductionDfStaticZeroOneTFTrueNotSolvedByKinductionPlainCorrectCputime\else\edef\PdrInvKinductionDfStaticZeroOneTFTrueNotSolvedByKinductionPlainCorrectCputime{0}\fi
\ifdefined\PdrInvKinductionDfStaticZeroOneTFTrueNotSolvedByKinductionPlainCorrectCputimeAvg\else\edef\PdrInvKinductionDfStaticZeroOneTFTrueNotSolvedByKinductionPlainCorrectCputimeAvg{None}\fi
\ifdefined\PdrInvKinductionDfStaticZeroOneTFTrueNotSolvedByKinductionPlainCorrectWalltime\else\edef\PdrInvKinductionDfStaticZeroOneTFTrueNotSolvedByKinductionPlainCorrectWalltime{0}\fi
\ifdefined\PdrInvKinductionDfStaticZeroOneTFTrueNotSolvedByKinductionPlainCorrectWalltimeAvg\else\edef\PdrInvKinductionDfStaticZeroOneTFTrueNotSolvedByKinductionPlainCorrectWalltimeAvg{None}\fi
\ifdefined\PdrInvKinductionDfStaticZeroTwoTTTrueNotSolvedByKinductionPlainTotalCount\else\edef\PdrInvKinductionDfStaticZeroTwoTTTrueNotSolvedByKinductionPlainTotalCount{0}\fi
\ifdefined\PdrInvKinductionDfStaticZeroTwoTTTrueNotSolvedByKinductionPlainCorrectCount\else\edef\PdrInvKinductionDfStaticZeroTwoTTTrueNotSolvedByKinductionPlainCorrectCount{0}\fi
\ifdefined\PdrInvKinductionDfStaticZeroTwoTTTrueNotSolvedByKinductionPlainCorrectTrueCount\else\edef\PdrInvKinductionDfStaticZeroTwoTTTrueNotSolvedByKinductionPlainCorrectTrueCount{0}\fi
\ifdefined\PdrInvKinductionDfStaticZeroTwoTTTrueNotSolvedByKinductionPlainCorrectFalseCount\else\edef\PdrInvKinductionDfStaticZeroTwoTTTrueNotSolvedByKinductionPlainCorrectFalseCount{0}\fi
\ifdefined\PdrInvKinductionDfStaticZeroTwoTTTrueNotSolvedByKinductionPlainWrongTrueCount\else\edef\PdrInvKinductionDfStaticZeroTwoTTTrueNotSolvedByKinductionPlainWrongTrueCount{0}\fi
\ifdefined\PdrInvKinductionDfStaticZeroTwoTTTrueNotSolvedByKinductionPlainWrongFalseCount\else\edef\PdrInvKinductionDfStaticZeroTwoTTTrueNotSolvedByKinductionPlainWrongFalseCount{0}\fi
\ifdefined\PdrInvKinductionDfStaticZeroTwoTTTrueNotSolvedByKinductionPlainErrorTimeoutCount\else\edef\PdrInvKinductionDfStaticZeroTwoTTTrueNotSolvedByKinductionPlainErrorTimeoutCount{0}\fi
\ifdefined\PdrInvKinductionDfStaticZeroTwoTTTrueNotSolvedByKinductionPlainErrorOutOfMemoryCount\else\edef\PdrInvKinductionDfStaticZeroTwoTTTrueNotSolvedByKinductionPlainErrorOutOfMemoryCount{0}\fi
\ifdefined\PdrInvKinductionDfStaticZeroTwoTTTrueNotSolvedByKinductionPlainCorrectCputime\else\edef\PdrInvKinductionDfStaticZeroTwoTTTrueNotSolvedByKinductionPlainCorrectCputime{0}\fi
\ifdefined\PdrInvKinductionDfStaticZeroTwoTTTrueNotSolvedByKinductionPlainCorrectCputimeAvg\else\edef\PdrInvKinductionDfStaticZeroTwoTTTrueNotSolvedByKinductionPlainCorrectCputimeAvg{None}\fi
\ifdefined\PdrInvKinductionDfStaticZeroTwoTTTrueNotSolvedByKinductionPlainCorrectWalltime\else\edef\PdrInvKinductionDfStaticZeroTwoTTTrueNotSolvedByKinductionPlainCorrectWalltime{0}\fi
\ifdefined\PdrInvKinductionDfStaticZeroTwoTTTrueNotSolvedByKinductionPlainCorrectWalltimeAvg\else\edef\PdrInvKinductionDfStaticZeroTwoTTTrueNotSolvedByKinductionPlainCorrectWalltimeAvg{None}\fi
\ifdefined\PdrInvKinductionDfStaticZeroTwoTFTrueNotSolvedByKinductionPlainTotalCount\else\edef\PdrInvKinductionDfStaticZeroTwoTFTrueNotSolvedByKinductionPlainTotalCount{0}\fi
\ifdefined\PdrInvKinductionDfStaticZeroTwoTFTrueNotSolvedByKinductionPlainCorrectCount\else\edef\PdrInvKinductionDfStaticZeroTwoTFTrueNotSolvedByKinductionPlainCorrectCount{0}\fi
\ifdefined\PdrInvKinductionDfStaticZeroTwoTFTrueNotSolvedByKinductionPlainCorrectTrueCount\else\edef\PdrInvKinductionDfStaticZeroTwoTFTrueNotSolvedByKinductionPlainCorrectTrueCount{0}\fi
\ifdefined\PdrInvKinductionDfStaticZeroTwoTFTrueNotSolvedByKinductionPlainCorrectFalseCount\else\edef\PdrInvKinductionDfStaticZeroTwoTFTrueNotSolvedByKinductionPlainCorrectFalseCount{0}\fi
\ifdefined\PdrInvKinductionDfStaticZeroTwoTFTrueNotSolvedByKinductionPlainWrongTrueCount\else\edef\PdrInvKinductionDfStaticZeroTwoTFTrueNotSolvedByKinductionPlainWrongTrueCount{0}\fi
\ifdefined\PdrInvKinductionDfStaticZeroTwoTFTrueNotSolvedByKinductionPlainWrongFalseCount\else\edef\PdrInvKinductionDfStaticZeroTwoTFTrueNotSolvedByKinductionPlainWrongFalseCount{0}\fi
\ifdefined\PdrInvKinductionDfStaticZeroTwoTFTrueNotSolvedByKinductionPlainErrorTimeoutCount\else\edef\PdrInvKinductionDfStaticZeroTwoTFTrueNotSolvedByKinductionPlainErrorTimeoutCount{0}\fi
\ifdefined\PdrInvKinductionDfStaticZeroTwoTFTrueNotSolvedByKinductionPlainErrorOutOfMemoryCount\else\edef\PdrInvKinductionDfStaticZeroTwoTFTrueNotSolvedByKinductionPlainErrorOutOfMemoryCount{0}\fi
\ifdefined\PdrInvKinductionDfStaticZeroTwoTFTrueNotSolvedByKinductionPlainCorrectCputime\else\edef\PdrInvKinductionDfStaticZeroTwoTFTrueNotSolvedByKinductionPlainCorrectCputime{0}\fi
\ifdefined\PdrInvKinductionDfStaticZeroTwoTFTrueNotSolvedByKinductionPlainCorrectCputimeAvg\else\edef\PdrInvKinductionDfStaticZeroTwoTFTrueNotSolvedByKinductionPlainCorrectCputimeAvg{None}\fi
\ifdefined\PdrInvKinductionDfStaticZeroTwoTFTrueNotSolvedByKinductionPlainCorrectWalltime\else\edef\PdrInvKinductionDfStaticZeroTwoTFTrueNotSolvedByKinductionPlainCorrectWalltime{0}\fi
\ifdefined\PdrInvKinductionDfStaticZeroTwoTFTrueNotSolvedByKinductionPlainCorrectWalltimeAvg\else\edef\PdrInvKinductionDfStaticZeroTwoTFTrueNotSolvedByKinductionPlainCorrectWalltimeAvg{None}\fi
\ifdefined\PdrInvKinductionDfStaticEightTwoTTrueNotSolvedByKinductionPlainTotalCount\else\edef\PdrInvKinductionDfStaticEightTwoTTrueNotSolvedByKinductionPlainTotalCount{0}\fi
\ifdefined\PdrInvKinductionDfStaticEightTwoTTrueNotSolvedByKinductionPlainCorrectCount\else\edef\PdrInvKinductionDfStaticEightTwoTTrueNotSolvedByKinductionPlainCorrectCount{0}\fi
\ifdefined\PdrInvKinductionDfStaticEightTwoTTrueNotSolvedByKinductionPlainCorrectTrueCount\else\edef\PdrInvKinductionDfStaticEightTwoTTrueNotSolvedByKinductionPlainCorrectTrueCount{0}\fi
\ifdefined\PdrInvKinductionDfStaticEightTwoTTrueNotSolvedByKinductionPlainCorrectFalseCount\else\edef\PdrInvKinductionDfStaticEightTwoTTrueNotSolvedByKinductionPlainCorrectFalseCount{0}\fi
\ifdefined\PdrInvKinductionDfStaticEightTwoTTrueNotSolvedByKinductionPlainWrongTrueCount\else\edef\PdrInvKinductionDfStaticEightTwoTTrueNotSolvedByKinductionPlainWrongTrueCount{0}\fi
\ifdefined\PdrInvKinductionDfStaticEightTwoTTrueNotSolvedByKinductionPlainWrongFalseCount\else\edef\PdrInvKinductionDfStaticEightTwoTTrueNotSolvedByKinductionPlainWrongFalseCount{0}\fi
\ifdefined\PdrInvKinductionDfStaticEightTwoTTrueNotSolvedByKinductionPlainErrorTimeoutCount\else\edef\PdrInvKinductionDfStaticEightTwoTTrueNotSolvedByKinductionPlainErrorTimeoutCount{0}\fi
\ifdefined\PdrInvKinductionDfStaticEightTwoTTrueNotSolvedByKinductionPlainErrorOutOfMemoryCount\else\edef\PdrInvKinductionDfStaticEightTwoTTrueNotSolvedByKinductionPlainErrorOutOfMemoryCount{0}\fi
\ifdefined\PdrInvKinductionDfStaticEightTwoTTrueNotSolvedByKinductionPlainCorrectCputime\else\edef\PdrInvKinductionDfStaticEightTwoTTrueNotSolvedByKinductionPlainCorrectCputime{0}\fi
\ifdefined\PdrInvKinductionDfStaticEightTwoTTrueNotSolvedByKinductionPlainCorrectCputimeAvg\else\edef\PdrInvKinductionDfStaticEightTwoTTrueNotSolvedByKinductionPlainCorrectCputimeAvg{None}\fi
\ifdefined\PdrInvKinductionDfStaticEightTwoTTrueNotSolvedByKinductionPlainCorrectWalltime\else\edef\PdrInvKinductionDfStaticEightTwoTTrueNotSolvedByKinductionPlainCorrectWalltime{0}\fi
\ifdefined\PdrInvKinductionDfStaticEightTwoTTrueNotSolvedByKinductionPlainCorrectWalltimeAvg\else\edef\PdrInvKinductionDfStaticEightTwoTTrueNotSolvedByKinductionPlainCorrectWalltimeAvg{None}\fi
\ifdefined\PdrInvKinductionDfStaticSixteenTwoTTrueNotSolvedByKinductionPlainTotalCount\else\edef\PdrInvKinductionDfStaticSixteenTwoTTrueNotSolvedByKinductionPlainTotalCount{0}\fi
\ifdefined\PdrInvKinductionDfStaticSixteenTwoTTrueNotSolvedByKinductionPlainCorrectCount\else\edef\PdrInvKinductionDfStaticSixteenTwoTTrueNotSolvedByKinductionPlainCorrectCount{0}\fi
\ifdefined\PdrInvKinductionDfStaticSixteenTwoTTrueNotSolvedByKinductionPlainCorrectTrueCount\else\edef\PdrInvKinductionDfStaticSixteenTwoTTrueNotSolvedByKinductionPlainCorrectTrueCount{0}\fi
\ifdefined\PdrInvKinductionDfStaticSixteenTwoTTrueNotSolvedByKinductionPlainCorrectFalseCount\else\edef\PdrInvKinductionDfStaticSixteenTwoTTrueNotSolvedByKinductionPlainCorrectFalseCount{0}\fi
\ifdefined\PdrInvKinductionDfStaticSixteenTwoTTrueNotSolvedByKinductionPlainWrongTrueCount\else\edef\PdrInvKinductionDfStaticSixteenTwoTTrueNotSolvedByKinductionPlainWrongTrueCount{0}\fi
\ifdefined\PdrInvKinductionDfStaticSixteenTwoTTrueNotSolvedByKinductionPlainWrongFalseCount\else\edef\PdrInvKinductionDfStaticSixteenTwoTTrueNotSolvedByKinductionPlainWrongFalseCount{0}\fi
\ifdefined\PdrInvKinductionDfStaticSixteenTwoTTrueNotSolvedByKinductionPlainErrorTimeoutCount\else\edef\PdrInvKinductionDfStaticSixteenTwoTTrueNotSolvedByKinductionPlainErrorTimeoutCount{0}\fi
\ifdefined\PdrInvKinductionDfStaticSixteenTwoTTrueNotSolvedByKinductionPlainErrorOutOfMemoryCount\else\edef\PdrInvKinductionDfStaticSixteenTwoTTrueNotSolvedByKinductionPlainErrorOutOfMemoryCount{0}\fi
\ifdefined\PdrInvKinductionDfStaticSixteenTwoTTrueNotSolvedByKinductionPlainCorrectCputime\else\edef\PdrInvKinductionDfStaticSixteenTwoTTrueNotSolvedByKinductionPlainCorrectCputime{0}\fi
\ifdefined\PdrInvKinductionDfStaticSixteenTwoTTrueNotSolvedByKinductionPlainCorrectCputimeAvg\else\edef\PdrInvKinductionDfStaticSixteenTwoTTrueNotSolvedByKinductionPlainCorrectCputimeAvg{None}\fi
\ifdefined\PdrInvKinductionDfStaticSixteenTwoTTrueNotSolvedByKinductionPlainCorrectWalltime\else\edef\PdrInvKinductionDfStaticSixteenTwoTTrueNotSolvedByKinductionPlainCorrectWalltime{0}\fi
\ifdefined\PdrInvKinductionDfStaticSixteenTwoTTrueNotSolvedByKinductionPlainCorrectWalltimeAvg\else\edef\PdrInvKinductionDfStaticSixteenTwoTTrueNotSolvedByKinductionPlainCorrectWalltimeAvg{None}\fi
\ifdefined\PdrInvKinductionDfStaticSixteenTwoFTrueNotSolvedByKinductionPlainTotalCount\else\edef\PdrInvKinductionDfStaticSixteenTwoFTrueNotSolvedByKinductionPlainTotalCount{0}\fi
\ifdefined\PdrInvKinductionDfStaticSixteenTwoFTrueNotSolvedByKinductionPlainCorrectCount\else\edef\PdrInvKinductionDfStaticSixteenTwoFTrueNotSolvedByKinductionPlainCorrectCount{0}\fi
\ifdefined\PdrInvKinductionDfStaticSixteenTwoFTrueNotSolvedByKinductionPlainCorrectTrueCount\else\edef\PdrInvKinductionDfStaticSixteenTwoFTrueNotSolvedByKinductionPlainCorrectTrueCount{0}\fi
\ifdefined\PdrInvKinductionDfStaticSixteenTwoFTrueNotSolvedByKinductionPlainCorrectFalseCount\else\edef\PdrInvKinductionDfStaticSixteenTwoFTrueNotSolvedByKinductionPlainCorrectFalseCount{0}\fi
\ifdefined\PdrInvKinductionDfStaticSixteenTwoFTrueNotSolvedByKinductionPlainWrongTrueCount\else\edef\PdrInvKinductionDfStaticSixteenTwoFTrueNotSolvedByKinductionPlainWrongTrueCount{0}\fi
\ifdefined\PdrInvKinductionDfStaticSixteenTwoFTrueNotSolvedByKinductionPlainWrongFalseCount\else\edef\PdrInvKinductionDfStaticSixteenTwoFTrueNotSolvedByKinductionPlainWrongFalseCount{0}\fi
\ifdefined\PdrInvKinductionDfStaticSixteenTwoFTrueNotSolvedByKinductionPlainErrorTimeoutCount\else\edef\PdrInvKinductionDfStaticSixteenTwoFTrueNotSolvedByKinductionPlainErrorTimeoutCount{0}\fi
\ifdefined\PdrInvKinductionDfStaticSixteenTwoFTrueNotSolvedByKinductionPlainErrorOutOfMemoryCount\else\edef\PdrInvKinductionDfStaticSixteenTwoFTrueNotSolvedByKinductionPlainErrorOutOfMemoryCount{0}\fi
\ifdefined\PdrInvKinductionDfStaticSixteenTwoFTrueNotSolvedByKinductionPlainCorrectCputime\else\edef\PdrInvKinductionDfStaticSixteenTwoFTrueNotSolvedByKinductionPlainCorrectCputime{0}\fi
\ifdefined\PdrInvKinductionDfStaticSixteenTwoFTrueNotSolvedByKinductionPlainCorrectCputimeAvg\else\edef\PdrInvKinductionDfStaticSixteenTwoFTrueNotSolvedByKinductionPlainCorrectCputimeAvg{None}\fi
\ifdefined\PdrInvKinductionDfStaticSixteenTwoFTrueNotSolvedByKinductionPlainCorrectWalltime\else\edef\PdrInvKinductionDfStaticSixteenTwoFTrueNotSolvedByKinductionPlainCorrectWalltime{0}\fi
\ifdefined\PdrInvKinductionDfStaticSixteenTwoFTrueNotSolvedByKinductionPlainCorrectWalltimeAvg\else\edef\PdrInvKinductionDfStaticSixteenTwoFTrueNotSolvedByKinductionPlainCorrectWalltimeAvg{None}\fi
\ifdefined\PdrInvKinductionDfTrueNotSolvedByKinductionPlainTotalCount\else\edef\PdrInvKinductionDfTrueNotSolvedByKinductionPlainTotalCount{0}\fi
\ifdefined\PdrInvKinductionDfTrueNotSolvedByKinductionPlainCorrectCount\else\edef\PdrInvKinductionDfTrueNotSolvedByKinductionPlainCorrectCount{0}\fi
\ifdefined\PdrInvKinductionDfTrueNotSolvedByKinductionPlainCorrectTrueCount\else\edef\PdrInvKinductionDfTrueNotSolvedByKinductionPlainCorrectTrueCount{0}\fi
\ifdefined\PdrInvKinductionDfTrueNotSolvedByKinductionPlainCorrectFalseCount\else\edef\PdrInvKinductionDfTrueNotSolvedByKinductionPlainCorrectFalseCount{0}\fi
\ifdefined\PdrInvKinductionDfTrueNotSolvedByKinductionPlainWrongTrueCount\else\edef\PdrInvKinductionDfTrueNotSolvedByKinductionPlainWrongTrueCount{0}\fi
\ifdefined\PdrInvKinductionDfTrueNotSolvedByKinductionPlainWrongFalseCount\else\edef\PdrInvKinductionDfTrueNotSolvedByKinductionPlainWrongFalseCount{0}\fi
\ifdefined\PdrInvKinductionDfTrueNotSolvedByKinductionPlainErrorTimeoutCount\else\edef\PdrInvKinductionDfTrueNotSolvedByKinductionPlainErrorTimeoutCount{0}\fi
\ifdefined\PdrInvKinductionDfTrueNotSolvedByKinductionPlainErrorOutOfMemoryCount\else\edef\PdrInvKinductionDfTrueNotSolvedByKinductionPlainErrorOutOfMemoryCount{0}\fi
\ifdefined\PdrInvKinductionDfTrueNotSolvedByKinductionPlainCorrectCputime\else\edef\PdrInvKinductionDfTrueNotSolvedByKinductionPlainCorrectCputime{0}\fi
\ifdefined\PdrInvKinductionDfTrueNotSolvedByKinductionPlainCorrectCputimeAvg\else\edef\PdrInvKinductionDfTrueNotSolvedByKinductionPlainCorrectCputimeAvg{None}\fi
\ifdefined\PdrInvKinductionDfTrueNotSolvedByKinductionPlainCorrectWalltime\else\edef\PdrInvKinductionDfTrueNotSolvedByKinductionPlainCorrectWalltime{0}\fi
\ifdefined\PdrInvKinductionDfTrueNotSolvedByKinductionPlainCorrectWalltimeAvg\else\edef\PdrInvKinductionDfTrueNotSolvedByKinductionPlainCorrectWalltimeAvg{None}\fi
\ifdefined\PdrInvKinductionKipdrTrueNotSolvedByKinductionPlainTotalCount\else\edef\PdrInvKinductionKipdrTrueNotSolvedByKinductionPlainTotalCount{0}\fi
\ifdefined\PdrInvKinductionKipdrTrueNotSolvedByKinductionPlainCorrectCount\else\edef\PdrInvKinductionKipdrTrueNotSolvedByKinductionPlainCorrectCount{0}\fi
\ifdefined\PdrInvKinductionKipdrTrueNotSolvedByKinductionPlainCorrectTrueCount\else\edef\PdrInvKinductionKipdrTrueNotSolvedByKinductionPlainCorrectTrueCount{0}\fi
\ifdefined\PdrInvKinductionKipdrTrueNotSolvedByKinductionPlainCorrectFalseCount\else\edef\PdrInvKinductionKipdrTrueNotSolvedByKinductionPlainCorrectFalseCount{0}\fi
\ifdefined\PdrInvKinductionKipdrTrueNotSolvedByKinductionPlainWrongTrueCount\else\edef\PdrInvKinductionKipdrTrueNotSolvedByKinductionPlainWrongTrueCount{0}\fi
\ifdefined\PdrInvKinductionKipdrTrueNotSolvedByKinductionPlainWrongFalseCount\else\edef\PdrInvKinductionKipdrTrueNotSolvedByKinductionPlainWrongFalseCount{0}\fi
\ifdefined\PdrInvKinductionKipdrTrueNotSolvedByKinductionPlainErrorTimeoutCount\else\edef\PdrInvKinductionKipdrTrueNotSolvedByKinductionPlainErrorTimeoutCount{0}\fi
\ifdefined\PdrInvKinductionKipdrTrueNotSolvedByKinductionPlainErrorOutOfMemoryCount\else\edef\PdrInvKinductionKipdrTrueNotSolvedByKinductionPlainErrorOutOfMemoryCount{0}\fi
\ifdefined\PdrInvKinductionKipdrTrueNotSolvedByKinductionPlainCorrectCputime\else\edef\PdrInvKinductionKipdrTrueNotSolvedByKinductionPlainCorrectCputime{0}\fi
\ifdefined\PdrInvKinductionKipdrTrueNotSolvedByKinductionPlainCorrectCputimeAvg\else\edef\PdrInvKinductionKipdrTrueNotSolvedByKinductionPlainCorrectCputimeAvg{None}\fi
\ifdefined\PdrInvKinductionKipdrTrueNotSolvedByKinductionPlainCorrectWalltime\else\edef\PdrInvKinductionKipdrTrueNotSolvedByKinductionPlainCorrectWalltime{0}\fi
\ifdefined\PdrInvKinductionKipdrTrueNotSolvedByKinductionPlainCorrectWalltimeAvg\else\edef\PdrInvKinductionKipdrTrueNotSolvedByKinductionPlainCorrectWalltimeAvg{None}\fi
\ifdefined\PdrInvKinductionKipdrdfTrueNotSolvedByKinductionPlainTotalCount\else\edef\PdrInvKinductionKipdrdfTrueNotSolvedByKinductionPlainTotalCount{0}\fi
\ifdefined\PdrInvKinductionKipdrdfTrueNotSolvedByKinductionPlainCorrectCount\else\edef\PdrInvKinductionKipdrdfTrueNotSolvedByKinductionPlainCorrectCount{0}\fi
\ifdefined\PdrInvKinductionKipdrdfTrueNotSolvedByKinductionPlainCorrectTrueCount\else\edef\PdrInvKinductionKipdrdfTrueNotSolvedByKinductionPlainCorrectTrueCount{0}\fi
\ifdefined\PdrInvKinductionKipdrdfTrueNotSolvedByKinductionPlainCorrectFalseCount\else\edef\PdrInvKinductionKipdrdfTrueNotSolvedByKinductionPlainCorrectFalseCount{0}\fi
\ifdefined\PdrInvKinductionKipdrdfTrueNotSolvedByKinductionPlainWrongTrueCount\else\edef\PdrInvKinductionKipdrdfTrueNotSolvedByKinductionPlainWrongTrueCount{0}\fi
\ifdefined\PdrInvKinductionKipdrdfTrueNotSolvedByKinductionPlainWrongFalseCount\else\edef\PdrInvKinductionKipdrdfTrueNotSolvedByKinductionPlainWrongFalseCount{0}\fi
\ifdefined\PdrInvKinductionKipdrdfTrueNotSolvedByKinductionPlainErrorTimeoutCount\else\edef\PdrInvKinductionKipdrdfTrueNotSolvedByKinductionPlainErrorTimeoutCount{0}\fi
\ifdefined\PdrInvKinductionKipdrdfTrueNotSolvedByKinductionPlainErrorOutOfMemoryCount\else\edef\PdrInvKinductionKipdrdfTrueNotSolvedByKinductionPlainErrorOutOfMemoryCount{0}\fi
\ifdefined\PdrInvKinductionKipdrdfTrueNotSolvedByKinductionPlainCorrectCputime\else\edef\PdrInvKinductionKipdrdfTrueNotSolvedByKinductionPlainCorrectCputime{0}\fi
\ifdefined\PdrInvKinductionKipdrdfTrueNotSolvedByKinductionPlainCorrectCputimeAvg\else\edef\PdrInvKinductionKipdrdfTrueNotSolvedByKinductionPlainCorrectCputimeAvg{None}\fi
\ifdefined\PdrInvKinductionKipdrdfTrueNotSolvedByKinductionPlainCorrectWalltime\else\edef\PdrInvKinductionKipdrdfTrueNotSolvedByKinductionPlainCorrectWalltime{0}\fi
\ifdefined\PdrInvKinductionKipdrdfTrueNotSolvedByKinductionPlainCorrectWalltimeAvg\else\edef\PdrInvKinductionKipdrdfTrueNotSolvedByKinductionPlainCorrectWalltimeAvg{None}\fi
\ifdefined\PdrInvPdrTrueNotSolvedByKinductionPlainTotalCount\else\edef\PdrInvPdrTrueNotSolvedByKinductionPlainTotalCount{0}\fi
\ifdefined\PdrInvPdrTrueNotSolvedByKinductionPlainCorrectCount\else\edef\PdrInvPdrTrueNotSolvedByKinductionPlainCorrectCount{0}\fi
\ifdefined\PdrInvPdrTrueNotSolvedByKinductionPlainCorrectTrueCount\else\edef\PdrInvPdrTrueNotSolvedByKinductionPlainCorrectTrueCount{0}\fi
\ifdefined\PdrInvPdrTrueNotSolvedByKinductionPlainCorrectFalseCount\else\edef\PdrInvPdrTrueNotSolvedByKinductionPlainCorrectFalseCount{0}\fi
\ifdefined\PdrInvPdrTrueNotSolvedByKinductionPlainWrongTrueCount\else\edef\PdrInvPdrTrueNotSolvedByKinductionPlainWrongTrueCount{0}\fi
\ifdefined\PdrInvPdrTrueNotSolvedByKinductionPlainWrongFalseCount\else\edef\PdrInvPdrTrueNotSolvedByKinductionPlainWrongFalseCount{0}\fi
\ifdefined\PdrInvPdrTrueNotSolvedByKinductionPlainErrorTimeoutCount\else\edef\PdrInvPdrTrueNotSolvedByKinductionPlainErrorTimeoutCount{0}\fi
\ifdefined\PdrInvPdrTrueNotSolvedByKinductionPlainErrorOutOfMemoryCount\else\edef\PdrInvPdrTrueNotSolvedByKinductionPlainErrorOutOfMemoryCount{0}\fi
\ifdefined\PdrInvPdrTrueNotSolvedByKinductionPlainCorrectCputime\else\edef\PdrInvPdrTrueNotSolvedByKinductionPlainCorrectCputime{0}\fi
\ifdefined\PdrInvPdrTrueNotSolvedByKinductionPlainCorrectCputimeAvg\else\edef\PdrInvPdrTrueNotSolvedByKinductionPlainCorrectCputimeAvg{None}\fi
\ifdefined\PdrInvPdrTrueNotSolvedByKinductionPlainCorrectWalltime\else\edef\PdrInvPdrTrueNotSolvedByKinductionPlainCorrectWalltime{0}\fi
\ifdefined\PdrInvPdrTrueNotSolvedByKinductionPlainCorrectWalltimeAvg\else\edef\PdrInvPdrTrueNotSolvedByKinductionPlainCorrectWalltimeAvg{None}\fi
\ifdefined\PdrInvOracleTrueNotSolvedByKinductionPlainTotalCount\else\edef\PdrInvOracleTrueNotSolvedByKinductionPlainTotalCount{0}\fi
\ifdefined\PdrInvOracleTrueNotSolvedByKinductionPlainCorrectCount\else\edef\PdrInvOracleTrueNotSolvedByKinductionPlainCorrectCount{0}\fi
\ifdefined\PdrInvOracleTrueNotSolvedByKinductionPlainCorrectTrueCount\else\edef\PdrInvOracleTrueNotSolvedByKinductionPlainCorrectTrueCount{0}\fi
\ifdefined\PdrInvOracleTrueNotSolvedByKinductionPlainCorrectFalseCount\else\edef\PdrInvOracleTrueNotSolvedByKinductionPlainCorrectFalseCount{0}\fi
\ifdefined\PdrInvOracleTrueNotSolvedByKinductionPlainWrongTrueCount\else\edef\PdrInvOracleTrueNotSolvedByKinductionPlainWrongTrueCount{0}\fi
\ifdefined\PdrInvOracleTrueNotSolvedByKinductionPlainWrongFalseCount\else\edef\PdrInvOracleTrueNotSolvedByKinductionPlainWrongFalseCount{0}\fi
\ifdefined\PdrInvOracleTrueNotSolvedByKinductionPlainErrorTimeoutCount\else\edef\PdrInvOracleTrueNotSolvedByKinductionPlainErrorTimeoutCount{0}\fi
\ifdefined\PdrInvOracleTrueNotSolvedByKinductionPlainErrorOutOfMemoryCount\else\edef\PdrInvOracleTrueNotSolvedByKinductionPlainErrorOutOfMemoryCount{0}\fi
\ifdefined\PdrInvOracleTrueNotSolvedByKinductionPlainCorrectCputime\else\edef\PdrInvOracleTrueNotSolvedByKinductionPlainCorrectCputime{0}\fi
\ifdefined\PdrInvOracleTrueNotSolvedByKinductionPlainCorrectCputimeAvg\else\edef\PdrInvOracleTrueNotSolvedByKinductionPlainCorrectCputimeAvg{None}\fi
\ifdefined\PdrInvOracleTrueNotSolvedByKinductionPlainCorrectWalltime\else\edef\PdrInvOracleTrueNotSolvedByKinductionPlainCorrectWalltime{0}\fi
\ifdefined\PdrInvOracleTrueNotSolvedByKinductionPlainCorrectWalltimeAvg\else\edef\PdrInvOracleTrueNotSolvedByKinductionPlainCorrectWalltimeAvg{None}\fi
\ifdefined\SeahornSeahornTrueNotSolvedByKinductionPlainTotalCount\else\edef\SeahornSeahornTrueNotSolvedByKinductionPlainTotalCount{0}\fi
\ifdefined\SeahornSeahornTrueNotSolvedByKinductionPlainCorrectCount\else\edef\SeahornSeahornTrueNotSolvedByKinductionPlainCorrectCount{0}\fi
\ifdefined\SeahornSeahornTrueNotSolvedByKinductionPlainCorrectTrueCount\else\edef\SeahornSeahornTrueNotSolvedByKinductionPlainCorrectTrueCount{0}\fi
\ifdefined\SeahornSeahornTrueNotSolvedByKinductionPlainCorrectFalseCount\else\edef\SeahornSeahornTrueNotSolvedByKinductionPlainCorrectFalseCount{0}\fi
\ifdefined\SeahornSeahornTrueNotSolvedByKinductionPlainWrongTrueCount\else\edef\SeahornSeahornTrueNotSolvedByKinductionPlainWrongTrueCount{0}\fi
\ifdefined\SeahornSeahornTrueNotSolvedByKinductionPlainWrongFalseCount\else\edef\SeahornSeahornTrueNotSolvedByKinductionPlainWrongFalseCount{0}\fi
\ifdefined\SeahornSeahornTrueNotSolvedByKinductionPlainErrorTimeoutCount\else\edef\SeahornSeahornTrueNotSolvedByKinductionPlainErrorTimeoutCount{0}\fi
\ifdefined\SeahornSeahornTrueNotSolvedByKinductionPlainErrorOutOfMemoryCount\else\edef\SeahornSeahornTrueNotSolvedByKinductionPlainErrorOutOfMemoryCount{0}\fi
\ifdefined\SeahornSeahornTrueNotSolvedByKinductionPlainCorrectCputime\else\edef\SeahornSeahornTrueNotSolvedByKinductionPlainCorrectCputime{0}\fi
\ifdefined\SeahornSeahornTrueNotSolvedByKinductionPlainCorrectCputimeAvg\else\edef\SeahornSeahornTrueNotSolvedByKinductionPlainCorrectCputimeAvg{None}\fi
\ifdefined\SeahornSeahornTrueNotSolvedByKinductionPlainCorrectWalltime\else\edef\SeahornSeahornTrueNotSolvedByKinductionPlainCorrectWalltime{0}\fi
\ifdefined\SeahornSeahornTrueNotSolvedByKinductionPlainCorrectWalltimeAvg\else\edef\SeahornSeahornTrueNotSolvedByKinductionPlainCorrectWalltimeAvg{None}\fi
\ifdefined\VvtCtigarTrueNotSolvedByKinductionPlainTotalCount\else\edef\VvtCtigarTrueNotSolvedByKinductionPlainTotalCount{0}\fi
\ifdefined\VvtCtigarTrueNotSolvedByKinductionPlainCorrectCount\else\edef\VvtCtigarTrueNotSolvedByKinductionPlainCorrectCount{0}\fi
\ifdefined\VvtCtigarTrueNotSolvedByKinductionPlainCorrectTrueCount\else\edef\VvtCtigarTrueNotSolvedByKinductionPlainCorrectTrueCount{0}\fi
\ifdefined\VvtCtigarTrueNotSolvedByKinductionPlainCorrectFalseCount\else\edef\VvtCtigarTrueNotSolvedByKinductionPlainCorrectFalseCount{0}\fi
\ifdefined\VvtCtigarTrueNotSolvedByKinductionPlainWrongTrueCount\else\edef\VvtCtigarTrueNotSolvedByKinductionPlainWrongTrueCount{0}\fi
\ifdefined\VvtCtigarTrueNotSolvedByKinductionPlainWrongFalseCount\else\edef\VvtCtigarTrueNotSolvedByKinductionPlainWrongFalseCount{0}\fi
\ifdefined\VvtCtigarTrueNotSolvedByKinductionPlainErrorTimeoutCount\else\edef\VvtCtigarTrueNotSolvedByKinductionPlainErrorTimeoutCount{0}\fi
\ifdefined\VvtCtigarTrueNotSolvedByKinductionPlainErrorOutOfMemoryCount\else\edef\VvtCtigarTrueNotSolvedByKinductionPlainErrorOutOfMemoryCount{0}\fi
\ifdefined\VvtCtigarTrueNotSolvedByKinductionPlainCorrectCputime\else\edef\VvtCtigarTrueNotSolvedByKinductionPlainCorrectCputime{0}\fi
\ifdefined\VvtCtigarTrueNotSolvedByKinductionPlainCorrectCputimeAvg\else\edef\VvtCtigarTrueNotSolvedByKinductionPlainCorrectCputimeAvg{None}\fi
\ifdefined\VvtCtigarTrueNotSolvedByKinductionPlainCorrectWalltime\else\edef\VvtCtigarTrueNotSolvedByKinductionPlainCorrectWalltime{0}\fi
\ifdefined\VvtCtigarTrueNotSolvedByKinductionPlainCorrectWalltimeAvg\else\edef\VvtCtigarTrueNotSolvedByKinductionPlainCorrectWalltimeAvg{None}\fi
\ifdefined\VvtPortfolioTrueNotSolvedByKinductionPlainTotalCount\else\edef\VvtPortfolioTrueNotSolvedByKinductionPlainTotalCount{0}\fi
\ifdefined\VvtPortfolioTrueNotSolvedByKinductionPlainCorrectCount\else\edef\VvtPortfolioTrueNotSolvedByKinductionPlainCorrectCount{0}\fi
\ifdefined\VvtPortfolioTrueNotSolvedByKinductionPlainCorrectTrueCount\else\edef\VvtPortfolioTrueNotSolvedByKinductionPlainCorrectTrueCount{0}\fi
\ifdefined\VvtPortfolioTrueNotSolvedByKinductionPlainCorrectFalseCount\else\edef\VvtPortfolioTrueNotSolvedByKinductionPlainCorrectFalseCount{0}\fi
\ifdefined\VvtPortfolioTrueNotSolvedByKinductionPlainWrongTrueCount\else\edef\VvtPortfolioTrueNotSolvedByKinductionPlainWrongTrueCount{0}\fi
\ifdefined\VvtPortfolioTrueNotSolvedByKinductionPlainWrongFalseCount\else\edef\VvtPortfolioTrueNotSolvedByKinductionPlainWrongFalseCount{0}\fi
\ifdefined\VvtPortfolioTrueNotSolvedByKinductionPlainErrorTimeoutCount\else\edef\VvtPortfolioTrueNotSolvedByKinductionPlainErrorTimeoutCount{0}\fi
\ifdefined\VvtPortfolioTrueNotSolvedByKinductionPlainErrorOutOfMemoryCount\else\edef\VvtPortfolioTrueNotSolvedByKinductionPlainErrorOutOfMemoryCount{0}\fi
\ifdefined\VvtPortfolioTrueNotSolvedByKinductionPlainCorrectCputime\else\edef\VvtPortfolioTrueNotSolvedByKinductionPlainCorrectCputime{0}\fi
\ifdefined\VvtPortfolioTrueNotSolvedByKinductionPlainCorrectCputimeAvg\else\edef\VvtPortfolioTrueNotSolvedByKinductionPlainCorrectCputimeAvg{None}\fi
\ifdefined\VvtPortfolioTrueNotSolvedByKinductionPlainCorrectWalltime\else\edef\VvtPortfolioTrueNotSolvedByKinductionPlainCorrectWalltime{0}\fi
\ifdefined\VvtPortfolioTrueNotSolvedByKinductionPlainCorrectWalltimeAvg\else\edef\VvtPortfolioTrueNotSolvedByKinductionPlainCorrectWalltimeAvg{None}\fi
\ifdefined\PdrInvKinductionPlainTrueNotSolvedByKinductionPlainButKipdrTotalCount\else\edef\PdrInvKinductionPlainTrueNotSolvedByKinductionPlainButKipdrTotalCount{0}\fi
\ifdefined\PdrInvKinductionPlainTrueNotSolvedByKinductionPlainButKipdrCorrectCount\else\edef\PdrInvKinductionPlainTrueNotSolvedByKinductionPlainButKipdrCorrectCount{0}\fi
\ifdefined\PdrInvKinductionPlainTrueNotSolvedByKinductionPlainButKipdrCorrectTrueCount\else\edef\PdrInvKinductionPlainTrueNotSolvedByKinductionPlainButKipdrCorrectTrueCount{0}\fi
\ifdefined\PdrInvKinductionPlainTrueNotSolvedByKinductionPlainButKipdrCorrectFalseCount\else\edef\PdrInvKinductionPlainTrueNotSolvedByKinductionPlainButKipdrCorrectFalseCount{0}\fi
\ifdefined\PdrInvKinductionPlainTrueNotSolvedByKinductionPlainButKipdrWrongTrueCount\else\edef\PdrInvKinductionPlainTrueNotSolvedByKinductionPlainButKipdrWrongTrueCount{0}\fi
\ifdefined\PdrInvKinductionPlainTrueNotSolvedByKinductionPlainButKipdrWrongFalseCount\else\edef\PdrInvKinductionPlainTrueNotSolvedByKinductionPlainButKipdrWrongFalseCount{0}\fi
\ifdefined\PdrInvKinductionPlainTrueNotSolvedByKinductionPlainButKipdrErrorTimeoutCount\else\edef\PdrInvKinductionPlainTrueNotSolvedByKinductionPlainButKipdrErrorTimeoutCount{0}\fi
\ifdefined\PdrInvKinductionPlainTrueNotSolvedByKinductionPlainButKipdrErrorOutOfMemoryCount\else\edef\PdrInvKinductionPlainTrueNotSolvedByKinductionPlainButKipdrErrorOutOfMemoryCount{0}\fi
\ifdefined\PdrInvKinductionPlainTrueNotSolvedByKinductionPlainButKipdrCorrectCputime\else\edef\PdrInvKinductionPlainTrueNotSolvedByKinductionPlainButKipdrCorrectCputime{0}\fi
\ifdefined\PdrInvKinductionPlainTrueNotSolvedByKinductionPlainButKipdrCorrectCputimeAvg\else\edef\PdrInvKinductionPlainTrueNotSolvedByKinductionPlainButKipdrCorrectCputimeAvg{None}\fi
\ifdefined\PdrInvKinductionPlainTrueNotSolvedByKinductionPlainButKipdrCorrectWalltime\else\edef\PdrInvKinductionPlainTrueNotSolvedByKinductionPlainButKipdrCorrectWalltime{0}\fi
\ifdefined\PdrInvKinductionPlainTrueNotSolvedByKinductionPlainButKipdrCorrectWalltimeAvg\else\edef\PdrInvKinductionPlainTrueNotSolvedByKinductionPlainButKipdrCorrectWalltimeAvg{None}\fi
\ifdefined\PdrInvKinductionDfStaticZeroZeroTTrueNotSolvedByKinductionPlainButKipdrTotalCount\else\edef\PdrInvKinductionDfStaticZeroZeroTTrueNotSolvedByKinductionPlainButKipdrTotalCount{0}\fi
\ifdefined\PdrInvKinductionDfStaticZeroZeroTTrueNotSolvedByKinductionPlainButKipdrCorrectCount\else\edef\PdrInvKinductionDfStaticZeroZeroTTrueNotSolvedByKinductionPlainButKipdrCorrectCount{0}\fi
\ifdefined\PdrInvKinductionDfStaticZeroZeroTTrueNotSolvedByKinductionPlainButKipdrCorrectTrueCount\else\edef\PdrInvKinductionDfStaticZeroZeroTTrueNotSolvedByKinductionPlainButKipdrCorrectTrueCount{0}\fi
\ifdefined\PdrInvKinductionDfStaticZeroZeroTTrueNotSolvedByKinductionPlainButKipdrCorrectFalseCount\else\edef\PdrInvKinductionDfStaticZeroZeroTTrueNotSolvedByKinductionPlainButKipdrCorrectFalseCount{0}\fi
\ifdefined\PdrInvKinductionDfStaticZeroZeroTTrueNotSolvedByKinductionPlainButKipdrWrongTrueCount\else\edef\PdrInvKinductionDfStaticZeroZeroTTrueNotSolvedByKinductionPlainButKipdrWrongTrueCount{0}\fi
\ifdefined\PdrInvKinductionDfStaticZeroZeroTTrueNotSolvedByKinductionPlainButKipdrWrongFalseCount\else\edef\PdrInvKinductionDfStaticZeroZeroTTrueNotSolvedByKinductionPlainButKipdrWrongFalseCount{0}\fi
\ifdefined\PdrInvKinductionDfStaticZeroZeroTTrueNotSolvedByKinductionPlainButKipdrErrorTimeoutCount\else\edef\PdrInvKinductionDfStaticZeroZeroTTrueNotSolvedByKinductionPlainButKipdrErrorTimeoutCount{0}\fi
\ifdefined\PdrInvKinductionDfStaticZeroZeroTTrueNotSolvedByKinductionPlainButKipdrErrorOutOfMemoryCount\else\edef\PdrInvKinductionDfStaticZeroZeroTTrueNotSolvedByKinductionPlainButKipdrErrorOutOfMemoryCount{0}\fi
\ifdefined\PdrInvKinductionDfStaticZeroZeroTTrueNotSolvedByKinductionPlainButKipdrCorrectCputime\else\edef\PdrInvKinductionDfStaticZeroZeroTTrueNotSolvedByKinductionPlainButKipdrCorrectCputime{0}\fi
\ifdefined\PdrInvKinductionDfStaticZeroZeroTTrueNotSolvedByKinductionPlainButKipdrCorrectCputimeAvg\else\edef\PdrInvKinductionDfStaticZeroZeroTTrueNotSolvedByKinductionPlainButKipdrCorrectCputimeAvg{None}\fi
\ifdefined\PdrInvKinductionDfStaticZeroZeroTTrueNotSolvedByKinductionPlainButKipdrCorrectWalltime\else\edef\PdrInvKinductionDfStaticZeroZeroTTrueNotSolvedByKinductionPlainButKipdrCorrectWalltime{0}\fi
\ifdefined\PdrInvKinductionDfStaticZeroZeroTTrueNotSolvedByKinductionPlainButKipdrCorrectWalltimeAvg\else\edef\PdrInvKinductionDfStaticZeroZeroTTrueNotSolvedByKinductionPlainButKipdrCorrectWalltimeAvg{None}\fi
\ifdefined\PdrInvKinductionDfStaticZeroOneTTTrueNotSolvedByKinductionPlainButKipdrTotalCount\else\edef\PdrInvKinductionDfStaticZeroOneTTTrueNotSolvedByKinductionPlainButKipdrTotalCount{0}\fi
\ifdefined\PdrInvKinductionDfStaticZeroOneTTTrueNotSolvedByKinductionPlainButKipdrCorrectCount\else\edef\PdrInvKinductionDfStaticZeroOneTTTrueNotSolvedByKinductionPlainButKipdrCorrectCount{0}\fi
\ifdefined\PdrInvKinductionDfStaticZeroOneTTTrueNotSolvedByKinductionPlainButKipdrCorrectTrueCount\else\edef\PdrInvKinductionDfStaticZeroOneTTTrueNotSolvedByKinductionPlainButKipdrCorrectTrueCount{0}\fi
\ifdefined\PdrInvKinductionDfStaticZeroOneTTTrueNotSolvedByKinductionPlainButKipdrCorrectFalseCount\else\edef\PdrInvKinductionDfStaticZeroOneTTTrueNotSolvedByKinductionPlainButKipdrCorrectFalseCount{0}\fi
\ifdefined\PdrInvKinductionDfStaticZeroOneTTTrueNotSolvedByKinductionPlainButKipdrWrongTrueCount\else\edef\PdrInvKinductionDfStaticZeroOneTTTrueNotSolvedByKinductionPlainButKipdrWrongTrueCount{0}\fi
\ifdefined\PdrInvKinductionDfStaticZeroOneTTTrueNotSolvedByKinductionPlainButKipdrWrongFalseCount\else\edef\PdrInvKinductionDfStaticZeroOneTTTrueNotSolvedByKinductionPlainButKipdrWrongFalseCount{0}\fi
\ifdefined\PdrInvKinductionDfStaticZeroOneTTTrueNotSolvedByKinductionPlainButKipdrErrorTimeoutCount\else\edef\PdrInvKinductionDfStaticZeroOneTTTrueNotSolvedByKinductionPlainButKipdrErrorTimeoutCount{0}\fi
\ifdefined\PdrInvKinductionDfStaticZeroOneTTTrueNotSolvedByKinductionPlainButKipdrErrorOutOfMemoryCount\else\edef\PdrInvKinductionDfStaticZeroOneTTTrueNotSolvedByKinductionPlainButKipdrErrorOutOfMemoryCount{0}\fi
\ifdefined\PdrInvKinductionDfStaticZeroOneTTTrueNotSolvedByKinductionPlainButKipdrCorrectCputime\else\edef\PdrInvKinductionDfStaticZeroOneTTTrueNotSolvedByKinductionPlainButKipdrCorrectCputime{0}\fi
\ifdefined\PdrInvKinductionDfStaticZeroOneTTTrueNotSolvedByKinductionPlainButKipdrCorrectCputimeAvg\else\edef\PdrInvKinductionDfStaticZeroOneTTTrueNotSolvedByKinductionPlainButKipdrCorrectCputimeAvg{None}\fi
\ifdefined\PdrInvKinductionDfStaticZeroOneTTTrueNotSolvedByKinductionPlainButKipdrCorrectWalltime\else\edef\PdrInvKinductionDfStaticZeroOneTTTrueNotSolvedByKinductionPlainButKipdrCorrectWalltime{0}\fi
\ifdefined\PdrInvKinductionDfStaticZeroOneTTTrueNotSolvedByKinductionPlainButKipdrCorrectWalltimeAvg\else\edef\PdrInvKinductionDfStaticZeroOneTTTrueNotSolvedByKinductionPlainButKipdrCorrectWalltimeAvg{None}\fi
\ifdefined\PdrInvKinductionDfStaticZeroOneTFTrueNotSolvedByKinductionPlainButKipdrTotalCount\else\edef\PdrInvKinductionDfStaticZeroOneTFTrueNotSolvedByKinductionPlainButKipdrTotalCount{0}\fi
\ifdefined\PdrInvKinductionDfStaticZeroOneTFTrueNotSolvedByKinductionPlainButKipdrCorrectCount\else\edef\PdrInvKinductionDfStaticZeroOneTFTrueNotSolvedByKinductionPlainButKipdrCorrectCount{0}\fi
\ifdefined\PdrInvKinductionDfStaticZeroOneTFTrueNotSolvedByKinductionPlainButKipdrCorrectTrueCount\else\edef\PdrInvKinductionDfStaticZeroOneTFTrueNotSolvedByKinductionPlainButKipdrCorrectTrueCount{0}\fi
\ifdefined\PdrInvKinductionDfStaticZeroOneTFTrueNotSolvedByKinductionPlainButKipdrCorrectFalseCount\else\edef\PdrInvKinductionDfStaticZeroOneTFTrueNotSolvedByKinductionPlainButKipdrCorrectFalseCount{0}\fi
\ifdefined\PdrInvKinductionDfStaticZeroOneTFTrueNotSolvedByKinductionPlainButKipdrWrongTrueCount\else\edef\PdrInvKinductionDfStaticZeroOneTFTrueNotSolvedByKinductionPlainButKipdrWrongTrueCount{0}\fi
\ifdefined\PdrInvKinductionDfStaticZeroOneTFTrueNotSolvedByKinductionPlainButKipdrWrongFalseCount\else\edef\PdrInvKinductionDfStaticZeroOneTFTrueNotSolvedByKinductionPlainButKipdrWrongFalseCount{0}\fi
\ifdefined\PdrInvKinductionDfStaticZeroOneTFTrueNotSolvedByKinductionPlainButKipdrErrorTimeoutCount\else\edef\PdrInvKinductionDfStaticZeroOneTFTrueNotSolvedByKinductionPlainButKipdrErrorTimeoutCount{0}\fi
\ifdefined\PdrInvKinductionDfStaticZeroOneTFTrueNotSolvedByKinductionPlainButKipdrErrorOutOfMemoryCount\else\edef\PdrInvKinductionDfStaticZeroOneTFTrueNotSolvedByKinductionPlainButKipdrErrorOutOfMemoryCount{0}\fi
\ifdefined\PdrInvKinductionDfStaticZeroOneTFTrueNotSolvedByKinductionPlainButKipdrCorrectCputime\else\edef\PdrInvKinductionDfStaticZeroOneTFTrueNotSolvedByKinductionPlainButKipdrCorrectCputime{0}\fi
\ifdefined\PdrInvKinductionDfStaticZeroOneTFTrueNotSolvedByKinductionPlainButKipdrCorrectCputimeAvg\else\edef\PdrInvKinductionDfStaticZeroOneTFTrueNotSolvedByKinductionPlainButKipdrCorrectCputimeAvg{None}\fi
\ifdefined\PdrInvKinductionDfStaticZeroOneTFTrueNotSolvedByKinductionPlainButKipdrCorrectWalltime\else\edef\PdrInvKinductionDfStaticZeroOneTFTrueNotSolvedByKinductionPlainButKipdrCorrectWalltime{0}\fi
\ifdefined\PdrInvKinductionDfStaticZeroOneTFTrueNotSolvedByKinductionPlainButKipdrCorrectWalltimeAvg\else\edef\PdrInvKinductionDfStaticZeroOneTFTrueNotSolvedByKinductionPlainButKipdrCorrectWalltimeAvg{None}\fi
\ifdefined\PdrInvKinductionDfStaticZeroTwoTTTrueNotSolvedByKinductionPlainButKipdrTotalCount\else\edef\PdrInvKinductionDfStaticZeroTwoTTTrueNotSolvedByKinductionPlainButKipdrTotalCount{0}\fi
\ifdefined\PdrInvKinductionDfStaticZeroTwoTTTrueNotSolvedByKinductionPlainButKipdrCorrectCount\else\edef\PdrInvKinductionDfStaticZeroTwoTTTrueNotSolvedByKinductionPlainButKipdrCorrectCount{0}\fi
\ifdefined\PdrInvKinductionDfStaticZeroTwoTTTrueNotSolvedByKinductionPlainButKipdrCorrectTrueCount\else\edef\PdrInvKinductionDfStaticZeroTwoTTTrueNotSolvedByKinductionPlainButKipdrCorrectTrueCount{0}\fi
\ifdefined\PdrInvKinductionDfStaticZeroTwoTTTrueNotSolvedByKinductionPlainButKipdrCorrectFalseCount\else\edef\PdrInvKinductionDfStaticZeroTwoTTTrueNotSolvedByKinductionPlainButKipdrCorrectFalseCount{0}\fi
\ifdefined\PdrInvKinductionDfStaticZeroTwoTTTrueNotSolvedByKinductionPlainButKipdrWrongTrueCount\else\edef\PdrInvKinductionDfStaticZeroTwoTTTrueNotSolvedByKinductionPlainButKipdrWrongTrueCount{0}\fi
\ifdefined\PdrInvKinductionDfStaticZeroTwoTTTrueNotSolvedByKinductionPlainButKipdrWrongFalseCount\else\edef\PdrInvKinductionDfStaticZeroTwoTTTrueNotSolvedByKinductionPlainButKipdrWrongFalseCount{0}\fi
\ifdefined\PdrInvKinductionDfStaticZeroTwoTTTrueNotSolvedByKinductionPlainButKipdrErrorTimeoutCount\else\edef\PdrInvKinductionDfStaticZeroTwoTTTrueNotSolvedByKinductionPlainButKipdrErrorTimeoutCount{0}\fi
\ifdefined\PdrInvKinductionDfStaticZeroTwoTTTrueNotSolvedByKinductionPlainButKipdrErrorOutOfMemoryCount\else\edef\PdrInvKinductionDfStaticZeroTwoTTTrueNotSolvedByKinductionPlainButKipdrErrorOutOfMemoryCount{0}\fi
\ifdefined\PdrInvKinductionDfStaticZeroTwoTTTrueNotSolvedByKinductionPlainButKipdrCorrectCputime\else\edef\PdrInvKinductionDfStaticZeroTwoTTTrueNotSolvedByKinductionPlainButKipdrCorrectCputime{0}\fi
\ifdefined\PdrInvKinductionDfStaticZeroTwoTTTrueNotSolvedByKinductionPlainButKipdrCorrectCputimeAvg\else\edef\PdrInvKinductionDfStaticZeroTwoTTTrueNotSolvedByKinductionPlainButKipdrCorrectCputimeAvg{None}\fi
\ifdefined\PdrInvKinductionDfStaticZeroTwoTTTrueNotSolvedByKinductionPlainButKipdrCorrectWalltime\else\edef\PdrInvKinductionDfStaticZeroTwoTTTrueNotSolvedByKinductionPlainButKipdrCorrectWalltime{0}\fi
\ifdefined\PdrInvKinductionDfStaticZeroTwoTTTrueNotSolvedByKinductionPlainButKipdrCorrectWalltimeAvg\else\edef\PdrInvKinductionDfStaticZeroTwoTTTrueNotSolvedByKinductionPlainButKipdrCorrectWalltimeAvg{None}\fi
\ifdefined\PdrInvKinductionDfStaticZeroTwoTFTrueNotSolvedByKinductionPlainButKipdrTotalCount\else\edef\PdrInvKinductionDfStaticZeroTwoTFTrueNotSolvedByKinductionPlainButKipdrTotalCount{0}\fi
\ifdefined\PdrInvKinductionDfStaticZeroTwoTFTrueNotSolvedByKinductionPlainButKipdrCorrectCount\else\edef\PdrInvKinductionDfStaticZeroTwoTFTrueNotSolvedByKinductionPlainButKipdrCorrectCount{0}\fi
\ifdefined\PdrInvKinductionDfStaticZeroTwoTFTrueNotSolvedByKinductionPlainButKipdrCorrectTrueCount\else\edef\PdrInvKinductionDfStaticZeroTwoTFTrueNotSolvedByKinductionPlainButKipdrCorrectTrueCount{0}\fi
\ifdefined\PdrInvKinductionDfStaticZeroTwoTFTrueNotSolvedByKinductionPlainButKipdrCorrectFalseCount\else\edef\PdrInvKinductionDfStaticZeroTwoTFTrueNotSolvedByKinductionPlainButKipdrCorrectFalseCount{0}\fi
\ifdefined\PdrInvKinductionDfStaticZeroTwoTFTrueNotSolvedByKinductionPlainButKipdrWrongTrueCount\else\edef\PdrInvKinductionDfStaticZeroTwoTFTrueNotSolvedByKinductionPlainButKipdrWrongTrueCount{0}\fi
\ifdefined\PdrInvKinductionDfStaticZeroTwoTFTrueNotSolvedByKinductionPlainButKipdrWrongFalseCount\else\edef\PdrInvKinductionDfStaticZeroTwoTFTrueNotSolvedByKinductionPlainButKipdrWrongFalseCount{0}\fi
\ifdefined\PdrInvKinductionDfStaticZeroTwoTFTrueNotSolvedByKinductionPlainButKipdrErrorTimeoutCount\else\edef\PdrInvKinductionDfStaticZeroTwoTFTrueNotSolvedByKinductionPlainButKipdrErrorTimeoutCount{0}\fi
\ifdefined\PdrInvKinductionDfStaticZeroTwoTFTrueNotSolvedByKinductionPlainButKipdrErrorOutOfMemoryCount\else\edef\PdrInvKinductionDfStaticZeroTwoTFTrueNotSolvedByKinductionPlainButKipdrErrorOutOfMemoryCount{0}\fi
\ifdefined\PdrInvKinductionDfStaticZeroTwoTFTrueNotSolvedByKinductionPlainButKipdrCorrectCputime\else\edef\PdrInvKinductionDfStaticZeroTwoTFTrueNotSolvedByKinductionPlainButKipdrCorrectCputime{0}\fi
\ifdefined\PdrInvKinductionDfStaticZeroTwoTFTrueNotSolvedByKinductionPlainButKipdrCorrectCputimeAvg\else\edef\PdrInvKinductionDfStaticZeroTwoTFTrueNotSolvedByKinductionPlainButKipdrCorrectCputimeAvg{None}\fi
\ifdefined\PdrInvKinductionDfStaticZeroTwoTFTrueNotSolvedByKinductionPlainButKipdrCorrectWalltime\else\edef\PdrInvKinductionDfStaticZeroTwoTFTrueNotSolvedByKinductionPlainButKipdrCorrectWalltime{0}\fi
\ifdefined\PdrInvKinductionDfStaticZeroTwoTFTrueNotSolvedByKinductionPlainButKipdrCorrectWalltimeAvg\else\edef\PdrInvKinductionDfStaticZeroTwoTFTrueNotSolvedByKinductionPlainButKipdrCorrectWalltimeAvg{None}\fi
\ifdefined\PdrInvKinductionDfStaticEightTwoTTrueNotSolvedByKinductionPlainButKipdrTotalCount\else\edef\PdrInvKinductionDfStaticEightTwoTTrueNotSolvedByKinductionPlainButKipdrTotalCount{0}\fi
\ifdefined\PdrInvKinductionDfStaticEightTwoTTrueNotSolvedByKinductionPlainButKipdrCorrectCount\else\edef\PdrInvKinductionDfStaticEightTwoTTrueNotSolvedByKinductionPlainButKipdrCorrectCount{0}\fi
\ifdefined\PdrInvKinductionDfStaticEightTwoTTrueNotSolvedByKinductionPlainButKipdrCorrectTrueCount\else\edef\PdrInvKinductionDfStaticEightTwoTTrueNotSolvedByKinductionPlainButKipdrCorrectTrueCount{0}\fi
\ifdefined\PdrInvKinductionDfStaticEightTwoTTrueNotSolvedByKinductionPlainButKipdrCorrectFalseCount\else\edef\PdrInvKinductionDfStaticEightTwoTTrueNotSolvedByKinductionPlainButKipdrCorrectFalseCount{0}\fi
\ifdefined\PdrInvKinductionDfStaticEightTwoTTrueNotSolvedByKinductionPlainButKipdrWrongTrueCount\else\edef\PdrInvKinductionDfStaticEightTwoTTrueNotSolvedByKinductionPlainButKipdrWrongTrueCount{0}\fi
\ifdefined\PdrInvKinductionDfStaticEightTwoTTrueNotSolvedByKinductionPlainButKipdrWrongFalseCount\else\edef\PdrInvKinductionDfStaticEightTwoTTrueNotSolvedByKinductionPlainButKipdrWrongFalseCount{0}\fi
\ifdefined\PdrInvKinductionDfStaticEightTwoTTrueNotSolvedByKinductionPlainButKipdrErrorTimeoutCount\else\edef\PdrInvKinductionDfStaticEightTwoTTrueNotSolvedByKinductionPlainButKipdrErrorTimeoutCount{0}\fi
\ifdefined\PdrInvKinductionDfStaticEightTwoTTrueNotSolvedByKinductionPlainButKipdrErrorOutOfMemoryCount\else\edef\PdrInvKinductionDfStaticEightTwoTTrueNotSolvedByKinductionPlainButKipdrErrorOutOfMemoryCount{0}\fi
\ifdefined\PdrInvKinductionDfStaticEightTwoTTrueNotSolvedByKinductionPlainButKipdrCorrectCputime\else\edef\PdrInvKinductionDfStaticEightTwoTTrueNotSolvedByKinductionPlainButKipdrCorrectCputime{0}\fi
\ifdefined\PdrInvKinductionDfStaticEightTwoTTrueNotSolvedByKinductionPlainButKipdrCorrectCputimeAvg\else\edef\PdrInvKinductionDfStaticEightTwoTTrueNotSolvedByKinductionPlainButKipdrCorrectCputimeAvg{None}\fi
\ifdefined\PdrInvKinductionDfStaticEightTwoTTrueNotSolvedByKinductionPlainButKipdrCorrectWalltime\else\edef\PdrInvKinductionDfStaticEightTwoTTrueNotSolvedByKinductionPlainButKipdrCorrectWalltime{0}\fi
\ifdefined\PdrInvKinductionDfStaticEightTwoTTrueNotSolvedByKinductionPlainButKipdrCorrectWalltimeAvg\else\edef\PdrInvKinductionDfStaticEightTwoTTrueNotSolvedByKinductionPlainButKipdrCorrectWalltimeAvg{None}\fi
\ifdefined\PdrInvKinductionDfStaticSixteenTwoTTrueNotSolvedByKinductionPlainButKipdrTotalCount\else\edef\PdrInvKinductionDfStaticSixteenTwoTTrueNotSolvedByKinductionPlainButKipdrTotalCount{0}\fi
\ifdefined\PdrInvKinductionDfStaticSixteenTwoTTrueNotSolvedByKinductionPlainButKipdrCorrectCount\else\edef\PdrInvKinductionDfStaticSixteenTwoTTrueNotSolvedByKinductionPlainButKipdrCorrectCount{0}\fi
\ifdefined\PdrInvKinductionDfStaticSixteenTwoTTrueNotSolvedByKinductionPlainButKipdrCorrectTrueCount\else\edef\PdrInvKinductionDfStaticSixteenTwoTTrueNotSolvedByKinductionPlainButKipdrCorrectTrueCount{0}\fi
\ifdefined\PdrInvKinductionDfStaticSixteenTwoTTrueNotSolvedByKinductionPlainButKipdrCorrectFalseCount\else\edef\PdrInvKinductionDfStaticSixteenTwoTTrueNotSolvedByKinductionPlainButKipdrCorrectFalseCount{0}\fi
\ifdefined\PdrInvKinductionDfStaticSixteenTwoTTrueNotSolvedByKinductionPlainButKipdrWrongTrueCount\else\edef\PdrInvKinductionDfStaticSixteenTwoTTrueNotSolvedByKinductionPlainButKipdrWrongTrueCount{0}\fi
\ifdefined\PdrInvKinductionDfStaticSixteenTwoTTrueNotSolvedByKinductionPlainButKipdrWrongFalseCount\else\edef\PdrInvKinductionDfStaticSixteenTwoTTrueNotSolvedByKinductionPlainButKipdrWrongFalseCount{0}\fi
\ifdefined\PdrInvKinductionDfStaticSixteenTwoTTrueNotSolvedByKinductionPlainButKipdrErrorTimeoutCount\else\edef\PdrInvKinductionDfStaticSixteenTwoTTrueNotSolvedByKinductionPlainButKipdrErrorTimeoutCount{0}\fi
\ifdefined\PdrInvKinductionDfStaticSixteenTwoTTrueNotSolvedByKinductionPlainButKipdrErrorOutOfMemoryCount\else\edef\PdrInvKinductionDfStaticSixteenTwoTTrueNotSolvedByKinductionPlainButKipdrErrorOutOfMemoryCount{0}\fi
\ifdefined\PdrInvKinductionDfStaticSixteenTwoTTrueNotSolvedByKinductionPlainButKipdrCorrectCputime\else\edef\PdrInvKinductionDfStaticSixteenTwoTTrueNotSolvedByKinductionPlainButKipdrCorrectCputime{0}\fi
\ifdefined\PdrInvKinductionDfStaticSixteenTwoTTrueNotSolvedByKinductionPlainButKipdrCorrectCputimeAvg\else\edef\PdrInvKinductionDfStaticSixteenTwoTTrueNotSolvedByKinductionPlainButKipdrCorrectCputimeAvg{None}\fi
\ifdefined\PdrInvKinductionDfStaticSixteenTwoTTrueNotSolvedByKinductionPlainButKipdrCorrectWalltime\else\edef\PdrInvKinductionDfStaticSixteenTwoTTrueNotSolvedByKinductionPlainButKipdrCorrectWalltime{0}\fi
\ifdefined\PdrInvKinductionDfStaticSixteenTwoTTrueNotSolvedByKinductionPlainButKipdrCorrectWalltimeAvg\else\edef\PdrInvKinductionDfStaticSixteenTwoTTrueNotSolvedByKinductionPlainButKipdrCorrectWalltimeAvg{None}\fi
\ifdefined\PdrInvKinductionDfStaticSixteenTwoFTrueNotSolvedByKinductionPlainButKipdrTotalCount\else\edef\PdrInvKinductionDfStaticSixteenTwoFTrueNotSolvedByKinductionPlainButKipdrTotalCount{0}\fi
\ifdefined\PdrInvKinductionDfStaticSixteenTwoFTrueNotSolvedByKinductionPlainButKipdrCorrectCount\else\edef\PdrInvKinductionDfStaticSixteenTwoFTrueNotSolvedByKinductionPlainButKipdrCorrectCount{0}\fi
\ifdefined\PdrInvKinductionDfStaticSixteenTwoFTrueNotSolvedByKinductionPlainButKipdrCorrectTrueCount\else\edef\PdrInvKinductionDfStaticSixteenTwoFTrueNotSolvedByKinductionPlainButKipdrCorrectTrueCount{0}\fi
\ifdefined\PdrInvKinductionDfStaticSixteenTwoFTrueNotSolvedByKinductionPlainButKipdrCorrectFalseCount\else\edef\PdrInvKinductionDfStaticSixteenTwoFTrueNotSolvedByKinductionPlainButKipdrCorrectFalseCount{0}\fi
\ifdefined\PdrInvKinductionDfStaticSixteenTwoFTrueNotSolvedByKinductionPlainButKipdrWrongTrueCount\else\edef\PdrInvKinductionDfStaticSixteenTwoFTrueNotSolvedByKinductionPlainButKipdrWrongTrueCount{0}\fi
\ifdefined\PdrInvKinductionDfStaticSixteenTwoFTrueNotSolvedByKinductionPlainButKipdrWrongFalseCount\else\edef\PdrInvKinductionDfStaticSixteenTwoFTrueNotSolvedByKinductionPlainButKipdrWrongFalseCount{0}\fi
\ifdefined\PdrInvKinductionDfStaticSixteenTwoFTrueNotSolvedByKinductionPlainButKipdrErrorTimeoutCount\else\edef\PdrInvKinductionDfStaticSixteenTwoFTrueNotSolvedByKinductionPlainButKipdrErrorTimeoutCount{0}\fi
\ifdefined\PdrInvKinductionDfStaticSixteenTwoFTrueNotSolvedByKinductionPlainButKipdrErrorOutOfMemoryCount\else\edef\PdrInvKinductionDfStaticSixteenTwoFTrueNotSolvedByKinductionPlainButKipdrErrorOutOfMemoryCount{0}\fi
\ifdefined\PdrInvKinductionDfStaticSixteenTwoFTrueNotSolvedByKinductionPlainButKipdrCorrectCputime\else\edef\PdrInvKinductionDfStaticSixteenTwoFTrueNotSolvedByKinductionPlainButKipdrCorrectCputime{0}\fi
\ifdefined\PdrInvKinductionDfStaticSixteenTwoFTrueNotSolvedByKinductionPlainButKipdrCorrectCputimeAvg\else\edef\PdrInvKinductionDfStaticSixteenTwoFTrueNotSolvedByKinductionPlainButKipdrCorrectCputimeAvg{None}\fi
\ifdefined\PdrInvKinductionDfStaticSixteenTwoFTrueNotSolvedByKinductionPlainButKipdrCorrectWalltime\else\edef\PdrInvKinductionDfStaticSixteenTwoFTrueNotSolvedByKinductionPlainButKipdrCorrectWalltime{0}\fi
\ifdefined\PdrInvKinductionDfStaticSixteenTwoFTrueNotSolvedByKinductionPlainButKipdrCorrectWalltimeAvg\else\edef\PdrInvKinductionDfStaticSixteenTwoFTrueNotSolvedByKinductionPlainButKipdrCorrectWalltimeAvg{None}\fi
\ifdefined\PdrInvKinductionDfTrueNotSolvedByKinductionPlainButKipdrTotalCount\else\edef\PdrInvKinductionDfTrueNotSolvedByKinductionPlainButKipdrTotalCount{0}\fi
\ifdefined\PdrInvKinductionDfTrueNotSolvedByKinductionPlainButKipdrCorrectCount\else\edef\PdrInvKinductionDfTrueNotSolvedByKinductionPlainButKipdrCorrectCount{0}\fi
\ifdefined\PdrInvKinductionDfTrueNotSolvedByKinductionPlainButKipdrCorrectTrueCount\else\edef\PdrInvKinductionDfTrueNotSolvedByKinductionPlainButKipdrCorrectTrueCount{0}\fi
\ifdefined\PdrInvKinductionDfTrueNotSolvedByKinductionPlainButKipdrCorrectFalseCount\else\edef\PdrInvKinductionDfTrueNotSolvedByKinductionPlainButKipdrCorrectFalseCount{0}\fi
\ifdefined\PdrInvKinductionDfTrueNotSolvedByKinductionPlainButKipdrWrongTrueCount\else\edef\PdrInvKinductionDfTrueNotSolvedByKinductionPlainButKipdrWrongTrueCount{0}\fi
\ifdefined\PdrInvKinductionDfTrueNotSolvedByKinductionPlainButKipdrWrongFalseCount\else\edef\PdrInvKinductionDfTrueNotSolvedByKinductionPlainButKipdrWrongFalseCount{0}\fi
\ifdefined\PdrInvKinductionDfTrueNotSolvedByKinductionPlainButKipdrErrorTimeoutCount\else\edef\PdrInvKinductionDfTrueNotSolvedByKinductionPlainButKipdrErrorTimeoutCount{0}\fi
\ifdefined\PdrInvKinductionDfTrueNotSolvedByKinductionPlainButKipdrErrorOutOfMemoryCount\else\edef\PdrInvKinductionDfTrueNotSolvedByKinductionPlainButKipdrErrorOutOfMemoryCount{0}\fi
\ifdefined\PdrInvKinductionDfTrueNotSolvedByKinductionPlainButKipdrCorrectCputime\else\edef\PdrInvKinductionDfTrueNotSolvedByKinductionPlainButKipdrCorrectCputime{0}\fi
\ifdefined\PdrInvKinductionDfTrueNotSolvedByKinductionPlainButKipdrCorrectCputimeAvg\else\edef\PdrInvKinductionDfTrueNotSolvedByKinductionPlainButKipdrCorrectCputimeAvg{None}\fi
\ifdefined\PdrInvKinductionDfTrueNotSolvedByKinductionPlainButKipdrCorrectWalltime\else\edef\PdrInvKinductionDfTrueNotSolvedByKinductionPlainButKipdrCorrectWalltime{0}\fi
\ifdefined\PdrInvKinductionDfTrueNotSolvedByKinductionPlainButKipdrCorrectWalltimeAvg\else\edef\PdrInvKinductionDfTrueNotSolvedByKinductionPlainButKipdrCorrectWalltimeAvg{None}\fi
\ifdefined\PdrInvKinductionKipdrTrueNotSolvedByKinductionPlainButKipdrTotalCount\else\edef\PdrInvKinductionKipdrTrueNotSolvedByKinductionPlainButKipdrTotalCount{0}\fi
\ifdefined\PdrInvKinductionKipdrTrueNotSolvedByKinductionPlainButKipdrCorrectCount\else\edef\PdrInvKinductionKipdrTrueNotSolvedByKinductionPlainButKipdrCorrectCount{0}\fi
\ifdefined\PdrInvKinductionKipdrTrueNotSolvedByKinductionPlainButKipdrCorrectTrueCount\else\edef\PdrInvKinductionKipdrTrueNotSolvedByKinductionPlainButKipdrCorrectTrueCount{0}\fi
\ifdefined\PdrInvKinductionKipdrTrueNotSolvedByKinductionPlainButKipdrCorrectFalseCount\else\edef\PdrInvKinductionKipdrTrueNotSolvedByKinductionPlainButKipdrCorrectFalseCount{0}\fi
\ifdefined\PdrInvKinductionKipdrTrueNotSolvedByKinductionPlainButKipdrWrongTrueCount\else\edef\PdrInvKinductionKipdrTrueNotSolvedByKinductionPlainButKipdrWrongTrueCount{0}\fi
\ifdefined\PdrInvKinductionKipdrTrueNotSolvedByKinductionPlainButKipdrWrongFalseCount\else\edef\PdrInvKinductionKipdrTrueNotSolvedByKinductionPlainButKipdrWrongFalseCount{0}\fi
\ifdefined\PdrInvKinductionKipdrTrueNotSolvedByKinductionPlainButKipdrErrorTimeoutCount\else\edef\PdrInvKinductionKipdrTrueNotSolvedByKinductionPlainButKipdrErrorTimeoutCount{0}\fi
\ifdefined\PdrInvKinductionKipdrTrueNotSolvedByKinductionPlainButKipdrErrorOutOfMemoryCount\else\edef\PdrInvKinductionKipdrTrueNotSolvedByKinductionPlainButKipdrErrorOutOfMemoryCount{0}\fi
\ifdefined\PdrInvKinductionKipdrTrueNotSolvedByKinductionPlainButKipdrCorrectCputime\else\edef\PdrInvKinductionKipdrTrueNotSolvedByKinductionPlainButKipdrCorrectCputime{0}\fi
\ifdefined\PdrInvKinductionKipdrTrueNotSolvedByKinductionPlainButKipdrCorrectCputimeAvg\else\edef\PdrInvKinductionKipdrTrueNotSolvedByKinductionPlainButKipdrCorrectCputimeAvg{None}\fi
\ifdefined\PdrInvKinductionKipdrTrueNotSolvedByKinductionPlainButKipdrCorrectWalltime\else\edef\PdrInvKinductionKipdrTrueNotSolvedByKinductionPlainButKipdrCorrectWalltime{0}\fi
\ifdefined\PdrInvKinductionKipdrTrueNotSolvedByKinductionPlainButKipdrCorrectWalltimeAvg\else\edef\PdrInvKinductionKipdrTrueNotSolvedByKinductionPlainButKipdrCorrectWalltimeAvg{None}\fi
\ifdefined\PdrInvKinductionKipdrdfTrueNotSolvedByKinductionPlainButKipdrTotalCount\else\edef\PdrInvKinductionKipdrdfTrueNotSolvedByKinductionPlainButKipdrTotalCount{0}\fi
\ifdefined\PdrInvKinductionKipdrdfTrueNotSolvedByKinductionPlainButKipdrCorrectCount\else\edef\PdrInvKinductionKipdrdfTrueNotSolvedByKinductionPlainButKipdrCorrectCount{0}\fi
\ifdefined\PdrInvKinductionKipdrdfTrueNotSolvedByKinductionPlainButKipdrCorrectTrueCount\else\edef\PdrInvKinductionKipdrdfTrueNotSolvedByKinductionPlainButKipdrCorrectTrueCount{0}\fi
\ifdefined\PdrInvKinductionKipdrdfTrueNotSolvedByKinductionPlainButKipdrCorrectFalseCount\else\edef\PdrInvKinductionKipdrdfTrueNotSolvedByKinductionPlainButKipdrCorrectFalseCount{0}\fi
\ifdefined\PdrInvKinductionKipdrdfTrueNotSolvedByKinductionPlainButKipdrWrongTrueCount\else\edef\PdrInvKinductionKipdrdfTrueNotSolvedByKinductionPlainButKipdrWrongTrueCount{0}\fi
\ifdefined\PdrInvKinductionKipdrdfTrueNotSolvedByKinductionPlainButKipdrWrongFalseCount\else\edef\PdrInvKinductionKipdrdfTrueNotSolvedByKinductionPlainButKipdrWrongFalseCount{0}\fi
\ifdefined\PdrInvKinductionKipdrdfTrueNotSolvedByKinductionPlainButKipdrErrorTimeoutCount\else\edef\PdrInvKinductionKipdrdfTrueNotSolvedByKinductionPlainButKipdrErrorTimeoutCount{0}\fi
\ifdefined\PdrInvKinductionKipdrdfTrueNotSolvedByKinductionPlainButKipdrErrorOutOfMemoryCount\else\edef\PdrInvKinductionKipdrdfTrueNotSolvedByKinductionPlainButKipdrErrorOutOfMemoryCount{0}\fi
\ifdefined\PdrInvKinductionKipdrdfTrueNotSolvedByKinductionPlainButKipdrCorrectCputime\else\edef\PdrInvKinductionKipdrdfTrueNotSolvedByKinductionPlainButKipdrCorrectCputime{0}\fi
\ifdefined\PdrInvKinductionKipdrdfTrueNotSolvedByKinductionPlainButKipdrCorrectCputimeAvg\else\edef\PdrInvKinductionKipdrdfTrueNotSolvedByKinductionPlainButKipdrCorrectCputimeAvg{None}\fi
\ifdefined\PdrInvKinductionKipdrdfTrueNotSolvedByKinductionPlainButKipdrCorrectWalltime\else\edef\PdrInvKinductionKipdrdfTrueNotSolvedByKinductionPlainButKipdrCorrectWalltime{0}\fi
\ifdefined\PdrInvKinductionKipdrdfTrueNotSolvedByKinductionPlainButKipdrCorrectWalltimeAvg\else\edef\PdrInvKinductionKipdrdfTrueNotSolvedByKinductionPlainButKipdrCorrectWalltimeAvg{None}\fi
\ifdefined\PdrInvPdrTrueNotSolvedByKinductionPlainButKipdrTotalCount\else\edef\PdrInvPdrTrueNotSolvedByKinductionPlainButKipdrTotalCount{0}\fi
\ifdefined\PdrInvPdrTrueNotSolvedByKinductionPlainButKipdrCorrectCount\else\edef\PdrInvPdrTrueNotSolvedByKinductionPlainButKipdrCorrectCount{0}\fi
\ifdefined\PdrInvPdrTrueNotSolvedByKinductionPlainButKipdrCorrectTrueCount\else\edef\PdrInvPdrTrueNotSolvedByKinductionPlainButKipdrCorrectTrueCount{0}\fi
\ifdefined\PdrInvPdrTrueNotSolvedByKinductionPlainButKipdrCorrectFalseCount\else\edef\PdrInvPdrTrueNotSolvedByKinductionPlainButKipdrCorrectFalseCount{0}\fi
\ifdefined\PdrInvPdrTrueNotSolvedByKinductionPlainButKipdrWrongTrueCount\else\edef\PdrInvPdrTrueNotSolvedByKinductionPlainButKipdrWrongTrueCount{0}\fi
\ifdefined\PdrInvPdrTrueNotSolvedByKinductionPlainButKipdrWrongFalseCount\else\edef\PdrInvPdrTrueNotSolvedByKinductionPlainButKipdrWrongFalseCount{0}\fi
\ifdefined\PdrInvPdrTrueNotSolvedByKinductionPlainButKipdrErrorTimeoutCount\else\edef\PdrInvPdrTrueNotSolvedByKinductionPlainButKipdrErrorTimeoutCount{0}\fi
\ifdefined\PdrInvPdrTrueNotSolvedByKinductionPlainButKipdrErrorOutOfMemoryCount\else\edef\PdrInvPdrTrueNotSolvedByKinductionPlainButKipdrErrorOutOfMemoryCount{0}\fi
\ifdefined\PdrInvPdrTrueNotSolvedByKinductionPlainButKipdrCorrectCputime\else\edef\PdrInvPdrTrueNotSolvedByKinductionPlainButKipdrCorrectCputime{0}\fi
\ifdefined\PdrInvPdrTrueNotSolvedByKinductionPlainButKipdrCorrectCputimeAvg\else\edef\PdrInvPdrTrueNotSolvedByKinductionPlainButKipdrCorrectCputimeAvg{None}\fi
\ifdefined\PdrInvPdrTrueNotSolvedByKinductionPlainButKipdrCorrectWalltime\else\edef\PdrInvPdrTrueNotSolvedByKinductionPlainButKipdrCorrectWalltime{0}\fi
\ifdefined\PdrInvPdrTrueNotSolvedByKinductionPlainButKipdrCorrectWalltimeAvg\else\edef\PdrInvPdrTrueNotSolvedByKinductionPlainButKipdrCorrectWalltimeAvg{None}\fi
\ifdefined\PdrInvOracleTrueNotSolvedByKinductionPlainButKipdrTotalCount\else\edef\PdrInvOracleTrueNotSolvedByKinductionPlainButKipdrTotalCount{0}\fi
\ifdefined\PdrInvOracleTrueNotSolvedByKinductionPlainButKipdrCorrectCount\else\edef\PdrInvOracleTrueNotSolvedByKinductionPlainButKipdrCorrectCount{0}\fi
\ifdefined\PdrInvOracleTrueNotSolvedByKinductionPlainButKipdrCorrectTrueCount\else\edef\PdrInvOracleTrueNotSolvedByKinductionPlainButKipdrCorrectTrueCount{0}\fi
\ifdefined\PdrInvOracleTrueNotSolvedByKinductionPlainButKipdrCorrectFalseCount\else\edef\PdrInvOracleTrueNotSolvedByKinductionPlainButKipdrCorrectFalseCount{0}\fi
\ifdefined\PdrInvOracleTrueNotSolvedByKinductionPlainButKipdrWrongTrueCount\else\edef\PdrInvOracleTrueNotSolvedByKinductionPlainButKipdrWrongTrueCount{0}\fi
\ifdefined\PdrInvOracleTrueNotSolvedByKinductionPlainButKipdrWrongFalseCount\else\edef\PdrInvOracleTrueNotSolvedByKinductionPlainButKipdrWrongFalseCount{0}\fi
\ifdefined\PdrInvOracleTrueNotSolvedByKinductionPlainButKipdrErrorTimeoutCount\else\edef\PdrInvOracleTrueNotSolvedByKinductionPlainButKipdrErrorTimeoutCount{0}\fi
\ifdefined\PdrInvOracleTrueNotSolvedByKinductionPlainButKipdrErrorOutOfMemoryCount\else\edef\PdrInvOracleTrueNotSolvedByKinductionPlainButKipdrErrorOutOfMemoryCount{0}\fi
\ifdefined\PdrInvOracleTrueNotSolvedByKinductionPlainButKipdrCorrectCputime\else\edef\PdrInvOracleTrueNotSolvedByKinductionPlainButKipdrCorrectCputime{0}\fi
\ifdefined\PdrInvOracleTrueNotSolvedByKinductionPlainButKipdrCorrectCputimeAvg\else\edef\PdrInvOracleTrueNotSolvedByKinductionPlainButKipdrCorrectCputimeAvg{None}\fi
\ifdefined\PdrInvOracleTrueNotSolvedByKinductionPlainButKipdrCorrectWalltime\else\edef\PdrInvOracleTrueNotSolvedByKinductionPlainButKipdrCorrectWalltime{0}\fi
\ifdefined\PdrInvOracleTrueNotSolvedByKinductionPlainButKipdrCorrectWalltimeAvg\else\edef\PdrInvOracleTrueNotSolvedByKinductionPlainButKipdrCorrectWalltimeAvg{None}\fi
\ifdefined\SeahornSeahornTrueNotSolvedByKinductionPlainButKipdrTotalCount\else\edef\SeahornSeahornTrueNotSolvedByKinductionPlainButKipdrTotalCount{0}\fi
\ifdefined\SeahornSeahornTrueNotSolvedByKinductionPlainButKipdrCorrectCount\else\edef\SeahornSeahornTrueNotSolvedByKinductionPlainButKipdrCorrectCount{0}\fi
\ifdefined\SeahornSeahornTrueNotSolvedByKinductionPlainButKipdrCorrectTrueCount\else\edef\SeahornSeahornTrueNotSolvedByKinductionPlainButKipdrCorrectTrueCount{0}\fi
\ifdefined\SeahornSeahornTrueNotSolvedByKinductionPlainButKipdrCorrectFalseCount\else\edef\SeahornSeahornTrueNotSolvedByKinductionPlainButKipdrCorrectFalseCount{0}\fi
\ifdefined\SeahornSeahornTrueNotSolvedByKinductionPlainButKipdrWrongTrueCount\else\edef\SeahornSeahornTrueNotSolvedByKinductionPlainButKipdrWrongTrueCount{0}\fi
\ifdefined\SeahornSeahornTrueNotSolvedByKinductionPlainButKipdrWrongFalseCount\else\edef\SeahornSeahornTrueNotSolvedByKinductionPlainButKipdrWrongFalseCount{0}\fi
\ifdefined\SeahornSeahornTrueNotSolvedByKinductionPlainButKipdrErrorTimeoutCount\else\edef\SeahornSeahornTrueNotSolvedByKinductionPlainButKipdrErrorTimeoutCount{0}\fi
\ifdefined\SeahornSeahornTrueNotSolvedByKinductionPlainButKipdrErrorOutOfMemoryCount\else\edef\SeahornSeahornTrueNotSolvedByKinductionPlainButKipdrErrorOutOfMemoryCount{0}\fi
\ifdefined\SeahornSeahornTrueNotSolvedByKinductionPlainButKipdrCorrectCputime\else\edef\SeahornSeahornTrueNotSolvedByKinductionPlainButKipdrCorrectCputime{0}\fi
\ifdefined\SeahornSeahornTrueNotSolvedByKinductionPlainButKipdrCorrectCputimeAvg\else\edef\SeahornSeahornTrueNotSolvedByKinductionPlainButKipdrCorrectCputimeAvg{None}\fi
\ifdefined\SeahornSeahornTrueNotSolvedByKinductionPlainButKipdrCorrectWalltime\else\edef\SeahornSeahornTrueNotSolvedByKinductionPlainButKipdrCorrectWalltime{0}\fi
\ifdefined\SeahornSeahornTrueNotSolvedByKinductionPlainButKipdrCorrectWalltimeAvg\else\edef\SeahornSeahornTrueNotSolvedByKinductionPlainButKipdrCorrectWalltimeAvg{None}\fi
\ifdefined\VvtCtigarTrueNotSolvedByKinductionPlainButKipdrTotalCount\else\edef\VvtCtigarTrueNotSolvedByKinductionPlainButKipdrTotalCount{0}\fi
\ifdefined\VvtCtigarTrueNotSolvedByKinductionPlainButKipdrCorrectCount\else\edef\VvtCtigarTrueNotSolvedByKinductionPlainButKipdrCorrectCount{0}\fi
\ifdefined\VvtCtigarTrueNotSolvedByKinductionPlainButKipdrCorrectTrueCount\else\edef\VvtCtigarTrueNotSolvedByKinductionPlainButKipdrCorrectTrueCount{0}\fi
\ifdefined\VvtCtigarTrueNotSolvedByKinductionPlainButKipdrCorrectFalseCount\else\edef\VvtCtigarTrueNotSolvedByKinductionPlainButKipdrCorrectFalseCount{0}\fi
\ifdefined\VvtCtigarTrueNotSolvedByKinductionPlainButKipdrWrongTrueCount\else\edef\VvtCtigarTrueNotSolvedByKinductionPlainButKipdrWrongTrueCount{0}\fi
\ifdefined\VvtCtigarTrueNotSolvedByKinductionPlainButKipdrWrongFalseCount\else\edef\VvtCtigarTrueNotSolvedByKinductionPlainButKipdrWrongFalseCount{0}\fi
\ifdefined\VvtCtigarTrueNotSolvedByKinductionPlainButKipdrErrorTimeoutCount\else\edef\VvtCtigarTrueNotSolvedByKinductionPlainButKipdrErrorTimeoutCount{0}\fi
\ifdefined\VvtCtigarTrueNotSolvedByKinductionPlainButKipdrErrorOutOfMemoryCount\else\edef\VvtCtigarTrueNotSolvedByKinductionPlainButKipdrErrorOutOfMemoryCount{0}\fi
\ifdefined\VvtCtigarTrueNotSolvedByKinductionPlainButKipdrCorrectCputime\else\edef\VvtCtigarTrueNotSolvedByKinductionPlainButKipdrCorrectCputime{0}\fi
\ifdefined\VvtCtigarTrueNotSolvedByKinductionPlainButKipdrCorrectCputimeAvg\else\edef\VvtCtigarTrueNotSolvedByKinductionPlainButKipdrCorrectCputimeAvg{None}\fi
\ifdefined\VvtCtigarTrueNotSolvedByKinductionPlainButKipdrCorrectWalltime\else\edef\VvtCtigarTrueNotSolvedByKinductionPlainButKipdrCorrectWalltime{0}\fi
\ifdefined\VvtCtigarTrueNotSolvedByKinductionPlainButKipdrCorrectWalltimeAvg\else\edef\VvtCtigarTrueNotSolvedByKinductionPlainButKipdrCorrectWalltimeAvg{None}\fi
\ifdefined\VvtPortfolioTrueNotSolvedByKinductionPlainButKipdrTotalCount\else\edef\VvtPortfolioTrueNotSolvedByKinductionPlainButKipdrTotalCount{0}\fi
\ifdefined\VvtPortfolioTrueNotSolvedByKinductionPlainButKipdrCorrectCount\else\edef\VvtPortfolioTrueNotSolvedByKinductionPlainButKipdrCorrectCount{0}\fi
\ifdefined\VvtPortfolioTrueNotSolvedByKinductionPlainButKipdrCorrectTrueCount\else\edef\VvtPortfolioTrueNotSolvedByKinductionPlainButKipdrCorrectTrueCount{0}\fi
\ifdefined\VvtPortfolioTrueNotSolvedByKinductionPlainButKipdrCorrectFalseCount\else\edef\VvtPortfolioTrueNotSolvedByKinductionPlainButKipdrCorrectFalseCount{0}\fi
\ifdefined\VvtPortfolioTrueNotSolvedByKinductionPlainButKipdrWrongTrueCount\else\edef\VvtPortfolioTrueNotSolvedByKinductionPlainButKipdrWrongTrueCount{0}\fi
\ifdefined\VvtPortfolioTrueNotSolvedByKinductionPlainButKipdrWrongFalseCount\else\edef\VvtPortfolioTrueNotSolvedByKinductionPlainButKipdrWrongFalseCount{0}\fi
\ifdefined\VvtPortfolioTrueNotSolvedByKinductionPlainButKipdrErrorTimeoutCount\else\edef\VvtPortfolioTrueNotSolvedByKinductionPlainButKipdrErrorTimeoutCount{0}\fi
\ifdefined\VvtPortfolioTrueNotSolvedByKinductionPlainButKipdrErrorOutOfMemoryCount\else\edef\VvtPortfolioTrueNotSolvedByKinductionPlainButKipdrErrorOutOfMemoryCount{0}\fi
\ifdefined\VvtPortfolioTrueNotSolvedByKinductionPlainButKipdrCorrectCputime\else\edef\VvtPortfolioTrueNotSolvedByKinductionPlainButKipdrCorrectCputime{0}\fi
\ifdefined\VvtPortfolioTrueNotSolvedByKinductionPlainButKipdrCorrectCputimeAvg\else\edef\VvtPortfolioTrueNotSolvedByKinductionPlainButKipdrCorrectCputimeAvg{None}\fi
\ifdefined\VvtPortfolioTrueNotSolvedByKinductionPlainButKipdrCorrectWalltime\else\edef\VvtPortfolioTrueNotSolvedByKinductionPlainButKipdrCorrectWalltime{0}\fi
\ifdefined\VvtPortfolioTrueNotSolvedByKinductionPlainButKipdrCorrectWalltimeAvg\else\edef\VvtPortfolioTrueNotSolvedByKinductionPlainButKipdrCorrectWalltimeAvg{None}\fi
\edef\PdrInvKinductionPlainErrorOtherInconclusiveCount{\the\numexpr \PdrInvKinductionPlainTotalCount - \PdrInvKinductionPlainCorrectCount - \PdrInvKinductionPlainWrongTrueCount - \PdrInvKinductionPlainWrongFalseCount - \PdrInvKinductionPlainErrorTimeoutCount - \PdrInvKinductionPlainErrorOutOfMemoryCount \relax}
\edef\PdrInvKinductionDfStaticZeroZeroTErrorOtherInconclusiveCount{\the\numexpr \PdrInvKinductionDfStaticZeroZeroTTotalCount - \PdrInvKinductionDfStaticZeroZeroTCorrectCount - \PdrInvKinductionDfStaticZeroZeroTWrongTrueCount - \PdrInvKinductionDfStaticZeroZeroTWrongFalseCount - \PdrInvKinductionDfStaticZeroZeroTErrorTimeoutCount - \PdrInvKinductionDfStaticZeroZeroTErrorOutOfMemoryCount \relax}
\edef\PdrInvKinductionDfStaticZeroOneTTErrorOtherInconclusiveCount{\the\numexpr \PdrInvKinductionDfStaticZeroOneTTTotalCount - \PdrInvKinductionDfStaticZeroOneTTCorrectCount - \PdrInvKinductionDfStaticZeroOneTTWrongTrueCount - \PdrInvKinductionDfStaticZeroOneTTWrongFalseCount - \PdrInvKinductionDfStaticZeroOneTTErrorTimeoutCount - \PdrInvKinductionDfStaticZeroOneTTErrorOutOfMemoryCount \relax}
\edef\PdrInvKinductionDfStaticZeroOneTFErrorOtherInconclusiveCount{\the\numexpr \PdrInvKinductionDfStaticZeroOneTFTotalCount - \PdrInvKinductionDfStaticZeroOneTFCorrectCount - \PdrInvKinductionDfStaticZeroOneTFWrongTrueCount - \PdrInvKinductionDfStaticZeroOneTFWrongFalseCount - \PdrInvKinductionDfStaticZeroOneTFErrorTimeoutCount - \PdrInvKinductionDfStaticZeroOneTFErrorOutOfMemoryCount \relax}
\edef\PdrInvKinductionDfStaticZeroTwoTTErrorOtherInconclusiveCount{\the\numexpr \PdrInvKinductionDfStaticZeroTwoTTTotalCount - \PdrInvKinductionDfStaticZeroTwoTTCorrectCount - \PdrInvKinductionDfStaticZeroTwoTTWrongTrueCount - \PdrInvKinductionDfStaticZeroTwoTTWrongFalseCount - \PdrInvKinductionDfStaticZeroTwoTTErrorTimeoutCount - \PdrInvKinductionDfStaticZeroTwoTTErrorOutOfMemoryCount \relax}
\edef\PdrInvKinductionDfStaticZeroTwoTFErrorOtherInconclusiveCount{\the\numexpr \PdrInvKinductionDfStaticZeroTwoTFTotalCount - \PdrInvKinductionDfStaticZeroTwoTFCorrectCount - \PdrInvKinductionDfStaticZeroTwoTFWrongTrueCount - \PdrInvKinductionDfStaticZeroTwoTFWrongFalseCount - \PdrInvKinductionDfStaticZeroTwoTFErrorTimeoutCount - \PdrInvKinductionDfStaticZeroTwoTFErrorOutOfMemoryCount \relax}
\edef\PdrInvKinductionDfStaticEightTwoTErrorOtherInconclusiveCount{\the\numexpr \PdrInvKinductionDfStaticEightTwoTTotalCount - \PdrInvKinductionDfStaticEightTwoTCorrectCount - \PdrInvKinductionDfStaticEightTwoTWrongTrueCount - \PdrInvKinductionDfStaticEightTwoTWrongFalseCount - \PdrInvKinductionDfStaticEightTwoTErrorTimeoutCount - \PdrInvKinductionDfStaticEightTwoTErrorOutOfMemoryCount \relax}
\edef\PdrInvKinductionDfStaticSixteenTwoTErrorOtherInconclusiveCount{\the\numexpr \PdrInvKinductionDfStaticSixteenTwoTTotalCount - \PdrInvKinductionDfStaticSixteenTwoTCorrectCount - \PdrInvKinductionDfStaticSixteenTwoTWrongTrueCount - \PdrInvKinductionDfStaticSixteenTwoTWrongFalseCount - \PdrInvKinductionDfStaticSixteenTwoTErrorTimeoutCount - \PdrInvKinductionDfStaticSixteenTwoTErrorOutOfMemoryCount \relax}
\edef\PdrInvKinductionDfStaticSixteenTwoFErrorOtherInconclusiveCount{\the\numexpr \PdrInvKinductionDfStaticSixteenTwoFTotalCount - \PdrInvKinductionDfStaticSixteenTwoFCorrectCount - \PdrInvKinductionDfStaticSixteenTwoFWrongTrueCount - \PdrInvKinductionDfStaticSixteenTwoFWrongFalseCount - \PdrInvKinductionDfStaticSixteenTwoFErrorTimeoutCount - \PdrInvKinductionDfStaticSixteenTwoFErrorOutOfMemoryCount \relax}
\edef\PdrInvKinductionDfErrorOtherInconclusiveCount{\the\numexpr \PdrInvKinductionDfTotalCount - \PdrInvKinductionDfCorrectCount - \PdrInvKinductionDfWrongTrueCount - \PdrInvKinductionDfWrongFalseCount - \PdrInvKinductionDfErrorTimeoutCount - \PdrInvKinductionDfErrorOutOfMemoryCount \relax}
\edef\PdrInvKinductionKipdrErrorOtherInconclusiveCount{\the\numexpr \PdrInvKinductionKipdrTotalCount - \PdrInvKinductionKipdrCorrectCount - \PdrInvKinductionKipdrWrongTrueCount - \PdrInvKinductionKipdrWrongFalseCount - \PdrInvKinductionKipdrErrorTimeoutCount - \PdrInvKinductionKipdrErrorOutOfMemoryCount \relax}
\edef\PdrInvKinductionKipdrdfErrorOtherInconclusiveCount{\the\numexpr \PdrInvKinductionKipdrdfTotalCount - \PdrInvKinductionKipdrdfCorrectCount - \PdrInvKinductionKipdrdfWrongTrueCount - \PdrInvKinductionKipdrdfWrongFalseCount - \PdrInvKinductionKipdrdfErrorTimeoutCount - \PdrInvKinductionKipdrdfErrorOutOfMemoryCount \relax}
\edef\PdrInvPdrErrorOtherInconclusiveCount{\the\numexpr \PdrInvPdrTotalCount - \PdrInvPdrCorrectCount - \PdrInvPdrWrongTrueCount - \PdrInvPdrWrongFalseCount - \PdrInvPdrErrorTimeoutCount - \PdrInvPdrErrorOutOfMemoryCount \relax}
\edef\PdrInvOracleErrorOtherInconclusiveCount{\the\numexpr \PdrInvOracleTotalCount - \PdrInvOracleCorrectCount - \PdrInvOracleWrongTrueCount - \PdrInvOracleWrongFalseCount - \PdrInvOracleErrorTimeoutCount - \PdrInvOracleErrorOutOfMemoryCount \relax}
\edef\SeahornSeahornErrorOtherInconclusiveCount{\the\numexpr \SeahornSeahornTotalCount - \SeahornSeahornCorrectCount - \SeahornSeahornWrongTrueCount - \SeahornSeahornWrongFalseCount - \SeahornSeahornErrorTimeoutCount - \SeahornSeahornErrorOutOfMemoryCount \relax}
\edef\VvtCtigarErrorOtherInconclusiveCount{\the\numexpr \VvtCtigarTotalCount - \VvtCtigarCorrectCount - \VvtCtigarWrongTrueCount - \VvtCtigarWrongFalseCount - \VvtCtigarErrorTimeoutCount - \VvtCtigarErrorOutOfMemoryCount \relax}
\edef\VvtPortfolioErrorOtherInconclusiveCount{\the\numexpr \VvtPortfolioTotalCount - \VvtPortfolioCorrectCount - \VvtPortfolioWrongTrueCount - \VvtPortfolioWrongFalseCount - \VvtPortfolioErrorTimeoutCount - \VvtPortfolioErrorOutOfMemoryCount \relax}
\edef\PdrInvKinductionPlainTrueNotSolvedByKinductionPlainErrorOtherInconclusiveCount{\the\numexpr \PdrInvKinductionPlainTrueNotSolvedByKinductionPlainTotalCount - \PdrInvKinductionPlainTrueNotSolvedByKinductionPlainCorrectCount - \PdrInvKinductionPlainTrueNotSolvedByKinductionPlainWrongTrueCount - \PdrInvKinductionPlainTrueNotSolvedByKinductionPlainWrongFalseCount - \PdrInvKinductionPlainTrueNotSolvedByKinductionPlainErrorTimeoutCount - \PdrInvKinductionPlainTrueNotSolvedByKinductionPlainErrorOutOfMemoryCount \relax}
\edef\PdrInvKinductionDfStaticZeroZeroTTrueNotSolvedByKinductionPlainErrorOtherInconclusiveCount{\the\numexpr \PdrInvKinductionDfStaticZeroZeroTTrueNotSolvedByKinductionPlainTotalCount - \PdrInvKinductionDfStaticZeroZeroTTrueNotSolvedByKinductionPlainCorrectCount - \PdrInvKinductionDfStaticZeroZeroTTrueNotSolvedByKinductionPlainWrongTrueCount - \PdrInvKinductionDfStaticZeroZeroTTrueNotSolvedByKinductionPlainWrongFalseCount - \PdrInvKinductionDfStaticZeroZeroTTrueNotSolvedByKinductionPlainErrorTimeoutCount - \PdrInvKinductionDfStaticZeroZeroTTrueNotSolvedByKinductionPlainErrorOutOfMemoryCount \relax}
\edef\PdrInvKinductionDfStaticZeroOneTTTrueNotSolvedByKinductionPlainErrorOtherInconclusiveCount{\the\numexpr \PdrInvKinductionDfStaticZeroOneTTTrueNotSolvedByKinductionPlainTotalCount - \PdrInvKinductionDfStaticZeroOneTTTrueNotSolvedByKinductionPlainCorrectCount - \PdrInvKinductionDfStaticZeroOneTTTrueNotSolvedByKinductionPlainWrongTrueCount - \PdrInvKinductionDfStaticZeroOneTTTrueNotSolvedByKinductionPlainWrongFalseCount - \PdrInvKinductionDfStaticZeroOneTTTrueNotSolvedByKinductionPlainErrorTimeoutCount - \PdrInvKinductionDfStaticZeroOneTTTrueNotSolvedByKinductionPlainErrorOutOfMemoryCount \relax}
\edef\PdrInvKinductionDfStaticZeroOneTFTrueNotSolvedByKinductionPlainErrorOtherInconclusiveCount{\the\numexpr \PdrInvKinductionDfStaticZeroOneTFTrueNotSolvedByKinductionPlainTotalCount - \PdrInvKinductionDfStaticZeroOneTFTrueNotSolvedByKinductionPlainCorrectCount - \PdrInvKinductionDfStaticZeroOneTFTrueNotSolvedByKinductionPlainWrongTrueCount - \PdrInvKinductionDfStaticZeroOneTFTrueNotSolvedByKinductionPlainWrongFalseCount - \PdrInvKinductionDfStaticZeroOneTFTrueNotSolvedByKinductionPlainErrorTimeoutCount - \PdrInvKinductionDfStaticZeroOneTFTrueNotSolvedByKinductionPlainErrorOutOfMemoryCount \relax}
\edef\PdrInvKinductionDfStaticZeroTwoTTTrueNotSolvedByKinductionPlainErrorOtherInconclusiveCount{\the\numexpr \PdrInvKinductionDfStaticZeroTwoTTTrueNotSolvedByKinductionPlainTotalCount - \PdrInvKinductionDfStaticZeroTwoTTTrueNotSolvedByKinductionPlainCorrectCount - \PdrInvKinductionDfStaticZeroTwoTTTrueNotSolvedByKinductionPlainWrongTrueCount - \PdrInvKinductionDfStaticZeroTwoTTTrueNotSolvedByKinductionPlainWrongFalseCount - \PdrInvKinductionDfStaticZeroTwoTTTrueNotSolvedByKinductionPlainErrorTimeoutCount - \PdrInvKinductionDfStaticZeroTwoTTTrueNotSolvedByKinductionPlainErrorOutOfMemoryCount \relax}
\edef\PdrInvKinductionDfStaticZeroTwoTFTrueNotSolvedByKinductionPlainErrorOtherInconclusiveCount{\the\numexpr \PdrInvKinductionDfStaticZeroTwoTFTrueNotSolvedByKinductionPlainTotalCount - \PdrInvKinductionDfStaticZeroTwoTFTrueNotSolvedByKinductionPlainCorrectCount - \PdrInvKinductionDfStaticZeroTwoTFTrueNotSolvedByKinductionPlainWrongTrueCount - \PdrInvKinductionDfStaticZeroTwoTFTrueNotSolvedByKinductionPlainWrongFalseCount - \PdrInvKinductionDfStaticZeroTwoTFTrueNotSolvedByKinductionPlainErrorTimeoutCount - \PdrInvKinductionDfStaticZeroTwoTFTrueNotSolvedByKinductionPlainErrorOutOfMemoryCount \relax}
\edef\PdrInvKinductionDfStaticEightTwoTTrueNotSolvedByKinductionPlainErrorOtherInconclusiveCount{\the\numexpr \PdrInvKinductionDfStaticEightTwoTTrueNotSolvedByKinductionPlainTotalCount - \PdrInvKinductionDfStaticEightTwoTTrueNotSolvedByKinductionPlainCorrectCount - \PdrInvKinductionDfStaticEightTwoTTrueNotSolvedByKinductionPlainWrongTrueCount - \PdrInvKinductionDfStaticEightTwoTTrueNotSolvedByKinductionPlainWrongFalseCount - \PdrInvKinductionDfStaticEightTwoTTrueNotSolvedByKinductionPlainErrorTimeoutCount - \PdrInvKinductionDfStaticEightTwoTTrueNotSolvedByKinductionPlainErrorOutOfMemoryCount \relax}
\edef\PdrInvKinductionDfStaticSixteenTwoTTrueNotSolvedByKinductionPlainErrorOtherInconclusiveCount{\the\numexpr \PdrInvKinductionDfStaticSixteenTwoTTrueNotSolvedByKinductionPlainTotalCount - \PdrInvKinductionDfStaticSixteenTwoTTrueNotSolvedByKinductionPlainCorrectCount - \PdrInvKinductionDfStaticSixteenTwoTTrueNotSolvedByKinductionPlainWrongTrueCount - \PdrInvKinductionDfStaticSixteenTwoTTrueNotSolvedByKinductionPlainWrongFalseCount - \PdrInvKinductionDfStaticSixteenTwoTTrueNotSolvedByKinductionPlainErrorTimeoutCount - \PdrInvKinductionDfStaticSixteenTwoTTrueNotSolvedByKinductionPlainErrorOutOfMemoryCount \relax}
\edef\PdrInvKinductionDfStaticSixteenTwoFTrueNotSolvedByKinductionPlainErrorOtherInconclusiveCount{\the\numexpr \PdrInvKinductionDfStaticSixteenTwoFTrueNotSolvedByKinductionPlainTotalCount - \PdrInvKinductionDfStaticSixteenTwoFTrueNotSolvedByKinductionPlainCorrectCount - \PdrInvKinductionDfStaticSixteenTwoFTrueNotSolvedByKinductionPlainWrongTrueCount - \PdrInvKinductionDfStaticSixteenTwoFTrueNotSolvedByKinductionPlainWrongFalseCount - \PdrInvKinductionDfStaticSixteenTwoFTrueNotSolvedByKinductionPlainErrorTimeoutCount - \PdrInvKinductionDfStaticSixteenTwoFTrueNotSolvedByKinductionPlainErrorOutOfMemoryCount \relax}
\edef\PdrInvKinductionDfTrueNotSolvedByKinductionPlainErrorOtherInconclusiveCount{\the\numexpr \PdrInvKinductionDfTrueNotSolvedByKinductionPlainTotalCount - \PdrInvKinductionDfTrueNotSolvedByKinductionPlainCorrectCount - \PdrInvKinductionDfTrueNotSolvedByKinductionPlainWrongTrueCount - \PdrInvKinductionDfTrueNotSolvedByKinductionPlainWrongFalseCount - \PdrInvKinductionDfTrueNotSolvedByKinductionPlainErrorTimeoutCount - \PdrInvKinductionDfTrueNotSolvedByKinductionPlainErrorOutOfMemoryCount \relax}
\edef\PdrInvKinductionKipdrTrueNotSolvedByKinductionPlainErrorOtherInconclusiveCount{\the\numexpr \PdrInvKinductionKipdrTrueNotSolvedByKinductionPlainTotalCount - \PdrInvKinductionKipdrTrueNotSolvedByKinductionPlainCorrectCount - \PdrInvKinductionKipdrTrueNotSolvedByKinductionPlainWrongTrueCount - \PdrInvKinductionKipdrTrueNotSolvedByKinductionPlainWrongFalseCount - \PdrInvKinductionKipdrTrueNotSolvedByKinductionPlainErrorTimeoutCount - \PdrInvKinductionKipdrTrueNotSolvedByKinductionPlainErrorOutOfMemoryCount \relax}
\edef\PdrInvKinductionKipdrdfTrueNotSolvedByKinductionPlainErrorOtherInconclusiveCount{\the\numexpr \PdrInvKinductionKipdrdfTrueNotSolvedByKinductionPlainTotalCount - \PdrInvKinductionKipdrdfTrueNotSolvedByKinductionPlainCorrectCount - \PdrInvKinductionKipdrdfTrueNotSolvedByKinductionPlainWrongTrueCount - \PdrInvKinductionKipdrdfTrueNotSolvedByKinductionPlainWrongFalseCount - \PdrInvKinductionKipdrdfTrueNotSolvedByKinductionPlainErrorTimeoutCount - \PdrInvKinductionKipdrdfTrueNotSolvedByKinductionPlainErrorOutOfMemoryCount \relax}
\edef\PdrInvPdrTrueNotSolvedByKinductionPlainErrorOtherInconclusiveCount{\the\numexpr \PdrInvPdrTrueNotSolvedByKinductionPlainTotalCount - \PdrInvPdrTrueNotSolvedByKinductionPlainCorrectCount - \PdrInvPdrTrueNotSolvedByKinductionPlainWrongTrueCount - \PdrInvPdrTrueNotSolvedByKinductionPlainWrongFalseCount - \PdrInvPdrTrueNotSolvedByKinductionPlainErrorTimeoutCount - \PdrInvPdrTrueNotSolvedByKinductionPlainErrorOutOfMemoryCount \relax}
\edef\PdrInvOracleTrueNotSolvedByKinductionPlainErrorOtherInconclusiveCount{\the\numexpr \PdrInvOracleTrueNotSolvedByKinductionPlainTotalCount - \PdrInvOracleTrueNotSolvedByKinductionPlainCorrectCount - \PdrInvOracleTrueNotSolvedByKinductionPlainWrongTrueCount - \PdrInvOracleTrueNotSolvedByKinductionPlainWrongFalseCount - \PdrInvOracleTrueNotSolvedByKinductionPlainErrorTimeoutCount - \PdrInvOracleTrueNotSolvedByKinductionPlainErrorOutOfMemoryCount \relax}
\edef\SeahornSeahornTrueNotSolvedByKinductionPlainErrorOtherInconclusiveCount{\the\numexpr \SeahornSeahornTrueNotSolvedByKinductionPlainTotalCount - \SeahornSeahornTrueNotSolvedByKinductionPlainCorrectCount - \SeahornSeahornTrueNotSolvedByKinductionPlainWrongTrueCount - \SeahornSeahornTrueNotSolvedByKinductionPlainWrongFalseCount - \SeahornSeahornTrueNotSolvedByKinductionPlainErrorTimeoutCount - \SeahornSeahornTrueNotSolvedByKinductionPlainErrorOutOfMemoryCount \relax}
\edef\VvtCtigarTrueNotSolvedByKinductionPlainErrorOtherInconclusiveCount{\the\numexpr \VvtCtigarTrueNotSolvedByKinductionPlainTotalCount - \VvtCtigarTrueNotSolvedByKinductionPlainCorrectCount - \VvtCtigarTrueNotSolvedByKinductionPlainWrongTrueCount - \VvtCtigarTrueNotSolvedByKinductionPlainWrongFalseCount - \VvtCtigarTrueNotSolvedByKinductionPlainErrorTimeoutCount - \VvtCtigarTrueNotSolvedByKinductionPlainErrorOutOfMemoryCount \relax}
\edef\VvtPortfolioTrueNotSolvedByKinductionPlainErrorOtherInconclusiveCount{\the\numexpr \VvtPortfolioTrueNotSolvedByKinductionPlainTotalCount - \VvtPortfolioTrueNotSolvedByKinductionPlainCorrectCount - \VvtPortfolioTrueNotSolvedByKinductionPlainWrongTrueCount - \VvtPortfolioTrueNotSolvedByKinductionPlainWrongFalseCount - \VvtPortfolioTrueNotSolvedByKinductionPlainErrorTimeoutCount - \VvtPortfolioTrueNotSolvedByKinductionPlainErrorOutOfMemoryCount \relax}
\edef\PdrInvKinductionPlainTrueNotSolvedByKinductionPlainButKipdrErrorOtherInconclusiveCount{\the\numexpr \PdrInvKinductionPlainTrueNotSolvedByKinductionPlainButKipdrTotalCount - \PdrInvKinductionPlainTrueNotSolvedByKinductionPlainButKipdrCorrectCount - \PdrInvKinductionPlainTrueNotSolvedByKinductionPlainButKipdrWrongTrueCount - \PdrInvKinductionPlainTrueNotSolvedByKinductionPlainButKipdrWrongFalseCount - \PdrInvKinductionPlainTrueNotSolvedByKinductionPlainButKipdrErrorTimeoutCount - \PdrInvKinductionPlainTrueNotSolvedByKinductionPlainButKipdrErrorOutOfMemoryCount \relax}
\edef\PdrInvKinductionDfStaticZeroZeroTTrueNotSolvedByKinductionPlainButKipdrErrorOtherInconclusiveCount{\the\numexpr \PdrInvKinductionDfStaticZeroZeroTTrueNotSolvedByKinductionPlainButKipdrTotalCount - \PdrInvKinductionDfStaticZeroZeroTTrueNotSolvedByKinductionPlainButKipdrCorrectCount - \PdrInvKinductionDfStaticZeroZeroTTrueNotSolvedByKinductionPlainButKipdrWrongTrueCount - \PdrInvKinductionDfStaticZeroZeroTTrueNotSolvedByKinductionPlainButKipdrWrongFalseCount - \PdrInvKinductionDfStaticZeroZeroTTrueNotSolvedByKinductionPlainButKipdrErrorTimeoutCount - \PdrInvKinductionDfStaticZeroZeroTTrueNotSolvedByKinductionPlainButKipdrErrorOutOfMemoryCount \relax}
\edef\PdrInvKinductionDfStaticZeroOneTTTrueNotSolvedByKinductionPlainButKipdrErrorOtherInconclusiveCount{\the\numexpr \PdrInvKinductionDfStaticZeroOneTTTrueNotSolvedByKinductionPlainButKipdrTotalCount - \PdrInvKinductionDfStaticZeroOneTTTrueNotSolvedByKinductionPlainButKipdrCorrectCount - \PdrInvKinductionDfStaticZeroOneTTTrueNotSolvedByKinductionPlainButKipdrWrongTrueCount - \PdrInvKinductionDfStaticZeroOneTTTrueNotSolvedByKinductionPlainButKipdrWrongFalseCount - \PdrInvKinductionDfStaticZeroOneTTTrueNotSolvedByKinductionPlainButKipdrErrorTimeoutCount - \PdrInvKinductionDfStaticZeroOneTTTrueNotSolvedByKinductionPlainButKipdrErrorOutOfMemoryCount \relax}
\edef\PdrInvKinductionDfStaticZeroOneTFTrueNotSolvedByKinductionPlainButKipdrErrorOtherInconclusiveCount{\the\numexpr \PdrInvKinductionDfStaticZeroOneTFTrueNotSolvedByKinductionPlainButKipdrTotalCount - \PdrInvKinductionDfStaticZeroOneTFTrueNotSolvedByKinductionPlainButKipdrCorrectCount - \PdrInvKinductionDfStaticZeroOneTFTrueNotSolvedByKinductionPlainButKipdrWrongTrueCount - \PdrInvKinductionDfStaticZeroOneTFTrueNotSolvedByKinductionPlainButKipdrWrongFalseCount - \PdrInvKinductionDfStaticZeroOneTFTrueNotSolvedByKinductionPlainButKipdrErrorTimeoutCount - \PdrInvKinductionDfStaticZeroOneTFTrueNotSolvedByKinductionPlainButKipdrErrorOutOfMemoryCount \relax}
\edef\PdrInvKinductionDfStaticZeroTwoTTTrueNotSolvedByKinductionPlainButKipdrErrorOtherInconclusiveCount{\the\numexpr \PdrInvKinductionDfStaticZeroTwoTTTrueNotSolvedByKinductionPlainButKipdrTotalCount - \PdrInvKinductionDfStaticZeroTwoTTTrueNotSolvedByKinductionPlainButKipdrCorrectCount - \PdrInvKinductionDfStaticZeroTwoTTTrueNotSolvedByKinductionPlainButKipdrWrongTrueCount - \PdrInvKinductionDfStaticZeroTwoTTTrueNotSolvedByKinductionPlainButKipdrWrongFalseCount - \PdrInvKinductionDfStaticZeroTwoTTTrueNotSolvedByKinductionPlainButKipdrErrorTimeoutCount - \PdrInvKinductionDfStaticZeroTwoTTTrueNotSolvedByKinductionPlainButKipdrErrorOutOfMemoryCount \relax}
\edef\PdrInvKinductionDfStaticZeroTwoTFTrueNotSolvedByKinductionPlainButKipdrErrorOtherInconclusiveCount{\the\numexpr \PdrInvKinductionDfStaticZeroTwoTFTrueNotSolvedByKinductionPlainButKipdrTotalCount - \PdrInvKinductionDfStaticZeroTwoTFTrueNotSolvedByKinductionPlainButKipdrCorrectCount - \PdrInvKinductionDfStaticZeroTwoTFTrueNotSolvedByKinductionPlainButKipdrWrongTrueCount - \PdrInvKinductionDfStaticZeroTwoTFTrueNotSolvedByKinductionPlainButKipdrWrongFalseCount - \PdrInvKinductionDfStaticZeroTwoTFTrueNotSolvedByKinductionPlainButKipdrErrorTimeoutCount - \PdrInvKinductionDfStaticZeroTwoTFTrueNotSolvedByKinductionPlainButKipdrErrorOutOfMemoryCount \relax}
\edef\PdrInvKinductionDfStaticEightTwoTTrueNotSolvedByKinductionPlainButKipdrErrorOtherInconclusiveCount{\the\numexpr \PdrInvKinductionDfStaticEightTwoTTrueNotSolvedByKinductionPlainButKipdrTotalCount - \PdrInvKinductionDfStaticEightTwoTTrueNotSolvedByKinductionPlainButKipdrCorrectCount - \PdrInvKinductionDfStaticEightTwoTTrueNotSolvedByKinductionPlainButKipdrWrongTrueCount - \PdrInvKinductionDfStaticEightTwoTTrueNotSolvedByKinductionPlainButKipdrWrongFalseCount - \PdrInvKinductionDfStaticEightTwoTTrueNotSolvedByKinductionPlainButKipdrErrorTimeoutCount - \PdrInvKinductionDfStaticEightTwoTTrueNotSolvedByKinductionPlainButKipdrErrorOutOfMemoryCount \relax}
\edef\PdrInvKinductionDfStaticSixteenTwoTTrueNotSolvedByKinductionPlainButKipdrErrorOtherInconclusiveCount{\the\numexpr \PdrInvKinductionDfStaticSixteenTwoTTrueNotSolvedByKinductionPlainButKipdrTotalCount - \PdrInvKinductionDfStaticSixteenTwoTTrueNotSolvedByKinductionPlainButKipdrCorrectCount - \PdrInvKinductionDfStaticSixteenTwoTTrueNotSolvedByKinductionPlainButKipdrWrongTrueCount - \PdrInvKinductionDfStaticSixteenTwoTTrueNotSolvedByKinductionPlainButKipdrWrongFalseCount - \PdrInvKinductionDfStaticSixteenTwoTTrueNotSolvedByKinductionPlainButKipdrErrorTimeoutCount - \PdrInvKinductionDfStaticSixteenTwoTTrueNotSolvedByKinductionPlainButKipdrErrorOutOfMemoryCount \relax}
\edef\PdrInvKinductionDfStaticSixteenTwoFTrueNotSolvedByKinductionPlainButKipdrErrorOtherInconclusiveCount{\the\numexpr \PdrInvKinductionDfStaticSixteenTwoFTrueNotSolvedByKinductionPlainButKipdrTotalCount - \PdrInvKinductionDfStaticSixteenTwoFTrueNotSolvedByKinductionPlainButKipdrCorrectCount - \PdrInvKinductionDfStaticSixteenTwoFTrueNotSolvedByKinductionPlainButKipdrWrongTrueCount - \PdrInvKinductionDfStaticSixteenTwoFTrueNotSolvedByKinductionPlainButKipdrWrongFalseCount - \PdrInvKinductionDfStaticSixteenTwoFTrueNotSolvedByKinductionPlainButKipdrErrorTimeoutCount - \PdrInvKinductionDfStaticSixteenTwoFTrueNotSolvedByKinductionPlainButKipdrErrorOutOfMemoryCount \relax}
\edef\PdrInvKinductionDfTrueNotSolvedByKinductionPlainButKipdrErrorOtherInconclusiveCount{\the\numexpr \PdrInvKinductionDfTrueNotSolvedByKinductionPlainButKipdrTotalCount - \PdrInvKinductionDfTrueNotSolvedByKinductionPlainButKipdrCorrectCount - \PdrInvKinductionDfTrueNotSolvedByKinductionPlainButKipdrWrongTrueCount - \PdrInvKinductionDfTrueNotSolvedByKinductionPlainButKipdrWrongFalseCount - \PdrInvKinductionDfTrueNotSolvedByKinductionPlainButKipdrErrorTimeoutCount - \PdrInvKinductionDfTrueNotSolvedByKinductionPlainButKipdrErrorOutOfMemoryCount \relax}
\edef\PdrInvKinductionKipdrTrueNotSolvedByKinductionPlainButKipdrErrorOtherInconclusiveCount{\the\numexpr \PdrInvKinductionKipdrTrueNotSolvedByKinductionPlainButKipdrTotalCount - \PdrInvKinductionKipdrTrueNotSolvedByKinductionPlainButKipdrCorrectCount - \PdrInvKinductionKipdrTrueNotSolvedByKinductionPlainButKipdrWrongTrueCount - \PdrInvKinductionKipdrTrueNotSolvedByKinductionPlainButKipdrWrongFalseCount - \PdrInvKinductionKipdrTrueNotSolvedByKinductionPlainButKipdrErrorTimeoutCount - \PdrInvKinductionKipdrTrueNotSolvedByKinductionPlainButKipdrErrorOutOfMemoryCount \relax}
\edef\PdrInvKinductionKipdrdfTrueNotSolvedByKinductionPlainButKipdrErrorOtherInconclusiveCount{\the\numexpr \PdrInvKinductionKipdrdfTrueNotSolvedByKinductionPlainButKipdrTotalCount - \PdrInvKinductionKipdrdfTrueNotSolvedByKinductionPlainButKipdrCorrectCount - \PdrInvKinductionKipdrdfTrueNotSolvedByKinductionPlainButKipdrWrongTrueCount - \PdrInvKinductionKipdrdfTrueNotSolvedByKinductionPlainButKipdrWrongFalseCount - \PdrInvKinductionKipdrdfTrueNotSolvedByKinductionPlainButKipdrErrorTimeoutCount - \PdrInvKinductionKipdrdfTrueNotSolvedByKinductionPlainButKipdrErrorOutOfMemoryCount \relax}
\edef\PdrInvPdrTrueNotSolvedByKinductionPlainButKipdrErrorOtherInconclusiveCount{\the\numexpr \PdrInvPdrTrueNotSolvedByKinductionPlainButKipdrTotalCount - \PdrInvPdrTrueNotSolvedByKinductionPlainButKipdrCorrectCount - \PdrInvPdrTrueNotSolvedByKinductionPlainButKipdrWrongTrueCount - \PdrInvPdrTrueNotSolvedByKinductionPlainButKipdrWrongFalseCount - \PdrInvPdrTrueNotSolvedByKinductionPlainButKipdrErrorTimeoutCount - \PdrInvPdrTrueNotSolvedByKinductionPlainButKipdrErrorOutOfMemoryCount \relax}
\edef\PdrInvOracleTrueNotSolvedByKinductionPlainButKipdrErrorOtherInconclusiveCount{\the\numexpr \PdrInvOracleTrueNotSolvedByKinductionPlainButKipdrTotalCount - \PdrInvOracleTrueNotSolvedByKinductionPlainButKipdrCorrectCount - \PdrInvOracleTrueNotSolvedByKinductionPlainButKipdrWrongTrueCount - \PdrInvOracleTrueNotSolvedByKinductionPlainButKipdrWrongFalseCount - \PdrInvOracleTrueNotSolvedByKinductionPlainButKipdrErrorTimeoutCount - \PdrInvOracleTrueNotSolvedByKinductionPlainButKipdrErrorOutOfMemoryCount \relax}
\edef\SeahornSeahornTrueNotSolvedByKinductionPlainButKipdrErrorOtherInconclusiveCount{\the\numexpr \SeahornSeahornTrueNotSolvedByKinductionPlainButKipdrTotalCount - \SeahornSeahornTrueNotSolvedByKinductionPlainButKipdrCorrectCount - \SeahornSeahornTrueNotSolvedByKinductionPlainButKipdrWrongTrueCount - \SeahornSeahornTrueNotSolvedByKinductionPlainButKipdrWrongFalseCount - \SeahornSeahornTrueNotSolvedByKinductionPlainButKipdrErrorTimeoutCount - \SeahornSeahornTrueNotSolvedByKinductionPlainButKipdrErrorOutOfMemoryCount \relax}
\edef\VvtCtigarTrueNotSolvedByKinductionPlainButKipdrErrorOtherInconclusiveCount{\the\numexpr \VvtCtigarTrueNotSolvedByKinductionPlainButKipdrTotalCount - \VvtCtigarTrueNotSolvedByKinductionPlainButKipdrCorrectCount - \VvtCtigarTrueNotSolvedByKinductionPlainButKipdrWrongTrueCount - \VvtCtigarTrueNotSolvedByKinductionPlainButKipdrWrongFalseCount - \VvtCtigarTrueNotSolvedByKinductionPlainButKipdrErrorTimeoutCount - \VvtCtigarTrueNotSolvedByKinductionPlainButKipdrErrorOutOfMemoryCount \relax}
\edef\VvtPortfolioTrueNotSolvedByKinductionPlainButKipdrErrorOtherInconclusiveCount{\the\numexpr \VvtPortfolioTrueNotSolvedByKinductionPlainButKipdrTotalCount - \VvtPortfolioTrueNotSolvedByKinductionPlainButKipdrCorrectCount - \VvtPortfolioTrueNotSolvedByKinductionPlainButKipdrWrongTrueCount - \VvtPortfolioTrueNotSolvedByKinductionPlainButKipdrWrongFalseCount - \VvtPortfolioTrueNotSolvedByKinductionPlainButKipdrErrorTimeoutCount - \VvtPortfolioTrueNotSolvedByKinductionPlainButKipdrErrorOutOfMemoryCount \relax}
\edef\PdrInvKinductionPlainScore{\the\numexpr (2 * \PdrInvKinductionPlainCorrectTrueCount) + \PdrInvKinductionPlainCorrectFalseCount - (32 * \PdrInvKinductionPlainWrongTrueCount) - (16 * \PdrInvKinductionPlainWrongFalseCount) \relax}
\edef\PdrInvKinductionDfStaticZeroZeroTScore{\the\numexpr (2 * \PdrInvKinductionDfStaticZeroZeroTCorrectTrueCount) + \PdrInvKinductionDfStaticZeroZeroTCorrectFalseCount - (32 * \PdrInvKinductionDfStaticZeroZeroTWrongTrueCount) - (16 * \PdrInvKinductionDfStaticZeroZeroTWrongFalseCount) \relax}
\edef\PdrInvKinductionDfStaticZeroOneTTScore{\the\numexpr (2 * \PdrInvKinductionDfStaticZeroOneTTCorrectTrueCount) + \PdrInvKinductionDfStaticZeroOneTTCorrectFalseCount - (32 * \PdrInvKinductionDfStaticZeroOneTTWrongTrueCount) - (16 * \PdrInvKinductionDfStaticZeroOneTTWrongFalseCount) \relax}
\edef\PdrInvKinductionDfStaticZeroOneTFScore{\the\numexpr (2 * \PdrInvKinductionDfStaticZeroOneTFCorrectTrueCount) + \PdrInvKinductionDfStaticZeroOneTFCorrectFalseCount - (32 * \PdrInvKinductionDfStaticZeroOneTFWrongTrueCount) - (16 * \PdrInvKinductionDfStaticZeroOneTFWrongFalseCount) \relax}
\edef\PdrInvKinductionDfStaticZeroTwoTTScore{\the\numexpr (2 * \PdrInvKinductionDfStaticZeroTwoTTCorrectTrueCount) + \PdrInvKinductionDfStaticZeroTwoTTCorrectFalseCount - (32 * \PdrInvKinductionDfStaticZeroTwoTTWrongTrueCount) - (16 * \PdrInvKinductionDfStaticZeroTwoTTWrongFalseCount) \relax}
\edef\PdrInvKinductionDfStaticZeroTwoTFScore{\the\numexpr (2 * \PdrInvKinductionDfStaticZeroTwoTFCorrectTrueCount) + \PdrInvKinductionDfStaticZeroTwoTFCorrectFalseCount - (32 * \PdrInvKinductionDfStaticZeroTwoTFWrongTrueCount) - (16 * \PdrInvKinductionDfStaticZeroTwoTFWrongFalseCount) \relax}
\edef\PdrInvKinductionDfStaticEightTwoTScore{\the\numexpr (2 * \PdrInvKinductionDfStaticEightTwoTCorrectTrueCount) + \PdrInvKinductionDfStaticEightTwoTCorrectFalseCount - (32 * \PdrInvKinductionDfStaticEightTwoTWrongTrueCount) - (16 * \PdrInvKinductionDfStaticEightTwoTWrongFalseCount) \relax}
\edef\PdrInvKinductionDfStaticSixteenTwoTScore{\the\numexpr (2 * \PdrInvKinductionDfStaticSixteenTwoTCorrectTrueCount) + \PdrInvKinductionDfStaticSixteenTwoTCorrectFalseCount - (32 * \PdrInvKinductionDfStaticSixteenTwoTWrongTrueCount) - (16 * \PdrInvKinductionDfStaticSixteenTwoTWrongFalseCount) \relax}
\edef\PdrInvKinductionDfStaticSixteenTwoFScore{\the\numexpr (2 * \PdrInvKinductionDfStaticSixteenTwoFCorrectTrueCount) + \PdrInvKinductionDfStaticSixteenTwoFCorrectFalseCount - (32 * \PdrInvKinductionDfStaticSixteenTwoFWrongTrueCount) - (16 * \PdrInvKinductionDfStaticSixteenTwoFWrongFalseCount) \relax}
\edef\PdrInvKinductionDfScore{\the\numexpr (2 * \PdrInvKinductionDfCorrectTrueCount) + \PdrInvKinductionDfCorrectFalseCount - (32 * \PdrInvKinductionDfWrongTrueCount) - (16 * \PdrInvKinductionDfWrongFalseCount) \relax}
\edef\PdrInvKinductionKipdrScore{\the\numexpr (2 * \PdrInvKinductionKipdrCorrectTrueCount) + \PdrInvKinductionKipdrCorrectFalseCount - (32 * \PdrInvKinductionKipdrWrongTrueCount) - (16 * \PdrInvKinductionKipdrWrongFalseCount) \relax}
\edef\PdrInvKinductionKipdrdfScore{\the\numexpr (2 * \PdrInvKinductionKipdrdfCorrectTrueCount) + \PdrInvKinductionKipdrdfCorrectFalseCount - (32 * \PdrInvKinductionKipdrdfWrongTrueCount) - (16 * \PdrInvKinductionKipdrdfWrongFalseCount) \relax}
\edef\PdrInvPdrScore{\the\numexpr (2 * \PdrInvPdrCorrectTrueCount) + \PdrInvPdrCorrectFalseCount - (32 * \PdrInvPdrWrongTrueCount) - (16 * \PdrInvPdrWrongFalseCount) \relax}
\edef\PdrInvOracleScore{\the\numexpr (2 * \PdrInvOracleCorrectTrueCount) + \PdrInvOracleCorrectFalseCount - (32 * \PdrInvOracleWrongTrueCount) - (16 * \PdrInvOracleWrongFalseCount) \relax}
\edef\SeahornSeahornScore{\the\numexpr (2 * \SeahornSeahornCorrectTrueCount) + \SeahornSeahornCorrectFalseCount - (32 * \SeahornSeahornWrongTrueCount) - (16 * \SeahornSeahornWrongFalseCount) \relax}
\edef\VvtCtigarScore{\the\numexpr (2 * \VvtCtigarCorrectTrueCount) + \VvtCtigarCorrectFalseCount - (32 * \VvtCtigarWrongTrueCount) - (16 * \VvtCtigarWrongFalseCount) \relax}
\edef\VvtPortfolioScore{\the\numexpr (2 * \VvtPortfolioCorrectTrueCount) + \VvtPortfolioCorrectFalseCount - (32 * \VvtPortfolioWrongTrueCount) - (16 * \VvtPortfolioWrongFalseCount) \relax}
\edef\PdrInvKinductionPlainTrueNotSolvedByKinductionPlainScore{\the\numexpr (2 * \PdrInvKinductionPlainTrueNotSolvedByKinductionPlainCorrectTrueCount) + \PdrInvKinductionPlainTrueNotSolvedByKinductionPlainCorrectFalseCount - (32 * \PdrInvKinductionPlainTrueNotSolvedByKinductionPlainWrongTrueCount) - (16 * \PdrInvKinductionPlainTrueNotSolvedByKinductionPlainWrongFalseCount) \relax}
\edef\PdrInvKinductionDfStaticZeroZeroTTrueNotSolvedByKinductionPlainScore{\the\numexpr (2 * \PdrInvKinductionDfStaticZeroZeroTTrueNotSolvedByKinductionPlainCorrectTrueCount) + \PdrInvKinductionDfStaticZeroZeroTTrueNotSolvedByKinductionPlainCorrectFalseCount - (32 * \PdrInvKinductionDfStaticZeroZeroTTrueNotSolvedByKinductionPlainWrongTrueCount) - (16 * \PdrInvKinductionDfStaticZeroZeroTTrueNotSolvedByKinductionPlainWrongFalseCount) \relax}
\edef\PdrInvKinductionDfStaticZeroOneTTTrueNotSolvedByKinductionPlainScore{\the\numexpr (2 * \PdrInvKinductionDfStaticZeroOneTTTrueNotSolvedByKinductionPlainCorrectTrueCount) + \PdrInvKinductionDfStaticZeroOneTTTrueNotSolvedByKinductionPlainCorrectFalseCount - (32 * \PdrInvKinductionDfStaticZeroOneTTTrueNotSolvedByKinductionPlainWrongTrueCount) - (16 * \PdrInvKinductionDfStaticZeroOneTTTrueNotSolvedByKinductionPlainWrongFalseCount) \relax}
\edef\PdrInvKinductionDfStaticZeroOneTFTrueNotSolvedByKinductionPlainScore{\the\numexpr (2 * \PdrInvKinductionDfStaticZeroOneTFTrueNotSolvedByKinductionPlainCorrectTrueCount) + \PdrInvKinductionDfStaticZeroOneTFTrueNotSolvedByKinductionPlainCorrectFalseCount - (32 * \PdrInvKinductionDfStaticZeroOneTFTrueNotSolvedByKinductionPlainWrongTrueCount) - (16 * \PdrInvKinductionDfStaticZeroOneTFTrueNotSolvedByKinductionPlainWrongFalseCount) \relax}
\edef\PdrInvKinductionDfStaticZeroTwoTTTrueNotSolvedByKinductionPlainScore{\the\numexpr (2 * \PdrInvKinductionDfStaticZeroTwoTTTrueNotSolvedByKinductionPlainCorrectTrueCount) + \PdrInvKinductionDfStaticZeroTwoTTTrueNotSolvedByKinductionPlainCorrectFalseCount - (32 * \PdrInvKinductionDfStaticZeroTwoTTTrueNotSolvedByKinductionPlainWrongTrueCount) - (16 * \PdrInvKinductionDfStaticZeroTwoTTTrueNotSolvedByKinductionPlainWrongFalseCount) \relax}
\edef\PdrInvKinductionDfStaticZeroTwoTFTrueNotSolvedByKinductionPlainScore{\the\numexpr (2 * \PdrInvKinductionDfStaticZeroTwoTFTrueNotSolvedByKinductionPlainCorrectTrueCount) + \PdrInvKinductionDfStaticZeroTwoTFTrueNotSolvedByKinductionPlainCorrectFalseCount - (32 * \PdrInvKinductionDfStaticZeroTwoTFTrueNotSolvedByKinductionPlainWrongTrueCount) - (16 * \PdrInvKinductionDfStaticZeroTwoTFTrueNotSolvedByKinductionPlainWrongFalseCount) \relax}
\edef\PdrInvKinductionDfStaticEightTwoTTrueNotSolvedByKinductionPlainScore{\the\numexpr (2 * \PdrInvKinductionDfStaticEightTwoTTrueNotSolvedByKinductionPlainCorrectTrueCount) + \PdrInvKinductionDfStaticEightTwoTTrueNotSolvedByKinductionPlainCorrectFalseCount - (32 * \PdrInvKinductionDfStaticEightTwoTTrueNotSolvedByKinductionPlainWrongTrueCount) - (16 * \PdrInvKinductionDfStaticEightTwoTTrueNotSolvedByKinductionPlainWrongFalseCount) \relax}
\edef\PdrInvKinductionDfStaticSixteenTwoTTrueNotSolvedByKinductionPlainScore{\the\numexpr (2 * \PdrInvKinductionDfStaticSixteenTwoTTrueNotSolvedByKinductionPlainCorrectTrueCount) + \PdrInvKinductionDfStaticSixteenTwoTTrueNotSolvedByKinductionPlainCorrectFalseCount - (32 * \PdrInvKinductionDfStaticSixteenTwoTTrueNotSolvedByKinductionPlainWrongTrueCount) - (16 * \PdrInvKinductionDfStaticSixteenTwoTTrueNotSolvedByKinductionPlainWrongFalseCount) \relax}
\edef\PdrInvKinductionDfStaticSixteenTwoFTrueNotSolvedByKinductionPlainScore{\the\numexpr (2 * \PdrInvKinductionDfStaticSixteenTwoFTrueNotSolvedByKinductionPlainCorrectTrueCount) + \PdrInvKinductionDfStaticSixteenTwoFTrueNotSolvedByKinductionPlainCorrectFalseCount - (32 * \PdrInvKinductionDfStaticSixteenTwoFTrueNotSolvedByKinductionPlainWrongTrueCount) - (16 * \PdrInvKinductionDfStaticSixteenTwoFTrueNotSolvedByKinductionPlainWrongFalseCount) \relax}
\edef\PdrInvKinductionDfTrueNotSolvedByKinductionPlainScore{\the\numexpr (2 * \PdrInvKinductionDfTrueNotSolvedByKinductionPlainCorrectTrueCount) + \PdrInvKinductionDfTrueNotSolvedByKinductionPlainCorrectFalseCount - (32 * \PdrInvKinductionDfTrueNotSolvedByKinductionPlainWrongTrueCount) - (16 * \PdrInvKinductionDfTrueNotSolvedByKinductionPlainWrongFalseCount) \relax}
\edef\PdrInvKinductionKipdrTrueNotSolvedByKinductionPlainScore{\the\numexpr (2 * \PdrInvKinductionKipdrTrueNotSolvedByKinductionPlainCorrectTrueCount) + \PdrInvKinductionKipdrTrueNotSolvedByKinductionPlainCorrectFalseCount - (32 * \PdrInvKinductionKipdrTrueNotSolvedByKinductionPlainWrongTrueCount) - (16 * \PdrInvKinductionKipdrTrueNotSolvedByKinductionPlainWrongFalseCount) \relax}
\edef\PdrInvKinductionKipdrdfTrueNotSolvedByKinductionPlainScore{\the\numexpr (2 * \PdrInvKinductionKipdrdfTrueNotSolvedByKinductionPlainCorrectTrueCount) + \PdrInvKinductionKipdrdfTrueNotSolvedByKinductionPlainCorrectFalseCount - (32 * \PdrInvKinductionKipdrdfTrueNotSolvedByKinductionPlainWrongTrueCount) - (16 * \PdrInvKinductionKipdrdfTrueNotSolvedByKinductionPlainWrongFalseCount) \relax}
\edef\PdrInvPdrTrueNotSolvedByKinductionPlainScore{\the\numexpr (2 * \PdrInvPdrTrueNotSolvedByKinductionPlainCorrectTrueCount) + \PdrInvPdrTrueNotSolvedByKinductionPlainCorrectFalseCount - (32 * \PdrInvPdrTrueNotSolvedByKinductionPlainWrongTrueCount) - (16 * \PdrInvPdrTrueNotSolvedByKinductionPlainWrongFalseCount) \relax}
\edef\PdrInvOracleTrueNotSolvedByKinductionPlainScore{\the\numexpr (2 * \PdrInvOracleTrueNotSolvedByKinductionPlainCorrectTrueCount) + \PdrInvOracleTrueNotSolvedByKinductionPlainCorrectFalseCount - (32 * \PdrInvOracleTrueNotSolvedByKinductionPlainWrongTrueCount) - (16 * \PdrInvOracleTrueNotSolvedByKinductionPlainWrongFalseCount) \relax}
\edef\SeahornSeahornTrueNotSolvedByKinductionPlainScore{\the\numexpr (2 * \SeahornSeahornTrueNotSolvedByKinductionPlainCorrectTrueCount) + \SeahornSeahornTrueNotSolvedByKinductionPlainCorrectFalseCount - (32 * \SeahornSeahornTrueNotSolvedByKinductionPlainWrongTrueCount) - (16 * \SeahornSeahornTrueNotSolvedByKinductionPlainWrongFalseCount) \relax}
\edef\VvtCtigarTrueNotSolvedByKinductionPlainScore{\the\numexpr (2 * \VvtCtigarTrueNotSolvedByKinductionPlainCorrectTrueCount) + \VvtCtigarTrueNotSolvedByKinductionPlainCorrectFalseCount - (32 * \VvtCtigarTrueNotSolvedByKinductionPlainWrongTrueCount) - (16 * \VvtCtigarTrueNotSolvedByKinductionPlainWrongFalseCount) \relax}
\edef\VvtPortfolioTrueNotSolvedByKinductionPlainScore{\the\numexpr (2 * \VvtPortfolioTrueNotSolvedByKinductionPlainCorrectTrueCount) + \VvtPortfolioTrueNotSolvedByKinductionPlainCorrectFalseCount - (32 * \VvtPortfolioTrueNotSolvedByKinductionPlainWrongTrueCount) - (16 * \VvtPortfolioTrueNotSolvedByKinductionPlainWrongFalseCount) \relax}
\edef\PdrInvKinductionPlainTrueNotSolvedByKinductionPlainButKipdrScore{\the\numexpr (2 * \PdrInvKinductionPlainTrueNotSolvedByKinductionPlainButKipdrCorrectTrueCount) + \PdrInvKinductionPlainTrueNotSolvedByKinductionPlainButKipdrCorrectFalseCount - (32 * \PdrInvKinductionPlainTrueNotSolvedByKinductionPlainButKipdrWrongTrueCount) - (16 * \PdrInvKinductionPlainTrueNotSolvedByKinductionPlainButKipdrWrongFalseCount) \relax}
\edef\PdrInvKinductionDfStaticZeroZeroTTrueNotSolvedByKinductionPlainButKipdrScore{\the\numexpr (2 * \PdrInvKinductionDfStaticZeroZeroTTrueNotSolvedByKinductionPlainButKipdrCorrectTrueCount) + \PdrInvKinductionDfStaticZeroZeroTTrueNotSolvedByKinductionPlainButKipdrCorrectFalseCount - (32 * \PdrInvKinductionDfStaticZeroZeroTTrueNotSolvedByKinductionPlainButKipdrWrongTrueCount) - (16 * \PdrInvKinductionDfStaticZeroZeroTTrueNotSolvedByKinductionPlainButKipdrWrongFalseCount) \relax}
\edef\PdrInvKinductionDfStaticZeroOneTTTrueNotSolvedByKinductionPlainButKipdrScore{\the\numexpr (2 * \PdrInvKinductionDfStaticZeroOneTTTrueNotSolvedByKinductionPlainButKipdrCorrectTrueCount) + \PdrInvKinductionDfStaticZeroOneTTTrueNotSolvedByKinductionPlainButKipdrCorrectFalseCount - (32 * \PdrInvKinductionDfStaticZeroOneTTTrueNotSolvedByKinductionPlainButKipdrWrongTrueCount) - (16 * \PdrInvKinductionDfStaticZeroOneTTTrueNotSolvedByKinductionPlainButKipdrWrongFalseCount) \relax}
\edef\PdrInvKinductionDfStaticZeroOneTFTrueNotSolvedByKinductionPlainButKipdrScore{\the\numexpr (2 * \PdrInvKinductionDfStaticZeroOneTFTrueNotSolvedByKinductionPlainButKipdrCorrectTrueCount) + \PdrInvKinductionDfStaticZeroOneTFTrueNotSolvedByKinductionPlainButKipdrCorrectFalseCount - (32 * \PdrInvKinductionDfStaticZeroOneTFTrueNotSolvedByKinductionPlainButKipdrWrongTrueCount) - (16 * \PdrInvKinductionDfStaticZeroOneTFTrueNotSolvedByKinductionPlainButKipdrWrongFalseCount) \relax}
\edef\PdrInvKinductionDfStaticZeroTwoTTTrueNotSolvedByKinductionPlainButKipdrScore{\the\numexpr (2 * \PdrInvKinductionDfStaticZeroTwoTTTrueNotSolvedByKinductionPlainButKipdrCorrectTrueCount) + \PdrInvKinductionDfStaticZeroTwoTTTrueNotSolvedByKinductionPlainButKipdrCorrectFalseCount - (32 * \PdrInvKinductionDfStaticZeroTwoTTTrueNotSolvedByKinductionPlainButKipdrWrongTrueCount) - (16 * \PdrInvKinductionDfStaticZeroTwoTTTrueNotSolvedByKinductionPlainButKipdrWrongFalseCount) \relax}
\edef\PdrInvKinductionDfStaticZeroTwoTFTrueNotSolvedByKinductionPlainButKipdrScore{\the\numexpr (2 * \PdrInvKinductionDfStaticZeroTwoTFTrueNotSolvedByKinductionPlainButKipdrCorrectTrueCount) + \PdrInvKinductionDfStaticZeroTwoTFTrueNotSolvedByKinductionPlainButKipdrCorrectFalseCount - (32 * \PdrInvKinductionDfStaticZeroTwoTFTrueNotSolvedByKinductionPlainButKipdrWrongTrueCount) - (16 * \PdrInvKinductionDfStaticZeroTwoTFTrueNotSolvedByKinductionPlainButKipdrWrongFalseCount) \relax}
\edef\PdrInvKinductionDfStaticEightTwoTTrueNotSolvedByKinductionPlainButKipdrScore{\the\numexpr (2 * \PdrInvKinductionDfStaticEightTwoTTrueNotSolvedByKinductionPlainButKipdrCorrectTrueCount) + \PdrInvKinductionDfStaticEightTwoTTrueNotSolvedByKinductionPlainButKipdrCorrectFalseCount - (32 * \PdrInvKinductionDfStaticEightTwoTTrueNotSolvedByKinductionPlainButKipdrWrongTrueCount) - (16 * \PdrInvKinductionDfStaticEightTwoTTrueNotSolvedByKinductionPlainButKipdrWrongFalseCount) \relax}
\edef\PdrInvKinductionDfStaticSixteenTwoTTrueNotSolvedByKinductionPlainButKipdrScore{\the\numexpr (2 * \PdrInvKinductionDfStaticSixteenTwoTTrueNotSolvedByKinductionPlainButKipdrCorrectTrueCount) + \PdrInvKinductionDfStaticSixteenTwoTTrueNotSolvedByKinductionPlainButKipdrCorrectFalseCount - (32 * \PdrInvKinductionDfStaticSixteenTwoTTrueNotSolvedByKinductionPlainButKipdrWrongTrueCount) - (16 * \PdrInvKinductionDfStaticSixteenTwoTTrueNotSolvedByKinductionPlainButKipdrWrongFalseCount) \relax}
\edef\PdrInvKinductionDfStaticSixteenTwoFTrueNotSolvedByKinductionPlainButKipdrScore{\the\numexpr (2 * \PdrInvKinductionDfStaticSixteenTwoFTrueNotSolvedByKinductionPlainButKipdrCorrectTrueCount) + \PdrInvKinductionDfStaticSixteenTwoFTrueNotSolvedByKinductionPlainButKipdrCorrectFalseCount - (32 * \PdrInvKinductionDfStaticSixteenTwoFTrueNotSolvedByKinductionPlainButKipdrWrongTrueCount) - (16 * \PdrInvKinductionDfStaticSixteenTwoFTrueNotSolvedByKinductionPlainButKipdrWrongFalseCount) \relax}
\edef\PdrInvKinductionDfTrueNotSolvedByKinductionPlainButKipdrScore{\the\numexpr (2 * \PdrInvKinductionDfTrueNotSolvedByKinductionPlainButKipdrCorrectTrueCount) + \PdrInvKinductionDfTrueNotSolvedByKinductionPlainButKipdrCorrectFalseCount - (32 * \PdrInvKinductionDfTrueNotSolvedByKinductionPlainButKipdrWrongTrueCount) - (16 * \PdrInvKinductionDfTrueNotSolvedByKinductionPlainButKipdrWrongFalseCount) \relax}
\edef\PdrInvKinductionKipdrTrueNotSolvedByKinductionPlainButKipdrScore{\the\numexpr (2 * \PdrInvKinductionKipdrTrueNotSolvedByKinductionPlainButKipdrCorrectTrueCount) + \PdrInvKinductionKipdrTrueNotSolvedByKinductionPlainButKipdrCorrectFalseCount - (32 * \PdrInvKinductionKipdrTrueNotSolvedByKinductionPlainButKipdrWrongTrueCount) - (16 * \PdrInvKinductionKipdrTrueNotSolvedByKinductionPlainButKipdrWrongFalseCount) \relax}
\edef\PdrInvKinductionKipdrdfTrueNotSolvedByKinductionPlainButKipdrScore{\the\numexpr (2 * \PdrInvKinductionKipdrdfTrueNotSolvedByKinductionPlainButKipdrCorrectTrueCount) + \PdrInvKinductionKipdrdfTrueNotSolvedByKinductionPlainButKipdrCorrectFalseCount - (32 * \PdrInvKinductionKipdrdfTrueNotSolvedByKinductionPlainButKipdrWrongTrueCount) - (16 * \PdrInvKinductionKipdrdfTrueNotSolvedByKinductionPlainButKipdrWrongFalseCount) \relax}
\edef\PdrInvPdrTrueNotSolvedByKinductionPlainButKipdrScore{\the\numexpr (2 * \PdrInvPdrTrueNotSolvedByKinductionPlainButKipdrCorrectTrueCount) + \PdrInvPdrTrueNotSolvedByKinductionPlainButKipdrCorrectFalseCount - (32 * \PdrInvPdrTrueNotSolvedByKinductionPlainButKipdrWrongTrueCount) - (16 * \PdrInvPdrTrueNotSolvedByKinductionPlainButKipdrWrongFalseCount) \relax}
\edef\PdrInvOracleTrueNotSolvedByKinductionPlainButKipdrScore{\the\numexpr (2 * \PdrInvOracleTrueNotSolvedByKinductionPlainButKipdrCorrectTrueCount) + \PdrInvOracleTrueNotSolvedByKinductionPlainButKipdrCorrectFalseCount - (32 * \PdrInvOracleTrueNotSolvedByKinductionPlainButKipdrWrongTrueCount) - (16 * \PdrInvOracleTrueNotSolvedByKinductionPlainButKipdrWrongFalseCount) \relax}
\edef\SeahornSeahornTrueNotSolvedByKinductionPlainButKipdrScore{\the\numexpr (2 * \SeahornSeahornTrueNotSolvedByKinductionPlainButKipdrCorrectTrueCount) + \SeahornSeahornTrueNotSolvedByKinductionPlainButKipdrCorrectFalseCount - (32 * \SeahornSeahornTrueNotSolvedByKinductionPlainButKipdrWrongTrueCount) - (16 * \SeahornSeahornTrueNotSolvedByKinductionPlainButKipdrWrongFalseCount) \relax}
\edef\VvtCtigarTrueNotSolvedByKinductionPlainButKipdrScore{\the\numexpr (2 * \VvtCtigarTrueNotSolvedByKinductionPlainButKipdrCorrectTrueCount) + \VvtCtigarTrueNotSolvedByKinductionPlainButKipdrCorrectFalseCount - (32 * \VvtCtigarTrueNotSolvedByKinductionPlainButKipdrWrongTrueCount) - (16 * \VvtCtigarTrueNotSolvedByKinductionPlainButKipdrWrongFalseCount) \relax}
\edef\VvtPortfolioTrueNotSolvedByKinductionPlainButKipdrScore{\the\numexpr (2 * \VvtPortfolioTrueNotSolvedByKinductionPlainButKipdrCorrectTrueCount) + \VvtPortfolioTrueNotSolvedByKinductionPlainButKipdrCorrectFalseCount - (32 * \VvtPortfolioTrueNotSolvedByKinductionPlainButKipdrWrongTrueCount) - (16 * \VvtPortfolioTrueNotSolvedByKinductionPlainButKipdrWrongFalseCount) \relax}
\edef\PdrInvKinductionPlainTrueScore{\the\numexpr (2 * \PdrInvKinductionPlainCorrectTrueCount) - (32 * \PdrInvKinductionPlainWrongTrueCount) \relax}
\edef\PdrInvKinductionDfStaticZeroZeroTTrueScore{\the\numexpr (2 * \PdrInvKinductionDfStaticZeroZeroTCorrectTrueCount) - (32 * \PdrInvKinductionDfStaticZeroZeroTWrongTrueCount) \relax}
\edef\PdrInvKinductionDfStaticZeroOneTTTrueScore{\the\numexpr (2 * \PdrInvKinductionDfStaticZeroOneTTCorrectTrueCount) - (32 * \PdrInvKinductionDfStaticZeroOneTTWrongTrueCount) \relax}
\edef\PdrInvKinductionDfStaticZeroOneTFTrueScore{\the\numexpr (2 * \PdrInvKinductionDfStaticZeroOneTFCorrectTrueCount) - (32 * \PdrInvKinductionDfStaticZeroOneTFWrongTrueCount) \relax}
\edef\PdrInvKinductionDfStaticZeroTwoTTTrueScore{\the\numexpr (2 * \PdrInvKinductionDfStaticZeroTwoTTCorrectTrueCount) - (32 * \PdrInvKinductionDfStaticZeroTwoTTWrongTrueCount) \relax}
\edef\PdrInvKinductionDfStaticZeroTwoTFTrueScore{\the\numexpr (2 * \PdrInvKinductionDfStaticZeroTwoTFCorrectTrueCount) - (32 * \PdrInvKinductionDfStaticZeroTwoTFWrongTrueCount) \relax}
\edef\PdrInvKinductionDfStaticEightTwoTTrueScore{\the\numexpr (2 * \PdrInvKinductionDfStaticEightTwoTCorrectTrueCount) - (32 * \PdrInvKinductionDfStaticEightTwoTWrongTrueCount) \relax}
\edef\PdrInvKinductionDfStaticSixteenTwoTTrueScore{\the\numexpr (2 * \PdrInvKinductionDfStaticSixteenTwoTCorrectTrueCount) - (32 * \PdrInvKinductionDfStaticSixteenTwoTWrongTrueCount) \relax}
\edef\PdrInvKinductionDfStaticSixteenTwoFTrueScore{\the\numexpr (2 * \PdrInvKinductionDfStaticSixteenTwoFCorrectTrueCount) - (32 * \PdrInvKinductionDfStaticSixteenTwoFWrongTrueCount) \relax}
\edef\PdrInvKinductionDfTrueScore{\the\numexpr (2 * \PdrInvKinductionDfCorrectTrueCount) - (32 * \PdrInvKinductionDfWrongTrueCount) \relax}
\edef\PdrInvKinductionKipdrTrueScore{\the\numexpr (2 * \PdrInvKinductionKipdrCorrectTrueCount) - (32 * \PdrInvKinductionKipdrWrongTrueCount) \relax}
\edef\PdrInvKinductionKipdrdfTrueScore{\the\numexpr (2 * \PdrInvKinductionKipdrdfCorrectTrueCount) - (32 * \PdrInvKinductionKipdrdfWrongTrueCount) \relax}
\edef\PdrInvPdrTrueScore{\the\numexpr (2 * \PdrInvPdrCorrectTrueCount) - (32 * \PdrInvPdrWrongTrueCount) \relax}
\edef\PdrInvOracleTrueScore{\the\numexpr (2 * \PdrInvOracleCorrectTrueCount) - (32 * \PdrInvOracleWrongTrueCount) \relax}
\edef\SeahornSeahornTrueScore{\the\numexpr (2 * \SeahornSeahornCorrectTrueCount) - (32 * \SeahornSeahornWrongTrueCount) \relax}
\edef\VvtCtigarTrueScore{\the\numexpr (2 * \VvtCtigarCorrectTrueCount) - (32 * \VvtCtigarWrongTrueCount) \relax}
\edef\VvtPortfolioTrueScore{\the\numexpr (2 * \VvtPortfolioCorrectTrueCount) - (32 * \VvtPortfolioWrongTrueCount) \relax}
\edef\PdrInvKinductionPlainTrueNotSolvedByKinductionPlainTrueScore{\the\numexpr (2 * \PdrInvKinductionPlainTrueNotSolvedByKinductionPlainCorrectTrueCount) - (32 * \PdrInvKinductionPlainTrueNotSolvedByKinductionPlainWrongTrueCount) \relax}
\edef\PdrInvKinductionDfStaticZeroZeroTTrueNotSolvedByKinductionPlainTrueScore{\the\numexpr (2 * \PdrInvKinductionDfStaticZeroZeroTTrueNotSolvedByKinductionPlainCorrectTrueCount) - (32 * \PdrInvKinductionDfStaticZeroZeroTTrueNotSolvedByKinductionPlainWrongTrueCount) \relax}
\edef\PdrInvKinductionDfStaticZeroOneTTTrueNotSolvedByKinductionPlainTrueScore{\the\numexpr (2 * \PdrInvKinductionDfStaticZeroOneTTTrueNotSolvedByKinductionPlainCorrectTrueCount) - (32 * \PdrInvKinductionDfStaticZeroOneTTTrueNotSolvedByKinductionPlainWrongTrueCount) \relax}
\edef\PdrInvKinductionDfStaticZeroOneTFTrueNotSolvedByKinductionPlainTrueScore{\the\numexpr (2 * \PdrInvKinductionDfStaticZeroOneTFTrueNotSolvedByKinductionPlainCorrectTrueCount) - (32 * \PdrInvKinductionDfStaticZeroOneTFTrueNotSolvedByKinductionPlainWrongTrueCount) \relax}
\edef\PdrInvKinductionDfStaticZeroTwoTTTrueNotSolvedByKinductionPlainTrueScore{\the\numexpr (2 * \PdrInvKinductionDfStaticZeroTwoTTTrueNotSolvedByKinductionPlainCorrectTrueCount) - (32 * \PdrInvKinductionDfStaticZeroTwoTTTrueNotSolvedByKinductionPlainWrongTrueCount) \relax}
\edef\PdrInvKinductionDfStaticZeroTwoTFTrueNotSolvedByKinductionPlainTrueScore{\the\numexpr (2 * \PdrInvKinductionDfStaticZeroTwoTFTrueNotSolvedByKinductionPlainCorrectTrueCount) - (32 * \PdrInvKinductionDfStaticZeroTwoTFTrueNotSolvedByKinductionPlainWrongTrueCount) \relax}
\edef\PdrInvKinductionDfStaticEightTwoTTrueNotSolvedByKinductionPlainTrueScore{\the\numexpr (2 * \PdrInvKinductionDfStaticEightTwoTTrueNotSolvedByKinductionPlainCorrectTrueCount) - (32 * \PdrInvKinductionDfStaticEightTwoTTrueNotSolvedByKinductionPlainWrongTrueCount) \relax}
\edef\PdrInvKinductionDfStaticSixteenTwoTTrueNotSolvedByKinductionPlainTrueScore{\the\numexpr (2 * \PdrInvKinductionDfStaticSixteenTwoTTrueNotSolvedByKinductionPlainCorrectTrueCount) - (32 * \PdrInvKinductionDfStaticSixteenTwoTTrueNotSolvedByKinductionPlainWrongTrueCount) \relax}
\edef\PdrInvKinductionDfStaticSixteenTwoFTrueNotSolvedByKinductionPlainTrueScore{\the\numexpr (2 * \PdrInvKinductionDfStaticSixteenTwoFTrueNotSolvedByKinductionPlainCorrectTrueCount) - (32 * \PdrInvKinductionDfStaticSixteenTwoFTrueNotSolvedByKinductionPlainWrongTrueCount) \relax}
\edef\PdrInvKinductionDfTrueNotSolvedByKinductionPlainTrueScore{\the\numexpr (2 * \PdrInvKinductionDfTrueNotSolvedByKinductionPlainCorrectTrueCount) - (32 * \PdrInvKinductionDfTrueNotSolvedByKinductionPlainWrongTrueCount) \relax}
\edef\PdrInvKinductionKipdrTrueNotSolvedByKinductionPlainTrueScore{\the\numexpr (2 * \PdrInvKinductionKipdrTrueNotSolvedByKinductionPlainCorrectTrueCount) - (32 * \PdrInvKinductionKipdrTrueNotSolvedByKinductionPlainWrongTrueCount) \relax}
\edef\PdrInvKinductionKipdrdfTrueNotSolvedByKinductionPlainTrueScore{\the\numexpr (2 * \PdrInvKinductionKipdrdfTrueNotSolvedByKinductionPlainCorrectTrueCount) - (32 * \PdrInvKinductionKipdrdfTrueNotSolvedByKinductionPlainWrongTrueCount) \relax}
\edef\PdrInvPdrTrueNotSolvedByKinductionPlainTrueScore{\the\numexpr (2 * \PdrInvPdrTrueNotSolvedByKinductionPlainCorrectTrueCount) - (32 * \PdrInvPdrTrueNotSolvedByKinductionPlainWrongTrueCount) \relax}
\edef\PdrInvOracleTrueNotSolvedByKinductionPlainTrueScore{\the\numexpr (2 * \PdrInvOracleTrueNotSolvedByKinductionPlainCorrectTrueCount) - (32 * \PdrInvOracleTrueNotSolvedByKinductionPlainWrongTrueCount) \relax}
\edef\SeahornSeahornTrueNotSolvedByKinductionPlainTrueScore{\the\numexpr (2 * \SeahornSeahornTrueNotSolvedByKinductionPlainCorrectTrueCount) - (32 * \SeahornSeahornTrueNotSolvedByKinductionPlainWrongTrueCount) \relax}
\edef\VvtCtigarTrueNotSolvedByKinductionPlainTrueScore{\the\numexpr (2 * \VvtCtigarTrueNotSolvedByKinductionPlainCorrectTrueCount) - (32 * \VvtCtigarTrueNotSolvedByKinductionPlainWrongTrueCount) \relax}
\edef\VvtPortfolioTrueNotSolvedByKinductionPlainTrueScore{\the\numexpr (2 * \VvtPortfolioTrueNotSolvedByKinductionPlainCorrectTrueCount) - (32 * \VvtPortfolioTrueNotSolvedByKinductionPlainWrongTrueCount) \relax}
\edef\PdrInvKinductionPlainTrueNotSolvedByKinductionPlainButKipdrTrueScore{\the\numexpr (2 * \PdrInvKinductionPlainTrueNotSolvedByKinductionPlainButKipdrCorrectTrueCount) - (32 * \PdrInvKinductionPlainTrueNotSolvedByKinductionPlainButKipdrWrongTrueCount) \relax}
\edef\PdrInvKinductionDfStaticZeroZeroTTrueNotSolvedByKinductionPlainButKipdrTrueScore{\the\numexpr (2 * \PdrInvKinductionDfStaticZeroZeroTTrueNotSolvedByKinductionPlainButKipdrCorrectTrueCount) - (32 * \PdrInvKinductionDfStaticZeroZeroTTrueNotSolvedByKinductionPlainButKipdrWrongTrueCount) \relax}
\edef\PdrInvKinductionDfStaticZeroOneTTTrueNotSolvedByKinductionPlainButKipdrTrueScore{\the\numexpr (2 * \PdrInvKinductionDfStaticZeroOneTTTrueNotSolvedByKinductionPlainButKipdrCorrectTrueCount) - (32 * \PdrInvKinductionDfStaticZeroOneTTTrueNotSolvedByKinductionPlainButKipdrWrongTrueCount) \relax}
\edef\PdrInvKinductionDfStaticZeroOneTFTrueNotSolvedByKinductionPlainButKipdrTrueScore{\the\numexpr (2 * \PdrInvKinductionDfStaticZeroOneTFTrueNotSolvedByKinductionPlainButKipdrCorrectTrueCount) - (32 * \PdrInvKinductionDfStaticZeroOneTFTrueNotSolvedByKinductionPlainButKipdrWrongTrueCount) \relax}
\edef\PdrInvKinductionDfStaticZeroTwoTTTrueNotSolvedByKinductionPlainButKipdrTrueScore{\the\numexpr (2 * \PdrInvKinductionDfStaticZeroTwoTTTrueNotSolvedByKinductionPlainButKipdrCorrectTrueCount) - (32 * \PdrInvKinductionDfStaticZeroTwoTTTrueNotSolvedByKinductionPlainButKipdrWrongTrueCount) \relax}
\edef\PdrInvKinductionDfStaticZeroTwoTFTrueNotSolvedByKinductionPlainButKipdrTrueScore{\the\numexpr (2 * \PdrInvKinductionDfStaticZeroTwoTFTrueNotSolvedByKinductionPlainButKipdrCorrectTrueCount) - (32 * \PdrInvKinductionDfStaticZeroTwoTFTrueNotSolvedByKinductionPlainButKipdrWrongTrueCount) \relax}
\edef\PdrInvKinductionDfStaticEightTwoTTrueNotSolvedByKinductionPlainButKipdrTrueScore{\the\numexpr (2 * \PdrInvKinductionDfStaticEightTwoTTrueNotSolvedByKinductionPlainButKipdrCorrectTrueCount) - (32 * \PdrInvKinductionDfStaticEightTwoTTrueNotSolvedByKinductionPlainButKipdrWrongTrueCount) \relax}
\edef\PdrInvKinductionDfStaticSixteenTwoTTrueNotSolvedByKinductionPlainButKipdrTrueScore{\the\numexpr (2 * \PdrInvKinductionDfStaticSixteenTwoTTrueNotSolvedByKinductionPlainButKipdrCorrectTrueCount) - (32 * \PdrInvKinductionDfStaticSixteenTwoTTrueNotSolvedByKinductionPlainButKipdrWrongTrueCount) \relax}
\edef\PdrInvKinductionDfStaticSixteenTwoFTrueNotSolvedByKinductionPlainButKipdrTrueScore{\the\numexpr (2 * \PdrInvKinductionDfStaticSixteenTwoFTrueNotSolvedByKinductionPlainButKipdrCorrectTrueCount) - (32 * \PdrInvKinductionDfStaticSixteenTwoFTrueNotSolvedByKinductionPlainButKipdrWrongTrueCount) \relax}
\edef\PdrInvKinductionDfTrueNotSolvedByKinductionPlainButKipdrTrueScore{\the\numexpr (2 * \PdrInvKinductionDfTrueNotSolvedByKinductionPlainButKipdrCorrectTrueCount) - (32 * \PdrInvKinductionDfTrueNotSolvedByKinductionPlainButKipdrWrongTrueCount) \relax}
\edef\PdrInvKinductionKipdrTrueNotSolvedByKinductionPlainButKipdrTrueScore{\the\numexpr (2 * \PdrInvKinductionKipdrTrueNotSolvedByKinductionPlainButKipdrCorrectTrueCount) - (32 * \PdrInvKinductionKipdrTrueNotSolvedByKinductionPlainButKipdrWrongTrueCount) \relax}
\edef\PdrInvKinductionKipdrdfTrueNotSolvedByKinductionPlainButKipdrTrueScore{\the\numexpr (2 * \PdrInvKinductionKipdrdfTrueNotSolvedByKinductionPlainButKipdrCorrectTrueCount) - (32 * \PdrInvKinductionKipdrdfTrueNotSolvedByKinductionPlainButKipdrWrongTrueCount) \relax}
\edef\PdrInvPdrTrueNotSolvedByKinductionPlainButKipdrTrueScore{\the\numexpr (2 * \PdrInvPdrTrueNotSolvedByKinductionPlainButKipdrCorrectTrueCount) - (32 * \PdrInvPdrTrueNotSolvedByKinductionPlainButKipdrWrongTrueCount) \relax}
\edef\PdrInvOracleTrueNotSolvedByKinductionPlainButKipdrTrueScore{\the\numexpr (2 * \PdrInvOracleTrueNotSolvedByKinductionPlainButKipdrCorrectTrueCount) - (32 * \PdrInvOracleTrueNotSolvedByKinductionPlainButKipdrWrongTrueCount) \relax}
\edef\SeahornSeahornTrueNotSolvedByKinductionPlainButKipdrTrueScore{\the\numexpr (2 * \SeahornSeahornTrueNotSolvedByKinductionPlainButKipdrCorrectTrueCount) - (32 * \SeahornSeahornTrueNotSolvedByKinductionPlainButKipdrWrongTrueCount) \relax}
\edef\VvtCtigarTrueNotSolvedByKinductionPlainButKipdrTrueScore{\the\numexpr (2 * \VvtCtigarTrueNotSolvedByKinductionPlainButKipdrCorrectTrueCount) - (32 * \VvtCtigarTrueNotSolvedByKinductionPlainButKipdrWrongTrueCount) \relax}
\edef\VvtPortfolioTrueNotSolvedByKinductionPlainButKipdrTrueScore{\the\numexpr (2 * \VvtPortfolioTrueNotSolvedByKinductionPlainButKipdrCorrectTrueCount) - (32 * \VvtPortfolioTrueNotSolvedByKinductionPlainButKipdrWrongTrueCount) \relax}
\newcommand{\ntasks}{\num{5591}}
\newcommand{\ntruetasks}{\num{4134}}
\newcommand{\nfalsetasks}{\num{1457}}
\newcommand{\vvtctigarXForYLessThanOneSecond}{505}
\newcommand{\vvtportfolioXForYLessThanOneSecond}{499}
\newcommand{\seahornXForYLessThanOneSecond}{1808}
\newcommand{\vvtctigarScoreXForYLessThanOneSecond}{483}
\newcommand{\vvtportfolioScoreXForYLessThanOneSecond}{191}
\newcommand{\seahornScoreXForYLessThanOneSecond}{1}

%% file: evaluation/new/data-commands.tex
\providecommand\StoreBenchExecResult[7]{\expandafter\newcommand\csname#1#2#3#4#5#6\endcsname{#7}}%
\StoreBenchExecResult{SvcompNineteenCpaSeq}{SvCompPropReachsafetyReachsafetyLoops}{Total}{}{Count}{}{208}%
\StoreBenchExecResult{SvcompNineteenCpaSeq}{SvCompPropReachsafetyReachsafetyLoops}{Total}{}{Cputime}{}{74365.958562103}%
\StoreBenchExecResult{SvcompNineteenCpaSeq}{SvCompPropReachsafetyReachsafetyLoops}{Total}{}{Cputime}{Avg}{357.5286469331875}%
\StoreBenchExecResult{SvcompNineteenCpaSeq}{SvCompPropReachsafetyReachsafetyLoops}{Total}{}{Cputime}{Median}{57.950741370}%
\StoreBenchExecResult{SvcompNineteenCpaSeq}{SvCompPropReachsafetyReachsafetyLoops}{Total}{}{Cputime}{Min}{2.679883697}%
\StoreBenchExecResult{SvcompNineteenCpaSeq}{SvCompPropReachsafetyReachsafetyLoops}{Total}{}{Cputime}{Max}{968.261604986}%
\StoreBenchExecResult{SvcompNineteenCpaSeq}{SvCompPropReachsafetyReachsafetyLoops}{Total}{}{Cputime}{Stdev}{428.3528738023139583818347889}%
\StoreBenchExecResult{SvcompNineteenCpaSeq}{SvCompPropReachsafetyReachsafetyLoops}{Total}{}{Walltime}{}{49192.97356366922}%
\StoreBenchExecResult{SvcompNineteenCpaSeq}{SvCompPropReachsafetyReachsafetyLoops}{Total}{}{Walltime}{Avg}{236.5046805945635576923076923}%
\StoreBenchExecResult{SvcompNineteenCpaSeq}{SvCompPropReachsafetyReachsafetyLoops}{Total}{}{Walltime}{Median}{33.03025138375}%
\StoreBenchExecResult{SvcompNineteenCpaSeq}{SvCompPropReachsafetyReachsafetyLoops}{Total}{}{Walltime}{Min}{1.16090703011}%
\StoreBenchExecResult{SvcompNineteenCpaSeq}{SvCompPropReachsafetyReachsafetyLoops}{Total}{}{Walltime}{Max}{889.649432898}%
\StoreBenchExecResult{SvcompNineteenCpaSeq}{SvCompPropReachsafetyReachsafetyLoops}{Total}{}{Walltime}{Stdev}{282.8994011066922728192630208}%
\StoreBenchExecResult{SvcompNineteenCpaSeq}{SvCompPropReachsafetyReachsafetyLoops}{Correct}{}{Count}{}{138}%
\StoreBenchExecResult{SvcompNineteenCpaSeq}{SvCompPropReachsafetyReachsafetyLoops}{Correct}{}{Cputime}{}{7975.434018784}%
\StoreBenchExecResult{SvcompNineteenCpaSeq}{SvCompPropReachsafetyReachsafetyLoops}{Correct}{}{Cputime}{Avg}{57.79300013611594202898550725}%
\StoreBenchExecResult{SvcompNineteenCpaSeq}{SvCompPropReachsafetyReachsafetyLoops}{Correct}{}{Cputime}{Median}{6.0889731375}%
\StoreBenchExecResult{SvcompNineteenCpaSeq}{SvCompPropReachsafetyReachsafetyLoops}{Correct}{}{Cputime}{Min}{2.679883697}%
\StoreBenchExecResult{SvcompNineteenCpaSeq}{SvCompPropReachsafetyReachsafetyLoops}{Correct}{}{Cputime}{Max}{354.725600663}%
\StoreBenchExecResult{SvcompNineteenCpaSeq}{SvCompPropReachsafetyReachsafetyLoops}{Correct}{}{Cputime}{Stdev}{96.79454365539337375781236095}%
\StoreBenchExecResult{SvcompNineteenCpaSeq}{SvCompPropReachsafetyReachsafetyLoops}{Correct}{}{Walltime}{}{6393.09706139522}%
\StoreBenchExecResult{SvcompNineteenCpaSeq}{SvCompPropReachsafetyReachsafetyLoops}{Correct}{}{Walltime}{Avg}{46.32679029996536231884057971}%
\StoreBenchExecResult{SvcompNineteenCpaSeq}{SvCompPropReachsafetyReachsafetyLoops}{Correct}{}{Walltime}{Median}{2.382009983065}%
\StoreBenchExecResult{SvcompNineteenCpaSeq}{SvCompPropReachsafetyReachsafetyLoops}{Correct}{}{Walltime}{Min}{1.16090703011}%
\StoreBenchExecResult{SvcompNineteenCpaSeq}{SvCompPropReachsafetyReachsafetyLoops}{Correct}{}{Walltime}{Max}{335.368287086}%
\StoreBenchExecResult{SvcompNineteenCpaSeq}{SvCompPropReachsafetyReachsafetyLoops}{Correct}{}{Walltime}{Stdev}{86.32756562220665836021220219}%
\StoreBenchExecResult{SvcompNineteenCpaSeq}{SvCompPropReachsafetyReachsafetyLoops}{Correct}{False}{Count}{}{41}%
\StoreBenchExecResult{SvcompNineteenCpaSeq}{SvCompPropReachsafetyReachsafetyLoops}{Correct}{False}{Cputime}{}{880.626608870}%
\StoreBenchExecResult{SvcompNineteenCpaSeq}{SvCompPropReachsafetyReachsafetyLoops}{Correct}{False}{Cputime}{Avg}{21.47869777731707317073170732}%
\StoreBenchExecResult{SvcompNineteenCpaSeq}{SvCompPropReachsafetyReachsafetyLoops}{Correct}{False}{Cputime}{Median}{4.429281215}%
\StoreBenchExecResult{SvcompNineteenCpaSeq}{SvCompPropReachsafetyReachsafetyLoops}{Correct}{False}{Cputime}{Min}{3.338786131}%
\StoreBenchExecResult{SvcompNineteenCpaSeq}{SvCompPropReachsafetyReachsafetyLoops}{Correct}{False}{Cputime}{Max}{227.124716141}%
\StoreBenchExecResult{SvcompNineteenCpaSeq}{SvCompPropReachsafetyReachsafetyLoops}{Correct}{False}{Cputime}{Stdev}{48.95841931169601386852089643}%
\StoreBenchExecResult{SvcompNineteenCpaSeq}{SvCompPropReachsafetyReachsafetyLoops}{Correct}{False}{Walltime}{}{580.98698282258}%
\StoreBenchExecResult{SvcompNineteenCpaSeq}{SvCompPropReachsafetyReachsafetyLoops}{Correct}{False}{Walltime}{Avg}{14.17041421518487804878048780}%
\StoreBenchExecResult{SvcompNineteenCpaSeq}{SvCompPropReachsafetyReachsafetyLoops}{Correct}{False}{Walltime}{Median}{1.78107714653}%
\StoreBenchExecResult{SvcompNineteenCpaSeq}{SvCompPropReachsafetyReachsafetyLoops}{Correct}{False}{Walltime}{Min}{1.44648504257}%
\StoreBenchExecResult{SvcompNineteenCpaSeq}{SvCompPropReachsafetyReachsafetyLoops}{Correct}{False}{Walltime}{Max}{186.446482182}%
\StoreBenchExecResult{SvcompNineteenCpaSeq}{SvCompPropReachsafetyReachsafetyLoops}{Correct}{False}{Walltime}{Stdev}{40.40251776171038065642131031}%
\StoreBenchExecResult{SvcompNineteenCpaSeq}{SvCompPropReachsafetyReachsafetyLoops}{Wrong}{False}{Count}{}{0}%
\StoreBenchExecResult{SvcompNineteenCpaSeq}{SvCompPropReachsafetyReachsafetyLoops}{Wrong}{False}{Cputime}{}{0}%
\StoreBenchExecResult{SvcompNineteenCpaSeq}{SvCompPropReachsafetyReachsafetyLoops}{Wrong}{False}{Cputime}{Avg}{None}%
\StoreBenchExecResult{SvcompNineteenCpaSeq}{SvCompPropReachsafetyReachsafetyLoops}{Wrong}{False}{Cputime}{Median}{None}%
\StoreBenchExecResult{SvcompNineteenCpaSeq}{SvCompPropReachsafetyReachsafetyLoops}{Wrong}{False}{Cputime}{Min}{None}%
\StoreBenchExecResult{SvcompNineteenCpaSeq}{SvCompPropReachsafetyReachsafetyLoops}{Wrong}{False}{Cputime}{Max}{None}%
\StoreBenchExecResult{SvcompNineteenCpaSeq}{SvCompPropReachsafetyReachsafetyLoops}{Wrong}{False}{Cputime}{Stdev}{None}%
\StoreBenchExecResult{SvcompNineteenCpaSeq}{SvCompPropReachsafetyReachsafetyLoops}{Wrong}{False}{Walltime}{}{0}%
\StoreBenchExecResult{SvcompNineteenCpaSeq}{SvCompPropReachsafetyReachsafetyLoops}{Wrong}{False}{Walltime}{Avg}{None}%
\StoreBenchExecResult{SvcompNineteenCpaSeq}{SvCompPropReachsafetyReachsafetyLoops}{Wrong}{False}{Walltime}{Median}{None}%
\StoreBenchExecResult{SvcompNineteenCpaSeq}{SvCompPropReachsafetyReachsafetyLoops}{Wrong}{False}{Walltime}{Min}{None}%
\StoreBenchExecResult{SvcompNineteenCpaSeq}{SvCompPropReachsafetyReachsafetyLoops}{Wrong}{False}{Walltime}{Max}{None}%
\StoreBenchExecResult{SvcompNineteenCpaSeq}{SvCompPropReachsafetyReachsafetyLoops}{Wrong}{False}{Walltime}{Stdev}{None}%
\StoreBenchExecResult{SvcompNineteenCpaSeq}{SvCompPropReachsafetyReachsafetyLoops}{Correct}{True}{Count}{}{97}%
\StoreBenchExecResult{SvcompNineteenCpaSeq}{SvCompPropReachsafetyReachsafetyLoops}{Correct}{True}{Cputime}{}{7094.807409914}%
\StoreBenchExecResult{SvcompNineteenCpaSeq}{SvCompPropReachsafetyReachsafetyLoops}{Correct}{True}{Cputime}{Avg}{73.14234443210309278350515464}%
\StoreBenchExecResult{SvcompNineteenCpaSeq}{SvCompPropReachsafetyReachsafetyLoops}{Correct}{True}{Cputime}{Median}{10.544024715}%
\StoreBenchExecResult{SvcompNineteenCpaSeq}{SvCompPropReachsafetyReachsafetyLoops}{Correct}{True}{Cputime}{Min}{2.679883697}%
\StoreBenchExecResult{SvcompNineteenCpaSeq}{SvCompPropReachsafetyReachsafetyLoops}{Correct}{True}{Cputime}{Max}{354.725600663}%
\StoreBenchExecResult{SvcompNineteenCpaSeq}{SvCompPropReachsafetyReachsafetyLoops}{Correct}{True}{Cputime}{Stdev}{107.3462481230200712524185159}%
\StoreBenchExecResult{SvcompNineteenCpaSeq}{SvCompPropReachsafetyReachsafetyLoops}{Correct}{True}{Walltime}{}{5812.11007857264}%
\StoreBenchExecResult{SvcompNineteenCpaSeq}{SvCompPropReachsafetyReachsafetyLoops}{Correct}{True}{Walltime}{Avg}{59.91866060384164948453608247}%
\StoreBenchExecResult{SvcompNineteenCpaSeq}{SvCompPropReachsafetyReachsafetyLoops}{Correct}{True}{Walltime}{Median}{4.6923160553}%
\StoreBenchExecResult{SvcompNineteenCpaSeq}{SvCompPropReachsafetyReachsafetyLoops}{Correct}{True}{Walltime}{Min}{1.16090703011}%
\StoreBenchExecResult{SvcompNineteenCpaSeq}{SvCompPropReachsafetyReachsafetyLoops}{Correct}{True}{Walltime}{Max}{335.368287086}%
\StoreBenchExecResult{SvcompNineteenCpaSeq}{SvCompPropReachsafetyReachsafetyLoops}{Correct}{True}{Walltime}{Stdev}{96.38817533151535981884744195}%
\StoreBenchExecResult{SvcompNineteenCpaSeq}{SvCompPropReachsafetyReachsafetyLoops}{Wrong}{True}{Count}{}{0}%
\StoreBenchExecResult{SvcompNineteenCpaSeq}{SvCompPropReachsafetyReachsafetyLoops}{Wrong}{True}{Cputime}{}{0}%
\StoreBenchExecResult{SvcompNineteenCpaSeq}{SvCompPropReachsafetyReachsafetyLoops}{Wrong}{True}{Cputime}{Avg}{None}%
\StoreBenchExecResult{SvcompNineteenCpaSeq}{SvCompPropReachsafetyReachsafetyLoops}{Wrong}{True}{Cputime}{Median}{None}%
\StoreBenchExecResult{SvcompNineteenCpaSeq}{SvCompPropReachsafetyReachsafetyLoops}{Wrong}{True}{Cputime}{Min}{None}%
\StoreBenchExecResult{SvcompNineteenCpaSeq}{SvCompPropReachsafetyReachsafetyLoops}{Wrong}{True}{Cputime}{Max}{None}%
\StoreBenchExecResult{SvcompNineteenCpaSeq}{SvCompPropReachsafetyReachsafetyLoops}{Wrong}{True}{Cputime}{Stdev}{None}%
\StoreBenchExecResult{SvcompNineteenCpaSeq}{SvCompPropReachsafetyReachsafetyLoops}{Wrong}{True}{Walltime}{}{0}%
\StoreBenchExecResult{SvcompNineteenCpaSeq}{SvCompPropReachsafetyReachsafetyLoops}{Wrong}{True}{Walltime}{Avg}{None}%
\StoreBenchExecResult{SvcompNineteenCpaSeq}{SvCompPropReachsafetyReachsafetyLoops}{Wrong}{True}{Walltime}{Median}{None}%
\StoreBenchExecResult{SvcompNineteenCpaSeq}{SvCompPropReachsafetyReachsafetyLoops}{Wrong}{True}{Walltime}{Min}{None}%
\StoreBenchExecResult{SvcompNineteenCpaSeq}{SvCompPropReachsafetyReachsafetyLoops}{Wrong}{True}{Walltime}{Max}{None}%
\StoreBenchExecResult{SvcompNineteenCpaSeq}{SvCompPropReachsafetyReachsafetyLoops}{Wrong}{True}{Walltime}{Stdev}{None}%
\StoreBenchExecResult{SvcompNineteenCpaSeq}{SvCompPropReachsafetyReachsafetyLoops}{Error}{}{Count}{}{70}%
\StoreBenchExecResult{SvcompNineteenCpaSeq}{SvCompPropReachsafetyReachsafetyLoops}{Error}{}{Cputime}{}{66390.524543319}%
\StoreBenchExecResult{SvcompNineteenCpaSeq}{SvCompPropReachsafetyReachsafetyLoops}{Error}{}{Cputime}{Avg}{948.4360649045571428571428571}%
\StoreBenchExecResult{SvcompNineteenCpaSeq}{SvCompPropReachsafetyReachsafetyLoops}{Error}{}{Cputime}{Median}{960.5014380335}%
\StoreBenchExecResult{SvcompNineteenCpaSeq}{SvCompPropReachsafetyReachsafetyLoops}{Error}{}{Cputime}{Min}{902.711084404}%
\StoreBenchExecResult{SvcompNineteenCpaSeq}{SvCompPropReachsafetyReachsafetyLoops}{Error}{}{Cputime}{Max}{968.261604986}%
\StoreBenchExecResult{SvcompNineteenCpaSeq}{SvCompPropReachsafetyReachsafetyLoops}{Error}{}{Cputime}{Stdev}{21.39675052333134778070845257}%
\StoreBenchExecResult{SvcompNineteenCpaSeq}{SvCompPropReachsafetyReachsafetyLoops}{Error}{}{Walltime}{}{42799.876502274}%
\StoreBenchExecResult{SvcompNineteenCpaSeq}{SvCompPropReachsafetyReachsafetyLoops}{Error}{}{Walltime}{Avg}{611.4268071753428571428571429}%
\StoreBenchExecResult{SvcompNineteenCpaSeq}{SvCompPropReachsafetyReachsafetyLoops}{Error}{}{Walltime}{Median}{602.672841072}%
\StoreBenchExecResult{SvcompNineteenCpaSeq}{SvCompPropReachsafetyReachsafetyLoops}{Error}{}{Walltime}{Min}{452.466502905}%
\StoreBenchExecResult{SvcompNineteenCpaSeq}{SvCompPropReachsafetyReachsafetyLoops}{Error}{}{Walltime}{Max}{889.649432898}%
\StoreBenchExecResult{SvcompNineteenCpaSeq}{SvCompPropReachsafetyReachsafetyLoops}{Error}{}{Walltime}{Stdev}{106.0618343589715861606660099}%
\StoreBenchExecResult{SvcompNineteenCpaSeq}{SvCompPropReachsafetyReachsafetyLoops}{Error}{Timeout}{Count}{}{70}%
\StoreBenchExecResult{SvcompNineteenCpaSeq}{SvCompPropReachsafetyReachsafetyLoops}{Error}{Timeout}{Cputime}{}{66390.524543319}%
\StoreBenchExecResult{SvcompNineteenCpaSeq}{SvCompPropReachsafetyReachsafetyLoops}{Error}{Timeout}{Cputime}{Avg}{948.4360649045571428571428571}%
\StoreBenchExecResult{SvcompNineteenCpaSeq}{SvCompPropReachsafetyReachsafetyLoops}{Error}{Timeout}{Cputime}{Median}{960.5014380335}%
\StoreBenchExecResult{SvcompNineteenCpaSeq}{SvCompPropReachsafetyReachsafetyLoops}{Error}{Timeout}{Cputime}{Min}{902.711084404}%
\StoreBenchExecResult{SvcompNineteenCpaSeq}{SvCompPropReachsafetyReachsafetyLoops}{Error}{Timeout}{Cputime}{Max}{968.261604986}%
\StoreBenchExecResult{SvcompNineteenCpaSeq}{SvCompPropReachsafetyReachsafetyLoops}{Error}{Timeout}{Cputime}{Stdev}{21.39675052333134778070845257}%
\StoreBenchExecResult{SvcompNineteenCpaSeq}{SvCompPropReachsafetyReachsafetyLoops}{Error}{Timeout}{Walltime}{}{42799.876502274}%
\StoreBenchExecResult{SvcompNineteenCpaSeq}{SvCompPropReachsafetyReachsafetyLoops}{Error}{Timeout}{Walltime}{Avg}{611.4268071753428571428571429}%
\StoreBenchExecResult{SvcompNineteenCpaSeq}{SvCompPropReachsafetyReachsafetyLoops}{Error}{Timeout}{Walltime}{Median}{602.672841072}%
\StoreBenchExecResult{SvcompNineteenCpaSeq}{SvCompPropReachsafetyReachsafetyLoops}{Error}{Timeout}{Walltime}{Min}{452.466502905}%
\StoreBenchExecResult{SvcompNineteenCpaSeq}{SvCompPropReachsafetyReachsafetyLoops}{Error}{Timeout}{Walltime}{Max}{889.649432898}%
\StoreBenchExecResult{SvcompNineteenCpaSeq}{SvCompPropReachsafetyReachsafetyLoops}{Error}{Timeout}{Walltime}{Stdev}{106.0618343589715861606660099}%
\providecommand\StoreBenchExecResult[7]{\expandafter\newcommand\csname#1#2#3#4#5#6\endcsname{#7}}%
\StoreBenchExecResult{SvcompNineteenPdrInv}{KinductionDfBoxesEqModReachsafetyLoops}{Total}{}{Count}{}{208}%
\StoreBenchExecResult{SvcompNineteenPdrInv}{KinductionDfBoxesEqModReachsafetyLoops}{Total}{}{Cputime}{}{80403.930307667}%
\StoreBenchExecResult{SvcompNineteenPdrInv}{KinductionDfBoxesEqModReachsafetyLoops}{Total}{}{Cputime}{Avg}{386.5573572483990384615384615}%
\StoreBenchExecResult{SvcompNineteenPdrInv}{KinductionDfBoxesEqModReachsafetyLoops}{Total}{}{Cputime}{Median}{23.593262464}%
\StoreBenchExecResult{SvcompNineteenPdrInv}{KinductionDfBoxesEqModReachsafetyLoops}{Total}{}{Cputime}{Min}{4.066413271}%
\StoreBenchExecResult{SvcompNineteenPdrInv}{KinductionDfBoxesEqModReachsafetyLoops}{Total}{}{Cputime}{Max}{930.835396438}%
\StoreBenchExecResult{SvcompNineteenPdrInv}{KinductionDfBoxesEqModReachsafetyLoops}{Total}{}{Cputime}{Stdev}{437.7524946695465612673059377}%
\StoreBenchExecResult{SvcompNineteenPdrInv}{KinductionDfBoxesEqModReachsafetyLoops}{Total}{}{Walltime}{}{77163.1280155730346226}%
\StoreBenchExecResult{SvcompNineteenPdrInv}{KinductionDfBoxesEqModReachsafetyLoops}{Total}{}{Walltime}{Avg}{370.9765769979472818394230769}%
\StoreBenchExecResult{SvcompNineteenPdrInv}{KinductionDfBoxesEqModReachsafetyLoops}{Total}{}{Walltime}{Median}{14.371967222999956}%
\StoreBenchExecResult{SvcompNineteenPdrInv}{KinductionDfBoxesEqModReachsafetyLoops}{Total}{}{Walltime}{Min}{2.1501494709955296}%
\StoreBenchExecResult{SvcompNineteenPdrInv}{KinductionDfBoxesEqModReachsafetyLoops}{Total}{}{Walltime}{Max}{907.2582092200028}%
\StoreBenchExecResult{SvcompNineteenPdrInv}{KinductionDfBoxesEqModReachsafetyLoops}{Total}{}{Walltime}{Stdev}{425.2025573121333131358477347}%
\StoreBenchExecResult{SvcompNineteenPdrInv}{KinductionDfBoxesEqModReachsafetyLoops}{Correct}{}{Count}{}{114}%
\StoreBenchExecResult{SvcompNineteenPdrInv}{KinductionDfBoxesEqModReachsafetyLoops}{Correct}{}{Cputime}{}{1661.941283645}%
\StoreBenchExecResult{SvcompNineteenPdrInv}{KinductionDfBoxesEqModReachsafetyLoops}{Correct}{}{Cputime}{Avg}{14.57843231267543859649122807}%
\StoreBenchExecResult{SvcompNineteenPdrInv}{KinductionDfBoxesEqModReachsafetyLoops}{Correct}{}{Cputime}{Median}{5.346966384}%
\StoreBenchExecResult{SvcompNineteenPdrInv}{KinductionDfBoxesEqModReachsafetyLoops}{Correct}{}{Cputime}{Min}{4.066413271}%
\StoreBenchExecResult{SvcompNineteenPdrInv}{KinductionDfBoxesEqModReachsafetyLoops}{Correct}{}{Cputime}{Max}{527.221964746}%
\StoreBenchExecResult{SvcompNineteenPdrInv}{KinductionDfBoxesEqModReachsafetyLoops}{Correct}{}{Cputime}{Stdev}{49.25934406168683594846712382}%
\StoreBenchExecResult{SvcompNineteenPdrInv}{KinductionDfBoxesEqModReachsafetyLoops}{Correct}{}{Walltime}{}{1185.8075379359943301}%
\StoreBenchExecResult{SvcompNineteenPdrInv}{KinductionDfBoxesEqModReachsafetyLoops}{Correct}{}{Walltime}{Avg}{10.40182050821047657982456140}%
\StoreBenchExecResult{SvcompNineteenPdrInv}{KinductionDfBoxesEqModReachsafetyLoops}{Correct}{}{Walltime}{Median}{2.8554765095032053}%
\StoreBenchExecResult{SvcompNineteenPdrInv}{KinductionDfBoxesEqModReachsafetyLoops}{Correct}{}{Walltime}{Min}{2.1501494709955296}%
\StoreBenchExecResult{SvcompNineteenPdrInv}{KinductionDfBoxesEqModReachsafetyLoops}{Correct}{}{Walltime}{Max}{505.04579130800266}%
\StoreBenchExecResult{SvcompNineteenPdrInv}{KinductionDfBoxesEqModReachsafetyLoops}{Correct}{}{Walltime}{Stdev}{47.06286059595872297383318415}%
\StoreBenchExecResult{SvcompNineteenPdrInv}{KinductionDfBoxesEqModReachsafetyLoops}{Correct}{False}{Count}{}{37}%
\StoreBenchExecResult{SvcompNineteenPdrInv}{KinductionDfBoxesEqModReachsafetyLoops}{Correct}{False}{Cputime}{}{334.009267860}%
\StoreBenchExecResult{SvcompNineteenPdrInv}{KinductionDfBoxesEqModReachsafetyLoops}{Correct}{False}{Cputime}{Avg}{9.027277509729729729729729730}%
\StoreBenchExecResult{SvcompNineteenPdrInv}{KinductionDfBoxesEqModReachsafetyLoops}{Correct}{False}{Cputime}{Median}{5.346694713}%
\StoreBenchExecResult{SvcompNineteenPdrInv}{KinductionDfBoxesEqModReachsafetyLoops}{Correct}{False}{Cputime}{Min}{4.188246525}%
\StoreBenchExecResult{SvcompNineteenPdrInv}{KinductionDfBoxesEqModReachsafetyLoops}{Correct}{False}{Cputime}{Max}{36.236669294}%
\StoreBenchExecResult{SvcompNineteenPdrInv}{KinductionDfBoxesEqModReachsafetyLoops}{Correct}{False}{Cputime}{Stdev}{8.498595350420289838638756657}%
\StoreBenchExecResult{SvcompNineteenPdrInv}{KinductionDfBoxesEqModReachsafetyLoops}{Correct}{False}{Walltime}{}{197.3464839750085952}%
\StoreBenchExecResult{SvcompNineteenPdrInv}{KinductionDfBoxesEqModReachsafetyLoops}{Correct}{False}{Walltime}{Avg}{5.333688756081313383783783784}%
\StoreBenchExecResult{SvcompNineteenPdrInv}{KinductionDfBoxesEqModReachsafetyLoops}{Correct}{False}{Walltime}{Median}{2.857629789999919}%
\StoreBenchExecResult{SvcompNineteenPdrInv}{KinductionDfBoxesEqModReachsafetyLoops}{Correct}{False}{Walltime}{Min}{2.254211919993395}%
\StoreBenchExecResult{SvcompNineteenPdrInv}{KinductionDfBoxesEqModReachsafetyLoops}{Correct}{False}{Walltime}{Max}{24.414077905006707}%
\StoreBenchExecResult{SvcompNineteenPdrInv}{KinductionDfBoxesEqModReachsafetyLoops}{Correct}{False}{Walltime}{Stdev}{5.847352552579505234717030473}%
\StoreBenchExecResult{SvcompNineteenPdrInv}{KinductionDfBoxesEqModReachsafetyLoops}{Wrong}{False}{Count}{}{0}%
\StoreBenchExecResult{SvcompNineteenPdrInv}{KinductionDfBoxesEqModReachsafetyLoops}{Wrong}{False}{Cputime}{}{0}%
\StoreBenchExecResult{SvcompNineteenPdrInv}{KinductionDfBoxesEqModReachsafetyLoops}{Wrong}{False}{Cputime}{Avg}{None}%
\StoreBenchExecResult{SvcompNineteenPdrInv}{KinductionDfBoxesEqModReachsafetyLoops}{Wrong}{False}{Cputime}{Median}{None}%
\StoreBenchExecResult{SvcompNineteenPdrInv}{KinductionDfBoxesEqModReachsafetyLoops}{Wrong}{False}{Cputime}{Min}{None}%
\StoreBenchExecResult{SvcompNineteenPdrInv}{KinductionDfBoxesEqModReachsafetyLoops}{Wrong}{False}{Cputime}{Max}{None}%
\StoreBenchExecResult{SvcompNineteenPdrInv}{KinductionDfBoxesEqModReachsafetyLoops}{Wrong}{False}{Cputime}{Stdev}{None}%
\StoreBenchExecResult{SvcompNineteenPdrInv}{KinductionDfBoxesEqModReachsafetyLoops}{Wrong}{False}{Walltime}{}{0}%
\StoreBenchExecResult{SvcompNineteenPdrInv}{KinductionDfBoxesEqModReachsafetyLoops}{Wrong}{False}{Walltime}{Avg}{None}%
\StoreBenchExecResult{SvcompNineteenPdrInv}{KinductionDfBoxesEqModReachsafetyLoops}{Wrong}{False}{Walltime}{Median}{None}%
\StoreBenchExecResult{SvcompNineteenPdrInv}{KinductionDfBoxesEqModReachsafetyLoops}{Wrong}{False}{Walltime}{Min}{None}%
\StoreBenchExecResult{SvcompNineteenPdrInv}{KinductionDfBoxesEqModReachsafetyLoops}{Wrong}{False}{Walltime}{Max}{None}%
\StoreBenchExecResult{SvcompNineteenPdrInv}{KinductionDfBoxesEqModReachsafetyLoops}{Wrong}{False}{Walltime}{Stdev}{None}%
\StoreBenchExecResult{SvcompNineteenPdrInv}{KinductionDfBoxesEqModReachsafetyLoops}{Correct}{True}{Count}{}{77}%
\StoreBenchExecResult{SvcompNineteenPdrInv}{KinductionDfBoxesEqModReachsafetyLoops}{Correct}{True}{Cputime}{}{1327.932015785}%
\StoreBenchExecResult{SvcompNineteenPdrInv}{KinductionDfBoxesEqModReachsafetyLoops}{Correct}{True}{Cputime}{Avg}{17.24587033487012987012987013}%
\StoreBenchExecResult{SvcompNineteenPdrInv}{KinductionDfBoxesEqModReachsafetyLoops}{Correct}{True}{Cputime}{Median}{5.347238055}%
\StoreBenchExecResult{SvcompNineteenPdrInv}{KinductionDfBoxesEqModReachsafetyLoops}{Correct}{True}{Cputime}{Min}{4.066413271}%
\StoreBenchExecResult{SvcompNineteenPdrInv}{KinductionDfBoxesEqModReachsafetyLoops}{Correct}{True}{Cputime}{Max}{527.221964746}%
\StoreBenchExecResult{SvcompNineteenPdrInv}{KinductionDfBoxesEqModReachsafetyLoops}{Correct}{True}{Cputime}{Stdev}{59.46281755105663967262219871}%
\StoreBenchExecResult{SvcompNineteenPdrInv}{KinductionDfBoxesEqModReachsafetyLoops}{Correct}{True}{Walltime}{}{988.4610539609857349}%
\StoreBenchExecResult{SvcompNineteenPdrInv}{KinductionDfBoxesEqModReachsafetyLoops}{Correct}{True}{Walltime}{Avg}{12.83715654494786668701298701}%
\StoreBenchExecResult{SvcompNineteenPdrInv}{KinductionDfBoxesEqModReachsafetyLoops}{Correct}{True}{Walltime}{Median}{2.8533232290064916}%
\StoreBenchExecResult{SvcompNineteenPdrInv}{KinductionDfBoxesEqModReachsafetyLoops}{Correct}{True}{Walltime}{Min}{2.1501494709955296}%
\StoreBenchExecResult{SvcompNineteenPdrInv}{KinductionDfBoxesEqModReachsafetyLoops}{Correct}{True}{Walltime}{Max}{505.04579130800266}%
\StoreBenchExecResult{SvcompNineteenPdrInv}{KinductionDfBoxesEqModReachsafetyLoops}{Correct}{True}{Walltime}{Stdev}{56.96067464223034515604653697}%
\StoreBenchExecResult{SvcompNineteenPdrInv}{KinductionDfBoxesEqModReachsafetyLoops}{Wrong}{True}{Count}{}{0}%
\StoreBenchExecResult{SvcompNineteenPdrInv}{KinductionDfBoxesEqModReachsafetyLoops}{Wrong}{True}{Cputime}{}{0}%
\StoreBenchExecResult{SvcompNineteenPdrInv}{KinductionDfBoxesEqModReachsafetyLoops}{Wrong}{True}{Cputime}{Avg}{None}%
\StoreBenchExecResult{SvcompNineteenPdrInv}{KinductionDfBoxesEqModReachsafetyLoops}{Wrong}{True}{Cputime}{Median}{None}%
\StoreBenchExecResult{SvcompNineteenPdrInv}{KinductionDfBoxesEqModReachsafetyLoops}{Wrong}{True}{Cputime}{Min}{None}%
\StoreBenchExecResult{SvcompNineteenPdrInv}{KinductionDfBoxesEqModReachsafetyLoops}{Wrong}{True}{Cputime}{Max}{None}%
\StoreBenchExecResult{SvcompNineteenPdrInv}{KinductionDfBoxesEqModReachsafetyLoops}{Wrong}{True}{Cputime}{Stdev}{None}%
\StoreBenchExecResult{SvcompNineteenPdrInv}{KinductionDfBoxesEqModReachsafetyLoops}{Wrong}{True}{Walltime}{}{0}%
\StoreBenchExecResult{SvcompNineteenPdrInv}{KinductionDfBoxesEqModReachsafetyLoops}{Wrong}{True}{Walltime}{Avg}{None}%
\StoreBenchExecResult{SvcompNineteenPdrInv}{KinductionDfBoxesEqModReachsafetyLoops}{Wrong}{True}{Walltime}{Median}{None}%
\StoreBenchExecResult{SvcompNineteenPdrInv}{KinductionDfBoxesEqModReachsafetyLoops}{Wrong}{True}{Walltime}{Min}{None}%
\StoreBenchExecResult{SvcompNineteenPdrInv}{KinductionDfBoxesEqModReachsafetyLoops}{Wrong}{True}{Walltime}{Max}{None}%
\StoreBenchExecResult{SvcompNineteenPdrInv}{KinductionDfBoxesEqModReachsafetyLoops}{Wrong}{True}{Walltime}{Stdev}{None}%
\StoreBenchExecResult{SvcompNineteenPdrInv}{KinductionDfBoxesEqModReachsafetyLoops}{Error}{}{Count}{}{94}%
\StoreBenchExecResult{SvcompNineteenPdrInv}{KinductionDfBoxesEqModReachsafetyLoops}{Error}{}{Cputime}{}{78741.989024022}%
\StoreBenchExecResult{SvcompNineteenPdrInv}{KinductionDfBoxesEqModReachsafetyLoops}{Error}{}{Cputime}{Avg}{837.6807342981063829787234043}%
\StoreBenchExecResult{SvcompNineteenPdrInv}{KinductionDfBoxesEqModReachsafetyLoops}{Error}{}{Cputime}{Median}{903.4126658645}%
\StoreBenchExecResult{SvcompNineteenPdrInv}{KinductionDfBoxesEqModReachsafetyLoops}{Error}{}{Cputime}{Min}{7.757978743}%
\StoreBenchExecResult{SvcompNineteenPdrInv}{KinductionDfBoxesEqModReachsafetyLoops}{Error}{}{Cputime}{Max}{930.835396438}%
\StoreBenchExecResult{SvcompNineteenPdrInv}{KinductionDfBoxesEqModReachsafetyLoops}{Error}{}{Cputime}{Stdev}{223.0757756726602031915287418}%
\StoreBenchExecResult{SvcompNineteenPdrInv}{KinductionDfBoxesEqModReachsafetyLoops}{Error}{}{Walltime}{}{75977.3204776370402925}%
\StoreBenchExecResult{SvcompNineteenPdrInv}{KinductionDfBoxesEqModReachsafetyLoops}{Error}{}{Walltime}{Avg}{808.2693667833727690691489362}%
\StoreBenchExecResult{SvcompNineteenPdrInv}{KinductionDfBoxesEqModReachsafetyLoops}{Error}{}{Walltime}{Median}{884.6063170054986}%
\StoreBenchExecResult{SvcompNineteenPdrInv}{KinductionDfBoxesEqModReachsafetyLoops}{Error}{}{Walltime}{Min}{4.0902234530076385}%
\StoreBenchExecResult{SvcompNineteenPdrInv}{KinductionDfBoxesEqModReachsafetyLoops}{Error}{}{Walltime}{Max}{907.2582092200028}%
\StoreBenchExecResult{SvcompNineteenPdrInv}{KinductionDfBoxesEqModReachsafetyLoops}{Error}{}{Walltime}{Stdev}{220.1681763979489351458462001}%
\StoreBenchExecResult{SvcompNineteenPdrInv}{KinductionDfBoxesEqModReachsafetyLoops}{Error}{Error}{Count}{}{5}%
\StoreBenchExecResult{SvcompNineteenPdrInv}{KinductionDfBoxesEqModReachsafetyLoops}{Error}{Error}{Cputime}{}{40.999137704}%
\StoreBenchExecResult{SvcompNineteenPdrInv}{KinductionDfBoxesEqModReachsafetyLoops}{Error}{Error}{Cputime}{Avg}{8.1998275408}%
\StoreBenchExecResult{SvcompNineteenPdrInv}{KinductionDfBoxesEqModReachsafetyLoops}{Error}{Error}{Cputime}{Median}{8.100824508}%
\StoreBenchExecResult{SvcompNineteenPdrInv}{KinductionDfBoxesEqModReachsafetyLoops}{Error}{Error}{Cputime}{Min}{7.757978743}%
\StoreBenchExecResult{SvcompNineteenPdrInv}{KinductionDfBoxesEqModReachsafetyLoops}{Error}{Error}{Cputime}{Max}{9.031081433}%
\StoreBenchExecResult{SvcompNineteenPdrInv}{KinductionDfBoxesEqModReachsafetyLoops}{Error}{Error}{Cputime}{Stdev}{0.4371019442375521419593240014}%
\StoreBenchExecResult{SvcompNineteenPdrInv}{KinductionDfBoxesEqModReachsafetyLoops}{Error}{Error}{Walltime}{}{21.4500404480058925}%
\StoreBenchExecResult{SvcompNineteenPdrInv}{KinductionDfBoxesEqModReachsafetyLoops}{Error}{Error}{Walltime}{Avg}{4.2900080896011785}%
\StoreBenchExecResult{SvcompNineteenPdrInv}{KinductionDfBoxesEqModReachsafetyLoops}{Error}{Error}{Walltime}{Median}{4.2497341910057}%
\StoreBenchExecResult{SvcompNineteenPdrInv}{KinductionDfBoxesEqModReachsafetyLoops}{Error}{Error}{Walltime}{Min}{4.0902234530076385}%
\StoreBenchExecResult{SvcompNineteenPdrInv}{KinductionDfBoxesEqModReachsafetyLoops}{Error}{Error}{Walltime}{Max}{4.709829411993269}%
\StoreBenchExecResult{SvcompNineteenPdrInv}{KinductionDfBoxesEqModReachsafetyLoops}{Error}{Error}{Walltime}{Stdev}{0.2189805165078343994103620851}%
\StoreBenchExecResult{SvcompNineteenPdrInv}{KinductionDfBoxesEqModReachsafetyLoops}{Error}{OutOfMemory}{Count}{}{10}%
\StoreBenchExecResult{SvcompNineteenPdrInv}{KinductionDfBoxesEqModReachsafetyLoops}{Error}{OutOfMemory}{Cputime}{}{7101.019708811}%
\StoreBenchExecResult{SvcompNineteenPdrInv}{KinductionDfBoxesEqModReachsafetyLoops}{Error}{OutOfMemory}{Cputime}{Avg}{710.1019708811}%
\StoreBenchExecResult{SvcompNineteenPdrInv}{KinductionDfBoxesEqModReachsafetyLoops}{Error}{OutOfMemory}{Cputime}{Median}{838.3914564525}%
\StoreBenchExecResult{SvcompNineteenPdrInv}{KinductionDfBoxesEqModReachsafetyLoops}{Error}{OutOfMemory}{Cputime}{Min}{175.73474717}%
\StoreBenchExecResult{SvcompNineteenPdrInv}{KinductionDfBoxesEqModReachsafetyLoops}{Error}{OutOfMemory}{Cputime}{Max}{885.897111722}%
\StoreBenchExecResult{SvcompNineteenPdrInv}{KinductionDfBoxesEqModReachsafetyLoops}{Error}{OutOfMemory}{Cputime}{Stdev}{264.5996682857255054286708809}%
\StoreBenchExecResult{SvcompNineteenPdrInv}{KinductionDfBoxesEqModReachsafetyLoops}{Error}{OutOfMemory}{Walltime}{}{6907.36267882400710}%
\StoreBenchExecResult{SvcompNineteenPdrInv}{KinductionDfBoxesEqModReachsafetyLoops}{Error}{OutOfMemory}{Walltime}{Avg}{690.73626788240071}%
\StoreBenchExecResult{SvcompNineteenPdrInv}{KinductionDfBoxesEqModReachsafetyLoops}{Error}{OutOfMemory}{Walltime}{Median}{817.4401663210010}%
\StoreBenchExecResult{SvcompNineteenPdrInv}{KinductionDfBoxesEqModReachsafetyLoops}{Error}{OutOfMemory}{Walltime}{Min}{165.32216841200716}%
\StoreBenchExecResult{SvcompNineteenPdrInv}{KinductionDfBoxesEqModReachsafetyLoops}{Error}{OutOfMemory}{Walltime}{Max}{862.7137118800019}%
\StoreBenchExecResult{SvcompNineteenPdrInv}{KinductionDfBoxesEqModReachsafetyLoops}{Error}{OutOfMemory}{Walltime}{Stdev}{260.0564997312307402181818412}%
\StoreBenchExecResult{SvcompNineteenPdrInv}{KinductionDfBoxesEqModReachsafetyLoops}{Error}{Timeout}{Count}{}{79}%
\StoreBenchExecResult{SvcompNineteenPdrInv}{KinductionDfBoxesEqModReachsafetyLoops}{Error}{Timeout}{Cputime}{}{71599.970177507}%
\StoreBenchExecResult{SvcompNineteenPdrInv}{KinductionDfBoxesEqModReachsafetyLoops}{Error}{Timeout}{Cputime}{Avg}{906.3287364241392405063291139}%
\StoreBenchExecResult{SvcompNineteenPdrInv}{KinductionDfBoxesEqModReachsafetyLoops}{Error}{Timeout}{Cputime}{Median}{904.234445108}%
\StoreBenchExecResult{SvcompNineteenPdrInv}{KinductionDfBoxesEqModReachsafetyLoops}{Error}{Timeout}{Cputime}{Min}{901.37807087}%
\StoreBenchExecResult{SvcompNineteenPdrInv}{KinductionDfBoxesEqModReachsafetyLoops}{Error}{Timeout}{Cputime}{Max}{930.835396438}%
\StoreBenchExecResult{SvcompNineteenPdrInv}{KinductionDfBoxesEqModReachsafetyLoops}{Error}{Timeout}{Cputime}{Stdev}{5.424941910590997140420754333}%
\StoreBenchExecResult{SvcompNineteenPdrInv}{KinductionDfBoxesEqModReachsafetyLoops}{Error}{Timeout}{Walltime}{}{69048.5077583650273}%
\StoreBenchExecResult{SvcompNineteenPdrInv}{KinductionDfBoxesEqModReachsafetyLoops}{Error}{Timeout}{Walltime}{Avg}{874.0317437767724974683544304}%
\StoreBenchExecResult{SvcompNineteenPdrInv}{KinductionDfBoxesEqModReachsafetyLoops}{Error}{Timeout}{Walltime}{Median}{885.8105239459983}%
\StoreBenchExecResult{SvcompNineteenPdrInv}{KinductionDfBoxesEqModReachsafetyLoops}{Error}{Timeout}{Walltime}{Min}{632.4179816800024}%
\StoreBenchExecResult{SvcompNineteenPdrInv}{KinductionDfBoxesEqModReachsafetyLoops}{Error}{Timeout}{Walltime}{Max}{907.2582092200028}%
\StoreBenchExecResult{SvcompNineteenPdrInv}{KinductionDfBoxesEqModReachsafetyLoops}{Error}{Timeout}{Walltime}{Stdev}{46.19160655352794710974224789}%
\providecommand\StoreBenchExecResult[7]{\expandafter\newcommand\csname#1#2#3#4#5#6\endcsname{#7}}%
\StoreBenchExecResult{SvcompNineteenPdrInv}{KinductionDfBoxesEqReachsafetyLoops}{Total}{}{Count}{}{208}%
\StoreBenchExecResult{SvcompNineteenPdrInv}{KinductionDfBoxesEqReachsafetyLoops}{Total}{}{Cputime}{}{85902.776759269}%
\StoreBenchExecResult{SvcompNineteenPdrInv}{KinductionDfBoxesEqReachsafetyLoops}{Total}{}{Cputime}{Avg}{412.9941190349471153846153846}%
\StoreBenchExecResult{SvcompNineteenPdrInv}{KinductionDfBoxesEqReachsafetyLoops}{Total}{}{Cputime}{Median}{31.997857109}%
\StoreBenchExecResult{SvcompNineteenPdrInv}{KinductionDfBoxesEqReachsafetyLoops}{Total}{}{Cputime}{Min}{4.120633785}%
\StoreBenchExecResult{SvcompNineteenPdrInv}{KinductionDfBoxesEqReachsafetyLoops}{Total}{}{Cputime}{Max}{940.707496472}%
\StoreBenchExecResult{SvcompNineteenPdrInv}{KinductionDfBoxesEqReachsafetyLoops}{Total}{}{Cputime}{Stdev}{441.6962509137787604097246326}%
\StoreBenchExecResult{SvcompNineteenPdrInv}{KinductionDfBoxesEqReachsafetyLoops}{Total}{}{Walltime}{}{82581.6757002659722348}%
\StoreBenchExecResult{SvcompNineteenPdrInv}{KinductionDfBoxesEqReachsafetyLoops}{Total}{}{Walltime}{Avg}{397.0272870205094818980769231}%
\StoreBenchExecResult{SvcompNineteenPdrInv}{KinductionDfBoxesEqReachsafetyLoops}{Total}{}{Walltime}{Median}{19.0139500215009325}%
\StoreBenchExecResult{SvcompNineteenPdrInv}{KinductionDfBoxesEqReachsafetyLoops}{Total}{}{Walltime}{Min}{2.214154396991944}%
\StoreBenchExecResult{SvcompNineteenPdrInv}{KinductionDfBoxesEqReachsafetyLoops}{Total}{}{Walltime}{Max}{915.5737941320112}%
\StoreBenchExecResult{SvcompNineteenPdrInv}{KinductionDfBoxesEqReachsafetyLoops}{Total}{}{Walltime}{Stdev}{429.4303881181196322526052596}%
\StoreBenchExecResult{SvcompNineteenPdrInv}{KinductionDfBoxesEqReachsafetyLoops}{Correct}{}{Count}{}{108}%
\StoreBenchExecResult{SvcompNineteenPdrInv}{KinductionDfBoxesEqReachsafetyLoops}{Correct}{}{Cputime}{}{1626.258783397}%
\StoreBenchExecResult{SvcompNineteenPdrInv}{KinductionDfBoxesEqReachsafetyLoops}{Correct}{}{Cputime}{Avg}{15.05795169812037037037037037}%
\StoreBenchExecResult{SvcompNineteenPdrInv}{KinductionDfBoxesEqReachsafetyLoops}{Correct}{}{Cputime}{Median}{5.459482723}%
\StoreBenchExecResult{SvcompNineteenPdrInv}{KinductionDfBoxesEqReachsafetyLoops}{Correct}{}{Cputime}{Min}{4.120633785}%
\StoreBenchExecResult{SvcompNineteenPdrInv}{KinductionDfBoxesEqReachsafetyLoops}{Correct}{}{Cputime}{Max}{526.675767489}%
\StoreBenchExecResult{SvcompNineteenPdrInv}{KinductionDfBoxesEqReachsafetyLoops}{Correct}{}{Cputime}{Stdev}{50.48239272680603772629328231}%
\StoreBenchExecResult{SvcompNineteenPdrInv}{KinductionDfBoxesEqReachsafetyLoops}{Correct}{}{Walltime}{}{1169.7543016089912508}%
\StoreBenchExecResult{SvcompNineteenPdrInv}{KinductionDfBoxesEqReachsafetyLoops}{Correct}{}{Walltime}{Avg}{10.83105834823140047037037037}%
\StoreBenchExecResult{SvcompNineteenPdrInv}{KinductionDfBoxesEqReachsafetyLoops}{Correct}{}{Walltime}{Median}{2.9121364310049102}%
\StoreBenchExecResult{SvcompNineteenPdrInv}{KinductionDfBoxesEqReachsafetyLoops}{Correct}{}{Walltime}{Min}{2.214154396991944}%
\StoreBenchExecResult{SvcompNineteenPdrInv}{KinductionDfBoxesEqReachsafetyLoops}{Correct}{}{Walltime}{Max}{504.7137087799929}%
\StoreBenchExecResult{SvcompNineteenPdrInv}{KinductionDfBoxesEqReachsafetyLoops}{Correct}{}{Walltime}{Stdev}{48.27824102559682685986761927}%
\StoreBenchExecResult{SvcompNineteenPdrInv}{KinductionDfBoxesEqReachsafetyLoops}{Correct}{False}{Count}{}{37}%
\StoreBenchExecResult{SvcompNineteenPdrInv}{KinductionDfBoxesEqReachsafetyLoops}{Correct}{False}{Cputime}{}{330.642588287}%
\StoreBenchExecResult{SvcompNineteenPdrInv}{KinductionDfBoxesEqReachsafetyLoops}{Correct}{False}{Cputime}{Avg}{8.936286169918918918918918919}%
\StoreBenchExecResult{SvcompNineteenPdrInv}{KinductionDfBoxesEqReachsafetyLoops}{Correct}{False}{Cputime}{Median}{5.039301792}%
\StoreBenchExecResult{SvcompNineteenPdrInv}{KinductionDfBoxesEqReachsafetyLoops}{Correct}{False}{Cputime}{Min}{4.238328599}%
\StoreBenchExecResult{SvcompNineteenPdrInv}{KinductionDfBoxesEqReachsafetyLoops}{Correct}{False}{Cputime}{Max}{36.380274732}%
\StoreBenchExecResult{SvcompNineteenPdrInv}{KinductionDfBoxesEqReachsafetyLoops}{Correct}{False}{Cputime}{Stdev}{8.575057880508487181069380008}%
\StoreBenchExecResult{SvcompNineteenPdrInv}{KinductionDfBoxesEqReachsafetyLoops}{Correct}{False}{Walltime}{}{196.9890340779820715}%
\StoreBenchExecResult{SvcompNineteenPdrInv}{KinductionDfBoxesEqReachsafetyLoops}{Correct}{False}{Walltime}{Avg}{5.3240279480535695}%
\StoreBenchExecResult{SvcompNineteenPdrInv}{KinductionDfBoxesEqReachsafetyLoops}{Correct}{False}{Walltime}{Median}{2.698206286993809}%
\StoreBenchExecResult{SvcompNineteenPdrInv}{KinductionDfBoxesEqReachsafetyLoops}{Correct}{False}{Walltime}{Min}{2.274272816008306}%
\StoreBenchExecResult{SvcompNineteenPdrInv}{KinductionDfBoxesEqReachsafetyLoops}{Correct}{False}{Walltime}{Max}{24.56515063099505}%
\StoreBenchExecResult{SvcompNineteenPdrInv}{KinductionDfBoxesEqReachsafetyLoops}{Correct}{False}{Walltime}{Stdev}{5.980578632494233216170339560}%
\StoreBenchExecResult{SvcompNineteenPdrInv}{KinductionDfBoxesEqReachsafetyLoops}{Wrong}{False}{Count}{}{0}%
\StoreBenchExecResult{SvcompNineteenPdrInv}{KinductionDfBoxesEqReachsafetyLoops}{Wrong}{False}{Cputime}{}{0}%
\StoreBenchExecResult{SvcompNineteenPdrInv}{KinductionDfBoxesEqReachsafetyLoops}{Wrong}{False}{Cputime}{Avg}{None}%
\StoreBenchExecResult{SvcompNineteenPdrInv}{KinductionDfBoxesEqReachsafetyLoops}{Wrong}{False}{Cputime}{Median}{None}%
\StoreBenchExecResult{SvcompNineteenPdrInv}{KinductionDfBoxesEqReachsafetyLoops}{Wrong}{False}{Cputime}{Min}{None}%
\StoreBenchExecResult{SvcompNineteenPdrInv}{KinductionDfBoxesEqReachsafetyLoops}{Wrong}{False}{Cputime}{Max}{None}%
\StoreBenchExecResult{SvcompNineteenPdrInv}{KinductionDfBoxesEqReachsafetyLoops}{Wrong}{False}{Cputime}{Stdev}{None}%
\StoreBenchExecResult{SvcompNineteenPdrInv}{KinductionDfBoxesEqReachsafetyLoops}{Wrong}{False}{Walltime}{}{0}%
\StoreBenchExecResult{SvcompNineteenPdrInv}{KinductionDfBoxesEqReachsafetyLoops}{Wrong}{False}{Walltime}{Avg}{None}%
\StoreBenchExecResult{SvcompNineteenPdrInv}{KinductionDfBoxesEqReachsafetyLoops}{Wrong}{False}{Walltime}{Median}{None}%
\StoreBenchExecResult{SvcompNineteenPdrInv}{KinductionDfBoxesEqReachsafetyLoops}{Wrong}{False}{Walltime}{Min}{None}%
\StoreBenchExecResult{SvcompNineteenPdrInv}{KinductionDfBoxesEqReachsafetyLoops}{Wrong}{False}{Walltime}{Max}{None}%
\StoreBenchExecResult{SvcompNineteenPdrInv}{KinductionDfBoxesEqReachsafetyLoops}{Wrong}{False}{Walltime}{Stdev}{None}%
\StoreBenchExecResult{SvcompNineteenPdrInv}{KinductionDfBoxesEqReachsafetyLoops}{Correct}{True}{Count}{}{71}%
\StoreBenchExecResult{SvcompNineteenPdrInv}{KinductionDfBoxesEqReachsafetyLoops}{Correct}{True}{Cputime}{}{1295.616195110}%
\StoreBenchExecResult{SvcompNineteenPdrInv}{KinductionDfBoxesEqReachsafetyLoops}{Correct}{True}{Cputime}{Avg}{18.24811542408450704225352113}%
\StoreBenchExecResult{SvcompNineteenPdrInv}{KinductionDfBoxesEqReachsafetyLoops}{Correct}{True}{Cputime}{Median}{5.770069675}%
\StoreBenchExecResult{SvcompNineteenPdrInv}{KinductionDfBoxesEqReachsafetyLoops}{Correct}{True}{Cputime}{Min}{4.120633785}%
\StoreBenchExecResult{SvcompNineteenPdrInv}{KinductionDfBoxesEqReachsafetyLoops}{Correct}{True}{Cputime}{Max}{526.675767489}%
\StoreBenchExecResult{SvcompNineteenPdrInv}{KinductionDfBoxesEqReachsafetyLoops}{Correct}{True}{Cputime}{Stdev}{61.71323493989286601688768214}%
\StoreBenchExecResult{SvcompNineteenPdrInv}{KinductionDfBoxesEqReachsafetyLoops}{Correct}{True}{Walltime}{}{972.7652675310091793}%
\StoreBenchExecResult{SvcompNineteenPdrInv}{KinductionDfBoxesEqReachsafetyLoops}{Correct}{True}{Walltime}{Avg}{13.70091926100012928591549296}%
\StoreBenchExecResult{SvcompNineteenPdrInv}{KinductionDfBoxesEqReachsafetyLoops}{Correct}{True}{Walltime}{Median}{3.045757558007608}%
\StoreBenchExecResult{SvcompNineteenPdrInv}{KinductionDfBoxesEqReachsafetyLoops}{Correct}{True}{Walltime}{Min}{2.214154396991944}%
\StoreBenchExecResult{SvcompNineteenPdrInv}{KinductionDfBoxesEqReachsafetyLoops}{Correct}{True}{Walltime}{Max}{504.7137087799929}%
\StoreBenchExecResult{SvcompNineteenPdrInv}{KinductionDfBoxesEqReachsafetyLoops}{Correct}{True}{Walltime}{Stdev}{59.18399298446035764757919632}%
\StoreBenchExecResult{SvcompNineteenPdrInv}{KinductionDfBoxesEqReachsafetyLoops}{Wrong}{True}{Count}{}{0}%
\StoreBenchExecResult{SvcompNineteenPdrInv}{KinductionDfBoxesEqReachsafetyLoops}{Wrong}{True}{Cputime}{}{0}%
\StoreBenchExecResult{SvcompNineteenPdrInv}{KinductionDfBoxesEqReachsafetyLoops}{Wrong}{True}{Cputime}{Avg}{None}%
\StoreBenchExecResult{SvcompNineteenPdrInv}{KinductionDfBoxesEqReachsafetyLoops}{Wrong}{True}{Cputime}{Median}{None}%
\StoreBenchExecResult{SvcompNineteenPdrInv}{KinductionDfBoxesEqReachsafetyLoops}{Wrong}{True}{Cputime}{Min}{None}%
\StoreBenchExecResult{SvcompNineteenPdrInv}{KinductionDfBoxesEqReachsafetyLoops}{Wrong}{True}{Cputime}{Max}{None}%
\StoreBenchExecResult{SvcompNineteenPdrInv}{KinductionDfBoxesEqReachsafetyLoops}{Wrong}{True}{Cputime}{Stdev}{None}%
\StoreBenchExecResult{SvcompNineteenPdrInv}{KinductionDfBoxesEqReachsafetyLoops}{Wrong}{True}{Walltime}{}{0}%
\StoreBenchExecResult{SvcompNineteenPdrInv}{KinductionDfBoxesEqReachsafetyLoops}{Wrong}{True}{Walltime}{Avg}{None}%
\StoreBenchExecResult{SvcompNineteenPdrInv}{KinductionDfBoxesEqReachsafetyLoops}{Wrong}{True}{Walltime}{Median}{None}%
\StoreBenchExecResult{SvcompNineteenPdrInv}{KinductionDfBoxesEqReachsafetyLoops}{Wrong}{True}{Walltime}{Min}{None}%
\StoreBenchExecResult{SvcompNineteenPdrInv}{KinductionDfBoxesEqReachsafetyLoops}{Wrong}{True}{Walltime}{Max}{None}%
\StoreBenchExecResult{SvcompNineteenPdrInv}{KinductionDfBoxesEqReachsafetyLoops}{Wrong}{True}{Walltime}{Stdev}{None}%
\StoreBenchExecResult{SvcompNineteenPdrInv}{KinductionDfBoxesEqReachsafetyLoops}{Error}{}{Count}{}{100}%
\StoreBenchExecResult{SvcompNineteenPdrInv}{KinductionDfBoxesEqReachsafetyLoops}{Error}{}{Cputime}{}{84276.517975872}%
\StoreBenchExecResult{SvcompNineteenPdrInv}{KinductionDfBoxesEqReachsafetyLoops}{Error}{}{Cputime}{Avg}{842.76517975872}%
\StoreBenchExecResult{SvcompNineteenPdrInv}{KinductionDfBoxesEqReachsafetyLoops}{Error}{}{Cputime}{Median}{904.2008815065}%
\StoreBenchExecResult{SvcompNineteenPdrInv}{KinductionDfBoxesEqReachsafetyLoops}{Error}{}{Cputime}{Min}{7.749711907}%
\StoreBenchExecResult{SvcompNineteenPdrInv}{KinductionDfBoxesEqReachsafetyLoops}{Error}{}{Cputime}{Max}{940.707496472}%
\StoreBenchExecResult{SvcompNineteenPdrInv}{KinductionDfBoxesEqReachsafetyLoops}{Error}{}{Cputime}{Stdev}{217.5358344790199862109332158}%
\StoreBenchExecResult{SvcompNineteenPdrInv}{KinductionDfBoxesEqReachsafetyLoops}{Error}{}{Walltime}{}{81411.921398656980984}%
\StoreBenchExecResult{SvcompNineteenPdrInv}{KinductionDfBoxesEqReachsafetyLoops}{Error}{}{Walltime}{Avg}{814.11921398656980984}%
\StoreBenchExecResult{SvcompNineteenPdrInv}{KinductionDfBoxesEqReachsafetyLoops}{Error}{}{Walltime}{Median}{884.8259822434993}%
\StoreBenchExecResult{SvcompNineteenPdrInv}{KinductionDfBoxesEqReachsafetyLoops}{Error}{}{Walltime}{Min}{4.046134623000398}%
\StoreBenchExecResult{SvcompNineteenPdrInv}{KinductionDfBoxesEqReachsafetyLoops}{Error}{}{Walltime}{Max}{915.5737941320112}%
\StoreBenchExecResult{SvcompNineteenPdrInv}{KinductionDfBoxesEqReachsafetyLoops}{Error}{}{Walltime}{Stdev}{214.5029061631688458570086486}%
\StoreBenchExecResult{SvcompNineteenPdrInv}{KinductionDfBoxesEqReachsafetyLoops}{Error}{Error}{Count}{}{5}%
\StoreBenchExecResult{SvcompNineteenPdrInv}{KinductionDfBoxesEqReachsafetyLoops}{Error}{Error}{Cputime}{}{42.028607577}%
\StoreBenchExecResult{SvcompNineteenPdrInv}{KinductionDfBoxesEqReachsafetyLoops}{Error}{Error}{Cputime}{Avg}{8.4057215154}%
\StoreBenchExecResult{SvcompNineteenPdrInv}{KinductionDfBoxesEqReachsafetyLoops}{Error}{Error}{Cputime}{Median}{8.753708037}%
\StoreBenchExecResult{SvcompNineteenPdrInv}{KinductionDfBoxesEqReachsafetyLoops}{Error}{Error}{Cputime}{Min}{7.749711907}%
\StoreBenchExecResult{SvcompNineteenPdrInv}{KinductionDfBoxesEqReachsafetyLoops}{Error}{Error}{Cputime}{Max}{8.897375434}%
\StoreBenchExecResult{SvcompNineteenPdrInv}{KinductionDfBoxesEqReachsafetyLoops}{Error}{Error}{Cputime}{Stdev}{0.5023314192099025878330944133}%
\StoreBenchExecResult{SvcompNineteenPdrInv}{KinductionDfBoxesEqReachsafetyLoops}{Error}{Error}{Walltime}{}{21.985260852001374}%
\StoreBenchExecResult{SvcompNineteenPdrInv}{KinductionDfBoxesEqReachsafetyLoops}{Error}{Error}{Walltime}{Avg}{4.3970521704002748}%
\StoreBenchExecResult{SvcompNineteenPdrInv}{KinductionDfBoxesEqReachsafetyLoops}{Error}{Error}{Walltime}{Median}{4.57799210200028}%
\StoreBenchExecResult{SvcompNineteenPdrInv}{KinductionDfBoxesEqReachsafetyLoops}{Error}{Error}{Walltime}{Min}{4.046134623000398}%
\StoreBenchExecResult{SvcompNineteenPdrInv}{KinductionDfBoxesEqReachsafetyLoops}{Error}{Error}{Walltime}{Max}{4.63377774800756}%
\StoreBenchExecResult{SvcompNineteenPdrInv}{KinductionDfBoxesEqReachsafetyLoops}{Error}{Error}{Walltime}{Stdev}{0.2559109433430099935219534629}%
\StoreBenchExecResult{SvcompNineteenPdrInv}{KinductionDfBoxesEqReachsafetyLoops}{Error}{OutOfMemory}{Count}{}{8}%
\StoreBenchExecResult{SvcompNineteenPdrInv}{KinductionDfBoxesEqReachsafetyLoops}{Error}{OutOfMemory}{Cputime}{}{5317.067232753}%
\StoreBenchExecResult{SvcompNineteenPdrInv}{KinductionDfBoxesEqReachsafetyLoops}{Error}{OutOfMemory}{Cputime}{Avg}{664.633404094125}%
\StoreBenchExecResult{SvcompNineteenPdrInv}{KinductionDfBoxesEqReachsafetyLoops}{Error}{OutOfMemory}{Cputime}{Median}{828.4829367775}%
\StoreBenchExecResult{SvcompNineteenPdrInv}{KinductionDfBoxesEqReachsafetyLoops}{Error}{OutOfMemory}{Cputime}{Min}{180.725181661}%
\StoreBenchExecResult{SvcompNineteenPdrInv}{KinductionDfBoxesEqReachsafetyLoops}{Error}{OutOfMemory}{Cputime}{Max}{883.773112119}%
\StoreBenchExecResult{SvcompNineteenPdrInv}{KinductionDfBoxesEqReachsafetyLoops}{Error}{OutOfMemory}{Cputime}{Stdev}{281.6073481619359866725607344}%
\StoreBenchExecResult{SvcompNineteenPdrInv}{KinductionDfBoxesEqReachsafetyLoops}{Error}{OutOfMemory}{Walltime}{}{5168.63376005999451}%
\StoreBenchExecResult{SvcompNineteenPdrInv}{KinductionDfBoxesEqReachsafetyLoops}{Error}{OutOfMemory}{Walltime}{Avg}{646.07922000749931375}%
\StoreBenchExecResult{SvcompNineteenPdrInv}{KinductionDfBoxesEqReachsafetyLoops}{Error}{OutOfMemory}{Walltime}{Median}{807.79627197800435}%
\StoreBenchExecResult{SvcompNineteenPdrInv}{KinductionDfBoxesEqReachsafetyLoops}{Error}{OutOfMemory}{Walltime}{Min}{172.4982765820023}%
\StoreBenchExecResult{SvcompNineteenPdrInv}{KinductionDfBoxesEqReachsafetyLoops}{Error}{OutOfMemory}{Walltime}{Max}{861.505236062003}%
\StoreBenchExecResult{SvcompNineteenPdrInv}{KinductionDfBoxesEqReachsafetyLoops}{Error}{OutOfMemory}{Walltime}{Stdev}{276.3514708024896455789812467}%
\StoreBenchExecResult{SvcompNineteenPdrInv}{KinductionDfBoxesEqReachsafetyLoops}{Error}{Timeout}{Count}{}{87}%
\StoreBenchExecResult{SvcompNineteenPdrInv}{KinductionDfBoxesEqReachsafetyLoops}{Error}{Timeout}{Cputime}{}{78917.422135542}%
\StoreBenchExecResult{SvcompNineteenPdrInv}{KinductionDfBoxesEqReachsafetyLoops}{Error}{Timeout}{Cputime}{Avg}{907.0968061556551724137931034}%
\StoreBenchExecResult{SvcompNineteenPdrInv}{KinductionDfBoxesEqReachsafetyLoops}{Error}{Timeout}{Cputime}{Median}{905.353073669}%
\StoreBenchExecResult{SvcompNineteenPdrInv}{KinductionDfBoxesEqReachsafetyLoops}{Error}{Timeout}{Cputime}{Min}{901.318430015}%
\StoreBenchExecResult{SvcompNineteenPdrInv}{KinductionDfBoxesEqReachsafetyLoops}{Error}{Timeout}{Cputime}{Max}{940.707496472}%
\StoreBenchExecResult{SvcompNineteenPdrInv}{KinductionDfBoxesEqReachsafetyLoops}{Error}{Timeout}{Cputime}{Stdev}{5.950519119424240547630423856}%
\StoreBenchExecResult{SvcompNineteenPdrInv}{KinductionDfBoxesEqReachsafetyLoops}{Error}{Timeout}{Walltime}{}{76221.3023777449851}%
\StoreBenchExecResult{SvcompNineteenPdrInv}{KinductionDfBoxesEqReachsafetyLoops}{Error}{Timeout}{Walltime}{Avg}{876.1069238821262655172413793}%
\StoreBenchExecResult{SvcompNineteenPdrInv}{KinductionDfBoxesEqReachsafetyLoops}{Error}{Timeout}{Walltime}{Median}{885.5097497810057}%
\StoreBenchExecResult{SvcompNineteenPdrInv}{KinductionDfBoxesEqReachsafetyLoops}{Error}{Timeout}{Walltime}{Min}{648.0461929839948}%
\StoreBenchExecResult{SvcompNineteenPdrInv}{KinductionDfBoxesEqReachsafetyLoops}{Error}{Timeout}{Walltime}{Max}{915.5737941320112}%
\StoreBenchExecResult{SvcompNineteenPdrInv}{KinductionDfBoxesEqReachsafetyLoops}{Error}{Timeout}{Walltime}{Stdev}{41.76334856980808613392930685}%
\providecommand\StoreBenchExecResult[7]{\expandafter\newcommand\csname#1#2#3#4#5#6\endcsname{#7}}%
\StoreBenchExecResult{SvcompNineteenPdrInv}{KinductionDfBoxesReachsafetyLoops}{Total}{}{Count}{}{208}%
\StoreBenchExecResult{SvcompNineteenPdrInv}{KinductionDfBoxesReachsafetyLoops}{Total}{}{Cputime}{}{90069.295853134}%
\StoreBenchExecResult{SvcompNineteenPdrInv}{KinductionDfBoxesReachsafetyLoops}{Total}{}{Cputime}{Avg}{433.025460832375}%
\StoreBenchExecResult{SvcompNineteenPdrInv}{KinductionDfBoxesReachsafetyLoops}{Total}{}{Cputime}{Median}{39.3475040715}%
\StoreBenchExecResult{SvcompNineteenPdrInv}{KinductionDfBoxesReachsafetyLoops}{Total}{}{Cputime}{Min}{4.050657157}%
\StoreBenchExecResult{SvcompNineteenPdrInv}{KinductionDfBoxesReachsafetyLoops}{Total}{}{Cputime}{Max}{982.460711955}%
\StoreBenchExecResult{SvcompNineteenPdrInv}{KinductionDfBoxesReachsafetyLoops}{Total}{}{Cputime}{Stdev}{444.9542546705409174291024924}%
\StoreBenchExecResult{SvcompNineteenPdrInv}{KinductionDfBoxesReachsafetyLoops}{Total}{}{Walltime}{}{86546.0014429149883519}%
\StoreBenchExecResult{SvcompNineteenPdrInv}{KinductionDfBoxesReachsafetyLoops}{Total}{}{Walltime}{Avg}{416.0865453986297516918269231}%
\StoreBenchExecResult{SvcompNineteenPdrInv}{KinductionDfBoxesReachsafetyLoops}{Total}{}{Walltime}{Median}{28.248418410999875}%
\StoreBenchExecResult{SvcompNineteenPdrInv}{KinductionDfBoxesReachsafetyLoops}{Total}{}{Walltime}{Min}{2.1701883140049176}%
\StoreBenchExecResult{SvcompNineteenPdrInv}{KinductionDfBoxesReachsafetyLoops}{Total}{}{Walltime}{Max}{976.3781876449939}%
\StoreBenchExecResult{SvcompNineteenPdrInv}{KinductionDfBoxesReachsafetyLoops}{Total}{}{Walltime}{Stdev}{432.7126850951775309712586245}%
\StoreBenchExecResult{SvcompNineteenPdrInv}{KinductionDfBoxesReachsafetyLoops}{Correct}{}{Count}{}{103}%
\StoreBenchExecResult{SvcompNineteenPdrInv}{KinductionDfBoxesReachsafetyLoops}{Correct}{}{Cputime}{}{1096.880119720}%
\StoreBenchExecResult{SvcompNineteenPdrInv}{KinductionDfBoxesReachsafetyLoops}{Correct}{}{Cputime}{Avg}{10.64932155067961165048543689}%
\StoreBenchExecResult{SvcompNineteenPdrInv}{KinductionDfBoxesReachsafetyLoops}{Correct}{}{Cputime}{Median}{5.654818925}%
\StoreBenchExecResult{SvcompNineteenPdrInv}{KinductionDfBoxesReachsafetyLoops}{Correct}{}{Cputime}{Min}{4.050657157}%
\StoreBenchExecResult{SvcompNineteenPdrInv}{KinductionDfBoxesReachsafetyLoops}{Correct}{}{Cputime}{Max}{47.160062179}%
\StoreBenchExecResult{SvcompNineteenPdrInv}{KinductionDfBoxesReachsafetyLoops}{Correct}{}{Cputime}{Stdev}{10.39118205470785840542892533}%
\StoreBenchExecResult{SvcompNineteenPdrInv}{KinductionDfBoxesReachsafetyLoops}{Correct}{}{Walltime}{}{661.9190334469167209}%
\StoreBenchExecResult{SvcompNineteenPdrInv}{KinductionDfBoxesReachsafetyLoops}{Correct}{}{Walltime}{Avg}{6.426398382979773989320388350}%
\StoreBenchExecResult{SvcompNineteenPdrInv}{KinductionDfBoxesReachsafetyLoops}{Correct}{}{Walltime}{Median}{3.013719273993047}%
\StoreBenchExecResult{SvcompNineteenPdrInv}{KinductionDfBoxesReachsafetyLoops}{Correct}{}{Walltime}{Min}{2.1701883140049176}%
\StoreBenchExecResult{SvcompNineteenPdrInv}{KinductionDfBoxesReachsafetyLoops}{Correct}{}{Walltime}{Max}{33.95817912100756}%
\StoreBenchExecResult{SvcompNineteenPdrInv}{KinductionDfBoxesReachsafetyLoops}{Correct}{}{Walltime}{Stdev}{7.359133152496425250533900548}%
\StoreBenchExecResult{SvcompNineteenPdrInv}{KinductionDfBoxesReachsafetyLoops}{Correct}{False}{Count}{}{37}%
\StoreBenchExecResult{SvcompNineteenPdrInv}{KinductionDfBoxesReachsafetyLoops}{Correct}{False}{Cputime}{}{334.095274125}%
\StoreBenchExecResult{SvcompNineteenPdrInv}{KinductionDfBoxesReachsafetyLoops}{Correct}{False}{Cputime}{Avg}{9.029602003378378378378378378}%
\StoreBenchExecResult{SvcompNineteenPdrInv}{KinductionDfBoxesReachsafetyLoops}{Correct}{False}{Cputime}{Median}{5.435698158}%
\StoreBenchExecResult{SvcompNineteenPdrInv}{KinductionDfBoxesReachsafetyLoops}{Correct}{False}{Cputime}{Min}{4.263957307}%
\StoreBenchExecResult{SvcompNineteenPdrInv}{KinductionDfBoxesReachsafetyLoops}{Correct}{False}{Cputime}{Max}{47.160062179}%
\StoreBenchExecResult{SvcompNineteenPdrInv}{KinductionDfBoxesReachsafetyLoops}{Correct}{False}{Cputime}{Stdev}{9.055691711420555079401620084}%
\StoreBenchExecResult{SvcompNineteenPdrInv}{KinductionDfBoxesReachsafetyLoops}{Correct}{False}{Walltime}{}{199.9563174259383202}%
\StoreBenchExecResult{SvcompNineteenPdrInv}{KinductionDfBoxesReachsafetyLoops}{Correct}{False}{Walltime}{Avg}{5.404224795295630275675675676}%
\StoreBenchExecResult{SvcompNineteenPdrInv}{KinductionDfBoxesReachsafetyLoops}{Correct}{False}{Walltime}{Median}{2.902230542997131}%
\StoreBenchExecResult{SvcompNineteenPdrInv}{KinductionDfBoxesReachsafetyLoops}{Correct}{False}{Walltime}{Min}{2.3022143589914776}%
\StoreBenchExecResult{SvcompNineteenPdrInv}{KinductionDfBoxesReachsafetyLoops}{Correct}{False}{Walltime}{Max}{33.84182383099687}%
\StoreBenchExecResult{SvcompNineteenPdrInv}{KinductionDfBoxesReachsafetyLoops}{Correct}{False}{Walltime}{Stdev}{6.493654779318735113536138391}%
\StoreBenchExecResult{SvcompNineteenPdrInv}{KinductionDfBoxesReachsafetyLoops}{Wrong}{False}{Count}{}{0}%
\StoreBenchExecResult{SvcompNineteenPdrInv}{KinductionDfBoxesReachsafetyLoops}{Wrong}{False}{Cputime}{}{0}%
\StoreBenchExecResult{SvcompNineteenPdrInv}{KinductionDfBoxesReachsafetyLoops}{Wrong}{False}{Cputime}{Avg}{None}%
\StoreBenchExecResult{SvcompNineteenPdrInv}{KinductionDfBoxesReachsafetyLoops}{Wrong}{False}{Cputime}{Median}{None}%
\StoreBenchExecResult{SvcompNineteenPdrInv}{KinductionDfBoxesReachsafetyLoops}{Wrong}{False}{Cputime}{Min}{None}%
\StoreBenchExecResult{SvcompNineteenPdrInv}{KinductionDfBoxesReachsafetyLoops}{Wrong}{False}{Cputime}{Max}{None}%
\StoreBenchExecResult{SvcompNineteenPdrInv}{KinductionDfBoxesReachsafetyLoops}{Wrong}{False}{Cputime}{Stdev}{None}%
\StoreBenchExecResult{SvcompNineteenPdrInv}{KinductionDfBoxesReachsafetyLoops}{Wrong}{False}{Walltime}{}{0}%
\StoreBenchExecResult{SvcompNineteenPdrInv}{KinductionDfBoxesReachsafetyLoops}{Wrong}{False}{Walltime}{Avg}{None}%
\StoreBenchExecResult{SvcompNineteenPdrInv}{KinductionDfBoxesReachsafetyLoops}{Wrong}{False}{Walltime}{Median}{None}%
\StoreBenchExecResult{SvcompNineteenPdrInv}{KinductionDfBoxesReachsafetyLoops}{Wrong}{False}{Walltime}{Min}{None}%
\StoreBenchExecResult{SvcompNineteenPdrInv}{KinductionDfBoxesReachsafetyLoops}{Wrong}{False}{Walltime}{Max}{None}%
\StoreBenchExecResult{SvcompNineteenPdrInv}{KinductionDfBoxesReachsafetyLoops}{Wrong}{False}{Walltime}{Stdev}{None}%
\StoreBenchExecResult{SvcompNineteenPdrInv}{KinductionDfBoxesReachsafetyLoops}{Correct}{True}{Count}{}{66}%
\StoreBenchExecResult{SvcompNineteenPdrInv}{KinductionDfBoxesReachsafetyLoops}{Correct}{True}{Cputime}{}{762.784845595}%
\StoreBenchExecResult{SvcompNineteenPdrInv}{KinductionDfBoxesReachsafetyLoops}{Correct}{True}{Cputime}{Avg}{11.55734614537878787878787879}%
\StoreBenchExecResult{SvcompNineteenPdrInv}{KinductionDfBoxesReachsafetyLoops}{Correct}{True}{Cputime}{Median}{5.902431385}%
\StoreBenchExecResult{SvcompNineteenPdrInv}{KinductionDfBoxesReachsafetyLoops}{Correct}{True}{Cputime}{Min}{4.050657157}%
\StoreBenchExecResult{SvcompNineteenPdrInv}{KinductionDfBoxesReachsafetyLoops}{Correct}{True}{Cputime}{Max}{44.049856775}%
\StoreBenchExecResult{SvcompNineteenPdrInv}{KinductionDfBoxesReachsafetyLoops}{Correct}{True}{Cputime}{Stdev}{10.96544451904232825670875558}%
\StoreBenchExecResult{SvcompNineteenPdrInv}{KinductionDfBoxesReachsafetyLoops}{Correct}{True}{Walltime}{}{461.9627160209784007}%
\StoreBenchExecResult{SvcompNineteenPdrInv}{KinductionDfBoxesReachsafetyLoops}{Correct}{True}{Walltime}{Avg}{6.999435091226945465151515152}%
\StoreBenchExecResult{SvcompNineteenPdrInv}{KinductionDfBoxesReachsafetyLoops}{Correct}{True}{Walltime}{Median}{3.115708031502436}%
\StoreBenchExecResult{SvcompNineteenPdrInv}{KinductionDfBoxesReachsafetyLoops}{Correct}{True}{Walltime}{Min}{2.1701883140049176}%
\StoreBenchExecResult{SvcompNineteenPdrInv}{KinductionDfBoxesReachsafetyLoops}{Correct}{True}{Walltime}{Max}{33.95817912100756}%
\StoreBenchExecResult{SvcompNineteenPdrInv}{KinductionDfBoxesReachsafetyLoops}{Correct}{True}{Walltime}{Stdev}{7.743642193308958178550264381}%
\StoreBenchExecResult{SvcompNineteenPdrInv}{KinductionDfBoxesReachsafetyLoops}{Wrong}{True}{Count}{}{0}%
\StoreBenchExecResult{SvcompNineteenPdrInv}{KinductionDfBoxesReachsafetyLoops}{Wrong}{True}{Cputime}{}{0}%
\StoreBenchExecResult{SvcompNineteenPdrInv}{KinductionDfBoxesReachsafetyLoops}{Wrong}{True}{Cputime}{Avg}{None}%
\StoreBenchExecResult{SvcompNineteenPdrInv}{KinductionDfBoxesReachsafetyLoops}{Wrong}{True}{Cputime}{Median}{None}%
\StoreBenchExecResult{SvcompNineteenPdrInv}{KinductionDfBoxesReachsafetyLoops}{Wrong}{True}{Cputime}{Min}{None}%
\StoreBenchExecResult{SvcompNineteenPdrInv}{KinductionDfBoxesReachsafetyLoops}{Wrong}{True}{Cputime}{Max}{None}%
\StoreBenchExecResult{SvcompNineteenPdrInv}{KinductionDfBoxesReachsafetyLoops}{Wrong}{True}{Cputime}{Stdev}{None}%
\StoreBenchExecResult{SvcompNineteenPdrInv}{KinductionDfBoxesReachsafetyLoops}{Wrong}{True}{Walltime}{}{0}%
\StoreBenchExecResult{SvcompNineteenPdrInv}{KinductionDfBoxesReachsafetyLoops}{Wrong}{True}{Walltime}{Avg}{None}%
\StoreBenchExecResult{SvcompNineteenPdrInv}{KinductionDfBoxesReachsafetyLoops}{Wrong}{True}{Walltime}{Median}{None}%
\StoreBenchExecResult{SvcompNineteenPdrInv}{KinductionDfBoxesReachsafetyLoops}{Wrong}{True}{Walltime}{Min}{None}%
\StoreBenchExecResult{SvcompNineteenPdrInv}{KinductionDfBoxesReachsafetyLoops}{Wrong}{True}{Walltime}{Max}{None}%
\StoreBenchExecResult{SvcompNineteenPdrInv}{KinductionDfBoxesReachsafetyLoops}{Wrong}{True}{Walltime}{Stdev}{None}%
\StoreBenchExecResult{SvcompNineteenPdrInv}{KinductionDfBoxesReachsafetyLoops}{Error}{}{Count}{}{105}%
\StoreBenchExecResult{SvcompNineteenPdrInv}{KinductionDfBoxesReachsafetyLoops}{Error}{}{Cputime}{}{88972.415733414}%
\StoreBenchExecResult{SvcompNineteenPdrInv}{KinductionDfBoxesReachsafetyLoops}{Error}{}{Cputime}{Avg}{847.3563403182285714285714286}%
\StoreBenchExecResult{SvcompNineteenPdrInv}{KinductionDfBoxesReachsafetyLoops}{Error}{}{Cputime}{Median}{904.267062911}%
\StoreBenchExecResult{SvcompNineteenPdrInv}{KinductionDfBoxesReachsafetyLoops}{Error}{}{Cputime}{Min}{8.404178375}%
\StoreBenchExecResult{SvcompNineteenPdrInv}{KinductionDfBoxesReachsafetyLoops}{Error}{}{Cputime}{Max}{982.460711955}%
\StoreBenchExecResult{SvcompNineteenPdrInv}{KinductionDfBoxesReachsafetyLoops}{Error}{}{Cputime}{Stdev}{213.1149297098928110911572791}%
\StoreBenchExecResult{SvcompNineteenPdrInv}{KinductionDfBoxesReachsafetyLoops}{Error}{}{Walltime}{}{85884.082409468071631}%
\StoreBenchExecResult{SvcompNineteenPdrInv}{KinductionDfBoxesReachsafetyLoops}{Error}{}{Walltime}{Avg}{817.9436419949340155333333333}%
\StoreBenchExecResult{SvcompNineteenPdrInv}{KinductionDfBoxesReachsafetyLoops}{Error}{}{Walltime}{Median}{884.7016637820052}%
\StoreBenchExecResult{SvcompNineteenPdrInv}{KinductionDfBoxesReachsafetyLoops}{Error}{}{Walltime}{Min}{4.386079442003393}%
\StoreBenchExecResult{SvcompNineteenPdrInv}{KinductionDfBoxesReachsafetyLoops}{Error}{}{Walltime}{Max}{976.3781876449939}%
\StoreBenchExecResult{SvcompNineteenPdrInv}{KinductionDfBoxesReachsafetyLoops}{Error}{}{Walltime}{Stdev}{211.5347861575073832594228079}%
\StoreBenchExecResult{SvcompNineteenPdrInv}{KinductionDfBoxesReachsafetyLoops}{Error}{Error}{Count}{}{5}%
\StoreBenchExecResult{SvcompNineteenPdrInv}{KinductionDfBoxesReachsafetyLoops}{Error}{Error}{Cputime}{}{45.042130420}%
\StoreBenchExecResult{SvcompNineteenPdrInv}{KinductionDfBoxesReachsafetyLoops}{Error}{Error}{Cputime}{Avg}{9.008426084}%
\StoreBenchExecResult{SvcompNineteenPdrInv}{KinductionDfBoxesReachsafetyLoops}{Error}{Error}{Cputime}{Median}{9.136781275}%
\StoreBenchExecResult{SvcompNineteenPdrInv}{KinductionDfBoxesReachsafetyLoops}{Error}{Error}{Cputime}{Min}{8.404178375}%
\StoreBenchExecResult{SvcompNineteenPdrInv}{KinductionDfBoxesReachsafetyLoops}{Error}{Error}{Cputime}{Max}{9.483872158}%
\StoreBenchExecResult{SvcompNineteenPdrInv}{KinductionDfBoxesReachsafetyLoops}{Error}{Error}{Cputime}{Stdev}{0.4521767503141462946749386225}%
\StoreBenchExecResult{SvcompNineteenPdrInv}{KinductionDfBoxesReachsafetyLoops}{Error}{Error}{Walltime}{}{23.424156356020831}%
\StoreBenchExecResult{SvcompNineteenPdrInv}{KinductionDfBoxesReachsafetyLoops}{Error}{Error}{Walltime}{Avg}{4.6848312712041662}%
\StoreBenchExecResult{SvcompNineteenPdrInv}{KinductionDfBoxesReachsafetyLoops}{Error}{Error}{Walltime}{Median}{4.729753541003447}%
\StoreBenchExecResult{SvcompNineteenPdrInv}{KinductionDfBoxesReachsafetyLoops}{Error}{Error}{Walltime}{Min}{4.386079442003393}%
\StoreBenchExecResult{SvcompNineteenPdrInv}{KinductionDfBoxesReachsafetyLoops}{Error}{Error}{Walltime}{Max}{4.916885338010616}%
\StoreBenchExecResult{SvcompNineteenPdrInv}{KinductionDfBoxesReachsafetyLoops}{Error}{Error}{Walltime}{Stdev}{0.2176879658619019929297554224}%
\StoreBenchExecResult{SvcompNineteenPdrInv}{KinductionDfBoxesReachsafetyLoops}{Error}{OutOfMemory}{Count}{}{9}%
\StoreBenchExecResult{SvcompNineteenPdrInv}{KinductionDfBoxesReachsafetyLoops}{Error}{OutOfMemory}{Cputime}{}{6270.302009383}%
\StoreBenchExecResult{SvcompNineteenPdrInv}{KinductionDfBoxesReachsafetyLoops}{Error}{OutOfMemory}{Cputime}{Avg}{696.7002232647777777777777778}%
\StoreBenchExecResult{SvcompNineteenPdrInv}{KinductionDfBoxesReachsafetyLoops}{Error}{OutOfMemory}{Cputime}{Median}{848.897379668}%
\StoreBenchExecResult{SvcompNineteenPdrInv}{KinductionDfBoxesReachsafetyLoops}{Error}{OutOfMemory}{Cputime}{Min}{174.789176257}%
\StoreBenchExecResult{SvcompNineteenPdrInv}{KinductionDfBoxesReachsafetyLoops}{Error}{OutOfMemory}{Cputime}{Max}{891.601728537}%
\StoreBenchExecResult{SvcompNineteenPdrInv}{KinductionDfBoxesReachsafetyLoops}{Error}{OutOfMemory}{Cputime}{Stdev}{279.3997718278862987088275115}%
\StoreBenchExecResult{SvcompNineteenPdrInv}{KinductionDfBoxesReachsafetyLoops}{Error}{OutOfMemory}{Walltime}{}{6092.56607323196660}%
\StoreBenchExecResult{SvcompNineteenPdrInv}{KinductionDfBoxesReachsafetyLoops}{Error}{OutOfMemory}{Walltime}{Avg}{676.9517859146629555555555556}%
\StoreBenchExecResult{SvcompNineteenPdrInv}{KinductionDfBoxesReachsafetyLoops}{Error}{OutOfMemory}{Walltime}{Median}{826.8620144029992}%
\StoreBenchExecResult{SvcompNineteenPdrInv}{KinductionDfBoxesReachsafetyLoops}{Error}{OutOfMemory}{Walltime}{Min}{166.24599786498584}%
\StoreBenchExecResult{SvcompNineteenPdrInv}{KinductionDfBoxesReachsafetyLoops}{Error}{OutOfMemory}{Walltime}{Max}{870.322107222004}%
\StoreBenchExecResult{SvcompNineteenPdrInv}{KinductionDfBoxesReachsafetyLoops}{Error}{OutOfMemory}{Walltime}{Stdev}{274.1058210926441096421736201}%
\StoreBenchExecResult{SvcompNineteenPdrInv}{KinductionDfBoxesReachsafetyLoops}{Error}{Timeout}{Count}{}{91}%
\StoreBenchExecResult{SvcompNineteenPdrInv}{KinductionDfBoxesReachsafetyLoops}{Error}{Timeout}{Cputime}{}{82657.071593611}%
\StoreBenchExecResult{SvcompNineteenPdrInv}{KinductionDfBoxesReachsafetyLoops}{Error}{Timeout}{Cputime}{Avg}{908.3194680616593406593406593}%
\StoreBenchExecResult{SvcompNineteenPdrInv}{KinductionDfBoxesReachsafetyLoops}{Error}{Timeout}{Cputime}{Median}{905.173503887}%
\StoreBenchExecResult{SvcompNineteenPdrInv}{KinductionDfBoxesReachsafetyLoops}{Error}{Timeout}{Cputime}{Min}{901.323274163}%
\StoreBenchExecResult{SvcompNineteenPdrInv}{KinductionDfBoxesReachsafetyLoops}{Error}{Timeout}{Cputime}{Max}{982.460711955}%
\StoreBenchExecResult{SvcompNineteenPdrInv}{KinductionDfBoxesReachsafetyLoops}{Error}{Timeout}{Cputime}{Stdev}{10.32163691993408929215500438}%
\StoreBenchExecResult{SvcompNineteenPdrInv}{KinductionDfBoxesReachsafetyLoops}{Error}{Timeout}{Walltime}{}{79768.0921798800842}%
\StoreBenchExecResult{SvcompNineteenPdrInv}{KinductionDfBoxesReachsafetyLoops}{Error}{Timeout}{Walltime}{Avg}{876.5724415371437824175824176}%
\StoreBenchExecResult{SvcompNineteenPdrInv}{KinductionDfBoxesReachsafetyLoops}{Error}{Timeout}{Walltime}{Median}{885.7339431299915}%
\StoreBenchExecResult{SvcompNineteenPdrInv}{KinductionDfBoxesReachsafetyLoops}{Error}{Timeout}{Walltime}{Min}{535.7662470190116}%
\StoreBenchExecResult{SvcompNineteenPdrInv}{KinductionDfBoxesReachsafetyLoops}{Error}{Timeout}{Walltime}{Max}{976.3781876449939}%
\StoreBenchExecResult{SvcompNineteenPdrInv}{KinductionDfBoxesReachsafetyLoops}{Error}{Timeout}{Walltime}{Stdev}{49.56622510881264181639824934}%
\providecommand\StoreBenchExecResult[7]{\expandafter\newcommand\csname#1#2#3#4#5#6\endcsname{#7}}%
\StoreBenchExecResult{SvcompNineteenPdrInv}{KinductionDfkipdrReachsafetyLoops}{Total}{}{Count}{}{208}%
\StoreBenchExecResult{SvcompNineteenPdrInv}{KinductionDfkipdrReachsafetyLoops}{Total}{}{Cputime}{}{73257.458564221}%
\StoreBenchExecResult{SvcompNineteenPdrInv}{KinductionDfkipdrReachsafetyLoops}{Total}{}{Cputime}{Avg}{352.1993200202932692307692308}%
\StoreBenchExecResult{SvcompNineteenPdrInv}{KinductionDfkipdrReachsafetyLoops}{Total}{}{Cputime}{Median}{23.686420936}%
\StoreBenchExecResult{SvcompNineteenPdrInv}{KinductionDfkipdrReachsafetyLoops}{Total}{}{Cputime}{Min}{4.164587888}%
\StoreBenchExecResult{SvcompNineteenPdrInv}{KinductionDfkipdrReachsafetyLoops}{Total}{}{Cputime}{Max}{928.897507159}%
\StoreBenchExecResult{SvcompNineteenPdrInv}{KinductionDfkipdrReachsafetyLoops}{Total}{}{Cputime}{Stdev}{425.7399033482480437626884359}%
\StoreBenchExecResult{SvcompNineteenPdrInv}{KinductionDfkipdrReachsafetyLoops}{Total}{}{Walltime}{}{36783.1845390700765062}%
\StoreBenchExecResult{SvcompNineteenPdrInv}{KinductionDfkipdrReachsafetyLoops}{Total}{}{Walltime}{Avg}{176.8422333609138293567307692}%
\StoreBenchExecResult{SvcompNineteenPdrInv}{KinductionDfkipdrReachsafetyLoops}{Total}{}{Walltime}{Median}{12.049948033003602}%
\StoreBenchExecResult{SvcompNineteenPdrInv}{KinductionDfkipdrReachsafetyLoops}{Total}{}{Walltime}{Min}{2.2418148500000825}%
\StoreBenchExecResult{SvcompNineteenPdrInv}{KinductionDfkipdrReachsafetyLoops}{Total}{}{Walltime}{Max}{477.6377339800092}%
\StoreBenchExecResult{SvcompNineteenPdrInv}{KinductionDfkipdrReachsafetyLoops}{Total}{}{Walltime}{Stdev}{213.5866006516365890521604670}%
\StoreBenchExecResult{SvcompNineteenPdrInv}{KinductionDfkipdrReachsafetyLoops}{Correct}{}{Count}{}{126}%
\StoreBenchExecResult{SvcompNineteenPdrInv}{KinductionDfkipdrReachsafetyLoops}{Correct}{}{Cputime}{}{4241.523149586}%
\StoreBenchExecResult{SvcompNineteenPdrInv}{KinductionDfkipdrReachsafetyLoops}{Correct}{}{Cputime}{Avg}{33.66288213957142857142857143}%
\StoreBenchExecResult{SvcompNineteenPdrInv}{KinductionDfkipdrReachsafetyLoops}{Correct}{}{Cputime}{Median}{6.155897468}%
\StoreBenchExecResult{SvcompNineteenPdrInv}{KinductionDfkipdrReachsafetyLoops}{Correct}{}{Cputime}{Min}{4.164587888}%
\StoreBenchExecResult{SvcompNineteenPdrInv}{KinductionDfkipdrReachsafetyLoops}{Correct}{}{Cputime}{Max}{764.118502657}%
\StoreBenchExecResult{SvcompNineteenPdrInv}{KinductionDfkipdrReachsafetyLoops}{Correct}{}{Cputime}{Stdev}{96.33543947829384455490872284}%
\StoreBenchExecResult{SvcompNineteenPdrInv}{KinductionDfkipdrReachsafetyLoops}{Correct}{}{Walltime}{}{2146.4297341300407252}%
\StoreBenchExecResult{SvcompNineteenPdrInv}{KinductionDfkipdrReachsafetyLoops}{Correct}{}{Walltime}{Avg}{17.03515662007968829523809524}%
\StoreBenchExecResult{SvcompNineteenPdrInv}{KinductionDfkipdrReachsafetyLoops}{Correct}{}{Walltime}{Median}{3.24594968200108265}%
\StoreBenchExecResult{SvcompNineteenPdrInv}{KinductionDfkipdrReachsafetyLoops}{Correct}{}{Walltime}{Min}{2.2418148500000825}%
\StoreBenchExecResult{SvcompNineteenPdrInv}{KinductionDfkipdrReachsafetyLoops}{Correct}{}{Walltime}{Max}{382.7099578359921}%
\StoreBenchExecResult{SvcompNineteenPdrInv}{KinductionDfkipdrReachsafetyLoops}{Correct}{}{Walltime}{Stdev}{48.25041776598824005142679389}%
\StoreBenchExecResult{SvcompNineteenPdrInv}{KinductionDfkipdrReachsafetyLoops}{Correct}{False}{Count}{}{37}%
\StoreBenchExecResult{SvcompNineteenPdrInv}{KinductionDfkipdrReachsafetyLoops}{Correct}{False}{Cputime}{}{579.005923219}%
\StoreBenchExecResult{SvcompNineteenPdrInv}{KinductionDfkipdrReachsafetyLoops}{Correct}{False}{Cputime}{Avg}{15.64880873564864864864864865}%
\StoreBenchExecResult{SvcompNineteenPdrInv}{KinductionDfkipdrReachsafetyLoops}{Correct}{False}{Cputime}{Median}{6.283672784}%
\StoreBenchExecResult{SvcompNineteenPdrInv}{KinductionDfkipdrReachsafetyLoops}{Correct}{False}{Cputime}{Min}{4.496091063}%
\StoreBenchExecResult{SvcompNineteenPdrInv}{KinductionDfkipdrReachsafetyLoops}{Correct}{False}{Cputime}{Max}{84.062649514}%
\StoreBenchExecResult{SvcompNineteenPdrInv}{KinductionDfkipdrReachsafetyLoops}{Correct}{False}{Cputime}{Stdev}{22.10400802136588150989673236}%
\StoreBenchExecResult{SvcompNineteenPdrInv}{KinductionDfkipdrReachsafetyLoops}{Correct}{False}{Walltime}{}{296.3480303640535655}%
\StoreBenchExecResult{SvcompNineteenPdrInv}{KinductionDfkipdrReachsafetyLoops}{Correct}{False}{Walltime}{Avg}{8.009406226055501770270270270}%
\StoreBenchExecResult{SvcompNineteenPdrInv}{KinductionDfkipdrReachsafetyLoops}{Correct}{False}{Walltime}{Median}{3.3102400290081277}%
\StoreBenchExecResult{SvcompNineteenPdrInv}{KinductionDfkipdrReachsafetyLoops}{Correct}{False}{Walltime}{Min}{2.3858039519982412}%
\StoreBenchExecResult{SvcompNineteenPdrInv}{KinductionDfkipdrReachsafetyLoops}{Correct}{False}{Walltime}{Max}{42.35810972900072}%
\StoreBenchExecResult{SvcompNineteenPdrInv}{KinductionDfkipdrReachsafetyLoops}{Correct}{False}{Walltime}{Stdev}{11.09040515893096911196280971}%
\StoreBenchExecResult{SvcompNineteenPdrInv}{KinductionDfkipdrReachsafetyLoops}{Wrong}{False}{Count}{}{0}%
\StoreBenchExecResult{SvcompNineteenPdrInv}{KinductionDfkipdrReachsafetyLoops}{Wrong}{False}{Cputime}{}{0}%
\StoreBenchExecResult{SvcompNineteenPdrInv}{KinductionDfkipdrReachsafetyLoops}{Wrong}{False}{Cputime}{Avg}{None}%
\StoreBenchExecResult{SvcompNineteenPdrInv}{KinductionDfkipdrReachsafetyLoops}{Wrong}{False}{Cputime}{Median}{None}%
\StoreBenchExecResult{SvcompNineteenPdrInv}{KinductionDfkipdrReachsafetyLoops}{Wrong}{False}{Cputime}{Min}{None}%
\StoreBenchExecResult{SvcompNineteenPdrInv}{KinductionDfkipdrReachsafetyLoops}{Wrong}{False}{Cputime}{Max}{None}%
\StoreBenchExecResult{SvcompNineteenPdrInv}{KinductionDfkipdrReachsafetyLoops}{Wrong}{False}{Cputime}{Stdev}{None}%
\StoreBenchExecResult{SvcompNineteenPdrInv}{KinductionDfkipdrReachsafetyLoops}{Wrong}{False}{Walltime}{}{0}%
\StoreBenchExecResult{SvcompNineteenPdrInv}{KinductionDfkipdrReachsafetyLoops}{Wrong}{False}{Walltime}{Avg}{None}%
\StoreBenchExecResult{SvcompNineteenPdrInv}{KinductionDfkipdrReachsafetyLoops}{Wrong}{False}{Walltime}{Median}{None}%
\StoreBenchExecResult{SvcompNineteenPdrInv}{KinductionDfkipdrReachsafetyLoops}{Wrong}{False}{Walltime}{Min}{None}%
\StoreBenchExecResult{SvcompNineteenPdrInv}{KinductionDfkipdrReachsafetyLoops}{Wrong}{False}{Walltime}{Max}{None}%
\StoreBenchExecResult{SvcompNineteenPdrInv}{KinductionDfkipdrReachsafetyLoops}{Wrong}{False}{Walltime}{Stdev}{None}%
\StoreBenchExecResult{SvcompNineteenPdrInv}{KinductionDfkipdrReachsafetyLoops}{Correct}{True}{Count}{}{89}%
\StoreBenchExecResult{SvcompNineteenPdrInv}{KinductionDfkipdrReachsafetyLoops}{Correct}{True}{Cputime}{}{3662.517226367}%
\StoreBenchExecResult{SvcompNineteenPdrInv}{KinductionDfkipdrReachsafetyLoops}{Correct}{True}{Cputime}{Avg}{41.15187894794382022471910112}%
\StoreBenchExecResult{SvcompNineteenPdrInv}{KinductionDfkipdrReachsafetyLoops}{Correct}{True}{Cputime}{Median}{6.100553162}%
\StoreBenchExecResult{SvcompNineteenPdrInv}{KinductionDfkipdrReachsafetyLoops}{Correct}{True}{Cputime}{Min}{4.164587888}%
\StoreBenchExecResult{SvcompNineteenPdrInv}{KinductionDfkipdrReachsafetyLoops}{Correct}{True}{Cputime}{Max}{764.118502657}%
\StoreBenchExecResult{SvcompNineteenPdrInv}{KinductionDfkipdrReachsafetyLoops}{Correct}{True}{Cputime}{Stdev}{112.8919678179378149617217853}%
\StoreBenchExecResult{SvcompNineteenPdrInv}{KinductionDfkipdrReachsafetyLoops}{Correct}{True}{Walltime}{}{1850.0817037659871597}%
\StoreBenchExecResult{SvcompNineteenPdrInv}{KinductionDfkipdrReachsafetyLoops}{Correct}{True}{Walltime}{Avg}{20.78743487377513662584269663}%
\StoreBenchExecResult{SvcompNineteenPdrInv}{KinductionDfkipdrReachsafetyLoops}{Correct}{True}{Walltime}{Median}{3.209783915008302}%
\StoreBenchExecResult{SvcompNineteenPdrInv}{KinductionDfkipdrReachsafetyLoops}{Correct}{True}{Walltime}{Min}{2.2418148500000825}%
\StoreBenchExecResult{SvcompNineteenPdrInv}{KinductionDfkipdrReachsafetyLoops}{Correct}{True}{Walltime}{Max}{382.7099578359921}%
\StoreBenchExecResult{SvcompNineteenPdrInv}{KinductionDfkipdrReachsafetyLoops}{Correct}{True}{Walltime}{Stdev}{56.54100673996947234710979406}%
\StoreBenchExecResult{SvcompNineteenPdrInv}{KinductionDfkipdrReachsafetyLoops}{Wrong}{True}{Count}{}{0}%
\StoreBenchExecResult{SvcompNineteenPdrInv}{KinductionDfkipdrReachsafetyLoops}{Wrong}{True}{Cputime}{}{0}%
\StoreBenchExecResult{SvcompNineteenPdrInv}{KinductionDfkipdrReachsafetyLoops}{Wrong}{True}{Cputime}{Avg}{None}%
\StoreBenchExecResult{SvcompNineteenPdrInv}{KinductionDfkipdrReachsafetyLoops}{Wrong}{True}{Cputime}{Median}{None}%
\StoreBenchExecResult{SvcompNineteenPdrInv}{KinductionDfkipdrReachsafetyLoops}{Wrong}{True}{Cputime}{Min}{None}%
\StoreBenchExecResult{SvcompNineteenPdrInv}{KinductionDfkipdrReachsafetyLoops}{Wrong}{True}{Cputime}{Max}{None}%
\StoreBenchExecResult{SvcompNineteenPdrInv}{KinductionDfkipdrReachsafetyLoops}{Wrong}{True}{Cputime}{Stdev}{None}%
\StoreBenchExecResult{SvcompNineteenPdrInv}{KinductionDfkipdrReachsafetyLoops}{Wrong}{True}{Walltime}{}{0}%
\StoreBenchExecResult{SvcompNineteenPdrInv}{KinductionDfkipdrReachsafetyLoops}{Wrong}{True}{Walltime}{Avg}{None}%
\StoreBenchExecResult{SvcompNineteenPdrInv}{KinductionDfkipdrReachsafetyLoops}{Wrong}{True}{Walltime}{Median}{None}%
\StoreBenchExecResult{SvcompNineteenPdrInv}{KinductionDfkipdrReachsafetyLoops}{Wrong}{True}{Walltime}{Min}{None}%
\StoreBenchExecResult{SvcompNineteenPdrInv}{KinductionDfkipdrReachsafetyLoops}{Wrong}{True}{Walltime}{Max}{None}%
\StoreBenchExecResult{SvcompNineteenPdrInv}{KinductionDfkipdrReachsafetyLoops}{Wrong}{True}{Walltime}{Stdev}{None}%
\StoreBenchExecResult{SvcompNineteenPdrInv}{KinductionDfkipdrReachsafetyLoops}{Error}{}{Count}{}{82}%
\StoreBenchExecResult{SvcompNineteenPdrInv}{KinductionDfkipdrReachsafetyLoops}{Error}{}{Cputime}{}{69015.935414635}%
\StoreBenchExecResult{SvcompNineteenPdrInv}{KinductionDfkipdrReachsafetyLoops}{Error}{}{Cputime}{Avg}{841.6577489589634146341463415}%
\StoreBenchExecResult{SvcompNineteenPdrInv}{KinductionDfkipdrReachsafetyLoops}{Error}{}{Cputime}{Median}{902.946215753}%
\StoreBenchExecResult{SvcompNineteenPdrInv}{KinductionDfkipdrReachsafetyLoops}{Error}{}{Cputime}{Min}{7.784846374}%
\StoreBenchExecResult{SvcompNineteenPdrInv}{KinductionDfkipdrReachsafetyLoops}{Error}{}{Cputime}{Max}{928.897507159}%
\StoreBenchExecResult{SvcompNineteenPdrInv}{KinductionDfkipdrReachsafetyLoops}{Error}{}{Cputime}{Stdev}{223.6676543980723824781103932}%
\StoreBenchExecResult{SvcompNineteenPdrInv}{KinductionDfkipdrReachsafetyLoops}{Error}{}{Walltime}{}{34636.754804940035781}%
\StoreBenchExecResult{SvcompNineteenPdrInv}{KinductionDfkipdrReachsafetyLoops}{Error}{}{Walltime}{Avg}{422.3994488407321436707317073}%
\StoreBenchExecResult{SvcompNineteenPdrInv}{KinductionDfkipdrReachsafetyLoops}{Error}{}{Walltime}{Median}{452.56462709099287}%
\StoreBenchExecResult{SvcompNineteenPdrInv}{KinductionDfkipdrReachsafetyLoops}{Error}{}{Walltime}{Min}{4.078142907994334}%
\StoreBenchExecResult{SvcompNineteenPdrInv}{KinductionDfkipdrReachsafetyLoops}{Error}{}{Walltime}{Max}{477.6377339800092}%
\StoreBenchExecResult{SvcompNineteenPdrInv}{KinductionDfkipdrReachsafetyLoops}{Error}{}{Walltime}{Stdev}{112.2480771116578484376268671}%
\StoreBenchExecResult{SvcompNineteenPdrInv}{KinductionDfkipdrReachsafetyLoops}{Error}{Error}{Count}{}{5}%
\StoreBenchExecResult{SvcompNineteenPdrInv}{KinductionDfkipdrReachsafetyLoops}{Error}{Error}{Cputime}{}{40.484258861}%
\StoreBenchExecResult{SvcompNineteenPdrInv}{KinductionDfkipdrReachsafetyLoops}{Error}{Error}{Cputime}{Avg}{8.0968517722}%
\StoreBenchExecResult{SvcompNineteenPdrInv}{KinductionDfkipdrReachsafetyLoops}{Error}{Error}{Cputime}{Median}{7.856856964}%
\StoreBenchExecResult{SvcompNineteenPdrInv}{KinductionDfkipdrReachsafetyLoops}{Error}{Error}{Cputime}{Min}{7.784846374}%
\StoreBenchExecResult{SvcompNineteenPdrInv}{KinductionDfkipdrReachsafetyLoops}{Error}{Error}{Cputime}{Max}{8.779949152}%
\StoreBenchExecResult{SvcompNineteenPdrInv}{KinductionDfkipdrReachsafetyLoops}{Error}{Error}{Cputime}{Stdev}{0.3863438474057328835825626947}%
\StoreBenchExecResult{SvcompNineteenPdrInv}{KinductionDfkipdrReachsafetyLoops}{Error}{Error}{Walltime}{}{21.178274624020561}%
\StoreBenchExecResult{SvcompNineteenPdrInv}{KinductionDfkipdrReachsafetyLoops}{Error}{Error}{Walltime}{Avg}{4.2356549248041122}%
\StoreBenchExecResult{SvcompNineteenPdrInv}{KinductionDfkipdrReachsafetyLoops}{Error}{Error}{Walltime}{Median}{4.098300837009447}%
\StoreBenchExecResult{SvcompNineteenPdrInv}{KinductionDfkipdrReachsafetyLoops}{Error}{Error}{Walltime}{Min}{4.078142907994334}%
\StoreBenchExecResult{SvcompNineteenPdrInv}{KinductionDfkipdrReachsafetyLoops}{Error}{Error}{Walltime}{Max}{4.585838975006482}%
\StoreBenchExecResult{SvcompNineteenPdrInv}{KinductionDfkipdrReachsafetyLoops}{Error}{Error}{Walltime}{Stdev}{0.1977803972480699670903746150}%
\StoreBenchExecResult{SvcompNineteenPdrInv}{KinductionDfkipdrReachsafetyLoops}{Error}{OutOfMemory}{Count}{}{1}%
\StoreBenchExecResult{SvcompNineteenPdrInv}{KinductionDfkipdrReachsafetyLoops}{Error}{OutOfMemory}{Cputime}{}{266.331312474}%
\StoreBenchExecResult{SvcompNineteenPdrInv}{KinductionDfkipdrReachsafetyLoops}{Error}{OutOfMemory}{Cputime}{Avg}{266.331312474}%
\StoreBenchExecResult{SvcompNineteenPdrInv}{KinductionDfkipdrReachsafetyLoops}{Error}{OutOfMemory}{Cputime}{Median}{266.331312474}%
\StoreBenchExecResult{SvcompNineteenPdrInv}{KinductionDfkipdrReachsafetyLoops}{Error}{OutOfMemory}{Cputime}{Min}{266.331312474}%
\StoreBenchExecResult{SvcompNineteenPdrInv}{KinductionDfkipdrReachsafetyLoops}{Error}{OutOfMemory}{Cputime}{Max}{266.331312474}%
\StoreBenchExecResult{SvcompNineteenPdrInv}{KinductionDfkipdrReachsafetyLoops}{Error}{OutOfMemory}{Cputime}{Stdev}{0E-9}%
\StoreBenchExecResult{SvcompNineteenPdrInv}{KinductionDfkipdrReachsafetyLoops}{Error}{OutOfMemory}{Walltime}{}{133.69831287200213}%
\StoreBenchExecResult{SvcompNineteenPdrInv}{KinductionDfkipdrReachsafetyLoops}{Error}{OutOfMemory}{Walltime}{Avg}{133.69831287200213}%
\StoreBenchExecResult{SvcompNineteenPdrInv}{KinductionDfkipdrReachsafetyLoops}{Error}{OutOfMemory}{Walltime}{Median}{133.69831287200213}%
\StoreBenchExecResult{SvcompNineteenPdrInv}{KinductionDfkipdrReachsafetyLoops}{Error}{OutOfMemory}{Walltime}{Min}{133.69831287200213}%
\StoreBenchExecResult{SvcompNineteenPdrInv}{KinductionDfkipdrReachsafetyLoops}{Error}{OutOfMemory}{Walltime}{Max}{133.69831287200213}%
\StoreBenchExecResult{SvcompNineteenPdrInv}{KinductionDfkipdrReachsafetyLoops}{Error}{OutOfMemory}{Walltime}{Stdev}{0E-14}%
\StoreBenchExecResult{SvcompNineteenPdrInv}{KinductionDfkipdrReachsafetyLoops}{Error}{Timeout}{Count}{}{76}%
\StoreBenchExecResult{SvcompNineteenPdrInv}{KinductionDfkipdrReachsafetyLoops}{Error}{Timeout}{Cputime}{}{68709.119843300}%
\StoreBenchExecResult{SvcompNineteenPdrInv}{KinductionDfkipdrReachsafetyLoops}{Error}{Timeout}{Cputime}{Avg}{904.0673663592105263157894737}%
\StoreBenchExecResult{SvcompNineteenPdrInv}{KinductionDfkipdrReachsafetyLoops}{Error}{Timeout}{Cputime}{Median}{903.127348608}%
\StoreBenchExecResult{SvcompNineteenPdrInv}{KinductionDfkipdrReachsafetyLoops}{Error}{Timeout}{Cputime}{Min}{901.370448238}%
\StoreBenchExecResult{SvcompNineteenPdrInv}{KinductionDfkipdrReachsafetyLoops}{Error}{Timeout}{Cputime}{Max}{928.897507159}%
\StoreBenchExecResult{SvcompNineteenPdrInv}{KinductionDfkipdrReachsafetyLoops}{Error}{Timeout}{Cputime}{Stdev}{3.795637376160078303432905639}%
\StoreBenchExecResult{SvcompNineteenPdrInv}{KinductionDfkipdrReachsafetyLoops}{Error}{Timeout}{Walltime}{}{34481.87821744401309}%
\StoreBenchExecResult{SvcompNineteenPdrInv}{KinductionDfkipdrReachsafetyLoops}{Error}{Timeout}{Walltime}{Avg}{453.7089239137370143421052632}%
\StoreBenchExecResult{SvcompNineteenPdrInv}{KinductionDfkipdrReachsafetyLoops}{Error}{Timeout}{Walltime}{Median}{452.5962240584995}%
\StoreBenchExecResult{SvcompNineteenPdrInv}{KinductionDfkipdrReachsafetyLoops}{Error}{Timeout}{Walltime}{Min}{451.3388211050042}%
\StoreBenchExecResult{SvcompNineteenPdrInv}{KinductionDfkipdrReachsafetyLoops}{Error}{Timeout}{Walltime}{Max}{477.6377339800092}%
\StoreBenchExecResult{SvcompNineteenPdrInv}{KinductionDfkipdrReachsafetyLoops}{Error}{Timeout}{Walltime}{Stdev}{3.654246296122719742785339145}%
\providecommand\StoreBenchExecResult[7]{\expandafter\newcommand\csname#1#2#3#4#5#6\endcsname{#7}}%
\StoreBenchExecResult{SvcompNineteenPdrInv}{KinductionDfReachsafetyLoops}{Total}{}{Count}{}{208}%
\StoreBenchExecResult{SvcompNineteenPdrInv}{KinductionDfReachsafetyLoops}{Total}{}{Cputime}{}{75458.592366226}%
\StoreBenchExecResult{SvcompNineteenPdrInv}{KinductionDfReachsafetyLoops}{Total}{}{Cputime}{Avg}{362.7816940683942307692307692}%
\StoreBenchExecResult{SvcompNineteenPdrInv}{KinductionDfReachsafetyLoops}{Total}{}{Cputime}{Median}{37.422541005}%
\StoreBenchExecResult{SvcompNineteenPdrInv}{KinductionDfReachsafetyLoops}{Total}{}{Cputime}{Min}{3.998420555}%
\StoreBenchExecResult{SvcompNineteenPdrInv}{KinductionDfReachsafetyLoops}{Total}{}{Cputime}{Max}{934.61590394}%
\StoreBenchExecResult{SvcompNineteenPdrInv}{KinductionDfReachsafetyLoops}{Total}{}{Cputime}{Stdev}{428.4849360975917388855809572}%
\StoreBenchExecResult{SvcompNineteenPdrInv}{KinductionDfReachsafetyLoops}{Total}{}{Walltime}{}{42055.9623150139086856}%
\StoreBenchExecResult{SvcompNineteenPdrInv}{KinductionDfReachsafetyLoops}{Total}{}{Walltime}{Avg}{202.1921265144899456038461538}%
\StoreBenchExecResult{SvcompNineteenPdrInv}{KinductionDfReachsafetyLoops}{Total}{}{Walltime}{Median}{19.229974232504901}%
\StoreBenchExecResult{SvcompNineteenPdrInv}{KinductionDfReachsafetyLoops}{Total}{}{Walltime}{Min}{2.146309845993528}%
\StoreBenchExecResult{SvcompNineteenPdrInv}{KinductionDfReachsafetyLoops}{Total}{}{Walltime}{Max}{879.2505336030008}%
\StoreBenchExecResult{SvcompNineteenPdrInv}{KinductionDfReachsafetyLoops}{Total}{}{Walltime}{Stdev}{252.7238460630081744470765609}%
\StoreBenchExecResult{SvcompNineteenPdrInv}{KinductionDfReachsafetyLoops}{Correct}{}{Count}{}{123}%
\StoreBenchExecResult{SvcompNineteenPdrInv}{KinductionDfReachsafetyLoops}{Correct}{}{Cputime}{}{3820.249067018}%
\StoreBenchExecResult{SvcompNineteenPdrInv}{KinductionDfReachsafetyLoops}{Correct}{}{Cputime}{Avg}{31.05893550421138211382113821}%
\StoreBenchExecResult{SvcompNineteenPdrInv}{KinductionDfReachsafetyLoops}{Correct}{}{Cputime}{Median}{5.778490019}%
\StoreBenchExecResult{SvcompNineteenPdrInv}{KinductionDfReachsafetyLoops}{Correct}{}{Cputime}{Min}{3.998420555}%
\StoreBenchExecResult{SvcompNineteenPdrInv}{KinductionDfReachsafetyLoops}{Correct}{}{Cputime}{Max}{668.58913311}%
\StoreBenchExecResult{SvcompNineteenPdrInv}{KinductionDfReachsafetyLoops}{Correct}{}{Cputime}{Stdev}{83.65021237822979440608244868}%
\StoreBenchExecResult{SvcompNineteenPdrInv}{KinductionDfReachsafetyLoops}{Correct}{}{Walltime}{}{1945.3075525859603931}%
\StoreBenchExecResult{SvcompNineteenPdrInv}{KinductionDfReachsafetyLoops}{Correct}{}{Walltime}{Avg}{15.81550855760943409024390244}%
\StoreBenchExecResult{SvcompNineteenPdrInv}{KinductionDfReachsafetyLoops}{Correct}{}{Walltime}{Median}{3.0578412859904347}%
\StoreBenchExecResult{SvcompNineteenPdrInv}{KinductionDfReachsafetyLoops}{Correct}{}{Walltime}{Min}{2.146309845993528}%
\StoreBenchExecResult{SvcompNineteenPdrInv}{KinductionDfReachsafetyLoops}{Correct}{}{Walltime}{Max}{335.3262249339896}%
\StoreBenchExecResult{SvcompNineteenPdrInv}{KinductionDfReachsafetyLoops}{Correct}{}{Walltime}{Stdev}{41.94498730172792381258528525}%
\StoreBenchExecResult{SvcompNineteenPdrInv}{KinductionDfReachsafetyLoops}{Correct}{False}{Count}{}{37}%
\StoreBenchExecResult{SvcompNineteenPdrInv}{KinductionDfReachsafetyLoops}{Correct}{False}{Cputime}{}{546.237996934}%
\StoreBenchExecResult{SvcompNineteenPdrInv}{KinductionDfReachsafetyLoops}{Correct}{False}{Cputime}{Avg}{14.76318910632432432432432432}%
\StoreBenchExecResult{SvcompNineteenPdrInv}{KinductionDfReachsafetyLoops}{Correct}{False}{Cputime}{Median}{5.819558802}%
\StoreBenchExecResult{SvcompNineteenPdrInv}{KinductionDfReachsafetyLoops}{Correct}{False}{Cputime}{Min}{4.310051095}%
\StoreBenchExecResult{SvcompNineteenPdrInv}{KinductionDfReachsafetyLoops}{Correct}{False}{Cputime}{Max}{96.528545089}%
\StoreBenchExecResult{SvcompNineteenPdrInv}{KinductionDfReachsafetyLoops}{Correct}{False}{Cputime}{Stdev}{21.03132639025557059757913862}%
\StoreBenchExecResult{SvcompNineteenPdrInv}{KinductionDfReachsafetyLoops}{Correct}{False}{Walltime}{}{279.9081650440057247}%
\StoreBenchExecResult{SvcompNineteenPdrInv}{KinductionDfReachsafetyLoops}{Correct}{False}{Walltime}{Avg}{7.565085541729884451351351351}%
\StoreBenchExecResult{SvcompNineteenPdrInv}{KinductionDfReachsafetyLoops}{Correct}{False}{Walltime}{Median}{3.078150872999686}%
\StoreBenchExecResult{SvcompNineteenPdrInv}{KinductionDfReachsafetyLoops}{Correct}{False}{Walltime}{Min}{2.306225372987683}%
\StoreBenchExecResult{SvcompNineteenPdrInv}{KinductionDfReachsafetyLoops}{Correct}{False}{Walltime}{Max}{48.586182525992626}%
\StoreBenchExecResult{SvcompNineteenPdrInv}{KinductionDfReachsafetyLoops}{Correct}{False}{Walltime}{Stdev}{10.55535914844450018100544852}%
\StoreBenchExecResult{SvcompNineteenPdrInv}{KinductionDfReachsafetyLoops}{Wrong}{False}{Count}{}{0}%
\StoreBenchExecResult{SvcompNineteenPdrInv}{KinductionDfReachsafetyLoops}{Wrong}{False}{Cputime}{}{0}%
\StoreBenchExecResult{SvcompNineteenPdrInv}{KinductionDfReachsafetyLoops}{Wrong}{False}{Cputime}{Avg}{None}%
\StoreBenchExecResult{SvcompNineteenPdrInv}{KinductionDfReachsafetyLoops}{Wrong}{False}{Cputime}{Median}{None}%
\StoreBenchExecResult{SvcompNineteenPdrInv}{KinductionDfReachsafetyLoops}{Wrong}{False}{Cputime}{Min}{None}%
\StoreBenchExecResult{SvcompNineteenPdrInv}{KinductionDfReachsafetyLoops}{Wrong}{False}{Cputime}{Max}{None}%
\StoreBenchExecResult{SvcompNineteenPdrInv}{KinductionDfReachsafetyLoops}{Wrong}{False}{Cputime}{Stdev}{None}%
\StoreBenchExecResult{SvcompNineteenPdrInv}{KinductionDfReachsafetyLoops}{Wrong}{False}{Walltime}{}{0}%
\StoreBenchExecResult{SvcompNineteenPdrInv}{KinductionDfReachsafetyLoops}{Wrong}{False}{Walltime}{Avg}{None}%
\StoreBenchExecResult{SvcompNineteenPdrInv}{KinductionDfReachsafetyLoops}{Wrong}{False}{Walltime}{Median}{None}%
\StoreBenchExecResult{SvcompNineteenPdrInv}{KinductionDfReachsafetyLoops}{Wrong}{False}{Walltime}{Min}{None}%
\StoreBenchExecResult{SvcompNineteenPdrInv}{KinductionDfReachsafetyLoops}{Wrong}{False}{Walltime}{Max}{None}%
\StoreBenchExecResult{SvcompNineteenPdrInv}{KinductionDfReachsafetyLoops}{Wrong}{False}{Walltime}{Stdev}{None}%
\StoreBenchExecResult{SvcompNineteenPdrInv}{KinductionDfReachsafetyLoops}{Correct}{True}{Count}{}{86}%
\StoreBenchExecResult{SvcompNineteenPdrInv}{KinductionDfReachsafetyLoops}{Correct}{True}{Cputime}{}{3274.011070084}%
\StoreBenchExecResult{SvcompNineteenPdrInv}{KinductionDfReachsafetyLoops}{Correct}{True}{Cputime}{Avg}{38.06989616376744186046511628}%
\StoreBenchExecResult{SvcompNineteenPdrInv}{KinductionDfReachsafetyLoops}{Correct}{True}{Cputime}{Median}{5.732203659}%
\StoreBenchExecResult{SvcompNineteenPdrInv}{KinductionDfReachsafetyLoops}{Correct}{True}{Cputime}{Min}{3.998420555}%
\StoreBenchExecResult{SvcompNineteenPdrInv}{KinductionDfReachsafetyLoops}{Correct}{True}{Cputime}{Max}{668.58913311}%
\StoreBenchExecResult{SvcompNineteenPdrInv}{KinductionDfReachsafetyLoops}{Correct}{True}{Cputime}{Stdev}{98.25552315045592013561562129}%
\StoreBenchExecResult{SvcompNineteenPdrInv}{KinductionDfReachsafetyLoops}{Correct}{True}{Walltime}{}{1665.3993875419546684}%
\StoreBenchExecResult{SvcompNineteenPdrInv}{KinductionDfReachsafetyLoops}{Correct}{True}{Walltime}{Avg}{19.36510915746458916744186047}%
\StoreBenchExecResult{SvcompNineteenPdrInv}{KinductionDfReachsafetyLoops}{Correct}{True}{Walltime}{Median}{3.03648480449191985}%
\StoreBenchExecResult{SvcompNineteenPdrInv}{KinductionDfReachsafetyLoops}{Correct}{True}{Walltime}{Min}{2.146309845993528}%
\StoreBenchExecResult{SvcompNineteenPdrInv}{KinductionDfReachsafetyLoops}{Correct}{True}{Walltime}{Max}{335.3262249339896}%
\StoreBenchExecResult{SvcompNineteenPdrInv}{KinductionDfReachsafetyLoops}{Correct}{True}{Walltime}{Stdev}{49.25957134011559250406738859}%
\StoreBenchExecResult{SvcompNineteenPdrInv}{KinductionDfReachsafetyLoops}{Wrong}{True}{Count}{}{0}%
\StoreBenchExecResult{SvcompNineteenPdrInv}{KinductionDfReachsafetyLoops}{Wrong}{True}{Cputime}{}{0}%
\StoreBenchExecResult{SvcompNineteenPdrInv}{KinductionDfReachsafetyLoops}{Wrong}{True}{Cputime}{Avg}{None}%
\StoreBenchExecResult{SvcompNineteenPdrInv}{KinductionDfReachsafetyLoops}{Wrong}{True}{Cputime}{Median}{None}%
\StoreBenchExecResult{SvcompNineteenPdrInv}{KinductionDfReachsafetyLoops}{Wrong}{True}{Cputime}{Min}{None}%
\StoreBenchExecResult{SvcompNineteenPdrInv}{KinductionDfReachsafetyLoops}{Wrong}{True}{Cputime}{Max}{None}%
\StoreBenchExecResult{SvcompNineteenPdrInv}{KinductionDfReachsafetyLoops}{Wrong}{True}{Cputime}{Stdev}{None}%
\StoreBenchExecResult{SvcompNineteenPdrInv}{KinductionDfReachsafetyLoops}{Wrong}{True}{Walltime}{}{0}%
\StoreBenchExecResult{SvcompNineteenPdrInv}{KinductionDfReachsafetyLoops}{Wrong}{True}{Walltime}{Avg}{None}%
\StoreBenchExecResult{SvcompNineteenPdrInv}{KinductionDfReachsafetyLoops}{Wrong}{True}{Walltime}{Median}{None}%
\StoreBenchExecResult{SvcompNineteenPdrInv}{KinductionDfReachsafetyLoops}{Wrong}{True}{Walltime}{Min}{None}%
\StoreBenchExecResult{SvcompNineteenPdrInv}{KinductionDfReachsafetyLoops}{Wrong}{True}{Walltime}{Max}{None}%
\StoreBenchExecResult{SvcompNineteenPdrInv}{KinductionDfReachsafetyLoops}{Wrong}{True}{Walltime}{Stdev}{None}%
\StoreBenchExecResult{SvcompNineteenPdrInv}{KinductionDfReachsafetyLoops}{Error}{}{Count}{}{85}%
\StoreBenchExecResult{SvcompNineteenPdrInv}{KinductionDfReachsafetyLoops}{Error}{}{Cputime}{}{71638.343299208}%
\StoreBenchExecResult{SvcompNineteenPdrInv}{KinductionDfReachsafetyLoops}{Error}{}{Cputime}{Avg}{842.8040388142117647058823529}%
\StoreBenchExecResult{SvcompNineteenPdrInv}{KinductionDfReachsafetyLoops}{Error}{}{Cputime}{Median}{902.938315895}%
\StoreBenchExecResult{SvcompNineteenPdrInv}{KinductionDfReachsafetyLoops}{Error}{}{Cputime}{Min}{7.72365526}%
\StoreBenchExecResult{SvcompNineteenPdrInv}{KinductionDfReachsafetyLoops}{Error}{}{Cputime}{Max}{934.61590394}%
\StoreBenchExecResult{SvcompNineteenPdrInv}{KinductionDfReachsafetyLoops}{Error}{}{Cputime}{Stdev}{222.4792340112614122542124682}%
\StoreBenchExecResult{SvcompNineteenPdrInv}{KinductionDfReachsafetyLoops}{Error}{}{Walltime}{}{40110.6547624279482925}%
\StoreBenchExecResult{SvcompNineteenPdrInv}{KinductionDfReachsafetyLoops}{Error}{}{Walltime}{Avg}{471.8900560285640975588235294}%
\StoreBenchExecResult{SvcompNineteenPdrInv}{KinductionDfReachsafetyLoops}{Error}{}{Walltime}{Median}{453.13015432799875}%
\StoreBenchExecResult{SvcompNineteenPdrInv}{KinductionDfReachsafetyLoops}{Error}{}{Walltime}{Min}{4.042247050994774}%
\StoreBenchExecResult{SvcompNineteenPdrInv}{KinductionDfReachsafetyLoops}{Error}{}{Walltime}{Max}{879.2505336030008}%
\StoreBenchExecResult{SvcompNineteenPdrInv}{KinductionDfReachsafetyLoops}{Error}{}{Walltime}{Stdev}{175.3389501822064974883570286}%
\StoreBenchExecResult{SvcompNineteenPdrInv}{KinductionDfReachsafetyLoops}{Error}{Error}{Count}{}{5}%
\StoreBenchExecResult{SvcompNineteenPdrInv}{KinductionDfReachsafetyLoops}{Error}{Error}{Cputime}{}{42.970628905}%
\StoreBenchExecResult{SvcompNineteenPdrInv}{KinductionDfReachsafetyLoops}{Error}{Error}{Cputime}{Avg}{8.594125781}%
\StoreBenchExecResult{SvcompNineteenPdrInv}{KinductionDfReachsafetyLoops}{Error}{Error}{Cputime}{Median}{8.826928744}%
\StoreBenchExecResult{SvcompNineteenPdrInv}{KinductionDfReachsafetyLoops}{Error}{Error}{Cputime}{Min}{7.72365526}%
\StoreBenchExecResult{SvcompNineteenPdrInv}{KinductionDfReachsafetyLoops}{Error}{Error}{Cputime}{Max}{9.33780541}%
\StoreBenchExecResult{SvcompNineteenPdrInv}{KinductionDfReachsafetyLoops}{Error}{Error}{Cputime}{Stdev}{0.5829865487319477226293157379}%
\StoreBenchExecResult{SvcompNineteenPdrInv}{KinductionDfReachsafetyLoops}{Error}{Error}{Walltime}{}{22.4371902829880125}%
\StoreBenchExecResult{SvcompNineteenPdrInv}{KinductionDfReachsafetyLoops}{Error}{Error}{Walltime}{Avg}{4.4874380565976025}%
\StoreBenchExecResult{SvcompNineteenPdrInv}{KinductionDfReachsafetyLoops}{Error}{Error}{Walltime}{Median}{4.609468067996204}%
\StoreBenchExecResult{SvcompNineteenPdrInv}{KinductionDfReachsafetyLoops}{Error}{Error}{Walltime}{Min}{4.042247050994774}%
\StoreBenchExecResult{SvcompNineteenPdrInv}{KinductionDfReachsafetyLoops}{Error}{Error}{Walltime}{Max}{4.8618512870016275}%
\StoreBenchExecResult{SvcompNineteenPdrInv}{KinductionDfReachsafetyLoops}{Error}{Error}{Walltime}{Stdev}{0.2992931240937007344205506998}%
\StoreBenchExecResult{SvcompNineteenPdrInv}{KinductionDfReachsafetyLoops}{Error}{OutOfMemory}{Count}{}{1}%
\StoreBenchExecResult{SvcompNineteenPdrInv}{KinductionDfReachsafetyLoops}{Error}{OutOfMemory}{Cputime}{}{185.945066399}%
\StoreBenchExecResult{SvcompNineteenPdrInv}{KinductionDfReachsafetyLoops}{Error}{OutOfMemory}{Cputime}{Avg}{185.945066399}%
\StoreBenchExecResult{SvcompNineteenPdrInv}{KinductionDfReachsafetyLoops}{Error}{OutOfMemory}{Cputime}{Median}{185.945066399}%
\StoreBenchExecResult{SvcompNineteenPdrInv}{KinductionDfReachsafetyLoops}{Error}{OutOfMemory}{Cputime}{Min}{185.945066399}%
\StoreBenchExecResult{SvcompNineteenPdrInv}{KinductionDfReachsafetyLoops}{Error}{OutOfMemory}{Cputime}{Max}{185.945066399}%
\StoreBenchExecResult{SvcompNineteenPdrInv}{KinductionDfReachsafetyLoops}{Error}{OutOfMemory}{Cputime}{Stdev}{0E-9}%
\StoreBenchExecResult{SvcompNineteenPdrInv}{KinductionDfReachsafetyLoops}{Error}{OutOfMemory}{Walltime}{}{174.4939333319926}%
\StoreBenchExecResult{SvcompNineteenPdrInv}{KinductionDfReachsafetyLoops}{Error}{OutOfMemory}{Walltime}{Avg}{174.4939333319926}%
\StoreBenchExecResult{SvcompNineteenPdrInv}{KinductionDfReachsafetyLoops}{Error}{OutOfMemory}{Walltime}{Median}{174.4939333319926}%
\StoreBenchExecResult{SvcompNineteenPdrInv}{KinductionDfReachsafetyLoops}{Error}{OutOfMemory}{Walltime}{Min}{174.4939333319926}%
\StoreBenchExecResult{SvcompNineteenPdrInv}{KinductionDfReachsafetyLoops}{Error}{OutOfMemory}{Walltime}{Max}{174.4939333319926}%
\StoreBenchExecResult{SvcompNineteenPdrInv}{KinductionDfReachsafetyLoops}{Error}{OutOfMemory}{Walltime}{Stdev}{0E-13}%
\StoreBenchExecResult{SvcompNineteenPdrInv}{KinductionDfReachsafetyLoops}{Error}{Timeout}{Count}{}{79}%
\StoreBenchExecResult{SvcompNineteenPdrInv}{KinductionDfReachsafetyLoops}{Error}{Timeout}{Cputime}{}{71409.427603904}%
\StoreBenchExecResult{SvcompNineteenPdrInv}{KinductionDfReachsafetyLoops}{Error}{Timeout}{Cputime}{Avg}{903.9168051127088607594936709}%
\StoreBenchExecResult{SvcompNineteenPdrInv}{KinductionDfReachsafetyLoops}{Error}{Timeout}{Cputime}{Median}{903.117745576}%
\StoreBenchExecResult{SvcompNineteenPdrInv}{KinductionDfReachsafetyLoops}{Error}{Timeout}{Cputime}{Min}{901.182823917}%
\StoreBenchExecResult{SvcompNineteenPdrInv}{KinductionDfReachsafetyLoops}{Error}{Timeout}{Cputime}{Max}{934.61590394}%
\StoreBenchExecResult{SvcompNineteenPdrInv}{KinductionDfReachsafetyLoops}{Error}{Timeout}{Cputime}{Stdev}{3.901740830451098368191828637}%
\StoreBenchExecResult{SvcompNineteenPdrInv}{KinductionDfReachsafetyLoops}{Error}{Timeout}{Walltime}{}{39913.72363881296768}%
\StoreBenchExecResult{SvcompNineteenPdrInv}{KinductionDfReachsafetyLoops}{Error}{Timeout}{Walltime}{Avg}{505.2370080862400972151898734}%
\StoreBenchExecResult{SvcompNineteenPdrInv}{KinductionDfReachsafetyLoops}{Error}{Timeout}{Walltime}{Median}{453.5618239809992}%
\StoreBenchExecResult{SvcompNineteenPdrInv}{KinductionDfReachsafetyLoops}{Error}{Timeout}{Walltime}{Min}{451.190239019008}%
\StoreBenchExecResult{SvcompNineteenPdrInv}{KinductionDfReachsafetyLoops}{Error}{Timeout}{Walltime}{Max}{879.2505336030008}%
\StoreBenchExecResult{SvcompNineteenPdrInv}{KinductionDfReachsafetyLoops}{Error}{Timeout}{Walltime}{Stdev}{130.4615990598684345444069370}%
\providecommand\StoreBenchExecResult[7]{\expandafter\newcommand\csname#1#2#3#4#5#6\endcsname{#7}}%
\StoreBenchExecResult{SvcompNineteenPdrInv}{KinductionKipdrReachsafetyLoops}{Total}{}{Count}{}{208}%
\StoreBenchExecResult{SvcompNineteenPdrInv}{KinductionKipdrReachsafetyLoops}{Total}{}{Cputime}{}{79097.727394688}%
\StoreBenchExecResult{SvcompNineteenPdrInv}{KinductionKipdrReachsafetyLoops}{Total}{}{Cputime}{Avg}{380.2775355513846153846153846}%
\StoreBenchExecResult{SvcompNineteenPdrInv}{KinductionKipdrReachsafetyLoops}{Total}{}{Cputime}{Median}{32.1324920725}%
\StoreBenchExecResult{SvcompNineteenPdrInv}{KinductionKipdrReachsafetyLoops}{Total}{}{Cputime}{Min}{4.091353234}%
\StoreBenchExecResult{SvcompNineteenPdrInv}{KinductionKipdrReachsafetyLoops}{Total}{}{Cputime}{Max}{921.503616529}%
\StoreBenchExecResult{SvcompNineteenPdrInv}{KinductionKipdrReachsafetyLoops}{Total}{}{Cputime}{Stdev}{437.2989939399087700657702632}%
\StoreBenchExecResult{SvcompNineteenPdrInv}{KinductionKipdrReachsafetyLoops}{Total}{}{Walltime}{}{39941.4423969879135778}%
\StoreBenchExecResult{SvcompNineteenPdrInv}{KinductionKipdrReachsafetyLoops}{Total}{}{Walltime}{Avg}{192.0261653701341998932692308}%
\StoreBenchExecResult{SvcompNineteenPdrInv}{KinductionKipdrReachsafetyLoops}{Total}{}{Walltime}{Median}{16.3221484785026405}%
\StoreBenchExecResult{SvcompNineteenPdrInv}{KinductionKipdrReachsafetyLoops}{Total}{}{Walltime}{Min}{2.186167938009021}%
\StoreBenchExecResult{SvcompNineteenPdrInv}{KinductionKipdrReachsafetyLoops}{Total}{}{Walltime}{Max}{716.646359025006}%
\StoreBenchExecResult{SvcompNineteenPdrInv}{KinductionKipdrReachsafetyLoops}{Total}{}{Walltime}{Stdev}{221.3592527221838420031012964}%
\StoreBenchExecResult{SvcompNineteenPdrInv}{KinductionKipdrReachsafetyLoops}{Correct}{}{Count}{}{114}%
\StoreBenchExecResult{SvcompNineteenPdrInv}{KinductionKipdrReachsafetyLoops}{Correct}{}{Cputime}{}{1600.917340279}%
\StoreBenchExecResult{SvcompNineteenPdrInv}{KinductionKipdrReachsafetyLoops}{Correct}{}{Cputime}{Avg}{14.04313456385087719298245614}%
\StoreBenchExecResult{SvcompNineteenPdrInv}{KinductionKipdrReachsafetyLoops}{Correct}{}{Cputime}{Median}{5.528112843}%
\StoreBenchExecResult{SvcompNineteenPdrInv}{KinductionKipdrReachsafetyLoops}{Correct}{}{Cputime}{Min}{4.091353234}%
\StoreBenchExecResult{SvcompNineteenPdrInv}{KinductionKipdrReachsafetyLoops}{Correct}{}{Cputime}{Max}{110.015123084}%
\StoreBenchExecResult{SvcompNineteenPdrInv}{KinductionKipdrReachsafetyLoops}{Correct}{}{Cputime}{Stdev}{19.87136794527928237462585727}%
\StoreBenchExecResult{SvcompNineteenPdrInv}{KinductionKipdrReachsafetyLoops}{Correct}{}{Walltime}{}{829.8410576200112654}%
\StoreBenchExecResult{SvcompNineteenPdrInv}{KinductionKipdrReachsafetyLoops}{Correct}{}{Walltime}{Avg}{7.279307522982554959649122807}%
\StoreBenchExecResult{SvcompNineteenPdrInv}{KinductionKipdrReachsafetyLoops}{Correct}{}{Walltime}{Median}{2.91611005400045535}%
\StoreBenchExecResult{SvcompNineteenPdrInv}{KinductionKipdrReachsafetyLoops}{Correct}{}{Walltime}{Min}{2.186167938009021}%
\StoreBenchExecResult{SvcompNineteenPdrInv}{KinductionKipdrReachsafetyLoops}{Correct}{}{Walltime}{Max}{55.24620384699665}%
\StoreBenchExecResult{SvcompNineteenPdrInv}{KinductionKipdrReachsafetyLoops}{Correct}{}{Walltime}{Stdev}{10.14930910758983663774946180}%
\StoreBenchExecResult{SvcompNineteenPdrInv}{KinductionKipdrReachsafetyLoops}{Correct}{False}{Count}{}{36}%
\StoreBenchExecResult{SvcompNineteenPdrInv}{KinductionKipdrReachsafetyLoops}{Correct}{False}{Cputime}{}{384.836274398}%
\StoreBenchExecResult{SvcompNineteenPdrInv}{KinductionKipdrReachsafetyLoops}{Correct}{False}{Cputime}{Avg}{10.68989651105555555555555556}%
\StoreBenchExecResult{SvcompNineteenPdrInv}{KinductionKipdrReachsafetyLoops}{Correct}{False}{Cputime}{Median}{5.243468615}%
\StoreBenchExecResult{SvcompNineteenPdrInv}{KinductionKipdrReachsafetyLoops}{Correct}{False}{Cputime}{Min}{4.218034337}%
\StoreBenchExecResult{SvcompNineteenPdrInv}{KinductionKipdrReachsafetyLoops}{Correct}{False}{Cputime}{Max}{64.856799181}%
\StoreBenchExecResult{SvcompNineteenPdrInv}{KinductionKipdrReachsafetyLoops}{Correct}{False}{Cputime}{Stdev}{14.16274745742660872394886428}%
\StoreBenchExecResult{SvcompNineteenPdrInv}{KinductionKipdrReachsafetyLoops}{Correct}{False}{Walltime}{}{206.0652520259318420}%
\StoreBenchExecResult{SvcompNineteenPdrInv}{KinductionKipdrReachsafetyLoops}{Correct}{False}{Walltime}{Avg}{5.724034778498106722222222222}%
\StoreBenchExecResult{SvcompNineteenPdrInv}{KinductionKipdrReachsafetyLoops}{Correct}{False}{Walltime}{Median}{2.7774848174958606}%
\StoreBenchExecResult{SvcompNineteenPdrInv}{KinductionKipdrReachsafetyLoops}{Correct}{False}{Walltime}{Min}{2.2463047199998982}%
\StoreBenchExecResult{SvcompNineteenPdrInv}{KinductionKipdrReachsafetyLoops}{Correct}{False}{Walltime}{Max}{39.16983112799062}%
\StoreBenchExecResult{SvcompNineteenPdrInv}{KinductionKipdrReachsafetyLoops}{Correct}{False}{Walltime}{Stdev}{7.912358324773184347699738966}%
\StoreBenchExecResult{SvcompNineteenPdrInv}{KinductionKipdrReachsafetyLoops}{Wrong}{False}{Count}{}{0}%
\StoreBenchExecResult{SvcompNineteenPdrInv}{KinductionKipdrReachsafetyLoops}{Wrong}{False}{Cputime}{}{0}%
\StoreBenchExecResult{SvcompNineteenPdrInv}{KinductionKipdrReachsafetyLoops}{Wrong}{False}{Cputime}{Avg}{None}%
\StoreBenchExecResult{SvcompNineteenPdrInv}{KinductionKipdrReachsafetyLoops}{Wrong}{False}{Cputime}{Median}{None}%
\StoreBenchExecResult{SvcompNineteenPdrInv}{KinductionKipdrReachsafetyLoops}{Wrong}{False}{Cputime}{Min}{None}%
\StoreBenchExecResult{SvcompNineteenPdrInv}{KinductionKipdrReachsafetyLoops}{Wrong}{False}{Cputime}{Max}{None}%
\StoreBenchExecResult{SvcompNineteenPdrInv}{KinductionKipdrReachsafetyLoops}{Wrong}{False}{Cputime}{Stdev}{None}%
\StoreBenchExecResult{SvcompNineteenPdrInv}{KinductionKipdrReachsafetyLoops}{Wrong}{False}{Walltime}{}{0}%
\StoreBenchExecResult{SvcompNineteenPdrInv}{KinductionKipdrReachsafetyLoops}{Wrong}{False}{Walltime}{Avg}{None}%
\StoreBenchExecResult{SvcompNineteenPdrInv}{KinductionKipdrReachsafetyLoops}{Wrong}{False}{Walltime}{Median}{None}%
\StoreBenchExecResult{SvcompNineteenPdrInv}{KinductionKipdrReachsafetyLoops}{Wrong}{False}{Walltime}{Min}{None}%
\StoreBenchExecResult{SvcompNineteenPdrInv}{KinductionKipdrReachsafetyLoops}{Wrong}{False}{Walltime}{Max}{None}%
\StoreBenchExecResult{SvcompNineteenPdrInv}{KinductionKipdrReachsafetyLoops}{Wrong}{False}{Walltime}{Stdev}{None}%
\StoreBenchExecResult{SvcompNineteenPdrInv}{KinductionKipdrReachsafetyLoops}{Correct}{True}{Count}{}{78}%
\StoreBenchExecResult{SvcompNineteenPdrInv}{KinductionKipdrReachsafetyLoops}{Correct}{True}{Cputime}{}{1216.081065881}%
\StoreBenchExecResult{SvcompNineteenPdrInv}{KinductionKipdrReachsafetyLoops}{Correct}{True}{Cputime}{Avg}{15.59078289591025641025641026}%
\StoreBenchExecResult{SvcompNineteenPdrInv}{KinductionKipdrReachsafetyLoops}{Correct}{True}{Cputime}{Median}{5.693433932}%
\StoreBenchExecResult{SvcompNineteenPdrInv}{KinductionKipdrReachsafetyLoops}{Correct}{True}{Cputime}{Min}{4.091353234}%
\StoreBenchExecResult{SvcompNineteenPdrInv}{KinductionKipdrReachsafetyLoops}{Correct}{True}{Cputime}{Max}{110.015123084}%
\StoreBenchExecResult{SvcompNineteenPdrInv}{KinductionKipdrReachsafetyLoops}{Correct}{True}{Cputime}{Stdev}{21.83936194255154709816985673}%
\StoreBenchExecResult{SvcompNineteenPdrInv}{KinductionKipdrReachsafetyLoops}{Correct}{True}{Walltime}{}{623.7758055940794234}%
\StoreBenchExecResult{SvcompNineteenPdrInv}{KinductionKipdrReachsafetyLoops}{Correct}{True}{Walltime}{Avg}{7.997125712744607992307692308}%
\StoreBenchExecResult{SvcompNineteenPdrInv}{KinductionKipdrReachsafetyLoops}{Correct}{True}{Walltime}{Median}{3.01201051900716265}%
\StoreBenchExecResult{SvcompNineteenPdrInv}{KinductionKipdrReachsafetyLoops}{Correct}{True}{Walltime}{Min}{2.186167938009021}%
\StoreBenchExecResult{SvcompNineteenPdrInv}{KinductionKipdrReachsafetyLoops}{Correct}{True}{Walltime}{Max}{55.24620384699665}%
\StoreBenchExecResult{SvcompNineteenPdrInv}{KinductionKipdrReachsafetyLoops}{Correct}{True}{Walltime}{Stdev}{10.95556370088683224876139810}%
\StoreBenchExecResult{SvcompNineteenPdrInv}{KinductionKipdrReachsafetyLoops}{Wrong}{True}{Count}{}{0}%
\StoreBenchExecResult{SvcompNineteenPdrInv}{KinductionKipdrReachsafetyLoops}{Wrong}{True}{Cputime}{}{0}%
\StoreBenchExecResult{SvcompNineteenPdrInv}{KinductionKipdrReachsafetyLoops}{Wrong}{True}{Cputime}{Avg}{None}%
\StoreBenchExecResult{SvcompNineteenPdrInv}{KinductionKipdrReachsafetyLoops}{Wrong}{True}{Cputime}{Median}{None}%
\StoreBenchExecResult{SvcompNineteenPdrInv}{KinductionKipdrReachsafetyLoops}{Wrong}{True}{Cputime}{Min}{None}%
\StoreBenchExecResult{SvcompNineteenPdrInv}{KinductionKipdrReachsafetyLoops}{Wrong}{True}{Cputime}{Max}{None}%
\StoreBenchExecResult{SvcompNineteenPdrInv}{KinductionKipdrReachsafetyLoops}{Wrong}{True}{Cputime}{Stdev}{None}%
\StoreBenchExecResult{SvcompNineteenPdrInv}{KinductionKipdrReachsafetyLoops}{Wrong}{True}{Walltime}{}{0}%
\StoreBenchExecResult{SvcompNineteenPdrInv}{KinductionKipdrReachsafetyLoops}{Wrong}{True}{Walltime}{Avg}{None}%
\StoreBenchExecResult{SvcompNineteenPdrInv}{KinductionKipdrReachsafetyLoops}{Wrong}{True}{Walltime}{Median}{None}%
\StoreBenchExecResult{SvcompNineteenPdrInv}{KinductionKipdrReachsafetyLoops}{Wrong}{True}{Walltime}{Min}{None}%
\StoreBenchExecResult{SvcompNineteenPdrInv}{KinductionKipdrReachsafetyLoops}{Wrong}{True}{Walltime}{Max}{None}%
\StoreBenchExecResult{SvcompNineteenPdrInv}{KinductionKipdrReachsafetyLoops}{Wrong}{True}{Walltime}{Stdev}{None}%
\StoreBenchExecResult{SvcompNineteenPdrInv}{KinductionKipdrReachsafetyLoops}{Error}{}{Count}{}{94}%
\StoreBenchExecResult{SvcompNineteenPdrInv}{KinductionKipdrReachsafetyLoops}{Error}{}{Cputime}{}{77496.810054409}%
\StoreBenchExecResult{SvcompNineteenPdrInv}{KinductionKipdrReachsafetyLoops}{Error}{}{Cputime}{Avg}{824.4341495149893617021276596}%
\StoreBenchExecResult{SvcompNineteenPdrInv}{KinductionKipdrReachsafetyLoops}{Error}{}{Cputime}{Median}{903.754299653}%
\StoreBenchExecResult{SvcompNineteenPdrInv}{KinductionKipdrReachsafetyLoops}{Error}{}{Cputime}{Min}{6.475513298}%
\StoreBenchExecResult{SvcompNineteenPdrInv}{KinductionKipdrReachsafetyLoops}{Error}{}{Cputime}{Max}{921.503616529}%
\StoreBenchExecResult{SvcompNineteenPdrInv}{KinductionKipdrReachsafetyLoops}{Error}{}{Cputime}{Stdev}{250.4570340117586128080647090}%
\StoreBenchExecResult{SvcompNineteenPdrInv}{KinductionKipdrReachsafetyLoops}{Error}{}{Walltime}{}{39111.6013393679023124}%
\StoreBenchExecResult{SvcompNineteenPdrInv}{KinductionKipdrReachsafetyLoops}{Error}{}{Walltime}{Avg}{416.0808653124244926851063830}%
\StoreBenchExecResult{SvcompNineteenPdrInv}{KinductionKipdrReachsafetyLoops}{Error}{}{Walltime}{Median}{452.606171987499685}%
\StoreBenchExecResult{SvcompNineteenPdrInv}{KinductionKipdrReachsafetyLoops}{Error}{}{Walltime}{Min}{3.4060808659996837}%
\StoreBenchExecResult{SvcompNineteenPdrInv}{KinductionKipdrReachsafetyLoops}{Error}{}{Walltime}{Max}{716.646359025006}%
\StoreBenchExecResult{SvcompNineteenPdrInv}{KinductionKipdrReachsafetyLoops}{Error}{}{Walltime}{Stdev}{129.2536909482714632097465842}%
\StoreBenchExecResult{SvcompNineteenPdrInv}{KinductionKipdrReachsafetyLoops}{Error}{Error}{Count}{}{5}%
\StoreBenchExecResult{SvcompNineteenPdrInv}{KinductionKipdrReachsafetyLoops}{Error}{Error}{Cputime}{}{39.464952922}%
\StoreBenchExecResult{SvcompNineteenPdrInv}{KinductionKipdrReachsafetyLoops}{Error}{Error}{Cputime}{Avg}{7.8929905844}%
\StoreBenchExecResult{SvcompNineteenPdrInv}{KinductionKipdrReachsafetyLoops}{Error}{Error}{Cputime}{Median}{7.719147481}%
\StoreBenchExecResult{SvcompNineteenPdrInv}{KinductionKipdrReachsafetyLoops}{Error}{Error}{Cputime}{Min}{7.128662005}%
\StoreBenchExecResult{SvcompNineteenPdrInv}{KinductionKipdrReachsafetyLoops}{Error}{Error}{Cputime}{Max}{8.982107397}%
\StoreBenchExecResult{SvcompNineteenPdrInv}{KinductionKipdrReachsafetyLoops}{Error}{Error}{Cputime}{Stdev}{0.6120255226650207435285251149}%
\StoreBenchExecResult{SvcompNineteenPdrInv}{KinductionKipdrReachsafetyLoops}{Error}{Error}{Walltime}{}{20.6179368739976770}%
\StoreBenchExecResult{SvcompNineteenPdrInv}{KinductionKipdrReachsafetyLoops}{Error}{Error}{Walltime}{Avg}{4.1235873747995354}%
\StoreBenchExecResult{SvcompNineteenPdrInv}{KinductionKipdrReachsafetyLoops}{Error}{Error}{Walltime}{Median}{4.029691502000787}%
\StoreBenchExecResult{SvcompNineteenPdrInv}{KinductionKipdrReachsafetyLoops}{Error}{Error}{Walltime}{Min}{3.7502993419911945}%
\StoreBenchExecResult{SvcompNineteenPdrInv}{KinductionKipdrReachsafetyLoops}{Error}{Error}{Walltime}{Max}{4.669572308004717}%
\StoreBenchExecResult{SvcompNineteenPdrInv}{KinductionKipdrReachsafetyLoops}{Error}{Error}{Walltime}{Stdev}{0.3042435651970596154701042843}%
\StoreBenchExecResult{SvcompNineteenPdrInv}{KinductionKipdrReachsafetyLoops}{Error}{OutOfMemory}{Count}{}{2}%
\StoreBenchExecResult{SvcompNineteenPdrInv}{KinductionKipdrReachsafetyLoops}{Error}{OutOfMemory}{Cputime}{}{492.721369582}%
\StoreBenchExecResult{SvcompNineteenPdrInv}{KinductionKipdrReachsafetyLoops}{Error}{OutOfMemory}{Cputime}{Avg}{246.360684791}%
\StoreBenchExecResult{SvcompNineteenPdrInv}{KinductionKipdrReachsafetyLoops}{Error}{OutOfMemory}{Cputime}{Median}{246.360684791}%
\StoreBenchExecResult{SvcompNineteenPdrInv}{KinductionKipdrReachsafetyLoops}{Error}{OutOfMemory}{Cputime}{Min}{238.333625179}%
\StoreBenchExecResult{SvcompNineteenPdrInv}{KinductionKipdrReachsafetyLoops}{Error}{OutOfMemory}{Cputime}{Max}{254.387744403}%
\StoreBenchExecResult{SvcompNineteenPdrInv}{KinductionKipdrReachsafetyLoops}{Error}{OutOfMemory}{Cputime}{Stdev}{8.027059612}%
\StoreBenchExecResult{SvcompNineteenPdrInv}{KinductionKipdrReachsafetyLoops}{Error}{OutOfMemory}{Walltime}{}{247.28456520198961}%
\StoreBenchExecResult{SvcompNineteenPdrInv}{KinductionKipdrReachsafetyLoops}{Error}{OutOfMemory}{Walltime}{Avg}{123.642282600994805}%
\StoreBenchExecResult{SvcompNineteenPdrInv}{KinductionKipdrReachsafetyLoops}{Error}{OutOfMemory}{Walltime}{Median}{123.642282600994805}%
\StoreBenchExecResult{SvcompNineteenPdrInv}{KinductionKipdrReachsafetyLoops}{Error}{OutOfMemory}{Walltime}{Min}{119.64219574299932}%
\StoreBenchExecResult{SvcompNineteenPdrInv}{KinductionKipdrReachsafetyLoops}{Error}{OutOfMemory}{Walltime}{Max}{127.64236945899029}%
\StoreBenchExecResult{SvcompNineteenPdrInv}{KinductionKipdrReachsafetyLoops}{Error}{OutOfMemory}{Walltime}{Stdev}{4.000086857995485000000000001}%
\StoreBenchExecResult{SvcompNineteenPdrInv}{KinductionKipdrReachsafetyLoops}{Error}{SegmentationFault}{Count}{}{2}%
\StoreBenchExecResult{SvcompNineteenPdrInv}{KinductionKipdrReachsafetyLoops}{Error}{SegmentationFault}{Cputime}{}{13.467177221}%
\StoreBenchExecResult{SvcompNineteenPdrInv}{KinductionKipdrReachsafetyLoops}{Error}{SegmentationFault}{Cputime}{Avg}{6.7335886105}%
\StoreBenchExecResult{SvcompNineteenPdrInv}{KinductionKipdrReachsafetyLoops}{Error}{SegmentationFault}{Cputime}{Median}{6.7335886105}%
\StoreBenchExecResult{SvcompNineteenPdrInv}{KinductionKipdrReachsafetyLoops}{Error}{SegmentationFault}{Cputime}{Min}{6.475513298}%
\StoreBenchExecResult{SvcompNineteenPdrInv}{KinductionKipdrReachsafetyLoops}{Error}{SegmentationFault}{Cputime}{Max}{6.991663923}%
\StoreBenchExecResult{SvcompNineteenPdrInv}{KinductionKipdrReachsafetyLoops}{Error}{SegmentationFault}{Cputime}{Stdev}{0.2580753125}%
\StoreBenchExecResult{SvcompNineteenPdrInv}{KinductionKipdrReachsafetyLoops}{Error}{SegmentationFault}{Walltime}{}{7.0562976700020954}%
\StoreBenchExecResult{SvcompNineteenPdrInv}{KinductionKipdrReachsafetyLoops}{Error}{SegmentationFault}{Walltime}{Avg}{3.5281488350010477}%
\StoreBenchExecResult{SvcompNineteenPdrInv}{KinductionKipdrReachsafetyLoops}{Error}{SegmentationFault}{Walltime}{Median}{3.5281488350010477}%
\StoreBenchExecResult{SvcompNineteenPdrInv}{KinductionKipdrReachsafetyLoops}{Error}{SegmentationFault}{Walltime}{Min}{3.4060808659996837}%
\StoreBenchExecResult{SvcompNineteenPdrInv}{KinductionKipdrReachsafetyLoops}{Error}{SegmentationFault}{Walltime}{Max}{3.6502168040024117}%
\StoreBenchExecResult{SvcompNineteenPdrInv}{KinductionKipdrReachsafetyLoops}{Error}{SegmentationFault}{Walltime}{Stdev}{0.1220679690013640000000000000}%
\StoreBenchExecResult{SvcompNineteenPdrInv}{KinductionKipdrReachsafetyLoops}{Error}{Timeout}{Count}{}{85}%
\StoreBenchExecResult{SvcompNineteenPdrInv}{KinductionKipdrReachsafetyLoops}{Error}{Timeout}{Cputime}{}{76951.156554684}%
\StoreBenchExecResult{SvcompNineteenPdrInv}{KinductionKipdrReachsafetyLoops}{Error}{Timeout}{Cputime}{Avg}{905.3077241727529411764705882}%
\StoreBenchExecResult{SvcompNineteenPdrInv}{KinductionKipdrReachsafetyLoops}{Error}{Timeout}{Cputime}{Median}{903.946185976}%
\StoreBenchExecResult{SvcompNineteenPdrInv}{KinductionKipdrReachsafetyLoops}{Error}{Timeout}{Cputime}{Min}{901.487323184}%
\StoreBenchExecResult{SvcompNineteenPdrInv}{KinductionKipdrReachsafetyLoops}{Error}{Timeout}{Cputime}{Max}{921.503616529}%
\StoreBenchExecResult{SvcompNineteenPdrInv}{KinductionKipdrReachsafetyLoops}{Error}{Timeout}{Cputime}{Stdev}{3.630396545771910336945047857}%
\StoreBenchExecResult{SvcompNineteenPdrInv}{KinductionKipdrReachsafetyLoops}{Error}{Timeout}{Walltime}{}{38836.64253962191293}%
\StoreBenchExecResult{SvcompNineteenPdrInv}{KinductionKipdrReachsafetyLoops}{Error}{Timeout}{Walltime}{Avg}{456.9016769367283874117647059}%
\StoreBenchExecResult{SvcompNineteenPdrInv}{KinductionKipdrReachsafetyLoops}{Error}{Timeout}{Walltime}{Median}{452.96576805198856}%
\StoreBenchExecResult{SvcompNineteenPdrInv}{KinductionKipdrReachsafetyLoops}{Error}{Timeout}{Walltime}{Min}{451.2513497409964}%
\StoreBenchExecResult{SvcompNineteenPdrInv}{KinductionKipdrReachsafetyLoops}{Error}{Timeout}{Walltime}{Max}{716.646359025006}%
\StoreBenchExecResult{SvcompNineteenPdrInv}{KinductionKipdrReachsafetyLoops}{Error}{Timeout}{Walltime}{Stdev}{28.44130200710534878589995256}%
\providecommand\StoreBenchExecResult[7]{\expandafter\newcommand\csname#1#2#3#4#5#6\endcsname{#7}}%
\StoreBenchExecResult{SvcompNineteenPdrInv}{KinductionMaxkKipdrMaxkReachsafetyLoops}{Total}{}{Count}{}{208}%
\StoreBenchExecResult{SvcompNineteenPdrInv}{KinductionMaxkKipdrMaxkReachsafetyLoops}{Total}{}{Cputime}{}{10857.408195942}%
\StoreBenchExecResult{SvcompNineteenPdrInv}{KinductionMaxkKipdrMaxkReachsafetyLoops}{Total}{}{Cputime}{Avg}{52.19907786510576923076923077}%
\StoreBenchExecResult{SvcompNineteenPdrInv}{KinductionMaxkKipdrMaxkReachsafetyLoops}{Total}{}{Cputime}{Median}{5.5678229805}%
\StoreBenchExecResult{SvcompNineteenPdrInv}{KinductionMaxkKipdrMaxkReachsafetyLoops}{Total}{}{Cputime}{Min}{4.103004202}%
\StoreBenchExecResult{SvcompNineteenPdrInv}{KinductionMaxkKipdrMaxkReachsafetyLoops}{Total}{}{Cputime}{Max}{905.582207567}%
\StoreBenchExecResult{SvcompNineteenPdrInv}{KinductionMaxkKipdrMaxkReachsafetyLoops}{Total}{}{Cputime}{Stdev}{193.0309891586687184441266600}%
\StoreBenchExecResult{SvcompNineteenPdrInv}{KinductionMaxkKipdrMaxkReachsafetyLoops}{Total}{}{Walltime}{}{5473.6374551351181895}%
\StoreBenchExecResult{SvcompNineteenPdrInv}{KinductionMaxkKipdrMaxkReachsafetyLoops}{Total}{}{Walltime}{Avg}{26.31556468814960668028846154}%
\StoreBenchExecResult{SvcompNineteenPdrInv}{KinductionMaxkKipdrMaxkReachsafetyLoops}{Total}{}{Walltime}{Median}{2.9597571045014775}%
\StoreBenchExecResult{SvcompNineteenPdrInv}{KinductionMaxkKipdrMaxkReachsafetyLoops}{Total}{}{Walltime}{Min}{2.1782093760120915}%
\StoreBenchExecResult{SvcompNineteenPdrInv}{KinductionMaxkKipdrMaxkReachsafetyLoops}{Total}{}{Walltime}{Max}{455.86239988899615}%
\StoreBenchExecResult{SvcompNineteenPdrInv}{KinductionMaxkKipdrMaxkReachsafetyLoops}{Total}{}{Walltime}{Stdev}{96.70346777728394907624010460}%
\StoreBenchExecResult{SvcompNineteenPdrInv}{KinductionMaxkKipdrMaxkReachsafetyLoops}{Correct}{}{Count}{}{65}%
\StoreBenchExecResult{SvcompNineteenPdrInv}{KinductionMaxkKipdrMaxkReachsafetyLoops}{Correct}{}{Cputime}{}{422.270114270}%
\StoreBenchExecResult{SvcompNineteenPdrInv}{KinductionMaxkKipdrMaxkReachsafetyLoops}{Correct}{}{Cputime}{Avg}{6.496463296461538461538461538}%
\StoreBenchExecResult{SvcompNineteenPdrInv}{KinductionMaxkKipdrMaxkReachsafetyLoops}{Correct}{}{Cputime}{Median}{4.955638738}%
\StoreBenchExecResult{SvcompNineteenPdrInv}{KinductionMaxkKipdrMaxkReachsafetyLoops}{Correct}{}{Cputime}{Min}{4.103004202}%
\StoreBenchExecResult{SvcompNineteenPdrInv}{KinductionMaxkKipdrMaxkReachsafetyLoops}{Correct}{}{Cputime}{Max}{38.849768044}%
\StoreBenchExecResult{SvcompNineteenPdrInv}{KinductionMaxkKipdrMaxkReachsafetyLoops}{Correct}{}{Cputime}{Stdev}{5.561099581415369934499870096}%
\StoreBenchExecResult{SvcompNineteenPdrInv}{KinductionMaxkKipdrMaxkReachsafetyLoops}{Correct}{}{Walltime}{}{222.4703679800150006}%
\StoreBenchExecResult{SvcompNineteenPdrInv}{KinductionMaxkKipdrMaxkReachsafetyLoops}{Correct}{}{Walltime}{Avg}{3.422621045846384624615384615}%
\StoreBenchExecResult{SvcompNineteenPdrInv}{KinductionMaxkKipdrMaxkReachsafetyLoops}{Correct}{}{Walltime}{Median}{2.6499554100009846}%
\StoreBenchExecResult{SvcompNineteenPdrInv}{KinductionMaxkKipdrMaxkReachsafetyLoops}{Correct}{}{Walltime}{Min}{2.1782093760120915}%
\StoreBenchExecResult{SvcompNineteenPdrInv}{KinductionMaxkKipdrMaxkReachsafetyLoops}{Correct}{}{Walltime}{Max}{19.6571699979977}%
\StoreBenchExecResult{SvcompNineteenPdrInv}{KinductionMaxkKipdrMaxkReachsafetyLoops}{Correct}{}{Walltime}{Stdev}{2.802322714398554015698179969}%
\StoreBenchExecResult{SvcompNineteenPdrInv}{KinductionMaxkKipdrMaxkReachsafetyLoops}{Correct}{False}{Count}{}{22}%
\StoreBenchExecResult{SvcompNineteenPdrInv}{KinductionMaxkKipdrMaxkReachsafetyLoops}{Correct}{False}{Cputime}{}{112.183896237}%
\StoreBenchExecResult{SvcompNineteenPdrInv}{KinductionMaxkKipdrMaxkReachsafetyLoops}{Correct}{False}{Cputime}{Avg}{5.099268010772727272727272727}%
\StoreBenchExecResult{SvcompNineteenPdrInv}{KinductionMaxkKipdrMaxkReachsafetyLoops}{Correct}{False}{Cputime}{Median}{4.7081338275}%
\StoreBenchExecResult{SvcompNineteenPdrInv}{KinductionMaxkKipdrMaxkReachsafetyLoops}{Correct}{False}{Cputime}{Min}{4.151785154}%
\StoreBenchExecResult{SvcompNineteenPdrInv}{KinductionMaxkKipdrMaxkReachsafetyLoops}{Correct}{False}{Cputime}{Max}{8.453906758}%
\StoreBenchExecResult{SvcompNineteenPdrInv}{KinductionMaxkKipdrMaxkReachsafetyLoops}{Correct}{False}{Cputime}{Stdev}{1.072951645471726960617724449}%
\StoreBenchExecResult{SvcompNineteenPdrInv}{KinductionMaxkKipdrMaxkReachsafetyLoops}{Correct}{False}{Walltime}{}{59.5689330959867220}%
\StoreBenchExecResult{SvcompNineteenPdrInv}{KinductionMaxkKipdrMaxkReachsafetyLoops}{Correct}{False}{Walltime}{Avg}{2.707678777090305545454545455}%
\StoreBenchExecResult{SvcompNineteenPdrInv}{KinductionMaxkKipdrMaxkReachsafetyLoops}{Correct}{False}{Walltime}{Median}{2.5247777924960247}%
\StoreBenchExecResult{SvcompNineteenPdrInv}{KinductionMaxkKipdrMaxkReachsafetyLoops}{Correct}{False}{Walltime}{Min}{2.2062276910000946}%
\StoreBenchExecResult{SvcompNineteenPdrInv}{KinductionMaxkKipdrMaxkReachsafetyLoops}{Correct}{False}{Walltime}{Max}{4.378346280005644}%
\StoreBenchExecResult{SvcompNineteenPdrInv}{KinductionMaxkKipdrMaxkReachsafetyLoops}{Correct}{False}{Walltime}{Stdev}{0.5412363624835912082461413296}%
\StoreBenchExecResult{SvcompNineteenPdrInv}{KinductionMaxkKipdrMaxkReachsafetyLoops}{Wrong}{False}{Count}{}{0}%
\StoreBenchExecResult{SvcompNineteenPdrInv}{KinductionMaxkKipdrMaxkReachsafetyLoops}{Wrong}{False}{Cputime}{}{0}%
\StoreBenchExecResult{SvcompNineteenPdrInv}{KinductionMaxkKipdrMaxkReachsafetyLoops}{Wrong}{False}{Cputime}{Avg}{None}%
\StoreBenchExecResult{SvcompNineteenPdrInv}{KinductionMaxkKipdrMaxkReachsafetyLoops}{Wrong}{False}{Cputime}{Median}{None}%
\StoreBenchExecResult{SvcompNineteenPdrInv}{KinductionMaxkKipdrMaxkReachsafetyLoops}{Wrong}{False}{Cputime}{Min}{None}%
\StoreBenchExecResult{SvcompNineteenPdrInv}{KinductionMaxkKipdrMaxkReachsafetyLoops}{Wrong}{False}{Cputime}{Max}{None}%
\StoreBenchExecResult{SvcompNineteenPdrInv}{KinductionMaxkKipdrMaxkReachsafetyLoops}{Wrong}{False}{Cputime}{Stdev}{None}%
\StoreBenchExecResult{SvcompNineteenPdrInv}{KinductionMaxkKipdrMaxkReachsafetyLoops}{Wrong}{False}{Walltime}{}{0}%
\StoreBenchExecResult{SvcompNineteenPdrInv}{KinductionMaxkKipdrMaxkReachsafetyLoops}{Wrong}{False}{Walltime}{Avg}{None}%
\StoreBenchExecResult{SvcompNineteenPdrInv}{KinductionMaxkKipdrMaxkReachsafetyLoops}{Wrong}{False}{Walltime}{Median}{None}%
\StoreBenchExecResult{SvcompNineteenPdrInv}{KinductionMaxkKipdrMaxkReachsafetyLoops}{Wrong}{False}{Walltime}{Min}{None}%
\StoreBenchExecResult{SvcompNineteenPdrInv}{KinductionMaxkKipdrMaxkReachsafetyLoops}{Wrong}{False}{Walltime}{Max}{None}%
\StoreBenchExecResult{SvcompNineteenPdrInv}{KinductionMaxkKipdrMaxkReachsafetyLoops}{Wrong}{False}{Walltime}{Stdev}{None}%
\StoreBenchExecResult{SvcompNineteenPdrInv}{KinductionMaxkKipdrMaxkReachsafetyLoops}{Correct}{True}{Count}{}{43}%
\StoreBenchExecResult{SvcompNineteenPdrInv}{KinductionMaxkKipdrMaxkReachsafetyLoops}{Correct}{True}{Cputime}{}{310.086218033}%
\StoreBenchExecResult{SvcompNineteenPdrInv}{KinductionMaxkKipdrMaxkReachsafetyLoops}{Correct}{True}{Cputime}{Avg}{7.211307396116279069767441860}%
\StoreBenchExecResult{SvcompNineteenPdrInv}{KinductionMaxkKipdrMaxkReachsafetyLoops}{Correct}{True}{Cputime}{Median}{5.102181066}%
\StoreBenchExecResult{SvcompNineteenPdrInv}{KinductionMaxkKipdrMaxkReachsafetyLoops}{Correct}{True}{Cputime}{Min}{4.103004202}%
\StoreBenchExecResult{SvcompNineteenPdrInv}{KinductionMaxkKipdrMaxkReachsafetyLoops}{Correct}{True}{Cputime}{Max}{38.849768044}%
\StoreBenchExecResult{SvcompNineteenPdrInv}{KinductionMaxkKipdrMaxkReachsafetyLoops}{Correct}{True}{Cputime}{Stdev}{6.682033196542849781777128603}%
\StoreBenchExecResult{SvcompNineteenPdrInv}{KinductionMaxkKipdrMaxkReachsafetyLoops}{Correct}{True}{Walltime}{}{162.9014348840282786}%
\StoreBenchExecResult{SvcompNineteenPdrInv}{KinductionMaxkKipdrMaxkReachsafetyLoops}{Correct}{True}{Walltime}{Avg}{3.788405462419262293023255814}%
\StoreBenchExecResult{SvcompNineteenPdrInv}{KinductionMaxkKipdrMaxkReachsafetyLoops}{Correct}{True}{Walltime}{Median}{2.710208563003107}%
\StoreBenchExecResult{SvcompNineteenPdrInv}{KinductionMaxkKipdrMaxkReachsafetyLoops}{Correct}{True}{Walltime}{Min}{2.1782093760120915}%
\StoreBenchExecResult{SvcompNineteenPdrInv}{KinductionMaxkKipdrMaxkReachsafetyLoops}{Correct}{True}{Walltime}{Max}{19.6571699979977}%
\StoreBenchExecResult{SvcompNineteenPdrInv}{KinductionMaxkKipdrMaxkReachsafetyLoops}{Correct}{True}{Walltime}{Stdev}{3.365359619926820686961438623}%
\StoreBenchExecResult{SvcompNineteenPdrInv}{KinductionMaxkKipdrMaxkReachsafetyLoops}{Wrong}{True}{Count}{}{0}%
\StoreBenchExecResult{SvcompNineteenPdrInv}{KinductionMaxkKipdrMaxkReachsafetyLoops}{Wrong}{True}{Cputime}{}{0}%
\StoreBenchExecResult{SvcompNineteenPdrInv}{KinductionMaxkKipdrMaxkReachsafetyLoops}{Wrong}{True}{Cputime}{Avg}{None}%
\StoreBenchExecResult{SvcompNineteenPdrInv}{KinductionMaxkKipdrMaxkReachsafetyLoops}{Wrong}{True}{Cputime}{Median}{None}%
\StoreBenchExecResult{SvcompNineteenPdrInv}{KinductionMaxkKipdrMaxkReachsafetyLoops}{Wrong}{True}{Cputime}{Min}{None}%
\StoreBenchExecResult{SvcompNineteenPdrInv}{KinductionMaxkKipdrMaxkReachsafetyLoops}{Wrong}{True}{Cputime}{Max}{None}%
\StoreBenchExecResult{SvcompNineteenPdrInv}{KinductionMaxkKipdrMaxkReachsafetyLoops}{Wrong}{True}{Cputime}{Stdev}{None}%
\StoreBenchExecResult{SvcompNineteenPdrInv}{KinductionMaxkKipdrMaxkReachsafetyLoops}{Wrong}{True}{Walltime}{}{0}%
\StoreBenchExecResult{SvcompNineteenPdrInv}{KinductionMaxkKipdrMaxkReachsafetyLoops}{Wrong}{True}{Walltime}{Avg}{None}%
\StoreBenchExecResult{SvcompNineteenPdrInv}{KinductionMaxkKipdrMaxkReachsafetyLoops}{Wrong}{True}{Walltime}{Median}{None}%
\StoreBenchExecResult{SvcompNineteenPdrInv}{KinductionMaxkKipdrMaxkReachsafetyLoops}{Wrong}{True}{Walltime}{Min}{None}%
\StoreBenchExecResult{SvcompNineteenPdrInv}{KinductionMaxkKipdrMaxkReachsafetyLoops}{Wrong}{True}{Walltime}{Max}{None}%
\StoreBenchExecResult{SvcompNineteenPdrInv}{KinductionMaxkKipdrMaxkReachsafetyLoops}{Wrong}{True}{Walltime}{Stdev}{None}%
\StoreBenchExecResult{SvcompNineteenPdrInv}{KinductionMaxkKipdrMaxkReachsafetyLoops}{Error}{}{Count}{}{19}%
\StoreBenchExecResult{SvcompNineteenPdrInv}{KinductionMaxkKipdrMaxkReachsafetyLoops}{Error}{}{Cputime}{}{9632.237455172}%
\StoreBenchExecResult{SvcompNineteenPdrInv}{KinductionMaxkKipdrMaxkReachsafetyLoops}{Error}{}{Cputime}{Avg}{506.9598660616842105263157895}%
\StoreBenchExecResult{SvcompNineteenPdrInv}{KinductionMaxkKipdrMaxkReachsafetyLoops}{Error}{}{Cputime}{Median}{901.179614139}%
\StoreBenchExecResult{SvcompNineteenPdrInv}{KinductionMaxkKipdrMaxkReachsafetyLoops}{Error}{}{Cputime}{Min}{5.697316356}%
\StoreBenchExecResult{SvcompNineteenPdrInv}{KinductionMaxkKipdrMaxkReachsafetyLoops}{Error}{}{Cputime}{Max}{905.582207567}%
\StoreBenchExecResult{SvcompNineteenPdrInv}{KinductionMaxkKipdrMaxkReachsafetyLoops}{Error}{}{Cputime}{Stdev}{424.4370531308998162410736376}%
\StoreBenchExecResult{SvcompNineteenPdrInv}{KinductionMaxkKipdrMaxkReachsafetyLoops}{Error}{}{Walltime}{}{4828.8529965190245663}%
\StoreBenchExecResult{SvcompNineteenPdrInv}{KinductionMaxkKipdrMaxkReachsafetyLoops}{Error}{}{Walltime}{Avg}{254.1501577115276087526315789}%
\StoreBenchExecResult{SvcompNineteenPdrInv}{KinductionMaxkKipdrMaxkReachsafetyLoops}{Error}{}{Walltime}{Median}{451.270210445}%
\StoreBenchExecResult{SvcompNineteenPdrInv}{KinductionMaxkKipdrMaxkReachsafetyLoops}{Error}{}{Walltime}{Min}{3.03022934900946}%
\StoreBenchExecResult{SvcompNineteenPdrInv}{KinductionMaxkKipdrMaxkReachsafetyLoops}{Error}{}{Walltime}{Max}{455.86239988899615}%
\StoreBenchExecResult{SvcompNineteenPdrInv}{KinductionMaxkKipdrMaxkReachsafetyLoops}{Error}{}{Walltime}{Stdev}{212.6175117420590312593228378}%
\StoreBenchExecResult{SvcompNineteenPdrInv}{KinductionMaxkKipdrMaxkReachsafetyLoops}{Error}{Error}{Count}{}{5}%
\StoreBenchExecResult{SvcompNineteenPdrInv}{KinductionMaxkKipdrMaxkReachsafetyLoops}{Error}{Error}{Cputime}{}{38.807053592}%
\StoreBenchExecResult{SvcompNineteenPdrInv}{KinductionMaxkKipdrMaxkReachsafetyLoops}{Error}{Error}{Cputime}{Avg}{7.7614107184}%
\StoreBenchExecResult{SvcompNineteenPdrInv}{KinductionMaxkKipdrMaxkReachsafetyLoops}{Error}{Error}{Cputime}{Median}{7.73147497}%
\StoreBenchExecResult{SvcompNineteenPdrInv}{KinductionMaxkKipdrMaxkReachsafetyLoops}{Error}{Error}{Cputime}{Min}{7.360436164}%
\StoreBenchExecResult{SvcompNineteenPdrInv}{KinductionMaxkKipdrMaxkReachsafetyLoops}{Error}{Error}{Cputime}{Max}{8.177425713}%
\StoreBenchExecResult{SvcompNineteenPdrInv}{KinductionMaxkKipdrMaxkReachsafetyLoops}{Error}{Error}{Cputime}{Stdev}{0.3034142543801000082473448722}%
\StoreBenchExecResult{SvcompNineteenPdrInv}{KinductionMaxkKipdrMaxkReachsafetyLoops}{Error}{Error}{Walltime}{}{20.3489461209828727}%
\StoreBenchExecResult{SvcompNineteenPdrInv}{KinductionMaxkKipdrMaxkReachsafetyLoops}{Error}{Error}{Walltime}{Avg}{4.06978922419657454}%
\StoreBenchExecResult{SvcompNineteenPdrInv}{KinductionMaxkKipdrMaxkReachsafetyLoops}{Error}{Error}{Walltime}{Median}{4.034079543998814}%
\StoreBenchExecResult{SvcompNineteenPdrInv}{KinductionMaxkKipdrMaxkReachsafetyLoops}{Error}{Error}{Walltime}{Min}{3.852745544005302}%
\StoreBenchExecResult{SvcompNineteenPdrInv}{KinductionMaxkKipdrMaxkReachsafetyLoops}{Error}{Error}{Walltime}{Max}{4.309755242997198}%
\StoreBenchExecResult{SvcompNineteenPdrInv}{KinductionMaxkKipdrMaxkReachsafetyLoops}{Error}{Error}{Walltime}{Stdev}{0.1613923286469425982793243962}%
\StoreBenchExecResult{SvcompNineteenPdrInv}{KinductionMaxkKipdrMaxkReachsafetyLoops}{Error}{OutOfMemory}{Count}{}{2}%
\StoreBenchExecResult{SvcompNineteenPdrInv}{KinductionMaxkKipdrMaxkReachsafetyLoops}{Error}{OutOfMemory}{Cputime}{}{551.537221335}%
\StoreBenchExecResult{SvcompNineteenPdrInv}{KinductionMaxkKipdrMaxkReachsafetyLoops}{Error}{OutOfMemory}{Cputime}{Avg}{275.7686106675}%
\StoreBenchExecResult{SvcompNineteenPdrInv}{KinductionMaxkKipdrMaxkReachsafetyLoops}{Error}{OutOfMemory}{Cputime}{Median}{275.7686106675}%
\StoreBenchExecResult{SvcompNineteenPdrInv}{KinductionMaxkKipdrMaxkReachsafetyLoops}{Error}{OutOfMemory}{Cputime}{Min}{271.025867649}%
\StoreBenchExecResult{SvcompNineteenPdrInv}{KinductionMaxkKipdrMaxkReachsafetyLoops}{Error}{OutOfMemory}{Cputime}{Max}{280.511353686}%
\StoreBenchExecResult{SvcompNineteenPdrInv}{KinductionMaxkKipdrMaxkReachsafetyLoops}{Error}{OutOfMemory}{Cputime}{Stdev}{4.7427430185}%
\StoreBenchExecResult{SvcompNineteenPdrInv}{KinductionMaxkKipdrMaxkReachsafetyLoops}{Error}{OutOfMemory}{Walltime}{}{276.87543502199696}%
\StoreBenchExecResult{SvcompNineteenPdrInv}{KinductionMaxkKipdrMaxkReachsafetyLoops}{Error}{OutOfMemory}{Walltime}{Avg}{138.43771751099848}%
\StoreBenchExecResult{SvcompNineteenPdrInv}{KinductionMaxkKipdrMaxkReachsafetyLoops}{Error}{OutOfMemory}{Walltime}{Median}{138.43771751099848}%
\StoreBenchExecResult{SvcompNineteenPdrInv}{KinductionMaxkKipdrMaxkReachsafetyLoops}{Error}{OutOfMemory}{Walltime}{Min}{136.03322649399342}%
\StoreBenchExecResult{SvcompNineteenPdrInv}{KinductionMaxkKipdrMaxkReachsafetyLoops}{Error}{OutOfMemory}{Walltime}{Max}{140.84220852800354}%
\StoreBenchExecResult{SvcompNineteenPdrInv}{KinductionMaxkKipdrMaxkReachsafetyLoops}{Error}{OutOfMemory}{Walltime}{Stdev}{2.404491017005060000000000000}%
\StoreBenchExecResult{SvcompNineteenPdrInv}{KinductionMaxkKipdrMaxkReachsafetyLoops}{Error}{SegmentationFault}{Count}{}{2}%
\StoreBenchExecResult{SvcompNineteenPdrInv}{KinductionMaxkKipdrMaxkReachsafetyLoops}{Error}{SegmentationFault}{Cputime}{}{12.270513198}%
\StoreBenchExecResult{SvcompNineteenPdrInv}{KinductionMaxkKipdrMaxkReachsafetyLoops}{Error}{SegmentationFault}{Cputime}{Avg}{6.135256599}%
\StoreBenchExecResult{SvcompNineteenPdrInv}{KinductionMaxkKipdrMaxkReachsafetyLoops}{Error}{SegmentationFault}{Cputime}{Median}{6.135256599}%
\StoreBenchExecResult{SvcompNineteenPdrInv}{KinductionMaxkKipdrMaxkReachsafetyLoops}{Error}{SegmentationFault}{Cputime}{Min}{5.697316356}%
\StoreBenchExecResult{SvcompNineteenPdrInv}{KinductionMaxkKipdrMaxkReachsafetyLoops}{Error}{SegmentationFault}{Cputime}{Max}{6.573196842}%
\StoreBenchExecResult{SvcompNineteenPdrInv}{KinductionMaxkKipdrMaxkReachsafetyLoops}{Error}{SegmentationFault}{Cputime}{Stdev}{0.437940243}%
\StoreBenchExecResult{SvcompNineteenPdrInv}{KinductionMaxkKipdrMaxkReachsafetyLoops}{Error}{SegmentationFault}{Walltime}{}{6.4644874000077836}%
\StoreBenchExecResult{SvcompNineteenPdrInv}{KinductionMaxkKipdrMaxkReachsafetyLoops}{Error}{SegmentationFault}{Walltime}{Avg}{3.2322437000038918}%
\StoreBenchExecResult{SvcompNineteenPdrInv}{KinductionMaxkKipdrMaxkReachsafetyLoops}{Error}{SegmentationFault}{Walltime}{Median}{3.2322437000038918}%
\StoreBenchExecResult{SvcompNineteenPdrInv}{KinductionMaxkKipdrMaxkReachsafetyLoops}{Error}{SegmentationFault}{Walltime}{Min}{3.03022934900946}%
\StoreBenchExecResult{SvcompNineteenPdrInv}{KinductionMaxkKipdrMaxkReachsafetyLoops}{Error}{SegmentationFault}{Walltime}{Max}{3.4342580509983236}%
\StoreBenchExecResult{SvcompNineteenPdrInv}{KinductionMaxkKipdrMaxkReachsafetyLoops}{Error}{SegmentationFault}{Walltime}{Stdev}{0.2020143509944318000000000000}%
\StoreBenchExecResult{SvcompNineteenPdrInv}{KinductionMaxkKipdrMaxkReachsafetyLoops}{Error}{Timeout}{Count}{}{10}%
\StoreBenchExecResult{SvcompNineteenPdrInv}{KinductionMaxkKipdrMaxkReachsafetyLoops}{Error}{Timeout}{Cputime}{}{9029.622667047}%
\StoreBenchExecResult{SvcompNineteenPdrInv}{KinductionMaxkKipdrMaxkReachsafetyLoops}{Error}{Timeout}{Cputime}{Avg}{902.9622667047}%
\StoreBenchExecResult{SvcompNineteenPdrInv}{KinductionMaxkKipdrMaxkReachsafetyLoops}{Error}{Timeout}{Cputime}{Median}{902.6869368665}%
\StoreBenchExecResult{SvcompNineteenPdrInv}{KinductionMaxkKipdrMaxkReachsafetyLoops}{Error}{Timeout}{Cputime}{Min}{901.179614139}%
\StoreBenchExecResult{SvcompNineteenPdrInv}{KinductionMaxkKipdrMaxkReachsafetyLoops}{Error}{Timeout}{Cputime}{Max}{905.582207567}%
\StoreBenchExecResult{SvcompNineteenPdrInv}{KinductionMaxkKipdrMaxkReachsafetyLoops}{Error}{Timeout}{Cputime}{Stdev}{1.404420416534626662791326985}%
\StoreBenchExecResult{SvcompNineteenPdrInv}{KinductionMaxkKipdrMaxkReachsafetyLoops}{Error}{Timeout}{Walltime}{}{4525.16412797603695}%
\StoreBenchExecResult{SvcompNineteenPdrInv}{KinductionMaxkKipdrMaxkReachsafetyLoops}{Error}{Timeout}{Walltime}{Avg}{452.516412797603695}%
\StoreBenchExecResult{SvcompNineteenPdrInv}{KinductionMaxkKipdrMaxkReachsafetyLoops}{Error}{Timeout}{Walltime}{Median}{452.311703796498475}%
\StoreBenchExecResult{SvcompNineteenPdrInv}{KinductionMaxkKipdrMaxkReachsafetyLoops}{Error}{Timeout}{Walltime}{Min}{451.270210445}%
\StoreBenchExecResult{SvcompNineteenPdrInv}{KinductionMaxkKipdrMaxkReachsafetyLoops}{Error}{Timeout}{Walltime}{Max}{455.86239988899615}%
\StoreBenchExecResult{SvcompNineteenPdrInv}{KinductionMaxkKipdrMaxkReachsafetyLoops}{Error}{Timeout}{Walltime}{Stdev}{1.230561519868755400980196220}%
\StoreBenchExecResult{SvcompNineteenPdrInv}{KinductionMaxkKipdrMaxkReachsafetyLoops}{Unknown}{}{Count}{}{124}%
\StoreBenchExecResult{SvcompNineteenPdrInv}{KinductionMaxkKipdrMaxkReachsafetyLoops}{Unknown}{}{Cputime}{}{802.900626500}%
\StoreBenchExecResult{SvcompNineteenPdrInv}{KinductionMaxkKipdrMaxkReachsafetyLoops}{Unknown}{}{Cputime}{Avg}{6.475005052419354838709677419}%
\StoreBenchExecResult{SvcompNineteenPdrInv}{KinductionMaxkKipdrMaxkReachsafetyLoops}{Unknown}{}{Cputime}{Median}{5.6040180765}%
\StoreBenchExecResult{SvcompNineteenPdrInv}{KinductionMaxkKipdrMaxkReachsafetyLoops}{Unknown}{}{Cputime}{Min}{4.339028991}%
\StoreBenchExecResult{SvcompNineteenPdrInv}{KinductionMaxkKipdrMaxkReachsafetyLoops}{Unknown}{}{Cputime}{Max}{25.104137117}%
\StoreBenchExecResult{SvcompNineteenPdrInv}{KinductionMaxkKipdrMaxkReachsafetyLoops}{Unknown}{}{Cputime}{Stdev}{3.015399853281926518474319553}%
\StoreBenchExecResult{SvcompNineteenPdrInv}{KinductionMaxkKipdrMaxkReachsafetyLoops}{Unknown}{}{Walltime}{}{422.3140906360786226}%
\StoreBenchExecResult{SvcompNineteenPdrInv}{KinductionMaxkKipdrMaxkReachsafetyLoops}{Unknown}{}{Walltime}{Avg}{3.405758795452246956451612903}%
\StoreBenchExecResult{SvcompNineteenPdrInv}{KinductionMaxkKipdrMaxkReachsafetyLoops}{Unknown}{}{Walltime}{Median}{2.97578099350357665}%
\StoreBenchExecResult{SvcompNineteenPdrInv}{KinductionMaxkKipdrMaxkReachsafetyLoops}{Unknown}{}{Walltime}{Min}{2.338161215011496}%
\StoreBenchExecResult{SvcompNineteenPdrInv}{KinductionMaxkKipdrMaxkReachsafetyLoops}{Unknown}{}{Walltime}{Max}{12.725751108999248}%
\StoreBenchExecResult{SvcompNineteenPdrInv}{KinductionMaxkKipdrMaxkReachsafetyLoops}{Unknown}{}{Walltime}{Stdev}{1.515069921398158714875202300}%
\StoreBenchExecResult{SvcompNineteenPdrInv}{KinductionMaxkKipdrMaxkReachsafetyLoops}{Unknown}{Unknown}{Count}{}{124}%
\StoreBenchExecResult{SvcompNineteenPdrInv}{KinductionMaxkKipdrMaxkReachsafetyLoops}{Unknown}{Unknown}{Cputime}{}{802.900626500}%
\StoreBenchExecResult{SvcompNineteenPdrInv}{KinductionMaxkKipdrMaxkReachsafetyLoops}{Unknown}{Unknown}{Cputime}{Avg}{6.475005052419354838709677419}%
\StoreBenchExecResult{SvcompNineteenPdrInv}{KinductionMaxkKipdrMaxkReachsafetyLoops}{Unknown}{Unknown}{Cputime}{Median}{5.6040180765}%
\StoreBenchExecResult{SvcompNineteenPdrInv}{KinductionMaxkKipdrMaxkReachsafetyLoops}{Unknown}{Unknown}{Cputime}{Min}{4.339028991}%
\StoreBenchExecResult{SvcompNineteenPdrInv}{KinductionMaxkKipdrMaxkReachsafetyLoops}{Unknown}{Unknown}{Cputime}{Max}{25.104137117}%
\StoreBenchExecResult{SvcompNineteenPdrInv}{KinductionMaxkKipdrMaxkReachsafetyLoops}{Unknown}{Unknown}{Cputime}{Stdev}{3.015399853281926518474319553}%
\StoreBenchExecResult{SvcompNineteenPdrInv}{KinductionMaxkKipdrMaxkReachsafetyLoops}{Unknown}{Unknown}{Walltime}{}{422.3140906360786226}%
\StoreBenchExecResult{SvcompNineteenPdrInv}{KinductionMaxkKipdrMaxkReachsafetyLoops}{Unknown}{Unknown}{Walltime}{Avg}{3.405758795452246956451612903}%
\StoreBenchExecResult{SvcompNineteenPdrInv}{KinductionMaxkKipdrMaxkReachsafetyLoops}{Unknown}{Unknown}{Walltime}{Median}{2.97578099350357665}%
\StoreBenchExecResult{SvcompNineteenPdrInv}{KinductionMaxkKipdrMaxkReachsafetyLoops}{Unknown}{Unknown}{Walltime}{Min}{2.338161215011496}%
\StoreBenchExecResult{SvcompNineteenPdrInv}{KinductionMaxkKipdrMaxkReachsafetyLoops}{Unknown}{Unknown}{Walltime}{Max}{12.725751108999248}%
\StoreBenchExecResult{SvcompNineteenPdrInv}{KinductionMaxkKipdrMaxkReachsafetyLoops}{Unknown}{Unknown}{Walltime}{Stdev}{1.515069921398158714875202300}%
\providecommand\StoreBenchExecResult[7]{\expandafter\newcommand\csname#1#2#3#4#5#6\endcsname{#7}}%
\StoreBenchExecResult{SvcompNineteenPdrInv}{KinductionMaxkKipdrReachsafetyLoops}{Total}{}{Count}{}{208}%
\StoreBenchExecResult{SvcompNineteenPdrInv}{KinductionMaxkKipdrReachsafetyLoops}{Total}{}{Cputime}{}{10806.557671004}%
\StoreBenchExecResult{SvcompNineteenPdrInv}{KinductionMaxkKipdrReachsafetyLoops}{Total}{}{Cputime}{Avg}{51.95460418751923076923076923}%
\StoreBenchExecResult{SvcompNineteenPdrInv}{KinductionMaxkKipdrReachsafetyLoops}{Total}{}{Cputime}{Median}{5.3612072595}%
\StoreBenchExecResult{SvcompNineteenPdrInv}{KinductionMaxkKipdrReachsafetyLoops}{Total}{}{Cputime}{Min}{4.037544963}%
\StoreBenchExecResult{SvcompNineteenPdrInv}{KinductionMaxkKipdrReachsafetyLoops}{Total}{}{Cputime}{Max}{906.229586286}%
\StoreBenchExecResult{SvcompNineteenPdrInv}{KinductionMaxkKipdrReachsafetyLoops}{Total}{}{Cputime}{Stdev}{192.9065076225289012980160488}%
\StoreBenchExecResult{SvcompNineteenPdrInv}{KinductionMaxkKipdrReachsafetyLoops}{Total}{}{Walltime}{}{5447.1999982401030484}%
\StoreBenchExecResult{SvcompNineteenPdrInv}{KinductionMaxkKipdrReachsafetyLoops}{Total}{}{Walltime}{Avg}{26.188461530000495425}%
\StoreBenchExecResult{SvcompNineteenPdrInv}{KinductionMaxkKipdrReachsafetyLoops}{Total}{}{Walltime}{Median}{2.85207816950423875}%
\StoreBenchExecResult{SvcompNineteenPdrInv}{KinductionMaxkKipdrReachsafetyLoops}{Total}{}{Walltime}{Min}{2.1662614660017425}%
\StoreBenchExecResult{SvcompNineteenPdrInv}{KinductionMaxkKipdrReachsafetyLoops}{Total}{}{Walltime}{Max}{455.44576951800263}%
\StoreBenchExecResult{SvcompNineteenPdrInv}{KinductionMaxkKipdrReachsafetyLoops}{Total}{}{Walltime}{Stdev}{96.62790212716429895342665655}%
\StoreBenchExecResult{SvcompNineteenPdrInv}{KinductionMaxkKipdrReachsafetyLoops}{Correct}{}{Count}{}{65}%
\StoreBenchExecResult{SvcompNineteenPdrInv}{KinductionMaxkKipdrReachsafetyLoops}{Correct}{}{Cputime}{}{422.528423861}%
\StoreBenchExecResult{SvcompNineteenPdrInv}{KinductionMaxkKipdrReachsafetyLoops}{Correct}{}{Cputime}{Avg}{6.500437290169230769230769231}%
\StoreBenchExecResult{SvcompNineteenPdrInv}{KinductionMaxkKipdrReachsafetyLoops}{Correct}{}{Cputime}{Median}{4.896394003}%
\StoreBenchExecResult{SvcompNineteenPdrInv}{KinductionMaxkKipdrReachsafetyLoops}{Correct}{}{Cputime}{Min}{4.037544963}%
\StoreBenchExecResult{SvcompNineteenPdrInv}{KinductionMaxkKipdrReachsafetyLoops}{Correct}{}{Cputime}{Max}{40.688569312}%
\StoreBenchExecResult{SvcompNineteenPdrInv}{KinductionMaxkKipdrReachsafetyLoops}{Correct}{}{Cputime}{Stdev}{5.735156200228870674658727147}%
\StoreBenchExecResult{SvcompNineteenPdrInv}{KinductionMaxkKipdrReachsafetyLoops}{Correct}{}{Walltime}{}{222.4853124480578127}%
\StoreBenchExecResult{SvcompNineteenPdrInv}{KinductionMaxkKipdrReachsafetyLoops}{Correct}{}{Walltime}{Avg}{3.422850960739350964615384615}%
\StoreBenchExecResult{SvcompNineteenPdrInv}{KinductionMaxkKipdrReachsafetyLoops}{Correct}{}{Walltime}{Median}{2.6062156350089936}%
\StoreBenchExecResult{SvcompNineteenPdrInv}{KinductionMaxkKipdrReachsafetyLoops}{Correct}{}{Walltime}{Min}{2.1662614660017425}%
\StoreBenchExecResult{SvcompNineteenPdrInv}{KinductionMaxkKipdrReachsafetyLoops}{Correct}{}{Walltime}{Max}{20.613732853991678}%
\StoreBenchExecResult{SvcompNineteenPdrInv}{KinductionMaxkKipdrReachsafetyLoops}{Correct}{}{Walltime}{Stdev}{2.893840988541019889078539896}%
\StoreBenchExecResult{SvcompNineteenPdrInv}{KinductionMaxkKipdrReachsafetyLoops}{Correct}{False}{Count}{}{22}%
\StoreBenchExecResult{SvcompNineteenPdrInv}{KinductionMaxkKipdrReachsafetyLoops}{Correct}{False}{Cputime}{}{108.568763557}%
\StoreBenchExecResult{SvcompNineteenPdrInv}{KinductionMaxkKipdrReachsafetyLoops}{Correct}{False}{Cputime}{Avg}{4.934943798045454545454545455}%
\StoreBenchExecResult{SvcompNineteenPdrInv}{KinductionMaxkKipdrReachsafetyLoops}{Correct}{False}{Cputime}{Median}{4.533614276}%
\StoreBenchExecResult{SvcompNineteenPdrInv}{KinductionMaxkKipdrReachsafetyLoops}{Correct}{False}{Cputime}{Min}{4.037544963}%
\StoreBenchExecResult{SvcompNineteenPdrInv}{KinductionMaxkKipdrReachsafetyLoops}{Correct}{False}{Cputime}{Max}{8.535485069}%
\StoreBenchExecResult{SvcompNineteenPdrInv}{KinductionMaxkKipdrReachsafetyLoops}{Correct}{False}{Cputime}{Stdev}{1.049856893656380245502577336}%
\StoreBenchExecResult{SvcompNineteenPdrInv}{KinductionMaxkKipdrReachsafetyLoops}{Correct}{False}{Walltime}{}{57.7129735280177562}%
\StoreBenchExecResult{SvcompNineteenPdrInv}{KinductionMaxkKipdrReachsafetyLoops}{Correct}{False}{Walltime}{Avg}{2.623316978546261645454545455}%
\StoreBenchExecResult{SvcompNineteenPdrInv}{KinductionMaxkKipdrReachsafetyLoops}{Correct}{False}{Walltime}{Median}{2.41814367699407735}%
\StoreBenchExecResult{SvcompNineteenPdrInv}{KinductionMaxkKipdrReachsafetyLoops}{Correct}{False}{Walltime}{Min}{2.1662614660017425}%
\StoreBenchExecResult{SvcompNineteenPdrInv}{KinductionMaxkKipdrReachsafetyLoops}{Correct}{False}{Walltime}{Max}{4.445943440005067}%
\StoreBenchExecResult{SvcompNineteenPdrInv}{KinductionMaxkKipdrReachsafetyLoops}{Correct}{False}{Walltime}{Stdev}{0.5343723888684248104980399268}%
\StoreBenchExecResult{SvcompNineteenPdrInv}{KinductionMaxkKipdrReachsafetyLoops}{Wrong}{False}{Count}{}{0}%
\StoreBenchExecResult{SvcompNineteenPdrInv}{KinductionMaxkKipdrReachsafetyLoops}{Wrong}{False}{Cputime}{}{0}%
\StoreBenchExecResult{SvcompNineteenPdrInv}{KinductionMaxkKipdrReachsafetyLoops}{Wrong}{False}{Cputime}{Avg}{None}%
\StoreBenchExecResult{SvcompNineteenPdrInv}{KinductionMaxkKipdrReachsafetyLoops}{Wrong}{False}{Cputime}{Median}{None}%
\StoreBenchExecResult{SvcompNineteenPdrInv}{KinductionMaxkKipdrReachsafetyLoops}{Wrong}{False}{Cputime}{Min}{None}%
\StoreBenchExecResult{SvcompNineteenPdrInv}{KinductionMaxkKipdrReachsafetyLoops}{Wrong}{False}{Cputime}{Max}{None}%
\StoreBenchExecResult{SvcompNineteenPdrInv}{KinductionMaxkKipdrReachsafetyLoops}{Wrong}{False}{Cputime}{Stdev}{None}%
\StoreBenchExecResult{SvcompNineteenPdrInv}{KinductionMaxkKipdrReachsafetyLoops}{Wrong}{False}{Walltime}{}{0}%
\StoreBenchExecResult{SvcompNineteenPdrInv}{KinductionMaxkKipdrReachsafetyLoops}{Wrong}{False}{Walltime}{Avg}{None}%
\StoreBenchExecResult{SvcompNineteenPdrInv}{KinductionMaxkKipdrReachsafetyLoops}{Wrong}{False}{Walltime}{Median}{None}%
\StoreBenchExecResult{SvcompNineteenPdrInv}{KinductionMaxkKipdrReachsafetyLoops}{Wrong}{False}{Walltime}{Min}{None}%
\StoreBenchExecResult{SvcompNineteenPdrInv}{KinductionMaxkKipdrReachsafetyLoops}{Wrong}{False}{Walltime}{Max}{None}%
\StoreBenchExecResult{SvcompNineteenPdrInv}{KinductionMaxkKipdrReachsafetyLoops}{Wrong}{False}{Walltime}{Stdev}{None}%
\StoreBenchExecResult{SvcompNineteenPdrInv}{KinductionMaxkKipdrReachsafetyLoops}{Correct}{True}{Count}{}{43}%
\StoreBenchExecResult{SvcompNineteenPdrInv}{KinductionMaxkKipdrReachsafetyLoops}{Correct}{True}{Cputime}{}{313.959660304}%
\StoreBenchExecResult{SvcompNineteenPdrInv}{KinductionMaxkKipdrReachsafetyLoops}{Correct}{True}{Cputime}{Avg}{7.301387448930232558139534884}%
\StoreBenchExecResult{SvcompNineteenPdrInv}{KinductionMaxkKipdrReachsafetyLoops}{Correct}{True}{Cputime}{Median}{5.164155292}%
\StoreBenchExecResult{SvcompNineteenPdrInv}{KinductionMaxkKipdrReachsafetyLoops}{Correct}{True}{Cputime}{Min}{4.173680291}%
\StoreBenchExecResult{SvcompNineteenPdrInv}{KinductionMaxkKipdrReachsafetyLoops}{Correct}{True}{Cputime}{Max}{40.688569312}%
\StoreBenchExecResult{SvcompNineteenPdrInv}{KinductionMaxkKipdrReachsafetyLoops}{Correct}{True}{Cputime}{Stdev}{6.874676052850637920109310268}%
\StoreBenchExecResult{SvcompNineteenPdrInv}{KinductionMaxkKipdrReachsafetyLoops}{Correct}{True}{Walltime}{}{164.7723389200400565}%
\StoreBenchExecResult{SvcompNineteenPdrInv}{KinductionMaxkKipdrReachsafetyLoops}{Correct}{True}{Walltime}{Avg}{3.831914858605582709302325581}%
\StoreBenchExecResult{SvcompNineteenPdrInv}{KinductionMaxkKipdrReachsafetyLoops}{Correct}{True}{Walltime}{Median}{2.7461751560040284}%
\StoreBenchExecResult{SvcompNineteenPdrInv}{KinductionMaxkKipdrReachsafetyLoops}{Correct}{True}{Walltime}{Min}{2.2459683010092704}%
\StoreBenchExecResult{SvcompNineteenPdrInv}{KinductionMaxkKipdrReachsafetyLoops}{Correct}{True}{Walltime}{Max}{20.613732853991678}%
\StoreBenchExecResult{SvcompNineteenPdrInv}{KinductionMaxkKipdrReachsafetyLoops}{Correct}{True}{Walltime}{Stdev}{3.466750378328748947238153724}%
\StoreBenchExecResult{SvcompNineteenPdrInv}{KinductionMaxkKipdrReachsafetyLoops}{Wrong}{True}{Count}{}{0}%
\StoreBenchExecResult{SvcompNineteenPdrInv}{KinductionMaxkKipdrReachsafetyLoops}{Wrong}{True}{Cputime}{}{0}%
\StoreBenchExecResult{SvcompNineteenPdrInv}{KinductionMaxkKipdrReachsafetyLoops}{Wrong}{True}{Cputime}{Avg}{None}%
\StoreBenchExecResult{SvcompNineteenPdrInv}{KinductionMaxkKipdrReachsafetyLoops}{Wrong}{True}{Cputime}{Median}{None}%
\StoreBenchExecResult{SvcompNineteenPdrInv}{KinductionMaxkKipdrReachsafetyLoops}{Wrong}{True}{Cputime}{Min}{None}%
\StoreBenchExecResult{SvcompNineteenPdrInv}{KinductionMaxkKipdrReachsafetyLoops}{Wrong}{True}{Cputime}{Max}{None}%
\StoreBenchExecResult{SvcompNineteenPdrInv}{KinductionMaxkKipdrReachsafetyLoops}{Wrong}{True}{Cputime}{Stdev}{None}%
\StoreBenchExecResult{SvcompNineteenPdrInv}{KinductionMaxkKipdrReachsafetyLoops}{Wrong}{True}{Walltime}{}{0}%
\StoreBenchExecResult{SvcompNineteenPdrInv}{KinductionMaxkKipdrReachsafetyLoops}{Wrong}{True}{Walltime}{Avg}{None}%
\StoreBenchExecResult{SvcompNineteenPdrInv}{KinductionMaxkKipdrReachsafetyLoops}{Wrong}{True}{Walltime}{Median}{None}%
\StoreBenchExecResult{SvcompNineteenPdrInv}{KinductionMaxkKipdrReachsafetyLoops}{Wrong}{True}{Walltime}{Min}{None}%
\StoreBenchExecResult{SvcompNineteenPdrInv}{KinductionMaxkKipdrReachsafetyLoops}{Wrong}{True}{Walltime}{Max}{None}%
\StoreBenchExecResult{SvcompNineteenPdrInv}{KinductionMaxkKipdrReachsafetyLoops}{Wrong}{True}{Walltime}{Stdev}{None}%
\StoreBenchExecResult{SvcompNineteenPdrInv}{KinductionMaxkKipdrReachsafetyLoops}{Error}{}{Count}{}{19}%
\StoreBenchExecResult{SvcompNineteenPdrInv}{KinductionMaxkKipdrReachsafetyLoops}{Error}{}{Cputime}{}{9600.123795064}%
\StoreBenchExecResult{SvcompNineteenPdrInv}{KinductionMaxkKipdrReachsafetyLoops}{Error}{}{Cputime}{Avg}{505.2696734244210526315789474}%
\StoreBenchExecResult{SvcompNineteenPdrInv}{KinductionMaxkKipdrReachsafetyLoops}{Error}{}{Cputime}{Median}{901.818155243}%
\StoreBenchExecResult{SvcompNineteenPdrInv}{KinductionMaxkKipdrReachsafetyLoops}{Error}{}{Cputime}{Min}{5.853028312}%
\StoreBenchExecResult{SvcompNineteenPdrInv}{KinductionMaxkKipdrReachsafetyLoops}{Error}{}{Cputime}{Max}{906.229586286}%
\StoreBenchExecResult{SvcompNineteenPdrInv}{KinductionMaxkKipdrReachsafetyLoops}{Error}{}{Cputime}{Stdev}{425.5251424144295785432988461}%
\StoreBenchExecResult{SvcompNineteenPdrInv}{KinductionMaxkKipdrReachsafetyLoops}{Error}{}{Walltime}{}{4811.9949443570222118}%
\StoreBenchExecResult{SvcompNineteenPdrInv}{KinductionMaxkKipdrReachsafetyLoops}{Error}{}{Walltime}{Avg}{253.2628918082643269368421053}%
\StoreBenchExecResult{SvcompNineteenPdrInv}{KinductionMaxkKipdrReachsafetyLoops}{Error}{}{Walltime}{Median}{451.32176349799556}%
\StoreBenchExecResult{SvcompNineteenPdrInv}{KinductionMaxkKipdrReachsafetyLoops}{Error}{}{Walltime}{Min}{3.0622444650070975}%
\StoreBenchExecResult{SvcompNineteenPdrInv}{KinductionMaxkKipdrReachsafetyLoops}{Error}{}{Walltime}{Max}{455.44576951800263}%
\StoreBenchExecResult{SvcompNineteenPdrInv}{KinductionMaxkKipdrReachsafetyLoops}{Error}{}{Walltime}{Stdev}{213.1389901134799842051121639}%
\StoreBenchExecResult{SvcompNineteenPdrInv}{KinductionMaxkKipdrReachsafetyLoops}{Error}{Error}{Count}{}{5}%
\StoreBenchExecResult{SvcompNineteenPdrInv}{KinductionMaxkKipdrReachsafetyLoops}{Error}{Error}{Cputime}{}{42.221019524}%
\StoreBenchExecResult{SvcompNineteenPdrInv}{KinductionMaxkKipdrReachsafetyLoops}{Error}{Error}{Cputime}{Avg}{8.4442039048}%
\StoreBenchExecResult{SvcompNineteenPdrInv}{KinductionMaxkKipdrReachsafetyLoops}{Error}{Error}{Cputime}{Median}{8.724153061}%
\StoreBenchExecResult{SvcompNineteenPdrInv}{KinductionMaxkKipdrReachsafetyLoops}{Error}{Error}{Cputime}{Min}{7.631698138}%
\StoreBenchExecResult{SvcompNineteenPdrInv}{KinductionMaxkKipdrReachsafetyLoops}{Error}{Error}{Cputime}{Max}{9.289803678}%
\StoreBenchExecResult{SvcompNineteenPdrInv}{KinductionMaxkKipdrReachsafetyLoops}{Error}{Error}{Cputime}{Stdev}{0.6279470633142145301038450984}%
\StoreBenchExecResult{SvcompNineteenPdrInv}{KinductionMaxkKipdrReachsafetyLoops}{Error}{Error}{Walltime}{}{21.946071856000344}%
\StoreBenchExecResult{SvcompNineteenPdrInv}{KinductionMaxkKipdrReachsafetyLoops}{Error}{Error}{Walltime}{Avg}{4.3892143712000688}%
\StoreBenchExecResult{SvcompNineteenPdrInv}{KinductionMaxkKipdrReachsafetyLoops}{Error}{Error}{Walltime}{Median}{4.542144835999352}%
\StoreBenchExecResult{SvcompNineteenPdrInv}{KinductionMaxkKipdrReachsafetyLoops}{Error}{Error}{Walltime}{Min}{3.96620453200012}%
\StoreBenchExecResult{SvcompNineteenPdrInv}{KinductionMaxkKipdrReachsafetyLoops}{Error}{Error}{Walltime}{Max}{4.825725217000581}%
\StoreBenchExecResult{SvcompNineteenPdrInv}{KinductionMaxkKipdrReachsafetyLoops}{Error}{Error}{Walltime}{Stdev}{0.3291511029851553760627140721}%
\StoreBenchExecResult{SvcompNineteenPdrInv}{KinductionMaxkKipdrReachsafetyLoops}{Error}{OutOfMemory}{Count}{}{2}%
\StoreBenchExecResult{SvcompNineteenPdrInv}{KinductionMaxkKipdrReachsafetyLoops}{Error}{OutOfMemory}{Cputime}{}{512.689623221}%
\StoreBenchExecResult{SvcompNineteenPdrInv}{KinductionMaxkKipdrReachsafetyLoops}{Error}{OutOfMemory}{Cputime}{Avg}{256.3448116105}%
\StoreBenchExecResult{SvcompNineteenPdrInv}{KinductionMaxkKipdrReachsafetyLoops}{Error}{OutOfMemory}{Cputime}{Median}{256.3448116105}%
\StoreBenchExecResult{SvcompNineteenPdrInv}{KinductionMaxkKipdrReachsafetyLoops}{Error}{OutOfMemory}{Cputime}{Min}{251.004011259}%
\StoreBenchExecResult{SvcompNineteenPdrInv}{KinductionMaxkKipdrReachsafetyLoops}{Error}{OutOfMemory}{Cputime}{Max}{261.685611962}%
\StoreBenchExecResult{SvcompNineteenPdrInv}{KinductionMaxkKipdrReachsafetyLoops}{Error}{OutOfMemory}{Cputime}{Stdev}{5.3408003515}%
\StoreBenchExecResult{SvcompNineteenPdrInv}{KinductionMaxkKipdrReachsafetyLoops}{Error}{OutOfMemory}{Walltime}{}{257.42503741399560}%
\StoreBenchExecResult{SvcompNineteenPdrInv}{KinductionMaxkKipdrReachsafetyLoops}{Error}{OutOfMemory}{Walltime}{Avg}{128.71251870699780}%
\StoreBenchExecResult{SvcompNineteenPdrInv}{KinductionMaxkKipdrReachsafetyLoops}{Error}{OutOfMemory}{Walltime}{Median}{128.71251870699780}%
\StoreBenchExecResult{SvcompNineteenPdrInv}{KinductionMaxkKipdrReachsafetyLoops}{Error}{OutOfMemory}{Walltime}{Min}{126.04679099400528}%
\StoreBenchExecResult{SvcompNineteenPdrInv}{KinductionMaxkKipdrReachsafetyLoops}{Error}{OutOfMemory}{Walltime}{Max}{131.37824641999032}%
\StoreBenchExecResult{SvcompNineteenPdrInv}{KinductionMaxkKipdrReachsafetyLoops}{Error}{OutOfMemory}{Walltime}{Stdev}{2.665727712992520000000000000}%
\StoreBenchExecResult{SvcompNineteenPdrInv}{KinductionMaxkKipdrReachsafetyLoops}{Error}{SegmentationFault}{Count}{}{2}%
\StoreBenchExecResult{SvcompNineteenPdrInv}{KinductionMaxkKipdrReachsafetyLoops}{Error}{SegmentationFault}{Cputime}{}{12.454593145}%
\StoreBenchExecResult{SvcompNineteenPdrInv}{KinductionMaxkKipdrReachsafetyLoops}{Error}{SegmentationFault}{Cputime}{Avg}{6.2272965725}%
\StoreBenchExecResult{SvcompNineteenPdrInv}{KinductionMaxkKipdrReachsafetyLoops}{Error}{SegmentationFault}{Cputime}{Median}{6.2272965725}%
\StoreBenchExecResult{SvcompNineteenPdrInv}{KinductionMaxkKipdrReachsafetyLoops}{Error}{SegmentationFault}{Cputime}{Min}{5.853028312}%
\StoreBenchExecResult{SvcompNineteenPdrInv}{KinductionMaxkKipdrReachsafetyLoops}{Error}{SegmentationFault}{Cputime}{Max}{6.601564833}%
\StoreBenchExecResult{SvcompNineteenPdrInv}{KinductionMaxkKipdrReachsafetyLoops}{Error}{SegmentationFault}{Cputime}{Stdev}{0.3742682605}%
\StoreBenchExecResult{SvcompNineteenPdrInv}{KinductionMaxkKipdrReachsafetyLoops}{Error}{SegmentationFault}{Walltime}{}{6.5244985760073178}%
\StoreBenchExecResult{SvcompNineteenPdrInv}{KinductionMaxkKipdrReachsafetyLoops}{Error}{SegmentationFault}{Walltime}{Avg}{3.2622492880036589}%
\StoreBenchExecResult{SvcompNineteenPdrInv}{KinductionMaxkKipdrReachsafetyLoops}{Error}{SegmentationFault}{Walltime}{Median}{3.2622492880036589}%
\StoreBenchExecResult{SvcompNineteenPdrInv}{KinductionMaxkKipdrReachsafetyLoops}{Error}{SegmentationFault}{Walltime}{Min}{3.0622444650070975}%
\StoreBenchExecResult{SvcompNineteenPdrInv}{KinductionMaxkKipdrReachsafetyLoops}{Error}{SegmentationFault}{Walltime}{Max}{3.4622541110002203}%
\StoreBenchExecResult{SvcompNineteenPdrInv}{KinductionMaxkKipdrReachsafetyLoops}{Error}{SegmentationFault}{Walltime}{Stdev}{0.2000048229965614000000000000}%
\StoreBenchExecResult{SvcompNineteenPdrInv}{KinductionMaxkKipdrReachsafetyLoops}{Error}{Timeout}{Count}{}{10}%
\StoreBenchExecResult{SvcompNineteenPdrInv}{KinductionMaxkKipdrReachsafetyLoops}{Error}{Timeout}{Cputime}{}{9032.758559174}%
\StoreBenchExecResult{SvcompNineteenPdrInv}{KinductionMaxkKipdrReachsafetyLoops}{Error}{Timeout}{Cputime}{Avg}{903.2758559174}%
\StoreBenchExecResult{SvcompNineteenPdrInv}{KinductionMaxkKipdrReachsafetyLoops}{Error}{Timeout}{Cputime}{Median}{902.5870419305}%
\StoreBenchExecResult{SvcompNineteenPdrInv}{KinductionMaxkKipdrReachsafetyLoops}{Error}{Timeout}{Cputime}{Min}{901.818155243}%
\StoreBenchExecResult{SvcompNineteenPdrInv}{KinductionMaxkKipdrReachsafetyLoops}{Error}{Timeout}{Cputime}{Max}{906.229586286}%
\StoreBenchExecResult{SvcompNineteenPdrInv}{KinductionMaxkKipdrReachsafetyLoops}{Error}{Timeout}{Cputime}{Stdev}{1.458745429311742247366047024}%
\StoreBenchExecResult{SvcompNineteenPdrInv}{KinductionMaxkKipdrReachsafetyLoops}{Error}{Timeout}{Walltime}{}{4526.09933651101895}%
\StoreBenchExecResult{SvcompNineteenPdrInv}{KinductionMaxkKipdrReachsafetyLoops}{Error}{Timeout}{Walltime}{Avg}{452.609933651101895}%
\StoreBenchExecResult{SvcompNineteenPdrInv}{KinductionMaxkKipdrReachsafetyLoops}{Error}{Timeout}{Walltime}{Median}{452.342141364002605}%
\StoreBenchExecResult{SvcompNineteenPdrInv}{KinductionMaxkKipdrReachsafetyLoops}{Error}{Timeout}{Walltime}{Min}{451.32176349799556}%
\StoreBenchExecResult{SvcompNineteenPdrInv}{KinductionMaxkKipdrReachsafetyLoops}{Error}{Timeout}{Walltime}{Max}{455.44576951800263}%
\StoreBenchExecResult{SvcompNineteenPdrInv}{KinductionMaxkKipdrReachsafetyLoops}{Error}{Timeout}{Walltime}{Stdev}{1.109873423035666906866645463}%
\StoreBenchExecResult{SvcompNineteenPdrInv}{KinductionMaxkKipdrReachsafetyLoops}{Unknown}{}{Count}{}{124}%
\StoreBenchExecResult{SvcompNineteenPdrInv}{KinductionMaxkKipdrReachsafetyLoops}{Unknown}{}{Cputime}{}{783.905452079}%
\StoreBenchExecResult{SvcompNineteenPdrInv}{KinductionMaxkKipdrReachsafetyLoops}{Unknown}{}{Cputime}{Avg}{6.321818161927419354838709677}%
\StoreBenchExecResult{SvcompNineteenPdrInv}{KinductionMaxkKipdrReachsafetyLoops}{Unknown}{}{Cputime}{Median}{5.4022587925}%
\StoreBenchExecResult{SvcompNineteenPdrInv}{KinductionMaxkKipdrReachsafetyLoops}{Unknown}{}{Cputime}{Min}{4.40142309}%
\StoreBenchExecResult{SvcompNineteenPdrInv}{KinductionMaxkKipdrReachsafetyLoops}{Unknown}{}{Cputime}{Max}{22.024999884}%
\StoreBenchExecResult{SvcompNineteenPdrInv}{KinductionMaxkKipdrReachsafetyLoops}{Unknown}{}{Cputime}{Stdev}{2.671407312561528673210839776}%
\StoreBenchExecResult{SvcompNineteenPdrInv}{KinductionMaxkKipdrReachsafetyLoops}{Unknown}{}{Walltime}{}{412.7197414350230239}%
\StoreBenchExecResult{SvcompNineteenPdrInv}{KinductionMaxkKipdrReachsafetyLoops}{Unknown}{}{Walltime}{Avg}{3.328385011572766321774193548}%
\StoreBenchExecResult{SvcompNineteenPdrInv}{KinductionMaxkKipdrReachsafetyLoops}{Unknown}{}{Walltime}{Median}{2.8679398025051341}%
\StoreBenchExecResult{SvcompNineteenPdrInv}{KinductionMaxkKipdrReachsafetyLoops}{Unknown}{}{Walltime}{Min}{2.370203449987457}%
\StoreBenchExecResult{SvcompNineteenPdrInv}{KinductionMaxkKipdrReachsafetyLoops}{Unknown}{}{Walltime}{Max}{11.214260701002786}%
\StoreBenchExecResult{SvcompNineteenPdrInv}{KinductionMaxkKipdrReachsafetyLoops}{Unknown}{}{Walltime}{Stdev}{1.344134421048809341824508134}%
\StoreBenchExecResult{SvcompNineteenPdrInv}{KinductionMaxkKipdrReachsafetyLoops}{Unknown}{Unknown}{Count}{}{124}%
\StoreBenchExecResult{SvcompNineteenPdrInv}{KinductionMaxkKipdrReachsafetyLoops}{Unknown}{Unknown}{Cputime}{}{783.905452079}%
\StoreBenchExecResult{SvcompNineteenPdrInv}{KinductionMaxkKipdrReachsafetyLoops}{Unknown}{Unknown}{Cputime}{Avg}{6.321818161927419354838709677}%
\StoreBenchExecResult{SvcompNineteenPdrInv}{KinductionMaxkKipdrReachsafetyLoops}{Unknown}{Unknown}{Cputime}{Median}{5.4022587925}%
\StoreBenchExecResult{SvcompNineteenPdrInv}{KinductionMaxkKipdrReachsafetyLoops}{Unknown}{Unknown}{Cputime}{Min}{4.40142309}%
\StoreBenchExecResult{SvcompNineteenPdrInv}{KinductionMaxkKipdrReachsafetyLoops}{Unknown}{Unknown}{Cputime}{Max}{22.024999884}%
\StoreBenchExecResult{SvcompNineteenPdrInv}{KinductionMaxkKipdrReachsafetyLoops}{Unknown}{Unknown}{Cputime}{Stdev}{2.671407312561528673210839776}%
\StoreBenchExecResult{SvcompNineteenPdrInv}{KinductionMaxkKipdrReachsafetyLoops}{Unknown}{Unknown}{Walltime}{}{412.7197414350230239}%
\StoreBenchExecResult{SvcompNineteenPdrInv}{KinductionMaxkKipdrReachsafetyLoops}{Unknown}{Unknown}{Walltime}{Avg}{3.328385011572766321774193548}%
\StoreBenchExecResult{SvcompNineteenPdrInv}{KinductionMaxkKipdrReachsafetyLoops}{Unknown}{Unknown}{Walltime}{Median}{2.8679398025051341}%
\StoreBenchExecResult{SvcompNineteenPdrInv}{KinductionMaxkKipdrReachsafetyLoops}{Unknown}{Unknown}{Walltime}{Min}{2.370203449987457}%
\StoreBenchExecResult{SvcompNineteenPdrInv}{KinductionMaxkKipdrReachsafetyLoops}{Unknown}{Unknown}{Walltime}{Max}{11.214260701002786}%
\StoreBenchExecResult{SvcompNineteenPdrInv}{KinductionMaxkKipdrReachsafetyLoops}{Unknown}{Unknown}{Walltime}{Stdev}{1.344134421048809341824508134}%
\providecommand\StoreBenchExecResult[7]{\expandafter\newcommand\csname#1#2#3#4#5#6\endcsname{#7}}%
\StoreBenchExecResult{SvcompNineteenPdrInv}{KinductionPlainReachsafetyLoops}{Total}{}{Count}{}{208}%
\StoreBenchExecResult{SvcompNineteenPdrInv}{KinductionPlainReachsafetyLoops}{Total}{}{Cputime}{}{91993.030741249}%
\StoreBenchExecResult{SvcompNineteenPdrInv}{KinductionPlainReachsafetyLoops}{Total}{}{Cputime}{Avg}{442.2741862560048076923076923}%
\StoreBenchExecResult{SvcompNineteenPdrInv}{KinductionPlainReachsafetyLoops}{Total}{}{Cputime}{Median}{43.9471823125}%
\StoreBenchExecResult{SvcompNineteenPdrInv}{KinductionPlainReachsafetyLoops}{Total}{}{Cputime}{Min}{3.700584376}%
\StoreBenchExecResult{SvcompNineteenPdrInv}{KinductionPlainReachsafetyLoops}{Total}{}{Cputime}{Max}{913.15335514}%
\StoreBenchExecResult{SvcompNineteenPdrInv}{KinductionPlainReachsafetyLoops}{Total}{}{Cputime}{Stdev}{443.0996879197507935468679721}%
\StoreBenchExecResult{SvcompNineteenPdrInv}{KinductionPlainReachsafetyLoops}{Total}{}{Walltime}{}{88684.9986353929702879}%
\StoreBenchExecResult{SvcompNineteenPdrInv}{KinductionPlainReachsafetyLoops}{Total}{}{Walltime}{Avg}{426.3701857470815879225961538}%
\StoreBenchExecResult{SvcompNineteenPdrInv}{KinductionPlainReachsafetyLoops}{Total}{}{Walltime}{Median}{33.0507409549973075}%
\StoreBenchExecResult{SvcompNineteenPdrInv}{KinductionPlainReachsafetyLoops}{Total}{}{Walltime}{Min}{1.9822720649972325}%
\StoreBenchExecResult{SvcompNineteenPdrInv}{KinductionPlainReachsafetyLoops}{Total}{}{Walltime}{Max}{906.6696613560052}%
\StoreBenchExecResult{SvcompNineteenPdrInv}{KinductionPlainReachsafetyLoops}{Total}{}{Walltime}{Stdev}{431.4704520487473705864292390}%
\StoreBenchExecResult{SvcompNineteenPdrInv}{KinductionPlainReachsafetyLoops}{Correct}{}{Count}{}{101}%
\StoreBenchExecResult{SvcompNineteenPdrInv}{KinductionPlainReachsafetyLoops}{Correct}{}{Cputime}{}{1627.799308354}%
\StoreBenchExecResult{SvcompNineteenPdrInv}{KinductionPlainReachsafetyLoops}{Correct}{}{Cputime}{Avg}{16.11682483518811881188118812}%
\StoreBenchExecResult{SvcompNineteenPdrInv}{KinductionPlainReachsafetyLoops}{Correct}{}{Cputime}{Median}{5.181980245}%
\StoreBenchExecResult{SvcompNineteenPdrInv}{KinductionPlainReachsafetyLoops}{Correct}{}{Cputime}{Min}{3.700584376}%
\StoreBenchExecResult{SvcompNineteenPdrInv}{KinductionPlainReachsafetyLoops}{Correct}{}{Cputime}{Max}{653.211812908}%
\StoreBenchExecResult{SvcompNineteenPdrInv}{KinductionPlainReachsafetyLoops}{Correct}{}{Cputime}{Stdev}{64.40572847840422157791855083}%
\StoreBenchExecResult{SvcompNineteenPdrInv}{KinductionPlainReachsafetyLoops}{Correct}{}{Walltime}{}{1237.0552919119799778}%
\StoreBenchExecResult{SvcompNineteenPdrInv}{KinductionPlainReachsafetyLoops}{Correct}{}{Walltime}{Avg}{12.24807219714831661188118812}%
\StoreBenchExecResult{SvcompNineteenPdrInv}{KinductionPlainReachsafetyLoops}{Correct}{}{Walltime}{Median}{2.721737344007124}%
\StoreBenchExecResult{SvcompNineteenPdrInv}{KinductionPlainReachsafetyLoops}{Correct}{}{Walltime}{Min}{1.9822720649972325}%
\StoreBenchExecResult{SvcompNineteenPdrInv}{KinductionPlainReachsafetyLoops}{Correct}{}{Walltime}{Max}{635.4222192799934}%
\StoreBenchExecResult{SvcompNineteenPdrInv}{KinductionPlainReachsafetyLoops}{Correct}{}{Walltime}{Stdev}{62.70152708561306653090719261}%
\StoreBenchExecResult{SvcompNineteenPdrInv}{KinductionPlainReachsafetyLoops}{Correct}{False}{Count}{}{37}%
\StoreBenchExecResult{SvcompNineteenPdrInv}{KinductionPlainReachsafetyLoops}{Correct}{False}{Cputime}{}{302.256499167}%
\StoreBenchExecResult{SvcompNineteenPdrInv}{KinductionPlainReachsafetyLoops}{Correct}{False}{Cputime}{Avg}{8.169094572081081081081081081}%
\StoreBenchExecResult{SvcompNineteenPdrInv}{KinductionPlainReachsafetyLoops}{Correct}{False}{Cputime}{Median}{4.928590924}%
\StoreBenchExecResult{SvcompNineteenPdrInv}{KinductionPlainReachsafetyLoops}{Correct}{False}{Cputime}{Min}{3.950322417}%
\StoreBenchExecResult{SvcompNineteenPdrInv}{KinductionPlainReachsafetyLoops}{Correct}{False}{Cputime}{Max}{35.94280372}%
\StoreBenchExecResult{SvcompNineteenPdrInv}{KinductionPlainReachsafetyLoops}{Correct}{False}{Cputime}{Stdev}{7.613642592095126515797315116}%
\StoreBenchExecResult{SvcompNineteenPdrInv}{KinductionPlainReachsafetyLoops}{Correct}{False}{Walltime}{}{182.0173294260021052}%
\StoreBenchExecResult{SvcompNineteenPdrInv}{KinductionPlainReachsafetyLoops}{Correct}{False}{Walltime}{Avg}{4.919387281783840681081081081}%
\StoreBenchExecResult{SvcompNineteenPdrInv}{KinductionPlainReachsafetyLoops}{Correct}{False}{Walltime}{Median}{2.6058429909899132}%
\StoreBenchExecResult{SvcompNineteenPdrInv}{KinductionPlainReachsafetyLoops}{Correct}{False}{Walltime}{Min}{2.1102478160028113}%
\StoreBenchExecResult{SvcompNineteenPdrInv}{KinductionPlainReachsafetyLoops}{Correct}{False}{Walltime}{Max}{26.782584966989816}%
\StoreBenchExecResult{SvcompNineteenPdrInv}{KinductionPlainReachsafetyLoops}{Correct}{False}{Walltime}{Stdev}{5.612746816351497363469431968}%
\StoreBenchExecResult{SvcompNineteenPdrInv}{KinductionPlainReachsafetyLoops}{Wrong}{False}{Count}{}{0}%
\StoreBenchExecResult{SvcompNineteenPdrInv}{KinductionPlainReachsafetyLoops}{Wrong}{False}{Cputime}{}{0}%
\StoreBenchExecResult{SvcompNineteenPdrInv}{KinductionPlainReachsafetyLoops}{Wrong}{False}{Cputime}{Avg}{None}%
\StoreBenchExecResult{SvcompNineteenPdrInv}{KinductionPlainReachsafetyLoops}{Wrong}{False}{Cputime}{Median}{None}%
\StoreBenchExecResult{SvcompNineteenPdrInv}{KinductionPlainReachsafetyLoops}{Wrong}{False}{Cputime}{Min}{None}%
\StoreBenchExecResult{SvcompNineteenPdrInv}{KinductionPlainReachsafetyLoops}{Wrong}{False}{Cputime}{Max}{None}%
\StoreBenchExecResult{SvcompNineteenPdrInv}{KinductionPlainReachsafetyLoops}{Wrong}{False}{Cputime}{Stdev}{None}%
\StoreBenchExecResult{SvcompNineteenPdrInv}{KinductionPlainReachsafetyLoops}{Wrong}{False}{Walltime}{}{0}%
\StoreBenchExecResult{SvcompNineteenPdrInv}{KinductionPlainReachsafetyLoops}{Wrong}{False}{Walltime}{Avg}{None}%
\StoreBenchExecResult{SvcompNineteenPdrInv}{KinductionPlainReachsafetyLoops}{Wrong}{False}{Walltime}{Median}{None}%
\StoreBenchExecResult{SvcompNineteenPdrInv}{KinductionPlainReachsafetyLoops}{Wrong}{False}{Walltime}{Min}{None}%
\StoreBenchExecResult{SvcompNineteenPdrInv}{KinductionPlainReachsafetyLoops}{Wrong}{False}{Walltime}{Max}{None}%
\StoreBenchExecResult{SvcompNineteenPdrInv}{KinductionPlainReachsafetyLoops}{Wrong}{False}{Walltime}{Stdev}{None}%
\StoreBenchExecResult{SvcompNineteenPdrInv}{KinductionPlainReachsafetyLoops}{Correct}{True}{Count}{}{64}%
\StoreBenchExecResult{SvcompNineteenPdrInv}{KinductionPlainReachsafetyLoops}{Correct}{True}{Cputime}{}{1325.542809187}%
\StoreBenchExecResult{SvcompNineteenPdrInv}{KinductionPlainReachsafetyLoops}{Correct}{True}{Cputime}{Avg}{20.711606393546875}%
\StoreBenchExecResult{SvcompNineteenPdrInv}{KinductionPlainReachsafetyLoops}{Correct}{True}{Cputime}{Median}{5.457414365}%
\StoreBenchExecResult{SvcompNineteenPdrInv}{KinductionPlainReachsafetyLoops}{Correct}{True}{Cputime}{Min}{3.700584376}%
\StoreBenchExecResult{SvcompNineteenPdrInv}{KinductionPlainReachsafetyLoops}{Correct}{True}{Cputime}{Max}{653.211812908}%
\StoreBenchExecResult{SvcompNineteenPdrInv}{KinductionPlainReachsafetyLoops}{Correct}{True}{Cputime}{Stdev}{80.34347728805948568504205860}%
\StoreBenchExecResult{SvcompNineteenPdrInv}{KinductionPlainReachsafetyLoops}{Correct}{True}{Walltime}{}{1055.0379624859778726}%
\StoreBenchExecResult{SvcompNineteenPdrInv}{KinductionPlainReachsafetyLoops}{Correct}{True}{Walltime}{Avg}{16.484968163843404259375}%
\StoreBenchExecResult{SvcompNineteenPdrInv}{KinductionPlainReachsafetyLoops}{Correct}{True}{Walltime}{Median}{2.90377774149965265}%
\StoreBenchExecResult{SvcompNineteenPdrInv}{KinductionPlainReachsafetyLoops}{Correct}{True}{Walltime}{Min}{1.9822720649972325}%
\StoreBenchExecResult{SvcompNineteenPdrInv}{KinductionPlainReachsafetyLoops}{Correct}{True}{Walltime}{Max}{635.4222192799934}%
\StoreBenchExecResult{SvcompNineteenPdrInv}{KinductionPlainReachsafetyLoops}{Correct}{True}{Walltime}{Stdev}{78.33999264640493079294889573}%
\StoreBenchExecResult{SvcompNineteenPdrInv}{KinductionPlainReachsafetyLoops}{Wrong}{True}{Count}{}{0}%
\StoreBenchExecResult{SvcompNineteenPdrInv}{KinductionPlainReachsafetyLoops}{Wrong}{True}{Cputime}{}{0}%
\StoreBenchExecResult{SvcompNineteenPdrInv}{KinductionPlainReachsafetyLoops}{Wrong}{True}{Cputime}{Avg}{None}%
\StoreBenchExecResult{SvcompNineteenPdrInv}{KinductionPlainReachsafetyLoops}{Wrong}{True}{Cputime}{Median}{None}%
\StoreBenchExecResult{SvcompNineteenPdrInv}{KinductionPlainReachsafetyLoops}{Wrong}{True}{Cputime}{Min}{None}%
\StoreBenchExecResult{SvcompNineteenPdrInv}{KinductionPlainReachsafetyLoops}{Wrong}{True}{Cputime}{Max}{None}%
\StoreBenchExecResult{SvcompNineteenPdrInv}{KinductionPlainReachsafetyLoops}{Wrong}{True}{Cputime}{Stdev}{None}%
\StoreBenchExecResult{SvcompNineteenPdrInv}{KinductionPlainReachsafetyLoops}{Wrong}{True}{Walltime}{}{0}%
\StoreBenchExecResult{SvcompNineteenPdrInv}{KinductionPlainReachsafetyLoops}{Wrong}{True}{Walltime}{Avg}{None}%
\StoreBenchExecResult{SvcompNineteenPdrInv}{KinductionPlainReachsafetyLoops}{Wrong}{True}{Walltime}{Median}{None}%
\StoreBenchExecResult{SvcompNineteenPdrInv}{KinductionPlainReachsafetyLoops}{Wrong}{True}{Walltime}{Min}{None}%
\StoreBenchExecResult{SvcompNineteenPdrInv}{KinductionPlainReachsafetyLoops}{Wrong}{True}{Walltime}{Max}{None}%
\StoreBenchExecResult{SvcompNineteenPdrInv}{KinductionPlainReachsafetyLoops}{Wrong}{True}{Walltime}{Stdev}{None}%
\StoreBenchExecResult{SvcompNineteenPdrInv}{KinductionPlainReachsafetyLoops}{Error}{}{Count}{}{107}%
\StoreBenchExecResult{SvcompNineteenPdrInv}{KinductionPlainReachsafetyLoops}{Error}{}{Cputime}{}{90365.231432895}%
\StoreBenchExecResult{SvcompNineteenPdrInv}{KinductionPlainReachsafetyLoops}{Error}{}{Cputime}{Avg}{844.5348732046261682242990654}%
\StoreBenchExecResult{SvcompNineteenPdrInv}{KinductionPlainReachsafetyLoops}{Error}{}{Cputime}{Median}{903.892771637}%
\StoreBenchExecResult{SvcompNineteenPdrInv}{KinductionPlainReachsafetyLoops}{Error}{}{Cputime}{Min}{6.406826619}%
\StoreBenchExecResult{SvcompNineteenPdrInv}{KinductionPlainReachsafetyLoops}{Error}{}{Cputime}{Max}{913.15335514}%
\StoreBenchExecResult{SvcompNineteenPdrInv}{KinductionPlainReachsafetyLoops}{Error}{}{Cputime}{Stdev}{210.9729587170591403609363145}%
\StoreBenchExecResult{SvcompNineteenPdrInv}{KinductionPlainReachsafetyLoops}{Error}{}{Walltime}{}{87447.9433434809903101}%
\StoreBenchExecResult{SvcompNineteenPdrInv}{KinductionPlainReachsafetyLoops}{Error}{}{Walltime}{Avg}{817.2704985372055169168224299}%
\StoreBenchExecResult{SvcompNineteenPdrInv}{KinductionPlainReachsafetyLoops}{Error}{}{Walltime}{Median}{886.0537621870026}%
\StoreBenchExecResult{SvcompNineteenPdrInv}{KinductionPlainReachsafetyLoops}{Error}{}{Walltime}{Min}{3.3741149400011636}%
\StoreBenchExecResult{SvcompNineteenPdrInv}{KinductionPlainReachsafetyLoops}{Error}{}{Walltime}{Max}{906.6696613560052}%
\StoreBenchExecResult{SvcompNineteenPdrInv}{KinductionPlainReachsafetyLoops}{Error}{}{Walltime}{Stdev}{208.5658219018502784991827855}%
\StoreBenchExecResult{SvcompNineteenPdrInv}{KinductionPlainReachsafetyLoops}{Error}{Error}{Count}{}{5}%
\StoreBenchExecResult{SvcompNineteenPdrInv}{KinductionPlainReachsafetyLoops}{Error}{Error}{Cputime}{}{34.190734850}%
\StoreBenchExecResult{SvcompNineteenPdrInv}{KinductionPlainReachsafetyLoops}{Error}{Error}{Cputime}{Avg}{6.838146970}%
\StoreBenchExecResult{SvcompNineteenPdrInv}{KinductionPlainReachsafetyLoops}{Error}{Error}{Cputime}{Median}{6.517755081}%
\StoreBenchExecResult{SvcompNineteenPdrInv}{KinductionPlainReachsafetyLoops}{Error}{Error}{Cputime}{Min}{6.406826619}%
\StoreBenchExecResult{SvcompNineteenPdrInv}{KinductionPlainReachsafetyLoops}{Error}{Error}{Cputime}{Max}{7.468300648}%
\StoreBenchExecResult{SvcompNineteenPdrInv}{KinductionPlainReachsafetyLoops}{Error}{Error}{Cputime}{Stdev}{0.4823912136364484937196952164}%
\StoreBenchExecResult{SvcompNineteenPdrInv}{KinductionPlainReachsafetyLoops}{Error}{Error}{Walltime}{}{17.9423556260007901}%
\StoreBenchExecResult{SvcompNineteenPdrInv}{KinductionPlainReachsafetyLoops}{Error}{Error}{Walltime}{Avg}{3.58847112520015802}%
\StoreBenchExecResult{SvcompNineteenPdrInv}{KinductionPlainReachsafetyLoops}{Error}{Error}{Walltime}{Median}{3.4022751260054065}%
\StoreBenchExecResult{SvcompNineteenPdrInv}{KinductionPlainReachsafetyLoops}{Error}{Error}{Walltime}{Min}{3.3741149400011636}%
\StoreBenchExecResult{SvcompNineteenPdrInv}{KinductionPlainReachsafetyLoops}{Error}{Error}{Walltime}{Max}{3.897819391990197}%
\StoreBenchExecResult{SvcompNineteenPdrInv}{KinductionPlainReachsafetyLoops}{Error}{Error}{Walltime}{Stdev}{0.2415567084795149196671504879}%
\StoreBenchExecResult{SvcompNineteenPdrInv}{KinductionPlainReachsafetyLoops}{Error}{OutOfMemory}{Count}{}{12}%
\StoreBenchExecResult{SvcompNineteenPdrInv}{KinductionPlainReachsafetyLoops}{Error}{OutOfMemory}{Cputime}{}{8770.661514067}%
\StoreBenchExecResult{SvcompNineteenPdrInv}{KinductionPlainReachsafetyLoops}{Error}{OutOfMemory}{Cputime}{Avg}{730.8884595055833333333333333}%
\StoreBenchExecResult{SvcompNineteenPdrInv}{KinductionPlainReachsafetyLoops}{Error}{OutOfMemory}{Cputime}{Median}{849.063699364}%
\StoreBenchExecResult{SvcompNineteenPdrInv}{KinductionPlainReachsafetyLoops}{Error}{OutOfMemory}{Cputime}{Min}{178.887943814}%
\StoreBenchExecResult{SvcompNineteenPdrInv}{KinductionPlainReachsafetyLoops}{Error}{OutOfMemory}{Cputime}{Max}{898.813397265}%
\StoreBenchExecResult{SvcompNineteenPdrInv}{KinductionPlainReachsafetyLoops}{Error}{OutOfMemory}{Cputime}{Stdev}{250.7727728084776618658552904}%
\StoreBenchExecResult{SvcompNineteenPdrInv}{KinductionPlainReachsafetyLoops}{Error}{OutOfMemory}{Walltime}{}{8554.37023485498492}%
\StoreBenchExecResult{SvcompNineteenPdrInv}{KinductionPlainReachsafetyLoops}{Error}{OutOfMemory}{Walltime}{Avg}{712.86418623791541}%
\StoreBenchExecResult{SvcompNineteenPdrInv}{KinductionPlainReachsafetyLoops}{Error}{OutOfMemory}{Walltime}{Median}{828.91003617899695}%
\StoreBenchExecResult{SvcompNineteenPdrInv}{KinductionPlainReachsafetyLoops}{Error}{OutOfMemory}{Walltime}{Min}{171.95480914300424}%
\StoreBenchExecResult{SvcompNineteenPdrInv}{KinductionPlainReachsafetyLoops}{Error}{OutOfMemory}{Walltime}{Max}{879.614286652999}%
\StoreBenchExecResult{SvcompNineteenPdrInv}{KinductionPlainReachsafetyLoops}{Error}{OutOfMemory}{Walltime}{Stdev}{245.8558897857167672484379466}%
\StoreBenchExecResult{SvcompNineteenPdrInv}{KinductionPlainReachsafetyLoops}{Error}{Timeout}{Count}{}{90}%
\StoreBenchExecResult{SvcompNineteenPdrInv}{KinductionPlainReachsafetyLoops}{Error}{Timeout}{Cputime}{}{81560.379183978}%
\StoreBenchExecResult{SvcompNineteenPdrInv}{KinductionPlainReachsafetyLoops}{Error}{Timeout}{Cputime}{Avg}{906.2264353775333333333333333}%
\StoreBenchExecResult{SvcompNineteenPdrInv}{KinductionPlainReachsafetyLoops}{Error}{Timeout}{Cputime}{Median}{904.5912823035}%
\StoreBenchExecResult{SvcompNineteenPdrInv}{KinductionPlainReachsafetyLoops}{Error}{Timeout}{Cputime}{Min}{900.043792419}%
\StoreBenchExecResult{SvcompNineteenPdrInv}{KinductionPlainReachsafetyLoops}{Error}{Timeout}{Cputime}{Max}{913.15335514}%
\StoreBenchExecResult{SvcompNineteenPdrInv}{KinductionPlainReachsafetyLoops}{Error}{Timeout}{Cputime}{Stdev}{4.332794684254140527668245386}%
\StoreBenchExecResult{SvcompNineteenPdrInv}{KinductionPlainReachsafetyLoops}{Error}{Timeout}{Walltime}{}{78875.6307530000046}%
\StoreBenchExecResult{SvcompNineteenPdrInv}{KinductionPlainReachsafetyLoops}{Error}{Timeout}{Walltime}{Avg}{876.3958972555556066666666667}%
\StoreBenchExecResult{SvcompNineteenPdrInv}{KinductionPlainReachsafetyLoops}{Error}{Timeout}{Walltime}{Median}{886.9042014785009}%
\StoreBenchExecResult{SvcompNineteenPdrInv}{KinductionPlainReachsafetyLoops}{Error}{Timeout}{Walltime}{Min}{614.9176561509958}%
\StoreBenchExecResult{SvcompNineteenPdrInv}{KinductionPlainReachsafetyLoops}{Error}{Timeout}{Walltime}{Max}{906.6696613560052}%
\StoreBenchExecResult{SvcompNineteenPdrInv}{KinductionPlainReachsafetyLoops}{Error}{Timeout}{Walltime}{Stdev}{43.88152128127502938669680706}%
\providecommand\StoreBenchExecResult[7]{\expandafter\newcommand\csname#1#2#3#4#5#6\endcsname{#7}}%
\StoreBenchExecResult{SvcompNineteenPdrInv}{KipdrReachsafetyLoops}{Total}{}{Count}{}{208}%
\StoreBenchExecResult{SvcompNineteenPdrInv}{KipdrReachsafetyLoops}{Total}{}{Cputime}{}{89153.394994794}%
\StoreBenchExecResult{SvcompNineteenPdrInv}{KipdrReachsafetyLoops}{Total}{}{Cputime}{Avg}{428.622091321125}%
\StoreBenchExecResult{SvcompNineteenPdrInv}{KipdrReachsafetyLoops}{Total}{}{Cputime}{Median}{47.5746390375}%
\StoreBenchExecResult{SvcompNineteenPdrInv}{KipdrReachsafetyLoops}{Total}{}{Cputime}{Min}{3.780974072}%
\StoreBenchExecResult{SvcompNineteenPdrInv}{KipdrReachsafetyLoops}{Total}{}{Cputime}{Max}{914.094549427}%
\StoreBenchExecResult{SvcompNineteenPdrInv}{KipdrReachsafetyLoops}{Total}{}{Cputime}{Stdev}{444.2291204080376012724172518}%
\StoreBenchExecResult{SvcompNineteenPdrInv}{KipdrReachsafetyLoops}{Total}{}{Walltime}{}{86586.7717947030027856}%
\StoreBenchExecResult{SvcompNineteenPdrInv}{KipdrReachsafetyLoops}{Total}{}{Walltime}{Avg}{416.2825567053028980076923077}%
\StoreBenchExecResult{SvcompNineteenPdrInv}{KipdrReachsafetyLoops}{Total}{}{Walltime}{Median}{37.794282779505011}%
\StoreBenchExecResult{SvcompNineteenPdrInv}{KipdrReachsafetyLoops}{Total}{}{Walltime}{Min}{2.0260269490099745}%
\StoreBenchExecResult{SvcompNineteenPdrInv}{KipdrReachsafetyLoops}{Total}{}{Walltime}{Max}{900.0657982929988}%
\StoreBenchExecResult{SvcompNineteenPdrInv}{KipdrReachsafetyLoops}{Total}{}{Walltime}{Stdev}{435.5210058069567462007118530}%
\StoreBenchExecResult{SvcompNineteenPdrInv}{KipdrReachsafetyLoops}{Correct}{}{Count}{}{103}%
\StoreBenchExecResult{SvcompNineteenPdrInv}{KipdrReachsafetyLoops}{Correct}{}{Cputime}{}{2001.140529958}%
\StoreBenchExecResult{SvcompNineteenPdrInv}{KipdrReachsafetyLoops}{Correct}{}{Cputime}{Avg}{19.42854883454368932038834951}%
\StoreBenchExecResult{SvcompNineteenPdrInv}{KipdrReachsafetyLoops}{Correct}{}{Cputime}{Median}{5.146735184}%
\StoreBenchExecResult{SvcompNineteenPdrInv}{KipdrReachsafetyLoops}{Correct}{}{Cputime}{Min}{3.780974072}%
\StoreBenchExecResult{SvcompNineteenPdrInv}{KipdrReachsafetyLoops}{Correct}{}{Cputime}{Max}{873.632759332}%
\StoreBenchExecResult{SvcompNineteenPdrInv}{KipdrReachsafetyLoops}{Correct}{}{Cputime}{Stdev}{86.37426590698974380758415335}%
\StoreBenchExecResult{SvcompNineteenPdrInv}{KipdrReachsafetyLoops}{Correct}{}{Walltime}{}{1642.1783220590150727}%
\StoreBenchExecResult{SvcompNineteenPdrInv}{KipdrReachsafetyLoops}{Correct}{}{Walltime}{Avg}{15.94347885494189390970873786}%
\StoreBenchExecResult{SvcompNineteenPdrInv}{KipdrReachsafetyLoops}{Correct}{}{Walltime}{Median}{2.745816278998973}%
\StoreBenchExecResult{SvcompNineteenPdrInv}{KipdrReachsafetyLoops}{Correct}{}{Walltime}{Min}{2.0260269490099745}%
\StoreBenchExecResult{SvcompNineteenPdrInv}{KipdrReachsafetyLoops}{Correct}{}{Walltime}{Max}{867.6820184899989}%
\StoreBenchExecResult{SvcompNineteenPdrInv}{KipdrReachsafetyLoops}{Correct}{}{Walltime}{Stdev}{85.70405152707022687572045957}%
\StoreBenchExecResult{SvcompNineteenPdrInv}{KipdrReachsafetyLoops}{Correct}{False}{Count}{}{33}%
\StoreBenchExecResult{SvcompNineteenPdrInv}{KipdrReachsafetyLoops}{Correct}{False}{Cputime}{}{354.172753864}%
\StoreBenchExecResult{SvcompNineteenPdrInv}{KipdrReachsafetyLoops}{Correct}{False}{Cputime}{Avg}{10.73250769284848484848484848}%
\StoreBenchExecResult{SvcompNineteenPdrInv}{KipdrReachsafetyLoops}{Correct}{False}{Cputime}{Median}{5.041272365}%
\StoreBenchExecResult{SvcompNineteenPdrInv}{KipdrReachsafetyLoops}{Correct}{False}{Cputime}{Min}{3.967238392}%
\StoreBenchExecResult{SvcompNineteenPdrInv}{KipdrReachsafetyLoops}{Correct}{False}{Cputime}{Max}{67.296555394}%
\StoreBenchExecResult{SvcompNineteenPdrInv}{KipdrReachsafetyLoops}{Correct}{False}{Cputime}{Stdev}{13.57329982774520032255126322}%
\StoreBenchExecResult{SvcompNineteenPdrInv}{KipdrReachsafetyLoops}{Correct}{False}{Walltime}{}{244.5009334069909558}%
\StoreBenchExecResult{SvcompNineteenPdrInv}{KipdrReachsafetyLoops}{Correct}{False}{Walltime}{Avg}{7.409119194151241084848484848}%
\StoreBenchExecResult{SvcompNineteenPdrInv}{KipdrReachsafetyLoops}{Correct}{False}{Walltime}{Median}{2.697783490002621}%
\StoreBenchExecResult{SvcompNineteenPdrInv}{KipdrReachsafetyLoops}{Correct}{False}{Walltime}{Min}{2.1341965129977325}%
\StoreBenchExecResult{SvcompNineteenPdrInv}{KipdrReachsafetyLoops}{Correct}{False}{Walltime}{Max}{56.98991096299142}%
\StoreBenchExecResult{SvcompNineteenPdrInv}{KipdrReachsafetyLoops}{Correct}{False}{Walltime}{Stdev}{11.95858635916589263396181823}%
\StoreBenchExecResult{SvcompNineteenPdrInv}{KipdrReachsafetyLoops}{Wrong}{False}{Count}{}{0}%
\StoreBenchExecResult{SvcompNineteenPdrInv}{KipdrReachsafetyLoops}{Wrong}{False}{Cputime}{}{0}%
\StoreBenchExecResult{SvcompNineteenPdrInv}{KipdrReachsafetyLoops}{Wrong}{False}{Cputime}{Avg}{None}%
\StoreBenchExecResult{SvcompNineteenPdrInv}{KipdrReachsafetyLoops}{Wrong}{False}{Cputime}{Median}{None}%
\StoreBenchExecResult{SvcompNineteenPdrInv}{KipdrReachsafetyLoops}{Wrong}{False}{Cputime}{Min}{None}%
\StoreBenchExecResult{SvcompNineteenPdrInv}{KipdrReachsafetyLoops}{Wrong}{False}{Cputime}{Max}{None}%
\StoreBenchExecResult{SvcompNineteenPdrInv}{KipdrReachsafetyLoops}{Wrong}{False}{Cputime}{Stdev}{None}%
\StoreBenchExecResult{SvcompNineteenPdrInv}{KipdrReachsafetyLoops}{Wrong}{False}{Walltime}{}{0}%
\StoreBenchExecResult{SvcompNineteenPdrInv}{KipdrReachsafetyLoops}{Wrong}{False}{Walltime}{Avg}{None}%
\StoreBenchExecResult{SvcompNineteenPdrInv}{KipdrReachsafetyLoops}{Wrong}{False}{Walltime}{Median}{None}%
\StoreBenchExecResult{SvcompNineteenPdrInv}{KipdrReachsafetyLoops}{Wrong}{False}{Walltime}{Min}{None}%
\StoreBenchExecResult{SvcompNineteenPdrInv}{KipdrReachsafetyLoops}{Wrong}{False}{Walltime}{Max}{None}%
\StoreBenchExecResult{SvcompNineteenPdrInv}{KipdrReachsafetyLoops}{Wrong}{False}{Walltime}{Stdev}{None}%
\StoreBenchExecResult{SvcompNineteenPdrInv}{KipdrReachsafetyLoops}{Correct}{True}{Count}{}{70}%
\StoreBenchExecResult{SvcompNineteenPdrInv}{KipdrReachsafetyLoops}{Correct}{True}{Cputime}{}{1646.967776094}%
\StoreBenchExecResult{SvcompNineteenPdrInv}{KipdrReachsafetyLoops}{Correct}{True}{Cputime}{Avg}{23.52811108705714285714285714}%
\StoreBenchExecResult{SvcompNineteenPdrInv}{KipdrReachsafetyLoops}{Correct}{True}{Cputime}{Median}{5.1746539495}%
\StoreBenchExecResult{SvcompNineteenPdrInv}{KipdrReachsafetyLoops}{Correct}{True}{Cputime}{Min}{3.780974072}%
\StoreBenchExecResult{SvcompNineteenPdrInv}{KipdrReachsafetyLoops}{Correct}{True}{Cputime}{Max}{873.632759332}%
\StoreBenchExecResult{SvcompNineteenPdrInv}{KipdrReachsafetyLoops}{Correct}{True}{Cputime}{Stdev}{104.1071727316121777114029646}%
\StoreBenchExecResult{SvcompNineteenPdrInv}{KipdrReachsafetyLoops}{Correct}{True}{Walltime}{}{1397.6773886520241169}%
\StoreBenchExecResult{SvcompNineteenPdrInv}{KipdrReachsafetyLoops}{Correct}{True}{Walltime}{Avg}{19.96681983788605881285714286}%
\StoreBenchExecResult{SvcompNineteenPdrInv}{KipdrReachsafetyLoops}{Correct}{True}{Walltime}{Median}{2.7558872904992313}%
\StoreBenchExecResult{SvcompNineteenPdrInv}{KipdrReachsafetyLoops}{Correct}{True}{Walltime}{Min}{2.0260269490099745}%
\StoreBenchExecResult{SvcompNineteenPdrInv}{KipdrReachsafetyLoops}{Correct}{True}{Walltime}{Max}{867.6820184899989}%
\StoreBenchExecResult{SvcompNineteenPdrInv}{KipdrReachsafetyLoops}{Correct}{True}{Walltime}{Stdev}{103.3923226874851151673313880}%
\StoreBenchExecResult{SvcompNineteenPdrInv}{KipdrReachsafetyLoops}{Wrong}{True}{Count}{}{0}%
\StoreBenchExecResult{SvcompNineteenPdrInv}{KipdrReachsafetyLoops}{Wrong}{True}{Cputime}{}{0}%
\StoreBenchExecResult{SvcompNineteenPdrInv}{KipdrReachsafetyLoops}{Wrong}{True}{Cputime}{Avg}{None}%
\StoreBenchExecResult{SvcompNineteenPdrInv}{KipdrReachsafetyLoops}{Wrong}{True}{Cputime}{Median}{None}%
\StoreBenchExecResult{SvcompNineteenPdrInv}{KipdrReachsafetyLoops}{Wrong}{True}{Cputime}{Min}{None}%
\StoreBenchExecResult{SvcompNineteenPdrInv}{KipdrReachsafetyLoops}{Wrong}{True}{Cputime}{Max}{None}%
\StoreBenchExecResult{SvcompNineteenPdrInv}{KipdrReachsafetyLoops}{Wrong}{True}{Cputime}{Stdev}{None}%
\StoreBenchExecResult{SvcompNineteenPdrInv}{KipdrReachsafetyLoops}{Wrong}{True}{Walltime}{}{0}%
\StoreBenchExecResult{SvcompNineteenPdrInv}{KipdrReachsafetyLoops}{Wrong}{True}{Walltime}{Avg}{None}%
\StoreBenchExecResult{SvcompNineteenPdrInv}{KipdrReachsafetyLoops}{Wrong}{True}{Walltime}{Median}{None}%
\StoreBenchExecResult{SvcompNineteenPdrInv}{KipdrReachsafetyLoops}{Wrong}{True}{Walltime}{Min}{None}%
\StoreBenchExecResult{SvcompNineteenPdrInv}{KipdrReachsafetyLoops}{Wrong}{True}{Walltime}{Max}{None}%
\StoreBenchExecResult{SvcompNineteenPdrInv}{KipdrReachsafetyLoops}{Wrong}{True}{Walltime}{Stdev}{None}%
\StoreBenchExecResult{SvcompNineteenPdrInv}{KipdrReachsafetyLoops}{Error}{}{Count}{}{105}%
\StoreBenchExecResult{SvcompNineteenPdrInv}{KipdrReachsafetyLoops}{Error}{}{Cputime}{}{87152.254464836}%
\StoreBenchExecResult{SvcompNineteenPdrInv}{KipdrReachsafetyLoops}{Error}{}{Cputime}{Avg}{830.0214710936761904761904762}%
\StoreBenchExecResult{SvcompNineteenPdrInv}{KipdrReachsafetyLoops}{Error}{}{Cputime}{Median}{902.386247526}%
\StoreBenchExecResult{SvcompNineteenPdrInv}{KipdrReachsafetyLoops}{Error}{}{Cputime}{Min}{5.236251169}%
\StoreBenchExecResult{SvcompNineteenPdrInv}{KipdrReachsafetyLoops}{Error}{}{Cputime}{Max}{914.094549427}%
\StoreBenchExecResult{SvcompNineteenPdrInv}{KipdrReachsafetyLoops}{Error}{}{Cputime}{Stdev}{241.3094900156614792285913944}%
\StoreBenchExecResult{SvcompNineteenPdrInv}{KipdrReachsafetyLoops}{Error}{}{Walltime}{}{84944.5934726439877129}%
\StoreBenchExecResult{SvcompNineteenPdrInv}{KipdrReachsafetyLoops}{Error}{}{Walltime}{Avg}{808.9961283108951210752380952}%
\StoreBenchExecResult{SvcompNineteenPdrInv}{KipdrReachsafetyLoops}{Error}{}{Walltime}{Median}{890.6096780819935}%
\StoreBenchExecResult{SvcompNineteenPdrInv}{KipdrReachsafetyLoops}{Error}{}{Walltime}{Min}{2.930289707001066}%
\StoreBenchExecResult{SvcompNineteenPdrInv}{KipdrReachsafetyLoops}{Error}{}{Walltime}{Max}{900.0657982929988}%
\StoreBenchExecResult{SvcompNineteenPdrInv}{KipdrReachsafetyLoops}{Error}{}{Walltime}{Stdev}{238.9484331208485067835052352}%
\StoreBenchExecResult{SvcompNineteenPdrInv}{KipdrReachsafetyLoops}{Error}{Error}{Count}{}{5}%
\StoreBenchExecResult{SvcompNineteenPdrInv}{KipdrReachsafetyLoops}{Error}{Error}{Cputime}{}{35.840210913}%
\StoreBenchExecResult{SvcompNineteenPdrInv}{KipdrReachsafetyLoops}{Error}{Error}{Cputime}{Avg}{7.1680421826}%
\StoreBenchExecResult{SvcompNineteenPdrInv}{KipdrReachsafetyLoops}{Error}{Error}{Cputime}{Median}{7.556258064}%
\StoreBenchExecResult{SvcompNineteenPdrInv}{KipdrReachsafetyLoops}{Error}{Error}{Cputime}{Min}{6.419385375}%
\StoreBenchExecResult{SvcompNineteenPdrInv}{KipdrReachsafetyLoops}{Error}{Error}{Cputime}{Max}{7.607837587}%
\StoreBenchExecResult{SvcompNineteenPdrInv}{KipdrReachsafetyLoops}{Error}{Error}{Cputime}{Stdev}{0.5056886459839748019967836029}%
\StoreBenchExecResult{SvcompNineteenPdrInv}{KipdrReachsafetyLoops}{Error}{Error}{Walltime}{}{18.8451899900246645}%
\StoreBenchExecResult{SvcompNineteenPdrInv}{KipdrReachsafetyLoops}{Error}{Error}{Walltime}{Avg}{3.7690379980049329}%
\StoreBenchExecResult{SvcompNineteenPdrInv}{KipdrReachsafetyLoops}{Error}{Error}{Walltime}{Median}{3.969759965009871}%
\StoreBenchExecResult{SvcompNineteenPdrInv}{KipdrReachsafetyLoops}{Error}{Error}{Walltime}{Min}{3.405665042999317}%
\StoreBenchExecResult{SvcompNineteenPdrInv}{KipdrReachsafetyLoops}{Error}{Error}{Walltime}{Max}{3.9856656920019304}%
\StoreBenchExecResult{SvcompNineteenPdrInv}{KipdrReachsafetyLoops}{Error}{Error}{Walltime}{Stdev}{0.2547796764257207457166545049}%
\StoreBenchExecResult{SvcompNineteenPdrInv}{KipdrReachsafetyLoops}{Error}{OutOfMemory}{Count}{}{2}%
\StoreBenchExecResult{SvcompNineteenPdrInv}{KipdrReachsafetyLoops}{Error}{OutOfMemory}{Cputime}{}{358.049159722}%
\StoreBenchExecResult{SvcompNineteenPdrInv}{KipdrReachsafetyLoops}{Error}{OutOfMemory}{Cputime}{Avg}{179.024579861}%
\StoreBenchExecResult{SvcompNineteenPdrInv}{KipdrReachsafetyLoops}{Error}{OutOfMemory}{Cputime}{Median}{179.024579861}%
\StoreBenchExecResult{SvcompNineteenPdrInv}{KipdrReachsafetyLoops}{Error}{OutOfMemory}{Cputime}{Min}{173.053521197}%
\StoreBenchExecResult{SvcompNineteenPdrInv}{KipdrReachsafetyLoops}{Error}{OutOfMemory}{Cputime}{Max}{184.995638525}%
\StoreBenchExecResult{SvcompNineteenPdrInv}{KipdrReachsafetyLoops}{Error}{OutOfMemory}{Cputime}{Stdev}{5.971058664}%
\StoreBenchExecResult{SvcompNineteenPdrInv}{KipdrReachsafetyLoops}{Error}{OutOfMemory}{Walltime}{}{343.73986024397891}%
\StoreBenchExecResult{SvcompNineteenPdrInv}{KipdrReachsafetyLoops}{Error}{OutOfMemory}{Walltime}{Avg}{171.869930121989455}%
\StoreBenchExecResult{SvcompNineteenPdrInv}{KipdrReachsafetyLoops}{Error}{OutOfMemory}{Walltime}{Median}{171.869930121989455}%
\StoreBenchExecResult{SvcompNineteenPdrInv}{KipdrReachsafetyLoops}{Error}{OutOfMemory}{Walltime}{Min}{166.35414471798867}%
\StoreBenchExecResult{SvcompNineteenPdrInv}{KipdrReachsafetyLoops}{Error}{OutOfMemory}{Walltime}{Max}{177.38571552599024}%
\StoreBenchExecResult{SvcompNineteenPdrInv}{KipdrReachsafetyLoops}{Error}{OutOfMemory}{Walltime}{Stdev}{5.515785404000785000000000000}%
\StoreBenchExecResult{SvcompNineteenPdrInv}{KipdrReachsafetyLoops}{Error}{SegmentationFault}{Count}{}{2}%
\StoreBenchExecResult{SvcompNineteenPdrInv}{KipdrReachsafetyLoops}{Error}{SegmentationFault}{Cputime}{}{10.726518174}%
\StoreBenchExecResult{SvcompNineteenPdrInv}{KipdrReachsafetyLoops}{Error}{SegmentationFault}{Cputime}{Avg}{5.363259087}%
\StoreBenchExecResult{SvcompNineteenPdrInv}{KipdrReachsafetyLoops}{Error}{SegmentationFault}{Cputime}{Median}{5.363259087}%
\StoreBenchExecResult{SvcompNineteenPdrInv}{KipdrReachsafetyLoops}{Error}{SegmentationFault}{Cputime}{Min}{5.236251169}%
\StoreBenchExecResult{SvcompNineteenPdrInv}{KipdrReachsafetyLoops}{Error}{SegmentationFault}{Cputime}{Max}{5.490267005}%
\StoreBenchExecResult{SvcompNineteenPdrInv}{KipdrReachsafetyLoops}{Error}{SegmentationFault}{Cputime}{Stdev}{0.127007918}%
\StoreBenchExecResult{SvcompNineteenPdrInv}{KipdrReachsafetyLoops}{Error}{SegmentationFault}{Walltime}{}{5.8765792699996384}%
\StoreBenchExecResult{SvcompNineteenPdrInv}{KipdrReachsafetyLoops}{Error}{SegmentationFault}{Walltime}{Avg}{2.9382896349998192}%
\StoreBenchExecResult{SvcompNineteenPdrInv}{KipdrReachsafetyLoops}{Error}{SegmentationFault}{Walltime}{Median}{2.9382896349998192}%
\StoreBenchExecResult{SvcompNineteenPdrInv}{KipdrReachsafetyLoops}{Error}{SegmentationFault}{Walltime}{Min}{2.930289707001066}%
\StoreBenchExecResult{SvcompNineteenPdrInv}{KipdrReachsafetyLoops}{Error}{SegmentationFault}{Walltime}{Max}{2.9462895629985724}%
\StoreBenchExecResult{SvcompNineteenPdrInv}{KipdrReachsafetyLoops}{Error}{SegmentationFault}{Walltime}{Stdev}{0.007999927998753200000000000001}%
\StoreBenchExecResult{SvcompNineteenPdrInv}{KipdrReachsafetyLoops}{Error}{Timeout}{Count}{}{96}%
\StoreBenchExecResult{SvcompNineteenPdrInv}{KipdrReachsafetyLoops}{Error}{Timeout}{Cputime}{}{86747.638576027}%
\StoreBenchExecResult{SvcompNineteenPdrInv}{KipdrReachsafetyLoops}{Error}{Timeout}{Cputime}{Avg}{903.6212351669479166666666667}%
\StoreBenchExecResult{SvcompNineteenPdrInv}{KipdrReachsafetyLoops}{Error}{Timeout}{Cputime}{Median}{902.576873401}%
\StoreBenchExecResult{SvcompNineteenPdrInv}{KipdrReachsafetyLoops}{Error}{Timeout}{Cputime}{Min}{901.181788202}%
\StoreBenchExecResult{SvcompNineteenPdrInv}{KipdrReachsafetyLoops}{Error}{Timeout}{Cputime}{Max}{914.094549427}%
\StoreBenchExecResult{SvcompNineteenPdrInv}{KipdrReachsafetyLoops}{Error}{Timeout}{Cputime}{Stdev}{3.105501820207186729081616024}%
\StoreBenchExecResult{SvcompNineteenPdrInv}{KipdrReachsafetyLoops}{Error}{Timeout}{Walltime}{}{84576.1318431399845}%
\StoreBenchExecResult{SvcompNineteenPdrInv}{KipdrReachsafetyLoops}{Error}{Timeout}{Walltime}{Avg}{881.0013733660415052083333333}%
\StoreBenchExecResult{SvcompNineteenPdrInv}{KipdrReachsafetyLoops}{Error}{Timeout}{Walltime}{Median}{891.02995344450755}%
\StoreBenchExecResult{SvcompNineteenPdrInv}{KipdrReachsafetyLoops}{Error}{Timeout}{Walltime}{Min}{644.7454644949903}%
\StoreBenchExecResult{SvcompNineteenPdrInv}{KipdrReachsafetyLoops}{Error}{Timeout}{Walltime}{Max}{900.0657982929988}%
\StoreBenchExecResult{SvcompNineteenPdrInv}{KipdrReachsafetyLoops}{Error}{Timeout}{Walltime}{Stdev}{38.73629549401359369727141614}%
\providecommand\StoreBenchExecResult[7]{\expandafter\newcommand\csname#1#2#3#4#5#6\endcsname{#7}}%
\StoreBenchExecResult{SvcompNineteenPesco}{SvCompPropReachsafetyReachsafetyLoops}{Total}{}{Count}{}{208}%
\StoreBenchExecResult{SvcompNineteenPesco}{SvCompPropReachsafetyReachsafetyLoops}{Total}{}{Cputime}{}{70456.113051686}%
\StoreBenchExecResult{SvcompNineteenPesco}{SvCompPropReachsafetyReachsafetyLoops}{Total}{}{Cputime}{Avg}{338.7313127484903846153846154}%
\StoreBenchExecResult{SvcompNineteenPesco}{SvCompPropReachsafetyReachsafetyLoops}{Total}{}{Cputime}{Median}{22.973791851}%
\StoreBenchExecResult{SvcompNineteenPesco}{SvCompPropReachsafetyReachsafetyLoops}{Total}{}{Cputime}{Min}{10.836018164}%
\StoreBenchExecResult{SvcompNineteenPesco}{SvCompPropReachsafetyReachsafetyLoops}{Total}{}{Cputime}{Max}{968.199552885}%
\StoreBenchExecResult{SvcompNineteenPesco}{SvCompPropReachsafetyReachsafetyLoops}{Total}{}{Cputime}{Stdev}{430.9909917763056957304652017}%
\StoreBenchExecResult{SvcompNineteenPesco}{SvCompPropReachsafetyReachsafetyLoops}{Total}{}{Walltime}{}{37372.71061921083}%
\StoreBenchExecResult{SvcompNineteenPesco}{SvCompPropReachsafetyReachsafetyLoops}{Total}{}{Walltime}{Avg}{179.6764933615905288461538462}%
\StoreBenchExecResult{SvcompNineteenPesco}{SvCompPropReachsafetyReachsafetyLoops}{Total}{}{Walltime}{Median}{8.11294758320}%
\StoreBenchExecResult{SvcompNineteenPesco}{SvCompPropReachsafetyReachsafetyLoops}{Total}{}{Walltime}{Min}{3.3984670639}%
\StoreBenchExecResult{SvcompNineteenPesco}{SvCompPropReachsafetyReachsafetyLoops}{Total}{}{Walltime}{Max}{886.269155025}%
\StoreBenchExecResult{SvcompNineteenPesco}{SvCompPropReachsafetyReachsafetyLoops}{Total}{}{Walltime}{Stdev}{242.9221371374295477108981943}%
\StoreBenchExecResult{SvcompNineteenPesco}{SvCompPropReachsafetyReachsafetyLoops}{Correct}{}{Count}{}{138}%
\StoreBenchExecResult{SvcompNineteenPesco}{SvCompPropReachsafetyReachsafetyLoops}{Correct}{}{Cputime}{}{6267.576871079}%
\StoreBenchExecResult{SvcompNineteenPesco}{SvCompPropReachsafetyReachsafetyLoops}{Correct}{}{Cputime}{Avg}{45.41722370347101449275362319}%
\StoreBenchExecResult{SvcompNineteenPesco}{SvCompPropReachsafetyReachsafetyLoops}{Correct}{}{Cputime}{Median}{14.846619309}%
\StoreBenchExecResult{SvcompNineteenPesco}{SvCompPropReachsafetyReachsafetyLoops}{Correct}{}{Cputime}{Min}{10.836018164}%
\StoreBenchExecResult{SvcompNineteenPesco}{SvCompPropReachsafetyReachsafetyLoops}{Correct}{}{Cputime}{Max}{833.145903519}%
\StoreBenchExecResult{SvcompNineteenPesco}{SvCompPropReachsafetyReachsafetyLoops}{Correct}{}{Cputime}{Stdev}{115.3112711551231497556901512}%
\StoreBenchExecResult{SvcompNineteenPesco}{SvCompPropReachsafetyReachsafetyLoops}{Correct}{}{Walltime}{}{2623.32415127813}%
\StoreBenchExecResult{SvcompNineteenPesco}{SvCompPropReachsafetyReachsafetyLoops}{Correct}{}{Walltime}{Avg}{19.00959529911688405797101449}%
\StoreBenchExecResult{SvcompNineteenPesco}{SvCompPropReachsafetyReachsafetyLoops}{Correct}{}{Walltime}{Median}{4.20904695988}%
\StoreBenchExecResult{SvcompNineteenPesco}{SvCompPropReachsafetyReachsafetyLoops}{Correct}{}{Walltime}{Min}{3.3984670639}%
\StoreBenchExecResult{SvcompNineteenPesco}{SvCompPropReachsafetyReachsafetyLoops}{Correct}{}{Walltime}{Max}{340.811495066}%
\StoreBenchExecResult{SvcompNineteenPesco}{SvCompPropReachsafetyReachsafetyLoops}{Correct}{}{Walltime}{Stdev}{54.93368736581610159505042930}%
\StoreBenchExecResult{SvcompNineteenPesco}{SvCompPropReachsafetyReachsafetyLoops}{Correct}{False}{Count}{}{40}%
\StoreBenchExecResult{SvcompNineteenPesco}{SvCompPropReachsafetyReachsafetyLoops}{Correct}{False}{Cputime}{}{1103.480304924}%
\StoreBenchExecResult{SvcompNineteenPesco}{SvCompPropReachsafetyReachsafetyLoops}{Correct}{False}{Cputime}{Avg}{27.5870076231}%
\StoreBenchExecResult{SvcompNineteenPesco}{SvCompPropReachsafetyReachsafetyLoops}{Correct}{False}{Cputime}{Median}{14.053559592}%
\StoreBenchExecResult{SvcompNineteenPesco}{SvCompPropReachsafetyReachsafetyLoops}{Correct}{False}{Cputime}{Min}{11.360098395}%
\StoreBenchExecResult{SvcompNineteenPesco}{SvCompPropReachsafetyReachsafetyLoops}{Correct}{False}{Cputime}{Max}{372.173194426}%
\StoreBenchExecResult{SvcompNineteenPesco}{SvCompPropReachsafetyReachsafetyLoops}{Correct}{False}{Cputime}{Stdev}{56.43982037388606261571220622}%
\StoreBenchExecResult{SvcompNineteenPesco}{SvCompPropReachsafetyReachsafetyLoops}{Correct}{False}{Walltime}{}{521.39588975909}%
\StoreBenchExecResult{SvcompNineteenPesco}{SvCompPropReachsafetyReachsafetyLoops}{Correct}{False}{Walltime}{Avg}{13.03489724397725}%
\StoreBenchExecResult{SvcompNineteenPesco}{SvCompPropReachsafetyReachsafetyLoops}{Correct}{False}{Walltime}{Median}{4.084911108015}%
\StoreBenchExecResult{SvcompNineteenPesco}{SvCompPropReachsafetyReachsafetyLoops}{Correct}{False}{Walltime}{Min}{3.62637400627}%
\StoreBenchExecResult{SvcompNineteenPesco}{SvCompPropReachsafetyReachsafetyLoops}{Correct}{False}{Walltime}{Max}{308.448894024}%
\StoreBenchExecResult{SvcompNineteenPesco}{SvCompPropReachsafetyReachsafetyLoops}{Correct}{False}{Walltime}{Stdev}{47.43545073009715931016167268}%
\StoreBenchExecResult{SvcompNineteenPesco}{SvCompPropReachsafetyReachsafetyLoops}{Wrong}{False}{Count}{}{0}%
\StoreBenchExecResult{SvcompNineteenPesco}{SvCompPropReachsafetyReachsafetyLoops}{Wrong}{False}{Cputime}{}{0}%
\StoreBenchExecResult{SvcompNineteenPesco}{SvCompPropReachsafetyReachsafetyLoops}{Wrong}{False}{Cputime}{Avg}{None}%
\StoreBenchExecResult{SvcompNineteenPesco}{SvCompPropReachsafetyReachsafetyLoops}{Wrong}{False}{Cputime}{Median}{None}%
\StoreBenchExecResult{SvcompNineteenPesco}{SvCompPropReachsafetyReachsafetyLoops}{Wrong}{False}{Cputime}{Min}{None}%
\StoreBenchExecResult{SvcompNineteenPesco}{SvCompPropReachsafetyReachsafetyLoops}{Wrong}{False}{Cputime}{Max}{None}%
\StoreBenchExecResult{SvcompNineteenPesco}{SvCompPropReachsafetyReachsafetyLoops}{Wrong}{False}{Cputime}{Stdev}{None}%
\StoreBenchExecResult{SvcompNineteenPesco}{SvCompPropReachsafetyReachsafetyLoops}{Wrong}{False}{Walltime}{}{0}%
\StoreBenchExecResult{SvcompNineteenPesco}{SvCompPropReachsafetyReachsafetyLoops}{Wrong}{False}{Walltime}{Avg}{None}%
\StoreBenchExecResult{SvcompNineteenPesco}{SvCompPropReachsafetyReachsafetyLoops}{Wrong}{False}{Walltime}{Median}{None}%
\StoreBenchExecResult{SvcompNineteenPesco}{SvCompPropReachsafetyReachsafetyLoops}{Wrong}{False}{Walltime}{Min}{None}%
\StoreBenchExecResult{SvcompNineteenPesco}{SvCompPropReachsafetyReachsafetyLoops}{Wrong}{False}{Walltime}{Max}{None}%
\StoreBenchExecResult{SvcompNineteenPesco}{SvCompPropReachsafetyReachsafetyLoops}{Wrong}{False}{Walltime}{Stdev}{None}%
\StoreBenchExecResult{SvcompNineteenPesco}{SvCompPropReachsafetyReachsafetyLoops}{Correct}{True}{Count}{}{98}%
\StoreBenchExecResult{SvcompNineteenPesco}{SvCompPropReachsafetyReachsafetyLoops}{Correct}{True}{Cputime}{}{5164.096566155}%
\StoreBenchExecResult{SvcompNineteenPesco}{SvCompPropReachsafetyReachsafetyLoops}{Correct}{True}{Cputime}{Avg}{52.69486291994897959183673469}%
\StoreBenchExecResult{SvcompNineteenPesco}{SvCompPropReachsafetyReachsafetyLoops}{Correct}{True}{Cputime}{Median}{15.3529259285}%
\StoreBenchExecResult{SvcompNineteenPesco}{SvCompPropReachsafetyReachsafetyLoops}{Correct}{True}{Cputime}{Min}{10.836018164}%
\StoreBenchExecResult{SvcompNineteenPesco}{SvCompPropReachsafetyReachsafetyLoops}{Correct}{True}{Cputime}{Max}{833.145903519}%
\StoreBenchExecResult{SvcompNineteenPesco}{SvCompPropReachsafetyReachsafetyLoops}{Correct}{True}{Cputime}{Stdev}{131.3049824537214351281299455}%
\StoreBenchExecResult{SvcompNineteenPesco}{SvCompPropReachsafetyReachsafetyLoops}{Correct}{True}{Walltime}{}{2101.92826151904}%
\StoreBenchExecResult{SvcompNineteenPesco}{SvCompPropReachsafetyReachsafetyLoops}{Correct}{True}{Walltime}{Avg}{21.44824756652081632653061224}%
\StoreBenchExecResult{SvcompNineteenPesco}{SvCompPropReachsafetyReachsafetyLoops}{Correct}{True}{Walltime}{Median}{4.33695256710}%
\StoreBenchExecResult{SvcompNineteenPesco}{SvCompPropReachsafetyReachsafetyLoops}{Correct}{True}{Walltime}{Min}{3.3984670639}%
\StoreBenchExecResult{SvcompNineteenPesco}{SvCompPropReachsafetyReachsafetyLoops}{Correct}{True}{Walltime}{Max}{340.811495066}%
\StoreBenchExecResult{SvcompNineteenPesco}{SvCompPropReachsafetyReachsafetyLoops}{Correct}{True}{Walltime}{Stdev}{57.53689255674326989434168303}%
\StoreBenchExecResult{SvcompNineteenPesco}{SvCompPropReachsafetyReachsafetyLoops}{Wrong}{True}{Count}{}{0}%
\StoreBenchExecResult{SvcompNineteenPesco}{SvCompPropReachsafetyReachsafetyLoops}{Wrong}{True}{Cputime}{}{0}%
\StoreBenchExecResult{SvcompNineteenPesco}{SvCompPropReachsafetyReachsafetyLoops}{Wrong}{True}{Cputime}{Avg}{None}%
\StoreBenchExecResult{SvcompNineteenPesco}{SvCompPropReachsafetyReachsafetyLoops}{Wrong}{True}{Cputime}{Median}{None}%
\StoreBenchExecResult{SvcompNineteenPesco}{SvCompPropReachsafetyReachsafetyLoops}{Wrong}{True}{Cputime}{Min}{None}%
\StoreBenchExecResult{SvcompNineteenPesco}{SvCompPropReachsafetyReachsafetyLoops}{Wrong}{True}{Cputime}{Max}{None}%
\StoreBenchExecResult{SvcompNineteenPesco}{SvCompPropReachsafetyReachsafetyLoops}{Wrong}{True}{Cputime}{Stdev}{None}%
\StoreBenchExecResult{SvcompNineteenPesco}{SvCompPropReachsafetyReachsafetyLoops}{Wrong}{True}{Walltime}{}{0}%
\StoreBenchExecResult{SvcompNineteenPesco}{SvCompPropReachsafetyReachsafetyLoops}{Wrong}{True}{Walltime}{Avg}{None}%
\StoreBenchExecResult{SvcompNineteenPesco}{SvCompPropReachsafetyReachsafetyLoops}{Wrong}{True}{Walltime}{Median}{None}%
\StoreBenchExecResult{SvcompNineteenPesco}{SvCompPropReachsafetyReachsafetyLoops}{Wrong}{True}{Walltime}{Min}{None}%
\StoreBenchExecResult{SvcompNineteenPesco}{SvCompPropReachsafetyReachsafetyLoops}{Wrong}{True}{Walltime}{Max}{None}%
\StoreBenchExecResult{SvcompNineteenPesco}{SvCompPropReachsafetyReachsafetyLoops}{Wrong}{True}{Walltime}{Stdev}{None}%
\StoreBenchExecResult{SvcompNineteenPesco}{SvCompPropReachsafetyReachsafetyLoops}{Error}{}{Count}{}{68}%
\StoreBenchExecResult{SvcompNineteenPesco}{SvCompPropReachsafetyReachsafetyLoops}{Error}{}{Cputime}{}{63984.193785832}%
\StoreBenchExecResult{SvcompNineteenPesco}{SvCompPropReachsafetyReachsafetyLoops}{Error}{}{Cputime}{Avg}{940.9440262622352941176470588}%
\StoreBenchExecResult{SvcompNineteenPesco}{SvCompPropReachsafetyReachsafetyLoops}{Error}{}{Cputime}{Median}{960.357043663}%
\StoreBenchExecResult{SvcompNineteenPesco}{SvCompPropReachsafetyReachsafetyLoops}{Error}{}{Cputime}{Min}{594.31700413}%
\StoreBenchExecResult{SvcompNineteenPesco}{SvCompPropReachsafetyReachsafetyLoops}{Error}{}{Cputime}{Max}{968.199552885}%
\StoreBenchExecResult{SvcompNineteenPesco}{SvCompPropReachsafetyReachsafetyLoops}{Error}{}{Cputime}{Stdev}{47.94949554456452145023495861}%
\StoreBenchExecResult{SvcompNineteenPesco}{SvCompPropReachsafetyReachsafetyLoops}{Error}{}{Walltime}{}{34598.228471755}%
\StoreBenchExecResult{SvcompNineteenPesco}{SvCompPropReachsafetyReachsafetyLoops}{Error}{}{Walltime}{Avg}{508.7974775258088235294117647}%
\StoreBenchExecResult{SvcompNineteenPesco}{SvCompPropReachsafetyReachsafetyLoops}{Error}{}{Walltime}{Median}{478.6927084925}%
\StoreBenchExecResult{SvcompNineteenPesco}{SvCompPropReachsafetyReachsafetyLoops}{Error}{}{Walltime}{Min}{403.778012037}%
\StoreBenchExecResult{SvcompNineteenPesco}{SvCompPropReachsafetyReachsafetyLoops}{Error}{}{Walltime}{Max}{886.269155025}%
\StoreBenchExecResult{SvcompNineteenPesco}{SvCompPropReachsafetyReachsafetyLoops}{Error}{}{Walltime}{Stdev}{115.5448362680987554676385545}%
\StoreBenchExecResult{SvcompNineteenPesco}{SvCompPropReachsafetyReachsafetyLoops}{Error}{OutOfMemory}{Count}{}{1}%
\StoreBenchExecResult{SvcompNineteenPesco}{SvCompPropReachsafetyReachsafetyLoops}{Error}{OutOfMemory}{Cputime}{}{594.31700413}%
\StoreBenchExecResult{SvcompNineteenPesco}{SvCompPropReachsafetyReachsafetyLoops}{Error}{OutOfMemory}{Cputime}{Avg}{594.31700413}%
\StoreBenchExecResult{SvcompNineteenPesco}{SvCompPropReachsafetyReachsafetyLoops}{Error}{OutOfMemory}{Cputime}{Median}{594.31700413}%
\StoreBenchExecResult{SvcompNineteenPesco}{SvCompPropReachsafetyReachsafetyLoops}{Error}{OutOfMemory}{Cputime}{Min}{594.31700413}%
\StoreBenchExecResult{SvcompNineteenPesco}{SvCompPropReachsafetyReachsafetyLoops}{Error}{OutOfMemory}{Cputime}{Max}{594.31700413}%
\StoreBenchExecResult{SvcompNineteenPesco}{SvCompPropReachsafetyReachsafetyLoops}{Error}{OutOfMemory}{Cputime}{Stdev}{0E-8}%
\StoreBenchExecResult{SvcompNineteenPesco}{SvCompPropReachsafetyReachsafetyLoops}{Error}{OutOfMemory}{Walltime}{}{555.2966609}%
\StoreBenchExecResult{SvcompNineteenPesco}{SvCompPropReachsafetyReachsafetyLoops}{Error}{OutOfMemory}{Walltime}{Avg}{555.2966609}%
\StoreBenchExecResult{SvcompNineteenPesco}{SvCompPropReachsafetyReachsafetyLoops}{Error}{OutOfMemory}{Walltime}{Median}{555.2966609}%
\StoreBenchExecResult{SvcompNineteenPesco}{SvCompPropReachsafetyReachsafetyLoops}{Error}{OutOfMemory}{Walltime}{Min}{555.2966609}%
\StoreBenchExecResult{SvcompNineteenPesco}{SvCompPropReachsafetyReachsafetyLoops}{Error}{OutOfMemory}{Walltime}{Max}{555.2966609}%
\StoreBenchExecResult{SvcompNineteenPesco}{SvCompPropReachsafetyReachsafetyLoops}{Error}{OutOfMemory}{Walltime}{Stdev}{0E-7}%
\StoreBenchExecResult{SvcompNineteenPesco}{SvCompPropReachsafetyReachsafetyLoops}{Error}{Timeout}{Count}{}{67}%
\StoreBenchExecResult{SvcompNineteenPesco}{SvCompPropReachsafetyReachsafetyLoops}{Error}{Timeout}{Cputime}{}{63389.876781702}%
\StoreBenchExecResult{SvcompNineteenPesco}{SvCompPropReachsafetyReachsafetyLoops}{Error}{Timeout}{Cputime}{Avg}{946.117563906}%
\StoreBenchExecResult{SvcompNineteenPesco}{SvCompPropReachsafetyReachsafetyLoops}{Error}{Timeout}{Cputime}{Median}{960.375905223}%
\StoreBenchExecResult{SvcompNineteenPesco}{SvCompPropReachsafetyReachsafetyLoops}{Error}{Timeout}{Cputime}{Min}{902.651286326}%
\StoreBenchExecResult{SvcompNineteenPesco}{SvCompPropReachsafetyReachsafetyLoops}{Error}{Timeout}{Cputime}{Max}{968.199552885}%
\StoreBenchExecResult{SvcompNineteenPesco}{SvCompPropReachsafetyReachsafetyLoops}{Error}{Timeout}{Cputime}{Stdev}{22.65869407221785553305800780}%
\StoreBenchExecResult{SvcompNineteenPesco}{SvCompPropReachsafetyReachsafetyLoops}{Error}{Timeout}{Walltime}{}{34042.931810855}%
\StoreBenchExecResult{SvcompNineteenPesco}{SvCompPropReachsafetyReachsafetyLoops}{Error}{Timeout}{Walltime}{Avg}{508.1034598635074626865671642}%
\StoreBenchExecResult{SvcompNineteenPesco}{SvCompPropReachsafetyReachsafetyLoops}{Error}{Timeout}{Walltime}{Median}{478.025187016}%
\StoreBenchExecResult{SvcompNineteenPesco}{SvCompPropReachsafetyReachsafetyLoops}{Error}{Timeout}{Walltime}{Min}{403.778012037}%
\StoreBenchExecResult{SvcompNineteenPesco}{SvCompPropReachsafetyReachsafetyLoops}{Error}{Timeout}{Walltime}{Max}{886.269155025}%
\StoreBenchExecResult{SvcompNineteenPesco}{SvCompPropReachsafetyReachsafetyLoops}{Error}{Timeout}{Walltime}{Stdev}{116.2631459098999568186646279}%
\StoreBenchExecResult{SvcompNineteenPesco}{SvCompPropReachsafetyReachsafetyLoops}{Unknown}{}{Count}{}{2}%
\StoreBenchExecResult{SvcompNineteenPesco}{SvCompPropReachsafetyReachsafetyLoops}{Unknown}{}{Cputime}{}{204.342394775}%
\StoreBenchExecResult{SvcompNineteenPesco}{SvCompPropReachsafetyReachsafetyLoops}{Unknown}{}{Cputime}{Avg}{102.1711973875}%
\StoreBenchExecResult{SvcompNineteenPesco}{SvCompPropReachsafetyReachsafetyLoops}{Unknown}{}{Cputime}{Median}{102.1711973875}%
\StoreBenchExecResult{SvcompNineteenPesco}{SvCompPropReachsafetyReachsafetyLoops}{Unknown}{}{Cputime}{Min}{101.871213833}%
\StoreBenchExecResult{SvcompNineteenPesco}{SvCompPropReachsafetyReachsafetyLoops}{Unknown}{}{Cputime}{Max}{102.471180942}%
\StoreBenchExecResult{SvcompNineteenPesco}{SvCompPropReachsafetyReachsafetyLoops}{Unknown}{}{Cputime}{Stdev}{0.2999835545}%
\StoreBenchExecResult{SvcompNineteenPesco}{SvCompPropReachsafetyReachsafetyLoops}{Unknown}{}{Walltime}{}{151.1579961777}%
\StoreBenchExecResult{SvcompNineteenPesco}{SvCompPropReachsafetyReachsafetyLoops}{Unknown}{}{Walltime}{Avg}{75.57899808885}%
\StoreBenchExecResult{SvcompNineteenPesco}{SvCompPropReachsafetyReachsafetyLoops}{Unknown}{}{Walltime}{Median}{75.57899808885}%
\StoreBenchExecResult{SvcompNineteenPesco}{SvCompPropReachsafetyReachsafetyLoops}{Unknown}{}{Walltime}{Min}{64.3903441429}%
\StoreBenchExecResult{SvcompNineteenPesco}{SvCompPropReachsafetyReachsafetyLoops}{Unknown}{}{Walltime}{Max}{86.7676520348}%
\StoreBenchExecResult{SvcompNineteenPesco}{SvCompPropReachsafetyReachsafetyLoops}{Unknown}{}{Walltime}{Stdev}{11.18865394595}%
\StoreBenchExecResult{SvcompNineteenPesco}{SvCompPropReachsafetyReachsafetyLoops}{Unknown}{Unknown}{Count}{}{2}%
\StoreBenchExecResult{SvcompNineteenPesco}{SvCompPropReachsafetyReachsafetyLoops}{Unknown}{Unknown}{Cputime}{}{204.342394775}%
\StoreBenchExecResult{SvcompNineteenPesco}{SvCompPropReachsafetyReachsafetyLoops}{Unknown}{Unknown}{Cputime}{Avg}{102.1711973875}%
\StoreBenchExecResult{SvcompNineteenPesco}{SvCompPropReachsafetyReachsafetyLoops}{Unknown}{Unknown}{Cputime}{Median}{102.1711973875}%
\StoreBenchExecResult{SvcompNineteenPesco}{SvCompPropReachsafetyReachsafetyLoops}{Unknown}{Unknown}{Cputime}{Min}{101.871213833}%
\StoreBenchExecResult{SvcompNineteenPesco}{SvCompPropReachsafetyReachsafetyLoops}{Unknown}{Unknown}{Cputime}{Max}{102.471180942}%
\StoreBenchExecResult{SvcompNineteenPesco}{SvCompPropReachsafetyReachsafetyLoops}{Unknown}{Unknown}{Cputime}{Stdev}{0.2999835545}%
\StoreBenchExecResult{SvcompNineteenPesco}{SvCompPropReachsafetyReachsafetyLoops}{Unknown}{Unknown}{Walltime}{}{151.1579961777}%
\StoreBenchExecResult{SvcompNineteenPesco}{SvCompPropReachsafetyReachsafetyLoops}{Unknown}{Unknown}{Walltime}{Avg}{75.57899808885}%
\StoreBenchExecResult{SvcompNineteenPesco}{SvCompPropReachsafetyReachsafetyLoops}{Unknown}{Unknown}{Walltime}{Median}{75.57899808885}%
\StoreBenchExecResult{SvcompNineteenPesco}{SvCompPropReachsafetyReachsafetyLoops}{Unknown}{Unknown}{Walltime}{Min}{64.3903441429}%
\StoreBenchExecResult{SvcompNineteenPesco}{SvCompPropReachsafetyReachsafetyLoops}{Unknown}{Unknown}{Walltime}{Max}{86.7676520348}%
\StoreBenchExecResult{SvcompNineteenPesco}{SvCompPropReachsafetyReachsafetyLoops}{Unknown}{Unknown}{Walltime}{Stdev}{11.18865394595}%
\providecommand\StoreBenchExecResult[7]{\expandafter\newcommand\csname#1#2#3#4#5#6\endcsname{#7}}%
\StoreBenchExecResult{SvcompNineteenSkink}{SvCompPropReachsafetyReachsafetyLoops}{Total}{}{Count}{}{208}%
\StoreBenchExecResult{SvcompNineteenSkink}{SvCompPropReachsafetyReachsafetyLoops}{Total}{}{Cputime}{}{18524.356983786}%
\StoreBenchExecResult{SvcompNineteenSkink}{SvCompPropReachsafetyReachsafetyLoops}{Total}{}{Cputime}{Avg}{89.05940857589423076923076923}%
\StoreBenchExecResult{SvcompNineteenSkink}{SvCompPropReachsafetyReachsafetyLoops}{Total}{}{Cputime}{Median}{10.2151298635}%
\StoreBenchExecResult{SvcompNineteenSkink}{SvCompPropReachsafetyReachsafetyLoops}{Total}{}{Cputime}{Min}{2.588134092}%
\StoreBenchExecResult{SvcompNineteenSkink}{SvCompPropReachsafetyReachsafetyLoops}{Total}{}{Cputime}{Max}{900.926218928}%
\StoreBenchExecResult{SvcompNineteenSkink}{SvCompPropReachsafetyReachsafetyLoops}{Total}{}{Cputime}{Stdev}{189.0302266803930026118712004}%
\StoreBenchExecResult{SvcompNineteenSkink}{SvCompPropReachsafetyReachsafetyLoops}{Total}{}{Walltime}{}{15475.02471900324}%
\StoreBenchExecResult{SvcompNineteenSkink}{SvCompPropReachsafetyReachsafetyLoops}{Total}{}{Walltime}{Avg}{74.39915730290019230769230769}%
\StoreBenchExecResult{SvcompNineteenSkink}{SvCompPropReachsafetyReachsafetyLoops}{Total}{}{Walltime}{Median}{4.132669568065}%
\StoreBenchExecResult{SvcompNineteenSkink}{SvCompPropReachsafetyReachsafetyLoops}{Total}{}{Walltime}{Min}{1.40682601929}%
\StoreBenchExecResult{SvcompNineteenSkink}{SvCompPropReachsafetyReachsafetyLoops}{Total}{}{Walltime}{Max}{876.589123964}%
\StoreBenchExecResult{SvcompNineteenSkink}{SvCompPropReachsafetyReachsafetyLoops}{Total}{}{Walltime}{Stdev}{175.2763944203405739846298172}%
\StoreBenchExecResult{SvcompNineteenSkink}{SvCompPropReachsafetyReachsafetyLoops}{Correct}{}{Count}{}{120}%
\StoreBenchExecResult{SvcompNineteenSkink}{SvCompPropReachsafetyReachsafetyLoops}{Correct}{}{Cputime}{}{1924.274850144}%
\StoreBenchExecResult{SvcompNineteenSkink}{SvCompPropReachsafetyReachsafetyLoops}{Correct}{}{Cputime}{Avg}{16.0356237512}%
\StoreBenchExecResult{SvcompNineteenSkink}{SvCompPropReachsafetyReachsafetyLoops}{Correct}{}{Cputime}{Median}{5.1796046455}%
\StoreBenchExecResult{SvcompNineteenSkink}{SvCompPropReachsafetyReachsafetyLoops}{Correct}{}{Cputime}{Min}{3.334866554}%
\StoreBenchExecResult{SvcompNineteenSkink}{SvCompPropReachsafetyReachsafetyLoops}{Correct}{}{Cputime}{Max}{292.610540875}%
\StoreBenchExecResult{SvcompNineteenSkink}{SvCompPropReachsafetyReachsafetyLoops}{Correct}{}{Cputime}{Stdev}{31.18300064259024829361162810}%
\StoreBenchExecResult{SvcompNineteenSkink}{SvCompPropReachsafetyReachsafetyLoops}{Correct}{}{Walltime}{}{1313.89790129641}%
\StoreBenchExecResult{SvcompNineteenSkink}{SvCompPropReachsafetyReachsafetyLoops}{Correct}{}{Walltime}{Avg}{10.94914917747008333333333333}%
\StoreBenchExecResult{SvcompNineteenSkink}{SvCompPropReachsafetyReachsafetyLoops}{Correct}{}{Walltime}{Median}{2.15304863453}%
\StoreBenchExecResult{SvcompNineteenSkink}{SvCompPropReachsafetyReachsafetyLoops}{Correct}{}{Walltime}{Min}{1.70069384575}%
\StoreBenchExecResult{SvcompNineteenSkink}{SvCompPropReachsafetyReachsafetyLoops}{Correct}{}{Walltime}{Max}{274.077408075}%
\StoreBenchExecResult{SvcompNineteenSkink}{SvCompPropReachsafetyReachsafetyLoops}{Correct}{}{Walltime}{Stdev}{28.21473980818222352229668995}%
\StoreBenchExecResult{SvcompNineteenSkink}{SvCompPropReachsafetyReachsafetyLoops}{Correct}{False}{Count}{}{35}%
\StoreBenchExecResult{SvcompNineteenSkink}{SvCompPropReachsafetyReachsafetyLoops}{Correct}{False}{Cputime}{}{247.227530777}%
\StoreBenchExecResult{SvcompNineteenSkink}{SvCompPropReachsafetyReachsafetyLoops}{Correct}{False}{Cputime}{Avg}{7.063643736485714285714285714}%
\StoreBenchExecResult{SvcompNineteenSkink}{SvCompPropReachsafetyReachsafetyLoops}{Correct}{False}{Cputime}{Median}{5.45958233}%
\StoreBenchExecResult{SvcompNineteenSkink}{SvCompPropReachsafetyReachsafetyLoops}{Correct}{False}{Cputime}{Min}{4.644882904}%
\StoreBenchExecResult{SvcompNineteenSkink}{SvCompPropReachsafetyReachsafetyLoops}{Correct}{False}{Cputime}{Max}{24.99745517}%
\StoreBenchExecResult{SvcompNineteenSkink}{SvCompPropReachsafetyReachsafetyLoops}{Correct}{False}{Cputime}{Stdev}{4.164703016839303603560232446}%
\StoreBenchExecResult{SvcompNineteenSkink}{SvCompPropReachsafetyReachsafetyLoops}{Correct}{False}{Walltime}{}{102.30581116680}%
\StoreBenchExecResult{SvcompNineteenSkink}{SvCompPropReachsafetyReachsafetyLoops}{Correct}{False}{Walltime}{Avg}{2.923023176194285714285714286}%
\StoreBenchExecResult{SvcompNineteenSkink}{SvCompPropReachsafetyReachsafetyLoops}{Correct}{False}{Walltime}{Median}{2.21888804436}%
\StoreBenchExecResult{SvcompNineteenSkink}{SvCompPropReachsafetyReachsafetyLoops}{Correct}{False}{Walltime}{Min}{2.09272408485}%
\StoreBenchExecResult{SvcompNineteenSkink}{SvCompPropReachsafetyReachsafetyLoops}{Correct}{False}{Walltime}{Max}{13.4272937775}%
\StoreBenchExecResult{SvcompNineteenSkink}{SvCompPropReachsafetyReachsafetyLoops}{Correct}{False}{Walltime}{Stdev}{2.086842627537168235996408101}%
\StoreBenchExecResult{SvcompNineteenSkink}{SvCompPropReachsafetyReachsafetyLoops}{Wrong}{False}{Count}{}{0}%
\StoreBenchExecResult{SvcompNineteenSkink}{SvCompPropReachsafetyReachsafetyLoops}{Wrong}{False}{Cputime}{}{0}%
\StoreBenchExecResult{SvcompNineteenSkink}{SvCompPropReachsafetyReachsafetyLoops}{Wrong}{False}{Cputime}{Avg}{None}%
\StoreBenchExecResult{SvcompNineteenSkink}{SvCompPropReachsafetyReachsafetyLoops}{Wrong}{False}{Cputime}{Median}{None}%
\StoreBenchExecResult{SvcompNineteenSkink}{SvCompPropReachsafetyReachsafetyLoops}{Wrong}{False}{Cputime}{Min}{None}%
\StoreBenchExecResult{SvcompNineteenSkink}{SvCompPropReachsafetyReachsafetyLoops}{Wrong}{False}{Cputime}{Max}{None}%
\StoreBenchExecResult{SvcompNineteenSkink}{SvCompPropReachsafetyReachsafetyLoops}{Wrong}{False}{Cputime}{Stdev}{None}%
\StoreBenchExecResult{SvcompNineteenSkink}{SvCompPropReachsafetyReachsafetyLoops}{Wrong}{False}{Walltime}{}{0}%
\StoreBenchExecResult{SvcompNineteenSkink}{SvCompPropReachsafetyReachsafetyLoops}{Wrong}{False}{Walltime}{Avg}{None}%
\StoreBenchExecResult{SvcompNineteenSkink}{SvCompPropReachsafetyReachsafetyLoops}{Wrong}{False}{Walltime}{Median}{None}%
\StoreBenchExecResult{SvcompNineteenSkink}{SvCompPropReachsafetyReachsafetyLoops}{Wrong}{False}{Walltime}{Min}{None}%
\StoreBenchExecResult{SvcompNineteenSkink}{SvCompPropReachsafetyReachsafetyLoops}{Wrong}{False}{Walltime}{Max}{None}%
\StoreBenchExecResult{SvcompNineteenSkink}{SvCompPropReachsafetyReachsafetyLoops}{Wrong}{False}{Walltime}{Stdev}{None}%
\StoreBenchExecResult{SvcompNineteenSkink}{SvCompPropReachsafetyReachsafetyLoops}{Correct}{True}{Count}{}{85}%
\StoreBenchExecResult{SvcompNineteenSkink}{SvCompPropReachsafetyReachsafetyLoops}{Correct}{True}{Cputime}{}{1677.047319367}%
\StoreBenchExecResult{SvcompNineteenSkink}{SvCompPropReachsafetyReachsafetyLoops}{Correct}{True}{Cputime}{Avg}{19.72996846314117647058823529}%
\StoreBenchExecResult{SvcompNineteenSkink}{SvCompPropReachsafetyReachsafetyLoops}{Correct}{True}{Cputime}{Median}{4.115406256}%
\StoreBenchExecResult{SvcompNineteenSkink}{SvCompPropReachsafetyReachsafetyLoops}{Correct}{True}{Cputime}{Min}{3.334866554}%
\StoreBenchExecResult{SvcompNineteenSkink}{SvCompPropReachsafetyReachsafetyLoops}{Correct}{True}{Cputime}{Max}{292.610540875}%
\StoreBenchExecResult{SvcompNineteenSkink}{SvCompPropReachsafetyReachsafetyLoops}{Correct}{True}{Cputime}{Stdev}{36.31577307058145788569712073}%
\StoreBenchExecResult{SvcompNineteenSkink}{SvCompPropReachsafetyReachsafetyLoops}{Correct}{True}{Walltime}{}{1211.59209012961}%
\StoreBenchExecResult{SvcompNineteenSkink}{SvCompPropReachsafetyReachsafetyLoops}{Correct}{True}{Walltime}{Avg}{14.25402458976011764705882353}%
\StoreBenchExecResult{SvcompNineteenSkink}{SvCompPropReachsafetyReachsafetyLoops}{Correct}{True}{Walltime}{Median}{1.86625599861}%
\StoreBenchExecResult{SvcompNineteenSkink}{SvCompPropReachsafetyReachsafetyLoops}{Correct}{True}{Walltime}{Min}{1.70069384575}%
\StoreBenchExecResult{SvcompNineteenSkink}{SvCompPropReachsafetyReachsafetyLoops}{Correct}{True}{Walltime}{Max}{274.077408075}%
\StoreBenchExecResult{SvcompNineteenSkink}{SvCompPropReachsafetyReachsafetyLoops}{Correct}{True}{Walltime}{Stdev}{32.93364480116309239068835328}%
\StoreBenchExecResult{SvcompNineteenSkink}{SvCompPropReachsafetyReachsafetyLoops}{Wrong}{True}{Count}{}{0}%
\StoreBenchExecResult{SvcompNineteenSkink}{SvCompPropReachsafetyReachsafetyLoops}{Wrong}{True}{Cputime}{}{0}%
\StoreBenchExecResult{SvcompNineteenSkink}{SvCompPropReachsafetyReachsafetyLoops}{Wrong}{True}{Cputime}{Avg}{None}%
\StoreBenchExecResult{SvcompNineteenSkink}{SvCompPropReachsafetyReachsafetyLoops}{Wrong}{True}{Cputime}{Median}{None}%
\StoreBenchExecResult{SvcompNineteenSkink}{SvCompPropReachsafetyReachsafetyLoops}{Wrong}{True}{Cputime}{Min}{None}%
\StoreBenchExecResult{SvcompNineteenSkink}{SvCompPropReachsafetyReachsafetyLoops}{Wrong}{True}{Cputime}{Max}{None}%
\StoreBenchExecResult{SvcompNineteenSkink}{SvCompPropReachsafetyReachsafetyLoops}{Wrong}{True}{Cputime}{Stdev}{None}%
\StoreBenchExecResult{SvcompNineteenSkink}{SvCompPropReachsafetyReachsafetyLoops}{Wrong}{True}{Walltime}{}{0}%
\StoreBenchExecResult{SvcompNineteenSkink}{SvCompPropReachsafetyReachsafetyLoops}{Wrong}{True}{Walltime}{Avg}{None}%
\StoreBenchExecResult{SvcompNineteenSkink}{SvCompPropReachsafetyReachsafetyLoops}{Wrong}{True}{Walltime}{Median}{None}%
\StoreBenchExecResult{SvcompNineteenSkink}{SvCompPropReachsafetyReachsafetyLoops}{Wrong}{True}{Walltime}{Min}{None}%
\StoreBenchExecResult{SvcompNineteenSkink}{SvCompPropReachsafetyReachsafetyLoops}{Wrong}{True}{Walltime}{Max}{None}%
\StoreBenchExecResult{SvcompNineteenSkink}{SvCompPropReachsafetyReachsafetyLoops}{Wrong}{True}{Walltime}{Stdev}{None}%
\StoreBenchExecResult{SvcompNineteenSkink}{SvCompPropReachsafetyReachsafetyLoops}{Error}{}{Count}{}{8}%
\StoreBenchExecResult{SvcompNineteenSkink}{SvCompPropReachsafetyReachsafetyLoops}{Error}{}{Cputime}{}{7203.790116870}%
\StoreBenchExecResult{SvcompNineteenSkink}{SvCompPropReachsafetyReachsafetyLoops}{Error}{}{Cputime}{Avg}{900.47376460875}%
\StoreBenchExecResult{SvcompNineteenSkink}{SvCompPropReachsafetyReachsafetyLoops}{Error}{}{Cputime}{Median}{900.3896637005}%
\StoreBenchExecResult{SvcompNineteenSkink}{SvCompPropReachsafetyReachsafetyLoops}{Error}{}{Cputime}{Min}{900.157085781}%
\StoreBenchExecResult{SvcompNineteenSkink}{SvCompPropReachsafetyReachsafetyLoops}{Error}{}{Cputime}{Max}{900.926218928}%
\StoreBenchExecResult{SvcompNineteenSkink}{SvCompPropReachsafetyReachsafetyLoops}{Error}{}{Cputime}{Stdev}{0.2712553548082073861743956968}%
\StoreBenchExecResult{SvcompNineteenSkink}{SvCompPropReachsafetyReachsafetyLoops}{Error}{}{Walltime}{}{6723.149426222}%
\StoreBenchExecResult{SvcompNineteenSkink}{SvCompPropReachsafetyReachsafetyLoops}{Error}{}{Walltime}{Avg}{840.39367827775}%
\StoreBenchExecResult{SvcompNineteenSkink}{SvCompPropReachsafetyReachsafetyLoops}{Error}{}{Walltime}{Median}{843.6162080765}%
\StoreBenchExecResult{SvcompNineteenSkink}{SvCompPropReachsafetyReachsafetyLoops}{Error}{}{Walltime}{Min}{797.886917114}%
\StoreBenchExecResult{SvcompNineteenSkink}{SvCompPropReachsafetyReachsafetyLoops}{Error}{}{Walltime}{Max}{876.589123964}%
\StoreBenchExecResult{SvcompNineteenSkink}{SvCompPropReachsafetyReachsafetyLoops}{Error}{}{Walltime}{Stdev}{22.76437227530552247621429649}%
\StoreBenchExecResult{SvcompNineteenSkink}{SvCompPropReachsafetyReachsafetyLoops}{Error}{Timeout}{Count}{}{8}%
\StoreBenchExecResult{SvcompNineteenSkink}{SvCompPropReachsafetyReachsafetyLoops}{Error}{Timeout}{Cputime}{}{7203.790116870}%
\StoreBenchExecResult{SvcompNineteenSkink}{SvCompPropReachsafetyReachsafetyLoops}{Error}{Timeout}{Cputime}{Avg}{900.47376460875}%
\StoreBenchExecResult{SvcompNineteenSkink}{SvCompPropReachsafetyReachsafetyLoops}{Error}{Timeout}{Cputime}{Median}{900.3896637005}%
\StoreBenchExecResult{SvcompNineteenSkink}{SvCompPropReachsafetyReachsafetyLoops}{Error}{Timeout}{Cputime}{Min}{900.157085781}%
\StoreBenchExecResult{SvcompNineteenSkink}{SvCompPropReachsafetyReachsafetyLoops}{Error}{Timeout}{Cputime}{Max}{900.926218928}%
\StoreBenchExecResult{SvcompNineteenSkink}{SvCompPropReachsafetyReachsafetyLoops}{Error}{Timeout}{Cputime}{Stdev}{0.2712553548082073861743956968}%
\StoreBenchExecResult{SvcompNineteenSkink}{SvCompPropReachsafetyReachsafetyLoops}{Error}{Timeout}{Walltime}{}{6723.149426222}%
\StoreBenchExecResult{SvcompNineteenSkink}{SvCompPropReachsafetyReachsafetyLoops}{Error}{Timeout}{Walltime}{Avg}{840.39367827775}%
\StoreBenchExecResult{SvcompNineteenSkink}{SvCompPropReachsafetyReachsafetyLoops}{Error}{Timeout}{Walltime}{Median}{843.6162080765}%
\StoreBenchExecResult{SvcompNineteenSkink}{SvCompPropReachsafetyReachsafetyLoops}{Error}{Timeout}{Walltime}{Min}{797.886917114}%
\StoreBenchExecResult{SvcompNineteenSkink}{SvCompPropReachsafetyReachsafetyLoops}{Error}{Timeout}{Walltime}{Max}{876.589123964}%
\StoreBenchExecResult{SvcompNineteenSkink}{SvCompPropReachsafetyReachsafetyLoops}{Error}{Timeout}{Walltime}{Stdev}{22.76437227530552247621429649}%
\StoreBenchExecResult{SvcompNineteenSkink}{SvCompPropReachsafetyReachsafetyLoops}{Unknown}{}{Count}{}{80}%
\StoreBenchExecResult{SvcompNineteenSkink}{SvCompPropReachsafetyReachsafetyLoops}{Unknown}{}{Cputime}{}{9396.292016772}%
\StoreBenchExecResult{SvcompNineteenSkink}{SvCompPropReachsafetyReachsafetyLoops}{Unknown}{}{Cputime}{Avg}{117.45365020965}%
\StoreBenchExecResult{SvcompNineteenSkink}{SvCompPropReachsafetyReachsafetyLoops}{Unknown}{}{Cputime}{Median}{70.476075657}%
\StoreBenchExecResult{SvcompNineteenSkink}{SvCompPropReachsafetyReachsafetyLoops}{Unknown}{}{Cputime}{Min}{2.588134092}%
\StoreBenchExecResult{SvcompNineteenSkink}{SvCompPropReachsafetyReachsafetyLoops}{Unknown}{}{Cputime}{Max}{644.4203508}%
\StoreBenchExecResult{SvcompNineteenSkink}{SvCompPropReachsafetyReachsafetyLoops}{Unknown}{}{Cputime}{Stdev}{129.6204617514656861918127261}%
\StoreBenchExecResult{SvcompNineteenSkink}{SvCompPropReachsafetyReachsafetyLoops}{Unknown}{}{Walltime}{}{7437.97739148483}%
\StoreBenchExecResult{SvcompNineteenSkink}{SvCompPropReachsafetyReachsafetyLoops}{Unknown}{}{Walltime}{Avg}{92.974717393560375}%
\StoreBenchExecResult{SvcompNineteenSkink}{SvCompPropReachsafetyReachsafetyLoops}{Unknown}{}{Walltime}{Median}{44.68006539345}%
\StoreBenchExecResult{SvcompNineteenSkink}{SvCompPropReachsafetyReachsafetyLoops}{Unknown}{}{Walltime}{Min}{1.40682601929}%
\StoreBenchExecResult{SvcompNineteenSkink}{SvCompPropReachsafetyReachsafetyLoops}{Unknown}{}{Walltime}{Max}{620.64004612}%
\StoreBenchExecResult{SvcompNineteenSkink}{SvCompPropReachsafetyReachsafetyLoops}{Unknown}{}{Walltime}{Stdev}{116.4994473064402855532924976}%
\StoreBenchExecResult{SvcompNineteenSkink}{SvCompPropReachsafetyReachsafetyLoops}{Unknown}{Unknown}{Count}{}{80}%
\StoreBenchExecResult{SvcompNineteenSkink}{SvCompPropReachsafetyReachsafetyLoops}{Unknown}{Unknown}{Cputime}{}{9396.292016772}%
\StoreBenchExecResult{SvcompNineteenSkink}{SvCompPropReachsafetyReachsafetyLoops}{Unknown}{Unknown}{Cputime}{Avg}{117.45365020965}%
\StoreBenchExecResult{SvcompNineteenSkink}{SvCompPropReachsafetyReachsafetyLoops}{Unknown}{Unknown}{Cputime}{Median}{70.476075657}%
\StoreBenchExecResult{SvcompNineteenSkink}{SvCompPropReachsafetyReachsafetyLoops}{Unknown}{Unknown}{Cputime}{Min}{2.588134092}%
\StoreBenchExecResult{SvcompNineteenSkink}{SvCompPropReachsafetyReachsafetyLoops}{Unknown}{Unknown}{Cputime}{Max}{644.4203508}%
\StoreBenchExecResult{SvcompNineteenSkink}{SvCompPropReachsafetyReachsafetyLoops}{Unknown}{Unknown}{Cputime}{Stdev}{129.6204617514656861918127261}%
\StoreBenchExecResult{SvcompNineteenSkink}{SvCompPropReachsafetyReachsafetyLoops}{Unknown}{Unknown}{Walltime}{}{7437.97739148483}%
\StoreBenchExecResult{SvcompNineteenSkink}{SvCompPropReachsafetyReachsafetyLoops}{Unknown}{Unknown}{Walltime}{Avg}{92.974717393560375}%
\StoreBenchExecResult{SvcompNineteenSkink}{SvCompPropReachsafetyReachsafetyLoops}{Unknown}{Unknown}{Walltime}{Median}{44.68006539345}%
\StoreBenchExecResult{SvcompNineteenSkink}{SvCompPropReachsafetyReachsafetyLoops}{Unknown}{Unknown}{Walltime}{Min}{1.40682601929}%
\StoreBenchExecResult{SvcompNineteenSkink}{SvCompPropReachsafetyReachsafetyLoops}{Unknown}{Unknown}{Walltime}{Max}{620.64004612}%
\StoreBenchExecResult{SvcompNineteenSkink}{SvCompPropReachsafetyReachsafetyLoops}{Unknown}{Unknown}{Walltime}{Stdev}{116.4994473064402855532924976}%
\providecommand\StoreBenchExecResult[7]{\expandafter\newcommand\csname#1#2#3#4#5#6\endcsname{#7}}%
\StoreBenchExecResult{SvcompNineteenUautomizer}{SvCompPropReachsafetyReachsafetyLoops}{Total}{}{Count}{}{208}%
\StoreBenchExecResult{SvcompNineteenUautomizer}{SvCompPropReachsafetyReachsafetyLoops}{Total}{}{Cputime}{}{54683.314026542}%
\StoreBenchExecResult{SvcompNineteenUautomizer}{SvCompPropReachsafetyReachsafetyLoops}{Total}{}{Cputime}{Avg}{262.9005482045288461538461538}%
\StoreBenchExecResult{SvcompNineteenUautomizer}{SvCompPropReachsafetyReachsafetyLoops}{Total}{}{Cputime}{Median}{11.5668570225}%
\StoreBenchExecResult{SvcompNineteenUautomizer}{SvCompPropReachsafetyReachsafetyLoops}{Total}{}{Cputime}{Min}{6.767440839}%
\StoreBenchExecResult{SvcompNineteenUautomizer}{SvCompPropReachsafetyReachsafetyLoops}{Total}{}{Cputime}{Max}{908.439713054}%
\StoreBenchExecResult{SvcompNineteenUautomizer}{SvCompPropReachsafetyReachsafetyLoops}{Total}{}{Cputime}{Stdev}{392.8901281218812072591676163}%
\StoreBenchExecResult{SvcompNineteenUautomizer}{SvCompPropReachsafetyReachsafetyLoops}{Total}{}{Walltime}{}{49041.38606095154}%
\StoreBenchExecResult{SvcompNineteenUautomizer}{SvCompPropReachsafetyReachsafetyLoops}{Total}{}{Walltime}{Avg}{235.7758945238054807692307692}%
\StoreBenchExecResult{SvcompNineteenUautomizer}{SvCompPropReachsafetyReachsafetyLoops}{Total}{}{Walltime}{Median}{4.274193048475}%
\StoreBenchExecResult{SvcompNineteenUautomizer}{SvCompPropReachsafetyReachsafetyLoops}{Total}{}{Walltime}{Min}{2.56650590897}%
\StoreBenchExecResult{SvcompNineteenUautomizer}{SvCompPropReachsafetyReachsafetyLoops}{Total}{}{Walltime}{Max}{883.707408905}%
\StoreBenchExecResult{SvcompNineteenUautomizer}{SvCompPropReachsafetyReachsafetyLoops}{Total}{}{Walltime}{Stdev}{367.2816587116427603857074217}%
\StoreBenchExecResult{SvcompNineteenUautomizer}{SvCompPropReachsafetyReachsafetyLoops}{Correct}{}{Count}{}{144}%
\StoreBenchExecResult{SvcompNineteenUautomizer}{SvCompPropReachsafetyReachsafetyLoops}{Correct}{}{Cputime}{}{4081.355570697}%
\StoreBenchExecResult{SvcompNineteenUautomizer}{SvCompPropReachsafetyReachsafetyLoops}{Correct}{}{Cputime}{Avg}{28.34274701872916666666666667}%
\StoreBenchExecResult{SvcompNineteenUautomizer}{SvCompPropReachsafetyReachsafetyLoops}{Correct}{}{Cputime}{Median}{9.6523869495}%
\StoreBenchExecResult{SvcompNineteenUautomizer}{SvCompPropReachsafetyReachsafetyLoops}{Correct}{}{Cputime}{Min}{6.767440839}%
\StoreBenchExecResult{SvcompNineteenUautomizer}{SvCompPropReachsafetyReachsafetyLoops}{Correct}{}{Cputime}{Max}{778.428379959}%
\StoreBenchExecResult{SvcompNineteenUautomizer}{SvCompPropReachsafetyReachsafetyLoops}{Correct}{}{Cputime}{Stdev}{79.90860668741459257357147252}%
\StoreBenchExecResult{SvcompNineteenUautomizer}{SvCompPropReachsafetyReachsafetyLoops}{Correct}{}{Walltime}{}{2763.71196675198}%
\StoreBenchExecResult{SvcompNineteenUautomizer}{SvCompPropReachsafetyReachsafetyLoops}{Correct}{}{Walltime}{Avg}{19.19244421355541666666666667}%
\StoreBenchExecResult{SvcompNineteenUautomizer}{SvCompPropReachsafetyReachsafetyLoops}{Correct}{}{Walltime}{Median}{3.424664974215}%
\StoreBenchExecResult{SvcompNineteenUautomizer}{SvCompPropReachsafetyReachsafetyLoops}{Correct}{}{Walltime}{Min}{2.56650590897}%
\StoreBenchExecResult{SvcompNineteenUautomizer}{SvCompPropReachsafetyReachsafetyLoops}{Correct}{}{Walltime}{Max}{741.794635057}%
\StoreBenchExecResult{SvcompNineteenUautomizer}{SvCompPropReachsafetyReachsafetyLoops}{Correct}{}{Walltime}{Stdev}{75.01184474646850042283079632}%
\StoreBenchExecResult{SvcompNineteenUautomizer}{SvCompPropReachsafetyReachsafetyLoops}{Correct}{False}{Count}{}{35}%
\StoreBenchExecResult{SvcompNineteenUautomizer}{SvCompPropReachsafetyReachsafetyLoops}{Correct}{False}{Cputime}{}{1667.001428005}%
\StoreBenchExecResult{SvcompNineteenUautomizer}{SvCompPropReachsafetyReachsafetyLoops}{Correct}{False}{Cputime}{Avg}{47.62861222871428571428571429}%
\StoreBenchExecResult{SvcompNineteenUautomizer}{SvCompPropReachsafetyReachsafetyLoops}{Correct}{False}{Cputime}{Median}{8.562734072}%
\StoreBenchExecResult{SvcompNineteenUautomizer}{SvCompPropReachsafetyReachsafetyLoops}{Correct}{False}{Cputime}{Min}{6.767440839}%
\StoreBenchExecResult{SvcompNineteenUautomizer}{SvCompPropReachsafetyReachsafetyLoops}{Correct}{False}{Cputime}{Max}{778.428379959}%
\StoreBenchExecResult{SvcompNineteenUautomizer}{SvCompPropReachsafetyReachsafetyLoops}{Correct}{False}{Cputime}{Stdev}{142.1378148606669382646595056}%
\StoreBenchExecResult{SvcompNineteenUautomizer}{SvCompPropReachsafetyReachsafetyLoops}{Correct}{False}{Walltime}{}{1376.12725639281}%
\StoreBenchExecResult{SvcompNineteenUautomizer}{SvCompPropReachsafetyReachsafetyLoops}{Correct}{False}{Walltime}{Avg}{39.31792161122314285714285714}%
\StoreBenchExecResult{SvcompNineteenUautomizer}{SvCompPropReachsafetyReachsafetyLoops}{Correct}{False}{Walltime}{Median}{3.0852458477}%
\StoreBenchExecResult{SvcompNineteenUautomizer}{SvCompPropReachsafetyReachsafetyLoops}{Correct}{False}{Walltime}{Min}{2.56650590897}%
\StoreBenchExecResult{SvcompNineteenUautomizer}{SvCompPropReachsafetyReachsafetyLoops}{Correct}{False}{Walltime}{Max}{741.794635057}%
\StoreBenchExecResult{SvcompNineteenUautomizer}{SvCompPropReachsafetyReachsafetyLoops}{Correct}{False}{Walltime}{Stdev}{135.0307939949358544134252027}%
\StoreBenchExecResult{SvcompNineteenUautomizer}{SvCompPropReachsafetyReachsafetyLoops}{Wrong}{False}{Count}{}{0}%
\StoreBenchExecResult{SvcompNineteenUautomizer}{SvCompPropReachsafetyReachsafetyLoops}{Wrong}{False}{Cputime}{}{0}%
\StoreBenchExecResult{SvcompNineteenUautomizer}{SvCompPropReachsafetyReachsafetyLoops}{Wrong}{False}{Cputime}{Avg}{None}%
\StoreBenchExecResult{SvcompNineteenUautomizer}{SvCompPropReachsafetyReachsafetyLoops}{Wrong}{False}{Cputime}{Median}{None}%
\StoreBenchExecResult{SvcompNineteenUautomizer}{SvCompPropReachsafetyReachsafetyLoops}{Wrong}{False}{Cputime}{Min}{None}%
\StoreBenchExecResult{SvcompNineteenUautomizer}{SvCompPropReachsafetyReachsafetyLoops}{Wrong}{False}{Cputime}{Max}{None}%
\StoreBenchExecResult{SvcompNineteenUautomizer}{SvCompPropReachsafetyReachsafetyLoops}{Wrong}{False}{Cputime}{Stdev}{None}%
\StoreBenchExecResult{SvcompNineteenUautomizer}{SvCompPropReachsafetyReachsafetyLoops}{Wrong}{False}{Walltime}{}{0}%
\StoreBenchExecResult{SvcompNineteenUautomizer}{SvCompPropReachsafetyReachsafetyLoops}{Wrong}{False}{Walltime}{Avg}{None}%
\StoreBenchExecResult{SvcompNineteenUautomizer}{SvCompPropReachsafetyReachsafetyLoops}{Wrong}{False}{Walltime}{Median}{None}%
\StoreBenchExecResult{SvcompNineteenUautomizer}{SvCompPropReachsafetyReachsafetyLoops}{Wrong}{False}{Walltime}{Min}{None}%
\StoreBenchExecResult{SvcompNineteenUautomizer}{SvCompPropReachsafetyReachsafetyLoops}{Wrong}{False}{Walltime}{Max}{None}%
\StoreBenchExecResult{SvcompNineteenUautomizer}{SvCompPropReachsafetyReachsafetyLoops}{Wrong}{False}{Walltime}{Stdev}{None}%
\StoreBenchExecResult{SvcompNineteenUautomizer}{SvCompPropReachsafetyReachsafetyLoops}{Correct}{True}{Count}{}{109}%
\StoreBenchExecResult{SvcompNineteenUautomizer}{SvCompPropReachsafetyReachsafetyLoops}{Correct}{True}{Cputime}{}{2414.354142692}%
\StoreBenchExecResult{SvcompNineteenUautomizer}{SvCompPropReachsafetyReachsafetyLoops}{Correct}{True}{Cputime}{Avg}{22.15003800634862385321100917}%
\StoreBenchExecResult{SvcompNineteenUautomizer}{SvCompPropReachsafetyReachsafetyLoops}{Correct}{True}{Cputime}{Median}{10.075581072}%
\StoreBenchExecResult{SvcompNineteenUautomizer}{SvCompPropReachsafetyReachsafetyLoops}{Correct}{True}{Cputime}{Min}{7.375673513}%
\StoreBenchExecResult{SvcompNineteenUautomizer}{SvCompPropReachsafetyReachsafetyLoops}{Correct}{True}{Cputime}{Max}{288.067036533}%
\StoreBenchExecResult{SvcompNineteenUautomizer}{SvCompPropReachsafetyReachsafetyLoops}{Correct}{True}{Cputime}{Stdev}{42.31671587403245146036076203}%
\StoreBenchExecResult{SvcompNineteenUautomizer}{SvCompPropReachsafetyReachsafetyLoops}{Correct}{True}{Walltime}{}{1387.58471035917}%
\StoreBenchExecResult{SvcompNineteenUautomizer}{SvCompPropReachsafetyReachsafetyLoops}{Correct}{True}{Walltime}{Avg}{12.73013495742357798165137615}%
\StoreBenchExecResult{SvcompNineteenUautomizer}{SvCompPropReachsafetyReachsafetyLoops}{Correct}{True}{Walltime}{Median}{3.56152796745}%
\StoreBenchExecResult{SvcompNineteenUautomizer}{SvCompPropReachsafetyReachsafetyLoops}{Correct}{True}{Walltime}{Min}{2.66832995415}%
\StoreBenchExecResult{SvcompNineteenUautomizer}{SvCompPropReachsafetyReachsafetyLoops}{Correct}{True}{Walltime}{Max}{268.866928816}%
\StoreBenchExecResult{SvcompNineteenUautomizer}{SvCompPropReachsafetyReachsafetyLoops}{Correct}{True}{Walltime}{Stdev}{37.50982914701981067858078505}%
\StoreBenchExecResult{SvcompNineteenUautomizer}{SvCompPropReachsafetyReachsafetyLoops}{Error}{}{Count}{}{63}%
\StoreBenchExecResult{SvcompNineteenUautomizer}{SvCompPropReachsafetyReachsafetyLoops}{Error}{}{Cputime}{}{50562.392398838}%
\StoreBenchExecResult{SvcompNineteenUautomizer}{SvCompPropReachsafetyReachsafetyLoops}{Error}{}{Cputime}{Avg}{802.5776571244126984126984127}%
\StoreBenchExecResult{SvcompNineteenUautomizer}{SvCompPropReachsafetyReachsafetyLoops}{Error}{}{Cputime}{Median}{900.754491938}%
\StoreBenchExecResult{SvcompNineteenUautomizer}{SvCompPropReachsafetyReachsafetyLoops}{Error}{}{Cputime}{Min}{10.872953767}%
\StoreBenchExecResult{SvcompNineteenUautomizer}{SvCompPropReachsafetyReachsafetyLoops}{Error}{}{Cputime}{Max}{908.439713054}%
\StoreBenchExecResult{SvcompNineteenUautomizer}{SvCompPropReachsafetyReachsafetyLoops}{Error}{}{Cputime}{Stdev}{277.9380078280354894813724849}%
\StoreBenchExecResult{SvcompNineteenUautomizer}{SvCompPropReachsafetyReachsafetyLoops}{Error}{}{Walltime}{}{46251.88049220986}%
\StoreBenchExecResult{SvcompNineteenUautomizer}{SvCompPropReachsafetyReachsafetyLoops}{Error}{}{Walltime}{Avg}{734.1568332096803174603174603}%
\StoreBenchExecResult{SvcompNineteenUautomizer}{SvCompPropReachsafetyReachsafetyLoops}{Error}{}{Walltime}{Median}{860.113828897}%
\StoreBenchExecResult{SvcompNineteenUautomizer}{SvCompPropReachsafetyReachsafetyLoops}{Error}{}{Walltime}{Min}{3.73326301575}%
\StoreBenchExecResult{SvcompNineteenUautomizer}{SvCompPropReachsafetyReachsafetyLoops}{Error}{}{Walltime}{Max}{883.707408905}%
\StoreBenchExecResult{SvcompNineteenUautomizer}{SvCompPropReachsafetyReachsafetyLoops}{Error}{}{Walltime}{Stdev}{276.0552305701352719481473300}%
\StoreBenchExecResult{SvcompNineteenUautomizer}{SvCompPropReachsafetyReachsafetyLoops}{Error}{Error}{Count}{}{7}%
\StoreBenchExecResult{SvcompNineteenUautomizer}{SvCompPropReachsafetyReachsafetyLoops}{Error}{Error}{Cputime}{}{115.432900564}%
\StoreBenchExecResult{SvcompNineteenUautomizer}{SvCompPropReachsafetyReachsafetyLoops}{Error}{Error}{Cputime}{Avg}{16.49041436628571428571428571}%
\StoreBenchExecResult{SvcompNineteenUautomizer}{SvCompPropReachsafetyReachsafetyLoops}{Error}{Error}{Cputime}{Median}{11.329751282}%
\StoreBenchExecResult{SvcompNineteenUautomizer}{SvCompPropReachsafetyReachsafetyLoops}{Error}{Error}{Cputime}{Min}{10.872953767}%
\StoreBenchExecResult{SvcompNineteenUautomizer}{SvCompPropReachsafetyReachsafetyLoops}{Error}{Error}{Cputime}{Max}{30.114591642}%
\StoreBenchExecResult{SvcompNineteenUautomizer}{SvCompPropReachsafetyReachsafetyLoops}{Error}{Error}{Cputime}{Stdev}{7.854189830364496146042516314}%
\StoreBenchExecResult{SvcompNineteenUautomizer}{SvCompPropReachsafetyReachsafetyLoops}{Error}{Error}{Walltime}{}{57.76225352286}%
\StoreBenchExecResult{SvcompNineteenUautomizer}{SvCompPropReachsafetyReachsafetyLoops}{Error}{Error}{Walltime}{Avg}{8.251750503265714285714285714}%
\StoreBenchExecResult{SvcompNineteenUautomizer}{SvCompPropReachsafetyReachsafetyLoops}{Error}{Error}{Walltime}{Median}{3.86558699608}%
\StoreBenchExecResult{SvcompNineteenUautomizer}{SvCompPropReachsafetyReachsafetyLoops}{Error}{Error}{Walltime}{Min}{3.73326301575}%
\StoreBenchExecResult{SvcompNineteenUautomizer}{SvCompPropReachsafetyReachsafetyLoops}{Error}{Error}{Walltime}{Max}{20.3489379883}%
\StoreBenchExecResult{SvcompNineteenUautomizer}{SvCompPropReachsafetyReachsafetyLoops}{Error}{Error}{Walltime}{Stdev}{6.950979463739041840132392902}%
\StoreBenchExecResult{SvcompNineteenUautomizer}{SvCompPropReachsafetyReachsafetyLoops}{Error}{Timeout}{Count}{}{56}%
\StoreBenchExecResult{SvcompNineteenUautomizer}{SvCompPropReachsafetyReachsafetyLoops}{Error}{Timeout}{Cputime}{}{50446.959498274}%
\StoreBenchExecResult{SvcompNineteenUautomizer}{SvCompPropReachsafetyReachsafetyLoops}{Error}{Timeout}{Cputime}{Avg}{900.8385624691785714285714286}%
\StoreBenchExecResult{SvcompNineteenUautomizer}{SvCompPropReachsafetyReachsafetyLoops}{Error}{Timeout}{Cputime}{Median}{900.834767230}%
\StoreBenchExecResult{SvcompNineteenUautomizer}{SvCompPropReachsafetyReachsafetyLoops}{Error}{Timeout}{Cputime}{Min}{900.054394963}%
\StoreBenchExecResult{SvcompNineteenUautomizer}{SvCompPropReachsafetyReachsafetyLoops}{Error}{Timeout}{Cputime}{Max}{908.439713054}%
\StoreBenchExecResult{SvcompNineteenUautomizer}{SvCompPropReachsafetyReachsafetyLoops}{Error}{Timeout}{Cputime}{Stdev}{1.080513790048480627113139595}%
\StoreBenchExecResult{SvcompNineteenUautomizer}{SvCompPropReachsafetyReachsafetyLoops}{Error}{Timeout}{Walltime}{}{46194.118238687}%
\StoreBenchExecResult{SvcompNineteenUautomizer}{SvCompPropReachsafetyReachsafetyLoops}{Error}{Timeout}{Walltime}{Avg}{824.8949685479821428571428571}%
\StoreBenchExecResult{SvcompNineteenUautomizer}{SvCompPropReachsafetyReachsafetyLoops}{Error}{Timeout}{Walltime}{Median}{865.581650972}%
\StoreBenchExecResult{SvcompNineteenUautomizer}{SvCompPropReachsafetyReachsafetyLoops}{Error}{Timeout}{Walltime}{Min}{141.088304043}%
\StoreBenchExecResult{SvcompNineteenUautomizer}{SvCompPropReachsafetyReachsafetyLoops}{Error}{Timeout}{Walltime}{Max}{883.707408905}%
\StoreBenchExecResult{SvcompNineteenUautomizer}{SvCompPropReachsafetyReachsafetyLoops}{Error}{Timeout}{Walltime}{Stdev}{107.8219794813928191141177246}%
\StoreBenchExecResult{SvcompNineteenUautomizer}{SvCompPropReachsafetyReachsafetyLoops}{Wrong}{}{Count}{}{1}%
\StoreBenchExecResult{SvcompNineteenUautomizer}{SvCompPropReachsafetyReachsafetyLoops}{Wrong}{}{Cputime}{}{39.566057007}%
\StoreBenchExecResult{SvcompNineteenUautomizer}{SvCompPropReachsafetyReachsafetyLoops}{Wrong}{}{Cputime}{Avg}{39.566057007}%
\StoreBenchExecResult{SvcompNineteenUautomizer}{SvCompPropReachsafetyReachsafetyLoops}{Wrong}{}{Cputime}{Median}{39.566057007}%
\StoreBenchExecResult{SvcompNineteenUautomizer}{SvCompPropReachsafetyReachsafetyLoops}{Wrong}{}{Cputime}{Min}{39.566057007}%
\StoreBenchExecResult{SvcompNineteenUautomizer}{SvCompPropReachsafetyReachsafetyLoops}{Wrong}{}{Cputime}{Max}{39.566057007}%
\StoreBenchExecResult{SvcompNineteenUautomizer}{SvCompPropReachsafetyReachsafetyLoops}{Wrong}{}{Cputime}{Stdev}{0E-9}%
\StoreBenchExecResult{SvcompNineteenUautomizer}{SvCompPropReachsafetyReachsafetyLoops}{Wrong}{}{Walltime}{}{25.7936019897}%
\StoreBenchExecResult{SvcompNineteenUautomizer}{SvCompPropReachsafetyReachsafetyLoops}{Wrong}{}{Walltime}{Avg}{25.7936019897}%
\StoreBenchExecResult{SvcompNineteenUautomizer}{SvCompPropReachsafetyReachsafetyLoops}{Wrong}{}{Walltime}{Median}{25.7936019897}%
\StoreBenchExecResult{SvcompNineteenUautomizer}{SvCompPropReachsafetyReachsafetyLoops}{Wrong}{}{Walltime}{Min}{25.7936019897}%
\StoreBenchExecResult{SvcompNineteenUautomizer}{SvCompPropReachsafetyReachsafetyLoops}{Wrong}{}{Walltime}{Max}{25.7936019897}%
\StoreBenchExecResult{SvcompNineteenUautomizer}{SvCompPropReachsafetyReachsafetyLoops}{Wrong}{}{Walltime}{Stdev}{0E-10}%
\StoreBenchExecResult{SvcompNineteenUautomizer}{SvCompPropReachsafetyReachsafetyLoops}{Wrong}{True}{Count}{}{1}%
\StoreBenchExecResult{SvcompNineteenUautomizer}{SvCompPropReachsafetyReachsafetyLoops}{Wrong}{True}{Cputime}{}{39.566057007}%
\StoreBenchExecResult{SvcompNineteenUautomizer}{SvCompPropReachsafetyReachsafetyLoops}{Wrong}{True}{Cputime}{Avg}{39.566057007}%
\StoreBenchExecResult{SvcompNineteenUautomizer}{SvCompPropReachsafetyReachsafetyLoops}{Wrong}{True}{Cputime}{Median}{39.566057007}%
\StoreBenchExecResult{SvcompNineteenUautomizer}{SvCompPropReachsafetyReachsafetyLoops}{Wrong}{True}{Cputime}{Min}{39.566057007}%
\StoreBenchExecResult{SvcompNineteenUautomizer}{SvCompPropReachsafetyReachsafetyLoops}{Wrong}{True}{Cputime}{Max}{39.566057007}%
\StoreBenchExecResult{SvcompNineteenUautomizer}{SvCompPropReachsafetyReachsafetyLoops}{Wrong}{True}{Cputime}{Stdev}{0E-9}%
\StoreBenchExecResult{SvcompNineteenUautomizer}{SvCompPropReachsafetyReachsafetyLoops}{Wrong}{True}{Walltime}{}{25.7936019897}%
\StoreBenchExecResult{SvcompNineteenUautomizer}{SvCompPropReachsafetyReachsafetyLoops}{Wrong}{True}{Walltime}{Avg}{25.7936019897}%
\StoreBenchExecResult{SvcompNineteenUautomizer}{SvCompPropReachsafetyReachsafetyLoops}{Wrong}{True}{Walltime}{Median}{25.7936019897}%
\StoreBenchExecResult{SvcompNineteenUautomizer}{SvCompPropReachsafetyReachsafetyLoops}{Wrong}{True}{Walltime}{Min}{25.7936019897}%
\StoreBenchExecResult{SvcompNineteenUautomizer}{SvCompPropReachsafetyReachsafetyLoops}{Wrong}{True}{Walltime}{Max}{25.7936019897}%
\StoreBenchExecResult{SvcompNineteenUautomizer}{SvCompPropReachsafetyReachsafetyLoops}{Wrong}{True}{Walltime}{Stdev}{0E-10}%
\providecommand\StoreBenchExecResult[7]{\expandafter\newcommand\csname#1#2#3#4#5#6\endcsname{#7}}%
\StoreBenchExecResult{SvcompNineteenUkojak}{SvCompPropReachsafetyReachsafetyLoops}{Total}{}{Count}{}{208}%
\StoreBenchExecResult{SvcompNineteenUkojak}{SvCompPropReachsafetyReachsafetyLoops}{Total}{}{Cputime}{}{68703.545553382}%
\StoreBenchExecResult{SvcompNineteenUkojak}{SvCompPropReachsafetyReachsafetyLoops}{Total}{}{Cputime}{Avg}{330.3055074681826923076923077}%
\StoreBenchExecResult{SvcompNineteenUkojak}{SvCompPropReachsafetyReachsafetyLoops}{Total}{}{Cputime}{Median}{17.450206873}%
\StoreBenchExecResult{SvcompNineteenUkojak}{SvCompPropReachsafetyReachsafetyLoops}{Total}{}{Cputime}{Min}{6.712709165}%
\StoreBenchExecResult{SvcompNineteenUkojak}{SvCompPropReachsafetyReachsafetyLoops}{Total}{}{Cputime}{Max}{901.584548426}%
\StoreBenchExecResult{SvcompNineteenUkojak}{SvCompPropReachsafetyReachsafetyLoops}{Total}{}{Cputime}{Stdev}{412.5073952128869416889187985}%
\StoreBenchExecResult{SvcompNineteenUkojak}{SvCompPropReachsafetyReachsafetyLoops}{Total}{}{Walltime}{}{58082.95056557712}%
\StoreBenchExecResult{SvcompNineteenUkojak}{SvCompPropReachsafetyReachsafetyLoops}{Total}{}{Walltime}{Avg}{279.2449546421976923076923077}%
\StoreBenchExecResult{SvcompNineteenUkojak}{SvCompPropReachsafetyReachsafetyLoops}{Total}{}{Walltime}{Median}{6.604523539545}%
\StoreBenchExecResult{SvcompNineteenUkojak}{SvCompPropReachsafetyReachsafetyLoops}{Total}{}{Walltime}{Min}{2.47493290901}%
\StoreBenchExecResult{SvcompNineteenUkojak}{SvCompPropReachsafetyReachsafetyLoops}{Total}{}{Walltime}{Max}{890.61344409}%
\StoreBenchExecResult{SvcompNineteenUkojak}{SvCompPropReachsafetyReachsafetyLoops}{Total}{}{Walltime}{Stdev}{358.1929519941166861437412510}%
\StoreBenchExecResult{SvcompNineteenUkojak}{SvCompPropReachsafetyReachsafetyLoops}{Correct}{}{Count}{}{132}%
\StoreBenchExecResult{SvcompNineteenUkojak}{SvCompPropReachsafetyReachsafetyLoops}{Correct}{}{Cputime}{}{5533.376254569}%
\StoreBenchExecResult{SvcompNineteenUkojak}{SvCompPropReachsafetyReachsafetyLoops}{Correct}{}{Cputime}{Avg}{41.91951708006818181818181818}%
\StoreBenchExecResult{SvcompNineteenUkojak}{SvCompPropReachsafetyReachsafetyLoops}{Correct}{}{Cputime}{Median}{9.948928160}%
\StoreBenchExecResult{SvcompNineteenUkojak}{SvCompPropReachsafetyReachsafetyLoops}{Correct}{}{Cputime}{Min}{6.712709165}%
\StoreBenchExecResult{SvcompNineteenUkojak}{SvCompPropReachsafetyReachsafetyLoops}{Correct}{}{Cputime}{Max}{485.582305311}%
\StoreBenchExecResult{SvcompNineteenUkojak}{SvCompPropReachsafetyReachsafetyLoops}{Correct}{}{Cputime}{Stdev}{89.69769109752424585986619936}%
\StoreBenchExecResult{SvcompNineteenUkojak}{SvCompPropReachsafetyReachsafetyLoops}{Correct}{}{Walltime}{}{3939.59568143006}%
\StoreBenchExecResult{SvcompNineteenUkojak}{SvCompPropReachsafetyReachsafetyLoops}{Correct}{}{Walltime}{Avg}{29.84542182901560606060606061}%
\StoreBenchExecResult{SvcompNineteenUkojak}{SvCompPropReachsafetyReachsafetyLoops}{Correct}{}{Walltime}{Median}{3.523302435875}%
\StoreBenchExecResult{SvcompNineteenUkojak}{SvCompPropReachsafetyReachsafetyLoops}{Correct}{}{Walltime}{Min}{2.47493290901}%
\StoreBenchExecResult{SvcompNineteenUkojak}{SvCompPropReachsafetyReachsafetyLoops}{Correct}{}{Walltime}{Max}{406.325536013}%
\StoreBenchExecResult{SvcompNineteenUkojak}{SvCompPropReachsafetyReachsafetyLoops}{Correct}{}{Walltime}{Stdev}{78.10263040289962082951404859}%
\StoreBenchExecResult{SvcompNineteenUkojak}{SvCompPropReachsafetyReachsafetyLoops}{Correct}{False}{Count}{}{36}%
\StoreBenchExecResult{SvcompNineteenUkojak}{SvCompPropReachsafetyReachsafetyLoops}{Correct}{False}{Cputime}{}{658.611875195}%
\StoreBenchExecResult{SvcompNineteenUkojak}{SvCompPropReachsafetyReachsafetyLoops}{Correct}{False}{Cputime}{Avg}{18.29477431097222222222222222}%
\StoreBenchExecResult{SvcompNineteenUkojak}{SvCompPropReachsafetyReachsafetyLoops}{Correct}{False}{Cputime}{Median}{9.107349338}%
\StoreBenchExecResult{SvcompNineteenUkojak}{SvCompPropReachsafetyReachsafetyLoops}{Correct}{False}{Cputime}{Min}{6.712709165}%
\StoreBenchExecResult{SvcompNineteenUkojak}{SvCompPropReachsafetyReachsafetyLoops}{Correct}{False}{Cputime}{Max}{158.137458566}%
\StoreBenchExecResult{SvcompNineteenUkojak}{SvCompPropReachsafetyReachsafetyLoops}{Correct}{False}{Cputime}{Stdev}{28.93971020768529058220161029}%
\StoreBenchExecResult{SvcompNineteenUkojak}{SvCompPropReachsafetyReachsafetyLoops}{Correct}{False}{Walltime}{}{352.58139324224}%
\StoreBenchExecResult{SvcompNineteenUkojak}{SvCompPropReachsafetyReachsafetyLoops}{Correct}{False}{Walltime}{Avg}{9.793927590062222222222222222}%
\StoreBenchExecResult{SvcompNineteenUkojak}{SvCompPropReachsafetyReachsafetyLoops}{Correct}{False}{Walltime}{Median}{3.17149198055}%
\StoreBenchExecResult{SvcompNineteenUkojak}{SvCompPropReachsafetyReachsafetyLoops}{Correct}{False}{Walltime}{Min}{2.5337729454}%
\StoreBenchExecResult{SvcompNineteenUkojak}{SvCompPropReachsafetyReachsafetyLoops}{Correct}{False}{Walltime}{Max}{119.206636906}%
\StoreBenchExecResult{SvcompNineteenUkojak}{SvCompPropReachsafetyReachsafetyLoops}{Correct}{False}{Walltime}{Stdev}{21.70891608315737394156845981}%
\StoreBenchExecResult{SvcompNineteenUkojak}{SvCompPropReachsafetyReachsafetyLoops}{Wrong}{False}{Count}{}{0}%
\StoreBenchExecResult{SvcompNineteenUkojak}{SvCompPropReachsafetyReachsafetyLoops}{Wrong}{False}{Cputime}{}{0}%
\StoreBenchExecResult{SvcompNineteenUkojak}{SvCompPropReachsafetyReachsafetyLoops}{Wrong}{False}{Cputime}{Avg}{None}%
\StoreBenchExecResult{SvcompNineteenUkojak}{SvCompPropReachsafetyReachsafetyLoops}{Wrong}{False}{Cputime}{Median}{None}%
\StoreBenchExecResult{SvcompNineteenUkojak}{SvCompPropReachsafetyReachsafetyLoops}{Wrong}{False}{Cputime}{Min}{None}%
\StoreBenchExecResult{SvcompNineteenUkojak}{SvCompPropReachsafetyReachsafetyLoops}{Wrong}{False}{Cputime}{Max}{None}%
\StoreBenchExecResult{SvcompNineteenUkojak}{SvCompPropReachsafetyReachsafetyLoops}{Wrong}{False}{Cputime}{Stdev}{None}%
\StoreBenchExecResult{SvcompNineteenUkojak}{SvCompPropReachsafetyReachsafetyLoops}{Wrong}{False}{Walltime}{}{0}%
\StoreBenchExecResult{SvcompNineteenUkojak}{SvCompPropReachsafetyReachsafetyLoops}{Wrong}{False}{Walltime}{Avg}{None}%
\StoreBenchExecResult{SvcompNineteenUkojak}{SvCompPropReachsafetyReachsafetyLoops}{Wrong}{False}{Walltime}{Median}{None}%
\StoreBenchExecResult{SvcompNineteenUkojak}{SvCompPropReachsafetyReachsafetyLoops}{Wrong}{False}{Walltime}{Min}{None}%
\StoreBenchExecResult{SvcompNineteenUkojak}{SvCompPropReachsafetyReachsafetyLoops}{Wrong}{False}{Walltime}{Max}{None}%
\StoreBenchExecResult{SvcompNineteenUkojak}{SvCompPropReachsafetyReachsafetyLoops}{Wrong}{False}{Walltime}{Stdev}{None}%
\StoreBenchExecResult{SvcompNineteenUkojak}{SvCompPropReachsafetyReachsafetyLoops}{Correct}{True}{Count}{}{96}%
\StoreBenchExecResult{SvcompNineteenUkojak}{SvCompPropReachsafetyReachsafetyLoops}{Correct}{True}{Cputime}{}{4874.764379374}%
\StoreBenchExecResult{SvcompNineteenUkojak}{SvCompPropReachsafetyReachsafetyLoops}{Correct}{True}{Cputime}{Avg}{50.77879561847916666666666667}%
\StoreBenchExecResult{SvcompNineteenUkojak}{SvCompPropReachsafetyReachsafetyLoops}{Correct}{True}{Cputime}{Median}{10.449611158}%
\StoreBenchExecResult{SvcompNineteenUkojak}{SvCompPropReachsafetyReachsafetyLoops}{Correct}{True}{Cputime}{Min}{6.812146972}%
\StoreBenchExecResult{SvcompNineteenUkojak}{SvCompPropReachsafetyReachsafetyLoops}{Correct}{True}{Cputime}{Max}{485.582305311}%
\StoreBenchExecResult{SvcompNineteenUkojak}{SvCompPropReachsafetyReachsafetyLoops}{Correct}{True}{Cputime}{Stdev}{102.2788059957280567612066957}%
\StoreBenchExecResult{SvcompNineteenUkojak}{SvCompPropReachsafetyReachsafetyLoops}{Correct}{True}{Walltime}{}{3587.01428818782}%
\StoreBenchExecResult{SvcompNineteenUkojak}{SvCompPropReachsafetyReachsafetyLoops}{Correct}{True}{Walltime}{Avg}{37.364732168623125}%
\StoreBenchExecResult{SvcompNineteenUkojak}{SvCompPropReachsafetyReachsafetyLoops}{Correct}{True}{Walltime}{Median}{3.638766050335}%
\StoreBenchExecResult{SvcompNineteenUkojak}{SvCompPropReachsafetyReachsafetyLoops}{Correct}{True}{Walltime}{Min}{2.47493290901}%
\StoreBenchExecResult{SvcompNineteenUkojak}{SvCompPropReachsafetyReachsafetyLoops}{Correct}{True}{Walltime}{Max}{406.325536013}%
\StoreBenchExecResult{SvcompNineteenUkojak}{SvCompPropReachsafetyReachsafetyLoops}{Correct}{True}{Walltime}{Stdev}{89.46220642982688084402282383}%
\StoreBenchExecResult{SvcompNineteenUkojak}{SvCompPropReachsafetyReachsafetyLoops}{Wrong}{True}{Count}{}{0}%
\StoreBenchExecResult{SvcompNineteenUkojak}{SvCompPropReachsafetyReachsafetyLoops}{Wrong}{True}{Cputime}{}{0}%
\StoreBenchExecResult{SvcompNineteenUkojak}{SvCompPropReachsafetyReachsafetyLoops}{Wrong}{True}{Cputime}{Avg}{None}%
\StoreBenchExecResult{SvcompNineteenUkojak}{SvCompPropReachsafetyReachsafetyLoops}{Wrong}{True}{Cputime}{Median}{None}%
\StoreBenchExecResult{SvcompNineteenUkojak}{SvCompPropReachsafetyReachsafetyLoops}{Wrong}{True}{Cputime}{Min}{None}%
\StoreBenchExecResult{SvcompNineteenUkojak}{SvCompPropReachsafetyReachsafetyLoops}{Wrong}{True}{Cputime}{Max}{None}%
\StoreBenchExecResult{SvcompNineteenUkojak}{SvCompPropReachsafetyReachsafetyLoops}{Wrong}{True}{Cputime}{Stdev}{None}%
\StoreBenchExecResult{SvcompNineteenUkojak}{SvCompPropReachsafetyReachsafetyLoops}{Wrong}{True}{Walltime}{}{0}%
\StoreBenchExecResult{SvcompNineteenUkojak}{SvCompPropReachsafetyReachsafetyLoops}{Wrong}{True}{Walltime}{Avg}{None}%
\StoreBenchExecResult{SvcompNineteenUkojak}{SvCompPropReachsafetyReachsafetyLoops}{Wrong}{True}{Walltime}{Median}{None}%
\StoreBenchExecResult{SvcompNineteenUkojak}{SvCompPropReachsafetyReachsafetyLoops}{Wrong}{True}{Walltime}{Min}{None}%
\StoreBenchExecResult{SvcompNineteenUkojak}{SvCompPropReachsafetyReachsafetyLoops}{Wrong}{True}{Walltime}{Max}{None}%
\StoreBenchExecResult{SvcompNineteenUkojak}{SvCompPropReachsafetyReachsafetyLoops}{Wrong}{True}{Walltime}{Stdev}{None}%
\StoreBenchExecResult{SvcompNineteenUkojak}{SvCompPropReachsafetyReachsafetyLoops}{Error}{}{Count}{}{76}%
\StoreBenchExecResult{SvcompNineteenUkojak}{SvCompPropReachsafetyReachsafetyLoops}{Error}{}{Cputime}{}{63170.169298813}%
\StoreBenchExecResult{SvcompNineteenUkojak}{SvCompPropReachsafetyReachsafetyLoops}{Error}{}{Cputime}{Avg}{831.1864381422763157894736842}%
\StoreBenchExecResult{SvcompNineteenUkojak}{SvCompPropReachsafetyReachsafetyLoops}{Error}{}{Cputime}{Median}{900.653031939}%
\StoreBenchExecResult{SvcompNineteenUkojak}{SvCompPropReachsafetyReachsafetyLoops}{Error}{}{Cputime}{Min}{10.774221952}%
\StoreBenchExecResult{SvcompNineteenUkojak}{SvCompPropReachsafetyReachsafetyLoops}{Error}{}{Cputime}{Max}{901.584548426}%
\StoreBenchExecResult{SvcompNineteenUkojak}{SvCompPropReachsafetyReachsafetyLoops}{Error}{}{Cputime}{Stdev}{237.4966711122728711762459926}%
\StoreBenchExecResult{SvcompNineteenUkojak}{SvCompPropReachsafetyReachsafetyLoops}{Error}{}{Walltime}{}{54143.35488414706}%
\StoreBenchExecResult{SvcompNineteenUkojak}{SvCompPropReachsafetyReachsafetyLoops}{Error}{}{Walltime}{Avg}{712.4125642650928947368421053}%
\StoreBenchExecResult{SvcompNineteenUkojak}{SvCompPropReachsafetyReachsafetyLoops}{Error}{}{Walltime}{Median}{759.3853514195}%
\StoreBenchExecResult{SvcompNineteenUkojak}{SvCompPropReachsafetyReachsafetyLoops}{Error}{}{Walltime}{Min}{3.82274198532}%
\StoreBenchExecResult{SvcompNineteenUkojak}{SvCompPropReachsafetyReachsafetyLoops}{Error}{}{Walltime}{Max}{890.61344409}%
\StoreBenchExecResult{SvcompNineteenUkojak}{SvCompPropReachsafetyReachsafetyLoops}{Error}{}{Walltime}{Stdev}{211.8539297300498824780311888}%
\StoreBenchExecResult{SvcompNineteenUkojak}{SvCompPropReachsafetyReachsafetyLoops}{Error}{Error}{Count}{}{6}%
\StoreBenchExecResult{SvcompNineteenUkojak}{SvCompPropReachsafetyReachsafetyLoops}{Error}{Error}{Cputime}{}{121.123063805}%
\StoreBenchExecResult{SvcompNineteenUkojak}{SvCompPropReachsafetyReachsafetyLoops}{Error}{Error}{Cputime}{Avg}{20.18717730083333333333333333}%
\StoreBenchExecResult{SvcompNineteenUkojak}{SvCompPropReachsafetyReachsafetyLoops}{Error}{Error}{Cputime}{Median}{11.5068203895}%
\StoreBenchExecResult{SvcompNineteenUkojak}{SvCompPropReachsafetyReachsafetyLoops}{Error}{Error}{Cputime}{Min}{10.774221952}%
\StoreBenchExecResult{SvcompNineteenUkojak}{SvCompPropReachsafetyReachsafetyLoops}{Error}{Error}{Cputime}{Max}{62.684064014}%
\StoreBenchExecResult{SvcompNineteenUkojak}{SvCompPropReachsafetyReachsafetyLoops}{Error}{Error}{Cputime}{Stdev}{19.02513454488440264583138570}%
\StoreBenchExecResult{SvcompNineteenUkojak}{SvCompPropReachsafetyReachsafetyLoops}{Error}{Error}{Walltime}{}{71.00238204006}%
\StoreBenchExecResult{SvcompNineteenUkojak}{SvCompPropReachsafetyReachsafetyLoops}{Error}{Error}{Walltime}{Avg}{11.83373034001}%
\StoreBenchExecResult{SvcompNineteenUkojak}{SvCompPropReachsafetyReachsafetyLoops}{Error}{Error}{Walltime}{Median}{4.151527523995}%
\StoreBenchExecResult{SvcompNineteenUkojak}{SvCompPropReachsafetyReachsafetyLoops}{Error}{Error}{Walltime}{Min}{3.82274198532}%
\StoreBenchExecResult{SvcompNineteenUkojak}{SvCompPropReachsafetyReachsafetyLoops}{Error}{Error}{Walltime}{Max}{50.8143088818}%
\StoreBenchExecResult{SvcompNineteenUkojak}{SvCompPropReachsafetyReachsafetyLoops}{Error}{Error}{Walltime}{Stdev}{17.43324589579832734160508649}%
\StoreBenchExecResult{SvcompNineteenUkojak}{SvCompPropReachsafetyReachsafetyLoops}{Error}{Timeout}{Count}{}{70}%
\StoreBenchExecResult{SvcompNineteenUkojak}{SvCompPropReachsafetyReachsafetyLoops}{Error}{Timeout}{Cputime}{}{63049.046235008}%
\StoreBenchExecResult{SvcompNineteenUkojak}{SvCompPropReachsafetyReachsafetyLoops}{Error}{Timeout}{Cputime}{Avg}{900.7006605001142857142857143}%
\StoreBenchExecResult{SvcompNineteenUkojak}{SvCompPropReachsafetyReachsafetyLoops}{Error}{Timeout}{Cputime}{Median}{900.7025083585}%
\StoreBenchExecResult{SvcompNineteenUkojak}{SvCompPropReachsafetyReachsafetyLoops}{Error}{Timeout}{Cputime}{Min}{900.039421913}%
\StoreBenchExecResult{SvcompNineteenUkojak}{SvCompPropReachsafetyReachsafetyLoops}{Error}{Timeout}{Cputime}{Max}{901.584548426}%
\StoreBenchExecResult{SvcompNineteenUkojak}{SvCompPropReachsafetyReachsafetyLoops}{Error}{Timeout}{Cputime}{Stdev}{0.3461897764310064360290156479}%
\StoreBenchExecResult{SvcompNineteenUkojak}{SvCompPropReachsafetyReachsafetyLoops}{Error}{Timeout}{Walltime}{}{54072.352502107}%
\StoreBenchExecResult{SvcompNineteenUkojak}{SvCompPropReachsafetyReachsafetyLoops}{Error}{Timeout}{Walltime}{Avg}{772.4621786015285714285714286}%
\StoreBenchExecResult{SvcompNineteenUkojak}{SvCompPropReachsafetyReachsafetyLoops}{Error}{Timeout}{Walltime}{Median}{760.380630970}%
\StoreBenchExecResult{SvcompNineteenUkojak}{SvCompPropReachsafetyReachsafetyLoops}{Error}{Timeout}{Walltime}{Min}{514.021726131}%
\StoreBenchExecResult{SvcompNineteenUkojak}{SvCompPropReachsafetyReachsafetyLoops}{Error}{Timeout}{Walltime}{Max}{890.61344409}%
\StoreBenchExecResult{SvcompNineteenUkojak}{SvCompPropReachsafetyReachsafetyLoops}{Error}{Timeout}{Walltime}{Stdev}{55.02389050252284350164432027}%
\providecommand\StoreBenchExecResult[7]{\expandafter\newcommand\csname#1#2#3#4#5#6\endcsname{#7}}%
\StoreBenchExecResult{SvcompNineteenUtaipan}{SvCompPropReachsafetyReachsafetyLoops}{Total}{}{Count}{}{208}%
\StoreBenchExecResult{SvcompNineteenUtaipan}{SvCompPropReachsafetyReachsafetyLoops}{Total}{}{Cputime}{}{50935.304825385}%
\StoreBenchExecResult{SvcompNineteenUtaipan}{SvCompPropReachsafetyReachsafetyLoops}{Total}{}{Cputime}{Avg}{244.8812731989663461538461538}%
\StoreBenchExecResult{SvcompNineteenUtaipan}{SvCompPropReachsafetyReachsafetyLoops}{Total}{}{Cputime}{Median}{12.149301121}%
\StoreBenchExecResult{SvcompNineteenUtaipan}{SvCompPropReachsafetyReachsafetyLoops}{Total}{}{Cputime}{Min}{6.837490405}%
\StoreBenchExecResult{SvcompNineteenUtaipan}{SvCompPropReachsafetyReachsafetyLoops}{Total}{}{Cputime}{Max}{901.074818567}%
\StoreBenchExecResult{SvcompNineteenUtaipan}{SvCompPropReachsafetyReachsafetyLoops}{Total}{}{Cputime}{Stdev}{382.7137906931626827771295371}%
\StoreBenchExecResult{SvcompNineteenUtaipan}{SvCompPropReachsafetyReachsafetyLoops}{Total}{}{Walltime}{}{45715.75471472729}%
\StoreBenchExecResult{SvcompNineteenUtaipan}{SvCompPropReachsafetyReachsafetyLoops}{Total}{}{Walltime}{Avg}{219.7872822823427403846153846}%
\StoreBenchExecResult{SvcompNineteenUtaipan}{SvCompPropReachsafetyReachsafetyLoops}{Total}{}{Walltime}{Median}{4.48053443432}%
\StoreBenchExecResult{SvcompNineteenUtaipan}{SvCompPropReachsafetyReachsafetyLoops}{Total}{}{Walltime}{Min}{2.52050876617}%
\StoreBenchExecResult{SvcompNineteenUtaipan}{SvCompPropReachsafetyReachsafetyLoops}{Total}{}{Walltime}{Max}{882.650352955}%
\StoreBenchExecResult{SvcompNineteenUtaipan}{SvCompPropReachsafetyReachsafetyLoops}{Total}{}{Walltime}{Stdev}{356.2811916604326307437221608}%
\StoreBenchExecResult{SvcompNineteenUtaipan}{SvCompPropReachsafetyReachsafetyLoops}{Correct}{}{Count}{}{150}%
\StoreBenchExecResult{SvcompNineteenUtaipan}{SvCompPropReachsafetyReachsafetyLoops}{Correct}{}{Cputime}{}{3975.319152554}%
\StoreBenchExecResult{SvcompNineteenUtaipan}{SvCompPropReachsafetyReachsafetyLoops}{Correct}{}{Cputime}{Avg}{26.50212768369333333333333333}%
\StoreBenchExecResult{SvcompNineteenUtaipan}{SvCompPropReachsafetyReachsafetyLoops}{Correct}{}{Cputime}{Median}{10.3913923855}%
\StoreBenchExecResult{SvcompNineteenUtaipan}{SvCompPropReachsafetyReachsafetyLoops}{Correct}{}{Cputime}{Min}{6.837490405}%
\StoreBenchExecResult{SvcompNineteenUtaipan}{SvCompPropReachsafetyReachsafetyLoops}{Correct}{}{Cputime}{Max}{681.14413899}%
\StoreBenchExecResult{SvcompNineteenUtaipan}{SvCompPropReachsafetyReachsafetyLoops}{Correct}{}{Cputime}{Stdev}{66.37963758471477091123499189}%
\StoreBenchExecResult{SvcompNineteenUtaipan}{SvCompPropReachsafetyReachsafetyLoops}{Correct}{}{Walltime}{}{2577.03907871322}%
\StoreBenchExecResult{SvcompNineteenUtaipan}{SvCompPropReachsafetyReachsafetyLoops}{Correct}{}{Walltime}{Avg}{17.1802605247548}%
\StoreBenchExecResult{SvcompNineteenUtaipan}{SvCompPropReachsafetyReachsafetyLoops}{Correct}{}{Walltime}{Median}{3.68777000904}%
\StoreBenchExecResult{SvcompNineteenUtaipan}{SvCompPropReachsafetyReachsafetyLoops}{Correct}{}{Walltime}{Min}{2.52050876617}%
\StoreBenchExecResult{SvcompNineteenUtaipan}{SvCompPropReachsafetyReachsafetyLoops}{Correct}{}{Walltime}{Max}{624.835301876}%
\StoreBenchExecResult{SvcompNineteenUtaipan}{SvCompPropReachsafetyReachsafetyLoops}{Correct}{}{Walltime}{Stdev}{60.64159345821072720783476188}%
\StoreBenchExecResult{SvcompNineteenUtaipan}{SvCompPropReachsafetyReachsafetyLoops}{Correct}{False}{Count}{}{34}%
\StoreBenchExecResult{SvcompNineteenUtaipan}{SvCompPropReachsafetyReachsafetyLoops}{Correct}{False}{Cputime}{}{1164.408302071}%
\StoreBenchExecResult{SvcompNineteenUtaipan}{SvCompPropReachsafetyReachsafetyLoops}{Correct}{False}{Cputime}{Avg}{34.24730300208823529411764706}%
\StoreBenchExecResult{SvcompNineteenUtaipan}{SvCompPropReachsafetyReachsafetyLoops}{Correct}{False}{Cputime}{Median}{8.463489162}%
\StoreBenchExecResult{SvcompNineteenUtaipan}{SvCompPropReachsafetyReachsafetyLoops}{Correct}{False}{Cputime}{Min}{6.837490405}%
\StoreBenchExecResult{SvcompNineteenUtaipan}{SvCompPropReachsafetyReachsafetyLoops}{Correct}{False}{Cputime}{Max}{681.14413899}%
\StoreBenchExecResult{SvcompNineteenUtaipan}{SvCompPropReachsafetyReachsafetyLoops}{Correct}{False}{Cputime}{Stdev}{115.6168107409001380004084490}%
\StoreBenchExecResult{SvcompNineteenUtaipan}{SvCompPropReachsafetyReachsafetyLoops}{Correct}{False}{Walltime}{}{897.04094934475}%
\StoreBenchExecResult{SvcompNineteenUtaipan}{SvCompPropReachsafetyReachsafetyLoops}{Correct}{False}{Walltime}{Avg}{26.38355733366911764705882353}%
\StoreBenchExecResult{SvcompNineteenUtaipan}{SvCompPropReachsafetyReachsafetyLoops}{Correct}{False}{Walltime}{Median}{3.18187189102}%
\StoreBenchExecResult{SvcompNineteenUtaipan}{SvCompPropReachsafetyReachsafetyLoops}{Correct}{False}{Walltime}{Min}{2.52050876617}%
\StoreBenchExecResult{SvcompNineteenUtaipan}{SvCompPropReachsafetyReachsafetyLoops}{Correct}{False}{Walltime}{Max}{624.835301876}%
\StoreBenchExecResult{SvcompNineteenUtaipan}{SvCompPropReachsafetyReachsafetyLoops}{Correct}{False}{Walltime}{Stdev}{107.2063316707634499575725667}%
\StoreBenchExecResult{SvcompNineteenUtaipan}{SvCompPropReachsafetyReachsafetyLoops}{Wrong}{False}{Count}{}{0}%
\StoreBenchExecResult{SvcompNineteenUtaipan}{SvCompPropReachsafetyReachsafetyLoops}{Wrong}{False}{Cputime}{}{0}%
\StoreBenchExecResult{SvcompNineteenUtaipan}{SvCompPropReachsafetyReachsafetyLoops}{Wrong}{False}{Cputime}{Avg}{None}%
\StoreBenchExecResult{SvcompNineteenUtaipan}{SvCompPropReachsafetyReachsafetyLoops}{Wrong}{False}{Cputime}{Median}{None}%
\StoreBenchExecResult{SvcompNineteenUtaipan}{SvCompPropReachsafetyReachsafetyLoops}{Wrong}{False}{Cputime}{Min}{None}%
\StoreBenchExecResult{SvcompNineteenUtaipan}{SvCompPropReachsafetyReachsafetyLoops}{Wrong}{False}{Cputime}{Max}{None}%
\StoreBenchExecResult{SvcompNineteenUtaipan}{SvCompPropReachsafetyReachsafetyLoops}{Wrong}{False}{Cputime}{Stdev}{None}%
\StoreBenchExecResult{SvcompNineteenUtaipan}{SvCompPropReachsafetyReachsafetyLoops}{Wrong}{False}{Walltime}{}{0}%
\StoreBenchExecResult{SvcompNineteenUtaipan}{SvCompPropReachsafetyReachsafetyLoops}{Wrong}{False}{Walltime}{Avg}{None}%
\StoreBenchExecResult{SvcompNineteenUtaipan}{SvCompPropReachsafetyReachsafetyLoops}{Wrong}{False}{Walltime}{Median}{None}%
\StoreBenchExecResult{SvcompNineteenUtaipan}{SvCompPropReachsafetyReachsafetyLoops}{Wrong}{False}{Walltime}{Min}{None}%
\StoreBenchExecResult{SvcompNineteenUtaipan}{SvCompPropReachsafetyReachsafetyLoops}{Wrong}{False}{Walltime}{Max}{None}%
\StoreBenchExecResult{SvcompNineteenUtaipan}{SvCompPropReachsafetyReachsafetyLoops}{Wrong}{False}{Walltime}{Stdev}{None}%
\StoreBenchExecResult{SvcompNineteenUtaipan}{SvCompPropReachsafetyReachsafetyLoops}{Correct}{True}{Count}{}{116}%
\StoreBenchExecResult{SvcompNineteenUtaipan}{SvCompPropReachsafetyReachsafetyLoops}{Correct}{True}{Cputime}{}{2810.910850483}%
\StoreBenchExecResult{SvcompNineteenUtaipan}{SvCompPropReachsafetyReachsafetyLoops}{Correct}{True}{Cputime}{Avg}{24.23199009037068965517241379}%
\StoreBenchExecResult{SvcompNineteenUtaipan}{SvCompPropReachsafetyReachsafetyLoops}{Correct}{True}{Cputime}{Median}{10.8887585815}%
\StoreBenchExecResult{SvcompNineteenUtaipan}{SvCompPropReachsafetyReachsafetyLoops}{Correct}{True}{Cputime}{Min}{7.082960827}%
\StoreBenchExecResult{SvcompNineteenUtaipan}{SvCompPropReachsafetyReachsafetyLoops}{Correct}{True}{Cputime}{Max}{278.768860203}%
\StoreBenchExecResult{SvcompNineteenUtaipan}{SvCompPropReachsafetyReachsafetyLoops}{Correct}{True}{Cputime}{Stdev}{41.91685688325154285546965276}%
\StoreBenchExecResult{SvcompNineteenUtaipan}{SvCompPropReachsafetyReachsafetyLoops}{Correct}{True}{Walltime}{}{1679.99812936847}%
\StoreBenchExecResult{SvcompNineteenUtaipan}{SvCompPropReachsafetyReachsafetyLoops}{Correct}{True}{Walltime}{Avg}{14.48274249455577586206896552}%
\StoreBenchExecResult{SvcompNineteenUtaipan}{SvCompPropReachsafetyReachsafetyLoops}{Correct}{True}{Walltime}{Median}{3.85159945488}%
\StoreBenchExecResult{SvcompNineteenUtaipan}{SvCompPropReachsafetyReachsafetyLoops}{Correct}{True}{Walltime}{Min}{2.66930294037}%
\StoreBenchExecResult{SvcompNineteenUtaipan}{SvCompPropReachsafetyReachsafetyLoops}{Correct}{True}{Walltime}{Max}{258.520524025}%
\StoreBenchExecResult{SvcompNineteenUtaipan}{SvCompPropReachsafetyReachsafetyLoops}{Correct}{True}{Walltime}{Stdev}{36.80304089297031868470619234}%
\StoreBenchExecResult{SvcompNineteenUtaipan}{SvCompPropReachsafetyReachsafetyLoops}{Error}{}{Count}{}{57}%
\StoreBenchExecResult{SvcompNineteenUtaipan}{SvCompPropReachsafetyReachsafetyLoops}{Error}{}{Cputime}{}{46915.953070275}%
\StoreBenchExecResult{SvcompNineteenUtaipan}{SvCompPropReachsafetyReachsafetyLoops}{Error}{}{Cputime}{Avg}{823.0868959697368421052631579}%
\StoreBenchExecResult{SvcompNineteenUtaipan}{SvCompPropReachsafetyReachsafetyLoops}{Error}{}{Cputime}{Median}{900.47688021}%
\StoreBenchExecResult{SvcompNineteenUtaipan}{SvCompPropReachsafetyReachsafetyLoops}{Error}{}{Cputime}{Min}{10.8287895}%
\StoreBenchExecResult{SvcompNineteenUtaipan}{SvCompPropReachsafetyReachsafetyLoops}{Error}{}{Cputime}{Max}{901.074818567}%
\StoreBenchExecResult{SvcompNineteenUtaipan}{SvCompPropReachsafetyReachsafetyLoops}{Error}{}{Cputime}{Stdev}{249.7258941406954674852320570}%
\StoreBenchExecResult{SvcompNineteenUtaipan}{SvCompPropReachsafetyReachsafetyLoops}{Error}{}{Walltime}{}{43112.60528612047}%
\StoreBenchExecResult{SvcompNineteenUtaipan}{SvCompPropReachsafetyReachsafetyLoops}{Error}{}{Walltime}{Avg}{756.3614962477275438596491228}%
\StoreBenchExecResult{SvcompNineteenUtaipan}{SvCompPropReachsafetyReachsafetyLoops}{Error}{}{Walltime}{Median}{845.190462828}%
\StoreBenchExecResult{SvcompNineteenUtaipan}{SvCompPropReachsafetyReachsafetyLoops}{Error}{}{Walltime}{Min}{3.7459499836}%
\StoreBenchExecResult{SvcompNineteenUtaipan}{SvCompPropReachsafetyReachsafetyLoops}{Error}{}{Walltime}{Max}{882.650352955}%
\StoreBenchExecResult{SvcompNineteenUtaipan}{SvCompPropReachsafetyReachsafetyLoops}{Error}{}{Walltime}{Stdev}{238.6073729259223895327118082}%
\StoreBenchExecResult{SvcompNineteenUtaipan}{SvCompPropReachsafetyReachsafetyLoops}{Error}{Error}{Count}{}{5}%
\StoreBenchExecResult{SvcompNineteenUtaipan}{SvCompPropReachsafetyReachsafetyLoops}{Error}{Error}{Cputime}{}{89.222926537}%
\StoreBenchExecResult{SvcompNineteenUtaipan}{SvCompPropReachsafetyReachsafetyLoops}{Error}{Error}{Cputime}{Avg}{17.8445853074}%
\StoreBenchExecResult{SvcompNineteenUtaipan}{SvCompPropReachsafetyReachsafetyLoops}{Error}{Error}{Cputime}{Median}{11.266396228}%
\StoreBenchExecResult{SvcompNineteenUtaipan}{SvCompPropReachsafetyReachsafetyLoops}{Error}{Error}{Cputime}{Min}{10.8287895}%
\StoreBenchExecResult{SvcompNineteenUtaipan}{SvCompPropReachsafetyReachsafetyLoops}{Error}{Error}{Cputime}{Max}{44.268983984}%
\StoreBenchExecResult{SvcompNineteenUtaipan}{SvCompPropReachsafetyReachsafetyLoops}{Error}{Error}{Cputime}{Stdev}{13.21467828494518534446625748}%
\StoreBenchExecResult{SvcompNineteenUtaipan}{SvCompPropReachsafetyReachsafetyLoops}{Error}{Error}{Walltime}{}{29.74106168747}%
\StoreBenchExecResult{SvcompNineteenUtaipan}{SvCompPropReachsafetyReachsafetyLoops}{Error}{Error}{Walltime}{Avg}{5.948212337494}%
\StoreBenchExecResult{SvcompNineteenUtaipan}{SvCompPropReachsafetyReachsafetyLoops}{Error}{Error}{Walltime}{Median}{4.12565112114}%
\StoreBenchExecResult{SvcompNineteenUtaipan}{SvCompPropReachsafetyReachsafetyLoops}{Error}{Error}{Walltime}{Min}{3.7459499836}%
\StoreBenchExecResult{SvcompNineteenUtaipan}{SvCompPropReachsafetyReachsafetyLoops}{Error}{Error}{Walltime}{Max}{13.8516178131}%
\StoreBenchExecResult{SvcompNineteenUtaipan}{SvCompPropReachsafetyReachsafetyLoops}{Error}{Error}{Walltime}{Stdev}{3.954603462956254204910899597}%
\StoreBenchExecResult{SvcompNineteenUtaipan}{SvCompPropReachsafetyReachsafetyLoops}{Error}{Timeout}{Count}{}{52}%
\StoreBenchExecResult{SvcompNineteenUtaipan}{SvCompPropReachsafetyReachsafetyLoops}{Error}{Timeout}{Cputime}{}{46826.730143738}%
\StoreBenchExecResult{SvcompNineteenUtaipan}{SvCompPropReachsafetyReachsafetyLoops}{Error}{Timeout}{Cputime}{Avg}{900.5140412257307692307692308}%
\StoreBenchExecResult{SvcompNineteenUtaipan}{SvCompPropReachsafetyReachsafetyLoops}{Error}{Timeout}{Cputime}{Median}{900.5392543895}%
\StoreBenchExecResult{SvcompNineteenUtaipan}{SvCompPropReachsafetyReachsafetyLoops}{Error}{Timeout}{Cputime}{Min}{900.055151799}%
\StoreBenchExecResult{SvcompNineteenUtaipan}{SvCompPropReachsafetyReachsafetyLoops}{Error}{Timeout}{Cputime}{Max}{901.074818567}%
\StoreBenchExecResult{SvcompNineteenUtaipan}{SvCompPropReachsafetyReachsafetyLoops}{Error}{Timeout}{Cputime}{Stdev}{0.3151881149562624353509336477}%
\StoreBenchExecResult{SvcompNineteenUtaipan}{SvCompPropReachsafetyReachsafetyLoops}{Error}{Timeout}{Walltime}{}{43082.864224433}%
\StoreBenchExecResult{SvcompNineteenUtaipan}{SvCompPropReachsafetyReachsafetyLoops}{Error}{Timeout}{Walltime}{Avg}{828.5166197006346153846153846}%
\StoreBenchExecResult{SvcompNineteenUtaipan}{SvCompPropReachsafetyReachsafetyLoops}{Error}{Timeout}{Walltime}{Median}{851.331788182}%
\StoreBenchExecResult{SvcompNineteenUtaipan}{SvCompPropReachsafetyReachsafetyLoops}{Error}{Timeout}{Walltime}{Min}{712.051084042}%
\StoreBenchExecResult{SvcompNineteenUtaipan}{SvCompPropReachsafetyReachsafetyLoops}{Error}{Timeout}{Walltime}{Max}{882.650352955}%
\StoreBenchExecResult{SvcompNineteenUtaipan}{SvCompPropReachsafetyReachsafetyLoops}{Error}{Timeout}{Walltime}{Stdev}{55.26140260670015215119214699}%
\StoreBenchExecResult{SvcompNineteenUtaipan}{SvCompPropReachsafetyReachsafetyLoops}{Wrong}{}{Count}{}{1}%
\StoreBenchExecResult{SvcompNineteenUtaipan}{SvCompPropReachsafetyReachsafetyLoops}{Wrong}{}{Cputime}{}{44.032602556}%
\StoreBenchExecResult{SvcompNineteenUtaipan}{SvCompPropReachsafetyReachsafetyLoops}{Wrong}{}{Cputime}{Avg}{44.032602556}%
\StoreBenchExecResult{SvcompNineteenUtaipan}{SvCompPropReachsafetyReachsafetyLoops}{Wrong}{}{Cputime}{Median}{44.032602556}%
\StoreBenchExecResult{SvcompNineteenUtaipan}{SvCompPropReachsafetyReachsafetyLoops}{Wrong}{}{Cputime}{Min}{44.032602556}%
\StoreBenchExecResult{SvcompNineteenUtaipan}{SvCompPropReachsafetyReachsafetyLoops}{Wrong}{}{Cputime}{Max}{44.032602556}%
\StoreBenchExecResult{SvcompNineteenUtaipan}{SvCompPropReachsafetyReachsafetyLoops}{Wrong}{}{Cputime}{Stdev}{0E-9}%
\StoreBenchExecResult{SvcompNineteenUtaipan}{SvCompPropReachsafetyReachsafetyLoops}{Wrong}{}{Walltime}{}{26.1103498936}%
\StoreBenchExecResult{SvcompNineteenUtaipan}{SvCompPropReachsafetyReachsafetyLoops}{Wrong}{}{Walltime}{Avg}{26.1103498936}%
\StoreBenchExecResult{SvcompNineteenUtaipan}{SvCompPropReachsafetyReachsafetyLoops}{Wrong}{}{Walltime}{Median}{26.1103498936}%
\StoreBenchExecResult{SvcompNineteenUtaipan}{SvCompPropReachsafetyReachsafetyLoops}{Wrong}{}{Walltime}{Min}{26.1103498936}%
\StoreBenchExecResult{SvcompNineteenUtaipan}{SvCompPropReachsafetyReachsafetyLoops}{Wrong}{}{Walltime}{Max}{26.1103498936}%
\StoreBenchExecResult{SvcompNineteenUtaipan}{SvCompPropReachsafetyReachsafetyLoops}{Wrong}{}{Walltime}{Stdev}{0E-10}%
\StoreBenchExecResult{SvcompNineteenUtaipan}{SvCompPropReachsafetyReachsafetyLoops}{Wrong}{True}{Count}{}{1}%
\StoreBenchExecResult{SvcompNineteenUtaipan}{SvCompPropReachsafetyReachsafetyLoops}{Wrong}{True}{Cputime}{}{44.032602556}%
\StoreBenchExecResult{SvcompNineteenUtaipan}{SvCompPropReachsafetyReachsafetyLoops}{Wrong}{True}{Cputime}{Avg}{44.032602556}%
\StoreBenchExecResult{SvcompNineteenUtaipan}{SvCompPropReachsafetyReachsafetyLoops}{Wrong}{True}{Cputime}{Median}{44.032602556}%
\StoreBenchExecResult{SvcompNineteenUtaipan}{SvCompPropReachsafetyReachsafetyLoops}{Wrong}{True}{Cputime}{Min}{44.032602556}%
\StoreBenchExecResult{SvcompNineteenUtaipan}{SvCompPropReachsafetyReachsafetyLoops}{Wrong}{True}{Cputime}{Max}{44.032602556}%
\StoreBenchExecResult{SvcompNineteenUtaipan}{SvCompPropReachsafetyReachsafetyLoops}{Wrong}{True}{Cputime}{Stdev}{0E-9}%
\StoreBenchExecResult{SvcompNineteenUtaipan}{SvCompPropReachsafetyReachsafetyLoops}{Wrong}{True}{Walltime}{}{26.1103498936}%
\StoreBenchExecResult{SvcompNineteenUtaipan}{SvCompPropReachsafetyReachsafetyLoops}{Wrong}{True}{Walltime}{Avg}{26.1103498936}%
\StoreBenchExecResult{SvcompNineteenUtaipan}{SvCompPropReachsafetyReachsafetyLoops}{Wrong}{True}{Walltime}{Median}{26.1103498936}%
\StoreBenchExecResult{SvcompNineteenUtaipan}{SvCompPropReachsafetyReachsafetyLoops}{Wrong}{True}{Walltime}{Min}{26.1103498936}%
\StoreBenchExecResult{SvcompNineteenUtaipan}{SvCompPropReachsafetyReachsafetyLoops}{Wrong}{True}{Walltime}{Max}{26.1103498936}%
\StoreBenchExecResult{SvcompNineteenUtaipan}{SvCompPropReachsafetyReachsafetyLoops}{Wrong}{True}{Walltime}{Stdev}{0E-10}%
\providecommand\StoreBenchExecResult[7]{\expandafter\newcommand\csname#1#2#3#4#5#6\endcsname{#7}}%
\StoreBenchExecResult{SvcompNineteenVeriabs}{SvCompPropReachsafetyReachsafetyLoops}{Total}{}{Count}{}{208}%
\StoreBenchExecResult{SvcompNineteenVeriabs}{SvCompPropReachsafetyReachsafetyLoops}{Total}{}{Cputime}{}{37769.617066955}%
\StoreBenchExecResult{SvcompNineteenVeriabs}{SvCompPropReachsafetyReachsafetyLoops}{Total}{}{Cputime}{Avg}{181.5846974372836538461538462}%
\StoreBenchExecResult{SvcompNineteenVeriabs}{SvCompPropReachsafetyReachsafetyLoops}{Total}{}{Cputime}{Median}{31.1441511075}%
\StoreBenchExecResult{SvcompNineteenVeriabs}{SvCompPropReachsafetyReachsafetyLoops}{Total}{}{Cputime}{Min}{2.971585435}%
\StoreBenchExecResult{SvcompNineteenVeriabs}{SvCompPropReachsafetyReachsafetyLoops}{Total}{}{Cputime}{Max}{901.106487171}%
\StoreBenchExecResult{SvcompNineteenVeriabs}{SvCompPropReachsafetyReachsafetyLoops}{Total}{}{Cputime}{Stdev}{279.6682305515304784180530277}%
\StoreBenchExecResult{SvcompNineteenVeriabs}{SvCompPropReachsafetyReachsafetyLoops}{Total}{}{Walltime}{}{33351.49895215141}%
\StoreBenchExecResult{SvcompNineteenVeriabs}{SvCompPropReachsafetyReachsafetyLoops}{Total}{}{Walltime}{Avg}{160.3437449622663942307692308}%
\StoreBenchExecResult{SvcompNineteenVeriabs}{SvCompPropReachsafetyReachsafetyLoops}{Total}{}{Walltime}{Median}{19.1055465937}%
\StoreBenchExecResult{SvcompNineteenVeriabs}{SvCompPropReachsafetyReachsafetyLoops}{Total}{}{Walltime}{Min}{1.22559690475}%
\StoreBenchExecResult{SvcompNineteenVeriabs}{SvCompPropReachsafetyReachsafetyLoops}{Total}{}{Walltime}{Max}{891.379369974}%
\StoreBenchExecResult{SvcompNineteenVeriabs}{SvCompPropReachsafetyReachsafetyLoops}{Total}{}{Walltime}{Stdev}{278.2386181491780658518560989}%
\StoreBenchExecResult{SvcompNineteenVeriabs}{SvCompPropReachsafetyReachsafetyLoops}{Correct}{}{Count}{}{182}%
\StoreBenchExecResult{SvcompNineteenVeriabs}{SvCompPropReachsafetyReachsafetyLoops}{Correct}{}{Cputime}{}{17679.548638922}%
\StoreBenchExecResult{SvcompNineteenVeriabs}{SvCompPropReachsafetyReachsafetyLoops}{Correct}{}{Cputime}{Avg}{97.14037713693406593406593407}%
\StoreBenchExecResult{SvcompNineteenVeriabs}{SvCompPropReachsafetyReachsafetyLoops}{Correct}{}{Cputime}{Median}{27.0723548685}%
\StoreBenchExecResult{SvcompNineteenVeriabs}{SvCompPropReachsafetyReachsafetyLoops}{Correct}{}{Cputime}{Min}{2.971585435}%
\StoreBenchExecResult{SvcompNineteenVeriabs}{SvCompPropReachsafetyReachsafetyLoops}{Correct}{}{Cputime}{Max}{824.625909549}%
\StoreBenchExecResult{SvcompNineteenVeriabs}{SvCompPropReachsafetyReachsafetyLoops}{Correct}{}{Cputime}{Stdev}{147.6748453552110584690451253}%
\StoreBenchExecResult{SvcompNineteenVeriabs}{SvCompPropReachsafetyReachsafetyLoops}{Correct}{}{Walltime}{}{13678.84736132731}%
\StoreBenchExecResult{SvcompNineteenVeriabs}{SvCompPropReachsafetyReachsafetyLoops}{Correct}{}{Walltime}{Avg}{75.15850198531489010989010989}%
\StoreBenchExecResult{SvcompNineteenVeriabs}{SvCompPropReachsafetyReachsafetyLoops}{Correct}{}{Walltime}{Median}{10.96807301045}%
\StoreBenchExecResult{SvcompNineteenVeriabs}{SvCompPropReachsafetyReachsafetyLoops}{Correct}{}{Walltime}{Min}{1.22559690475}%
\StoreBenchExecResult{SvcompNineteenVeriabs}{SvCompPropReachsafetyReachsafetyLoops}{Correct}{}{Walltime}{Max}{800.405259132}%
\StoreBenchExecResult{SvcompNineteenVeriabs}{SvCompPropReachsafetyReachsafetyLoops}{Correct}{}{Walltime}{Stdev}{141.5634996074967304679724092}%
\StoreBenchExecResult{SvcompNineteenVeriabs}{SvCompPropReachsafetyReachsafetyLoops}{Correct}{False}{Count}{}{57}%
\StoreBenchExecResult{SvcompNineteenVeriabs}{SvCompPropReachsafetyReachsafetyLoops}{Correct}{False}{Cputime}{}{7434.908996489}%
\StoreBenchExecResult{SvcompNineteenVeriabs}{SvCompPropReachsafetyReachsafetyLoops}{Correct}{False}{Cputime}{Avg}{130.4369999384035087719298246}%
\StoreBenchExecResult{SvcompNineteenVeriabs}{SvCompPropReachsafetyReachsafetyLoops}{Correct}{False}{Cputime}{Median}{17.842475275}%
\StoreBenchExecResult{SvcompNineteenVeriabs}{SvCompPropReachsafetyReachsafetyLoops}{Correct}{False}{Cputime}{Min}{2.971585435}%
\StoreBenchExecResult{SvcompNineteenVeriabs}{SvCompPropReachsafetyReachsafetyLoops}{Correct}{False}{Cputime}{Max}{641.462232979}%
\StoreBenchExecResult{SvcompNineteenVeriabs}{SvCompPropReachsafetyReachsafetyLoops}{Correct}{False}{Cputime}{Stdev}{176.4308909437953762716396444}%
\StoreBenchExecResult{SvcompNineteenVeriabs}{SvCompPropReachsafetyReachsafetyLoops}{Correct}{False}{Walltime}{}{6734.80822420220}%
\StoreBenchExecResult{SvcompNineteenVeriabs}{SvCompPropReachsafetyReachsafetyLoops}{Correct}{False}{Walltime}{Avg}{118.1545302491614035087719298}%
\StoreBenchExecResult{SvcompNineteenVeriabs}{SvCompPropReachsafetyReachsafetyLoops}{Correct}{False}{Walltime}{Median}{6.52734899521}%
\StoreBenchExecResult{SvcompNineteenVeriabs}{SvCompPropReachsafetyReachsafetyLoops}{Correct}{False}{Walltime}{Min}{1.22559690475}%
\StoreBenchExecResult{SvcompNineteenVeriabs}{SvCompPropReachsafetyReachsafetyLoops}{Correct}{False}{Walltime}{Max}{621.834995985}%
\StoreBenchExecResult{SvcompNineteenVeriabs}{SvCompPropReachsafetyReachsafetyLoops}{Correct}{False}{Walltime}{Stdev}{172.2114426894122551070019411}%
\StoreBenchExecResult{SvcompNineteenVeriabs}{SvCompPropReachsafetyReachsafetyLoops}{Wrong}{False}{Count}{}{0}%
\StoreBenchExecResult{SvcompNineteenVeriabs}{SvCompPropReachsafetyReachsafetyLoops}{Wrong}{False}{Cputime}{}{0}%
\StoreBenchExecResult{SvcompNineteenVeriabs}{SvCompPropReachsafetyReachsafetyLoops}{Wrong}{False}{Cputime}{Avg}{None}%
\StoreBenchExecResult{SvcompNineteenVeriabs}{SvCompPropReachsafetyReachsafetyLoops}{Wrong}{False}{Cputime}{Median}{None}%
\StoreBenchExecResult{SvcompNineteenVeriabs}{SvCompPropReachsafetyReachsafetyLoops}{Wrong}{False}{Cputime}{Min}{None}%
\StoreBenchExecResult{SvcompNineteenVeriabs}{SvCompPropReachsafetyReachsafetyLoops}{Wrong}{False}{Cputime}{Max}{None}%
\StoreBenchExecResult{SvcompNineteenVeriabs}{SvCompPropReachsafetyReachsafetyLoops}{Wrong}{False}{Cputime}{Stdev}{None}%
\StoreBenchExecResult{SvcompNineteenVeriabs}{SvCompPropReachsafetyReachsafetyLoops}{Wrong}{False}{Walltime}{}{0}%
\StoreBenchExecResult{SvcompNineteenVeriabs}{SvCompPropReachsafetyReachsafetyLoops}{Wrong}{False}{Walltime}{Avg}{None}%
\StoreBenchExecResult{SvcompNineteenVeriabs}{SvCompPropReachsafetyReachsafetyLoops}{Wrong}{False}{Walltime}{Median}{None}%
\StoreBenchExecResult{SvcompNineteenVeriabs}{SvCompPropReachsafetyReachsafetyLoops}{Wrong}{False}{Walltime}{Min}{None}%
\StoreBenchExecResult{SvcompNineteenVeriabs}{SvCompPropReachsafetyReachsafetyLoops}{Wrong}{False}{Walltime}{Max}{None}%
\StoreBenchExecResult{SvcompNineteenVeriabs}{SvCompPropReachsafetyReachsafetyLoops}{Wrong}{False}{Walltime}{Stdev}{None}%
\StoreBenchExecResult{SvcompNineteenVeriabs}{SvCompPropReachsafetyReachsafetyLoops}{Correct}{True}{Count}{}{125}%
\StoreBenchExecResult{SvcompNineteenVeriabs}{SvCompPropReachsafetyReachsafetyLoops}{Correct}{True}{Cputime}{}{10244.639642433}%
\StoreBenchExecResult{SvcompNineteenVeriabs}{SvCompPropReachsafetyReachsafetyLoops}{Correct}{True}{Cputime}{Avg}{81.957117139464}%
\StoreBenchExecResult{SvcompNineteenVeriabs}{SvCompPropReachsafetyReachsafetyLoops}{Correct}{True}{Cputime}{Median}{27.420784753}%
\StoreBenchExecResult{SvcompNineteenVeriabs}{SvCompPropReachsafetyReachsafetyLoops}{Correct}{True}{Cputime}{Min}{3.128185104}%
\StoreBenchExecResult{SvcompNineteenVeriabs}{SvCompPropReachsafetyReachsafetyLoops}{Correct}{True}{Cputime}{Max}{824.625909549}%
\StoreBenchExecResult{SvcompNineteenVeriabs}{SvCompPropReachsafetyReachsafetyLoops}{Correct}{True}{Cputime}{Stdev}{129.6991040512229134836877741}%
\StoreBenchExecResult{SvcompNineteenVeriabs}{SvCompPropReachsafetyReachsafetyLoops}{Correct}{True}{Walltime}{}{6944.03913712511}%
\StoreBenchExecResult{SvcompNineteenVeriabs}{SvCompPropReachsafetyReachsafetyLoops}{Correct}{True}{Walltime}{Avg}{55.55231309700088}%
\StoreBenchExecResult{SvcompNineteenVeriabs}{SvCompPropReachsafetyReachsafetyLoops}{Correct}{True}{Walltime}{Median}{11.1442070007}%
\StoreBenchExecResult{SvcompNineteenVeriabs}{SvCompPropReachsafetyReachsafetyLoops}{Correct}{True}{Walltime}{Min}{1.36005187035}%
\StoreBenchExecResult{SvcompNineteenVeriabs}{SvCompPropReachsafetyReachsafetyLoops}{Correct}{True}{Walltime}{Max}{800.405259132}%
\StoreBenchExecResult{SvcompNineteenVeriabs}{SvCompPropReachsafetyReachsafetyLoops}{Correct}{True}{Walltime}{Stdev}{120.1152934890298929562801015}%
\StoreBenchExecResult{SvcompNineteenVeriabs}{SvCompPropReachsafetyReachsafetyLoops}{Wrong}{True}{Count}{}{0}%
\StoreBenchExecResult{SvcompNineteenVeriabs}{SvCompPropReachsafetyReachsafetyLoops}{Wrong}{True}{Cputime}{}{0}%
\StoreBenchExecResult{SvcompNineteenVeriabs}{SvCompPropReachsafetyReachsafetyLoops}{Wrong}{True}{Cputime}{Avg}{None}%
\StoreBenchExecResult{SvcompNineteenVeriabs}{SvCompPropReachsafetyReachsafetyLoops}{Wrong}{True}{Cputime}{Median}{None}%
\StoreBenchExecResult{SvcompNineteenVeriabs}{SvCompPropReachsafetyReachsafetyLoops}{Wrong}{True}{Cputime}{Min}{None}%
\StoreBenchExecResult{SvcompNineteenVeriabs}{SvCompPropReachsafetyReachsafetyLoops}{Wrong}{True}{Cputime}{Max}{None}%
\StoreBenchExecResult{SvcompNineteenVeriabs}{SvCompPropReachsafetyReachsafetyLoops}{Wrong}{True}{Cputime}{Stdev}{None}%
\StoreBenchExecResult{SvcompNineteenVeriabs}{SvCompPropReachsafetyReachsafetyLoops}{Wrong}{True}{Walltime}{}{0}%
\StoreBenchExecResult{SvcompNineteenVeriabs}{SvCompPropReachsafetyReachsafetyLoops}{Wrong}{True}{Walltime}{Avg}{None}%
\StoreBenchExecResult{SvcompNineteenVeriabs}{SvCompPropReachsafetyReachsafetyLoops}{Wrong}{True}{Walltime}{Median}{None}%
\StoreBenchExecResult{SvcompNineteenVeriabs}{SvCompPropReachsafetyReachsafetyLoops}{Wrong}{True}{Walltime}{Min}{None}%
\StoreBenchExecResult{SvcompNineteenVeriabs}{SvCompPropReachsafetyReachsafetyLoops}{Wrong}{True}{Walltime}{Max}{None}%
\StoreBenchExecResult{SvcompNineteenVeriabs}{SvCompPropReachsafetyReachsafetyLoops}{Wrong}{True}{Walltime}{Stdev}{None}%
\StoreBenchExecResult{SvcompNineteenVeriabs}{SvCompPropReachsafetyReachsafetyLoops}{Error}{}{Count}{}{20}%
\StoreBenchExecResult{SvcompNineteenVeriabs}{SvCompPropReachsafetyReachsafetyLoops}{Error}{}{Cputime}{}{16313.462417245}%
\StoreBenchExecResult{SvcompNineteenVeriabs}{SvCompPropReachsafetyReachsafetyLoops}{Error}{}{Cputime}{Avg}{815.67312086225}%
\StoreBenchExecResult{SvcompNineteenVeriabs}{SvCompPropReachsafetyReachsafetyLoops}{Error}{}{Cputime}{Median}{900.8279615725}%
\StoreBenchExecResult{SvcompNineteenVeriabs}{SvCompPropReachsafetyReachsafetyLoops}{Error}{}{Cputime}{Min}{48.937238342}%
\StoreBenchExecResult{SvcompNineteenVeriabs}{SvCompPropReachsafetyReachsafetyLoops}{Error}{}{Cputime}{Max}{901.106487171}%
\StoreBenchExecResult{SvcompNineteenVeriabs}{SvCompPropReachsafetyReachsafetyLoops}{Error}{}{Cputime}{Stdev}{255.5050342879982237744155565}%
\StoreBenchExecResult{SvcompNineteenVeriabs}{SvCompPropReachsafetyReachsafetyLoops}{Error}{}{Walltime}{}{15997.1992347247}%
\StoreBenchExecResult{SvcompNineteenVeriabs}{SvCompPropReachsafetyReachsafetyLoops}{Error}{}{Walltime}{Avg}{799.859961736235}%
\StoreBenchExecResult{SvcompNineteenVeriabs}{SvCompPropReachsafetyReachsafetyLoops}{Error}{}{Walltime}{Median}{885.232069373}%
\StoreBenchExecResult{SvcompNineteenVeriabs}{SvCompPropReachsafetyReachsafetyLoops}{Error}{}{Walltime}{Min}{41.7469789982}%
\StoreBenchExecResult{SvcompNineteenVeriabs}{SvCompPropReachsafetyReachsafetyLoops}{Error}{}{Walltime}{Max}{891.379369974}%
\StoreBenchExecResult{SvcompNineteenVeriabs}{SvCompPropReachsafetyReachsafetyLoops}{Error}{}{Walltime}{Stdev}{252.7398287538653134113354669}%
\StoreBenchExecResult{SvcompNineteenVeriabs}{SvCompPropReachsafetyReachsafetyLoops}{Error}{OutOfMemory}{Count}{}{2}%
\StoreBenchExecResult{SvcompNineteenVeriabs}{SvCompPropReachsafetyReachsafetyLoops}{Error}{OutOfMemory}{Cputime}{}{98.316226320}%
\StoreBenchExecResult{SvcompNineteenVeriabs}{SvCompPropReachsafetyReachsafetyLoops}{Error}{OutOfMemory}{Cputime}{Avg}{49.158113160}%
\StoreBenchExecResult{SvcompNineteenVeriabs}{SvCompPropReachsafetyReachsafetyLoops}{Error}{OutOfMemory}{Cputime}{Median}{49.158113160}%
\StoreBenchExecResult{SvcompNineteenVeriabs}{SvCompPropReachsafetyReachsafetyLoops}{Error}{OutOfMemory}{Cputime}{Min}{48.937238342}%
\StoreBenchExecResult{SvcompNineteenVeriabs}{SvCompPropReachsafetyReachsafetyLoops}{Error}{OutOfMemory}{Cputime}{Max}{49.378987978}%
\StoreBenchExecResult{SvcompNineteenVeriabs}{SvCompPropReachsafetyReachsafetyLoops}{Error}{OutOfMemory}{Cputime}{Stdev}{0.220874818}%
\StoreBenchExecResult{SvcompNineteenVeriabs}{SvCompPropReachsafetyReachsafetyLoops}{Error}{OutOfMemory}{Walltime}{}{83.5150198937}%
\StoreBenchExecResult{SvcompNineteenVeriabs}{SvCompPropReachsafetyReachsafetyLoops}{Error}{OutOfMemory}{Walltime}{Avg}{41.75750994685}%
\StoreBenchExecResult{SvcompNineteenVeriabs}{SvCompPropReachsafetyReachsafetyLoops}{Error}{OutOfMemory}{Walltime}{Median}{41.75750994685}%
\StoreBenchExecResult{SvcompNineteenVeriabs}{SvCompPropReachsafetyReachsafetyLoops}{Error}{OutOfMemory}{Walltime}{Min}{41.7469789982}%
\StoreBenchExecResult{SvcompNineteenVeriabs}{SvCompPropReachsafetyReachsafetyLoops}{Error}{OutOfMemory}{Walltime}{Max}{41.7680408955}%
\StoreBenchExecResult{SvcompNineteenVeriabs}{SvCompPropReachsafetyReachsafetyLoops}{Error}{OutOfMemory}{Walltime}{Stdev}{0.01053094865}%
\StoreBenchExecResult{SvcompNineteenVeriabs}{SvCompPropReachsafetyReachsafetyLoops}{Error}{Timeout}{Count}{}{18}%
\StoreBenchExecResult{SvcompNineteenVeriabs}{SvCompPropReachsafetyReachsafetyLoops}{Error}{Timeout}{Cputime}{}{16215.146190925}%
\StoreBenchExecResult{SvcompNineteenVeriabs}{SvCompPropReachsafetyReachsafetyLoops}{Error}{Timeout}{Cputime}{Avg}{900.8414550513888888888888889}%
\StoreBenchExecResult{SvcompNineteenVeriabs}{SvCompPropReachsafetyReachsafetyLoops}{Error}{Timeout}{Cputime}{Median}{900.84733321}%
\StoreBenchExecResult{SvcompNineteenVeriabs}{SvCompPropReachsafetyReachsafetyLoops}{Error}{Timeout}{Cputime}{Min}{900.710296457}%
\StoreBenchExecResult{SvcompNineteenVeriabs}{SvCompPropReachsafetyReachsafetyLoops}{Error}{Timeout}{Cputime}{Max}{901.106487171}%
\StoreBenchExecResult{SvcompNineteenVeriabs}{SvCompPropReachsafetyReachsafetyLoops}{Error}{Timeout}{Cputime}{Stdev}{0.1122050059683391023123569098}%
\StoreBenchExecResult{SvcompNineteenVeriabs}{SvCompPropReachsafetyReachsafetyLoops}{Error}{Timeout}{Walltime}{}{15913.684214831}%
\StoreBenchExecResult{SvcompNineteenVeriabs}{SvCompPropReachsafetyReachsafetyLoops}{Error}{Timeout}{Walltime}{Avg}{884.0935674906111111111111111}%
\StoreBenchExecResult{SvcompNineteenVeriabs}{SvCompPropReachsafetyReachsafetyLoops}{Error}{Timeout}{Walltime}{Median}{885.5380539895}%
\StoreBenchExecResult{SvcompNineteenVeriabs}{SvCompPropReachsafetyReachsafetyLoops}{Error}{Timeout}{Walltime}{Min}{873.657593966}%
\StoreBenchExecResult{SvcompNineteenVeriabs}{SvCompPropReachsafetyReachsafetyLoops}{Error}{Timeout}{Walltime}{Max}{891.379369974}%
\StoreBenchExecResult{SvcompNineteenVeriabs}{SvCompPropReachsafetyReachsafetyLoops}{Error}{Timeout}{Walltime}{Stdev}{4.680691068409216015989369537}%
\StoreBenchExecResult{SvcompNineteenVeriabs}{SvCompPropReachsafetyReachsafetyLoops}{Unknown}{}{Count}{}{6}%
\StoreBenchExecResult{SvcompNineteenVeriabs}{SvCompPropReachsafetyReachsafetyLoops}{Unknown}{}{Cputime}{}{3776.606010788}%
\StoreBenchExecResult{SvcompNineteenVeriabs}{SvCompPropReachsafetyReachsafetyLoops}{Unknown}{}{Cputime}{Avg}{629.4343351313333333333333333}%
\StoreBenchExecResult{SvcompNineteenVeriabs}{SvCompPropReachsafetyReachsafetyLoops}{Unknown}{}{Cputime}{Median}{756.6552243795}%
\StoreBenchExecResult{SvcompNineteenVeriabs}{SvCompPropReachsafetyReachsafetyLoops}{Unknown}{}{Cputime}{Min}{41.049367753}%
\StoreBenchExecResult{SvcompNineteenVeriabs}{SvCompPropReachsafetyReachsafetyLoops}{Unknown}{}{Cputime}{Max}{844.728390616}%
\StoreBenchExecResult{SvcompNineteenVeriabs}{SvCompPropReachsafetyReachsafetyLoops}{Unknown}{}{Cputime}{Stdev}{274.2564484447750536235803905}%
\StoreBenchExecResult{SvcompNineteenVeriabs}{SvCompPropReachsafetyReachsafetyLoops}{Unknown}{}{Walltime}{}{3675.4523560994}%
\StoreBenchExecResult{SvcompNineteenVeriabs}{SvCompPropReachsafetyReachsafetyLoops}{Unknown}{}{Walltime}{Avg}{612.5753926832333333333333333}%
\StoreBenchExecResult{SvcompNineteenVeriabs}{SvCompPropReachsafetyReachsafetyLoops}{Unknown}{}{Walltime}{Median}{742.781126499}%
\StoreBenchExecResult{SvcompNineteenVeriabs}{SvCompPropReachsafetyReachsafetyLoops}{Unknown}{}{Walltime}{Min}{25.8356699944}%
\StoreBenchExecResult{SvcompNineteenVeriabs}{SvCompPropReachsafetyReachsafetyLoops}{Unknown}{}{Walltime}{Max}{829.863758087}%
\StoreBenchExecResult{SvcompNineteenVeriabs}{SvCompPropReachsafetyReachsafetyLoops}{Unknown}{}{Walltime}{Stdev}{273.9896643319813464280608613}%
\StoreBenchExecResult{SvcompNineteenVeriabs}{SvCompPropReachsafetyReachsafetyLoops}{Unknown}{Unknown}{Count}{}{6}%
\StoreBenchExecResult{SvcompNineteenVeriabs}{SvCompPropReachsafetyReachsafetyLoops}{Unknown}{Unknown}{Cputime}{}{3776.606010788}%
\StoreBenchExecResult{SvcompNineteenVeriabs}{SvCompPropReachsafetyReachsafetyLoops}{Unknown}{Unknown}{Cputime}{Avg}{629.4343351313333333333333333}%
\StoreBenchExecResult{SvcompNineteenVeriabs}{SvCompPropReachsafetyReachsafetyLoops}{Unknown}{Unknown}{Cputime}{Median}{756.6552243795}%
\StoreBenchExecResult{SvcompNineteenVeriabs}{SvCompPropReachsafetyReachsafetyLoops}{Unknown}{Unknown}{Cputime}{Min}{41.049367753}%
\StoreBenchExecResult{SvcompNineteenVeriabs}{SvCompPropReachsafetyReachsafetyLoops}{Unknown}{Unknown}{Cputime}{Max}{844.728390616}%
\StoreBenchExecResult{SvcompNineteenVeriabs}{SvCompPropReachsafetyReachsafetyLoops}{Unknown}{Unknown}{Cputime}{Stdev}{274.2564484447750536235803905}%
\StoreBenchExecResult{SvcompNineteenVeriabs}{SvCompPropReachsafetyReachsafetyLoops}{Unknown}{Unknown}{Walltime}{}{3675.4523560994}%
\StoreBenchExecResult{SvcompNineteenVeriabs}{SvCompPropReachsafetyReachsafetyLoops}{Unknown}{Unknown}{Walltime}{Avg}{612.5753926832333333333333333}%
\StoreBenchExecResult{SvcompNineteenVeriabs}{SvCompPropReachsafetyReachsafetyLoops}{Unknown}{Unknown}{Walltime}{Median}{742.781126499}%
\StoreBenchExecResult{SvcompNineteenVeriabs}{SvCompPropReachsafetyReachsafetyLoops}{Unknown}{Unknown}{Walltime}{Min}{25.8356699944}%
\StoreBenchExecResult{SvcompNineteenVeriabs}{SvCompPropReachsafetyReachsafetyLoops}{Unknown}{Unknown}{Walltime}{Max}{829.863758087}%
\StoreBenchExecResult{SvcompNineteenVeriabs}{SvCompPropReachsafetyReachsafetyLoops}{Unknown}{Unknown}{Walltime}{Stdev}{273.9896643319813464280608613}%
\providecommand\StoreBenchExecResult[7]{\expandafter\newcommand\csname#1#2#3#4#5#6\endcsname{#7}}%
\StoreBenchExecResult{SvcompNineteenViap}{SvCompPropReachsafetyReachsafetyLoops}{Total}{}{Count}{}{208}%
\StoreBenchExecResult{SvcompNineteenViap}{SvCompPropReachsafetyReachsafetyLoops}{Total}{}{Cputime}{}{17565.330193665}%
\StoreBenchExecResult{SvcompNineteenViap}{SvCompPropReachsafetyReachsafetyLoops}{Total}{}{Cputime}{Avg}{84.44870285415865384615384615}%
\StoreBenchExecResult{SvcompNineteenViap}{SvCompPropReachsafetyReachsafetyLoops}{Total}{}{Cputime}{Median}{7.877640262}%
\StoreBenchExecResult{SvcompNineteenViap}{SvCompPropReachsafetyReachsafetyLoops}{Total}{}{Cputime}{Min}{2.386299718}%
\StoreBenchExecResult{SvcompNineteenViap}{SvCompPropReachsafetyReachsafetyLoops}{Total}{}{Cputime}{Max}{900.621247091}%
\StoreBenchExecResult{SvcompNineteenViap}{SvCompPropReachsafetyReachsafetyLoops}{Total}{}{Cputime}{Stdev}{187.6433561431482581498199252}%
\StoreBenchExecResult{SvcompNineteenViap}{SvCompPropReachsafetyReachsafetyLoops}{Total}{}{Walltime}{}{17444.46097731728}%
\StoreBenchExecResult{SvcompNineteenViap}{SvCompPropReachsafetyReachsafetyLoops}{Total}{}{Walltime}{Avg}{83.86760085248692307692307692}%
\StoreBenchExecResult{SvcompNineteenViap}{SvCompPropReachsafetyReachsafetyLoops}{Total}{}{Walltime}{Median}{7.951982975005}%
\StoreBenchExecResult{SvcompNineteenViap}{SvCompPropReachsafetyReachsafetyLoops}{Total}{}{Walltime}{Min}{2.38919401169}%
\StoreBenchExecResult{SvcompNineteenViap}{SvCompPropReachsafetyReachsafetyLoops}{Total}{}{Walltime}{Max}{900.750031948}%
\StoreBenchExecResult{SvcompNineteenViap}{SvCompPropReachsafetyReachsafetyLoops}{Total}{}{Walltime}{Stdev}{187.6314971414078894217567056}%
\StoreBenchExecResult{SvcompNineteenViap}{SvCompPropReachsafetyReachsafetyLoops}{Correct}{}{Count}{}{124}%
\StoreBenchExecResult{SvcompNineteenViap}{SvCompPropReachsafetyReachsafetyLoops}{Correct}{}{Cputime}{}{4484.338328725}%
\StoreBenchExecResult{SvcompNineteenViap}{SvCompPropReachsafetyReachsafetyLoops}{Correct}{}{Cputime}{Avg}{36.16401878004032258064516129}%
\StoreBenchExecResult{SvcompNineteenViap}{SvCompPropReachsafetyReachsafetyLoops}{Correct}{}{Cputime}{Median}{8.025255248}%
\StoreBenchExecResult{SvcompNineteenViap}{SvCompPropReachsafetyReachsafetyLoops}{Correct}{}{Cputime}{Min}{4.35589619}%
\StoreBenchExecResult{SvcompNineteenViap}{SvCompPropReachsafetyReachsafetyLoops}{Correct}{}{Cputime}{Max}{306.758221271}%
\StoreBenchExecResult{SvcompNineteenViap}{SvCompPropReachsafetyReachsafetyLoops}{Correct}{}{Cputime}{Stdev}{60.91859398277514749025338838}%
\StoreBenchExecResult{SvcompNineteenViap}{SvCompPropReachsafetyReachsafetyLoops}{Correct}{}{Walltime}{}{4496.95529770902}%
\StoreBenchExecResult{SvcompNineteenViap}{SvCompPropReachsafetyReachsafetyLoops}{Correct}{}{Walltime}{Avg}{36.26576852991145161290322581}%
\StoreBenchExecResult{SvcompNineteenViap}{SvCompPropReachsafetyReachsafetyLoops}{Correct}{}{Walltime}{Median}{8.097622513775}%
\StoreBenchExecResult{SvcompNineteenViap}{SvCompPropReachsafetyReachsafetyLoops}{Correct}{}{Walltime}{Min}{4.4462211132}%
\StoreBenchExecResult{SvcompNineteenViap}{SvCompPropReachsafetyReachsafetyLoops}{Correct}{}{Walltime}{Max}{306.862892866}%
\StoreBenchExecResult{SvcompNineteenViap}{SvCompPropReachsafetyReachsafetyLoops}{Correct}{}{Walltime}{Stdev}{60.95560507632146869506727195}%
\StoreBenchExecResult{SvcompNineteenViap}{SvCompPropReachsafetyReachsafetyLoops}{Correct}{False}{Count}{}{36}%
\StoreBenchExecResult{SvcompNineteenViap}{SvCompPropReachsafetyReachsafetyLoops}{Correct}{False}{Cputime}{}{377.467215058}%
\StoreBenchExecResult{SvcompNineteenViap}{SvCompPropReachsafetyReachsafetyLoops}{Correct}{False}{Cputime}{Avg}{10.48520041827777777777777778}%
\StoreBenchExecResult{SvcompNineteenViap}{SvCompPropReachsafetyReachsafetyLoops}{Correct}{False}{Cputime}{Median}{5.6777372835}%
\StoreBenchExecResult{SvcompNineteenViap}{SvCompPropReachsafetyReachsafetyLoops}{Correct}{False}{Cputime}{Min}{4.671753868}%
\StoreBenchExecResult{SvcompNineteenViap}{SvCompPropReachsafetyReachsafetyLoops}{Correct}{False}{Cputime}{Max}{63.628251052}%
\StoreBenchExecResult{SvcompNineteenViap}{SvCompPropReachsafetyReachsafetyLoops}{Correct}{False}{Cputime}{Stdev}{13.92716901612960271010899951}%
\StoreBenchExecResult{SvcompNineteenViap}{SvCompPropReachsafetyReachsafetyLoops}{Correct}{False}{Walltime}{}{376.29400277140}%
\StoreBenchExecResult{SvcompNineteenViap}{SvCompPropReachsafetyReachsafetyLoops}{Correct}{False}{Walltime}{Avg}{10.45261118809444444444444444}%
\StoreBenchExecResult{SvcompNineteenViap}{SvCompPropReachsafetyReachsafetyLoops}{Correct}{False}{Walltime}{Median}{5.67969954014}%
\StoreBenchExecResult{SvcompNineteenViap}{SvCompPropReachsafetyReachsafetyLoops}{Correct}{False}{Walltime}{Min}{4.67577815056}%
\StoreBenchExecResult{SvcompNineteenViap}{SvCompPropReachsafetyReachsafetyLoops}{Correct}{False}{Walltime}{Max}{62.8327150345}%
\StoreBenchExecResult{SvcompNineteenViap}{SvCompPropReachsafetyReachsafetyLoops}{Correct}{False}{Walltime}{Stdev}{13.75159650335940776212423787}%
\StoreBenchExecResult{SvcompNineteenViap}{SvCompPropReachsafetyReachsafetyLoops}{Wrong}{False}{Count}{}{0}%
\StoreBenchExecResult{SvcompNineteenViap}{SvCompPropReachsafetyReachsafetyLoops}{Wrong}{False}{Cputime}{}{0}%
\StoreBenchExecResult{SvcompNineteenViap}{SvCompPropReachsafetyReachsafetyLoops}{Wrong}{False}{Cputime}{Avg}{None}%
\StoreBenchExecResult{SvcompNineteenViap}{SvCompPropReachsafetyReachsafetyLoops}{Wrong}{False}{Cputime}{Median}{None}%
\StoreBenchExecResult{SvcompNineteenViap}{SvCompPropReachsafetyReachsafetyLoops}{Wrong}{False}{Cputime}{Min}{None}%
\StoreBenchExecResult{SvcompNineteenViap}{SvCompPropReachsafetyReachsafetyLoops}{Wrong}{False}{Cputime}{Max}{None}%
\StoreBenchExecResult{SvcompNineteenViap}{SvCompPropReachsafetyReachsafetyLoops}{Wrong}{False}{Cputime}{Stdev}{None}%
\StoreBenchExecResult{SvcompNineteenViap}{SvCompPropReachsafetyReachsafetyLoops}{Wrong}{False}{Walltime}{}{0}%
\StoreBenchExecResult{SvcompNineteenViap}{SvCompPropReachsafetyReachsafetyLoops}{Wrong}{False}{Walltime}{Avg}{None}%
\StoreBenchExecResult{SvcompNineteenViap}{SvCompPropReachsafetyReachsafetyLoops}{Wrong}{False}{Walltime}{Median}{None}%
\StoreBenchExecResult{SvcompNineteenViap}{SvCompPropReachsafetyReachsafetyLoops}{Wrong}{False}{Walltime}{Min}{None}%
\StoreBenchExecResult{SvcompNineteenViap}{SvCompPropReachsafetyReachsafetyLoops}{Wrong}{False}{Walltime}{Max}{None}%
\StoreBenchExecResult{SvcompNineteenViap}{SvCompPropReachsafetyReachsafetyLoops}{Wrong}{False}{Walltime}{Stdev}{None}%
\StoreBenchExecResult{SvcompNineteenViap}{SvCompPropReachsafetyReachsafetyLoops}{Correct}{True}{Count}{}{88}%
\StoreBenchExecResult{SvcompNineteenViap}{SvCompPropReachsafetyReachsafetyLoops}{Correct}{True}{Cputime}{}{4106.871113667}%
\StoreBenchExecResult{SvcompNineteenViap}{SvCompPropReachsafetyReachsafetyLoops}{Correct}{True}{Cputime}{Avg}{46.66898992803409090909090909}%
\StoreBenchExecResult{SvcompNineteenViap}{SvCompPropReachsafetyReachsafetyLoops}{Correct}{True}{Cputime}{Median}{10.901746776}%
\StoreBenchExecResult{SvcompNineteenViap}{SvCompPropReachsafetyReachsafetyLoops}{Correct}{True}{Cputime}{Min}{4.35589619}%
\StoreBenchExecResult{SvcompNineteenViap}{SvCompPropReachsafetyReachsafetyLoops}{Correct}{True}{Cputime}{Max}{306.758221271}%
\StoreBenchExecResult{SvcompNineteenViap}{SvCompPropReachsafetyReachsafetyLoops}{Correct}{True}{Cputime}{Stdev}{69.06361393033315257071772480}%
\StoreBenchExecResult{SvcompNineteenViap}{SvCompPropReachsafetyReachsafetyLoops}{Correct}{True}{Walltime}{}{4120.66129493762}%
\StoreBenchExecResult{SvcompNineteenViap}{SvCompPropReachsafetyReachsafetyLoops}{Correct}{True}{Walltime}{Avg}{46.82569653338204545454545455}%
\StoreBenchExecResult{SvcompNineteenViap}{SvCompPropReachsafetyReachsafetyLoops}{Correct}{True}{Walltime}{Median}{11.02876102925}%
\StoreBenchExecResult{SvcompNineteenViap}{SvCompPropReachsafetyReachsafetyLoops}{Correct}{True}{Walltime}{Min}{4.4462211132}%
\StoreBenchExecResult{SvcompNineteenViap}{SvCompPropReachsafetyReachsafetyLoops}{Correct}{True}{Walltime}{Max}{306.862892866}%
\StoreBenchExecResult{SvcompNineteenViap}{SvCompPropReachsafetyReachsafetyLoops}{Correct}{True}{Walltime}{Stdev}{69.09514660763953017024855749}%
\StoreBenchExecResult{SvcompNineteenViap}{SvCompPropReachsafetyReachsafetyLoops}{Wrong}{True}{Count}{}{0}%
\StoreBenchExecResult{SvcompNineteenViap}{SvCompPropReachsafetyReachsafetyLoops}{Wrong}{True}{Cputime}{}{0}%
\StoreBenchExecResult{SvcompNineteenViap}{SvCompPropReachsafetyReachsafetyLoops}{Wrong}{True}{Cputime}{Avg}{None}%
\StoreBenchExecResult{SvcompNineteenViap}{SvCompPropReachsafetyReachsafetyLoops}{Wrong}{True}{Cputime}{Median}{None}%
\StoreBenchExecResult{SvcompNineteenViap}{SvCompPropReachsafetyReachsafetyLoops}{Wrong}{True}{Cputime}{Min}{None}%
\StoreBenchExecResult{SvcompNineteenViap}{SvCompPropReachsafetyReachsafetyLoops}{Wrong}{True}{Cputime}{Max}{None}%
\StoreBenchExecResult{SvcompNineteenViap}{SvCompPropReachsafetyReachsafetyLoops}{Wrong}{True}{Cputime}{Stdev}{None}%
\StoreBenchExecResult{SvcompNineteenViap}{SvCompPropReachsafetyReachsafetyLoops}{Wrong}{True}{Walltime}{}{0}%
\StoreBenchExecResult{SvcompNineteenViap}{SvCompPropReachsafetyReachsafetyLoops}{Wrong}{True}{Walltime}{Avg}{None}%
\StoreBenchExecResult{SvcompNineteenViap}{SvCompPropReachsafetyReachsafetyLoops}{Wrong}{True}{Walltime}{Median}{None}%
\StoreBenchExecResult{SvcompNineteenViap}{SvCompPropReachsafetyReachsafetyLoops}{Wrong}{True}{Walltime}{Min}{None}%
\StoreBenchExecResult{SvcompNineteenViap}{SvCompPropReachsafetyReachsafetyLoops}{Wrong}{True}{Walltime}{Max}{None}%
\StoreBenchExecResult{SvcompNineteenViap}{SvCompPropReachsafetyReachsafetyLoops}{Wrong}{True}{Walltime}{Stdev}{None}%
\StoreBenchExecResult{SvcompNineteenViap}{SvCompPropReachsafetyReachsafetyLoops}{Error}{}{Count}{}{19}%
\StoreBenchExecResult{SvcompNineteenViap}{SvCompPropReachsafetyReachsafetyLoops}{Error}{}{Cputime}{}{8791.830223089}%
\StoreBenchExecResult{SvcompNineteenViap}{SvCompPropReachsafetyReachsafetyLoops}{Error}{}{Cputime}{Avg}{462.7279064783684210526315789}%
\StoreBenchExecResult{SvcompNineteenViap}{SvCompPropReachsafetyReachsafetyLoops}{Error}{}{Cputime}{Median}{589.672547593}%
\StoreBenchExecResult{SvcompNineteenViap}{SvCompPropReachsafetyReachsafetyLoops}{Error}{}{Cputime}{Min}{4.088264796}%
\StoreBenchExecResult{SvcompNineteenViap}{SvCompPropReachsafetyReachsafetyLoops}{Error}{}{Cputime}{Max}{900.621247091}%
\StoreBenchExecResult{SvcompNineteenViap}{SvCompPropReachsafetyReachsafetyLoops}{Error}{}{Cputime}{Stdev}{394.4918968866540613785802822}%
\StoreBenchExecResult{SvcompNineteenViap}{SvCompPropReachsafetyReachsafetyLoops}{Error}{}{Walltime}{}{8733.41831970238}%
\StoreBenchExecResult{SvcompNineteenViap}{SvCompPropReachsafetyReachsafetyLoops}{Error}{}{Walltime}{Avg}{459.6535957738094736842105263}%
\StoreBenchExecResult{SvcompNineteenViap}{SvCompPropReachsafetyReachsafetyLoops}{Error}{}{Walltime}{Median}{589.688944101}%
\StoreBenchExecResult{SvcompNineteenViap}{SvCompPropReachsafetyReachsafetyLoops}{Error}{}{Walltime}{Min}{4.17580294609}%
\StoreBenchExecResult{SvcompNineteenViap}{SvCompPropReachsafetyReachsafetyLoops}{Error}{}{Walltime}{Max}{900.750031948}%
\StoreBenchExecResult{SvcompNineteenViap}{SvCompPropReachsafetyReachsafetyLoops}{Error}{}{Walltime}{Stdev}{397.4516384474170399156664024}%
\StoreBenchExecResult{SvcompNineteenViap}{SvCompPropReachsafetyReachsafetyLoops}{Error}{Error}{Count}{}{4}%
\StoreBenchExecResult{SvcompNineteenViap}{SvCompPropReachsafetyReachsafetyLoops}{Error}{Error}{Cputime}{}{172.735089248}%
\StoreBenchExecResult{SvcompNineteenViap}{SvCompPropReachsafetyReachsafetyLoops}{Error}{Error}{Cputime}{Avg}{43.183772312}%
\StoreBenchExecResult{SvcompNineteenViap}{SvCompPropReachsafetyReachsafetyLoops}{Error}{Error}{Cputime}{Median}{4.699564602}%
\StoreBenchExecResult{SvcompNineteenViap}{SvCompPropReachsafetyReachsafetyLoops}{Error}{Error}{Cputime}{Min}{4.088264796}%
\StoreBenchExecResult{SvcompNineteenViap}{SvCompPropReachsafetyReachsafetyLoops}{Error}{Error}{Cputime}{Max}{159.247695248}%
\StoreBenchExecResult{SvcompNineteenViap}{SvCompPropReachsafetyReachsafetyLoops}{Error}{Error}{Cputime}{Stdev}{67.01006164120438948037542386}%
\StoreBenchExecResult{SvcompNineteenViap}{SvCompPropReachsafetyReachsafetyLoops}{Error}{Error}{Walltime}{}{172.90521192513}%
\StoreBenchExecResult{SvcompNineteenViap}{SvCompPropReachsafetyReachsafetyLoops}{Error}{Error}{Walltime}{Avg}{43.2263029812825}%
\StoreBenchExecResult{SvcompNineteenViap}{SvCompPropReachsafetyReachsafetyLoops}{Error}{Error}{Walltime}{Median}{4.73306894302}%
\StoreBenchExecResult{SvcompNineteenViap}{SvCompPropReachsafetyReachsafetyLoops}{Error}{Error}{Walltime}{Min}{4.17580294609}%
\StoreBenchExecResult{SvcompNineteenViap}{SvCompPropReachsafetyReachsafetyLoops}{Error}{Error}{Walltime}{Max}{159.263271093}%
\StoreBenchExecResult{SvcompNineteenViap}{SvCompPropReachsafetyReachsafetyLoops}{Error}{Error}{Walltime}{Stdev}{66.99442768834933612499074084}%
\StoreBenchExecResult{SvcompNineteenViap}{SvCompPropReachsafetyReachsafetyLoops}{Error}{OutOfMemory}{Count}{}{8}%
\StoreBenchExecResult{SvcompNineteenViap}{SvCompPropReachsafetyReachsafetyLoops}{Error}{OutOfMemory}{Cputime}{}{3209.007488301}%
\StoreBenchExecResult{SvcompNineteenViap}{SvCompPropReachsafetyReachsafetyLoops}{Error}{OutOfMemory}{Cputime}{Avg}{401.125936037625}%
\StoreBenchExecResult{SvcompNineteenViap}{SvCompPropReachsafetyReachsafetyLoops}{Error}{OutOfMemory}{Cputime}{Median}{343.2834904885}%
\StoreBenchExecResult{SvcompNineteenViap}{SvCompPropReachsafetyReachsafetyLoops}{Error}{OutOfMemory}{Cputime}{Min}{39.452537188}%
\StoreBenchExecResult{SvcompNineteenViap}{SvCompPropReachsafetyReachsafetyLoops}{Error}{OutOfMemory}{Cputime}{Max}{819.818409719}%
\StoreBenchExecResult{SvcompNineteenViap}{SvCompPropReachsafetyReachsafetyLoops}{Error}{OutOfMemory}{Cputime}{Stdev}{325.1807575091416383302407044}%
\StoreBenchExecResult{SvcompNineteenViap}{SvCompPropReachsafetyReachsafetyLoops}{Error}{OutOfMemory}{Walltime}{}{3150.6463730341}%
\StoreBenchExecResult{SvcompNineteenViap}{SvCompPropReachsafetyReachsafetyLoops}{Error}{OutOfMemory}{Walltime}{Avg}{393.8307966292625}%
\StoreBenchExecResult{SvcompNineteenViap}{SvCompPropReachsafetyReachsafetyLoops}{Error}{OutOfMemory}{Walltime}{Median}{335.5094325541}%
\StoreBenchExecResult{SvcompNineteenViap}{SvCompPropReachsafetyReachsafetyLoops}{Error}{OutOfMemory}{Walltime}{Min}{27.718611002}%
\StoreBenchExecResult{SvcompNineteenViap}{SvCompPropReachsafetyReachsafetyLoops}{Error}{OutOfMemory}{Walltime}{Max}{819.836544991}%
\StoreBenchExecResult{SvcompNineteenViap}{SvCompPropReachsafetyReachsafetyLoops}{Error}{OutOfMemory}{Walltime}{Stdev}{332.3020600618857460342878505}%
\StoreBenchExecResult{SvcompNineteenViap}{SvCompPropReachsafetyReachsafetyLoops}{Error}{Timeout}{Count}{}{6}%
\StoreBenchExecResult{SvcompNineteenViap}{SvCompPropReachsafetyReachsafetyLoops}{Error}{Timeout}{Cputime}{}{5403.587217035}%
\StoreBenchExecResult{SvcompNineteenViap}{SvCompPropReachsafetyReachsafetyLoops}{Error}{Timeout}{Cputime}{Avg}{900.5978695058333333333333333}%
\StoreBenchExecResult{SvcompNineteenViap}{SvCompPropReachsafetyReachsafetyLoops}{Error}{Timeout}{Cputime}{Median}{900.593506608}%
\StoreBenchExecResult{SvcompNineteenViap}{SvCompPropReachsafetyReachsafetyLoops}{Error}{Timeout}{Cputime}{Min}{900.592007709}%
\StoreBenchExecResult{SvcompNineteenViap}{SvCompPropReachsafetyReachsafetyLoops}{Error}{Timeout}{Cputime}{Max}{900.621247091}%
\StoreBenchExecResult{SvcompNineteenViap}{SvCompPropReachsafetyReachsafetyLoops}{Error}{Timeout}{Cputime}{Stdev}{0.01047158020871077217233227785}%
\StoreBenchExecResult{SvcompNineteenViap}{SvCompPropReachsafetyReachsafetyLoops}{Error}{Timeout}{Walltime}{}{5403.358414650}%
\StoreBenchExecResult{SvcompNineteenViap}{SvCompPropReachsafetyReachsafetyLoops}{Error}{Timeout}{Walltime}{Avg}{900.559735775}%
\StoreBenchExecResult{SvcompNineteenViap}{SvCompPropReachsafetyReachsafetyLoops}{Error}{Timeout}{Walltime}{Median}{900.587189436}%
\StoreBenchExecResult{SvcompNineteenViap}{SvCompPropReachsafetyReachsafetyLoops}{Error}{Timeout}{Walltime}{Min}{900.255277872}%
\StoreBenchExecResult{SvcompNineteenViap}{SvCompPropReachsafetyReachsafetyLoops}{Error}{Timeout}{Walltime}{Max}{900.750031948}%
\StoreBenchExecResult{SvcompNineteenViap}{SvCompPropReachsafetyReachsafetyLoops}{Error}{Timeout}{Walltime}{Stdev}{0.1485519049568780813171233543}%
\StoreBenchExecResult{SvcompNineteenViap}{SvCompPropReachsafetyReachsafetyLoops}{Error}{WitnessInvalid}{Count}{}{1}%
\StoreBenchExecResult{SvcompNineteenViap}{SvCompPropReachsafetyReachsafetyLoops}{Error}{WitnessInvalid}{Cputime}{}{6.500428505}%
\StoreBenchExecResult{SvcompNineteenViap}{SvCompPropReachsafetyReachsafetyLoops}{Error}{WitnessInvalid}{Cputime}{Avg}{6.500428505}%
\StoreBenchExecResult{SvcompNineteenViap}{SvCompPropReachsafetyReachsafetyLoops}{Error}{WitnessInvalid}{Cputime}{Median}{6.500428505}%
\StoreBenchExecResult{SvcompNineteenViap}{SvCompPropReachsafetyReachsafetyLoops}{Error}{WitnessInvalid}{Cputime}{Min}{6.500428505}%
\StoreBenchExecResult{SvcompNineteenViap}{SvCompPropReachsafetyReachsafetyLoops}{Error}{WitnessInvalid}{Cputime}{Max}{6.500428505}%
\StoreBenchExecResult{SvcompNineteenViap}{SvCompPropReachsafetyReachsafetyLoops}{Error}{WitnessInvalid}{Cputime}{Stdev}{0E-9}%
\StoreBenchExecResult{SvcompNineteenViap}{SvCompPropReachsafetyReachsafetyLoops}{Error}{WitnessInvalid}{Walltime}{}{6.50832009315}%
\StoreBenchExecResult{SvcompNineteenViap}{SvCompPropReachsafetyReachsafetyLoops}{Error}{WitnessInvalid}{Walltime}{Avg}{6.50832009315}%
\StoreBenchExecResult{SvcompNineteenViap}{SvCompPropReachsafetyReachsafetyLoops}{Error}{WitnessInvalid}{Walltime}{Median}{6.50832009315}%
\StoreBenchExecResult{SvcompNineteenViap}{SvCompPropReachsafetyReachsafetyLoops}{Error}{WitnessInvalid}{Walltime}{Min}{6.50832009315}%
\StoreBenchExecResult{SvcompNineteenViap}{SvCompPropReachsafetyReachsafetyLoops}{Error}{WitnessInvalid}{Walltime}{Max}{6.50832009315}%
\StoreBenchExecResult{SvcompNineteenViap}{SvCompPropReachsafetyReachsafetyLoops}{Error}{WitnessInvalid}{Walltime}{Stdev}{0E-11}%
\StoreBenchExecResult{SvcompNineteenViap}{SvCompPropReachsafetyReachsafetyLoops}{Unknown}{}{Count}{}{65}%
\StoreBenchExecResult{SvcompNineteenViap}{SvCompPropReachsafetyReachsafetyLoops}{Unknown}{}{Cputime}{}{4289.161641851}%
\StoreBenchExecResult{SvcompNineteenViap}{SvCompPropReachsafetyReachsafetyLoops}{Unknown}{}{Cputime}{Avg}{65.98710218232307692307692308}%
\StoreBenchExecResult{SvcompNineteenViap}{SvCompPropReachsafetyReachsafetyLoops}{Unknown}{}{Cputime}{Median}{2.581800723}%
\StoreBenchExecResult{SvcompNineteenViap}{SvCompPropReachsafetyReachsafetyLoops}{Unknown}{}{Cputime}{Min}{2.386299718}%
\StoreBenchExecResult{SvcompNineteenViap}{SvCompPropReachsafetyReachsafetyLoops}{Unknown}{}{Cputime}{Max}{503.598877367}%
\StoreBenchExecResult{SvcompNineteenViap}{SvCompPropReachsafetyReachsafetyLoops}{Unknown}{}{Cputime}{Stdev}{116.1300715395828834269637117}%
\StoreBenchExecResult{SvcompNineteenViap}{SvCompPropReachsafetyReachsafetyLoops}{Unknown}{}{Walltime}{}{4214.08735990588}%
\StoreBenchExecResult{SvcompNineteenViap}{SvCompPropReachsafetyReachsafetyLoops}{Unknown}{}{Walltime}{Avg}{64.83211322932123076923076923}%
\StoreBenchExecResult{SvcompNineteenViap}{SvCompPropReachsafetyReachsafetyLoops}{Unknown}{}{Walltime}{Median}{2.582903862}%
\StoreBenchExecResult{SvcompNineteenViap}{SvCompPropReachsafetyReachsafetyLoops}{Unknown}{}{Walltime}{Min}{2.38919401169}%
\StoreBenchExecResult{SvcompNineteenViap}{SvCompPropReachsafetyReachsafetyLoops}{Unknown}{}{Walltime}{Max}{503.665392876}%
\StoreBenchExecResult{SvcompNineteenViap}{SvCompPropReachsafetyReachsafetyLoops}{Unknown}{}{Walltime}{Stdev}{115.8927667471787265596491462}%
\StoreBenchExecResult{SvcompNineteenViap}{SvCompPropReachsafetyReachsafetyLoops}{Unknown}{Unknown}{Count}{}{65}%
\StoreBenchExecResult{SvcompNineteenViap}{SvCompPropReachsafetyReachsafetyLoops}{Unknown}{Unknown}{Cputime}{}{4289.161641851}%
\StoreBenchExecResult{SvcompNineteenViap}{SvCompPropReachsafetyReachsafetyLoops}{Unknown}{Unknown}{Cputime}{Avg}{65.98710218232307692307692308}%
\StoreBenchExecResult{SvcompNineteenViap}{SvCompPropReachsafetyReachsafetyLoops}{Unknown}{Unknown}{Cputime}{Median}{2.581800723}%
\StoreBenchExecResult{SvcompNineteenViap}{SvCompPropReachsafetyReachsafetyLoops}{Unknown}{Unknown}{Cputime}{Min}{2.386299718}%
\StoreBenchExecResult{SvcompNineteenViap}{SvCompPropReachsafetyReachsafetyLoops}{Unknown}{Unknown}{Cputime}{Max}{503.598877367}%
\StoreBenchExecResult{SvcompNineteenViap}{SvCompPropReachsafetyReachsafetyLoops}{Unknown}{Unknown}{Cputime}{Stdev}{116.1300715395828834269637117}%
\StoreBenchExecResult{SvcompNineteenViap}{SvCompPropReachsafetyReachsafetyLoops}{Unknown}{Unknown}{Walltime}{}{4214.08735990588}%
\StoreBenchExecResult{SvcompNineteenViap}{SvCompPropReachsafetyReachsafetyLoops}{Unknown}{Unknown}{Walltime}{Avg}{64.83211322932123076923076923}%
\StoreBenchExecResult{SvcompNineteenViap}{SvCompPropReachsafetyReachsafetyLoops}{Unknown}{Unknown}{Walltime}{Median}{2.582903862}%
\StoreBenchExecResult{SvcompNineteenViap}{SvCompPropReachsafetyReachsafetyLoops}{Unknown}{Unknown}{Walltime}{Min}{2.38919401169}%
\StoreBenchExecResult{SvcompNineteenViap}{SvCompPropReachsafetyReachsafetyLoops}{Unknown}{Unknown}{Walltime}{Max}{503.665392876}%
\StoreBenchExecResult{SvcompNineteenViap}{SvCompPropReachsafetyReachsafetyLoops}{Unknown}{Unknown}{Walltime}{Stdev}{115.8927667471787265596491462}%
\ifdefined\SvcompNineteenPdrInvKinductionPlainReachsafetyLoopsTotalCount\else\edef\SvcompNineteenPdrInvKinductionPlainReachsafetyLoopsTotalCount{0}\fi
\ifdefined\SvcompNineteenPdrInvKinductionPlainReachsafetyLoopsCorrectCount\else\edef\SvcompNineteenPdrInvKinductionPlainReachsafetyLoopsCorrectCount{0}\fi
\ifdefined\SvcompNineteenPdrInvKinductionPlainReachsafetyLoopsCorrectTrueCount\else\edef\SvcompNineteenPdrInvKinductionPlainReachsafetyLoopsCorrectTrueCount{0}\fi
\ifdefined\SvcompNineteenPdrInvKinductionPlainReachsafetyLoopsCorrectFalseCount\else\edef\SvcompNineteenPdrInvKinductionPlainReachsafetyLoopsCorrectFalseCount{0}\fi
\ifdefined\SvcompNineteenPdrInvKinductionPlainReachsafetyLoopsWrongTrueCount\else\edef\SvcompNineteenPdrInvKinductionPlainReachsafetyLoopsWrongTrueCount{0}\fi
\ifdefined\SvcompNineteenPdrInvKinductionPlainReachsafetyLoopsWrongFalseCount\else\edef\SvcompNineteenPdrInvKinductionPlainReachsafetyLoopsWrongFalseCount{0}\fi
\ifdefined\SvcompNineteenPdrInvKinductionPlainReachsafetyLoopsErrorTimeoutCount\else\edef\SvcompNineteenPdrInvKinductionPlainReachsafetyLoopsErrorTimeoutCount{0}\fi
\ifdefined\SvcompNineteenPdrInvKinductionPlainReachsafetyLoopsErrorOutOfMemoryCount\else\edef\SvcompNineteenPdrInvKinductionPlainReachsafetyLoopsErrorOutOfMemoryCount{0}\fi
\ifdefined\SvcompNineteenPdrInvKinductionPlainReachsafetyLoopsCorrectCputime\else\edef\SvcompNineteenPdrInvKinductionPlainReachsafetyLoopsCorrectCputime{0}\fi
\ifdefined\SvcompNineteenPdrInvKinductionPlainReachsafetyLoopsCorrectCputimeAvg\else\edef\SvcompNineteenPdrInvKinductionPlainReachsafetyLoopsCorrectCputimeAvg{None}\fi
\ifdefined\SvcompNineteenPdrInvKinductionPlainReachsafetyLoopsCorrectWalltime\else\edef\SvcompNineteenPdrInvKinductionPlainReachsafetyLoopsCorrectWalltime{0}\fi
\ifdefined\SvcompNineteenPdrInvKinductionPlainReachsafetyLoopsCorrectWalltimeAvg\else\edef\SvcompNineteenPdrInvKinductionPlainReachsafetyLoopsCorrectWalltimeAvg{None}\fi
\ifdefined\SvcompNineteenPdrInvKinductionDfReachsafetyLoopsTotalCount\else\edef\SvcompNineteenPdrInvKinductionDfReachsafetyLoopsTotalCount{0}\fi
\ifdefined\SvcompNineteenPdrInvKinductionDfReachsafetyLoopsCorrectCount\else\edef\SvcompNineteenPdrInvKinductionDfReachsafetyLoopsCorrectCount{0}\fi
\ifdefined\SvcompNineteenPdrInvKinductionDfReachsafetyLoopsCorrectTrueCount\else\edef\SvcompNineteenPdrInvKinductionDfReachsafetyLoopsCorrectTrueCount{0}\fi
\ifdefined\SvcompNineteenPdrInvKinductionDfReachsafetyLoopsCorrectFalseCount\else\edef\SvcompNineteenPdrInvKinductionDfReachsafetyLoopsCorrectFalseCount{0}\fi
\ifdefined\SvcompNineteenPdrInvKinductionDfReachsafetyLoopsWrongTrueCount\else\edef\SvcompNineteenPdrInvKinductionDfReachsafetyLoopsWrongTrueCount{0}\fi
\ifdefined\SvcompNineteenPdrInvKinductionDfReachsafetyLoopsWrongFalseCount\else\edef\SvcompNineteenPdrInvKinductionDfReachsafetyLoopsWrongFalseCount{0}\fi
\ifdefined\SvcompNineteenPdrInvKinductionDfReachsafetyLoopsErrorTimeoutCount\else\edef\SvcompNineteenPdrInvKinductionDfReachsafetyLoopsErrorTimeoutCount{0}\fi
\ifdefined\SvcompNineteenPdrInvKinductionDfReachsafetyLoopsErrorOutOfMemoryCount\else\edef\SvcompNineteenPdrInvKinductionDfReachsafetyLoopsErrorOutOfMemoryCount{0}\fi
\ifdefined\SvcompNineteenPdrInvKinductionDfReachsafetyLoopsCorrectCputime\else\edef\SvcompNineteenPdrInvKinductionDfReachsafetyLoopsCorrectCputime{0}\fi
\ifdefined\SvcompNineteenPdrInvKinductionDfReachsafetyLoopsCorrectCputimeAvg\else\edef\SvcompNineteenPdrInvKinductionDfReachsafetyLoopsCorrectCputimeAvg{None}\fi
\ifdefined\SvcompNineteenPdrInvKinductionDfReachsafetyLoopsCorrectWalltime\else\edef\SvcompNineteenPdrInvKinductionDfReachsafetyLoopsCorrectWalltime{0}\fi
\ifdefined\SvcompNineteenPdrInvKinductionDfReachsafetyLoopsCorrectWalltimeAvg\else\edef\SvcompNineteenPdrInvKinductionDfReachsafetyLoopsCorrectWalltimeAvg{None}\fi
\ifdefined\SvcompNineteenPdrInvKinductionKipdrReachsafetyLoopsTotalCount\else\edef\SvcompNineteenPdrInvKinductionKipdrReachsafetyLoopsTotalCount{0}\fi
\ifdefined\SvcompNineteenPdrInvKinductionKipdrReachsafetyLoopsCorrectCount\else\edef\SvcompNineteenPdrInvKinductionKipdrReachsafetyLoopsCorrectCount{0}\fi
\ifdefined\SvcompNineteenPdrInvKinductionKipdrReachsafetyLoopsCorrectTrueCount\else\edef\SvcompNineteenPdrInvKinductionKipdrReachsafetyLoopsCorrectTrueCount{0}\fi
\ifdefined\SvcompNineteenPdrInvKinductionKipdrReachsafetyLoopsCorrectFalseCount\else\edef\SvcompNineteenPdrInvKinductionKipdrReachsafetyLoopsCorrectFalseCount{0}\fi
\ifdefined\SvcompNineteenPdrInvKinductionKipdrReachsafetyLoopsWrongTrueCount\else\edef\SvcompNineteenPdrInvKinductionKipdrReachsafetyLoopsWrongTrueCount{0}\fi
\ifdefined\SvcompNineteenPdrInvKinductionKipdrReachsafetyLoopsWrongFalseCount\else\edef\SvcompNineteenPdrInvKinductionKipdrReachsafetyLoopsWrongFalseCount{0}\fi
\ifdefined\SvcompNineteenPdrInvKinductionKipdrReachsafetyLoopsErrorTimeoutCount\else\edef\SvcompNineteenPdrInvKinductionKipdrReachsafetyLoopsErrorTimeoutCount{0}\fi
\ifdefined\SvcompNineteenPdrInvKinductionKipdrReachsafetyLoopsErrorOutOfMemoryCount\else\edef\SvcompNineteenPdrInvKinductionKipdrReachsafetyLoopsErrorOutOfMemoryCount{0}\fi
\ifdefined\SvcompNineteenPdrInvKinductionKipdrReachsafetyLoopsCorrectCputime\else\edef\SvcompNineteenPdrInvKinductionKipdrReachsafetyLoopsCorrectCputime{0}\fi
\ifdefined\SvcompNineteenPdrInvKinductionKipdrReachsafetyLoopsCorrectCputimeAvg\else\edef\SvcompNineteenPdrInvKinductionKipdrReachsafetyLoopsCorrectCputimeAvg{None}\fi
\ifdefined\SvcompNineteenPdrInvKinductionKipdrReachsafetyLoopsCorrectWalltime\else\edef\SvcompNineteenPdrInvKinductionKipdrReachsafetyLoopsCorrectWalltime{0}\fi
\ifdefined\SvcompNineteenPdrInvKinductionKipdrReachsafetyLoopsCorrectWalltimeAvg\else\edef\SvcompNineteenPdrInvKinductionKipdrReachsafetyLoopsCorrectWalltimeAvg{None}\fi
\ifdefined\SvcompNineteenPdrInvKinductionDfkipdrReachsafetyLoopsTotalCount\else\edef\SvcompNineteenPdrInvKinductionDfkipdrReachsafetyLoopsTotalCount{0}\fi
\ifdefined\SvcompNineteenPdrInvKinductionDfkipdrReachsafetyLoopsCorrectCount\else\edef\SvcompNineteenPdrInvKinductionDfkipdrReachsafetyLoopsCorrectCount{0}\fi
\ifdefined\SvcompNineteenPdrInvKinductionDfkipdrReachsafetyLoopsCorrectTrueCount\else\edef\SvcompNineteenPdrInvKinductionDfkipdrReachsafetyLoopsCorrectTrueCount{0}\fi
\ifdefined\SvcompNineteenPdrInvKinductionDfkipdrReachsafetyLoopsCorrectFalseCount\else\edef\SvcompNineteenPdrInvKinductionDfkipdrReachsafetyLoopsCorrectFalseCount{0}\fi
\ifdefined\SvcompNineteenPdrInvKinductionDfkipdrReachsafetyLoopsWrongTrueCount\else\edef\SvcompNineteenPdrInvKinductionDfkipdrReachsafetyLoopsWrongTrueCount{0}\fi
\ifdefined\SvcompNineteenPdrInvKinductionDfkipdrReachsafetyLoopsWrongFalseCount\else\edef\SvcompNineteenPdrInvKinductionDfkipdrReachsafetyLoopsWrongFalseCount{0}\fi
\ifdefined\SvcompNineteenPdrInvKinductionDfkipdrReachsafetyLoopsErrorTimeoutCount\else\edef\SvcompNineteenPdrInvKinductionDfkipdrReachsafetyLoopsErrorTimeoutCount{0}\fi
\ifdefined\SvcompNineteenPdrInvKinductionDfkipdrReachsafetyLoopsErrorOutOfMemoryCount\else\edef\SvcompNineteenPdrInvKinductionDfkipdrReachsafetyLoopsErrorOutOfMemoryCount{0}\fi
\ifdefined\SvcompNineteenPdrInvKinductionDfkipdrReachsafetyLoopsCorrectCputime\else\edef\SvcompNineteenPdrInvKinductionDfkipdrReachsafetyLoopsCorrectCputime{0}\fi
\ifdefined\SvcompNineteenPdrInvKinductionDfkipdrReachsafetyLoopsCorrectCputimeAvg\else\edef\SvcompNineteenPdrInvKinductionDfkipdrReachsafetyLoopsCorrectCputimeAvg{None}\fi
\ifdefined\SvcompNineteenPdrInvKinductionDfkipdrReachsafetyLoopsCorrectWalltime\else\edef\SvcompNineteenPdrInvKinductionDfkipdrReachsafetyLoopsCorrectWalltime{0}\fi
\ifdefined\SvcompNineteenPdrInvKinductionDfkipdrReachsafetyLoopsCorrectWalltimeAvg\else\edef\SvcompNineteenPdrInvKinductionDfkipdrReachsafetyLoopsCorrectWalltimeAvg{None}\fi
\ifdefined\SvcompNineteenCpaSeqSvCompPropReachsafetyReachsafetyLoopsTotalCount\else\edef\SvcompNineteenCpaSeqSvCompPropReachsafetyReachsafetyLoopsTotalCount{0}\fi
\ifdefined\SvcompNineteenCpaSeqSvCompPropReachsafetyReachsafetyLoopsCorrectCount\else\edef\SvcompNineteenCpaSeqSvCompPropReachsafetyReachsafetyLoopsCorrectCount{0}\fi
\ifdefined\SvcompNineteenCpaSeqSvCompPropReachsafetyReachsafetyLoopsCorrectTrueCount\else\edef\SvcompNineteenCpaSeqSvCompPropReachsafetyReachsafetyLoopsCorrectTrueCount{0}\fi
\ifdefined\SvcompNineteenCpaSeqSvCompPropReachsafetyReachsafetyLoopsCorrectFalseCount\else\edef\SvcompNineteenCpaSeqSvCompPropReachsafetyReachsafetyLoopsCorrectFalseCount{0}\fi
\ifdefined\SvcompNineteenCpaSeqSvCompPropReachsafetyReachsafetyLoopsWrongTrueCount\else\edef\SvcompNineteenCpaSeqSvCompPropReachsafetyReachsafetyLoopsWrongTrueCount{0}\fi
\ifdefined\SvcompNineteenCpaSeqSvCompPropReachsafetyReachsafetyLoopsWrongFalseCount\else\edef\SvcompNineteenCpaSeqSvCompPropReachsafetyReachsafetyLoopsWrongFalseCount{0}\fi
\ifdefined\SvcompNineteenCpaSeqSvCompPropReachsafetyReachsafetyLoopsErrorTimeoutCount\else\edef\SvcompNineteenCpaSeqSvCompPropReachsafetyReachsafetyLoopsErrorTimeoutCount{0}\fi
\ifdefined\SvcompNineteenCpaSeqSvCompPropReachsafetyReachsafetyLoopsErrorOutOfMemoryCount\else\edef\SvcompNineteenCpaSeqSvCompPropReachsafetyReachsafetyLoopsErrorOutOfMemoryCount{0}\fi
\ifdefined\SvcompNineteenCpaSeqSvCompPropReachsafetyReachsafetyLoopsCorrectCputime\else\edef\SvcompNineteenCpaSeqSvCompPropReachsafetyReachsafetyLoopsCorrectCputime{0}\fi
\ifdefined\SvcompNineteenCpaSeqSvCompPropReachsafetyReachsafetyLoopsCorrectCputimeAvg\else\edef\SvcompNineteenCpaSeqSvCompPropReachsafetyReachsafetyLoopsCorrectCputimeAvg{None}\fi
\ifdefined\SvcompNineteenCpaSeqSvCompPropReachsafetyReachsafetyLoopsCorrectWalltime\else\edef\SvcompNineteenCpaSeqSvCompPropReachsafetyReachsafetyLoopsCorrectWalltime{0}\fi
\ifdefined\SvcompNineteenCpaSeqSvCompPropReachsafetyReachsafetyLoopsCorrectWalltimeAvg\else\edef\SvcompNineteenCpaSeqSvCompPropReachsafetyReachsafetyLoopsCorrectWalltimeAvg{None}\fi
\ifdefined\SvcompNineteenPescoSvCompPropReachsafetyReachsafetyLoopsTotalCount\else\edef\SvcompNineteenPescoSvCompPropReachsafetyReachsafetyLoopsTotalCount{0}\fi
\ifdefined\SvcompNineteenPescoSvCompPropReachsafetyReachsafetyLoopsCorrectCount\else\edef\SvcompNineteenPescoSvCompPropReachsafetyReachsafetyLoopsCorrectCount{0}\fi
\ifdefined\SvcompNineteenPescoSvCompPropReachsafetyReachsafetyLoopsCorrectTrueCount\else\edef\SvcompNineteenPescoSvCompPropReachsafetyReachsafetyLoopsCorrectTrueCount{0}\fi
\ifdefined\SvcompNineteenPescoSvCompPropReachsafetyReachsafetyLoopsCorrectFalseCount\else\edef\SvcompNineteenPescoSvCompPropReachsafetyReachsafetyLoopsCorrectFalseCount{0}\fi
\ifdefined\SvcompNineteenPescoSvCompPropReachsafetyReachsafetyLoopsWrongTrueCount\else\edef\SvcompNineteenPescoSvCompPropReachsafetyReachsafetyLoopsWrongTrueCount{0}\fi
\ifdefined\SvcompNineteenPescoSvCompPropReachsafetyReachsafetyLoopsWrongFalseCount\else\edef\SvcompNineteenPescoSvCompPropReachsafetyReachsafetyLoopsWrongFalseCount{0}\fi
\ifdefined\SvcompNineteenPescoSvCompPropReachsafetyReachsafetyLoopsErrorTimeoutCount\else\edef\SvcompNineteenPescoSvCompPropReachsafetyReachsafetyLoopsErrorTimeoutCount{0}\fi
\ifdefined\SvcompNineteenPescoSvCompPropReachsafetyReachsafetyLoopsErrorOutOfMemoryCount\else\edef\SvcompNineteenPescoSvCompPropReachsafetyReachsafetyLoopsErrorOutOfMemoryCount{0}\fi
\ifdefined\SvcompNineteenPescoSvCompPropReachsafetyReachsafetyLoopsCorrectCputime\else\edef\SvcompNineteenPescoSvCompPropReachsafetyReachsafetyLoopsCorrectCputime{0}\fi
\ifdefined\SvcompNineteenPescoSvCompPropReachsafetyReachsafetyLoopsCorrectCputimeAvg\else\edef\SvcompNineteenPescoSvCompPropReachsafetyReachsafetyLoopsCorrectCputimeAvg{None}\fi
\ifdefined\SvcompNineteenPescoSvCompPropReachsafetyReachsafetyLoopsCorrectWalltime\else\edef\SvcompNineteenPescoSvCompPropReachsafetyReachsafetyLoopsCorrectWalltime{0}\fi
\ifdefined\SvcompNineteenPescoSvCompPropReachsafetyReachsafetyLoopsCorrectWalltimeAvg\else\edef\SvcompNineteenPescoSvCompPropReachsafetyReachsafetyLoopsCorrectWalltimeAvg{None}\fi
\ifdefined\SvcompNineteenSkinkSvCompPropReachsafetyReachsafetyLoopsTotalCount\else\edef\SvcompNineteenSkinkSvCompPropReachsafetyReachsafetyLoopsTotalCount{0}\fi
\ifdefined\SvcompNineteenSkinkSvCompPropReachsafetyReachsafetyLoopsCorrectCount\else\edef\SvcompNineteenSkinkSvCompPropReachsafetyReachsafetyLoopsCorrectCount{0}\fi
\ifdefined\SvcompNineteenSkinkSvCompPropReachsafetyReachsafetyLoopsCorrectTrueCount\else\edef\SvcompNineteenSkinkSvCompPropReachsafetyReachsafetyLoopsCorrectTrueCount{0}\fi
\ifdefined\SvcompNineteenSkinkSvCompPropReachsafetyReachsafetyLoopsCorrectFalseCount\else\edef\SvcompNineteenSkinkSvCompPropReachsafetyReachsafetyLoopsCorrectFalseCount{0}\fi
\ifdefined\SvcompNineteenSkinkSvCompPropReachsafetyReachsafetyLoopsWrongTrueCount\else\edef\SvcompNineteenSkinkSvCompPropReachsafetyReachsafetyLoopsWrongTrueCount{0}\fi
\ifdefined\SvcompNineteenSkinkSvCompPropReachsafetyReachsafetyLoopsWrongFalseCount\else\edef\SvcompNineteenSkinkSvCompPropReachsafetyReachsafetyLoopsWrongFalseCount{0}\fi
\ifdefined\SvcompNineteenSkinkSvCompPropReachsafetyReachsafetyLoopsErrorTimeoutCount\else\edef\SvcompNineteenSkinkSvCompPropReachsafetyReachsafetyLoopsErrorTimeoutCount{0}\fi
\ifdefined\SvcompNineteenSkinkSvCompPropReachsafetyReachsafetyLoopsErrorOutOfMemoryCount\else\edef\SvcompNineteenSkinkSvCompPropReachsafetyReachsafetyLoopsErrorOutOfMemoryCount{0}\fi
\ifdefined\SvcompNineteenSkinkSvCompPropReachsafetyReachsafetyLoopsCorrectCputime\else\edef\SvcompNineteenSkinkSvCompPropReachsafetyReachsafetyLoopsCorrectCputime{0}\fi
\ifdefined\SvcompNineteenSkinkSvCompPropReachsafetyReachsafetyLoopsCorrectCputimeAvg\else\edef\SvcompNineteenSkinkSvCompPropReachsafetyReachsafetyLoopsCorrectCputimeAvg{None}\fi
\ifdefined\SvcompNineteenSkinkSvCompPropReachsafetyReachsafetyLoopsCorrectWalltime\else\edef\SvcompNineteenSkinkSvCompPropReachsafetyReachsafetyLoopsCorrectWalltime{0}\fi
\ifdefined\SvcompNineteenSkinkSvCompPropReachsafetyReachsafetyLoopsCorrectWalltimeAvg\else\edef\SvcompNineteenSkinkSvCompPropReachsafetyReachsafetyLoopsCorrectWalltimeAvg{None}\fi
\ifdefined\SvcompNineteenUautomizerSvCompPropReachsafetyReachsafetyLoopsTotalCount\else\edef\SvcompNineteenUautomizerSvCompPropReachsafetyReachsafetyLoopsTotalCount{0}\fi
\ifdefined\SvcompNineteenUautomizerSvCompPropReachsafetyReachsafetyLoopsCorrectCount\else\edef\SvcompNineteenUautomizerSvCompPropReachsafetyReachsafetyLoopsCorrectCount{0}\fi
\ifdefined\SvcompNineteenUautomizerSvCompPropReachsafetyReachsafetyLoopsCorrectTrueCount\else\edef\SvcompNineteenUautomizerSvCompPropReachsafetyReachsafetyLoopsCorrectTrueCount{0}\fi
\ifdefined\SvcompNineteenUautomizerSvCompPropReachsafetyReachsafetyLoopsCorrectFalseCount\else\edef\SvcompNineteenUautomizerSvCompPropReachsafetyReachsafetyLoopsCorrectFalseCount{0}\fi
\ifdefined\SvcompNineteenUautomizerSvCompPropReachsafetyReachsafetyLoopsWrongTrueCount\else\edef\SvcompNineteenUautomizerSvCompPropReachsafetyReachsafetyLoopsWrongTrueCount{0}\fi
\ifdefined\SvcompNineteenUautomizerSvCompPropReachsafetyReachsafetyLoopsWrongFalseCount\else\edef\SvcompNineteenUautomizerSvCompPropReachsafetyReachsafetyLoopsWrongFalseCount{0}\fi
\ifdefined\SvcompNineteenUautomizerSvCompPropReachsafetyReachsafetyLoopsErrorTimeoutCount\else\edef\SvcompNineteenUautomizerSvCompPropReachsafetyReachsafetyLoopsErrorTimeoutCount{0}\fi
\ifdefined\SvcompNineteenUautomizerSvCompPropReachsafetyReachsafetyLoopsErrorOutOfMemoryCount\else\edef\SvcompNineteenUautomizerSvCompPropReachsafetyReachsafetyLoopsErrorOutOfMemoryCount{0}\fi
\ifdefined\SvcompNineteenUautomizerSvCompPropReachsafetyReachsafetyLoopsCorrectCputime\else\edef\SvcompNineteenUautomizerSvCompPropReachsafetyReachsafetyLoopsCorrectCputime{0}\fi
\ifdefined\SvcompNineteenUautomizerSvCompPropReachsafetyReachsafetyLoopsCorrectCputimeAvg\else\edef\SvcompNineteenUautomizerSvCompPropReachsafetyReachsafetyLoopsCorrectCputimeAvg{None}\fi
\ifdefined\SvcompNineteenUautomizerSvCompPropReachsafetyReachsafetyLoopsCorrectWalltime\else\edef\SvcompNineteenUautomizerSvCompPropReachsafetyReachsafetyLoopsCorrectWalltime{0}\fi
\ifdefined\SvcompNineteenUautomizerSvCompPropReachsafetyReachsafetyLoopsCorrectWalltimeAvg\else\edef\SvcompNineteenUautomizerSvCompPropReachsafetyReachsafetyLoopsCorrectWalltimeAvg{None}\fi
\ifdefined\SvcompNineteenUkojakSvCompPropReachsafetyReachsafetyLoopsTotalCount\else\edef\SvcompNineteenUkojakSvCompPropReachsafetyReachsafetyLoopsTotalCount{0}\fi
\ifdefined\SvcompNineteenUkojakSvCompPropReachsafetyReachsafetyLoopsCorrectCount\else\edef\SvcompNineteenUkojakSvCompPropReachsafetyReachsafetyLoopsCorrectCount{0}\fi
\ifdefined\SvcompNineteenUkojakSvCompPropReachsafetyReachsafetyLoopsCorrectTrueCount\else\edef\SvcompNineteenUkojakSvCompPropReachsafetyReachsafetyLoopsCorrectTrueCount{0}\fi
\ifdefined\SvcompNineteenUkojakSvCompPropReachsafetyReachsafetyLoopsCorrectFalseCount\else\edef\SvcompNineteenUkojakSvCompPropReachsafetyReachsafetyLoopsCorrectFalseCount{0}\fi
\ifdefined\SvcompNineteenUkojakSvCompPropReachsafetyReachsafetyLoopsWrongTrueCount\else\edef\SvcompNineteenUkojakSvCompPropReachsafetyReachsafetyLoopsWrongTrueCount{0}\fi
\ifdefined\SvcompNineteenUkojakSvCompPropReachsafetyReachsafetyLoopsWrongFalseCount\else\edef\SvcompNineteenUkojakSvCompPropReachsafetyReachsafetyLoopsWrongFalseCount{0}\fi
\ifdefined\SvcompNineteenUkojakSvCompPropReachsafetyReachsafetyLoopsErrorTimeoutCount\else\edef\SvcompNineteenUkojakSvCompPropReachsafetyReachsafetyLoopsErrorTimeoutCount{0}\fi
\ifdefined\SvcompNineteenUkojakSvCompPropReachsafetyReachsafetyLoopsErrorOutOfMemoryCount\else\edef\SvcompNineteenUkojakSvCompPropReachsafetyReachsafetyLoopsErrorOutOfMemoryCount{0}\fi
\ifdefined\SvcompNineteenUkojakSvCompPropReachsafetyReachsafetyLoopsCorrectCputime\else\edef\SvcompNineteenUkojakSvCompPropReachsafetyReachsafetyLoopsCorrectCputime{0}\fi
\ifdefined\SvcompNineteenUkojakSvCompPropReachsafetyReachsafetyLoopsCorrectCputimeAvg\else\edef\SvcompNineteenUkojakSvCompPropReachsafetyReachsafetyLoopsCorrectCputimeAvg{None}\fi
\ifdefined\SvcompNineteenUkojakSvCompPropReachsafetyReachsafetyLoopsCorrectWalltime\else\edef\SvcompNineteenUkojakSvCompPropReachsafetyReachsafetyLoopsCorrectWalltime{0}\fi
\ifdefined\SvcompNineteenUkojakSvCompPropReachsafetyReachsafetyLoopsCorrectWalltimeAvg\else\edef\SvcompNineteenUkojakSvCompPropReachsafetyReachsafetyLoopsCorrectWalltimeAvg{None}\fi
\ifdefined\SvcompNineteenUtaipanSvCompPropReachsafetyReachsafetyLoopsTotalCount\else\edef\SvcompNineteenUtaipanSvCompPropReachsafetyReachsafetyLoopsTotalCount{0}\fi
\ifdefined\SvcompNineteenUtaipanSvCompPropReachsafetyReachsafetyLoopsCorrectCount\else\edef\SvcompNineteenUtaipanSvCompPropReachsafetyReachsafetyLoopsCorrectCount{0}\fi
\ifdefined\SvcompNineteenUtaipanSvCompPropReachsafetyReachsafetyLoopsCorrectTrueCount\else\edef\SvcompNineteenUtaipanSvCompPropReachsafetyReachsafetyLoopsCorrectTrueCount{0}\fi
\ifdefined\SvcompNineteenUtaipanSvCompPropReachsafetyReachsafetyLoopsCorrectFalseCount\else\edef\SvcompNineteenUtaipanSvCompPropReachsafetyReachsafetyLoopsCorrectFalseCount{0}\fi
\ifdefined\SvcompNineteenUtaipanSvCompPropReachsafetyReachsafetyLoopsWrongTrueCount\else\edef\SvcompNineteenUtaipanSvCompPropReachsafetyReachsafetyLoopsWrongTrueCount{0}\fi
\ifdefined\SvcompNineteenUtaipanSvCompPropReachsafetyReachsafetyLoopsWrongFalseCount\else\edef\SvcompNineteenUtaipanSvCompPropReachsafetyReachsafetyLoopsWrongFalseCount{0}\fi
\ifdefined\SvcompNineteenUtaipanSvCompPropReachsafetyReachsafetyLoopsErrorTimeoutCount\else\edef\SvcompNineteenUtaipanSvCompPropReachsafetyReachsafetyLoopsErrorTimeoutCount{0}\fi
\ifdefined\SvcompNineteenUtaipanSvCompPropReachsafetyReachsafetyLoopsErrorOutOfMemoryCount\else\edef\SvcompNineteenUtaipanSvCompPropReachsafetyReachsafetyLoopsErrorOutOfMemoryCount{0}\fi
\ifdefined\SvcompNineteenUtaipanSvCompPropReachsafetyReachsafetyLoopsCorrectCputime\else\edef\SvcompNineteenUtaipanSvCompPropReachsafetyReachsafetyLoopsCorrectCputime{0}\fi
\ifdefined\SvcompNineteenUtaipanSvCompPropReachsafetyReachsafetyLoopsCorrectCputimeAvg\else\edef\SvcompNineteenUtaipanSvCompPropReachsafetyReachsafetyLoopsCorrectCputimeAvg{None}\fi
\ifdefined\SvcompNineteenUtaipanSvCompPropReachsafetyReachsafetyLoopsCorrectWalltime\else\edef\SvcompNineteenUtaipanSvCompPropReachsafetyReachsafetyLoopsCorrectWalltime{0}\fi
\ifdefined\SvcompNineteenUtaipanSvCompPropReachsafetyReachsafetyLoopsCorrectWalltimeAvg\else\edef\SvcompNineteenUtaipanSvCompPropReachsafetyReachsafetyLoopsCorrectWalltimeAvg{None}\fi
\ifdefined\SvcompNineteenVeriabsSvCompPropReachsafetyReachsafetyLoopsTotalCount\else\edef\SvcompNineteenVeriabsSvCompPropReachsafetyReachsafetyLoopsTotalCount{0}\fi
\ifdefined\SvcompNineteenVeriabsSvCompPropReachsafetyReachsafetyLoopsCorrectCount\else\edef\SvcompNineteenVeriabsSvCompPropReachsafetyReachsafetyLoopsCorrectCount{0}\fi
\ifdefined\SvcompNineteenVeriabsSvCompPropReachsafetyReachsafetyLoopsCorrectTrueCount\else\edef\SvcompNineteenVeriabsSvCompPropReachsafetyReachsafetyLoopsCorrectTrueCount{0}\fi
\ifdefined\SvcompNineteenVeriabsSvCompPropReachsafetyReachsafetyLoopsCorrectFalseCount\else\edef\SvcompNineteenVeriabsSvCompPropReachsafetyReachsafetyLoopsCorrectFalseCount{0}\fi
\ifdefined\SvcompNineteenVeriabsSvCompPropReachsafetyReachsafetyLoopsWrongTrueCount\else\edef\SvcompNineteenVeriabsSvCompPropReachsafetyReachsafetyLoopsWrongTrueCount{0}\fi
\ifdefined\SvcompNineteenVeriabsSvCompPropReachsafetyReachsafetyLoopsWrongFalseCount\else\edef\SvcompNineteenVeriabsSvCompPropReachsafetyReachsafetyLoopsWrongFalseCount{0}\fi
\ifdefined\SvcompNineteenVeriabsSvCompPropReachsafetyReachsafetyLoopsErrorTimeoutCount\else\edef\SvcompNineteenVeriabsSvCompPropReachsafetyReachsafetyLoopsErrorTimeoutCount{0}\fi
\ifdefined\SvcompNineteenVeriabsSvCompPropReachsafetyReachsafetyLoopsErrorOutOfMemoryCount\else\edef\SvcompNineteenVeriabsSvCompPropReachsafetyReachsafetyLoopsErrorOutOfMemoryCount{0}\fi
\ifdefined\SvcompNineteenVeriabsSvCompPropReachsafetyReachsafetyLoopsCorrectCputime\else\edef\SvcompNineteenVeriabsSvCompPropReachsafetyReachsafetyLoopsCorrectCputime{0}\fi
\ifdefined\SvcompNineteenVeriabsSvCompPropReachsafetyReachsafetyLoopsCorrectCputimeAvg\else\edef\SvcompNineteenVeriabsSvCompPropReachsafetyReachsafetyLoopsCorrectCputimeAvg{None}\fi
\ifdefined\SvcompNineteenVeriabsSvCompPropReachsafetyReachsafetyLoopsCorrectWalltime\else\edef\SvcompNineteenVeriabsSvCompPropReachsafetyReachsafetyLoopsCorrectWalltime{0}\fi
\ifdefined\SvcompNineteenVeriabsSvCompPropReachsafetyReachsafetyLoopsCorrectWalltimeAvg\else\edef\SvcompNineteenVeriabsSvCompPropReachsafetyReachsafetyLoopsCorrectWalltimeAvg{None}\fi
\ifdefined\SvcompNineteenViapSvCompPropReachsafetyReachsafetyLoopsTotalCount\else\edef\SvcompNineteenViapSvCompPropReachsafetyReachsafetyLoopsTotalCount{0}\fi
\ifdefined\SvcompNineteenViapSvCompPropReachsafetyReachsafetyLoopsCorrectCount\else\edef\SvcompNineteenViapSvCompPropReachsafetyReachsafetyLoopsCorrectCount{0}\fi
\ifdefined\SvcompNineteenViapSvCompPropReachsafetyReachsafetyLoopsCorrectTrueCount\else\edef\SvcompNineteenViapSvCompPropReachsafetyReachsafetyLoopsCorrectTrueCount{0}\fi
\ifdefined\SvcompNineteenViapSvCompPropReachsafetyReachsafetyLoopsCorrectFalseCount\else\edef\SvcompNineteenViapSvCompPropReachsafetyReachsafetyLoopsCorrectFalseCount{0}\fi
\ifdefined\SvcompNineteenViapSvCompPropReachsafetyReachsafetyLoopsWrongTrueCount\else\edef\SvcompNineteenViapSvCompPropReachsafetyReachsafetyLoopsWrongTrueCount{0}\fi
\ifdefined\SvcompNineteenViapSvCompPropReachsafetyReachsafetyLoopsWrongFalseCount\else\edef\SvcompNineteenViapSvCompPropReachsafetyReachsafetyLoopsWrongFalseCount{0}\fi
\ifdefined\SvcompNineteenViapSvCompPropReachsafetyReachsafetyLoopsErrorTimeoutCount\else\edef\SvcompNineteenViapSvCompPropReachsafetyReachsafetyLoopsErrorTimeoutCount{0}\fi
\ifdefined\SvcompNineteenViapSvCompPropReachsafetyReachsafetyLoopsErrorOutOfMemoryCount\else\edef\SvcompNineteenViapSvCompPropReachsafetyReachsafetyLoopsErrorOutOfMemoryCount{0}\fi
\ifdefined\SvcompNineteenViapSvCompPropReachsafetyReachsafetyLoopsCorrectCputime\else\edef\SvcompNineteenViapSvCompPropReachsafetyReachsafetyLoopsCorrectCputime{0}\fi
\ifdefined\SvcompNineteenViapSvCompPropReachsafetyReachsafetyLoopsCorrectCputimeAvg\else\edef\SvcompNineteenViapSvCompPropReachsafetyReachsafetyLoopsCorrectCputimeAvg{None}\fi
\ifdefined\SvcompNineteenViapSvCompPropReachsafetyReachsafetyLoopsCorrectWalltime\else\edef\SvcompNineteenViapSvCompPropReachsafetyReachsafetyLoopsCorrectWalltime{0}\fi
\ifdefined\SvcompNineteenViapSvCompPropReachsafetyReachsafetyLoopsCorrectWalltimeAvg\else\edef\SvcompNineteenViapSvCompPropReachsafetyReachsafetyLoopsCorrectWalltimeAvg{None}\fi
\edef\SvcompNineteenPdrInvKinductionPlainReachsafetyLoopsErrorOtherInconclusiveCount{\the\numexpr \SvcompNineteenPdrInvKinductionPlainReachsafetyLoopsTotalCount - \SvcompNineteenPdrInvKinductionPlainReachsafetyLoopsCorrectCount - \SvcompNineteenPdrInvKinductionPlainReachsafetyLoopsWrongTrueCount - \SvcompNineteenPdrInvKinductionPlainReachsafetyLoopsWrongFalseCount - \SvcompNineteenPdrInvKinductionPlainReachsafetyLoopsErrorTimeoutCount - \SvcompNineteenPdrInvKinductionPlainReachsafetyLoopsErrorOutOfMemoryCount \relax}
\edef\SvcompNineteenPdrInvKinductionDfReachsafetyLoopsErrorOtherInconclusiveCount{\the\numexpr \SvcompNineteenPdrInvKinductionDfReachsafetyLoopsTotalCount - \SvcompNineteenPdrInvKinductionDfReachsafetyLoopsCorrectCount - \SvcompNineteenPdrInvKinductionDfReachsafetyLoopsWrongTrueCount - \SvcompNineteenPdrInvKinductionDfReachsafetyLoopsWrongFalseCount - \SvcompNineteenPdrInvKinductionDfReachsafetyLoopsErrorTimeoutCount - \SvcompNineteenPdrInvKinductionDfReachsafetyLoopsErrorOutOfMemoryCount \relax}
\edef\SvcompNineteenPdrInvKinductionKipdrReachsafetyLoopsErrorOtherInconclusiveCount{\the\numexpr \SvcompNineteenPdrInvKinductionKipdrReachsafetyLoopsTotalCount - \SvcompNineteenPdrInvKinductionKipdrReachsafetyLoopsCorrectCount - \SvcompNineteenPdrInvKinductionKipdrReachsafetyLoopsWrongTrueCount - \SvcompNineteenPdrInvKinductionKipdrReachsafetyLoopsWrongFalseCount - \SvcompNineteenPdrInvKinductionKipdrReachsafetyLoopsErrorTimeoutCount - \SvcompNineteenPdrInvKinductionKipdrReachsafetyLoopsErrorOutOfMemoryCount \relax}
\edef\SvcompNineteenPdrInvKinductionDfkipdrReachsafetyLoopsErrorOtherInconclusiveCount{\the\numexpr \SvcompNineteenPdrInvKinductionDfkipdrReachsafetyLoopsTotalCount - \SvcompNineteenPdrInvKinductionDfkipdrReachsafetyLoopsCorrectCount - \SvcompNineteenPdrInvKinductionDfkipdrReachsafetyLoopsWrongTrueCount - \SvcompNineteenPdrInvKinductionDfkipdrReachsafetyLoopsWrongFalseCount - \SvcompNineteenPdrInvKinductionDfkipdrReachsafetyLoopsErrorTimeoutCount - \SvcompNineteenPdrInvKinductionDfkipdrReachsafetyLoopsErrorOutOfMemoryCount \relax}
\edef\SvcompNineteenCpaSeqSvCompPropReachsafetyReachsafetyLoopsErrorOtherInconclusiveCount{\the\numexpr \SvcompNineteenCpaSeqSvCompPropReachsafetyReachsafetyLoopsTotalCount - \SvcompNineteenCpaSeqSvCompPropReachsafetyReachsafetyLoopsCorrectCount - \SvcompNineteenCpaSeqSvCompPropReachsafetyReachsafetyLoopsWrongTrueCount - \SvcompNineteenCpaSeqSvCompPropReachsafetyReachsafetyLoopsWrongFalseCount - \SvcompNineteenCpaSeqSvCompPropReachsafetyReachsafetyLoopsErrorTimeoutCount - \SvcompNineteenCpaSeqSvCompPropReachsafetyReachsafetyLoopsErrorOutOfMemoryCount \relax}
\edef\SvcompNineteenPescoSvCompPropReachsafetyReachsafetyLoopsErrorOtherInconclusiveCount{\the\numexpr \SvcompNineteenPescoSvCompPropReachsafetyReachsafetyLoopsTotalCount - \SvcompNineteenPescoSvCompPropReachsafetyReachsafetyLoopsCorrectCount - \SvcompNineteenPescoSvCompPropReachsafetyReachsafetyLoopsWrongTrueCount - \SvcompNineteenPescoSvCompPropReachsafetyReachsafetyLoopsWrongFalseCount - \SvcompNineteenPescoSvCompPropReachsafetyReachsafetyLoopsErrorTimeoutCount - \SvcompNineteenPescoSvCompPropReachsafetyReachsafetyLoopsErrorOutOfMemoryCount \relax}
\edef\SvcompNineteenSkinkSvCompPropReachsafetyReachsafetyLoopsErrorOtherInconclusiveCount{\the\numexpr \SvcompNineteenSkinkSvCompPropReachsafetyReachsafetyLoopsTotalCount - \SvcompNineteenSkinkSvCompPropReachsafetyReachsafetyLoopsCorrectCount - \SvcompNineteenSkinkSvCompPropReachsafetyReachsafetyLoopsWrongTrueCount - \SvcompNineteenSkinkSvCompPropReachsafetyReachsafetyLoopsWrongFalseCount - \SvcompNineteenSkinkSvCompPropReachsafetyReachsafetyLoopsErrorTimeoutCount - \SvcompNineteenSkinkSvCompPropReachsafetyReachsafetyLoopsErrorOutOfMemoryCount \relax}
\edef\SvcompNineteenUautomizerSvCompPropReachsafetyReachsafetyLoopsErrorOtherInconclusiveCount{\the\numexpr \SvcompNineteenUautomizerSvCompPropReachsafetyReachsafetyLoopsTotalCount - \SvcompNineteenUautomizerSvCompPropReachsafetyReachsafetyLoopsCorrectCount - \SvcompNineteenUautomizerSvCompPropReachsafetyReachsafetyLoopsWrongTrueCount - \SvcompNineteenUautomizerSvCompPropReachsafetyReachsafetyLoopsWrongFalseCount - \SvcompNineteenUautomizerSvCompPropReachsafetyReachsafetyLoopsErrorTimeoutCount - \SvcompNineteenUautomizerSvCompPropReachsafetyReachsafetyLoopsErrorOutOfMemoryCount \relax}
\edef\SvcompNineteenUkojakSvCompPropReachsafetyReachsafetyLoopsErrorOtherInconclusiveCount{\the\numexpr \SvcompNineteenUkojakSvCompPropReachsafetyReachsafetyLoopsTotalCount - \SvcompNineteenUkojakSvCompPropReachsafetyReachsafetyLoopsCorrectCount - \SvcompNineteenUkojakSvCompPropReachsafetyReachsafetyLoopsWrongTrueCount - \SvcompNineteenUkojakSvCompPropReachsafetyReachsafetyLoopsWrongFalseCount - \SvcompNineteenUkojakSvCompPropReachsafetyReachsafetyLoopsErrorTimeoutCount - \SvcompNineteenUkojakSvCompPropReachsafetyReachsafetyLoopsErrorOutOfMemoryCount \relax}
\edef\SvcompNineteenUtaipanSvCompPropReachsafetyReachsafetyLoopsErrorOtherInconclusiveCount{\the\numexpr \SvcompNineteenUtaipanSvCompPropReachsafetyReachsafetyLoopsTotalCount - \SvcompNineteenUtaipanSvCompPropReachsafetyReachsafetyLoopsCorrectCount - \SvcompNineteenUtaipanSvCompPropReachsafetyReachsafetyLoopsWrongTrueCount - \SvcompNineteenUtaipanSvCompPropReachsafetyReachsafetyLoopsWrongFalseCount - \SvcompNineteenUtaipanSvCompPropReachsafetyReachsafetyLoopsErrorTimeoutCount - \SvcompNineteenUtaipanSvCompPropReachsafetyReachsafetyLoopsErrorOutOfMemoryCount \relax}
\edef\SvcompNineteenVeriabsSvCompPropReachsafetyReachsafetyLoopsErrorOtherInconclusiveCount{\the\numexpr \SvcompNineteenVeriabsSvCompPropReachsafetyReachsafetyLoopsTotalCount - \SvcompNineteenVeriabsSvCompPropReachsafetyReachsafetyLoopsCorrectCount - \SvcompNineteenVeriabsSvCompPropReachsafetyReachsafetyLoopsWrongTrueCount - \SvcompNineteenVeriabsSvCompPropReachsafetyReachsafetyLoopsWrongFalseCount - \SvcompNineteenVeriabsSvCompPropReachsafetyReachsafetyLoopsErrorTimeoutCount - \SvcompNineteenVeriabsSvCompPropReachsafetyReachsafetyLoopsErrorOutOfMemoryCount \relax}
\edef\SvcompNineteenViapSvCompPropReachsafetyReachsafetyLoopsErrorOtherInconclusiveCount{\the\numexpr \SvcompNineteenViapSvCompPropReachsafetyReachsafetyLoopsTotalCount - \SvcompNineteenViapSvCompPropReachsafetyReachsafetyLoopsCorrectCount - \SvcompNineteenViapSvCompPropReachsafetyReachsafetyLoopsWrongTrueCount - \SvcompNineteenViapSvCompPropReachsafetyReachsafetyLoopsWrongFalseCount - \SvcompNineteenViapSvCompPropReachsafetyReachsafetyLoopsErrorTimeoutCount - \SvcompNineteenViapSvCompPropReachsafetyReachsafetyLoopsErrorOutOfMemoryCount \relax}
\edef\SvcompNineteenPdrInvKinductionPlainReachsafetyLoopsScore{\the\numexpr (2 * \SvcompNineteenPdrInvKinductionPlainReachsafetyLoopsCorrectTrueCount) + \SvcompNineteenPdrInvKinductionPlainReachsafetyLoopsCorrectFalseCount - (32 * \SvcompNineteenPdrInvKinductionPlainReachsafetyLoopsWrongTrueCount) - (16 * \SvcompNineteenPdrInvKinductionPlainReachsafetyLoopsWrongFalseCount) \relax}
\edef\SvcompNineteenPdrInvKinductionDfReachsafetyLoopsScore{\the\numexpr (2 * \SvcompNineteenPdrInvKinductionDfReachsafetyLoopsCorrectTrueCount) + \SvcompNineteenPdrInvKinductionDfReachsafetyLoopsCorrectFalseCount - (32 * \SvcompNineteenPdrInvKinductionDfReachsafetyLoopsWrongTrueCount) - (16 * \SvcompNineteenPdrInvKinductionDfReachsafetyLoopsWrongFalseCount) \relax}
\edef\SvcompNineteenPdrInvKinductionKipdrReachsafetyLoopsScore{\the\numexpr (2 * \SvcompNineteenPdrInvKinductionKipdrReachsafetyLoopsCorrectTrueCount) + \SvcompNineteenPdrInvKinductionKipdrReachsafetyLoopsCorrectFalseCount - (32 * \SvcompNineteenPdrInvKinductionKipdrReachsafetyLoopsWrongTrueCount) - (16 * \SvcompNineteenPdrInvKinductionKipdrReachsafetyLoopsWrongFalseCount) \relax}
\edef\SvcompNineteenPdrInvKinductionDfkipdrReachsafetyLoopsScore{\the\numexpr (2 * \SvcompNineteenPdrInvKinductionDfkipdrReachsafetyLoopsCorrectTrueCount) + \SvcompNineteenPdrInvKinductionDfkipdrReachsafetyLoopsCorrectFalseCount - (32 * \SvcompNineteenPdrInvKinductionDfkipdrReachsafetyLoopsWrongTrueCount) - (16 * \SvcompNineteenPdrInvKinductionDfkipdrReachsafetyLoopsWrongFalseCount) \relax}
\edef\SvcompNineteenCpaSeqSvCompPropReachsafetyReachsafetyLoopsScore{\the\numexpr (2 * \SvcompNineteenCpaSeqSvCompPropReachsafetyReachsafetyLoopsCorrectTrueCount) + \SvcompNineteenCpaSeqSvCompPropReachsafetyReachsafetyLoopsCorrectFalseCount - (32 * \SvcompNineteenCpaSeqSvCompPropReachsafetyReachsafetyLoopsWrongTrueCount) - (16 * \SvcompNineteenCpaSeqSvCompPropReachsafetyReachsafetyLoopsWrongFalseCount) \relax}
\edef\SvcompNineteenPescoSvCompPropReachsafetyReachsafetyLoopsScore{\the\numexpr (2 * \SvcompNineteenPescoSvCompPropReachsafetyReachsafetyLoopsCorrectTrueCount) + \SvcompNineteenPescoSvCompPropReachsafetyReachsafetyLoopsCorrectFalseCount - (32 * \SvcompNineteenPescoSvCompPropReachsafetyReachsafetyLoopsWrongTrueCount) - (16 * \SvcompNineteenPescoSvCompPropReachsafetyReachsafetyLoopsWrongFalseCount) \relax}
\edef\SvcompNineteenSkinkSvCompPropReachsafetyReachsafetyLoopsScore{\the\numexpr (2 * \SvcompNineteenSkinkSvCompPropReachsafetyReachsafetyLoopsCorrectTrueCount) + \SvcompNineteenSkinkSvCompPropReachsafetyReachsafetyLoopsCorrectFalseCount - (32 * \SvcompNineteenSkinkSvCompPropReachsafetyReachsafetyLoopsWrongTrueCount) - (16 * \SvcompNineteenSkinkSvCompPropReachsafetyReachsafetyLoopsWrongFalseCount) \relax}
\edef\SvcompNineteenUautomizerSvCompPropReachsafetyReachsafetyLoopsScore{\the\numexpr (2 * \SvcompNineteenUautomizerSvCompPropReachsafetyReachsafetyLoopsCorrectTrueCount) + \SvcompNineteenUautomizerSvCompPropReachsafetyReachsafetyLoopsCorrectFalseCount - (32 * \SvcompNineteenUautomizerSvCompPropReachsafetyReachsafetyLoopsWrongTrueCount) - (16 * \SvcompNineteenUautomizerSvCompPropReachsafetyReachsafetyLoopsWrongFalseCount) \relax}
\edef\SvcompNineteenUkojakSvCompPropReachsafetyReachsafetyLoopsScore{\the\numexpr (2 * \SvcompNineteenUkojakSvCompPropReachsafetyReachsafetyLoopsCorrectTrueCount) + \SvcompNineteenUkojakSvCompPropReachsafetyReachsafetyLoopsCorrectFalseCount - (32 * \SvcompNineteenUkojakSvCompPropReachsafetyReachsafetyLoopsWrongTrueCount) - (16 * \SvcompNineteenUkojakSvCompPropReachsafetyReachsafetyLoopsWrongFalseCount) \relax}
\edef\SvcompNineteenUtaipanSvCompPropReachsafetyReachsafetyLoopsScore{\the\numexpr (2 * \SvcompNineteenUtaipanSvCompPropReachsafetyReachsafetyLoopsCorrectTrueCount) + \SvcompNineteenUtaipanSvCompPropReachsafetyReachsafetyLoopsCorrectFalseCount - (32 * \SvcompNineteenUtaipanSvCompPropReachsafetyReachsafetyLoopsWrongTrueCount) - (16 * \SvcompNineteenUtaipanSvCompPropReachsafetyReachsafetyLoopsWrongFalseCount) \relax}
\edef\SvcompNineteenVeriabsSvCompPropReachsafetyReachsafetyLoopsScore{\the\numexpr (2 * \SvcompNineteenVeriabsSvCompPropReachsafetyReachsafetyLoopsCorrectTrueCount) + \SvcompNineteenVeriabsSvCompPropReachsafetyReachsafetyLoopsCorrectFalseCount - (32 * \SvcompNineteenVeriabsSvCompPropReachsafetyReachsafetyLoopsWrongTrueCount) - (16 * \SvcompNineteenVeriabsSvCompPropReachsafetyReachsafetyLoopsWrongFalseCount) \relax}
\edef\SvcompNineteenViapSvCompPropReachsafetyReachsafetyLoopsScore{\the\numexpr (2 * \SvcompNineteenViapSvCompPropReachsafetyReachsafetyLoopsCorrectTrueCount) + \SvcompNineteenViapSvCompPropReachsafetyReachsafetyLoopsCorrectFalseCount - (32 * \SvcompNineteenViapSvCompPropReachsafetyReachsafetyLoopsWrongTrueCount) - (16 * \SvcompNineteenViapSvCompPropReachsafetyReachsafetyLoopsWrongFalseCount) \relax}
\edef\SvcompNineteenPdrInvKinductionPlainReachsafetyLoopsTrueScore{\the\numexpr (2 * \SvcompNineteenPdrInvKinductionPlainReachsafetyLoopsCorrectTrueCount) - (32 * \SvcompNineteenPdrInvKinductionPlainReachsafetyLoopsWrongTrueCount) \relax}
\edef\SvcompNineteenPdrInvKinductionDfReachsafetyLoopsTrueScore{\the\numexpr (2 * \SvcompNineteenPdrInvKinductionDfReachsafetyLoopsCorrectTrueCount) - (32 * \SvcompNineteenPdrInvKinductionDfReachsafetyLoopsWrongTrueCount) \relax}
\edef\SvcompNineteenPdrInvKinductionKipdrReachsafetyLoopsTrueScore{\the\numexpr (2 * \SvcompNineteenPdrInvKinductionKipdrReachsafetyLoopsCorrectTrueCount) - (32 * \SvcompNineteenPdrInvKinductionKipdrReachsafetyLoopsWrongTrueCount) \relax}
\edef\SvcompNineteenPdrInvKinductionDfkipdrReachsafetyLoopsTrueScore{\the\numexpr (2 * \SvcompNineteenPdrInvKinductionDfkipdrReachsafetyLoopsCorrectTrueCount) - (32 * \SvcompNineteenPdrInvKinductionDfkipdrReachsafetyLoopsWrongTrueCount) \relax}
\edef\SvcompNineteenCpaSeqSvCompPropReachsafetyReachsafetyLoopsTrueScore{\the\numexpr (2 * \SvcompNineteenCpaSeqSvCompPropReachsafetyReachsafetyLoopsCorrectTrueCount) - (32 * \SvcompNineteenCpaSeqSvCompPropReachsafetyReachsafetyLoopsWrongTrueCount) \relax}
\edef\SvcompNineteenPescoSvCompPropReachsafetyReachsafetyLoopsTrueScore{\the\numexpr (2 * \SvcompNineteenPescoSvCompPropReachsafetyReachsafetyLoopsCorrectTrueCount) - (32 * \SvcompNineteenPescoSvCompPropReachsafetyReachsafetyLoopsWrongTrueCount) \relax}
\edef\SvcompNineteenSkinkSvCompPropReachsafetyReachsafetyLoopsTrueScore{\the\numexpr (2 * \SvcompNineteenSkinkSvCompPropReachsafetyReachsafetyLoopsCorrectTrueCount) - (32 * \SvcompNineteenSkinkSvCompPropReachsafetyReachsafetyLoopsWrongTrueCount) \relax}
\edef\SvcompNineteenUautomizerSvCompPropReachsafetyReachsafetyLoopsTrueScore{\the\numexpr (2 * \SvcompNineteenUautomizerSvCompPropReachsafetyReachsafetyLoopsCorrectTrueCount) - (32 * \SvcompNineteenUautomizerSvCompPropReachsafetyReachsafetyLoopsWrongTrueCount) \relax}
\edef\SvcompNineteenUkojakSvCompPropReachsafetyReachsafetyLoopsTrueScore{\the\numexpr (2 * \SvcompNineteenUkojakSvCompPropReachsafetyReachsafetyLoopsCorrectTrueCount) - (32 * \SvcompNineteenUkojakSvCompPropReachsafetyReachsafetyLoopsWrongTrueCount) \relax}
\edef\SvcompNineteenUtaipanSvCompPropReachsafetyReachsafetyLoopsTrueScore{\the\numexpr (2 * \SvcompNineteenUtaipanSvCompPropReachsafetyReachsafetyLoopsCorrectTrueCount) - (32 * \SvcompNineteenUtaipanSvCompPropReachsafetyReachsafetyLoopsWrongTrueCount) \relax}
\edef\SvcompNineteenVeriabsSvCompPropReachsafetyReachsafetyLoopsTrueScore{\the\numexpr (2 * \SvcompNineteenVeriabsSvCompPropReachsafetyReachsafetyLoopsCorrectTrueCount) - (32 * \SvcompNineteenVeriabsSvCompPropReachsafetyReachsafetyLoopsWrongTrueCount) \relax}
\edef\SvcompNineteenViapSvCompPropReachsafetyReachsafetyLoopsTrueScore{\the\numexpr (2 * \SvcompNineteenViapSvCompPropReachsafetyReachsafetyLoopsCorrectTrueCount) - (32 * \SvcompNineteenViapSvCompPropReachsafetyReachsafetyLoopsWrongTrueCount) \relax}
\newcommand{\nSvcompNineteenLoopsTasks}{\num{208}}
\newcommand{\nSvcompNineteenLoopsTruetasks}{\num{149}}
\newcommand{\nSvcompNineteenLoopsFalsetasks}{\num{59}}

%% file: evaluation/pathprograms/data-commands.tex
\providecommand\StoreBenchExecResult[7]{\expandafter\newcommand\csname#1#2#3#4#5#6\endcsname{#7}}%
\StoreBenchExecResult{PdrInvPathprograms}{KinductionDfStaticZeroZeroTTrueNotSolvedByKinductionPlain}{Total}{}{Count}{}{114}%
\StoreBenchExecResult{PdrInvPathprograms}{KinductionDfStaticZeroZeroTTrueNotSolvedByKinductionPlain}{Total}{}{Cputime}{}{62809.261009515}%
\StoreBenchExecResult{PdrInvPathprograms}{KinductionDfStaticZeroZeroTTrueNotSolvedByKinductionPlain}{Total}{}{Cputime}{Avg}{550.9584299080263157894736842}%
\StoreBenchExecResult{PdrInvPathprograms}{KinductionDfStaticZeroZeroTTrueNotSolvedByKinductionPlain}{Total}{}{Cputime}{Median}{827.6515852845}%
\StoreBenchExecResult{PdrInvPathprograms}{KinductionDfStaticZeroZeroTTrueNotSolvedByKinductionPlain}{Total}{}{Cputime}{Min}{3.197946868}%
\StoreBenchExecResult{PdrInvPathprograms}{KinductionDfStaticZeroZeroTTrueNotSolvedByKinductionPlain}{Total}{}{Cputime}{Max}{1000.76279416}%
\StoreBenchExecResult{PdrInvPathprograms}{KinductionDfStaticZeroZeroTTrueNotSolvedByKinductionPlain}{Total}{}{Cputime}{Stdev}{401.7666503385628152278191051}%
\StoreBenchExecResult{PdrInvPathprograms}{KinductionDfStaticZeroZeroTTrueNotSolvedByKinductionPlain}{Total}{}{Walltime}{}{56147.38730740658}%
\StoreBenchExecResult{PdrInvPathprograms}{KinductionDfStaticZeroZeroTTrueNotSolvedByKinductionPlain}{Total}{}{Walltime}{Avg}{492.5209412930401754385964912}%
\StoreBenchExecResult{PdrInvPathprograms}{KinductionDfStaticZeroZeroTTrueNotSolvedByKinductionPlain}{Total}{}{Walltime}{Median}{645.025067091}%
\StoreBenchExecResult{PdrInvPathprograms}{KinductionDfStaticZeroZeroTTrueNotSolvedByKinductionPlain}{Total}{}{Walltime}{Min}{1.76477193832}%
\StoreBenchExecResult{PdrInvPathprograms}{KinductionDfStaticZeroZeroTTrueNotSolvedByKinductionPlain}{Total}{}{Walltime}{Max}{898.954440117}%
\StoreBenchExecResult{PdrInvPathprograms}{KinductionDfStaticZeroZeroTTrueNotSolvedByKinductionPlain}{Total}{}{Walltime}{Stdev}{395.2144457324145355180975899}%
\StoreBenchExecResult{PdrInvPathprograms}{KinductionDfStaticZeroZeroTTrueNotSolvedByKinductionPlain}{Correct}{}{Count}{}{54}%
\StoreBenchExecResult{PdrInvPathprograms}{KinductionDfStaticZeroZeroTTrueNotSolvedByKinductionPlain}{Correct}{}{Cputime}{}{9285.765614906}%
\StoreBenchExecResult{PdrInvPathprograms}{KinductionDfStaticZeroZeroTTrueNotSolvedByKinductionPlain}{Correct}{}{Cputime}{Avg}{171.9586224982592592592592593}%
\StoreBenchExecResult{PdrInvPathprograms}{KinductionDfStaticZeroZeroTTrueNotSolvedByKinductionPlain}{Correct}{}{Cputime}{Median}{33.160734036}%
\StoreBenchExecResult{PdrInvPathprograms}{KinductionDfStaticZeroZeroTTrueNotSolvedByKinductionPlain}{Correct}{}{Cputime}{Min}{3.197946868}%
\StoreBenchExecResult{PdrInvPathprograms}{KinductionDfStaticZeroZeroTTrueNotSolvedByKinductionPlain}{Correct}{}{Cputime}{Max}{824.212066709}%
\StoreBenchExecResult{PdrInvPathprograms}{KinductionDfStaticZeroZeroTTrueNotSolvedByKinductionPlain}{Correct}{}{Cputime}{Stdev}{240.9712799865800036518560915}%
\StoreBenchExecResult{PdrInvPathprograms}{KinductionDfStaticZeroZeroTTrueNotSolvedByKinductionPlain}{Correct}{}{Walltime}{}{5374.98777460958}%
\StoreBenchExecResult{PdrInvPathprograms}{KinductionDfStaticZeroZeroTTrueNotSolvedByKinductionPlain}{Correct}{}{Walltime}{Avg}{99.53681064091814814814814815}%
\StoreBenchExecResult{PdrInvPathprograms}{KinductionDfStaticZeroZeroTTrueNotSolvedByKinductionPlain}{Correct}{}{Walltime}{Median}{16.8302990198}%
\StoreBenchExecResult{PdrInvPathprograms}{KinductionDfStaticZeroZeroTTrueNotSolvedByKinductionPlain}{Correct}{}{Walltime}{Min}{1.76477193832}%
\StoreBenchExecResult{PdrInvPathprograms}{KinductionDfStaticZeroZeroTTrueNotSolvedByKinductionPlain}{Correct}{}{Walltime}{Max}{612.673634052}%
\StoreBenchExecResult{PdrInvPathprograms}{KinductionDfStaticZeroZeroTTrueNotSolvedByKinductionPlain}{Correct}{}{Walltime}{Stdev}{145.5227444048886123919830606}%
\StoreBenchExecResult{PdrInvPathprograms}{KinductionDfStaticZeroZeroTTrueNotSolvedByKinductionPlain}{Correct}{True}{Count}{}{54}%
\StoreBenchExecResult{PdrInvPathprograms}{KinductionDfStaticZeroZeroTTrueNotSolvedByKinductionPlain}{Correct}{True}{Cputime}{}{9285.765614906}%
\StoreBenchExecResult{PdrInvPathprograms}{KinductionDfStaticZeroZeroTTrueNotSolvedByKinductionPlain}{Correct}{True}{Cputime}{Avg}{171.9586224982592592592592593}%
\StoreBenchExecResult{PdrInvPathprograms}{KinductionDfStaticZeroZeroTTrueNotSolvedByKinductionPlain}{Correct}{True}{Cputime}{Median}{33.160734036}%
\StoreBenchExecResult{PdrInvPathprograms}{KinductionDfStaticZeroZeroTTrueNotSolvedByKinductionPlain}{Correct}{True}{Cputime}{Min}{3.197946868}%
\StoreBenchExecResult{PdrInvPathprograms}{KinductionDfStaticZeroZeroTTrueNotSolvedByKinductionPlain}{Correct}{True}{Cputime}{Max}{824.212066709}%
\StoreBenchExecResult{PdrInvPathprograms}{KinductionDfStaticZeroZeroTTrueNotSolvedByKinductionPlain}{Correct}{True}{Cputime}{Stdev}{240.9712799865800036518560915}%
\StoreBenchExecResult{PdrInvPathprograms}{KinductionDfStaticZeroZeroTTrueNotSolvedByKinductionPlain}{Correct}{True}{Walltime}{}{5374.98777460958}%
\StoreBenchExecResult{PdrInvPathprograms}{KinductionDfStaticZeroZeroTTrueNotSolvedByKinductionPlain}{Correct}{True}{Walltime}{Avg}{99.53681064091814814814814815}%
\StoreBenchExecResult{PdrInvPathprograms}{KinductionDfStaticZeroZeroTTrueNotSolvedByKinductionPlain}{Correct}{True}{Walltime}{Median}{16.8302990198}%
\StoreBenchExecResult{PdrInvPathprograms}{KinductionDfStaticZeroZeroTTrueNotSolvedByKinductionPlain}{Correct}{True}{Walltime}{Min}{1.76477193832}%
\StoreBenchExecResult{PdrInvPathprograms}{KinductionDfStaticZeroZeroTTrueNotSolvedByKinductionPlain}{Correct}{True}{Walltime}{Max}{612.673634052}%
\StoreBenchExecResult{PdrInvPathprograms}{KinductionDfStaticZeroZeroTTrueNotSolvedByKinductionPlain}{Correct}{True}{Walltime}{Stdev}{145.5227444048886123919830606}%
\StoreBenchExecResult{PdrInvPathprograms}{KinductionDfStaticZeroZeroTTrueNotSolvedByKinductionPlain}{Wrong}{True}{Count}{}{0}%
\StoreBenchExecResult{PdrInvPathprograms}{KinductionDfStaticZeroZeroTTrueNotSolvedByKinductionPlain}{Wrong}{True}{Cputime}{}{0}%
\StoreBenchExecResult{PdrInvPathprograms}{KinductionDfStaticZeroZeroTTrueNotSolvedByKinductionPlain}{Wrong}{True}{Cputime}{Avg}{None}%
\StoreBenchExecResult{PdrInvPathprograms}{KinductionDfStaticZeroZeroTTrueNotSolvedByKinductionPlain}{Wrong}{True}{Cputime}{Median}{None}%
\StoreBenchExecResult{PdrInvPathprograms}{KinductionDfStaticZeroZeroTTrueNotSolvedByKinductionPlain}{Wrong}{True}{Cputime}{Min}{None}%
\StoreBenchExecResult{PdrInvPathprograms}{KinductionDfStaticZeroZeroTTrueNotSolvedByKinductionPlain}{Wrong}{True}{Cputime}{Max}{None}%
\StoreBenchExecResult{PdrInvPathprograms}{KinductionDfStaticZeroZeroTTrueNotSolvedByKinductionPlain}{Wrong}{True}{Cputime}{Stdev}{None}%
\StoreBenchExecResult{PdrInvPathprograms}{KinductionDfStaticZeroZeroTTrueNotSolvedByKinductionPlain}{Wrong}{True}{Walltime}{}{0}%
\StoreBenchExecResult{PdrInvPathprograms}{KinductionDfStaticZeroZeroTTrueNotSolvedByKinductionPlain}{Wrong}{True}{Walltime}{Avg}{None}%
\StoreBenchExecResult{PdrInvPathprograms}{KinductionDfStaticZeroZeroTTrueNotSolvedByKinductionPlain}{Wrong}{True}{Walltime}{Median}{None}%
\StoreBenchExecResult{PdrInvPathprograms}{KinductionDfStaticZeroZeroTTrueNotSolvedByKinductionPlain}{Wrong}{True}{Walltime}{Min}{None}%
\StoreBenchExecResult{PdrInvPathprograms}{KinductionDfStaticZeroZeroTTrueNotSolvedByKinductionPlain}{Wrong}{True}{Walltime}{Max}{None}%
\StoreBenchExecResult{PdrInvPathprograms}{KinductionDfStaticZeroZeroTTrueNotSolvedByKinductionPlain}{Wrong}{True}{Walltime}{Stdev}{None}%
\StoreBenchExecResult{PdrInvPathprograms}{KinductionDfStaticZeroZeroTTrueNotSolvedByKinductionPlain}{Error}{}{Count}{}{60}%
\StoreBenchExecResult{PdrInvPathprograms}{KinductionDfStaticZeroZeroTTrueNotSolvedByKinductionPlain}{Error}{}{Cputime}{}{53523.495394609}%
\StoreBenchExecResult{PdrInvPathprograms}{KinductionDfStaticZeroZeroTTrueNotSolvedByKinductionPlain}{Error}{}{Cputime}{Avg}{892.0582565768166666666666667}%
\StoreBenchExecResult{PdrInvPathprograms}{KinductionDfStaticZeroZeroTTrueNotSolvedByKinductionPlain}{Error}{}{Cputime}{Median}{903.244046556}%
\StoreBenchExecResult{PdrInvPathprograms}{KinductionDfStaticZeroZeroTTrueNotSolvedByKinductionPlain}{Error}{}{Cputime}{Min}{212.831758958}%
\StoreBenchExecResult{PdrInvPathprograms}{KinductionDfStaticZeroZeroTTrueNotSolvedByKinductionPlain}{Error}{}{Cputime}{Max}{1000.76279416}%
\StoreBenchExecResult{PdrInvPathprograms}{KinductionDfStaticZeroZeroTTrueNotSolvedByKinductionPlain}{Error}{}{Cputime}{Stdev}{93.83462126061011959890590529}%
\StoreBenchExecResult{PdrInvPathprograms}{KinductionDfStaticZeroZeroTTrueNotSolvedByKinductionPlain}{Error}{}{Walltime}{}{50772.399532797}%
\StoreBenchExecResult{PdrInvPathprograms}{KinductionDfStaticZeroZeroTTrueNotSolvedByKinductionPlain}{Error}{}{Walltime}{Avg}{846.20665887995}%
\StoreBenchExecResult{PdrInvPathprograms}{KinductionDfStaticZeroZeroTTrueNotSolvedByKinductionPlain}{Error}{}{Walltime}{Median}{887.519270420}%
\StoreBenchExecResult{PdrInvPathprograms}{KinductionDfStaticZeroZeroTTrueNotSolvedByKinductionPlain}{Error}{}{Walltime}{Min}{133.803637981}%
\StoreBenchExecResult{PdrInvPathprograms}{KinductionDfStaticZeroZeroTTrueNotSolvedByKinductionPlain}{Error}{}{Walltime}{Max}{898.954440117}%
\StoreBenchExecResult{PdrInvPathprograms}{KinductionDfStaticZeroZeroTTrueNotSolvedByKinductionPlain}{Error}{}{Walltime}{Stdev}{116.7211511228910938251545339}%
\StoreBenchExecResult{PdrInvPathprograms}{KinductionDfStaticZeroZeroTTrueNotSolvedByKinductionPlain}{Error}{OutOfJavaMemory}{Count}{}{1}%
\StoreBenchExecResult{PdrInvPathprograms}{KinductionDfStaticZeroZeroTTrueNotSolvedByKinductionPlain}{Error}{OutOfJavaMemory}{Cputime}{}{212.831758958}%
\StoreBenchExecResult{PdrInvPathprograms}{KinductionDfStaticZeroZeroTTrueNotSolvedByKinductionPlain}{Error}{OutOfJavaMemory}{Cputime}{Avg}{212.831758958}%
\StoreBenchExecResult{PdrInvPathprograms}{KinductionDfStaticZeroZeroTTrueNotSolvedByKinductionPlain}{Error}{OutOfJavaMemory}{Cputime}{Median}{212.831758958}%
\StoreBenchExecResult{PdrInvPathprograms}{KinductionDfStaticZeroZeroTTrueNotSolvedByKinductionPlain}{Error}{OutOfJavaMemory}{Cputime}{Min}{212.831758958}%
\StoreBenchExecResult{PdrInvPathprograms}{KinductionDfStaticZeroZeroTTrueNotSolvedByKinductionPlain}{Error}{OutOfJavaMemory}{Cputime}{Max}{212.831758958}%
\StoreBenchExecResult{PdrInvPathprograms}{KinductionDfStaticZeroZeroTTrueNotSolvedByKinductionPlain}{Error}{OutOfJavaMemory}{Cputime}{Stdev}{0E-9}%
\StoreBenchExecResult{PdrInvPathprograms}{KinductionDfStaticZeroZeroTTrueNotSolvedByKinductionPlain}{Error}{OutOfJavaMemory}{Walltime}{}{133.803637981}%
\StoreBenchExecResult{PdrInvPathprograms}{KinductionDfStaticZeroZeroTTrueNotSolvedByKinductionPlain}{Error}{OutOfJavaMemory}{Walltime}{Avg}{133.803637981}%
\StoreBenchExecResult{PdrInvPathprograms}{KinductionDfStaticZeroZeroTTrueNotSolvedByKinductionPlain}{Error}{OutOfJavaMemory}{Walltime}{Median}{133.803637981}%
\StoreBenchExecResult{PdrInvPathprograms}{KinductionDfStaticZeroZeroTTrueNotSolvedByKinductionPlain}{Error}{OutOfJavaMemory}{Walltime}{Min}{133.803637981}%
\StoreBenchExecResult{PdrInvPathprograms}{KinductionDfStaticZeroZeroTTrueNotSolvedByKinductionPlain}{Error}{OutOfJavaMemory}{Walltime}{Max}{133.803637981}%
\StoreBenchExecResult{PdrInvPathprograms}{KinductionDfStaticZeroZeroTTrueNotSolvedByKinductionPlain}{Error}{OutOfJavaMemory}{Walltime}{Stdev}{0E-9}%
\StoreBenchExecResult{PdrInvPathprograms}{KinductionDfStaticZeroZeroTTrueNotSolvedByKinductionPlain}{Error}{OutOfMemory}{Count}{}{4}%
\StoreBenchExecResult{PdrInvPathprograms}{KinductionDfStaticZeroZeroTTrueNotSolvedByKinductionPlain}{Error}{OutOfMemory}{Cputime}{}{3271.670550001}%
\StoreBenchExecResult{PdrInvPathprograms}{KinductionDfStaticZeroZeroTTrueNotSolvedByKinductionPlain}{Error}{OutOfMemory}{Cputime}{Avg}{817.91763750025}%
\StoreBenchExecResult{PdrInvPathprograms}{KinductionDfStaticZeroZeroTTrueNotSolvedByKinductionPlain}{Error}{OutOfMemory}{Cputime}{Median}{813.973233431}%
\StoreBenchExecResult{PdrInvPathprograms}{KinductionDfStaticZeroZeroTTrueNotSolvedByKinductionPlain}{Error}{OutOfMemory}{Cputime}{Min}{793.421967169}%
\StoreBenchExecResult{PdrInvPathprograms}{KinductionDfStaticZeroZeroTTrueNotSolvedByKinductionPlain}{Error}{OutOfMemory}{Cputime}{Max}{850.30211597}%
\StoreBenchExecResult{PdrInvPathprograms}{KinductionDfStaticZeroZeroTTrueNotSolvedByKinductionPlain}{Error}{OutOfMemory}{Cputime}{Stdev}{23.80100803383167630697836785}%
\StoreBenchExecResult{PdrInvPathprograms}{KinductionDfStaticZeroZeroTTrueNotSolvedByKinductionPlain}{Error}{OutOfMemory}{Walltime}{}{3206.224534035}%
\StoreBenchExecResult{PdrInvPathprograms}{KinductionDfStaticZeroZeroTTrueNotSolvedByKinductionPlain}{Error}{OutOfMemory}{Walltime}{Avg}{801.55613350875}%
\StoreBenchExecResult{PdrInvPathprograms}{KinductionDfStaticZeroZeroTTrueNotSolvedByKinductionPlain}{Error}{OutOfMemory}{Walltime}{Median}{798.862776637}%
\StoreBenchExecResult{PdrInvPathprograms}{KinductionDfStaticZeroZeroTTrueNotSolvedByKinductionPlain}{Error}{OutOfMemory}{Walltime}{Min}{776.4936409}%
\StoreBenchExecResult{PdrInvPathprograms}{KinductionDfStaticZeroZeroTTrueNotSolvedByKinductionPlain}{Error}{OutOfMemory}{Walltime}{Max}{832.005339861}%
\StoreBenchExecResult{PdrInvPathprograms}{KinductionDfStaticZeroZeroTTrueNotSolvedByKinductionPlain}{Error}{OutOfMemory}{Walltime}{Stdev}{23.09260564244950163147088329}%
\StoreBenchExecResult{PdrInvPathprograms}{KinductionDfStaticZeroZeroTTrueNotSolvedByKinductionPlain}{Error}{Timeout}{Count}{}{55}%
\StoreBenchExecResult{PdrInvPathprograms}{KinductionDfStaticZeroZeroTTrueNotSolvedByKinductionPlain}{Error}{Timeout}{Cputime}{}{50038.993085650}%
\StoreBenchExecResult{PdrInvPathprograms}{KinductionDfStaticZeroZeroTTrueNotSolvedByKinductionPlain}{Error}{Timeout}{Cputime}{Avg}{909.7998742845454545454545455}%
\StoreBenchExecResult{PdrInvPathprograms}{KinductionDfStaticZeroZeroTTrueNotSolvedByKinductionPlain}{Error}{Timeout}{Cputime}{Median}{903.543912582}%
\StoreBenchExecResult{PdrInvPathprograms}{KinductionDfStaticZeroZeroTTrueNotSolvedByKinductionPlain}{Error}{Timeout}{Cputime}{Min}{901.049511544}%
\StoreBenchExecResult{PdrInvPathprograms}{KinductionDfStaticZeroZeroTTrueNotSolvedByKinductionPlain}{Error}{Timeout}{Cputime}{Max}{1000.76279416}%
\StoreBenchExecResult{PdrInvPathprograms}{KinductionDfStaticZeroZeroTTrueNotSolvedByKinductionPlain}{Error}{Timeout}{Cputime}{Stdev}{21.48241619180090350773157245}%
\StoreBenchExecResult{PdrInvPathprograms}{KinductionDfStaticZeroZeroTTrueNotSolvedByKinductionPlain}{Error}{Timeout}{Walltime}{}{47432.371360781}%
\StoreBenchExecResult{PdrInvPathprograms}{KinductionDfStaticZeroZeroTTrueNotSolvedByKinductionPlain}{Error}{Timeout}{Walltime}{Avg}{862.4067520142}%
\StoreBenchExecResult{PdrInvPathprograms}{KinductionDfStaticZeroZeroTTrueNotSolvedByKinductionPlain}{Error}{Timeout}{Walltime}{Median}{887.822458029}%
\StoreBenchExecResult{PdrInvPathprograms}{KinductionDfStaticZeroZeroTTrueNotSolvedByKinductionPlain}{Error}{Timeout}{Walltime}{Min}{551.987067938}%
\StoreBenchExecResult{PdrInvPathprograms}{KinductionDfStaticZeroZeroTTrueNotSolvedByKinductionPlain}{Error}{Timeout}{Walltime}{Max}{898.954440117}%
\StoreBenchExecResult{PdrInvPathprograms}{KinductionDfStaticZeroZeroTTrueNotSolvedByKinductionPlain}{Error}{Timeout}{Walltime}{Stdev}{72.03148877216244357388896877}%
\providecommand\StoreBenchExecResult[7]{\expandafter\newcommand\csname#1#2#3#4#5#6\endcsname{#7}}%
\StoreBenchExecResult{PdrInvPathprograms}{KinductionDfStaticZeroOneTFTrueNotSolvedByKinductionPlain}{Total}{}{Count}{}{114}%
\StoreBenchExecResult{PdrInvPathprograms}{KinductionDfStaticZeroOneTFTrueNotSolvedByKinductionPlain}{Total}{}{Cputime}{}{62573.522186263}%
\StoreBenchExecResult{PdrInvPathprograms}{KinductionDfStaticZeroOneTFTrueNotSolvedByKinductionPlain}{Total}{}{Cputime}{Avg}{548.8905454935350877192982456}%
\StoreBenchExecResult{PdrInvPathprograms}{KinductionDfStaticZeroOneTFTrueNotSolvedByKinductionPlain}{Total}{}{Cputime}{Median}{885.341494683}%
\StoreBenchExecResult{PdrInvPathprograms}{KinductionDfStaticZeroOneTFTrueNotSolvedByKinductionPlain}{Total}{}{Cputime}{Min}{3.229539062}%
\StoreBenchExecResult{PdrInvPathprograms}{KinductionDfStaticZeroOneTFTrueNotSolvedByKinductionPlain}{Total}{}{Cputime}{Max}{923.06364486}%
\StoreBenchExecResult{PdrInvPathprograms}{KinductionDfStaticZeroOneTFTrueNotSolvedByKinductionPlain}{Total}{}{Cputime}{Stdev}{397.8880182626475416736998761}%
\StoreBenchExecResult{PdrInvPathprograms}{KinductionDfStaticZeroOneTFTrueNotSolvedByKinductionPlain}{Total}{}{Walltime}{}{54826.84929394840}%
\StoreBenchExecResult{PdrInvPathprograms}{KinductionDfStaticZeroOneTFTrueNotSolvedByKinductionPlain}{Total}{}{Walltime}{Avg}{480.9372745083192982456140351}%
\StoreBenchExecResult{PdrInvPathprograms}{KinductionDfStaticZeroOneTFTrueNotSolvedByKinductionPlain}{Total}{}{Walltime}{Median}{458.158100963}%
\StoreBenchExecResult{PdrInvPathprograms}{KinductionDfStaticZeroOneTFTrueNotSolvedByKinductionPlain}{Total}{}{Walltime}{Min}{1.76596307755}%
\StoreBenchExecResult{PdrInvPathprograms}{KinductionDfStaticZeroOneTFTrueNotSolvedByKinductionPlain}{Total}{}{Walltime}{Max}{907.747550011}%
\StoreBenchExecResult{PdrInvPathprograms}{KinductionDfStaticZeroOneTFTrueNotSolvedByKinductionPlain}{Total}{}{Walltime}{Stdev}{391.9833455111248539202385054}%
\StoreBenchExecResult{PdrInvPathprograms}{KinductionDfStaticZeroOneTFTrueNotSolvedByKinductionPlain}{Correct}{}{Count}{}{54}%
\StoreBenchExecResult{PdrInvPathprograms}{KinductionDfStaticZeroOneTFTrueNotSolvedByKinductionPlain}{Correct}{}{Cputime}{}{9384.089778303}%
\StoreBenchExecResult{PdrInvPathprograms}{KinductionDfStaticZeroOneTFTrueNotSolvedByKinductionPlain}{Correct}{}{Cputime}{Avg}{173.7794403389444444444444444}%
\StoreBenchExecResult{PdrInvPathprograms}{KinductionDfStaticZeroOneTFTrueNotSolvedByKinductionPlain}{Correct}{}{Cputime}{Median}{33.1220371265}%
\StoreBenchExecResult{PdrInvPathprograms}{KinductionDfStaticZeroOneTFTrueNotSolvedByKinductionPlain}{Correct}{}{Cputime}{Min}{3.229539062}%
\StoreBenchExecResult{PdrInvPathprograms}{KinductionDfStaticZeroOneTFTrueNotSolvedByKinductionPlain}{Correct}{}{Cputime}{Max}{869.696698055}%
\StoreBenchExecResult{PdrInvPathprograms}{KinductionDfStaticZeroOneTFTrueNotSolvedByKinductionPlain}{Correct}{}{Cputime}{Stdev}{237.8683490144561142699405214}%
\StoreBenchExecResult{PdrInvPathprograms}{KinductionDfStaticZeroOneTFTrueNotSolvedByKinductionPlain}{Correct}{}{Walltime}{}{5471.19169569040}%
\StoreBenchExecResult{PdrInvPathprograms}{KinductionDfStaticZeroOneTFTrueNotSolvedByKinductionPlain}{Correct}{}{Walltime}{Avg}{101.3183647350074074074074074}%
\StoreBenchExecResult{PdrInvPathprograms}{KinductionDfStaticZeroOneTFTrueNotSolvedByKinductionPlain}{Correct}{}{Walltime}{Median}{16.78567910195}%
\StoreBenchExecResult{PdrInvPathprograms}{KinductionDfStaticZeroOneTFTrueNotSolvedByKinductionPlain}{Correct}{}{Walltime}{Min}{1.76596307755}%
\StoreBenchExecResult{PdrInvPathprograms}{KinductionDfStaticZeroOneTFTrueNotSolvedByKinductionPlain}{Correct}{}{Walltime}{Max}{669.033270836}%
\StoreBenchExecResult{PdrInvPathprograms}{KinductionDfStaticZeroOneTFTrueNotSolvedByKinductionPlain}{Correct}{}{Walltime}{Stdev}{146.9996278940373511529106148}%
\StoreBenchExecResult{PdrInvPathprograms}{KinductionDfStaticZeroOneTFTrueNotSolvedByKinductionPlain}{Correct}{True}{Count}{}{54}%
\StoreBenchExecResult{PdrInvPathprograms}{KinductionDfStaticZeroOneTFTrueNotSolvedByKinductionPlain}{Correct}{True}{Cputime}{}{9384.089778303}%
\StoreBenchExecResult{PdrInvPathprograms}{KinductionDfStaticZeroOneTFTrueNotSolvedByKinductionPlain}{Correct}{True}{Cputime}{Avg}{173.7794403389444444444444444}%
\StoreBenchExecResult{PdrInvPathprograms}{KinductionDfStaticZeroOneTFTrueNotSolvedByKinductionPlain}{Correct}{True}{Cputime}{Median}{33.1220371265}%
\StoreBenchExecResult{PdrInvPathprograms}{KinductionDfStaticZeroOneTFTrueNotSolvedByKinductionPlain}{Correct}{True}{Cputime}{Min}{3.229539062}%
\StoreBenchExecResult{PdrInvPathprograms}{KinductionDfStaticZeroOneTFTrueNotSolvedByKinductionPlain}{Correct}{True}{Cputime}{Max}{869.696698055}%
\StoreBenchExecResult{PdrInvPathprograms}{KinductionDfStaticZeroOneTFTrueNotSolvedByKinductionPlain}{Correct}{True}{Cputime}{Stdev}{237.8683490144561142699405214}%
\StoreBenchExecResult{PdrInvPathprograms}{KinductionDfStaticZeroOneTFTrueNotSolvedByKinductionPlain}{Correct}{True}{Walltime}{}{5471.19169569040}%
\StoreBenchExecResult{PdrInvPathprograms}{KinductionDfStaticZeroOneTFTrueNotSolvedByKinductionPlain}{Correct}{True}{Walltime}{Avg}{101.3183647350074074074074074}%
\StoreBenchExecResult{PdrInvPathprograms}{KinductionDfStaticZeroOneTFTrueNotSolvedByKinductionPlain}{Correct}{True}{Walltime}{Median}{16.78567910195}%
\StoreBenchExecResult{PdrInvPathprograms}{KinductionDfStaticZeroOneTFTrueNotSolvedByKinductionPlain}{Correct}{True}{Walltime}{Min}{1.76596307755}%
\StoreBenchExecResult{PdrInvPathprograms}{KinductionDfStaticZeroOneTFTrueNotSolvedByKinductionPlain}{Correct}{True}{Walltime}{Max}{669.033270836}%
\StoreBenchExecResult{PdrInvPathprograms}{KinductionDfStaticZeroOneTFTrueNotSolvedByKinductionPlain}{Correct}{True}{Walltime}{Stdev}{146.9996278940373511529106148}%
\StoreBenchExecResult{PdrInvPathprograms}{KinductionDfStaticZeroOneTFTrueNotSolvedByKinductionPlain}{Wrong}{True}{Count}{}{0}%
\StoreBenchExecResult{PdrInvPathprograms}{KinductionDfStaticZeroOneTFTrueNotSolvedByKinductionPlain}{Wrong}{True}{Cputime}{}{0}%
\StoreBenchExecResult{PdrInvPathprograms}{KinductionDfStaticZeroOneTFTrueNotSolvedByKinductionPlain}{Wrong}{True}{Cputime}{Avg}{None}%
\StoreBenchExecResult{PdrInvPathprograms}{KinductionDfStaticZeroOneTFTrueNotSolvedByKinductionPlain}{Wrong}{True}{Cputime}{Median}{None}%
\StoreBenchExecResult{PdrInvPathprograms}{KinductionDfStaticZeroOneTFTrueNotSolvedByKinductionPlain}{Wrong}{True}{Cputime}{Min}{None}%
\StoreBenchExecResult{PdrInvPathprograms}{KinductionDfStaticZeroOneTFTrueNotSolvedByKinductionPlain}{Wrong}{True}{Cputime}{Max}{None}%
\StoreBenchExecResult{PdrInvPathprograms}{KinductionDfStaticZeroOneTFTrueNotSolvedByKinductionPlain}{Wrong}{True}{Cputime}{Stdev}{None}%
\StoreBenchExecResult{PdrInvPathprograms}{KinductionDfStaticZeroOneTFTrueNotSolvedByKinductionPlain}{Wrong}{True}{Walltime}{}{0}%
\StoreBenchExecResult{PdrInvPathprograms}{KinductionDfStaticZeroOneTFTrueNotSolvedByKinductionPlain}{Wrong}{True}{Walltime}{Avg}{None}%
\StoreBenchExecResult{PdrInvPathprograms}{KinductionDfStaticZeroOneTFTrueNotSolvedByKinductionPlain}{Wrong}{True}{Walltime}{Median}{None}%
\StoreBenchExecResult{PdrInvPathprograms}{KinductionDfStaticZeroOneTFTrueNotSolvedByKinductionPlain}{Wrong}{True}{Walltime}{Min}{None}%
\StoreBenchExecResult{PdrInvPathprograms}{KinductionDfStaticZeroOneTFTrueNotSolvedByKinductionPlain}{Wrong}{True}{Walltime}{Max}{None}%
\StoreBenchExecResult{PdrInvPathprograms}{KinductionDfStaticZeroOneTFTrueNotSolvedByKinductionPlain}{Wrong}{True}{Walltime}{Stdev}{None}%
\StoreBenchExecResult{PdrInvPathprograms}{KinductionDfStaticZeroOneTFTrueNotSolvedByKinductionPlain}{Error}{}{Count}{}{60}%
\StoreBenchExecResult{PdrInvPathprograms}{KinductionDfStaticZeroOneTFTrueNotSolvedByKinductionPlain}{Error}{}{Cputime}{}{53189.432407960}%
\StoreBenchExecResult{PdrInvPathprograms}{KinductionDfStaticZeroOneTFTrueNotSolvedByKinductionPlain}{Error}{}{Cputime}{Avg}{886.4905401326666666666666667}%
\StoreBenchExecResult{PdrInvPathprograms}{KinductionDfStaticZeroOneTFTrueNotSolvedByKinductionPlain}{Error}{}{Cputime}{Median}{903.2029294235}%
\StoreBenchExecResult{PdrInvPathprograms}{KinductionDfStaticZeroOneTFTrueNotSolvedByKinductionPlain}{Error}{}{Cputime}{Min}{249.03118587}%
\StoreBenchExecResult{PdrInvPathprograms}{KinductionDfStaticZeroOneTFTrueNotSolvedByKinductionPlain}{Error}{}{Cputime}{Max}{923.06364486}%
\StoreBenchExecResult{PdrInvPathprograms}{KinductionDfStaticZeroOneTFTrueNotSolvedByKinductionPlain}{Error}{}{Cputime}{Stdev}{96.24854731407224913592640801}%
\StoreBenchExecResult{PdrInvPathprograms}{KinductionDfStaticZeroOneTFTrueNotSolvedByKinductionPlain}{Error}{}{Walltime}{}{49355.657598258}%
\StoreBenchExecResult{PdrInvPathprograms}{KinductionDfStaticZeroOneTFTrueNotSolvedByKinductionPlain}{Error}{}{Walltime}{Avg}{822.5942933043}%
\StoreBenchExecResult{PdrInvPathprograms}{KinductionDfStaticZeroOneTFTrueNotSolvedByKinductionPlain}{Error}{}{Walltime}{Median}{888.074846506}%
\StoreBenchExecResult{PdrInvPathprograms}{KinductionDfStaticZeroOneTFTrueNotSolvedByKinductionPlain}{Error}{}{Walltime}{Min}{152.069455862}%
\StoreBenchExecResult{PdrInvPathprograms}{KinductionDfStaticZeroOneTFTrueNotSolvedByKinductionPlain}{Error}{}{Walltime}{Max}{907.747550011}%
\StoreBenchExecResult{PdrInvPathprograms}{KinductionDfStaticZeroOneTFTrueNotSolvedByKinductionPlain}{Error}{}{Walltime}{Stdev}{161.4305019976220256117186223}%
\StoreBenchExecResult{PdrInvPathprograms}{KinductionDfStaticZeroOneTFTrueNotSolvedByKinductionPlain}{Error}{OutOfJavaMemory}{Count}{}{2}%
\StoreBenchExecResult{PdrInvPathprograms}{KinductionDfStaticZeroOneTFTrueNotSolvedByKinductionPlain}{Error}{OutOfJavaMemory}{Cputime}{}{782.643199003}%
\StoreBenchExecResult{PdrInvPathprograms}{KinductionDfStaticZeroOneTFTrueNotSolvedByKinductionPlain}{Error}{OutOfJavaMemory}{Cputime}{Avg}{391.3215995015}%
\StoreBenchExecResult{PdrInvPathprograms}{KinductionDfStaticZeroOneTFTrueNotSolvedByKinductionPlain}{Error}{OutOfJavaMemory}{Cputime}{Median}{391.3215995015}%
\StoreBenchExecResult{PdrInvPathprograms}{KinductionDfStaticZeroOneTFTrueNotSolvedByKinductionPlain}{Error}{OutOfJavaMemory}{Cputime}{Min}{249.03118587}%
\StoreBenchExecResult{PdrInvPathprograms}{KinductionDfStaticZeroOneTFTrueNotSolvedByKinductionPlain}{Error}{OutOfJavaMemory}{Cputime}{Max}{533.612013133}%
\StoreBenchExecResult{PdrInvPathprograms}{KinductionDfStaticZeroOneTFTrueNotSolvedByKinductionPlain}{Error}{OutOfJavaMemory}{Cputime}{Stdev}{142.2904136315}%
\StoreBenchExecResult{PdrInvPathprograms}{KinductionDfStaticZeroOneTFTrueNotSolvedByKinductionPlain}{Error}{OutOfJavaMemory}{Walltime}{}{425.709290981}%
\StoreBenchExecResult{PdrInvPathprograms}{KinductionDfStaticZeroOneTFTrueNotSolvedByKinductionPlain}{Error}{OutOfJavaMemory}{Walltime}{Avg}{212.8546454905}%
\StoreBenchExecResult{PdrInvPathprograms}{KinductionDfStaticZeroOneTFTrueNotSolvedByKinductionPlain}{Error}{OutOfJavaMemory}{Walltime}{Median}{212.8546454905}%
\StoreBenchExecResult{PdrInvPathprograms}{KinductionDfStaticZeroOneTFTrueNotSolvedByKinductionPlain}{Error}{OutOfJavaMemory}{Walltime}{Min}{152.069455862}%
\StoreBenchExecResult{PdrInvPathprograms}{KinductionDfStaticZeroOneTFTrueNotSolvedByKinductionPlain}{Error}{OutOfJavaMemory}{Walltime}{Max}{273.639835119}%
\StoreBenchExecResult{PdrInvPathprograms}{KinductionDfStaticZeroOneTFTrueNotSolvedByKinductionPlain}{Error}{OutOfJavaMemory}{Walltime}{Stdev}{60.7851896285}%
\StoreBenchExecResult{PdrInvPathprograms}{KinductionDfStaticZeroOneTFTrueNotSolvedByKinductionPlain}{Error}{OutOfMemory}{Count}{}{1}%
\StoreBenchExecResult{PdrInvPathprograms}{KinductionDfStaticZeroOneTFTrueNotSolvedByKinductionPlain}{Error}{OutOfMemory}{Cputime}{}{821.233264578}%
\StoreBenchExecResult{PdrInvPathprograms}{KinductionDfStaticZeroOneTFTrueNotSolvedByKinductionPlain}{Error}{OutOfMemory}{Cputime}{Avg}{821.233264578}%
\StoreBenchExecResult{PdrInvPathprograms}{KinductionDfStaticZeroOneTFTrueNotSolvedByKinductionPlain}{Error}{OutOfMemory}{Cputime}{Median}{821.233264578}%
\StoreBenchExecResult{PdrInvPathprograms}{KinductionDfStaticZeroOneTFTrueNotSolvedByKinductionPlain}{Error}{OutOfMemory}{Cputime}{Min}{821.233264578}%
\StoreBenchExecResult{PdrInvPathprograms}{KinductionDfStaticZeroOneTFTrueNotSolvedByKinductionPlain}{Error}{OutOfMemory}{Cputime}{Max}{821.233264578}%
\StoreBenchExecResult{PdrInvPathprograms}{KinductionDfStaticZeroOneTFTrueNotSolvedByKinductionPlain}{Error}{OutOfMemory}{Cputime}{Stdev}{0E-9}%
\StoreBenchExecResult{PdrInvPathprograms}{KinductionDfStaticZeroOneTFTrueNotSolvedByKinductionPlain}{Error}{OutOfMemory}{Walltime}{}{806.828974962}%
\StoreBenchExecResult{PdrInvPathprograms}{KinductionDfStaticZeroOneTFTrueNotSolvedByKinductionPlain}{Error}{OutOfMemory}{Walltime}{Avg}{806.828974962}%
\StoreBenchExecResult{PdrInvPathprograms}{KinductionDfStaticZeroOneTFTrueNotSolvedByKinductionPlain}{Error}{OutOfMemory}{Walltime}{Median}{806.828974962}%
\StoreBenchExecResult{PdrInvPathprograms}{KinductionDfStaticZeroOneTFTrueNotSolvedByKinductionPlain}{Error}{OutOfMemory}{Walltime}{Min}{806.828974962}%
\StoreBenchExecResult{PdrInvPathprograms}{KinductionDfStaticZeroOneTFTrueNotSolvedByKinductionPlain}{Error}{OutOfMemory}{Walltime}{Max}{806.828974962}%
\StoreBenchExecResult{PdrInvPathprograms}{KinductionDfStaticZeroOneTFTrueNotSolvedByKinductionPlain}{Error}{OutOfMemory}{Walltime}{Stdev}{0E-9}%
\StoreBenchExecResult{PdrInvPathprograms}{KinductionDfStaticZeroOneTFTrueNotSolvedByKinductionPlain}{Error}{Timeout}{Count}{}{57}%
\StoreBenchExecResult{PdrInvPathprograms}{KinductionDfStaticZeroOneTFTrueNotSolvedByKinductionPlain}{Error}{Timeout}{Cputime}{}{51585.555944379}%
\StoreBenchExecResult{PdrInvPathprograms}{KinductionDfStaticZeroOneTFTrueNotSolvedByKinductionPlain}{Error}{Timeout}{Cputime}{Avg}{905.0097534101578947368421053}%
\StoreBenchExecResult{PdrInvPathprograms}{KinductionDfStaticZeroOneTFTrueNotSolvedByKinductionPlain}{Error}{Timeout}{Cputime}{Median}{903.527506903}%
\StoreBenchExecResult{PdrInvPathprograms}{KinductionDfStaticZeroOneTFTrueNotSolvedByKinductionPlain}{Error}{Timeout}{Cputime}{Min}{900.986291311}%
\StoreBenchExecResult{PdrInvPathprograms}{KinductionDfStaticZeroOneTFTrueNotSolvedByKinductionPlain}{Error}{Timeout}{Cputime}{Max}{923.06364486}%
\StoreBenchExecResult{PdrInvPathprograms}{KinductionDfStaticZeroOneTFTrueNotSolvedByKinductionPlain}{Error}{Timeout}{Cputime}{Stdev}{4.476026968553235933003574338}%
\StoreBenchExecResult{PdrInvPathprograms}{KinductionDfStaticZeroOneTFTrueNotSolvedByKinductionPlain}{Error}{Timeout}{Walltime}{}{48123.119332315}%
\StoreBenchExecResult{PdrInvPathprograms}{KinductionDfStaticZeroOneTFTrueNotSolvedByKinductionPlain}{Error}{Timeout}{Walltime}{Avg}{844.2652514441228070175438596}%
\StoreBenchExecResult{PdrInvPathprograms}{KinductionDfStaticZeroOneTFTrueNotSolvedByKinductionPlain}{Error}{Timeout}{Walltime}{Median}{888.284183979}%
\StoreBenchExecResult{PdrInvPathprograms}{KinductionDfStaticZeroOneTFTrueNotSolvedByKinductionPlain}{Error}{Timeout}{Walltime}{Min}{452.436802149}%
\StoreBenchExecResult{PdrInvPathprograms}{KinductionDfStaticZeroOneTFTrueNotSolvedByKinductionPlain}{Error}{Timeout}{Walltime}{Max}{907.747550011}%
\StoreBenchExecResult{PdrInvPathprograms}{KinductionDfStaticZeroOneTFTrueNotSolvedByKinductionPlain}{Error}{Timeout}{Walltime}{Stdev}{117.3999300740622662331248081}%
\providecommand\StoreBenchExecResult[7]{\expandafter\newcommand\csname#1#2#3#4#5#6\endcsname{#7}}%
\StoreBenchExecResult{PdrInvPathprograms}{KinductionDfStaticZeroOneTTTrueNotSolvedByKinductionPlain}{Total}{}{Count}{}{114}%
\StoreBenchExecResult{PdrInvPathprograms}{KinductionDfStaticZeroOneTTTrueNotSolvedByKinductionPlain}{Total}{}{Cputime}{}{59119.481402288}%
\StoreBenchExecResult{PdrInvPathprograms}{KinductionDfStaticZeroOneTTTrueNotSolvedByKinductionPlain}{Total}{}{Cputime}{Avg}{518.5919421253333333333333333}%
\StoreBenchExecResult{PdrInvPathprograms}{KinductionDfStaticZeroOneTTTrueNotSolvedByKinductionPlain}{Total}{}{Cputime}{Median}{661.304196262}%
\StoreBenchExecResult{PdrInvPathprograms}{KinductionDfStaticZeroOneTTTrueNotSolvedByKinductionPlain}{Total}{}{Cputime}{Min}{3.235300754}%
\StoreBenchExecResult{PdrInvPathprograms}{KinductionDfStaticZeroOneTTTrueNotSolvedByKinductionPlain}{Total}{}{Cputime}{Max}{925.230062222}%
\StoreBenchExecResult{PdrInvPathprograms}{KinductionDfStaticZeroOneTTTrueNotSolvedByKinductionPlain}{Total}{}{Cputime}{Stdev}{404.8653091502951805363279309}%
\StoreBenchExecResult{PdrInvPathprograms}{KinductionDfStaticZeroOneTTTrueNotSolvedByKinductionPlain}{Total}{}{Walltime}{}{51360.06629657774}%
\StoreBenchExecResult{PdrInvPathprograms}{KinductionDfStaticZeroOneTTTrueNotSolvedByKinductionPlain}{Total}{}{Walltime}{Avg}{450.5268973384012280701754386}%
\StoreBenchExecResult{PdrInvPathprograms}{KinductionDfStaticZeroOneTTTrueNotSolvedByKinductionPlain}{Total}{}{Walltime}{Median}{416.040808916}%
\StoreBenchExecResult{PdrInvPathprograms}{KinductionDfStaticZeroOneTTTrueNotSolvedByKinductionPlain}{Total}{}{Walltime}{Min}{1.78474998474}%
\StoreBenchExecResult{PdrInvPathprograms}{KinductionDfStaticZeroOneTTTrueNotSolvedByKinductionPlain}{Total}{}{Walltime}{Max}{901.014378071}%
\StoreBenchExecResult{PdrInvPathprograms}{KinductionDfStaticZeroOneTTTrueNotSolvedByKinductionPlain}{Total}{}{Walltime}{Stdev}{393.3238620249297958871134663}%
\StoreBenchExecResult{PdrInvPathprograms}{KinductionDfStaticZeroOneTTTrueNotSolvedByKinductionPlain}{Correct}{}{Count}{}{57}%
\StoreBenchExecResult{PdrInvPathprograms}{KinductionDfStaticZeroOneTTTrueNotSolvedByKinductionPlain}{Correct}{}{Cputime}{}{8591.515383151}%
\StoreBenchExecResult{PdrInvPathprograms}{KinductionDfStaticZeroOneTTTrueNotSolvedByKinductionPlain}{Correct}{}{Cputime}{Avg}{150.7283400552807017543859649}%
\StoreBenchExecResult{PdrInvPathprograms}{KinductionDfStaticZeroOneTTTrueNotSolvedByKinductionPlain}{Correct}{}{Cputime}{Median}{22.576946887}%
\StoreBenchExecResult{PdrInvPathprograms}{KinductionDfStaticZeroOneTTTrueNotSolvedByKinductionPlain}{Correct}{}{Cputime}{Min}{3.235300754}%
\StoreBenchExecResult{PdrInvPathprograms}{KinductionDfStaticZeroOneTTTrueNotSolvedByKinductionPlain}{Correct}{}{Cputime}{Max}{899.345297569}%
\StoreBenchExecResult{PdrInvPathprograms}{KinductionDfStaticZeroOneTTTrueNotSolvedByKinductionPlain}{Correct}{}{Cputime}{Stdev}{219.0767349914196701726891414}%
\StoreBenchExecResult{PdrInvPathprograms}{KinductionDfStaticZeroOneTTTrueNotSolvedByKinductionPlain}{Correct}{}{Walltime}{}{5085.44161391274}%
\StoreBenchExecResult{PdrInvPathprograms}{KinductionDfStaticZeroOneTTTrueNotSolvedByKinductionPlain}{Correct}{}{Walltime}{Avg}{89.21827392829368421052631579}%
\StoreBenchExecResult{PdrInvPathprograms}{KinductionDfStaticZeroOneTTTrueNotSolvedByKinductionPlain}{Correct}{}{Walltime}{Median}{11.500025034}%
\StoreBenchExecResult{PdrInvPathprograms}{KinductionDfStaticZeroOneTTTrueNotSolvedByKinductionPlain}{Correct}{}{Walltime}{Min}{1.78474998474}%
\StoreBenchExecResult{PdrInvPathprograms}{KinductionDfStaticZeroOneTTTrueNotSolvedByKinductionPlain}{Correct}{}{Walltime}{Max}{687.831533909}%
\StoreBenchExecResult{PdrInvPathprograms}{KinductionDfStaticZeroOneTTTrueNotSolvedByKinductionPlain}{Correct}{}{Walltime}{Stdev}{139.7267341353487106284029688}%
\StoreBenchExecResult{PdrInvPathprograms}{KinductionDfStaticZeroOneTTTrueNotSolvedByKinductionPlain}{Correct}{True}{Count}{}{57}%
\StoreBenchExecResult{PdrInvPathprograms}{KinductionDfStaticZeroOneTTTrueNotSolvedByKinductionPlain}{Correct}{True}{Cputime}{}{8591.515383151}%
\StoreBenchExecResult{PdrInvPathprograms}{KinductionDfStaticZeroOneTTTrueNotSolvedByKinductionPlain}{Correct}{True}{Cputime}{Avg}{150.7283400552807017543859649}%
\StoreBenchExecResult{PdrInvPathprograms}{KinductionDfStaticZeroOneTTTrueNotSolvedByKinductionPlain}{Correct}{True}{Cputime}{Median}{22.576946887}%
\StoreBenchExecResult{PdrInvPathprograms}{KinductionDfStaticZeroOneTTTrueNotSolvedByKinductionPlain}{Correct}{True}{Cputime}{Min}{3.235300754}%
\StoreBenchExecResult{PdrInvPathprograms}{KinductionDfStaticZeroOneTTTrueNotSolvedByKinductionPlain}{Correct}{True}{Cputime}{Max}{899.345297569}%
\StoreBenchExecResult{PdrInvPathprograms}{KinductionDfStaticZeroOneTTTrueNotSolvedByKinductionPlain}{Correct}{True}{Cputime}{Stdev}{219.0767349914196701726891414}%
\StoreBenchExecResult{PdrInvPathprograms}{KinductionDfStaticZeroOneTTTrueNotSolvedByKinductionPlain}{Correct}{True}{Walltime}{}{5085.44161391274}%
\StoreBenchExecResult{PdrInvPathprograms}{KinductionDfStaticZeroOneTTTrueNotSolvedByKinductionPlain}{Correct}{True}{Walltime}{Avg}{89.21827392829368421052631579}%
\StoreBenchExecResult{PdrInvPathprograms}{KinductionDfStaticZeroOneTTTrueNotSolvedByKinductionPlain}{Correct}{True}{Walltime}{Median}{11.500025034}%
\StoreBenchExecResult{PdrInvPathprograms}{KinductionDfStaticZeroOneTTTrueNotSolvedByKinductionPlain}{Correct}{True}{Walltime}{Min}{1.78474998474}%
\StoreBenchExecResult{PdrInvPathprograms}{KinductionDfStaticZeroOneTTTrueNotSolvedByKinductionPlain}{Correct}{True}{Walltime}{Max}{687.831533909}%
\StoreBenchExecResult{PdrInvPathprograms}{KinductionDfStaticZeroOneTTTrueNotSolvedByKinductionPlain}{Correct}{True}{Walltime}{Stdev}{139.7267341353487106284029688}%
\StoreBenchExecResult{PdrInvPathprograms}{KinductionDfStaticZeroOneTTTrueNotSolvedByKinductionPlain}{Wrong}{True}{Count}{}{0}%
\StoreBenchExecResult{PdrInvPathprograms}{KinductionDfStaticZeroOneTTTrueNotSolvedByKinductionPlain}{Wrong}{True}{Cputime}{}{0}%
\StoreBenchExecResult{PdrInvPathprograms}{KinductionDfStaticZeroOneTTTrueNotSolvedByKinductionPlain}{Wrong}{True}{Cputime}{Avg}{None}%
\StoreBenchExecResult{PdrInvPathprograms}{KinductionDfStaticZeroOneTTTrueNotSolvedByKinductionPlain}{Wrong}{True}{Cputime}{Median}{None}%
\StoreBenchExecResult{PdrInvPathprograms}{KinductionDfStaticZeroOneTTTrueNotSolvedByKinductionPlain}{Wrong}{True}{Cputime}{Min}{None}%
\StoreBenchExecResult{PdrInvPathprograms}{KinductionDfStaticZeroOneTTTrueNotSolvedByKinductionPlain}{Wrong}{True}{Cputime}{Max}{None}%
\StoreBenchExecResult{PdrInvPathprograms}{KinductionDfStaticZeroOneTTTrueNotSolvedByKinductionPlain}{Wrong}{True}{Cputime}{Stdev}{None}%
\StoreBenchExecResult{PdrInvPathprograms}{KinductionDfStaticZeroOneTTTrueNotSolvedByKinductionPlain}{Wrong}{True}{Walltime}{}{0}%
\StoreBenchExecResult{PdrInvPathprograms}{KinductionDfStaticZeroOneTTTrueNotSolvedByKinductionPlain}{Wrong}{True}{Walltime}{Avg}{None}%
\StoreBenchExecResult{PdrInvPathprograms}{KinductionDfStaticZeroOneTTTrueNotSolvedByKinductionPlain}{Wrong}{True}{Walltime}{Median}{None}%
\StoreBenchExecResult{PdrInvPathprograms}{KinductionDfStaticZeroOneTTTrueNotSolvedByKinductionPlain}{Wrong}{True}{Walltime}{Min}{None}%
\StoreBenchExecResult{PdrInvPathprograms}{KinductionDfStaticZeroOneTTTrueNotSolvedByKinductionPlain}{Wrong}{True}{Walltime}{Max}{None}%
\StoreBenchExecResult{PdrInvPathprograms}{KinductionDfStaticZeroOneTTTrueNotSolvedByKinductionPlain}{Wrong}{True}{Walltime}{Stdev}{None}%
\StoreBenchExecResult{PdrInvPathprograms}{KinductionDfStaticZeroOneTTTrueNotSolvedByKinductionPlain}{Error}{}{Count}{}{57}%
\StoreBenchExecResult{PdrInvPathprograms}{KinductionDfStaticZeroOneTTTrueNotSolvedByKinductionPlain}{Error}{}{Cputime}{}{50527.966019137}%
\StoreBenchExecResult{PdrInvPathprograms}{KinductionDfStaticZeroOneTTTrueNotSolvedByKinductionPlain}{Error}{}{Cputime}{Avg}{886.4555441953859649122807018}%
\StoreBenchExecResult{PdrInvPathprograms}{KinductionDfStaticZeroOneTTTrueNotSolvedByKinductionPlain}{Error}{}{Cputime}{Median}{903.323183726}%
\StoreBenchExecResult{PdrInvPathprograms}{KinductionDfStaticZeroOneTTTrueNotSolvedByKinductionPlain}{Error}{}{Cputime}{Min}{255.489342379}%
\StoreBenchExecResult{PdrInvPathprograms}{KinductionDfStaticZeroOneTTTrueNotSolvedByKinductionPlain}{Error}{}{Cputime}{Max}{925.230062222}%
\StoreBenchExecResult{PdrInvPathprograms}{KinductionDfStaticZeroOneTTTrueNotSolvedByKinductionPlain}{Error}{}{Cputime}{Stdev}{95.86428863930194329261261275}%
\StoreBenchExecResult{PdrInvPathprograms}{KinductionDfStaticZeroOneTTTrueNotSolvedByKinductionPlain}{Error}{}{Walltime}{}{46274.624682665}%
\StoreBenchExecResult{PdrInvPathprograms}{KinductionDfStaticZeroOneTTTrueNotSolvedByKinductionPlain}{Error}{}{Walltime}{Avg}{811.8355207485087719298245614}%
\StoreBenchExecResult{PdrInvPathprograms}{KinductionDfStaticZeroOneTTTrueNotSolvedByKinductionPlain}{Error}{}{Walltime}{Median}{887.547874928}%
\StoreBenchExecResult{PdrInvPathprograms}{KinductionDfStaticZeroOneTTTrueNotSolvedByKinductionPlain}{Error}{}{Walltime}{Min}{155.164362907}%
\StoreBenchExecResult{PdrInvPathprograms}{KinductionDfStaticZeroOneTTTrueNotSolvedByKinductionPlain}{Error}{}{Walltime}{Max}{901.014378071}%
\StoreBenchExecResult{PdrInvPathprograms}{KinductionDfStaticZeroOneTTTrueNotSolvedByKinductionPlain}{Error}{}{Walltime}{Stdev}{169.6936001835890499459134925}%
\StoreBenchExecResult{PdrInvPathprograms}{KinductionDfStaticZeroOneTTTrueNotSolvedByKinductionPlain}{Error}{OutOfJavaMemory}{Count}{}{2}%
\StoreBenchExecResult{PdrInvPathprograms}{KinductionDfStaticZeroOneTTTrueNotSolvedByKinductionPlain}{Error}{OutOfJavaMemory}{Cputime}{}{816.083683080}%
\StoreBenchExecResult{PdrInvPathprograms}{KinductionDfStaticZeroOneTTTrueNotSolvedByKinductionPlain}{Error}{OutOfJavaMemory}{Cputime}{Avg}{408.041841540}%
\StoreBenchExecResult{PdrInvPathprograms}{KinductionDfStaticZeroOneTTTrueNotSolvedByKinductionPlain}{Error}{OutOfJavaMemory}{Cputime}{Median}{408.041841540}%
\StoreBenchExecResult{PdrInvPathprograms}{KinductionDfStaticZeroOneTTTrueNotSolvedByKinductionPlain}{Error}{OutOfJavaMemory}{Cputime}{Min}{255.489342379}%
\StoreBenchExecResult{PdrInvPathprograms}{KinductionDfStaticZeroOneTTTrueNotSolvedByKinductionPlain}{Error}{OutOfJavaMemory}{Cputime}{Max}{560.594340701}%
\StoreBenchExecResult{PdrInvPathprograms}{KinductionDfStaticZeroOneTTTrueNotSolvedByKinductionPlain}{Error}{OutOfJavaMemory}{Cputime}{Stdev}{152.552499161}%
\StoreBenchExecResult{PdrInvPathprograms}{KinductionDfStaticZeroOneTTTrueNotSolvedByKinductionPlain}{Error}{OutOfJavaMemory}{Walltime}{}{442.774129867}%
\StoreBenchExecResult{PdrInvPathprograms}{KinductionDfStaticZeroOneTTTrueNotSolvedByKinductionPlain}{Error}{OutOfJavaMemory}{Walltime}{Avg}{221.3870649335}%
\StoreBenchExecResult{PdrInvPathprograms}{KinductionDfStaticZeroOneTTTrueNotSolvedByKinductionPlain}{Error}{OutOfJavaMemory}{Walltime}{Median}{221.3870649335}%
\StoreBenchExecResult{PdrInvPathprograms}{KinductionDfStaticZeroOneTTTrueNotSolvedByKinductionPlain}{Error}{OutOfJavaMemory}{Walltime}{Min}{155.164362907}%
\StoreBenchExecResult{PdrInvPathprograms}{KinductionDfStaticZeroOneTTTrueNotSolvedByKinductionPlain}{Error}{OutOfJavaMemory}{Walltime}{Max}{287.60976696}%
\StoreBenchExecResult{PdrInvPathprograms}{KinductionDfStaticZeroOneTTTrueNotSolvedByKinductionPlain}{Error}{OutOfJavaMemory}{Walltime}{Stdev}{66.2227020265}%
\StoreBenchExecResult{PdrInvPathprograms}{KinductionDfStaticZeroOneTTTrueNotSolvedByKinductionPlain}{Error}{OutOfMemory}{Count}{}{1}%
\StoreBenchExecResult{PdrInvPathprograms}{KinductionDfStaticZeroOneTTTrueNotSolvedByKinductionPlain}{Error}{OutOfMemory}{Cputime}{}{859.58239486}%
\StoreBenchExecResult{PdrInvPathprograms}{KinductionDfStaticZeroOneTTTrueNotSolvedByKinductionPlain}{Error}{OutOfMemory}{Cputime}{Avg}{859.58239486}%
\StoreBenchExecResult{PdrInvPathprograms}{KinductionDfStaticZeroOneTTTrueNotSolvedByKinductionPlain}{Error}{OutOfMemory}{Cputime}{Median}{859.58239486}%
\StoreBenchExecResult{PdrInvPathprograms}{KinductionDfStaticZeroOneTTTrueNotSolvedByKinductionPlain}{Error}{OutOfMemory}{Cputime}{Min}{859.58239486}%
\StoreBenchExecResult{PdrInvPathprograms}{KinductionDfStaticZeroOneTTTrueNotSolvedByKinductionPlain}{Error}{OutOfMemory}{Cputime}{Max}{859.58239486}%
\StoreBenchExecResult{PdrInvPathprograms}{KinductionDfStaticZeroOneTTTrueNotSolvedByKinductionPlain}{Error}{OutOfMemory}{Cputime}{Stdev}{0E-8}%
\StoreBenchExecResult{PdrInvPathprograms}{KinductionDfStaticZeroOneTTTrueNotSolvedByKinductionPlain}{Error}{OutOfMemory}{Walltime}{}{841.881070137}%
\StoreBenchExecResult{PdrInvPathprograms}{KinductionDfStaticZeroOneTTTrueNotSolvedByKinductionPlain}{Error}{OutOfMemory}{Walltime}{Avg}{841.881070137}%
\StoreBenchExecResult{PdrInvPathprograms}{KinductionDfStaticZeroOneTTTrueNotSolvedByKinductionPlain}{Error}{OutOfMemory}{Walltime}{Median}{841.881070137}%
\StoreBenchExecResult{PdrInvPathprograms}{KinductionDfStaticZeroOneTTTrueNotSolvedByKinductionPlain}{Error}{OutOfMemory}{Walltime}{Min}{841.881070137}%
\StoreBenchExecResult{PdrInvPathprograms}{KinductionDfStaticZeroOneTTTrueNotSolvedByKinductionPlain}{Error}{OutOfMemory}{Walltime}{Max}{841.881070137}%
\StoreBenchExecResult{PdrInvPathprograms}{KinductionDfStaticZeroOneTTTrueNotSolvedByKinductionPlain}{Error}{OutOfMemory}{Walltime}{Stdev}{0E-9}%
\StoreBenchExecResult{PdrInvPathprograms}{KinductionDfStaticZeroOneTTTrueNotSolvedByKinductionPlain}{Error}{Timeout}{Count}{}{54}%
\StoreBenchExecResult{PdrInvPathprograms}{KinductionDfStaticZeroOneTTTrueNotSolvedByKinductionPlain}{Error}{Timeout}{Cputime}{}{48852.299941197}%
\StoreBenchExecResult{PdrInvPathprograms}{KinductionDfStaticZeroOneTTTrueNotSolvedByKinductionPlain}{Error}{Timeout}{Cputime}{Avg}{904.6722211332777777777777778}%
\StoreBenchExecResult{PdrInvPathprograms}{KinductionDfStaticZeroOneTTTrueNotSolvedByKinductionPlain}{Error}{Timeout}{Cputime}{Median}{903.8159143175}%
\StoreBenchExecResult{PdrInvPathprograms}{KinductionDfStaticZeroOneTTTrueNotSolvedByKinductionPlain}{Error}{Timeout}{Cputime}{Min}{901.018388933}%
\StoreBenchExecResult{PdrInvPathprograms}{KinductionDfStaticZeroOneTTTrueNotSolvedByKinductionPlain}{Error}{Timeout}{Cputime}{Max}{925.230062222}%
\StoreBenchExecResult{PdrInvPathprograms}{KinductionDfStaticZeroOneTTTrueNotSolvedByKinductionPlain}{Error}{Timeout}{Cputime}{Stdev}{4.041524121160103536271484543}%
\StoreBenchExecResult{PdrInvPathprograms}{KinductionDfStaticZeroOneTTTrueNotSolvedByKinductionPlain}{Error}{Timeout}{Walltime}{}{44989.969482661}%
\StoreBenchExecResult{PdrInvPathprograms}{KinductionDfStaticZeroOneTTTrueNotSolvedByKinductionPlain}{Error}{Timeout}{Walltime}{Avg}{833.1475830122407407407407407}%
\StoreBenchExecResult{PdrInvPathprograms}{KinductionDfStaticZeroOneTTTrueNotSolvedByKinductionPlain}{Error}{Timeout}{Walltime}{Median}{887.6305850745}%
\StoreBenchExecResult{PdrInvPathprograms}{KinductionDfStaticZeroOneTTTrueNotSolvedByKinductionPlain}{Error}{Timeout}{Walltime}{Min}{452.403551817}%
\StoreBenchExecResult{PdrInvPathprograms}{KinductionDfStaticZeroOneTTTrueNotSolvedByKinductionPlain}{Error}{Timeout}{Walltime}{Max}{901.014378071}%
\StoreBenchExecResult{PdrInvPathprograms}{KinductionDfStaticZeroOneTTTrueNotSolvedByKinductionPlain}{Error}{Timeout}{Walltime}{Stdev}{129.8081148924126051035492246}%
\providecommand\StoreBenchExecResult[7]{\expandafter\newcommand\csname#1#2#3#4#5#6\endcsname{#7}}%
\StoreBenchExecResult{PdrInvPathprograms}{KinductionDfStaticZeroTwoTFTrueNotSolvedByKinductionPlain}{Total}{}{Count}{}{114}%
\StoreBenchExecResult{PdrInvPathprograms}{KinductionDfStaticZeroTwoTFTrueNotSolvedByKinductionPlain}{Total}{}{Cputime}{}{62126.733679630}%
\StoreBenchExecResult{PdrInvPathprograms}{KinductionDfStaticZeroTwoTFTrueNotSolvedByKinductionPlain}{Total}{}{Cputime}{Avg}{544.9713480669298245614035088}%
\StoreBenchExecResult{PdrInvPathprograms}{KinductionDfStaticZeroTwoTFTrueNotSolvedByKinductionPlain}{Total}{}{Cputime}{Median}{890.349993096}%
\StoreBenchExecResult{PdrInvPathprograms}{KinductionDfStaticZeroTwoTFTrueNotSolvedByKinductionPlain}{Total}{}{Cputime}{Min}{3.113819648}%
\StoreBenchExecResult{PdrInvPathprograms}{KinductionDfStaticZeroTwoTFTrueNotSolvedByKinductionPlain}{Total}{}{Cputime}{Max}{915.033246713}%
\StoreBenchExecResult{PdrInvPathprograms}{KinductionDfStaticZeroTwoTFTrueNotSolvedByKinductionPlain}{Total}{}{Cputime}{Stdev}{398.4821729090036789768860352}%
\StoreBenchExecResult{PdrInvPathprograms}{KinductionDfStaticZeroTwoTFTrueNotSolvedByKinductionPlain}{Total}{}{Walltime}{}{54560.51193714261}%
\StoreBenchExecResult{PdrInvPathprograms}{KinductionDfStaticZeroTwoTFTrueNotSolvedByKinductionPlain}{Total}{}{Walltime}{Avg}{478.6009819047597368421052632}%
\StoreBenchExecResult{PdrInvPathprograms}{KinductionDfStaticZeroTwoTFTrueNotSolvedByKinductionPlain}{Total}{}{Walltime}{Median}{458.4202769995}%
\StoreBenchExecResult{PdrInvPathprograms}{KinductionDfStaticZeroTwoTFTrueNotSolvedByKinductionPlain}{Total}{}{Walltime}{Min}{1.7338218689}%
\StoreBenchExecResult{PdrInvPathprograms}{KinductionDfStaticZeroTwoTFTrueNotSolvedByKinductionPlain}{Total}{}{Walltime}{Max}{900.796468019}%
\StoreBenchExecResult{PdrInvPathprograms}{KinductionDfStaticZeroTwoTFTrueNotSolvedByKinductionPlain}{Total}{}{Walltime}{Stdev}{392.6521730968776226597445238}%
\StoreBenchExecResult{PdrInvPathprograms}{KinductionDfStaticZeroTwoTFTrueNotSolvedByKinductionPlain}{Correct}{}{Count}{}{53}%
\StoreBenchExecResult{PdrInvPathprograms}{KinductionDfStaticZeroTwoTFTrueNotSolvedByKinductionPlain}{Correct}{}{Cputime}{}{8135.537758952}%
\StoreBenchExecResult{PdrInvPathprograms}{KinductionDfStaticZeroTwoTFTrueNotSolvedByKinductionPlain}{Correct}{}{Cputime}{Avg}{153.5007124330566037735849057}%
\StoreBenchExecResult{PdrInvPathprograms}{KinductionDfStaticZeroTwoTFTrueNotSolvedByKinductionPlain}{Correct}{}{Cputime}{Median}{32.679810931}%
\StoreBenchExecResult{PdrInvPathprograms}{KinductionDfStaticZeroTwoTFTrueNotSolvedByKinductionPlain}{Correct}{}{Cputime}{Min}{3.113819648}%
\StoreBenchExecResult{PdrInvPathprograms}{KinductionDfStaticZeroTwoTFTrueNotSolvedByKinductionPlain}{Correct}{}{Cputime}{Max}{873.045055426}%
\StoreBenchExecResult{PdrInvPathprograms}{KinductionDfStaticZeroTwoTFTrueNotSolvedByKinductionPlain}{Correct}{}{Cputime}{Stdev}{210.9445743881895777413078012}%
\StoreBenchExecResult{PdrInvPathprograms}{KinductionDfStaticZeroTwoTFTrueNotSolvedByKinductionPlain}{Correct}{}{Walltime}{}{4810.34620141961}%
\StoreBenchExecResult{PdrInvPathprograms}{KinductionDfStaticZeroTwoTFTrueNotSolvedByKinductionPlain}{Correct}{}{Walltime}{Avg}{90.76124908338886792452830189}%
\StoreBenchExecResult{PdrInvPathprograms}{KinductionDfStaticZeroTwoTFTrueNotSolvedByKinductionPlain}{Correct}{}{Walltime}{Median}{16.581346035}%
\StoreBenchExecResult{PdrInvPathprograms}{KinductionDfStaticZeroTwoTFTrueNotSolvedByKinductionPlain}{Correct}{}{Walltime}{Min}{1.7338218689}%
\StoreBenchExecResult{PdrInvPathprograms}{KinductionDfStaticZeroTwoTFTrueNotSolvedByKinductionPlain}{Correct}{}{Walltime}{Max}{669.999941111}%
\StoreBenchExecResult{PdrInvPathprograms}{KinductionDfStaticZeroTwoTFTrueNotSolvedByKinductionPlain}{Correct}{}{Walltime}{Stdev}{136.4442796981906818135370406}%
\StoreBenchExecResult{PdrInvPathprograms}{KinductionDfStaticZeroTwoTFTrueNotSolvedByKinductionPlain}{Correct}{True}{Count}{}{53}%
\StoreBenchExecResult{PdrInvPathprograms}{KinductionDfStaticZeroTwoTFTrueNotSolvedByKinductionPlain}{Correct}{True}{Cputime}{}{8135.537758952}%
\StoreBenchExecResult{PdrInvPathprograms}{KinductionDfStaticZeroTwoTFTrueNotSolvedByKinductionPlain}{Correct}{True}{Cputime}{Avg}{153.5007124330566037735849057}%
\StoreBenchExecResult{PdrInvPathprograms}{KinductionDfStaticZeroTwoTFTrueNotSolvedByKinductionPlain}{Correct}{True}{Cputime}{Median}{32.679810931}%
\StoreBenchExecResult{PdrInvPathprograms}{KinductionDfStaticZeroTwoTFTrueNotSolvedByKinductionPlain}{Correct}{True}{Cputime}{Min}{3.113819648}%
\StoreBenchExecResult{PdrInvPathprograms}{KinductionDfStaticZeroTwoTFTrueNotSolvedByKinductionPlain}{Correct}{True}{Cputime}{Max}{873.045055426}%
\StoreBenchExecResult{PdrInvPathprograms}{KinductionDfStaticZeroTwoTFTrueNotSolvedByKinductionPlain}{Correct}{True}{Cputime}{Stdev}{210.9445743881895777413078012}%
\StoreBenchExecResult{PdrInvPathprograms}{KinductionDfStaticZeroTwoTFTrueNotSolvedByKinductionPlain}{Correct}{True}{Walltime}{}{4810.34620141961}%
\StoreBenchExecResult{PdrInvPathprograms}{KinductionDfStaticZeroTwoTFTrueNotSolvedByKinductionPlain}{Correct}{True}{Walltime}{Avg}{90.76124908338886792452830189}%
\StoreBenchExecResult{PdrInvPathprograms}{KinductionDfStaticZeroTwoTFTrueNotSolvedByKinductionPlain}{Correct}{True}{Walltime}{Median}{16.581346035}%
\StoreBenchExecResult{PdrInvPathprograms}{KinductionDfStaticZeroTwoTFTrueNotSolvedByKinductionPlain}{Correct}{True}{Walltime}{Min}{1.7338218689}%
\StoreBenchExecResult{PdrInvPathprograms}{KinductionDfStaticZeroTwoTFTrueNotSolvedByKinductionPlain}{Correct}{True}{Walltime}{Max}{669.999941111}%
\StoreBenchExecResult{PdrInvPathprograms}{KinductionDfStaticZeroTwoTFTrueNotSolvedByKinductionPlain}{Correct}{True}{Walltime}{Stdev}{136.4442796981906818135370406}%
\StoreBenchExecResult{PdrInvPathprograms}{KinductionDfStaticZeroTwoTFTrueNotSolvedByKinductionPlain}{Wrong}{True}{Count}{}{0}%
\StoreBenchExecResult{PdrInvPathprograms}{KinductionDfStaticZeroTwoTFTrueNotSolvedByKinductionPlain}{Wrong}{True}{Cputime}{}{0}%
\StoreBenchExecResult{PdrInvPathprograms}{KinductionDfStaticZeroTwoTFTrueNotSolvedByKinductionPlain}{Wrong}{True}{Cputime}{Avg}{None}%
\StoreBenchExecResult{PdrInvPathprograms}{KinductionDfStaticZeroTwoTFTrueNotSolvedByKinductionPlain}{Wrong}{True}{Cputime}{Median}{None}%
\StoreBenchExecResult{PdrInvPathprograms}{KinductionDfStaticZeroTwoTFTrueNotSolvedByKinductionPlain}{Wrong}{True}{Cputime}{Min}{None}%
\StoreBenchExecResult{PdrInvPathprograms}{KinductionDfStaticZeroTwoTFTrueNotSolvedByKinductionPlain}{Wrong}{True}{Cputime}{Max}{None}%
\StoreBenchExecResult{PdrInvPathprograms}{KinductionDfStaticZeroTwoTFTrueNotSolvedByKinductionPlain}{Wrong}{True}{Cputime}{Stdev}{None}%
\StoreBenchExecResult{PdrInvPathprograms}{KinductionDfStaticZeroTwoTFTrueNotSolvedByKinductionPlain}{Wrong}{True}{Walltime}{}{0}%
\StoreBenchExecResult{PdrInvPathprograms}{KinductionDfStaticZeroTwoTFTrueNotSolvedByKinductionPlain}{Wrong}{True}{Walltime}{Avg}{None}%
\StoreBenchExecResult{PdrInvPathprograms}{KinductionDfStaticZeroTwoTFTrueNotSolvedByKinductionPlain}{Wrong}{True}{Walltime}{Median}{None}%
\StoreBenchExecResult{PdrInvPathprograms}{KinductionDfStaticZeroTwoTFTrueNotSolvedByKinductionPlain}{Wrong}{True}{Walltime}{Min}{None}%
\StoreBenchExecResult{PdrInvPathprograms}{KinductionDfStaticZeroTwoTFTrueNotSolvedByKinductionPlain}{Wrong}{True}{Walltime}{Max}{None}%
\StoreBenchExecResult{PdrInvPathprograms}{KinductionDfStaticZeroTwoTFTrueNotSolvedByKinductionPlain}{Wrong}{True}{Walltime}{Stdev}{None}%
\StoreBenchExecResult{PdrInvPathprograms}{KinductionDfStaticZeroTwoTFTrueNotSolvedByKinductionPlain}{Error}{}{Count}{}{61}%
\StoreBenchExecResult{PdrInvPathprograms}{KinductionDfStaticZeroTwoTFTrueNotSolvedByKinductionPlain}{Error}{}{Cputime}{}{53991.195920678}%
\StoreBenchExecResult{PdrInvPathprograms}{KinductionDfStaticZeroTwoTFTrueNotSolvedByKinductionPlain}{Error}{}{Cputime}{Avg}{885.1015724701311475409836066}%
\StoreBenchExecResult{PdrInvPathprograms}{KinductionDfStaticZeroTwoTFTrueNotSolvedByKinductionPlain}{Error}{}{Cputime}{Median}{903.193692078}%
\StoreBenchExecResult{PdrInvPathprograms}{KinductionDfStaticZeroTwoTFTrueNotSolvedByKinductionPlain}{Error}{}{Cputime}{Min}{262.518296654}%
\StoreBenchExecResult{PdrInvPathprograms}{KinductionDfStaticZeroTwoTFTrueNotSolvedByKinductionPlain}{Error}{}{Cputime}{Max}{915.033246713}%
\StoreBenchExecResult{PdrInvPathprograms}{KinductionDfStaticZeroTwoTFTrueNotSolvedByKinductionPlain}{Error}{}{Cputime}{Stdev}{96.17690686291439165319938144}%
\StoreBenchExecResult{PdrInvPathprograms}{KinductionDfStaticZeroTwoTFTrueNotSolvedByKinductionPlain}{Error}{}{Walltime}{}{49750.165735723}%
\StoreBenchExecResult{PdrInvPathprograms}{KinductionDfStaticZeroTwoTFTrueNotSolvedByKinductionPlain}{Error}{}{Walltime}{Avg}{815.5764874708688524590163934}%
\StoreBenchExecResult{PdrInvPathprograms}{KinductionDfStaticZeroTwoTFTrueNotSolvedByKinductionPlain}{Error}{}{Walltime}{Median}{887.75613904}%
\StoreBenchExecResult{PdrInvPathprograms}{KinductionDfStaticZeroTwoTFTrueNotSolvedByKinductionPlain}{Error}{}{Walltime}{Min}{161.511502028}%
\StoreBenchExecResult{PdrInvPathprograms}{KinductionDfStaticZeroTwoTFTrueNotSolvedByKinductionPlain}{Error}{}{Walltime}{Max}{900.796468019}%
\StoreBenchExecResult{PdrInvPathprograms}{KinductionDfStaticZeroTwoTFTrueNotSolvedByKinductionPlain}{Error}{}{Walltime}{Stdev}{166.4669537203421092592141498}%
\StoreBenchExecResult{PdrInvPathprograms}{KinductionDfStaticZeroTwoTFTrueNotSolvedByKinductionPlain}{Error}{OutOfJavaMemory}{Count}{}{2}%
\StoreBenchExecResult{PdrInvPathprograms}{KinductionDfStaticZeroTwoTFTrueNotSolvedByKinductionPlain}{Error}{OutOfJavaMemory}{Cputime}{}{765.508348079}%
\StoreBenchExecResult{PdrInvPathprograms}{KinductionDfStaticZeroTwoTFTrueNotSolvedByKinductionPlain}{Error}{OutOfJavaMemory}{Cputime}{Avg}{382.7541740395}%
\StoreBenchExecResult{PdrInvPathprograms}{KinductionDfStaticZeroTwoTFTrueNotSolvedByKinductionPlain}{Error}{OutOfJavaMemory}{Cputime}{Median}{382.7541740395}%
\StoreBenchExecResult{PdrInvPathprograms}{KinductionDfStaticZeroTwoTFTrueNotSolvedByKinductionPlain}{Error}{OutOfJavaMemory}{Cputime}{Min}{262.518296654}%
\StoreBenchExecResult{PdrInvPathprograms}{KinductionDfStaticZeroTwoTFTrueNotSolvedByKinductionPlain}{Error}{OutOfJavaMemory}{Cputime}{Max}{502.990051425}%
\StoreBenchExecResult{PdrInvPathprograms}{KinductionDfStaticZeroTwoTFTrueNotSolvedByKinductionPlain}{Error}{OutOfJavaMemory}{Cputime}{Stdev}{120.2358773855}%
\StoreBenchExecResult{PdrInvPathprograms}{KinductionDfStaticZeroTwoTFTrueNotSolvedByKinductionPlain}{Error}{OutOfJavaMemory}{Walltime}{}{418.905094147}%
\StoreBenchExecResult{PdrInvPathprograms}{KinductionDfStaticZeroTwoTFTrueNotSolvedByKinductionPlain}{Error}{OutOfJavaMemory}{Walltime}{Avg}{209.4525470735}%
\StoreBenchExecResult{PdrInvPathprograms}{KinductionDfStaticZeroTwoTFTrueNotSolvedByKinductionPlain}{Error}{OutOfJavaMemory}{Walltime}{Median}{209.4525470735}%
\StoreBenchExecResult{PdrInvPathprograms}{KinductionDfStaticZeroTwoTFTrueNotSolvedByKinductionPlain}{Error}{OutOfJavaMemory}{Walltime}{Min}{161.511502028}%
\StoreBenchExecResult{PdrInvPathprograms}{KinductionDfStaticZeroTwoTFTrueNotSolvedByKinductionPlain}{Error}{OutOfJavaMemory}{Walltime}{Max}{257.393592119}%
\StoreBenchExecResult{PdrInvPathprograms}{KinductionDfStaticZeroTwoTFTrueNotSolvedByKinductionPlain}{Error}{OutOfJavaMemory}{Walltime}{Stdev}{47.9410450455}%
\StoreBenchExecResult{PdrInvPathprograms}{KinductionDfStaticZeroTwoTFTrueNotSolvedByKinductionPlain}{Error}{OutOfMemory}{Count}{}{2}%
\StoreBenchExecResult{PdrInvPathprograms}{KinductionDfStaticZeroTwoTFTrueNotSolvedByKinductionPlain}{Error}{OutOfMemory}{Cputime}{}{1671.654924537}%
\StoreBenchExecResult{PdrInvPathprograms}{KinductionDfStaticZeroTwoTFTrueNotSolvedByKinductionPlain}{Error}{OutOfMemory}{Cputime}{Avg}{835.8274622685}%
\StoreBenchExecResult{PdrInvPathprograms}{KinductionDfStaticZeroTwoTFTrueNotSolvedByKinductionPlain}{Error}{OutOfMemory}{Cputime}{Median}{835.8274622685}%
\StoreBenchExecResult{PdrInvPathprograms}{KinductionDfStaticZeroTwoTFTrueNotSolvedByKinductionPlain}{Error}{OutOfMemory}{Cputime}{Min}{791.951949067}%
\StoreBenchExecResult{PdrInvPathprograms}{KinductionDfStaticZeroTwoTFTrueNotSolvedByKinductionPlain}{Error}{OutOfMemory}{Cputime}{Max}{879.70297547}%
\StoreBenchExecResult{PdrInvPathprograms}{KinductionDfStaticZeroTwoTFTrueNotSolvedByKinductionPlain}{Error}{OutOfMemory}{Cputime}{Stdev}{43.8755132015}%
\StoreBenchExecResult{PdrInvPathprograms}{KinductionDfStaticZeroTwoTFTrueNotSolvedByKinductionPlain}{Error}{OutOfMemory}{Walltime}{}{1642.462596893}%
\StoreBenchExecResult{PdrInvPathprograms}{KinductionDfStaticZeroTwoTFTrueNotSolvedByKinductionPlain}{Error}{OutOfMemory}{Walltime}{Avg}{821.2312984465}%
\StoreBenchExecResult{PdrInvPathprograms}{KinductionDfStaticZeroTwoTFTrueNotSolvedByKinductionPlain}{Error}{OutOfMemory}{Walltime}{Median}{821.2312984465}%
\StoreBenchExecResult{PdrInvPathprograms}{KinductionDfStaticZeroTwoTFTrueNotSolvedByKinductionPlain}{Error}{OutOfMemory}{Walltime}{Min}{777.53171587}%
\StoreBenchExecResult{PdrInvPathprograms}{KinductionDfStaticZeroTwoTFTrueNotSolvedByKinductionPlain}{Error}{OutOfMemory}{Walltime}{Max}{864.930881023}%
\StoreBenchExecResult{PdrInvPathprograms}{KinductionDfStaticZeroTwoTFTrueNotSolvedByKinductionPlain}{Error}{OutOfMemory}{Walltime}{Stdev}{43.6995825765}%
\StoreBenchExecResult{PdrInvPathprograms}{KinductionDfStaticZeroTwoTFTrueNotSolvedByKinductionPlain}{Error}{Timeout}{Count}{}{57}%
\StoreBenchExecResult{PdrInvPathprograms}{KinductionDfStaticZeroTwoTFTrueNotSolvedByKinductionPlain}{Error}{Timeout}{Cputime}{}{51554.032648062}%
\StoreBenchExecResult{PdrInvPathprograms}{KinductionDfStaticZeroTwoTFTrueNotSolvedByKinductionPlain}{Error}{Timeout}{Cputime}{Avg}{904.4567131238947368421052632}%
\StoreBenchExecResult{PdrInvPathprograms}{KinductionDfStaticZeroTwoTFTrueNotSolvedByKinductionPlain}{Error}{Timeout}{Cputime}{Median}{903.227462456}%
\StoreBenchExecResult{PdrInvPathprograms}{KinductionDfStaticZeroTwoTFTrueNotSolvedByKinductionPlain}{Error}{Timeout}{Cputime}{Min}{900.997010722}%
\StoreBenchExecResult{PdrInvPathprograms}{KinductionDfStaticZeroTwoTFTrueNotSolvedByKinductionPlain}{Error}{Timeout}{Cputime}{Max}{915.033246713}%
\StoreBenchExecResult{PdrInvPathprograms}{KinductionDfStaticZeroTwoTFTrueNotSolvedByKinductionPlain}{Error}{Timeout}{Cputime}{Stdev}{3.165802554103391478647645797}%
\StoreBenchExecResult{PdrInvPathprograms}{KinductionDfStaticZeroTwoTFTrueNotSolvedByKinductionPlain}{Error}{Timeout}{Walltime}{}{47688.798044683}%
\StoreBenchExecResult{PdrInvPathprograms}{KinductionDfStaticZeroTwoTFTrueNotSolvedByKinductionPlain}{Error}{Timeout}{Walltime}{Avg}{836.6455797312807017543859649}%
\StoreBenchExecResult{PdrInvPathprograms}{KinductionDfStaticZeroTwoTFTrueNotSolvedByKinductionPlain}{Error}{Timeout}{Walltime}{Median}{887.850430012}%
\StoreBenchExecResult{PdrInvPathprograms}{KinductionDfStaticZeroTwoTFTrueNotSolvedByKinductionPlain}{Error}{Timeout}{Walltime}{Min}{452.320882082}%
\StoreBenchExecResult{PdrInvPathprograms}{KinductionDfStaticZeroTwoTFTrueNotSolvedByKinductionPlain}{Error}{Timeout}{Walltime}{Max}{900.796468019}%
\StoreBenchExecResult{PdrInvPathprograms}{KinductionDfStaticZeroTwoTFTrueNotSolvedByKinductionPlain}{Error}{Timeout}{Walltime}{Stdev}{127.1710381562641613686862458}%
\providecommand\StoreBenchExecResult[7]{\expandafter\newcommand\csname#1#2#3#4#5#6\endcsname{#7}}%
\StoreBenchExecResult{PdrInvPathprograms}{KinductionDfStaticZeroTwoTTTrueNotSolvedByKinductionPlain}{Total}{}{Count}{}{114}%
\StoreBenchExecResult{PdrInvPathprograms}{KinductionDfStaticZeroTwoTTTrueNotSolvedByKinductionPlain}{Total}{}{Cputime}{}{58762.472287053}%
\StoreBenchExecResult{PdrInvPathprograms}{KinductionDfStaticZeroTwoTTTrueNotSolvedByKinductionPlain}{Total}{}{Cputime}{Avg}{515.4602832197631578947368421}%
\StoreBenchExecResult{PdrInvPathprograms}{KinductionDfStaticZeroTwoTTTrueNotSolvedByKinductionPlain}{Total}{}{Cputime}{Median}{607.486606330}%
\StoreBenchExecResult{PdrInvPathprograms}{KinductionDfStaticZeroTwoTTTrueNotSolvedByKinductionPlain}{Total}{}{Cputime}{Min}{3.171097586}%
\StoreBenchExecResult{PdrInvPathprograms}{KinductionDfStaticZeroTwoTTTrueNotSolvedByKinductionPlain}{Total}{}{Cputime}{Max}{915.506980113}%
\StoreBenchExecResult{PdrInvPathprograms}{KinductionDfStaticZeroTwoTTTrueNotSolvedByKinductionPlain}{Total}{}{Cputime}{Stdev}{404.1102870542328612128146794}%
\StoreBenchExecResult{PdrInvPathprograms}{KinductionDfStaticZeroTwoTTTrueNotSolvedByKinductionPlain}{Total}{}{Walltime}{}{51170.92638325661}%
\StoreBenchExecResult{PdrInvPathprograms}{KinductionDfStaticZeroTwoTTTrueNotSolvedByKinductionPlain}{Total}{}{Walltime}{Avg}{448.8677752917246491228070175}%
\StoreBenchExecResult{PdrInvPathprograms}{KinductionDfStaticZeroTwoTTTrueNotSolvedByKinductionPlain}{Total}{}{Walltime}{Median}{381.1539975405}%
\StoreBenchExecResult{PdrInvPathprograms}{KinductionDfStaticZeroTwoTTTrueNotSolvedByKinductionPlain}{Total}{}{Walltime}{Min}{1.75759005547}%
\StoreBenchExecResult{PdrInvPathprograms}{KinductionDfStaticZeroTwoTTTrueNotSolvedByKinductionPlain}{Total}{}{Walltime}{Max}{900.985891104}%
\StoreBenchExecResult{PdrInvPathprograms}{KinductionDfStaticZeroTwoTTTrueNotSolvedByKinductionPlain}{Total}{}{Walltime}{Stdev}{393.7556635949067486807289642}%
\StoreBenchExecResult{PdrInvPathprograms}{KinductionDfStaticZeroTwoTTTrueNotSolvedByKinductionPlain}{Correct}{}{Count}{}{58}%
\StoreBenchExecResult{PdrInvPathprograms}{KinductionDfStaticZeroTwoTTTrueNotSolvedByKinductionPlain}{Correct}{}{Cputime}{}{9226.219201241}%
\StoreBenchExecResult{PdrInvPathprograms}{KinductionDfStaticZeroTwoTTTrueNotSolvedByKinductionPlain}{Correct}{}{Cputime}{Avg}{159.0727448489827586206896552}%
\StoreBenchExecResult{PdrInvPathprograms}{KinductionDfStaticZeroTwoTTTrueNotSolvedByKinductionPlain}{Correct}{}{Cputime}{Median}{25.381801055}%
\StoreBenchExecResult{PdrInvPathprograms}{KinductionDfStaticZeroTwoTTTrueNotSolvedByKinductionPlain}{Correct}{}{Cputime}{Min}{3.171097586}%
\StoreBenchExecResult{PdrInvPathprograms}{KinductionDfStaticZeroTwoTTTrueNotSolvedByKinductionPlain}{Correct}{}{Cputime}{Max}{893.664252456}%
\StoreBenchExecResult{PdrInvPathprograms}{KinductionDfStaticZeroTwoTTTrueNotSolvedByKinductionPlain}{Correct}{}{Cputime}{Stdev}{229.9026619594938551183489038}%
\StoreBenchExecResult{PdrInvPathprograms}{KinductionDfStaticZeroTwoTTTrueNotSolvedByKinductionPlain}{Correct}{}{Walltime}{}{5391.39205074261}%
\StoreBenchExecResult{PdrInvPathprograms}{KinductionDfStaticZeroTwoTTTrueNotSolvedByKinductionPlain}{Correct}{}{Walltime}{Avg}{92.95503535763120689655172414}%
\StoreBenchExecResult{PdrInvPathprograms}{KinductionDfStaticZeroTwoTTTrueNotSolvedByKinductionPlain}{Correct}{}{Walltime}{Median}{13.1424765587}%
\StoreBenchExecResult{PdrInvPathprograms}{KinductionDfStaticZeroTwoTTTrueNotSolvedByKinductionPlain}{Correct}{}{Walltime}{Min}{1.75759005547}%
\StoreBenchExecResult{PdrInvPathprograms}{KinductionDfStaticZeroTwoTTTrueNotSolvedByKinductionPlain}{Correct}{}{Walltime}{Max}{685.111480951}%
\StoreBenchExecResult{PdrInvPathprograms}{KinductionDfStaticZeroTwoTTTrueNotSolvedByKinductionPlain}{Correct}{}{Walltime}{Stdev}{142.9640554358894899938707376}%
\StoreBenchExecResult{PdrInvPathprograms}{KinductionDfStaticZeroTwoTTTrueNotSolvedByKinductionPlain}{Correct}{True}{Count}{}{58}%
\StoreBenchExecResult{PdrInvPathprograms}{KinductionDfStaticZeroTwoTTTrueNotSolvedByKinductionPlain}{Correct}{True}{Cputime}{}{9226.219201241}%
\StoreBenchExecResult{PdrInvPathprograms}{KinductionDfStaticZeroTwoTTTrueNotSolvedByKinductionPlain}{Correct}{True}{Cputime}{Avg}{159.0727448489827586206896552}%
\StoreBenchExecResult{PdrInvPathprograms}{KinductionDfStaticZeroTwoTTTrueNotSolvedByKinductionPlain}{Correct}{True}{Cputime}{Median}{25.381801055}%
\StoreBenchExecResult{PdrInvPathprograms}{KinductionDfStaticZeroTwoTTTrueNotSolvedByKinductionPlain}{Correct}{True}{Cputime}{Min}{3.171097586}%
\StoreBenchExecResult{PdrInvPathprograms}{KinductionDfStaticZeroTwoTTTrueNotSolvedByKinductionPlain}{Correct}{True}{Cputime}{Max}{893.664252456}%
\StoreBenchExecResult{PdrInvPathprograms}{KinductionDfStaticZeroTwoTTTrueNotSolvedByKinductionPlain}{Correct}{True}{Cputime}{Stdev}{229.9026619594938551183489038}%
\StoreBenchExecResult{PdrInvPathprograms}{KinductionDfStaticZeroTwoTTTrueNotSolvedByKinductionPlain}{Correct}{True}{Walltime}{}{5391.39205074261}%
\StoreBenchExecResult{PdrInvPathprograms}{KinductionDfStaticZeroTwoTTTrueNotSolvedByKinductionPlain}{Correct}{True}{Walltime}{Avg}{92.95503535763120689655172414}%
\StoreBenchExecResult{PdrInvPathprograms}{KinductionDfStaticZeroTwoTTTrueNotSolvedByKinductionPlain}{Correct}{True}{Walltime}{Median}{13.1424765587}%
\StoreBenchExecResult{PdrInvPathprograms}{KinductionDfStaticZeroTwoTTTrueNotSolvedByKinductionPlain}{Correct}{True}{Walltime}{Min}{1.75759005547}%
\StoreBenchExecResult{PdrInvPathprograms}{KinductionDfStaticZeroTwoTTTrueNotSolvedByKinductionPlain}{Correct}{True}{Walltime}{Max}{685.111480951}%
\StoreBenchExecResult{PdrInvPathprograms}{KinductionDfStaticZeroTwoTTTrueNotSolvedByKinductionPlain}{Correct}{True}{Walltime}{Stdev}{142.9640554358894899938707376}%
\StoreBenchExecResult{PdrInvPathprograms}{KinductionDfStaticZeroTwoTTTrueNotSolvedByKinductionPlain}{Wrong}{True}{Count}{}{0}%
\StoreBenchExecResult{PdrInvPathprograms}{KinductionDfStaticZeroTwoTTTrueNotSolvedByKinductionPlain}{Wrong}{True}{Cputime}{}{0}%
\StoreBenchExecResult{PdrInvPathprograms}{KinductionDfStaticZeroTwoTTTrueNotSolvedByKinductionPlain}{Wrong}{True}{Cputime}{Avg}{None}%
\StoreBenchExecResult{PdrInvPathprograms}{KinductionDfStaticZeroTwoTTTrueNotSolvedByKinductionPlain}{Wrong}{True}{Cputime}{Median}{None}%
\StoreBenchExecResult{PdrInvPathprograms}{KinductionDfStaticZeroTwoTTTrueNotSolvedByKinductionPlain}{Wrong}{True}{Cputime}{Min}{None}%
\StoreBenchExecResult{PdrInvPathprograms}{KinductionDfStaticZeroTwoTTTrueNotSolvedByKinductionPlain}{Wrong}{True}{Cputime}{Max}{None}%
\StoreBenchExecResult{PdrInvPathprograms}{KinductionDfStaticZeroTwoTTTrueNotSolvedByKinductionPlain}{Wrong}{True}{Cputime}{Stdev}{None}%
\StoreBenchExecResult{PdrInvPathprograms}{KinductionDfStaticZeroTwoTTTrueNotSolvedByKinductionPlain}{Wrong}{True}{Walltime}{}{0}%
\StoreBenchExecResult{PdrInvPathprograms}{KinductionDfStaticZeroTwoTTTrueNotSolvedByKinductionPlain}{Wrong}{True}{Walltime}{Avg}{None}%
\StoreBenchExecResult{PdrInvPathprograms}{KinductionDfStaticZeroTwoTTTrueNotSolvedByKinductionPlain}{Wrong}{True}{Walltime}{Median}{None}%
\StoreBenchExecResult{PdrInvPathprograms}{KinductionDfStaticZeroTwoTTTrueNotSolvedByKinductionPlain}{Wrong}{True}{Walltime}{Min}{None}%
\StoreBenchExecResult{PdrInvPathprograms}{KinductionDfStaticZeroTwoTTTrueNotSolvedByKinductionPlain}{Wrong}{True}{Walltime}{Max}{None}%
\StoreBenchExecResult{PdrInvPathprograms}{KinductionDfStaticZeroTwoTTTrueNotSolvedByKinductionPlain}{Wrong}{True}{Walltime}{Stdev}{None}%
\StoreBenchExecResult{PdrInvPathprograms}{KinductionDfStaticZeroTwoTTTrueNotSolvedByKinductionPlain}{Error}{}{Count}{}{56}%
\StoreBenchExecResult{PdrInvPathprograms}{KinductionDfStaticZeroTwoTTTrueNotSolvedByKinductionPlain}{Error}{}{Cputime}{}{49536.253085812}%
\StoreBenchExecResult{PdrInvPathprograms}{KinductionDfStaticZeroTwoTTTrueNotSolvedByKinductionPlain}{Error}{}{Cputime}{Avg}{884.5759479609285714285714286}%
\StoreBenchExecResult{PdrInvPathprograms}{KinductionDfStaticZeroTwoTTTrueNotSolvedByKinductionPlain}{Error}{}{Cputime}{Median}{903.1670066845}%
\StoreBenchExecResult{PdrInvPathprograms}{KinductionDfStaticZeroTwoTTTrueNotSolvedByKinductionPlain}{Error}{}{Cputime}{Min}{246.452079879}%
\StoreBenchExecResult{PdrInvPathprograms}{KinductionDfStaticZeroTwoTTTrueNotSolvedByKinductionPlain}{Error}{}{Cputime}{Max}{915.506980113}%
\StoreBenchExecResult{PdrInvPathprograms}{KinductionDfStaticZeroTwoTTTrueNotSolvedByKinductionPlain}{Error}{}{Cputime}{Stdev}{99.52414819853489320619219936}%
\StoreBenchExecResult{PdrInvPathprograms}{KinductionDfStaticZeroTwoTTTrueNotSolvedByKinductionPlain}{Error}{}{Walltime}{}{45779.534332514}%
\StoreBenchExecResult{PdrInvPathprograms}{KinductionDfStaticZeroTwoTTTrueNotSolvedByKinductionPlain}{Error}{}{Walltime}{Avg}{817.4916845091785714285714286}%
\StoreBenchExecResult{PdrInvPathprograms}{KinductionDfStaticZeroTwoTTTrueNotSolvedByKinductionPlain}{Error}{}{Walltime}{Median}{887.274067402}%
\StoreBenchExecResult{PdrInvPathprograms}{KinductionDfStaticZeroTwoTTTrueNotSolvedByKinductionPlain}{Error}{}{Walltime}{Min}{150.216253042}%
\StoreBenchExecResult{PdrInvPathprograms}{KinductionDfStaticZeroTwoTTTrueNotSolvedByKinductionPlain}{Error}{}{Walltime}{Max}{900.985891104}%
\StoreBenchExecResult{PdrInvPathprograms}{KinductionDfStaticZeroTwoTTTrueNotSolvedByKinductionPlain}{Error}{}{Walltime}{Stdev}{165.4512207146920179125781706}%
\StoreBenchExecResult{PdrInvPathprograms}{KinductionDfStaticZeroTwoTTTrueNotSolvedByKinductionPlain}{Error}{OutOfJavaMemory}{Count}{}{2}%
\StoreBenchExecResult{PdrInvPathprograms}{KinductionDfStaticZeroTwoTTTrueNotSolvedByKinductionPlain}{Error}{OutOfJavaMemory}{Cputime}{}{782.572637048}%
\StoreBenchExecResult{PdrInvPathprograms}{KinductionDfStaticZeroTwoTTTrueNotSolvedByKinductionPlain}{Error}{OutOfJavaMemory}{Cputime}{Avg}{391.286318524}%
\StoreBenchExecResult{PdrInvPathprograms}{KinductionDfStaticZeroTwoTTTrueNotSolvedByKinductionPlain}{Error}{OutOfJavaMemory}{Cputime}{Median}{391.286318524}%
\StoreBenchExecResult{PdrInvPathprograms}{KinductionDfStaticZeroTwoTTTrueNotSolvedByKinductionPlain}{Error}{OutOfJavaMemory}{Cputime}{Min}{246.452079879}%
\StoreBenchExecResult{PdrInvPathprograms}{KinductionDfStaticZeroTwoTTTrueNotSolvedByKinductionPlain}{Error}{OutOfJavaMemory}{Cputime}{Max}{536.120557169}%
\StoreBenchExecResult{PdrInvPathprograms}{KinductionDfStaticZeroTwoTTTrueNotSolvedByKinductionPlain}{Error}{OutOfJavaMemory}{Cputime}{Stdev}{144.834238645}%
\StoreBenchExecResult{PdrInvPathprograms}{KinductionDfStaticZeroTwoTTTrueNotSolvedByKinductionPlain}{Error}{OutOfJavaMemory}{Walltime}{}{424.922064066}%
\StoreBenchExecResult{PdrInvPathprograms}{KinductionDfStaticZeroTwoTTTrueNotSolvedByKinductionPlain}{Error}{OutOfJavaMemory}{Walltime}{Avg}{212.461032033}%
\StoreBenchExecResult{PdrInvPathprograms}{KinductionDfStaticZeroTwoTTTrueNotSolvedByKinductionPlain}{Error}{OutOfJavaMemory}{Walltime}{Median}{212.461032033}%
\StoreBenchExecResult{PdrInvPathprograms}{KinductionDfStaticZeroTwoTTTrueNotSolvedByKinductionPlain}{Error}{OutOfJavaMemory}{Walltime}{Min}{150.216253042}%
\StoreBenchExecResult{PdrInvPathprograms}{KinductionDfStaticZeroTwoTTTrueNotSolvedByKinductionPlain}{Error}{OutOfJavaMemory}{Walltime}{Max}{274.705811024}%
\StoreBenchExecResult{PdrInvPathprograms}{KinductionDfStaticZeroTwoTTTrueNotSolvedByKinductionPlain}{Error}{OutOfJavaMemory}{Walltime}{Stdev}{62.244778991}%
\StoreBenchExecResult{PdrInvPathprograms}{KinductionDfStaticZeroTwoTTTrueNotSolvedByKinductionPlain}{Error}{OutOfMemory}{Count}{}{1}%
\StoreBenchExecResult{PdrInvPathprograms}{KinductionDfStaticZeroTwoTTTrueNotSolvedByKinductionPlain}{Error}{OutOfMemory}{Cputime}{}{817.435108748}%
\StoreBenchExecResult{PdrInvPathprograms}{KinductionDfStaticZeroTwoTTTrueNotSolvedByKinductionPlain}{Error}{OutOfMemory}{Cputime}{Avg}{817.435108748}%
\StoreBenchExecResult{PdrInvPathprograms}{KinductionDfStaticZeroTwoTTTrueNotSolvedByKinductionPlain}{Error}{OutOfMemory}{Cputime}{Median}{817.435108748}%
\StoreBenchExecResult{PdrInvPathprograms}{KinductionDfStaticZeroTwoTTTrueNotSolvedByKinductionPlain}{Error}{OutOfMemory}{Cputime}{Min}{817.435108748}%
\StoreBenchExecResult{PdrInvPathprograms}{KinductionDfStaticZeroTwoTTTrueNotSolvedByKinductionPlain}{Error}{OutOfMemory}{Cputime}{Max}{817.435108748}%
\StoreBenchExecResult{PdrInvPathprograms}{KinductionDfStaticZeroTwoTTTrueNotSolvedByKinductionPlain}{Error}{OutOfMemory}{Cputime}{Stdev}{0E-9}%
\StoreBenchExecResult{PdrInvPathprograms}{KinductionDfStaticZeroTwoTTTrueNotSolvedByKinductionPlain}{Error}{OutOfMemory}{Walltime}{}{802.653079987}%
\StoreBenchExecResult{PdrInvPathprograms}{KinductionDfStaticZeroTwoTTTrueNotSolvedByKinductionPlain}{Error}{OutOfMemory}{Walltime}{Avg}{802.653079987}%
\StoreBenchExecResult{PdrInvPathprograms}{KinductionDfStaticZeroTwoTTTrueNotSolvedByKinductionPlain}{Error}{OutOfMemory}{Walltime}{Median}{802.653079987}%
\StoreBenchExecResult{PdrInvPathprograms}{KinductionDfStaticZeroTwoTTTrueNotSolvedByKinductionPlain}{Error}{OutOfMemory}{Walltime}{Min}{802.653079987}%
\StoreBenchExecResult{PdrInvPathprograms}{KinductionDfStaticZeroTwoTTTrueNotSolvedByKinductionPlain}{Error}{OutOfMemory}{Walltime}{Max}{802.653079987}%
\StoreBenchExecResult{PdrInvPathprograms}{KinductionDfStaticZeroTwoTTTrueNotSolvedByKinductionPlain}{Error}{OutOfMemory}{Walltime}{Stdev}{0E-9}%
\StoreBenchExecResult{PdrInvPathprograms}{KinductionDfStaticZeroTwoTTTrueNotSolvedByKinductionPlain}{Error}{Timeout}{Count}{}{53}%
\StoreBenchExecResult{PdrInvPathprograms}{KinductionDfStaticZeroTwoTTTrueNotSolvedByKinductionPlain}{Error}{Timeout}{Cputime}{}{47936.245340016}%
\StoreBenchExecResult{PdrInvPathprograms}{KinductionDfStaticZeroTwoTTTrueNotSolvedByKinductionPlain}{Error}{Timeout}{Cputime}{Avg}{904.4574592455849056603773585}%
\StoreBenchExecResult{PdrInvPathprograms}{KinductionDfStaticZeroTwoTTTrueNotSolvedByKinductionPlain}{Error}{Timeout}{Cputime}{Median}{903.26082264}%
\StoreBenchExecResult{PdrInvPathprograms}{KinductionDfStaticZeroTwoTTTrueNotSolvedByKinductionPlain}{Error}{Timeout}{Cputime}{Min}{900.953046204}%
\StoreBenchExecResult{PdrInvPathprograms}{KinductionDfStaticZeroTwoTTTrueNotSolvedByKinductionPlain}{Error}{Timeout}{Cputime}{Max}{915.506980113}%
\StoreBenchExecResult{PdrInvPathprograms}{KinductionDfStaticZeroTwoTTTrueNotSolvedByKinductionPlain}{Error}{Timeout}{Cputime}{Stdev}{3.371549895719411581145250399}%
\StoreBenchExecResult{PdrInvPathprograms}{KinductionDfStaticZeroTwoTTTrueNotSolvedByKinductionPlain}{Error}{Timeout}{Walltime}{}{44551.959188461}%
\StoreBenchExecResult{PdrInvPathprograms}{KinductionDfStaticZeroTwoTTTrueNotSolvedByKinductionPlain}{Error}{Timeout}{Walltime}{Avg}{840.6030035558679245283018868}%
\StoreBenchExecResult{PdrInvPathprograms}{KinductionDfStaticZeroTwoTTTrueNotSolvedByKinductionPlain}{Error}{Timeout}{Walltime}{Median}{887.58338809}%
\StoreBenchExecResult{PdrInvPathprograms}{KinductionDfStaticZeroTwoTTTrueNotSolvedByKinductionPlain}{Error}{Timeout}{Walltime}{Min}{452.554687977}%
\StoreBenchExecResult{PdrInvPathprograms}{KinductionDfStaticZeroTwoTTTrueNotSolvedByKinductionPlain}{Error}{Timeout}{Walltime}{Max}{900.985891104}%
\StoreBenchExecResult{PdrInvPathprograms}{KinductionDfStaticZeroTwoTTTrueNotSolvedByKinductionPlain}{Error}{Timeout}{Walltime}{Stdev}{120.1059059179672580528233601}%
\providecommand\StoreBenchExecResult[7]{\expandafter\newcommand\csname#1#2#3#4#5#6\endcsname{#7}}%
\StoreBenchExecResult{PdrInvPathprograms}{KinductionDfStaticSixteenTwoFTrueNotSolvedByKinductionPlain}{Total}{}{Count}{}{114}%
\StoreBenchExecResult{PdrInvPathprograms}{KinductionDfStaticSixteenTwoFTrueNotSolvedByKinductionPlain}{Total}{}{Cputime}{}{49370.262843292}%
\StoreBenchExecResult{PdrInvPathprograms}{KinductionDfStaticSixteenTwoFTrueNotSolvedByKinductionPlain}{Total}{}{Cputime}{Avg}{433.0724810815087719298245614}%
\StoreBenchExecResult{PdrInvPathprograms}{KinductionDfStaticSixteenTwoFTrueNotSolvedByKinductionPlain}{Total}{}{Cputime}{Median}{286.090658745}%
\StoreBenchExecResult{PdrInvPathprograms}{KinductionDfStaticSixteenTwoFTrueNotSolvedByKinductionPlain}{Total}{}{Cputime}{Min}{3.140401466}%
\StoreBenchExecResult{PdrInvPathprograms}{KinductionDfStaticSixteenTwoFTrueNotSolvedByKinductionPlain}{Total}{}{Cputime}{Max}{923.481149277}%
\StoreBenchExecResult{PdrInvPathprograms}{KinductionDfStaticSixteenTwoFTrueNotSolvedByKinductionPlain}{Total}{}{Cputime}{Stdev}{406.7562557583777101803374661}%
\StoreBenchExecResult{PdrInvPathprograms}{KinductionDfStaticSixteenTwoFTrueNotSolvedByKinductionPlain}{Total}{}{Walltime}{}{25597.04405761091}%
\StoreBenchExecResult{PdrInvPathprograms}{KinductionDfStaticSixteenTwoFTrueNotSolvedByKinductionPlain}{Total}{}{Walltime}{Avg}{224.5354741895693859649122807}%
\StoreBenchExecResult{PdrInvPathprograms}{KinductionDfStaticSixteenTwoFTrueNotSolvedByKinductionPlain}{Total}{}{Walltime}{Median}{158.585852027}%
\StoreBenchExecResult{PdrInvPathprograms}{KinductionDfStaticSixteenTwoFTrueNotSolvedByKinductionPlain}{Total}{}{Walltime}{Min}{1.730052948}%
\StoreBenchExecResult{PdrInvPathprograms}{KinductionDfStaticSixteenTwoFTrueNotSolvedByKinductionPlain}{Total}{}{Walltime}{Max}{641.675709963}%
\StoreBenchExecResult{PdrInvPathprograms}{KinductionDfStaticSixteenTwoFTrueNotSolvedByKinductionPlain}{Total}{}{Walltime}{Stdev}{208.8621580750765738508442641}%
\StoreBenchExecResult{PdrInvPathprograms}{KinductionDfStaticSixteenTwoFTrueNotSolvedByKinductionPlain}{Correct}{}{Count}{}{65}%
\StoreBenchExecResult{PdrInvPathprograms}{KinductionDfStaticSixteenTwoFTrueNotSolvedByKinductionPlain}{Correct}{}{Cputime}{}{7253.881752785}%
\StoreBenchExecResult{PdrInvPathprograms}{KinductionDfStaticSixteenTwoFTrueNotSolvedByKinductionPlain}{Correct}{}{Cputime}{Avg}{111.5981808120769230769230769}%
\StoreBenchExecResult{PdrInvPathprograms}{KinductionDfStaticSixteenTwoFTrueNotSolvedByKinductionPlain}{Correct}{}{Cputime}{Median}{36.534204835}%
\StoreBenchExecResult{PdrInvPathprograms}{KinductionDfStaticSixteenTwoFTrueNotSolvedByKinductionPlain}{Correct}{}{Cputime}{Min}{3.140401466}%
\StoreBenchExecResult{PdrInvPathprograms}{KinductionDfStaticSixteenTwoFTrueNotSolvedByKinductionPlain}{Correct}{}{Cputime}{Max}{824.299716021}%
\StoreBenchExecResult{PdrInvPathprograms}{KinductionDfStaticSixteenTwoFTrueNotSolvedByKinductionPlain}{Correct}{}{Cputime}{Stdev}{180.9456286328689874995821532}%
\StoreBenchExecResult{PdrInvPathprograms}{KinductionDfStaticSixteenTwoFTrueNotSolvedByKinductionPlain}{Correct}{}{Walltime}{}{4352.64344668491}%
\StoreBenchExecResult{PdrInvPathprograms}{KinductionDfStaticSixteenTwoFTrueNotSolvedByKinductionPlain}{Correct}{}{Walltime}{Avg}{66.963745333614}%
\StoreBenchExecResult{PdrInvPathprograms}{KinductionDfStaticSixteenTwoFTrueNotSolvedByKinductionPlain}{Correct}{}{Walltime}{Median}{18.6862211227}%
\StoreBenchExecResult{PdrInvPathprograms}{KinductionDfStaticSixteenTwoFTrueNotSolvedByKinductionPlain}{Correct}{}{Walltime}{Min}{1.730052948}%
\StoreBenchExecResult{PdrInvPathprograms}{KinductionDfStaticSixteenTwoFTrueNotSolvedByKinductionPlain}{Correct}{}{Walltime}{Max}{641.675709963}%
\StoreBenchExecResult{PdrInvPathprograms}{KinductionDfStaticSixteenTwoFTrueNotSolvedByKinductionPlain}{Correct}{}{Walltime}{Stdev}{120.5156039318774493090411149}%
\StoreBenchExecResult{PdrInvPathprograms}{KinductionDfStaticSixteenTwoFTrueNotSolvedByKinductionPlain}{Correct}{True}{Count}{}{65}%
\StoreBenchExecResult{PdrInvPathprograms}{KinductionDfStaticSixteenTwoFTrueNotSolvedByKinductionPlain}{Correct}{True}{Cputime}{}{7253.881752785}%
\StoreBenchExecResult{PdrInvPathprograms}{KinductionDfStaticSixteenTwoFTrueNotSolvedByKinductionPlain}{Correct}{True}{Cputime}{Avg}{111.5981808120769230769230769}%
\StoreBenchExecResult{PdrInvPathprograms}{KinductionDfStaticSixteenTwoFTrueNotSolvedByKinductionPlain}{Correct}{True}{Cputime}{Median}{36.534204835}%
\StoreBenchExecResult{PdrInvPathprograms}{KinductionDfStaticSixteenTwoFTrueNotSolvedByKinductionPlain}{Correct}{True}{Cputime}{Min}{3.140401466}%
\StoreBenchExecResult{PdrInvPathprograms}{KinductionDfStaticSixteenTwoFTrueNotSolvedByKinductionPlain}{Correct}{True}{Cputime}{Max}{824.299716021}%
\StoreBenchExecResult{PdrInvPathprograms}{KinductionDfStaticSixteenTwoFTrueNotSolvedByKinductionPlain}{Correct}{True}{Cputime}{Stdev}{180.9456286328689874995821532}%
\StoreBenchExecResult{PdrInvPathprograms}{KinductionDfStaticSixteenTwoFTrueNotSolvedByKinductionPlain}{Correct}{True}{Walltime}{}{4352.64344668491}%
\StoreBenchExecResult{PdrInvPathprograms}{KinductionDfStaticSixteenTwoFTrueNotSolvedByKinductionPlain}{Correct}{True}{Walltime}{Avg}{66.963745333614}%
\StoreBenchExecResult{PdrInvPathprograms}{KinductionDfStaticSixteenTwoFTrueNotSolvedByKinductionPlain}{Correct}{True}{Walltime}{Median}{18.6862211227}%
\StoreBenchExecResult{PdrInvPathprograms}{KinductionDfStaticSixteenTwoFTrueNotSolvedByKinductionPlain}{Correct}{True}{Walltime}{Min}{1.730052948}%
\StoreBenchExecResult{PdrInvPathprograms}{KinductionDfStaticSixteenTwoFTrueNotSolvedByKinductionPlain}{Correct}{True}{Walltime}{Max}{641.675709963}%
\StoreBenchExecResult{PdrInvPathprograms}{KinductionDfStaticSixteenTwoFTrueNotSolvedByKinductionPlain}{Correct}{True}{Walltime}{Stdev}{120.5156039318774493090411149}%
\StoreBenchExecResult{PdrInvPathprograms}{KinductionDfStaticSixteenTwoFTrueNotSolvedByKinductionPlain}{Wrong}{True}{Count}{}{0}%
\StoreBenchExecResult{PdrInvPathprograms}{KinductionDfStaticSixteenTwoFTrueNotSolvedByKinductionPlain}{Wrong}{True}{Cputime}{}{0}%
\StoreBenchExecResult{PdrInvPathprograms}{KinductionDfStaticSixteenTwoFTrueNotSolvedByKinductionPlain}{Wrong}{True}{Cputime}{Avg}{None}%
\StoreBenchExecResult{PdrInvPathprograms}{KinductionDfStaticSixteenTwoFTrueNotSolvedByKinductionPlain}{Wrong}{True}{Cputime}{Median}{None}%
\StoreBenchExecResult{PdrInvPathprograms}{KinductionDfStaticSixteenTwoFTrueNotSolvedByKinductionPlain}{Wrong}{True}{Cputime}{Min}{None}%
\StoreBenchExecResult{PdrInvPathprograms}{KinductionDfStaticSixteenTwoFTrueNotSolvedByKinductionPlain}{Wrong}{True}{Cputime}{Max}{None}%
\StoreBenchExecResult{PdrInvPathprograms}{KinductionDfStaticSixteenTwoFTrueNotSolvedByKinductionPlain}{Wrong}{True}{Cputime}{Stdev}{None}%
\StoreBenchExecResult{PdrInvPathprograms}{KinductionDfStaticSixteenTwoFTrueNotSolvedByKinductionPlain}{Wrong}{True}{Walltime}{}{0}%
\StoreBenchExecResult{PdrInvPathprograms}{KinductionDfStaticSixteenTwoFTrueNotSolvedByKinductionPlain}{Wrong}{True}{Walltime}{Avg}{None}%
\StoreBenchExecResult{PdrInvPathprograms}{KinductionDfStaticSixteenTwoFTrueNotSolvedByKinductionPlain}{Wrong}{True}{Walltime}{Median}{None}%
\StoreBenchExecResult{PdrInvPathprograms}{KinductionDfStaticSixteenTwoFTrueNotSolvedByKinductionPlain}{Wrong}{True}{Walltime}{Min}{None}%
\StoreBenchExecResult{PdrInvPathprograms}{KinductionDfStaticSixteenTwoFTrueNotSolvedByKinductionPlain}{Wrong}{True}{Walltime}{Max}{None}%
\StoreBenchExecResult{PdrInvPathprograms}{KinductionDfStaticSixteenTwoFTrueNotSolvedByKinductionPlain}{Wrong}{True}{Walltime}{Stdev}{None}%
\StoreBenchExecResult{PdrInvPathprograms}{KinductionDfStaticSixteenTwoFTrueNotSolvedByKinductionPlain}{Error}{}{Count}{}{49}%
\StoreBenchExecResult{PdrInvPathprograms}{KinductionDfStaticSixteenTwoFTrueNotSolvedByKinductionPlain}{Error}{}{Cputime}{}{42116.381090507}%
\StoreBenchExecResult{PdrInvPathprograms}{KinductionDfStaticSixteenTwoFTrueNotSolvedByKinductionPlain}{Error}{}{Cputime}{Avg}{859.5179814389183673469387755}%
\StoreBenchExecResult{PdrInvPathprograms}{KinductionDfStaticSixteenTwoFTrueNotSolvedByKinductionPlain}{Error}{}{Cputime}{Median}{901.30809491}%
\StoreBenchExecResult{PdrInvPathprograms}{KinductionDfStaticSixteenTwoFTrueNotSolvedByKinductionPlain}{Error}{}{Cputime}{Min}{238.665273596}%
\StoreBenchExecResult{PdrInvPathprograms}{KinductionDfStaticSixteenTwoFTrueNotSolvedByKinductionPlain}{Error}{}{Cputime}{Max}{923.481149277}%
\StoreBenchExecResult{PdrInvPathprograms}{KinductionDfStaticSixteenTwoFTrueNotSolvedByKinductionPlain}{Error}{}{Cputime}{Stdev}{150.1552802852333329666618244}%
\StoreBenchExecResult{PdrInvPathprograms}{KinductionDfStaticSixteenTwoFTrueNotSolvedByKinductionPlain}{Error}{}{Walltime}{}{21244.400610926}%
\StoreBenchExecResult{PdrInvPathprograms}{KinductionDfStaticSixteenTwoFTrueNotSolvedByKinductionPlain}{Error}{}{Walltime}{Avg}{433.5591961413469387755102041}%
\StoreBenchExecResult{PdrInvPathprograms}{KinductionDfStaticSixteenTwoFTrueNotSolvedByKinductionPlain}{Error}{}{Walltime}{Median}{452.370603085}%
\StoreBenchExecResult{PdrInvPathprograms}{KinductionDfStaticSixteenTwoFTrueNotSolvedByKinductionPlain}{Error}{}{Walltime}{Min}{124.706032991}%
\StoreBenchExecResult{PdrInvPathprograms}{KinductionDfStaticSixteenTwoFTrueNotSolvedByKinductionPlain}{Error}{}{Walltime}{Max}{490.974872112}%
\StoreBenchExecResult{PdrInvPathprograms}{KinductionDfStaticSixteenTwoFTrueNotSolvedByKinductionPlain}{Error}{}{Walltime}{Stdev}{74.81635855673469266929521641}%
\StoreBenchExecResult{PdrInvPathprograms}{KinductionDfStaticSixteenTwoFTrueNotSolvedByKinductionPlain}{Error}{OutOfJavaMemory}{Count}{}{4}%
\StoreBenchExecResult{PdrInvPathprograms}{KinductionDfStaticSixteenTwoFTrueNotSolvedByKinductionPlain}{Error}{OutOfJavaMemory}{Cputime}{}{1476.994056821}%
\StoreBenchExecResult{PdrInvPathprograms}{KinductionDfStaticSixteenTwoFTrueNotSolvedByKinductionPlain}{Error}{OutOfJavaMemory}{Cputime}{Avg}{369.24851420525}%
\StoreBenchExecResult{PdrInvPathprograms}{KinductionDfStaticSixteenTwoFTrueNotSolvedByKinductionPlain}{Error}{OutOfJavaMemory}{Cputime}{Median}{360.547576043}%
\StoreBenchExecResult{PdrInvPathprograms}{KinductionDfStaticSixteenTwoFTrueNotSolvedByKinductionPlain}{Error}{OutOfJavaMemory}{Cputime}{Min}{238.665273596}%
\StoreBenchExecResult{PdrInvPathprograms}{KinductionDfStaticSixteenTwoFTrueNotSolvedByKinductionPlain}{Error}{OutOfJavaMemory}{Cputime}{Max}{517.233631139}%
\StoreBenchExecResult{PdrInvPathprograms}{KinductionDfStaticSixteenTwoFTrueNotSolvedByKinductionPlain}{Error}{OutOfJavaMemory}{Cputime}{Stdev}{119.2974086207675350337844513}%
\StoreBenchExecResult{PdrInvPathprograms}{KinductionDfStaticSixteenTwoFTrueNotSolvedByKinductionPlain}{Error}{OutOfJavaMemory}{Walltime}{}{763.208463192}%
\StoreBenchExecResult{PdrInvPathprograms}{KinductionDfStaticSixteenTwoFTrueNotSolvedByKinductionPlain}{Error}{OutOfJavaMemory}{Walltime}{Avg}{190.802115798}%
\StoreBenchExecResult{PdrInvPathprograms}{KinductionDfStaticSixteenTwoFTrueNotSolvedByKinductionPlain}{Error}{OutOfJavaMemory}{Walltime}{Median}{187.198547602}%
\StoreBenchExecResult{PdrInvPathprograms}{KinductionDfStaticSixteenTwoFTrueNotSolvedByKinductionPlain}{Error}{OutOfJavaMemory}{Walltime}{Min}{124.706032991}%
\StoreBenchExecResult{PdrInvPathprograms}{KinductionDfStaticSixteenTwoFTrueNotSolvedByKinductionPlain}{Error}{OutOfJavaMemory}{Walltime}{Max}{264.105334997}%
\StoreBenchExecResult{PdrInvPathprograms}{KinductionDfStaticSixteenTwoFTrueNotSolvedByKinductionPlain}{Error}{OutOfJavaMemory}{Walltime}{Stdev}{61.41229381140544460867309139}%
\StoreBenchExecResult{PdrInvPathprograms}{KinductionDfStaticSixteenTwoFTrueNotSolvedByKinductionPlain}{Error}{Timeout}{Count}{}{45}%
\StoreBenchExecResult{PdrInvPathprograms}{KinductionDfStaticSixteenTwoFTrueNotSolvedByKinductionPlain}{Error}{Timeout}{Cputime}{}{40639.387033686}%
\StoreBenchExecResult{PdrInvPathprograms}{KinductionDfStaticSixteenTwoFTrueNotSolvedByKinductionPlain}{Error}{Timeout}{Cputime}{Avg}{903.0974896374666666666666667}%
\StoreBenchExecResult{PdrInvPathprograms}{KinductionDfStaticSixteenTwoFTrueNotSolvedByKinductionPlain}{Error}{Timeout}{Cputime}{Median}{901.699804567}%
\StoreBenchExecResult{PdrInvPathprograms}{KinductionDfStaticSixteenTwoFTrueNotSolvedByKinductionPlain}{Error}{Timeout}{Cputime}{Min}{900.827926904}%
\StoreBenchExecResult{PdrInvPathprograms}{KinductionDfStaticSixteenTwoFTrueNotSolvedByKinductionPlain}{Error}{Timeout}{Cputime}{Max}{923.481149277}%
\StoreBenchExecResult{PdrInvPathprograms}{KinductionDfStaticSixteenTwoFTrueNotSolvedByKinductionPlain}{Error}{Timeout}{Cputime}{Stdev}{4.562903050618069467536339505}%
\StoreBenchExecResult{PdrInvPathprograms}{KinductionDfStaticSixteenTwoFTrueNotSolvedByKinductionPlain}{Error}{Timeout}{Walltime}{}{20481.192147734}%
\StoreBenchExecResult{PdrInvPathprograms}{KinductionDfStaticSixteenTwoFTrueNotSolvedByKinductionPlain}{Error}{Timeout}{Walltime}{Avg}{455.1376032829777777777777778}%
\StoreBenchExecResult{PdrInvPathprograms}{KinductionDfStaticSixteenTwoFTrueNotSolvedByKinductionPlain}{Error}{Timeout}{Walltime}{Median}{452.41753602}%
\StoreBenchExecResult{PdrInvPathprograms}{KinductionDfStaticSixteenTwoFTrueNotSolvedByKinductionPlain}{Error}{Timeout}{Walltime}{Min}{451.232038975}%
\StoreBenchExecResult{PdrInvPathprograms}{KinductionDfStaticSixteenTwoFTrueNotSolvedByKinductionPlain}{Error}{Timeout}{Walltime}{Max}{490.974872112}%
\StoreBenchExecResult{PdrInvPathprograms}{KinductionDfStaticSixteenTwoFTrueNotSolvedByKinductionPlain}{Error}{Timeout}{Walltime}{Stdev}{7.474049713117553667826631507}%
\providecommand\StoreBenchExecResult[7]{\expandafter\newcommand\csname#1#2#3#4#5#6\endcsname{#7}}%
\StoreBenchExecResult{PdrInvPathprograms}{KinductionDfStaticSixteenTwoTTrueNotSolvedByKinductionPlain}{Total}{}{Count}{}{114}%
\StoreBenchExecResult{PdrInvPathprograms}{KinductionDfStaticSixteenTwoTTrueNotSolvedByKinductionPlain}{Total}{}{Cputime}{}{46072.570951415}%
\StoreBenchExecResult{PdrInvPathprograms}{KinductionDfStaticSixteenTwoTTrueNotSolvedByKinductionPlain}{Total}{}{Cputime}{Avg}{404.1453592229385964912280702}%
\StoreBenchExecResult{PdrInvPathprograms}{KinductionDfStaticSixteenTwoTTrueNotSolvedByKinductionPlain}{Total}{}{Cputime}{Median}{201.650918367}%
\StoreBenchExecResult{PdrInvPathprograms}{KinductionDfStaticSixteenTwoTTrueNotSolvedByKinductionPlain}{Total}{}{Cputime}{Min}{3.213554809}%
\StoreBenchExecResult{PdrInvPathprograms}{KinductionDfStaticSixteenTwoTTrueNotSolvedByKinductionPlain}{Total}{}{Cputime}{Max}{918.601201373}%
\StoreBenchExecResult{PdrInvPathprograms}{KinductionDfStaticSixteenTwoTTrueNotSolvedByKinductionPlain}{Total}{}{Cputime}{Stdev}{386.6532475599080075521261604}%
\StoreBenchExecResult{PdrInvPathprograms}{KinductionDfStaticSixteenTwoTTrueNotSolvedByKinductionPlain}{Total}{}{Walltime}{}{28095.23683309602}%
\StoreBenchExecResult{PdrInvPathprograms}{KinductionDfStaticSixteenTwoTTrueNotSolvedByKinductionPlain}{Total}{}{Walltime}{Avg}{246.4494459043510526315789474}%
\StoreBenchExecResult{PdrInvPathprograms}{KinductionDfStaticSixteenTwoTTrueNotSolvedByKinductionPlain}{Total}{}{Walltime}{Median}{101.44143307215}%
\StoreBenchExecResult{PdrInvPathprograms}{KinductionDfStaticSixteenTwoTTrueNotSolvedByKinductionPlain}{Total}{}{Walltime}{Min}{1.77312803268}%
\StoreBenchExecResult{PdrInvPathprograms}{KinductionDfStaticSixteenTwoTTrueNotSolvedByKinductionPlain}{Total}{}{Walltime}{Max}{898.312685966}%
\StoreBenchExecResult{PdrInvPathprograms}{KinductionDfStaticSixteenTwoTTrueNotSolvedByKinductionPlain}{Total}{}{Walltime}{Stdev}{265.6173626376659351992348006}%
\StoreBenchExecResult{PdrInvPathprograms}{KinductionDfStaticSixteenTwoTTrueNotSolvedByKinductionPlain}{Correct}{}{Count}{}{74}%
\StoreBenchExecResult{PdrInvPathprograms}{KinductionDfStaticSixteenTwoTTrueNotSolvedByKinductionPlain}{Correct}{}{Cputime}{}{10973.828822546}%
\StoreBenchExecResult{PdrInvPathprograms}{KinductionDfStaticSixteenTwoTTrueNotSolvedByKinductionPlain}{Correct}{}{Cputime}{Avg}{148.2949840884594594594594595}%
\StoreBenchExecResult{PdrInvPathprograms}{KinductionDfStaticSixteenTwoTTrueNotSolvedByKinductionPlain}{Correct}{}{Cputime}{Median}{77.5897263095}%
\StoreBenchExecResult{PdrInvPathprograms}{KinductionDfStaticSixteenTwoTTrueNotSolvedByKinductionPlain}{Correct}{}{Cputime}{Min}{3.213554809}%
\StoreBenchExecResult{PdrInvPathprograms}{KinductionDfStaticSixteenTwoTTrueNotSolvedByKinductionPlain}{Correct}{}{Cputime}{Max}{831.582240838}%
\StoreBenchExecResult{PdrInvPathprograms}{KinductionDfStaticSixteenTwoTTrueNotSolvedByKinductionPlain}{Correct}{}{Cputime}{Stdev}{192.9199340462793923004816620}%
\StoreBenchExecResult{PdrInvPathprograms}{KinductionDfStaticSixteenTwoTTrueNotSolvedByKinductionPlain}{Correct}{}{Walltime}{}{6210.78783226002}%
\StoreBenchExecResult{PdrInvPathprograms}{KinductionDfStaticSixteenTwoTTrueNotSolvedByKinductionPlain}{Correct}{}{Walltime}{Avg}{83.92956530081108108108108108}%
\StoreBenchExecResult{PdrInvPathprograms}{KinductionDfStaticSixteenTwoTTrueNotSolvedByKinductionPlain}{Correct}{}{Walltime}{Median}{39.98029303555}%
\StoreBenchExecResult{PdrInvPathprograms}{KinductionDfStaticSixteenTwoTTrueNotSolvedByKinductionPlain}{Correct}{}{Walltime}{Min}{1.77312803268}%
\StoreBenchExecResult{PdrInvPathprograms}{KinductionDfStaticSixteenTwoTTrueNotSolvedByKinductionPlain}{Correct}{}{Walltime}{Max}{637.909646988}%
\StoreBenchExecResult{PdrInvPathprograms}{KinductionDfStaticSixteenTwoTTrueNotSolvedByKinductionPlain}{Correct}{}{Walltime}{Stdev}{119.8927129455105861115187060}%
\StoreBenchExecResult{PdrInvPathprograms}{KinductionDfStaticSixteenTwoTTrueNotSolvedByKinductionPlain}{Correct}{True}{Count}{}{74}%
\StoreBenchExecResult{PdrInvPathprograms}{KinductionDfStaticSixteenTwoTTrueNotSolvedByKinductionPlain}{Correct}{True}{Cputime}{}{10973.828822546}%
\StoreBenchExecResult{PdrInvPathprograms}{KinductionDfStaticSixteenTwoTTrueNotSolvedByKinductionPlain}{Correct}{True}{Cputime}{Avg}{148.2949840884594594594594595}%
\StoreBenchExecResult{PdrInvPathprograms}{KinductionDfStaticSixteenTwoTTrueNotSolvedByKinductionPlain}{Correct}{True}{Cputime}{Median}{77.5897263095}%
\StoreBenchExecResult{PdrInvPathprograms}{KinductionDfStaticSixteenTwoTTrueNotSolvedByKinductionPlain}{Correct}{True}{Cputime}{Min}{3.213554809}%
\StoreBenchExecResult{PdrInvPathprograms}{KinductionDfStaticSixteenTwoTTrueNotSolvedByKinductionPlain}{Correct}{True}{Cputime}{Max}{831.582240838}%
\StoreBenchExecResult{PdrInvPathprograms}{KinductionDfStaticSixteenTwoTTrueNotSolvedByKinductionPlain}{Correct}{True}{Cputime}{Stdev}{192.9199340462793923004816620}%
\StoreBenchExecResult{PdrInvPathprograms}{KinductionDfStaticSixteenTwoTTrueNotSolvedByKinductionPlain}{Correct}{True}{Walltime}{}{6210.78783226002}%
\StoreBenchExecResult{PdrInvPathprograms}{KinductionDfStaticSixteenTwoTTrueNotSolvedByKinductionPlain}{Correct}{True}{Walltime}{Avg}{83.92956530081108108108108108}%
\StoreBenchExecResult{PdrInvPathprograms}{KinductionDfStaticSixteenTwoTTrueNotSolvedByKinductionPlain}{Correct}{True}{Walltime}{Median}{39.98029303555}%
\StoreBenchExecResult{PdrInvPathprograms}{KinductionDfStaticSixteenTwoTTrueNotSolvedByKinductionPlain}{Correct}{True}{Walltime}{Min}{1.77312803268}%
\StoreBenchExecResult{PdrInvPathprograms}{KinductionDfStaticSixteenTwoTTrueNotSolvedByKinductionPlain}{Correct}{True}{Walltime}{Max}{637.909646988}%
\StoreBenchExecResult{PdrInvPathprograms}{KinductionDfStaticSixteenTwoTTrueNotSolvedByKinductionPlain}{Correct}{True}{Walltime}{Stdev}{119.8927129455105861115187060}%
\StoreBenchExecResult{PdrInvPathprograms}{KinductionDfStaticSixteenTwoTTrueNotSolvedByKinductionPlain}{Wrong}{True}{Count}{}{0}%
\StoreBenchExecResult{PdrInvPathprograms}{KinductionDfStaticSixteenTwoTTrueNotSolvedByKinductionPlain}{Wrong}{True}{Cputime}{}{0}%
\StoreBenchExecResult{PdrInvPathprograms}{KinductionDfStaticSixteenTwoTTrueNotSolvedByKinductionPlain}{Wrong}{True}{Cputime}{Avg}{None}%
\StoreBenchExecResult{PdrInvPathprograms}{KinductionDfStaticSixteenTwoTTrueNotSolvedByKinductionPlain}{Wrong}{True}{Cputime}{Median}{None}%
\StoreBenchExecResult{PdrInvPathprograms}{KinductionDfStaticSixteenTwoTTrueNotSolvedByKinductionPlain}{Wrong}{True}{Cputime}{Min}{None}%
\StoreBenchExecResult{PdrInvPathprograms}{KinductionDfStaticSixteenTwoTTrueNotSolvedByKinductionPlain}{Wrong}{True}{Cputime}{Max}{None}%
\StoreBenchExecResult{PdrInvPathprograms}{KinductionDfStaticSixteenTwoTTrueNotSolvedByKinductionPlain}{Wrong}{True}{Cputime}{Stdev}{None}%
\StoreBenchExecResult{PdrInvPathprograms}{KinductionDfStaticSixteenTwoTTrueNotSolvedByKinductionPlain}{Wrong}{True}{Walltime}{}{0}%
\StoreBenchExecResult{PdrInvPathprograms}{KinductionDfStaticSixteenTwoTTrueNotSolvedByKinductionPlain}{Wrong}{True}{Walltime}{Avg}{None}%
\StoreBenchExecResult{PdrInvPathprograms}{KinductionDfStaticSixteenTwoTTrueNotSolvedByKinductionPlain}{Wrong}{True}{Walltime}{Median}{None}%
\StoreBenchExecResult{PdrInvPathprograms}{KinductionDfStaticSixteenTwoTTrueNotSolvedByKinductionPlain}{Wrong}{True}{Walltime}{Min}{None}%
\StoreBenchExecResult{PdrInvPathprograms}{KinductionDfStaticSixteenTwoTTrueNotSolvedByKinductionPlain}{Wrong}{True}{Walltime}{Max}{None}%
\StoreBenchExecResult{PdrInvPathprograms}{KinductionDfStaticSixteenTwoTTrueNotSolvedByKinductionPlain}{Wrong}{True}{Walltime}{Stdev}{None}%
\StoreBenchExecResult{PdrInvPathprograms}{KinductionDfStaticSixteenTwoTTrueNotSolvedByKinductionPlain}{Error}{}{Count}{}{40}%
\StoreBenchExecResult{PdrInvPathprograms}{KinductionDfStaticSixteenTwoTTrueNotSolvedByKinductionPlain}{Error}{}{Cputime}{}{35098.742128869}%
\StoreBenchExecResult{PdrInvPathprograms}{KinductionDfStaticSixteenTwoTTrueNotSolvedByKinductionPlain}{Error}{}{Cputime}{Avg}{877.468553221725}%
\StoreBenchExecResult{PdrInvPathprograms}{KinductionDfStaticSixteenTwoTTrueNotSolvedByKinductionPlain}{Error}{}{Cputime}{Median}{902.046677284}%
\StoreBenchExecResult{PdrInvPathprograms}{KinductionDfStaticSixteenTwoTTrueNotSolvedByKinductionPlain}{Error}{}{Cputime}{Min}{281.625429389}%
\StoreBenchExecResult{PdrInvPathprograms}{KinductionDfStaticSixteenTwoTTrueNotSolvedByKinductionPlain}{Error}{}{Cputime}{Max}{918.601201373}%
\StoreBenchExecResult{PdrInvPathprograms}{KinductionDfStaticSixteenTwoTTrueNotSolvedByKinductionPlain}{Error}{}{Cputime}{Stdev}{109.9492700519034787060320103}%
\StoreBenchExecResult{PdrInvPathprograms}{KinductionDfStaticSixteenTwoTTrueNotSolvedByKinductionPlain}{Error}{}{Walltime}{}{21884.449000836}%
\StoreBenchExecResult{PdrInvPathprograms}{KinductionDfStaticSixteenTwoTTrueNotSolvedByKinductionPlain}{Error}{}{Walltime}{Avg}{547.1112250209}%
\StoreBenchExecResult{PdrInvPathprograms}{KinductionDfStaticSixteenTwoTTrueNotSolvedByKinductionPlain}{Error}{}{Walltime}{Median}{453.696599603}%
\StoreBenchExecResult{PdrInvPathprograms}{KinductionDfStaticSixteenTwoTTrueNotSolvedByKinductionPlain}{Error}{}{Walltime}{Min}{170.200577974}%
\StoreBenchExecResult{PdrInvPathprograms}{KinductionDfStaticSixteenTwoTTrueNotSolvedByKinductionPlain}{Error}{}{Walltime}{Max}{898.312685966}%
\StoreBenchExecResult{PdrInvPathprograms}{KinductionDfStaticSixteenTwoTTrueNotSolvedByKinductionPlain}{Error}{}{Walltime}{Stdev}{187.6737989057813855104598077}%
\StoreBenchExecResult{PdrInvPathprograms}{KinductionDfStaticSixteenTwoTTrueNotSolvedByKinductionPlain}{Error}{OutOfJavaMemory}{Count}{}{2}%
\StoreBenchExecResult{PdrInvPathprograms}{KinductionDfStaticSixteenTwoTTrueNotSolvedByKinductionPlain}{Error}{OutOfJavaMemory}{Cputime}{}{845.143957640}%
\StoreBenchExecResult{PdrInvPathprograms}{KinductionDfStaticSixteenTwoTTrueNotSolvedByKinductionPlain}{Error}{OutOfJavaMemory}{Cputime}{Avg}{422.571978820}%
\StoreBenchExecResult{PdrInvPathprograms}{KinductionDfStaticSixteenTwoTTrueNotSolvedByKinductionPlain}{Error}{OutOfJavaMemory}{Cputime}{Median}{422.571978820}%
\StoreBenchExecResult{PdrInvPathprograms}{KinductionDfStaticSixteenTwoTTrueNotSolvedByKinductionPlain}{Error}{OutOfJavaMemory}{Cputime}{Min}{281.625429389}%
\StoreBenchExecResult{PdrInvPathprograms}{KinductionDfStaticSixteenTwoTTrueNotSolvedByKinductionPlain}{Error}{OutOfJavaMemory}{Cputime}{Max}{563.518528251}%
\StoreBenchExecResult{PdrInvPathprograms}{KinductionDfStaticSixteenTwoTTrueNotSolvedByKinductionPlain}{Error}{OutOfJavaMemory}{Cputime}{Stdev}{140.946549431}%
\StoreBenchExecResult{PdrInvPathprograms}{KinductionDfStaticSixteenTwoTTrueNotSolvedByKinductionPlain}{Error}{OutOfJavaMemory}{Walltime}{}{457.510756969}%
\StoreBenchExecResult{PdrInvPathprograms}{KinductionDfStaticSixteenTwoTTrueNotSolvedByKinductionPlain}{Error}{OutOfJavaMemory}{Walltime}{Avg}{228.7553784845}%
\StoreBenchExecResult{PdrInvPathprograms}{KinductionDfStaticSixteenTwoTTrueNotSolvedByKinductionPlain}{Error}{OutOfJavaMemory}{Walltime}{Median}{228.7553784845}%
\StoreBenchExecResult{PdrInvPathprograms}{KinductionDfStaticSixteenTwoTTrueNotSolvedByKinductionPlain}{Error}{OutOfJavaMemory}{Walltime}{Min}{170.200577974}%
\StoreBenchExecResult{PdrInvPathprograms}{KinductionDfStaticSixteenTwoTTrueNotSolvedByKinductionPlain}{Error}{OutOfJavaMemory}{Walltime}{Max}{287.310178995}%
\StoreBenchExecResult{PdrInvPathprograms}{KinductionDfStaticSixteenTwoTTrueNotSolvedByKinductionPlain}{Error}{OutOfJavaMemory}{Walltime}{Stdev}{58.5548005105}%
\StoreBenchExecResult{PdrInvPathprograms}{KinductionDfStaticSixteenTwoTTrueNotSolvedByKinductionPlain}{Error}{OutOfMemory}{Count}{}{1}%
\StoreBenchExecResult{PdrInvPathprograms}{KinductionDfStaticSixteenTwoTTrueNotSolvedByKinductionPlain}{Error}{OutOfMemory}{Cputime}{}{815.376880425}%
\StoreBenchExecResult{PdrInvPathprograms}{KinductionDfStaticSixteenTwoTTrueNotSolvedByKinductionPlain}{Error}{OutOfMemory}{Cputime}{Avg}{815.376880425}%
\StoreBenchExecResult{PdrInvPathprograms}{KinductionDfStaticSixteenTwoTTrueNotSolvedByKinductionPlain}{Error}{OutOfMemory}{Cputime}{Median}{815.376880425}%
\StoreBenchExecResult{PdrInvPathprograms}{KinductionDfStaticSixteenTwoTTrueNotSolvedByKinductionPlain}{Error}{OutOfMemory}{Cputime}{Min}{815.376880425}%
\StoreBenchExecResult{PdrInvPathprograms}{KinductionDfStaticSixteenTwoTTrueNotSolvedByKinductionPlain}{Error}{OutOfMemory}{Cputime}{Max}{815.376880425}%
\StoreBenchExecResult{PdrInvPathprograms}{KinductionDfStaticSixteenTwoTTrueNotSolvedByKinductionPlain}{Error}{OutOfMemory}{Cputime}{Stdev}{0E-9}%
\StoreBenchExecResult{PdrInvPathprograms}{KinductionDfStaticSixteenTwoTTrueNotSolvedByKinductionPlain}{Error}{OutOfMemory}{Walltime}{}{800.159305096}%
\StoreBenchExecResult{PdrInvPathprograms}{KinductionDfStaticSixteenTwoTTrueNotSolvedByKinductionPlain}{Error}{OutOfMemory}{Walltime}{Avg}{800.159305096}%
\StoreBenchExecResult{PdrInvPathprograms}{KinductionDfStaticSixteenTwoTTrueNotSolvedByKinductionPlain}{Error}{OutOfMemory}{Walltime}{Median}{800.159305096}%
\StoreBenchExecResult{PdrInvPathprograms}{KinductionDfStaticSixteenTwoTTrueNotSolvedByKinductionPlain}{Error}{OutOfMemory}{Walltime}{Min}{800.159305096}%
\StoreBenchExecResult{PdrInvPathprograms}{KinductionDfStaticSixteenTwoTTrueNotSolvedByKinductionPlain}{Error}{OutOfMemory}{Walltime}{Max}{800.159305096}%
\StoreBenchExecResult{PdrInvPathprograms}{KinductionDfStaticSixteenTwoTTrueNotSolvedByKinductionPlain}{Error}{OutOfMemory}{Walltime}{Stdev}{0E-9}%
\StoreBenchExecResult{PdrInvPathprograms}{KinductionDfStaticSixteenTwoTTrueNotSolvedByKinductionPlain}{Error}{Timeout}{Count}{}{37}%
\StoreBenchExecResult{PdrInvPathprograms}{KinductionDfStaticSixteenTwoTTrueNotSolvedByKinductionPlain}{Error}{Timeout}{Cputime}{}{33438.221290804}%
\StoreBenchExecResult{PdrInvPathprograms}{KinductionDfStaticSixteenTwoTTrueNotSolvedByKinductionPlain}{Error}{Timeout}{Cputime}{Avg}{903.7357105622702702702702703}%
\StoreBenchExecResult{PdrInvPathprograms}{KinductionDfStaticSixteenTwoTTrueNotSolvedByKinductionPlain}{Error}{Timeout}{Cputime}{Median}{902.136993894}%
\StoreBenchExecResult{PdrInvPathprograms}{KinductionDfStaticSixteenTwoTTrueNotSolvedByKinductionPlain}{Error}{Timeout}{Cputime}{Min}{900.878051541}%
\StoreBenchExecResult{PdrInvPathprograms}{KinductionDfStaticSixteenTwoTTrueNotSolvedByKinductionPlain}{Error}{Timeout}{Cputime}{Max}{918.601201373}%
\StoreBenchExecResult{PdrInvPathprograms}{KinductionDfStaticSixteenTwoTTrueNotSolvedByKinductionPlain}{Error}{Timeout}{Cputime}{Stdev}{3.945595023950781210563899246}%
\StoreBenchExecResult{PdrInvPathprograms}{KinductionDfStaticSixteenTwoTTrueNotSolvedByKinductionPlain}{Error}{Timeout}{Walltime}{}{20626.778938771}%
\StoreBenchExecResult{PdrInvPathprograms}{KinductionDfStaticSixteenTwoTTrueNotSolvedByKinductionPlain}{Error}{Timeout}{Walltime}{Avg}{557.4805118586756756756756757}%
\StoreBenchExecResult{PdrInvPathprograms}{KinductionDfStaticSixteenTwoTTrueNotSolvedByKinductionPlain}{Error}{Timeout}{Walltime}{Median}{454.142408133}%
\StoreBenchExecResult{PdrInvPathprograms}{KinductionDfStaticSixteenTwoTTrueNotSolvedByKinductionPlain}{Error}{Timeout}{Walltime}{Min}{451.102282047}%
\StoreBenchExecResult{PdrInvPathprograms}{KinductionDfStaticSixteenTwoTTrueNotSolvedByKinductionPlain}{Error}{Timeout}{Walltime}{Max}{898.312685966}%
\StoreBenchExecResult{PdrInvPathprograms}{KinductionDfStaticSixteenTwoTTrueNotSolvedByKinductionPlain}{Error}{Timeout}{Walltime}{Stdev}{174.8581141857095335939384303}%
\providecommand\StoreBenchExecResult[7]{\expandafter\newcommand\csname#1#2#3#4#5#6\endcsname{#7}}%
\StoreBenchExecResult{PdrInvPathprograms}{KinductionDfStaticEightTwoTTrueNotSolvedByKinductionPlain}{Total}{}{Count}{}{114}%
\StoreBenchExecResult{PdrInvPathprograms}{KinductionDfStaticEightTwoTTrueNotSolvedByKinductionPlain}{Total}{}{Cputime}{}{37657.884626913}%
\StoreBenchExecResult{PdrInvPathprograms}{KinductionDfStaticEightTwoTTrueNotSolvedByKinductionPlain}{Total}{}{Cputime}{Avg}{330.3323212887105263157894737}%
\StoreBenchExecResult{PdrInvPathprograms}{KinductionDfStaticEightTwoTTrueNotSolvedByKinductionPlain}{Total}{}{Cputime}{Median}{116.5932438875}%
\StoreBenchExecResult{PdrInvPathprograms}{KinductionDfStaticEightTwoTTrueNotSolvedByKinductionPlain}{Total}{}{Cputime}{Min}{3.2741362}%
\StoreBenchExecResult{PdrInvPathprograms}{KinductionDfStaticEightTwoTTrueNotSolvedByKinductionPlain}{Total}{}{Cputime}{Max}{1000.74474958}%
\StoreBenchExecResult{PdrInvPathprograms}{KinductionDfStaticEightTwoTTrueNotSolvedByKinductionPlain}{Total}{}{Cputime}{Stdev}{376.8926205798792362210125133}%
\StoreBenchExecResult{PdrInvPathprograms}{KinductionDfStaticEightTwoTTrueNotSolvedByKinductionPlain}{Total}{}{Walltime}{}{25869.53749298827}%
\StoreBenchExecResult{PdrInvPathprograms}{KinductionDfStaticEightTwoTTrueNotSolvedByKinductionPlain}{Total}{}{Walltime}{Avg}{226.9257674823532456140350877}%
\StoreBenchExecResult{PdrInvPathprograms}{KinductionDfStaticEightTwoTTrueNotSolvedByKinductionPlain}{Total}{}{Walltime}{Median}{74.81367206575}%
\StoreBenchExecResult{PdrInvPathprograms}{KinductionDfStaticEightTwoTTrueNotSolvedByKinductionPlain}{Total}{}{Walltime}{Min}{1.79608201981}%
\StoreBenchExecResult{PdrInvPathprograms}{KinductionDfStaticEightTwoTTrueNotSolvedByKinductionPlain}{Total}{}{Walltime}{Max}{898.08901906}%
\StoreBenchExecResult{PdrInvPathprograms}{KinductionDfStaticEightTwoTTrueNotSolvedByKinductionPlain}{Total}{}{Walltime}{Stdev}{295.1341538397902946402250215}%
\StoreBenchExecResult{PdrInvPathprograms}{KinductionDfStaticEightTwoTTrueNotSolvedByKinductionPlain}{Correct}{}{Count}{}{83}%
\StoreBenchExecResult{PdrInvPathprograms}{KinductionDfStaticEightTwoTTrueNotSolvedByKinductionPlain}{Correct}{}{Cputime}{}{10654.602968223}%
\StoreBenchExecResult{PdrInvPathprograms}{KinductionDfStaticEightTwoTTrueNotSolvedByKinductionPlain}{Correct}{}{Cputime}{Avg}{128.3687104605180722891566265}%
\StoreBenchExecResult{PdrInvPathprograms}{KinductionDfStaticEightTwoTTrueNotSolvedByKinductionPlain}{Correct}{}{Cputime}{Median}{39.809312868}%
\StoreBenchExecResult{PdrInvPathprograms}{KinductionDfStaticEightTwoTTrueNotSolvedByKinductionPlain}{Correct}{}{Cputime}{Min}{3.2741362}%
\StoreBenchExecResult{PdrInvPathprograms}{KinductionDfStaticEightTwoTTrueNotSolvedByKinductionPlain}{Correct}{}{Cputime}{Max}{898.051709025}%
\StoreBenchExecResult{PdrInvPathprograms}{KinductionDfStaticEightTwoTTrueNotSolvedByKinductionPlain}{Correct}{}{Cputime}{Stdev}{196.0085401852064763373261141}%
\StoreBenchExecResult{PdrInvPathprograms}{KinductionDfStaticEightTwoTTrueNotSolvedByKinductionPlain}{Correct}{}{Walltime}{}{6128.85636758827}%
\StoreBenchExecResult{PdrInvPathprograms}{KinductionDfStaticEightTwoTTrueNotSolvedByKinductionPlain}{Correct}{}{Walltime}{Avg}{73.84164298299120481927710843}%
\StoreBenchExecResult{PdrInvPathprograms}{KinductionDfStaticEightTwoTTrueNotSolvedByKinductionPlain}{Correct}{}{Walltime}{Median}{20.2034180164}%
\StoreBenchExecResult{PdrInvPathprograms}{KinductionDfStaticEightTwoTTrueNotSolvedByKinductionPlain}{Correct}{}{Walltime}{Min}{1.79608201981}%
\StoreBenchExecResult{PdrInvPathprograms}{KinductionDfStaticEightTwoTTrueNotSolvedByKinductionPlain}{Correct}{}{Walltime}{Max}{695.898958206}%
\StoreBenchExecResult{PdrInvPathprograms}{KinductionDfStaticEightTwoTTrueNotSolvedByKinductionPlain}{Correct}{}{Walltime}{Stdev}{123.6438010218835952549116862}%
\StoreBenchExecResult{PdrInvPathprograms}{KinductionDfStaticEightTwoTTrueNotSolvedByKinductionPlain}{Correct}{True}{Count}{}{83}%
\StoreBenchExecResult{PdrInvPathprograms}{KinductionDfStaticEightTwoTTrueNotSolvedByKinductionPlain}{Correct}{True}{Cputime}{}{10654.602968223}%
\StoreBenchExecResult{PdrInvPathprograms}{KinductionDfStaticEightTwoTTrueNotSolvedByKinductionPlain}{Correct}{True}{Cputime}{Avg}{128.3687104605180722891566265}%
\StoreBenchExecResult{PdrInvPathprograms}{KinductionDfStaticEightTwoTTrueNotSolvedByKinductionPlain}{Correct}{True}{Cputime}{Median}{39.809312868}%
\StoreBenchExecResult{PdrInvPathprograms}{KinductionDfStaticEightTwoTTrueNotSolvedByKinductionPlain}{Correct}{True}{Cputime}{Min}{3.2741362}%
\StoreBenchExecResult{PdrInvPathprograms}{KinductionDfStaticEightTwoTTrueNotSolvedByKinductionPlain}{Correct}{True}{Cputime}{Max}{898.051709025}%
\StoreBenchExecResult{PdrInvPathprograms}{KinductionDfStaticEightTwoTTrueNotSolvedByKinductionPlain}{Correct}{True}{Cputime}{Stdev}{196.0085401852064763373261141}%
\StoreBenchExecResult{PdrInvPathprograms}{KinductionDfStaticEightTwoTTrueNotSolvedByKinductionPlain}{Correct}{True}{Walltime}{}{6128.85636758827}%
\StoreBenchExecResult{PdrInvPathprograms}{KinductionDfStaticEightTwoTTrueNotSolvedByKinductionPlain}{Correct}{True}{Walltime}{Avg}{73.84164298299120481927710843}%
\StoreBenchExecResult{PdrInvPathprograms}{KinductionDfStaticEightTwoTTrueNotSolvedByKinductionPlain}{Correct}{True}{Walltime}{Median}{20.2034180164}%
\StoreBenchExecResult{PdrInvPathprograms}{KinductionDfStaticEightTwoTTrueNotSolvedByKinductionPlain}{Correct}{True}{Walltime}{Min}{1.79608201981}%
\StoreBenchExecResult{PdrInvPathprograms}{KinductionDfStaticEightTwoTTrueNotSolvedByKinductionPlain}{Correct}{True}{Walltime}{Max}{695.898958206}%
\StoreBenchExecResult{PdrInvPathprograms}{KinductionDfStaticEightTwoTTrueNotSolvedByKinductionPlain}{Correct}{True}{Walltime}{Stdev}{123.6438010218835952549116862}%
\StoreBenchExecResult{PdrInvPathprograms}{KinductionDfStaticEightTwoTTrueNotSolvedByKinductionPlain}{Wrong}{True}{Count}{}{0}%
\StoreBenchExecResult{PdrInvPathprograms}{KinductionDfStaticEightTwoTTrueNotSolvedByKinductionPlain}{Wrong}{True}{Cputime}{}{0}%
\StoreBenchExecResult{PdrInvPathprograms}{KinductionDfStaticEightTwoTTrueNotSolvedByKinductionPlain}{Wrong}{True}{Cputime}{Avg}{None}%
\StoreBenchExecResult{PdrInvPathprograms}{KinductionDfStaticEightTwoTTrueNotSolvedByKinductionPlain}{Wrong}{True}{Cputime}{Median}{None}%
\StoreBenchExecResult{PdrInvPathprograms}{KinductionDfStaticEightTwoTTrueNotSolvedByKinductionPlain}{Wrong}{True}{Cputime}{Min}{None}%
\StoreBenchExecResult{PdrInvPathprograms}{KinductionDfStaticEightTwoTTrueNotSolvedByKinductionPlain}{Wrong}{True}{Cputime}{Max}{None}%
\StoreBenchExecResult{PdrInvPathprograms}{KinductionDfStaticEightTwoTTrueNotSolvedByKinductionPlain}{Wrong}{True}{Cputime}{Stdev}{None}%
\StoreBenchExecResult{PdrInvPathprograms}{KinductionDfStaticEightTwoTTrueNotSolvedByKinductionPlain}{Wrong}{True}{Walltime}{}{0}%
\StoreBenchExecResult{PdrInvPathprograms}{KinductionDfStaticEightTwoTTrueNotSolvedByKinductionPlain}{Wrong}{True}{Walltime}{Avg}{None}%
\StoreBenchExecResult{PdrInvPathprograms}{KinductionDfStaticEightTwoTTrueNotSolvedByKinductionPlain}{Wrong}{True}{Walltime}{Median}{None}%
\StoreBenchExecResult{PdrInvPathprograms}{KinductionDfStaticEightTwoTTrueNotSolvedByKinductionPlain}{Wrong}{True}{Walltime}{Min}{None}%
\StoreBenchExecResult{PdrInvPathprograms}{KinductionDfStaticEightTwoTTrueNotSolvedByKinductionPlain}{Wrong}{True}{Walltime}{Max}{None}%
\StoreBenchExecResult{PdrInvPathprograms}{KinductionDfStaticEightTwoTTrueNotSolvedByKinductionPlain}{Wrong}{True}{Walltime}{Stdev}{None}%
\StoreBenchExecResult{PdrInvPathprograms}{KinductionDfStaticEightTwoTTrueNotSolvedByKinductionPlain}{Error}{}{Count}{}{31}%
\StoreBenchExecResult{PdrInvPathprograms}{KinductionDfStaticEightTwoTTrueNotSolvedByKinductionPlain}{Error}{}{Cputime}{}{27003.281658690}%
\StoreBenchExecResult{PdrInvPathprograms}{KinductionDfStaticEightTwoTTrueNotSolvedByKinductionPlain}{Error}{}{Cputime}{Avg}{871.0736018932258064516129032}%
\StoreBenchExecResult{PdrInvPathprograms}{KinductionDfStaticEightTwoTTrueNotSolvedByKinductionPlain}{Error}{}{Cputime}{Median}{902.942360903}%
\StoreBenchExecResult{PdrInvPathprograms}{KinductionDfStaticEightTwoTTrueNotSolvedByKinductionPlain}{Error}{}{Cputime}{Min}{235.819950611}%
\StoreBenchExecResult{PdrInvPathprograms}{KinductionDfStaticEightTwoTTrueNotSolvedByKinductionPlain}{Error}{}{Cputime}{Max}{1000.74474958}%
\StoreBenchExecResult{PdrInvPathprograms}{KinductionDfStaticEightTwoTTrueNotSolvedByKinductionPlain}{Error}{}{Cputime}{Stdev}{133.7697253783802163454155637}%
\StoreBenchExecResult{PdrInvPathprograms}{KinductionDfStaticEightTwoTTrueNotSolvedByKinductionPlain}{Error}{}{Walltime}{}{19740.681125400}%
\StoreBenchExecResult{PdrInvPathprograms}{KinductionDfStaticEightTwoTTrueNotSolvedByKinductionPlain}{Error}{}{Walltime}{Avg}{636.7961653354838709677419355}%
\StoreBenchExecResult{PdrInvPathprograms}{KinductionDfStaticEightTwoTTrueNotSolvedByKinductionPlain}{Error}{}{Walltime}{Median}{681.305367947}%
\StoreBenchExecResult{PdrInvPathprograms}{KinductionDfStaticEightTwoTTrueNotSolvedByKinductionPlain}{Error}{}{Walltime}{Min}{144.522119045}%
\StoreBenchExecResult{PdrInvPathprograms}{KinductionDfStaticEightTwoTTrueNotSolvedByKinductionPlain}{Error}{}{Walltime}{Max}{898.08901906}%
\StoreBenchExecResult{PdrInvPathprograms}{KinductionDfStaticEightTwoTTrueNotSolvedByKinductionPlain}{Error}{}{Walltime}{Stdev}{220.5636834612379094228146535}%
\StoreBenchExecResult{PdrInvPathprograms}{KinductionDfStaticEightTwoTTrueNotSolvedByKinductionPlain}{Error}{OutOfJavaMemory}{Count}{}{2}%
\StoreBenchExecResult{PdrInvPathprograms}{KinductionDfStaticEightTwoTTrueNotSolvedByKinductionPlain}{Error}{OutOfJavaMemory}{Cputime}{}{783.510348824}%
\StoreBenchExecResult{PdrInvPathprograms}{KinductionDfStaticEightTwoTTrueNotSolvedByKinductionPlain}{Error}{OutOfJavaMemory}{Cputime}{Avg}{391.755174412}%
\StoreBenchExecResult{PdrInvPathprograms}{KinductionDfStaticEightTwoTTrueNotSolvedByKinductionPlain}{Error}{OutOfJavaMemory}{Cputime}{Median}{391.755174412}%
\StoreBenchExecResult{PdrInvPathprograms}{KinductionDfStaticEightTwoTTrueNotSolvedByKinductionPlain}{Error}{OutOfJavaMemory}{Cputime}{Min}{235.819950611}%
\StoreBenchExecResult{PdrInvPathprograms}{KinductionDfStaticEightTwoTTrueNotSolvedByKinductionPlain}{Error}{OutOfJavaMemory}{Cputime}{Max}{547.690398213}%
\StoreBenchExecResult{PdrInvPathprograms}{KinductionDfStaticEightTwoTTrueNotSolvedByKinductionPlain}{Error}{OutOfJavaMemory}{Cputime}{Stdev}{155.935223801}%
\StoreBenchExecResult{PdrInvPathprograms}{KinductionDfStaticEightTwoTTrueNotSolvedByKinductionPlain}{Error}{OutOfJavaMemory}{Walltime}{}{423.194770097}%
\StoreBenchExecResult{PdrInvPathprograms}{KinductionDfStaticEightTwoTTrueNotSolvedByKinductionPlain}{Error}{OutOfJavaMemory}{Walltime}{Avg}{211.5973850485}%
\StoreBenchExecResult{PdrInvPathprograms}{KinductionDfStaticEightTwoTTrueNotSolvedByKinductionPlain}{Error}{OutOfJavaMemory}{Walltime}{Median}{211.5973850485}%
\StoreBenchExecResult{PdrInvPathprograms}{KinductionDfStaticEightTwoTTrueNotSolvedByKinductionPlain}{Error}{OutOfJavaMemory}{Walltime}{Min}{144.522119045}%
\StoreBenchExecResult{PdrInvPathprograms}{KinductionDfStaticEightTwoTTrueNotSolvedByKinductionPlain}{Error}{OutOfJavaMemory}{Walltime}{Max}{278.672651052}%
\StoreBenchExecResult{PdrInvPathprograms}{KinductionDfStaticEightTwoTTrueNotSolvedByKinductionPlain}{Error}{OutOfJavaMemory}{Walltime}{Stdev}{67.0752660035}%
\StoreBenchExecResult{PdrInvPathprograms}{KinductionDfStaticEightTwoTTrueNotSolvedByKinductionPlain}{Error}{OutOfMemory}{Count}{}{2}%
\StoreBenchExecResult{PdrInvPathprograms}{KinductionDfStaticEightTwoTTrueNotSolvedByKinductionPlain}{Error}{OutOfMemory}{Cputime}{}{1706.755596622}%
\StoreBenchExecResult{PdrInvPathprograms}{KinductionDfStaticEightTwoTTrueNotSolvedByKinductionPlain}{Error}{OutOfMemory}{Cputime}{Avg}{853.377798311}%
\StoreBenchExecResult{PdrInvPathprograms}{KinductionDfStaticEightTwoTTrueNotSolvedByKinductionPlain}{Error}{OutOfMemory}{Cputime}{Median}{853.377798311}%
\StoreBenchExecResult{PdrInvPathprograms}{KinductionDfStaticEightTwoTTrueNotSolvedByKinductionPlain}{Error}{OutOfMemory}{Cputime}{Min}{852.079705363}%
\StoreBenchExecResult{PdrInvPathprograms}{KinductionDfStaticEightTwoTTrueNotSolvedByKinductionPlain}{Error}{OutOfMemory}{Cputime}{Max}{854.675891259}%
\StoreBenchExecResult{PdrInvPathprograms}{KinductionDfStaticEightTwoTTrueNotSolvedByKinductionPlain}{Error}{OutOfMemory}{Cputime}{Stdev}{1.298092948}%
\StoreBenchExecResult{PdrInvPathprograms}{KinductionDfStaticEightTwoTTrueNotSolvedByKinductionPlain}{Error}{OutOfMemory}{Walltime}{}{1674.055427789}%
\StoreBenchExecResult{PdrInvPathprograms}{KinductionDfStaticEightTwoTTrueNotSolvedByKinductionPlain}{Error}{OutOfMemory}{Walltime}{Avg}{837.0277138945}%
\StoreBenchExecResult{PdrInvPathprograms}{KinductionDfStaticEightTwoTTrueNotSolvedByKinductionPlain}{Error}{OutOfMemory}{Walltime}{Median}{837.0277138945}%
\StoreBenchExecResult{PdrInvPathprograms}{KinductionDfStaticEightTwoTTrueNotSolvedByKinductionPlain}{Error}{OutOfMemory}{Walltime}{Min}{835.730414867}%
\StoreBenchExecResult{PdrInvPathprograms}{KinductionDfStaticEightTwoTTrueNotSolvedByKinductionPlain}{Error}{OutOfMemory}{Walltime}{Max}{838.325012922}%
\StoreBenchExecResult{PdrInvPathprograms}{KinductionDfStaticEightTwoTTrueNotSolvedByKinductionPlain}{Error}{OutOfMemory}{Walltime}{Stdev}{1.2972990275}%
\StoreBenchExecResult{PdrInvPathprograms}{KinductionDfStaticEightTwoTTrueNotSolvedByKinductionPlain}{Error}{Timeout}{Count}{}{27}%
\StoreBenchExecResult{PdrInvPathprograms}{KinductionDfStaticEightTwoTTrueNotSolvedByKinductionPlain}{Error}{Timeout}{Cputime}{}{24513.015713244}%
\StoreBenchExecResult{PdrInvPathprograms}{KinductionDfStaticEightTwoTTrueNotSolvedByKinductionPlain}{Error}{Timeout}{Cputime}{Avg}{907.8894708608888888888888889}%
\StoreBenchExecResult{PdrInvPathprograms}{KinductionDfStaticEightTwoTTrueNotSolvedByKinductionPlain}{Error}{Timeout}{Cputime}{Median}{903.024274154}%
\StoreBenchExecResult{PdrInvPathprograms}{KinductionDfStaticEightTwoTTrueNotSolvedByKinductionPlain}{Error}{Timeout}{Cputime}{Min}{900.937191228}%
\StoreBenchExecResult{PdrInvPathprograms}{KinductionDfStaticEightTwoTTrueNotSolvedByKinductionPlain}{Error}{Timeout}{Cputime}{Max}{1000.74474958}%
\StoreBenchExecResult{PdrInvPathprograms}{KinductionDfStaticEightTwoTTrueNotSolvedByKinductionPlain}{Error}{Timeout}{Cputime}{Stdev}{18.63384986453860440780298865}%
\StoreBenchExecResult{PdrInvPathprograms}{KinductionDfStaticEightTwoTTrueNotSolvedByKinductionPlain}{Error}{Timeout}{Walltime}{}{17643.430927514}%
\StoreBenchExecResult{PdrInvPathprograms}{KinductionDfStaticEightTwoTTrueNotSolvedByKinductionPlain}{Error}{Timeout}{Walltime}{Avg}{653.4604047227407407407407407}%
\StoreBenchExecResult{PdrInvPathprograms}{KinductionDfStaticEightTwoTTrueNotSolvedByKinductionPlain}{Error}{Timeout}{Walltime}{Median}{681.305367947}%
\StoreBenchExecResult{PdrInvPathprograms}{KinductionDfStaticEightTwoTTrueNotSolvedByKinductionPlain}{Error}{Timeout}{Walltime}{Min}{451.166331053}%
\StoreBenchExecResult{PdrInvPathprograms}{KinductionDfStaticEightTwoTTrueNotSolvedByKinductionPlain}{Error}{Timeout}{Walltime}{Max}{898.08901906}%
\StoreBenchExecResult{PdrInvPathprograms}{KinductionDfStaticEightTwoTTrueNotSolvedByKinductionPlain}{Error}{Timeout}{Walltime}{Stdev}{197.1862981241890484629243161}%
\providecommand\StoreBenchExecResult[7]{\expandafter\newcommand\csname#1#2#3#4#5#6\endcsname{#7}}%
\StoreBenchExecResult{PdrInvPathprograms}{KinductionDfTrueNotSolvedByKinductionPlain}{Total}{}{Count}{}{114}%
\StoreBenchExecResult{PdrInvPathprograms}{KinductionDfTrueNotSolvedByKinductionPlain}{Total}{}{Cputime}{}{24518.346937238}%
\StoreBenchExecResult{PdrInvPathprograms}{KinductionDfTrueNotSolvedByKinductionPlain}{Total}{}{Cputime}{Avg}{215.0732187477017543859649123}%
\StoreBenchExecResult{PdrInvPathprograms}{KinductionDfTrueNotSolvedByKinductionPlain}{Total}{}{Cputime}{Median}{35.0728247395}%
\StoreBenchExecResult{PdrInvPathprograms}{KinductionDfTrueNotSolvedByKinductionPlain}{Total}{}{Cputime}{Min}{3.291779731}%
\StoreBenchExecResult{PdrInvPathprograms}{KinductionDfTrueNotSolvedByKinductionPlain}{Total}{}{Cputime}{Max}{905.693468505}%
\StoreBenchExecResult{PdrInvPathprograms}{KinductionDfTrueNotSolvedByKinductionPlain}{Total}{}{Cputime}{Stdev}{312.6603975762021073467330447}%
\StoreBenchExecResult{PdrInvPathprograms}{KinductionDfTrueNotSolvedByKinductionPlain}{Total}{}{Walltime}{}{13150.66718054044}%
\StoreBenchExecResult{PdrInvPathprograms}{KinductionDfTrueNotSolvedByKinductionPlain}{Total}{}{Walltime}{Avg}{115.3567296538635087719298246}%
\StoreBenchExecResult{PdrInvPathprograms}{KinductionDfTrueNotSolvedByKinductionPlain}{Total}{}{Walltime}{Median}{17.7938940525}%
\StoreBenchExecResult{PdrInvPathprograms}{KinductionDfTrueNotSolvedByKinductionPlain}{Total}{}{Walltime}{Min}{1.8098859787}%
\StoreBenchExecResult{PdrInvPathprograms}{KinductionDfTrueNotSolvedByKinductionPlain}{Total}{}{Walltime}{Max}{694.941759109}%
\StoreBenchExecResult{PdrInvPathprograms}{KinductionDfTrueNotSolvedByKinductionPlain}{Total}{}{Walltime}{Stdev}{168.5263941674911859261782480}%
\StoreBenchExecResult{PdrInvPathprograms}{KinductionDfTrueNotSolvedByKinductionPlain}{Correct}{}{Count}{}{99}%
\StoreBenchExecResult{PdrInvPathprograms}{KinductionDfTrueNotSolvedByKinductionPlain}{Correct}{}{Cputime}{}{11376.605482415}%
\StoreBenchExecResult{PdrInvPathprograms}{KinductionDfTrueNotSolvedByKinductionPlain}{Correct}{}{Cputime}{Avg}{114.9152068930808080808080808}%
\StoreBenchExecResult{PdrInvPathprograms}{KinductionDfTrueNotSolvedByKinductionPlain}{Correct}{}{Cputime}{Median}{25.510950194}%
\StoreBenchExecResult{PdrInvPathprograms}{KinductionDfTrueNotSolvedByKinductionPlain}{Correct}{}{Cputime}{Min}{3.291779731}%
\StoreBenchExecResult{PdrInvPathprograms}{KinductionDfTrueNotSolvedByKinductionPlain}{Correct}{}{Cputime}{Max}{848.648438629}%
\StoreBenchExecResult{PdrInvPathprograms}{KinductionDfTrueNotSolvedByKinductionPlain}{Correct}{}{Cputime}{Stdev}{186.6271377220947712901169616}%
\StoreBenchExecResult{PdrInvPathprograms}{KinductionDfTrueNotSolvedByKinductionPlain}{Correct}{}{Walltime}{}{6291.62521481644}%
\StoreBenchExecResult{PdrInvPathprograms}{KinductionDfTrueNotSolvedByKinductionPlain}{Correct}{}{Walltime}{Avg}{63.55176984663070707070707071}%
\StoreBenchExecResult{PdrInvPathprograms}{KinductionDfTrueNotSolvedByKinductionPlain}{Correct}{}{Walltime}{Median}{13.0032818317}%
\StoreBenchExecResult{PdrInvPathprograms}{KinductionDfTrueNotSolvedByKinductionPlain}{Correct}{}{Walltime}{Min}{1.8098859787}%
\StoreBenchExecResult{PdrInvPathprograms}{KinductionDfTrueNotSolvedByKinductionPlain}{Correct}{}{Walltime}{Max}{506.149466991}%
\StoreBenchExecResult{PdrInvPathprograms}{KinductionDfTrueNotSolvedByKinductionPlain}{Correct}{}{Walltime}{Stdev}{106.4658150652412100943847934}%
\StoreBenchExecResult{PdrInvPathprograms}{KinductionDfTrueNotSolvedByKinductionPlain}{Correct}{True}{Count}{}{99}%
\StoreBenchExecResult{PdrInvPathprograms}{KinductionDfTrueNotSolvedByKinductionPlain}{Correct}{True}{Cputime}{}{11376.605482415}%
\StoreBenchExecResult{PdrInvPathprograms}{KinductionDfTrueNotSolvedByKinductionPlain}{Correct}{True}{Cputime}{Avg}{114.9152068930808080808080808}%
\StoreBenchExecResult{PdrInvPathprograms}{KinductionDfTrueNotSolvedByKinductionPlain}{Correct}{True}{Cputime}{Median}{25.510950194}%
\StoreBenchExecResult{PdrInvPathprograms}{KinductionDfTrueNotSolvedByKinductionPlain}{Correct}{True}{Cputime}{Min}{3.291779731}%
\StoreBenchExecResult{PdrInvPathprograms}{KinductionDfTrueNotSolvedByKinductionPlain}{Correct}{True}{Cputime}{Max}{848.648438629}%
\StoreBenchExecResult{PdrInvPathprograms}{KinductionDfTrueNotSolvedByKinductionPlain}{Correct}{True}{Cputime}{Stdev}{186.6271377220947712901169616}%
\StoreBenchExecResult{PdrInvPathprograms}{KinductionDfTrueNotSolvedByKinductionPlain}{Correct}{True}{Walltime}{}{6291.62521481644}%
\StoreBenchExecResult{PdrInvPathprograms}{KinductionDfTrueNotSolvedByKinductionPlain}{Correct}{True}{Walltime}{Avg}{63.55176984663070707070707071}%
\StoreBenchExecResult{PdrInvPathprograms}{KinductionDfTrueNotSolvedByKinductionPlain}{Correct}{True}{Walltime}{Median}{13.0032818317}%
\StoreBenchExecResult{PdrInvPathprograms}{KinductionDfTrueNotSolvedByKinductionPlain}{Correct}{True}{Walltime}{Min}{1.8098859787}%
\StoreBenchExecResult{PdrInvPathprograms}{KinductionDfTrueNotSolvedByKinductionPlain}{Correct}{True}{Walltime}{Max}{506.149466991}%
\StoreBenchExecResult{PdrInvPathprograms}{KinductionDfTrueNotSolvedByKinductionPlain}{Correct}{True}{Walltime}{Stdev}{106.4658150652412100943847934}%
\StoreBenchExecResult{PdrInvPathprograms}{KinductionDfTrueNotSolvedByKinductionPlain}{Wrong}{True}{Count}{}{0}%
\StoreBenchExecResult{PdrInvPathprograms}{KinductionDfTrueNotSolvedByKinductionPlain}{Wrong}{True}{Cputime}{}{0}%
\StoreBenchExecResult{PdrInvPathprograms}{KinductionDfTrueNotSolvedByKinductionPlain}{Wrong}{True}{Cputime}{Avg}{None}%
\StoreBenchExecResult{PdrInvPathprograms}{KinductionDfTrueNotSolvedByKinductionPlain}{Wrong}{True}{Cputime}{Median}{None}%
\StoreBenchExecResult{PdrInvPathprograms}{KinductionDfTrueNotSolvedByKinductionPlain}{Wrong}{True}{Cputime}{Min}{None}%
\StoreBenchExecResult{PdrInvPathprograms}{KinductionDfTrueNotSolvedByKinductionPlain}{Wrong}{True}{Cputime}{Max}{None}%
\StoreBenchExecResult{PdrInvPathprograms}{KinductionDfTrueNotSolvedByKinductionPlain}{Wrong}{True}{Cputime}{Stdev}{None}%
\StoreBenchExecResult{PdrInvPathprograms}{KinductionDfTrueNotSolvedByKinductionPlain}{Wrong}{True}{Walltime}{}{0}%
\StoreBenchExecResult{PdrInvPathprograms}{KinductionDfTrueNotSolvedByKinductionPlain}{Wrong}{True}{Walltime}{Avg}{None}%
\StoreBenchExecResult{PdrInvPathprograms}{KinductionDfTrueNotSolvedByKinductionPlain}{Wrong}{True}{Walltime}{Median}{None}%
\StoreBenchExecResult{PdrInvPathprograms}{KinductionDfTrueNotSolvedByKinductionPlain}{Wrong}{True}{Walltime}{Min}{None}%
\StoreBenchExecResult{PdrInvPathprograms}{KinductionDfTrueNotSolvedByKinductionPlain}{Wrong}{True}{Walltime}{Max}{None}%
\StoreBenchExecResult{PdrInvPathprograms}{KinductionDfTrueNotSolvedByKinductionPlain}{Wrong}{True}{Walltime}{Stdev}{None}%
\StoreBenchExecResult{PdrInvPathprograms}{KinductionDfTrueNotSolvedByKinductionPlain}{Error}{}{Count}{}{15}%
\StoreBenchExecResult{PdrInvPathprograms}{KinductionDfTrueNotSolvedByKinductionPlain}{Error}{}{Cputime}{}{13141.741454823}%
\StoreBenchExecResult{PdrInvPathprograms}{KinductionDfTrueNotSolvedByKinductionPlain}{Error}{}{Cputime}{Avg}{876.1160969882}%
\StoreBenchExecResult{PdrInvPathprograms}{KinductionDfTrueNotSolvedByKinductionPlain}{Error}{}{Cputime}{Median}{902.081636279}%
\StoreBenchExecResult{PdrInvPathprograms}{KinductionDfTrueNotSolvedByKinductionPlain}{Error}{}{Cputime}{Min}{504.107534998}%
\StoreBenchExecResult{PdrInvPathprograms}{KinductionDfTrueNotSolvedByKinductionPlain}{Error}{}{Cputime}{Max}{905.693468505}%
\StoreBenchExecResult{PdrInvPathprograms}{KinductionDfTrueNotSolvedByKinductionPlain}{Error}{}{Cputime}{Stdev}{99.43446274156896181633101646}%
\StoreBenchExecResult{PdrInvPathprograms}{KinductionDfTrueNotSolvedByKinductionPlain}{Error}{}{Walltime}{}{6859.041965724}%
\StoreBenchExecResult{PdrInvPathprograms}{KinductionDfTrueNotSolvedByKinductionPlain}{Error}{}{Walltime}{Avg}{457.2694643816}%
\StoreBenchExecResult{PdrInvPathprograms}{KinductionDfTrueNotSolvedByKinductionPlain}{Error}{}{Walltime}{Median}{453.353525877}%
\StoreBenchExecResult{PdrInvPathprograms}{KinductionDfTrueNotSolvedByKinductionPlain}{Error}{}{Walltime}{Min}{258.119006872}%
\StoreBenchExecResult{PdrInvPathprograms}{KinductionDfTrueNotSolvedByKinductionPlain}{Error}{}{Walltime}{Max}{694.941759109}%
\StoreBenchExecResult{PdrInvPathprograms}{KinductionDfTrueNotSolvedByKinductionPlain}{Error}{}{Walltime}{Stdev}{80.12996852390953662495974103}%
\StoreBenchExecResult{PdrInvPathprograms}{KinductionDfTrueNotSolvedByKinductionPlain}{Error}{OutOfJavaMemory}{Count}{}{1}%
\StoreBenchExecResult{PdrInvPathprograms}{KinductionDfTrueNotSolvedByKinductionPlain}{Error}{OutOfJavaMemory}{Cputime}{}{504.107534998}%
\StoreBenchExecResult{PdrInvPathprograms}{KinductionDfTrueNotSolvedByKinductionPlain}{Error}{OutOfJavaMemory}{Cputime}{Avg}{504.107534998}%
\StoreBenchExecResult{PdrInvPathprograms}{KinductionDfTrueNotSolvedByKinductionPlain}{Error}{OutOfJavaMemory}{Cputime}{Median}{504.107534998}%
\StoreBenchExecResult{PdrInvPathprograms}{KinductionDfTrueNotSolvedByKinductionPlain}{Error}{OutOfJavaMemory}{Cputime}{Min}{504.107534998}%
\StoreBenchExecResult{PdrInvPathprograms}{KinductionDfTrueNotSolvedByKinductionPlain}{Error}{OutOfJavaMemory}{Cputime}{Max}{504.107534998}%
\StoreBenchExecResult{PdrInvPathprograms}{KinductionDfTrueNotSolvedByKinductionPlain}{Error}{OutOfJavaMemory}{Cputime}{Stdev}{0E-9}%
\StoreBenchExecResult{PdrInvPathprograms}{KinductionDfTrueNotSolvedByKinductionPlain}{Error}{OutOfJavaMemory}{Walltime}{}{258.119006872}%
\StoreBenchExecResult{PdrInvPathprograms}{KinductionDfTrueNotSolvedByKinductionPlain}{Error}{OutOfJavaMemory}{Walltime}{Avg}{258.119006872}%
\StoreBenchExecResult{PdrInvPathprograms}{KinductionDfTrueNotSolvedByKinductionPlain}{Error}{OutOfJavaMemory}{Walltime}{Median}{258.119006872}%
\StoreBenchExecResult{PdrInvPathprograms}{KinductionDfTrueNotSolvedByKinductionPlain}{Error}{OutOfJavaMemory}{Walltime}{Min}{258.119006872}%
\StoreBenchExecResult{PdrInvPathprograms}{KinductionDfTrueNotSolvedByKinductionPlain}{Error}{OutOfJavaMemory}{Walltime}{Max}{258.119006872}%
\StoreBenchExecResult{PdrInvPathprograms}{KinductionDfTrueNotSolvedByKinductionPlain}{Error}{OutOfJavaMemory}{Walltime}{Stdev}{0E-9}%
\StoreBenchExecResult{PdrInvPathprograms}{KinductionDfTrueNotSolvedByKinductionPlain}{Error}{Timeout}{Count}{}{14}%
\StoreBenchExecResult{PdrInvPathprograms}{KinductionDfTrueNotSolvedByKinductionPlain}{Error}{Timeout}{Cputime}{}{12637.633919825}%
\StoreBenchExecResult{PdrInvPathprograms}{KinductionDfTrueNotSolvedByKinductionPlain}{Error}{Timeout}{Cputime}{Avg}{902.6881371303571428571428571}%
\StoreBenchExecResult{PdrInvPathprograms}{KinductionDfTrueNotSolvedByKinductionPlain}{Error}{Timeout}{Cputime}{Median}{902.1912542035}%
\StoreBenchExecResult{PdrInvPathprograms}{KinductionDfTrueNotSolvedByKinductionPlain}{Error}{Timeout}{Cputime}{Min}{900.993297887}%
\StoreBenchExecResult{PdrInvPathprograms}{KinductionDfTrueNotSolvedByKinductionPlain}{Error}{Timeout}{Cputime}{Max}{905.693468505}%
\StoreBenchExecResult{PdrInvPathprograms}{KinductionDfTrueNotSolvedByKinductionPlain}{Error}{Timeout}{Cputime}{Stdev}{1.530384501656980605078940702}%
\StoreBenchExecResult{PdrInvPathprograms}{KinductionDfTrueNotSolvedByKinductionPlain}{Error}{Timeout}{Walltime}{}{6600.922958852}%
\StoreBenchExecResult{PdrInvPathprograms}{KinductionDfTrueNotSolvedByKinductionPlain}{Error}{Timeout}{Walltime}{Avg}{471.4944970608571428571428571}%
\StoreBenchExecResult{PdrInvPathprograms}{KinductionDfTrueNotSolvedByKinductionPlain}{Error}{Timeout}{Walltime}{Median}{453.464579940}%
\StoreBenchExecResult{PdrInvPathprograms}{KinductionDfTrueNotSolvedByKinductionPlain}{Error}{Timeout}{Walltime}{Min}{452.436374903}%
\StoreBenchExecResult{PdrInvPathprograms}{KinductionDfTrueNotSolvedByKinductionPlain}{Error}{Timeout}{Walltime}{Max}{694.941759109}%
\StoreBenchExecResult{PdrInvPathprograms}{KinductionDfTrueNotSolvedByKinductionPlain}{Error}{Timeout}{Walltime}{Stdev}{62.00135444499518013783893070}%
\providecommand\StoreBenchExecResult[7]{\expandafter\newcommand\csname#1#2#3#4#5#6\endcsname{#7}}%
\StoreBenchExecResult{PdrInvPathprograms}{KinductionKipdrdfTrueNotSolvedByKinductionPlain}{Total}{}{Count}{}{114}%
\StoreBenchExecResult{PdrInvPathprograms}{KinductionKipdrdfTrueNotSolvedByKinductionPlain}{Total}{}{Cputime}{}{24349.854041886}%
\StoreBenchExecResult{PdrInvPathprograms}{KinductionKipdrdfTrueNotSolvedByKinductionPlain}{Total}{}{Cputime}{Avg}{213.5952108937368421052631579}%
\StoreBenchExecResult{PdrInvPathprograms}{KinductionKipdrdfTrueNotSolvedByKinductionPlain}{Total}{}{Cputime}{Median}{35.725623178}%
\StoreBenchExecResult{PdrInvPathprograms}{KinductionKipdrdfTrueNotSolvedByKinductionPlain}{Total}{}{Cputime}{Min}{3.286148479}%
\StoreBenchExecResult{PdrInvPathprograms}{KinductionKipdrdfTrueNotSolvedByKinductionPlain}{Total}{}{Cputime}{Max}{905.602447968}%
\StoreBenchExecResult{PdrInvPathprograms}{KinductionKipdrdfTrueNotSolvedByKinductionPlain}{Total}{}{Cputime}{Stdev}{310.1485506223835998226070589}%
\StoreBenchExecResult{PdrInvPathprograms}{KinductionKipdrdfTrueNotSolvedByKinductionPlain}{Total}{}{Walltime}{}{13035.54750752507}%
\StoreBenchExecResult{PdrInvPathprograms}{KinductionKipdrdfTrueNotSolvedByKinductionPlain}{Total}{}{Walltime}{Avg}{114.3469079607462280701754386}%
\StoreBenchExecResult{PdrInvPathprograms}{KinductionKipdrdfTrueNotSolvedByKinductionPlain}{Total}{}{Walltime}{Median}{18.10811793805}%
\StoreBenchExecResult{PdrInvPathprograms}{KinductionKipdrdfTrueNotSolvedByKinductionPlain}{Total}{}{Walltime}{Min}{1.80929398537}%
\StoreBenchExecResult{PdrInvPathprograms}{KinductionKipdrdfTrueNotSolvedByKinductionPlain}{Total}{}{Walltime}{Max}{670.088958025}%
\StoreBenchExecResult{PdrInvPathprograms}{KinductionKipdrdfTrueNotSolvedByKinductionPlain}{Total}{}{Walltime}{Stdev}{166.7742286543851050553802981}%
\StoreBenchExecResult{PdrInvPathprograms}{KinductionKipdrdfTrueNotSolvedByKinductionPlain}{Correct}{}{Count}{}{100}%
\StoreBenchExecResult{PdrInvPathprograms}{KinductionKipdrdfTrueNotSolvedByKinductionPlain}{Correct}{}{Cputime}{}{12093.068106020}%
\StoreBenchExecResult{PdrInvPathprograms}{KinductionKipdrdfTrueNotSolvedByKinductionPlain}{Correct}{}{Cputime}{Avg}{120.9306810602}%
\StoreBenchExecResult{PdrInvPathprograms}{KinductionKipdrdfTrueNotSolvedByKinductionPlain}{Correct}{}{Cputime}{Median}{27.190704853}%
\StoreBenchExecResult{PdrInvPathprograms}{KinductionKipdrdfTrueNotSolvedByKinductionPlain}{Correct}{}{Cputime}{Min}{3.286148479}%
\StoreBenchExecResult{PdrInvPathprograms}{KinductionKipdrdfTrueNotSolvedByKinductionPlain}{Correct}{}{Cputime}{Max}{870.224613672}%
\StoreBenchExecResult{PdrInvPathprograms}{KinductionKipdrdfTrueNotSolvedByKinductionPlain}{Correct}{}{Cputime}{Stdev}{196.0063447579497993864006810}%
\StoreBenchExecResult{PdrInvPathprograms}{KinductionKipdrdfTrueNotSolvedByKinductionPlain}{Correct}{}{Walltime}{}{6864.59471178007}%
\StoreBenchExecResult{PdrInvPathprograms}{KinductionKipdrdfTrueNotSolvedByKinductionPlain}{Correct}{}{Walltime}{Avg}{68.6459471178007}%
\StoreBenchExecResult{PdrInvPathprograms}{KinductionKipdrdfTrueNotSolvedByKinductionPlain}{Correct}{}{Walltime}{Median}{14.0381740332}%
\StoreBenchExecResult{PdrInvPathprograms}{KinductionKipdrdfTrueNotSolvedByKinductionPlain}{Correct}{}{Walltime}{Min}{1.80929398537}%
\StoreBenchExecResult{PdrInvPathprograms}{KinductionKipdrdfTrueNotSolvedByKinductionPlain}{Correct}{}{Walltime}{Max}{670.088958025}%
\StoreBenchExecResult{PdrInvPathprograms}{KinductionKipdrdfTrueNotSolvedByKinductionPlain}{Correct}{}{Walltime}{Stdev}{119.9336339418239129019263755}%
\StoreBenchExecResult{PdrInvPathprograms}{KinductionKipdrdfTrueNotSolvedByKinductionPlain}{Correct}{True}{Count}{}{100}%
\StoreBenchExecResult{PdrInvPathprograms}{KinductionKipdrdfTrueNotSolvedByKinductionPlain}{Correct}{True}{Cputime}{}{12093.068106020}%
\StoreBenchExecResult{PdrInvPathprograms}{KinductionKipdrdfTrueNotSolvedByKinductionPlain}{Correct}{True}{Cputime}{Avg}{120.9306810602}%
\StoreBenchExecResult{PdrInvPathprograms}{KinductionKipdrdfTrueNotSolvedByKinductionPlain}{Correct}{True}{Cputime}{Median}{27.190704853}%
\StoreBenchExecResult{PdrInvPathprograms}{KinductionKipdrdfTrueNotSolvedByKinductionPlain}{Correct}{True}{Cputime}{Min}{3.286148479}%
\StoreBenchExecResult{PdrInvPathprograms}{KinductionKipdrdfTrueNotSolvedByKinductionPlain}{Correct}{True}{Cputime}{Max}{870.224613672}%
\StoreBenchExecResult{PdrInvPathprograms}{KinductionKipdrdfTrueNotSolvedByKinductionPlain}{Correct}{True}{Cputime}{Stdev}{196.0063447579497993864006810}%
\StoreBenchExecResult{PdrInvPathprograms}{KinductionKipdrdfTrueNotSolvedByKinductionPlain}{Correct}{True}{Walltime}{}{6864.59471178007}%
\StoreBenchExecResult{PdrInvPathprograms}{KinductionKipdrdfTrueNotSolvedByKinductionPlain}{Correct}{True}{Walltime}{Avg}{68.6459471178007}%
\StoreBenchExecResult{PdrInvPathprograms}{KinductionKipdrdfTrueNotSolvedByKinductionPlain}{Correct}{True}{Walltime}{Median}{14.0381740332}%
\StoreBenchExecResult{PdrInvPathprograms}{KinductionKipdrdfTrueNotSolvedByKinductionPlain}{Correct}{True}{Walltime}{Min}{1.80929398537}%
\StoreBenchExecResult{PdrInvPathprograms}{KinductionKipdrdfTrueNotSolvedByKinductionPlain}{Correct}{True}{Walltime}{Max}{670.088958025}%
\StoreBenchExecResult{PdrInvPathprograms}{KinductionKipdrdfTrueNotSolvedByKinductionPlain}{Correct}{True}{Walltime}{Stdev}{119.9336339418239129019263755}%
\StoreBenchExecResult{PdrInvPathprograms}{KinductionKipdrdfTrueNotSolvedByKinductionPlain}{Wrong}{True}{Count}{}{0}%
\StoreBenchExecResult{PdrInvPathprograms}{KinductionKipdrdfTrueNotSolvedByKinductionPlain}{Wrong}{True}{Cputime}{}{0}%
\StoreBenchExecResult{PdrInvPathprograms}{KinductionKipdrdfTrueNotSolvedByKinductionPlain}{Wrong}{True}{Cputime}{Avg}{None}%
\StoreBenchExecResult{PdrInvPathprograms}{KinductionKipdrdfTrueNotSolvedByKinductionPlain}{Wrong}{True}{Cputime}{Median}{None}%
\StoreBenchExecResult{PdrInvPathprograms}{KinductionKipdrdfTrueNotSolvedByKinductionPlain}{Wrong}{True}{Cputime}{Min}{None}%
\StoreBenchExecResult{PdrInvPathprograms}{KinductionKipdrdfTrueNotSolvedByKinductionPlain}{Wrong}{True}{Cputime}{Max}{None}%
\StoreBenchExecResult{PdrInvPathprograms}{KinductionKipdrdfTrueNotSolvedByKinductionPlain}{Wrong}{True}{Cputime}{Stdev}{None}%
\StoreBenchExecResult{PdrInvPathprograms}{KinductionKipdrdfTrueNotSolvedByKinductionPlain}{Wrong}{True}{Walltime}{}{0}%
\StoreBenchExecResult{PdrInvPathprograms}{KinductionKipdrdfTrueNotSolvedByKinductionPlain}{Wrong}{True}{Walltime}{Avg}{None}%
\StoreBenchExecResult{PdrInvPathprograms}{KinductionKipdrdfTrueNotSolvedByKinductionPlain}{Wrong}{True}{Walltime}{Median}{None}%
\StoreBenchExecResult{PdrInvPathprograms}{KinductionKipdrdfTrueNotSolvedByKinductionPlain}{Wrong}{True}{Walltime}{Min}{None}%
\StoreBenchExecResult{PdrInvPathprograms}{KinductionKipdrdfTrueNotSolvedByKinductionPlain}{Wrong}{True}{Walltime}{Max}{None}%
\StoreBenchExecResult{PdrInvPathprograms}{KinductionKipdrdfTrueNotSolvedByKinductionPlain}{Wrong}{True}{Walltime}{Stdev}{None}%
\StoreBenchExecResult{PdrInvPathprograms}{KinductionKipdrdfTrueNotSolvedByKinductionPlain}{Error}{}{Count}{}{14}%
\StoreBenchExecResult{PdrInvPathprograms}{KinductionKipdrdfTrueNotSolvedByKinductionPlain}{Error}{}{Cputime}{}{12256.785935866}%
\StoreBenchExecResult{PdrInvPathprograms}{KinductionKipdrdfTrueNotSolvedByKinductionPlain}{Error}{}{Cputime}{Avg}{875.4847097047142857142857143}%
\StoreBenchExecResult{PdrInvPathprograms}{KinductionKipdrdfTrueNotSolvedByKinductionPlain}{Error}{}{Cputime}{Median}{901.9126290765}%
\StoreBenchExecResult{PdrInvPathprograms}{KinductionKipdrdfTrueNotSolvedByKinductionPlain}{Error}{}{Cputime}{Min}{525.399517777}%
\StoreBenchExecResult{PdrInvPathprograms}{KinductionKipdrdfTrueNotSolvedByKinductionPlain}{Error}{}{Cputime}{Max}{905.602447968}%
\StoreBenchExecResult{PdrInvPathprograms}{KinductionKipdrdfTrueNotSolvedByKinductionPlain}{Error}{}{Cputime}{Stdev}{97.10595291782706858549995675}%
\StoreBenchExecResult{PdrInvPathprograms}{KinductionKipdrdfTrueNotSolvedByKinductionPlain}{Error}{}{Walltime}{}{6170.952795745}%
\StoreBenchExecResult{PdrInvPathprograms}{KinductionKipdrdfTrueNotSolvedByKinductionPlain}{Error}{}{Walltime}{Avg}{440.7823425532142857142857143}%
\StoreBenchExecResult{PdrInvPathprograms}{KinductionKipdrdfTrueNotSolvedByKinductionPlain}{Error}{}{Walltime}{Median}{452.973734975}%
\StoreBenchExecResult{PdrInvPathprograms}{KinductionKipdrdfTrueNotSolvedByKinductionPlain}{Error}{}{Walltime}{Min}{269.522426844}%
\StoreBenchExecResult{PdrInvPathprograms}{KinductionKipdrdfTrueNotSolvedByKinductionPlain}{Error}{}{Walltime}{Max}{459.128829956}%
\StoreBenchExecResult{PdrInvPathprograms}{KinductionKipdrdfTrueNotSolvedByKinductionPlain}{Error}{}{Walltime}{Stdev}{47.54591951863086692793060846}%
\StoreBenchExecResult{PdrInvPathprograms}{KinductionKipdrdfTrueNotSolvedByKinductionPlain}{Error}{OutOfJavaMemory}{Count}{}{1}%
\StoreBenchExecResult{PdrInvPathprograms}{KinductionKipdrdfTrueNotSolvedByKinductionPlain}{Error}{OutOfJavaMemory}{Cputime}{}{525.399517777}%
\StoreBenchExecResult{PdrInvPathprograms}{KinductionKipdrdfTrueNotSolvedByKinductionPlain}{Error}{OutOfJavaMemory}{Cputime}{Avg}{525.399517777}%
\StoreBenchExecResult{PdrInvPathprograms}{KinductionKipdrdfTrueNotSolvedByKinductionPlain}{Error}{OutOfJavaMemory}{Cputime}{Median}{525.399517777}%
\StoreBenchExecResult{PdrInvPathprograms}{KinductionKipdrdfTrueNotSolvedByKinductionPlain}{Error}{OutOfJavaMemory}{Cputime}{Min}{525.399517777}%
\StoreBenchExecResult{PdrInvPathprograms}{KinductionKipdrdfTrueNotSolvedByKinductionPlain}{Error}{OutOfJavaMemory}{Cputime}{Max}{525.399517777}%
\StoreBenchExecResult{PdrInvPathprograms}{KinductionKipdrdfTrueNotSolvedByKinductionPlain}{Error}{OutOfJavaMemory}{Cputime}{Stdev}{0E-9}%
\StoreBenchExecResult{PdrInvPathprograms}{KinductionKipdrdfTrueNotSolvedByKinductionPlain}{Error}{OutOfJavaMemory}{Walltime}{}{269.522426844}%
\StoreBenchExecResult{PdrInvPathprograms}{KinductionKipdrdfTrueNotSolvedByKinductionPlain}{Error}{OutOfJavaMemory}{Walltime}{Avg}{269.522426844}%
\StoreBenchExecResult{PdrInvPathprograms}{KinductionKipdrdfTrueNotSolvedByKinductionPlain}{Error}{OutOfJavaMemory}{Walltime}{Median}{269.522426844}%
\StoreBenchExecResult{PdrInvPathprograms}{KinductionKipdrdfTrueNotSolvedByKinductionPlain}{Error}{OutOfJavaMemory}{Walltime}{Min}{269.522426844}%
\StoreBenchExecResult{PdrInvPathprograms}{KinductionKipdrdfTrueNotSolvedByKinductionPlain}{Error}{OutOfJavaMemory}{Walltime}{Max}{269.522426844}%
\StoreBenchExecResult{PdrInvPathprograms}{KinductionKipdrdfTrueNotSolvedByKinductionPlain}{Error}{OutOfJavaMemory}{Walltime}{Stdev}{0E-9}%
\StoreBenchExecResult{PdrInvPathprograms}{KinductionKipdrdfTrueNotSolvedByKinductionPlain}{Error}{Timeout}{Count}{}{13}%
\StoreBenchExecResult{PdrInvPathprograms}{KinductionKipdrdfTrueNotSolvedByKinductionPlain}{Error}{Timeout}{Cputime}{}{11731.386418089}%
\StoreBenchExecResult{PdrInvPathprograms}{KinductionKipdrdfTrueNotSolvedByKinductionPlain}{Error}{Timeout}{Cputime}{Avg}{902.414339853}%
\StoreBenchExecResult{PdrInvPathprograms}{KinductionKipdrdfTrueNotSolvedByKinductionPlain}{Error}{Timeout}{Cputime}{Median}{901.993194977}%
\StoreBenchExecResult{PdrInvPathprograms}{KinductionKipdrdfTrueNotSolvedByKinductionPlain}{Error}{Timeout}{Cputime}{Min}{900.98805237}%
\StoreBenchExecResult{PdrInvPathprograms}{KinductionKipdrdfTrueNotSolvedByKinductionPlain}{Error}{Timeout}{Cputime}{Max}{905.602447968}%
\StoreBenchExecResult{PdrInvPathprograms}{KinductionKipdrdfTrueNotSolvedByKinductionPlain}{Error}{Timeout}{Cputime}{Stdev}{1.430947589212513884107605523}%
\StoreBenchExecResult{PdrInvPathprograms}{KinductionKipdrdfTrueNotSolvedByKinductionPlain}{Error}{Timeout}{Walltime}{}{5901.430368901}%
\StoreBenchExecResult{PdrInvPathprograms}{KinductionKipdrdfTrueNotSolvedByKinductionPlain}{Error}{Timeout}{Walltime}{Avg}{453.9561822231538461538461538}%
\StoreBenchExecResult{PdrInvPathprograms}{KinductionKipdrdfTrueNotSolvedByKinductionPlain}{Error}{Timeout}{Walltime}{Median}{453.267354012}%
\StoreBenchExecResult{PdrInvPathprograms}{KinductionKipdrdfTrueNotSolvedByKinductionPlain}{Error}{Timeout}{Walltime}{Min}{451.667281866}%
\StoreBenchExecResult{PdrInvPathprograms}{KinductionKipdrdfTrueNotSolvedByKinductionPlain}{Error}{Timeout}{Walltime}{Max}{459.128829956}%
\StoreBenchExecResult{PdrInvPathprograms}{KinductionKipdrdfTrueNotSolvedByKinductionPlain}{Error}{Timeout}{Walltime}{Stdev}{2.192523660814291568239124110}%
\providecommand\StoreBenchExecResult[7]{\expandafter\newcommand\csname#1#2#3#4#5#6\endcsname{#7}}%
\StoreBenchExecResult{PdrInvPathprograms}{KinductionKipdrTrueNotSolvedByKinductionPlain}{Total}{}{Count}{}{114}%
\StoreBenchExecResult{PdrInvPathprograms}{KinductionKipdrTrueNotSolvedByKinductionPlain}{Total}{}{Cputime}{}{86678.810449876}%
\StoreBenchExecResult{PdrInvPathprograms}{KinductionKipdrTrueNotSolvedByKinductionPlain}{Total}{}{Cputime}{Avg}{760.3404425427719298245614035}%
\StoreBenchExecResult{PdrInvPathprograms}{KinductionKipdrTrueNotSolvedByKinductionPlain}{Total}{}{Cputime}{Median}{902.3387082625}%
\StoreBenchExecResult{PdrInvPathprograms}{KinductionKipdrTrueNotSolvedByKinductionPlain}{Total}{}{Cputime}{Min}{3.930682161}%
\StoreBenchExecResult{PdrInvPathprograms}{KinductionKipdrTrueNotSolvedByKinductionPlain}{Total}{}{Cputime}{Max}{1000.66668317}%
\StoreBenchExecResult{PdrInvPathprograms}{KinductionKipdrTrueNotSolvedByKinductionPlain}{Total}{}{Cputime}{Stdev}{309.5056350312842344197956811}%
\StoreBenchExecResult{PdrInvPathprograms}{KinductionKipdrTrueNotSolvedByKinductionPlain}{Total}{}{Walltime}{}{43986.34993410065}%
\StoreBenchExecResult{PdrInvPathprograms}{KinductionKipdrTrueNotSolvedByKinductionPlain}{Total}{}{Walltime}{Avg}{385.8451748605320175438596491}%
\StoreBenchExecResult{PdrInvPathprograms}{KinductionKipdrTrueNotSolvedByKinductionPlain}{Total}{}{Walltime}{Median}{452.1061639785}%
\StoreBenchExecResult{PdrInvPathprograms}{KinductionKipdrTrueNotSolvedByKinductionPlain}{Total}{}{Walltime}{Min}{2.16458201408}%
\StoreBenchExecResult{PdrInvPathprograms}{KinductionKipdrTrueNotSolvedByKinductionPlain}{Total}{}{Walltime}{Max}{546.541609049}%
\StoreBenchExecResult{PdrInvPathprograms}{KinductionKipdrTrueNotSolvedByKinductionPlain}{Total}{}{Walltime}{Stdev}{152.7644172108835944347505034}%
\StoreBenchExecResult{PdrInvPathprograms}{KinductionKipdrTrueNotSolvedByKinductionPlain}{Correct}{}{Count}{}{10}%
\StoreBenchExecResult{PdrInvPathprograms}{KinductionKipdrTrueNotSolvedByKinductionPlain}{Correct}{}{Cputime}{}{444.823119940}%
\StoreBenchExecResult{PdrInvPathprograms}{KinductionKipdrTrueNotSolvedByKinductionPlain}{Correct}{}{Cputime}{Avg}{44.482311994}%
\StoreBenchExecResult{PdrInvPathprograms}{KinductionKipdrTrueNotSolvedByKinductionPlain}{Correct}{}{Cputime}{Median}{5.3520158005}%
\StoreBenchExecResult{PdrInvPathprograms}{KinductionKipdrTrueNotSolvedByKinductionPlain}{Correct}{}{Cputime}{Min}{3.930682161}%
\StoreBenchExecResult{PdrInvPathprograms}{KinductionKipdrTrueNotSolvedByKinductionPlain}{Correct}{}{Cputime}{Max}{354.440635031}%
\StoreBenchExecResult{PdrInvPathprograms}{KinductionKipdrTrueNotSolvedByKinductionPlain}{Correct}{}{Cputime}{Stdev}{103.8511688648563336514295253}%
\StoreBenchExecResult{PdrInvPathprograms}{KinductionKipdrTrueNotSolvedByKinductionPlain}{Correct}{}{Walltime}{}{227.36772418005}%
\StoreBenchExecResult{PdrInvPathprograms}{KinductionKipdrTrueNotSolvedByKinductionPlain}{Correct}{}{Walltime}{Avg}{22.736772418005}%
\StoreBenchExecResult{PdrInvPathprograms}{KinductionKipdrTrueNotSolvedByKinductionPlain}{Correct}{}{Walltime}{Median}{2.91957008839}%
\StoreBenchExecResult{PdrInvPathprograms}{KinductionKipdrTrueNotSolvedByKinductionPlain}{Correct}{}{Walltime}{Min}{2.16458201408}%
\StoreBenchExecResult{PdrInvPathprograms}{KinductionKipdrTrueNotSolvedByKinductionPlain}{Correct}{}{Walltime}{Max}{179.861695051}%
\StoreBenchExecResult{PdrInvPathprograms}{KinductionKipdrTrueNotSolvedByKinductionPlain}{Correct}{}{Walltime}{Stdev}{52.64742245826541199468189167}%
\StoreBenchExecResult{PdrInvPathprograms}{KinductionKipdrTrueNotSolvedByKinductionPlain}{Correct}{True}{Count}{}{10}%
\StoreBenchExecResult{PdrInvPathprograms}{KinductionKipdrTrueNotSolvedByKinductionPlain}{Correct}{True}{Cputime}{}{444.823119940}%
\StoreBenchExecResult{PdrInvPathprograms}{KinductionKipdrTrueNotSolvedByKinductionPlain}{Correct}{True}{Cputime}{Avg}{44.482311994}%
\StoreBenchExecResult{PdrInvPathprograms}{KinductionKipdrTrueNotSolvedByKinductionPlain}{Correct}{True}{Cputime}{Median}{5.3520158005}%
\StoreBenchExecResult{PdrInvPathprograms}{KinductionKipdrTrueNotSolvedByKinductionPlain}{Correct}{True}{Cputime}{Min}{3.930682161}%
\StoreBenchExecResult{PdrInvPathprograms}{KinductionKipdrTrueNotSolvedByKinductionPlain}{Correct}{True}{Cputime}{Max}{354.440635031}%
\StoreBenchExecResult{PdrInvPathprograms}{KinductionKipdrTrueNotSolvedByKinductionPlain}{Correct}{True}{Cputime}{Stdev}{103.8511688648563336514295253}%
\StoreBenchExecResult{PdrInvPathprograms}{KinductionKipdrTrueNotSolvedByKinductionPlain}{Correct}{True}{Walltime}{}{227.36772418005}%
\StoreBenchExecResult{PdrInvPathprograms}{KinductionKipdrTrueNotSolvedByKinductionPlain}{Correct}{True}{Walltime}{Avg}{22.736772418005}%
\StoreBenchExecResult{PdrInvPathprograms}{KinductionKipdrTrueNotSolvedByKinductionPlain}{Correct}{True}{Walltime}{Median}{2.91957008839}%
\StoreBenchExecResult{PdrInvPathprograms}{KinductionKipdrTrueNotSolvedByKinductionPlain}{Correct}{True}{Walltime}{Min}{2.16458201408}%
\StoreBenchExecResult{PdrInvPathprograms}{KinductionKipdrTrueNotSolvedByKinductionPlain}{Correct}{True}{Walltime}{Max}{179.861695051}%
\StoreBenchExecResult{PdrInvPathprograms}{KinductionKipdrTrueNotSolvedByKinductionPlain}{Correct}{True}{Walltime}{Stdev}{52.64742245826541199468189167}%
\StoreBenchExecResult{PdrInvPathprograms}{KinductionKipdrTrueNotSolvedByKinductionPlain}{Wrong}{True}{Count}{}{0}%
\StoreBenchExecResult{PdrInvPathprograms}{KinductionKipdrTrueNotSolvedByKinductionPlain}{Wrong}{True}{Cputime}{}{0}%
\StoreBenchExecResult{PdrInvPathprograms}{KinductionKipdrTrueNotSolvedByKinductionPlain}{Wrong}{True}{Cputime}{Avg}{None}%
\StoreBenchExecResult{PdrInvPathprograms}{KinductionKipdrTrueNotSolvedByKinductionPlain}{Wrong}{True}{Cputime}{Median}{None}%
\StoreBenchExecResult{PdrInvPathprograms}{KinductionKipdrTrueNotSolvedByKinductionPlain}{Wrong}{True}{Cputime}{Min}{None}%
\StoreBenchExecResult{PdrInvPathprograms}{KinductionKipdrTrueNotSolvedByKinductionPlain}{Wrong}{True}{Cputime}{Max}{None}%
\StoreBenchExecResult{PdrInvPathprograms}{KinductionKipdrTrueNotSolvedByKinductionPlain}{Wrong}{True}{Cputime}{Stdev}{None}%
\StoreBenchExecResult{PdrInvPathprograms}{KinductionKipdrTrueNotSolvedByKinductionPlain}{Wrong}{True}{Walltime}{}{0}%
\StoreBenchExecResult{PdrInvPathprograms}{KinductionKipdrTrueNotSolvedByKinductionPlain}{Wrong}{True}{Walltime}{Avg}{None}%
\StoreBenchExecResult{PdrInvPathprograms}{KinductionKipdrTrueNotSolvedByKinductionPlain}{Wrong}{True}{Walltime}{Median}{None}%
\StoreBenchExecResult{PdrInvPathprograms}{KinductionKipdrTrueNotSolvedByKinductionPlain}{Wrong}{True}{Walltime}{Min}{None}%
\StoreBenchExecResult{PdrInvPathprograms}{KinductionKipdrTrueNotSolvedByKinductionPlain}{Wrong}{True}{Walltime}{Max}{None}%
\StoreBenchExecResult{PdrInvPathprograms}{KinductionKipdrTrueNotSolvedByKinductionPlain}{Wrong}{True}{Walltime}{Stdev}{None}%
\StoreBenchExecResult{PdrInvPathprograms}{KinductionKipdrTrueNotSolvedByKinductionPlain}{Error}{}{Count}{}{104}%
\StoreBenchExecResult{PdrInvPathprograms}{KinductionKipdrTrueNotSolvedByKinductionPlain}{Error}{}{Cputime}{}{86233.987329936}%
\StoreBenchExecResult{PdrInvPathprograms}{KinductionKipdrTrueNotSolvedByKinductionPlain}{Error}{}{Cputime}{Avg}{829.1729550955384615384615385}%
\StoreBenchExecResult{PdrInvPathprograms}{KinductionKipdrTrueNotSolvedByKinductionPlain}{Error}{}{Cputime}{Median}{902.657465653}%
\StoreBenchExecResult{PdrInvPathprograms}{KinductionKipdrTrueNotSolvedByKinductionPlain}{Error}{}{Cputime}{Min}{5.072465065}%
\StoreBenchExecResult{PdrInvPathprograms}{KinductionKipdrTrueNotSolvedByKinductionPlain}{Error}{}{Cputime}{Max}{1000.66668317}%
\StoreBenchExecResult{PdrInvPathprograms}{KinductionKipdrTrueNotSolvedByKinductionPlain}{Error}{}{Cputime}{Stdev}{223.5070925607054175545368983}%
\StoreBenchExecResult{PdrInvPathprograms}{KinductionKipdrTrueNotSolvedByKinductionPlain}{Error}{}{Walltime}{}{43758.98220992060}%
\StoreBenchExecResult{PdrInvPathprograms}{KinductionKipdrTrueNotSolvedByKinductionPlain}{Error}{}{Walltime}{Avg}{420.7594443261596153846153846}%
\StoreBenchExecResult{PdrInvPathprograms}{KinductionKipdrTrueNotSolvedByKinductionPlain}{Error}{}{Walltime}{Median}{452.408055067}%
\StoreBenchExecResult{PdrInvPathprograms}{KinductionKipdrTrueNotSolvedByKinductionPlain}{Error}{}{Walltime}{Min}{2.75841403008}%
\StoreBenchExecResult{PdrInvPathprograms}{KinductionKipdrTrueNotSolvedByKinductionPlain}{Error}{}{Walltime}{Max}{546.541609049}%
\StoreBenchExecResult{PdrInvPathprograms}{KinductionKipdrTrueNotSolvedByKinductionPlain}{Error}{}{Walltime}{Stdev}{106.8537363309192547315887770}%
\StoreBenchExecResult{PdrInvPathprograms}{KinductionKipdrTrueNotSolvedByKinductionPlain}{Error}{Error}{Count}{}{5}%
\StoreBenchExecResult{PdrInvPathprograms}{KinductionKipdrTrueNotSolvedByKinductionPlain}{Error}{Error}{Cputime}{}{943.334091609}%
\StoreBenchExecResult{PdrInvPathprograms}{KinductionKipdrTrueNotSolvedByKinductionPlain}{Error}{Error}{Cputime}{Avg}{188.6668183218}%
\StoreBenchExecResult{PdrInvPathprograms}{KinductionKipdrTrueNotSolvedByKinductionPlain}{Error}{Error}{Cputime}{Median}{131.125702909}%
\StoreBenchExecResult{PdrInvPathprograms}{KinductionKipdrTrueNotSolvedByKinductionPlain}{Error}{Error}{Cputime}{Min}{109.535690626}%
\StoreBenchExecResult{PdrInvPathprograms}{KinductionKipdrTrueNotSolvedByKinductionPlain}{Error}{Error}{Cputime}{Max}{445.155296816}%
\StoreBenchExecResult{PdrInvPathprograms}{KinductionKipdrTrueNotSolvedByKinductionPlain}{Error}{Error}{Cputime}{Stdev}{128.8210202064423038674625274}%
\StoreBenchExecResult{PdrInvPathprograms}{KinductionKipdrTrueNotSolvedByKinductionPlain}{Error}{Error}{Walltime}{}{796.9146015647}%
\StoreBenchExecResult{PdrInvPathprograms}{KinductionKipdrTrueNotSolvedByKinductionPlain}{Error}{Error}{Walltime}{Avg}{159.38292031294}%
\StoreBenchExecResult{PdrInvPathprograms}{KinductionKipdrTrueNotSolvedByKinductionPlain}{Error}{Error}{Walltime}{Median}{99.0471858978}%
\StoreBenchExecResult{PdrInvPathprograms}{KinductionKipdrTrueNotSolvedByKinductionPlain}{Error}{Error}{Walltime}{Min}{81.0172679424}%
\StoreBenchExecResult{PdrInvPathprograms}{KinductionKipdrTrueNotSolvedByKinductionPlain}{Error}{Error}{Walltime}{Max}{418.556979895}%
\StoreBenchExecResult{PdrInvPathprograms}{KinductionKipdrTrueNotSolvedByKinductionPlain}{Error}{Error}{Walltime}{Stdev}{130.1780786272533479408758973}%
\StoreBenchExecResult{PdrInvPathprograms}{KinductionKipdrTrueNotSolvedByKinductionPlain}{Error}{OutOfJavaMemory}{Count}{}{4}%
\StoreBenchExecResult{PdrInvPathprograms}{KinductionKipdrTrueNotSolvedByKinductionPlain}{Error}{OutOfJavaMemory}{Cputime}{}{1985.309934181}%
\StoreBenchExecResult{PdrInvPathprograms}{KinductionKipdrTrueNotSolvedByKinductionPlain}{Error}{OutOfJavaMemory}{Cputime}{Avg}{496.32748354525}%
\StoreBenchExecResult{PdrInvPathprograms}{KinductionKipdrTrueNotSolvedByKinductionPlain}{Error}{OutOfJavaMemory}{Cputime}{Median}{456.143516939}%
\StoreBenchExecResult{PdrInvPathprograms}{KinductionKipdrTrueNotSolvedByKinductionPlain}{Error}{OutOfJavaMemory}{Cputime}{Min}{399.368695766}%
\StoreBenchExecResult{PdrInvPathprograms}{KinductionKipdrTrueNotSolvedByKinductionPlain}{Error}{OutOfJavaMemory}{Cputime}{Max}{673.654204537}%
\StoreBenchExecResult{PdrInvPathprograms}{KinductionKipdrTrueNotSolvedByKinductionPlain}{Error}{OutOfJavaMemory}{Cputime}{Stdev}{109.6290369365102197869973778}%
\StoreBenchExecResult{PdrInvPathprograms}{KinductionKipdrTrueNotSolvedByKinductionPlain}{Error}{OutOfJavaMemory}{Walltime}{}{1035.638725042}%
\StoreBenchExecResult{PdrInvPathprograms}{KinductionKipdrTrueNotSolvedByKinductionPlain}{Error}{OutOfJavaMemory}{Walltime}{Avg}{258.9096812605}%
\StoreBenchExecResult{PdrInvPathprograms}{KinductionKipdrTrueNotSolvedByKinductionPlain}{Error}{OutOfJavaMemory}{Walltime}{Median}{237.972959399}%
\StoreBenchExecResult{PdrInvPathprograms}{KinductionKipdrTrueNotSolvedByKinductionPlain}{Error}{OutOfJavaMemory}{Walltime}{Min}{211.779084206}%
\StoreBenchExecResult{PdrInvPathprograms}{KinductionKipdrTrueNotSolvedByKinductionPlain}{Error}{OutOfJavaMemory}{Walltime}{Max}{347.913722038}%
\StoreBenchExecResult{PdrInvPathprograms}{KinductionKipdrTrueNotSolvedByKinductionPlain}{Error}{OutOfJavaMemory}{Walltime}{Stdev}{54.33321409655694819182798349}%
\StoreBenchExecResult{PdrInvPathprograms}{KinductionKipdrTrueNotSolvedByKinductionPlain}{Error}{SegmentationFault}{Count}{}{3}%
\StoreBenchExecResult{PdrInvPathprograms}{KinductionKipdrTrueNotSolvedByKinductionPlain}{Error}{SegmentationFault}{Cputime}{}{22.445000716}%
\StoreBenchExecResult{PdrInvPathprograms}{KinductionKipdrTrueNotSolvedByKinductionPlain}{Error}{SegmentationFault}{Cputime}{Avg}{7.481666905333333333333333333}%
\StoreBenchExecResult{PdrInvPathprograms}{KinductionKipdrTrueNotSolvedByKinductionPlain}{Error}{SegmentationFault}{Cputime}{Median}{6.898001669}%
\StoreBenchExecResult{PdrInvPathprograms}{KinductionKipdrTrueNotSolvedByKinductionPlain}{Error}{SegmentationFault}{Cputime}{Min}{5.072465065}%
\StoreBenchExecResult{PdrInvPathprograms}{KinductionKipdrTrueNotSolvedByKinductionPlain}{Error}{SegmentationFault}{Cputime}{Max}{10.474533982}%
\StoreBenchExecResult{PdrInvPathprograms}{KinductionKipdrTrueNotSolvedByKinductionPlain}{Error}{SegmentationFault}{Cputime}{Stdev}{2.243670501222260221972946155}%
\StoreBenchExecResult{PdrInvPathprograms}{KinductionKipdrTrueNotSolvedByKinductionPlain}{Error}{SegmentationFault}{Walltime}{}{11.93145203590}%
\StoreBenchExecResult{PdrInvPathprograms}{KinductionKipdrTrueNotSolvedByKinductionPlain}{Error}{SegmentationFault}{Walltime}{Avg}{3.977150678633333333333333333}%
\StoreBenchExecResult{PdrInvPathprograms}{KinductionKipdrTrueNotSolvedByKinductionPlain}{Error}{SegmentationFault}{Walltime}{Median}{3.66410708427}%
\StoreBenchExecResult{PdrInvPathprograms}{KinductionKipdrTrueNotSolvedByKinductionPlain}{Error}{SegmentationFault}{Walltime}{Min}{2.75841403008}%
\StoreBenchExecResult{PdrInvPathprograms}{KinductionKipdrTrueNotSolvedByKinductionPlain}{Error}{SegmentationFault}{Walltime}{Max}{5.50893092155}%
\StoreBenchExecResult{PdrInvPathprograms}{KinductionKipdrTrueNotSolvedByKinductionPlain}{Error}{SegmentationFault}{Walltime}{Stdev}{1.144503680362331942675887443}%
\StoreBenchExecResult{PdrInvPathprograms}{KinductionKipdrTrueNotSolvedByKinductionPlain}{Error}{Timeout}{Count}{}{92}%
\StoreBenchExecResult{PdrInvPathprograms}{KinductionKipdrTrueNotSolvedByKinductionPlain}{Error}{Timeout}{Cputime}{}{83282.898303430}%
\StoreBenchExecResult{PdrInvPathprograms}{KinductionKipdrTrueNotSolvedByKinductionPlain}{Error}{Timeout}{Cputime}{Avg}{905.2488946025}%
\StoreBenchExecResult{PdrInvPathprograms}{KinductionKipdrTrueNotSolvedByKinductionPlain}{Error}{Timeout}{Cputime}{Median}{902.825918228}%
\StoreBenchExecResult{PdrInvPathprograms}{KinductionKipdrTrueNotSolvedByKinductionPlain}{Error}{Timeout}{Cputime}{Min}{900.795123117}%
\StoreBenchExecResult{PdrInvPathprograms}{KinductionKipdrTrueNotSolvedByKinductionPlain}{Error}{Timeout}{Cputime}{Max}{1000.66668317}%
\StoreBenchExecResult{PdrInvPathprograms}{KinductionKipdrTrueNotSolvedByKinductionPlain}{Error}{Timeout}{Cputime}{Stdev}{11.38795464090762305110787211}%
\StoreBenchExecResult{PdrInvPathprograms}{KinductionKipdrTrueNotSolvedByKinductionPlain}{Error}{Timeout}{Walltime}{}{41914.497431278}%
\StoreBenchExecResult{PdrInvPathprograms}{KinductionKipdrTrueNotSolvedByKinductionPlain}{Error}{Timeout}{Walltime}{Avg}{455.5923633834565217391304348}%
\StoreBenchExecResult{PdrInvPathprograms}{KinductionKipdrTrueNotSolvedByKinductionPlain}{Error}{Timeout}{Walltime}{Median}{452.6569219825}%
\StoreBenchExecResult{PdrInvPathprograms}{KinductionKipdrTrueNotSolvedByKinductionPlain}{Error}{Timeout}{Walltime}{Min}{451.306558132}%
\StoreBenchExecResult{PdrInvPathprograms}{KinductionKipdrTrueNotSolvedByKinductionPlain}{Error}{Timeout}{Walltime}{Max}{546.541609049}%
\StoreBenchExecResult{PdrInvPathprograms}{KinductionKipdrTrueNotSolvedByKinductionPlain}{Error}{Timeout}{Walltime}{Stdev}{11.31571145081013244045374991}%
\providecommand\StoreBenchExecResult[7]{\expandafter\newcommand\csname#1#2#3#4#5#6\endcsname{#7}}%
\StoreBenchExecResult{PdrInvPathprograms}{KinductionPlainTrueNotSolvedByKinductionPlain}{Total}{}{Count}{}{114}%
\StoreBenchExecResult{PdrInvPathprograms}{KinductionPlainTrueNotSolvedByKinductionPlain}{Total}{}{Cputime}{}{96336.091440619}%
\StoreBenchExecResult{PdrInvPathprograms}{KinductionPlainTrueNotSolvedByKinductionPlain}{Total}{}{Cputime}{Avg}{845.0534336896403508771929825}%
\StoreBenchExecResult{PdrInvPathprograms}{KinductionPlainTrueNotSolvedByKinductionPlain}{Total}{}{Cputime}{Median}{902.002942506}%
\StoreBenchExecResult{PdrInvPathprograms}{KinductionPlainTrueNotSolvedByKinductionPlain}{Total}{}{Cputime}{Min}{116.813024209}%
\StoreBenchExecResult{PdrInvPathprograms}{KinductionPlainTrueNotSolvedByKinductionPlain}{Total}{}{Cputime}{Max}{1000.54078193}%
\StoreBenchExecResult{PdrInvPathprograms}{KinductionPlainTrueNotSolvedByKinductionPlain}{Total}{}{Cputime}{Stdev}{193.4702209804432051324986843}%
\StoreBenchExecResult{PdrInvPathprograms}{KinductionPlainTrueNotSolvedByKinductionPlain}{Total}{}{Walltime}{}{91653.3571939505}%
\StoreBenchExecResult{PdrInvPathprograms}{KinductionPlainTrueNotSolvedByKinductionPlain}{Total}{}{Walltime}{Avg}{803.9768174907938596491228070}%
\StoreBenchExecResult{PdrInvPathprograms}{KinductionPlainTrueNotSolvedByKinductionPlain}{Total}{}{Walltime}{Median}{885.859370947}%
\StoreBenchExecResult{PdrInvPathprograms}{KinductionPlainTrueNotSolvedByKinductionPlain}{Total}{}{Walltime}{Min}{86.2834601402}%
\StoreBenchExecResult{PdrInvPathprograms}{KinductionPlainTrueNotSolvedByKinductionPlain}{Total}{}{Walltime}{Max}{944.178749084}%
\StoreBenchExecResult{PdrInvPathprograms}{KinductionPlainTrueNotSolvedByKinductionPlain}{Total}{}{Walltime}{Stdev}{204.3858612456389961979367229}%
\StoreBenchExecResult{PdrInvPathprograms}{KinductionPlainTrueNotSolvedByKinductionPlain}{Error}{}{Count}{}{114}%
\StoreBenchExecResult{PdrInvPathprograms}{KinductionPlainTrueNotSolvedByKinductionPlain}{Error}{}{Cputime}{}{96336.091440619}%
\StoreBenchExecResult{PdrInvPathprograms}{KinductionPlainTrueNotSolvedByKinductionPlain}{Error}{}{Cputime}{Avg}{845.0534336896403508771929825}%
\StoreBenchExecResult{PdrInvPathprograms}{KinductionPlainTrueNotSolvedByKinductionPlain}{Error}{}{Cputime}{Median}{902.002942506}%
\StoreBenchExecResult{PdrInvPathprograms}{KinductionPlainTrueNotSolvedByKinductionPlain}{Error}{}{Cputime}{Min}{116.813024209}%
\StoreBenchExecResult{PdrInvPathprograms}{KinductionPlainTrueNotSolvedByKinductionPlain}{Error}{}{Cputime}{Max}{1000.54078193}%
\StoreBenchExecResult{PdrInvPathprograms}{KinductionPlainTrueNotSolvedByKinductionPlain}{Error}{}{Cputime}{Stdev}{193.4702209804432051324986843}%
\StoreBenchExecResult{PdrInvPathprograms}{KinductionPlainTrueNotSolvedByKinductionPlain}{Error}{}{Walltime}{}{91653.3571939505}%
\StoreBenchExecResult{PdrInvPathprograms}{KinductionPlainTrueNotSolvedByKinductionPlain}{Error}{}{Walltime}{Avg}{803.9768174907938596491228070}%
\StoreBenchExecResult{PdrInvPathprograms}{KinductionPlainTrueNotSolvedByKinductionPlain}{Error}{}{Walltime}{Median}{885.859370947}%
\StoreBenchExecResult{PdrInvPathprograms}{KinductionPlainTrueNotSolvedByKinductionPlain}{Error}{}{Walltime}{Min}{86.2834601402}%
\StoreBenchExecResult{PdrInvPathprograms}{KinductionPlainTrueNotSolvedByKinductionPlain}{Error}{}{Walltime}{Max}{944.178749084}%
\StoreBenchExecResult{PdrInvPathprograms}{KinductionPlainTrueNotSolvedByKinductionPlain}{Error}{}{Walltime}{Stdev}{204.3858612456389961979367229}%
\StoreBenchExecResult{PdrInvPathprograms}{KinductionPlainTrueNotSolvedByKinductionPlain}{Error}{Error}{Count}{}{5}%
\StoreBenchExecResult{PdrInvPathprograms}{KinductionPlainTrueNotSolvedByKinductionPlain}{Error}{Error}{Cputime}{}{982.418801883}%
\StoreBenchExecResult{PdrInvPathprograms}{KinductionPlainTrueNotSolvedByKinductionPlain}{Error}{Error}{Cputime}{Avg}{196.4837603766}%
\StoreBenchExecResult{PdrInvPathprograms}{KinductionPlainTrueNotSolvedByKinductionPlain}{Error}{Error}{Cputime}{Median}{120.593235354}%
\StoreBenchExecResult{PdrInvPathprograms}{KinductionPlainTrueNotSolvedByKinductionPlain}{Error}{Error}{Cputime}{Min}{116.813024209}%
\StoreBenchExecResult{PdrInvPathprograms}{KinductionPlainTrueNotSolvedByKinductionPlain}{Error}{Error}{Cputime}{Max}{492.36271057}%
\StoreBenchExecResult{PdrInvPathprograms}{KinductionPlainTrueNotSolvedByKinductionPlain}{Error}{Error}{Cputime}{Stdev}{148.0974343335334908191010605}%
\StoreBenchExecResult{PdrInvPathprograms}{KinductionPlainTrueNotSolvedByKinductionPlain}{Error}{Error}{Walltime}{}{837.6751639845}%
\StoreBenchExecResult{PdrInvPathprograms}{KinductionPlainTrueNotSolvedByKinductionPlain}{Error}{Error}{Walltime}{Avg}{167.5350327969}%
\StoreBenchExecResult{PdrInvPathprograms}{KinductionPlainTrueNotSolvedByKinductionPlain}{Error}{Error}{Walltime}{Median}{93.5712780952}%
\StoreBenchExecResult{PdrInvPathprograms}{KinductionPlainTrueNotSolvedByKinductionPlain}{Error}{Error}{Walltime}{Min}{86.2834601402}%
\StoreBenchExecResult{PdrInvPathprograms}{KinductionPlainTrueNotSolvedByKinductionPlain}{Error}{Error}{Walltime}{Max}{460.645820856}%
\StoreBenchExecResult{PdrInvPathprograms}{KinductionPlainTrueNotSolvedByKinductionPlain}{Error}{Error}{Walltime}{Stdev}{146.6701456211216979679601609}%
\StoreBenchExecResult{PdrInvPathprograms}{KinductionPlainTrueNotSolvedByKinductionPlain}{Error}{OutOfJavaMemory}{Count}{}{5}%
\StoreBenchExecResult{PdrInvPathprograms}{KinductionPlainTrueNotSolvedByKinductionPlain}{Error}{OutOfJavaMemory}{Cputime}{}{1610.742452090}%
\StoreBenchExecResult{PdrInvPathprograms}{KinductionPlainTrueNotSolvedByKinductionPlain}{Error}{OutOfJavaMemory}{Cputime}{Avg}{322.148490418}%
\StoreBenchExecResult{PdrInvPathprograms}{KinductionPlainTrueNotSolvedByKinductionPlain}{Error}{OutOfJavaMemory}{Cputime}{Median}{184.890148498}%
\StoreBenchExecResult{PdrInvPathprograms}{KinductionPlainTrueNotSolvedByKinductionPlain}{Error}{OutOfJavaMemory}{Cputime}{Min}{155.594367288}%
\StoreBenchExecResult{PdrInvPathprograms}{KinductionPlainTrueNotSolvedByKinductionPlain}{Error}{OutOfJavaMemory}{Cputime}{Max}{806.271756232}%
\StoreBenchExecResult{PdrInvPathprograms}{KinductionPlainTrueNotSolvedByKinductionPlain}{Error}{OutOfJavaMemory}{Cputime}{Stdev}{245.9527685431288978962005426}%
\StoreBenchExecResult{PdrInvPathprograms}{KinductionPlainTrueNotSolvedByKinductionPlain}{Error}{OutOfJavaMemory}{Walltime}{}{997.312891245}%
\StoreBenchExecResult{PdrInvPathprograms}{KinductionPlainTrueNotSolvedByKinductionPlain}{Error}{OutOfJavaMemory}{Walltime}{Avg}{199.462578249}%
\StoreBenchExecResult{PdrInvPathprograms}{KinductionPlainTrueNotSolvedByKinductionPlain}{Error}{OutOfJavaMemory}{Walltime}{Median}{119.467287064}%
\StoreBenchExecResult{PdrInvPathprograms}{KinductionPlainTrueNotSolvedByKinductionPlain}{Error}{OutOfJavaMemory}{Walltime}{Min}{101.580156088}%
\StoreBenchExecResult{PdrInvPathprograms}{KinductionPlainTrueNotSolvedByKinductionPlain}{Error}{OutOfJavaMemory}{Walltime}{Max}{473.32741785}%
\StoreBenchExecResult{PdrInvPathprograms}{KinductionPlainTrueNotSolvedByKinductionPlain}{Error}{OutOfJavaMemory}{Walltime}{Stdev}{139.8138721662483206020354543}%
\StoreBenchExecResult{PdrInvPathprograms}{KinductionPlainTrueNotSolvedByKinductionPlain}{Error}{OutOfMemory}{Count}{}{4}%
\StoreBenchExecResult{PdrInvPathprograms}{KinductionPlainTrueNotSolvedByKinductionPlain}{Error}{OutOfMemory}{Cputime}{}{3241.367197416}%
\StoreBenchExecResult{PdrInvPathprograms}{KinductionPlainTrueNotSolvedByKinductionPlain}{Error}{OutOfMemory}{Cputime}{Avg}{810.341799354}%
\StoreBenchExecResult{PdrInvPathprograms}{KinductionPlainTrueNotSolvedByKinductionPlain}{Error}{OutOfMemory}{Cputime}{Median}{796.996441908}%
\StoreBenchExecResult{PdrInvPathprograms}{KinductionPlainTrueNotSolvedByKinductionPlain}{Error}{OutOfMemory}{Cputime}{Min}{760.628865776}%
\StoreBenchExecResult{PdrInvPathprograms}{KinductionPlainTrueNotSolvedByKinductionPlain}{Error}{OutOfMemory}{Cputime}{Max}{886.745447824}%
\StoreBenchExecResult{PdrInvPathprograms}{KinductionPlainTrueNotSolvedByKinductionPlain}{Error}{OutOfMemory}{Cputime}{Stdev}{47.54224480225450639301130771}%
\StoreBenchExecResult{PdrInvPathprograms}{KinductionPlainTrueNotSolvedByKinductionPlain}{Error}{OutOfMemory}{Walltime}{}{3181.345288039}%
\StoreBenchExecResult{PdrInvPathprograms}{KinductionPlainTrueNotSolvedByKinductionPlain}{Error}{OutOfMemory}{Walltime}{Avg}{795.33632200975}%
\StoreBenchExecResult{PdrInvPathprograms}{KinductionPlainTrueNotSolvedByKinductionPlain}{Error}{OutOfMemory}{Walltime}{Median}{782.7712146045}%
\StoreBenchExecResult{PdrInvPathprograms}{KinductionPlainTrueNotSolvedByKinductionPlain}{Error}{OutOfMemory}{Walltime}{Min}{744.673719883}%
\StoreBenchExecResult{PdrInvPathprograms}{KinductionPlainTrueNotSolvedByKinductionPlain}{Error}{OutOfMemory}{Walltime}{Max}{871.129138947}%
\StoreBenchExecResult{PdrInvPathprograms}{KinductionPlainTrueNotSolvedByKinductionPlain}{Error}{OutOfMemory}{Walltime}{Stdev}{47.39657896343965756193109276}%
\StoreBenchExecResult{PdrInvPathprograms}{KinductionPlainTrueNotSolvedByKinductionPlain}{Error}{Timeout}{Count}{}{100}%
\StoreBenchExecResult{PdrInvPathprograms}{KinductionPlainTrueNotSolvedByKinductionPlain}{Error}{Timeout}{Cputime}{}{90501.562989230}%
\StoreBenchExecResult{PdrInvPathprograms}{KinductionPlainTrueNotSolvedByKinductionPlain}{Error}{Timeout}{Cputime}{Avg}{905.0156298923}%
\StoreBenchExecResult{PdrInvPathprograms}{KinductionPlainTrueNotSolvedByKinductionPlain}{Error}{Timeout}{Cputime}{Median}{902.3272660095}%
\StoreBenchExecResult{PdrInvPathprograms}{KinductionPlainTrueNotSolvedByKinductionPlain}{Error}{Timeout}{Cputime}{Min}{900.996856015}%
\StoreBenchExecResult{PdrInvPathprograms}{KinductionPlainTrueNotSolvedByKinductionPlain}{Error}{Timeout}{Cputime}{Max}{1000.54078193}%
\StoreBenchExecResult{PdrInvPathprograms}{KinductionPlainTrueNotSolvedByKinductionPlain}{Error}{Timeout}{Cputime}{Stdev}{10.58611653097657671063896261}%
\StoreBenchExecResult{PdrInvPathprograms}{KinductionPlainTrueNotSolvedByKinductionPlain}{Error}{Timeout}{Walltime}{}{86637.023850682}%
\StoreBenchExecResult{PdrInvPathprograms}{KinductionPlainTrueNotSolvedByKinductionPlain}{Error}{Timeout}{Walltime}{Avg}{866.37023850682}%
\StoreBenchExecResult{PdrInvPathprograms}{KinductionPlainTrueNotSolvedByKinductionPlain}{Error}{Timeout}{Walltime}{Median}{888.422053933}%
\StoreBenchExecResult{PdrInvPathprograms}{KinductionPlainTrueNotSolvedByKinductionPlain}{Error}{Timeout}{Walltime}{Min}{650.421828032}%
\StoreBenchExecResult{PdrInvPathprograms}{KinductionPlainTrueNotSolvedByKinductionPlain}{Error}{Timeout}{Walltime}{Max}{944.178749084}%
\StoreBenchExecResult{PdrInvPathprograms}{KinductionPlainTrueNotSolvedByKinductionPlain}{Error}{Timeout}{Walltime}{Stdev}{55.30204078149356193467890384}%
\ifdefined\PdrInvPathprogramsKinductionPlainTotalCount\else\edef\PdrInvPathprogramsKinductionPlainTotalCount{0}\fi
\ifdefined\PdrInvPathprogramsKinductionPlainCorrectCount\else\edef\PdrInvPathprogramsKinductionPlainCorrectCount{0}\fi
\ifdefined\PdrInvPathprogramsKinductionPlainCorrectTrueCount\else\edef\PdrInvPathprogramsKinductionPlainCorrectTrueCount{0}\fi
\ifdefined\PdrInvPathprogramsKinductionPlainCorrectFalseCount\else\edef\PdrInvPathprogramsKinductionPlainCorrectFalseCount{0}\fi
\ifdefined\PdrInvPathprogramsKinductionPlainWrongTrueCount\else\edef\PdrInvPathprogramsKinductionPlainWrongTrueCount{0}\fi
\ifdefined\PdrInvPathprogramsKinductionPlainWrongFalseCount\else\edef\PdrInvPathprogramsKinductionPlainWrongFalseCount{0}\fi
\ifdefined\PdrInvPathprogramsKinductionPlainErrorTimeoutCount\else\edef\PdrInvPathprogramsKinductionPlainErrorTimeoutCount{0}\fi
\ifdefined\PdrInvPathprogramsKinductionPlainErrorOutOfMemoryCount\else\edef\PdrInvPathprogramsKinductionPlainErrorOutOfMemoryCount{0}\fi
\ifdefined\PdrInvPathprogramsKinductionPlainCorrectCputime\else\edef\PdrInvPathprogramsKinductionPlainCorrectCputime{0}\fi
\ifdefined\PdrInvPathprogramsKinductionPlainCorrectCputimeAvg\else\edef\PdrInvPathprogramsKinductionPlainCorrectCputimeAvg{None}\fi
\ifdefined\PdrInvPathprogramsKinductionPlainCorrectWalltime\else\edef\PdrInvPathprogramsKinductionPlainCorrectWalltime{0}\fi
\ifdefined\PdrInvPathprogramsKinductionPlainCorrectWalltimeAvg\else\edef\PdrInvPathprogramsKinductionPlainCorrectWalltimeAvg{None}\fi
\ifdefined\PdrInvPathprogramsKinductionDfStaticZeroZeroTTotalCount\else\edef\PdrInvPathprogramsKinductionDfStaticZeroZeroTTotalCount{0}\fi
\ifdefined\PdrInvPathprogramsKinductionDfStaticZeroZeroTCorrectCount\else\edef\PdrInvPathprogramsKinductionDfStaticZeroZeroTCorrectCount{0}\fi
\ifdefined\PdrInvPathprogramsKinductionDfStaticZeroZeroTCorrectTrueCount\else\edef\PdrInvPathprogramsKinductionDfStaticZeroZeroTCorrectTrueCount{0}\fi
\ifdefined\PdrInvPathprogramsKinductionDfStaticZeroZeroTCorrectFalseCount\else\edef\PdrInvPathprogramsKinductionDfStaticZeroZeroTCorrectFalseCount{0}\fi
\ifdefined\PdrInvPathprogramsKinductionDfStaticZeroZeroTWrongTrueCount\else\edef\PdrInvPathprogramsKinductionDfStaticZeroZeroTWrongTrueCount{0}\fi
\ifdefined\PdrInvPathprogramsKinductionDfStaticZeroZeroTWrongFalseCount\else\edef\PdrInvPathprogramsKinductionDfStaticZeroZeroTWrongFalseCount{0}\fi
\ifdefined\PdrInvPathprogramsKinductionDfStaticZeroZeroTErrorTimeoutCount\else\edef\PdrInvPathprogramsKinductionDfStaticZeroZeroTErrorTimeoutCount{0}\fi
\ifdefined\PdrInvPathprogramsKinductionDfStaticZeroZeroTErrorOutOfMemoryCount\else\edef\PdrInvPathprogramsKinductionDfStaticZeroZeroTErrorOutOfMemoryCount{0}\fi
\ifdefined\PdrInvPathprogramsKinductionDfStaticZeroZeroTCorrectCputime\else\edef\PdrInvPathprogramsKinductionDfStaticZeroZeroTCorrectCputime{0}\fi
\ifdefined\PdrInvPathprogramsKinductionDfStaticZeroZeroTCorrectCputimeAvg\else\edef\PdrInvPathprogramsKinductionDfStaticZeroZeroTCorrectCputimeAvg{None}\fi
\ifdefined\PdrInvPathprogramsKinductionDfStaticZeroZeroTCorrectWalltime\else\edef\PdrInvPathprogramsKinductionDfStaticZeroZeroTCorrectWalltime{0}\fi
\ifdefined\PdrInvPathprogramsKinductionDfStaticZeroZeroTCorrectWalltimeAvg\else\edef\PdrInvPathprogramsKinductionDfStaticZeroZeroTCorrectWalltimeAvg{None}\fi
\ifdefined\PdrInvPathprogramsKinductionDfStaticZeroOneTTTotalCount\else\edef\PdrInvPathprogramsKinductionDfStaticZeroOneTTTotalCount{0}\fi
\ifdefined\PdrInvPathprogramsKinductionDfStaticZeroOneTTCorrectCount\else\edef\PdrInvPathprogramsKinductionDfStaticZeroOneTTCorrectCount{0}\fi
\ifdefined\PdrInvPathprogramsKinductionDfStaticZeroOneTTCorrectTrueCount\else\edef\PdrInvPathprogramsKinductionDfStaticZeroOneTTCorrectTrueCount{0}\fi
\ifdefined\PdrInvPathprogramsKinductionDfStaticZeroOneTTCorrectFalseCount\else\edef\PdrInvPathprogramsKinductionDfStaticZeroOneTTCorrectFalseCount{0}\fi
\ifdefined\PdrInvPathprogramsKinductionDfStaticZeroOneTTWrongTrueCount\else\edef\PdrInvPathprogramsKinductionDfStaticZeroOneTTWrongTrueCount{0}\fi
\ifdefined\PdrInvPathprogramsKinductionDfStaticZeroOneTTWrongFalseCount\else\edef\PdrInvPathprogramsKinductionDfStaticZeroOneTTWrongFalseCount{0}\fi
\ifdefined\PdrInvPathprogramsKinductionDfStaticZeroOneTTErrorTimeoutCount\else\edef\PdrInvPathprogramsKinductionDfStaticZeroOneTTErrorTimeoutCount{0}\fi
\ifdefined\PdrInvPathprogramsKinductionDfStaticZeroOneTTErrorOutOfMemoryCount\else\edef\PdrInvPathprogramsKinductionDfStaticZeroOneTTErrorOutOfMemoryCount{0}\fi
\ifdefined\PdrInvPathprogramsKinductionDfStaticZeroOneTTCorrectCputime\else\edef\PdrInvPathprogramsKinductionDfStaticZeroOneTTCorrectCputime{0}\fi
\ifdefined\PdrInvPathprogramsKinductionDfStaticZeroOneTTCorrectCputimeAvg\else\edef\PdrInvPathprogramsKinductionDfStaticZeroOneTTCorrectCputimeAvg{None}\fi
\ifdefined\PdrInvPathprogramsKinductionDfStaticZeroOneTTCorrectWalltime\else\edef\PdrInvPathprogramsKinductionDfStaticZeroOneTTCorrectWalltime{0}\fi
\ifdefined\PdrInvPathprogramsKinductionDfStaticZeroOneTTCorrectWalltimeAvg\else\edef\PdrInvPathprogramsKinductionDfStaticZeroOneTTCorrectWalltimeAvg{None}\fi
\ifdefined\PdrInvPathprogramsKinductionDfStaticZeroOneTFTotalCount\else\edef\PdrInvPathprogramsKinductionDfStaticZeroOneTFTotalCount{0}\fi
\ifdefined\PdrInvPathprogramsKinductionDfStaticZeroOneTFCorrectCount\else\edef\PdrInvPathprogramsKinductionDfStaticZeroOneTFCorrectCount{0}\fi
\ifdefined\PdrInvPathprogramsKinductionDfStaticZeroOneTFCorrectTrueCount\else\edef\PdrInvPathprogramsKinductionDfStaticZeroOneTFCorrectTrueCount{0}\fi
\ifdefined\PdrInvPathprogramsKinductionDfStaticZeroOneTFCorrectFalseCount\else\edef\PdrInvPathprogramsKinductionDfStaticZeroOneTFCorrectFalseCount{0}\fi
\ifdefined\PdrInvPathprogramsKinductionDfStaticZeroOneTFWrongTrueCount\else\edef\PdrInvPathprogramsKinductionDfStaticZeroOneTFWrongTrueCount{0}\fi
\ifdefined\PdrInvPathprogramsKinductionDfStaticZeroOneTFWrongFalseCount\else\edef\PdrInvPathprogramsKinductionDfStaticZeroOneTFWrongFalseCount{0}\fi
\ifdefined\PdrInvPathprogramsKinductionDfStaticZeroOneTFErrorTimeoutCount\else\edef\PdrInvPathprogramsKinductionDfStaticZeroOneTFErrorTimeoutCount{0}\fi
\ifdefined\PdrInvPathprogramsKinductionDfStaticZeroOneTFErrorOutOfMemoryCount\else\edef\PdrInvPathprogramsKinductionDfStaticZeroOneTFErrorOutOfMemoryCount{0}\fi
\ifdefined\PdrInvPathprogramsKinductionDfStaticZeroOneTFCorrectCputime\else\edef\PdrInvPathprogramsKinductionDfStaticZeroOneTFCorrectCputime{0}\fi
\ifdefined\PdrInvPathprogramsKinductionDfStaticZeroOneTFCorrectCputimeAvg\else\edef\PdrInvPathprogramsKinductionDfStaticZeroOneTFCorrectCputimeAvg{None}\fi
\ifdefined\PdrInvPathprogramsKinductionDfStaticZeroOneTFCorrectWalltime\else\edef\PdrInvPathprogramsKinductionDfStaticZeroOneTFCorrectWalltime{0}\fi
\ifdefined\PdrInvPathprogramsKinductionDfStaticZeroOneTFCorrectWalltimeAvg\else\edef\PdrInvPathprogramsKinductionDfStaticZeroOneTFCorrectWalltimeAvg{None}\fi
\ifdefined\PdrInvPathprogramsKinductionDfStaticZeroTwoTTTotalCount\else\edef\PdrInvPathprogramsKinductionDfStaticZeroTwoTTTotalCount{0}\fi
\ifdefined\PdrInvPathprogramsKinductionDfStaticZeroTwoTTCorrectCount\else\edef\PdrInvPathprogramsKinductionDfStaticZeroTwoTTCorrectCount{0}\fi
\ifdefined\PdrInvPathprogramsKinductionDfStaticZeroTwoTTCorrectTrueCount\else\edef\PdrInvPathprogramsKinductionDfStaticZeroTwoTTCorrectTrueCount{0}\fi
\ifdefined\PdrInvPathprogramsKinductionDfStaticZeroTwoTTCorrectFalseCount\else\edef\PdrInvPathprogramsKinductionDfStaticZeroTwoTTCorrectFalseCount{0}\fi
\ifdefined\PdrInvPathprogramsKinductionDfStaticZeroTwoTTWrongTrueCount\else\edef\PdrInvPathprogramsKinductionDfStaticZeroTwoTTWrongTrueCount{0}\fi
\ifdefined\PdrInvPathprogramsKinductionDfStaticZeroTwoTTWrongFalseCount\else\edef\PdrInvPathprogramsKinductionDfStaticZeroTwoTTWrongFalseCount{0}\fi
\ifdefined\PdrInvPathprogramsKinductionDfStaticZeroTwoTTErrorTimeoutCount\else\edef\PdrInvPathprogramsKinductionDfStaticZeroTwoTTErrorTimeoutCount{0}\fi
\ifdefined\PdrInvPathprogramsKinductionDfStaticZeroTwoTTErrorOutOfMemoryCount\else\edef\PdrInvPathprogramsKinductionDfStaticZeroTwoTTErrorOutOfMemoryCount{0}\fi
\ifdefined\PdrInvPathprogramsKinductionDfStaticZeroTwoTTCorrectCputime\else\edef\PdrInvPathprogramsKinductionDfStaticZeroTwoTTCorrectCputime{0}\fi
\ifdefined\PdrInvPathprogramsKinductionDfStaticZeroTwoTTCorrectCputimeAvg\else\edef\PdrInvPathprogramsKinductionDfStaticZeroTwoTTCorrectCputimeAvg{None}\fi
\ifdefined\PdrInvPathprogramsKinductionDfStaticZeroTwoTTCorrectWalltime\else\edef\PdrInvPathprogramsKinductionDfStaticZeroTwoTTCorrectWalltime{0}\fi
\ifdefined\PdrInvPathprogramsKinductionDfStaticZeroTwoTTCorrectWalltimeAvg\else\edef\PdrInvPathprogramsKinductionDfStaticZeroTwoTTCorrectWalltimeAvg{None}\fi
\ifdefined\PdrInvPathprogramsKinductionDfStaticZeroTwoTFTotalCount\else\edef\PdrInvPathprogramsKinductionDfStaticZeroTwoTFTotalCount{0}\fi
\ifdefined\PdrInvPathprogramsKinductionDfStaticZeroTwoTFCorrectCount\else\edef\PdrInvPathprogramsKinductionDfStaticZeroTwoTFCorrectCount{0}\fi
\ifdefined\PdrInvPathprogramsKinductionDfStaticZeroTwoTFCorrectTrueCount\else\edef\PdrInvPathprogramsKinductionDfStaticZeroTwoTFCorrectTrueCount{0}\fi
\ifdefined\PdrInvPathprogramsKinductionDfStaticZeroTwoTFCorrectFalseCount\else\edef\PdrInvPathprogramsKinductionDfStaticZeroTwoTFCorrectFalseCount{0}\fi
\ifdefined\PdrInvPathprogramsKinductionDfStaticZeroTwoTFWrongTrueCount\else\edef\PdrInvPathprogramsKinductionDfStaticZeroTwoTFWrongTrueCount{0}\fi
\ifdefined\PdrInvPathprogramsKinductionDfStaticZeroTwoTFWrongFalseCount\else\edef\PdrInvPathprogramsKinductionDfStaticZeroTwoTFWrongFalseCount{0}\fi
\ifdefined\PdrInvPathprogramsKinductionDfStaticZeroTwoTFErrorTimeoutCount\else\edef\PdrInvPathprogramsKinductionDfStaticZeroTwoTFErrorTimeoutCount{0}\fi
\ifdefined\PdrInvPathprogramsKinductionDfStaticZeroTwoTFErrorOutOfMemoryCount\else\edef\PdrInvPathprogramsKinductionDfStaticZeroTwoTFErrorOutOfMemoryCount{0}\fi
\ifdefined\PdrInvPathprogramsKinductionDfStaticZeroTwoTFCorrectCputime\else\edef\PdrInvPathprogramsKinductionDfStaticZeroTwoTFCorrectCputime{0}\fi
\ifdefined\PdrInvPathprogramsKinductionDfStaticZeroTwoTFCorrectCputimeAvg\else\edef\PdrInvPathprogramsKinductionDfStaticZeroTwoTFCorrectCputimeAvg{None}\fi
\ifdefined\PdrInvPathprogramsKinductionDfStaticZeroTwoTFCorrectWalltime\else\edef\PdrInvPathprogramsKinductionDfStaticZeroTwoTFCorrectWalltime{0}\fi
\ifdefined\PdrInvPathprogramsKinductionDfStaticZeroTwoTFCorrectWalltimeAvg\else\edef\PdrInvPathprogramsKinductionDfStaticZeroTwoTFCorrectWalltimeAvg{None}\fi
\ifdefined\PdrInvPathprogramsKinductionDfStaticEightTwoTTotalCount\else\edef\PdrInvPathprogramsKinductionDfStaticEightTwoTTotalCount{0}\fi
\ifdefined\PdrInvPathprogramsKinductionDfStaticEightTwoTCorrectCount\else\edef\PdrInvPathprogramsKinductionDfStaticEightTwoTCorrectCount{0}\fi
\ifdefined\PdrInvPathprogramsKinductionDfStaticEightTwoTCorrectTrueCount\else\edef\PdrInvPathprogramsKinductionDfStaticEightTwoTCorrectTrueCount{0}\fi
\ifdefined\PdrInvPathprogramsKinductionDfStaticEightTwoTCorrectFalseCount\else\edef\PdrInvPathprogramsKinductionDfStaticEightTwoTCorrectFalseCount{0}\fi
\ifdefined\PdrInvPathprogramsKinductionDfStaticEightTwoTWrongTrueCount\else\edef\PdrInvPathprogramsKinductionDfStaticEightTwoTWrongTrueCount{0}\fi
\ifdefined\PdrInvPathprogramsKinductionDfStaticEightTwoTWrongFalseCount\else\edef\PdrInvPathprogramsKinductionDfStaticEightTwoTWrongFalseCount{0}\fi
\ifdefined\PdrInvPathprogramsKinductionDfStaticEightTwoTErrorTimeoutCount\else\edef\PdrInvPathprogramsKinductionDfStaticEightTwoTErrorTimeoutCount{0}\fi
\ifdefined\PdrInvPathprogramsKinductionDfStaticEightTwoTErrorOutOfMemoryCount\else\edef\PdrInvPathprogramsKinductionDfStaticEightTwoTErrorOutOfMemoryCount{0}\fi
\ifdefined\PdrInvPathprogramsKinductionDfStaticEightTwoTCorrectCputime\else\edef\PdrInvPathprogramsKinductionDfStaticEightTwoTCorrectCputime{0}\fi
\ifdefined\PdrInvPathprogramsKinductionDfStaticEightTwoTCorrectCputimeAvg\else\edef\PdrInvPathprogramsKinductionDfStaticEightTwoTCorrectCputimeAvg{None}\fi
\ifdefined\PdrInvPathprogramsKinductionDfStaticEightTwoTCorrectWalltime\else\edef\PdrInvPathprogramsKinductionDfStaticEightTwoTCorrectWalltime{0}\fi
\ifdefined\PdrInvPathprogramsKinductionDfStaticEightTwoTCorrectWalltimeAvg\else\edef\PdrInvPathprogramsKinductionDfStaticEightTwoTCorrectWalltimeAvg{None}\fi
\ifdefined\PdrInvPathprogramsKinductionDfStaticSixteenTwoTTotalCount\else\edef\PdrInvPathprogramsKinductionDfStaticSixteenTwoTTotalCount{0}\fi
\ifdefined\PdrInvPathprogramsKinductionDfStaticSixteenTwoTCorrectCount\else\edef\PdrInvPathprogramsKinductionDfStaticSixteenTwoTCorrectCount{0}\fi
\ifdefined\PdrInvPathprogramsKinductionDfStaticSixteenTwoTCorrectTrueCount\else\edef\PdrInvPathprogramsKinductionDfStaticSixteenTwoTCorrectTrueCount{0}\fi
\ifdefined\PdrInvPathprogramsKinductionDfStaticSixteenTwoTCorrectFalseCount\else\edef\PdrInvPathprogramsKinductionDfStaticSixteenTwoTCorrectFalseCount{0}\fi
\ifdefined\PdrInvPathprogramsKinductionDfStaticSixteenTwoTWrongTrueCount\else\edef\PdrInvPathprogramsKinductionDfStaticSixteenTwoTWrongTrueCount{0}\fi
\ifdefined\PdrInvPathprogramsKinductionDfStaticSixteenTwoTWrongFalseCount\else\edef\PdrInvPathprogramsKinductionDfStaticSixteenTwoTWrongFalseCount{0}\fi
\ifdefined\PdrInvPathprogramsKinductionDfStaticSixteenTwoTErrorTimeoutCount\else\edef\PdrInvPathprogramsKinductionDfStaticSixteenTwoTErrorTimeoutCount{0}\fi
\ifdefined\PdrInvPathprogramsKinductionDfStaticSixteenTwoTErrorOutOfMemoryCount\else\edef\PdrInvPathprogramsKinductionDfStaticSixteenTwoTErrorOutOfMemoryCount{0}\fi
\ifdefined\PdrInvPathprogramsKinductionDfStaticSixteenTwoTCorrectCputime\else\edef\PdrInvPathprogramsKinductionDfStaticSixteenTwoTCorrectCputime{0}\fi
\ifdefined\PdrInvPathprogramsKinductionDfStaticSixteenTwoTCorrectCputimeAvg\else\edef\PdrInvPathprogramsKinductionDfStaticSixteenTwoTCorrectCputimeAvg{None}\fi
\ifdefined\PdrInvPathprogramsKinductionDfStaticSixteenTwoTCorrectWalltime\else\edef\PdrInvPathprogramsKinductionDfStaticSixteenTwoTCorrectWalltime{0}\fi
\ifdefined\PdrInvPathprogramsKinductionDfStaticSixteenTwoTCorrectWalltimeAvg\else\edef\PdrInvPathprogramsKinductionDfStaticSixteenTwoTCorrectWalltimeAvg{None}\fi
\ifdefined\PdrInvPathprogramsKinductionDfStaticSixteenTwoFTotalCount\else\edef\PdrInvPathprogramsKinductionDfStaticSixteenTwoFTotalCount{0}\fi
\ifdefined\PdrInvPathprogramsKinductionDfStaticSixteenTwoFCorrectCount\else\edef\PdrInvPathprogramsKinductionDfStaticSixteenTwoFCorrectCount{0}\fi
\ifdefined\PdrInvPathprogramsKinductionDfStaticSixteenTwoFCorrectTrueCount\else\edef\PdrInvPathprogramsKinductionDfStaticSixteenTwoFCorrectTrueCount{0}\fi
\ifdefined\PdrInvPathprogramsKinductionDfStaticSixteenTwoFCorrectFalseCount\else\edef\PdrInvPathprogramsKinductionDfStaticSixteenTwoFCorrectFalseCount{0}\fi
\ifdefined\PdrInvPathprogramsKinductionDfStaticSixteenTwoFWrongTrueCount\else\edef\PdrInvPathprogramsKinductionDfStaticSixteenTwoFWrongTrueCount{0}\fi
\ifdefined\PdrInvPathprogramsKinductionDfStaticSixteenTwoFWrongFalseCount\else\edef\PdrInvPathprogramsKinductionDfStaticSixteenTwoFWrongFalseCount{0}\fi
\ifdefined\PdrInvPathprogramsKinductionDfStaticSixteenTwoFErrorTimeoutCount\else\edef\PdrInvPathprogramsKinductionDfStaticSixteenTwoFErrorTimeoutCount{0}\fi
\ifdefined\PdrInvPathprogramsKinductionDfStaticSixteenTwoFErrorOutOfMemoryCount\else\edef\PdrInvPathprogramsKinductionDfStaticSixteenTwoFErrorOutOfMemoryCount{0}\fi
\ifdefined\PdrInvPathprogramsKinductionDfStaticSixteenTwoFCorrectCputime\else\edef\PdrInvPathprogramsKinductionDfStaticSixteenTwoFCorrectCputime{0}\fi
\ifdefined\PdrInvPathprogramsKinductionDfStaticSixteenTwoFCorrectCputimeAvg\else\edef\PdrInvPathprogramsKinductionDfStaticSixteenTwoFCorrectCputimeAvg{None}\fi
\ifdefined\PdrInvPathprogramsKinductionDfStaticSixteenTwoFCorrectWalltime\else\edef\PdrInvPathprogramsKinductionDfStaticSixteenTwoFCorrectWalltime{0}\fi
\ifdefined\PdrInvPathprogramsKinductionDfStaticSixteenTwoFCorrectWalltimeAvg\else\edef\PdrInvPathprogramsKinductionDfStaticSixteenTwoFCorrectWalltimeAvg{None}\fi
\ifdefined\PdrInvPathprogramsKinductionDfTotalCount\else\edef\PdrInvPathprogramsKinductionDfTotalCount{0}\fi
\ifdefined\PdrInvPathprogramsKinductionDfCorrectCount\else\edef\PdrInvPathprogramsKinductionDfCorrectCount{0}\fi
\ifdefined\PdrInvPathprogramsKinductionDfCorrectTrueCount\else\edef\PdrInvPathprogramsKinductionDfCorrectTrueCount{0}\fi
\ifdefined\PdrInvPathprogramsKinductionDfCorrectFalseCount\else\edef\PdrInvPathprogramsKinductionDfCorrectFalseCount{0}\fi
\ifdefined\PdrInvPathprogramsKinductionDfWrongTrueCount\else\edef\PdrInvPathprogramsKinductionDfWrongTrueCount{0}\fi
\ifdefined\PdrInvPathprogramsKinductionDfWrongFalseCount\else\edef\PdrInvPathprogramsKinductionDfWrongFalseCount{0}\fi
\ifdefined\PdrInvPathprogramsKinductionDfErrorTimeoutCount\else\edef\PdrInvPathprogramsKinductionDfErrorTimeoutCount{0}\fi
\ifdefined\PdrInvPathprogramsKinductionDfErrorOutOfMemoryCount\else\edef\PdrInvPathprogramsKinductionDfErrorOutOfMemoryCount{0}\fi
\ifdefined\PdrInvPathprogramsKinductionDfCorrectCputime\else\edef\PdrInvPathprogramsKinductionDfCorrectCputime{0}\fi
\ifdefined\PdrInvPathprogramsKinductionDfCorrectCputimeAvg\else\edef\PdrInvPathprogramsKinductionDfCorrectCputimeAvg{None}\fi
\ifdefined\PdrInvPathprogramsKinductionDfCorrectWalltime\else\edef\PdrInvPathprogramsKinductionDfCorrectWalltime{0}\fi
\ifdefined\PdrInvPathprogramsKinductionDfCorrectWalltimeAvg\else\edef\PdrInvPathprogramsKinductionDfCorrectWalltimeAvg{None}\fi
\ifdefined\PdrInvPathprogramsKinductionKipdrTotalCount\else\edef\PdrInvPathprogramsKinductionKipdrTotalCount{0}\fi
\ifdefined\PdrInvPathprogramsKinductionKipdrCorrectCount\else\edef\PdrInvPathprogramsKinductionKipdrCorrectCount{0}\fi
\ifdefined\PdrInvPathprogramsKinductionKipdrCorrectTrueCount\else\edef\PdrInvPathprogramsKinductionKipdrCorrectTrueCount{0}\fi
\ifdefined\PdrInvPathprogramsKinductionKipdrCorrectFalseCount\else\edef\PdrInvPathprogramsKinductionKipdrCorrectFalseCount{0}\fi
\ifdefined\PdrInvPathprogramsKinductionKipdrWrongTrueCount\else\edef\PdrInvPathprogramsKinductionKipdrWrongTrueCount{0}\fi
\ifdefined\PdrInvPathprogramsKinductionKipdrWrongFalseCount\else\edef\PdrInvPathprogramsKinductionKipdrWrongFalseCount{0}\fi
\ifdefined\PdrInvPathprogramsKinductionKipdrErrorTimeoutCount\else\edef\PdrInvPathprogramsKinductionKipdrErrorTimeoutCount{0}\fi
\ifdefined\PdrInvPathprogramsKinductionKipdrErrorOutOfMemoryCount\else\edef\PdrInvPathprogramsKinductionKipdrErrorOutOfMemoryCount{0}\fi
\ifdefined\PdrInvPathprogramsKinductionKipdrCorrectCputime\else\edef\PdrInvPathprogramsKinductionKipdrCorrectCputime{0}\fi
\ifdefined\PdrInvPathprogramsKinductionKipdrCorrectCputimeAvg\else\edef\PdrInvPathprogramsKinductionKipdrCorrectCputimeAvg{None}\fi
\ifdefined\PdrInvPathprogramsKinductionKipdrCorrectWalltime\else\edef\PdrInvPathprogramsKinductionKipdrCorrectWalltime{0}\fi
\ifdefined\PdrInvPathprogramsKinductionKipdrCorrectWalltimeAvg\else\edef\PdrInvPathprogramsKinductionKipdrCorrectWalltimeAvg{None}\fi
\ifdefined\PdrInvPathprogramsKinductionKipdrdfTotalCount\else\edef\PdrInvPathprogramsKinductionKipdrdfTotalCount{0}\fi
\ifdefined\PdrInvPathprogramsKinductionKipdrdfCorrectCount\else\edef\PdrInvPathprogramsKinductionKipdrdfCorrectCount{0}\fi
\ifdefined\PdrInvPathprogramsKinductionKipdrdfCorrectTrueCount\else\edef\PdrInvPathprogramsKinductionKipdrdfCorrectTrueCount{0}\fi
\ifdefined\PdrInvPathprogramsKinductionKipdrdfCorrectFalseCount\else\edef\PdrInvPathprogramsKinductionKipdrdfCorrectFalseCount{0}\fi
\ifdefined\PdrInvPathprogramsKinductionKipdrdfWrongTrueCount\else\edef\PdrInvPathprogramsKinductionKipdrdfWrongTrueCount{0}\fi
\ifdefined\PdrInvPathprogramsKinductionKipdrdfWrongFalseCount\else\edef\PdrInvPathprogramsKinductionKipdrdfWrongFalseCount{0}\fi
\ifdefined\PdrInvPathprogramsKinductionKipdrdfErrorTimeoutCount\else\edef\PdrInvPathprogramsKinductionKipdrdfErrorTimeoutCount{0}\fi
\ifdefined\PdrInvPathprogramsKinductionKipdrdfErrorOutOfMemoryCount\else\edef\PdrInvPathprogramsKinductionKipdrdfErrorOutOfMemoryCount{0}\fi
\ifdefined\PdrInvPathprogramsKinductionKipdrdfCorrectCputime\else\edef\PdrInvPathprogramsKinductionKipdrdfCorrectCputime{0}\fi
\ifdefined\PdrInvPathprogramsKinductionKipdrdfCorrectCputimeAvg\else\edef\PdrInvPathprogramsKinductionKipdrdfCorrectCputimeAvg{None}\fi
\ifdefined\PdrInvPathprogramsKinductionKipdrdfCorrectWalltime\else\edef\PdrInvPathprogramsKinductionKipdrdfCorrectWalltime{0}\fi
\ifdefined\PdrInvPathprogramsKinductionKipdrdfCorrectWalltimeAvg\else\edef\PdrInvPathprogramsKinductionKipdrdfCorrectWalltimeAvg{None}\fi
\ifdefined\PdrInvPathprogramsKinductionPlainTrueNotSolvedByKinductionPlainTotalCount\else\edef\PdrInvPathprogramsKinductionPlainTrueNotSolvedByKinductionPlainTotalCount{0}\fi
\ifdefined\PdrInvPathprogramsKinductionPlainTrueNotSolvedByKinductionPlainCorrectCount\else\edef\PdrInvPathprogramsKinductionPlainTrueNotSolvedByKinductionPlainCorrectCount{0}\fi
\ifdefined\PdrInvPathprogramsKinductionPlainTrueNotSolvedByKinductionPlainCorrectTrueCount\else\edef\PdrInvPathprogramsKinductionPlainTrueNotSolvedByKinductionPlainCorrectTrueCount{0}\fi
\ifdefined\PdrInvPathprogramsKinductionPlainTrueNotSolvedByKinductionPlainCorrectFalseCount\else\edef\PdrInvPathprogramsKinductionPlainTrueNotSolvedByKinductionPlainCorrectFalseCount{0}\fi
\ifdefined\PdrInvPathprogramsKinductionPlainTrueNotSolvedByKinductionPlainWrongTrueCount\else\edef\PdrInvPathprogramsKinductionPlainTrueNotSolvedByKinductionPlainWrongTrueCount{0}\fi
\ifdefined\PdrInvPathprogramsKinductionPlainTrueNotSolvedByKinductionPlainWrongFalseCount\else\edef\PdrInvPathprogramsKinductionPlainTrueNotSolvedByKinductionPlainWrongFalseCount{0}\fi
\ifdefined\PdrInvPathprogramsKinductionPlainTrueNotSolvedByKinductionPlainErrorTimeoutCount\else\edef\PdrInvPathprogramsKinductionPlainTrueNotSolvedByKinductionPlainErrorTimeoutCount{0}\fi
\ifdefined\PdrInvPathprogramsKinductionPlainTrueNotSolvedByKinductionPlainErrorOutOfMemoryCount\else\edef\PdrInvPathprogramsKinductionPlainTrueNotSolvedByKinductionPlainErrorOutOfMemoryCount{0}\fi
\ifdefined\PdrInvPathprogramsKinductionPlainTrueNotSolvedByKinductionPlainCorrectCputime\else\edef\PdrInvPathprogramsKinductionPlainTrueNotSolvedByKinductionPlainCorrectCputime{0}\fi
\ifdefined\PdrInvPathprogramsKinductionPlainTrueNotSolvedByKinductionPlainCorrectCputimeAvg\else\edef\PdrInvPathprogramsKinductionPlainTrueNotSolvedByKinductionPlainCorrectCputimeAvg{None}\fi
\ifdefined\PdrInvPathprogramsKinductionPlainTrueNotSolvedByKinductionPlainCorrectWalltime\else\edef\PdrInvPathprogramsKinductionPlainTrueNotSolvedByKinductionPlainCorrectWalltime{0}\fi
\ifdefined\PdrInvPathprogramsKinductionPlainTrueNotSolvedByKinductionPlainCorrectWalltimeAvg\else\edef\PdrInvPathprogramsKinductionPlainTrueNotSolvedByKinductionPlainCorrectWalltimeAvg{None}\fi
\ifdefined\PdrInvPathprogramsKinductionDfStaticZeroZeroTTrueNotSolvedByKinductionPlainTotalCount\else\edef\PdrInvPathprogramsKinductionDfStaticZeroZeroTTrueNotSolvedByKinductionPlainTotalCount{0}\fi
\ifdefined\PdrInvPathprogramsKinductionDfStaticZeroZeroTTrueNotSolvedByKinductionPlainCorrectCount\else\edef\PdrInvPathprogramsKinductionDfStaticZeroZeroTTrueNotSolvedByKinductionPlainCorrectCount{0}\fi
\ifdefined\PdrInvPathprogramsKinductionDfStaticZeroZeroTTrueNotSolvedByKinductionPlainCorrectTrueCount\else\edef\PdrInvPathprogramsKinductionDfStaticZeroZeroTTrueNotSolvedByKinductionPlainCorrectTrueCount{0}\fi
\ifdefined\PdrInvPathprogramsKinductionDfStaticZeroZeroTTrueNotSolvedByKinductionPlainCorrectFalseCount\else\edef\PdrInvPathprogramsKinductionDfStaticZeroZeroTTrueNotSolvedByKinductionPlainCorrectFalseCount{0}\fi
\ifdefined\PdrInvPathprogramsKinductionDfStaticZeroZeroTTrueNotSolvedByKinductionPlainWrongTrueCount\else\edef\PdrInvPathprogramsKinductionDfStaticZeroZeroTTrueNotSolvedByKinductionPlainWrongTrueCount{0}\fi
\ifdefined\PdrInvPathprogramsKinductionDfStaticZeroZeroTTrueNotSolvedByKinductionPlainWrongFalseCount\else\edef\PdrInvPathprogramsKinductionDfStaticZeroZeroTTrueNotSolvedByKinductionPlainWrongFalseCount{0}\fi
\ifdefined\PdrInvPathprogramsKinductionDfStaticZeroZeroTTrueNotSolvedByKinductionPlainErrorTimeoutCount\else\edef\PdrInvPathprogramsKinductionDfStaticZeroZeroTTrueNotSolvedByKinductionPlainErrorTimeoutCount{0}\fi
\ifdefined\PdrInvPathprogramsKinductionDfStaticZeroZeroTTrueNotSolvedByKinductionPlainErrorOutOfMemoryCount\else\edef\PdrInvPathprogramsKinductionDfStaticZeroZeroTTrueNotSolvedByKinductionPlainErrorOutOfMemoryCount{0}\fi
\ifdefined\PdrInvPathprogramsKinductionDfStaticZeroZeroTTrueNotSolvedByKinductionPlainCorrectCputime\else\edef\PdrInvPathprogramsKinductionDfStaticZeroZeroTTrueNotSolvedByKinductionPlainCorrectCputime{0}\fi
\ifdefined\PdrInvPathprogramsKinductionDfStaticZeroZeroTTrueNotSolvedByKinductionPlainCorrectCputimeAvg\else\edef\PdrInvPathprogramsKinductionDfStaticZeroZeroTTrueNotSolvedByKinductionPlainCorrectCputimeAvg{None}\fi
\ifdefined\PdrInvPathprogramsKinductionDfStaticZeroZeroTTrueNotSolvedByKinductionPlainCorrectWalltime\else\edef\PdrInvPathprogramsKinductionDfStaticZeroZeroTTrueNotSolvedByKinductionPlainCorrectWalltime{0}\fi
\ifdefined\PdrInvPathprogramsKinductionDfStaticZeroZeroTTrueNotSolvedByKinductionPlainCorrectWalltimeAvg\else\edef\PdrInvPathprogramsKinductionDfStaticZeroZeroTTrueNotSolvedByKinductionPlainCorrectWalltimeAvg{None}\fi
\ifdefined\PdrInvPathprogramsKinductionDfStaticZeroOneTTTrueNotSolvedByKinductionPlainTotalCount\else\edef\PdrInvPathprogramsKinductionDfStaticZeroOneTTTrueNotSolvedByKinductionPlainTotalCount{0}\fi
\ifdefined\PdrInvPathprogramsKinductionDfStaticZeroOneTTTrueNotSolvedByKinductionPlainCorrectCount\else\edef\PdrInvPathprogramsKinductionDfStaticZeroOneTTTrueNotSolvedByKinductionPlainCorrectCount{0}\fi
\ifdefined\PdrInvPathprogramsKinductionDfStaticZeroOneTTTrueNotSolvedByKinductionPlainCorrectTrueCount\else\edef\PdrInvPathprogramsKinductionDfStaticZeroOneTTTrueNotSolvedByKinductionPlainCorrectTrueCount{0}\fi
\ifdefined\PdrInvPathprogramsKinductionDfStaticZeroOneTTTrueNotSolvedByKinductionPlainCorrectFalseCount\else\edef\PdrInvPathprogramsKinductionDfStaticZeroOneTTTrueNotSolvedByKinductionPlainCorrectFalseCount{0}\fi
\ifdefined\PdrInvPathprogramsKinductionDfStaticZeroOneTTTrueNotSolvedByKinductionPlainWrongTrueCount\else\edef\PdrInvPathprogramsKinductionDfStaticZeroOneTTTrueNotSolvedByKinductionPlainWrongTrueCount{0}\fi
\ifdefined\PdrInvPathprogramsKinductionDfStaticZeroOneTTTrueNotSolvedByKinductionPlainWrongFalseCount\else\edef\PdrInvPathprogramsKinductionDfStaticZeroOneTTTrueNotSolvedByKinductionPlainWrongFalseCount{0}\fi
\ifdefined\PdrInvPathprogramsKinductionDfStaticZeroOneTTTrueNotSolvedByKinductionPlainErrorTimeoutCount\else\edef\PdrInvPathprogramsKinductionDfStaticZeroOneTTTrueNotSolvedByKinductionPlainErrorTimeoutCount{0}\fi
\ifdefined\PdrInvPathprogramsKinductionDfStaticZeroOneTTTrueNotSolvedByKinductionPlainErrorOutOfMemoryCount\else\edef\PdrInvPathprogramsKinductionDfStaticZeroOneTTTrueNotSolvedByKinductionPlainErrorOutOfMemoryCount{0}\fi
\ifdefined\PdrInvPathprogramsKinductionDfStaticZeroOneTTTrueNotSolvedByKinductionPlainCorrectCputime\else\edef\PdrInvPathprogramsKinductionDfStaticZeroOneTTTrueNotSolvedByKinductionPlainCorrectCputime{0}\fi
\ifdefined\PdrInvPathprogramsKinductionDfStaticZeroOneTTTrueNotSolvedByKinductionPlainCorrectCputimeAvg\else\edef\PdrInvPathprogramsKinductionDfStaticZeroOneTTTrueNotSolvedByKinductionPlainCorrectCputimeAvg{None}\fi
\ifdefined\PdrInvPathprogramsKinductionDfStaticZeroOneTTTrueNotSolvedByKinductionPlainCorrectWalltime\else\edef\PdrInvPathprogramsKinductionDfStaticZeroOneTTTrueNotSolvedByKinductionPlainCorrectWalltime{0}\fi
\ifdefined\PdrInvPathprogramsKinductionDfStaticZeroOneTTTrueNotSolvedByKinductionPlainCorrectWalltimeAvg\else\edef\PdrInvPathprogramsKinductionDfStaticZeroOneTTTrueNotSolvedByKinductionPlainCorrectWalltimeAvg{None}\fi
\ifdefined\PdrInvPathprogramsKinductionDfStaticZeroOneTFTrueNotSolvedByKinductionPlainTotalCount\else\edef\PdrInvPathprogramsKinductionDfStaticZeroOneTFTrueNotSolvedByKinductionPlainTotalCount{0}\fi
\ifdefined\PdrInvPathprogramsKinductionDfStaticZeroOneTFTrueNotSolvedByKinductionPlainCorrectCount\else\edef\PdrInvPathprogramsKinductionDfStaticZeroOneTFTrueNotSolvedByKinductionPlainCorrectCount{0}\fi
\ifdefined\PdrInvPathprogramsKinductionDfStaticZeroOneTFTrueNotSolvedByKinductionPlainCorrectTrueCount\else\edef\PdrInvPathprogramsKinductionDfStaticZeroOneTFTrueNotSolvedByKinductionPlainCorrectTrueCount{0}\fi
\ifdefined\PdrInvPathprogramsKinductionDfStaticZeroOneTFTrueNotSolvedByKinductionPlainCorrectFalseCount\else\edef\PdrInvPathprogramsKinductionDfStaticZeroOneTFTrueNotSolvedByKinductionPlainCorrectFalseCount{0}\fi
\ifdefined\PdrInvPathprogramsKinductionDfStaticZeroOneTFTrueNotSolvedByKinductionPlainWrongTrueCount\else\edef\PdrInvPathprogramsKinductionDfStaticZeroOneTFTrueNotSolvedByKinductionPlainWrongTrueCount{0}\fi
\ifdefined\PdrInvPathprogramsKinductionDfStaticZeroOneTFTrueNotSolvedByKinductionPlainWrongFalseCount\else\edef\PdrInvPathprogramsKinductionDfStaticZeroOneTFTrueNotSolvedByKinductionPlainWrongFalseCount{0}\fi
\ifdefined\PdrInvPathprogramsKinductionDfStaticZeroOneTFTrueNotSolvedByKinductionPlainErrorTimeoutCount\else\edef\PdrInvPathprogramsKinductionDfStaticZeroOneTFTrueNotSolvedByKinductionPlainErrorTimeoutCount{0}\fi
\ifdefined\PdrInvPathprogramsKinductionDfStaticZeroOneTFTrueNotSolvedByKinductionPlainErrorOutOfMemoryCount\else\edef\PdrInvPathprogramsKinductionDfStaticZeroOneTFTrueNotSolvedByKinductionPlainErrorOutOfMemoryCount{0}\fi
\ifdefined\PdrInvPathprogramsKinductionDfStaticZeroOneTFTrueNotSolvedByKinductionPlainCorrectCputime\else\edef\PdrInvPathprogramsKinductionDfStaticZeroOneTFTrueNotSolvedByKinductionPlainCorrectCputime{0}\fi
\ifdefined\PdrInvPathprogramsKinductionDfStaticZeroOneTFTrueNotSolvedByKinductionPlainCorrectCputimeAvg\else\edef\PdrInvPathprogramsKinductionDfStaticZeroOneTFTrueNotSolvedByKinductionPlainCorrectCputimeAvg{None}\fi
\ifdefined\PdrInvPathprogramsKinductionDfStaticZeroOneTFTrueNotSolvedByKinductionPlainCorrectWalltime\else\edef\PdrInvPathprogramsKinductionDfStaticZeroOneTFTrueNotSolvedByKinductionPlainCorrectWalltime{0}\fi
\ifdefined\PdrInvPathprogramsKinductionDfStaticZeroOneTFTrueNotSolvedByKinductionPlainCorrectWalltimeAvg\else\edef\PdrInvPathprogramsKinductionDfStaticZeroOneTFTrueNotSolvedByKinductionPlainCorrectWalltimeAvg{None}\fi
\ifdefined\PdrInvPathprogramsKinductionDfStaticZeroTwoTTTrueNotSolvedByKinductionPlainTotalCount\else\edef\PdrInvPathprogramsKinductionDfStaticZeroTwoTTTrueNotSolvedByKinductionPlainTotalCount{0}\fi
\ifdefined\PdrInvPathprogramsKinductionDfStaticZeroTwoTTTrueNotSolvedByKinductionPlainCorrectCount\else\edef\PdrInvPathprogramsKinductionDfStaticZeroTwoTTTrueNotSolvedByKinductionPlainCorrectCount{0}\fi
\ifdefined\PdrInvPathprogramsKinductionDfStaticZeroTwoTTTrueNotSolvedByKinductionPlainCorrectTrueCount\else\edef\PdrInvPathprogramsKinductionDfStaticZeroTwoTTTrueNotSolvedByKinductionPlainCorrectTrueCount{0}\fi
\ifdefined\PdrInvPathprogramsKinductionDfStaticZeroTwoTTTrueNotSolvedByKinductionPlainCorrectFalseCount\else\edef\PdrInvPathprogramsKinductionDfStaticZeroTwoTTTrueNotSolvedByKinductionPlainCorrectFalseCount{0}\fi
\ifdefined\PdrInvPathprogramsKinductionDfStaticZeroTwoTTTrueNotSolvedByKinductionPlainWrongTrueCount\else\edef\PdrInvPathprogramsKinductionDfStaticZeroTwoTTTrueNotSolvedByKinductionPlainWrongTrueCount{0}\fi
\ifdefined\PdrInvPathprogramsKinductionDfStaticZeroTwoTTTrueNotSolvedByKinductionPlainWrongFalseCount\else\edef\PdrInvPathprogramsKinductionDfStaticZeroTwoTTTrueNotSolvedByKinductionPlainWrongFalseCount{0}\fi
\ifdefined\PdrInvPathprogramsKinductionDfStaticZeroTwoTTTrueNotSolvedByKinductionPlainErrorTimeoutCount\else\edef\PdrInvPathprogramsKinductionDfStaticZeroTwoTTTrueNotSolvedByKinductionPlainErrorTimeoutCount{0}\fi
\ifdefined\PdrInvPathprogramsKinductionDfStaticZeroTwoTTTrueNotSolvedByKinductionPlainErrorOutOfMemoryCount\else\edef\PdrInvPathprogramsKinductionDfStaticZeroTwoTTTrueNotSolvedByKinductionPlainErrorOutOfMemoryCount{0}\fi
\ifdefined\PdrInvPathprogramsKinductionDfStaticZeroTwoTTTrueNotSolvedByKinductionPlainCorrectCputime\else\edef\PdrInvPathprogramsKinductionDfStaticZeroTwoTTTrueNotSolvedByKinductionPlainCorrectCputime{0}\fi
\ifdefined\PdrInvPathprogramsKinductionDfStaticZeroTwoTTTrueNotSolvedByKinductionPlainCorrectCputimeAvg\else\edef\PdrInvPathprogramsKinductionDfStaticZeroTwoTTTrueNotSolvedByKinductionPlainCorrectCputimeAvg{None}\fi
\ifdefined\PdrInvPathprogramsKinductionDfStaticZeroTwoTTTrueNotSolvedByKinductionPlainCorrectWalltime\else\edef\PdrInvPathprogramsKinductionDfStaticZeroTwoTTTrueNotSolvedByKinductionPlainCorrectWalltime{0}\fi
\ifdefined\PdrInvPathprogramsKinductionDfStaticZeroTwoTTTrueNotSolvedByKinductionPlainCorrectWalltimeAvg\else\edef\PdrInvPathprogramsKinductionDfStaticZeroTwoTTTrueNotSolvedByKinductionPlainCorrectWalltimeAvg{None}\fi
\ifdefined\PdrInvPathprogramsKinductionDfStaticZeroTwoTFTrueNotSolvedByKinductionPlainTotalCount\else\edef\PdrInvPathprogramsKinductionDfStaticZeroTwoTFTrueNotSolvedByKinductionPlainTotalCount{0}\fi
\ifdefined\PdrInvPathprogramsKinductionDfStaticZeroTwoTFTrueNotSolvedByKinductionPlainCorrectCount\else\edef\PdrInvPathprogramsKinductionDfStaticZeroTwoTFTrueNotSolvedByKinductionPlainCorrectCount{0}\fi
\ifdefined\PdrInvPathprogramsKinductionDfStaticZeroTwoTFTrueNotSolvedByKinductionPlainCorrectTrueCount\else\edef\PdrInvPathprogramsKinductionDfStaticZeroTwoTFTrueNotSolvedByKinductionPlainCorrectTrueCount{0}\fi
\ifdefined\PdrInvPathprogramsKinductionDfStaticZeroTwoTFTrueNotSolvedByKinductionPlainCorrectFalseCount\else\edef\PdrInvPathprogramsKinductionDfStaticZeroTwoTFTrueNotSolvedByKinductionPlainCorrectFalseCount{0}\fi
\ifdefined\PdrInvPathprogramsKinductionDfStaticZeroTwoTFTrueNotSolvedByKinductionPlainWrongTrueCount\else\edef\PdrInvPathprogramsKinductionDfStaticZeroTwoTFTrueNotSolvedByKinductionPlainWrongTrueCount{0}\fi
\ifdefined\PdrInvPathprogramsKinductionDfStaticZeroTwoTFTrueNotSolvedByKinductionPlainWrongFalseCount\else\edef\PdrInvPathprogramsKinductionDfStaticZeroTwoTFTrueNotSolvedByKinductionPlainWrongFalseCount{0}\fi
\ifdefined\PdrInvPathprogramsKinductionDfStaticZeroTwoTFTrueNotSolvedByKinductionPlainErrorTimeoutCount\else\edef\PdrInvPathprogramsKinductionDfStaticZeroTwoTFTrueNotSolvedByKinductionPlainErrorTimeoutCount{0}\fi
\ifdefined\PdrInvPathprogramsKinductionDfStaticZeroTwoTFTrueNotSolvedByKinductionPlainErrorOutOfMemoryCount\else\edef\PdrInvPathprogramsKinductionDfStaticZeroTwoTFTrueNotSolvedByKinductionPlainErrorOutOfMemoryCount{0}\fi
\ifdefined\PdrInvPathprogramsKinductionDfStaticZeroTwoTFTrueNotSolvedByKinductionPlainCorrectCputime\else\edef\PdrInvPathprogramsKinductionDfStaticZeroTwoTFTrueNotSolvedByKinductionPlainCorrectCputime{0}\fi
\ifdefined\PdrInvPathprogramsKinductionDfStaticZeroTwoTFTrueNotSolvedByKinductionPlainCorrectCputimeAvg\else\edef\PdrInvPathprogramsKinductionDfStaticZeroTwoTFTrueNotSolvedByKinductionPlainCorrectCputimeAvg{None}\fi
\ifdefined\PdrInvPathprogramsKinductionDfStaticZeroTwoTFTrueNotSolvedByKinductionPlainCorrectWalltime\else\edef\PdrInvPathprogramsKinductionDfStaticZeroTwoTFTrueNotSolvedByKinductionPlainCorrectWalltime{0}\fi
\ifdefined\PdrInvPathprogramsKinductionDfStaticZeroTwoTFTrueNotSolvedByKinductionPlainCorrectWalltimeAvg\else\edef\PdrInvPathprogramsKinductionDfStaticZeroTwoTFTrueNotSolvedByKinductionPlainCorrectWalltimeAvg{None}\fi
\ifdefined\PdrInvPathprogramsKinductionDfStaticEightTwoTTrueNotSolvedByKinductionPlainTotalCount\else\edef\PdrInvPathprogramsKinductionDfStaticEightTwoTTrueNotSolvedByKinductionPlainTotalCount{0}\fi
\ifdefined\PdrInvPathprogramsKinductionDfStaticEightTwoTTrueNotSolvedByKinductionPlainCorrectCount\else\edef\PdrInvPathprogramsKinductionDfStaticEightTwoTTrueNotSolvedByKinductionPlainCorrectCount{0}\fi
\ifdefined\PdrInvPathprogramsKinductionDfStaticEightTwoTTrueNotSolvedByKinductionPlainCorrectTrueCount\else\edef\PdrInvPathprogramsKinductionDfStaticEightTwoTTrueNotSolvedByKinductionPlainCorrectTrueCount{0}\fi
\ifdefined\PdrInvPathprogramsKinductionDfStaticEightTwoTTrueNotSolvedByKinductionPlainCorrectFalseCount\else\edef\PdrInvPathprogramsKinductionDfStaticEightTwoTTrueNotSolvedByKinductionPlainCorrectFalseCount{0}\fi
\ifdefined\PdrInvPathprogramsKinductionDfStaticEightTwoTTrueNotSolvedByKinductionPlainWrongTrueCount\else\edef\PdrInvPathprogramsKinductionDfStaticEightTwoTTrueNotSolvedByKinductionPlainWrongTrueCount{0}\fi
\ifdefined\PdrInvPathprogramsKinductionDfStaticEightTwoTTrueNotSolvedByKinductionPlainWrongFalseCount\else\edef\PdrInvPathprogramsKinductionDfStaticEightTwoTTrueNotSolvedByKinductionPlainWrongFalseCount{0}\fi
\ifdefined\PdrInvPathprogramsKinductionDfStaticEightTwoTTrueNotSolvedByKinductionPlainErrorTimeoutCount\else\edef\PdrInvPathprogramsKinductionDfStaticEightTwoTTrueNotSolvedByKinductionPlainErrorTimeoutCount{0}\fi
\ifdefined\PdrInvPathprogramsKinductionDfStaticEightTwoTTrueNotSolvedByKinductionPlainErrorOutOfMemoryCount\else\edef\PdrInvPathprogramsKinductionDfStaticEightTwoTTrueNotSolvedByKinductionPlainErrorOutOfMemoryCount{0}\fi
\ifdefined\PdrInvPathprogramsKinductionDfStaticEightTwoTTrueNotSolvedByKinductionPlainCorrectCputime\else\edef\PdrInvPathprogramsKinductionDfStaticEightTwoTTrueNotSolvedByKinductionPlainCorrectCputime{0}\fi
\ifdefined\PdrInvPathprogramsKinductionDfStaticEightTwoTTrueNotSolvedByKinductionPlainCorrectCputimeAvg\else\edef\PdrInvPathprogramsKinductionDfStaticEightTwoTTrueNotSolvedByKinductionPlainCorrectCputimeAvg{None}\fi
\ifdefined\PdrInvPathprogramsKinductionDfStaticEightTwoTTrueNotSolvedByKinductionPlainCorrectWalltime\else\edef\PdrInvPathprogramsKinductionDfStaticEightTwoTTrueNotSolvedByKinductionPlainCorrectWalltime{0}\fi
\ifdefined\PdrInvPathprogramsKinductionDfStaticEightTwoTTrueNotSolvedByKinductionPlainCorrectWalltimeAvg\else\edef\PdrInvPathprogramsKinductionDfStaticEightTwoTTrueNotSolvedByKinductionPlainCorrectWalltimeAvg{None}\fi
\ifdefined\PdrInvPathprogramsKinductionDfStaticSixteenTwoTTrueNotSolvedByKinductionPlainTotalCount\else\edef\PdrInvPathprogramsKinductionDfStaticSixteenTwoTTrueNotSolvedByKinductionPlainTotalCount{0}\fi
\ifdefined\PdrInvPathprogramsKinductionDfStaticSixteenTwoTTrueNotSolvedByKinductionPlainCorrectCount\else\edef\PdrInvPathprogramsKinductionDfStaticSixteenTwoTTrueNotSolvedByKinductionPlainCorrectCount{0}\fi
\ifdefined\PdrInvPathprogramsKinductionDfStaticSixteenTwoTTrueNotSolvedByKinductionPlainCorrectTrueCount\else\edef\PdrInvPathprogramsKinductionDfStaticSixteenTwoTTrueNotSolvedByKinductionPlainCorrectTrueCount{0}\fi
\ifdefined\PdrInvPathprogramsKinductionDfStaticSixteenTwoTTrueNotSolvedByKinductionPlainCorrectFalseCount\else\edef\PdrInvPathprogramsKinductionDfStaticSixteenTwoTTrueNotSolvedByKinductionPlainCorrectFalseCount{0}\fi
\ifdefined\PdrInvPathprogramsKinductionDfStaticSixteenTwoTTrueNotSolvedByKinductionPlainWrongTrueCount\else\edef\PdrInvPathprogramsKinductionDfStaticSixteenTwoTTrueNotSolvedByKinductionPlainWrongTrueCount{0}\fi
\ifdefined\PdrInvPathprogramsKinductionDfStaticSixteenTwoTTrueNotSolvedByKinductionPlainWrongFalseCount\else\edef\PdrInvPathprogramsKinductionDfStaticSixteenTwoTTrueNotSolvedByKinductionPlainWrongFalseCount{0}\fi
\ifdefined\PdrInvPathprogramsKinductionDfStaticSixteenTwoTTrueNotSolvedByKinductionPlainErrorTimeoutCount\else\edef\PdrInvPathprogramsKinductionDfStaticSixteenTwoTTrueNotSolvedByKinductionPlainErrorTimeoutCount{0}\fi
\ifdefined\PdrInvPathprogramsKinductionDfStaticSixteenTwoTTrueNotSolvedByKinductionPlainErrorOutOfMemoryCount\else\edef\PdrInvPathprogramsKinductionDfStaticSixteenTwoTTrueNotSolvedByKinductionPlainErrorOutOfMemoryCount{0}\fi
\ifdefined\PdrInvPathprogramsKinductionDfStaticSixteenTwoTTrueNotSolvedByKinductionPlainCorrectCputime\else\edef\PdrInvPathprogramsKinductionDfStaticSixteenTwoTTrueNotSolvedByKinductionPlainCorrectCputime{0}\fi
\ifdefined\PdrInvPathprogramsKinductionDfStaticSixteenTwoTTrueNotSolvedByKinductionPlainCorrectCputimeAvg\else\edef\PdrInvPathprogramsKinductionDfStaticSixteenTwoTTrueNotSolvedByKinductionPlainCorrectCputimeAvg{None}\fi
\ifdefined\PdrInvPathprogramsKinductionDfStaticSixteenTwoTTrueNotSolvedByKinductionPlainCorrectWalltime\else\edef\PdrInvPathprogramsKinductionDfStaticSixteenTwoTTrueNotSolvedByKinductionPlainCorrectWalltime{0}\fi
\ifdefined\PdrInvPathprogramsKinductionDfStaticSixteenTwoTTrueNotSolvedByKinductionPlainCorrectWalltimeAvg\else\edef\PdrInvPathprogramsKinductionDfStaticSixteenTwoTTrueNotSolvedByKinductionPlainCorrectWalltimeAvg{None}\fi
\ifdefined\PdrInvPathprogramsKinductionDfStaticSixteenTwoFTrueNotSolvedByKinductionPlainTotalCount\else\edef\PdrInvPathprogramsKinductionDfStaticSixteenTwoFTrueNotSolvedByKinductionPlainTotalCount{0}\fi
\ifdefined\PdrInvPathprogramsKinductionDfStaticSixteenTwoFTrueNotSolvedByKinductionPlainCorrectCount\else\edef\PdrInvPathprogramsKinductionDfStaticSixteenTwoFTrueNotSolvedByKinductionPlainCorrectCount{0}\fi
\ifdefined\PdrInvPathprogramsKinductionDfStaticSixteenTwoFTrueNotSolvedByKinductionPlainCorrectTrueCount\else\edef\PdrInvPathprogramsKinductionDfStaticSixteenTwoFTrueNotSolvedByKinductionPlainCorrectTrueCount{0}\fi
\ifdefined\PdrInvPathprogramsKinductionDfStaticSixteenTwoFTrueNotSolvedByKinductionPlainCorrectFalseCount\else\edef\PdrInvPathprogramsKinductionDfStaticSixteenTwoFTrueNotSolvedByKinductionPlainCorrectFalseCount{0}\fi
\ifdefined\PdrInvPathprogramsKinductionDfStaticSixteenTwoFTrueNotSolvedByKinductionPlainWrongTrueCount\else\edef\PdrInvPathprogramsKinductionDfStaticSixteenTwoFTrueNotSolvedByKinductionPlainWrongTrueCount{0}\fi
\ifdefined\PdrInvPathprogramsKinductionDfStaticSixteenTwoFTrueNotSolvedByKinductionPlainWrongFalseCount\else\edef\PdrInvPathprogramsKinductionDfStaticSixteenTwoFTrueNotSolvedByKinductionPlainWrongFalseCount{0}\fi
\ifdefined\PdrInvPathprogramsKinductionDfStaticSixteenTwoFTrueNotSolvedByKinductionPlainErrorTimeoutCount\else\edef\PdrInvPathprogramsKinductionDfStaticSixteenTwoFTrueNotSolvedByKinductionPlainErrorTimeoutCount{0}\fi
\ifdefined\PdrInvPathprogramsKinductionDfStaticSixteenTwoFTrueNotSolvedByKinductionPlainErrorOutOfMemoryCount\else\edef\PdrInvPathprogramsKinductionDfStaticSixteenTwoFTrueNotSolvedByKinductionPlainErrorOutOfMemoryCount{0}\fi
\ifdefined\PdrInvPathprogramsKinductionDfStaticSixteenTwoFTrueNotSolvedByKinductionPlainCorrectCputime\else\edef\PdrInvPathprogramsKinductionDfStaticSixteenTwoFTrueNotSolvedByKinductionPlainCorrectCputime{0}\fi
\ifdefined\PdrInvPathprogramsKinductionDfStaticSixteenTwoFTrueNotSolvedByKinductionPlainCorrectCputimeAvg\else\edef\PdrInvPathprogramsKinductionDfStaticSixteenTwoFTrueNotSolvedByKinductionPlainCorrectCputimeAvg{None}\fi
\ifdefined\PdrInvPathprogramsKinductionDfStaticSixteenTwoFTrueNotSolvedByKinductionPlainCorrectWalltime\else\edef\PdrInvPathprogramsKinductionDfStaticSixteenTwoFTrueNotSolvedByKinductionPlainCorrectWalltime{0}\fi
\ifdefined\PdrInvPathprogramsKinductionDfStaticSixteenTwoFTrueNotSolvedByKinductionPlainCorrectWalltimeAvg\else\edef\PdrInvPathprogramsKinductionDfStaticSixteenTwoFTrueNotSolvedByKinductionPlainCorrectWalltimeAvg{None}\fi
\ifdefined\PdrInvPathprogramsKinductionDfTrueNotSolvedByKinductionPlainTotalCount\else\edef\PdrInvPathprogramsKinductionDfTrueNotSolvedByKinductionPlainTotalCount{0}\fi
\ifdefined\PdrInvPathprogramsKinductionDfTrueNotSolvedByKinductionPlainCorrectCount\else\edef\PdrInvPathprogramsKinductionDfTrueNotSolvedByKinductionPlainCorrectCount{0}\fi
\ifdefined\PdrInvPathprogramsKinductionDfTrueNotSolvedByKinductionPlainCorrectTrueCount\else\edef\PdrInvPathprogramsKinductionDfTrueNotSolvedByKinductionPlainCorrectTrueCount{0}\fi
\ifdefined\PdrInvPathprogramsKinductionDfTrueNotSolvedByKinductionPlainCorrectFalseCount\else\edef\PdrInvPathprogramsKinductionDfTrueNotSolvedByKinductionPlainCorrectFalseCount{0}\fi
\ifdefined\PdrInvPathprogramsKinductionDfTrueNotSolvedByKinductionPlainWrongTrueCount\else\edef\PdrInvPathprogramsKinductionDfTrueNotSolvedByKinductionPlainWrongTrueCount{0}\fi
\ifdefined\PdrInvPathprogramsKinductionDfTrueNotSolvedByKinductionPlainWrongFalseCount\else\edef\PdrInvPathprogramsKinductionDfTrueNotSolvedByKinductionPlainWrongFalseCount{0}\fi
\ifdefined\PdrInvPathprogramsKinductionDfTrueNotSolvedByKinductionPlainErrorTimeoutCount\else\edef\PdrInvPathprogramsKinductionDfTrueNotSolvedByKinductionPlainErrorTimeoutCount{0}\fi
\ifdefined\PdrInvPathprogramsKinductionDfTrueNotSolvedByKinductionPlainErrorOutOfMemoryCount\else\edef\PdrInvPathprogramsKinductionDfTrueNotSolvedByKinductionPlainErrorOutOfMemoryCount{0}\fi
\ifdefined\PdrInvPathprogramsKinductionDfTrueNotSolvedByKinductionPlainCorrectCputime\else\edef\PdrInvPathprogramsKinductionDfTrueNotSolvedByKinductionPlainCorrectCputime{0}\fi
\ifdefined\PdrInvPathprogramsKinductionDfTrueNotSolvedByKinductionPlainCorrectCputimeAvg\else\edef\PdrInvPathprogramsKinductionDfTrueNotSolvedByKinductionPlainCorrectCputimeAvg{None}\fi
\ifdefined\PdrInvPathprogramsKinductionDfTrueNotSolvedByKinductionPlainCorrectWalltime\else\edef\PdrInvPathprogramsKinductionDfTrueNotSolvedByKinductionPlainCorrectWalltime{0}\fi
\ifdefined\PdrInvPathprogramsKinductionDfTrueNotSolvedByKinductionPlainCorrectWalltimeAvg\else\edef\PdrInvPathprogramsKinductionDfTrueNotSolvedByKinductionPlainCorrectWalltimeAvg{None}\fi
\ifdefined\PdrInvPathprogramsKinductionKipdrTrueNotSolvedByKinductionPlainTotalCount\else\edef\PdrInvPathprogramsKinductionKipdrTrueNotSolvedByKinductionPlainTotalCount{0}\fi
\ifdefined\PdrInvPathprogramsKinductionKipdrTrueNotSolvedByKinductionPlainCorrectCount\else\edef\PdrInvPathprogramsKinductionKipdrTrueNotSolvedByKinductionPlainCorrectCount{0}\fi
\ifdefined\PdrInvPathprogramsKinductionKipdrTrueNotSolvedByKinductionPlainCorrectTrueCount\else\edef\PdrInvPathprogramsKinductionKipdrTrueNotSolvedByKinductionPlainCorrectTrueCount{0}\fi
\ifdefined\PdrInvPathprogramsKinductionKipdrTrueNotSolvedByKinductionPlainCorrectFalseCount\else\edef\PdrInvPathprogramsKinductionKipdrTrueNotSolvedByKinductionPlainCorrectFalseCount{0}\fi
\ifdefined\PdrInvPathprogramsKinductionKipdrTrueNotSolvedByKinductionPlainWrongTrueCount\else\edef\PdrInvPathprogramsKinductionKipdrTrueNotSolvedByKinductionPlainWrongTrueCount{0}\fi
\ifdefined\PdrInvPathprogramsKinductionKipdrTrueNotSolvedByKinductionPlainWrongFalseCount\else\edef\PdrInvPathprogramsKinductionKipdrTrueNotSolvedByKinductionPlainWrongFalseCount{0}\fi
\ifdefined\PdrInvPathprogramsKinductionKipdrTrueNotSolvedByKinductionPlainErrorTimeoutCount\else\edef\PdrInvPathprogramsKinductionKipdrTrueNotSolvedByKinductionPlainErrorTimeoutCount{0}\fi
\ifdefined\PdrInvPathprogramsKinductionKipdrTrueNotSolvedByKinductionPlainErrorOutOfMemoryCount\else\edef\PdrInvPathprogramsKinductionKipdrTrueNotSolvedByKinductionPlainErrorOutOfMemoryCount{0}\fi
\ifdefined\PdrInvPathprogramsKinductionKipdrTrueNotSolvedByKinductionPlainCorrectCputime\else\edef\PdrInvPathprogramsKinductionKipdrTrueNotSolvedByKinductionPlainCorrectCputime{0}\fi
\ifdefined\PdrInvPathprogramsKinductionKipdrTrueNotSolvedByKinductionPlainCorrectCputimeAvg\else\edef\PdrInvPathprogramsKinductionKipdrTrueNotSolvedByKinductionPlainCorrectCputimeAvg{None}\fi
\ifdefined\PdrInvPathprogramsKinductionKipdrTrueNotSolvedByKinductionPlainCorrectWalltime\else\edef\PdrInvPathprogramsKinductionKipdrTrueNotSolvedByKinductionPlainCorrectWalltime{0}\fi
\ifdefined\PdrInvPathprogramsKinductionKipdrTrueNotSolvedByKinductionPlainCorrectWalltimeAvg\else\edef\PdrInvPathprogramsKinductionKipdrTrueNotSolvedByKinductionPlainCorrectWalltimeAvg{None}\fi
\ifdefined\PdrInvPathprogramsKinductionKipdrdfTrueNotSolvedByKinductionPlainTotalCount\else\edef\PdrInvPathprogramsKinductionKipdrdfTrueNotSolvedByKinductionPlainTotalCount{0}\fi
\ifdefined\PdrInvPathprogramsKinductionKipdrdfTrueNotSolvedByKinductionPlainCorrectCount\else\edef\PdrInvPathprogramsKinductionKipdrdfTrueNotSolvedByKinductionPlainCorrectCount{0}\fi
\ifdefined\PdrInvPathprogramsKinductionKipdrdfTrueNotSolvedByKinductionPlainCorrectTrueCount\else\edef\PdrInvPathprogramsKinductionKipdrdfTrueNotSolvedByKinductionPlainCorrectTrueCount{0}\fi
\ifdefined\PdrInvPathprogramsKinductionKipdrdfTrueNotSolvedByKinductionPlainCorrectFalseCount\else\edef\PdrInvPathprogramsKinductionKipdrdfTrueNotSolvedByKinductionPlainCorrectFalseCount{0}\fi
\ifdefined\PdrInvPathprogramsKinductionKipdrdfTrueNotSolvedByKinductionPlainWrongTrueCount\else\edef\PdrInvPathprogramsKinductionKipdrdfTrueNotSolvedByKinductionPlainWrongTrueCount{0}\fi
\ifdefined\PdrInvPathprogramsKinductionKipdrdfTrueNotSolvedByKinductionPlainWrongFalseCount\else\edef\PdrInvPathprogramsKinductionKipdrdfTrueNotSolvedByKinductionPlainWrongFalseCount{0}\fi
\ifdefined\PdrInvPathprogramsKinductionKipdrdfTrueNotSolvedByKinductionPlainErrorTimeoutCount\else\edef\PdrInvPathprogramsKinductionKipdrdfTrueNotSolvedByKinductionPlainErrorTimeoutCount{0}\fi
\ifdefined\PdrInvPathprogramsKinductionKipdrdfTrueNotSolvedByKinductionPlainErrorOutOfMemoryCount\else\edef\PdrInvPathprogramsKinductionKipdrdfTrueNotSolvedByKinductionPlainErrorOutOfMemoryCount{0}\fi
\ifdefined\PdrInvPathprogramsKinductionKipdrdfTrueNotSolvedByKinductionPlainCorrectCputime\else\edef\PdrInvPathprogramsKinductionKipdrdfTrueNotSolvedByKinductionPlainCorrectCputime{0}\fi
\ifdefined\PdrInvPathprogramsKinductionKipdrdfTrueNotSolvedByKinductionPlainCorrectCputimeAvg\else\edef\PdrInvPathprogramsKinductionKipdrdfTrueNotSolvedByKinductionPlainCorrectCputimeAvg{None}\fi
\ifdefined\PdrInvPathprogramsKinductionKipdrdfTrueNotSolvedByKinductionPlainCorrectWalltime\else\edef\PdrInvPathprogramsKinductionKipdrdfTrueNotSolvedByKinductionPlainCorrectWalltime{0}\fi
\ifdefined\PdrInvPathprogramsKinductionKipdrdfTrueNotSolvedByKinductionPlainCorrectWalltimeAvg\else\edef\PdrInvPathprogramsKinductionKipdrdfTrueNotSolvedByKinductionPlainCorrectWalltimeAvg{None}\fi
\edef\PdrInvPathprogramsKinductionPlainErrorOtherInconclusiveCount{\the\numexpr \PdrInvPathprogramsKinductionPlainTotalCount - \PdrInvPathprogramsKinductionPlainCorrectCount - \PdrInvPathprogramsKinductionPlainWrongTrueCount - \PdrInvPathprogramsKinductionPlainWrongFalseCount - \PdrInvPathprogramsKinductionPlainErrorTimeoutCount - \PdrInvPathprogramsKinductionPlainErrorOutOfMemoryCount \relax}
\edef\PdrInvPathprogramsKinductionDfStaticZeroZeroTErrorOtherInconclusiveCount{\the\numexpr \PdrInvPathprogramsKinductionDfStaticZeroZeroTTotalCount - \PdrInvPathprogramsKinductionDfStaticZeroZeroTCorrectCount - \PdrInvPathprogramsKinductionDfStaticZeroZeroTWrongTrueCount - \PdrInvPathprogramsKinductionDfStaticZeroZeroTWrongFalseCount - \PdrInvPathprogramsKinductionDfStaticZeroZeroTErrorTimeoutCount - \PdrInvPathprogramsKinductionDfStaticZeroZeroTErrorOutOfMemoryCount \relax}
\edef\PdrInvPathprogramsKinductionDfStaticZeroOneTTErrorOtherInconclusiveCount{\the\numexpr \PdrInvPathprogramsKinductionDfStaticZeroOneTTTotalCount - \PdrInvPathprogramsKinductionDfStaticZeroOneTTCorrectCount - \PdrInvPathprogramsKinductionDfStaticZeroOneTTWrongTrueCount - \PdrInvPathprogramsKinductionDfStaticZeroOneTTWrongFalseCount - \PdrInvPathprogramsKinductionDfStaticZeroOneTTErrorTimeoutCount - \PdrInvPathprogramsKinductionDfStaticZeroOneTTErrorOutOfMemoryCount \relax}
\edef\PdrInvPathprogramsKinductionDfStaticZeroOneTFErrorOtherInconclusiveCount{\the\numexpr \PdrInvPathprogramsKinductionDfStaticZeroOneTFTotalCount - \PdrInvPathprogramsKinductionDfStaticZeroOneTFCorrectCount - \PdrInvPathprogramsKinductionDfStaticZeroOneTFWrongTrueCount - \PdrInvPathprogramsKinductionDfStaticZeroOneTFWrongFalseCount - \PdrInvPathprogramsKinductionDfStaticZeroOneTFErrorTimeoutCount - \PdrInvPathprogramsKinductionDfStaticZeroOneTFErrorOutOfMemoryCount \relax}
\edef\PdrInvPathprogramsKinductionDfStaticZeroTwoTTErrorOtherInconclusiveCount{\the\numexpr \PdrInvPathprogramsKinductionDfStaticZeroTwoTTTotalCount - \PdrInvPathprogramsKinductionDfStaticZeroTwoTTCorrectCount - \PdrInvPathprogramsKinductionDfStaticZeroTwoTTWrongTrueCount - \PdrInvPathprogramsKinductionDfStaticZeroTwoTTWrongFalseCount - \PdrInvPathprogramsKinductionDfStaticZeroTwoTTErrorTimeoutCount - \PdrInvPathprogramsKinductionDfStaticZeroTwoTTErrorOutOfMemoryCount \relax}
\edef\PdrInvPathprogramsKinductionDfStaticZeroTwoTFErrorOtherInconclusiveCount{\the\numexpr \PdrInvPathprogramsKinductionDfStaticZeroTwoTFTotalCount - \PdrInvPathprogramsKinductionDfStaticZeroTwoTFCorrectCount - \PdrInvPathprogramsKinductionDfStaticZeroTwoTFWrongTrueCount - \PdrInvPathprogramsKinductionDfStaticZeroTwoTFWrongFalseCount - \PdrInvPathprogramsKinductionDfStaticZeroTwoTFErrorTimeoutCount - \PdrInvPathprogramsKinductionDfStaticZeroTwoTFErrorOutOfMemoryCount \relax}
\edef\PdrInvPathprogramsKinductionDfStaticEightTwoTErrorOtherInconclusiveCount{\the\numexpr \PdrInvPathprogramsKinductionDfStaticEightTwoTTotalCount - \PdrInvPathprogramsKinductionDfStaticEightTwoTCorrectCount - \PdrInvPathprogramsKinductionDfStaticEightTwoTWrongTrueCount - \PdrInvPathprogramsKinductionDfStaticEightTwoTWrongFalseCount - \PdrInvPathprogramsKinductionDfStaticEightTwoTErrorTimeoutCount - \PdrInvPathprogramsKinductionDfStaticEightTwoTErrorOutOfMemoryCount \relax}
\edef\PdrInvPathprogramsKinductionDfStaticSixteenTwoTErrorOtherInconclusiveCount{\the\numexpr \PdrInvPathprogramsKinductionDfStaticSixteenTwoTTotalCount - \PdrInvPathprogramsKinductionDfStaticSixteenTwoTCorrectCount - \PdrInvPathprogramsKinductionDfStaticSixteenTwoTWrongTrueCount - \PdrInvPathprogramsKinductionDfStaticSixteenTwoTWrongFalseCount - \PdrInvPathprogramsKinductionDfStaticSixteenTwoTErrorTimeoutCount - \PdrInvPathprogramsKinductionDfStaticSixteenTwoTErrorOutOfMemoryCount \relax}
\edef\PdrInvPathprogramsKinductionDfStaticSixteenTwoFErrorOtherInconclusiveCount{\the\numexpr \PdrInvPathprogramsKinductionDfStaticSixteenTwoFTotalCount - \PdrInvPathprogramsKinductionDfStaticSixteenTwoFCorrectCount - \PdrInvPathprogramsKinductionDfStaticSixteenTwoFWrongTrueCount - \PdrInvPathprogramsKinductionDfStaticSixteenTwoFWrongFalseCount - \PdrInvPathprogramsKinductionDfStaticSixteenTwoFErrorTimeoutCount - \PdrInvPathprogramsKinductionDfStaticSixteenTwoFErrorOutOfMemoryCount \relax}
\edef\PdrInvPathprogramsKinductionDfErrorOtherInconclusiveCount{\the\numexpr \PdrInvPathprogramsKinductionDfTotalCount - \PdrInvPathprogramsKinductionDfCorrectCount - \PdrInvPathprogramsKinductionDfWrongTrueCount - \PdrInvPathprogramsKinductionDfWrongFalseCount - \PdrInvPathprogramsKinductionDfErrorTimeoutCount - \PdrInvPathprogramsKinductionDfErrorOutOfMemoryCount \relax}
\edef\PdrInvPathprogramsKinductionKipdrErrorOtherInconclusiveCount{\the\numexpr \PdrInvPathprogramsKinductionKipdrTotalCount - \PdrInvPathprogramsKinductionKipdrCorrectCount - \PdrInvPathprogramsKinductionKipdrWrongTrueCount - \PdrInvPathprogramsKinductionKipdrWrongFalseCount - \PdrInvPathprogramsKinductionKipdrErrorTimeoutCount - \PdrInvPathprogramsKinductionKipdrErrorOutOfMemoryCount \relax}
\edef\PdrInvPathprogramsKinductionKipdrdfErrorOtherInconclusiveCount{\the\numexpr \PdrInvPathprogramsKinductionKipdrdfTotalCount - \PdrInvPathprogramsKinductionKipdrdfCorrectCount - \PdrInvPathprogramsKinductionKipdrdfWrongTrueCount - \PdrInvPathprogramsKinductionKipdrdfWrongFalseCount - \PdrInvPathprogramsKinductionKipdrdfErrorTimeoutCount - \PdrInvPathprogramsKinductionKipdrdfErrorOutOfMemoryCount \relax}
\edef\PdrInvPathprogramsKinductionPlainTrueNotSolvedByKinductionPlainErrorOtherInconclusiveCount{\the\numexpr \PdrInvPathprogramsKinductionPlainTrueNotSolvedByKinductionPlainTotalCount - \PdrInvPathprogramsKinductionPlainTrueNotSolvedByKinductionPlainCorrectCount - \PdrInvPathprogramsKinductionPlainTrueNotSolvedByKinductionPlainWrongTrueCount - \PdrInvPathprogramsKinductionPlainTrueNotSolvedByKinductionPlainWrongFalseCount - \PdrInvPathprogramsKinductionPlainTrueNotSolvedByKinductionPlainErrorTimeoutCount - \PdrInvPathprogramsKinductionPlainTrueNotSolvedByKinductionPlainErrorOutOfMemoryCount \relax}
\edef\PdrInvPathprogramsKinductionDfStaticZeroZeroTTrueNotSolvedByKinductionPlainErrorOtherInconclusiveCount{\the\numexpr \PdrInvPathprogramsKinductionDfStaticZeroZeroTTrueNotSolvedByKinductionPlainTotalCount - \PdrInvPathprogramsKinductionDfStaticZeroZeroTTrueNotSolvedByKinductionPlainCorrectCount - \PdrInvPathprogramsKinductionDfStaticZeroZeroTTrueNotSolvedByKinductionPlainWrongTrueCount - \PdrInvPathprogramsKinductionDfStaticZeroZeroTTrueNotSolvedByKinductionPlainWrongFalseCount - \PdrInvPathprogramsKinductionDfStaticZeroZeroTTrueNotSolvedByKinductionPlainErrorTimeoutCount - \PdrInvPathprogramsKinductionDfStaticZeroZeroTTrueNotSolvedByKinductionPlainErrorOutOfMemoryCount \relax}
\edef\PdrInvPathprogramsKinductionDfStaticZeroOneTTTrueNotSolvedByKinductionPlainErrorOtherInconclusiveCount{\the\numexpr \PdrInvPathprogramsKinductionDfStaticZeroOneTTTrueNotSolvedByKinductionPlainTotalCount - \PdrInvPathprogramsKinductionDfStaticZeroOneTTTrueNotSolvedByKinductionPlainCorrectCount - \PdrInvPathprogramsKinductionDfStaticZeroOneTTTrueNotSolvedByKinductionPlainWrongTrueCount - \PdrInvPathprogramsKinductionDfStaticZeroOneTTTrueNotSolvedByKinductionPlainWrongFalseCount - \PdrInvPathprogramsKinductionDfStaticZeroOneTTTrueNotSolvedByKinductionPlainErrorTimeoutCount - \PdrInvPathprogramsKinductionDfStaticZeroOneTTTrueNotSolvedByKinductionPlainErrorOutOfMemoryCount \relax}
\edef\PdrInvPathprogramsKinductionDfStaticZeroOneTFTrueNotSolvedByKinductionPlainErrorOtherInconclusiveCount{\the\numexpr \PdrInvPathprogramsKinductionDfStaticZeroOneTFTrueNotSolvedByKinductionPlainTotalCount - \PdrInvPathprogramsKinductionDfStaticZeroOneTFTrueNotSolvedByKinductionPlainCorrectCount - \PdrInvPathprogramsKinductionDfStaticZeroOneTFTrueNotSolvedByKinductionPlainWrongTrueCount - \PdrInvPathprogramsKinductionDfStaticZeroOneTFTrueNotSolvedByKinductionPlainWrongFalseCount - \PdrInvPathprogramsKinductionDfStaticZeroOneTFTrueNotSolvedByKinductionPlainErrorTimeoutCount - \PdrInvPathprogramsKinductionDfStaticZeroOneTFTrueNotSolvedByKinductionPlainErrorOutOfMemoryCount \relax}
\edef\PdrInvPathprogramsKinductionDfStaticZeroTwoTTTrueNotSolvedByKinductionPlainErrorOtherInconclusiveCount{\the\numexpr \PdrInvPathprogramsKinductionDfStaticZeroTwoTTTrueNotSolvedByKinductionPlainTotalCount - \PdrInvPathprogramsKinductionDfStaticZeroTwoTTTrueNotSolvedByKinductionPlainCorrectCount - \PdrInvPathprogramsKinductionDfStaticZeroTwoTTTrueNotSolvedByKinductionPlainWrongTrueCount - \PdrInvPathprogramsKinductionDfStaticZeroTwoTTTrueNotSolvedByKinductionPlainWrongFalseCount - \PdrInvPathprogramsKinductionDfStaticZeroTwoTTTrueNotSolvedByKinductionPlainErrorTimeoutCount - \PdrInvPathprogramsKinductionDfStaticZeroTwoTTTrueNotSolvedByKinductionPlainErrorOutOfMemoryCount \relax}
\edef\PdrInvPathprogramsKinductionDfStaticZeroTwoTFTrueNotSolvedByKinductionPlainErrorOtherInconclusiveCount{\the\numexpr \PdrInvPathprogramsKinductionDfStaticZeroTwoTFTrueNotSolvedByKinductionPlainTotalCount - \PdrInvPathprogramsKinductionDfStaticZeroTwoTFTrueNotSolvedByKinductionPlainCorrectCount - \PdrInvPathprogramsKinductionDfStaticZeroTwoTFTrueNotSolvedByKinductionPlainWrongTrueCount - \PdrInvPathprogramsKinductionDfStaticZeroTwoTFTrueNotSolvedByKinductionPlainWrongFalseCount - \PdrInvPathprogramsKinductionDfStaticZeroTwoTFTrueNotSolvedByKinductionPlainErrorTimeoutCount - \PdrInvPathprogramsKinductionDfStaticZeroTwoTFTrueNotSolvedByKinductionPlainErrorOutOfMemoryCount \relax}
\edef\PdrInvPathprogramsKinductionDfStaticEightTwoTTrueNotSolvedByKinductionPlainErrorOtherInconclusiveCount{\the\numexpr \PdrInvPathprogramsKinductionDfStaticEightTwoTTrueNotSolvedByKinductionPlainTotalCount - \PdrInvPathprogramsKinductionDfStaticEightTwoTTrueNotSolvedByKinductionPlainCorrectCount - \PdrInvPathprogramsKinductionDfStaticEightTwoTTrueNotSolvedByKinductionPlainWrongTrueCount - \PdrInvPathprogramsKinductionDfStaticEightTwoTTrueNotSolvedByKinductionPlainWrongFalseCount - \PdrInvPathprogramsKinductionDfStaticEightTwoTTrueNotSolvedByKinductionPlainErrorTimeoutCount - \PdrInvPathprogramsKinductionDfStaticEightTwoTTrueNotSolvedByKinductionPlainErrorOutOfMemoryCount \relax}
\edef\PdrInvPathprogramsKinductionDfStaticSixteenTwoTTrueNotSolvedByKinductionPlainErrorOtherInconclusiveCount{\the\numexpr \PdrInvPathprogramsKinductionDfStaticSixteenTwoTTrueNotSolvedByKinductionPlainTotalCount - \PdrInvPathprogramsKinductionDfStaticSixteenTwoTTrueNotSolvedByKinductionPlainCorrectCount - \PdrInvPathprogramsKinductionDfStaticSixteenTwoTTrueNotSolvedByKinductionPlainWrongTrueCount - \PdrInvPathprogramsKinductionDfStaticSixteenTwoTTrueNotSolvedByKinductionPlainWrongFalseCount - \PdrInvPathprogramsKinductionDfStaticSixteenTwoTTrueNotSolvedByKinductionPlainErrorTimeoutCount - \PdrInvPathprogramsKinductionDfStaticSixteenTwoTTrueNotSolvedByKinductionPlainErrorOutOfMemoryCount \relax}
\edef\PdrInvPathprogramsKinductionDfStaticSixteenTwoFTrueNotSolvedByKinductionPlainErrorOtherInconclusiveCount{\the\numexpr \PdrInvPathprogramsKinductionDfStaticSixteenTwoFTrueNotSolvedByKinductionPlainTotalCount - \PdrInvPathprogramsKinductionDfStaticSixteenTwoFTrueNotSolvedByKinductionPlainCorrectCount - \PdrInvPathprogramsKinductionDfStaticSixteenTwoFTrueNotSolvedByKinductionPlainWrongTrueCount - \PdrInvPathprogramsKinductionDfStaticSixteenTwoFTrueNotSolvedByKinductionPlainWrongFalseCount - \PdrInvPathprogramsKinductionDfStaticSixteenTwoFTrueNotSolvedByKinductionPlainErrorTimeoutCount - \PdrInvPathprogramsKinductionDfStaticSixteenTwoFTrueNotSolvedByKinductionPlainErrorOutOfMemoryCount \relax}
\edef\PdrInvPathprogramsKinductionDfTrueNotSolvedByKinductionPlainErrorOtherInconclusiveCount{\the\numexpr \PdrInvPathprogramsKinductionDfTrueNotSolvedByKinductionPlainTotalCount - \PdrInvPathprogramsKinductionDfTrueNotSolvedByKinductionPlainCorrectCount - \PdrInvPathprogramsKinductionDfTrueNotSolvedByKinductionPlainWrongTrueCount - \PdrInvPathprogramsKinductionDfTrueNotSolvedByKinductionPlainWrongFalseCount - \PdrInvPathprogramsKinductionDfTrueNotSolvedByKinductionPlainErrorTimeoutCount - \PdrInvPathprogramsKinductionDfTrueNotSolvedByKinductionPlainErrorOutOfMemoryCount \relax}
\edef\PdrInvPathprogramsKinductionKipdrTrueNotSolvedByKinductionPlainErrorOtherInconclusiveCount{\the\numexpr \PdrInvPathprogramsKinductionKipdrTrueNotSolvedByKinductionPlainTotalCount - \PdrInvPathprogramsKinductionKipdrTrueNotSolvedByKinductionPlainCorrectCount - \PdrInvPathprogramsKinductionKipdrTrueNotSolvedByKinductionPlainWrongTrueCount - \PdrInvPathprogramsKinductionKipdrTrueNotSolvedByKinductionPlainWrongFalseCount - \PdrInvPathprogramsKinductionKipdrTrueNotSolvedByKinductionPlainErrorTimeoutCount - \PdrInvPathprogramsKinductionKipdrTrueNotSolvedByKinductionPlainErrorOutOfMemoryCount \relax}
\edef\PdrInvPathprogramsKinductionKipdrdfTrueNotSolvedByKinductionPlainErrorOtherInconclusiveCount{\the\numexpr \PdrInvPathprogramsKinductionKipdrdfTrueNotSolvedByKinductionPlainTotalCount - \PdrInvPathprogramsKinductionKipdrdfTrueNotSolvedByKinductionPlainCorrectCount - \PdrInvPathprogramsKinductionKipdrdfTrueNotSolvedByKinductionPlainWrongTrueCount - \PdrInvPathprogramsKinductionKipdrdfTrueNotSolvedByKinductionPlainWrongFalseCount - \PdrInvPathprogramsKinductionKipdrdfTrueNotSolvedByKinductionPlainErrorTimeoutCount - \PdrInvPathprogramsKinductionKipdrdfTrueNotSolvedByKinductionPlainErrorOutOfMemoryCount \relax}
\edef\PdrInvPathprogramsKinductionPlainScore{\the\numexpr (2 * \PdrInvPathprogramsKinductionPlainCorrectTrueCount) + \PdrInvPathprogramsKinductionPlainCorrectFalseCount - (32 * \PdrInvPathprogramsKinductionPlainWrongTrueCount) - (16 * \PdrInvPathprogramsKinductionPlainWrongFalseCount) \relax}
\edef\PdrInvPathprogramsKinductionDfStaticZeroZeroTScore{\the\numexpr (2 * \PdrInvPathprogramsKinductionDfStaticZeroZeroTCorrectTrueCount) + \PdrInvPathprogramsKinductionDfStaticZeroZeroTCorrectFalseCount - (32 * \PdrInvPathprogramsKinductionDfStaticZeroZeroTWrongTrueCount) - (16 * \PdrInvPathprogramsKinductionDfStaticZeroZeroTWrongFalseCount) \relax}
\edef\PdrInvPathprogramsKinductionDfStaticZeroOneTTScore{\the\numexpr (2 * \PdrInvPathprogramsKinductionDfStaticZeroOneTTCorrectTrueCount) + \PdrInvPathprogramsKinductionDfStaticZeroOneTTCorrectFalseCount - (32 * \PdrInvPathprogramsKinductionDfStaticZeroOneTTWrongTrueCount) - (16 * \PdrInvPathprogramsKinductionDfStaticZeroOneTTWrongFalseCount) \relax}
\edef\PdrInvPathprogramsKinductionDfStaticZeroOneTFScore{\the\numexpr (2 * \PdrInvPathprogramsKinductionDfStaticZeroOneTFCorrectTrueCount) + \PdrInvPathprogramsKinductionDfStaticZeroOneTFCorrectFalseCount - (32 * \PdrInvPathprogramsKinductionDfStaticZeroOneTFWrongTrueCount) - (16 * \PdrInvPathprogramsKinductionDfStaticZeroOneTFWrongFalseCount) \relax}
\edef\PdrInvPathprogramsKinductionDfStaticZeroTwoTTScore{\the\numexpr (2 * \PdrInvPathprogramsKinductionDfStaticZeroTwoTTCorrectTrueCount) + \PdrInvPathprogramsKinductionDfStaticZeroTwoTTCorrectFalseCount - (32 * \PdrInvPathprogramsKinductionDfStaticZeroTwoTTWrongTrueCount) - (16 * \PdrInvPathprogramsKinductionDfStaticZeroTwoTTWrongFalseCount) \relax}
\edef\PdrInvPathprogramsKinductionDfStaticZeroTwoTFScore{\the\numexpr (2 * \PdrInvPathprogramsKinductionDfStaticZeroTwoTFCorrectTrueCount) + \PdrInvPathprogramsKinductionDfStaticZeroTwoTFCorrectFalseCount - (32 * \PdrInvPathprogramsKinductionDfStaticZeroTwoTFWrongTrueCount) - (16 * \PdrInvPathprogramsKinductionDfStaticZeroTwoTFWrongFalseCount) \relax}
\edef\PdrInvPathprogramsKinductionDfStaticEightTwoTScore{\the\numexpr (2 * \PdrInvPathprogramsKinductionDfStaticEightTwoTCorrectTrueCount) + \PdrInvPathprogramsKinductionDfStaticEightTwoTCorrectFalseCount - (32 * \PdrInvPathprogramsKinductionDfStaticEightTwoTWrongTrueCount) - (16 * \PdrInvPathprogramsKinductionDfStaticEightTwoTWrongFalseCount) \relax}
\edef\PdrInvPathprogramsKinductionDfStaticSixteenTwoTScore{\the\numexpr (2 * \PdrInvPathprogramsKinductionDfStaticSixteenTwoTCorrectTrueCount) + \PdrInvPathprogramsKinductionDfStaticSixteenTwoTCorrectFalseCount - (32 * \PdrInvPathprogramsKinductionDfStaticSixteenTwoTWrongTrueCount) - (16 * \PdrInvPathprogramsKinductionDfStaticSixteenTwoTWrongFalseCount) \relax}
\edef\PdrInvPathprogramsKinductionDfStaticSixteenTwoFScore{\the\numexpr (2 * \PdrInvPathprogramsKinductionDfStaticSixteenTwoFCorrectTrueCount) + \PdrInvPathprogramsKinductionDfStaticSixteenTwoFCorrectFalseCount - (32 * \PdrInvPathprogramsKinductionDfStaticSixteenTwoFWrongTrueCount) - (16 * \PdrInvPathprogramsKinductionDfStaticSixteenTwoFWrongFalseCount) \relax}
\edef\PdrInvPathprogramsKinductionDfScore{\the\numexpr (2 * \PdrInvPathprogramsKinductionDfCorrectTrueCount) + \PdrInvPathprogramsKinductionDfCorrectFalseCount - (32 * \PdrInvPathprogramsKinductionDfWrongTrueCount) - (16 * \PdrInvPathprogramsKinductionDfWrongFalseCount) \relax}
\edef\PdrInvPathprogramsKinductionKipdrScore{\the\numexpr (2 * \PdrInvPathprogramsKinductionKipdrCorrectTrueCount) + \PdrInvPathprogramsKinductionKipdrCorrectFalseCount - (32 * \PdrInvPathprogramsKinductionKipdrWrongTrueCount) - (16 * \PdrInvPathprogramsKinductionKipdrWrongFalseCount) \relax}
\edef\PdrInvPathprogramsKinductionKipdrdfScore{\the\numexpr (2 * \PdrInvPathprogramsKinductionKipdrdfCorrectTrueCount) + \PdrInvPathprogramsKinductionKipdrdfCorrectFalseCount - (32 * \PdrInvPathprogramsKinductionKipdrdfWrongTrueCount) - (16 * \PdrInvPathprogramsKinductionKipdrdfWrongFalseCount) \relax}
\edef\PdrInvPathprogramsKinductionPlainTrueNotSolvedByKinductionPlainScore{\the\numexpr (2 * \PdrInvPathprogramsKinductionPlainTrueNotSolvedByKinductionPlainCorrectTrueCount) + \PdrInvPathprogramsKinductionPlainTrueNotSolvedByKinductionPlainCorrectFalseCount - (32 * \PdrInvPathprogramsKinductionPlainTrueNotSolvedByKinductionPlainWrongTrueCount) - (16 * \PdrInvPathprogramsKinductionPlainTrueNotSolvedByKinductionPlainWrongFalseCount) \relax}
\edef\PdrInvPathprogramsKinductionDfStaticZeroZeroTTrueNotSolvedByKinductionPlainScore{\the\numexpr (2 * \PdrInvPathprogramsKinductionDfStaticZeroZeroTTrueNotSolvedByKinductionPlainCorrectTrueCount) + \PdrInvPathprogramsKinductionDfStaticZeroZeroTTrueNotSolvedByKinductionPlainCorrectFalseCount - (32 * \PdrInvPathprogramsKinductionDfStaticZeroZeroTTrueNotSolvedByKinductionPlainWrongTrueCount) - (16 * \PdrInvPathprogramsKinductionDfStaticZeroZeroTTrueNotSolvedByKinductionPlainWrongFalseCount) \relax}
\edef\PdrInvPathprogramsKinductionDfStaticZeroOneTTTrueNotSolvedByKinductionPlainScore{\the\numexpr (2 * \PdrInvPathprogramsKinductionDfStaticZeroOneTTTrueNotSolvedByKinductionPlainCorrectTrueCount) + \PdrInvPathprogramsKinductionDfStaticZeroOneTTTrueNotSolvedByKinductionPlainCorrectFalseCount - (32 * \PdrInvPathprogramsKinductionDfStaticZeroOneTTTrueNotSolvedByKinductionPlainWrongTrueCount) - (16 * \PdrInvPathprogramsKinductionDfStaticZeroOneTTTrueNotSolvedByKinductionPlainWrongFalseCount) \relax}
\edef\PdrInvPathprogramsKinductionDfStaticZeroOneTFTrueNotSolvedByKinductionPlainScore{\the\numexpr (2 * \PdrInvPathprogramsKinductionDfStaticZeroOneTFTrueNotSolvedByKinductionPlainCorrectTrueCount) + \PdrInvPathprogramsKinductionDfStaticZeroOneTFTrueNotSolvedByKinductionPlainCorrectFalseCount - (32 * \PdrInvPathprogramsKinductionDfStaticZeroOneTFTrueNotSolvedByKinductionPlainWrongTrueCount) - (16 * \PdrInvPathprogramsKinductionDfStaticZeroOneTFTrueNotSolvedByKinductionPlainWrongFalseCount) \relax}
\edef\PdrInvPathprogramsKinductionDfStaticZeroTwoTTTrueNotSolvedByKinductionPlainScore{\the\numexpr (2 * \PdrInvPathprogramsKinductionDfStaticZeroTwoTTTrueNotSolvedByKinductionPlainCorrectTrueCount) + \PdrInvPathprogramsKinductionDfStaticZeroTwoTTTrueNotSolvedByKinductionPlainCorrectFalseCount - (32 * \PdrInvPathprogramsKinductionDfStaticZeroTwoTTTrueNotSolvedByKinductionPlainWrongTrueCount) - (16 * \PdrInvPathprogramsKinductionDfStaticZeroTwoTTTrueNotSolvedByKinductionPlainWrongFalseCount) \relax}
\edef\PdrInvPathprogramsKinductionDfStaticZeroTwoTFTrueNotSolvedByKinductionPlainScore{\the\numexpr (2 * \PdrInvPathprogramsKinductionDfStaticZeroTwoTFTrueNotSolvedByKinductionPlainCorrectTrueCount) + \PdrInvPathprogramsKinductionDfStaticZeroTwoTFTrueNotSolvedByKinductionPlainCorrectFalseCount - (32 * \PdrInvPathprogramsKinductionDfStaticZeroTwoTFTrueNotSolvedByKinductionPlainWrongTrueCount) - (16 * \PdrInvPathprogramsKinductionDfStaticZeroTwoTFTrueNotSolvedByKinductionPlainWrongFalseCount) \relax}
\edef\PdrInvPathprogramsKinductionDfStaticEightTwoTTrueNotSolvedByKinductionPlainScore{\the\numexpr (2 * \PdrInvPathprogramsKinductionDfStaticEightTwoTTrueNotSolvedByKinductionPlainCorrectTrueCount) + \PdrInvPathprogramsKinductionDfStaticEightTwoTTrueNotSolvedByKinductionPlainCorrectFalseCount - (32 * \PdrInvPathprogramsKinductionDfStaticEightTwoTTrueNotSolvedByKinductionPlainWrongTrueCount) - (16 * \PdrInvPathprogramsKinductionDfStaticEightTwoTTrueNotSolvedByKinductionPlainWrongFalseCount) \relax}
\edef\PdrInvPathprogramsKinductionDfStaticSixteenTwoTTrueNotSolvedByKinductionPlainScore{\the\numexpr (2 * \PdrInvPathprogramsKinductionDfStaticSixteenTwoTTrueNotSolvedByKinductionPlainCorrectTrueCount) + \PdrInvPathprogramsKinductionDfStaticSixteenTwoTTrueNotSolvedByKinductionPlainCorrectFalseCount - (32 * \PdrInvPathprogramsKinductionDfStaticSixteenTwoTTrueNotSolvedByKinductionPlainWrongTrueCount) - (16 * \PdrInvPathprogramsKinductionDfStaticSixteenTwoTTrueNotSolvedByKinductionPlainWrongFalseCount) \relax}
\edef\PdrInvPathprogramsKinductionDfStaticSixteenTwoFTrueNotSolvedByKinductionPlainScore{\the\numexpr (2 * \PdrInvPathprogramsKinductionDfStaticSixteenTwoFTrueNotSolvedByKinductionPlainCorrectTrueCount) + \PdrInvPathprogramsKinductionDfStaticSixteenTwoFTrueNotSolvedByKinductionPlainCorrectFalseCount - (32 * \PdrInvPathprogramsKinductionDfStaticSixteenTwoFTrueNotSolvedByKinductionPlainWrongTrueCount) - (16 * \PdrInvPathprogramsKinductionDfStaticSixteenTwoFTrueNotSolvedByKinductionPlainWrongFalseCount) \relax}
\edef\PdrInvPathprogramsKinductionDfTrueNotSolvedByKinductionPlainScore{\the\numexpr (2 * \PdrInvPathprogramsKinductionDfTrueNotSolvedByKinductionPlainCorrectTrueCount) + \PdrInvPathprogramsKinductionDfTrueNotSolvedByKinductionPlainCorrectFalseCount - (32 * \PdrInvPathprogramsKinductionDfTrueNotSolvedByKinductionPlainWrongTrueCount) - (16 * \PdrInvPathprogramsKinductionDfTrueNotSolvedByKinductionPlainWrongFalseCount) \relax}
\edef\PdrInvPathprogramsKinductionKipdrTrueNotSolvedByKinductionPlainScore{\the\numexpr (2 * \PdrInvPathprogramsKinductionKipdrTrueNotSolvedByKinductionPlainCorrectTrueCount) + \PdrInvPathprogramsKinductionKipdrTrueNotSolvedByKinductionPlainCorrectFalseCount - (32 * \PdrInvPathprogramsKinductionKipdrTrueNotSolvedByKinductionPlainWrongTrueCount) - (16 * \PdrInvPathprogramsKinductionKipdrTrueNotSolvedByKinductionPlainWrongFalseCount) \relax}
\edef\PdrInvPathprogramsKinductionKipdrdfTrueNotSolvedByKinductionPlainScore{\the\numexpr (2 * \PdrInvPathprogramsKinductionKipdrdfTrueNotSolvedByKinductionPlainCorrectTrueCount) + \PdrInvPathprogramsKinductionKipdrdfTrueNotSolvedByKinductionPlainCorrectFalseCount - (32 * \PdrInvPathprogramsKinductionKipdrdfTrueNotSolvedByKinductionPlainWrongTrueCount) - (16 * \PdrInvPathprogramsKinductionKipdrdfTrueNotSolvedByKinductionPlainWrongFalseCount) \relax}
\edef\PdrInvPathprogramsKinductionPlainTrueScore{\the\numexpr (2 * \PdrInvPathprogramsKinductionPlainCorrectTrueCount) - (32 * \PdrInvPathprogramsKinductionPlainWrongTrueCount) \relax}
\edef\PdrInvPathprogramsKinductionDfStaticZeroZeroTTrueScore{\the\numexpr (2 * \PdrInvPathprogramsKinductionDfStaticZeroZeroTCorrectTrueCount) - (32 * \PdrInvPathprogramsKinductionDfStaticZeroZeroTWrongTrueCount) \relax}
\edef\PdrInvPathprogramsKinductionDfStaticZeroOneTTTrueScore{\the\numexpr (2 * \PdrInvPathprogramsKinductionDfStaticZeroOneTTCorrectTrueCount) - (32 * \PdrInvPathprogramsKinductionDfStaticZeroOneTTWrongTrueCount) \relax}
\edef\PdrInvPathprogramsKinductionDfStaticZeroOneTFTrueScore{\the\numexpr (2 * \PdrInvPathprogramsKinductionDfStaticZeroOneTFCorrectTrueCount) - (32 * \PdrInvPathprogramsKinductionDfStaticZeroOneTFWrongTrueCount) \relax}
\edef\PdrInvPathprogramsKinductionDfStaticZeroTwoTTTrueScore{\the\numexpr (2 * \PdrInvPathprogramsKinductionDfStaticZeroTwoTTCorrectTrueCount) - (32 * \PdrInvPathprogramsKinductionDfStaticZeroTwoTTWrongTrueCount) \relax}
\edef\PdrInvPathprogramsKinductionDfStaticZeroTwoTFTrueScore{\the\numexpr (2 * \PdrInvPathprogramsKinductionDfStaticZeroTwoTFCorrectTrueCount) - (32 * \PdrInvPathprogramsKinductionDfStaticZeroTwoTFWrongTrueCount) \relax}
\edef\PdrInvPathprogramsKinductionDfStaticEightTwoTTrueScore{\the\numexpr (2 * \PdrInvPathprogramsKinductionDfStaticEightTwoTCorrectTrueCount) - (32 * \PdrInvPathprogramsKinductionDfStaticEightTwoTWrongTrueCount) \relax}
\edef\PdrInvPathprogramsKinductionDfStaticSixteenTwoTTrueScore{\the\numexpr (2 * \PdrInvPathprogramsKinductionDfStaticSixteenTwoTCorrectTrueCount) - (32 * \PdrInvPathprogramsKinductionDfStaticSixteenTwoTWrongTrueCount) \relax}
\edef\PdrInvPathprogramsKinductionDfStaticSixteenTwoFTrueScore{\the\numexpr (2 * \PdrInvPathprogramsKinductionDfStaticSixteenTwoFCorrectTrueCount) - (32 * \PdrInvPathprogramsKinductionDfStaticSixteenTwoFWrongTrueCount) \relax}
\edef\PdrInvPathprogramsKinductionDfTrueScore{\the\numexpr (2 * \PdrInvPathprogramsKinductionDfCorrectTrueCount) - (32 * \PdrInvPathprogramsKinductionDfWrongTrueCount) \relax}
\edef\PdrInvPathprogramsKinductionKipdrTrueScore{\the\numexpr (2 * \PdrInvPathprogramsKinductionKipdrCorrectTrueCount) - (32 * \PdrInvPathprogramsKinductionKipdrWrongTrueCount) \relax}
\edef\PdrInvPathprogramsKinductionKipdrdfTrueScore{\the\numexpr (2 * \PdrInvPathprogramsKinductionKipdrdfCorrectTrueCount) - (32 * \PdrInvPathprogramsKinductionKipdrdfWrongTrueCount) \relax}
\edef\PdrInvPathprogramsKinductionPlainTrueNotSolvedByKinductionPlainTrueScore{\the\numexpr (2 * \PdrInvPathprogramsKinductionPlainTrueNotSolvedByKinductionPlainCorrectTrueCount) - (32 * \PdrInvPathprogramsKinductionPlainTrueNotSolvedByKinductionPlainWrongTrueCount) \relax}
\edef\PdrInvPathprogramsKinductionDfStaticZeroZeroTTrueNotSolvedByKinductionPlainTrueScore{\the\numexpr (2 * \PdrInvPathprogramsKinductionDfStaticZeroZeroTTrueNotSolvedByKinductionPlainCorrectTrueCount) - (32 * \PdrInvPathprogramsKinductionDfStaticZeroZeroTTrueNotSolvedByKinductionPlainWrongTrueCount) \relax}
\edef\PdrInvPathprogramsKinductionDfStaticZeroOneTTTrueNotSolvedByKinductionPlainTrueScore{\the\numexpr (2 * \PdrInvPathprogramsKinductionDfStaticZeroOneTTTrueNotSolvedByKinductionPlainCorrectTrueCount) - (32 * \PdrInvPathprogramsKinductionDfStaticZeroOneTTTrueNotSolvedByKinductionPlainWrongTrueCount) \relax}
\edef\PdrInvPathprogramsKinductionDfStaticZeroOneTFTrueNotSolvedByKinductionPlainTrueScore{\the\numexpr (2 * \PdrInvPathprogramsKinductionDfStaticZeroOneTFTrueNotSolvedByKinductionPlainCorrectTrueCount) - (32 * \PdrInvPathprogramsKinductionDfStaticZeroOneTFTrueNotSolvedByKinductionPlainWrongTrueCount) \relax}
\edef\PdrInvPathprogramsKinductionDfStaticZeroTwoTTTrueNotSolvedByKinductionPlainTrueScore{\the\numexpr (2 * \PdrInvPathprogramsKinductionDfStaticZeroTwoTTTrueNotSolvedByKinductionPlainCorrectTrueCount) - (32 * \PdrInvPathprogramsKinductionDfStaticZeroTwoTTTrueNotSolvedByKinductionPlainWrongTrueCount) \relax}
\edef\PdrInvPathprogramsKinductionDfStaticZeroTwoTFTrueNotSolvedByKinductionPlainTrueScore{\the\numexpr (2 * \PdrInvPathprogramsKinductionDfStaticZeroTwoTFTrueNotSolvedByKinductionPlainCorrectTrueCount) - (32 * \PdrInvPathprogramsKinductionDfStaticZeroTwoTFTrueNotSolvedByKinductionPlainWrongTrueCount) \relax}
\edef\PdrInvPathprogramsKinductionDfStaticEightTwoTTrueNotSolvedByKinductionPlainTrueScore{\the\numexpr (2 * \PdrInvPathprogramsKinductionDfStaticEightTwoTTrueNotSolvedByKinductionPlainCorrectTrueCount) - (32 * \PdrInvPathprogramsKinductionDfStaticEightTwoTTrueNotSolvedByKinductionPlainWrongTrueCount) \relax}
\edef\PdrInvPathprogramsKinductionDfStaticSixteenTwoTTrueNotSolvedByKinductionPlainTrueScore{\the\numexpr (2 * \PdrInvPathprogramsKinductionDfStaticSixteenTwoTTrueNotSolvedByKinductionPlainCorrectTrueCount) - (32 * \PdrInvPathprogramsKinductionDfStaticSixteenTwoTTrueNotSolvedByKinductionPlainWrongTrueCount) \relax}
\edef\PdrInvPathprogramsKinductionDfStaticSixteenTwoFTrueNotSolvedByKinductionPlainTrueScore{\the\numexpr (2 * \PdrInvPathprogramsKinductionDfStaticSixteenTwoFTrueNotSolvedByKinductionPlainCorrectTrueCount) - (32 * \PdrInvPathprogramsKinductionDfStaticSixteenTwoFTrueNotSolvedByKinductionPlainWrongTrueCount) \relax}
\edef\PdrInvPathprogramsKinductionDfTrueNotSolvedByKinductionPlainTrueScore{\the\numexpr (2 * \PdrInvPathprogramsKinductionDfTrueNotSolvedByKinductionPlainCorrectTrueCount) - (32 * \PdrInvPathprogramsKinductionDfTrueNotSolvedByKinductionPlainWrongTrueCount) \relax}
\edef\PdrInvPathprogramsKinductionKipdrTrueNotSolvedByKinductionPlainTrueScore{\the\numexpr (2 * \PdrInvPathprogramsKinductionKipdrTrueNotSolvedByKinductionPlainCorrectTrueCount) - (32 * \PdrInvPathprogramsKinductionKipdrTrueNotSolvedByKinductionPlainWrongTrueCount) \relax}
\edef\PdrInvPathprogramsKinductionKipdrdfTrueNotSolvedByKinductionPlainTrueScore{\the\numexpr (2 * \PdrInvPathprogramsKinductionKipdrdfTrueNotSolvedByKinductionPlainCorrectTrueCount) - (32 * \PdrInvPathprogramsKinductionKipdrdfTrueNotSolvedByKinductionPlainWrongTrueCount) \relax}
\edef\npathprogramstasksplain{1167}
\newcommand{\npathprogramstasks}{\num{\npathprogramstasksplain}}
\edef\npathprogramstruetasksplain{1167}
\newcommand{\npathprogramstruetasks}{\num{\npathprogramstruetasksplain}}

%% file: main-configs-table.tex
\begin{table*}[t]
  \centering
  \caption{Results for all~\ntasks~verification tasks,
           \nfalsetasks~of which contain bugs,
           while the other~\ntruetasks~are considered to be safe,
           for the two CTIGAR implementations \cpactigar and \vvtctigar,
           for a theoretical ``virtual best'' combination of both CTIGAR implementations where an oracle selects the best implementation for each task,
	   for \kinduction without auxiliary invariants (\kind),
	   and for the best configurations of each tool:
	   \cpachecker's \kidfkipdr,
	   \seahorn,
	   and \vvt as a portfolio verifier.
  }
  \label{tab:main-configs-results}
  \newcommand\stoh[1]{\fpeval{#1/3600}}
  \newcommand\precnum[1]{\tablenum[round-mode=off,table-format=4.1]{#1}}
  \newcommand\rndnum[1]{\tablenum[round-mode=figures,round-precision=2,table-format=4.1]{#1}}
  \newcommand\precnumnodec[1]{\tablenum[round-mode=off,table-format=4.0]{#1}}
  \newcommand\rndnumnodec[1]{\tablenum[round-mode=figures,round-precision=2,table-format=4.0]{#1}}

  \newcommand\virtualBestColor{black!70}

\scalebox{0.95}{
  \begin{tabular}{l|cc>{\color{\virtualBestColor}}c|c|ccc}
    \toprule
    \multicolumn{1}{l|}{\multirow{2}{*}{Verifier}}
        & \multicolumn{3}{c|}{CTIGAR}
	& \multicolumn{1}{c|}{\multirow{2}{*}{\kind}}
	& \multicolumn{3}{c}{Best of each tool} \\
          & \multicolumn{1}{c}{\cpachecker}
          & \multicolumn{1}{c}{\vvt}
          & \multicolumn{1}{c|}{\color{\virtualBestColor}\begin{minipage}{1cm}Virt. Best\end{minipage}}
	  &
	  & \multicolumn{1}{c}{\begin{minipage}{1.35cm}\smaller\kidf;\\\kipdr\end{minipage}}
          & \multicolumn{1}{c}{\seahorn}
          & \multicolumn{1}{c}{\begin{minipage}{1.35cm}\mbox{\vvt\smaller-}\\\mbox{Portfolio}\end{minipage}} \\
    \midrule
    Score
          & \precnum{\PdrInvPdrScore}
          & \precnum{\VvtCtigarScore}
          & \precnum{\PdrInvOracleScore}
	  & \precnum{\PdrInvKinductionPlainScore}
	  & \textbf{\precnum{\PdrInvKinductionKipdrdfScore}}
	  & \precnum{\SeahornSeahornScore}
          & \VvtPortfolioScore \\
    Correct results
          & \precnum{\PdrInvPdrCorrectCount}
          & \precnum{\VvtCtigarCorrectCount}
          & \precnum{\PdrInvOracleCorrectCount}
	  & \precnum{\PdrInvKinductionPlainCorrectCount}
	  & \precnum{\PdrInvKinductionKipdrdfCorrectCount}
	  & \precnum{\SeahornSeahornCorrectCount}
          & \precnum{\VvtPortfolioCorrectCount} \\
    \quad Correct proofs
          & \precnum{\PdrInvPdrCorrectTrueCount}
          & \precnum{\VvtCtigarCorrectTrueCount}
          & \precnum{\PdrInvOracleCorrectTrueCount}
	  & \precnum{\PdrInvKinductionPlainCorrectTrueCount}
	  & \precnum{\PdrInvKinductionKipdrdfCorrectTrueCount}
	  & \precnum{\SeahornSeahornCorrectTrueCount}
          & \precnum{\VvtPortfolioCorrectTrueCount} \\
    \quad Correct alarms
          & \precnum{\PdrInvPdrCorrectFalseCount}
          & \precnum{\VvtCtigarCorrectFalseCount}
          & \precnum{\PdrInvOracleCorrectFalseCount}
	  & \precnum{\PdrInvKinductionPlainCorrectFalseCount}
	  & \precnum{\PdrInvKinductionKipdrdfCorrectFalseCount}
	  & \precnum{\SeahornSeahornCorrectFalseCount}
          & \precnum{\VvtPortfolioCorrectFalseCount} \\
    Wrong proofs
          & \textbf{\precnum{\PdrInvPdrWrongTrueCount}}
          & \precnum{\VvtCtigarWrongTrueCount}
          & \precnum{\PdrInvOracleWrongTrueCount}
	  & \textbf{\precnum{\PdrInvKinductionPlainWrongTrueCount}}
	  & \textbf{\precnum{\PdrInvKinductionKipdrdfWrongTrueCount}}
	  & \precnum{\SeahornSeahornWrongTrueCount}
          & \precnum{\VvtPortfolioWrongTrueCount} \\
    Wrong alarms
          & \precnum{\PdrInvPdrWrongFalseCount}
          & \precnum{\VvtCtigarWrongFalseCount}
          & \precnum{\PdrInvOracleWrongFalseCount}
	  & \precnum{\PdrInvKinductionPlainWrongFalseCount}
	  & \precnum{\PdrInvKinductionKipdrdfWrongFalseCount}
	  & \precnum{\SeahornSeahornWrongFalseCount}
          & \precnum{\VvtPortfolioWrongFalseCount} \\
    Timeouts
          & \precnum{\PdrInvPdrErrorTimeoutCount}
          & \precnum{\VvtCtigarErrorTimeoutCount}
          & \precnum{\PdrInvOracleErrorTimeoutCount}
	  & \precnum{\PdrInvKinductionPlainErrorTimeoutCount}
	  & \precnum{\PdrInvKinductionKipdrdfErrorTimeoutCount}
	  & \precnum{\SeahornSeahornErrorTimeoutCount}
          & \precnum{\VvtPortfolioErrorTimeoutCount} \\
    Out of memory
          & \precnum{\PdrInvPdrErrorOutOfMemoryCount}
          & \precnum{\VvtCtigarErrorOutOfMemoryCount}
          & \precnum{\PdrInvOracleErrorOutOfMemoryCount}
	  & \precnum{\PdrInvKinductionPlainErrorOutOfMemoryCount}
	  & \precnum{\PdrInvKinductionKipdrdfErrorOutOfMemoryCount}
	  & \precnum{\SeahornSeahornErrorOutOfMemoryCount}
          & \precnum{\VvtPortfolioErrorOutOfMemoryCount} \\
    Other inconclusive
          & \precnum{\PdrInvPdrErrorOtherInconclusiveCount}
          & \precnum{\VvtCtigarErrorOtherInconclusiveCount}
          & \precnum{\PdrInvOracleErrorOtherInconclusiveCount}
	  & \precnum{\PdrInvKinductionPlainErrorOtherInconclusiveCount}
	  & \precnum{\PdrInvKinductionKipdrdfErrorOtherInconclusiveCount}
	  & \precnum{\SeahornSeahornErrorOtherInconclusiveCount}
          & \precnum{\VvtPortfolioErrorOtherInconclusiveCount} \\
    \midrule
    \multicolumn{4}{l}{Times for correct results} \\
    Total CPU Time (h)
         & \rndnum{\stoh{\PdrInvPdrCorrectCputime}}
         & \rndnum{\stoh{\VvtCtigarCorrectCputime}}
         & \rndnum{\stoh{\PdrInvOracleCorrectCputime}}
	 & \rndnum{\stoh{\PdrInvKinductionPlainCorrectCputime}}
	 & \rndnum{\stoh{\PdrInvKinductionKipdrdfCorrectCputime}}
	 & \rndnum{\stoh{\SeahornSeahornCorrectCputime}}
         & \rndnum{\stoh{\VvtPortfolioCorrectCputime}} \\
    Mean CPU Time (s)
         & \rndnum{\PdrInvPdrCorrectCputimeAvg}
         & \rndnum{\VvtCtigarCorrectCputimeAvg}
         & \rndnum{\PdrInvOracleCorrectCputimeAvg}
	 & \rndnum{\PdrInvKinductionPlainCorrectCputimeAvg}
	 & \rndnum{\PdrInvKinductionKipdrdfCorrectCputimeAvg}
	 & \rndnum{\SeahornSeahornCorrectCputimeAvg}
         & \rndnum{\VvtPortfolioCorrectCputimeAvg} \\
    Median CPU Time (s)
         & \rndnum{\PdrInvPdrCorrectCputimeMedian}
         & \rndnum{\VvtCtigarCorrectCputimeMedian}
         & \rndnum{\PdrInvOracleCorrectCputimeMedian}
	 & \rndnum{\PdrInvKinductionPlainCorrectCputimeMedian}
	 & \rndnum{\PdrInvKinductionKipdrdfCorrectCputimeMedian}
	 & \rndnum{\SeahornSeahornCorrectCputimeMedian}
         & \rndnum{\VvtPortfolioCorrectCputimeMedian} \\
    Total Wall Time (h)
         & \rndnum{\stoh{\PdrInvPdrCorrectWalltime}}
         & \rndnum{\stoh{\VvtCtigarCorrectWalltime}}
         & \rndnum{\stoh{\PdrInvOracleCorrectWalltime}}
	 & \rndnum{\stoh{\PdrInvKinductionPlainCorrectWalltime}}
	 & \rndnum{\stoh{\PdrInvKinductionKipdrdfCorrectWalltime}}
	 & \rndnum{\stoh{\SeahornSeahornCorrectWalltime}}
         & \rndnum{\stoh{\VvtPortfolioCorrectWalltime}} \\
    Mean Wall Time (s)
         & \rndnum{\PdrInvPdrCorrectWalltimeAvg}
         & \rndnum{\VvtCtigarCorrectWalltimeAvg}
         & \rndnum{\PdrInvOracleCorrectWalltimeAvg}
	 & \rndnum{\PdrInvKinductionPlainCorrectWalltimeAvg}
	 & \rndnum{\PdrInvKinductionKipdrdfCorrectWalltimeAvg}
	 & \rndnum{\SeahornSeahornCorrectWalltimeAvg}
         & \rndnum{\VvtPortfolioCorrectWalltimeAvg} \\
    Median Wall Time (s)
         & \rndnum{\PdrInvPdrCorrectWalltimeMedian}
         & \rndnum{\VvtCtigarCorrectWalltimeMedian}
         & \rndnum{\PdrInvOracleCorrectWalltimeMedian}
	 & \rndnum{\PdrInvKinductionPlainCorrectWalltimeMedian}
	 & \rndnum{\PdrInvKinductionKipdrdfCorrectWalltimeMedian}
	 & \rndnum{\SeahornSeahornCorrectWalltimeMedian}
         & \rndnum{\VvtPortfolioCorrectWalltimeMedian} \\
  \bottomrule
  \end{tabular}
}
\end{table*}

%% file: evaluation/ctigar.tex
\begin{pgfplot}
\setlength{\parindent}{0pt}
\begin{groupplot}[
    split quantile plot,
    y label style={inner sep=0pt},
    group style={group name=myplot,group size= 1 by 2,vertical sep=0pt},
]
\nextgroupplot[split quantile plot top,
    mark repeat=200,
    xmin=0,xmax=1200,
    ymode=log,
  ]
  \addgraph[violet,solid]{\cpactigar}{pdr.quantile.csv};
  \addgraph[
      mark phase=\offset{\vvtctigarXForYLessThanOneSecond}{200},
      cpacheckergreen,
      solid
    ]{\vvtctigar}{vvt-ctigar.quantile.csv};
\nextgroupplot[split quantile plot bottom,
    mark repeat=200,
    xmin=0,xmax=1200,
    xlabel=$n$-th fastest correct result (proof or alarm),
    ]
  \addgraph[violet,solid]{\cpactigar}{pdr.quantile.csv};
  \addgraph[
      cpacheckergreen,
      solid
    ]{\vvtctigar}{vvt-ctigar.quantile.csv};
  \legend{}
\end{groupplot}
\end{pgfplot}

%% file: k-induction-true-not-solved-by-kInduction-plain-table.tex
\begin{table}[t]
  \centering
  \caption{Results of \kinduction-based configurations in \cpachecker
           with different approaches for generating auxiliary invariants
           for~all~\ntasksTrueNotSolvedByKinductionPlain~verification tasks
	   that do not contain bugs
           and are not solved by \kinduction without auxiliary invariants.
	   }
  \label{tab:kind-results-true-not-solved-by-k-induction-plain}
  \newcommand\stoh[1]{\fpeval{#1/3600}}
  \newcommand\precnum[1]{\tablenum[round-mode=off,table-format=4.1]{#1}}
  \newcommand\rndnum[1]{\tablenum[round-mode=figures,round-precision=2,table-format=4.1]{#1}}
  \begin{tabular}{l|ccc}
    \toprule
    \multicolumn{1}{l|}{Approach}
        & \multicolumn{1}{c}{
              \kikipdr
          }
        & \multicolumn{1}{@{~~}c@{~~}}{
              \kidf
          }
        & \multicolumn{1}{c@{}}{
              \kidfkipdr
          } \\
    \midrule
    Correct proofs
	  & \precnum{\PdrInvKinductionKipdrTrueNotSolvedByKinductionPlainCorrectTrueCount}
	  & \precnum{\PdrInvKinductionDfTrueNotSolvedByKinductionPlainCorrectTrueCount}
	  & \bfseries\precnum{\PdrInvKinductionKipdrdfTrueNotSolvedByKinductionPlainCorrectTrueCount} \\
    Timeouts
	  & \precnum{\PdrInvKinductionKipdrTrueNotSolvedByKinductionPlainErrorTimeoutCount}
	  & \precnum{\PdrInvKinductionDfTrueNotSolvedByKinductionPlainErrorTimeoutCount}
	  & \precnum{\PdrInvKinductionKipdrdfTrueNotSolvedByKinductionPlainErrorTimeoutCount} \\
    Out of memory
	  & \precnum{\PdrInvKinductionKipdrTrueNotSolvedByKinductionPlainErrorOutOfMemoryCount}
	  & \precnum{\PdrInvKinductionDfTrueNotSolvedByKinductionPlainErrorOutOfMemoryCount}
	  & \precnum{\PdrInvKinductionKipdrdfTrueNotSolvedByKinductionPlainErrorOutOfMemoryCount} \\
    Other inconclusive
	  & \precnum{\PdrInvKinductionKipdrTrueNotSolvedByKinductionPlainErrorOtherInconclusiveCount}
	  & \precnum{\PdrInvKinductionDfTrueNotSolvedByKinductionPlainErrorOtherInconclusiveCount}
	  & \precnum{\PdrInvKinductionKipdrdfTrueNotSolvedByKinductionPlainErrorOtherInconclusiveCount} \\
    \midrule
    \multicolumn{4}{l}{Times for correct results} \\
    Total CPU Time (h)
         & \rndnum{\stoh{\PdrInvKinductionKipdrTrueNotSolvedByKinductionPlainCorrectCputime}}
         & \rndnum{\stoh{\PdrInvKinductionDfTrueNotSolvedByKinductionPlainCorrectCputime}}
         & \rndnum{\stoh{\PdrInvKinductionKipdrdfTrueNotSolvedByKinductionPlainCorrectCputime}}
\\
    Mean CPU Time (s)
         & \rndnum{\PdrInvKinductionKipdrTrueNotSolvedByKinductionPlainCorrectCputimeAvg}
         & \rndnum{\PdrInvKinductionDfTrueNotSolvedByKinductionPlainCorrectCputimeAvg}
         & \rndnum{\PdrInvKinductionKipdrdfTrueNotSolvedByKinductionPlainCorrectCputimeAvg}
\\
    Median CPU Time (s)
         & \rndnum{\PdrInvKinductionKipdrTrueNotSolvedByKinductionPlainCorrectCputimeMedian}
         & \rndnum{\PdrInvKinductionDfTrueNotSolvedByKinductionPlainCorrectCputimeMedian}
         & \rndnum{\PdrInvKinductionKipdrdfTrueNotSolvedByKinductionPlainCorrectCputimeMedian}
\\
    Total Wall Time (h)
         & \rndnum{\stoh{\PdrInvKinductionKipdrTrueNotSolvedByKinductionPlainCorrectWalltime}}
         & \rndnum{\stoh{\PdrInvKinductionDfTrueNotSolvedByKinductionPlainCorrectWalltime}}
         & \rndnum{\stoh{\PdrInvKinductionKipdrdfTrueNotSolvedByKinductionPlainCorrectWalltime}}
\\
    Mean Wall Time (s)
         & \rndnum{\PdrInvKinductionKipdrTrueNotSolvedByKinductionPlainCorrectWalltimeAvg}
         & \rndnum{\PdrInvKinductionDfTrueNotSolvedByKinductionPlainCorrectWalltimeAvg}
         & \rndnum{\PdrInvKinductionKipdrdfTrueNotSolvedByKinductionPlainCorrectWalltimeAvg}
\\
    Median Wall Time (s)
         & \rndnum{\PdrInvKinductionKipdrTrueNotSolvedByKinductionPlainCorrectWalltimeMedian}
         & \rndnum{\PdrInvKinductionDfTrueNotSolvedByKinductionPlainCorrectWalltimeMedian}
         & \rndnum{\PdrInvKinductionKipdrdfTrueNotSolvedByKinductionPlainCorrectWalltimeMedian}
\\
  \bottomrule
  \end{tabular}
  \vspace{-2mm}
\end{table}

%% file: evaluation/kInduction-true-not-solved-by-kInduction-plain.tex
\begin{pgfplot}
\begin{semilogyaxis}[quantile plot, mark repeat=200,xmin=0,xlabel=$n$-th fastest correct proof,
    legend style={at={(1,0)}, anchor=south east, outer xsep=4pt, outer ysep=4pt, inner ysep=0pt, fill=none, font={\smaller}}
  ]
  \addgraph[magenta,solid,mark=square]{\kikipdr}{kInduction-kipdr-true-not-solved-by-kInduction-plain.quantile-true.csv};
  \addgraph[red,solid,mark=o]{\kidf}{kInduction-df-true-not-solved-by-kInduction-plain.quantile-true.csv};
  \addgraph[blue,solid,mark=asterisk]{\kidfkipdr}{kInduction-kipdrdf-true-not-solved-by-kInduction-plain.quantile-true.csv};
\end{semilogyaxis}
\end{pgfplot}

%% file: pathprograms.k-induction-true-not-solved-by-kInduction-plain-table.tex
\begin{table}[t]
  \centering
  \caption{Results of \kinduction-based configurations in \cpachecker
           with different approaches for generating auxiliary invariants
           for~all~\npathprogramstasksTrueNotSolvedByKinductionPlain~path programs
	   that we assume do not contain bugs
           and that are not solved by \kinduction without auxiliary invariants.
	   }
  \label{tab:pathprograms.kind-results-true-not-solved-by-k-induction-plain}
  \newcommand\stoh[1]{\fpeval{#1/3600}}
  \newcommand\precnum[1]{\tablenum[round-mode=off,table-format=4.1]{#1}}
  \newcommand\rndnum[1]{\tablenum[round-mode=figures,round-precision=2,table-format=4.1]{#1}}
  \begin{tabular}{l|ccc}
    \toprule
    \multicolumn{1}{l|}{Approach}
        & \multicolumn{1}{c}{
              \kikipdr
          }
        & \multicolumn{1}{@{~~}c@{~~}}{
              \kidf
          }
        & \multicolumn{1}{c@{}}{
              \kidfkipdr
          } \\
    \midrule
    Correct proofs
	  & \precnum{\PdrInvPathprogramsKinductionKipdrTrueNotSolvedByKinductionPlainCorrectTrueCount}
	  & \precnum{\PdrInvPathprogramsKinductionDfTrueNotSolvedByKinductionPlainCorrectTrueCount}
	  & \bfseries\precnum{\PdrInvPathprogramsKinductionKipdrdfTrueNotSolvedByKinductionPlainCorrectTrueCount} \\
    Timeouts
	  & \precnum{\PdrInvPathprogramsKinductionKipdrTrueNotSolvedByKinductionPlainErrorTimeoutCount}
	  & \precnum{\PdrInvPathprogramsKinductionDfTrueNotSolvedByKinductionPlainErrorTimeoutCount}
	  & \precnum{\PdrInvPathprogramsKinductionKipdrdfTrueNotSolvedByKinductionPlainErrorTimeoutCount} \\
    Out of memory
	  & \precnum{\PdrInvPathprogramsKinductionKipdrTrueNotSolvedByKinductionPlainErrorOutOfMemoryCount}
	  & \precnum{\PdrInvPathprogramsKinductionDfTrueNotSolvedByKinductionPlainErrorOutOfMemoryCount}
	  & \precnum{\PdrInvPathprogramsKinductionKipdrdfTrueNotSolvedByKinductionPlainErrorOutOfMemoryCount} \\
    Other inconclusive
	  & \precnum{\PdrInvPathprogramsKinductionKipdrTrueNotSolvedByKinductionPlainErrorOtherInconclusiveCount}
	  & \precnum{\PdrInvPathprogramsKinductionDfTrueNotSolvedByKinductionPlainErrorOtherInconclusiveCount}
	  & \precnum{\PdrInvPathprogramsKinductionKipdrdfTrueNotSolvedByKinductionPlainErrorOtherInconclusiveCount} \\
    \midrule
    \multicolumn{4}{l}{Times for correct results} \\
    Total CPU Time (h)
         & \rndnum{\stoh{\PdrInvPathprogramsKinductionKipdrTrueNotSolvedByKinductionPlainCorrectCputime}}
         & \rndnum{\stoh{\PdrInvPathprogramsKinductionDfTrueNotSolvedByKinductionPlainCorrectCputime}}
         & \rndnum{\stoh{\PdrInvPathprogramsKinductionKipdrdfTrueNotSolvedByKinductionPlainCorrectCputime}}
\\
    Mean CPU Time (s)
         & \rndnum{\PdrInvPathprogramsKinductionKipdrTrueNotSolvedByKinductionPlainCorrectCputimeAvg}
         & \rndnum{\PdrInvPathprogramsKinductionDfTrueNotSolvedByKinductionPlainCorrectCputimeAvg}
         & \rndnum{\PdrInvPathprogramsKinductionKipdrdfTrueNotSolvedByKinductionPlainCorrectCputimeAvg}
\\
    Median CPU Time (s)
         & \rndnum{\PdrInvPathprogramsKinductionKipdrTrueNotSolvedByKinductionPlainCorrectCputimeMedian}
         & \rndnum{\PdrInvPathprogramsKinductionDfTrueNotSolvedByKinductionPlainCorrectCputimeMedian}
         & \rndnum{\PdrInvPathprogramsKinductionKipdrdfTrueNotSolvedByKinductionPlainCorrectCputimeMedian}
\\
    Total Wall Time (h)
         & \rndnum{\stoh{\PdrInvPathprogramsKinductionKipdrTrueNotSolvedByKinductionPlainCorrectWalltime}}
         & \rndnum{\stoh{\PdrInvPathprogramsKinductionDfTrueNotSolvedByKinductionPlainCorrectWalltime}}
         & \rndnum{\stoh{\PdrInvPathprogramsKinductionKipdrdfTrueNotSolvedByKinductionPlainCorrectWalltime}}
\\
    Mean Wall Time (s)
         & \rndnum{\PdrInvPathprogramsKinductionKipdrTrueNotSolvedByKinductionPlainCorrectWalltimeAvg}
         & \rndnum{\PdrInvPathprogramsKinductionDfTrueNotSolvedByKinductionPlainCorrectWalltimeAvg}
         & \rndnum{\PdrInvPathprogramsKinductionKipdrdfTrueNotSolvedByKinductionPlainCorrectWalltimeAvg}
\\
    Median Wall Time (s)
         & \rndnum{\PdrInvPathprogramsKinductionKipdrTrueNotSolvedByKinductionPlainCorrectWalltimeMedian}
         & \rndnum{\PdrInvPathprogramsKinductionDfTrueNotSolvedByKinductionPlainCorrectWalltimeMedian}
         & \rndnum{\PdrInvPathprogramsKinductionKipdrdfTrueNotSolvedByKinductionPlainCorrectWalltimeMedian}
\\
  \bottomrule
  \end{tabular}
\end{table}

%% file: invariants-table.tex
\begin{table}[t]
  \centering
  \caption{Results of four \kinduction-based configurations in \cpachecker
           with different approaches for generating auxiliary invariants
           for~all~\num{\PdrInvKinductionKipdrTrueNotSolvedByKinductionPlainButKipdrCorrectTrueCount}~verification tasks
	   that do not contain bugs
           and are not solved by \kinduction without auxiliary invariants,
	   but are solved by \kikipdr.
	   }
  \label{tab:kind-results-not-solved-by-k-induction-plain-but-kipdr}
  \newcommand\stoh[1]{\fpeval{#1/3600}}
  \newcommand\precnum[1]{\tablenum[round-mode=off,table-format=3.2]{#1}}
  \newcommand\rndnum[1]{\tablenum[round-mode=figures,round-precision=2,table-format=3.2]{#1}}
  \begin{tabular}{l|ccc|c}
    \toprule
    \multicolumn{1}{l|}{\multirow{2}{*}{Approach}}
        & \multicolumn{3}{c|}{\static}
        & \multicolumn{1}{c}{\multirow{2}{*}{\kikipdr}}
	\\
             & \multicolumn{1}{c}{Boxes}
	     & \multicolumn{1}{@{~~~}c@{~~~}}{\begin{minipage}{5ex}Boxes,\\$\mathsf{Eq}$\end{minipage}}
	     & \multicolumn{1}{@{~~}c@{~~~}|}{\begin{minipage}{5ex}Boxes,\\$\mathsf{Eq}$,\\$\mathsf{Mod2}$\end{minipage}} \\
    \midrule
    Correct proofs
	  & \precnum{\PdrInvKinductionDfStaticZeroZeroTTrueNotSolvedByKinductionPlainButKipdrCorrectTrueCount}
	  & \precnum{\PdrInvKinductionDfStaticZeroOneTFTrueNotSolvedByKinductionPlainButKipdrCorrectTrueCount}
	  & \precnum{\PdrInvKinductionDfStaticZeroOneTTTrueNotSolvedByKinductionPlainButKipdrCorrectTrueCount}
	  & \bfseries\precnum{\PdrInvKinductionKipdrTrueNotSolvedByKinductionPlainButKipdrCorrectTrueCount}
\\
    Timeouts
	  & \precnum{\PdrInvKinductionDfStaticZeroZeroTTrueNotSolvedByKinductionPlainButKipdrErrorTimeoutCount}
	  & \precnum{\PdrInvKinductionDfStaticZeroOneTFTrueNotSolvedByKinductionPlainButKipdrErrorTimeoutCount}
	  & \precnum{\PdrInvKinductionDfStaticZeroOneTTTrueNotSolvedByKinductionPlainButKipdrErrorTimeoutCount}
	  & \precnum{\PdrInvKinductionKipdrTrueNotSolvedByKinductionPlainButKipdrErrorTimeoutCount}
\\
    Out of memory
	  & \precnum{\PdrInvKinductionDfStaticZeroZeroTTrueNotSolvedByKinductionPlainButKipdrErrorOutOfMemoryCount}
	  & \precnum{\PdrInvKinductionDfStaticZeroOneTFTrueNotSolvedByKinductionPlainButKipdrErrorOutOfMemoryCount}
	  & \precnum{\PdrInvKinductionDfStaticZeroOneTTTrueNotSolvedByKinductionPlainButKipdrErrorOutOfMemoryCount}
	  & \precnum{\PdrInvKinductionKipdrTrueNotSolvedByKinductionPlainButKipdrErrorOutOfMemoryCount}
\\
    \midrule
    \multicolumn{5}{l}{Times for correct results} \\
    Total CPU Time (h)
         & \rndnum{\stoh{\PdrInvKinductionDfStaticZeroZeroTTrueNotSolvedByKinductionPlainButKipdrCorrectCputime}}
         & \rndnum{\stoh{\PdrInvKinductionDfStaticZeroOneTFTrueNotSolvedByKinductionPlainButKipdrCorrectCputime}}
         & \rndnum{\stoh{\PdrInvKinductionDfStaticZeroOneTTTrueNotSolvedByKinductionPlainButKipdrCorrectCputime}}
         & \rndnum{\stoh{\PdrInvKinductionKipdrTrueNotSolvedByKinductionPlainButKipdrCorrectCputime}}
\\
    Mean CPU Time (s)
         & \rndnum{\PdrInvKinductionDfStaticZeroZeroTTrueNotSolvedByKinductionPlainButKipdrCorrectCputimeAvg}
         & \rndnum{\PdrInvKinductionDfStaticZeroOneTFTrueNotSolvedByKinductionPlainButKipdrCorrectCputimeAvg}
         & \rndnum{\PdrInvKinductionDfStaticZeroOneTTTrueNotSolvedByKinductionPlainButKipdrCorrectCputimeAvg}
         & \rndnum{\PdrInvKinductionKipdrTrueNotSolvedByKinductionPlainButKipdrCorrectCputimeAvg}
\\
    Median CPU Time (s)
         & \rndnum{\PdrInvKinductionDfStaticZeroZeroTTrueNotSolvedByKinductionPlainButKipdrCorrectCputimeMedian}
         & \rndnum{\PdrInvKinductionDfStaticZeroOneTFTrueNotSolvedByKinductionPlainButKipdrCorrectCputimeMedian}
         & \rndnum{\PdrInvKinductionDfStaticZeroOneTTTrueNotSolvedByKinductionPlainButKipdrCorrectCputimeMedian}
         & \rndnum{\PdrInvKinductionKipdrTrueNotSolvedByKinductionPlainButKipdrCorrectCputimeMedian}
\\
    Total Wall Time (h)
         & \rndnum{\stoh{\PdrInvKinductionDfStaticZeroZeroTTrueNotSolvedByKinductionPlainButKipdrCorrectWalltime}}
         & \rndnum{\stoh{\PdrInvKinductionDfStaticZeroOneTFTrueNotSolvedByKinductionPlainButKipdrCorrectWalltime}}
         & \rndnum{\stoh{\PdrInvKinductionDfStaticZeroOneTTTrueNotSolvedByKinductionPlainButKipdrCorrectWalltime}}
         & \rndnum{\stoh{\PdrInvKinductionKipdrTrueNotSolvedByKinductionPlainButKipdrCorrectWalltime}}
\\
    Mean Wall Time (s)
         & \rndnum{\PdrInvKinductionDfStaticZeroZeroTTrueNotSolvedByKinductionPlainButKipdrCorrectWalltimeAvg}
         & \rndnum{\PdrInvKinductionDfStaticZeroOneTFTrueNotSolvedByKinductionPlainButKipdrCorrectWalltimeAvg}
         & \rndnum{\PdrInvKinductionDfStaticZeroOneTTTrueNotSolvedByKinductionPlainButKipdrCorrectWalltimeAvg}
         & \rndnum{\PdrInvKinductionKipdrTrueNotSolvedByKinductionPlainButKipdrCorrectWalltimeAvg}
\\
    Median Wall Time (s)
         & \rndnum{\PdrInvKinductionDfStaticZeroZeroTTrueNotSolvedByKinductionPlainButKipdrCorrectWalltimeMedian}
         & \rndnum{\PdrInvKinductionDfStaticZeroOneTFTrueNotSolvedByKinductionPlainButKipdrCorrectWalltimeMedian}
         & \rndnum{\PdrInvKinductionDfStaticZeroOneTTTrueNotSolvedByKinductionPlainButKipdrCorrectWalltimeMedian}
         & \rndnum{\PdrInvKinductionKipdrTrueNotSolvedByKinductionPlainButKipdrCorrectWalltimeMedian}
\\
  \bottomrule
  \end{tabular}
\end{table}

%% file: hand-crafted-examples.tex
\lstset{
    breaklines=true,
    postbreak=\raisebox{0ex}[0ex][0ex]{\hspace{-3em}\ensuremath{\color{red}\hookrightarrow\space}},
    breakatwhitespace=true,
}

\begin{figure}[t]
\begin{lstlisting}[style=C,firstline=9,firstnumber=9]
int main(void) {
  unsigned int s = 0;
  while (__VERIFIER_nondet_uint()) {
    if (s != 0) {
      ++s;
    }
    if (__VERIFIER_nondet_uint()) {
      __VERIFIER_assert(s == 0);
    }
  }
  return 0;
}
\end{lstlisting}
  \vspace{-5mm}
  \caption{Program \progconst}
  \label{fig:const.c}
  \vspace{-6mm}
\end{figure}

\begin{figure}[t]
\begin{lstlisting}[style=C,firstline=9,firstnumber=9]
int main(void) {
  unsigned int w = __VERIFIER_nondet_uint();
  unsigned int x = w;
  unsigned int y = __VERIFIER_nondet_uint();
  unsigned int z = y;
  while (__VERIFIER_nondet_uint()) {
    if (__VERIFIER_nondet_uint()) {
      ++w; ++x;
    } else {
      --y; --z;
    }
  }
  __VERIFIER_assert(w == x && y == z);
  return 0;
}
\end{lstlisting}
  \vspace{-5mm}
  \caption{Program \progeqone}
  \label{fig:eq1.c}
  \vspace{-6mm}
\end{figure}

\begin{figure}[t]
\begin{lstlisting}[style=C,firstline=9,firstnumber=9]
int main(void) {
  unsigned int x = 0;
  while (__VERIFIER_nondet_int()) {
    x += 2;
  }
  __VERIFIER_assert(!(x % 2));
  return 0;
}
\end{lstlisting}
  \vspace{-5mm}
  \caption{Program \progeven}
  \label{fig:even.c}
  \vspace{-6mm}
\end{figure}

\begin{figure}[t]
\begin{lstlisting}[style=C,firstline=9,firstnumber=9]
int main(void) {
  unsigned int x = 1;
  while (__VERIFIER_nondet_int()) {
    x += 2;
  }
  __VERIFIER_assert(x % 2);
  return 0;
}
\end{lstlisting}
  \vspace{-5mm}
  \caption{Program \progodd}
  \label{fig:odd.c}
  \vspace{-6mm}
\end{figure}

\begin{figure}[t]
\begin{lstlisting}[style=C,firstline=9,firstnumber=9]
int main(void) {
  unsigned int x = 0;
  while (__VERIFIER_nondet_int()) {
    x += 4;
  }
  __VERIFIER_assert(!(x % 4));
  return 0;
}
\end{lstlisting}
  \vspace{-5mm}
  \caption{Program \progmodfour}
  \label{fig:mod4.c}
  \vspace{-6mm}
\end{figure}

\begin{figure}[t]
\begin{lstlisting}[style=C,firstline=9,firstnumber=9]
int main(void) {
  unsigned int x = 5;
  while (__VERIFIER_nondet_int()) {
    x += 8;
  }
  __VERIFIER_assert((x & 5) == 5);
  return 0;
}
\end{lstlisting}
  \vspace{-6mm}
  \caption{Program \progbin}
  \label{fig:bin-suffix-5.c}
  \vspace{-5mm}
\end{figure}

%% file: evaluation/tools-score.tex
\begin{pgfplot}
\begin{groupplot}[split quantile plot,/pgfplots/table/col sep=space,
    group style={group name=myplot,group size= 1 by 2,vertical sep=0pt},
]
\nextgroupplot[split quantile plot top,
    mark repeat=200,
    xmin=-3500,xmax=5500,
    ymode=log,
    height=.60\linewidth,
  ]
  \addgraph[blue,solid,mark=asterisk]{\kidfkipdr}{kInduction-kipdrdf.quantile-score.csv};
  \addgraph[
      mark phase=\offset{\seahornScoreXForYLessThanOneSecond}{200},
      brown,solid,mark=triangle*
    ]{\seahorn}{seahorn.quantile-score.csv};
  \addgraph[
      mark phase=\offset{\vvtportfolioScoreXForYLessThanOneSecond}{200},
      green,solid,mark=halfsquare*
    ]{\vvt-Portfolio}{vvt-portfolio.quantile-score.csv};
\nextgroupplot[split quantile plot bottom,
    mark repeat=200,
    xmin=-3500,xmax=5500,
    xminorticks=true,
    minor xtick={-3000,-1000,1000,3000,5000},
    height=.08\linewidth,
    xlabel=Accumulated score,
  ]
  \addgraph[blue,solid,mark=asterisk]{\kidfkipdr}{kInduction-kipdrdf.quantile-score.csv};
  \addgraph[brown,solid,mark=triangle*]{\seahorn}{seahorn.quantile-score.csv};
  \addgraph[green,solid,mark=halfsquare*]{\vvt-Portfolio}{vvt-portfolio.quantile-score.csv};
  \legend{}
\end{groupplot}
\end{pgfplot}

%% file: svcomp19-loops-table.tex
\begin{table*}[t]
  \centering
  \caption{Results for all~\nSvcompNineteenLoopsTasks~verification tasks of the \svcomp~2019 subcategory \textit{ReachSafety-Loops},
           \nSvcompNineteenLoopsFalsetasks~of which contain bugs,
           while the other~\nSvcompNineteenLoopsTruetasks~are considered to be safe,
           for the best three verifiers in that category
           (\uautomizer, \utaipan, and \veriabs),
           as well as for {\smaller\kikipdr} and {\smaller\kidfkipdr}.
  }
  \label{tab:svcomp19-loops-results}
  \newcommand\stoh[1]{\fpeval{#1/3600}}
  \newcommand\precnum[1]{\tablenum[round-mode=off,table-format=4.1]{#1}}
  \newcommand\rndnum[1]{\tablenum[round-mode=figures,round-precision=2,table-format=4.1]{#1}}
  \newcommand\precnumnodec[1]{\tablenum[round-mode=off,table-format=4.0]{#1}}
  \newcommand\rndnumnodec[1]{\tablenum[round-mode=figures,round-precision=2,table-format=4.0]{#1}}
  \renewcommand{\uautomizer}{\tool{UAutomizer}\xspace}
  \renewcommand{\ukojak}{\tool{UKojak}\xspace}
  \renewcommand{\utaipan}{\tool{UTaipan}\xspace}

  \begin{tabular}{l|ccc|cc}
    \toprule
    \multicolumn{1}{l|}{\multirow{2}{*}{Verifier}}
        & \multicolumn{3}{c|}{\svcomp~2019}
        & \multicolumn{1}{c}{\multirow{2}{*}{\smaller\kikipdr}}
        & \multicolumn{1}{c}{\multirow{2}{*}{\begin{minipage}{1.35cm}\smaller\kidf;\\\kipdr\end{minipage}}} \\
        & \multicolumn{1}{c}{\uautomizer}
        & \multicolumn{1}{c}{\utaipan}
        & \multicolumn{1}{c|}{\veriabs}
        &
        & \\

    \midrule
    Score
          & \precnum{\SvcompNineteenUautomizerSvCompPropReachsafetyReachsafetyLoopsScore}
          & \precnum{\SvcompNineteenUtaipanSvCompPropReachsafetyReachsafetyLoopsScore}
          & \bfseries\precnum{\SvcompNineteenVeriabsSvCompPropReachsafetyReachsafetyLoopsScore}
          & \precnum{\SvcompNineteenPdrInvKinductionKipdrReachsafetyLoopsScore}
          & \SvcompNineteenPdrInvKinductionDfkipdrReachsafetyLoopsScore \\
    Correct results
          & \precnum{\SvcompNineteenUautomizerSvCompPropReachsafetyReachsafetyLoopsCorrectCount}
          & \precnum{\SvcompNineteenUtaipanSvCompPropReachsafetyReachsafetyLoopsCorrectCount}
          & \bfseries\precnum{\SvcompNineteenVeriabsSvCompPropReachsafetyReachsafetyLoopsCorrectCount}
          & \precnum{\SvcompNineteenPdrInvKinductionKipdrReachsafetyLoopsCorrectCount}
          & \precnum{\SvcompNineteenPdrInvKinductionDfkipdrReachsafetyLoopsCorrectCount} \\
    \quad Correct proofs
          & \precnum{\SvcompNineteenUautomizerSvCompPropReachsafetyReachsafetyLoopsCorrectTrueCount}
          & \precnum{\SvcompNineteenUtaipanSvCompPropReachsafetyReachsafetyLoopsCorrectTrueCount}
          & \bfseries\precnum{\SvcompNineteenVeriabsSvCompPropReachsafetyReachsafetyLoopsCorrectTrueCount}
          & \precnum{\SvcompNineteenPdrInvKinductionKipdrReachsafetyLoopsCorrectTrueCount}
          & \precnum{\SvcompNineteenPdrInvKinductionDfkipdrReachsafetyLoopsCorrectTrueCount} \\
    \quad Correct alarms
          & \precnum{\SvcompNineteenUautomizerSvCompPropReachsafetyReachsafetyLoopsCorrectFalseCount}
          & \precnum{\SvcompNineteenUtaipanSvCompPropReachsafetyReachsafetyLoopsCorrectFalseCount}
          & \bfseries\precnum{\SvcompNineteenVeriabsSvCompPropReachsafetyReachsafetyLoopsCorrectFalseCount}
          & \precnum{\SvcompNineteenPdrInvKinductionKipdrReachsafetyLoopsCorrectFalseCount}
          & \precnum{\SvcompNineteenPdrInvKinductionDfkipdrReachsafetyLoopsCorrectFalseCount} \\
    Wrong proofs
          & \precnum{\SvcompNineteenUautomizerSvCompPropReachsafetyReachsafetyLoopsWrongTrueCount}
          & \precnum{\SvcompNineteenUtaipanSvCompPropReachsafetyReachsafetyLoopsWrongTrueCount}
          & \bfseries\precnum{\SvcompNineteenVeriabsSvCompPropReachsafetyReachsafetyLoopsWrongTrueCount}
          & \bfseries\precnum{\SvcompNineteenPdrInvKinductionKipdrReachsafetyLoopsWrongTrueCount}
          & \bfseries\precnum{\SvcompNineteenPdrInvKinductionDfkipdrReachsafetyLoopsWrongTrueCount} \\
    Wrong alarms
          & \bfseries\precnum{\SvcompNineteenUautomizerSvCompPropReachsafetyReachsafetyLoopsWrongFalseCount}
          & \bfseries\precnum{\SvcompNineteenUtaipanSvCompPropReachsafetyReachsafetyLoopsWrongFalseCount}
          & \bfseries\precnum{\SvcompNineteenVeriabsSvCompPropReachsafetyReachsafetyLoopsWrongFalseCount}
          & \bfseries\precnum{\SvcompNineteenPdrInvKinductionKipdrReachsafetyLoopsWrongFalseCount}
          & \bfseries\precnum{\SvcompNineteenPdrInvKinductionDfkipdrReachsafetyLoopsWrongFalseCount} \\
    Timeouts
          & \precnum{\SvcompNineteenUautomizerSvCompPropReachsafetyReachsafetyLoopsErrorTimeoutCount}
          & \precnum{\SvcompNineteenUtaipanSvCompPropReachsafetyReachsafetyLoopsErrorTimeoutCount}
          & \precnum{\SvcompNineteenVeriabsSvCompPropReachsafetyReachsafetyLoopsErrorTimeoutCount}
          & \precnum{\SvcompNineteenPdrInvKinductionKipdrReachsafetyLoopsErrorTimeoutCount}
          & \precnum{\SvcompNineteenPdrInvKinductionDfkipdrReachsafetyLoopsErrorTimeoutCount} \\
    Out of memory
          & \precnum{\SvcompNineteenUautomizerSvCompPropReachsafetyReachsafetyLoopsErrorOutOfMemoryCount}
          & \precnum{\SvcompNineteenUtaipanSvCompPropReachsafetyReachsafetyLoopsErrorOutOfMemoryCount}
          & \precnum{\SvcompNineteenVeriabsSvCompPropReachsafetyReachsafetyLoopsErrorOutOfMemoryCount}
          & \precnum{\SvcompNineteenPdrInvKinductionKipdrReachsafetyLoopsErrorOutOfMemoryCount}
          & \precnum{\SvcompNineteenPdrInvKinductionDfkipdrReachsafetyLoopsErrorOutOfMemoryCount} \\
    Other inconclusive
          & \precnum{\SvcompNineteenUautomizerSvCompPropReachsafetyReachsafetyLoopsErrorOtherInconclusiveCount}
          & \precnum{\SvcompNineteenUtaipanSvCompPropReachsafetyReachsafetyLoopsErrorOtherInconclusiveCount}
          & \precnum{\SvcompNineteenVeriabsSvCompPropReachsafetyReachsafetyLoopsErrorOtherInconclusiveCount}
          & \precnum{\SvcompNineteenPdrInvKinductionKipdrReachsafetyLoopsErrorOtherInconclusiveCount}
          & \precnum{\SvcompNineteenPdrInvKinductionDfkipdrReachsafetyLoopsErrorOtherInconclusiveCount} \\
    \midrule
    \multicolumn{4}{l}{Times for correct results} \\
    Total CPU Time (h)
         & \rndnum{\stoh{\SvcompNineteenUautomizerSvCompPropReachsafetyReachsafetyLoopsCorrectCputime}}
         & \rndnum{\stoh{\SvcompNineteenUtaipanSvCompPropReachsafetyReachsafetyLoopsCorrectCputime}}
         & \rndnum{\stoh{\SvcompNineteenVeriabsSvCompPropReachsafetyReachsafetyLoopsCorrectCputime}}
         & \rndnum{\stoh{\SvcompNineteenPdrInvKinductionKipdrReachsafetyLoopsCorrectCputime}}
         & \rndnum{\stoh{\SvcompNineteenPdrInvKinductionDfkipdrReachsafetyLoopsCorrectCputime}} \\
    Mean CPU Time (s)
         & \rndnum{\SvcompNineteenUautomizerSvCompPropReachsafetyReachsafetyLoopsCorrectCputimeAvg}
         & \rndnum{\SvcompNineteenUtaipanSvCompPropReachsafetyReachsafetyLoopsCorrectCputimeAvg}
         & \rndnum{\SvcompNineteenVeriabsSvCompPropReachsafetyReachsafetyLoopsCorrectCputimeAvg}
         & \bfseries\rndnum{\SvcompNineteenPdrInvKinductionKipdrReachsafetyLoopsCorrectCputimeAvg}
         & \rndnum{\SvcompNineteenPdrInvKinductionDfkipdrReachsafetyLoopsCorrectCputimeAvg} \\
    Median CPU Time (s)
         & \rndnum{\SvcompNineteenUautomizerSvCompPropReachsafetyReachsafetyLoopsCorrectCputimeMedian}
         & \rndnum{\SvcompNineteenUtaipanSvCompPropReachsafetyReachsafetyLoopsCorrectCputimeMedian}
         & \rndnum{\SvcompNineteenVeriabsSvCompPropReachsafetyReachsafetyLoopsCorrectCputimeMedian}
         & \bfseries\rndnum{\SvcompNineteenPdrInvKinductionKipdrReachsafetyLoopsCorrectCputimeMedian}
         & \rndnum{\SvcompNineteenPdrInvKinductionDfkipdrReachsafetyLoopsCorrectCputimeMedian} \\
    Total Wall Time (h)
         & \rndnum{\stoh{\SvcompNineteenUautomizerSvCompPropReachsafetyReachsafetyLoopsCorrectWalltime}}
         & \rndnum{\stoh{\SvcompNineteenUtaipanSvCompPropReachsafetyReachsafetyLoopsCorrectWalltime}}
         & \rndnum{\stoh{\SvcompNineteenVeriabsSvCompPropReachsafetyReachsafetyLoopsCorrectWalltime}}
         & \rndnum{\stoh{\SvcompNineteenPdrInvKinductionKipdrReachsafetyLoopsCorrectWalltime}}
         & \rndnum{\stoh{\SvcompNineteenPdrInvKinductionDfkipdrReachsafetyLoopsCorrectWalltime}} \\
    Mean Wall Time (s)
         & \rndnum{\SvcompNineteenUautomizerSvCompPropReachsafetyReachsafetyLoopsCorrectWalltimeAvg}
         & \rndnum{\SvcompNineteenUtaipanSvCompPropReachsafetyReachsafetyLoopsCorrectWalltimeAvg}
         & \rndnum{\SvcompNineteenVeriabsSvCompPropReachsafetyReachsafetyLoopsCorrectWalltimeAvg}
         & \rndnum{\SvcompNineteenPdrInvKinductionKipdrReachsafetyLoopsCorrectWalltimeAvg}
         & \rndnum{\SvcompNineteenPdrInvKinductionDfkipdrReachsafetyLoopsCorrectWalltimeAvg} \\
    Median Wall Time (s)
         & \rndnum{\SvcompNineteenUautomizerSvCompPropReachsafetyReachsafetyLoopsCorrectWalltimeMedian}
         & \rndnum{\SvcompNineteenUtaipanSvCompPropReachsafetyReachsafetyLoopsCorrectWalltimeMedian}
         & \rndnum{\SvcompNineteenVeriabsSvCompPropReachsafetyReachsafetyLoopsCorrectWalltimeMedian}
         & \rndnum{\SvcompNineteenPdrInvKinductionKipdrReachsafetyLoopsCorrectWalltimeMedian}
         & \rndnum{\SvcompNineteenPdrInvKinductionDfkipdrReachsafetyLoopsCorrectWalltimeMedian} \\
  \bottomrule
  \end{tabular}
\end{table*}

%% file: evaluation/svcomp19-loops.scatter.tex
\begin{tikzpicture}
\begin{loglogaxis}[
    xlabel=CPU time for \veriabs (\second),
    ylabel=CPU time for \kikipdr (\second),
    xmin=0.01,
    xmax=1000,
    ymin=0.01,
    ymax=1000,
    domain=0.01:1001,
    clip mode=individual,
    axis equal image,
    ]
    \addplot+[blue, mark=+,only marks]
         table[
             header=false,
             skip first n=3, %
             x index=2, %
             y index=6  %
             ] {\plotpath/svcomp19-loops.scatter.csv};
    \addplot[gray] {x};
    \addplot[gray] {10*x};
    \addplot[gray] {x/10};
\end{loglogaxis}
\end{tikzpicture}

%% file: conclusion.tex
\section{Conclusion}
\vspace{-2mm}
Property-directed reachability (a.k.a. IC3) is a verification approach that
is popular and successful in some fields of formal verification
(e.g., hardware designs, Horn clauses).
Unfortunately, there is a large gap between this success story and the
applicability in practical software verification.
We are closing this gap by
(a)~providing a well-engineered implementation of one published
adaptation of \pdr to software verification,
(b)~designing and implementing an invariant generator based on the ideas of \pdr,
and
(c)~providing an evaluation of all applicable tools and approaches
on the largest available benchmark set of C verification tasks.
This provides a good foundation as baseline for ongoing research in this area.

The results of our comparative evaluation extend the knowledge about \pdr
for software verification in the following ways:
(1)~Our implementation outperforms the existing implementation of \pdr (\vvt)
and is more precise than the other software verifier that uses \pdr (\seahorn).
Thus, our implementation can serve as a reference implementation for further
research on \pdr for software verification.
(2)~On most of the programs in the widely used \emph{sv-benchmarks} collection of verification tasks,
other techniques are more effective (solve more problems) and more efficient (solve the problems faster).
(3)~\pdr can be an effective and efficient technique for computing invariants
that are difficult to obtain:
there are programs for which our \pdr-based approach
is more efficient than the best invariant generator from \svcomp
in the subcategory \emph{ReachSafety-Loops}.